\documentclass[11pt]{article}
\usepackage{iftex}
\ifPDFTeX
  \usepackage[utf8]{inputenc}
\fi
\usepackage[T1]{fontenc}
\usepackage{lmodern}
\usepackage{amsmath,amssymb,amsthm,mathtools,bm}
\usepackage[a4paper,left=0.8in,right=0.8in,top=1in,bottom=1in]{geometry}
\usepackage{booktabs}
\usepackage{graphicx}
\usepackage[final]{microtype}
\usepackage[font=small,labelfont=bf,labelsep=period,
  format=hang,justification=justified,singlelinecheck=true]{caption}
\usepackage{float,placeins}
\usepackage{xcolor,fancyhdr}
\usepackage[hidelinks,bookmarksnumbered,pdfdisplaydoctitle,
  linktoc=all,bookmarksdepth=2,pdfpagelabels]{hyperref}

\allowdisplaybreaks[1]
\numberwithin{equation}{section}
\theoremstyle{definition}
\newtheorem{definition}{Definition}[section]
\newtheorem{postulate}{Postulate}
\newtheorem{example}{Example}[section]
\theoremstyle{plain}
\newtheorem{proposition}{Proposition}[section]
\newtheorem{theorem}{Theorem}[section]
\theoremstyle{remark}

\newcommand{\R}{\mathbb R}
\newcommand{\C}{\mathbb C}
\newcommand{\ket}[1]{\lvert #1\rangle}
\newcommand{\bra}[1]{\langle #1\rvert}
\newcommand{\braket}[2]{\langle #1\mid #2\rangle}
\newcommand{\vct}[1]{\bm{#1}}
\newcommand{\Id}{\mathsf I}
\newcommand{\dd}{\mathrm d}
\DeclareMathOperator{\tr}{tr}
\DeclareMathOperator{\diag}{diag}
\definecolor{vlblue}{RGB}{0,0,0}
\definecolor{vlgray}{RGB}{85,85,85}
\definecolor{vlfGreen}{HTML}{05472A}
\definecolor{vlfCopper}{HTML}{98613C}
\definecolor{vlfBlue}{HTML}{426A7B}
\definecolor{vlfRed}{HTML}{873F48}
\definecolor{vlfDarkRed}{HTML}{762B38}
\newcommand{\vlStroke}[6]{%
  \put(#1,#2){\rotatebox{#3}{\rule[-#6pt]{#4pt}{#5pt}}}}
\ifPDFTeX
  \newcommand{\vlLiteral}[1]{\pdfliteral{#1}}
\fi
\ifLuaTeX
  \newcommand{\vlLiteral}[1]{\pdfextension literal {#1}}
\fi
\ifdefined\vlLiteral
\makeatletter
\renewcommand{\vlStroke}[6]{%
  \begingroup
  \ifcsname sin(\number#3)\endcsname
    \edef\vl@sin{\UseSin{#3}}%
  \else
    \dimen@=#3\p@\relax\TG@@sin
    \edef\vl@sin{\strip@pt\dimen@}%
  \fi
  \ifcsname cos(\number#3)\endcsname
    \edef\vl@cos{\UseCos{#3}}%
  \else
    \dimen@=90\p@\advance\dimen@-#3\p@\relax\TG@@sin
    \edef\vl@cos{\strip@pt\dimen@}%
  \fi
  \dimen@=#4pt\relax
  \@tempdima=\z@
  \ifdim\vl@cos\p@<\z@
    \@tempdima=\vl@cos\dimen@
  \fi
  \ifdim\vl@sin\p@>\z@
    \dimen@=\dimexpr #5pt-#6pt\relax
    \@tempdimb=\vl@sin\dimen@
    \advance\@tempdima-\@tempdimb
  \else
    \dimen@=#6pt\relax
    \@tempdimb=\vl@sin\dimen@
    \advance\@tempdima\@tempdimb
  \fi
  \@tempdima=-\@tempdima
  \edef\vl@shift{\strip@pt\dimexpr 0.99626401\@tempdima\relax}%
  \dimen@=-\vl@sin\p@
  \edef\vl@negsin{\strip@pt\dimen@}%
  \dimen@=#4pt\relax
  \edef\vl@width{\strip@pt\dimexpr 0.99626401\dimen@\relax}%
  \dimen@=#5pt\relax
  \edef\vl@thick{\strip@pt\dimexpr 0.99626401\dimen@\relax}%
  \dimen@=-#6pt\relax
  \edef\vl@depth{\strip@pt\dimexpr 0.99626401\dimen@\relax}%
  \edef\vl@data{q \vl@cos\space\vl@sin\space
    \vl@negsin\space\vl@cos\space\vl@shift\space0 cm
    0 \vl@depth\space\vl@width\space\vl@thick\space re f Q}%
  \put(#1,#2){\vlLiteral{\vl@data}}%
  \endgroup}
\makeatother
\fi
\newcommand{\Dt}[1]{\frac{\dd #1}{\dd t}}
\newcommand{\Dtt}[1]{\frac{\dd^2 #1}{\dd t^2}}
\newcommand{\Pdt}[1]{\frac{\partial #1}{\partial t}}
\newcommand{\Pdtt}[1]{\frac{\partial^2 #1}{\partial t^2}}
\renewcommand{\sectionmark}[1]{%
 \markboth{\thesection\enspace\ifcase\value{section}%
 \or Introduction\or Architecture\or Structure and Interactions%
 \or Viscoelastic Dynamics\or Complex Displacement%
 \or Operators and Observables\or Viral Phonons\or Conclusion\fi}{}}
\fancypagestyle{plain}{\fancyhf{}%
 \fancyfoot[L]{\footnotesize Lillian St. Kleess}%
 \fancyfoot[R]{\footnotesize\thepage}%
 }

\title{Viral Lattice Theory:\\A Biophysical Framework for Collective Virion Motion}
\author{Lillian St. Kleess\thanks{Correspondence:
\href{mailto:Lilliankleess@baruch.cuny.edu}{\texttt{Lilliankleess@baruch.cuny.edu}}.}\\
\small Department of Natural Sciences, Baruch College}
\date{Original submission: December 25, 2024\\[2pt]
\small Revised: October 31, 2026}
\hypersetup{
  pdftitle={Viral Lattice Theory: A Biophysical Framework for Collective Virion Motion},
  pdfauthor={Lillian St. Kleess},
  pdflang={en},
  pdfsubject={Theoretical biophysics of collective virion motion, material memory, particle tracking and structural stability},
  pdfkeywords={Viral Lattice Theory, collective virion motion, virus-like particles, human papillomavirus, lattice dynamics, viscoelasticity, particle tracking, biophysical inference, variable selection, structural stability, nonequilibrium dynamics, viral phonons}
}

\input{vlt-graphics.tex}
\begin{document}
\pagenumbering{roman}
\maketitle
\vspace{-2.5em}
\markboth{Abstract}{Abstract}

\begin{abstract}
The collective motion of virions and virus-like particles (VLPs) depends
on both particle interactions and the material around them. Viral Lattice
Theory connects these physical inputs to predictions of motion, mechanical
response and experimental records. Its reference system contains ideal
icosahedral particles in a cylindrical cavity. Interaction forces and
constraints determine collective displacement patterns; material variables
carry relaxation and preparation history. Eliminating those variables
gives memory-dependent forces and, under specified thermal-equilibrium
conditions, the matching fluctuations.
The framework combines established lattice dynamics, viscoelasticity,
stochastic mechanics and response theory to address four practical tasks:
interpreting particle tracking with finite camera exposure; separating
elastic coupling from shared-fluid motion; choosing complementary
measurements that distinguish hidden parameters; and testing which
variables are needed to predict a chosen record. A synthetic
response--fluctuation benchmark gives models with matching equilibrium
covariance and nearly matching response magnitude, but about $28^\circ$
phase difference and a $40\%$ spectral-density difference at a selected
crossover. A crystal-response example uses satellite tobacco mosaic virus
geometry and published acoustic summaries to calibrate a peak and
test a separately reserved linewidth. Human papillomavirus (HPV)-inspired ideal
VLPs provide a geometric reference for a worked guide, with collective
mechanical parameters assigned independently of capsid size.
A transport extension resolves normal, reversibly deformed and damaged
structural classes using classical probabilities. An assigned-rate
simulation predicts both recovery and delayed structural loss after
delivery. Separate path-survival and driven-cycle calculations distinguish
finite loading histories from maintained nonequilibrium steady states.
These comparisons formulate tests of structural stability; an HPV
application requires measured loading, transition kinetics and structural
readouts. A separate viral-phonon quantization postulate assigns
mode-energy increments $\hbar\omega_s$ and supports conditional quantum
models, whose biological use requires suitable preparation, damping,
thermal conditions and a distinguishing measurement. Results include
synthetic comparisons and a conditional test against published acoustic summaries. With independently
constrained interactions, material response and observation maps, they
make particular biophysical inference and variable-selection problems
explicit and testable.
\end{abstract}

\noindent{\small\textbf{Keywords:} virus-like particles; collective virion motion;
lattice dynamics; viscoelasticity; particle tracking; biophysical inference;
variable selection; structural stability; virus crystals; Brillouin scattering.}
\par\medskip
\noindent\small\textbf{Reading conventions.} Bold symbols denote vectors or
matrices. Hats distinguish quantized observables where needed; they also
mark Fourier transforms or unit directions when stated. Time derivatives use
$\dd/\dd t$, and field derivatives at fixed position use
$\partial/\partial t$. The fixed particle pool is counted by
$N_{\rm tot}$; $n$ is the retained array size in a given calculation,
and $n_{\mathcal C}(t)$ is the size of a selected cluster. The symbol
$N_s$ counts mode excitations, and $T_{\rm cap}$ denotes capsid triangulation.
\par\normalsize\medskip

\noindent{\small\textbf{Scope and reading guide.}
Section~\ref{sec:introduction} introduces the physical model and maps its
results. Section~\ref{sec:introduction-worked-guide} works through four
prediction tasks and a synthetic HPV-inspired transport simulation.
Biological applications require preparation-specific calibration;
quantized extensions add their own postulates.}

\clearpage
\pdfbookmark[0]{Contents}{contents}
\begingroup
\setlength{\parskip}{0pt}
\tableofcontents
\endgroup
\clearpage
\pagenumbering{arabic}

\section{Introduction}
\label{sec:introduction}

Viral Lattice Theory (VLT) describes the collective motion of virions
and virus-like particles (VLPs): how a disturbance of one particle spreads
through an assembly, changes particle separations and exchanges energy
with the surroundings. It is a theoretical framework that may potentially
be useful to some virologists, biophysicists and theoretical physicists
developing quantitative models or interpreting measurements of particle
motion and structural state.

The starting point is a finite population of ideal icosahedral particles
in a cylindrical medium. Each whole particle is one mechanical node.
Its mass, surface preparation and interactions determine how it responds
to forces from its neighbours and environment. Lennard--Jones and Coulomb
or explicitly screened electrostatic potentials supply the central pair
reference. Resolved surface patches add direction-dependent forces and
torques; ionic and configurational populations give a finer description
of charge and binding. Neutral, same-sign and mixed-sign charge
preparations are included. A chosen application specifies these inputs
for its physical preparation.

A viral lattice is a reference arrangement together with its interaction
network. Compatible cells join local contributions into larger finite
assemblies, including irregular ones. The enclosed particle pool
$N_{\rm tot}$ is fixed, while a selected cluster $\mathcal C(t)$ can
recruit, release or exchange members. Its size is
$n_{\mathcal C}(t)=|\mathcal C(t)|$. Thus the same construction describes
collective motion within an arrangement and continues the physical
preparation when the arrangement changes.

The energy supplies forces and, near a stable force-balanced reference,
the restoring matrix. Inertia and constraints turn that matrix into
collective displacement patterns. The surrounding material contributes
resistance and relaxing strain. Keeping that strain as a state variable
makes its history explicit; eliminating it gives a force that depends on
past motion and initial preparation. In thermal equilibrium, the same
coupling fixes the corresponding fluctuations. A measured signal then
follows from an observation model: a camera averages motion during
exposure, a marker can detect turning, and a structural assay can
distinguish intact and deformed shells.

A second geometric reference is the small T=1 shell of satellite tobacco
mosaic virus (STMV). Its internal protein contacts and the contacts
between whole particles in a crystal occupy different levels of the
construction. Section~\ref{sec:stmv-response-benchmark} connects a
documented orthorhombic packing to published acoustic line summaries.
It calibrates a response peak, predicts an unused linewidth, and gives
prospective directional records that separate otherwise compatible
contact networks. This comparison anchors a limited mechanical test to
measurements while retaining the assigned parts of its material and
observation model.

Four practical questions organize the development. How much of a tracked
correlation comes from the acquisition process? Does correlated motion
reflect an elastic link or shared fluid response? Which additional
measurement separates parameters that mechanical tracking leaves
ambiguous? Which particle, material or structural variables are needed
to predict what happens next? These are specific forms of the
biophysical inference and variable-selection problems. Inference asks
which model inputs a record distinguishes; variable selection asks which
evolving quantities carry the information required for a prediction.

Transport and delivery provide a further application of these questions.
Container acceleration, flow gradients and contacts can load particle
or shell coordinates. A structural-state extension follows a population
through normal, reversibly deformed and long-lived damaged classes.
Its synthetic example predicts both recovery and further structural
loss after unloading, because delivery preserves a prepared population.
The loading-to-transition law and structural readout are inputs to test
for an actual VLP preparation. The sealed reference cavity, driven flow
extensions and delivery protocol each have their stated boundary and
work accounts.

The calculation proceeds from physical state and interactions to dynamics,
then to measured records and comparisons between alternatives. One
equilibrium parameter set predicts both driven response and fluctuations.
The worked guide in Section~\ref{sec:introduction-worked-guide} shows
how to perform and interpret those comparisons for HPV-inspired ideal
VLPs, then continues the procedure through a finite delivery protocol.
This introduction supplies the conceptual entry point to the theory:
its objects, variables, principal results and application requirements.
The later sections provide the hypotheses, derivations, proofs and
controlled extensions behind that map.

\begin{center}
\fbox{\begin{minipage}{0.92\linewidth}
\textbf{The central mechanical question.}
Which particle and material variables are needed to predict collective
virion motion, and which measurements distinguish the interaction and
medium parameters that generate it?
\end{minipage}}
\end{center}

The sealed, reflecting \emph{ideal reference cavity} confines the pool.
Interparticle forces and declared supports establish force balance and
stability. Small displacements within a stable basin admit a harmonic
description; encounters, energy exchange and escape connect different
arrangements. Figure~\ref{fig:opening-reference-network} traces the
construction from particles and retained interactions to mechanical
operators.

\begin{figure}[H]
\centering\begingroup
\definecolor{vltTeal}{HTML}{176B64}
\definecolor{vltBlue}{HTML}{315D80}
\definecolor{vltGold}{HTML}{A87520}
\definecolor{vltPlum}{HTML}{725877}
\definecolor{vltCoral}{HTML}{A94D42}
\definecolor{vltInk}{HTML}{172027}
\definecolor{vltFog}{HTML}{EDF2F2}
\definecolor{vltRed}{HTML}{762B38}
\setlength{\unitlength}{1pt}
\scalebox{0.84}{%
\input{vlt-figures/figure-001.tex}%
}\endgroup
\caption{A four-virion reference and its matrix construction. (a) Labelled rigid bodies share five retained interactions inside a projected cylindrical medium. (b) The graph preserves their identities and positions; the teal vector is one centre displacement. The lower arrows map interaction energy to curvature and collective motion. Connectors identify pair-energy contributions, whose constitutive law supplies stiffness. Geometry and displacement gain are schematic. Particle hues register corresponding objects within a figure; facet tint supplies shape cues.}
\label{fig:opening-reference-network}
\end{figure}

\paragraph{Principal calculations and proposed uses.}
The construction developed here treats compatible finite assemblies,
retained material state, registered arrangement changes and calibrated
observations within one physical specification. Table~\ref{tab:introduction-contribution-map}
identifies calculations tied to four prediction tasks. Each supplies a
comparison between declared models or observation settings; applying
that comparison to virions requires a specified biological preparation.

\begin{table}[H]
\centering\small
\begin{tabular}{@{}p{3.0cm}p{8.0cm}p{4.3cm}@{}}
\toprule
Prediction task & Calculated comparison & Detailed construction\\
\midrule
\raggedright Interpret tracked motion
& \raggedright Joint frame--state statistics retain motion during exposure and its shared thermal noise. Acquisition settings can be varied while the mechanical law remains fixed.
& \raggedright Section~\ref{sec:operator-exact-camera-frames}\tabularnewline\addlinespace[4pt]
\raggedright Separate elastic and shared-fluid coupling
& \raggedright A trapped-pair reference distinguishes changes in equilibrium covariance from lagged correlations and common/relative response caused by shared mobility.
& \raggedright Section~\ref{sec:phonon-two-probe-application}\tabularnewline\addlinespace[4pt]
\raggedright Design complementary measurements
& \raggedright An explicit hidden-parameter family has identical mechanical records. A calibrated additional ionic or configurational readout distinguishes members of that family.
& \raggedright Proposition~\ref{prop:s7-mechanical-parameter-equivalence}\tabularnewline\addlinespace[4pt]
\raggedright Test retained variables
& \raggedright Linear closure criteria test exact sufficiency for specified histories; task-dependent error comparisons assess whether retained variables suffice for the chosen readout and window.
& \raggedright Proposition~\ref{prop:s6-history-closed-variables}; Section~\ref{sec:phonon-establishing-biophysical-variables}\tabularnewline
\bottomrule
\end{tabular}
\caption{Four prediction tasks, their calculated comparisons and the sections
giving the detailed assumptions and derivations.}
\label{tab:introduction-contribution-map}
\end{table}

\paragraph{Established foundations and the contribution developed here.}
Harmonic lattice mechanics, coarse-graining, linear response and
equilibrium stochastic dynamics provide the foundations
\cite{Misra2012,Mori1965,Noid2008,Kubo1957}. Particle-tracking statistics
and microrheology supply established connections between motion,
acquisition and material response~\cite{Berglund2010,Crocker2000}.
Their standard identities and component methods retain that provenance.
The contribution developed here is their organization into a reproducible
assembly-to-observation construction and the conditional comparisons
derived from it. These include compatible cell assembly and response
reduction, continuation of prepared material state, exposure-resolved
records, hidden-parameter equivalences and tests of predictive state.
Surface populations and structural classes extend the same construction
when the target prediction depends on them. The transport calculation
connects conditional mechanics and kinetics to delivery statistics and
path survival. Each comparison identifies the inputs it uses, the record
it predicts and the alternative that record can distinguish.

\paragraph{Reading route and status of the results.}
The four tasks in Table~\ref{tab:introduction-contribution-map} and the
structural transport example have classical formulations. Sections on
interactions and assembly define coordinates and forces; material and
response sections derive their evolution; observation sections determine
which information reaches a record. Later calculations treat confined
excitations, responsive surfaces and finite populations. The quantized
passages add a mode-quantization postulate and can be followed as a
separate development. Standard methods retain their cited provenance;
derived identities and synthetic comparisons state their assumptions.
Biological applications require preparation-specific inputs and tests
of the resulting predictions.

Relative displacement is the immediate mechanical link between the
picture and the equations. Two particles moving together can preserve
their separation; moving toward or away from one another changes it and
loads their interaction. Figure~\ref{fig:introduction-relative-spring-geometry}
isolates this distinction before the same construction is assembled over
many particles.

\begin{figure}[H]
\centering\begingroup
\definecolor{vltTeal}{HTML}{176B64}
\definecolor{vltBlue}{HTML}{315D80}
\definecolor{vltGold}{HTML}{A87520}
\definecolor{vltPlum}{HTML}{725877}
\definecolor{vltCoral}{HTML}{A94D42}
\definecolor{vltInk}{HTML}{172027}
\definecolor{vltFog}{HTML}{EDF2F2}
\definecolor{vltRed}{HTML}{762B38}
\setlength{\unitlength}{1pt}
\scalebox{0.90}{%
\input{vlt-figures/figure-002.tex}%
}\endgroup
\caption{Relative displacement of a free pair with $V(r)=k(r-r_\star)^2/2$, $k>0$. The ruler uses the common reference separation $r_\star$ and origin at the left reference centre; grey guides register these positions in every row. Teal arrows show displacement, and burgundy arrows below the bodies show restoring force. Nonzero shifts have magnitude $A$, giving separation changes $0,0,-2A,+2A$. The last two rows store $2kA^2$. All connectors retain seven turns, radius, leads and outline attachment, so pitch follows actual surface-to-surface span. Force and displacement use separate fixed arrow gains; the displayed amplitude is illustrative.}
\label{fig:introduction-relative-spring-geometry}
\end{figure}

The same relative-motion construction extends across rows and layers.
Figure~\ref{fig:introduction-assembly-motion-3d} compares two trial fields
on a 24-particle assembly: a common shift preserves all pair separations,
whereas opposite layer shifts load interactions between the layers.
The global restoring matrix collects these directional contributions.
Its size follows the retained coordinates, and its connectivity follows
the interaction model.

\begin{figure}[H]
\centering\begingroup
\definecolor{vltTeal}{HTML}{176B64}
\definecolor{vltBlue}{HTML}{315D80}
\definecolor{vltGold}{HTML}{A87520}
\definecolor{vltPlum}{HTML}{725877}
\definecolor{vltRed}{HTML}{762B38}
\definecolor{vltYellow}{HTML}{D5AD3C}
\definecolor{vltNeutral}{HTML}{777D83}
\setlength{\unitlength}{1pt}
\scalebox{0.84}{%
\input{vlt-figures/figure-003.tex}%
}\endgroup
\caption{Two trial displacement fields on the same $4\times3\times2$ assembly. Reference nearest-neighbour spacing is $a$, and layer heights are $z=\pm a/2$. All 24 ideal virions appear in both views; four colored sites register particle identity. Grey cages and dots locate the reference, and black arrows give the displayed shifts of the labelled sites. A common displacement preserves every pair separation. Opposite layer shifts deform the selected interlayer diagonals, with burgundy and blue marking increased and decreased spans. All coils retain five turns, radius and leads; their pitch follows the displayed surface-to-surface distance. The displacement gain is five in both panels, and particle sizes are schematic. Connectors show a retained interaction subset; supports and additional long-range terms enter the specified energy.}
\label{fig:introduction-assembly-motion-3d}
\end{figure}

Matrices supply the computational links: they assemble interactions,
resolve collective patterns, eliminate selected coordinates and update
response after a contact or material change. Their predictions include
trajectories, response phase, energy transfer and correlations. Connecting
such predictions to a biophysical measurement requires an explicit
observation model~\cite{Leake2016}.

Camera-based tracking gives one concrete application. A frame averages
motion during exposure, while thermal forcing continues to change the
state. The resulting frame statistics depend on both the mechanics and
the acquisition settings. Section~\ref{sec:operator-exact-camera-frames}
derives their joint evolution, including the shared fluctuation entering
an image and the next state. Other applications resolve material strain,
ionic populations, molecular configurations or membrane shape. The
synthetic benchmark in Section~\ref{sec:complete-benchmark} uses one
parameter set for complementary response and fluctuation tests.

\subsection{Mechanical resolution: particles, coordinates and observables}
\label{sec:measles}

Mechanical resolution specifies the objects and motions retained.
For whole-particle translation, $\vct r_i(t)=\vct R_i+\vct u_i(t)$:
$\vct R_i$ is the reference centre and $\vct u_i$ its displacement.
Stacking the retained components gives one real column $\vct u$;
unconstrained translation of $n$ particles has $3n$ components.
Orientation adds rigid turning, shape coordinates add deformation,
and internal relative coordinates add motion between components.
Figure~\ref{fig:opening-coordinate-hierarchy} identifies the distinct
motions and their readouts.

\begin{figure}[H]
\centering\begingroup
\definecolor{vltTeal}{HTML}{176B64}
\definecolor{vltBlue}{HTML}{315D80}
\definecolor{vltGold}{HTML}{A87520}
\definecolor{vltPlum}{HTML}{725877}
\definecolor{vltCoral}{HTML}{A94D42}
\definecolor{vltInk}{HTML}{172027}
\definecolor{vltFog}{HTML}{EDF2F2}
\definecolor{vltRed}{HTML}{762B38}
\setlength{\unitlength}{1pt}
\scalebox{0.84}{%
\input{vlt-figures/figure-004.tex}%
}\endgroup
\caption{Three coordinate choices and their readouts. (a) Rigid translation preserves shape relative to a registered outline. (b) A body direction defines an angle from the dashed reference ray. (c) Opposed component shifts change separation at fixed equal-weight centre. Upper arrows describe coordinate changes; lower arrows map each coordinate to a measured quantity. The tracking vector $\vct e$ is fixed and unit length. Orientation and internal outlines are auxiliary geometric models; internal motion is independent of the retained rigid coordinates.}
\label{fig:opening-coordinate-hierarchy}
\end{figure}

Viral structure motivates several useful resolutions. Measles particles
are enveloped and pleomorphic, with a packaged RNA--protein
ribonucleoprotein complex (RNP), membrane and associated proteins.
Cryo-electron tomography has resolved their organization in intact
particles and at cellular assembly sites~\cite{Liljeroos2011,Ke2018}.
These observations suggest whole-particle, membrane and internal relative
coordinates for different mechanical questions. The ideal icosahedron is
the framework's reference particle geometry; a measles-specific
realization assigns geometry and couplings appropriate to that preparation.

The readout selects which motion matters. Tracking can follow a centre;
a contact probe weights the structures it loads; a structural reporter
can resolve an internal separation. Single-particle acoustic spectroscopy
has resolved vibrations of chemically fixed, substrate-supported
lentiviral pseudovirions~\cite{Zhang2025}, motivating internal and contact
coordinates alongside translation. Each coordinate requires its own
inertia, force coupling and detector map. Introducing internal motion
counts rigid translation and rotation only once.

The construction also applies at other biological scales. Protein
elastic networks assign nodes to residues and edges to reference contacts;
their harmonic models predict fluctuations and
cross-correlations~\cite{Bahar1997}. A collective opening can resolve
relative motion between structural regions. In the whole-particle model,
nodes are ideal virions and edges encode interparticle mechanics. Physical
nodes, energy and measurement provide the common construction; each
resolution supplies its own parameters.

At centre-coordinate resolution, the same equations also describe other
interacting colloidal particles. A virion-specific prediction assigns
geometry, surface interactions, medium coupling, internal resolution and
readout for a declared preparation. Structural observations motivate
candidate coordinates; interaction, transport and detector measurements
constrain the inputs used to predict their motion. Comparing calibrated
records then tests that resolution over the chosen lengths and times.

\subsection{Virion geometry and the spherical approximation}
\label{sec:introduction-virion-geometry}

Shape determines force directions, contact geometry and coupling to the
medium. The regular icosahedral reference has twenty triangular faces
and two-, three- and fivefold rotation axes. Icosahedral organization
occurs in many protein capsids; other virions have helical, pleomorphic
or more complex architectures. Envelope, shell and surface proteins
can also have different structural organization within one particle.

The Caspar--Klug construction relates a triangular surface subdivision
to the subunit count of a class of quasi-equivalent
capsids~\cite{CasparKlug1962}:

\begin{equation}
 T_{\rm cap}=h_g^2+h_g k_g+k_g^2,\qquad
 N_{\rm sub}=60T_{\rm cap},\qquad h_g,k_g\in\mathbb N_0,\quad h_g+k_g>0.
 \label{eq:introduction-triangulation}
\end{equation}

The integers $h_g,k_g$ specify steps on the triangular construction, and
$N_{\rm sub}$ counts protein subunits in that quasi-equivalent construction.
HPV uses an all-pentameric $T_{\rm cap}=7d$ organization with 360 L1
subunits~\cite{Goetschius2021}; its composition is specified separately
in the worked guide. These are structural counts within one shell;
the particle pool and mode occupations have their own counts.
Figure~\ref{fig:introduction-triangulation-geometry} supplies the
geometric construction. Elasticity also shapes the shell: the balance of
stretching and bending can drive faceting or buckling~\cite{Lidmar2003}.

\begin{figure}[H]
\centering\begingroup
\definecolor{vltTeal}{HTML}{176B64}
\definecolor{vltBlue}{HTML}{315D80}
\definecolor{vltGold}{HTML}{A87520}
\definecolor{vltPlum}{HTML}{725877}
\definecolor{vltCoral}{HTML}{A94D42}
\definecolor{vltInk}{HTML}{172027}
\definecolor{vltFog}{HTML}{EDF2F2}
\definecolor{vltRed}{HTML}{762B38}
\setlength{\unitlength}{1pt}
\scalebox{0.80}{%
\input{vlt-figures/figure-005.tex}%
}\endgroup
\caption{Caspar--Klug surface construction at $(h_g,k_g)=(2,1)$. Equal triangular-lattice basis lengths enclose $60^\circ$. Two steps along $\vct a_1$ and one along $\vct a_2$ reach $A$; the resultant is burgundy. Equilateral triangle $OAB$ contains seven unit-triangle areas, giving $T_{\rm cap}=7$ and $N_{\rm sub}=420$ for the stated capsid class. The faceted shell supplies orientation context and does not locate individual proteins. These are subunits within one shell; whole-particle count is separate.}
\label{fig:introduction-triangulation-geometry}
\end{figure}

An effective radius simplifies translation when the force and resistance
laws support geometric averaging. Resolved orientation or deformation
becomes useful for facet-dependent contacts, directional drag or
shape-sensitive signals. Figure~\ref{fig:added-equal-volume-views}
compares distinct shapes sharing one volume-based size measure.

\begin{figure}[H]
\centering\begingroup
\definecolor{vltTeal}{HTML}{176B64}
\definecolor{vltBlue}{HTML}{315D80}
\definecolor{vltGold}{HTML}{A87520}
\definecolor{vltPlum}{HTML}{725877}
\definecolor{vltCoral}{HTML}{A94D42}
\definecolor{vltInk}{HTML}{172027}
\definecolor{vltFog}{HTML}{EDF2F2}
\definecolor{vltRed}{HTML}{762B38}
\setlength{\unitlength}{1pt}
\scalebox{0.80}{%
\input{vlt-figures/figure-006.tex}%
}\endgroup
\caption{Equal volume with different projected shapes. Solid teal outlines project an ellipsoid of semiaxes $(2b,b,b/2)$; dashed outlines project a sphere of radius $b$. All views use one orthographic length scale and have volume $4\pi b^3/3$. Printed values are projected areas. The middle ellipsoid view shares the sphere's area while retaining a different aspect ratio. These geometric comparisons do not prescribe a resistance law.}
\label{fig:added-equal-volume-views}
\end{figure}

An equal-volume radius characterizes size; a hydrodynamic radius
characterizes a specified resistance. An elongated particle instead
has a mobility tensor linking force and velocity along body directions,
which orientation places in the laboratory frame. Tensorial microrheology
connects translation and rotation to material
response~\cite{SquiresMason2010}. Comparing these motions tests a scalar
sphere approximation and helps separate particle shape from material
anisotropy.

Structural and mechanical measurements supply complementary inputs.
The mature Zika glycoprotein shell has icosahedral organization with
180 copies of E~\cite{Sirohi2016}. Atomic-force microscopy of
bacteriophage shells resolved a local elastic indentation
response~\cite{Ivanovska2004}; measurements on minute virus of mice
showed dependence on loading direction and packaged genome
state~\cite{Carrasco2006}. These examples identify concrete mechanical
variables: shell structure, orientation, packaged state, loading and
support. Figure~\ref{fig:final-capsid-loading-axes} identifies the ideal
icosahedral symmetry directions used to describe orientation.

\begin{figure}[H]
\centering\begingroup
\definecolor{vltTeal}{HTML}{176B64}
\definecolor{vltBlue}{HTML}{315D80}
\definecolor{vltGold}{HTML}{A87520}
\definecolor{vltPlum}{HTML}{725877}
\definecolor{vltCoral}{HTML}{A94D42}
\definecolor{vltInk}{HTML}{172027}
\definecolor{vltFog}{HTML}{EDF2F2}
\definecolor{vltRed}{HTML}{762B38}
\setlength{\unitlength}{1pt}
\scalebox{0.80}{%
\input{vlt-figures/figure-007.tex}%
}\endgroup
\caption{Symmetry directions of one regular icosahedron at a common circumscribed radius. Views pass through opposite edge midpoints, face centres and vertices. Labels give the smallest positive restoring rotation and the number of distinct axes of that type. The central $\odot$ points towards the observer. Visible facets and light tint convey depth; the marker specifies orientation rather than applied force.}
\label{fig:final-capsid-loading-axes}
\end{figure}

The symmetry of an assembly belongs to its complete arrangement of
particle centres, orientations, interactions and boundaries. An array
of icosahedral particles can have a periodic centre pattern whose symmetry
is lower than that of one body. Figure~\ref{fig:added-shell-centre-symmetry}
separates the particle geometry from the centre arrangement.
Section~\ref{sec:operator-geometric-extensions} extends the mechanical
coordinates to orientation and shape and tests their dynamical effects.

\begin{figure}[H]
\centering\begingroup
\definecolor{vltTeal}{HTML}{176B64}
\definecolor{vltBlue}{HTML}{315D80}
\definecolor{vltGold}{HTML}{A87520}
\definecolor{vltPlum}{HTML}{725877}
\definecolor{vltCoral}{HTML}{A94D42}
\definecolor{vltInk}{HTML}{172027}
\definecolor{vltFog}{HTML}{EDF2F2}
\definecolor{vltRed}{HTML}{762B38}
\setlength{\unitlength}{1pt}
\scalebox{0.80}{%
\input{vlt-figures/figure-008.tex}%
}\endgroup
\caption{Body geometry and centre-array translation. The right panel preserves all six reference positions and labels while replacing each body by its centre. Open arrows define translations of an auxiliary infinite rectangular extension, with one primitive repeat outlined. The finite patch retains its own boundary. Replacing bodies by centres or specifying a repeat leaves the physical particle identities unchanged.}
\label{fig:added-shell-centre-symmetry}
\end{figure}

\subsection{From energy storage to Hamiltonian motion}
\label{sec:introduction-hamiltonian}

Potential energy records the work stored by changing the particle
configuration. Its negative derivative gives force. Near a force-balanced
reference, the leading change in a twice continuously differentiable
potential is quadratic in the translational displacement column
$\vct u$:

\begin{equation}
 \begin{aligned}
 U(\vct u)&=U_0+\frac12\vct u^{\mathsf T}\mathsf K\vct u
                         +o(\|\vct u\|^2),\\
 K_{ab}&=\left.\frac{\partial^2U}{\partial u_a\partial u_b}
                                           \right|_{\vct u=0},\qquad
 \vct F_{\rm cons}=-\mathsf K\vct u+o(\|\vct u\|).
 \end{aligned}
 \label{eq:introduction-energy-curvature}
\end{equation}

Force balance removes the linear energy term. The Hessian $\mathsf K$
then collects restoring couplings: diagonal entries give curvature
along each coordinate, and off-diagonal entries couple displacements
to forces along other coordinates. Translational entries have units
$\mathrm{N\,m^{-1}}$. Positive curvature defines the stable vibrational
sector; unconstrained rigid motions may remain zero modes.

A pair interaction contributes both the curvature along its separation
and, when the pair carries a reference force, a transverse geometric
stiffness. This \emph{prestress} contribution makes the assembled
Hessian sensitive to force balance and geometry as well as to which
pairs interact. Sections~\ref{sec:lattice-architecture}
and~\ref{sec:structure-interactions} construct these contributions
and test the resulting collective stability.

Local curvature and an escape barrier describe different parts of the
same energy landscape. Curvature determines small-motion stiffness;
the landscape along a separation or rearrangement path determines the
barrier. Figure~\ref{fig:added-curvature-barriers} shows why equal local
stiffness can coexist with different escape costs.

\begin{figure}[H]
\centering\begingroup
\definecolor{vltTeal}{HTML}{176B64}
\definecolor{vltBlue}{HTML}{315D80}
\definecolor{vltGold}{HTML}{A87520}
\definecolor{vltPlum}{HTML}{725877}
\definecolor{vltCoral}{HTML}{A94D42}
\definecolor{vltInk}{HTML}{172027}
\definecolor{vltFog}{HTML}{EDF2F2}
\definecolor{vltRed}{HTML}{762B38}
\setlength{\unitlength}{1pt}
\scalebox{0.80}{%
\input{vlt-figures/figure-009.tex}%
}\endgroup
\caption{Shared local stiffness with different escape barriers. The curves are $h_1=X^2/2-X^4/4$ and $h_2=X^2/2-X^4/16$, with $X=u/\ell_\star$ and $h=U/(k_\star\ell_\star^2)$, $k_\star>0$. Both have $h''(0)=1$. Local magnification compares them with $X^2/2$; the wider view locates positive maxima of height $1/4$ and $1$. Burgundy identifies the higher barrier. The pale window relates the two axis ranges, and negative $X$ mirrors positive $X$. The polynomials describe local basins and escape directions; they are not global thermal confinement potentials.}
\label{fig:added-curvature-barriers}
\end{figure}

The same distinction applies to structural integrity. A shell can have
a well-defined small-deformation response while retaining a different
barrier to collapse or rearrangement. A resolved shell coordinate and
its landscape supply that transition problem; a translational pair
curvature supplies the local interparticle motion.

Capture requires an encounter and accommodation of approach and binding
energy. Escape depends on the path, loading and available time.
Section~\ref{sec:s23-transient-recruitment} connects these fixed-pool
processes to forces generated by prepared mode populations. The Hessian
governs small motion during residence, and transition dynamics governs
passage between arrangements.

For conservative motion, let $\mathsf M$ be the constant positive mass
matrix and $\vct p=\mathsf M\vct v$ the momentum, with
$\vct v=\dd\vct u/\dd t$. The harmonic Hamiltonian is the sum of
kinetic and elastic potential energies, after subtracting the reference
constant:

\begin{equation}
 \boxed{H_{\rm harm}(\vct u,\vct p)
    =\underbrace{\frac12\vct p^{\mathsf T}\mathsf M^{-1}\vct p}
                       _{\text{kinetic energy}}
      +\underbrace{\frac12\vct u^{\mathsf T}\mathsf K\vct u}
                       _{\text{elastic potential energy}}.}
 \label{eq:introduction-unfolded-hamiltonian}
\end{equation}

Positions and momenta specify the conservative state in \emph{phase
space}. At identical positions, different momenta give different future
motion. Hamilton's equations generate that evolution from the energy:

\begin{equation}
 \frac{\dd\vct u}{\dd t}=\frac{\partial H_{\rm harm}}{\partial\vct p}
       =\mathsf M^{-1}\vct p,\qquad
 \frac{\dd\vct p}{\dd t}=-\frac{\partial H_{\rm harm}}{\partial\vct u}
       =-\mathsf K\vct u.
 \label{eq:introduction-hamilton-equations}
\end{equation}

The first equation gives velocity and the second gives force.
Figure~\ref{fig:introduction-phase-space} connects this evolution to
the exchange of kinetic and potential energy in one mode.

\begin{figure}[H]
\centering\begingroup
\definecolor{vltTeal}{HTML}{176B64}
\definecolor{vltBlue}{HTML}{315D80}
\definecolor{vltGold}{HTML}{A87520}
\definecolor{vltPlum}{HTML}{725877}
\definecolor{vltCoral}{HTML}{A94D42}
\definecolor{vltInk}{HTML}{172027}
\definecolor{vltFog}{HTML}{EDF2F2}
\definecolor{vltRed}{HTML}{762B38}
\setlength{\unitlength}{1pt}
\scalebox{0.80}{%
\input{vlt-figures/figure-010.tex}%
}\endgroup
\caption{Energy exchange in one mass-normalized harmonic mode, $H=(\Pi^2+\omega^2Q^2)/2$, at $E_\star=\omega^2Q_\star^2/2$, with positive $\omega,Q_\star$. The potential and equal-scale phase-space circle share $X=Q/Q_\star$, $P=\Pi/(\omega Q_\star)$. Blue vectors have direction $(P,-X)$ and display gain $0.26$. Turning, mixed and crossing states partition the same energy into teal potential and burgundy kinetic fractions. The phase-space orbit describes a modal coordinate and momentum. The boxed result illustrated here is~\eqref{eq:introduction-unfolded-hamiltonian}.}
\label{fig:introduction-phase-space}
\end{figure}

The same regular constant-mass mechanics has a Lagrangian formulation:

\begin{equation}
 L(\vct u,\vct v)=\frac12\vct v^{\mathsf T}\mathsf M\vct v
                 -\frac12\vct u^{\mathsf T}\mathsf K\vct u,
 \qquad \vct p=\frac{\partial L}{\partial\vct v},\qquad
 H=\vct p^{\mathsf T}\vct v-L.
 \label{eq:introduction-legendre-link}
\end{equation}

The Lagrangian organizes configurations and constraints; the Hamiltonian
organizes phase space and energy~\cite{Deriglazov2010,FeynmanLectures1963}.
Their relation supplies the canonical coordinates used later for mode
quantization. If the restoring law is an equilibrium potential of mean
force, its curvature already includes the statistical effect of the
eliminated coordinates at the assigned temperature~\cite{Noid2008}.
The associated energy and thermal account is chosen consistently with
that reduction. Section~\ref{sec:hamiltonian-control} also includes
the work supplied when an external control changes the interaction.

Compatible cells make this energy construction scalable. Local mass and
stiffness terms are embedded in one set of global physical coordinates,
with each owned contribution counted once. Eliminating cell interiors
then gives an exact interface response for the stated linear problem;
retaining fewer interior patterns gives an approximation whose accuracy
can be checked by spectral residuals and response comparisons.
The reduction carries inertia, memory, preparation and thermal forcing
into the interface response. Retained long-range electrostatic and
shared-fluid couplings enter the same global construction.

\subsection{Biophysical scales: fluctuations and causal medium response}
\label{sec:biophysical-scales}

The medium transmits momentum and exchanges energy with the particles.
Deformation stores strain and produces a relaxing reaction force.
Retaining strain gives material state variables; eliminating it gives
history-dependent forces. Unresolved preparation also contributes
fluctuations, linking memory to reduced
dynamics~\cite{Mori1965}. Conservative surface forces and screening enter
the energy~\cite{Israelachvili2011}; resistance and relaxation specify
the dynamical material response.

Charge is one input to this mechanics. Each effective particle charge
$q_i$, measured in coulombs, is specified with its electrostatic law
and medium conditions. Resolved surface patches or ionic populations
give a finer charge description. Particle orientation then matters
because contact and charge patterns turn with the body. A ligand pool
shared by several particles has a finite free inventory; binding and
release couple that inventory to surface forces and fluctuations. An electrokinetic model connects
quantities such as electrophoretic mobility or slip-plane potential to
an effective interaction charge. These inputs set force strengths and
can also alter the stability and response of the assembled network.

An isolated sphere supplies useful scales for translation. For radius
$b$, mass $m$ and an unbounded Newtonian liquid of viscosity $\eta$,
the no-slip, quasisteady, low-Reynolds-number model
gives~\cite{Cotterill2002}

\begin{equation}
 c_{\rm drag}=6\pi\eta b,\qquad
 D_{\mathrm{tr}}=\frac{k_{\mathrm B}T}{c_{\rm drag}},\qquad
 \tau_{\mathrm I}=\frac{m}{c_{\rm drag}}.
 \label{eq:introduction-stokes-scales}
\end{equation}

The resistance is $c_{\rm drag}$, the free translational diffusion
coefficient is $D_{\rm tr}$, and the bare inertial time is $\tau_{\rm I}$.
For a uniform sphere of density $\rho_{\rm v}$,
$m=4\pi\rho_{\rm v}b^3/3$ reduces the inertial time to
$\tau_{\rm I}=2\rho_{\rm v}b^2/(9\eta)$.
At illustrative $b=100\,\mathrm{nm}$,
$\rho_{\rm v}=10^3\,\mathrm{kg\,m^{-3}}$,
$\eta=10^{-3}\,\mathrm{Pa\,s}$ and $T=300\,\mathrm K$, the scales give
$D_{\rm tr}\simeq2.20\,\mu\mathrm m^2\,\mathrm s^{-1}$ and
$\tau_{\rm I}\simeq2.22\,\mathrm{ns}$. Confinement and interactions
add wall effects, restoring forces and collective resistance.

The frequency range also matters. Vorticity spreads across the sphere
on the time $\tau_\nu=b^2/\nu$, where
$\nu=\eta/\rho_{\rm f}$ and $\rho_{\rm f}$ is liquid density.
Quasisteady drag requires $\omega\tau_\nu\ll1$ as well as small
Reynolds number. Fluid inertia and history become relevant outside
that range; Brownian experiments have resolved hydrodynamic memory in
trapped-particle spectra~\cite{Franosch2011}. In a scalar
constant-drag model, $m\Dtt{y}+c_{\rm drag}\Dt{y}+k_y y=0$
oscillates when $c_{\rm drag}^2<4mk_y$. For comparable particle and
liquid densities, using this estimate near oscillatory onset already
requires checking unsteady fluid effects. The full coupled response
determines whether a conservative mode appears as an oscillation or
a relaxation in the measured record.

Thermal fluctuations connect the restoring law to the spread of
positions. A fluctuation is a departure from the ensemble mean,
$\delta\vct u=\vct u-\langle\vct u\rangle$, for a specified
preparation. In classical equilibrium, a scalar harmonic coordinate
with energy $k_y y^2/2$ and $k_y>0$ has

\begin{equation}
 y_{\mathrm{rms}}=\sqrt{\frac{k_{\mathrm B}T}{k_y}}.
 \label{eq:introduction-thermal-excursion}
\end{equation}

A stiffer coordinate has a narrower thermal distribution. The full
covariance additionally describes which particles fluctuate together.
The harmonic approximation is useful while thermal and prepared
excursions remain small relative to the basin, force-variation and
clearance scales.

The simplest overdamped example separates fluctuation width from
relaxation time. With $c_{\rm loc}>0$ and zero-mean thermal force $f_T$,

\begin{equation}
 c_{\rm loc}\frac{\dd y}{\dd t}=-k_y y+f_T(t),\qquad
 \langle f_T(t)f_T(t')\rangle
       =2k_{\rm B}Tc_{\rm loc}\delta(t-t').
 \label{eq:opening-thermal-reduction}
\end{equation}

The noise and resistance are matched to the equilibrium
preparation~\cite{CallenWelton1951}. In the stationary Gibbs state,
$\langle y\rangle=0$ and $\operatorname{Var}y=k_{\rm B}T/k_y$.
For ideal instantaneous observations, the mean-square displacement is

\begin{equation}
 \mathrm{MSD}(\Delta t)
    =\big\langle[y(t+\Delta t)-y(t)]^2\big\rangle
    =\frac{2k_{\rm B}T}{k_y}
                      \left(1-e^{-\Delta t/\tau_y}\right),\qquad
 \tau_y=\frac{c_{\rm loc}}{k_y}.
 \label{eq:opening-thermal-msd}
\end{equation}

Curvature and temperature set the positional width; the ratio
$c_{\rm loc}/k_y$ sets the correlation time. At long lag, decorrelated
positions give twice the equilibrium variance.
Figure~\ref{fig:opening-thermal-width-time} separates these static and
dynamic scales. The constant-drag description applies beyond its
omitted inertial and material-memory transients.

\begin{figure}[H]
\centering\begingroup
\definecolor{vltTeal}{HTML}{176B64}
\definecolor{vltBlue}{HTML}{315D80}
\definecolor{vltGold}{HTML}{A87520}
\definecolor{vltPlum}{HTML}{725877}
\definecolor{vltRed}{HTML}{762B38}
\definecolor{vltYellow}{HTML}{D5AD3C}
\definecolor{vltNeutral}{HTML}{777D83}
\setlength{\unitlength}{1pt}
\scalebox{0.80}{%
\input{vlt-figures/figure-011.tex}%
}\endgroup
\caption{Static width and relaxation time in a classical overdamped harmonic model. (a) At fixed temperature, the labelled curvatures give normalized Gaussian position densities on shared axes. (b) At fixed $k_y=k_\star$, changing resistance changes $\tau_y=c_{\rm loc}/k_\star$ while preserving the MSD plateau. Dots locate the normalized value $1-e^{-1}$ at each correlation time. The declared scales are $\ell_\star=\sqrt{k_{\rm B}T/k_\star}$ and $\tau_\star=c_\star/k_\star$. Solid, dashed and dotted styles preserve the case identities in grayscale.}
\label{fig:opening-thermal-width-time}
\end{figure}

Section~\ref{sec:viscoelastic-dynamics} derives the more general causal
response. In equilibrium, the same eliminated coupling determines memory
and compatible thermal forcing. At fixed relaxed curvature and
temperature, different memory laws can redistribute spectral power while
preserving the static covariance. In the thermally isolated realization,
energy transferred to the medium stays within the complete system; an
isothermal reduction includes its thermal exchange. Prepared material
strain can return energy to the particles. That strain is retained through
a change of reference or cluster membership. Different stored strains
can therefore produce different futures from the same particle positions
and velocities.

Biophysical variables assign these mechanisms to concrete processes.
Redistribution among configurations softens relaxed curvature, with
transition rates setting the response time. Ionic transfer uses an
electrochemical affinity and reciprocal force derived from one free
energy. Sections~\ref{sec:phonon-ionic-extension}
and~\ref{sec:phonon-biochemical-extensions} develop finite populations,
gates and chemical resources. Prepared gradients and fuel can supply
work while depleting; sustained activity uses a continuing supply and
loss account as in Section~\ref{sec:self-excitation-passive-boundary}.

Complex viscosity $\eta^*(\omega)$ expresses material response in the
frequency domain. With the $e^{-i\omega t}$ convention,
$G^*(\omega)=-i\omega\eta^*(\omega)$ relates the strain modulus
to viscosity. Geometry and boundary coupling convert that material law,
in $\mathrm{Pa\,s}$, to a particle resistance or hydraulic impedance.
Section~\ref{sec:s5-cylindrical-fluid-context} uses pressure-driven
cylindrical flow as an externally driven extension: pressure difference
and volume flux supply an additional force--rate pair and a measurable
power input. Poiseuille and unsteady-flow limits provide benchmarks for
this extension, with pressure ports specified separately from the sealed
reference boundary.

\subsection{Collective modes and the viral-phonon hypothesis}
\label{sec:architecture-introduction}

A normal mode is a coordinated displacement pattern preserved by the
conservative restoring force and inertia. Its shape $\vct U_s$ and
frequency $\omega_s$ solve

\begin{equation}
 \boxed{\mathsf K\vct U_s=\omega_s^2\mathsf M\vct U_s.}
 \label{eq:introduction-mode-problem}
\end{equation}

In a positive-frequency mode, all participating coordinates oscillate at
the same frequency with fixed relative amplitudes. General conservative
harmonic motion is a superposition of these patterns~\cite{Misra2012,FeynmanLectures1963}.
The medium can then attenuate, relax or mix their amplitudes. The
conservative mode basis remains a useful representation even when the
full dissipative response couples its coordinates.

\paragraph{Superposition and what the coefficients mean.}
Several collective patterns can contribute to the same particle motion
at once. In the linear reference, their signed displacements add at
each particle. Relative phases determine where they reinforce or cancel;
a small signal at one location can coexist with substantial motion
elsewhere. This classical superposition underlies the mode
reconstruction, packet and interference calculations below. Section~\ref{sec:sup-mechanical}
states its mechanical conditions; Section~\ref{sec:complex-field-observables}
connects it to measured signals. Quantum occupation amplitudes and
classical structural probabilities have their separate normalization
and interpretation, summarized in Table~\ref{tab:sup-coefficient-meanings}.

For two equal masses on a line,
$\vct e_+=(1,1)^{\mathsf T}/\sqrt2$ describes common motion and
$\vct e_-=(1,-1)^{\mathsf T}/\sqrt2$ describes relative motion.
Writing $\vct u=a_+\vct e_++a_-\vct e_-$ resolves the physical
displacement into these two patterns. A free dimer with spring stiffness
$k$ has a common translational zero mode and a relative frequency
$\omega_-^2=2k/m$. Support can restore common translation as well.
Figure~\ref{fig:introduction-modes-waves} connects this pair to spatial
patterns in a larger assembly.

Pattern overlap gives the linear algebra its mechanical meaning.
The real symmetric matrix $\mathsf M^{-1/2}\mathsf K\mathsf M^{-1/2}$
has a Euclidean-orthonormal eigenbasis. Physical modes can consequently
be mass normalized. In a fixed positive-frequency sector,
\begin{equation}
 \vct U_s^{\mathsf T}\mathsf M\vct U_{s'}=\delta_{ss'},\qquad
 \vct u_{\rm vib}=\sum_s\vct U_sQ_s,\qquad
 Q_s=\vct U_s^{\mathsf T}\mathsf M\vct u.
 \label{eq:introduction-modal-reconstruction}
\end{equation}
Here $\vct u_{\rm vib}$ is the vibrational part of the displacement,
and $Q_s$ specifies how much of pattern $s$ is present;
its units are $\mathrm{kg^{1/2}\,m}$ for translational coordinates,
while $\vct U_s$ has units $\mathrm{kg^{-1/2}}$. For the equal-mass
pair, $\vct U_\pm=\vct e_\pm/\sqrt m$ and $Q_\pm=\sqrt m\,a_\pm$.
Any free rigid coordinates are treated separately from this vibrational
sector. Choosing a basis changes the coordinates of a motion while
preserving its physical displacement and energy.

The resulting finite-dimensional inner-product space is a Hilbert
space. Bra--ket notation records vectors, adjoints and
projections~\cite{Shankar1994}; complex coefficients record amplitude
and phase. Continuum self-adjointness additionally specifies operator
domains and boundary conditions~\cite{Teschl2009}.
Section~\ref{sec:complex-displacement} develops the corresponding
real-motion, energy and finite-record representations.

\begin{figure}[H]
\centering\begingroup
\definecolor{vltTeal}{HTML}{176B64}
\definecolor{vltBlue}{HTML}{315D80}
\definecolor{vltGold}{HTML}{A87520}
\definecolor{vltPlum}{HTML}{725877}
\definecolor{vltRed}{HTML}{762B38}
\definecolor{vltYellow}{HTML}{D5AD3C}
\definecolor{vltNeutral}{HTML}{777D83}
\setlength{\unitlength}{1pt}
\scalebox{0.90}{%
\input{vlt-figures/figure-012.tex}%
}\endgroup
\caption{Collective-coordinate geometry for an equal-mass pair and an auxiliary interval. (a) Equal-scale axes $X_i=u_i/\ell_\star$ carry orthonormal common and relative directions. Dashed sides reconstruct the displayed displacement; the two virion rows decode signs of the corresponding components. (b) Peak-normalized free-end interval patterns have rings at displacement nodes and short tangents at their zero-slope ends. Multiplying a pattern by its modal amplitude reconstructs motion. The interval boundary problem has its own spatial normalization. The boxed result illustrated here is~\eqref{eq:introduction-mode-problem}.}
\label{fig:introduction-modes-waves}
\end{figure}

The quantized extension assigns energy levels to each positive-frequency
mode. A \emph{viral phonon} is one excitation quantum, with energy
increment $\hbar\omega_s$. Mode occupation counts these quanta, while
$N_{\rm tot}$ counts particles and $n_{\mathcal C}(t)$ counts selected
members. A spatially localized excitation generally combines modes.
Sections~\ref{sec:lattice-architecture}
and~\ref{sec:viral-phonon-mechanics} develop excitation transfer,
scattering and wave packets from the mechanical couplings. Bulk
wave-vector labels describe an auxiliary repeated reference; the finite
cavity retains its own boundaries and mode spectrum. Mode quantization
is an additional physical postulate beyond this classical spectrum.

For an independent quantized harmonic mode in thermal equilibrium,

\begin{equation}
 x=\frac{\hbar\omega}{k_{\mathrm B}T}
   =\frac{hf}{k_{\mathrm B}T},\qquad
 \bar N=\frac{1}{e^x-1},\qquad f=\frac{\omega}{2\pi},
 \label{eq:introduction-quantum-thermal-scale}
\end{equation}

where $h=2\pi\hbar$ and $\bar N$ is mean occupation. The ratio $x$
compares a quantum with thermal energy. At $x\ll1$, many quanta are
occupied and equilibrium oscillator statistics approach the classical
limit. For illustration, $k_{\rm B}T/h\simeq6.25\,\mathrm{THz}$ at
$T=300\,\mathrm K$, giving $\bar N\simeq6.25\times10^3$ for an
assigned $1\,\mathrm{GHz}$ mode. Thermal occupation tends to zero
with temperature in the fixed harmonic model, while zero-point
fluctuations remain. Section~\ref{sec:phonon-temperature-comparison}
also treats temperature-dependent material properties.

The quantum extension specifies state preparation, coupling and
observables together. Section~\ref{sec:phonon-quantum-active} applies the
quantum van der Pol construction to one-quantum gain and ordinary and
two-quantum loss~\cite{LeeSadeghpour2013,WalterNunnenkampBruder2014}, with
an explicit resource and energy account. Its calculated occupation and
correlation statistics are predictions of those specified channels.
The classical mechanical and observation construction, including the
four tasks in Table~\ref{tab:introduction-contribution-map}, can be used
without the quantization postulate. A biological quantum interpretation
additionally requires an identified mode, preparation, coupling, damping
and thermal regime, together with a calibrated readout that distinguishes
the proposed statistics from relevant classical alternatives.
Section~\ref{sec:qfa-feasibility} makes this assessment explicit for the
selected HPV preparation, with calculated thermal scales and a
conditional preparation--storage--counting protocol.

\subsection{Collective response, controlled coupling and inference}
\label{sec:tunable-introduction}

A measurement selects part of the motion. Let $\vct w$ be a fixed,
real, dimensionless weighting column and let
$Y=\vct w^{\mathsf T}\vct u$ be its length-valued displacement
readout. The force pattern $F\vct w$ is conjugate to $Y$, so its
elementary work is $F\,\dd Y$. In linear response, with the
$e^{-i\omega t}$ convention,

\begin{equation}
 \boxed{Y(\omega)=\mathcal R_Y(\omega)F(\omega),\qquad
 \mathcal R_Y(\omega)=\vct w^{\mathsf T}
                        \mathcal R(\omega)\vct w.}
 \label{eq:introduction-readout-response}
\end{equation}

The matrix $\mathcal R$ maps force to displacement;
$\mathcal R_Y$ has units $\mathrm{m\,N^{-1}}$. Its magnitude is
displacement per force amplitude, and its argument is the phase relative
to that force. Tracking one centre, averaging a group and measuring
a separation assign different weights. Each mode contributes through
$\vct w^{\mathsf T}\vct U_s$. Material coupling determines the full
mechanical response; detector transfer and observation time determine
the recorded signal. Figure~\ref{fig:introduction-work-conjugate-readout}
compares common and relative channels on the same pair.

\begin{figure}[H]
\centering\begingroup
\definecolor{vltTeal}{HTML}{176B64}
\definecolor{vltBlue}{HTML}{315D80}
\definecolor{vltGold}{HTML}{A87520}
\definecolor{vltPlum}{HTML}{725877}
\definecolor{vltRed}{HTML}{762B38}
\definecolor{vltYellow}{HTML}{D5AD3C}
\definecolor{vltNeutral}{HTML}{777D83}
\setlength{\unitlength}{1pt}
\scalebox{0.90}{%
\input{vlt-figures/figure-013.tex}%
}\endgroup
\caption{Conjugate force and readout on one supported pair. Equal masses have identical supports $k_p$ and coupling $k_c$, shown at the reference positions. Matching coil spans and turn counts indicate geometry; labels specify stiffness. Equal-length arrows in (b,c) give the signed normalized force weights. The associated common and relative readouts each select one pattern and cancel the other. The conservative responses hold away from poles. Arrows encode force weights rather than displaced configurations; size and force gain are schematic. The boxed result illustrated here is~\eqref{eq:introduction-readout-response}.}
\label{fig:introduction-work-conjugate-readout}
\end{figure}

Camera frames make the observation step explicit. Let $Y(t)$ be the
instantaneous displacement channel and $Y_k$ its measured frame. Uniform
exposure of duration $\tau_e$, with frame-start interval $\Delta$, gives
\begin{equation}
 Y_k=\frac1{\tau_e}\int_{t_k}^{t_k+\tau_e}Y(t)\,\dd t
                  +\epsilon_k,\qquad
 t_k=k\Delta,\quad 0<\tau_e\leq\Delta.
 \label{eq:introduction-camera-frame}
\end{equation}
Here $\epsilon_k$ is calibrated localization error. For the stationary
scalar model in \eqref{eq:opening-thermal-reduction}, take $Y(t)=y(t)$,
write $\sigma_y^2=k_{\rm B}T/k_y$ for its instantaneous variance,
and set the localization error to zero. Exposure reduces the recorded
variance by a known factor,
\begin{equation}
 \frac{\operatorname{Var}Y_k}{\sigma_y^2}
     =\frac{2(a-1+e^{-a})}{a^2},\qquad a=\frac{\tau_e}{\tau_y}>0.
 \label{eq:introduction-camera-variance}
\end{equation}
Exposure suppresses the recorded variance even at fixed physical
curvature. Section~\ref{sec:introduction-worked-guide} evaluates this
bias in the shared example below.
Figure~\ref{fig:introduction-camera-observation} illustrates this
acquisition effect with the physical curvature held fixed. Camera blur
and localization noise have established roles in particle
tracking~\cite{Berglund2010,Calderon2016}. The exact joint frame--state
model in Section~\ref{sec:operator-exact-camera-frames} also retains
their lagged covariance, enabling a comparison across exposure settings
without assigning their effects to a changed interaction law.

\begin{figure}[H]
\centering\begingroup
\definecolor{vltTeal}{HTML}{176B64}
\definecolor{vltBlue}{HTML}{315D80}
\definecolor{vltGold}{HTML}{A87520}
\definecolor{vltPlum}{HTML}{725877}
\definecolor{vltRed}{HTML}{762B38}
\definecolor{vltYellow}{HTML}{D5AD3C}
\definecolor{vltNeutral}{HTML}{777D83}
\setlength{\unitlength}{1pt}
\scalebox{0.84}{%
\input{vlt-figures/figure-014.tex}%
}\endgroup
\caption{Acquisition changes the record of one overdamped harmonic coordinate. (a) Black shows the conditional mean after a prepared displacement, with time measured in $\tau_y$. Colored segments give its exact averages over exposures starting at zero; dashed guides mark their end times. The virion inset illustrates translation relative to a registered outline. (b) For a separate stationary thermal record of the same model, applying instantaneous equipartition directly to the frame variance gives the plotted apparent curvature. The marked exposure ratios are one and two. The physical curvature $k_y$, temperature and relaxation law stay fixed, with zero localization error. The curve follows \eqref{eq:introduction-camera-variance}; $k_{\rm app}/k_y$ tends to one for vanishing exposure.}
\label{fig:introduction-camera-observation}
\end{figure}

Direct particle interactions and a shared medium can both create
collective signals. Two-point microrheology uses correlated tracer motion
to probe the surrounding material~\cite{Crocker2000}. A particle's
local confinement can set its positional variance, while spatially shared
flow or deformation affects lagged cross-correlations and response phase.
Comparing separations, force patterns and frequency bands helps
distinguish retained elastic links from medium-mediated coupling.
These measurements give a concrete route from group motion to material
properties.

A controlled target interaction introduces another comparison. The
dimensionless parameter $\lambda\in[0,1]$ sets its strength, with
zero target coupling at $\lambda=0$. A stationary stable reference
at each value determines the corresponding susceptibility. Its variation
performs control work according to the energy derivative with respect
to $\lambda$. Binding, recruitment
and molecular transitions additionally require the physical coordinates
and kinetics governing those events. Across a registered rearrangement,
the theory continues particle positions, velocities and stored material
strain into the updated representation. Its causal response then depends
on both excitation and observation times, as well as on the specified
assembly history.

The \emph{forward problem} predicts readouts from those physical inputs.
The \emph{inverse problem} asks which inputs or parameter combinations
the observations determine. \emph{Identifiability} is the requirement
that the specified observations distinguish the quantity being inferred;
noise, record length and calibration add practical
uncertainty~\cite{Raue2009}. Section~\ref{sec:phonon-calculable-outcomes}
constructs distinct hidden storage, coupling and mobility parameters with
the same mechanical response and equilibrium fluctuations. An
independently calibrated ionic or configurational channel separates them.
The example makes the choice of a second measurement a physical part of
the inference problem.

Variable selection asks which retained quantities predict the target
observations, and over what window they form a closed state.
Section~\ref{sec:operator-framework} gives exact linear criteria for
target recovery and closure, including assembly events.
Section~\ref{sec:viral-phonon-mechanics} applies a specified preparation,
control, readout, time window and error tolerance to biophysical
examples. An added variable is justified by the predictive information
it supplies at that resolution.

Membrane fluctuations illustrate the distinct roles of state, material
law and observation. Shape amplitudes are state coordinates; bending
rigidity and tension determine their energy cost; an imaged contour
coefficient weights the shape through optical projection and exposure.
Contour analysis has used spatial fluctuations to infer bending and
tension~\cite{Pecreaux2004}. Time-dependent correlations additionally
test relaxation laws. Retained membrane, ionic or configurational
variables become useful when their dynamics change the selected
prediction. Figure~\ref{fig:introduction-biophysical-variable-roles}
organizes these roles, and
Section~\ref{sec:phonon-membrane-application} develops the membrane
example.

\begin{figure}[H]
\centering\begingroup
\definecolor{vltTeal}{HTML}{176B64}
\definecolor{vltBlue}{HTML}{315D80}
\definecolor{vltGold}{HTML}{A87520}
\definecolor{vltPlum}{HTML}{725877}
\definecolor{vltRed}{HTML}{762B38}
\definecolor{vltYellow}{HTML}{D5AD3C}
\definecolor{vltNeutral}{HTML}{777D83}
\setlength{\unitlength}{1pt}
\scalebox{0.90}{%
\input{vlt-figures/figure-015.tex}%
}\endgroup
\caption{State, material law and signal at three biophysical resolutions. A translating probe, a membrane surface and an internal separation provide different coordinates. The middle row identifies the retained state; subsequent rows identify constitutive inputs and the calibrated observation. A membrane image is a projection of surface shape, and the protein reporter measures the specified site separation. Coordinate-change arrows and registered outlines clarify the geometry. Each drawing uses an independent illustrative scale; the lobes have no atomistic assignment.}
\label{fig:introduction-biophysical-variable-roles}
\end{figure}

Figure~\ref{fig:introduction-calibration-linked-predictions} summarizes
the calibration-to-prediction sequence for a single passive coordinate.
The worked guide that follows applies the sequence to an HPV-inspired
pair, compares complementary motion records and then resolves
structural outcomes during transport and delivery.

\begin{figure}[H]
\centering\begingroup
\definecolor{vltTeal}{HTML}{176B64}
\definecolor{vltBlue}{HTML}{315D80}
\definecolor{vltGold}{HTML}{A87520}
\definecolor{vltPlum}{HTML}{725877}
\definecolor{vltRed}{HTML}{762B38}
\definecolor{vltYellow}{HTML}{D5AD3C}
\definecolor{vltNeutral}{HTML}{777D83}
\setlength{\unitlength}{1pt}
\scalebox{0.90}{%
\input{vlt-figures/figure-016.tex}%
}\endgroup
\caption{Two predictions from one calibrated overdamped coordinate with $k,c,T>0$. The upper spring is at reference geometry; force and displacement arrows identify separate physical quantities. The left response has $k|\mathcal R_Y|=[1+(\omega\tau)^2]^{-1/2}$, and the right normalized MSD is $1-e^{-\Delta t/\tau}$, with $\tau=c/k$. Both use the same restoring law, resistance and matched thermal forcing. Independent support and detector constraints precede comparison with the reserved observations.}
\label{fig:introduction-calibration-linked-predictions}
\end{figure}

\subsection{A worked guide: HPV-inspired motion and transport}
\label{sec:introduction-worked-guide}

This guide shows how to use the four tasks in
Table~\ref{tab:introduction-contribution-map}. Fix a physical preparation
and comparison models, calibrate inputs on designated records, then
predict complementary measurements using those same inputs. The first
calculation resolves two supported particle centres and a relaxing
material state. A subsequent kinetic example follows twelve structural
objects through delivery. Both use the same HPV-inspired empty-shell
identity, at different mechanical resolutions.

For the motion calculation, use two ideal, rigid HPV16-inspired L1 VLPs. HPV capsids have 72
pentameric capsomers, containing 360 L1 subunits, with $T_{\rm cap}=7d$
organization~\cite{Goetschius2021}. Hydrated HPV11/16 VLPs selected for
cryoTEM reconstruction had diameters $54\pm3\,\mathrm{nm}$
\cite{Zhao2014}; we choose a nominal diameter $55\,\mathrm{nm}$.
Each whole VLP is one mechanical node: capsid triangulation describes
its internal shell, whereas the lattice graph describes interactions
between particles. The chosen reference is an empty L1 shell, with no
genome or L2 coordinate retained.

\paragraph{Set up and calibrate the common model.}
Take $N_{\rm tot}=n=2$ particles, with axial reference centres at
$z=\pm75\,\mathrm{nm}$ in a cylinder of radius $500\,\mathrm{nm}$
and length $2\,\mu\mathrm m$. Independent supports retain their axial
displacements $u_1,u_2$; transverse displacement, orientation and capsid
deformation are omitted. Introduce the orthonormal coordinates
$q_\pm=(u_1\pm u_2)/\sqrt2$, with conjugate forces
$f_\pm=(f_1\pm f_2)/\sqrt2$. A relaxed interparticle curvature
$k_l$ and a material coordinate $h$ give
\begin{equation}
 E=\frac{k_pq_+^2}{2}+\frac{k_-q_-^2}{2}
       +\frac{\kappa_b}{2}(q_--h)^2,
 \qquad k_-=k_p+2k_l,\quad \kappa_b=2\Delta k.
 \label{eq:introduction-hpv-energy}
\end{equation}
Here $h$ is a length-valued relaxed material displacement, not a
capsomer position; it equals $y/\sqrt2$ in the full benchmark of
Section~\ref{sec:complete-benchmark}. For a specified Lennard--Jones
plus electrostatic pair law $U_{\rm pair}(r)$, the axial local input
is $k_l=U_{\rm pair}''(150\,\mathrm{nm})$, with support forces
balancing any nonzero first derivative. The guide prescribes this
curvature rather than inferring an HPV pair potential from capsid size.
Surface charge, screening and the harmonic range would be additional
preparation-specific calibration inputs. Intravirion shell stiffness
and disulfide bonds do not supply this intervirion curvature.

\begin{table}[H]
\centering\small
\begin{tabular}{@{}p{5.0cm}p{10.3cm}@{}}
\toprule
Input or scale & Assignment and status\\
\midrule
Nominal particle radius & $b_v=27.5\,\mathrm{nm}$; HPV-inspired geometric choice.\\
Temperature & $T=300\,\mathrm K$; prescribed isothermal bath.\\
Relaxed support and link & $k_p=k_l=3\times10^{-4}\,\mathrm{N\,m^{-1}}$; synthetic local curvatures.\\
Material branch & $\Delta k=6\times10^{-4}\,\mathrm{N\,m^{-1}}$, $\tau=0.5\,\mathrm{ms}$; synthetic stiffness and relaxation time.\\
Instantaneous resistance & $c_0=k_p\tau=1.5\times10^{-7}\,\mathrm{kg\,s^{-1}}$ per particle; prescribed diagonal resistance.\\
Observation & Calibrated axial positions, exposure, timing and localization covariance; declared separately from the mechanics.\\
\bottomrule
\end{tabular}
\caption{Inputs for the introductory guide. Structural information fixes
the particle identity and nominal size. The dynamical inputs are synthetic
and differ from the dimensional inputs of Section~\ref{sec:complete-benchmark},
while preserving its stiffness and resistance ratios. Neither a cavity
wall nor a particle diameter determines the supplied support or drag.}
\label{tab:introduction-hpv-inputs}
\end{table}

Use the overdamped regime over the selected observation band, with
independent instantaneous particle resistance and an equilibrated
isothermal bath. On $\vct X=(q_+,q_-,h)^{\mathsf T}$ the entire
linear stochastic calculation is
\begin{equation}
 \begin{aligned}
 \dd\vct X&=\mathsf A\vct X\,\dd t
       +\frac1{c_0}(f_+,f_-,0)^{\mathsf T}\dd t
       +\sqrt{\frac{2k_{\rm B}T}{c_0}}
                         \diag(1,1,1/2)\,\dd\vct W,\\
 \mathsf A&=\frac1\tau
       \begin{pmatrix}-1&0&0\\0&-7&4\\0&1&-1\end{pmatrix}.
 \end{aligned}
 \label{eq:introduction-hpv-drift}
\end{equation}
The three Wiener components are independent. Noise amplitudes follow
from the same mobilities as the drift. Before using this reduction for
data, check that inertia and unresolved fluid and shell dynamics lie
outside the band. Diagonal particle resistance is a comparison
assumption, tested below. Fix $k_p,k_l$ from relaxed force gradients;
holding $q_-=q_0$ after rapid loading gives excess force
$\kappa_bq_0e^{-t/\tau}$, fixing the branch amplitude and time.
At $q_0=1\,\mathrm{nm}$ its initial excess is $1.2\,\mathrm{pN}$.
A finite loading rise uses the correction in
Section~\ref{sec:complete-benchmark}. Reserve separate stationary and
driven records for the predictions that follow. The slowest drift
relaxation is $\tau/(4-\sqrt{13})=1.27\,\mathrm{ms}$, so stationary
comparisons require equilibration beyond the branch time $\tau$.

\paragraph{Turn the model into a calculation.}
The energy in~\eqref{eq:introduction-hpv-energy} fixes the restoring
coefficients; the mobilities and bath temperature fix both the drift
and noise in~\eqref{eq:introduction-hpv-drift}. Define
$\ell_T^2=k_{\rm B}T/k_p$. The stationary covariance of the retained
state and its exact sampled evolution are
\begin{equation}
 \begin{aligned}
 \mathsf C_X&=\ell_T^2
 \begin{pmatrix}
 1&0&0\\0&1/3&1/3\\0&1/3&7/12
 \end{pmatrix},\\
 \mathsf F_\Delta&=e^{\mathsf A\Delta},\qquad
 \mathsf Q_\Delta=\mathsf C_X-
          \mathsf F_\Delta\mathsf C_X\mathsf F_\Delta^{\mathsf T}.
 \end{aligned}
 \label{eq:introduction-hpv-computation}
\end{equation}
Here $\mathsf F_\Delta$ advances the conditional mean over a sampling
interval $\Delta$, and $\mathsf Q_\Delta$ is the covariance of the new
thermal increment. These follow from the matrix exponential and the
stationary Lyapunov equation. A deterministic applied force contributes
its propagated integral. To calculate a camera record, integrate the
same evolution over the exposure; Section~\ref{sec:operator-exact-camera-frames}
gives the joint frame--state law. This preserves the shared thermal
forcing of the frame and next state. A numerical implementation therefore
needs one drift matrix, a positive stationary covariance, a force input
and the observation timing, rather than separately fitted dynamics for
each measurement. The four tasks below use this model as a common
reference, with each alternative model and readout stated explicitly.

\paragraph{1. Interpret tracked motion through the camera model.}
Transform the calibrated particle records into common and relative
records and average the predicted motion over each exposure. For the
common coordinate, \eqref{eq:introduction-hpv-drift} gives relaxation
time $\tau$ and instantaneous variance $k_{\rm B}T/k_p$.
With exposure $\tau_e=\tau$ and zero localization error, applying
\eqref{eq:introduction-camera-variance} predicts a recorded variance
$2/e$ times that value. The instantaneous common-coordinate rms is $3.72\,\mathrm{nm}$;
the recorded rms becomes $3.19\,\mathrm{nm}$ in this exposure example.
The same physical confinement would appear $36\%$ stiffer if exposure
were ignored. For the relative coordinate,
use the full matrix drift and joint frame--state covariances of
Section~\ref{sec:operator-exact-camera-frames}, including the shared
thermal increments. Comparing several known exposures tests acquisition
effects at unchanged mechanics. Longer exposure reduces recorded
motion; it does not by itself establish a change in the pair potential.

\paragraph{2. Separate elastic coupling from shared fluid motion.}
First solve the stationary Lyapunov equation for
\eqref{eq:introduction-hpv-drift} and transform back to particle coordinates.
Equivalently, integrate the Gibbs law over $h$. Both routes give
\begin{equation}
 \mathsf C_u=\frac{k_{\rm B}T}{k_p}
       \begin{pmatrix}2/3&1/3\\1/3&2/3\end{pmatrix},
 \qquad \sqrt{\langle u_i^2\rangle}=3.03\,\mathrm{nm}.
 \label{eq:introduction-hpv-covariance}
\end{equation}
The particle correlation is $1/2$. Excursions are small relative to
geometric clearances; a physical pair law must additionally support
this harmonic range. Next declare a fluid-only comparison: remove the
elastic and material links, retain the independent supports, and use
constant mobility $\mu_0\begin{pmatrix}1&\epsilon_\mu\\
\epsilon_\mu&1\end{pmatrix}$ with $\mu_0=1/c_0$ and
$\epsilon_\mu=0.2$, together with its matched thermal noise.
Its equilibrium cross-covariance is zero, but
\eqref{eq:phonon-hydrodynamic-lagged-anticorrelation} gives
$C_{12}(\tau)=-0.0741\,k_{\rm B}T/k_p$ and common/relative
decay rates $1.2/\tau$ and $0.8/\tau$. These stipulated models
therefore separate an energy coupling from a mobility coupling using
covariance and lagged or driven records together. A real preparation
can contain both. Driving the two particles in phase measures the common
channel; driving with equal and opposite forces measures the relative
channel. The pair thus provides a control and a link-sensitive record
without changing the specimen. The cylinder requires its own mobility calibration;
the chosen off-diagonal mobility is not derived from its geometry.
Compare instantaneous model covariances or forward-filter every
candidate: exposure itself can alter measured equal-time correlations.

\paragraph{3. Design a complementary measurement.}
The same relative energy can be written
$E_-=(k_xx^2+2b_{xh}xh+a_hh^2)/2$, where $x=q_-$,
$k_x=k_-+\kappa_b$, $b_{xh}=-\kappa_b$, $a_h=\kappa_b$,
and the hidden mobility is $\Lambda=(\kappa_b\tau)^{-1}$.
At fixed physical coordinate units, replace
$(a_h,b_{xh},\Lambda)$ by $(4a_h,2b_{xh},\Lambda/4)$.
Proposition~\ref{prop:s7-mechanical-parameter-equivalence} shows that
the mechanical response and matched equilibrium particle records are
unchanged. Yet the conditional slope
$\dd\mathbb E[h\mid x]/\dd x=-b_{xh}/a_h$ halves, and
$\operatorname{Var}(h\mid x)=k_{\rm B}T/a_h$ quarters.
A calibrated hidden-coordinate channel can distinguish those models
when more mechanical averaging cannot. An HPV realization would need
to identify that coordinate and its readout. Complementary light-scattering
and electrophoretic-mobility measurements have already resolved distinct
aspects of HPV16 L1 VLP stabilization~\cite{Vega2016}; they motivate
additional channels without assigning a measured zeta potential to
this length-valued $h$ or to a universal particle charge.

\paragraph{4. Test retained state with linked response and fluctuations.}
Prepare identical particle positions but different mean material
displacements. For $q_-=0$, changing $h$ by $1\,\mathrm{nm}$ changes
the initial mean relative drift by
$(\kappa_b/c_0)\,1\,\mathrm{nm}=8\,\mathrm{nm\,ms^{-1}}$.
The particle positions alone cannot predict both conditional futures.
Retain $h$, or its equivalent causal memory with the preparation term.
This is a state-sufficiency question, distinct from the hidden-parameter
ambiguity in task 3.

For a stationary comparison, eliminate $h$ and calculate the relative
susceptibility using $e^{-i\omega t}$. With $r=\omega\tau$,
\begin{equation}
 \begin{aligned}
 \mathcal R_-(\omega)&=\frac1{k_p}
       \left[3-ir+4\frac{-ir}{1-ir}\right]^{-1},\\
 S_-(\omega)&=2k_{\rm B}T
       \left[c_0+\frac{\kappa_b\tau}{1+r^2}\right]
                              |\mathcal R_-(\omega)|^2.
 \end{aligned}
 \label{eq:introduction-hpv-response-spectrum}
\end{equation}
Here $S_-$ is the two-sided angular-frequency displacement spectrum.
Compare the full model with the memory-free Langevin model L having
the same $k_-$ and low-frequency relative drag
$c_0+\kappa_b\tau=5c_0$, with its matched white noise.
At $r=1$, or $f=1/(2\pi\tau)=318\,\mathrm{Hz}$,
$k_p\mathcal R_-=(5-3i)^{-1}$ and
$k_p\mathcal R_-^{\rm L}=(3-5i)^{-1}$. Consequently,
\begin{equation}
 \begin{array}{c|cc}
 &\text{full material state}&\text{memory-free L}\\ \hline
 \sqrt{\langle u_i^2\rangle}&3.03\,\mathrm{nm}&3.03\,\mathrm{nm}\\
 |\mathcal R_-|&0.572\,\mathrm{nm\,pN^{-1}}&0.572\,\mathrm{nm\,pN^{-1}}\\
 \arg\mathcal R_-&30.96^\circ&59.04^\circ\\
 k_pS_-/(k_{\rm B}T\tau)&3/17&5/17
 \end{array}
 \label{eq:introduction-hpv-comparison}
\end{equation}
The variances refer to instantaneous positions before observation filtering.
Positive phase denotes displacement lag. The equal crossover magnitudes
and integrated variances coexist with a $28.1^\circ$ phase difference
and a $40\%$ lower full-model spectral density at that frequency.
A $1\,\mathrm{pN}$ conjugate relative force gives a mean $q_-$ amplitude
$0.572\,\mathrm{nm}$ in either model; the individual forces are
$\pm1/\sqrt2\,\mathrm{pN}$ and particle amplitudes are
$0.404\,\mathrm{nm}$. The phase distinguishes the models. After
matching detector transfer, exposure and sampling, compare
held-out phase and spectra across the crossover rather than fitting
each record separately. Section~\ref{sec:complete-benchmark} supplies
the finite-inertia calculation and continued link-change test. The
present overdamped rescaling preserves its dimensionless discrimination;
it does not replace its dimensional parameter table.

\paragraph{Apply the guide to transport and delivery.}
Structural integrity adds another resolved quantity. Use the normal,
reversibly deformed and long-lived damaged classes
$\mathrm N,\mathrm R,\mathrm D$ of
Section~\ref{sec:s7transport-delivery}, with probability column
$\vct p=(p_N,p_R,p_D)^{\mathsf T}$. The corresponding ket
$\ket p=p_N\ket{\mathrm N}+p_R\ket{\mathrm R}+p_D\ket{\mathrm D}$
is a classical probability representation. A shell reaction coordinate,
barrier model or measured transition law connects mechanical loading to
these classes. The translational Hessian and the material displacement
$h$ alone do not determine their transition rates. The assigned
transport unit $\tau_T$ is independent of the pair's material time
$\tau=0.5\,\mathrm{ms}$.

Start every object in $\mathrm N$. During transport, the rate $a$ moves
$\mathrm N\to\mathrm R$, $b$ returns $\mathrm R\to\mathrm N$,
and $d$ moves $\mathrm R\to\mathrm D$. The generator in
\eqref{eq:s7transport-three-state-generator} conserves probability.
Run it for $T=2\tau_T$ with $(a,b,d)\tau_T=(1.2,0.4,0.1)$.
At delivery, retain the structural class and change the rates to
$(a,b,d)\tau_T=(0,0.6,0.3)$. The ensemble prediction is
\begin{equation}
 \vct p(T+u)=
 e^{u\mathsf W_{\rm off}}e^{T\mathsf W_{\rm on}}
       (1,0,0)^{\mathsf T},\qquad u\geq0.
 \label{eq:introduction-transport-propagation}
\end{equation}
The matrix product preserves both probability and the preparation at
delivery. Damage is absorbing over this observation window.
The holding rates are assigned inputs; in particular, the increase
in $d$ is not derived from unloading or release of the material branch.

\begin{table}[H]
\centering\small
\begin{tabular}{@{}p{8.0cm}rrr@{}}
\toprule
Observation or holding comparison & $p_N$ & $p_R$ & $p_D$\\
\midrule
Delivery, $T=2\tau_T$ & $0.26494$ & $0.63703$ & $0.09804$\\
Main holding law, $u=3\tau_T$ & $0.66108$ & $0.04281$ & $0.29611$\\
Hold with unchanged $d\tau_T=0.1$, same $u$ & $0.74410$ & $0.07801$ & $0.17790$\\
Reset $\mathrm R\to\mathrm N$ at delivery, then main law & $0.90196$ & $0$ & $0.09804$\\
\bottomrule
\end{tabular}
\caption{Exact synthetic delivery fractions and two diagnostic comparisons.
The unchanged-hazard comparison keeps $a=0$ and $b\tau_T=0.6$.
The reset comparison erases the delivered deformed population; the
continued-state models preserve it. Parameters and the time unit are
assigned, rather than fitted to HPV transport data.}
\label{tab:introduction-transport-comparisons}
\end{table}

After unloading, no new $\mathrm R$ particles are created in this model.
Those already in $\mathrm R$ wait at total rate $b+d$, then recover
with probability $b/(b+d)$ or become damaged with probability
$d/(b+d)$. For the main holding law these branching probabilities are
$2/3$ and $1/3$. Over $u=3\tau_T$, the delivered population contributes
$0.39614$ to recovery and $0.19807$ to new damage; $0.04281$ remains
deformed. Both the normal and damaged fractions can therefore rise
while deformation subsides. Keeping $d$ at its transport value still
adds $7.99$ percentage points of damage after delivery. The continued
deformed population produces delayed loss, and the post-delivery rates
control its magnitude. Resetting that population to normal would miss
the later loss. This is a concrete instance of the retained-state test
in task~4.

The executable ancillary file
\texttt{VLP\_transport\_state\_benchmark.py} computes the ensemble law
and samples exact continuous-time paths. At each event it draws an
exponential waiting time from the total outgoing rate, then chooses a
destination in proportion to its rate. A delivery boundary changes the
rates while keeping the current class. With seed $20261102$, the twelve
illustrative paths in Figure~\ref{fig:s7transport-delivery-states} contain
$(N,R,D)=(4,7,1)$ at delivery and $(9,0,3)$ after holding.
The corresponding expected damaged counts are $1.18$ and $3.55$.
An individual finite shipment fluctuates around the ensemble fractions;
the illustrated count is one realization. The script also reports the
two holding comparisons in Table~\ref{tab:introduction-transport-comparisons}.

To apply this procedure, record the loading history and specify how it
loads shell or contact coordinates. Calibrate the structural classifier
and a transition law on designated data, then predict held-out protocols
and several post-delivery waiting times. Centre tracking constrains
within-basin motion; an integrity-sensitive image, compliance or antigenic
readout constrains the structural classes. Correlated outcomes require
the joint population and shared environment rather than the independent
objects used here. The coupled generator and path-survival construction
in Section~\ref{sec:s7transport-delivery} give that more general model.
A finite delivery calculation carries an initial preparation through a
time-dependent protocol; maintained steady circulation additionally
requires a continuing work or chemical source.

The output is a reproducible model comparison for an HPV-inspired
particle resolution, not an experimental determination of HPV collective
mechanics. Section~\ref{sec:hpva-application} develops
a separate, preparation-specific literature application with a provenance
ledger, competing models and a reserved prediction. Its measured summaries
do not replace the assigned constants of this guide. Glycan-induced changes in HPV16 capsid size and indentation
response~\cite{Feng2024} illustrate why preparation and internal resolution
can matter; those measurements would enter a deformation model rather
than directly set the translational springs above. A specific application
now has a short list of inputs to constrain and complementary records
to predict. The four motion calculations use classical mechanics and thermal
stochastic dynamics; the delivery example adds a classical kinetic law
whose mechanical calibration is an independent part of the application.

The comparisons in Table~\ref{tab:introduction-contribution-map} give
specific routes through the detailed development. A generalized Brownian
model with the same mechanical inputs, memory, preparation and thermal
noise gives the same passive predictions. The framework makes those
inputs and their reductions explicit, so that the adequacy of a chosen
particle, material and observation description can be calculated and
tested as additional processes become observable.

Sections~\ref{sec:lattice-architecture}
and~\ref{sec:structure-interactions} develop interactions, compatible
cells, collective spectra and transient membership.
Section~\ref{sec:viscoelastic-dynamics} derives causal memory, energy
exchange and continuation through rearrangement.
Section~\ref{sec:complex-displacement} connects real motion to complex
response, fluid benchmarks and finite records.
Section~\ref{sec:operator-framework} organizes operators, states,
observables and inference, including exact finite-exposure frame dynamics,
and combines them in the stationary and continued-response benchmark of
Section~\ref{sec:complete-benchmark}.
Section~\ref{sec:viral-phonon-mechanics} develops confinement, excitation
transfer, responsive surfaces and finite biophysical populations.
Section~\ref{sec:s7transport-delivery} connects structural classes to
delivery and survival, distinguishing finite protocols from maintained
nonequilibrium steady operation. The final applications compare which
effective variables and readouts suffice for a chosen prediction.

\clearpage

\section{The Architecture of the Viral Lattice}
\label{sec:lattice-architecture}

Section~\ref{sec:introduction} introduced the physical system and its
measurement questions. Here we specify its configuration space, energy,
interaction network and prepared state, then derive the collective modes.
The fixed pool belongs to the complete cavity; a selected assembly can
contain some or all of its particles. Subsection~\ref{sec:s23-membership-architecture}
defines changing membership, and Subsection~\ref{sec:s2-assembly-cells}
constructs compatible assembly cells. The harmonic calculation retains
one coordinate set near a persistent reference.
Subsection~\ref{sec:s23-transient-recruitment} supplies the nonlinear
encounter and escape dynamics that connect such references.

Three specifications organize the construction. Geometry locates the
particles and their clearances. The interaction graph selects retained
pairs. Each pair's constitutive law assigns its energy, force and
stiffness. Inertia, support and medium response complete the equations
of motion. Irregular finite arrangements and periodic reference examples
use this same construction.

To interpret a local interaction, imagine giving one virion a small trial
displacement while registering the other reference centres. The resulting
force changes act on several coordinates at once. A diagram spring
summarizes this local derivative of the interaction energy; its direction
and any reference tension matter alongside its longitudinal curvature.
The energy and coordinate definitions below specify exactly what each
matrix entry measures.

Preparation then turns those force laws into a prediction. Positions
and, when resolved, orientations locate the assembly. An inertial state
also retains momenta; material coordinates and surface populations retain
storage and chemical preparation. Conservative normal modes organize
the selected small-motion mechanical sector. Surface relaxation and
medium evolution use their corresponding augmented dynamics. Initial values, forcing and observation determine which
patterns are excited and which become visible. Keeping these roles
explicit makes the same construction usable for a small cell or a larger
assembly.

A \emph{field} assigns a value to each point of a specified domain.
A spatial force field assigns a three-component force to a position;
the collective force assigns $3n$ components to a particle configuration.
The Hamiltonian vector field assigns position and momentum rates to a
phase-space state. Figure~\ref{fig:architecture-field-assignment-map}
distinguishes these domains and units.

The three assignments use the same mechanical energy, viewed through
different inputs and outputs.

\begin{figure}[H]
\centering\begingroup
\definecolor{vltTeal}{HTML}{176B64}
\definecolor{vltBlue}{HTML}{315D80}
\definecolor{vltGold}{HTML}{A87520}
\definecolor{vltPlum}{HTML}{725877}
\definecolor{vltRed}{HTML}{762B38}
\definecolor{vltYellow}{HTML}{D5AD3C}
\definecolor{vltNeutral}{HTML}{777D83}
\setlength{\unitlength}{1pt}
\scalebox{0.84}{%
\input{vlt-figures/figure-017.tex}%
}\endgroup
\caption{Three field assignments for the same conservative model. (a) One particle position varies while the others are fixed. (b) An entire configuration receives one force component for each retained coordinate. (c) A phase-space state receives position and momentum rates; the small graph is a schematic two-coordinate slice. The open-headed downward arrows denote assignments. The two Hamiltonian output blocks retain their different physical units.}
\label{fig:architecture-field-assignment-map}
\end{figure}

Table~\ref{tab:architecture-model-inputs} collects the inputs needed to
turn structural information into a dynamical prediction and a calibrated
signal~\cite{Leake2016}.

\begin{table}[H]
\centering\small
\caption{The physical specification of a whole-virion mechanical model.
Inputs must be constrained independently where possible; their adequacy
is judged against the stated observation band.}
\label{tab:architecture-model-inputs}
\begin{tabular}{p{0.17\linewidth}p{0.34\linewidth}p{0.39\linewidth}}
\toprule
Input & What it specifies & Role in a prediction\\
\midrule
Resolution and geometry & Retained particles, reference positions,
exclusion sizes and any orientation or shape coordinates & Defines the
motions represented and the available clearances\\[3pt]
Inertia & Particle masses; moments of inertia for resolved rotations &
Relates forces and torques to acceleration at the retained resolution\\[3pt]
Conservative energy & Pair and boundary interactions; their local
curvatures in $\mathrm{N\,m^{-1}}$ for translations & Supplies force
balance and restoring stiffness; combines with inertia to set mode
frequencies\\[3pt]
Medium dynamics & Frequency-dependent storage, loss and spatial coupling &
Describes relaxation, energy transfer and motion transmitted through the
medium\\[3pt]
Preparation & Initial positions and momenta; medium history or an
explicit equilibrium ensemble & Selects a trajectory or statistical
distribution for the stated observation\\[3pt]
Force and readout & Where a calibrated force acts and which displacement
combination is recorded & Connects the mechanical response to measured
signals and to the parameters those signals constrain\\
\bottomrule
\end{tabular}
\end{table}

Each physical contribution enters the energy account once. If an
effective inertia or interaction already incorporates an environmental
effect, the explicit medium model contains its remaining contributions.
The reductions below state this matching.

\subsection{The sealed domain, configuration and mechanical state}
\label{sec:domain}

The ideal reference cavity is a stationary cylinder,
\begin{equation}
 \mathcal V=\{(x,y,z)\in\R^3:x^2+y^2<R_c^2,\ 0<z<L_c\},
 \label{eq:cylinder}
\end{equation}
where $R_c$ and $L_c$ are the radius and length. The wall and medium
boundary conditions specify mechanical energy flux. Particle positions,
charges and material preparation determine which cylindrical symmetries
also belong to the complete dynamics.

Choose $n$ reference positions $\vct R_i$ and masses $m_i>0$. Position,
displacement and velocity are related by
\begin{equation}
 \vct r_i(t)=\vct R_i+\vct u_i(t),\qquad
 \vct v_i(t)=\Dt{\vct u_i}(t),\qquad i=1,\ldots,n.
 \label{eq:reference-position}
\end{equation}
The reference position is fixed during a calculation; $\vct u_i(t)$ is
the displacement of particle $i$ from that position. The baseline collects
the $3n$ centre translations in
$\vct u=(\vct u_1,\ldots,\vct u_n)^{\mathsf T}$. A displacement pattern
therefore says which particles move, in which directions, and by how much.

These coordinates track whole-particle centres. Membrane, capsid or
internal relative motion can be resolved by additional coordinates,
or represented through matched effective interactions and medium
response over the selected observation band.

An interpolated displacement field draws a continuous pattern through
the particle displacements. Its domain and interpolation rule are part
of the model specification.

\begin{definition}[Reference displacement interpolation]
\label{def:s2-reference-interpolation}
Let $\vct X$ label a point in a declared reference region
$\mathcal D_{\rm ref}$ containing the retained sites. A differentiable
reference displacement interpolation is a map satisfying
\begin{equation}
 \vct u_c:\mathcal D_{\rm ref}\times\R\longrightarrow\R^3,
 \qquad \vct u_c(\vct R_i,t)=\vct u_i(t),\qquad
 \vct v_i(t)=\left.\frac{\partial\vct u_c}{\partial t}
                                  \right|_{\vct X=\vct R_i}.
 \label{eq:architecture-displacement-field}
\end{equation}
\end{definition}
Because $\vct R_i$ is fixed, the partial time derivative gives the
particle velocity at each site. Between sites, the interpolation rule
determines field values and spatial derivatives. Differentiating a field
indexed by current position along a moving point uses the material derivative
$\partial_t+\vct v\cdot\nabla_{\vct x}$ along a moving point.
A fluid velocity field additionally requires the medium's own equations.

Figure~\ref{fig:architecture-interpolation-nonuniqueness} shows explicitly why an interpolated field and its spatial derivatives require more information than the finite displacement list.

\begin{figure}[H]
\centering\begingroup
\definecolor{vltTeal}{HTML}{176B64}
\definecolor{vltBlue}{HTML}{315D80}
\definecolor{vltGold}{HTML}{A87520}
\definecolor{vltPlum}{HTML}{725877}
\definecolor{vltRed}{HTML}{762B38}
\definecolor{vltYellow}{HTML}{D5AD3C}
\definecolor{vltNeutral}{HTML}{777D83}
\setlength{\unitlength}{1pt}
\scalebox{0.90}{%
\input{vlt-figures/figure-018.tex}%
}\endgroup
\caption{(a) One reference-coordinate ruler locates all five displacement samples, with one transverse-arrow gain. (b,c) Both interpolants match the samples, while their spatial derivatives differ. The common horizontal range is $0\le X/a\le4$. Spatial interpolation supplies information beyond the finite displacement column; no force law or mode condition is assigned to these auxiliary profiles.}
\label{fig:architecture-interpolation-nonuniqueness}
\end{figure}

A small finite example is useful for reading the field and network
diagrams. Set $R_c=2a$, $L_c=4a$, $b_i=0.1a$, where $a>0$ is a length.
Place six reference centres at $\vct R_i=(a c_{i1},a c_{i2},L_c/2)$, with
\begin{equation}
 \begin{aligned}
 (\vct c_1,\ldots,\vct c_6)={}&
 \bigl((-1.2,-0.7),(0,-0.9),(1.15,-0.6),\\
 &\hspace{14mm}(-0.9,0.65),(0.25,0.45),(1.2,0.95)\bigr),\\
 \mathcal E={}&\{\{1,2\},\{2,3\},\{1,4\},\{2,5\},
                  \{3,5\},\{3,6\},\{4,5\},\{5,6\}\}.
 \end{aligned}
 \label{eq:architecture-illustrative-graph}
\end{equation}
All centres have positive pair and wall clearance. The corresponding
adjacency matrix simply records inclusion of a pair:
\begin{equation}
 A^{\rm net}_{ij}=\begin{cases}1,&\{i,j\}\in\mathcal E,\\
                            0,&\text{otherwise},\end{cases}
 \qquad
 \mathsf A_{\rm net}=
 \begin{pmatrix}
 0&1&0&1&0&0\\1&0&1&0&1&0\\0&1&0&0&1&1\\
 1&0&0&0&1&0\\0&1&1&1&0&1\\0&0&1&0&1&0
 \end{pmatrix}.
 \label{eq:architecture-illustrative-adjacency}
\end{equation}
The dimensionless symmetric matrix has zero diagonal. Its sixteen
off-diagonal entries record eight unordered pairs, each carrying its
assigned energy law. Force balance and three-dimensional stability are
then checked on the completed mechanical model. The same sites can
sample an imposed affine displacement:
\begin{equation}
 \frac{\vct u_c(X,Y,L_c/2)}{a}
       =0.08\left(\frac{X}{a},-\frac{Y}{a},0\right).
 \label{eq:architecture-illustrative-displacement}
\end{equation}
This field extends $X$ and contracts $Y$. Its linearized dilation is
zero; its exact in-plane area ratio $(1+0.08)(1-0.08)$ includes a
second-order decrease. Figure~\ref{fig:architecture-network-fields}
uses one arrow gain to compare displacement magnitudes. This imposed
snapshot precedes the normal-mode calculation.

\begin{figure}[H]
\centering\begingroup
\definecolor{vltTeal}{HTML}{176B64}
\definecolor{vltBlue}{HTML}{315D80}
\definecolor{vltGold}{HTML}{A87520}
\definecolor{vltPlum}{HTML}{725877}
\definecolor{vltRed}{HTML}{762B38}
\definecolor{vltYellow}{HTML}{D5AD3C}
\definecolor{vltNeutral}{HTML}{777D83}
\setlength{\unitlength}{1pt}
\scalebox{0.90}{%
\input{vlt-figures/figure-019.tex}%
}\endgroup
\caption{The same six labelled virions and eight pair-energy terms in the plane $z=L_c/2$. (a) Reference geometry. (b) The imposed affine field expands $X$ and contracts $Y$. Springs attach to the displayed body outlines; each matched connector retains its turn count and radius as its span changes. Body positions use the actual displacement; black arrows start at the references and show four times that displacement. The field is a prescribed deformation, with equal spatial scales in both panels.}
\label{fig:architecture-network-fields}
\end{figure}

\begin{definition}[Admissible finite-radius configuration]
\label{def:s2-admissible-configuration}
Assign effective exclusion radii $b_i>0$. With
$\vct r_i=(x_i,y_i,z_i)$ and $r_{ij}=|\vct r_i-\vct r_j|$, the
configuration set of the retained particle centres is
\begin{equation}
 \begin{split}
 \mathcal Q=\{\vct r\in\mathcal V^n:\;&
       \sqrt{x_i^2+y_i^2}\le R_c-b_i,\quad
       b_i\le z_i\le L_c-b_i,\\
      &r_{ij}\ge b_i+b_j\quad(i\ne j)\}.
 \end{split}
 \label{eq:finite-radius-configuration}
\end{equation}
\end{definition}
The dimensions and radii are chosen so that this configuration set is
nonempty. The reference has positive wall and pair clearances. These
clearances set the available room for local motion, while the lengths
over which forces vary set the range of the harmonic approximation. A
hard wall acts at contact. Smooth confinement, when retained, contributes
local restoring stiffness through the curvature of its boundary energy.

Ordinary space $\mathcal V$ locates one point. Configuration space
$\mathcal Q$ records all particle positions simultaneously; its
collision-free interior has dimension $3n$. Adding conjugate momenta
specifies the particle mechanical state.

\begin{definition}[Particle phase state]
\label{def:s2-particle-phase-state}
For the smooth translational sector, a particle phase state is a point
in the classical phase space
\begin{equation}
 \mathcal P=T^*\mathcal Q^\circ
       \simeq\mathcal Q^\circ\times\R^{3n},\qquad
       (\vct r,\vct p)\in\mathcal P .
 \label{eq:architecture-phase-space}
\end{equation}
\end{definition}
Here $T^*$ denotes the cotangent bundle: each configuration carries its
conjugate momentum coordinates. In these Cartesian coordinates it is
simply the displayed position--momentum product, with $6n$ state
coordinates. This is the particle state of the conservative baseline.
If dynamical medium variables are resolved, they also belong to the state;
if they are eliminated, their relevant history or equivalent auxiliary
variables must still be supplied.

Figure~\ref{fig:more-configuration-space} shows a two-dimensional slice
of the configuration set. Two individually allowed positions can still
be incompatible because the particles overlap.

\begin{figure}[H]
\centering\begingroup
\definecolor{vltTeal}{HTML}{176B64}
\definecolor{vltBlue}{HTML}{315D80}
\definecolor{vltGold}{HTML}{A87520}
\definecolor{vltPlum}{HTML}{725877}
\definecolor{vltRed}{HTML}{762B38}
\definecolor{vltYellow}{HTML}{D5AD3C}
\definecolor{vltNeutral}{HTML}{777D83}
\setlength{\unitlength}{1pt}
\scalebox{0.90}{%
\input{vlt-figures/figure-020.tex}%
}\endgroup
\caption{(a) Two labelled virions within their radius-$b$ spherical exclusion envelopes. Spatial lengths share one scale; the first centre moves from $A$ to $B$ by $b/4$. (b) The corresponding configurations are two nearby points in an equal-scale axial configuration slice. Shaded triangles satisfy both wall clearance and nonoverlap; solid sloping edges are contact boundaries. This slice fixes all other coordinates. It shows admissibility rather than a force or an equilibrium probability.}
\label{fig:more-configuration-space}
\end{figure}

A structural snapshot constrains configuration. Time-resolved data also
constrain momentum and the resulting trajectory, as illustrated by the
phase-space construction in Section~\ref{sec:introduction-hamiltonian}.

Table~\ref{tab:architecture-field-domains} fixes the vocabulary for the
construction. Each field assigns a value to a specified domain: a spatial
force has three components, while the collective force has one
three-component entry per particle~\cite{FeynmanLectures1963}.

\begin{table}[H]
\centering\small
\caption{Fields and state functions at the declared mechanical resolution.
The domain, value and units determine what a field describes.}
\label{tab:architecture-field-domains}
\begin{tabular}{p{0.23\linewidth}p{0.28\linewidth}p{0.39\linewidth}}
\toprule
Quantity & Domain and value & Physical interpretation\\
\midrule
Discrete displacement $\vct u_i$ & Particle label $i$; three components,
$\mathrm m$ & Motion of the represented whole-particle centre\\[3pt]
Interpolated $\vct u_c(\vct X,t)$ & Declared reference region;
$\R^3$, $\mathrm m$ & An additional spatial representation consistent
with the retained samples\\[3pt]
Energy $U_\lambda(\vct r)$ & $\mathcal Q^\circ$;
$\R$, $\mathrm J$ & Total potential energy of the retained
particle configuration\\[3pt]
Collective force $\vct F(\vct r)$ & $\mathcal Q^\circ$;
$\R^{3n}$, $\mathrm N$ & One three-component force per particle\\[3pt]
Electric potential and field & Ordinary space;
$\R$ in $\mathrm V$, $\R^3$ in $\mathrm{V\,m^{-1}}$ &
Energy per unit test charge and force per unit test charge\\[3pt]
Hamiltonian field $X_H$ & Phase space; tangent rates &
Position rates in $\mathrm{m\,s^{-1}}$ and momentum rates in $\mathrm N$\\
\bottomrule
\end{tabular}
\end{table}

The ideal hard wall supplies an elastic impact rule.

\begin{definition}[Specular hard-wall impact]
\label{def:s2-specular-impact}
At a single smooth contact with a stationary wall, let the unit normal
$\vct n_w$ be evaluated at the contact surface for the finite-radius
centre. The specular impact map is
\begin{equation}
 \vct v_i^{+}=\vct v_i^{-}
       -2(\vct v_i^{-}\cdot\vct n_w)\vct n_w,
 \qquad |\vct v_i^{+}|^2=|\vct v_i^{-}|^2.
 \label{eq:reflection}
\end{equation}
\end{definition}
Splitting velocity into normal and tangential components proves the
speed identity: reflection reverses the former and preserves the latter.
The stationary wall exchanges momentum without doing work. Frictionless
elastic particle contacts conserve pair momentum and kinetic energy.
Simultaneous contacts require a specified impact rule; resolved contact
deformation requires additional coordinates. The smooth-basin Hessian
applies between impacts.

Figure~\ref{fig:architecture-specular-velocity-map} shows the finite-radius contact surface and the velocity-component reflection, with equal incident and outgoing speeds.

\begin{figure}[H]
\centering\begingroup
\definecolor{vltTeal}{HTML}{176B64}
\definecolor{vltBlue}{HTML}{315D80}
\definecolor{vltGold}{HTML}{A87520}
\definecolor{vltPlum}{HTML}{725877}
\definecolor{vltRed}{HTML}{762B38}
\definecolor{vltYellow}{HTML}{D5AD3C}
\definecolor{vltNeutral}{HTML}{777D83}
\setlength{\unitlength}{1pt}
\scalebox{0.90}{%
\input{vlt-figures/figure-021.tex}%
}\endgroup
\caption{Specular reflection at one smooth hard contact. (a) The spherical contact envelope has radius $b$; its centre reflects at the dashed surface one radius from the wall. The faceted glyph identifies the ideal virion inside that envelope. (b) Equal-scale velocity axes show normal reversal and the unchanged tangential component. Dashed projections and the common speed circle expose both invariants. Each view has one velocity gain. The ideal impulse transfers momentum without changing kinetic energy.}
\label{fig:architecture-specular-velocity-map}
\end{figure}

A resolved continuum medium has its own displacement or traction
condition at the boundary, chosen here to give zero outward mechanical
energy flux. Particle reflection and medium boundary conditions together
complete the sealed-domain description. Periodic boundary conditions
define a separate idealization used in bulk calculations.

\begin{definition}[Resolved whole-particle configuration]
\label{def:s2patch-body-configuration}
Give each retained particle a compact body $\mathcal P_i^0$ centred at
the origin, an orientation $\mathsf O_i\in\operatorname{SO}(3)$ and
body-frame surface sites $\vct a_{ia}$. Its occupied body and site
positions are
\begin{equation}
 \mathcal P_i=\vct r_i+\mathsf O_i\mathcal P_i^0,
 \qquad \vct x_{ia}=\vct r_i+\mathsf O_i\vct a_{ia}.
 \label{eq:s2patch-body-sites}
\end{equation}
The smooth admissible configuration manifold $\mathcal Q_{\rm b}^\circ$
contains the labelled $(\vct r_i,\mathsf O_i)_{i=1}^n$ whose bodies
have positive pair and wall clearance in $\mathcal V$. Its local
dimension is $6n$ before imposed constraints. The mechanical phase
space is $T^*\mathcal Q_{\rm b}^\circ$. Retained surface populations
$\boldsymbol\alpha$ and material coordinates $\vct z$ enlarge the
state beyond this mechanical phase space.
\end{definition}

The spherical-envelope set in~\eqref{eq:finite-radius-configuration}
provides a sufficient positive-clearance sector of this body description.
A faceted contact calculation instead uses the actual occupied bodies.
Neither the number of surface sites nor the 72 capsomers of an
HPV-inspired shell changes the number of whole-particle nodes. Resolving
capsid assembly from pentamers would change the conserved building
blocks and require their own interactions and coordinates.

A transport-related integrity variable can label a normal shell, a
reversibly deformed shell and a long-lived damaged morphology. Those
classes retain their own structural coordinates and conditional
mechanics; Section~\ref{sec:s7transport-delivery} develops their
force-driven transition and delivery statistics. Material conservation
persists across a change of description, while disassembly requires the
resolved capsomer or protein bookkeeping.

Orientations have local three-component tangent increments, rather
than one globally regular triple of angles. Labeled body axes may be
used computationally; orientations related by a symmetry of the complete
body, surface pattern and readout represent the same physical state.
An asymmetric ligand or reporter can reduce that symmetry. The local
translation--rotation--deformation split is developed in
Section~\ref{sec:operator-geometric-extensions}.

An inertial realization retains linear and angular momenta. An
overdamped realization retains configuration and its material/surface
state with a mobility, thermal law and preparation; it does not acquire
instantaneous velocities by differentiating Brownian paths. The
translational Hamiltonian and mode theorems below specialize to fixed
orientations, fixed charges and the stated conservative sector. This
keeps the central reference available while making the additional
physical information explicit. The specular centre-reflection rule above
belongs to its spherical-envelope translational reference; actual facet
contact requires a specified force--torque impulse and angular-momentum
map.

\subsection{Modeling postulates and their validity}
\label{sec:postulates}

The postulates fix one autonomous reference calculation. Their use for a
particular preparation is assessed through force balance, stability,
excursion size and the observation interval.

\begin{postulate}[Fixed population and passive motion]
\label{post:population}
The complete cavity contains a fixed pool of $N_{\rm tot}$ ideal virions.
A reference calculation retains a fixed array of $n$ particles, equal to
the pool or a specified subset with its environmental coupling retained.
Their centre translations respond passively to the declared forces.
The particle and medium preparations supply the initial state.
\end{postulate}

\begin{postulate}[A stationary local reference]
\label{post:reference}
For each selected interaction setting, the complete conservative force
vanishes at the chosen reference. Its interaction network is retained
within a smooth local basin. The force-constant matrix is nonnegative
and positive on the selected vibrational sector. Its kernel is identified,
including symmetry directions and any additional nonrigid mechanisms.
\end{postulate}

\begin{postulate}[Resolved interaction energy]
\label{post:interaction}
The conservative model has a real, twice-differentiable energy near the
reference. Pair interactions, confinement and any retained medium or
target coupling are specified at the chosen structural resolution.
Effective terms state which unresolved effects they already include.
\end{postulate}

\begin{postulate}[Complete energy accounting]
\label{post:energy}
The reference enclosure is mechanically and thermally isolated, with
stationary boundaries and fixed controls. Neither particles nor energy
cross its boundary, and the complete enclosed energy, including coupling
terms, is conserved. A resolved subsystem may exchange
energy with omitted coordinates; its energy need not be conserved alone.
\end{postulate}

\begin{postulate}[A declared control setting]
\label{post:control}
A dimensionless parameter $\lambda\in[0,1]$ multiplies a prescribed
conservative target interaction. At $\lambda=0$ that coupling vanishes.
Each autonomous closed evolution uses a fixed setting for which a
stationary, stable reference exists. A change of setting during evolution
has a separate work account.
\end{postulate}

These postulates specify the closed conservative reference. A fixed VLP
population and a fixed surface charge are different restrictions.
Resolved rotations add torque balance; equilibrated surface populations
add stationarity under their chemical constraints. An isothermal reduced
free energy uses its stated bath or large-reservoir approximation,
while imposed pH, salt or flow introduces its chemical or mechanical
ports. In a strictly closed realization those reservoir variables are
included in the enclosed state and conservation account. The reduced
free energy governs available work and equilibrium weighting; the
complete isolated Hamiltonian remains the conserved energy.

Force balance concerns the sum at each particle. Individual pair forces
can be nonzero, balanced by other interactions or smooth support.
Positive constrained curvature gives a strict local minimum on the
vibrational sector. Symmetry motions belong to the kernel; additional
nonrigid mechanisms require higher-order analysis or a physical constraint.

A fixed-network calculation follows motion within the same local
basin. If $\tau_{\rm net}$ characterizes network changes and
$\tau_{\rm obs}$ the observation time, this description requires
$\tau_{\rm obs}\ll\tau_{\rm net}$ together with small excursions.
Thermal escape depends on the barriers in the nonlinear energy landscape
and on the environment, while the local curvature determines the
restoring force within the basin. Changing $\lambda$ generally changes
both the reference and the spectrum; each selected setting therefore has
its own force-balance and stability calculation.

The boundary, energy law and preparation specify complementary properties.
Material sealing fixes the particle pool. Postulate~\ref{post:energy}
also imposes thermal and mechanical isolation. An autonomous conservative
model exchanges energy among retained and hidden coordinates while
conserving their total. Equilibrium additionally specifies a stationary
statistical preparation; a sealed system can be prepared away from it.

\begin{definition}[Metastable residence at a declared resolution]
\label{def:s2patch-metastable-residence}
A metastable assembly is an admissible basin whose typical residence
time is long compared with its local relaxation time under the declared
preparation and dynamics. A kinetic trap for a target is a basin that
suppresses reaching that target within a specified observation window.
Membership size, topology, orientation and stored surface/material state
are candidate descriptors; sufficiency is assessed by the future record
they predict. The local Hessian describes basin fluctuations, while
escape uses the nonlinear landscape and dynamics of
Section~\ref{sec:s23-transient-recruitment}.
\end{definition}

The local approximation is controlled by spatial and observational
scales. Particle excursions must be small compared with the pair and
wall clearances and with the distance over which the force law changes
appreciably. A steep short-range force can therefore require a tighter
displacement range than the available clearance suggests. The retained
coordinates and constitutive inputs must also describe the relevant time
and frequency band. Internal coordinates can contribute effective inertia
or memory within that band. The mode construction below carries these
scale assumptions into the collective dynamics.

\begin{definition}[Parameters and evolving state]
\label{def:s2-parameters-and-state}
Parameters specify the mechanical model; state variables describe its
current condition. Masses, radii, interaction coefficients and a fixed
$\lambda$ are inputs to one autonomous calculation. Positions and momenta
evolve according to that model. Comparing preparations requires recording
both the initial states and any changed effective parameters, so an
observed change in motion can be related to its physical cause. A
resolved protonation occupancy or retained material strain is state;
its intrinsic acidity, mobility or stiffness specifies its law.
Prescribed pH and ionic chemical potentials are reservoir descriptors
or controls. Fixed charge and screening remain parameters in the
central reference, with their physical-to-effective mapping stated.
\end{definition}

Figure~\ref{fig:architecture-clearance-harmonic-accuracy} separates the geometric clearance test from the accuracy of a harmonic restoring force in a declared scalar example.

\begin{figure}[H]
\centering\begingroup
\definecolor{vltTeal}{HTML}{176B64}
\definecolor{vltBlue}{HTML}{315D80}
\definecolor{vltGold}{HTML}{A87520}
\definecolor{vltPlum}{HTML}{725877}
\definecolor{vltRed}{HTML}{762B38}
\definecolor{vltYellow}{HTML}{D5AD3C}
\definecolor{vltNeutral}{HTML}{777D83}
\setlength{\unitlength}{1pt}
\scalebox{0.80}{%
\input{vlt-figures/figure-022.tex}%
}\endgroup
\caption{The same trial interval is shaded in both panels; dotted blue boundaries mark an independently imposed clearance. (a) The quadratic and full auxiliary energy have the same curvature at the origin. (b) Their relative restoring-force error has continuous value zero at the origin and maximum $0.176$ on $|X|\leq0.4$. The two tests quantify geometric admissibility and local force accuracy using the displayed normalization.}
\label{fig:architecture-clearance-harmonic-accuracy}
\end{figure}

\subsection{The ideal virion and its interaction energy}
\label{sec:ideal-virion}

\begin{definition}[Ideal virion]
\label{def:ideal-virion}
An \emph{ideal virion}, or \emph{ideal VLP}, is a whole-particle
mechanical representation with mass $m_i$, position $\vct r_i$ and
specified body geometry. The central reference uses a rigid icosahedral
body, spherical exclusion radius $b_i$, fixed effective interaction
charge $q_i$ and centre translation under radial pair forces.
Definition~\ref{def:s2patch-body-configuration} retains orientation,
resolved surface states and their additional physical laws when needed.
Internal deformation requires its own coordinates and constitutive law.
\end{definition}

The effective parameters summarize structure below the whole-particle
resolution established in Section~\ref{sec:measles}. Every point of the
rigid body shares its centre translation. Resolving shell components or
internal units instead assigns nodes and relative coordinates at that
smaller scale.

The radius $b_i$ sets geometric exclusion. A circumsphere provides a
sufficient nonoverlap rule for the faceted body; a fitted effective
sphere applies over its stated range. Hydrodynamic and contact radii
belong to their corresponding measurements. Particle symmetry and
assembly symmetry remain separate, as discussed in
Section~\ref{sec:introduction-virion-geometry}.
Section~\ref{sec:operator-geometric-extensions} develops resolved shape
and orientation.

Figure~\ref{fig:architecture-faceted-exclusion-map} shows how an ideal faceted particle, its exclusion envelope and its retained centre coordinate enter this description.

\begin{figure}[H]
\centering\begingroup
\definecolor{vltTeal}{HTML}{176B64}
\definecolor{vltBlue}{HTML}{315D80}
\definecolor{vltGold}{HTML}{A87520}
\definecolor{vltPlum}{HTML}{725877}
\definecolor{vltRed}{HTML}{762B38}
\definecolor{vltYellow}{HTML}{D5AD3C}
\definecolor{vltNeutral}{HTML}{777D83}
\setlength{\unitlength}{1pt}
\scalebox{0.84}{%
\input{vlt-figures/figure-023.tex}%
}\endgroup
\caption{(a) The dashed circumsphere encloses the faceted body. (b) A rigidly translated copy retains its shape and orientation. (c) The gold gap is measured between the spherical envelopes along the centre line; it is not the nearest facet-to-facet distance. Nonoverlapping circumspheres provide the displayed sufficient exclusion test. Particle sizes and shifts are schematic; no hydrodynamic or contact radius is inferred.}
\label{fig:architecture-faceted-exclusion-map}
\end{figure}

The mass $m_i$ sets the inertia of the centre coordinate. Fluid loading
can be retained in the medium response or incorporated consistently into
an effective inertia. The charge $q_i$ sets the amplitude and sign of the
particle's contribution to the chosen electrostatic law. It summarizes
surface and environmental structure at that law's resolution. Structural
images constrain particle geometry; mechanical or interaction data
constrain the forces and their derivatives. These inputs are interpreted
for the same declared preparation.

The charge pattern enters through the signed products $q_iq_j$ at
labelled sites. Reassigning those charges can therefore change the
mechanics at unchanged positions. The scalar law below uses monopole
resolution; a finer model can resolve surface-charge structure.

A concrete pair model combines a screened Coulomb interaction with a
Lennard--Jones interaction:
\begin{equation}
 V_{ij}(r)=\frac{q_iq_j}{4\pi\varepsilon_m}
                \frac{e^{-\kappa r}}{r}
       +4\epsilon_{ij}\left[
          \left(\frac{\sigma_{ij}}{r}\right)^{12}
          -\left(\frac{\sigma_{ij}}{r}\right)^6\right].
 \label{eq:pair-potential}
\end{equation}
Here $\varepsilon_m$ is absolute permittivity in
$\mathrm{F\,m^{-1}}$, $\kappa$ is inverse screening length,
$\epsilon_{ij}>0$ is an energy, and $\sigma_{ij}>0$ is an interaction
length. The unscreened limit is $\kappa=0$. These effective interaction
laws~\cite{Israelachvili2011} provide a tractable reference model whose
parameters and range of applicability are specified for the preparation.
Surface geometry can be included through orientation-dependent or
many-body terms when the observations require that resolution.

The screened term pushes like-sign charges apart and pulls opposite-sign
charges together. Its screening length controls how rapidly this
electrostatic contribution falls with separation. The Lennard--Jones term
supplies steep short-range repulsion and an attractive range. Together
they determine the pair force and its change as neighbouring centres
move. The domain in~\eqref{eq:finite-radius-configuration} separately
enforces hard exclusion at $r=b_i+b_j$; $\sigma_{ij}$ sets an interaction
scale within the allowed domain.

The connection between energy and force begins with reversible work.

\begin{definition}[Conservative force at fixed controls]
\label{def:s2-conservative-force}
For differentiable $U_\lambda$ in the smooth configuration region, the
conservative force is the negative configuration gradient. Its virtual
work under $\delta\vct r$ is
\begin{equation}
 \delta W_{\rm cons}=\sum_i\vct F_i\cdot\delta\vct r_i
                  =-\delta U_\lambda,\qquad
 \boxed{\vct F_i=-\frac{\partial U_\lambda}{\partial\vct r_i}.}
 \label{eq:architecture-work-force}
\end{equation}
\end{definition}

Figure~\ref{fig:architecture-pair-virtual-work} applies this work relation to one central harmonic pair, keeping its internal forces distinct from any support reactions.

\begin{figure}[H]
\centering\begingroup
\definecolor{vltTeal}{HTML}{176B64}
\definecolor{vltBlue}{HTML}{315D80}
\definecolor{vltGold}{HTML}{A87520}
\definecolor{vltPlum}{HTML}{725877}
\definecolor{vltRed}{HTML}{762B38}
\definecolor{vltYellow}{HTML}{D5AD3C}
\definecolor{vltNeutral}{HTML}{777D83}
\setlength{\unitlength}{1pt}
\scalebox{0.90}{%
\input{vlt-figures/figure-024.tex}%
}\endgroup
\caption{A central harmonic pair at two separations. The common ruler fixes particle $i$ at $z=0$, with the reference $j$ at $2\ell_\star$ and the extended $j$ at $2.4\ell_\star$. Equal body sizes and seven-turn coils retain their family as the free span increases by $0.4\ell_\star$. Teal arrows show attractive pair forces; the plum arrow is a virtual displacement. Panel (b) evaluates the same state, $X=0.4$. Force and virtual-motion arrows use separate gains; support reactions, if present, enter separately. The boxed result illustrated here is~\eqref{eq:architecture-work-force}.}
\label{fig:architecture-pair-virtual-work}
\end{figure}

Differentiating with respect to $\vct r_i$ holds the other positions
fixed and gives three force components, in $\mathrm{J\,m^{-1}}=\mathrm N$.
Every energy term contributes, including
$-\partial U_{\mathrm{vm}}/\partial\vct r_i$ when the medium coupling
is explicit. For a central law,
$\vct n_{ij}=(\vct r_i-\vct r_j)/r_{ij}$ points from $j$ to $i$ and
$\vct F_{ij}=-V'_{ij}(r_{ij})\vct n_{ij}$. The two pair contributions are
\begin{equation}
 \vct F_{ij}^{\mathrm{scr}}=
 \frac{q_iq_j}{4\pi\varepsilon_m}
 e^{-\kappa r_{ij}}
 \left(\frac{1}{r_{ij}^2}+\frac{\kappa}{r_{ij}}\right)\vct n_{ij},
 \label{eq:screened-force}
\end{equation}
\begin{equation}
 \vct F_{ij}^{\mathrm{LJ}}=
 \frac{24\epsilon_{ij}}{r_{ij}}
 \left[2\left(\frac{\sigma_{ij}}{r_{ij}}\right)^{12}
        -\left(\frac{\sigma_{ij}}{r_{ij}}\right)^6\right]\vct n_{ij}.
 \label{eq:lj-force}
\end{equation}
The equal and opposite force acts on particle $j$. Each prefactor has
units of force, and its sign selects pushing or pulling along the line
of centres. These vector forces are summed over the retained neighbours
to obtain each particle's total pair force.

Fixing one source provides a spatial view of the same electrostatic
interaction. Hold particle $j$ at $\vct R_j$ in a homogeneous bulk
medium, and define
\begin{equation}
 \varphi_{{\rm el},j}(\vct x)=
   \frac{q_j}{4\pi\varepsilon_m}
     \frac{e^{-\kappa|\vct x-\vct R_j|}}{|\vct x-\vct R_j|},
 \qquad \vct E_j=-\nabla_{\vct x}\varphi_{{\rm el},j},
 \qquad \vct F_i(\vct x;\vct R_j)=q_i\vct E_j(\vct x).
 \label{eq:architecture-conditional-electric-field}
\end{equation}
The potential has units of volts, the electric field of volts per metre,
and the force of newtons. Moving the test position through ordinary space
samples the force from this fixed source. In the full moving assembly,
$-\nabla_{\vct r}U_\lambda$ instead supplies a three-component force for
each particle, giving $3n$ components at every configuration. A spatial
arrow plot displays a selected slice of that collective force. The
displayed electrostatic expression uses homogeneous bulk screening with
prescribed parameters. Resolving dielectric boundaries or ion transport
requires the corresponding electrical boundary conditions and medium
variables in addition to the cylinder's mechanical reflection rule.

An effective screened interaction may be specified as a reversible model
energy or obtained by integrating out equilibrium environmental variables.
The resulting potential of mean force preserves the selected equilibrium
distribution at the chosen coarse-grained resolution~\cite{Noid2008}.
Predicting time-dependent motion additionally requires a specified
dynamical closure for the omitted variables.
Contributions already included by that averaging are counted once when
explicit medium coordinates are introduced. Subsection~\ref{sec:hamiltonian-control}
shows how to match these energy and medium descriptions.

For the Lennard--Jones component alone,
\begin{equation}
 r_{m,ij}=2^{1/6}\sigma_{ij},\qquad
 V_{ij}^{\rm LJ}(r_{m,ij})=-\epsilon_{ij},\qquad
 \left.\frac{\dd^2V_{ij}^{\rm LJ}}{\dd r^2}\right|_{r=r_{m,ij}}
                   =\frac{72\epsilon_{ij}}{r_{m,ij}^2}.
 \label{eq:architecture-lj-minimum}
\end{equation}
The depth $\epsilon_{ij}$ measures the energy below the separated-pair
level. The curvature measures how quickly the restoring force changes
near the minimum: it is the local radial spring stiffness. The minimum
is accessible when $r_{m,ij}>b_i+b_j$. In the complete lattice, all
retained interactions and supports jointly determine force balance, so
the reference separations are found from their combined forces.
To display the pair law as both energy and force, suppress pair indices
and set $x=r/r_m$. Then
\begin{equation}
 \begin{aligned}
 v(x)&=\frac{V_{\rm LJ}}{\epsilon}=x^{-12}-2x^{-6},&
 f(x)&=\frac{F_r r_m}{\epsilon}=12(x^{-13}-x^{-7})=-v'(x),\\
 v_h(x)&=-1+36(x-1)^2,& f_h(x)&=-72(x-1).
 \end{aligned}
 \label{eq:architecture-normalized-pair-law}
\end{equation}
The subscript $h$ denotes the local harmonic expansion: a parabolic
energy and a linear restoring force near the minimum. This is the
spring approximation used for small lattice displacements.
Figure~\ref{fig:architecture-lj-curvature} compares it with the full pair
law over a small displayed interval. Their departure measures the force
error of the approximation and helps set the displacement range for the
desired prediction accuracy.

\begin{figure}[H]
\centering\begingroup
\definecolor{vltTeal}{HTML}{176B64}
\definecolor{vltBlue}{HTML}{315D80}
\definecolor{vltGold}{HTML}{A87520}
\definecolor{vltPlum}{HTML}{725877}
\definecolor{vltRed}{HTML}{762B38}
\definecolor{vltYellow}{HTML}{D5AD3C}
\definecolor{vltNeutral}{HTML}{777D83}
\setlength{\unitlength}{1pt}
\scalebox{0.90}{%
\input{vlt-figures/figure-025.tex}%
}\endgroup
\caption{Exact Lennard--Jones energy and radial force share the minimum $x=1$. Dashed gold segments show the harmonic laws only on $0.95\le x\le1.05$. Pair sketches use one position scale to compare repulsive and attractive separations; force arrows use $2.5\,\mathrm{pt}$ per unit $f$. Positive $f$ points outward. Hard exclusion is specified independently of this interaction law.}
\label{fig:architecture-lj-curvature}
\end{figure}

The screened component gives a complementary spatial illustration.
For $\kappa>0$, write $\ell=\kappa^{-1}$ and
$A=q_iq_j/(4\pi\varepsilon_m)>0$. In the plane through a fixed source,
use dimensionless coordinates $(X,Y)=(x/\ell,y/\ell)$ with source at
the origin and $s=\sqrt{X^2+Y^2}$. The exact normalized force is
\begin{equation}
 \frac{\vct F}{A/\ell^2}
      =\frac{e^{-s}(1+s)}{s^3}(X,Y,0),\qquad
 g(s)=\frac{|\vct F|\ell^2}{A}
      =\frac{e^{-s}(1+s)}{s^2}.
 \label{eq:architecture-screened-field-slice}
\end{equation}
Both exponential screening and radial spreading reduce the force with
distance. Circles about the source are equipotentials; the force for a
like-sign test charge points outward, normal to them.
Figure~\ref{fig:architecture-screened-field} uses one global arrow scale
so arrow lengths show that decay, alongside the force-magnitude curve.

\begin{figure}[H]
\centering\begingroup
\definecolor{vltTeal}{HTML}{176B64}
\definecolor{vltBlue}{HTML}{315D80}
\definecolor{vltGold}{HTML}{A87520}
\definecolor{vltPlum}{HTML}{725877}
\definecolor{vltRed}{HTML}{762B38}
\definecolor{vltYellow}{HTML}{D5AD3C}
\definecolor{vltNeutral}{HTML}{777D83}
\setlength{\unitlength}{1pt}
\scalebox{0.84}{%
\input{vlt-figures/figure-026.tex}%
}\endgroup
\caption{The Cartesian sampling grid uses $0.75\le s\le1.65$. Arrows display $0.25\vct F/(A/\ell^2)$, preserving relative magnitudes; the circles $s=1,1.5$ are equipotentials. The pale disk $s<0.65$ is an illustrative excluded region, and the central virion identifies the held source. This plane is a bulk-law slice through that source. Cylindrical electrical boundary conditions require their own field solution.}
\label{fig:architecture-screened-field}
\end{figure}

The local force change supplies the linear restoring law. At a
stationary reference, define
\begin{equation}
 \mathsf K=\left.\nabla_{\vct r}^{2}U_\lambda\right|_{\vct r=\vct R},
 \qquad \delta\vct F=-\mathsf K\,\delta\vct r.
 \label{eq:architecture-force-tangent}
\end{equation}
Each Hessian entry is the negative derivative of a force component with
respect to a position component, in $\mathrm{N\,m^{-1}}$. The graph
selects direct pairs; their directions and energy derivatives supply
the matrix entries. The inertia then turns this restoring law into the
collective frequency problem.

\subsubsection{From a central reference to a specified colloidal interaction}
\label{sec:s2-dlvo-foundation}

The pair law in~\eqref{eq:pair-potential} assigns a tractable force to
each retained edge. A colloidal realization instead begins with the
particle geometry, electrolyte, surface condition and dispersion model.
These choices determine the reversible interaction whose derivatives
enter the same assembly calculation. Here DLVO denotes the combination
of a specified double-layer contribution and dispersion attraction,
completed by the short-range exclusion appropriate to the preparation.
It provides a comparison model whose accuracy is assessed over the
separations and observations being used.

\begin{definition}[Linear electrolyte reference]
\label{def:s2-linear-electrolyte}
Let $c_\nu^\infty$ be the bulk \emph{number density} of ionic species
$\nu$, of valence $z_\nu$, in a homogeneous solvent of absolute
permittivity $\varepsilon_m$. For an electroneutral bulk, the
Poisson--Boltzmann mean-field model and its linearization are
\begin{equation}
 -\varepsilon_m\nabla^2\psi
 =e\sum_\nu z_\nu c_\nu^\infty
       e^{-z_\nu e\psi/(k_BT)},\qquad
 (\nabla^2-\kappa^2)\psi=0,\qquad
 \kappa^2=\frac{e^2}{\varepsilon_m k_BT}
                \sum_\nu z_\nu^2c_\nu^\infty .
 \label{eq:s2-dh-reference}
\end{equation}
The second equation holds outside fixed charge where
$|z_\nu e\psi|/(k_BT)\ll1$ for the relevant species throughout the
accessible electrolyte. Surface charge supplies boundary data.
\end{definition}

Small potential controls the linearization; the underlying mean-field
description also requires an appropriate treatment of ion correlations,
size and solvent structure. Strongly correlated multivalent ions or
molecular contact need additional physics. Zeta potential describes an
electrokinetic measurement and is not automatically the electrostatic
potential at the capsid surface. The bulk concentrations above describe
an electrolyte reservoir or a sufficiently weakly depleted local
reference. A finite sealed ionic population instead needs its conserved
counts and self-consistent concentration fields.

\begin{proposition}[An isolated spherical monopole and its pair approximation]
\label{prop:s2-finite-sphere-monopole}
For a uniformly charged sphere of radius $b_i$ excluding mobile ions
from its interior, with prescribed total charge $Q_i$ and potential zero
at infinity, the isolated exterior linear solution is
\begin{equation}
 \psi_i(s)=\frac{Q_i}{4\pi\varepsilon_m}
       \frac{e^{-\kappa(s-b_i)}}{(1+\kappa b_i)s},\qquad s\ge b_i .
 \label{eq:s2-isolated-finite-sphere}
\end{equation}
Its effective Yukawa amplitude is
$q_i^{Y}=Q_i e^{\kappa b_i}/(1+\kappa b_i)$. A linear-superposition
pair reference consequently has
\begin{equation}
 U_{ij}^{\rm LSA}(r)
 =\frac{Q_iQ_j}{4\pi\varepsilon_m}
   \frac{e^{-\kappa h}}
   {(1+\kappa b_i)(1+\kappa b_j)r},\qquad
 h=r-b_i-b_j>0 .
 \label{eq:s2-finite-sphere-lsa}
\end{equation}
\end{proposition}

\begin{proof}
The decaying radial solution is $C e^{-\kappa s}/s$; applying the
normal-field condition
$-4\pi\varepsilon_m b_i^2\psi_i'(b_i)=Q_i$ gives the first formula.
Multiplying the corresponding Yukawa amplitudes gives the declared
superposition pair law.
\end{proof}

The isolated field is exact under its spherical linear boundary model;
the pair formula additionally neglects mutual distortion of the two
double layers. Its usefulness requires that distortion to be small.
Low-potential sphere--sphere calculations already distinguish interaction
laws and surface conditions~\cite{HoggHealyFuerstenau1966}. Thus the
$q_i$ in~\eqref{eq:pair-potential} can be a fitted Yukawa amplitude
rather than the bare total $Q_i$. Ion-permeable shells and ion-excluding
solid spheres have different fields; neither automatically represents
a hollow capsid with a permeable wall and dielectric contrast.

\begin{definition}[A nonretarded dispersion comparison]
\label{def:s2-hamaker-comparison}
For two equal, homogeneous solid spheres of radius $b$ in a specified
medium, the pairwise-additive nonretarded Hamaker model is
\begin{equation}
 \begin{aligned}
 U_{\rm H}(r)&=-\frac{A_{\rm H}}6
 \left[\frac{2b^2}{r^2-4b^2}+\frac{2b^2}{r^2}
       +\log\!\left(\frac{r^2-4b^2}{r^2}\right)\right],\quad r>2b,\\
 U_{\rm H}(2b+h)&\sim-\frac{A_{\rm H}b}{12h}\quad(h/b\to0).
 \end{aligned}
 \label{eq:s2-hamaker-equal-spheres}
\end{equation}
$A_{\rm H}>0$ is the attractive Hamaker constant for the declared
materials and intervening medium, in joules~\cite{Hamaker1937}.
\end{definition}

This continuum formula integrates molecular dispersion over solid
volumes. A hollow protein shell requires its shell geometry and material
response, rather than treating the entire enclosed volume as protein.
Nonretardation requires relevant gaps below the retardation scales set
by the dielectric spectrum; there is no universal distance cutoff for
all materials. At molecular gaps hydration, roughness and steric forces
complete the model. The bare $-1/h$ attraction diverges at contact, so
the condition $h>0$ alone does not ensure a finite equilibrium partition
function. Use a positive minimum molecular gap or a short-range
repulsion that dominates the attraction. Any regularization used for
Hessians must be twice differentiable on its sampled domain.

The centre-pair Lennard--Jones attraction and the integrated Hamaker
attraction are alternative specifications of the attractive component;
the Hamaker realization supplies its short-range repulsion separately. A fitted
Lennard--Jones attraction already accounting for dispersion is replaced,
not supplemented by the same dispersion contribution. Distinct hydration
or ligand-mediated terms can be added when their physical content is
specified independently. The same rule applies to an empirical potential
of mean force that has already averaged over surface or ionic states.

\begin{definition}[Surface boundary condition and regulation]
\label{def:s2-surface-boundary-condition}
Fixed charge prescribes the surface charge distribution; fixed potential
prescribes the surface potential and permits charge exchange with its
maintaining reservoir. Charge regulation instead determines the charge
from retained protonation or adsorption states and their coupling to the
local potential. Each choice uses the thermodynamic potential
appropriate to its electrical and chemical reservoirs.
\end{definition}

For regulated coordinates $\boldsymbol\alpha$, specify a surface free
energy including mixing and chemical work, together with the electrical
field problem. Differentiate that complete free energy at fixed
$\boldsymbol\alpha$ when the states are retained, or its consistently
reduced free energy when they equilibrate. Separation-dependent charge
must not be substituted into a fixed-charge pair formula while omitting
the charging work. Regulation can mix charge multipoles and alter their
response~\cite{BozicPodgornik2018}; finite relaxation also requires an
evolution law and preparation.

An orientation-resolved alternative assigns body-frame surface charge
patterns to each rigid particle. Exact permeable-shell linear theory
already treats arbitrary patterns, including icosahedral
ones~\cite{BozicPodgornik2013}. A screened sum over patch sites is a
useful declared approximation in homogeneous screening, but does not
solve ion exclusion or dielectric matching at two finite bodies.
Potential-matched anisotropic DLVO-like models provide a controlled
alternative; their finite multipole truncation and probe-particle pair
approximation must be assessed~\cite{Gnidovec2025}. Net-zero charge
removes the monopole alone. The calculation required here is to carry
one specified interaction, its allowed charge response and its
force--torque derivatives into the finite assembly and observation map.

Figure~\ref{fig:s2patch-dlvo-landscape} shows why the full interaction
landscape is inspected before selecting a reference configuration and
its local curvature.

\begin{figure}[H]
\centering\begingroup
\setlength{\unitlength}{1pt}
\scalebox{0.86}{%
\input{vlt-figures/figure-282.tex}%
}\endgroup
\caption{A regularized synthetic DLVO comparison. The upper drawing identifies the spherical-envelope gap; its dimensions are schematic. Curves use the finite-sphere reference~\eqref{eq:s2-finite-sphere-lsa} and equal-solid-sphere dispersion~\eqref{eq:s2-hamaker-equal-spheres}, with $\kappa b=5$, $A_H/(k_BT)=3$ and $Q^2/[4\pi\varepsilon_m b k_BT(1+\kappa b)^2]=7$. The added smooth core is $U_{\rm core}/(k_BT)=0.4(0.03/g)^{12}$. Black dots locate the total-energy minimum and barrier. The core prevents contact collapse; all curves share one energy scale. These assigned spherical laws illustrate the construction, without fitting a hollow HPV capsid.}
\label{fig:s2patch-dlvo-landscape}
\end{figure}

\subsubsection{Surface patches, torque and chemical preparation}
\label{sec:s2patch-surface-interactions}

An oriented surface converts the geometry in
Definition~\ref{def:s2patch-body-configuration} into a mechanical law.
A positive patch can approach a negative one while another orientation
brings like signs together. The resulting force can change the centre
separation and rotate both bodies. The physical input is a specified
interaction of those surfaces, including which ionic and molecular
coordinates have already been averaged out.

\begin{definition}[Resolved surface interaction]
\label{def:s2patch-surface-energy}
Write $\mathfrak q=(\vct r_i,\mathsf O_i)_{i=1}^n$ for a mechanical
configuration and $Q_{ia}(\boldsymbol\alpha_i)$ for the charge of
body-frame patch $a$ on particle $i$. One declared homogeneous-screening
approximation is
\begin{equation}
 U_{ij}^{\rm patch}
 =\sum_{a,b}\frac{Q_{ia}Q_{jb}}{4\pi\varepsilon_m}
                \frac{e^{-\kappa d_{ia,jb}}}{d_{ia,jb}},
 \qquad d_{ia,jb}=|\vct x_{ia}-\vct x_{jb}|.
 \label{eq:s2patch-site-yukawa}
\end{equation}
The interparticle distances stay positive on the sampled domain.
The complete retained free energy, at its stated thermodynamic controls,
is a specified function
\begin{equation}
 \mathcal F(\mathfrak q,\boldsymbol\alpha,\vct z)
 =U_{\rm excl}+U_{\rm disp}+U_{\rm el}
        +F_{\rm surf}+F_{\rm mat}+U_{\rm sup}+U_{\rm ctrl}.
 \label{eq:s2patch-free-energy-ledger}
\end{equation}
Each term states its physical content and reference. The site sum can
supply $U_{\rm el}$ only when its common-screening approximation is
the selected electrical model.
\end{definition}

The charges in~\eqref{eq:s2patch-site-yukawa} can be effective patch
amplitudes, distinct from chemically counted charges when a finite-body
matching procedure has been used. The electrical model specifies that
mapping. Dielectric interfaces, excluded ions, overlapping double layers
or strong ion correlations require the corresponding boundary solution
or a controlled reduction of it. Hydrophobic and hydration interactions
have their own surface laws; they do not follow from assigning charge.
An explicit surface model replaces contributions already included in a
fitted centre-pair potential of mean force. Physical self and intra-body
electrical terms, when charge states vary, belong to the same ledger;
singular point-site self-energies are not added as chemical energies.

\begin{proposition}[Work-conjugate surface force and torque]
\label{prop:s2patch-force-torque}
At fixed surface and material states, define forces and laboratory-frame
torques by the virtual-work identity
\begin{equation}
 \delta\mathcal F=-\sum_i\left(
       \vct F_i\cdot\delta\vct r_i
       +\vct T_i\cdot\delta\vct\theta_i\right),
 \qquad
 \delta\mathsf O_i=[\delta\vct\theta_i]_\times\mathsf O_i.
 \label{eq:s2patch-virtual-force-torque}
\end{equation}
For a site force $\vct f_{ia}=-\partial\mathcal F/\partial\vct x_{ia}$
in a site-position realization,
\begin{equation}
 \vct F_i=\sum_a\vct f_{ia},\qquad
 \vct T_i=\sum_a
        (\mathsf O_i\vct a_{ia})\times\vct f_{ia}.
 \label{eq:s2patch-site-force-torque}
\end{equation}
Direct centre, orientational or boundary contributions are added through
the same virtual-work identity.
\end{proposition}
\begin{proof}
Use $\delta\vct x_{ia}=\delta\vct r_i+
\delta\vct\theta_i\times\mathsf O_i\vct a_{ia}$ in
$\delta\mathcal F=-\sum_{ia}\vct f_{ia}\cdot\delta\vct x_{ia}$
and collect the independent translations and rotations.
\end{proof}
Torque has units of joules, conjugate to a dimensionless angle. For an
isolated pair with a common-rotation invariant energy, the sum of its
spin torques and the moment of its centre forces vanishes. Pair forces
can therefore be noncentral without violating angular-momentum balance.
The central monopole limit has zero intrinsic orientational torque.

Figure~\ref{fig:s2patch-orientation-torque} translates the site law into
an orientation-dependent energy and a torque along a specified rotation.
The same centre separation can support different rotational responses.

\begin{figure}[H]
\centering\begingroup
\setlength{\unitlength}{1pt}
\scalebox{0.92}{%
\input{vlt-figures/figure-283.tex}%
}\endgroup
\caption{Orientation changes a neutral pair interaction. Each rigid icosahedral body carries two opposite vertex charges $\pm q$; charge symmetry is lower than shape symmetry. Panel (a) uses one orthographic geometry and shows two right-body rotations about $z$ at fixed centres. Red/green marks positive/negative charge; filled/open markers distinguish near/far sites. The curved arrow gives the torque sign. Panel (b) evaluates the site law~\eqref{eq:s2patch-site-yukawa} at $d=3b$ and $\kappa b=1$, with $U_\star=q^2/(4\pi\varepsilon_m b)$ and $T_{jz}=-\partial_\phi U$. Open curve markers register the turned view. The two-site pattern and homogeneous screening are synthetic inputs.}
\label{fig:s2patch-orientation-torque}
\end{figure}

\begin{definition}[An ideal acidic surface population]
\label{def:s2patch-acidic-population}
Let patch $a$ contain $N_{ia}\geq1$ identical titratable acidic groups.
The deprotonated fraction $0<\alpha_{ia}<1$ gives chemical charge
$Q_{ia}^{\rm chem}=Q_{ia}^{0}-eN_{ia}\alpha_{ia}$, with $e>0$ the
elementary charge. For a reservoir at prescribed pH and temperature,
an ideal mixing contribution to the surface grand free energy is
\begin{equation}
 G_{\rm surf}=k_BT\sum_{ia}N_{ia}
 \left[\alpha_{ia}\ln\alpha_{ia}
 +(1-\alpha_{ia})\ln(1-\alpha_{ia})
 +\alpha_{ia}\ln(10)(pK_{a,ia}^{0}-\mathrm{pH})\right].
 \label{eq:s2patch-surface-mixing}
\end{equation}
Here $pK_{a,ia}^{0}$ is the intrinsic value consistent with the chosen
electrical self-energy reference. Finite groups can instead be treated
as discrete occupation states with reversible reaction rates.
\end{definition}

With the electrostatic and any nonideal surface terms included in
$\mathcal G$, stationarity in a free surface fraction gives
\begin{equation}
 \ln\frac{\alpha_{ia}}{1-\alpha_{ia}}
 =\ln(10)(\mathrm{pH}-pK_{a,ia}^{0})
       -\frac{1}{N_{ia}k_BT}
         \frac{\partial(\mathcal G-G_{\rm surf})}
              {\partial\alpha_{ia}}.
 \label{eq:s2patch-regulation-stationarity}
\end{equation}
For chemical charge entering only an electrical free energy with
$\partial F_{\rm el}/\partial Q_{ia}^{\rm chem}=\psi_{ia}$, the last
term is $e\psi_{ia}/(k_BT)$. At zero electrical and nonideal contribution this recovers the ideal
acid-equilibrium relation, while neighbouring particles can shift
protonation through their electrical contribution. Ideal mixing is a
specified starting model; site correlations, specific ion binding and
different basic groups need their own free energy and kinetics
\cite{NinhamParsegian1971,BozicPodgornik2018}.

The complete cavity retains $N_{\rm tot}$ whole VLPs through any such
change. For closed proton exchange, an illustrative conserved count is
\begin{equation}
 N_H^{\rm tot}=N_H^{\rm free}+N_H^{\rm buffer}
               +\sum_{ia}N_{ia}(1-\alpha_{ia}),
 \qquad \frac{\dd N_H^{\rm tot}}{\dd t}=0.
 \label{eq:s2patch-proton-pool}
\end{equation}
The buffer count includes its stated bound proton inventory. Mobile
ions and bound groups also enter the conserved total charge. A fixed-pH
reservoir supplies chemical exchange instead; its chemical potential,
work and ionic flux are part of the specification. Finite ionic states
and their dynamics are developed in Sections~\ref{sec:phonon-ionic-extension}
and~\ref{sec:phonon-biochemical-extensions}.

The three comparison levels are now explicit: a fixed-charge central
law, a frozen body-frame patch pattern with evolving orientation, and
responsive surface populations. A patch pattern can have zero total
charge and nonzero forces or torques. The isolated body's geometric
icosahedral symmetry constrains the charge pattern only when the
surface populations, bound species and electrical environment share
that symmetry. These distinctions identify which added state is being
tested by a mechanical or charge-sensitive record.

\begin{definition}[Coupled tangent force constants]
\label{def:s2patch-tangent-hessian}
At a force-, torque- and surface-balanced reference, use
$\mathsf O_i=\exp([\vct\theta_i]_\times)\mathsf O_i^0$,
where $[\vct\theta]_\times\vct b=\vct\theta\times\vct b$.
Set $\vct\zeta=(\vct u,\vct\theta)$ and write admissible independent
surface perturbations as $\delta\boldsymbol\alpha=\mathsf C_\alpha\vct\eta$.
The Hessian of this pulled-back free energy has blocks
\begin{equation}
 \mathcal F^{(2)}=
 \tfrac12\vct\zeta^{\mathsf T}\mathsf H_{\rm m}\vct\zeta
 +\vct\zeta^{\mathsf T}\mathsf B_\alpha\vct\eta
 +\tfrac12\vct\eta^{\mathsf T}\mathsf A_\alpha\vct\eta.
 \label{eq:s2patch-tangent-energy}
\end{equation}
\end{definition}
At stationarity this symmetric Hessian is a well-defined bilinear form
on the allowed tangent space. Angular entries have energy units,
translation--angle entries force units, and translation--translation
entries stiffness units. Its mechanical force and torque perturbations
are $-\mathsf H_{\rm m}\vct\zeta-\mathsf B_\alpha\vct\eta$.
Conservation removes surface directions before $\mathsf A_\alpha$ is
inverted. Occupancies remain near an interior reference; small integer
populations can instead require explicit jump states.

\subsubsection{Capsid architecture as an internal mechanical resolution}
\label{sec:cmd-capsid-architecture}

Caspar--Klug geometry provides a useful bridge between the shell of one
particle and the mechanics of an assembly. A structural model supplies
candidate contact locations and equivalent environments; the interaction
law assigns forces to them. At whole-particle resolution, these details
enter through surface maps, effective inertia and compliance. Resolving
the shell promotes some of those internal motions to explicit coordinates.
The same whole VLP then carries a centre, an orientation and a deformable
internal network.

\begin{definition}[Capsomer graph and whole-particle identity]
\label{def:cmd-capsomer-graph}
For particle $i$, a shell graph has reference sites $\vct X_{ia}$,
declared contact edges and, where needed, triangular faces or bending
hinges. A site represents a capsomer or another stated coarse-grained
unit. Write its position as
\begin{equation}
 \vct x_{ia}=\vct r_i+\mathsf R_i
       \left(\vct X_{ia}+\sum_\mu\vct v_{ia\mu}q_{i\mu}\right).
 \label{eq:cmd-shell-embedding}
\end{equation}
Here $\mathsf R_i\in SO(3)$ is the whole-body orientation and $q_{i\mu}$
are internal deformation coordinates with their declared length scales.
Their basis excludes rigid translations and rotations in the chosen
mass metric. The shell sites remain constituents of particle $i$;
they do not increase $N_{\rm tot}$.
\end{definition}

For a spherical triangular scaffold, Euler's identity and edge counting
give $V-E+F=2$, $3F=2E$ and $\sum_a(6-z_a)=12$. In a
Caspar--Klug icosadeltahedral scaffold,
\begin{equation}
 V=10T_{\rm cap}+2,\qquad E=30T_{\rm cap},\qquad
 F=20T_{\rm cap}.
 \label{eq:cmd-capsomer-topology}
\end{equation}
With only five- and six-coordinated sites there are twelve of the former.
Coordination counts neighbours in the scaffold; the number of proteins
inside a capsomer is a separate assignment. The usual quasi-equivalent
protein count is $60T_{\rm cap}$~\cite{CasparKlug1962}.

A synthetic CCMV-inspired $T_{\rm cap}=3$ reference has 32 capsomer
sites: twelve pentamers and twenty hexamers, comprising 180 proteins.
CCMV nanoindentation supplies an experimental precedent for studying
capsid elasticity and strength~\cite{Michel2006}. Here this alternative
serves as a geometric reference; its material parameters are assigned
separately from those of HPV. The HPV-inspired $T_{\rm cap}=7d$ shell
instead has 72 pentameric capsomers and 360 L1 proteins. Its twelve
five-coordinated and sixty six-coordinated environments do not convert
the latter capsomers into hexamers~\cite{Goetschius2021}. Tiling and
interaction-network descriptions accommodate such distinct bonding
environments~\cite{KeefTaorminaTwarock2005}. The scaffold's edges are
not automatically a map of molecular bonds.

A mechanically specified coarse-grained shell can use
\begin{equation}
 \mathcal U_{\rm sh}
  =\sum_{e}\phi_e(\ell_e;\chi,h)
   +\sum_{\langle f,g\rangle}\kappa_{fg}
                       (1-\vct n_f\cdot\vct n_g)
   +\mathcal U_{\rm other},\qquad \kappa_{fg}\geq0.
 \label{eq:cmd-shell-energy}
\end{equation}
Lengths $\ell_e$, adjacent face normals and the remaining electrostatic,
excluded-volume, support and surface terms specify the realization.
Reference curvature and prestress are retained when differentiating
this energy. Shell and interparticle contacts may have different
energies and kinetics. The C-terminal-arm and disulfide organization
of HPV motivates contact classes, but its elastic coefficients still
need physical assignment~\cite{Feng2024}.

For a continuum shell approximation the ratio
$\gamma_{\rm FvK}=Y_{2D}R^2/\kappa_b$ compares stretching with bending;
$Y_{2D}$ has force-per-length units and $\kappa_b$ has energy units.
Elastic-shell calculations relate this ratio to faceting
\cite{Lidmar2003}. It is a useful organizing parameter when a thin-shell
description applies. The finite contact network remains available for
site-dependent loading, chemistry and defects. No universal collapse
force follows from $T_{\rm cap}$ or $\gamma_{\rm FvK}$ alone.
Figure~\ref{fig:cmd-capsid-levels} distinguishes these structural levels
and compares the two synthetic architectures.

\begin{figure}[H]
\centering\begingroup
\setlength{\unitlength}{1pt}
\scalebox{0.84}{%
\input{vlt-figures/figure-306.tex}%
}\endgroup
\caption{Three levels of a synthetic capsid-to-assembly construction. (a) Twelve whole-body nodes retain their identities and three-dimensional coupling graph, including vertical links. (b,c) Radially projected triangular scaffolds have $(V,E,F)=(32,90,60)$ and $(72,210,140)$. Their degree counts are twelve five-coordinated sites and twenty or sixty six-coordinated sites. Green and blue mark those environments; rear sites and edges are pale. In (b), pentagonal and hexagonal marks represent the usual $T=3$ capsomer composition. In (c), every mark is pentagonal: the HPV-inspired sites are all pentamers, including those with six neighbours. The geometric realizations supply topology, not atomic coordinates or measured bond constants. The lower map is the physical assignment needed to turn a scaffold into mechanics. Whole-particle, capsomer and protein counts retain separate meanings.}
\label{fig:cmd-capsid-levels}
\end{figure}

\subsubsection{An STMV reference: protein contacts and whole-virion packing}
\label{sec:stmv-geometric-hierarchy}

Satellite tobacco mosaic virus (STMV) supplies a compact reference for
the two structural levels. Its $T_{\rm cap}=1$ shell contains sixty
coat proteins. The corresponding triangular capsomer scaffold has
$(V,E,F)=(12,30,20)$; its twelve sites describe one shell, whereas
each node of a crystal of intact particles describes one whole virion
\cite{LarsonSTMV2014}. The conventional Caspar--Klug construction uses
$T_{\rm cap}=h^2+hk+k^2$ with nonnegative integers $h,k$, not both zero.
The triangulation specifies candidate shell architecture. A contact
law and its kinetics determine the forces, rearrangements and persistence
of a realization.

\begin{definition}[Nested shell and assembly contact graphs]
\label{def:stmv-nested-graphs}
Index a resolved shell constituent by $(i,a)$, where $i$ is its parent
whole particle. The intraparticle edge set $\mathcal E_{\rm cap}$ contains
contacts with the same parent; $\mathcal E_{\rm xtal}$ contains contacts
with different parents. Use the embedding
\eqref{eq:cmd-shell-embedding} and an energy of the form
\begin{equation}
 \mathcal F=
 \sum_i\mathcal F_i^{\rm cap}
 +\sum_{\{(i,a),(j,b)\}\in\mathcal E_{\rm xtal}}
       \phi_{ia,jb}(\vct x_{ia},\vct x_{jb};\chi)
 +\mathcal F_{\rm other},\qquad i\ne j.
 \label{eq:stmv-nested-energy}
\end{equation}
The first sum contains the declared intraparticle contacts and shell
terms. Every interaction is assigned once at the chosen resolution.
An already reduced whole-particle potential replaces the contacts it
has integrated out. It is not added to those same contacts again.
\end{definition}

A twelve-site capsomer model, a sixty-protein model and a model of
protein dimers retain different internal detail. Their mass, force and
observation maps identify the same parent particle. The internal basis
is mass-orthogonal to its rigid translations and rotations, as in
Section~\ref{sec:cmd-capsid-architecture}; this avoids counting a rigid
motion again as shell deformation. Cargo and any co-moving material
have their own declared inertial assignment. A coordinate file supplies
geometry and possible interfaces, while interaction coefficients require
separate constraints. A scaffold edge is not automatically a molecular
bond.

For a crystal, let $\ell$ enumerate translations and $\alpha$ enumerate
whole virions in a conventional cell. Their centre references are
\begin{equation}
 \vct r_{\ell\alpha}^0=\mathsf A\ell+\vct d_\alpha,
 \qquad \ell\in\mathbb Z^3.
 \label{eq:stmv-crystal-basis}
\end{equation}
The room-temperature STMV reference in
Table~1 of Ref.~\cite{LarsonSTMV2014} has space group $I222$ and
conventional edge lengths
\begin{equation}
 (a,b,c)=(17.427,19.177,20.250)\ {\rm nm},\qquad
 \vct d_0=\vct0,\quad
 \vct d_1=\tfrac12(a,b,c)^{\mathsf T}.
 \label{eq:stmv-reference-cell}
\end{equation}
There are two whole particles per conventional cell. The origin has
been placed at one centre. Centring is a translation, so the two sites
have the same reference orientation; further crystallographic rotations
act on the complete particle. The nearest body-diagonal separation is
$\sqrt{a^2+b^2+c^2}/2\simeq16.443$~nm. The primitive volume is $abc/2$.
These packing quantities and the capsid's $T_{\rm cap}$ refer to
different lattices. Icosahedral constituents can form an orthorhombic
periodic assembly.

The geometry in \eqref{eq:stmv-reference-cell} is a documented reference,
not an identification of the particular crystals used in the acoustic
measurements below. Their preparation and orientation need their own
ledger. A periodic bulk calculation describes a region far from the
crystal boundary. A finite crystal in the cylindrical reference cavity
instead uses its actual truncation, supports and fluid boundary; a
clearance wall adds no artificial stiffness. Neither a macroscopic
crystal habit nor a virion silhouette determines its contact coefficients.
Figure~\ref{fig:stmv-nested-geometry} registers the two resolutions.

\begin{figure}[H]
\centering\begingroup
\setlength{\unitlength}{1pt}
\scalebox{0.84}{%
\input{vlt-figures/figure-315.tex}%
}\endgroup
\caption{Two nested geometric resolutions. Panel (a) puts three geometric protein marks in each triangular icosahedral face, giving sixty marks for one ideal T=1 shell; they are not atomic coordinates or an inferred contact map. Panel (b) retains 48 whole-particle centres in a three-by-two-by-two conventional-cell cutout of the body-centred orthorhombic reference, including its boundary centres. Body-diagonal interparticle links are shown; rear objects remain visible in the common oblique projection. Internal contacts belong to one parent and interparticle contacts join different parents, as in \eqref{eq:stmv-nested-energy}. The capsomer, protein, whole-particle and crystallographic counts identify different objects.}
\label{fig:stmv-nested-geometry}
\end{figure}

\subsection{Hamiltonian mechanics, medium reduction and control work}
\label{sec:hamiltonian-control}

The total particle energy collects the pair forces into one mechanical
system. For the translational baseline, write
\begin{equation}
 U_\lambda(\vct r)=
       \sum_{\{i,j\}\in\mathcal E}V_{ij}(r_{ij})
       +U_{\mathrm w}(\vct r)+\lambda U_{\mathrm T}(\vct r),
 \label{eq:controlled-potential}
\end{equation}
Each unordered edge in $\mathcal E$ contributes one pair energy.
The term $U_{\mathrm w}$ describes smooth confinement or, in the hard-wall
limit, the previously specified boundary rule. The term $U_{\mathrm T}$
describes a prescribed stationary mechanical target, such as a supported
object coupled to selected particles. The dimensionless control $\lambda$
scales that coupling. A movable target is included by adding its position,
inertia and reaction forces to the same mechanical system.

The particle Hamiltonian combines this configuration energy with
$E_{\rm kin}=\sum_i m_i|\vct v_i|^2/2$. For constant Cartesian
masses it generates the conservative motion from the initial positions
and velocities. The Hamiltonian and Lagrangian constructions introduced
in Section~\ref{sec:introduction-hamiltonian} now use the full nonlinear
energy~\eqref{eq:controlled-potential}.

Let $\mathsf M=\diag(m_1\Id_3,\ldots,m_n\Id_3)$. The Lagrangian is
kinetic energy minus potential energy, and its velocity derivative defines
canonical momentum:
\begin{equation}
 L(\vct r,\vct v)=\frac12\vct v^{\mathsf T}\mathsf M\vct v
                     -U_\lambda(\vct r),\qquad
 \vct p=\frac{\partial L}{\partial\vct v}=\mathsf M\vct v.
 \label{eq:viral-lagrangian-momentum}
\end{equation}
Positive masses make the velocity--momentum map invertible.

\begin{definition}[Particle Hamiltonian]
\label{def:s2-particle-hamiltonian}
For constant positive Cartesian masses, the particle Hamiltonian is the
Legendre transform of the declared Lagrangian:
\begin{equation}
 H_{\mathrm v}(\vct r,\vct p)
   =\vct p^{\mathsf T}\vct v-L
   =\frac12\vct p^{\mathsf T}\mathsf M^{-1}\vct p+U_\lambda(\vct r),
 \qquad\vct v=\mathsf M^{-1}\vct p.
 \label{eq:viral-legendre-transform}
\end{equation}
\end{definition}
The Euler--Lagrange equation
$\dd(\partial L/\partial\vct v)/\dd t=\partial L/\partial\vct r$
and Hamilton's equations
$\Dt{\vct r}=\partial H_{\mathrm v}/\partial\vct p$,
$\Dt{\vct p}=-\partial H_{\mathrm v}/\partial\vct r$
both give $\mathsf M\Dtt{\vct r}=-\nabla U_\lambda$ in the smooth
region~\cite{Deriglazov2010}: mass times acceleration equals total force.
The Lagrangian is useful for a variational derivation and constraints;
the Hamiltonian organizes position--momentum pairs and energy exchange.
Including the environment's energy then permits a consistent reduced
description of dissipative particle motion.

For the particle-only conservative evolution, the equations define an
explicit vector field in phase space,
\begin{equation}
 X_H(\vct r,\vct p)=
 \begin{pmatrix}\partial_{\vct p}H_{\mathrm v}\\
                   -\partial_{\vct r}H_{\mathrm v}\end{pmatrix}
 =\begin{pmatrix}\mathsf M^{-1}\vct p\\-\nabla U_\lambda\end{pmatrix},
 \qquad
 \begin{pmatrix}\Dt{\vct r}\\\Dt{\vct p}\end{pmatrix}=X_H.
 \label{eq:architecture-hamiltonian-field}
\end{equation}
The first block has velocity units and the second has force units.
Together they give the tangent to a phase-space trajectory. The chain
rule yields
$\dd H_{\mathrm v}/\dd t=
(\partial_{\vct r}H_{\mathrm v})^{\mathsf T}\partial_{\vct p}H_{\mathrm v}
-(\partial_{\vct p}H_{\mathrm v})^{\mathsf T}\partial_{\vct r}H_{\mathrm v}=0$.
Thus autonomous particle-only motion follows a constant-energy surface;
the impact map supplies the continuation at hard contact.

One harmonic coordinate unfolds this statement completely.

\begin{example}[Normalized harmonic phase flow]
\label{ex:s2-harmonic-phase-flow}
Let $u$ be a displacement with mass $m_\star>0$ and stiffness
$k_\star>0$. Choose a length $\ell_\star>0$ and define
\begin{equation}
 \begin{aligned}
 H_\star&=\frac{p^2}{2m_\star}+\frac{k_\star u^2}{2},&
 \omega_\star&=\sqrt{k_\star/m_\star},\\
 X&=u/\ell_\star,&P&=p/(m_\star\omega_\star\ell_\star),
 \quad\tau=\omega_\star t,\\
 \frac{H_\star}{E_\star}&=\frac{X^2+P^2}{2},&
 E_\star&=m_\star\omega_\star^2\ell_\star^2,\\
 \frac{\dd X}{\dd\tau}&=P,&\frac{\dd P}{\dd\tau}&=-X.
 \end{aligned}
 \label{eq:architecture-dimensionless-oscillator}
\end{equation}
\end{example}
At $P=0$, displacement is instantaneously stationary while force changes
momentum. At $X=0$, momentum carries the coordinate through its reference.
The normalized phase plane has circular energy contours, and $(P,-X)$
is tangent because $XP+P(-X)=0$.
Figure~\ref{fig:architecture-hamiltonian-flow} shows the field and one
exact orbit~\cite{FeynmanLectures1963}.

\begin{figure}[H]
\centering\begingroup
\definecolor{vltTeal}{HTML}{176B64}
\definecolor{vltBlue}{HTML}{315D80}
\definecolor{vltGold}{HTML}{A87520}
\definecolor{vltPlum}{HTML}{725877}
\definecolor{vltRed}{HTML}{762B38}
\definecolor{vltYellow}{HTML}{D5AD3C}
\definecolor{vltNeutral}{HTML}{777D83}
\setlength{\unitlength}{1pt}
\scalebox{0.80}{%
\input{vlt-figures/figure-027.tex}%
}\endgroup
\caption{(a) Equal phase-plane units give circular energy contours of radii $0.5,1,1.5$. Blue arrows sample $(P,-X)$ with gain $0.12$; the origin is a fixed point. (b) The initial state $(X,P)=(1,0)$ generates the two time traces. Here $H/E_\star=(X^2+P^2)/2$ and $E_\star=m_\star\omega_\star^2\ell_\star^2$. The vectors belong to nondimensional classical phase space; position and momentum regain different units under the stated scaling.}
\label{fig:architecture-hamiltonian-flow}
\end{figure}

The Hamiltonian also generates the rate of any differentiable classical
observable. For scalar functions $A(\vct r,\vct p,t)$ and
$B(\vct r,\vct p,t)$, define the Poisson bracket
\begin{equation}
 \{A,B\}=\sum_{a=1}^{3n}
 \left(\frac{\partial A}{\partial r_a}\frac{\partial B}{\partial p_a}
      -\frac{\partial A}{\partial p_a}\frac{\partial B}{\partial r_a}\right),
 \qquad \Dt{A}=\frac{\partial A}{\partial t}+\{A,H_{\mathrm v}\}.
 \label{eq:architecture-poisson-generator}
\end{equation}
The second identity follows from the chain rule and Hamilton's equations.
It gives a compact rule for predicting how any chosen observable changes
along the classical motion. In particular, $\{r_a,p_b\}=\delta_{ab}$
expresses the position--momentum pairing. The bracket has units
$[A][B]/\text{action}$, so bracketing with an energy produces a
rate~\cite{Deriglazov2010,NeumaierWestra2011}.

When medium coordinates $\vct\xi$ and conjugate momenta $\vct\pi$ are
retained, the full conservative energy can be written
\begin{equation}
 H_{\mathrm{tot}}=
       \sum_i\frac{|\vct p_i|^2}{2m_i}
       +U_\lambda(\vct r)
       +H_{\mathrm m}(\vct\xi,\vct\pi)
       +U_{\mathrm{vm}}(\vct r,\vct\xi),
 \qquad \vct p_i=m_i\vct v_i.
 \label{eq:total-hamiltonian}
\end{equation}
Here $H_{\mathrm m}$ stores the medium's own energy and
$U_{\mathrm{vm}}$ stores particle--medium coupling energy. These terms
must be matched to the effective inputs already included in $U_\lambda$.
For fixed controls the complete Hamiltonian is autonomous and conserved.
Particle motion can still lose or regain energy through the medium.

\begin{proposition}[Energy balance under prescribed control]
\label{prop:s2-control-energy-balance}
Let the complete particle--medium system follow Hamilton's equations
for~\eqref{eq:total-hamiltonian} on a smooth trajectory. Suppose its only
explicit time dependence is a differentiable prescribed $\lambda(t)$,
and that $U_{\mathrm T}$, $H_{\mathrm m}$ and $U_{\mathrm{vm}}$ have no
additional control dependence. Then
\begin{equation}
 \frac{\dd H_{\mathrm{tot}}}{\dd t}
       =\Dt{\lambda}\,\partial_\lambda H_{\mathrm{tot}}
       =\Dt{\lambda}\,U_{\mathrm T}(\vct r(t)),\qquad
 W_\lambda[0,t]=\int_0^t
       \frac{\dd\lambda(\tau)}{\dd\tau}U_{\mathrm T}(\vct r(\tau))\,\dd\tau .
 \label{eq:control-work}
\end{equation}
In particular, fixed controls conserve the complete enclosed energy.
\end{proposition}
\begin{proof}
In the derivative of $H_{\mathrm{tot}}$ along Hamiltonian flow, the
position and momentum terms cancel pairwise. The remaining explicit
partial derivative is $\dot\lambda\,\partial_\lambda H_{\mathrm{tot}}
=\dot\lambda U_{\mathrm T}$. Integrating it gives the stated work.
\end{proof}
The integrand is power: the dimensionless control rate multiplies an
energy. Changing a setting between separately prepared runs and changing
it during one trajectory are physically different operations. During a
protocol the external control performs the displayed work; after it is
fixed the autonomous complete-system energy is again conserved.

With $H_{\mathrm v}$ defined by~\eqref{eq:viral-legendre-transform},
the subsystem balance is
\begin{equation}
 \Dt{H_{\mathrm v}}
       =-\Dt{H_{\mathrm m}}-\Dt{U_{\mathrm{vm}}}
          +\Dt{\lambda}\,U_{\mathrm T}.
 \label{eq:subsystem-energy-balance}
\end{equation}
The particles exchange energy with both the medium and their coupling
to it. The displayed balance keeps all three contributions in the same
energy account. A resolved movable target or another added sector enters
with its own energy and exchange terms. Equation~\eqref{eq:control-work}
uses the stated sole-control assumption. A pH or salt protocol also
specifies chemical exchange and the explicit dependence of the retained
thermodynamic potential on its controls. A grand-potential term
$-N_\nu\dot\mu_\nu$ from varying a chemical potential is distinguished
from chemical transfer work $\mu_\nu\dot N_\nu$.
Sustained flow includes supplied pressure or wall work. The corresponding
ports replace the isolated setting rather than being assigned to an
unobserved particle force.

A surrounding material can accommodate particle displacement and thereby
change the force needed to hold it. To calculate this static relaxation,
expand the joint potential about a stationary reference, with medium
displacement $\vct y$:
\begin{equation}
 U^{(2)}_{\mathrm{joint}}=
 \frac12
 \begin{pmatrix}\vct u\\\vct y\end{pmatrix}^{\mathsf T}
 \begin{pmatrix}
   \mathsf K_{rr}&\mathsf K_{r\xi}\\
   \mathsf K_{\xi r}&\mathsf K_{\xi\xi}
 \end{pmatrix}
 \begin{pmatrix}\vct u\\\vct y\end{pmatrix}.
 \label{eq:joint-quadratic-potential}
\end{equation}
\begin{proposition}[Static relaxation and reduced curvature]
\label{prop:s2-static-medium-relaxation}
For~\eqref{eq:joint-quadratic-potential}, assume
$\mathsf K_{\xi\xi}\succ0$ on the retained medium sector and
$\mathsf K_{\xi r}=\mathsf K_{r\xi}^{\mathsf T}$. The clamped
particle curvature is $\mathsf K_{rr}$. The unique medium minimizer
at fixed $\vct u$ and the relaxed particle curvature are
\begin{equation}
 \vct y_{\mathrm{rel}}
     =-\mathsf K_{\xi\xi}^{-1}\mathsf K_{\xi r}\vct u,\qquad
 \mathsf K_{\mathrm{rel}}
     =\mathsf K_{rr}
       -\mathsf K_{r\xi}\mathsf K_{\xi\xi}^{-1}\mathsf K_{\xi r}.
 \label{eq:relaxed-medium-curvature}
\end{equation}
The joint energy separates as
\begin{equation}
 U^{(2)}_{\mathrm{joint}}
   =\frac12\vct u^{\mathsf T}\mathsf K_{\mathrm{rel}}\vct u
    +\frac12(\vct y-\vct y_{\mathrm{rel}})^{\mathsf T}
       \mathsf K_{\xi\xi}(\vct y-\vct y_{\mathrm{rel}}),\qquad
 \boxed{\mathsf K_{\mathrm{rel}}\preceq\mathsf K_{rr}.}
 \label{eq:architecture-completed-square}
\end{equation}
If the joint Hessian is nonnegative, so is $\mathsf K_{\mathrm{rel}}$.
\end{proposition}
\begin{proof}
Stationarity with respect to $\vct y$ gives
$\mathsf K_{\xi\xi}\vct y=-\mathsf K_{\xi r}\vct u$; positive
definiteness makes this the unique minimizer. Substitution, or direct
completion of the square, yields the two displayed identities. The
subtracted matrix is nonnegative because
$\vct u^{\mathsf T}\mathsf K_{r\xi}\mathsf K_{\xi\xi}^{-1}
\mathsf K_{\xi r}\vct u$
is a squared norm. Evaluating a nonnegative joint quadratic form at
its minimizing $\vct y$ proves the final assertion.
\end{proof}
The Schur complement gives the static stiffness after medium
accommodation. Releasing the medium lowers or preserves the quadratic
energy cost for every held particle pattern. The dynamics determine how
quickly that accommodation occurs.

A two-coordinate example makes the reduction concrete. Let $u$ and $y$
be particle and medium displacements with a common length unit and take
\begin{equation}
 \begin{aligned}
 U_{\rm ex}(u,y)&=\frac{k_\star}{2}(2u^2-2uy+y^2)
             =\frac{k_\star u^2}{2}+\frac{k_\star(y-u)^2}{2},\\
 \mathsf K_{\rm ex}&=k_\star
             \begin{pmatrix}2&-1\\-1&1\end{pmatrix},\qquad k_\star>0.
 \end{aligned}
 \label{eq:architecture-medium-example}
\end{equation}
The joint energy is positive for every nonzero $(u,y)$.
Clamping $y=0$ costs $k_\star u^2$ and gives stiffness $2k_\star$.
Letting the medium follow the held particle gives $y=u$, costs
$k_\star u^2/2$, and gives stiffness $k_\star$. Accommodation therefore
halves the stiffness in this example. Figure~\ref{fig:architecture-medium-relaxation}
displays both paths in the same landscape. Its valley joins the
minimum-energy medium positions at successive held values of $u$.
Normal-mode directions and trajectories are found separately by adding
inertia and, for a trajectory, initial data.

In a classical canonical preparation at $T>0$, this same quadratic model
also illustrates the connection to fluctuations. Its clamped distribution
has $\operatorname{Var}(u)=k_{\rm B}T/(2k_\star)$; integrating over the
medium gives $\operatorname{Var}(u)=k_{\rm B}T/k_\star$.
Each preparation has its own curvature. The equilibrating medium makes
particle displacement less costly, so the particle's equilibrium variance
is twice as large in this example. The Gaussian distribution gives that
width; a dynamical law supplies the time over which it is sampled.

\begin{figure}[H]
\centering\begingroup
\definecolor{vltTeal}{HTML}{176B64}
\definecolor{vltBlue}{HTML}{315D80}
\definecolor{vltGold}{HTML}{A87520}
\definecolor{vltPlum}{HTML}{725877}
\definecolor{vltRed}{HTML}{762B38}
\definecolor{vltYellow}{HTML}{D5AD3C}
\definecolor{vltNeutral}{HTML}{777D83}
\setlength{\unitlength}{1pt}
\scalebox{0.80}{%
\input{vlt-figures/figure-028.tex}%
}\endgroup
\caption{Contours in (a) have $h=0.25,0.5,0.9$. The blue construction holds $X=1$ and changes $Y$ from the clamped slice to its conditional minimum. The corresponding points in (b) differ by $1/2$ in scaled energy; pale shading compares the two energy functions. The valley $Y=X$ specifies a static minimization operation. It is not a normal-mode axis or a time-parametrized relaxation path. The boxed result illustrated here is~\eqref{eq:architecture-completed-square}.}
\label{fig:architecture-medium-relaxation}
\end{figure}

The corresponding dynamical elimination needs medium inertia as well.
\begin{proposition}[Frequency-dependent medium elimination]
\label{prop:s2-dynamic-medium-elimination}
Take the joint quadratic potential above, constant block-diagonal
positive inertia $(\mathsf M,\mathsf M_{\mathrm m})$,
$\mathsf K_{\xi\xi}\succ0$ on the retained medium sector, and forcing
only the particle sector. At complex frequency $z$ for which
$\mathsf K_{\xi\xi}-z^2\mathsf M_{\mathrm m}$ is invertible, the
particle dynamic stiffness after eliminating the medium is
\begin{equation}
 \begin{aligned}
 \mathsf A_{\rm eff}(z)={}&\mathsf K_{rr}-z^2\mathsf M\\
 &-\mathsf K_{r\xi}
   (\mathsf K_{\xi\xi}-z^2\mathsf M_{\mathrm m})^{-1}
                                  \mathsf K_{\xi r},\qquad
 \mathsf A_{\rm eff}(0)=\mathsf K_{\mathrm{rel}} .
 \end{aligned}
 \label{eq:architecture-dynamic-elimination}
\end{equation}
\end{proposition}
\begin{proof}
The unforced medium block gives
$\vct y=-(\mathsf K_{\xi\xi}-z^2\mathsf M_{\mathrm m})^{-1}
\mathsf K_{\xi r}\vct u$.
Substitute this expression into the particle block. At $z=0$ the
result reduces to Proposition~\ref{prop:s2-static-medium-relaxation}.
\end{proof}
With time dependence $e^{-i\omega t}$, the retarded limit is
$z=\omega+i0^+$, taken where the relevant inverse exists. Medium forcing
and preparation contribute to the effective right-hand side. Static
response gives the relaxed curvature; finite-frequency response retains
the medium's inertia and phase. A finite conservative medium stores and
returns energy through its discrete modes. Attenuation and thermal
forcing additionally require the dynamical and statistical reduction
specified below.

At thermal equilibrium, the medium samples a range of configurations.
Integrating their statistical weights gives the particle's effective
free energy~\cite{Noid2008}.

\begin{definition}[Conditional medium free energy]
\label{def:s2-conditional-medium-free-energy}
For $T>0$, take a joint coordinate potential
$U_{\mathrm{joint}}(\vct u,\vct y)$, coordinate-independent inertia,
and a medium integration measure and domain independent of $\vct u$.
When the integrals are finite and positive, define
\begin{equation}
 \begin{aligned}
 Z_{\mathrm m}(\vct u;T)&=
     \int\exp\!\left[-\frac{U_{\mathrm{joint}}(\vct u,\vct y)}
                              {k_{\mathrm B}T}\right]\dd\vct y,\\
 \mathcal F_{\rm eff}(\vct u;T)&=-k_{\mathrm B}T\,
       \ln\frac{Z_{\mathrm m}(\vct u;T)}{Z_{\mathrm m}(\vct0;T)}.
 \end{aligned}
 \label{eq:architecture-conditional-free-energy}
\end{equation}
\end{definition}
The normalized ratio makes the logarithm dimensionless. The conditional
mean $\langle\cdot\rangle_{\mathrm m|u,T}$ weights medium configurations
by their Boltzmann factor at the specified particle displacement. For
Cartesian indices $A,B$, write $\partial_A=\partial/\partial u_A$.

\begin{proposition}[Conditional mean force and curvature]
\label{prop:s2-conditional-free-energy-curvature}
Assume the defining integrals admit differentiation twice under the
integral sign. For the fixed measure and domain in
Definition~\ref{def:s2-conditional-medium-free-energy},
\begin{equation}
 \begin{aligned}
 \partial_A\mathcal F_{\rm eff}
    &=\langle\partial_AU_{\mathrm{joint}}\rangle_{\mathrm m|u,T},\\
 \partial_A\partial_B\mathcal F_{\rm eff}
    &=\langle\partial_A\partial_BU_{\mathrm{joint}}\rangle_{\mathrm m|u,T}
      -\frac{\operatorname{Cov}_{\mathrm m|u,T}
           (\partial_AU_{\mathrm{joint}},\partial_BU_{\mathrm{joint}})}
                         {k_{\mathrm B}T}.
 \end{aligned}
 \label{eq:architecture-mean-force-curvature}
\end{equation}
\end{proposition}
\begin{proof}
Differentiate the logarithm of $Z_{\mathrm m}$. For a differentiable
integrable quantity $A$, differentiating its normalized conditional
average gives
$\partial_B\langle A\rangle=
\langle\partial_B A\rangle-
\operatorname{Cov}(A,\partial_B U_{\mathrm{joint}})/(k_{\mathrm B}T)$.
Use $A=\partial_AU_{\mathrm{joint}}$ for the second derivative.
\end{proof}
Here $\operatorname{Cov}(a,b)=\langle ab\rangle-\langle a\rangle\langle b\rangle$.
The negative free-energy gradient is the conditional mean force. Its
curvature combines mean microscopic curvature and conditional force
fluctuations. For the constant-coefficient Gaussian potential above,
integration yields $\mathsf K_{\mathrm{rel}}$; coordinate-dependent
medium stiffness can add entropic curvature. Moving integration domains
or coordinate-dependent inertia also require their measure terms.

Figure~\ref{fig:architecture-gaussian-reduction} gives the conditional and marginal distributions of the same quadratic example, making the clamped-versus-integrated comparison explicit.

\begin{figure}[H]
\centering\begingroup
\definecolor{vltTeal}{HTML}{176B64}
\definecolor{vltBlue}{HTML}{315D80}
\definecolor{vltGold}{HTML}{A87520}
\definecolor{vltPlum}{HTML}{725877}
\definecolor{vltRed}{HTML}{762B38}
\definecolor{vltYellow}{HTML}{D5AD3C}
\definecolor{vltNeutral}{HTML}{777D83}
\setlength{\unitlength}{1pt}
\scalebox{0.80}{%
\input{vlt-figures/figure-029.tex}%
}\endgroup
\caption{(a) The conditional Gaussians have means $0$ and $1$ and the same unit variance. (b) Integrating over $Y$ gives the solid marginal; fixing $Y=0$ and renormalizing gives the dotted clamped density. The shaded interval $|X|\le1$ lies under the marginal and marks its one-standard-deviation width. All curves are normalized model densities in dimensionless coordinates. This equilibrium comparison uses the quadratic energy printed below the panels.}
\label{fig:architecture-gaussian-reduction}
\end{figure}

The covariance term measures how medium configurations at the same
particle displacement differ in the forces they exert. That spread
changes the slope of the conditional mean force. Its units are
$\mathrm{N^2/J}=\mathrm{N\,m^{-1}}$, as required for a stiffness.
For the stated fixed measure, the covariance matrix is nonnegative and
reduces the \emph{conditional mean} microscopic curvature. Both terms
can vary with displacement and temperature. A comparison with a fixed
clamped matrix, such as~\eqref{eq:architecture-completed-square},
therefore uses the additional constant-coefficient quadratic assumptions
of that example.

The reductions answer different questions. Static minimization finds
the medium position at held particle displacement. Dynamic elimination
finds its frequency-dependent response. Equilibrium integration weights
all thermally sampled configurations and includes entropic effects.
Preparation and readout determine the relevant comparison.

An equilibrium harmonic model expands the resulting free energy about
its own stationary reference. Explicit medium coordinates are matched
to the contributions already included in that free energy, counting each
relaxation once. The canonical ensemble used here specifies a temperature
and statistical preparation under equilibrium and reservoir
approximations~\cite{FeynmanLectures1963}. The enclosure specifies the
physical boundary; the ensemble specifies how states are populated.

\paragraph{Application: a scaffold's relaxed modulus.}
The same reduction connects a microscopic network to a measured tissue
or gel deformation. Under boundary shear, a passive filament scaffold
accommodates loading by stretching and bending filaments and moving its
internal cross-links. Motion beyond the uniform affine deformation is
called \emph{nonaffine}. Cross-link positions are state coordinates;
stretching and bending coefficients set the constitutive law; boundary
stress and imaged displacements provide observables. Semiflexible-network
models connect these quantities through the network architecture and its
allowed deformation mechanisms~\cite{HeadLevineMacKintosh2003}.

For a finite three-dimensional specimen of reference volume $V_s$, let
$\gamma$ be a small dimensionless shear and $\vct v$ its internal
displacements relative to the imposed affine motion. With zero reference
shear stress, a local quadratic energy can be written
\begin{equation}
 \begin{aligned}
 U_\gamma^{(2)}(\vct v)&=
   \tfrac12V_sG_{\rm aff}\gamma^2
   +\gamma\vct\Xi^{\mathsf T}\vct v
   +\tfrac12\vct v^{\mathsf T}\mathsf H_{\rm int}\vct v,\\
 G_{\rm rel}&=G_{\rm aff}
   -\frac{\vct\Xi^{\mathsf T}\mathsf H_{\rm int}^{-1}\vct\Xi}{V_s}.
 \end{aligned}
 \label{eq:architecture-nonaffine-scaffold-modulus}
\end{equation}
Here $\mathsf H_{\rm int}\succ0$ on the constrained internal sector and
$\vct\Xi$ has units of force. Minimizing over $\vct v$ gives the second
line, a direct instance of~\eqref{eq:relaxed-medium-curvature}.
Both moduli have pressure units. The affine value holds internal
accommodation fixed; the relaxed value allows it. Joint stability makes the
relaxed value nonnegative, while the subtracted term shows how coupled
internal motions soften this particular boundary response.

Structural imaging and bulk mechanics therefore probe complementary
parts of the same response. Images of nonaffine node motion reveal how
the scaffold accommodates loading; the shear modulus measures the
resulting boundary stiffness. Together, they help separate the
contribution of internal accommodation from changes in filament
stiffness. The displayed comparison holds the reference, boundaries and
constitutive law fixed while releasing internal motion. Its rate is
determined by the corresponding dynamical model.

\subsubsection{Surface relaxation, preparation and matched fluctuations}
\label{sec:s2patch-surface-relaxation}
For $\mathsf A_\alpha\succ0$, the existing Schur construction gives
$\mathsf H_{\rm rel}=\mathsf H_{\rm m}
-\mathsf B_\alpha\mathsf A_\alpha^{-1}\mathsf B_\alpha^{\mathsf T}$.
Thus releasing charge can create cross-couplings between translation
and rotation even when their clamped block vanishes. Stability concerns
the complete constrained quadratic form and its relaxed sector.

\begin{proposition}[Prepared linear surface reduction]
\label{prop:s2patch-surface-reduction}
Choose constant reciprocal mobility
$\mathsf L_\alpha=\mathsf L_\alpha^{\mathsf T}\succ0$, and thermal
surface dynamics
\begin{equation}
 \dd\vct\eta=-\mathsf L_\alpha
   (\mathsf A_\alpha\vct\eta+\mathsf B_\alpha^{\mathsf T}\vct\zeta)
       \dd t+\sqrt{2k_{\rm B}T}\,\mathsf L_\alpha^{1/2}\dd\vct W_\alpha.
 \label{eq:s2patch-surface-kinetics}
\end{equation}
This is a local Gaussian surface approximation about an interior
population; a global bounded population uses its specified chemical
transition or boundary law. Its steady harmonic mechanical stiffness,
for $e^{-i\omega t}$, is
\begin{equation}
 \mathsf H_{\rm eff}(\omega)=\mathsf H_{\rm m}
 -\mathsf B_\alpha(\mathsf A_\alpha-i\omega\mathsf L_\alpha^{-1})^{-1}
                      \mathsf B_\alpha^{\mathsf T}.
 \label{eq:s2patch-charge-dynamic-schur}
\end{equation}
No commutation of $\mathsf A_\alpha$ and $\mathsf L_\alpha$ is required.
\end{proposition}
\begin{proof}
Multiply the harmonic surface equation by
$\mathsf L_\alpha^{-1}$, solve for $\vct\eta$, and substitute in
$-\mathsf H_{\rm m}\vct\zeta-\mathsf B_\alpha\vct\eta$.
\end{proof}
At finite time, the same elimination retains
\begin{equation}
 \begin{aligned}
 \vct\eta(t)={}&e^{-\mathsf L_\alpha\mathsf A_\alpha t}\vct\eta(0)
 -\int_0^t e^{-\mathsf L_\alpha\mathsf A_\alpha(t-s)}
          \mathsf L_\alpha\mathsf B_\alpha^{\mathsf T}\vct\zeta(s)\,\dd s\\
 &+\sqrt{2k_{\rm B}T}\int_0^t
    e^{-\mathsf L_\alpha\mathsf A_\alpha(t-s)}
                 \mathsf L_\alpha^{1/2}\dd\vct W_\alpha(s).
 \end{aligned}
 \label{eq:s2patch-finite-prepared-charge}
\end{equation}
The mechanical force contains $-\mathsf B_\alpha$ times every term.
Conditional Gibbs preparation has mean
$-\mathsf A_\alpha^{-1}\mathsf B_\alpha^{\mathsf T}\vct\zeta(0)$
and covariance $k_{\rm B}T\mathsf A_\alpha^{-1}$.
For stationary elimination, with
$\mathsf G_\alpha=(\mathsf A_\alpha-i\omega\mathsf L_\alpha^{-1})^{-1}$,
the two-sided effective force spectrum is
\begin{equation}
 \mathsf S_{f_\alpha}=2k_{\rm B}T\mathsf B_\alpha\mathsf G_\alpha
  \mathsf L_\alpha^{-1}\mathsf G_\alpha^\dagger\mathsf B_\alpha^{\mathsf T}
 =\frac{2k_{\rm B}T}{\omega}
       [-\operatorname{Im}_{\rm H}\mathsf H_{\rm eff}],\quad\omega>0.
 \label{eq:s2patch-charge-noise}
\end{equation}
Here $\operatorname{Im}_{\rm H}\mathsf X=(\mathsf X-\mathsf X^\dagger)/(2i)$.
The inverse identity
$\operatorname{Im}_{\rm H}\mathsf G_\alpha
=\omega\mathsf G_\alpha\mathsf L_\alpha^{-1}\mathsf G_\alpha^\dagger$
proves positivity. Retaining a fitted phase law while resetting its
surface preparation or choosing independent noise changes the model.

\begin{example}[Surface relaxation creates a mechanical cross-response]
\label{ex:s2patch-induced-coupling}
Let $X=u/\ell_\star$ be a dimensionless separation perturbation,
$\Theta$ a small relative angle, and $\eta$ a dimensionless surface-state
perturbation about an interior population. For $E_\star,\ell_\star>0$,
take the synthetic local free energy
\begin{equation}
 \mathcal F_2=\frac{E_\star}{2}
       (4X^2+3\Theta^2+2\eta^2-2X\eta+2\Theta\eta).
 \label{eq:s2patch-toy-energy}
\end{equation}
Its Hessian divided by $E_\star$ has leading principal minors $4,12,17$,
so it is positive definite. Clamping $\eta=0$ leaves no $X$--$\Theta$
cross term. Surface relaxation gives
\begin{equation}
 \eta_{\rm rel}=\frac{X-\Theta}{2},\qquad
 \mathsf H_{\rm rel}=E_\star
       \begin{pmatrix}7/2&1/2\\1/2&5/2\end{pmatrix}.
 \label{eq:s2patch-toy-relaxation}
\end{equation}
Holding $X$ now exerts a torque $-E_\star X/2$ at $\Theta=0$.
The physical separation--angle curvature is $E_\star/(2\ell_\star)$,
with units of force. The induced cross term records a surface state
that responds to both motions.

For positive mobility $\ell_\eta$, use
$\dot\eta=-\ell_\eta\partial_\eta\mathcal F_2$ and
$\tau_\eta=(2\ell_\eta E_\star)^{-1}$. Harmonic elimination gives
\begin{equation}
 \frac{\mathsf H_{\rm eff}(\omega)}{E_\star}
 =\begin{pmatrix}4&0\\0&3\end{pmatrix}
 -\frac{1}{2(1-i\omega\tau_\eta)}
           \begin{pmatrix}1&-1\\-1&1\end{pmatrix}.
 \label{eq:s2patch-toy-dynamic}
\end{equation}
The low-frequency limit is~\eqref{eq:s2patch-toy-relaxation}; a rapidly
varying drive freezes the surface response. With $e^{-i\omega t}$,
the positive loss matrix is $-\operatorname{Im}\mathsf H_{\rm eff}$
for $\omega>0$. Equilibrium fluctuations require the matching surface
noise and preparation. These coefficients illustrate the calculation;
they are not measured HPV parameters or a derived molecular patch map.
The quadratic population approximation is used only over excursions
remaining within its physical occupancy range.
\end{example}

Figure~\ref{fig:s2patch-induced-cross-response} connects this example to
separation and rotation: the relaxed energy contours tilt, and the
induced cross-response varies with the driving timescale.

\begin{figure}[H]
\centering\begingroup
\setlength{\unitlength}{1pt}
\scalebox{0.92}{%
\input{vlt-figures/figure-284.tex}%
}\endgroup
\caption{The synthetic surface-relaxation example~\eqref{eq:s2patch-toy-energy} in physical and coordinate views. The VLP glyphs identify separation and relative orientation; they do not specify a molecular charge map. Both lower-left contours use equal coordinate scales and energy $\mathcal F/E_\star=1/2$, with clamped $\eta=0$ or minimized $\eta$. Relaxation rotates the contour through the induced cross term. Panel (b) plots the real and imaginary parts of the cross stiffness in~\eqref{eq:s2patch-toy-dynamic}; the imaginary cross component is a quadrature, not a scalar dissipation rate. The marked crossover has both components $1/4$.}
\label{fig:s2patch-induced-cross-response}
\end{figure}

\subsubsection{A material state that survives separation}
\label{sec:s2patch-persistent-material}
The same retained-state principle can be specified globally before
linearization. It supplies a useful input to the nonlinear medium
construction in Section~\ref{sec:viscoelastic-dynamics}.
\begin{definition}[Fixed-dimensional material attachment]
\label{def:s2patch-persistent-material-state}
For a smooth position/orientation configuration $\mathfrak q$, let
$\vct z\in\R^{p_h}$ have fixed dimension and measure, choose constant
$\mathsf A_h\succ0$, and prescribe a smooth physical map
$\vct h(\mathfrak q)$ with the same units as $\vct z$. Set
\begin{equation}
 \begin{aligned}
 \mathcal F_h&=\mathcal F_0(\mathfrak q)
  +\tfrac12(\vct z-\vct h)^{\mathsf T}\mathsf A_h(\vct z-\vct h),\\
 \vct f_h&=(D\vct h)^{\mathsf T}\mathsf A_h(\vct z-\vct h),
 \qquad \vct g_z=\mathsf A_h(\vct z-\vct h).
 \end{aligned}
 \label{eq:s2patch-persistent-material-energy}
\end{equation}
\end{definition}
Take $\vct h$ to vary near contact and approach zero far away. The
coordinate $\vct z$ remains after separation, with positive self-energy,
retaining old material preparation for its specified dynamics rather
than erasing it at a membership change. Positive material mobility
relaxes every material direction; a null mobility direction remains
frozen until another specified process acts.

\begin{proposition}[Equal equilibrium and passive material relaxation]
\label{prop:s2patch-material-marginal}
Integration over $\vct z\in\R^{p_h}$ leaves $\mathcal F_0$ plus a
configuration-independent constant, for $T>0$ and fixed controls.
With reciprocal mechanics and
$\dot{\vct z}=-\mathsf L_h\vct g_z$, constant $\mathsf L_h\succeq0$, its
deterministic free-energy rate is
\begin{equation}
 \dot{\mathcal F}_h
 =-\vct f_{\rm rev}\cdot\dot{\mathfrak q}
       -\vct g_z^{\mathsf T}\mathsf L_h\vct g_z,
 \qquad \vct f_{\rm rev}=-\partial_{\mathfrak q}\mathcal F_h.
 \label{eq:s2patch-material-free-energy-rate}
\end{equation}
The first pairing includes translational force and rotational torque.
If also $\dot{\mathfrak q}=\mathsf M_q\vct f_{\rm rev}$ with
$\mathsf M_q\succeq0$, then
$\dot{\mathcal F}_h=-\vct f_{\rm rev}^{\mathsf T}\mathsf M_q\vct f_{\rm rev}
-\vct g_z^{\mathsf T}\mathsf L_h\vct g_z\leq0$.
\end{proposition}
\begin{proof}
Translate the Gaussian integration variable to $\vct z-\vct h$.
Its integral is $(2\pi k_{\rm B}T)^{p_h/2}(\det\mathsf A_h)^{-1/2}$.
The chain rule and the declared gradient kinetics give the rate identity.
\end{proof}
Matched material noise has covariance intensity
$2k_{\rm B}T\mathsf L_h$;
equilibrium initialization uses
$\vct z|\mathfrak q\sim\mathcal N(\vct h,k_{\rm B}T\mathsf A_h^{-1})$.
Stochastic configuration dynamics uses its matched mobility, rotational
measure and thermal drift. This is an isothermal free-energy account;
an internal-energy ledger also specifies entropy and heat exchange.
Changing relaxation can therefore change escape or driven-cycle records
without changing the equilibrium configuration distribution. Its
low-frequency effective drag can also change; matching that drag is an
additional calibration condition, not a consequence of equal equilibrium. Such
predictions still require a separation landscape and calibrated kinetics;
passive equilibrium supplies no sustained cyclic work. Configuration-dependent $\mathsf A_h$, bounded populations or deleted coordinates
require their additional normalization and event account.

Figure~\ref{fig:s2patch-prepared-pair-futures} gives a local preparation
test. Two pairs start at the same positions but with opposite retained
material offsets; their subsequent mean separations distinguish the
prepared states.

\begin{figure}[H]
\centering\begingroup
\setlength{\unitlength}{1pt}
\scalebox{0.92}{%
\input{vlt-figures/figure-285.tex}%
}\endgroup
\caption{A retained material state distinguishes futures from the same particle positions. Use the local attachment model $\mathcal F=k_\star[q^2+(z-q)^2]/2$, $q=u_2-u_1$, and equal positive coordinate mobilities $\mu$, giving $\tau_\star=(\mu k_\star)^{-1}$. The centre of mass is fixed. Panel (a) prepares $q=0$, $z=\pm z_0$; arrows show equal and opposite site forces with one common gain. The material rulers use $z/z_0$; their dots are state markers. Panel (b) gives $\langle q\rangle/z_0=\pm[e^{-(3-\sqrt5)t/(2\tau_\star)}-e^{-(3+\sqrt5)t/(2\tau_\star)}]/\sqrt5$. A position-only, instantaneous-material reduction initialized at $q=0$ predicts zero mean. This is a local synthetic preparation test; matched zero-mean thermal forcing preserves the mean curves.}
\label{fig:s2patch-prepared-pair-futures}
\end{figure}

\subsubsection{Environmental resolution, crowding and mobility}
\label{sec:s2patch-environmental-resolution}

The conditional construction in
Definition~\ref{def:s2-conditional-medium-free-energy} also specifies
what an effective interaction means when ions, surface populations or
crowders are omitted. Discrete equilibrated occupations contribute a
sum over allowed states as well as any continuous integral. Finite
species counts restrict that sum; reservoirs contribute their
chemical-potential weights. Surface chemistry, particle geometry and
medium transport can therefore change different parts of the same
prediction.

\begin{definition}[A specified crowder reduction]
\label{def:s2patch-crowder-reduction}
Let $\vct c$ be explicit crowder coordinates, with their number and
admissible region declared. At fixed particle configuration
$\mathfrak q$, define $Z_c$ by integrating
$\exp[-(U_c+U_{\rm VLP-c})/(k_BT)]$ over their allowed configurations
with the stated measure. Here $U_c$ contains crowder and wall terms and
$U_{\rm VLP-c}$ contains VLP--crowder coupling; the separately retained
$U_{\rm VLP}$ is excluded from $Z_c$. Their conditional reduction gives
\begin{equation}
 \mathcal F_{\rm eff}(\mathfrak q)-\mathcal F_{\rm eff}(\mathfrak q_0)
 =U_{\rm VLP}(\mathfrak q)-U_{\rm VLP}(\mathfrak q_0)
       -k_BT\ln\frac{Z_c(\mathfrak q)}{Z_c(\mathfrak q_0)}.
 \label{eq:s2patch-crowder-pmf}
\end{equation}
For $N_c$ fixed, ideal noninteracting depletants with accessible
centre-volume $V_{\rm free}(\mathfrak q)>0$, their contribution is
\begin{equation}
 \Delta\mathcal F_c=-N_ck_BT
       \ln\frac{V_{\rm free}(\mathfrak q)}{V_{\rm free}(\mathfrak q_0)}.
 \label{eq:s2patch-fixed-crowder-free-volume}
\end{equation}
An ideal crowder reservoir at fixed osmotic pressure $\Pi_c$ instead
gives $\Delta\Omega_c=-\Pi_c\Delta V_{\rm free}$; its two-particle
overlap attraction is $-\Pi_cV_{\rm overlap}$.
\end{definition}
The fixed-count result follows from $Z_c\propto V_{\rm free}^{N_c}$;
the reservoir result follows by summing the ideal-gas number sectors
\cite{AsakuraOosawa1958}. Their different ensembles matter in a finite
sealed cavity. Exclusion can favor arrangements leaving more space to
crowders, while crowding can also slow encounter and rearrangement.
Wall and many-particle overlap terms generally prevent a universal
pairwise reduction. Explicit crowders already generate their depletion
contribution; adding it again as an effective attraction counts it twice.
With hard exclusion, moving accessible domains carry contact/boundary
terms. Applying~\eqref{eq:architecture-mean-force-curvature} requires
its fixed-domain differentiability hypotheses, or an appropriate smooth
regularization before the limit. Slow crowder relaxation requires its
state or matched memory in addition to this equilibrium free energy.

\begin{proposition}[Passive mobility and the equilibrium measure]
\label{prop:s2patch-passive-configuration-generator}
Use local coordinates $q^a$ on the retained continuous configuration
space, with physical measure $J(q)\,\dd q$ and fixed isothermal free
energy $\mathcal F$. Let $\mathsf\mu^{ab}(q)=\mathsf\mu^{ba}(q)$
be a smooth positive mobility mapping conjugate forces to coordinate
rates. A passive overdamped density $\rho$ relative to that measure
obeys
\begin{equation}
 \partial_t\rho=\frac1J\partial_a
 \left[J\mathsf\mu^{ab}
       (\rho\,\partial_b\mathcal F+k_BT\partial_b\rho)\right].
 \label{eq:s2patch-reversible-configuration-diffusion}
\end{equation}
Repeated coordinate indices are summed. With no-flux boundaries and a
finite normalization, $\rho_{\rm eq}\propto e^{-\mathcal F/(k_BT)}$
is stationary. Its local It\^o drift and noise covariance are
\begin{equation}
 b^a=-\mathsf\mu^{ab}\partial_b\mathcal F
       +\frac{k_BT}{J}\partial_b(J\mathsf\mu^{ab}),
 \qquad
 \langle\dd q^a\dd q^b\rangle=2k_BT\mathsf\mu^{ab}\dd t.
 \label{eq:s2patch-mobility-thermal-drift}
\end{equation}
The covariance concerns the stochastic increments.
\end{proposition}
\begin{proof}
Insert $\partial_b\rho_{\rm eq}=-\rho_{\rm eq}
\partial_b\mathcal F/(k_BT)$ to make the probability flux vanish.
Expanding the It\^o forward equation for coordinate density $J\rho$
with the displayed drift and covariance gives
\eqref{eq:s2patch-reversible-configuration-diffusion}.
\end{proof}
For orientations, the measure includes rotational Haar volume. The
mobility includes its translation--rotation, interparticle and wall
blocks, and the thermal increments are correspondingly correlated.
Configuration-dependent mobility therefore requires its thermal and
geometric drift; independent Cartesian kicks do not supply the enlarged
law~\cite{DelongUsabiagaDonev2015}. This configuration diffusion and
the inertial Hamiltonian are different dynamical resolutions of a stated
physical preparation.

A finite prepared fluid motion can relax within the enclosure. A
sustained prescribed flow instead specifies forcing, wall or inlet/outlet
conditions and supplied work. Its conservative interaction graph and
hydrodynamic mobility have different roles: one supplies an energy
derivative, the other transmits motion and dissipation. Driven flow or
active forcing adds its own generator and source account; the equilibrium
proposition alone does not describe that driving. These choices set the
inputs for the fluid construction in Section~\ref{sec:complex-displacement}
and the state/readout tests in Section~\ref{sec:operator-framework}.

\subsection{Interaction network and coordination shells}
\label{sec:network}

\subsubsection{The finite charged ideal-virion assembly}
\label{sec:charged-ideal-vlp-assembly}

\begin{definition}[Finite charged ideal-virion assembly]
\label{def:s2-charged-vlp-assembly}
An \emph{ideal virion} is the rigid icosahedral, whole-particle specialization
of Definition~\ref{def:ideal-virion}, with its centre as the retained
mechanical coordinate. A \emph{finite charged assembly} consists of a
fixed labelled population of these particles, their positive masses,
assigned effective charges, admissible reference positions and declared
interaction and environmental laws.
\end{definition}

Particle identities persist as their centres move through the medium.
In the translational baseline, fixed orientations leave $3n$ centre
coordinates. The surrounding material supplies the assigned dielectric
and mechanical response.

One can separate body geometry from the centre coordinate by writing
\begin{equation}
 \mathcal P_i(t)=\vct r_i(t)+\mathsf O_i\mathcal P_i^0,
 \qquad \mathsf O_i\in\operatorname{SO}(3),\qquad
 \mathcal P_i^0\subseteq\overline B_{b_i}(\vct0).
 \label{eq:vlp-rigid-body-centre-map}
\end{equation}
Here $\mathcal P_i^0$ is a centred regular icosahedral body and
$\mathsf O_i$ is its fixed orientation in the translational baseline.
The circumscribed radius $b_i$ gives the sufficient exclusion construction
already used in~\eqref{eq:finite-radius-configuration}. There are $3n$
translational coordinates before fixed constraints, independently of the
number of faces or vertices used to depict a particle. Retaining
$\mathsf O_i(t)$ or deforming the body would add rotational or internal
coordinates and their corresponding inertia. In particular, the
icosahedral point symmetry of one body does not specify the translation
symmetry of the set of centres.

The second variation of~\eqref{eq:pair-potential} supplies the effective
spring law. Its electrostatic term uses a radial monopole approximation;
an enclosing Gaussian surface alone does not imply a spherical local
field. Orientation-dependent energies can resolve surface anisotropy.

The finite population is the cardinality of the retained index set, not
the volume of a drawn coordination cube. For the cubic reference below,
an occupancy function $\nu_{\vct\ell}\in\{0,1\}$ records which sites
are present:
\begin{equation}
 \mathcal I=\{\vct\ell\in\mathbb Z^3:\nu_{\vct\ell}=1\},
 \qquad n=\sum_{\vct\ell}\nu_{\vct\ell},\qquad
 \vct R_*+a\vct\ell\in\mathcal V_{b}
       \quad(\vct\ell\in\mathcal I),
 \label{eq:vlp-finite-site-occupancy}
\end{equation}
where equal exclusion radii $b$ are assumed in the last condition and
$\mathcal V_b=\{\vct x\in\mathcal V:
\operatorname{dist}(\vct x,\partial\mathcal V)>b\}$.
The sum is finite. The cylinder's radius, length, population and spacing
are separate specifications subject to the admissibility conditions.
Occupancy records the reference sites; the particle labels persist as the
centres move away from those sites. The network's eigenmodes will
describe correlated centre displacements relative to one another within
this same medium.

Figure~\ref{fig:vlp-finite-assembly-coordinate-map} separates the fixed finite population from the centre displacements of its rigid particles.

\begin{figure}[H]
\centering\begingroup
\definecolor{vltTeal}{HTML}{176B64}
\definecolor{vltBlue}{HTML}{315D80}
\definecolor{vltGold}{HTML}{A87520}
\definecolor{vltPlum}{HTML}{725877}
\definecolor{vltRed}{HTML}{762B38}
\definecolor{vltYellow}{HTML}{D5AD3C}
\definecolor{vltNeutral}{HTML}{777D83}
\setlength{\unitlength}{1pt}
\scalebox{0.90}{%
\input{vlt-figures/figure-030.tex}%
}\endgroup
\caption{Six rigid ideal virions occupy a projected triangular prism in the cylindrical medium. Nine lines identify the selected pairs; a black coordinate triad fixes the viewing directions. Panel (b) retains shape and orientation under a centre translation. Each faceted body contributes one centre, not one coordinate per vertex. Positions and sizes are illustrative; force balance is a separate mechanical input. Figure~\ref{fig:vlp-three-dimensional-confined-network} shows a larger connector network.}
\label{fig:vlp-finite-assembly-coordinate-map}
\end{figure}

Figure~\ref{fig:vlp-three-dimensional-confined-network} supplies a spatial realization of a finite interaction network within the cavity.

\begin{figure}[H]
\centering\begingroup
\definecolor{vltTeal}{HTML}{176B64}
\definecolor{vltBlue}{HTML}{315D80}
\definecolor{vltGold}{HTML}{A87520}
\definecolor{vltPlum}{HTML}{725877}
\definecolor{vltRed}{HTML}{762B38}
\definecolor{vltYellow}{HTML}{D5AD3C}
\definecolor{vltNeutral}{HTML}{777D83}
\setlength{\unitlength}{1pt}
\scalebox{0.90}{%
\input{vlt-figures/figure-031.tex}%
}\endgroup
\caption{An illustrative $3\times2\times2$ occupied patch lies inside a cylinder whose axis is $x$. All twelve ideal virions are retained. The twenty axial pairs span the three spatial directions; four additional diagonals are explicitly selected. Each connector follows a three-dimensional helix between points on the rigid body surfaces. The body circumscribed radius is $0.12a$, giving positive wall and pair clearances for these coordinates. Coils are reference interaction glyphs; pair laws and support conditions complete the mechanical specification.}
\label{fig:vlp-three-dimensional-confined-network}
\end{figure}

The general reference is a finite set of positions and an interaction
graph. For a tractable geometric example, choose a subset of a simple
cubic arrangement inside the cylinder:
\begin{equation}
 \vct R_i=\vct R_*+a\vct\ell_i,\qquad
 \vct\ell_i\in\mathbb Z^3,\qquad a>0,
 \label{eq:cubic-lattice}
\end{equation}
The spacing $a$ and integer vectors $\vct\ell_i$ locate the sites relative
to $\vct R_*$. For equal exclusion radii $b$, $a>2b$ gives positive nearest
pair clearance; every retained site also needs positive wall clearance.
This is an imposed reference geometry. A chosen energy must separately
make it force-balanced and stable. Neither follows from drawing a grid.

For a central site $\alpha_i$, the coordination labels $\beta_i$,
$\gamma_i$ and $\Omega_i$ classify separations, as listed in
Table~\ref{tab:coordination-shells}.
\begin{table}[H]
\centering
\caption{Coordination shells of the one-site simple cubic reference.
Counts describe an interior site before finite-boundary truncation.}
\label{tab:coordination-shells}
\begin{tabular}{llll}
\toprule
Label & Relationship & Reference distance & Bulk count\\
\midrule
$\alpha_i$ & Central site & $0$ & $1$\\
$\beta_i$ & Axial nearest neighbors & $a$ & $6$\\
$\gamma_i$ & Face-diagonal neighbors & $\sqrt{2}a$ & $12$\\
$\Omega_i$ & Body-diagonal neighbors & $\sqrt{3}a$ & $8$\\
\bottomrule
\end{tabular}
\end{table}
The central site and shells form a 27-site interior coordination
neighborhood, called a \emph{viral cell} here. The shell labels classify
geometry. A primitive cell instead specifies the smallest periodic
repeat, which contains one site for this cubic reference. Finite
boundaries truncate the neighborhood.
The coordination diagrams consistently use teal, gold, blue and plum for
$\alpha,\beta,\gamma,\Omega$; printed labels retain the classification
in grayscale.

The shell counts follow from the integer offsets. Relative to an
interior central site, define
\begin{equation}
 \begin{aligned}
 \mathcal S_\beta&=\{\vct d=a\vct j:\vct j\in\{-1,0,1\}^3,
                                      |\vct j|^2=1\},\\
 \mathcal S_\gamma&=\{\vct d=a\vct j:\vct j\in\{-1,0,1\}^3,
                                      |\vct j|^2=2\},\\
 \mathcal S_\Omega&=\{\vct d=a\vct j:\vct j\in\{-1,0,1\}^3,
                                      |\vct j|^2=3\}.
 \end{aligned}
 \label{eq:architecture-shell-sets}
\end{equation}
For $\beta$, one of three components is nonzero, with two signs, giving
$3\times2=6$. For $\gamma$, choose two components and their signs,
giving $3\times4=12$. For $\Omega$, all three components are nonzero,
giving $2^3=8$. Figure~\ref{fig:architecture-coordination-layers}
shows all 27 sites as three parallel layers. A planar picture alone
would conceal the out-of-plane distances; retaining the layer label
makes the shell assignment exact~\cite{Misra2012}.

\begin{figure}[H]
\centering\begingroup
\definecolor{vltTeal}{HTML}{176B64}
\definecolor{vltBlue}{HTML}{315D80}
\definecolor{vltGold}{HTML}{A87520}
\definecolor{vltPlum}{HTML}{725877}
\definecolor{vltRed}{HTML}{762B38}
\definecolor{vltYellow}{HTML}{D5AD3C}
\definecolor{vltNeutral}{HTML}{777D83}
\setlength{\unitlength}{1pt}
\scalebox{0.90}{%
\input{vlt-figures/figure-032.tex}%
}\endgroup
\caption{All $27$ sites of $\{-1,0,1\}^3$ appear in three coordinate-labelled layers. Teal, gold, blue and plum identify $\alpha,\beta,\gamma,\Omega$, respectively, with explicit class labels. The rule $\Delta\ell_x^2+\Delta\ell_y^2+\Delta\ell_z^2=0,1,2,3$ gives the assignment and the counts. Each faceted icon represents one whole ideal virion. Particle radius and lattice spacing are independent inputs.}
\label{fig:architecture-coordination-layers}
\end{figure}

Figure~\ref{fig:added-three-dimensional-shells} gives complementary three-dimensional views of the three shell sets.

\begin{figure}[H]
\centering\begingroup
\definecolor{vltTeal}{HTML}{176B64}
\definecolor{vltBlue}{HTML}{315D80}
\definecolor{vltGold}{HTML}{A87520}
\definecolor{vltPlum}{HTML}{725877}
\definecolor{vltRed}{HTML}{8E3B46}
\definecolor{vltYellow}{HTML}{D5AD3C}
\definecolor{vltNeutral}{HTML}{777D83}
\setlength{\unitlength}{1pt}
\scalebox{0.90}{%
\input{vlt-figures/figure-033.tex}%
}\endgroup
\caption{The same $27$ ideal virions appear in every frame at one orthographic scale. The selected class alone is colored: teal $\alpha$, gold $\beta$, blue $\gamma$ or plum $\Omega$; the remainder is gray. Axis ticks are one lattice unit apart, and the central ring fixes the origin. Spokes identify centre-to-shell relationships and the printed distances. They specify geometry; the pair energy supplies the force law.}
\label{fig:added-three-dimensional-shells}
\end{figure}

\begin{definition}[Interaction graph and coordination neighborhood]
\label{def:s2-interaction-graph}
The \emph{interaction graph} $G=(\{1,\ldots,n\},\mathcal E)$ has one
vertex for each retained ideal virion and one undirected edge for each
unordered pair entering the declared conservative pair energy. A
\emph{coordination shell} groups reference displacements by a specified
geometric relation; the cubic shells are
\eqref{eq:architecture-shell-sets}. Their union with the central site is
the corresponding coordination neighborhood, called a viral cell here.
\end{definition}

A nearest-neighbor model retains $\beta$ edges; additional shells enlarge
that energy model. Range truncation is assessed through its effect on
force balance, stiffness and the selected readout. Longer-range forces
and eliminated medium coordinates can add nonlocal couplings.
Hydrodynamic coupling need not follow the conservative pair graph.

Figure~\ref{fig:added-boundary-coordination} displays the loss of incident edges at several boundaries of a finite cubic patch.

\begin{figure}[H]
\centering\begingroup
\definecolor{vltTeal}{HTML}{176B64}
\definecolor{vltBlue}{HTML}{315D80}
\definecolor{vltGold}{HTML}{A87520}
\definecolor{vltPlum}{HTML}{725877}
\definecolor{vltRed}{HTML}{8E3B46}
\definecolor{vltYellow}{HTML}{D5AD3C}
\definecolor{vltNeutral}{HTML}{777D83}
\setlength{\unitlength}{1pt}
\scalebox{0.90}{%
\input{vlt-figures/figure-034.tex}%
}\endgroup
\caption{The complete finite $3\times3\times3$ patch is retained in every view. The marked site and its axial neighbors are highlighted; the rest remain gray. Lines select only incident nearest-neighbor pairs within the patch, giving degrees $6,5,4,3$. No edge wraps across a boundary. The cube and coordinate triad use the same projection and scale throughout. These are finite cubic examples; a cylindrical cut has its own occupancy and boundary coordination.}
\label{fig:added-boundary-coordination}
\end{figure}

The three-dimensional construction in Figure~\ref{fig:vlp-three-dimensional-affine-pairs} makes the relation between a displacement field and its edge strains explicit.

\begin{figure}[H]
\centering\begingroup
\definecolor{vltTeal}{HTML}{176B64}
\definecolor{vltBlue}{HTML}{315D80}
\definecolor{vltGold}{HTML}{A87520}
\definecolor{vltPlum}{HTML}{725877}
\definecolor{vltRed}{HTML}{8E3B46}
\definecolor{vltYellow}{HTML}{D5AD3C}
\definecolor{vltNeutral}{HTML}{777D83}
\setlength{\unitlength}{1pt}
\scalebox{0.90}{%
\input{vlt-figures/figure-035.tex}%
}\endgroup
\caption{All eight particles and fourteen selected pairs are retained in both frames. Both assemblies share a projection and length scale; gray wireframes retain the reference cube and black triads give axis directions. The twelve axial connectors and the highlighted face and body diagonals use five turns, fixed coil radius and fixed leads in three dimensions. Their surface attachment points are recomputed after the centre shifts, so pitch follows the actual free span. The tabulated ratios describe centre separations. This is an imposed kinematic deformation; it does not assume that the field is a normal mode or an equilibrium.}
\label{fig:vlp-three-dimensional-affine-pairs}
\end{figure}

The degree $d_i$ counts incident edges and obeys
$\sum_i d_i=2|\mathcal E|$. A boundary site can have fewer neighbors
than an interior site. Removing a site removes its pairs and can change
force balance; the completed boundary reference must therefore be
checked. A reflecting cylinder retains those finite boundaries.

Shells organize repeated directions efficiently. An irregular assembly
can instead store each retained direction and separation directly.
Both representations answer the same mechanical question: which
relative displacements change which forces?

\paragraph{Charge assignments and complete pair bookkeeping.}
\begin{definition}[Fixed charge pattern]
\label{def:s2-fixed-charge-pattern}
The \emph{charge pattern} is the ordered list $(q_1,\ldots,q_n)$
assigned to the labelled particle centres. Its total charge and sign
populations are
\begin{equation}
 \begin{gathered}
 Q_{\rm virion}=\sum_{i=1}^n q_i,\qquad n_++n_-+n_0=n,\\
 n_+=\#\{i:q_i>0\},\quad n_-=\#\{i:q_i<0\},\quad
 n_0=\#\{i:q_i=0\}.
 \end{gathered}
 \label{eq:vlp-charge-population}
\end{equation}
\end{definition}
All-positive, all-negative and mixed-sign arrays are allowed assignments
of this finite model. The charges are fixed parameters during this central
mechanical evolution. Sections~\ref{sec:s2patch-surface-interactions}
and~\ref{sec:s2patch-surface-relaxation} specify a different realization
with resolved surface exchange or polarization. Its chemical preparation
and relaxation need not share the reversal symmetry below. A positive total charge does not imply that every
pair is repulsive: a mixed array can have $Q_{\rm virion}>0$ while still
containing attractive electrostatic pairs. The products $q_iq_j$, their
positions and the short-range law enter the force constants; the scalar
$Q_{\rm virion}$ alone does not determine them.

For a complete finite interaction model, the pair part of
\eqref{eq:controlled-potential} is
\begin{equation}
 U_{\rm pair}^{\rm full}(\vct r)
   =\sum_{1\le i<j\le n}V_{ij}(r_{ij}),\qquad
 V_{ij}=V_{ij}^{\rm el}+V_{ij}^{\rm LJ}.
 \label{eq:vlp-complete-finite-pair-energy}
\end{equation}
One adds the same declared wall, support and control energies used in the
Hamiltonian. Each unordered pair occurs once. With fixed screening and
homogeneous absolute permittivity, $V_{ij}^{\rm el}$ is the screened term
in~\eqref{eq:pair-potential}; $\kappa=0$ gives unscreened Coulomb.
For $\kappa>0$ its far field decays exponentially, whereas the
Lennard--Jones attractive tail is algebraic. Thus electrostatic and
short-range refer to the specified laws and length scales, rather than
a universal ordering of their magnitudes at every separation.

\begin{proposition}[Global charge-reversal symmetry]
\label{prop:s2-charge-reversal}
For the pair energy~\eqref{eq:vlp-complete-finite-pair-energy} with fixed
screening and permittivity, charge dependence only through $q_iq_j$ and
charge-independent Lennard--Jones parameters, reversing every charge
preserves the pair energy and its Hessian at a common admissible geometry:
\begin{equation}
 U_{\rm pair}^{\rm full}(\vct r;\{-q_i\})
      =U_{\rm pair}^{\rm full}(\vct r;\{q_i\}),\qquad
 \mathsf K_{\rm pair}(\{-q_i\})
      =\mathsf K_{\rm pair}(\{q_i\}).
 \label{eq:vlp-global-charge-reversal}
\end{equation}
The complete mechanical model has the same symmetry when its remaining
energies, inertia, constraints and environmental response are unchanged
by that reversal.
\end{proposition}
\begin{proof}
Every product $(-q_i)(-q_j)$ equals $q_iq_j$, so the two pair potentials
agree as functions of position. Their first and second spatial
derivatives consequently agree. Adding the same remaining terms and
using the same mechanical inputs preserves the full equations.
\end{proof}
An all-positive array and its all-negative counterpart therefore have
the same pair restoring matrix at the same geometry in this
specialization. An external electric field, charge-selective boundary or
charge-regulating environment requires its additional energy and can
remove that equivalence. Redistributing signs among sites generally
changes the pair matrix even at unchanged total charge.

Figure~\ref{fig:vlp-charge-pattern-pair-products} compares sign rearrangement with global reversal at a shared finite geometry.

\begin{figure}[H]
\centering\begingroup
\definecolor{vltTeal}{HTML}{176B64}
\definecolor{vltBlue}{HTML}{315D80}
\definecolor{vltGold}{HTML}{A87520}
\definecolor{vltPlum}{HTML}{725877}
\definecolor{vltRed}{HTML}{8E3B46}
\definecolor{vltYellow}{HTML}{D5AD3C}
\definecolor{vltNeutral}{HTML}{777D83}
\setlength{\unitlength}{1pt}
\scalebox{0.84}{%
\input{vlt-figures/figure-036.tex}%
}\endgroup
\caption{All three frames retain four positions and two charges of each sign, with $q_\star>0$. The thick line selects the same pair $\{1,2\}$: its product changes sign in (b), then returns to the value in (a) under global reversal in (c). Positive particles are yellow and negative particles blue, with explicit signs. Lines denote unordered pairs. Charge-independent inputs and the homogeneous pair law are held fixed in the reversal comparison.}
\label{fig:vlp-charge-pattern-pair-products}
\end{figure}

The shells in Table~\ref{tab:coordination-shells} organize geometry.
They do not remove farther interactions from
\eqref{eq:vlp-complete-finite-pair-energy}. A three-shell approximation
retains a declared subset of pairs, while a complete finite Coulomb
model retains all charged pairs. One may use different retained pair
sets for the two energy terms, provided the force and Hessian are
derived from that same energy. A distance-dependent switching function
also contributes its derivatives; discarding those derivatives changes
the mechanical law.

\begin{proposition}[Fixed-reference error from omitted interactions]
\label{prop:s2-omitted-interaction-error}
Compare complete and truncated conservative pair energies at the same
admissible reference, with the same masses and all other terms fixed.
Let $m_{\min}=\min_i m_i>0$, let $\mathcal E_{\rm om}$ be the omitted
unordered edges, and let their symmetric directional curvatures be
$\mathsf C_{ij}$ from~\eqref{eq:edge-stiffness}. The resulting force and
mass-normalized Hessian changes satisfy
\begin{equation}
 \begin{aligned}
 \|\Delta\vct f_i^{(0)}\|
    &\le\sum_{j:\{i,j\}\in\mathcal E_{\rm om}}
                         |V_{ij}'(R_{ij})|,\\
 \|\mathsf D_{\rm full}-\mathsf D_{\rm tr}\|_2
    &\le\frac{2}{m_{\min}}
       \max_i\sum_{j:\{i,j\}\in\mathcal E_{\rm om}}
                              \|\mathsf C_{ij}\|_2
       =:\delta_{\rm om}.
 \end{aligned}
 \label{eq:vlp-omitted-interaction-bounds}
\end{equation}
Every ordered eigenvalue of the two real symmetric dynamical matrices
differs by at most $\delta_{\rm om}$ at that reference.
\end{proposition}
\begin{proof}
The triangle inequality applied to the omitted radial forces gives the
first line. Form the scalar symmetric comparison matrix with entries
$\|\Delta\mathsf K_{ij}\|_2$. Each block-row sum is at most twice the
sum of the incident omitted curvature norms, which bounds
$\|\Delta\mathsf K\|_2$. Congruence by $\mathsf M^{-1/2}$ multiplies
this bound by at most $1/m_{\min}$. The eigenvalue estimate follows
from the min--max principle~\cite{HornJohnson2013}.
\end{proof}

The first bound measures changed reference force; $\delta_{\rm om}$
has units $\mathrm{s^{-2}}$ and controls squared modal frequencies when
the compared references are stable.
If only one energy component of a pair is omitted, the same bounds
apply with its own omitted force derivative and curvature, before
combining it with retained contributions.
A truncation can also destroy stationarity. Frequencies about a newly
relaxed reference require its newly evaluated Hessian; the displayed
fixed-geometry estimate does not include that additional change.

Finite particle charge need not sum to zero for the nonoverlapping
finite pair sum to exist. The dielectric and screening approximation
still specifies how the surrounding medium is represented. The
cylinder's mechanical wall is not, by itself, an electrostatic boundary
condition. Dielectric interfaces or imposed electric potentials require
the corresponding field energy or Green function.
A separately repeated \emph{unscreened} charged reference also needs a
neutrality or compensating-background prescription and an electrostatic
summation/boundary convention. It cannot be obtained by silently wrapping
the finite charged cylinder. These distinctions matter before adopting
a three-shell Fourier symbol or a local continuum tensor.

\paragraph{Application: choosing coordinates for a filament network.}
The geometry--graph--law distinction extends naturally to passive
cytoskeletal or extracellular-matrix scaffolds, provided the represented
object is changed explicitly. A node can describe a cross-link or a point
along a filament rather than a whole particle. The retained positions then
describe a material's deformation, while the graph records which segments
and attachments share mechanical interactions. A structural image can
constrain this architecture without fixing its deformation energy.

A filament segment can store energy by stretching and by bending. An
unstressed central spring between its endpoints retains only the first
mechanism at harmonic order: a small transverse endpoint motion has no
quadratic cost when $V'(R)=0$. A discretized bending law instead couples
successive segment directions and therefore generally contributes
multi-node Hessian terms. Cross-link rotations may also need separate
coordinates or a declared freely rotating constraint. Such choices determine
which relative motions can carry energy; they cannot be replaced by
changing the color or weight of an adjacency edge.

Head, Levine and MacKintosh's network model explicitly retains stretching
and bending and finds affine and nonaffine deformation regimes
~\cite{HeadLevineMacKintosh2003}. Its relevance here is architectural: a
model that omits bending can label a motion mechanically free even when a
filament with bending rigidity resists it. Calibrated segment geometry,
boundary loading and resolved transverse motion can distinguish these
descriptions. Effective spring constants remain tied to the chosen scale
and small-deformation range~\cite{Cotterill2002}. If finer bending
coordinates are eliminated, their static accommodation and dynamic memory
must be matched to the retained network. Thus establishing a displacement
variable includes establishing the deformation that it represents and the
energy terms that make it observable mechanically.

\subsubsection{A fixed particle pool and changing lattice membership}
\label{sec:s23-membership-architecture}

A selected lattice can contain any admissible subset of the fixed
particle pool. Recruitment and release change that subset; excitation
specifies the motion of its members.

\begin{definition}[Particle pool and cluster membership]
\label{def:s23-cluster-membership}
Let $\mathcal I_{\rm tot}=\{1,\ldots,N_{\rm tot}\}$ index the fixed
particle pool in the sealed cavity. A selected cluster is a time-dependent
subset $\mathcal C(t)\subseteq\mathcal I_{\rm tot}$, with
\begin{equation}
 n_{\mathcal C}(t)=|\mathcal C(t)|,\qquad
 0\le n_{\mathcal C}(t)\le N_{\rm tot},\qquad
 \sum_{\mathcal C\in\pi(t)}n_{\mathcal C}(t)=N_{\rm tot}.
 \label{eq:s23-fixed-pool-cluster-size}
\end{equation}
Here $\pi(t)$ is a partition into clusters, including singleton particles.
The full retained translational state has $3N_{\rm tot}$ position
coordinates before constraints; a conditional cluster calculation retains
its selected $3n_{\mathcal C}$ coordinates and states how the remaining
particles and medium influence them.
\end{definition}

Membership requires an operational rule. For example, an observed
association edge $h_{ij}$ can use separate entry and exit distances,
\begin{equation}
 h_{ij}(t^+)=
 \begin{cases}
 1,&r_{ij}(t)\le r_{\rm on},\\
 0,&r_{ij}(t)\ge r_{\rm off},\\
 h_{ij}(t^-),&r_{\rm on}<r_{ij}(t)<r_{\rm off},
 \end{cases}
 \qquad r_{\rm on}<r_{\rm off}.
 \label{eq:s23-hysteretic-membership}
\end{equation}
Connected components of this observed graph define clusters. A minimum
residence time can additionally exclude brief encounters. These thresholds
specify a measurement convention; changing them does not remove the
Lennard--Jones or electrostatic force from the Hamiltonian. A mechanically
useful cluster also persists long enough for its conditional reference and
collective response to be resolved. Metastable bound and separated sets
can be compared more directly through the probability of reaching one
before returning to the other, the committor of transition-path theory
\cite{EVandenEijnden2006}.

\begin{definition}[Engagement and excitation]
\label{def:s23-engagement-state}
An ideal virion is \emph{engaged} with a selected cluster when it satisfies
the declared membership and residence criteria. Its mechanical excitation
is specified separately by displacement, momentum or modal occupation.
An unengaged particle retains its mass, charge and interaction law.
If a distinct internal activation state changes that law, the state,
its energy and its transition mechanism are additional model variables.
\end{definition}

Practical independence is assessed through the selected observable.
A separated virion is effectively independent over an observation band
when its residual coupling changes the declared response by less than
the stated tolerance. The response-update bounds in
Subsection~\ref{sec:s3-phonon-scattering-updates} provide one way to test
this approximation. Distance and membership labels alone do not establish
that tolerance for a charged assembly.

Thus a cluster can grow from $n_{\mathcal C}=4$ toward
$n_{\mathcal C}=1000$ within a pool with $N_{\rm tot}\ge1000$, while
the complete particle and energy accounts remain closed. Individual
members can join, leave or exchange between clusters. The transient
extension in Subsection~\ref{sec:s23-transient-recruitment} supplies the
encounter, energy-transfer and escape mechanisms for those changes.

Figure~\ref{fig:s23-membership-snapshots} follows fixed particle identities through an illustrative recruitment and release event.

\begin{figure}[H]
\centering\begingroup
\definecolor{vltTeal}{HTML}{176B64}
\definecolor{vltBlue}{HTML}{315D80}
\definecolor{vltGold}{HTML}{A87520}
\definecolor{vltPlum}{HTML}{725877}
\definecolor{vltRed}{HTML}{8E3B46}
\definecolor{vltYellow}{HTML}{D5AD3C}
\definecolor{vltNeutral}{HTML}{777D83}
\setlength{\unitlength}{1pt}
\scalebox{0.90}{%
\input{vlt-figures/figure-037.tex}%
}\endgroup
\caption{All views retain the same eight numbered ideal virions, camera and length scale. Red displacement arrows give virion 5's approach and release directions, separately from the contact record below. The selected core is teal, its two new contacts are red, and particles outside $\mathcal C$ are gray. The lower step diagram records both displayed association edges during a residence interval; its horizontal coordinate orders the snapshots without assigning transition rates. Identical four-turn core coils preserve their free spans. The graph omits force tails only from the drawing. These kinematic and membership constructions do not prescribe a solved capture trajectory.}
\label{fig:s23-membership-snapshots}
\end{figure}

\subsubsection{Assembly cells as mechanical building blocks}
\label{sec:s2-assembly-cells}

Assembly cells organize larger populations by combining local energies
on shared physical coordinates and specifying cross-interface pairs.
Periodic repetition is one special placement of these building blocks.

\begin{definition}[Assembly cell and compatible placement]
\label{def:s2-assembly-cell}
An \emph{assembly cell} $c$ is a finite list of ideal virions, reference
positions and assigned internal interactions, together with the particle
sites through which it can couple to other cells. Its local particle
labels $a=1,\ldots,n_c$ are mapped injectively to physical labels by
$\iota_c(a)$. For a collection $\mathcal F_{\rm cell}$ of cells, define
\begin{equation}
 \mathcal I_{\rm asm}=\bigcup_{c\in\mathcal F_{\rm cell}}
       \iota_c(\{1,\ldots,n_c\}),\qquad
 n_{\rm asm}=|\mathcal I_{\rm asm}|.
 \label{eq:s2-cell-union-count}
\end{equation}
A \emph{compatible placement} assigns one reference position, mass and
charge specification to each physical label. In a resolved surface model
it also identifies that particle's orientation, patch map and occupancy
state. Shared material coordinates have one declared physical owner and
history. If two cell descriptions contain the same label, they describe
the same physical particle and state; its site chemistry and energy are
counted once.
\end{definition}

The label map makes the particle count explicit. Two four-particle
blocks with no shared particles contain eight particles. Two tetrahedral
blocks sharing the same three particles of one face contain five. Their
interface copies are useful for computation, but do not supply another
mass or charge. Thus 250 disjoint four-particle blocks contain 1000
particles, whereas a chain in which every new tetrahedron shares exactly
one existing face introduces one new particle per step and has
$n_{\rm asm}=N_{\rm cell}+3$. These are counting statements; admissible
placement, the retained cross-pairs and global force balance complete
the mechanical reference. In the sealed-system description,
$n_{\rm asm}\le N_{\rm tot}$ when the cells describe a selected assembly.

The 27-site viral cell already defined above is an overlapping
\emph{coordination neighborhood}. It is useful for inspecting the
surroundings of a central particle. An assembly cell instead specifies a
block for constructing the complete mechanical description. A
\emph{primitive unit cell} is the smallest repeat of a periodic reference,
including its particle basis and mechanical parameters. These three
uses have different roles: neighborhood analysis, finite assembly and
translation symmetry. The spatial volume cells used later to represent
fields in Definition~\ref{def:s3-cell-field-representation} are a fourth, purely
representational partition; they do not add particles or mechanical
connections.

For genuinely repeated copies, the reference takes the form
$\vct R_{\vct\ell\sigma}=\mathsf A\vct\ell+\vct\rho_\sigma$,
with a nonsingular translation matrix $\mathsf A$,
$\vct\ell\in\mathbb Z^3$ and a basis of $n_{\rm b}$ distinct physical
sites. Assigning each site to one half-open primitive cell avoids
counting a face, edge or corner repeatedly. A finite index set
$\mathcal J$ of complete cells then has
$n_{\rm asm}=|\mathcal J|n_{\rm b}$. A cylindrical cut instead uses
$n_{\rm asm}=\sum_{\vct\ell,\sigma}\nu_{\vct\ell\sigma}$ with its
declared zero--one occupancy. Masses, charges and interactions repeat
with the basis when periodic dynamics is claimed; the corresponding
Bloch construction is developed in
Section~\ref{sec:structure-charged-periodic-basis}.

\begin{definition}[Internal pairs and cross-cell pairs]
\label{def:s2-cross-cell-pairs}
Give every physical particle one cell owner, even when its coordinate is
copied into neighboring cell descriptions. Internal pairs have both
endpoints in one owner's set; a \emph{cross-cell pair} has endpoints with
different owners. Let $\mathcal E_\psi$ denote the retained cross-cell
pairs. Its pair law remains the declared $V_{ij}(r)$, with the same
charge, screening and short-range parameters as any other retained pair.
The tag $\psi$ specifies where a pair enters the assembly, while
$\beta,\gamma,\Omega$ specify its reference geometry.
\end{definition}

An axial neighbor can also be a cross-cell neighbor: cell ownership
changes bookkeeping, while the pair retains its physical law. The
$\psi$ notation identifies such cross-pairs. Their full central-force
tensor is
\begin{equation}
 \mathsf C_{ij}
 =V_{ij}^{\prime\prime}(R_{ij})\vct n_{ij}^0
                              (\vct n_{ij}^0)^{\mathsf T}
 +\frac{V_{ij}^{\prime}(R_{ij})}{R_{ij}}
       \bigl[\Id_3-\vct n_{ij}^0(\vct n_{ij}^0)^{\mathsf T}\bigr],
 \qquad \{i,j\}\in\mathcal E_\psi.
 \label{eq:s2-cross-cell-tensor}
\end{equation}
The first term measures longitudinal curvature; the second retains the
geometric contribution of an existing pair force. Which pairs belong
to $\mathcal E_\psi$ depends on the chosen cell ownership, but their
physical forces and their contribution to the full energy do not. A
different decomposition of the same assembly must therefore produce
the same global matrix.

Figure~\ref{fig:s23-compatible-cell-building} compares disjoint and shared-particle assemblies using explicit label maps.

\begin{figure}[H]
\centering\begingroup
\definecolor{vltTeal}{HTML}{176B64}
\definecolor{vltBlue}{HTML}{315D80}
\definecolor{vltGold}{HTML}{A87520}
\definecolor{vltPlum}{HTML}{725877}
\definecolor{vltRed}{HTML}{8E3B46}
\definecolor{vltYellow}{HTML}{D5AD3C}
\definecolor{vltNeutral}{HTML}{777D83}
\setlength{\unitlength}{1pt}
\scalebox{0.90}{%
\input{vlt-figures/figure-038.tex}%
}\endgroup
\caption{(a) A red $\psi$ pair joins disjoint four-particle lists; its endpoints retain their numbered identities. (b) Gold sites 1--3 belong to both regular tetrahedra, while apices 4 and 5 differ. The shaded face and its three pair energies occur once in the assembled model. (c) Two local displacement entries for site 1 map to the same physical virion. Compatibility equates their motion; bookkeeping retains one mass and charge. The same identification applies to shared sites 2 and 3. Numbers below the views are local-to-physical maps. The two views use different display scales; selected reference coils specify no equilibrium or additional boundary force.}
\label{fig:s23-compatible-cell-building}
\end{figure}

\subsection{Force constants, stability and energy transport}
\label{sec:hessian}

\begin{definition}[Force-constant matrix]
\label{def:s2-force-constant-matrix}
At a smooth stationary reference $\vct R$ of the particle-only potential,
$\nabla U_\lambda(\vct R)=0$, the \emph{force-constant matrix} is its
spatial Hessian:
\begin{equation}
 \mathsf K_{i\mu,j\nu}=
 \left.\frac{\partial^2U_\lambda}
 {\partial r_{i\mu}\partial r_{j\nu}}\right|_{\vct r=\vct R},
 \qquad \mu,\nu\in\{x,y,z\}.
 \label{eq:hessian}
\end{equation}
\end{definition}
The entry indexed by $i\mu,j\nu$ is the negative derivative of force
$F_{i\mu}$ with respect to $r_{j\nu}$:
$\delta F_{i\mu}=-\sum_{j\nu}\mathsf K_{i\mu,j\nu}\delta r_{j\nu}$.
Symmetry of mixed
derivatives gives $\mathsf K=\mathsf K^{\mathsf T}$. Translational entries
have units $\mathrm{N\,m^{-1}}$. In the local expansion,
$U_\lambda(\vct R+\vct u)=U_\lambda(\vct R)
+\tfrac12\vct u^{\mathsf T}\mathsf K\vct u+o(\|\vct u\|^2)$.
The linear term vanishes by force balance. A controlled cubic remainder
would need additional smoothness and derivative bounds.

For a resolved orientation/surface model,
Definition~\ref{def:s2patch-tangent-hessian} supplies the larger tangent
form and states which populations are clamped. Fast equilibration uses
the reduced curvature; finite-rate states use their augmented dynamics.
The central-pair tensor, incidence assembly and translational energy
currents below retain their stated radial specialization. General patch
or many-body energies contribute their full derivatives instead of being
inserted into a scalar radial curvature.

\paragraph{Reference force, harmonic accuracy and fast-coordinate reduction.}
With $\lambda$ held fixed, the harmonic construction applies the lattice-dynamics expansion to the
\emph{whole ideal virion coordinates}~\cite{Dove2011}. At a trial reference
with $\vct g=\nabla U_\lambda(\vct R)\ne\vct0$, the expansion includes
$\vct g^{\mathsf T}\vct u$ and the linearized equation is
$\mathsf M\Dtt{\vct u}+\mathsf K\vct u=-\vct g$.
Diagonalizing $\mathsf K$ remains an algebraic operation, but its
homogeneous modes then do not describe oscillations about a stationary
reference. The finite-cylinder force balance in
\eqref{eq:finite-reference-force-balance} is the physical condition
that removes this constant forcing, including the boundary sites.

\begin{proposition}[Controlled harmonic approximation]
\label{prop:s2-harmonic-remainder}
Suppose $U_\lambda$ is three times continuously differentiable on an
admissible convex neighborhood of the stationary reference $\vct R$,
and its third derivative has operator norm at most $B_3$. For
$\vct R+\vct u$ in that neighborhood, the energy and force remainders obey
\begin{equation}
 \begin{aligned}
 U_\lambda(\vct R+\vct u)-U_\lambda(\vct R)
       &=\tfrac12\vct u^{\mathsf T}\mathsf K\vct u
                              +\mathcal R_3(\vct u),
 &|\mathcal R_3(\vct u)|&\le\tfrac16B_3\|\vct u\|^3,\\
 -\nabla U_\lambda(\vct R+\vct u)
       &=-\mathsf K\vct u+\vct\rho_2(\vct u),
 &\|\vct\rho_2(\vct u)\|&\le\tfrac12B_3\|\vct u\|^2.
 \end{aligned}
 \label{eq:vlp-controlled-harmonic-remainders}
\end{equation}
On a fixed linear constrained subspace with Euclidean-orthonormal
coordinates and
$\mathsf K_{\rm c}\succeq k_{\rm c,min}\Id>0$, the same derivative bound
gives, for nonzero admissible $\vct q_{\rm c}$,
\begin{equation}
 \frac{\|\vct\rho_{2,\rm c}\|}
      {\|\mathsf K_{\rm c}\vct q_{\rm c}\|}
       \le\frac{B_3\|\vct q_{\rm c}\|}{2k_{\rm c,min}}.
 \label{eq:vlp-relative-harmonic-force-error}
\end{equation}
\end{proposition}
\begin{proof}
Taylor's integral remainder for the scalar energy gives the factor
$1/6$; applying it to its gradient gives $1/2$. Orthogonal restriction
does not increase the derivative bound. Finally,
$\|\mathsf K_{\rm c}\vct q_{\rm c}\|\ge
k_{\rm c,min}\|\vct q_{\rm c}\|$ gives the relative estimate.
\end{proof}

For translational coordinates $B_3$ has units $\mathrm{J\,m^{-3}}$.
The bound uses the same orthonormal restriction of the derivative tensor.
A soft restoring direction can therefore demand a smaller displacement
for the same relative force accuracy. This test and geometric clearance
are independent requirements. If the constrained stiffness has a kernel,
the relative estimate requires a specified positive sector; it does not
control force along a zero-curvature mechanism. The quartic example in
\eqref{eq:structure-transverse-quartic} illustrates that distinction.

The harmonic approximation truncates displacement dependence. In an
atomistic electronic calculation, the Born--Oppenheimer approximation
instead supplies an electronic adiabatic potential surface before its
harmonic expansion~\cite{Giustino2017}. Here the effective whole-particle
pair law is specified directly. Replacing solvent, polarization or
internal coordinates by an instantaneous law requires its own closure.
For a coupled relaxing hidden coordinate, a standard quasistatic regime
is $|\omega|\tau_{\rm fast}\ll1$, where $\tau_{\rm fast}$ is its
characteristic relaxation time, with coupling and response errors
also controlled over the chosen band. A large particle mass alone does
not establish this regime. Static, dynamical and equilibrium
reductions retain the different information derived in
\eqref{eq:architecture-completed-square}--\eqref{eq:architecture-mean-force-curvature}.
Their consistent choice precedes assigning the ideal virions their
effective lattice ``springs.''

Figure~\ref{fig:vlp-harmonic-and-relaxation-tests} distinguishes displacement linearization from the separate time-scale test for a relaxing hidden coordinate.

\begin{figure}[H]
\centering\begingroup
\definecolor{vltTeal}{HTML}{176B64}
\definecolor{vltBlue}{HTML}{315D80}
\definecolor{vltGold}{HTML}{A87520}
\definecolor{vltPlum}{HTML}{725877}
\definecolor{vltRed}{HTML}{8E3B46}
\definecolor{vltYellow}{HTML}{D5AD3C}
\definecolor{vltNeutral}{HTML}{777D83}
\setlength{\unitlength}{1pt}
\scalebox{0.80}{%
\input{vlt-figures/figure-039.tex}%
}\endgroup
\caption{(a) The curves plot $B_3\|\vct q_{\rm c}\|/(2k_{\rm c,min})$ for the three labelled positive minimum curvatures. (b) The logarithmic horizontal axis belongs to the separate one-pole model $\tau_{\rm fast}\dot y+y=f$, with the $e^{-i\omega t}$ convention. The marked $10\%$ complex-response error occurs at $|\omega|\tau_{\rm fast}=0.1/\sqrt{0.99}$. The force bounds and scalar closure error answer different approximation questions.}
\label{fig:vlp-harmonic-and-relaxation-tests}
\end{figure}

Retaining the quadratic term gives
\begin{equation}
 E_{\mathrm{harm}}=
 \tfrac12\Dt{\vct u}^{\mathsf T}\mathsf M\Dt{\vct u}
 +\tfrac12\vct u^{\mathsf T}\mathsf K\vct u,
 \qquad \mathsf M\Dtt{\vct u}+\mathsf K\vct u=0.
 \label{eq:harmonic-mechanics}
\end{equation}
The first term stores motion and the second stores reversible deformation
of the selected arrangement. The constant reference energy has been
subtracted. In an explicitly retained medium model, the joint Hessian
replaces this particle-only construction; an equilibrium reduced model
instead uses its consistently matched free-energy curvature. These
choices are stated before interpreting a fitted stiffness.

The complete Hessian couples coordinate directions and particles. Its
eigenvectors organize their collective motion; the diagonal blocks
alone describe clamped local curvature.

\begin{proposition}[Second variation of a central pair energy]
\label{prop:s2-central-pair-variation}
Let $\vct R_{ij}=\vct R_i-\vct R_j=R\vct n$ and
$\vct\delta=\vct u_i-\vct u_j$, with $R>0$ and
$\|\vct\delta\|/R\ll1$. The length expansion is
\begin{equation}
 |\vct R_{ij}+\vct\delta|
 =R+\vct n^{\mathsf T}\vct\delta
   +\frac{\|\vct\delta\|^2-(\vct n^{\mathsf T}\vct\delta)^2}{2R}
   +O(\|\vct\delta\|^3/R^2).
 \label{eq:architecture-separation-expansion}
\end{equation}
For twice continuously differentiable $V$, the pair energy has expansion
\begin{equation}
 \begin{aligned}
 V(|\vct R_{ij}+\vct\delta|)={}&V(R)
       +V'(R)\vct n^{\mathsf T}\vct\delta\\
   &+\frac12V''(R)(\vct n^{\mathsf T}\vct\delta)^2
       +\frac{V'(R)}{2R}\|\mathsf P_\perp\vct\delta\|^2
       +o(\|\vct\delta\|^2),\\
 \mathsf P_\perp={}&\Id_3-\vct n\vct n^{\mathsf T}.
 \end{aligned}
 \label{eq:architecture-pair-expansion}
\end{equation}
\end{proposition}
\begin{proof}
Expand the Euclidean norm at the nonzero vector $R\vct n$; its gradient
is $\vct n$ and its Hessian is $\mathsf P_\perp/R$. Substituting this
expansion into the second-order Taylor formula for $V$ gives the
longitudinal and transverse terms.
\end{proof}
Longitudinal displacement changes length at first order. A transverse
displacement changes it only at second order, by forming a longer diagonal.
The projector $\mathsf P_\perp$ removes the longitudinal component and
retains the two transverse components. The linear contribution of an
individual pair need not vanish. It is the complete first variation,
including all incident interactions and confinement, that vanishes at
a force-balanced reference. This is why the transverse quadratic term
cannot generally be discarded.

\begin{definition}[Directional edge curvature and prestress]
\label{def:s2-edge-curvature}
For a central pair interaction, the \emph{directional edge curvature} is
the Hessian with respect to its relative displacement:
\begin{equation}
 \boxed{\begin{aligned}
 \mathsf C_{ij}={}&
 V''_{ij}(R_{ij})\vct n_{ij}^{0}(\vct n_{ij}^{0})^{\mathsf T}\\
 &+\frac{V'_{ij}(R_{ij})}{R_{ij}}
   \left[\Id_3-\vct n_{ij}^{0}(\vct n_{ij}^{0})^{\mathsf T}\right],
 \end{aligned}}
 \label{eq:edge-stiffness}
\end{equation}
where $R_{ij}=|\vct R_i-\vct R_j|$ and
$\vct n_{ij}^{0}=(\vct R_i-\vct R_j)/R_{ij}$.
Its transverse term $V'_{ij}(R_{ij})/R_{ij}$ is the geometric
\emph{prestress} contribution to that tangent law.
\end{definition} The projector
$\vct n_{ij}^{0}(\vct n_{ij}^{0})^{\mathsf T}$ selects displacement along
the line of centres; its complement selects transverse displacement.
Writing $\Delta\vct u_{ij}=\vct u_i-\vct u_j$ unfolds the edge energy:
\begin{equation}
 \begin{aligned}
 \frac12\Delta\vct u_{ij}^{\mathsf T}\mathsf C_{ij}\Delta\vct u_{ij}
  ={}&\frac12V''_{ij}(R_{ij})
        [ (\vct n_{ij}^{0})^{\mathsf T}\Delta\vct u_{ij}]^2\\
   &+\frac{V'_{ij}(R_{ij})}{2R_{ij}}
       \bigl\|[\Id_3-\vct n_{ij}^{0}(\vct n_{ij}^{0})^{\mathsf T}]
                                  \Delta\vct u_{ij}\bigr\|^2.
 \end{aligned}
 \label{eq:architecture-edge-cost-components}
\end{equation}
Radial curvature gives the longitudinal coefficient. The transverse
coefficient $V'_{ij}/R_{ij}$ arises because transverse motion changes
the separation at second order. At an isolated minimum it vanishes;
a force-balanced assembly can still have nonzero individual pair forces
and must retain their prestress contributions.

For $V'(R)>0$, the existing attractive force gives positive transverse
curvature. For $V'(R)<0$, that contribution is negative even when
$V''(R)>0$. Figure~\ref{fig:architecture-edge-prestress} compares these
signed edge forms at common scales. Stability is assessed after all
edge and support contributions have been assembled.

\begin{figure}[H]
\centering\begingroup
\definecolor{vltTeal}{HTML}{176B64}
\definecolor{vltBlue}{HTML}{315D80}
\definecolor{vltGold}{HTML}{A87520}
\definecolor{vltPlum}{HTML}{725877}
\definecolor{vltRed}{HTML}{8E3B46}
\definecolor{vltYellow}{HTML}{D5AD3C}
\definecolor{vltNeutral}{HTML}{777D83}
\setlength{\unitlength}{1pt}
\scalebox{0.80}{%
\input{vlt-figures/figure-040.tex}%
}\endgroup
\caption{One longitudinal and one transverse relative coordinate are resolved at identical scales. Positive edge-energy contours change from ellipses to parallel lines to hyperbolas as $\rho$ changes. The dashed negative contour and pale negative region occur only in (c). Here $X=\Delta u_\parallel/\ell_\star$, $Y=\Delta u_\perp/\ell_\star$ and $h=E_{\rm edge}/(V^{\prime\prime}(R)\ell_\star^2)$, with $V^{\prime\prime}(R)>0$. These are signed edge contributions; the assembled Hessian determines network stability. The boxed result illustrated here is~\eqref{eq:edge-stiffness}.}
\label{fig:architecture-edge-prestress}
\end{figure}

Figure~\ref{fig:vlp-relative-and-common-motion} relates the directional coefficients to relative whole-particle motion and distinguishes it from a common rigid translation.

\begin{figure}[H]
\centering\begingroup
\definecolor{vltTeal}{HTML}{176B64}
\definecolor{vltBlue}{HTML}{315D80}
\definecolor{vltGold}{HTML}{A87520}
\definecolor{vltPlum}{HTML}{725877}
\definecolor{vltRed}{HTML}{8E3B46}
\definecolor{vltYellow}{HTML}{D5AD3C}
\definecolor{vltNeutral}{HTML}{777D83}
\setlength{\unitlength}{1pt}
\scalebox{0.90}{%
\input{vlt-figures/figure-041.tex}%
}\endgroup
\caption{In (a), $j$ is held while $i$ moves from its dashed reference outline. Longitudinal and transverse component arrows close to the total relative displacement. In (b), common translation preserves the pair separation and connector pitch. The seven-turn coils follow the displayed rigid body outlines. Both directional coefficients are derivatives of the same central energy; the pair-translation identity applies to that energy separately from one-body supports.}
\label{fig:vlp-relative-and-common-motion}
\end{figure}

For pair interactions with one-body wall and target terms,
\begin{equation}
 \mathsf K_{ij}=-\mathsf C_{ij}\quad(i\ne j),\qquad
 \mathsf K_{ii}=\sum_{j:\{i,j\}\in\mathcal E}\mathsf C_{ij}
                    +\mathsf K^{\mathrm{w,T}}_{ii},
 \label{eq:blocks}
\end{equation}
The off-diagonal block transmits a displacement's force effect to its
neighbor, while the diagonal block supplies the site's combined restoring
response. Nonlocal target or medium terms contribute their full Hessian
blocks. The diagonal block is therefore not automatically an intrinsic
capsid or envelope stiffness.

A force--indentation slope describes deformation under a particular probe
and support geometry. Relating it to a centre interaction requires that
geometry and its deformation coordinates. Translational stiffness has
units $\mathrm{N\,m^{-1}}$, a Young modulus has pressure units, and a
membrane bending modulus has energy units~\cite{Ivanovska2004,Landau1986}.

\paragraph{Application: what a local stiffness measures.}
A local loading measurement on a supported protein or polymer scaffold
samples the response of its mechanically connected surroundings.

\begin{proposition}[Relaxed stiffness of a calibrated displacement]
\label{prop:s2-relaxed-probe-stiffness}
Take a real symmetric positive definite constrained stiffness $\mathsf K$ and a
nonzero, real, dimensionless vector $\vct w$. Let
$Y=\vct w^{\mathsf T}\vct u$ be the calibrated displacement and apply
its conjugate force pattern $\vct wF$. Static equilibrium gives
$Y=(\vct w^{\mathsf T}\mathsf K^{-1}\vct w)F$. Equivalently,
\begin{equation}
 \min_{\vct w^{\mathsf T}\vct u=Y}
           \tfrac12\vct u^{\mathsf T}\mathsf K\vct u
       =\tfrac12 k_Y Y^2,
 \qquad
 k_Y=\bigl(\vct w^{\mathsf T}\mathsf K^{-1}\vct w\bigr)^{-1}.
 \label{eq:architecture-network-probe-stiffness}
\end{equation}
The inverse is restricted to the stated supported sector.
\end{proposition}
\begin{proof}
Stationarity of the constrained quadratic energy gives
$\mathsf K\vct u=\eta\vct w$, where $\eta$ is the multiplier.
The constraint fixes $\eta=Y/(\vct w^{\mathsf T}\mathsf K^{-1}\vct w)$.
Substitution gives the displayed minimum. Positive definiteness makes
this stationary point the unique constrained minimizer.
\end{proof}

The resulting $k_Y$ has units $\mathrm{N\,m^{-1}}$.
For a readout of one coordinate $a$,
$k_Y=1/(\mathsf K^{-1})_{aa}$ permits every other free coordinate to
relax. In contrast, $\mathsf K_{aa}$ is the slope with all those other
coordinates held fixed. The two loading questions generally differ even
though both answers are called a stiffness.

Figure~\ref{fig:architecture-probe-compliance-channels} compares these
operations in a supported two-coordinate example. Local, common and relative
force--readout channels interrogate different displacement combinations of
the same matrix; permitting the unmeasured coordinate to relax further
distinguishes a local compliance from a clamped slope.

Changing support, loading or readout weights changes $k_Y$ even when
individual constitutive coefficients remain fixed. Uncalibrated probe
and attachment compliance introduce further coordinates. The matched
energy or free-energy Hessian gives the appropriate static calculation.
Additional force--readout channels then constrain the spatial couplings
behind a local slope~\cite{Landau1986,HornJohnson2013}.

\begin{figure}[H]
\centering\begingroup
\definecolor{vltTeal}{HTML}{176B64}
\definecolor{vltBlue}{HTML}{315D80}
\definecolor{vltGold}{HTML}{A87520}
\definecolor{vltPlum}{HTML}{725877}
\definecolor{vltRed}{HTML}{762B38}
\definecolor{vltYellow}{HTML}{D5AD3C}
\definecolor{vltNeutral}{HTML}{777D83}
\setlength{\unitlength}{1pt}
\scalebox{0.90}{%
\input{vlt-figures/figure-042.tex}%
}\endgroup
\caption{Three equal reference connectors link two scalar centre coordinates to fixed outer ends. Every glyph has the same free span, five turns and transverse radius; no displaced state is assigned. The open arrows define positive coordinates. Color keys identify unit-normalized force--readout channels in the same matrix. The exact compliance bars are $2/3,1,1/3$ in units of $k_\star^{-1}$; the dashed $1/2$ level uses a different clamped boundary condition.}
\label{fig:architecture-probe-compliance-channels}
\end{figure}

\begin{definition}[Incidence assembly of pair stiffness]
\label{def:s2-incidence-assembly}
The \emph{block incidence matrix} $\mathsf B$ maps node displacements to
the differences $\vct u_i-\vct u_j$ along an arbitrary bookkeeping
orientation of each retained edge. The pair stiffness is the
matrix-weighted graph Laplacian
\begin{equation}
 \mathsf K_{\mathrm{pair}}
   =\mathsf B^{\mathsf T}\diag_{\{i,j\}\in\mathcal E}
       (\mathsf C_{ij})\mathsf B.
 \label{eq:graph-hessian}
\end{equation}
\end{definition}
The factorization forms relative edge motion, applies each curvature
and collects node forces. A finite-range graph preserves sparse
structure. Global positivity must still be checked: unstressed axial
central springs on a simple cubic network leave shear mechanisms, and
additional interactions are tested through the resulting
spectrum~\cite{Misra2012}.

Choose a bookkeeping orientation $i\to j$ for each edge. Its row of
$\mathsf B$ has blocks $+\Id_3$ at $i$, $-\Id_3$ at $j$, and zero
elsewhere, so $(\mathsf B\vct u)_{ij}=\vct u_i-\vct u_j$.
Reversing an edge changes the sign of that relative displacement, but
not its quadratic energy or the assembled stiffness. Thus an undirected
interaction graph can be assembled with oriented rows without assigning
a preferred physical direction of transmission. In the pair-only
model, equal translation of every node leaves all differences zero.
The resulting translational kernel can be removed by a justified
constraint or lifted by specified one-body supports; it is not removed
by mass weighting.

The matrix weights are the full $3\times3$ tensors $\mathsf C_{ij}$,
which can be signed under prestress.

\begin{proposition}[Finite acoustic sum rule]
\label{prop:s2-finite-acoustic-sum-rule}
For the unconstrained pair operator in~\eqref{eq:graph-hessian} and
$\mathsf D_{\rm pair}=\mathsf M^{-1/2}\mathsf K_{\rm pair}\mathsf M^{-1/2}$,
the three uniform translations satisfy
\begin{equation}
 \begin{aligned}
 \sum_j(\mathsf K_{\rm pair})_{i\mu,j\nu}&=0,\qquad
 \mathsf K_{\rm pair}\vct t_a=\vct0,\\
 \mathsf D_{\rm pair}\mathsf M^{1/2}\vct t_a&=\vct0,
 \qquad (\vct t_a)_i=\vct e_a,
 \quad a\in\{x,y,z\}.
 \end{aligned}
 \label{eq:vlp-finite-acoustic-sum-rule}
\end{equation}
The first equality sums all site blocks, including the diagonal one.
\end{proposition}
\begin{proof}
Uniform translation gives $\mathsf B\vct t_a=\vct0$, so the incidence
factorization annihilates $\vct t_a$. Taking individual coordinate
entries gives the block-row sum. Mass normalization then gives
$\mathsf D_{\rm pair}\mathsf M^{1/2}\vct t_a=\vct0$.
\end{proof}

This identity holds for arbitrary positive masses, finite geometry and
charge pattern. For unequal masses the null column of $\mathsf D$ has
components $\sqrt{m_i}\vct e_a$; an unweighted constant column is
generally the wrong test. This is the finite pair analogue of the bulk
acoustic sum rule. Smooth support changes the full operator, and fixed
constraints change the admissible space. The three translation
directions of the pair energy alone neither count all mechanisms nor
assign traveling acoustic branches to the finite cylinder.

Figure~\ref{fig:architecture-edge-hessian-assembly} gives an explicit scalar assembly, including its translation kernel, before the general energy balance.

\begin{figure}[H]
\centering\begingroup
\definecolor{vltTeal}{HTML}{176B64}
\definecolor{vltBlue}{HTML}{315D80}
\definecolor{vltGold}{HTML}{A87520}
\definecolor{vltPlum}{HTML}{725877}
\definecolor{vltRed}{HTML}{762B38}
\definecolor{vltYellow}{HTML}{D5AD3C}
\definecolor{vltNeutral}{HTML}{777D83}
\setlength{\unitlength}{1pt}
\scalebox{0.90}{%
\input{vlt-figures/figure-043.tex}%
}\endgroup
\caption{The free scalar chain has the two labelled pair curvatures and one displayed centre coordinate per ideal virion. Both connectors are reference glyphs; their equal pitch does not equate the labelled stiffnesses. Numbers give every signed entry of the assembled matrix; tint reinforces sign and magnitude. Each row sums to zero. The printed $\kappa_\pm$ are the nonzero stiffness eigenvalues. Masses are required to convert them into squared frequencies.}
\label{fig:architecture-edge-hessian-assembly}
\end{figure}

\begin{proposition}[Harmonic energy conservation]
\label{prop:energy}
For constant symmetric $\mathsf K$ and positive $\mathsf M$, solutions of
\eqref{eq:harmonic-mechanics} conserve $E_{\mathrm{harm}}$. If
$\mathsf K\succeq0$, that energy is nonnegative.
\end{proposition}
\begin{proof}
The power is
$\Dt{E_{\mathrm{harm}}}=\Dt{\vct u}^{\mathsf T}
(\mathsf M\Dtt{\vct u}+\mathsf K\vct u)=0$.
Nonnegativity follows from the two quadratic forms.
\end{proof}

Strictly positive curvature bounds displacement at fixed energy. A free
zero coordinate can drift, so a stability statement also specifies its
constraints, sector and preparation.

\begin{proposition}[Local harmonic energy balance]
\label{prop:s2-local-energy-balance}
In a pair-only harmonic network with scalar node masses $m_i>0$ and
constant symmetric edge tensors $\mathsf C_{ij}=\mathsf C_{ji}$, allocate
half of each edge's quadratic energy to each endpoint. Along solutions
of the unforced equations of motion, the resulting local energy and
oriented power obey
\begin{equation}
 \begin{split}
 e_i&=\tfrac12m_i|\Dt{\vct u_i}|^2
       +\tfrac14\sum_j(\vct u_i-\vct u_j)^{\mathsf T}
                 \mathsf C_{ij}(\vct u_i-\vct u_j),\\
 J_{i\to j}&=\tfrac12(\Dt{\vct u_i}+\Dt{\vct u_j})^{\mathsf T}
                    \mathsf C_{ij}(\vct u_i-\vct u_j),\qquad
 \Dt{e_i}=-\sum_jJ_{i\to j}.
 \end{split}
 \label{eq:edge-flux}
\end{equation}
The sums run over retained neighbors and $J_{j\to i}=-J_{i\to j}$.
\end{proposition}
\begin{proof}
Differentiate the kinetic term and substitute
$m_i\ddot{\vct u}_i=-\sum_j\mathsf C_{ij}(\vct u_i-\vct u_j)$.
Differentiating the half-edge storage contributes
$\tfrac12\sum_j(\dot{\vct u}_i-\dot{\vct u}_j)^{\mathsf T}
\mathsf C_{ij}(\vct u_i-\vct u_j)$.
Combining the two terms gives $-\sum_jJ_{i\to j}$. Reversing an
edge reverses its displacement difference and hence its current.
\end{proof}

Each $e_i$ is an assigned portion of quadratic perturbation energy;
$J_{i\to j}$ is power, positive for transfer from $i$ to $j$.
Antisymmetry cancels internal currents in the total energy.
The half-edge allocation is a convention, and a signed prestressed edge
term can make an assigned local contribution signed even when total
energy is nonnegative. One-body potentials add local storage; dynamical
medium and target coordinates add exchange channels. This energy
transport statement concerns mechanical motion, without an implied
biochemical or viral functional outcome.

The current depends on both displacement and velocity. A stretched
edge with endpoints instantaneously at rest stores energy but has zero
current. 
\begin{example}[Instantaneous current in a three-node chain]
\label{ex:s2-local-current}
For a scalar three-node chain with equal $m,k>0$,
choose $\omega_\star=\sqrt{k/m}$, $E_\star=k\ell_\star^2$ and
$\tau=\omega_\star t$. At one instant take
\begin{equation}
 \begin{aligned}
 \frac{\vct u}{\ell_\star}&=(0.4,-0.2,0.1)^{\mathsf T},&
 \frac{\Dt{\vct u}}{\omega_\star\ell_\star}
                         &=(0.2,0.1,-0.3)^{\mathsf T},\\
 \frac{(e_1,e_2,e_3)}{E_\star}&=(0.110,0.1175,0.0675),&
 \frac{(J_{1\to2},J_{2\to3})}{E_\star\omega_\star}&=(0.09,0.03),\\
 \frac{\dd}{\dd\tau}\frac{(e_1,e_2,e_3)}{E_\star}
                         &=(-0.09,0.06,0.03).
 \end{aligned}
 \label{eq:architecture-current-snapshot}
\end{equation}
There are only the two neighbor edges and no one-body terms in this
example. Node 1 loses energy; node 2 receives more than it sends onward;
node 3 gains energy. Their rates sum to zero. The numbers describe an
instantaneous harmonic preparation, not a sustained directed current or
an irreversible loss law. They explain why a network line, a displacement
arrow and an energy-current arrow are different pieces of information.
\end{example}

Figure~\ref{fig:architecture-local-energy-current} displays this snapshot as a local energy balance, using a common scale for the two power arrows.

\begin{figure}[H]
\centering\begingroup
\definecolor{vltTeal}{HTML}{176B64}
\definecolor{vltBlue}{HTML}{315D80}
\definecolor{vltGold}{HTML}{A87520}
\definecolor{vltPlum}{HTML}{725877}
\definecolor{vltRed}{HTML}{762B38}
\definecolor{vltYellow}{HTML}{D5AD3C}
\definecolor{vltNeutral}{HTML}{777D83}
\setlength{\unitlength}{1pt}
\scalebox{0.90}{%
\input{vlt-figures/figure-044.tex}%
}\endgroup
\caption{The body positions use $30$ points per $\ell_\star$; gray ticks retain the reference centres. Both seven-turn springs have fixed body radii and coil radius, so their span changes are $-0.6\ell_\star$ and $+0.3\ell_\star$. Black velocity arrows use a separate common gain. Plum power arrows have the exact $3:1$ length ratio. The node-energy rates use the half-edge allocation in~\eqref{eq:edge-flux}; their signed sum vanishes. This is the free-chain snapshot in~\eqref{eq:architecture-current-snapshot}, with no support springs.}
\label{fig:architecture-local-energy-current}
\end{figure}

\paragraph{Application: distinguishing stored deformation from power.}
A displacement image records deformation. Compliance requires its force
relation, and edge power also requires endpoint velocities and the signed coupling.
The same stretched configuration has zero current at a turning instant
and nonzero current during motion. A displacement correlation therefore
supplies only part of the information needed for~\eqref{eq:edge-flux}.

The externally accessible energy account has a direct mechanical readout.
For a calibrated conjugate pair, force $F(t)$ and displacement $Y(t)$,
the supplied power is $F(t)\Dt{Y}(t)$; the time average also depends on their
relative phase. A purely elastic channel stores and returns work over a
cycle, whereas a dissipative response has a positive net cycle work in a
passive stationary regime. Resolving this balance separates compliance from
loss without assigning a biological function to a resonance. The local
half-edge energies above are a useful consistent partition, rather than
separately unique microscopic observables~\cite{Dhar2008}.

These distinctions guide variable selection: a displacement image constrains
shape, a time-resolved signal adds velocity or phase, and a force calibration
connects that motion to work. If an omitted material coordinate stores and
returns appreciable energy, a particle-only power balance identifies the
need for its exchange channel; it does not by itself identify the molecular
mechanism. The medium realization below shows one controlled way to retain
such storage explicitly.

\subsection{Hilbert spaces, columns and bra--ket notation}
\label{sec:dirac-notation}

Mass weighting turns the unequal-inertia problem into a symmetric
eigenproblem:
\begin{equation}
 \vct x=\mathsf M^{1/2}\vct u,\qquad
 \mathsf D=\mathsf M^{-1/2}\mathsf K\mathsf M^{-1/2},\qquad
 \Dtt{\vct x}+\mathsf D\vct x=0.
 \label{eq:dynamical-matrix}
\end{equation}
The coordinate $\vct x$ has units $\mathrm{kg^{1/2}m}$ and
$\tfrac12\|\Dt{\vct x}\|^2$ is kinetic energy. The entries of
$\mathsf D$ have units $\mathrm{s^{-2}}$. For equal masses,
$\mathsf D=\mathsf K/m$. This transformation accounts for how particles
with equal velocity but different masses store different energy.

Mass weighting also fixes how physical displacement patterns are compared.
\begin{definition}[Mass metric]
\label{def:s2-mass-metric}
For positive particle masses, the mass inner product in displacement
coordinates and its mass-weighted representation are
\begin{equation}
 \langle\vct u,\vct v\rangle_M=\vct u^\dagger\mathsf M\vct v,
 \qquad
 \langle\vct u,\vct v\rangle_M
  =\langle\mathsf M^{1/2}\vct u,\mathsf M^{1/2}\vct v\rangle.
 \label{eq:architecture-mass-isometry}
\end{equation}
\end{definition}
This equality is an isometry: it preserves lengths and angles measured
with the declared mass metric. The squared-frequency operator in physical coordinates,
$\mathsf M^{-1}\mathsf K$, generally is not symmetric in the ordinary
Euclidean metric when masses differ. It is self-adjoint in the mass
metric because $\mathsf K=\mathsf K^\dagger$. The matrix $\mathsf D$
expresses the same problem in coordinates where the ordinary inner
product already includes the inertia. Geometry, inertia and the chosen
representation are thereby kept distinct.

\paragraph{Normalizing coupled coordinates with their physical inertia.}
Charge occupancies with overdamped kinetics do not acquire particle
mass. For the inertial mechanical tangent, let
$\mathsf G=\diag(\mathsf M,\mathsf J_1,\ldots,\mathsf J_n)$,
where $\mathsf J_i$ is the specified laboratory rotational inertia at
the resting reference. Choose a constant diagonal scale map
$\vct\zeta=\mathsf S\vct X$ so $\vct X$ is dimensionless. Then
\begin{equation}
 \mathsf G_X=\mathsf S^{\mathsf T}\mathsf G\mathsf S,\quad
 \mathsf H_X=\mathsf S^{\mathsf T}\mathsf H_{\rm m}\mathsf S,\quad
 \mathsf D_X=\mathsf G_X^{-1/2}\mathsf H_X\mathsf G_X^{-1/2}.
 \label{eq:s2patch-generalized-inertia}
\end{equation}
All entries of $\mathsf G_X$ have units $\mathrm{J\,s^2}$, those of
$\mathsf H_X$ energy units, and those of $\mathsf D_X$ inverse-time
squared. Congruence of inertia and stiffness, with the corresponding dual force
transformation and composed readout map, preserves generalized
frequencies and work. In particular,
$\vct f_X=\mathsf S^{\mathsf T}\vct f_\zeta$ and a scalar readout
$w_\zeta^{\mathsf T}\vct\zeta$ becomes
$w_\zeta^{\mathsf T}\mathsf S\vct X$. Compatible cell maps assemble these same quadratic forms with
shared identities and owned contributions; mixed units are transformed
with the coordinates. The purely translational central reference uses
$\mathsf K=\partial_{\vct u}^2U_\lambda$,
$\mathsf M=\diag(m_i\Id_3)$ and~\eqref{eq:dynamical-matrix}; any
retained material state is clamped or relaxed as specified. In the
attachment model, $\mathcal F_0$ is the relaxed/integrated free energy,
whereas clamping a material state generally adds its restoring curvature.
Additional orientation and surface relaxation uses the corresponding
reduced or augmented operator. The central tensor and edge-current formulas remain
the stated fixed-charge translational specialization.

\begin{figure}[H]
\centering\begingroup
\definecolor{vltTeal}{HTML}{176B64}
\definecolor{vltBlue}{HTML}{315D80}
\definecolor{vltGold}{HTML}{A87520}
\definecolor{vltPlum}{HTML}{725877}
\definecolor{vltRed}{HTML}{762B38}
\definecolor{vltYellow}{HTML}{D5AD3C}
\definecolor{vltNeutral}{HTML}{777D83}
\setlength{\unitlength}{1pt}
\scalebox{0.90}{%
\input{vlt-figures/figure-045.tex}%
}\endgroup
\caption{Mass weighting maps the unit circle to the physical displacement ellipse for $m_2/m_1=4$. Common and relative columns are $(1,2)/\sqrt5$ and $(2,-1)/\sqrt5$ in the mass-weighted plane; its right-angle mark uses that metric. The lower rulers locate reference centres at $0,a$. One displacement gain gives equal common shifts and opposite relative shifts with amplitude ratio $4:1$. Body sizes and spring turns remain fixed; assigned mass is independent of marker size.}
\label{fig:final-mass-weighted-modes}
\end{figure}

\begin{definition}[Mechanical Hilbert space]
\label{def:s2-mechanical-hilbert-space}
The complex coordinate representation of the retained translations is
$\mathcal H_{\mathrm{mech}}=\C^{3n}$ with
$\braket{x}{y}=\sum_{A=1}^{3n}x_A^*y_A$. The Euclidean inner product makes this finite-dimensional space a Hilbert
space; completeness is automatic.
\end{definition}
Complex coefficients conveniently record phase. Physical translational
coordinates are recovered as real coordinates under the specified
representation. A quantum probability interpretation, when used, belongs
to the separate occupation construction in Subsection~\ref{sec:viral-phonon}.

The inner product defines overlap and orthogonality. Self-adjointness
means equality with the adjoint in that metric; in an orthonormal basis
it is Hermitian matrix symmetry. Its real spectrum gives the mode
decomposition. Conservative displacement norm $\|\vct x(t)\|$ can
vary as kinetic and potential energies exchange.
Subsection~\ref{sec:complex-amplitudes} constructs the conserved energy
norm of a first-order representation.

In a chosen coordinate basis, the vector and its dual are
\begin{equation}
 \ket{x}=\begin{pmatrix}x_1\\x_2\\\vdots\\x_{3n}\end{pmatrix},
 \qquad
 \bra{x}=\begin{pmatrix}x_1^*&x_2^*&\cdots&x_{3n}^*\end{pmatrix},
 \qquad \bra{x}=(\ket{x})^\dagger.
 \label{eq:ket-column}
\end{equation}
In this orthonormal basis, the ket is a column and the bra its
conjugate-transpose row. An operator
is a linear map: $\mathsf D\ket{x}$ is matrix multiplication, while
$\bra{x}\mathsf D\ket{x}$ is a scalar quadratic form. For orthonormal
coordinate vectors $\ket{A}$,
\begin{equation}
 \ket{x}=\sum_A x_A\ket{A},\qquad
 x_A=\braket{A}{x},\qquad
 \sum_A\ket{A}\bra{A}=\Id.
 \label{eq:basis}
\end{equation}
The outer product $\ket{A}\bra{A}$ projects onto coordinate $A$.
This notation distinguishes an object from its coordinates and permits
the same displacement to be expressed in particle or mode bases
~\cite{Shankar1994,HornJohnson2013}.

\begin{theorem}[Finite conservative normal modes]
\label{thm:s2-finite-normal-modes}
Let $\mathsf M$ be real symmetric positive definite and $\mathsf K$ real
symmetric nonnegative on the retained coordinates. Then
$\mathsf D=\mathsf M^{-1/2}\mathsf K\mathsf M^{-1/2}$ has a complete
real orthonormal eigenbasis, including all zero modes, with
$\omega_s^2\ge0$, and
\begin{equation}
 \mathsf D\ket{s}=\omega_s^2\ket{s},\qquad
 \mathsf D=\sum_s\omega_s^2\ket{s}\bra{s},\qquad
 \braket{s}{t}=\delta_{st}.
 \label{eq:spectral-decomposition}
\end{equation}
The physical columns $\vct\phi_s=\mathsf M^{-1/2}\ket{s}$ form a
complete mass-orthonormal basis. The full real modal coordinate and
momentum transformation in~\eqref{eq:architecture-canonical-mode-map}
is canonical, including its zero-frequency coordinates.
\end{theorem}
\begin{proof}
The real spectral theorem applies because $\mathsf D$ is symmetric;
$\vct y^{\mathsf T}\mathsf D\vct y\ge0$ gives nonnegative eigenvalues.
Writing $\mathsf\Phi$ for the matrix of physical columns gives
$\mathsf\Phi^{\mathsf T}\mathsf M\mathsf\Phi=\Id$ and
$\mathsf\Phi\mathsf\Phi^{\mathsf T}=\mathsf M^{-1}$. Thus the modal
map is invertible and its particle Poisson brackets give
$\{Q_s,\Pi_t\}=\delta_{st}$, with both other bracket types zero.
\end{proof}

Figure~\ref{fig:architecture-mechanical-basis-projections} illustrates the mechanical eigenbasis and its geometric projections before translating the columns back to physical displacements.

\begin{figure}[H]
\centering\begingroup
\definecolor{vltTeal}{HTML}{176B64}
\definecolor{vltBlue}{HTML}{315D80}
\definecolor{vltGold}{HTML}{A87520}
\definecolor{vltPlum}{HTML}{725877}
\definecolor{vltRed}{HTML}{762B38}
\definecolor{vltYellow}{HTML}{D5AD3C}
\definecolor{vltNeutral}{HTML}{777D83}
\setlength{\unitlength}{1pt}
\scalebox{0.90}{%
\input{vlt-figures/figure-046.tex}%
}\endgroup
\caption{A free equal-mass three-site chain has orthonormal columns $(1,1,1)/\sqrt3$, $(1,0,-1)/\sqrt2$ and $(1,-2,1)/\sqrt6$. All rows use the same reference-position ruler and displacement gain. Rigid shifts and arrows reconstruct each column, while matched coils follow the endpoints. The stationary middle component in row $s=1$ is explicit. Panel (b) gives squared projections of a local displacement direction. These are displacement-norm shares; the zero mode has no quadratic potential cost.}
\label{fig:architecture-mechanical-basis-projections}
\end{figure}

A separate four-particle example in Figure~\ref{fig:vlp-three-dimensional-tetrahedral-mode} follows one mechanical column through three-dimensional particle motion.

\begin{figure}[H]
\centering\begingroup
\definecolor{vltTeal}{HTML}{176B64}
\definecolor{vltBlue}{HTML}{315D80}
\definecolor{vltGold}{HTML}{A87520}
\definecolor{vltPlum}{HTML}{725877}
\definecolor{vltRed}{HTML}{762B38}
\definecolor{vltYellow}{HTML}{D5AD3C}
\definecolor{vltNeutral}{HTML}{777D83}
\setlength{\unitlength}{1pt}
\scalebox{0.90}{%
\input{vlt-figures/figure-047.tex}%
}\endgroup
\caption{A regular tetrahedral reference has four equal masses, six equal reference separations $a$ and pair laws $V_{ij}(r)=k(r-a)^2/2$, $k>0$. The displayed breathing mode follows the printed equation with $\epsilon=0.16$ and $\sum_i\vct R_i=0$. All four rigid ideal virions and all six connectors appear in every frame at one scale. Their body radius, five-turn coils and transverse coil radius stay fixed. Triads give axis directions. The curve locates the three position snapshots; the middle frame has zero displacement, with nonzero velocity. This is a declared mechanical example, without fitted viral parameters.}
\label{fig:vlp-three-dimensional-tetrahedral-mode}
\end{figure}

The physical displacement columns satisfy the generalized eigenproblem
and mass orthogonality:
\begin{equation}
 \mathsf K\vct\phi_s=\omega_s^2\mathsf M\vct\phi_s,\qquad
 \vct\phi_s^\dagger\mathsf M\vct\phi_t=\delta_{st}.
 \label{eq:physical-mode-shapes}
\end{equation}
Orthogonality in physical coordinates uses the mass metric
$\mathsf M$, not an unweighted geometric dot product. The unit-normalized
mass-weighted columns are dimensionless; $\vct\phi_s$ has units
$\mathrm{kg^{-1/2}}$.

Writing $\vct u=\sum_s\vct\phi_sQ_s$ with $Q_s=\braket{s}{x}$ gives
$\Dtt{Q_s}+\omega_s^2Q_s=0$. Each positive mode has a fixed spatial
pattern and its own oscillation frequency; superposition gives the full
motion. A negative eigenvalue would give exponential instability rather
than oscillation. A zero eigenvalue gives no linear restoring force.
One may choose real eigenvectors for this real conservative model.

The mode coordinate $Q_s$ has units $\mathrm{kg^{1/2}m}$. In a real
mode basis its conjugate momentum is $\Pi_s=\Dt{Q_s}$, and the linear
canonical transformation is
\begin{equation}
 Q_s=\vct\phi_s^{\mathsf T}\mathsf M\vct u,
 \qquad \Pi_s=\vct\phi_s^{\mathsf T}\vct p,
 \qquad
 \vct u=\sum_s\vct\phi_sQ_s,\quad
 \vct p=\mathsf M\sum_s\vct\phi_s\Pi_s.
 \label{eq:architecture-canonical-mode-map}
\end{equation}
The mass orthogonality in~\eqref{eq:physical-mode-shapes} gives
$\{Q_s,\Pi_t\}=\delta_{st}$ from the particle Poisson brackets.
This establishes the classical meaning of the modal canonical pair
before the quantization postulate. For a positive mode its spatial
column remains fixed while $Q_s$ oscillates; a sum of such columns
describes a disturbance that can change shape as their phases separate
~\cite{FeynmanLectures1963}.

\begin{example}[Two coupled translations]
\label{ex:two-virions}
Two equal masses coupled along one coordinate by stiffness $k>0$ have
\begin{equation}
 \mathsf D=\frac{k}{m}\begin{pmatrix}1&-1\\-1&1\end{pmatrix},\qquad
 \ket{+}=\frac1{\sqrt2}\begin{pmatrix}1\\1\end{pmatrix},\qquad
 \ket{-}=\frac1{\sqrt2}\begin{pmatrix}1\\-1\end{pmatrix}.
 \label{eq:dimer}
\end{equation}
The common pattern has $\omega_+=0$ and the relative pattern has
$\omega_-^2=2k/m$. A displacement of the first mass alone is
$(1,0)^{\mathsf T}=(\ket{+}+\ket{-})/\sqrt2$: a localized disturbance
contains more than one collective pattern. For unequal masses the
relative coordinate has reduced mass
$\mu_{12}=m_1m_2/(m_1+m_2)$ and
$\omega_-^2=k/\mu_{12}$. Common physical translation remains equal
displacement; its mass-weighted column is proportional to
$(\sqrt{m_1},\sqrt{m_2})$.
\end{example}

Figure~\ref{fig:final-mass-weighted-modes} uses $m_2/m_1=4$ to compare
the mass-weighted and physical columns. Displacement measurements read
the physical columns; inertia enters their mass metric.

For the unequal-mass figure, the physical common and relative columns
have an ordinary dimensionless dot product $3/10$, yet their
mass-metric dot product is zero. The apparent angle in physical
coordinates therefore does not replace the inertia-weighted
orthogonality used in the calculation. In the relative mode,
$m_1u_1+m_2u_2=0$: the centre of mass remains fixed. At an instant of
nonzero modal velocity, the light and heavy particles store respectively
$4/5$ and $1/5$ of that mode's kinetic energy. A smaller displacement
of the heavy particle is a result of the coupled inertia and force law,
not a change in its visibility to a measurement.

A column order is part of the representation. Figure~\ref{fig:added-network-relabeling} follows one scalar network through an explicit permutation of that order.

\begin{figure}[H]
\centering\begingroup
\definecolor{vltTeal}{HTML}{176B64}
\definecolor{vltBlue}{HTML}{315D80}
\definecolor{vltGold}{HTML}{A87520}
\definecolor{vltPlum}{HTML}{725877}
\definecolor{vltRed}{HTML}{762B38}
\definecolor{vltYellow}{HTML}{D5AD3C}
\definecolor{vltNeutral}{HTML}{777D83}
\setlength{\unitlength}{1pt}
\scalebox{0.90}{%
\input{vlt-figures/figure-048.tex}%
}\endgroup
\caption{Physical identities and coordinate order for the same scalar weighted network. (a) Each edge penalizes a displacement difference, with $k_{12}=k_\star$, $k_{23}=2k_\star$ and $k_{13}=3k_\star$; there are no supports. Faceted glyphs identify the three ideal virions. The triangle is a graph layout, not a three-dimensional central-force geometry. (b,c) Listing the coordinates as $(u_3,u_1,u_2)$ permutes rows and columns together. The displayed orthogonal $\mathsf P$ preserves the quadratic energy, the two positive stiffness eigenvalues and the common-displacement kernel. Edge colors identify the same pair laws throughout; changing the listing moves no physical particle.}
\label{fig:added-network-relabeling}
\end{figure}

\subsubsection{Superposition of mechanical solutions}
\label{sec:sup-mechanical}

Superposition makes a mode basis useful: it reconstructs a physical
motion by adding contributions, rather than assigning each particle
to one mode. The same VLP can move with several patterns simultaneously.
Their sum determines its instantaneous displacement and velocity
\cite{FeynmanLectures1963,Misra2012}.

\begin{definition}[Mechanical superposition in one linear reference]
\label{def:sup-mechanical-reference}
Fix a reference, retained coordinates, homogeneous linear constraints
and the same matrices $\mathsf M>0$ and $\mathsf K\geq0$ for
$\mathsf M\ddot{\vct u}+\mathsf K\vct u=\vct f$.
If $\vct u^{(1)},\vct u^{(2)}$ solve this problem with their respective
forces and initial data, then
$a\vct u^{(1)}+b\vct u^{(2)}$, for real constants $a,b$, solves
the problem with the corresponding sums of forces and initial data.
The combined motion must remain in the reference's validity range.
\end{definition}

\begin{samepage}
\begin{proposition}[Reconstruction from a complete mode basis]
\leavevmode\label{prop:sup-complete-reconstruction}
With the full basis of Theorem~\ref{thm:s2-finite-normal-modes},
initial displacement $\vct u_0$ and velocity $\vct v_0$ give
\begin{equation}
 \begin{aligned}
 \vct u(t)=\sum_s\vct\phi_s\bigg[
 &Q_{s0}\cos(\omega_st)+\Pi_{s0}g_s(t)\\
 &+\int_0^t g_s(t-v)\vct\phi_s^{\mathsf T}\vct f(v)\,\dd v\bigg],\\
 Q_{s0}&=\vct\phi_s^{\mathsf T}\mathsf M\vct u_0,\qquad
 \Pi_{s0}=\vct\phi_s^{\mathsf T}\mathsf M\vct v_0.
 \end{aligned}
 \label{eq:sup-complete-reconstruction}
\end{equation}
Here $g_s(t)=\sin(\omega_st)/\omega_s$ for $\omega_s>0$ and
$g_s(t)=t$ for a zero mode. The sum includes the full retained basis.
\end{proposition}
\begin{proof}
Mass orthogonality projects the equation onto
$\ddot Q_s+\omega_s^2Q_s=\vct\phi_s^{\mathsf T}\vct f$.
The scalar initial-value solution gives the displayed bracket;
completeness reconstructs the original displacement. Direct substitution
also proves the linear-combination statement in the definition.
\end{proof}
\end{samepage}

The coefficients are modal coordinates with their physical units.
A local push generally excites several columns through their force
overlaps. Each contributes to motion across the assembly, including
different rows and layers. In an exactly degenerate eigenspace, any
orthonormal combination is an equally valid modal basis: the eigenspace
projector and reconstructed motion are invariant. The preparation and
readout determine which combination is visible.

Displacements add linearly, while their mechanical energy is quadratic.
For real $\vct u=\vct u^{(1)}+\vct u^{(2)}$ and
$\vct v=\vct v^{(1)}+\vct v^{(2)}$, the harmonic energy satisfies
\begin{equation}
 E_{\rm harm}[\vct u,\vct v]=E_1+E_2+
 (\vct v^{(1)})^{\mathsf T}\mathsf M\vct v^{(2)}+
 (\vct u^{(1)})^{\mathsf T}\mathsf K\vct u^{(2)}.
 \label{eq:sup-energy-cross-terms}
\end{equation}
For contributions in disjoint orthogonal normal modes, these global
cross terms vanish. A local displacement, bond deformation or detector
projection can nevertheless contain interference. Figure~\ref{fig:architecture-energy-readout}
already shows an energetic mode cancelling in a chosen readout.
The local cancellation changes what that channel sees; total storage
and source work are calculated from the full state.

\begin{table}[H]
\centering\small
\begin{tabular}{@{}p{3.1cm}p{5.8cm}p{6.4cm}@{}}
\toprule
Representation & Coefficients and normalization & Physical interpretation\\
\midrule
\raggedright Classical motion
& \raggedright Signed or complex modal amplitudes, with the stated displacement or energy units.
& \raggedright Components add to one motion; phase can change a local signal. Sections~\ref{sec:sup-mechanical} and~\ref{sec:complex-amplitudes}.\tabularnewline\addlinespace[4pt]
\raggedright Quantum occupation state
& \raggedright Complex amplitudes $c_{\vct N}$ with $\sum_{\vct N}|c_{\vct N}|^2=1$ for a pure state.
& \raggedright Amplitudes determine probabilities and coherences under Postulate~\ref{post:quantization}; mixtures use a density operator. Example~\ref{ex:sup-quantum-versus-mixture}.\tabularnewline\addlinespace[4pt]
\raggedright Classical structural population
& \raggedright Nonnegative probabilities $p_c$ with $\sum_c p_c=1$.
& \raggedright A distribution of normal, deformed and damaged objects. Combining preparations averages their populations. Section~\ref{sec:s7transport-delivery}.\tabularnewline
\bottomrule
\end{tabular}
\caption{Three uses of coefficient columns and bra--ket notation. The
normalization and physical readout identify the represented object.
A Hilbert-space notation alone supplies no quantum-state interpretation.}
\label{tab:sup-coefficient-meanings}
\end{table}

\subsection{Complex mode amplitudes and conserved energy}
\label{sec:complex-amplitudes}

The same classical motion admits an energy-based first-order
representation.
\begin{definition}[Positive-sector energy amplitude]
\label{def:s2-energy-amplitude}
For a fixed nonnegative conservative reference, let $\mathsf P_+$ project onto positive-frequency modes
and $\mathsf P_0$ onto the kernel. Set
$\ket{x_+}=\mathsf P_+\ket{x}$ and
$\mathsf W=\mathsf D^{1/2}$ on the positive sector. Define
\begin{equation}
 \ket{\psi}=\frac1{\sqrt2}
                  \bigl(\mathsf W\ket{x_+}+i\ket{\Dt{x_+}}\bigr),\qquad
 \braket{\psi}{\psi}=E_+.
 \label{eq:energy-amplitude}
\end{equation}
Here $E_+$ denotes the harmonic energy of the positive-frequency sector.
\end{definition}
The real projected coordinates and velocities form two quadratures,
combined as the real and imaginary parts of $\ket{\psi}$. Their squared
norm is the positive-sector harmonic energy $E_+$. The complete harmonic
energy also includes zero-sector kinetic energy:
$E_{\mathrm{harm}}=E_++\tfrac12\|\mathsf P_0\ket{\Dt{x}}\|^2$.
The components of $\ket{\psi}$ have units $\mathrm{J^{1/2}}$.

\begin{proposition}[Conservative energy-amplitude evolution]
\label{prop:s2-energy-amplitude-evolution}
For fixed $\mathsf M$, $\mathsf K$ and $\mathsf P_+$, autonomous
conservative motion satisfies
\begin{equation}
 i\frac{\dd}{\dd t}\ket{\psi}=\mathsf W\ket{\psi},\qquad
 \ket{\psi(t)}=e^{-it\mathsf W}\ket{\psi(0)}.
 \label{eq:first-order}
\end{equation}
The map between real positive-sector displacement--velocity pairs and
$\ket\psi$ is invertible, and this evolution preserves $E_+$.
\end{proposition}
\begin{proof}
Differentiate~\eqref{eq:energy-amplitude} and substitute
$\ket{\Dtt{x_+}}=-\mathsf W^2\ket{x_+}$ to obtain the first equation.
Its exponential solution is unitary because $\mathsf W$ is self-adjoint.
Its squared norm is the sum of the two quadratic energy terms, and
\eqref{eq:energy-ket-reconstruction} recovers both real quadratures.
\end{proof}

This unitary rewriting assumes fixed $\mathsf M$, $\mathsf K$ and
positive-sector projection throughout the autonomous evolution. A
time-dependent control or changing basis produces additional derivative
terms and its own work account. For each fixed positive mode,
\begin{equation}
 \begin{aligned}
 \psi_s&=\braket{s}{\psi}
       =\frac{\omega_sQ_s+i\Pi_s}{\sqrt2}
       =\sqrt{E_s}\,e^{-i(\omega_st+\theta_s)},\\
 Q_s(t)&=\frac{\sqrt{2E_s}}{\omega_s}
                              \cos(\omega_st+\theta_s),\qquad
 \Pi_s(t)=-\sqrt{2E_s}\sin(\omega_st+\theta_s).
 \end{aligned}
 \label{eq:architecture-modal-energy-phase}
\end{equation}
Its radius is $\sqrt{E_s}$. Increasing $\omega_st+\theta_s$ rotates
the complex argument clockwise and reconstructs an oscillating physical
pattern $\vct\phi_s$. At fixed energy, the displacement scale grows
as $1/\omega_s$. Multiplying the evolution by $\hbar$ gives a
Schr\"odinger form while preserving its classical energy norm.

Figure~\ref{fig:architecture-energy-phase-quadratures} displays the classical energy amplitude and its two reconstructed quadratures at the same phase.

\begin{figure}[H]
\centering\begingroup
\definecolor{vltTeal}{HTML}{176B64}
\definecolor{vltBlue}{HTML}{315D80}
\definecolor{vltGold}{HTML}{A87520}
\definecolor{vltPlum}{HTML}{725877}
\definecolor{vltRed}{HTML}{762B38}
\definecolor{vltYellow}{HTML}{D5AD3C}
\definecolor{vltNeutral}{HTML}{777D83}
\setlength{\unitlength}{1pt}
\scalebox{0.80}{%
\input{vlt-figures/figure-049.tex}%
}\endgroup
\caption{Clockwise phase evolution of the classical modal amplitude in \eqref{eq:architecture-modal-energy-phase}. A fixed positive-frequency mode with $E_s>0$ gives the normalized coordinate $A=\omega_sQ_s/\sqrt{2E_s}$ and momentum $B=\Pi_s/\sqrt{2E_s}$. (a) The unit-circle geometry uses equal axis scales; its projections are the two quadratures. (b) The same preparation traces $A=\cos\varphi_s$ and $B=-\sin\varphi_s$. The point at $\varphi_s=\pi/3$ is registered in both panels. Their common energy scale is retained in $\sqrt{E_s}$ before normalization. These are two components of one classical mode amplitude, rather than a probability density or two independently prepared modes.}
\label{fig:architecture-energy-phase-quadratures}
\end{figure}

In a fixed real coordinate basis the trajectory is recovered by
\begin{equation}
 \ket{x_+}=\sqrt2\,\mathsf W^{-1}\operatorname{Re}\ket{\psi},
 \qquad
 \ket{\Dt{x_+}}=\sqrt2\,\operatorname{Im}\ket{\psi}.
 \label{eq:energy-ket-reconstruction}
\end{equation}
The inverse exists only on the positive sector. Zero coordinates instead
obey
$\mathsf P_0\ket{x(t)}=\mathsf P_0\ket{x(0)}
+t\mathsf P_0\ket{\Dt{x}(0)}$
while the local description remains applicable. For $E_+>0$,
$w_s=|\braket{s}{\psi}|^2/E_+$ gives the fraction of positive-sector
energy in mode $s$. These weights sum to one but are not quantum
probabilities. Inside a degenerate eigenspace individual weights depend
on the chosen basis; their sum is the invariant weight of that frequency
subspace.

\begin{proposition}[Harmonic excursion and sufficient safe preparation]
\label{prop:s2-safe-harmonic-preparation}
Consider autonomous harmonic motion with a nonempty positive sector,
zero initial displacement and velocity in the zero sector, and fixed
particle masses. Set $m_{\min}=\min_i m_i$ and
$\omega_{\min,+}=\min_{s:\omega_s>0}\omega_s$. Then
\begin{equation}
 \|\vct u_+(t)\|^2
 \le\frac{\|\vct x_+(t)\|^2}{m_{\min}}
 \le\frac{2E_+}{m_{\min}\omega_{\min,+}^2},\qquad
 \vct u_+=\mathsf M^{-1/2}\vct x_+.
 \label{eq:architecture-energy-excursion}
\end{equation}
Choose $\ell_{\rm safe}$ smaller
than every reference wall clearance and half every pair clearance, with
the ball $\|\vct u\|<\ell_{\rm safe}$ inside the chosen harmonic validity
region. The sufficient condition is
\begin{equation}
 \boxed{E_+<\frac12m_{\min}\omega_{\min,+}^2\ell_{\rm safe}^2
       \quad\Longrightarrow\quad\|\vct u(t)\|<\ell_{\rm safe}.}
 \label{eq:architecture-safe-preparation}
\end{equation}
\end{proposition}
\begin{proof}
Since $\mathsf M\succeq m_{\min}\Id$,
$\|\vct u_+\|^2\le\|\vct x_+\|^2/m_{\min}$.
On the positive sector,
$E_+\ge\tfrac12\omega_{\min,+}^2\|\vct x_+\|^2$ at every time.
Energy conservation and the stated strict energy bound therefore keep
the trajectory within the chosen ball.
\end{proof}
A softer positive mode permits a larger excursion at the same energy,
so it tightens this condition. This is a statement about trajectories of
the harmonic reference, not a bound on all nonlinear preparations or on
the spatial support of a quantum state. Zero-sector drift, forcing and
additional coordinates need their own bounds.
Figure~\ref{fig:architecture-energy-safe-bound} shows how a prepared
energy region fits inside a declared safe displacement ball.

\begin{figure}[H]
\centering\begingroup
\definecolor{vltTeal}{HTML}{176B64}
\definecolor{vltBlue}{HTML}{315D80}
\definecolor{vltGold}{HTML}{A87520}
\definecolor{vltPlum}{HTML}{725877}
\definecolor{vltRed}{HTML}{762B38}
\definecolor{vltYellow}{HTML}{D5AD3C}
\definecolor{vltNeutral}{HTML}{777D83}
\setlength{\unitlength}{1pt}
\scalebox{0.80}{%
\input{vlt-figures/figure-050.tex}%
}\endgroup
\caption{The sufficient energy bound \eqref{eq:architecture-safe-preparation} in an auxiliary two-coordinate system. Both masses are $m_\star=m_{\min}$, the positive frequencies are $\omega_1=\omega_{\min,+}$ and $\omega_2=2\omega_1$, and no zero sector, force or medium exchange is present. (a) At $e=0.64$, nonnegative kinetic energy restricts positions to the shaded ellipse $X^2+4Y^2\leq0.64$, with $X=u_1/\ell_{\rm safe}$ and $Y=u_2/\ell_{\rm safe}$. This allowed-position region lies inside the unit safe circle; it is not a trajectory. (b) The conserved-energy estimate is $\|\vct u\|/\ell_{\rm safe}\leq\sqrt e$. The open point marks the strict boundary of the sufficient guarantee. Above it this estimate alone leaves escape undecided. The safe length is supplied by the physical clearances and harmonic validity region; \eqref{eq:architecture-energy-excursion} gives the general mass-weighted result.}
\label{fig:architecture-energy-safe-bound}
\end{figure}

The finite positive-mode density of states is
$g(\omega)=\sum_{s:\omega_s>0}\delta(\omega-\omega_s)$; mode shapes,
energy weights and edge currents describe complementary features of a
prepared motion.

A prepared mode's energy and its visibility in a displacement channel
are separate quantities.
\begin{definition}[Modal coupling to a conjugate displacement channel]
\label{def:s2-modal-channel-coupling}
On the positive sector, with zero-frequency coordinates constrained to
remain zero throughout the forced response, a real dimensionless weight $\vct w$ defines the readout, modal
couplings and conservative retarded response by
\begin{equation}
 Y=\vct w^{\mathsf T}\vct u=\sum_s g_sQ_s,
 \qquad g_s=\vct w^{\mathsf T}\vct\phi_s,\qquad
 \mathcal R_Y^{(0)}(\omega)
       =\sum_{s:\omega_s>0}
          \frac{|g_s|^2}{\omega_s^2-(\omega+i0^+)^2}.
 \label{eq:architecture-modal-readout-residues}
\end{equation}
The conjugate force pattern is $\vct wF$, so its work is $F\,\dd Y$.
\end{definition}
The final expression is the conservative retarded response on the
selected positive sector to the conjugate force pattern $\vct wF$.
The residue $|g_s|^2$ has units $\mathrm{kg^{-1}}$ and states how strongly
that force and readout couple to the mode. It is a property of the
mechanics and readout, whereas $E_s/E_+$ describes a particular
preparation. A large energy weight can coexist with $g_s=0$.

For a concrete finite example, take three scalar translations with equal
mass $m_\star$, nearest springs $k_\star>0$, and fixed end coordinates
$u_0=u_4=0$. Their stiffness and normalized columns are
\begin{equation}
 \frac{\mathsf K_{\rm ch}}{k_\star}=
  \begin{pmatrix}2&-1&0\\-1&2&-1\\0&-1&2\end{pmatrix},
 \quad e_{s,j}=\frac{\sin(sj\pi/4)}{\sqrt2},
 \quad \omega_s^2=\frac{k_\star}{m_\star}
                        [2-2\cos(s\pi/4)].
 \label{eq:architecture-readout-chain}
\end{equation}
Here $s,j=1,2,3$ and $\vct\phi_s=\vct e_s/\sqrt{m_\star}$.
For $Y=(u_1+u_3)/\sqrt2$, the three dimensionless residues are
$(1/2,0,1/2)$ after multiplication by $m_\star$.
The antisymmetric second mode has opposite end motions that cancel in
this channel. The relative readout $(u_1-u_3)/\sqrt2$ instead has
residues $(0,1,0)$. Figure~\ref{fig:architecture-energy-readout}
compares one prepared energy allocation with this readout selection.
The supports and one-dimensional coordinates are declared parts of this
illustrative model, not consequences of the cylindrical boundary.

For particle tracking, the same principle applies to a calibrated weighted
combination of centre displacements. A weak or absent spectral peak may
reflect the readout projection rather than absence of a mechanical mode.
Complementary centre and relative channels can reveal different patterns.
Interpreting that difference requires the coordinate resolution, weights
and observation band, rather than assigning an unseen mode a biological
function.

\begin{figure}[H]
\centering\begingroup
\definecolor{vltTeal}{HTML}{176B64}
\definecolor{vltBlue}{HTML}{315D80}
\definecolor{vltGold}{HTML}{A87520}
\definecolor{vltPlum}{HTML}{725877}
\definecolor{vltRed}{HTML}{762B38}
\definecolor{vltYellow}{HTML}{D5AD3C}
\definecolor{vltNeutral}{HTML}{777D83}
\setlength{\unitlength}{1pt}
\scalebox{0.90}{%
\input{vlt-figures/figure-051.tex}%
}\endgroup
\caption{The supported three-site chain has $\mathsf K/k_\star=\operatorname{tridiag}(-1,2,-1)$ and equal masses. The common ruler places reference centres at $0,a,2a$; the $s=2$ snapshot has shifts $(A,0,-A)$ at $14\,\mathrm{pt}$ per $A$. Dashed outlines and matched surface-attached coils retain the reference comparison. The separate multimode energy shares in (b) are $(0.1,0.7,0.2)$. For $Y=(u_1+u_3)/\sqrt2$, (c) has residues $(1/2,0,1/2)/m_\star$. The most energetic mode in this preparation cancels in this readout; the relative-end channel isolates it.}
\label{fig:architecture-energy-readout}
\end{figure}

\subsection{The viral phonon as a quantized excitation}
\label{sec:viral-phonon}

The finite mechanical spectrum describes the normal frequencies of the
classical particle assembly. Elementary energy quanta enter through an
additional physical postulate. This postulate assigns quantum canonical
operators and an occupation ladder to each positive-frequency mode.

\begin{postulate}[Quantized harmonic occupations]
\label{post:quantization}
The positive-frequency modes admit self-adjoint canonical operators
$\widehat Q_s,\widehat\Pi_s$ with
$[\widehat Q_s,\widehat\Pi_t]=i\hbar\delta_{st}$ and vanishing
coordinate--coordinate and momentum--momentum commutators, on a common
invariant domain. Their quadratic mechanical energy defines the
quantized harmonic Hamiltonian.
\end{postulate}

In a real normal-mode basis, the classical momentum is
$\Pi_s=\Dt{Q_s}$ and the positive-sector energy is
$\tfrac12\sum_s(\Pi_s^2+\omega_s^2Q_s^2)$. The standard oscillator
construction yields
\begin{equation}
 \widehat b_s=\sqrt{\frac{\omega_s}{2\hbar}}\widehat Q_s
             +\frac{i\widehat\Pi_s}{\sqrt{2\hbar\omega_s}},\qquad
 [\widehat b_s,\widehat b_t^\dagger]=\delta_{st},\qquad
 \widehat H_{\mathrm{harm}}
   =\sum_s\hbar\omega_s
         \left(\widehat b_s^\dagger\widehat b_s+\tfrac12\right).
 \label{eq:quantization}
\end{equation}
The dimensionless creation and annihilation operators raise and lower
occupation.

The ladder operators combine the two canonical quadratures into
dimensionless quantities, with the scale set by $\hbar$ and $\omega_s$.
Inverting the definitions gives
\begin{equation}
 \widehat Q_s=\sqrt{\frac{\hbar}{2\omega_s}}
                    (\widehat b_s+\widehat b_s^\dagger),\qquad
 \widehat\Pi_s=-i\sqrt{\frac{\hbar\omega_s}{2}}
                    (\widehat b_s-\widehat b_s^\dagger).
 \label{eq:architecture-ladder-quadratures}
\end{equation}
The first scale has units $\mathrm{kg^{1/2}m}$ and the second
$\mathrm{kg^{1/2}m\,s^{-1}}$, matching the classical modal coordinate
and momentum. Thus the operator construction quantizes a specified
collective coordinate, rather than assigning a quantum state to an
undefined whole-particle motion. This controlled harmonic construction does not assert an
exact quantization rule for every nonlinear mechanical observable
~\cite{Shankar1994,NeumaierWestra2011}.

\begin{definition}[Viral phonon]
\label{def:viral-phonon}
A viral phonon of mode $s$ is one elementary excitation of that mode in
the postulated quantized harmonic description. Raising occupation from
$N_s$ to $N_s+1$ increases its energy by $\hbar\omega_s$.
\end{definition}

Thus $E_s(N_s)=\hbar\omega_s(N_s+\tfrac12)$ and
$\widehat b_s^\dagger\ket{N_s}=\sqrt{N_s+1}\ket{N_s+1}$.
The integer $N_s$ counts excitation quanta. The retained particle count
is $n$, drawn from the fixed pool $N_{\rm tot}$.
\begin{definition}[Harmonic occupation space]
\label{def:s2-harmonic-occupation-space}
For $r\ge1$ positive modes under Postulate~\ref{post:quantization}, the
bosonic occupation space is
\begin{equation}
 \mathcal F_{\mathrm B}(\mathcal H_+)
       \cong\bigotimes_{s=1}^r\ell^2(\mathbb N_0),
 \qquad \mathcal H_+=\operatorname{ran}\mathsf P_+ .
 \label{eq:occupation-space}
\end{equation}
Each factor carries the square-summable occupation amplitudes of one mode.
\end{definition}
Each $\ell^2(\mathbb N_0)$ factor consists of square-summable amplitudes
over $N_s=0,1,\ldots$. This Fock space is infinite-dimensional despite a
finite number of mechanical modes. It differs from the coordinate space
$\C^{3n}$: a displacement vector, a classical energy-amplitude vector
and a quantum occupation state are distinct objects.

The oscillator polynomials and canonical commutators act on the common
invariant domain of finite linear combinations of occupation kets;
self-adjoint realizations specify the physical observables and evolution.
Section~\ref{sec:operator-framework} develops these domain issues.
Bosonic symmetrization concerns mode excitations and introduces no
exchange-statistics assumption about whole virions. For two modes,
$\ket{2,0}$ and $\ket{1,1}$ contain two excitations of the same fixed
particle population, with energies above vacuum
$2\hbar\omega_1$ and $\hbar(\omega_1+\omega_2)$, respectively. This
oscillator vacuum is an occupation state, separate from the physical
contents of the ideal reference cavity.

Occupation space organizes how energy is distributed among modes.
Figure~\ref{fig:architecture-occupation-ladder} places two-mode labels
on a grid and compares them with a single-mode energy ladder. That grid
is a representation of occupation numbers: moving to an adjacent point
adds one excitation in one mode, not a particle or a spatial displacement.
The zero-point energy is retained in the oscillator Hamiltonian, while
excitation energies are measured relative to its occupation vacuum.

\begin{figure}[H]
\centering\begingroup
\definecolor{vltTeal}{HTML}{176B64}
\definecolor{vltBlue}{HTML}{315D80}
\definecolor{vltGold}{HTML}{A87520}
\definecolor{vltPlum}{HTML}{725877}
\definecolor{vltRed}{HTML}{762B38}
\definecolor{vltYellow}{HTML}{D5AD3C}
\definecolor{vltNeutral}{HTML}{777D83}
\setlength{\unitlength}{1pt}
\scalebox{0.80}{%
\input{vlt-figures/figure-052.tex}%
}\endgroup
\caption{Occupation coordinates and their energy labels under the quantized harmonic postulate. (a) With $\omega_1=\omega_\star$ and $\omega_2=3\omega_\star/2$, each integer point labels $\ket{N_1,N_2}$; its adjacent number is excitation energy in units $\hbar\omega_\star$. The circled states each have two excitations, with energies $2$ and $5/2$ in those units. (b) A single oscillator instead shows full energies $N+1/2$, including its zero-point offset. The arrow spans one adjacent-level spacing. Both displays are finite crops of unbounded occupation ladders, with the mechanical particle population fixed. Integer occupation and discrete classical normal-mode frequency are different constructions.}
\label{fig:architecture-occupation-ladder}
\end{figure}

The unlimited occupation ladder belongs to the quadratic model. Its
physical use still requires small excursions.
\begin{proposition}[Excursion in an occupation-diagonal preparation]
\label{prop:s2-occupation-excursion}
Under Postulate~\ref{post:quantization}, a mixture diagonal in the joint
number basis, with finite first occupation moments, has
\begin{equation}
 \left\langle|\widehat{\vct u}_{i,+}|^2\right\rangle
    =\sum_{s=1}^r
       \frac{\hbar\bigl(2\langle\widehat N_s\rangle+1\bigr)}{2\omega_s}
       \|\vct\phi_{s,i}\|^2,\qquad
 \widehat N_s=\widehat b_s^\dagger\widehat b_s .
 \label{eq:occupation-displacement-scale}
\end{equation}
\end{proposition}
\begin{proof}
Expand the displacement as $\sum_s\vct\phi_{s,i}\widehat Q_s$.
Diagonal number-state averages make the intermode products and the
single-mode $\widehat b_s^2$ and
$(\widehat b_s^\dagger)^2$ terms vanish. The remaining identity
$\langle\widehat Q_s^2\rangle=
\hbar(2\langle\widehat N_s\rangle+1)/(2\omega_s)$ gives the result.
\end{proof}
Here $\vct\phi_{s,i}$ is particle $i$'s physical mode block. The factor
$1/\omega_s$ explains why a soft mode can contribute strongly to
mean-square displacement. General preparations also contain mean motion
and cross-mode covariances. Their mean-square excursion must be compared
with the reference clearances and force-variation scales. This rms test
does not guarantee strict confinement of every measurement outcome.

The same ladder calculation gives, for a single mode in a number state
$\ket{N}$,
\begin{equation}
 \langle\widehat Q_s\rangle_N=\langle\widehat\Pi_s\rangle_N=0,
 \qquad
 \langle\widehat Q_s^2\rangle_N
       =\frac{\hbar(N+1/2)}{\omega_s},\qquad
 \langle\widehat\Pi_s^2\rangle_N
       =\hbar\omega_s(N+1/2).
 \label{eq:architecture-number-quadrature-variance}
\end{equation}
Zero mean displacement therefore does not mean absence of motion-related
variance or excitation energy. Increasing $N$ raises both variances,
while lowering $\omega_s$ raises the displacement variance at fixed
occupation. The physical amplitude at a particle additionally contains
its mode block $\vct\phi_{s,i}$. These three elements---occupation,
frequency and spatial mode shape---must be kept together when connecting
the quantum postulate to a displacement observable. Thermal occupation
and its classical limit are developed in
Subsection~\ref{sec:phonon-temperature-comparison}, rather than inferred
from normalization of the classical energy-amplitude vector.

The quantized Hamiltonian refers to the conservative reference with its
canonical inertia. A damped pole subsequently tests whether that
reference supports a resolved oscillation in the actual medium; a
complex pole is not substituted as a level spacing in a self-adjoint
Hamiltonian. Section~\ref{sec:qfa-lifetimes} develops the conditional
single-mode bath dynamics, including the survival of single-phonon
statistics. Section~\ref{sec:qfa-feasibility} keeps the selected HPV
preparation's measured inputs distinct from the synthetic guide's
assigned scales and specifies the remaining preparation and detector
requirements.

Occupation measurements have integer outcomes, while expectations in
mixtures or superpositions can be noninteger.

\begin{definition}[One-phonon packet]
\label{def:s2-one-phonon-packet}
A normalized state in the single-excitation sector is
\begin{equation}
 \ket{\Phi}=\sum_s c_s\widehat b_s^\dagger\ket{0},\qquad
 \sum_s|c_s|^2=1,\qquad
 \langle E_{\mathrm{exc}}\rangle_\Phi
       =\sum_s\hbar\omega_s|c_s|^2,
 \label{eq:phonon-packet}
\end{equation}
Its energy expectation is measured above the occupation vacuum.
\end{definition}
Here $|c_s|^2$ is the probability of measuring the excitation in mode
$s$. Different mode frequencies give an excitation-energy distribution.
Removing the common vacuum phase yields
\begin{equation}
 \ket{\Phi(t)}
       =\sum_s c_s e^{-i\omega_s t}\ket{1_s},
 \qquad \ket{1_s}=\widehat b_s^\dagger\ket{0}.
 \label{eq:one-phonon-evolution}
\end{equation}
The mechanical eigenket $\ket{s}$ denotes a spatial pattern, whereas
$\ket{1_s}$ denotes an occupation state. A pure one-phonon packet has
zero mean displacement because the displacement operator connects
sectors differing by one excitation. Its spatial information appears in
transition amplitudes and correlations, as in
\eqref{eq:phonon-transition-amplitude}. An oscillating mean displacement
uses another preparation, such as
\eqref{eq:coherent-mean-energy-amplitude}.

Mode occupations are conserved by this harmonic Hamiltonian. Nonlinear
couplings can redistribute excitation and need a separately specified
energy-conserving model. The evidence for a mechanical resonance and the
evidence for elementary energy exchange therefore test different claims.
Classical mechanics and thermal fluctuations remain a complete baseline
without the occupation postulate.

\subsubsection{Energy exchange and the physical attributes of a viral phonon}
\label{sec:s2-phonon-attributes}

A source, a collective mechanical field and a receiver provide the
coupling architecture. The field is built from the retained
particle--medium coordinates; its couplings follow the assembled
operator~\cite{Misra2012,TongSolidState2017}. An isolated cluster mode
is supported on that cluster. Transfer between clusters requires the
declared intercluster interaction or a retained host field.

\begin{definition}[Fixed-charge phonon sector]
\label{def:s2-fixed-charge-phonon-sector}
For a fixed labelled particle population with fixed assigned charges,
the total virion charge acts on its mechanical occupation space as
\begin{equation}
 \widehat{\mathcal Q}_{\rm virion}
   =\left(\sum_iq_i\right)\Id,\qquad
 [\widehat{\mathcal Q}_{\rm virion},\widehat b_s^\dagger]=0.
 \label{eq:s2-phonon-charge-sector}
\end{equation}
Creating one phonon changes the mode energy by $\hbar\omega_s$ and
leaves this total charge and the material-particle population unchanged.
\end{definition}

An electrically active displacement can nevertheless change a dipole.
Let the declared electrical coupling have the linearization
$\delta\vct d=\sum_i\mathsf Z_i^{\rm em}\vct u_i$, where
$\mathsf Z_i^{\rm em}$ has charge units. Projection onto the real
positive-frequency modes gives
\begin{equation}
 \delta\widehat{\vct d}
     =\sum_s\vct d_s^{\rm em}
            (\widehat b_s+\widehat b_s^\dagger),\qquad
 \vct d_s^{\rm em}
     =\sqrt{\frac{\hbar}{2\omega_s}}
                \sum_i\mathsf Z_i^{\rm em}\vct\phi_{s,i}.
 \label{eq:s2-phonon-transition-dipole}
\end{equation}
The vector $\vct d_s^{\rm em}$ has units of electric dipole moment.
A uniform field couples through
$-\vct E\cdot\delta\widehat{\vct d}$ and therefore selects modes
by this overlap. In a specified point-charge electrical model,
$\mathsf Z_i^{\rm em}=q_i^{\rm em}\Id_3$. An effective charge fitted
to a screened pair potential need not equal this electrical coupling
charge without a model connecting the two measurements. For a charged
object the dipole also requires a declared origin and translational
frame. Motion-induced polarization and its associated current can thus
be present even though the phonon changes no total electric charge.
Coupling this polarization to a dynamical electromagnetic field
enlarges the normal-mode problem; mixed phonon--photon modes then
require both sectors and their coupling~\cite{HellmanKronik2026}.

\begin{proposition}[Internal modes and true particle momentum]
\label{prop:s2-positive-modes-zero-net-momentum}
For a free, stationary pair reference with positive masses and
$\mathsf K\vct t_a=0$ in each uniform translation direction, every
positive-frequency physical mode satisfies
\begin{equation}
 \sum_i m_i\vct\phi_{s,i}=\vct0,\qquad
 \vct P_{\rm virion}
       =\sum_i m_i\dot{\vct u}_i=\vct0
 \quad\hbox{for a pure internal mode.}
 \label{eq:s2-phonon-internal-momentum}
\end{equation}
\end{proposition}
\begin{proof}
Multiply $\mathsf K\vct\phi_s=\omega_s^2\mathsf M\vct\phi_s$
by $\vct t_a^{\mathsf T}$. Symmetry and
$\mathsf K\vct t_a=0$ make the left side zero. Since
$\omega_s>0$, the mass-weighted sum of each Cartesian component
vanishes. Multiplication by the modal velocity gives the momentum
identity.
\end{proof}

The wavevector of a periodic displacement pattern labels another
symmetry. A bulk phonon can carry crystal pseudomomentum
$\hbar\vct k$, defined modulo a reciprocal-lattice vector, while the
internal pattern above has no net first-order particle momentum.
Mechanical momentum transfer is evaluated from particle forces and the
recoil of the medium, boundary and source. In a finite reflecting cavity,
a standing mode generally has no unique traveling wavevector. These
distinctions are essential when transferring momentum arguments from
electrodynamics to waves in matter~\cite{Stone2000,TongSolidState2017}.

The particle masses in $\mathsf M$ remain material inputs. The linear
theory assigns no universal Newtonian rest mass to a phonon, and
$\hbar\omega_s/c_s^2$ is not a general rule for modifying that matrix.
A smooth bulk branch can instead have a \emph{dispersion mass tensor},
defined locally through its energy curvature,
\begin{equation}
 (\mathsf m_{\rm disp}^{-1})_{ab}
    :=\frac{1}{\hbar^2}
       \partial_{k_a}\partial_{k_b}(\hbar\omega_j)
     =\frac{1}{\hbar}\partial_{k_a}\partial_{k_b}\omega_j.
 \label{eq:s2-phonon-dispersion-mass}
\end{equation}
It describes the quadratic packet equation, can be anisotropic or
singular, and is not an added virion mass. Nonlinear acoustic waves can
also transport small, medium-dependent amounts of material; those
second-order mass-transport questions require the medium dynamics
beyond this linear mode account~\cite{EspositoKrichevskyNicolis2019,Soper2020}.

\subsubsection{A source, an absorber and a measurable exchange channel}
\label{sec:s2-phonon-source-absorber}

A statement that a particle emits an excitation becomes predictive when
the particle's energy source and its mechanical coupling are specified.
For example, retain two material states $\ket e,\ket g$ of a selected
source, separated by $\hbar\Omega_A>0$, and a Hermitian material-force
operator $\widehat{\vct F}_A$ evaluated at the reference.
The interaction energy and its single-mode coupling are
\begin{equation}
 \begin{aligned}
 \widehat H_{A{\rm m}}
    &=-\widehat{\vct F}_A^{\mathsf T}\widehat{\vct u},\\
 g_{As}
    &:=-\frac{1}{\hbar}\sqrt{\frac{\hbar}{2\omega_s}}
       \vct\phi_s^{\mathsf T}
                   \langle g|\widehat{\vct F}_A|e\rangle .
 \end{aligned}
 \label{eq:s2-phonon-transition-force-coupling}
\end{equation}
Here $g_{As}$ has units $\mathrm{s^{-1}}$. The source can represent
a declared internal, contact or chemical transition; its level spacing
and transition force are new physical inputs rather than consequences
of an icosahedral shape. Near resonance and at weak coupling, retaining
the slowly varying exchange terms gives
\begin{equation}
 \widehat H_{A{\rm m}}^{\rm exch}
  =\hbar\sum_s\left[
       g_{As}\widehat\sigma_-\widehat b_s^\dagger
       +g_{As}^*\widehat\sigma_+\widehat b_s\right],
 \qquad \widehat\sigma_-=\ket g\bra e.
 \label{eq:s2-phonon-exchange-hamiltonian}
\end{equation}
The first term lowers the source and raises a mode occupation; the
second reverses that transfer. A receiving transition has its own
frequency and force overlap $g_{Bs}$. Spatial proximity alone does not
fix that overlap. Force-mediated coupling to mechanical resonators
provides an established setting in which this source--mode account
is developed quantitatively~\cite{AspelmeyerKippenbergMarquardt2014}.

\begin{proposition}[Single-quantum exchange weights]
\label{prop:s2-phonon-emission-absorption}
For the interaction above, the squared transition matrix elements
from mode occupation $N_s$ are
\begin{equation}
 \begin{aligned}
 |\langle g,N_s+1|\widehat H^{\rm exch}|e,N_s\rangle|^2
       &=\hbar^2|g_{As}|^2(N_s+1),\\
 |\langle e,N_s-1|\widehat H^{\rm exch}|g,N_s\rangle|^2
       &=\hbar^2|g_{As}|^2N_s .
 \end{aligned}
 \label{eq:s2-phonon-exchange-weights}
\end{equation}
\end{proposition}
\begin{proof}
Apply $\widehat\sigma_-$ or $\widehat\sigma_+$ to the material state
and the corresponding ladder operator to the number state.
\end{proof}

For weak coupling to a dense mode spectrum, a golden-rule limit gives
\begin{equation}
 \begin{aligned}
 \Gamma_{\rm em}(\Omega_A)
    &=2\pi\sum_s|g_{As}|^2(N_s+1)
                            \delta(\Omega_A-\omega_s),\\
 \Gamma_{\rm abs}(\Omega_A)
    &=2\pi\sum_s|g_{As}|^2N_s
                            \delta(\Omega_A-\omega_s).
 \end{aligned}
 \label{eq:s2-phonon-golden-rule-rates}
\end{equation}
The delta function is in angular frequency and has time units. A physically
broadened spectrum replaces each delta function by its normalized spectral
line shape within the same weak-coupling approximation. A constant decay
rate requires the reservoir and coarse-time assumptions
of this limit. A single isolated resonant mode instead gives reversible
exchange, with transition probability
$\sin^2(|g_{As}|\sqrt{N_s+1}\,t)$ in the resonant exchange model.
For a thermal mode, the mean occupation
$\overline N_s=(e^{\hbar\omega_s/(k_{\rm B}T)}-1)^{-1}$ obeys
$\overline N_s/(\overline N_s+1)=e^{-\hbar\omega_s/(k_{\rm B}T)}$.
Averaging the two rates over this thermal state therefore recovers
detailed balance on resonance. Phonon exchange redistributes an accounted-for energy supply;
it supplies neither an independent source of work nor a new material
particle. The classical high-occupation description uses the same
mechanical couplings without requiring resolved quantum jumps.

An experimentally specified force can also be treated directly.
For steady harmonic response about a stable reference, use the
$e^{-i\omega t}$ convention and let
$\vct U=\mathsf R(\omega)\vct F$ and
$\mathsf R=[\mathsf K-\omega^2\mathsf M-i\omega\mathsf C]^{-1}$,
with passive $\mathsf C\succeq0$. If $\mathsf C_B\succeq0$ is the
receiving region's contribution to the total resistance, its mean
absorbed power is
\begin{equation}
 \overline P_B
   =\frac{\omega^2}{2}\vct U^\dagger\mathsf C_B\vct U\ge0 .
 \label{eq:s2-phonon-receiver-power}
\end{equation}
This quantity has power units and can be compared with a specified
incident energy flux to define an absorption cross section. It measures
net energy deposition in that channel; an equilibrium upward transition
rate or a displacement amplitude is a different observable. More general
passive memory uses the dissipative impedance in the same quadratic
power account, as developed in Section~\ref{sec:viscoelastic-dynamics}.

Figure~\ref{fig:s23-source-mode-receiver} aligns the material-state energy changes with a displacement realization of the coupled collective mode.

\begin{figure}[H]
\centering\begingroup
\definecolor{vltTeal}{HTML}{176B64}
\definecolor{vltBlue}{HTML}{315D80}
\definecolor{vltGold}{HTML}{A87520}
\definecolor{vltPlum}{HTML}{725877}
\definecolor{vltRed}{HTML}{762B38}
\definecolor{vltYellow}{HTML}{D5AD3C}
\definecolor{vltNeutral}{HTML}{777D83}
\setlength{\unitlength}{1pt}
\scalebox{0.90}{%
\input{vlt-figures/figure-053.tex}%
}\endgroup
\caption{(a) Vertical level arrows record source and receiver energy changes; horizontal arrows mark force-overlap couplings. Each boxed transition starts from its own occupation $N$. The declared transition-force patterns are parallel to $\vct\phi_s=\vct e_s/\sqrt m$, selecting this column of the mode basis. (b) An auxiliary equal-mass array has $\vct R_i/a\in\{-1,0,1\}\times\{-1,0,1\}\times\{-1/2,1/2\}$, 33 nearest-axis central springs of curvature $k_\star$ and isotropic onsite pinning $k_\star$. The unit mass-weighted pattern $e_{iz}=2(R_{iz}/a)/\sqrt{18}$, $e_{ix}=e_{iy}=0$, has $\omega_s^2=3k_\star/m$. All eighteen rigid virions move at one scale; nine blue vertical links stretch and the other 24 retain their lengths. Coils retain their radius and turns and attach to actual body surfaces. Pins are implicit. This displacement snapshot represents a classical or coherent preparation; a number state has zero mean displacement.}
\label{fig:s23-source-mode-receiver}
\end{figure}

\subsection{Acoustic and optical organization of the modes}
\label{sec:branches}

Acoustic and optical branches organize an additional translation-invariant
bulk reference. The finite particle model retains finitely many modes;
a bounded continuum can have an infinite discrete spectrum when its
operator and boundary conditions give a compact resolvent. Each spectrum
therefore belongs to a specified resolution and boundary problem.

\begin{definition}[Bloch displacement]
\label{def:s2-bloch-displacement}
In a translation-invariant bulk reference, a Bloch displacement has the form
$\vct u_\ell(t)=\vct e\exp[i(\vct k\cdot\vct R_\ell-\omega t)]$.
Its real part is the physical motion.
\end{definition}
The phase changes by
$\vct k\cdot\vct d$ between cells separated by $\vct d$; small wavevector
gives nearly in-phase neighbors, with wavelength $2\pi/|\vct k|$ for
$\vct k\ne0$. For one equal-mass site per primitive cell,
\begin{equation}
 \mathsf D(\vct k)=\frac1m\sum_{\vct d}
          \mathsf C(\vct d)\bigl(1-e^{i\vct k\cdot\vct d}\bigr),
 \label{eq:bulk-dispersion}
\end{equation}
The sum contains both directions of every retained separation, with
$\mathsf C(-\vct d)=\mathsf C(\vct d)$. Pairing them replaces complex
phases by real cosine terms and gives a real symmetric matrix. Its three
eigenvalues are the squared frequencies of the translational branches.

\begin{definition}[Acoustic and optical translational branches]
\label{def:s2-acoustic-optical-branches}
For a stable unpinned bulk reference with only translation-generated
zero modes at zero wavevector, the three branches generated by common
translation are acoustic:
$\omega\to0$ as $\vct k\to0$. A primitive basis with
$n_{\mathrm b}$ sites gives $3n_{\mathrm b}$ translational branches.
Three are acoustic, and $3n_{\mathrm b}-3$ are optical when internal
relative motion has finite restoring frequencies at $\vct k=0$.
\end{definition}
``Optical'' describes a branch type, not a requirement for visible-light
excitation~\cite{Misra2012}. A crystallographic basis specifies sites
that repeat together; a vector-space basis specifies representation
vectors. They count different things.

A repeated multi-site virion basis supports internal relative patterns.
Resolved particle-internal coordinates add their own modes and restoring
mechanisms. Coordination shells classify pair geometry; the repeated
particle basis determines the number of translational branches.

The bulk calculation is an organizing limit for finite-cylinder motion,
not a replacement of reflecting walls by periodic boundaries. Where a
smooth dispersion is a useful local description, group velocity
$\nabla_{\vct k}\omega_s$ describes wave-packet transport away from the
walls. In a finite or strongly irregular network, the actual mode shapes
and the specified drive/readout determine the measured response. A
single-particle deformation mode requires different retained coordinates
from a whole-particle collective translation.

An exact two-site bulk example gives the branch labels a concrete
meaning. Consider a separate one-dimensional chain with equal springs
$k_\star$, alternating masses $m_1=m_\star$, $m_2=4m_\star$, and
primitive repeat $a$. Consecutive particles are separated by $a/2$.

\begin{example}[Alternating-mass bulk chain]
\label{ex:s2-alternating-mass-chain}
For the declared masses, springs and repeat, cell phase $e^{i\ell q}$
with $q=ka$ gives the mass-weighted Bloch matrix
\begin{equation}
 \frac{\mathsf D_{\rm alt}(q)}{k_\star/m_\star}
  =\begin{pmatrix}
      2&-(1+e^{-iq})/2\\-(1+e^{iq})/2&1/2
    \end{pmatrix},\qquad
 \frac{\omega_\pm^2(q)}{k_\star/m_\star}
  =\frac54\pm\sqrt{\frac{25}{16}-\sin^2(q/2)}.
 \label{eq:architecture-alternating-dispersion}
\end{equation}
The off-diagonal entries are conjugates, so the Bloch matrix is
Hermitian. Its trace is $5/2$ and its determinant is $\sin^2(q/2)$,
which give the displayed two eigenvalues. At $q=0$, common physical
motion has zero restoring force and lies on the acoustic branch. The
optical pattern has $u_2=-u_1/4$, keeping the cell's centre of mass fixed,
and $\omega_+^2(0)=5k_\star/(2m_\star)$. Two adjacent springs act on
each particle in this bulk model; its relative frequency therefore
differs from that of the single-spring free pair in
Example~\ref{ex:two-virions}. The zone edge is $|q|=\pi$, set by the
primitive repeat, not by a new boundary condition in the cylinder.
\end{example}
Figure~\ref{fig:architecture-diatomic-branches} plots the exact curves
and the corresponding physical ratios at $q=0$~\cite{Misra2012}.

\begin{figure}[H]
\centering\begingroup
\definecolor{vltTeal}{HTML}{176B64}
\definecolor{vltBlue}{HTML}{315D80}
\definecolor{vltGold}{HTML}{A87520}
\definecolor{vltPlum}{HTML}{725877}
\definecolor{vltRed}{HTML}{762B38}
\definecolor{vltYellow}{HTML}{D5AD3C}
\definecolor{vltNeutral}{HTML}{777D83}
\setlength{\unitlength}{1pt}
\scalebox{0.90}{%
\input{vlt-figures/figure-054.tex}%
}\endgroup
\caption{The alternating masses $m_\star,4m_\star$ share one reference-position ruler; two adjacent spacings form the primitive repeat $a$. Identical reference coils encode geometry independently of stiffness or mass. The exact bulk branches use $q=ka$ and $\omega_\star=\sqrt{k_\star/m_\star}$. At $q=0$, frequencies are $0$ and $\sqrt{5/2}\,\omega_\star$. The cell snapshots have $u_2/u_1=1$ and $-1/4$, with one shift gain of $20\,\mathrm{pt}$ per $A=u_1>0$. Dashed outlines retain reference positions. This auxiliary periodic chain has its own mechanics; the finite cylinder retains boundary-adapted modes.}
\label{fig:architecture-diatomic-branches}
\end{figure}

Periodicity labels repeated phases by wavevector and organizes common
and internal motion. For a finite preparation these labels guide a
justified bulk approximation; its boundary-adapted Hessian and readout
still determine the response. Mode frequency, dispersion branch and
phonon occupation retain their separate definitions.

\subsection{Medium memory, preparation and linked observables}
\label{sec:viscoelastic}

A viscoelastic medium stores mechanical energy and transfers energy
among retained and hidden coordinates. Its reduced force can depend on
the history of motion. Memory kernels or explicit auxiliary variables
represent that dependence~\cite{Mori1965,Lakes2009}.

\begin{definition}[Prepared hereditary particle equation]
\label{def:s2-prepared-hereditary-equation}
For a specified causal kernel and medium preparation, the linear particle
equation is
\begin{equation}
 \mathsf M\Dtt{\vct u}(t)+\mathsf K\vct u(t)
   +\int_0^t\mathsf\Gamma(t-\tau)\frac{\dd\vct u(\tau)}{\dd\tau}\,\dd\tau
   =\vct f_{\mathrm m}(t),
 \label{eq:memory}
\end{equation}
The preparation term $\vct f_{\mathrm m}$ includes the force generated by
the initial medium state; it may also contain explicitly declared drives.
\end{definition}
The kernel weights earlier velocities and can couple different particles.
A regular translational kernel has units $\mathrm{N\,m^{-1}}$, so its
velocity--time integral is a force. The preparation term contains the
initial medium state. An ensemble of these states supplies the
corresponding fluctuating forcing.

Evolution requires initial particle positions and velocities together
with medium history or equivalent auxiliary variables. A transformed
kernel or complex viscosity specifies the constitutive law; the
auxiliary coordinates specify its current stored state.

Causality restricts force to past and present inputs. Passivity further
requires an energy-storage and loss law: once initial storage is
accounted for, the medium supplies no net resource work. A prepared
medium can return stored energy. Equilibrium preparation additionally
matches its dissipative response to thermal forcing.

In the deterministic unforced viscous limit,
$\mathsf M\Dtt{\vct u}+\mathsf C_{\mathrm d}\Dt{\vct u}
+\mathsf K\vct u=0$ with
$\mathsf C_{\mathrm d}=\mathsf C_{\mathrm d}^{\mathsf T}\succeq0$,
the energy account is particularly simple:
\begin{equation}
 \Dt{E_{\mathrm{harm}}}=-\Dt{\vct u}^{\mathsf T}
                  \mathsf C_{\mathrm d}\Dt{\vct u},\qquad
 \Dt{E_{\mathrm{rest}}}=\Dt{\vct u}^{\mathsf T}
                  \mathsf C_{\mathrm d}\Dt{\vct u},\qquad
 \Dt{E_{\mathrm{tot}}}=0.
 \label{eq:reduced-energy}
\end{equation}
The nonnegative quadratic power is removed from resolved motion and
recorded in $E_{\mathrm{rest}}$, the energy of the eliminated medium and
coupling contributions. A regular medium force adds
$\Dt{\vct u}^{\mathsf T}\vct f_{\mathrm m}$ to resolved power with
the opposite change in the remaining account. A thermal stochastic limit
requires its corresponding stochastic energy balance. Hereditary models
also retain stored energy and its return; their reduced energy need not
decrease monotonically. Complete-system conservation is compatible with
attenuation of a selected displacement channel.

One force-valued auxiliary variable makes memory and stored energy
explicit. For a scalar coordinate $u$, choose $m,k_0,k_m,\tau_m>0$ and
the causal exponential kernel
\begin{equation}
 \Gamma_m(t)=k_m e^{-t/\tau_m}\quad(t\ge0),\qquad
 m\Dtt{u}+k_0u+z=f,\qquad
 \Dt{z}=k_m\Dt{u}-\frac{z}{\tau_m}.
 \label{eq:architecture-scalar-memory-realization}
\end{equation}

Figure~\ref{fig:architecture-maxwell-memory-account} realizes the same auxiliary equation as a Maxwell branch and shows an exact clamped relaxation with its energy transfer accounted for.

Here $z$ is a force, not a displacement. Solving its first-order equation
gives
$z(t)=z(0)e^{-t/\tau_m}+
\int_0^t k_m e^{-(t-s)/\tau_m}\dd u(s)/\dd s\,\dd s$.
Thus it stores the exponentially weighted velocity history. A nonzero
initial $z(0)$ is part of the preparation and cannot be discarded when
starting the finite-time equation.

\begin{proposition}[Storage and loss in an exponential-memory realization]
\label{prop:s2-exponential-memory-storage}
For $m,k_0,k_m,\tau_m>0$ and sufficiently regular solutions of
\eqref{eq:architecture-scalar-memory-realization}, the auxiliary stores
recoverable energy and the combined power balance is
\begin{equation}
 E_m=\frac{z^2}{2k_m},\qquad
 \frac{\dd}{\dd t}\left(\frac{m}{2}\left(\Dt{u}\right)^2
                       +\frac{k_0u^2}{2}+E_m\right)
      =f\Dt{u}-\frac{z^2}{k_m\tau_m}.
 \label{eq:architecture-memory-storage-power}
\end{equation}
\end{proposition}
\begin{proof}
Multiplying the particle equation by $\Dt u$ gives the particle power
$f\Dt u-z\Dt u$. Differentiating $z^2/(2k_m)$ and substituting its
evolution gives $z\Dt u-z^2/(k_m\tau_m)$. Adding cancels the reversible
coupling power and leaves the stated nonnegative loss.
\end{proof}
The power $z\Dt{u}$ cancels between particle and auxiliary storage.
Their combined recoverable energy decreases in unforced motion, even
when the particle regains energy from the auxiliary. The remaining loss
belongs to still-eliminated coordinates. This constitutive realization
approximates those dynamics over a stated observation
band~\cite{Lakes2009}.

\begin{figure}[H]
\centering\begingroup
\definecolor{vltTeal}{HTML}{176B64}
\definecolor{vltBlue}{HTML}{315D80}
\definecolor{vltGold}{HTML}{A87520}
\definecolor{vltPlum}{HTML}{725877}
\definecolor{vltRed}{HTML}{762B38}
\definecolor{vltYellow}{HTML}{D5AD3C}
\definecolor{vltNeutral}{HTML}{777D83}
\setlength{\unitlength}{1pt}
\scalebox{0.90}{%
\input{vlt-figures/figure-055.tex}%
}\endgroup
\caption{The scalar material-memory realization \eqref{eq:architecture-scalar-memory-realization}. (a) Spring $k_0$ is in parallel with a series dashpot--spring branch. Only the left rail is fixed; the right rail follows the rigid particle coordinate $u$. The black internal dot has displacement $y$ and carries no extra virion. The two spring glyphs share their geometry and describe the reference topology, with no strain or stiffness scale. Eliminating $y$ gives the stated force-valued memory $z$. (b) Clamping $u$ with $z(0)=z_0\ne0$ gives $z/z_0=e^{-t/\tau_m}$, stored-energy fraction $e^{-2t/\tau_m}$ and dissipated fraction $1-e^{-2t/\tau_m}$. Their exact account is \eqref{eq:architecture-memory-storage-power}. The clamp does no work while material storage relaxes. This is a deterministic constitutive analogue; an equilibrium preparation also supplies matched noise.}
\label{fig:architecture-maxwell-memory-account}
\end{figure}

\paragraph{Application: transient attachments and material memory.}
A passive filament gel with reversible attachments illustrates why a
material's state can require more than its instantaneous displacement.
At the level of an effective element, the auxiliary force $z$ in
\eqref{eq:architecture-scalar-memory-realization} distinguishes two
preparations having the same $u$ and $\Dt{u}$ but different stored stress.
Those preparations exert different restoring forces and subsequently relax
differently. Here $z$ is an evolving internal state, whereas $k_m$ and
$\tau_m$ specify its constitutive law. The measured force and displacement
remain outputs of a stated loading geometry.

In a transient network, structural accommodation can involve many
attachment changes and filament deformation scales. Broedersz and
coauthors show that a single microscopic cross-link time can produce a
broad spectrum of collective relaxation times
~\cite{Broedersz2010}. Consequently a fitted single exponential is an
effective description over its successful frequency range, and its fitted
time need not equal a molecular attachment lifetime. A broad measured
response may require several auxiliary stresses or a continuous memory
kernel even when a static modulus is well reproduced.

A short-time elastic calculation conditions on persistent attachments.
At longer times, attachment states or a justified effective constitutive
law account for their changes. Static curvature, response phase and
relaxation records test whether the retained state has enough storage
channels. A fitted relaxation time is interpreted through that tested
resolution and need not identify a unique molecular process.

Stationary expressions use the equilibrium history or the long-time limit
of the prepared equation; finite-time transients retain their initial terms.
For stationary linear response, use the $e^{-i\omega t}$ convention and
the causal transform
$\widetilde{\mathsf\Gamma}(\omega)
=\int_0^\infty\mathsf\Gamma(s)e^{i\omega s}\,\dd s$.
After matching the static curvature and inertia to this medium model,
define
\begin{equation}
 \mathcal R(\omega)=
 [\mathsf K-\omega^2\mathsf M
            -i\omega\widetilde{\mathsf\Gamma}(\omega)]^{-1},\qquad
 Y=\vct w^{\mathsf T}\vct u,\qquad
 \mathcal R_Y=\vct w^{\mathsf T}\mathcal R\vct w .
 \label{eq:architecture-linked-response}
\end{equation}
The inverse is taken on the specified stable sector and where the
response exists. For real dimensionless readout weights $\vct w$,
$Y$ is a displacement. Its conjugate force has pattern $\vct w F$
and work $F\,\dd Y$, as introduced in
\eqref{eq:introduction-readout-response}. A response may be small even
at a mode frequency if that mode couples weakly to the force or readout.

For a consistently prepared classical equilibrium model at $T>0$, let
$\vct f_{\rm th}$ be the centred stationary thermal force. On stable
thermalized sectors its two-sided angular-frequency spectra satisfy
\begin{equation}
 \begin{aligned}
 \mathsf S_{\rm th}(\omega)
   &=k_{\mathrm B}T\bigl[
       \widetilde{\mathsf\Gamma}(\omega)
        +\widetilde{\mathsf\Gamma}(\omega)^\dagger\bigr],\\
 \mathsf S_u(\omega)&=
      \mathcal R(\omega)\mathsf S_{\rm th}(\omega)
                                  \mathcal R(\omega)^\dagger,\qquad
 S_Y(\omega)=\vct w^{\mathsf T}\mathsf S_u(\omega)\vct w .
 \end{aligned}
 \label{eq:architecture-linked-fluctuations}
\end{equation}
The convention is
$\mathsf S_u(\omega)=\int_{\R}e^{i\omega t}
\operatorname{Cov}[\vct u(t),\vct u(0)]\,\dd t$, with
equal-time covariance $\int_{\R}\mathsf S_u(\omega)\dd\omega/(2\pi)$.
Thus $\mathsf S_{\rm th}$ has units $\mathrm{N^2s}$ and
$\mathsf S_u$ has units $\mathrm{m^2s}$. The noise identity concerns
thermal forcing, not a prescribed external drive. The Hermitian part of
the transformed friction is nonnegative for the passive equilibrium
model. Undamped modes retain initial-condition spectral contributions;
they cannot be reconstructed by setting thermal noise to zero in this
formula. Unpinned zero coordinates need separate treatment rather than
an unrestricted inverse covariance. Section~\ref{sec:viscoelastic-dynamics}
develops the response and fluctuation assumptions in detail.

The matched equilibrium force law makes the response and fluctuations
testable through the same channel~\cite{CallenWelton1951}.
\begin{proposition}[Matched equilibrium readout relation]
\label{prop:s2-matched-readout-fdt}
Assume the stable reciprocal classical equilibrium model and thermal
force spectrum in~\eqref{eq:architecture-linked-fluctuations}, with
real conjugate readout weights. Set
$\mathsf G=[\widetilde{\mathsf\Gamma}+
\widetilde{\mathsf\Gamma}^{\dagger}]/2$.
Where the regular response exists and $\omega\ne0$,
\begin{equation}
 \frac{\mathcal R-\mathcal R^\dagger}{2i}
       =\omega\mathcal R\mathsf G\mathcal R^\dagger,
 \qquad
 \boxed{S_Y(\omega)=\frac{2k_{\rm B}T}{\omega}
                            \operatorname{Im}\mathcal R_Y(\omega).}
 \label{eq:architecture-readout-fdt}
\end{equation}
\end{proposition}
\begin{proof}
Write $\mathsf A=\mathcal R^{-1}$. Real symmetric
$\mathsf K,\mathsf M$ give
$\mathsf A^\dagger-\mathsf A=2i\omega\mathsf G$.
The identity
$\mathcal R-\mathcal R^\dagger=
\mathcal R(\mathsf A^\dagger-\mathsf A)\mathcal R^\dagger$
proves the first equality. Combining it with
$\mathsf S_{\rm th}=2k_{\rm B}T\mathsf G$ and projecting with
$\vct w$ proves the scalar spectrum relation.
\end{proof}
For scalar $\mathcal R_Y$, the imaginary part gives the displacement
response in quadrature with the force; the phase lag itself is
$\arg\mathcal R_Y$ under the stated Fourier convention. The same dissipative part weights its
equilibrium fluctuations; the factor $2$ follows from the stated
two-sided angular-frequency convention. The expression at zero
frequency is obtained by a continuous limit when one exists. Discrete
undamped contributions still require the equilibrium initial-state
account described above.

For the exponential example, the transform and scalar response are
\begin{equation}
 \begin{aligned}
 \widetilde\Gamma_m(\omega)&=\frac{k_m\tau_m}{1-i\omega\tau_m},\\
 \mathcal R_m(\omega)&=
  \left[k_0-m\omega^2-
        \frac{i\omega k_m\tau_m}{1-i\omega\tau_m}\right]^{-1},\\
 S_u(\omega)&=\frac{2k_{\rm B}T k_m\tau_m}
                         {1+\omega^2\tau_m^2}|\mathcal R_m(\omega)|^2 .
 \end{aligned}
 \label{eq:architecture-scalar-memory-spectrum}
\end{equation}
The real part of the transformed friction is
$k_m\tau_m/(1+\omega^2\tau_m^2)>0$. At low frequency the kernel
approaches a viscous coefficient $c=k_m\tau_m$; at finite frequency it
also adds storage. The regular scalar thermal-force covariance is
$\langle f_{\rm th}(t)f_{\rm th}(0)\rangle
=k_{\rm B}Tk_m e^{-|t|/\tau_m}$.
An instantaneous-drag comparison uses
$\mathcal R_d=[k_0-m\omega^2-i\omega c]^{-1}$ and
$S_{u,d}=2k_{\rm B}Tc|\mathcal R_d|^2$ with the same $m,k_0,T$.
These are different constitutive models with a matched low-frequency
drag and static curvature, not two independent fits to response and noise.

Introduce dimensionless variables only for plotting:
\begin{equation}
 \begin{aligned}
 \omega_\star&=\sqrt{k_0/m},\quad\nu=\omega/\omega_\star,
 \quad\rho=k_m/k_0,\quad\theta=\tau_m\omega_\star,\\
 \overline{\mathcal R}_m&=k_0\mathcal R_m
      =\left[1-\nu^2-
          \frac{i\nu\rho\theta}{1-i\nu\theta}\right]^{-1},\\
 \overline S_u&=\frac{k_0\omega_\star}{k_{\rm B}T}S_u
      =\frac{2\rho\theta}{1+\nu^2\theta^2}
                            |\overline{\mathcal R}_m|^2.
 \end{aligned}
 \label{eq:architecture-memory-plot-normalization}
\end{equation}
Figure~\ref{fig:architecture-linked-memory-response} sets
$(\rho,\theta)=(0.6,0.8)$ to display magnitude, phase and fluctuation
spectrum from the same specified parameters. Both models have
$\mathcal R(0)=1/k_0$ and the classical equilibrium variance
$\langle u^2\rangle=k_{\rm B}T/k_0$, obtained by integrating their
complete spectra. Their finite-frequency predictions differ. At fixed
mechanical coefficients, changing $T$ rescales classical fluctuation
strength; temperature-dependent coefficients are additional physical
inputs. This simple illustration isolates the value of memory before
the full benchmark in Subsection~\ref{sec:complete-benchmark}.

\begin{figure}[H]
\centering\begingroup
\definecolor{vltTeal}{HTML}{176B64}
\definecolor{vltBlue}{HTML}{315D80}
\definecolor{vltGold}{HTML}{A87520}
\definecolor{vltPlum}{HTML}{725877}
\definecolor{vltRed}{HTML}{762B38}
\definecolor{vltYellow}{HTML}{D5AD3C}
\definecolor{vltNeutral}{HTML}{777D83}
\setlength{\unitlength}{1pt}
\scalebox{0.80}{%
\input{vlt-figures/figure-056.tex}%
}\endgroup
\caption{Response, lag and fluctuations of the same scalar channel, using \eqref{eq:architecture-memory-plot-normalization}. Solid curves retain an exponential memory with $(\rho,\theta)=(0.6,0.8)$; dashed curves use the instantaneous drag $c=k_m\tau_m$ with the same $m,k_0,T$. The plots show response magnitude, positive displacement-lag phase under $e^{-i\omega t}$, and two-sided thermal spectrum $\overline S_u=k_0\omega_\star S_u/(k_{\rm B}T)$. Open points mark the common zero-frequency limits $(1,0,0.96)$. The boxed readout relation \eqref{eq:architecture-readout-fdt} links each response to its matched noise. Their static compliance and fully integrated equilibrium variance agree; their finite-frequency records differ. Only the nonnegative frequency band is displayed. These are declared synthetic models, with no fitted virion material parameters.}
\label{fig:architecture-linked-memory-response}
\end{figure}

Equations~\eqref{eq:architecture-linked-response} and
\eqref{eq:architecture-linked-fluctuations} state the central test of the
architecture: one specified mechanical model predicts both forced
response and spontaneous motion through the same calibrated channel.
Static covariance constrains curvature; phase and frequency dependence
can constrain dynamical environmental effects that static data leave
unresolved. Correlation alone does not establish a direct edge, because
shared medium forcing can also correlate particles. A comparison with a
simpler model must therefore use the same calibration, preparation and
observation band. This provides the physical specification for the
later benchmark, without making that synthetic calculation an empirical
validation of a viral preparation.

\subsubsection{Inverse questions and integration of biophysical variables}
\label{sec:s2-open-inference-variables}

The forward construction sharpens two biophysical questions: which
quantities the observations determine, and which variables predict their
future signals. Geometry, interactions, preparation, forcing and readout
define the forward map. Inference then asks what remains identifiable
among specifications producing the same record. Recovering a target
prediction can require less information than recovering every microscopic
input~\cite{ClermontZenker2015};
Definition~\ref{def:s6-identifiable-prediction} states the later criterion.

The charge symmetry gives an exact example within this lattice.
Proposition~\ref{prop:s2-charge-reversal} identifies globally reversed
charge patterns with the same mechanical law when the remaining inputs
are reversal invariant. With the same particle and medium preparation,
forcing and charge-independent displacement readout, the two assignments
therefore give identical mechanical predictions. Additional samples of
those same observables cannot select their common charge sign. They can
still constrain restoring curvatures or predict a specified displacement
response. A charge-sensitive observation or coupling would instead enter
through its explicitly stated measurement or energy law. This separates
an exact observational equivalence from uncertainty caused by finite
sampling or calibration.

The exponential-memory realization supplies a complementary test of
variable sufficiency. Consider two admissible preparations with the same
$u$, $\dot u$ and subsequent force $f$, but different auxiliary stresses
$z$. Equation~\eqref{eq:architecture-scalar-memory-realization} gives
$\Delta\ddot u=-\Delta z/m$: their particle motions immediately develop
different accelerations. Retaining $z$, or its equivalent prepared
history, distinguishes those futures. Here $u$, $\dot u$ and $z$ are
state variables; $k_m$ and $\tau_m$ are constitutive parameters; a
measured displacement is the output of a specified channel. Integrating
a new biophysical variable means assigning it one of these physical
roles and a consistent evolution or observation law. The exact linear
closure criterion in Proposition~\ref{prop:s4-state-exact-closure}
extends this same question to many coordinates.

For a patch-sensitive application, first constrain the body and surface
preparation; select fixed-charge, frozen-patch or regulated resolution;
then calculate force, torque and the appropriate clamped or relaxed
tangent. Supply mobility, material/surface history and matched thermal
forcing before predicting a record. Complementary translational,
orientational and chemical readouts can then test which of those
resolutions is sufficient. Their surface charges and kinetics are
synthetic inputs until independently constrained for the preparation.

A concrete research task is then to find the smallest physically
specified state that predicts the selected signals within a declared
error, time range and frequency band. Structural measurements constrain
geometry; calibrated static loading constrains relaxed stiffness;
transients, response phase and equilibrium fluctuations test the
additional storage and loss channels. Parameters constrained by one
measurement can be carried into another force--readout channel or a new
preparation to assess predictive sufficiency. Repeated disagreement can
motivate a resolved coordinate, a revised interaction or a richer medium
law, each with its own energy account. These comparisons make variable
selection and inverse prediction testable within the same finite
mechanical system.

The architecture now supplies the objects needed for computation:
geometry and energy determine the Hessian; inertia determines mode
frequencies; preparation and readout determine the visible motion.
Medium reduction adds storage, loss, history and matched fluctuations.
The occupation postulate adds a separate quantum state description.
Section~\ref{sec:structure-interactions} develops finite assembly and
controlled bulk approximations from these inputs.

\clearpage

\section{Viral Lattice Structure and Interactions}
\label{sec:structure-interactions}

The geometry and interactions specified in Section~\ref{sec:lattice-architecture}
now determine the mechanics of the finite assembly. In the central
reference, each rigid ideal VLP contributes its centre coordinates and
mass. Resolving its surface also retains orientation and rotational
inertia. A patch interaction then supplies both forces and torques;
surface populations and material preparation determine which restoring
law is held fixed, equilibrated or evolved during the observation.
Supports and constraints complete the mechanical model.

The central-force calculations below provide an explicit translational
baseline. Their resolved-body extension uses the same energy, inertia
and virtual-work construction on the larger tangent space. It connects
three calculations: the motion within an arrangement, the response after
surface or material relaxation, and the probability of changing
arrangements. All three refer to the same labelled particle pool and
specified nonlinear interaction.

The calculation proceeds through three tests: force balance at the
reference, quadratic energy on the admissible displacement space, and
restoring curvature relative to inertia. The resulting eigenproblem
gives collective motion and conservative frequencies. Preparation fixes
amplitudes and phases, and the force--readout map connects that motion
to a signal~\cite{Leake2016}.

The essential step is to turn a set of pair interactions into a law for
the whole assembly. A trial displacement can stretch one interaction,
shorten another and turn a third without appreciably changing its length.
Adding their energy changes determines the collective restoring force.
Uniform translation preserves pair separations; relative patterns load
the network according to its geometry and reference forces. This is why
the global matrix carries information that a list of isolated spring
constants leaves unresolved.

The same reasoning explains the computational role of a cell. Its
coordinates and interaction contributions are embedded in the larger
assembly, with shared objects registered consistently. Interior motion
can then be retained or eliminated according to the response being
predicted. The finite model supplies the reference calculation; bulk
dispersion and continuum fields are controlled comparisons over their
resolved wavelengths and observation times.

The section first derives finite-domain balances, directional tensors
and global assembly. Controlled bulk comparisons then describe
dispersion, internal relaxation and packet propagation. Compatible cells
provide computational building blocks for larger populations. The final
subsection returns to the nonlinear mechanics of recruitment and breakup,
with the total particle pool in the cylinder held fixed.

\subsection{Reference positions and finite-domain constraints}
\label{sec:reference-constraints}

\begin{definition}[Finite shell truncation]
\label{def:s3-finite-shell-truncation}
Let $\mathcal I$ be the finite set of integer triples retained in
\eqref{eq:cubic-lattice}, with $n=|\mathcal I|$. Its reference shell
offsets are
\begin{equation}
 \begin{split}
 \mathcal N_\beta&=a\{\pm\vct e_x,\pm\vct e_y,\pm\vct e_z\},\\
 \mathcal N_\gamma&=
   a\{\sigma_\mu\vct e_\mu+\sigma_\nu\vct e_\nu:
           \mu<\nu,\ \sigma_\mu,\sigma_\nu\in\{-1,1\}\},\\
 \mathcal N_\Omega&=
   a\{\sigma_x\vct e_x+\sigma_y\vct e_y+\sigma_z\vct e_z:
                \sigma_x,\sigma_y,\sigma_z\in\{-1,1\}\}.
 \end{split}
 \label{eq:shell-offsets}
\end{equation}
At site $\vct\ell\in\mathcal I$, an offset $\vct d$ is retained only
when $\vct\ell+\vct d/a\in\mathcal I$ and its pair belongs to the
declared interaction set.
\end{definition}
Here $\vct e_x,\vct e_y,\vct e_z$ are Cartesian unit vectors. An offset
joins two reference positions; a displacement measures motion from one
reference. The dimensionless signs $\sigma_\mu=\pm1$ enumerate
directions and are separate from the interaction lengths $\sigma_{ij}$.
The shells contain six axial, twelve face-diagonal and eight
body-diagonal offsets. Retaining only endpoints in $\mathcal I$ gives
the actual coordination at a finite boundary.

\begin{definition}[Reference clearance margins]
\label{def:s3-clearance-margins}
For exclusion radii $b_i$, positive pair and wall margins
$\delta_{ij}$ and $\delta_i^{\mathrm w}$ are specified by
\begin{equation}
 R_{ij}\ge b_i+b_j+\delta_{ij},\qquad
 \operatorname{dist}(\vct R_i,\partial\mathcal V)
                       \ge b_i+\delta_i^{\mathrm w}.
 \label{eq:clearance}
\end{equation}
\end{definition}
Distance to $\partial\mathcal V$ means the shortest distance to a side
wall or end cap. Thus the conditions constrain both separation and
confinement. For equal radii the nearest-pair condition is
$a\ge2b+\delta$. The exact population is the number of retained sites;
the cylinder volume $\pi R_c^2L_c$ does not impose a cubic-box counting
rule.

\begin{proposition}[Sufficient excursion margins]
\label{prop:s3-clearance-excursion}
For the margins in Definition~\ref{def:s3-clearance-margins}, the
conditions $|\vct u_i-\vct u_j|\le\delta_{ij}/2$ and
$|\vct u_i|\le\delta_i^{\mathrm w}/2$ preserve strictly positive
pair and wall clearance.
\end{proposition}
\begin{proof}
The reverse triangle inequality gives
$r_{ij}(t)\ge R_{ij}-|\vct u_i-\vct u_j|$.
Distance to a closed boundary is $1$-Lipschitz, so it decreases by at
most $|\vct u_i|$. The displacement bound also keeps the centre inside
the ball about its reference that lies within the cylinder. Each
remaining clearance is at least half its positive reference margin.
\end{proof}
These bounds test a prepared trajectory against its geometric margins;
the force-variation scale separately controls harmonic accuracy. At hard
contact, the impact rule~\eqref{eq:reflection} supplies the event dynamics.

Figure~\ref{fig:structure-finite-boundary-forces} separates admissible positions from the force balance needed to prepare a stationary finite reference.

\begin{definition}[Smooth stationary finite reference]
\label{def:s3-stationary-finite-reference}
For the unconstrained conservative baseline, a smooth stationary
reference is an admissible configuration at which every retained
site satisfies
\begin{equation}
 \vct f_i^{(0)}
   =-\sum_{j:\{i,j\}\in\mathcal E}
        V'_{ij}(R_{ij})\vct n_{ij}^{(0)}
     -\left.\nabla_i(U_{\mathrm w}+\lambda U_{\mathrm T})
                                            \right|_{\vct R}
   =\vct0,\qquad
 \vct n_{ij}^{(0)}=\frac{\vct R_i-\vct R_j}{R_{ij}}.
 \label{eq:finite-reference-force-balance}
\end{equation}
\end{definition}
Stationarity balances the vector sum of incident pair forces and smooth
confinement or target forces. Individual values of $V'_{ij}(R_{ij})$
may be nonzero. Opposite forces cancel in a homogeneous repeated bulk;
finite boundary sites require their actual incident forces and supports.
A positive-clearance hard wall is inactive in this balance.

\paragraph{Global force, torque and dilation tests.}
Equation~\eqref{eq:finite-reference-force-balance} states the
unconstrained, smooth-support equilibrium condition. Under the fixed
linear constraints of~\eqref{eq:structure-constrained-sector}, the
necessary stationarity condition is instead
$\mathsf U_{\rm c}^{\mathsf T}\nabla U_\lambda(\vct R)=\vct0$.
An ideal reaction can balance a nonzero full gradient while doing no
work on allowed displacements. A hard wall at positive clearance
provides no such reaction.

\begin{proposition}[Finite support identities]
\label{prop:s3-finite-support-identities}
Suppose all site forces balance, including any ideal constraint
reactions. Let $\vct F_i^{\rm sup,0}$ collect the non-pair reference
forces from the supports, target terms and those reactions.
For central pair interactions,
\begin{equation}
 \begin{aligned}
 \sum_i\vct F_i^{\rm sup,0}&=\vct0,
 &\sum_i\vct R_i\times\vct F_i^{\rm sup,0}&=\vct0,\\
 \sum_i\vct R_i\cdot\vct F_i^{\rm sup,0}
      &=\sum_{\{i,j\}\in\mathcal E}R_{ij}V'_{ij}(R_{ij}).
 \end{aligned}
 \label{eq:structure-finite-support-identities}
\end{equation}
\end{proposition}
\begin{proof}
Sum the site balances. Equal and opposite central pair forces cancel
their total force and torque. Dotting each balance with $\vct R_i$
makes the contribution of one unordered pair
$-V'_{ij}(R_{ij})(\vct R_i-\vct R_j)\cdot\vct n_{ij}^{(0)}
=-R_{ij}V'_{ij}(R_{ij})$, giving the last identity.
\end{proof}
The first two relations test total force and torque. The last is the
static scalar virial identity, obtained by a position-weighted force sum
or a formal uniform dilation of the pair energy. Each unordered edge
enters once, including any retained farther charged pairs. Both sides
have energy units; a pressure interpretation additionally specifies
volume and boundary work. When the total support force vanishes,
torque and virial are independent of the coordinate origin.

These global identities supplement the all-site equilibrium test.
They can expose omitted support forces, but cancellation across the
whole assembly can coexist with a residual force on an individual virion.
They describe the central reference; the resolved-body construction
below includes the intrinsic torques of oriented surfaces.

An unstressed reference places every retained pair at its rest length.
A prestressed reference instead balances its nonzero pair forces through
the complete geometry and declared supports, often after relaxation.
A residual force drives the reference itself. The following harmonic
calculation therefore starts from a stationary preparation.

Periodic face identification supplies the auxiliary bulk comparison.
The reflecting cylinder retains its finite boundary problem; comparing
their spectra isolates bulk behavior and finite-size effects.

\begin{figure}[H]
\centering\begingroup
\definecolor{vltTeal}{HTML}{176B64}
\definecolor{vltBlue}{HTML}{315D80}
\definecolor{vltGold}{HTML}{A87520}
\definecolor{vltPlum}{HTML}{725877}
\definecolor{vltRed}{HTML}{762B38}
\definecolor{vltYellow}{HTML}{D5AD3C}
\definecolor{vltNeutral}{HTML}{777D83}
\setlength{\unitlength}{1pt}
\scalebox{0.90}{%
\input{vlt-figures/figure-057.tex}%
}\endgroup
\caption{The same three virions test reference balance with different supports. Centres lie at $z_i=(i-1/2)a$ in a slice of length $3a$, with body radius $0.12a$ and scale $120\,\mathrm{pt}$ per $a$. Transverse coordinates are fixed. Adjacent pairs have $V^\prime(a)=F_0>0$; black incident-force arrows share the gain $24\,\mathrm{pt}/F_0$. Their middle contributions cancel, leaving opposite end residuals. In (a), positive-clearance hard caps are inactive. In (b), burgundy smooth-support forces balance those residuals, and support curvature also enters the Hessian. The shared ruler locates both preparations. For an unstressed reference, the corresponding pair-force arrows vanish.}
\label{fig:structure-finite-boundary-forces}
\end{figure}

\subsubsection{Force--torque balance and the resolved patch Hessian}
\label{sec:s3patch-body-tangent}

An aligned pair can resist separation and still turn under its patch
torques. A resting body reference therefore balances both kinds of
motion. Write $\vct b_{ia}=\mathsf O_i^0\vct a_{ia}$ for a reference
surface arm and use
$\mathsf O_i=\exp([\vct\theta_i]_\times)\mathsf O_i^0$ near that
reference. The tangent $\vct\zeta=(\vct u,\vct\theta)$ contains
translations and laboratory angular increments, as in
Definition~\ref{def:s2patch-tangent-hessian}.

\begin{definition}[Stationary resolved-body reference]
\label{def:s3patch-stationary-body}
At specified frozen surface/material states, an unconstrained resting
reference satisfies $\vct F_i^0=\vct0$ and $\vct T_i^0=\vct0$ for
every body. Any state to be equilibrated is stationary on its permitted
population subspace as well. The mechanical Hessian is the second
derivative of the complete retained free energy in this tangent chart.
\end{definition}

For smooth independent bilateral constraints $c_a(\vct\zeta)=0$,
choose reference multipliers with
$\nabla_\zeta\mathcal F+(Dc)^{\mathsf T}\vct\lambda=0$ and a
full-column-rank tangent map $\mathsf P$ with
$\operatorname{Ran}\mathsf P=\ker Dc$.
The second variation on the constrained manifold is
\begin{equation}
 \mathsf H_{\rm c}=\mathsf P^{\mathsf T}
  \left[\nabla_\zeta^2\mathcal F+
         \sum_a\lambda_a\nabla_\zeta^2c_a\right]\mathsf P,
 \qquad
 \mathsf G_{\rm c}=\mathsf P^{\mathsf T}\mathsf G\mathsf P .
 \label{eq:s3patch-constrained-body-hessian}
\end{equation}
This follows by differentiating $c_a(\vct\zeta(t))=0$ twice along
an allowed curve and using stationarity to eliminate its acceleration
term from the free-energy derivative. The reaction-curvature term
vanishes for the affine constraints used in
\eqref{eq:structure-constrained-sector}. Actual contact and changes of
the active constraint set retain their specified event dynamics.

\begin{proposition}[Patch-site second variation]
\label{prop:s3patch-site-second-variation}
For one frozen-charge site interaction $v(d)$, let
$\vct d^0=\vct R_i+\vct b_{ia}-\vct R_j-\vct b_{jb}$,
$d=|\vct d^0|>0$, $\vct n=\vct d^0/d$, and define
\begin{equation}
 \begin{aligned}
 \vct a_e(\vct\zeta)
  &=\vct u_i-\vct u_j+
      \vct\theta_i\times\vct b_{ia}
      -\vct\theta_j\times\vct b_{jb},\\
 \vct b_e(\vct\zeta)
  &=\vct\theta_i\times(\vct\theta_i\times\vct b_{ia})
     -\vct\theta_j\times(\vct\theta_j\times\vct b_{jb}),\\
 \mathsf C_e
  &=v''(d)\vct n\vct n^{\mathsf T}
     +\frac{v'(d)}d(\Id_3-\vct n\vct n^{\mathsf T}).
 \end{aligned}
 \label{eq:s3patch-site-kinematics}
\end{equation}
Along the dimensionless path $\varepsilon\vct\zeta$ in exponential
coordinates, the exact second derivative at $\varepsilon=0$ is
\begin{equation}
 \frac{\dd^2}{\dd\varepsilon^2}v(d(\varepsilon))\bigg|_{0}
  =\vct a_e^{\mathsf T}\mathsf C_e\vct a_e
       +v'(d)\vct n\cdot\vct b_e .
 \label{eq:s3patch-site-hessian-geometric}
\end{equation}
\end{proposition}
\begin{proof}
The relative site vector has derivatives $\vct a_e$ and $\vct b_e$.
The Cartesian gradient of $v(|\vct d|)$ is $v'(d)\vct n$, and its
Cartesian Hessian is $\mathsf C_e$. The second-order chain rule gives
the result.
\end{proof}

The first term turns site motion into Cartesian interaction curvature.
The second records the curvature of rotational motion of the arms.
Individual site interactions can remain loaded when their total force
and torque balance, so both terms enter the assembled Hessian. The
screened site sum in~\eqref{eq:s2patch-site-yukawa} uses this formula
term by term; dielectric, exclusion, surface and support energies supply
their own derivatives. Twice differentiating the complete pulled-back
energy also handles models that are not sums of site distances.

For translation- and rotation-invariant internal energy, a free
stationary body assembly has the rigid tangents
\begin{equation}
 \vct u_i=\vct a+\vct\psi\times\vct R_i,
 \qquad \vct\theta_i=\vct\psi .
 \label{eq:s3patch-global-rigid-tangents}
\end{equation}
They lie in its Hessian kernel: differentiate the stationary gradient
along the common rigid-motion orbit. Rotating the centres while holding
the surface orientations fixed generally changes the interaction.
The corresponding support test includes intrinsic torque,
$\sum_i\vct F_i^{\rm sup}=0$ and
$\sum_i(\vct R_i\times\vct F_i^{\rm sup}+\vct T_i^{\rm sup})=0$.
Confinement and the declared supports determine which of these free
symmetries survive. Mechanical normalization uses
\eqref{eq:s2patch-generalized-inertia}, so translational and angular
entries are compared through their physical inertia and coordinate
scales.

Figure~\ref{fig:s34patch-site-path-curvature} separates motion of a body centre from motion of a surface site. Its enlarged arc identifies the geometric contribution that enters the resolved patch Hessian.

\begin{figure}[H]
\centering\begingroup
\setlength{\unitlength}{1pt}
\scalebox{0.94}{%
\input{vlt-figures/figure-286.tex}%
}\endgroup
\caption{Body motion and site-distance curvature. Panel (a) is a top view of two ideal VLPs with one marked site on each; translation and rotation are separate tangent components. Panel (b) enlarges the right-body arm: black is the exact circular site path, green its first-order displacement and burgundy the second-order correction, with an open marker at the quadratic endpoint. The finite angle is exaggerated to separate these contributions. The lower identity is the exact second variation at the reference, as in~\eqref{eq:s3patch-site-hessian-geometric}; it retains the rotational path term even when the assembled forces and torques balance. Site markers identify interaction locations, not the net charge of either body.}
\label{fig:s34patch-site-path-curvature}
\end{figure}

\subsection{Discrete mechanics and the controlled continuum approximation}
\label{sec:continuum-bridge}

\paragraph{The finite lattice in virtual-power form.}
Before making a smooth-field approximation, the finite linearized
assembly has an exact work statement.

\begin{proposition}[Finite virtual-power identity]
\label{prop:s3-finite-virtual-power}
For constant symmetric edge tensors and smooth-support curvature
$\mathsf K_{\rm sup}$, write
$\Delta_{ij}\vct u=\vct u_i-\vct u_j$ and let $\vct f_i(t)$ be
the applied incremental force. On a fixed linear constrained sector,
the equations imply, for every admissible test velocity $\vct w$,
\begin{equation}
 \begin{aligned}
 &\sum_i m_i\vct w_i\cdot\Dtt{\vct u_i}
 +\sum_{\{i,j\}\in\mathcal E}
       (\Delta_{ij}\vct w)^{\mathsf T}
                  \mathsf C_{ij}\Delta_{ij}\vct u
 \\
 &+\vct w^{\mathsf T}\mathsf K_{\rm sup}\vct u
       =\sum_i\vct w_i\cdot\vct f_i.
 \end{aligned}
 \label{eq:structure-finite-virtual-power}
\end{equation}
\end{proposition}
\begin{proof}
Dot each site equation with its test velocity and sum. The two endpoint
terms of each pair combine into
$(\Delta_{ij}\vct w)^{\mathsf T}\mathsf C_{ij}\Delta_{ij}\vct u$.
Ideal fixed reactions pair to zero with admissible test velocities.
\end{proof}
Each term is power. Reversing an edge reverses both differences and
preserves their pairing. Symmetric pair curvatures follow from the
energy Hessian, including when prestress gives signed curvature.
Setting $\vct w=\Dt{\vct u}$ yields the quadratic energy balance:
the rate of kinetic-plus-potential energy equals applied power.
Ideal fixed reactions do zero work on admissible test velocities
~\cite{Dove2011}.

Summing the force equations over sites $S$ cancels their internal pairs.
The remaining interaction force is
$-\sum_{i\in S,j\notin S:\{i,j\}\in\mathcal E}
\mathsf C_{ij}(\vct u_i-\vct u_j)$, alongside applied and support
forces. A cut through the network therefore transmits the forces on its
crossing edges. Continuum volume and surface integrals must preserve
this work pairing, which leads to the traction defined below.

Figure~\ref{fig:structure-cut-and-virtual-power} identifies the crossing-edge forces and their endpoint power pairing before the continuum approximation.

\begin{figure}[H]
\centering\begingroup
\definecolor{vltTeal}{HTML}{176B64}
\definecolor{vltBlue}{HTML}{315D80}
\definecolor{vltGold}{HTML}{A87520}
\definecolor{vltPlum}{HTML}{725877}
\definecolor{vltRed}{HTML}{762B38}
\definecolor{vltYellow}{HTML}{D5AD3C}
\definecolor{vltNeutral}{HTML}{777D83}
\setlength{\unitlength}{1pt}
\scalebox{0.90}{%
\input{vlt-figures/figure-058.tex}%
}\endgroup
\caption{(a) Twelve ideal virions occupy $\vct R_i/a\in\{-1,0,1\}\times\{-1/2,1/2\}^2$. The set $S$ contains the eight sites at $x/a=-1,0$; four of the 20 nearest-axis pairs cross the dashed plane $x/a=1/2$. Each centre and connector uses one orthographic scale; body identities persist across the cut. Teal highlights membership and crossing pairs here. (b) A separate longitudinal edge test has reference centres $0,a$, $u_i=0.12a$, $u_j=0$ and unstressed curvature $k>0$. The shortened seven-turn connector follows the moved surfaces. Teal arrows denote displacement and blue arrows equal and opposite incremental forces, with separate gains. The lower work pairing holds for the general symmetric edge tensor, including signed prestress. Coils in (a) specify the retained network at an auxiliary unstressed reference; no force balance or rigidity of a loaded preparation is inferred from the drawing.}
\label{fig:structure-cut-and-virtual-power}
\end{figure}

Begin with identical masses in an unpinned, translation-invariant pair
reference. The linear force equation is a difference equation on sites:
\begin{equation}
 m\Pdtt{\vct u}(\vct R,t)=
 \sum_{\vct d}\mathsf C(\vct d)
      [\vct u(\vct R+\vct d,t)-\vct u(\vct R,t)].
 \label{eq:bulk-difference-equation}
\end{equation}
The sum includes both directions of each bulk offset, with
$\mathsf C(-\vct d)=\mathsf C(\vct d)$. Each relative displacement
produces a directional force; common translation gives zero.
The field values here are the displacements of the particle centres.

\begin{definition}[Long-wavelength comparison]
\label{def:s3-long-wavelength-comparison}
A long-wavelength comparison uses an interpolation reproducing the
site values and varying over a length $\ell$ large compared with the
retained interaction range. For the finite shell set this is
$\ell\gg a$; a Fourier component satisfies $|\vct k|a\ll1$.
\end{definition}

Long wavelength and small amplitude are independent requirements.
A small alternating displacement can satisfy harmonic mechanics while
varying too rapidly for a continuum approximation.

\begin{proposition}[Opposite-offset expansion]
\label{prop:s3-opposite-offset-expansion}
For a displacement interpolation with bounded sixth spatial
derivatives along the expanded line segments, opposite offsets
cancel odd Taylor terms and give
\begin{equation}
 \begin{aligned}
 \vct u(\vct R+\vct d,t)+\vct u(\vct R-\vct d,t)-2\vct u(\vct R,t)
   &=d_\mu d_\nu\partial_\mu\partial_\nu\vct u\\
   &\quad
      +\frac1{12}d_\mu d_\nu d_\alpha d_\beta
          \partial_\mu\partial_\nu\partial_\alpha\partial_\beta\vct u
      +\vct r_6.
 \end{aligned}
 \label{eq:structure-opposite-offset-expansion}
\end{equation}
Repeated Cartesian indices are summed and derivatives are evaluated at
$(\vct R,t)$. The remainder obeys $|\vct r_6|\le C_6|\vct d|^6$
for a bound $C_6$ on the sixth-order Taylor remainders.
\end{proposition}
\begin{proof}
Apply the one-dimensional Taylor formula along $\vct R+s\vct d$
to fifth order, with integral sixth-order remainder. Adding its
values at $s=1$ and $s=-1$ cancels odd terms. The surviving even
coefficients are $2/2!=1$ and $2/4!=1/12$.
\end{proof}
The constant $C_6$ has units of displacement per length to the sixth
power. The second derivative gives the leading continuum force, and
the fourth gives a dispersion correction. For a nonzero leading
acoustic curvature, their relative scale is $(a/\ell)^2$, with
model-dependent coefficients. A vanishing leading term requires a
higher-gradient analysis. The truncated expansion is used only within
its wavelength band; extrapolating it can introduce artificial instability.

A scalar spring chain makes the approximation quantitative.

\begin{example}[Exact scalar-chain continuum comparison]
\label{ex:s3-chain-continuum}
Take an infinite or auxiliary periodic chain with mass $m>0$, spacing
$a$ and nearest-neighbor longitudinal stiffness $k_{\rm ch}>0$. For scalar site motion $u_j$, its force law is
$m\Dtt{u_j}=k_{\rm ch}(u_{j+1}-2u_j+u_{j-1})$. Substituting
$u_j=A\exp[i(kja-\omega t)]$ gives
\begin{equation}
 \begin{aligned}
 \omega_{\rm ch}^2(k)&=\frac{4k_{\rm ch}}m\sin^2\frac{ka}{2},
       &c_{\rm ch}^2&=\frac{k_{\rm ch}a^2}m,\\
 \frac{\omega_{\rm ch}^2(k)}{c_{\rm ch}^2 k^2}
   &=\left[\frac{\sin(ka/2)}{ka/2}\right]^2
     =1-\frac{(ka)^2}{12}+O((ka)^4).
 \end{aligned}
 \label{eq:structure-chain-continuum-check}
\end{equation}
The ratio is defined by continuity at $k=0$. The leading wave equation
has $\omega^2=c_{\rm ch}^2k^2$, while the exact difference equation
retains the spacing-dependent dispersion~\cite{Misra2012}. The speed
has units $\mathrm{m\,s^{-1}}$; the ratio is dimensionless. This chain
is a check of the limiting argument, not a numerical material law for
virions.
\end{example}

Figure~\ref{fig:structure-discrete-continuum} shows the site-to-field
mapping and the departure of the two predictions as wavelength decreases.

\begin{figure}[H]
\centering\begingroup
\definecolor{vltTeal}{HTML}{176B64}
\definecolor{vltBlue}{HTML}{315D80}
\definecolor{vltGold}{HTML}{A87520}
\definecolor{vltPlum}{HTML}{725877}
\definecolor{vltRed}{HTML}{762B38}
\definecolor{vltYellow}{HTML}{D5AD3C}
\definecolor{vltNeutral}{HTML}{777D83}
\setlength{\unitlength}{1pt}
\scalebox{0.90}{%
\input{vlt-figures/figure-059.tex}%
}\endgroup
\caption{(a) The longitudinal snapshot $u_j/A=\sin(\pi j/4)$ uses nine rigid virions, one position scale and a displacement gain of $7\,\mathrm{pt}/A$. The ruler marks the fixed reference coordinates $X/a=j$; bodies and matched coils follow the displaced centres. (b) Dots record the same site amplitudes and the curve is their chosen interpolation. (c) The auxiliary scalar-chain squared-frequency ratio has a declared $5\%$ tolerance at $|k|a\simeq0.783$, or $\lambda/a\simeq8.03$. Shading identifies the accepted wavelength band. The amplitude can be chosen within the harmonic regime; the cylinder retains its finite boundary problem.}
\label{fig:structure-discrete-continuum}
\end{figure}

\begin{definition}[Leading homogenized particle network]
\label{def:s3-homogenized-particle-network}
For the homogeneous three-dimensional one-site comparison with identical
masses, primitive volume $a^3$ and inversion-paired offsets, set
$\rho=m/a^3$. Its leading gradient-stiffness tensor and continuum
equation are
\begin{equation}
 \mathcal A_{p\mu q\nu}=
   \frac{1}{2a^3}\sum_{\vct d}
        C_{pq}(\vct d)d_\mu d_\nu,
 \qquad
 \rho\Pdtt{u_p}=\partial_\mu
            (\mathcal A_{p\mu q\nu}\partial_\nu u_q).
\label{eq:continuum-tensor}
\end{equation}
\end{definition}
The factor $1/2$ accounts for the two directions of each offset. The
tensor maps a displacement gradient to stress and then, through its
divergence, to force per volume. Its units are pressure:
$(\mathrm{N\,m^{-1}})\mathrm{m^2}/\mathrm{m^3}=\mathrm{Pa}$.
It is obtained from the chosen interactions; isotropy has not been
assumed.

\begin{definition}[Directional acoustic matrix]
\label{def:s3-directional-acoustic-matrix}
For the leading homogenized network and a unit propagation direction
$\vct{\hat k}$, the acoustic matrix is
\begin{equation}
 \bigl[\mathsf G_{\rm ac}(\vct{\hat k})\bigr]_{pq}
       =\rho^{-1}\mathcal A_{p\mu q\nu}
                               \hat k_\mu\hat k_\nu .
 \label{eq:directional-acoustic-matrix}
\end{equation}
\end{definition}
The eigenvalues are squared directional acoustic speeds
$c_\nu(\vct{\hat k})^2$. The notation $\mathsf G_{\rm ac}$ separates
this matrix from the medium memory kernel $\mathsf\Gamma(t)$.
Positivity for every direction and retained polarization tests
long-wavelength restoring response. Stability at other wavevectors
and at finite boundaries is checked on the corresponding full operator.
Anisotropy can tilt both polarization and group velocity relative to
the wavevector.

These speeds describe collective motion of whole-particle centres.
The density $\rho=m/a^3$ is that of the uniformly occupied reference
network; medium inertia and internal particle elasticity enter through
their own retained coordinates. For the occupied region
$\mathcal V_{\mathrm{lat}}$, the bulk estimate
$mn\simeq\rho|\mathcal V_{\mathrm{lat}}|$ has boundary corrections,
while the finite mass is exactly $mn$. This homogenization holds at
specified spacing. A refinement limit changing $a$ must also specify
mass and stiffness scaling.

\begin{proposition}[Continuum boundary power]
\label{prop:s3-continuum-boundary-power}
For a sufficiently smooth solution of the unforced continuum equation
on a fixed material domain, with constant real coefficients satisfying
$\mathcal A_{p\mu q\nu}=\mathcal A_{q\nu p\mu}$, energy and traction obey
\begin{equation}
 \begin{split}
 E_{\mathrm c}&=\frac12\int_{\mathcal V_{\mathrm{lat}}}
       [\rho|\Pdt{\vct u}|^2+
        (\partial_\mu u_p)\mathcal A_{p\mu q\nu}
                                  (\partial_\nu u_q)]\,\dd^3r,\\
 t_p&=n_\mu\mathcal A_{p\mu q\nu}\partial_\nu u_q,
 \qquad
 \Dt{E_{\mathrm c}}=\int_{\partial\mathcal V_{\mathrm{lat}}}
                                  \Pdt{\vct u}\cdot\vct t\,\dd S.
 \end{split}
 \label{eq:continuum-boundary-power}
\end{equation}
\end{proposition}
\begin{proof}
Differentiate the two quadratic energy terms. Symmetry of
$\mathcal A$ combines the two gradient derivatives. Substitute the
field equation into the kinetic term and integrate by parts; the
volume terms cancel and leave the stated boundary pairing.
\end{proof}
The traction is force per area exerted at the material surface, and its
scalar product with velocity is boundary power per area. A fixed-displacement boundary has zero velocity; a free boundary
has zero traction. Each is lossless. A stationary slip boundary imposes
$\vct u\cdot\vct n=0$ and zero tangential traction, so its normal
reaction also does no work~\cite{Landau1986}.

A collision-free finite graph has a free material surface where bonds
end, subject to any specified smooth support forces. An engaged slip
boundary adds a normal displacement constraint. These domains describe
different preparations, each with its own boundary power; the finite
impact rule alone leaves that continuum choice open.

For unstressed central springs,
$\mathsf C(\vct d)=k(\vct d)\vct n\vct n^{\mathsf T}$,
$\vct n=\vct d/|\vct d|$. Only projected extension contributes, so the
elastic energy can be written using symmetric strain
$\epsilon_{pq}=(\partial_pu_q+\partial_qu_p)/2$ and
\begin{equation}
 \mathcal C_{pqrs}^{\mathrm{el}}
       =\frac1{2a^3}\sum_{\vct d}k(\vct d)|\vct d|^2
                     n_pn_qn_rn_s.
 \label{eq:unstressed-elastic-tensor}
\end{equation}
Symmetric strain measures local stretching and shear while removing
infinitesimal rigid rotation from this unstressed energy. Under prestress,
the full gradient tensor must retain the transverse terms of the discrete
Hessian and the corresponding initial-stress treatment. A leading
acoustic test and positivity of every admissible finite deformation are
different requirements. The continuum construction carries forward the
stated mechanical assumptions instead of assigning an isotropic
constitutive law from particle shape alone.

\paragraph{An explicit wavelength-error estimate.}
\begin{proposition}[Fourth-moment continuum error]
\label{prop:s3-fourth-moment-error}
For the identical-mass one-site comparison in
\eqref{eq:bulk-difference-equation}, with real symmetric
$\mathsf C(-\vct d)=\mathsf C(\vct d)$ and finite absolute fourth
interaction moment $M_4$, define its exact Fourier symbol
$\mathsf D(\vct k)=m^{-1}\sum_{\vct d}
\mathsf C(\vct d)[1-\cos(\vct k\cdot\vct d)]$
and its leading continuum symbol
$[\mathsf D_2(\vct k)]_{pq}
=\rho^{-1}\mathcal A_{p\mu q\nu}k_\mu k_\nu$.
The sum includes both directions. Then
\begin{equation}
 \begin{aligned}
 \|\mathsf D(\vct k)-\mathsf D_2(\vct k)\|_2
 &\le\frac1{24m}\sum_{\vct d}
       \|\mathsf C(\vct d)\|_2|\vct k\cdot\vct d|^4
 \le\frac{M_4}{24m}|\vct k|^4,\\
 M_4&:=\sum_{\vct d}\|\mathsf C(\vct d)\|_2|\vct d|^4.
 \end{aligned}
 \label{eq:structure-fourth-moment-symbol-bound}
\end{equation}
Every ordered eigenvalue difference between the two symmetric symbols
is bounded by the same right-hand side.
\end{proposition}
\begin{proof}
The scalar Taylor estimate
$|1-\cos x-x^2/2|\le |x|^4/24$, the triangle inequality and
$|\vct k\cdot\vct d|\le|\vct k||\vct d|$ give the operator bound.
Absolute convergence justifies the sum. The min--max principle then
bounds the ordered eigenvalues.
\end{proof}
This absolute bound permits signed prestress tensors. Its units are
$\mathrm{s^{-2}}$; the eigenvalue bounds are squared-frequency
differences for stable branches. If the
smallest directional acoustic speed has a positive lower bound
$c_{\min}$, division by $c_{\min}^2|\vct k|^2$ supplies a uniform
relative estimate against the continuum squared frequencies away from
$\vct k=\vct0$.
A vanishing leading speed cannot use that denominator.

Finite shells make $M_4$ finite. An infinite-range comparison requires
convergence of this absolute interaction moment before applying the
estimate; unscreened Coulomb interactions do not satisfy it merely
because the finite-cylinder pair energy exists. Charge cancellation
or a specified electrostatic boundary treatment needs its own limiting
analysis. The finite cylinder itself retains its site equations and
boundary conditions, rather than acquiring plane-wave eigenvectors
from this bulk error estimate.

\paragraph{A continuum comparison that retains body rotation.}
A slowly varying patch-locked assembly can require both a centre
displacement $\vct u(\vct r)$ and an independent small body rotation
$\vct\theta(\vct r)$. A micropolar comparison uses
\begin{equation}
 \mathsf E=\nabla\vct u-[\vct\theta]_\times,
 \qquad \mathsf K_\theta=\nabla\vct\theta,
 \qquad (\nabla\vct u)_{ij}=\partial_j u_i .
 \label{eq:s3patch-micropolar-strains}
\end{equation}
Both vanish under $\vct u=\vct a+\vct\psi\times\vct r$,
$\vct\theta=\vct\psi$. This is the rigid-motion structure of
micropolar elasticity~\cite{Eringen1966,Eringen1968NASA}, applied here
to whole-particle rotations. For a matched quadratic energy density
$w(\mathsf E,\mathsf K_\theta)$, including allowed cross terms,
define $\mathsf\sigma=\partial_{\mathsf E}w$ and
$\mathsf m=\partial_{\mathsf K_\theta}w$. Internal virtual power is
$\int[\mathsf\sigma:\dot{\mathsf E}
+\mathsf m:\dot{\mathsf K}_\theta]\,\dd V$; integration by parts
gives traction $\mathsf\sigma\vct n$ and couple traction
$\mathsf m\vct n$, paired with $\dot{\vct u}$ and
$\dot{\vct\theta}$. Stress has pressure units; couple stress has
force-per-length units. Its angular balance retains the skew stress.
Coefficients and spin inertia are matched to the actual cell energy
and its stated internal relaxation. A displacement-only approximation
then requires a controlled elimination of rotation. Long wavelengths
alone do not establish that elimination or isotropic coefficients.

\subsection{Directional matrices for the neighbor shells}
\label{sec:shell-matrices}

The shell classification now supplies explicit $3\times3$
force-constant blocks.

\begin{definition}[Homogeneous shell principal curvatures]
\label{def:s3-shell-principal-curvatures}
When the radial law is the same for every direction within a shell,
its longitudinal and transverse curvatures and directional tensor are
\begin{equation}
 \begin{aligned}
 &r_\beta=a,\qquad r_\gamma=\sqrt2a,\qquad r_\Omega=\sqrt3a,\\
 &L_\eta=V_\eta''(r_\eta),\qquad
 T_\eta=\frac{V_\eta'(r_\eta)}{r_\eta},\\
 &\mathsf C_\eta(\vct n)=T_\eta\Id_3+
                           (L_\eta-T_\eta)\vct n\vct n^{\mathsf T}.
 \end{aligned}
 \label{eq:shell-curvatures}
\end{equation}
\end{definition}
The coefficient $L_\eta$ is curvature along the pair separation and
$T_\eta$ is transverse curvature, both in $\mathrm{N\,m^{-1}}$.
Here $T_\eta$ denotes a stiffness coefficient, distinct from temperature
$T$. The matrix has eigenvalue $L_\eta$ along $\vct n$ and two eigenvalues
$T_\eta$ in the perpendicular plane. Thus its meaning is directional
restoring force, rather than three unrelated matrix entries. If every
pair uses the same radial potential, $V_\eta$ means that potential
evaluated at the shell distance; shell-specific laws are an additional
model choice.

Figure~\ref{fig:structure-mirrored-shell-forces} shows how mirrored bond directions cancel their cross-axis force components before the complete cubic-shell sum is assembled.

\begin{figure}[H]
\centering\begingroup
\definecolor{vltTeal}{HTML}{176B64}
\definecolor{vltBlue}{HTML}{315D80}
\definecolor{vltGold}{HTML}{A87520}
\definecolor{vltPlum}{HTML}{725877}
\definecolor{vltRed}{HTML}{762B38}
\definecolor{vltYellow}{HTML}{D5AD3C}
\definecolor{vltNeutral}{HTML}{777D83}
\setlength{\unitlength}{1pt}
\scalebox{0.90}{%
\input{vlt-figures/figure-060.tex}%
}\endgroup
\caption{Two mirrored face-diagonal directions $\vct n_\pm=(1,\pm1,0)/\sqrt2$ have $L=2k_\star$, $T=k_\star/2$, and fixed neighbors. A separate test displacement $\vct d=\ell_\star\vct e_x$ gives $\vct F_\pm=-\mathsf C_\pm\vct d$. The force panels share equal Cartesian scales of $30\,\mathrm{pt}/(k_\star\ell_\star)$. Their transverse components cancel; their sum is $(-5/2,0)k_\star\ell_\star$. The virion panel locates the reference directions, and the vector panels evaluate this incremental force test. A complete cubic-shell sum additionally uses all retained directions and their actual weights.}
\label{fig:structure-mirrored-shell-forces}
\end{figure}

For the illustrative law~\eqref{eq:pair-potential}, write
$A_{ij}=q_iq_j/(4\pi\varepsilon_m)$. Its screened derivatives are
\begin{equation}
 V_{ij}^{\mathrm{scr}\,\prime}(r)
     =-A_{ij}e^{-\kappa r}
                  \left(\frac\kappa r+\frac1{r^2}\right),\qquad
 V_{ij}^{\mathrm{scr}\,\prime\prime}(r)
     =A_{ij}e^{-\kappa r}
             \left(\frac{\kappa^2}r+\frac{2\kappa}{r^2}+
                                      \frac2{r^3}\right).
 \label{eq:screened-derivatives}
\end{equation}
The Lennard--Jones derivatives are
\begin{equation}
 \begin{split}
 V_{ij}^{\mathrm{LJ}\,\prime}(r)&=
       \frac{24\epsilon_{ij}}r
       \left[\left(\frac{\sigma_{ij}}r\right)^6
              -2\left(\frac{\sigma_{ij}}r\right)^{12}\right],\\
 V_{ij}^{\mathrm{LJ}\,\prime\prime}(r)&=
       \frac{24\epsilon_{ij}}{r^2}
       \left[26\left(\frac{\sigma_{ij}}r\right)^{12}
              -7\left(\frac{\sigma_{ij}}r\right)^6\right].
 \end{split}
 \label{eq:lj-derivatives}
\end{equation}
Adding the electrostatic and Lennard--Jones derivatives gives the two
principal curvatures used in assembly. First derivatives have force
units and enter reference balance; second derivatives have force-per-length
units. A common potential evaluated at several shell distances generally
produces a prestressed reference.

Prestress makes the transverse coefficient particularly informative.
For a compressed central spring, $V'(r)<0$ and hence $T_\eta<0$:
transverse motion can reduce its stored energy to quadratic order even
when its longitudinal curvature is positive. Conversely, at a rest
length $V'(r)=0$, transverse quadratic curvature vanishes. The full
network, including admissible motions and support, decides stability.

\paragraph{A signed electrostatic spring tensor.}
For the ideal virion monopole model, set
$A_{ij}=q_iq_j/(4\pi\varepsilon_m)$ and let $R=R_{ij}>0$.
Substituting~\eqref{eq:screened-derivatives} into the Cartesian edge law
gives
\begin{equation}
 \begin{aligned}
 \mathsf C_{ij}^{\rm el}
  &=\frac{A_{ij}e^{-\kappa R}}{R^3}
    \left[\bigl((\kappa R)^2+3\kappa R+3\bigr)
                  \vct n\vct n^{\mathsf T}
                  -(\kappa R+1)\Id_3\right],\\
 \kappa=0:\qquad
 \mathsf C_{ij}^{\rm C}
  &=\frac{A_{ij}}{R^3}(3\vct n\vct n^{\mathsf T}-\Id_3),
 \qquad L_{ij}^{\rm C}=\frac{2A_{ij}}{R^3},\quad
 T_{ij}^{\rm C}=-\frac{A_{ij}}{R^3},\quad
 \tr\mathsf C_{ij}^{\rm C}=0.
 \end{aligned}
 \label{eq:vlp-electrostatic-cartesian-curvature}
\end{equation}
For like-sign charges, the longitudinal Coulomb curvature is positive
while both transverse curvatures are negative; reversing the pair product
reverses those signs. Thus a repulsive force, an attractive force and a
positive scalar radial curvature are different statements. The complete
edge tensor is the sum of this electrostatic contribution and the
Lennard--Jones contribution derived from~\eqref{eq:lj-derivatives}.
The incremental forces are
$\delta\vct f_i^{(ij)}=-\mathsf C_{ij}(\vct u_i-\vct u_j)$ and
$\delta\vct f_j^{(ij)}=-\delta\vct f_i^{(ij)}$.
This is the precise Cartesian meaning of the connecting spring.

\begin{proposition}[Finite Coulomb trace obstruction]
\label{prop:s3-coulomb-trace-obstruction}
For the finite unscreened central-pair Coulomb operator on the full
three-dimensional translational space, with positive scalar node masses,
$\tr\mathsf K_{\rm C}=\tr\mathsf D_{\rm C}=0$. If this real symmetric
operator is nonzero, it has both positive and negative eigenvalues and
cannot be a nonnegative harmonic restoring operator.
\end{proposition}
\begin{proof}
Equation~\eqref{eq:vlp-electrostatic-cartesian-curvature} makes each
pair tensor traceless. A diagonal block of $\mathsf K_{\rm C}$ is
their sum; the corresponding block of $\mathsf D_{\rm C}$ is divided
by the positive scalar $m_i$. Both traces vanish. A nonzero symmetric
matrix with zero eigenvalue sum has eigenvalues of both signs.
\end{proof}
The stability test therefore assembles Coulomb, short-range and support
terms on the admissible coordinate space. The trace obstruction applies
to the full finite unscreened pair operator. Screening, electric boundary
terms or restricted motion change that operator and its premises.

Figure~\ref{fig:structure-coulomb-principal-curvatures} separates the longitudinal and two transverse Coulomb curvatures for either sign of the pair product.

\begin{figure}[H]
\centering\begingroup
\definecolor{vltTeal}{HTML}{176B64}
\definecolor{vltBlue}{HTML}{315D80}
\definecolor{vltGold}{HTML}{A87520}
\definecolor{vltPlum}{HTML}{725877}
\definecolor{vltRed}{HTML}{762B38}
\definecolor{vltYellow}{HTML}{D5AD3C}
\definecolor{vltNeutral}{HTML}{777D83}
\setlength{\unitlength}{1pt}
\scalebox{0.90}{%
\input{vlt-figures/figure-061.tex}%
}\endgroup
\caption{The three principal Cartesian curvatures of one unscreened Coulomb edge, \eqref{eq:vlp-electrostatic-cartesian-curvature}. Signs on the same-sized ideal-virion glyphs specify the charge product. The open arrow runs from $j$ to $i$ and defines the reference direction, with no assigned force magnitude. In units $k_{\rm C}=|q_iq_j|/(4\pi\varepsilon_mR^3)$, radial/transverse eigenvalues are $(2,-1,-1)$ for like signs and $(-2,1,1)$ for opposite signs. Equal bar axes, a black zero line and signed values expose the vanishing trace. $T_1,T_2$ are separate orthogonal transverse components. The complete assembly adds its short-range, support and constraint contributions to these signed edge curvatures.}
\label{fig:structure-coulomb-principal-curvatures}
\end{figure}

\begin{example}[Transverse quartic energy]
\label{ex:s3-transverse-quartic}
Consider the generic radial spring
$V_{\rm ex}(r)=k_{\rm ex}(r-a)^2/2$, with $k_{\rm ex}>0$.
Hold its longitudinal separation at $a$ and add a transverse relative
shift $\delta$. Its extension begins at second order:
\begin{equation}
 \Delta V_{\rm ex}
   =\frac{k_{\rm ex}}2\bigl(\sqrt{a^2+\delta^2}-a\bigr)^2
   =\frac{k_{\rm ex}\delta^4}{8a^2}
      +O\!\left(\frac{k_{\rm ex}\delta^6}{a^4}\right),
 \qquad |\delta|/a\ll1 .
 \label{eq:structure-transverse-quartic}
\end{equation}
The harmonic energy misses this quartic cost along the stated straight
displacement path. A finite rotation can also change the longitudinal
component to preserve length. Whether all bonds of a network permit
such a finite deformation is therefore a separate geometric question
from finding a zero Hessian vector.
\end{example}

\paragraph{Rigid rotation and the reference force.}
\begin{proposition}[Rotation, gradient and tangent curvature]
\label{prop:s3-rotation-gradient}
Let the twice continuously differentiable energy satisfy
$U(e^{\theta\mathsf A}\vct r)=U(\vct r)$ near the reference for the
tested rotations, where
$\mathsf A=\Id_n\otimes\mathsf\Omega$ and
$\mathsf\Omega^{\mathsf T}=-\mathsf\Omega$ is dimensionless.
For $\vct g=\nabla U(\vct R)$ and $\mathsf K=\nabla^2U(\vct R)$,

\begin{equation}
 \mathsf K\mathsf A\vct R=\mathsf A\vct g,
 \qquad
 (\mathsf A\vct R)^{\mathsf T}\mathsf K(\mathsf A\vct R)
       +\vct g^{\mathsf T}\mathsf A^2\vct R=0.
 \label{eq:structure-rotation-gradient-identity}
\end{equation}
\end{proposition}
\begin{proof}
Rotational invariance gives gradient covariance
$\nabla U(e^{\theta\mathsf A}\vct R)=e^{\theta\mathsf A}\vct g$.
Differentiate once at zero for the first identity. Differentiating the
constant scalar $U(e^{\theta\mathsf A}\vct R)$ twice gives the second.
\end{proof}
The second relation is the second derivative of the constant energy
$U(e^{\theta\mathsf A}\vct R)$. An exact rotation has both a
first-order tangent $\mathsf A\vct R$ and a second-order acceleration
$\mathsf A^2\vct R$ in Cartesian configuration space. At a
nonstationary reference, the gradient contribution from that acceleration
cancels the apparent quadratic cost of following only the straight
tangent. Dropping it would incorrectly interpret a change of direction
as a restoring force against a true rigid rotation.

For a free stationary assembly, $\vct g=0$ and every independent
rotational tangent is a Hessian null vector even when individual edges
carry nonzero tension. This cancellation is global; removing transverse
terms from the individual edge tensors would change other physical
motions. If external terms balance the pair forces, the pair-only
gradient need not vanish, and its Hessian alone need not annihilate a
rotation. The actual total energy must be tested for the symmetry in
question. An axially symmetric smooth cylinder can preserve rotation
about its axis while breaking rotations that tilt the array; an
asymmetric target can break the remaining symmetry. Fixed constraints
must also admit the whole symmetry path before it is counted as a free
rigid motion. Thus particle shape, freedom to rotate the array within
the available clearance, and rotational invariance of the declared
center-coordinate energy are separate physical statements.

Figure~\ref{fig:structure-rotation-tangent-and-path} compares a straight tangent displacement with the exact rotation path that supplies the second-order position change.

\begin{figure}[H]
\centering\begingroup
\definecolor{vltTeal}{HTML}{176B64}
\definecolor{vltBlue}{HTML}{315D80}
\definecolor{vltGold}{HTML}{A87520}
\definecolor{vltPlum}{HTML}{725877}
\definecolor{vltRed}{HTML}{762B38}
\definecolor{vltYellow}{HTML}{D5AD3C}
\definecolor{vltNeutral}{HTML}{777D83}
\setlength{\unitlength}{1pt}
\scalebox{0.90}{%
\input{vlt-figures/figure-062.tex}%
}\endgroup
\caption{A pair starts at centres $(0,0)$ and $(R,0)$. The straight tangent increment in (a) reaches $(R,R\theta)$, whereas the exact rotation in (b) follows $(R\cos\theta,R\sin\theta)$ at constant separation. Both use the same body size and position scale, with $\theta=0.5$ exaggerated for visibility. Derivatives $\vct v,\vct a$ are with respect to angle. For the radial energy, the gradient contribution from the curved path cancels its tangent-curvature contribution as in~\eqref{eq:structure-rotation-gradient-identity}. The straight path changes length by $R\theta^2/2+O(\theta^4)$; an admissible rigid rotation at a free stationary reference is a Hessian null direction.}
\label{fig:structure-rotation-tangent-and-path}
\end{figure}

For an axial direction,
\begin{equation}
 \mathsf C_\beta(\vct e_x)=
 \begin{pmatrix}L_\beta&0&0\\0&T_\beta&0\\0&0&T_\beta\end{pmatrix}.
 \label{eq:beta-matrix}
\end{equation}
For a face diagonal
$\vct n=(\vct e_x+\sigma\vct e_y)/\sqrt2$, $\sigma=\pm1$,
\begin{equation}
 \mathsf C_\gamma(\vct n)=
 \begin{pmatrix}
 (L_\gamma+T_\gamma)/2&\sigma(L_\gamma-T_\gamma)/2&0\\
 \sigma(L_\gamma-T_\gamma)/2&(L_\gamma+T_\gamma)/2&0\\
 0&0&T_\gamma
 \end{pmatrix}.
 \label{eq:gamma-matrix}
\end{equation}
For a body diagonal
$\vct n=(\sigma_x,\sigma_y,\sigma_z)^{\mathsf T}/\sqrt3$,
\begin{equation}
 \mathsf C_\Omega(\vct n)=\frac13
 \begin{pmatrix}
 L_\Omega+2T_\Omega&(L_\Omega-T_\Omega)\sigma_x\sigma_y&
                         (L_\Omega-T_\Omega)\sigma_x\sigma_z\\
 (L_\Omega-T_\Omega)\sigma_x\sigma_y&L_\Omega+2T_\Omega&
                         (L_\Omega-T_\Omega)\sigma_y\sigma_z\\
 (L_\Omega-T_\Omega)\sigma_x\sigma_z&
 (L_\Omega-T_\Omega)\sigma_y\sigma_z&L_\Omega+2T_\Omega
 \end{pmatrix}.
 \label{eq:omega-shell-matrix}
\end{equation}
Coordinate permutations give the remaining directions, and bond reversal
preserves the tensor. An oblique bond can turn motion along one laboratory
axis into force along another; its off-diagonal entries express that
geometry. Global assembly uses $\mathsf K_{ij}=-\mathsf C_{ij}$ and
$\mathsf D_{ij}=-\mathsf C_{ij}/\sqrt{m_im_j}$.

Figure~\ref{fig:structure-shell-stiffness-map} visualizes one directional shell matrix as an exact displacement-to-force map in laboratory coordinates.

\begin{figure}[H]
\centering\begingroup
\definecolor{vltTeal}{HTML}{176B64}
\definecolor{vltBlue}{HTML}{315D80}
\definecolor{vltGold}{HTML}{A87520}
\definecolor{vltPlum}{HTML}{725877}
\definecolor{vltRed}{HTML}{762B38}
\definecolor{vltYellow}{HTML}{D5AD3C}
\definecolor{vltNeutral}{HTML}{777D83}
\setlength{\unitlength}{1pt}
\scalebox{0.80}{%
\input{vlt-figures/figure-063.tex}%
}\endgroup
\caption{The directional edge map of \eqref{eq:shell-curvatures}, with $\vct n=(1,1,0)/\sqrt2$, longitudinal curvature $2k_\star$ and transverse curvature $k_\star/2$. (a) Unit relative displacements form the laboratory-component circle. (b) Their negative-gradient forces form an ellipse of principal radii $2$ along $\vct n$ and $1/2$ perpendicular to it. Both panels use the same normalized component gain and equal Cartesian scales. Matching arrow colors register the longitudinal and laboratory-$x$ tests; open heads denote the test displacement and filled heads its restoring force. The second transverse direction is outside this planar section. This is one edge contribution; full-network stability also depends on its assembled neighbours, supports and reference.}
\label{fig:structure-shell-stiffness-map}
\end{figure}

\paragraph{Charge heterogeneity and shell symmetry.}
The reduction to one pair of coefficients $(L_\eta,T_\eta)$ per shell
assumes a homogeneous interaction environment. In a mixed-charge array,
particles at the same distance can instead have different products
$q_iq_j$, hence different $L_{ij}$ and $T_{ij}$ even when their bodies
and short-range parameters are identical. The correct local sum is
over those actual pair tensors. For distinct Cartesian axes,
\begin{equation}
 (\mathsf K_{ii,\rm pair})_{\mu\nu}
   =\sum_{j:\{i,j\}\in\mathcal E}
         (L_{ij}-T_{ij})n_{ij,\mu}^{(0)}n_{ij,\nu}^{(0)},
 \qquad \mu\ne\nu.
 \label{eq:structure-charge-weighted-shell-symmetry}
\end{equation}
Mirrored directions cancel only when their curvature weights share the
required symmetry. The cubic reduction below applies when the charge-dependent
law and retained edges transform together under cubic rotations.
Missing neighbors or asymmetric charge patterns can break that symmetry.
Shell labels still specify geometric relationships; mass and charge
remain attributes of each physical virion.

\begin{proposition}[Complete homogeneous cubic local sum]
\label{prop:s3-cubic-local-sum}
For a complete cubic neighborhood with one common pair of coefficients
$(L_\eta,T_\eta)$ per shell, the pair contribution to the central
diagonal block is
\begin{equation}
 \mathsf K_{\alpha,\mathrm{pair}}=
 \left[2(L_\beta+2T_\beta)+4(L_\gamma+2T_\gamma)
                  +\frac83(L_\Omega+2T_\Omega)\right]\Id_3.
 \label{eq:central-shell-sum}
\end{equation}
\end{proposition}
\begin{proof}
For each complete shell of $N_\eta$ directions, mirrored directions
cancel off-diagonal dyads and cubic permutations equalize the
diagonals. Since each unit dyad has trace one,
$\sum_{\vct n}\vct n\vct n^{\mathsf T}=(N_\eta/3)\Id_3$.
Its tensor sum is $N_\eta(L_\eta+2T_\eta)\Id_3/3$.
Insert $N_\eta=6,12,8$.
\end{proof}
Cubic symmetry makes this central block isotropic for a displacement
with the neighbors held fixed. Collective motion instead samples the
correlations between moving sites, so its elastic response and frequencies
require the global matrix. Boundary sites retain their actual edge sums
and any smooth support curvature.

Imaging constrains contact directions; independent force or response
measurements constrain their curvatures. Together they specify the
whole-particle displacement model. Intrinsic capsid deformation or
orientation-dependent interactions enter through additional coordinates
and their joint energy.

\subsection{Global assembly, normal modes and spectral bounds}
\label{sec:global-assembly}

Directional force constants, inertia and supports determine the
collective normal modes. In each positive mode, particle displacement
ratios remain fixed while one normal coordinate oscillates. The same
assembly applies to regular references and irregular finite arrays.

Let $\mathsf E_{ij}$ extract relative displacement:
$\mathsf E_{ij}\vct u=\vct u_i-\vct u_j$. Each unordered edge enters
the global stiffness once:
\begin{equation}
 \boxed{\begin{aligned}
 \mathsf K&=\sum_{\{i,j\}\in\mathcal E}
      \mathsf E_{ij}^{\mathsf T}\mathsf C_{ij}\mathsf E_{ij}
                     +\mathsf K_{\mathrm{w,T}},\\
 \mathsf D&=\mathsf M^{-1/2}\mathsf K\mathsf M^{-1/2}.
 \end{aligned}}
 \label{eq:global-edge-assembly}
\end{equation}
The sum contains the pairwise baseline and smooth wall/target curvature.
Nonnegative edge tensors and support curvature give a sufficient route
to $\mathsf K\succeq0$, with its null mechanisms retained. Prestressed
signed tensors require a test of the assembled operator.

To see the assembly in physical coordinates, write
$\vct\delta_{ij}=\vct u_i-\vct u_j$. The energy and force of one edge are
\begin{equation}
 U_{ij}^{(2)}=\tfrac12\vct\delta_{ij}^{\mathsf T}
                         \mathsf C_{ij}\vct\delta_{ij},\qquad
 \vct f_i^{(ij)}=-\mathsf C_{ij}\vct\delta_{ij},\qquad
 \vct f_j^{(ij)}=-\vct f_i^{(ij)}.
 \label{eq:structure-edge-energy-force}
\end{equation}
The edge contributes $+\mathsf C_{ij}$ to both diagonal blocks and
$-\mathsf C_{ij}$ to both cross blocks. Its stiffness has units
$\mathrm{N\,m^{-1}}$, and $\mathsf D$ has units $\mathrm{s^{-2}}$;
for equal masses, $\mathsf D=\mathsf K/m$. A pair-only block row sums
to zero because common translation changes no separation. This is a
direct assembly check. Smooth supports can break the translation
symmetry; additional generalized coordinates require their own inertia
and work-conjugate forces.

Figure~\ref{fig:structure-particle-to-mode-map} connects the whole-particle geometry, edge-block assembly and inertial weighting in one construction.

\begin{figure}[H]
\centering\begingroup
\definecolor{vltTeal}{HTML}{176B64}
\definecolor{vltBlue}{HTML}{315D80}
\definecolor{vltGold}{HTML}{A87520}
\definecolor{vltPlum}{HTML}{725877}
\definecolor{vltRed}{HTML}{762B38}
\definecolor{vltYellow}{HTML}{D5AD3C}
\definecolor{vltNeutral}{HTML}{777D83}
\setlength{\unitlength}{1pt}
\scalebox{0.90}{%
\input{vlt-figures/figure-064.tex}%
}\endgroup
\caption{Particle identities, pair ownership and the eigenproblem in \eqref{eq:global-edge-assembly}. (a) Each rigid faceted symbol marks a whole ideal virion and its retained centre coordinates. The highlighted pair selects one quadratic energy. (b) Relative displacement gives $+\mathsf C_{ij}$ on each diagonal block and $-\mathsf C_{ij}$ on each cross block. These algebraic signs retain the actual signed curvatures of the edge. (c) After owned pairs, smooth wall and target terms are assembled, the declared positive inertia supplies the mass-weighted matrix. Its eigenvectors reconstruct physical displacement columns through $\mathsf M^{-1/2}$. Force balance and stability concern the assembled reference. Open connecting arrows indicate construction maps.}
\label{fig:structure-particle-to-mode-map}
\end{figure}

An equilibrium reduction can supply nonlocal or many-body curvature,
as in~\eqref{eq:relaxed-medium-curvature}. Those full blocks must be added
consistently, rather than automatically interpreted as three-shell
central pairs. The geometric graph and the reduced effective force law
remain different specifications.

\subsubsection{Rigidity, mechanisms and boundary conditions}
\label{sec:finite-rigidity}

\begin{proposition}[Unstressed compatibility and harmonic kernel]
\label{prop:s3-compatibility-kernel}
For unstressed central springs with $k_{ij}>0$ and no smooth support,
the scalar compatibility matrix and pair stiffness satisfy
\begin{equation}
 (\mathsf B_G\vct u)_{\{i,j\}}
       =(\vct n_{ij}^{(0)})^{\mathsf T}(\vct u_i-\vct u_j),
 \qquad
 \mathsf K=\mathsf B_G^{\mathsf T}
                    \diag(k_{ij})\mathsf B_G.
 \label{eq:rigidity-factorization}
\end{equation}
In particular, $\ker\mathsf K=\ker\mathsf B_G$.
\end{proposition}
\begin{proof}
Each row measures first-order bond extension. The quadratic form is
$\vct u^{\mathsf T}\mathsf K\vct u=
\sum_e k_e| (\mathsf B_G\vct u)_e|^2$; positivity of every $k_e$
makes it zero exactly when all extensions vanish. A nonnegative
symmetric matrix has zero quadratic form precisely on its kernel.
\end{proof}

This scalar compatibility matrix is distinct from
the block incidence matrix that records all three components of an
edge displacement. A connected graph can transmit force and still have
motions that extend none of its springs to first order.

\begin{definition}[Infinitesimal mechanisms and self-stress]
\label{def:s3-mechanisms-self-stress}
An \emph{infinitesimal mechanism} has zero first-order extension on
every retained edge. For a free noncollinear configuration, the six
rigid directions are $\vct u_i=\vct a+\vct\vartheta\times\vct R_i$;
$n_{\mathrm{mech}}$ counts additional independent mechanisms modulo
these rigid motions. A \emph{self-stress} is an edge-tension vector in
$\ker\mathsf B_G^{\mathsf T}$, and $n_{\mathrm{ss}}$ is that space's
dimension.
\end{definition}

\begin{proposition}[Finite Maxwell--Calladine count]
\label{prop:s3-maxwell-calladine}
For the free noncollinear framework in
Definition~\ref{def:s3-mechanisms-self-stress}, let
$N_{\mathrm e}=|\mathcal E|$ and
$q_G=\operatorname{rank}\mathsf B_G$. Then
\begin{equation}
 n_{\mathrm{mech}}=3n-6-q_G,\qquad
 n_{\mathrm{ss}}=N_{\mathrm e}-q_G,\qquad
 n_{\mathrm{mech}}-n_{\mathrm{ss}}=3n-6-N_{\mathrm e}.
 \label{eq:maxwell-calladine}
\end{equation}
\end{proposition}
\begin{proof}
The right nullity is $3n-q_G$. Rigid motions have zero first-order
extension, so six independent kernel directions are rigid. The left
nullity is $N_{\mathrm e}-q_G$.
Subtracting the two counts gives the final identity.
\end{proof}

A self-stress balances edge forces at every site~\cite{Lubensky2015}.
The nullity counts such tension patterns; the actual preparation sets
their amplitudes. Redundant constraints and internal mechanisms can coexist.

Thus $N_{\mathrm e}\ge3n-6$ is necessary for infinitesimal rigidity of
this free spring framework but not sufficient. Eight cube corners joined
only by twelve axial edge springs have twelve independent constraints,
leaving six internal mechanisms in addition to rigid motion. Additional
diagonals are assessed by their actual rank. Figure~\ref{fig:structure-rigidity-bracing}
shows the same distinction in an auxiliary planar frame.

\begin{figure}[H]
\centering\begingroup
\definecolor{vltTeal}{HTML}{176B64}
\definecolor{vltBlue}{HTML}{315D80}
\definecolor{vltGold}{HTML}{A87520}
\definecolor{vltPlum}{HTML}{725877}
\definecolor{vltRed}{HTML}{762B38}
\definecolor{vltYellow}{HTML}{D5AD3C}
\definecolor{vltNeutral}{HTML}{777D83}
\setlength{\unitlength}{1pt}
\scalebox{0.90}{%
\input{vlt-figures/figure-065.tex}%
}\endgroup
\caption{One planar shear tests a square with and without a positive diagonal spring. Four rigid virions retain two centre coordinates each; unstressed side springs have positive curvature and no onsite constraint. Both panels use $\vct u_i=(g_{\rm sh}y_i,0)$, exaggerated to $g_{\rm sh}=0.22$. Reference centres and edges are gray. The side compatibility rank is four, leaving three planar rigid motions and one mechanism; the brace raises the rank to five. Matched coils follow the actual endpoints. At finite shear, vertical sides elongate by $a[\sqrt{1+g_{\rm sh}^2}-1]$, horizontal lengths stay fixed, and the diagonal changes by $a[\sqrt{(1+g_{\rm sh})^2+1}-\sqrt2]$. Their different leading orders explain the harmonic test.}
\label{fig:structure-rigidity-bracing}
\end{figure}

This is a local, infinitesimal count. The quartic example
\eqref{eq:structure-transverse-quartic} explains why zero harmonic cost
does not settle finite-amplitude motion. Prestressed references also
retain the transverse curvature of~\eqref{eq:shell-curvatures}, so the
unstressed factorization alone no longer decides their stability. The
bulk has three translation modes at zero wavevector; an infinitesimal
rotation of an entire finite object is not a periodic Bloch field and
does not create three additional acoustic branches.

\paragraph{Compatibility and geometric stiffness under prestress.}
\begin{proposition}[Compatibility and geometric stiffness]
\label{prop:s3-prestressed-rigidity}
For a smooth central-pair reference, use the same edge directions in
the compatibility and relative-displacement maps. For each retained
unordered edge
$e=\{i,j\}$, write $L_e=V_e''(R_e)$,
$t_e=V_e'(R_e)$, $T_e=t_e/R_e$ and
$\mathsf P_e=\Id_3-\vct n_e\vct n_e^{\mathsf T}$.
Then the pair stiffness and its restriction to a compatibility mechanism are
\begin{equation}
 \begin{aligned}
 \mathsf K_{\rm pair}
  &=\mathsf B_G^{\mathsf T}\diag(L_e)\mathsf B_G
       +\sum_{e\in\mathcal E}T_e\,
          \mathsf E_e^{\mathsf T}\mathsf P_e\mathsf E_e,\\
 \mathsf B_G\vct u=0:\qquad
 \tfrac12\vct u^{\mathsf T}\mathsf K_{\rm pair}\vct u
  &=\tfrac12\sum_{e\in\mathcal E}T_e
               |\mathsf P_e\mathsf E_e\vct u|^2.
 \end{aligned}
 \label{eq:structure-prestressed-rigidity-decomposition}
\end{equation}
\end{proposition}
\begin{proof}
Insert $\mathsf C_e=L_e\vct n_e\vct n_e^{\mathsf T}+T_e\mathsf P_e$
into the edge assembly. The longitudinal part is the compatibility
quadratic form. It vanishes on $\ker\mathsf B_G$, leaving the second
displayed expression.
\end{proof}

Here $t_e$ is a signed force, whereas $L_e,T_e$ are stiffnesses.
The first term penalizes first-order extension. The geometric term
records the second-order length change of a loaded edge under transverse
motion. Tension and compression contribute opposite signs, so different
edges can reinforce or oppose one another. The total curvature on the
admissible space, including supports, distinguishes oscillatory,
marginal and unstable directions.

The reference tensions and the space of possible self-stresses must
also be distinguished. With
$\vct t=(t_e)_{e\in\mathcal E}$ and
$\vct g_{\rm w,T}=\nabla(U_{\rm w}+\lambda U_{\rm T})|_{\vct R}$,
stationarity with smooth supports and no ideal constraint reactions gives
\begin{equation}
 \mathsf B_G^{\mathsf T}\vct t+\vct g_{\rm w,T}=0.
 \label{eq:structure-prestress-force-balance}
\end{equation}
At free pair-only equilibrium, the realized $\vct t$ lies in
$\ker\mathsf B_G^{\mathsf T}$. Smooth supports can balance a nonzero
pair resultant. Fixed linear constraints instead impose
$\mathsf U_{\rm c}^{\mathsf T}
(\mathsf B_G^{\mathsf T}\vct t+\vct g_{\rm w,T})=\vct0$,
with ideal reactions balancing excluded components. The potential,
reference distances and this force balance determine actual tension
amplitudes. Thus the same edge geometry can have different mechanics
under different loading states~\cite{Lubensky2015}.

Positive-clearance hard walls leave rigid directions locally admissible
until contact. Smooth pinning changes the energy; ideal fixed constraints
restrict the allowed directions.

\begin{proposition}[Reduction to a fixed admissible sector]
\label{prop:s3-fixed-constraint-reduction}
For homogeneous linear constraints about the chosen equilibrium, let
$\mathsf U_{\rm c}$ be a constant, dimensionless, full-column-rank basis
of admissible displacements and let $\mathsf M\succ0$. The reduced
inertia, stiffness and dynamical matrix are
\begin{equation}
 \begin{aligned}
 \vct u&=\mathsf U_{\rm c}\vct q_{\rm c},
 &\mathsf M_{\rm c}&=\mathsf U_{\rm c}^{\mathsf T}
                              \mathsf M\mathsf U_{\rm c},\\
 \mathsf K_{\rm c}&=\mathsf U_{\rm c}^{\mathsf T}
                              \mathsf K\mathsf U_{\rm c},
 &\mathsf D_{\rm c}&=\mathsf M_{\rm c}^{-1/2}
                    \mathsf K_{\rm c}\mathsf M_{\rm c}^{-1/2}.
 \end{aligned}
 \label{eq:structure-constrained-sector}
\end{equation}
The restricted mass matrix is positive definite, including for a
nonorthonormal basis.
\end{proposition}
\begin{proof}
Substitution into kinetic and potential energy gives the two congruent
restrictions. For $\vct q_{\rm c}\ne0$, full column rank implies
$\mathsf U_{\rm c}\vct q_{\rm c}\ne0$, so its kinetic quadratic form
is positive. Virtual work by an ideal reaction vanishes on the same
admissible variations, giving the restricted equations of motion.
\end{proof}

Both inertia and stiffness are restricted before mass normalization.
The fixed map applies to linear constraints about the reference.
Moving supports and nonlinear constraint geometry supply additional
terms. The resulting admissible sector identifies the retained zero
directions and the equilibrium fluctuations that are well defined.

\paragraph{Rigidity of a patch-connected body assembly.}
\begin{definition}[Patch-distance compatibility]
\label{def:s3patch-compatibility}
For $p$ retained scalar site distances, let $\mathsf R_{\rm p}$ map
the body tangent to their first variations:
\begin{equation}
 (\mathsf R_{\rm p}\vct\zeta)_e
   =\vct n_e^{\mathsf T}\vct a_e(\vct\zeta),\qquad
 \mathsf H_{\rm pair}
   =\mathsf R_{\rm p}^{\mathsf T}\diag(v_e'')\mathsf R_{\rm p}
       +\sum_e v_e'\nabla_\zeta^2d_e .
 \label{eq:s3patch-distance-compatibility}
\end{equation}
Derivatives are evaluated at the same reference. The final term
contains both transverse distance curvature and rotational geometry.
\end{definition}

\begin{proposition}[Unstressed body-framework count]
\label{prop:s3patch-body-framework-count}
Suppose these are the only energy terms, every $v_e'=0$ and every
$v_e''>0$. After any imposed tangent restriction, use the corresponding
restricted compatibility map. On its domain of dimension $d$, let
$r=\operatorname{rank}\mathsf R_{\rm p}$ and let $r_0$ independent
allowed rigid directions lie in its kernel. Then
\begin{equation}
 \begin{aligned}
 \ker\mathsf H_{\rm pair}&=\ker\mathsf R_{\rm p},\\
 n_{\rm mech}&=d-r-r_0,\qquad n_{\rm ss}=p-r,\qquad
 n_{\rm mech}-n_{\rm ss}=d-r_0-p .
 \end{aligned}
 \label{eq:s3patch-body-framework-count}
\end{equation}
Here $n_{\rm ss}$ counts scalar site-tension combinations in
$\ker\mathsf R_{\rm p}^{\mathsf T}$. For free oriented bodies with
discrete body symmetry, $d=6n$ and $r_0=6$.
\end{proposition}
\begin{proof}
The positive weighted sum of squared distance variations vanishes
exactly on $\ker\mathsf R_{\rm p}$. Rank--nullity gives the right
and left nullities; subtracting the rigid directions gives the count.
\end{proof}

The number of particle-pair edges and the number of independent patch
constraints now differ. One scalar distance between two rigid bodies
leaves five relative infinitesimal mechanisms in the free unstressed
case. Several well-placed surface contacts can constrain translation
and rotation together; their Jacobian rank decides how many motions
remain. A centre-only tetrahedron can be rigid in translation while
leaving all four independent body orientations unrestrained. An
orientation-dependent interaction tests and can remove those extra
directions. Loaded electrostatic patches instead require the complete
geometric Hessian in~\eqref{eq:s3patch-distance-compatibility} and
the actual support terms. A rank count is a compatibility test for its
stated unstressed construction, while stability is a test of the full
assembled second variation.

Figure~\ref{fig:s34patch-body-rigidity-rank} evaluates the compatibility rank on an explicit two-body geometry. Multiple interactions between the same pair can restrict different relative translations and rotations.

\begin{figure}[H]
\centering\begingroup
\setlength{\unitlength}{1pt}
\scalebox{0.86}{%
\input{vlt-figures/figure-287.tex}%
}\endgroup
\caption{Two bodies, several independent patch restrictions. The displayed channels join distinct vertices of unit-circumradius icosahedra at centres $(\mp1.65,0,0)$; the right orientation is $\exp([\,(0.18,0.22,0.34)\,]_{\times})$. Line colours separate the overlaid distance channels. Direct evaluation of the relative six-column compatibility matrix gives rank one for any single displayed distance, rank three for any displayed triple and rank six for all six. Filled bar segments count the rank; grey segments count the remaining relative infinitesimal mechanisms. One body is held as the rigid reference, removing the six common rigid motions. Each distance is assigned a positive quadratic curvature and zero initial tension, so Proposition~\ref{prop:s3patch-body-framework-count} applies. The lines denote scalar interaction channels; loaded patch stability instead uses the full geometric Hessian.}
\label{fig:s34patch-body-rigidity-rank}
\end{figure}

\subsubsection{Assembly and spectral enclosures}
\label{sec:assembly-enclosures}

For computation, partition sites into disjoint groups
$I_1,\ldots,I_g$ and reorder coordinates:
\begin{equation}
 \mathsf D=
 \begin{pmatrix}
 \mathsf D_{11}&\mathsf D_{12}&\cdots&\mathsf D_{1g}\\
 \mathsf D_{21}&\mathsf D_{22}&\cdots&\mathsf D_{2g}\\
 \vdots&\vdots&\ddots&\vdots\\
 \mathsf D_{g1}&\mathsf D_{g2}&\cdots&\mathsf D_{gg}
 \end{pmatrix}.
 \label{eq:group-block-matrix}
\end{equation}
A bond crossing groups contributes both off-diagonal coupling and the
appropriate diagonal terms. Grouping creates no physical boundary.
Overlapping 27-site coordination neighborhoods describe local relations;
the global edge list prevents shared sites and bonds from being counted
repeatedly.

The assembled finite matrix gives the normal modes:
\begin{equation}
 \det(\mathsf D-\omega^2\Id)=0,
 \qquad \mathsf D\ket{s}=\omega_s^2\ket{s}.
 \label{eq:finite-characteristic}
\end{equation}
The eigenvalue is squared frequency, with units $\mathrm{s^{-2}}$.

\begin{proposition}[Rayleigh quotient of a finite mode]
\label{prop:s3-mode-rayleigh}
For real symmetric $\mathsf K$ and $\mathsf M\succ0$, a nonzero real
generalized eigenpattern $\vct\phi_s$ satisfies
\begin{equation}
 \boxed{\omega_s^2=
       \frac{\vct\phi_s^{\mathsf T}\mathsf K\vct\phi_s}
            {\vct\phi_s^{\mathsf T}\mathsf M\vct\phi_s}.}
 \label{eq:structure-mode-rayleigh}
\end{equation}
For any nonzero trial pattern, this quotient lies between the extremal
generalized eigenvalues on the stated coordinate space.
\end{proposition}
\begin{proof}
Multiplying $\mathsf K\vct\phi_s=\omega_s^2\mathsf M\vct\phi_s$
by $\vct\phi_s^{\mathsf T}$ gives the identity. For a trial pattern,
expand its mass-weighted column in the orthonormal eigenbasis; the
quotient is a convex combination of the eigenvalues.
\end{proof}

The numerator is restoring curvature and the denominator inertial
weight. A trial quotient evaluates one deformation; an eigenpattern
preserves its displacement ratios during free motion. Preparation then
sets its amplitude and phase.

Figure~\ref{fig:structure-rayleigh-patterns} makes the quotient explicit for the supported pair used below: its extremal trial patterns are individual eigenmodes.

\begin{figure}[H]
\centering\begingroup
\definecolor{vltTeal}{HTML}{176B64}
\definecolor{vltBlue}{HTML}{315D80}
\definecolor{vltGold}{HTML}{A87520}
\definecolor{vltPlum}{HTML}{725877}
\definecolor{vltRed}{HTML}{762B38}
\definecolor{vltYellow}{HTML}{D5AD3C}
\definecolor{vltNeutral}{HTML}{777D83}
\setlength{\unitlength}{1pt}
\scalebox{0.90}{%
\input{vlt-figures/figure-066.tex}%
}\endgroup
\caption{The supported equal-mass pair has three spring curvatures $k_\star>0$. Common $(A,A)$ and opposed $(A,-A)$ snapshots use one gain of $12\,\mathrm{pt}/A$. Rulers locate reference centres at $0,a$ in both rows. The pair span stays fixed under common motion and shortens by $2A$ under opposed motion; body sizes and coil families are unchanged. The trial quotient is $Q/\omega_\star^2=2-\sin(2\theta)$, with eigenmode extrema $1,3$. Intermediate trial directions combine the two patterns. Shading marks a squared-frequency interval. The boxed result illustrated here is~\eqref{eq:structure-mode-rayleigh}.}
\label{fig:structure-rayleigh-patterns}
\end{figure}

More explicitly, let $\mathsf D$ have a real orthonormal eigenbasis
$\vct e_s=\ket{s}$ and write
$\vct u=\sum_s\mathsf M^{-1/2}\vct e_s Q_s$.
The transformed equation $\mathsf M\Dtt{\vct u}+\mathsf K\vct u=0$
then becomes
\begin{equation}
 \Dtt{Q_s}+\omega_s^2Q_s=0,\qquad
 E_s=\tfrac12\bigl[(\Dt{Q_s})^2+\omega_s^2Q_s^2\bigr].
 \label{eq:structure-assembled-mode-energy}
\end{equation}
Here $Q_s$ has units $\mathrm{kg^{1/2}\,m}$: mass normalization absorbs
inertia into the coordinate. Each positive mode exchanges kinetic and
potential energy while conserving $E_s$. The eigenproblem supplies
patterns and frequencies; preparation and readout complete the prediction.
For fixed constraints, the same construction uses the reduced matrices
in~\eqref{eq:structure-constrained-sector}~\cite{Misra2012,Deriglazov2010}.

\paragraph{From a finite eigenpattern to particle and bond motion.}

The eigenvectors describe correlated motion of the same labelled particles
in the finite assembly. In the full Cartesian
space the physical mass-normalized patterns are
$\vct\phi_s=\mathsf M^{-1/2}\vct e_s$. For the fixed constraints of
\eqref{eq:structure-constrained-sector}, their embedded form is instead
\begin{equation}
 \vct\phi_s=\mathsf U_{\rm c}\mathsf M_{\rm c}^{-1/2}\vct e_{{\rm c},s},
 \qquad \mathsf D_{\rm c}\vct e_{{\rm c},s}
        =\omega_s^2\vct e_{{\rm c},s},\qquad
 \vct\phi_s^{\mathsf T}\mathsf M\vct\phi_t=\delta_{st}.
 \label{eq:structure-embedded-constrained-modes}
\end{equation}
The unconstrained construction is recovered with $\mathsf U_{\rm c}=\Id$.
Initial displacements and velocities must belong to the same admissible
space. The three entries $\vct\phi_{si}$ belonging to site $i$
give its displacement relative to the other sites in that same mode.
For the positive modes, an initial configuration and velocity determine
the entire conservative preparation:
\begin{equation}
 \begin{aligned}
 Q_s(0)&=\vct\phi_s^{\mathsf T}\mathsf M\vct u(0),&
 \Dt{Q_s}(0)&=\vct\phi_s^{\mathsf T}\mathsf M\Dt{\vct u}(0),\\
 Q_s(t)&=Q_s(0)\cos\omega_st+
       \frac{\Dt{Q_s}(0)}{\omega_s}\sin\omega_st,&
 \vct u_i(t)&=\sum_{s:\,\omega_s>0}\vct\phi_{si}Q_s(t).
 \end{aligned}
 \label{eq:structure-finite-mode-preparation}
\end{equation}
The final equality assumes that zero-sector displacement and velocity
have been excluded by the stated preparation or constraints. A retained
zero coordinate instead contributes its separately specified constant
and linear-in-time terms. Mass orthogonality is the normalization
appropriate to kinetic energy; physical displacement patterns need not
be orthogonal in an unweighted Cartesian dot product
~\cite{Dove2011,Misra2012}.

The same patterns determine which local separations change.
\begin{definition}[Modal bond-extension coefficient]
\label{def:s3-modal-bond-extension}
For one included edge with reference vector
$\vct R_{ij}=R_{ij}\vct n_{ij}^{(0)}$, its modal bond-extension
coefficient $b_{ij,s}$ is the first-order change of separation per unit
normal coordinate. For a positive-mode preparation,
\begin{equation}
 b_{ij,s}:=\vct n_{ij}^{(0)\mathsf T}
               (\vct\phi_{si}-\vct\phi_{sj}),\qquad
 r_{ij}(t)-R_{ij}
   =\sum_{s:\,\omega_s>0}b_{ij,s}Q_s(t)
     +O\!\left(\frac{|\vct u_i-\vct u_j|^2}{R_{ij}}\right).
 \label{eq:structure-mode-bond-extension}
\end{equation}
\end{definition}
Taylor expansion of the edge length gives the separation formula.
The coefficient $b_{ij,s}$ has units $\mathrm{kg^{-1/2}}$, so
$b_{ij,s}Q_s$ is a length. Both particles can move while their separation
is unchanged to first order. For an unstressed edge this removes its
quadratic stretching cost along that mode; a prestressed edge also
samples transverse relative motion through $\mathsf C_{ij}$.
The $\beta_i$, $\gamma_i$ and $\Omega_i$ labels select which physical
edges are examined in a repeated neighborhood.

Figure~\ref{fig:structure-mass-normalized-particle-modes} resolves physical mode columns and an edge-extension coefficient in an unequal-mass supported chain.

\begin{figure}[H]
\centering\begingroup
\definecolor{vltTeal}{HTML}{176B64}
\definecolor{vltBlue}{HTML}{315D80}
\definecolor{vltGold}{HTML}{A87520}
\definecolor{vltPlum}{HTML}{725877}
\definecolor{vltRed}{HTML}{762B38}
\definecolor{vltYellow}{HTML}{D5AD3C}
\definecolor{vltNeutral}{HTML}{777D83}
\setlength{\unitlength}{1pt}
\scalebox{0.90}{%
\input{vlt-figures/figure-067.tex}%
}\endgroup
\caption{The supported scalar chain uses the displayed $\overline{\mathsf K}$ and $\overline{\mathsf M}$, with twice the assigned mass at site 2 and equal body sizes. All rows share reference centres $x/a=(2,1,0)$, so positive scalar motion points left, consistently with the upper arrow. Each normalized column has one displacement gain of $20\,\mathrm{pt}$; a short tick denotes a zero component. The physical pattern is $\vct\phi_s=\vct v_s/\sqrt{m_\star}$ and $b_{12}=\phi_{s1}-\phi_{s2}$. Equal reference spans and fixed coil families give left-to-right length changes $(-u_1,u_1-u_2,u_2-u_3,u_3)$. Initial conditions can reverse or rescale these eigenpatterns.}
\label{fig:structure-mass-normalized-particle-modes}
\end{figure}

This reconstruction also separates an eigenpattern from its observation.
A displacement coordinate $Y=\vct\ell^{\mathsf T}\vct u$ has modal
weight $\vct\ell^{\mathsf T}\vct\phi_s$, whereas an edge-extension
coordinate has weight $b_{ij,s}$. The same finite mode can consequently
be prominent in one defined coordinate and absent in another. None of
these weights changes its eigenfrequency; they specify how a chosen
preparation or readout couples to that motion.

Self-adjointness gives real conservative eigenvalues. On a nonnegative
sector, positive modes oscillate and zero coordinates retain their
specified drift. This restoring spectrum remains useful when the medium
overdamps the observed motion. Thermal position correlations in colloidal
model crystals have, for example, constrained spring constants under
those conditions~\cite{Keim2004}. For the virion model, resolved peaks and
propagation also depend on the medium and channel overlaps developed
in Sections~\ref{sec:viscoelastic} and~\ref{sec:viscoelastic-dynamics}.

\begin{proposition}[Scalar and site-block spectral enclosures]
\label{prop:s3-spectral-enclosures}
Let $\mathsf D$ be real symmetric. With
$r_A=\sum_{B\ne A}|D_{AB}|$, every eigenvalue obeys
\begin{equation}
 \min_A(D_{AA}-r_A)\le\omega_s^2\le
                              \max_A(D_{AA}+r_A).
 \label{eq:gershgorin}
\end{equation}
Grouping each site's three components, set
$\mathcal R_i=\sum_{j\ne i}\|\mathsf D_{ij}\|_2$. Then also
\begin{equation}
 \ell_-:=\min_i[\lambda_{\min}(\mathsf D_{ii})-\mathcal R_i]
 \ \le\ \omega_s^2\ \le\
 \max_i[\lambda_{\max}(\mathsf D_{ii})+\mathcal R_i]=:\ell_+.
 \label{eq:block-spectral-bounds}
\end{equation}
\end{proposition}
\begin{proof}
Choose an eigenvector component of maximal absolute value. Its row
equation gives $|\omega_s^2-D_{AA}|\le r_A$ and hence the scalar
enclosure~\cite{HornJohnson2013}. For the block estimate, choose a site
block of maximal norm, take its inner product with that block's row
equation, and bound the coupling terms by
$\mathcal R_i$ times its squared norm. The diagonal quadratic form
lies between its two extremal eigenvalues, proving both bounds.
\end{proof}

These enclosures locate eigenvalues before diagonalization. Mode shapes
and measured resonances still require the mechanical and observation
maps. For shell tensors,
$\|\mathsf C_\eta\|_2=\max(|L_\eta|,|T_\eta|)$, so masses and actual
coordination counts provide immediate estimates.

A strictly positive lower bound proves a spectral gap for the matrix
being bounded. A free network has rigid zero directions, so a positive
gap belongs to an explicitly supported or restricted sector rather than
its full unconstrained space. A negative lower bound leaves stability
undecided; it does not prove a negative eigenvalue. Once stability is
established independently, the enclosed frequencies obey
$\sqrt{\max(0,\ell_-)}\le\omega_s\le\sqrt{\ell_+}$.
The following subsection distinguishes this collective calculation from
the local restoring response measured with neighboring coordinates held
fixed.

Figure~\ref{fig:structure-spectral-enclosures} compares the row enclosures with the actual positive eigenvalues of a supported three-site reference.

\begin{figure}[H]
\centering\begingroup
\definecolor{vltTeal}{HTML}{176B64}
\definecolor{vltBlue}{HTML}{315D80}
\definecolor{vltGold}{HTML}{A87520}
\definecolor{vltPlum}{HTML}{725877}
\definecolor{vltRed}{HTML}{762B38}
\definecolor{vltYellow}{HTML}{D5AD3C}
\definecolor{vltNeutral}{HTML}{777D83}
\setlength{\unitlength}{1pt}
\scalebox{0.80}{%
\begin{picture}(470,369)%
{\color{black}\put(235.00000,350.00000){\makebox(0,0)[c]{\fontfamily{lmss}\fontsize{11.000}{13.000}\selectfont \bfseries Row intervals enclose eigenvalues without specifying their patterns.}}}%
{\color{black}\put(42.00000,292.00000){\makebox(0,0)[r]{\fontfamily{lmss}\fontsize{10.500}{12.500}\selectfont row 1}}}%
{\color{vltBlue}\vlB{%
\vSab{151.000000}{292.000000}{0.000000}{172.000000}%
}}%
{\color{black}\vlB{%
\vSf{150.550000}{287.000000}{90.000000}{10.000000}%
\vSf{322.550000}{287.000000}{90.000000}{10.000000}%
}}%
{\color{black}\put(237.000000,292.000000){\circle*{5.000000}}}%
{\color{black}\put(42.00000,239.00000){\makebox(0,0)[r]{\fontfamily{lmss}\fontsize{10.500}{12.500}\selectfont row 2}}}%
{\color{vltBlue}\vlB{%
\vSab{65.000000}{239.000000}{0.000000}{344.000000}%
}}%
{\color{black}\vlB{%
\vSf{64.550000}{234.000000}{90.000000}{10.000000}%
\vSf{408.550000}{234.000000}{90.000000}{10.000000}%
}}%
{\color{black}\put(237.000000,239.000000){\circle*{5.000000}}}%
{\color{black}\put(42.00000,186.00000){\makebox(0,0)[r]{\fontfamily{lmss}\fontsize{10.500}{12.500}\selectfont row 3}}}%
{\color{vltBlue}\vlB{%
\vSab{151.000000}{186.000000}{0.000000}{172.000000}%
}}%
{\color{black}\vlB{%
\vSf{150.550000}{181.000000}{90.000000}{10.000000}%
\vSf{322.550000}{181.000000}{90.000000}{10.000000}%
}}%
{\color{black}\put(237.000000,186.000000){\circle*{5.000000}}}%
{\color{black}\put(42.00000,106.00000){\makebox(0,0)[r]{\fontfamily{lmss}\fontsize{10.500}{12.500}\selectfont spectrum}}}%
{\color{vltTeal}\vlB{%
\vSv{114.477634}{91.000000}{90.000000}{33.000000}%
}}%
{\color{black}\put(115.37763,141.00000){\makebox(0,0)[c]{\fontfamily{lmss}\fontsize{10.500}{12.500}\selectfont $2-\sqrt2$}}}%
{\color{vltTeal}\vlB{%
\vSv{236.100000}{91.000000}{90.000000}{33.000000}%
}}%
{\color{black}\put(237.00000,141.00000){\makebox(0,0)[c]{\fontfamily{lmss}\fontsize{10.500}{12.500}\selectfont $2$}}}%
{\color{vltTeal}\vlB{%
\vSv{357.722366}{91.000000}{90.000000}{33.000000}%
}}%
{\color{black}\put(358.62237,141.00000){\makebox(0,0)[c]{\fontfamily{lmss}\fontsize{10.500}{12.500}\selectfont $2+\sqrt2$}}}%
{\color{black}\vlB{%
\vSe{65.000000}{91.000000}{0.000000}{344.000000}%
\vSk{64.600000}{88.000000}{90.000000}{6.000000}%
}}%
{\color{black}\put(65.00000,75.00000){\makebox(0,0)[c]{\fontfamily{lmss}\fontsize{10.500}{12.500}\selectfont $0$}}}%
{\color{black}\vlB{%
\vSk{150.600000}{88.000000}{90.000000}{6.000000}%
}}%
{\color{black}\put(151.00000,75.00000){\makebox(0,0)[c]{\fontfamily{lmss}\fontsize{10.500}{12.500}\selectfont $1$}}}%
{\color{black}\vlB{%
\vSk{236.600000}{88.000000}{90.000000}{6.000000}%
}}%
{\color{black}\put(237.00000,75.00000){\makebox(0,0)[c]{\fontfamily{lmss}\fontsize{10.500}{12.500}\selectfont $2$}}}%
{\color{black}\vlB{%
\vSk{322.600000}{88.000000}{90.000000}{6.000000}%
}}%
{\color{black}\put(323.00000,75.00000){\makebox(0,0)[c]{\fontfamily{lmss}\fontsize{10.500}{12.500}\selectfont $3$}}}%
{\color{black}\vlB{%
\vSk{408.600000}{88.000000}{90.000000}{6.000000}%
}}%
{\color{black}\put(409.00000,75.00000){\makebox(0,0)[c]{\fontfamily{lmss}\fontsize{10.500}{12.500}\selectfont $4$}}}%
{\color{black}\put(235.00000,48.00000){\makebox(0,0)[c]{\fontfamily{lmss}\fontsize{10.500}{12.500}\selectfont $x=\omega^2/\omega_\star^2,\qquad\mathsf D/\omega_\star^2=\operatorname{tridiag}(-1,2,-1)$}}}%
{\color{black}\put(235.00000,20.00000){\makebox(0,0)[c]{\fontfamily{lmss}\fontsize{10.500}{12.500}\selectfont The enclosure reaches zero; the pinned system has a strictly positive gap.}}}%
\end{picture}%
%
}\endgroup
\caption{Gershgorin intervals and the exact spectrum of the equal-mass supported three-site chain. Its row intervals are $[1,3]$, $[0,4]$, $[1,3]$ in units $\omega_\star^2=k_\star/m_\star$, with center entries at two. The lower sticks locate $2-\sqrt2$, $2$, $2+\sqrt2$ on exactly the same horizontal scale. Every stick lies within the enclosure, while none lies at its zero lower endpoint. These are squared-frequency locations, rather than mode weights or a spectral density. End springs supply the declared support; all three particle displacements remain free variables.}
\label{fig:structure-spectral-enclosures}
\end{figure}

\subsubsection{Compatible cell assembly and interface reduction}
\label{sec:s3-compatible-cell-assembly}

The cell construction gives a reusable route from small particle blocks
to the operator of a large assembly. The particle coordinates remain
global, while a cell matrix records only its assigned contributions.
This is the energy-based assembly principle familiar from structural
matrix methods, applied here to the ideal-virion interaction
energy~\cite{RoylanceFEA2001}.

\begin{definition}[Cell restriction and energy ownership]
\label{def:s3-cell-restriction}
Let $\mathsf R_c$ be the dimensionless Boolean restriction from the
global laboratory displacement column to the particle coordinates used
by cell $c$: $\vct u^{(c)}=\mathsf R_c\vct u$. For the chosen
particle-coordinate potential, assign each retained
unordered pair, each one-body term and each other coupling term to
one cell contribution. Thus
\begin{equation}
 U(\vct r)=\sum_{c\in\mathcal F_{\rm cell}}U_c(\mathsf R_c\vct r),\qquad
 \mathsf K_c=\nabla^2 U_c(\mathsf R_c\vct R).
 \label{eq:s3-cell-energy-ownership}
\end{equation}
The coordinate list of a cell may include neighboring particles needed
to evaluate its owned terms. When a particle occurs in several lists,
choose mass fractions $\mu_{ci}\ge0$ satisfying
$\sum_{c:i\in c}\mu_{ci}=1$ and let
$\mathsf M_c=\operatorname{diag}_{i\in c}(\mu_{ci}m_i\Id_3)$.
\end{definition}

\begin{proposition}[Conforming assembly preserves the mechanics]
\label{prop:s3-conforming-cell-assembly}
Under the ownership and compatibility conditions above,
\begin{equation}
 \mathsf K=\sum_c\mathsf R_c^{\mathsf T}\mathsf K_c\mathsf R_c,
 \qquad
 \mathsf M=\sum_c\mathsf R_c^{\mathsf T}\mathsf M_c\mathsf R_c.
 \label{eq:s3-conforming-cell-assembly}
\end{equation}
The assembled kinetic and quadratic potential energies equal their
physical global values, $\mathsf M\succ0$ for positive physical masses,
and, for the unconstrained smooth baseline, force balance is the global condition
$\sum_c\mathsf R_c^{\mathsf T}\nabla U_c=\vct0$ at the reference.
\end{proposition}
\begin{proof}
The Hessian identity follows by differentiating
\eqref{eq:s3-cell-energy-ownership} twice. Substitution of
$\vct u^{(c)}=\mathsf R_c\vct u$ into each quadratic form proves energy
equality. The mass fractions restore exactly one $m_i\Id_3$ at each
physical site, so the mass form is positive. Differentiating the full
energy once gives the stated force balance. Local contributions need
not be separately stationary.
\end{proof}

Under the declared fixed linear constraints, this force-balance sum is
instead projected by $\mathsf U_{\rm c}^{\mathsf T}$ as
in~\eqref{eq:structure-constrained-sector}; ideal reactions can balance
its excluded components.

\paragraph{A shared surface state belongs to one physical VLP.}
The same ownership rule applies to orientation, surface populations and
material states. Cells sharing a particle share its one orientation and
its one retained population. Differentiating every owned energy after
pulling it into a common global position/orientation chart preserves
\eqref{eq:s3-conforming-cell-assembly}. For a nonlinear local chart
$\vct y_c=f_c(\vct y)$, the Hessian chain rule additionally contains
$\sum_a(\partial_{y_{c,a}}\mathcal F_c)\nabla_y^2 f_{c,a}$.
Local owned terms need not be separately stationary, so this term is
retained when changing charts. Rotated cells transform angular and
translational components together, with dual force and torque maps.

\begin{example}[Shared regulation couples two cell ports]
\label{ex:s3patch-shared-cell-state}
Let dimensionless port motions $X_1,X_2$ couple to one shared
dimensionless population $\eta$. Assign the cell energies
$\mathcal F_c=k_cX_c^2/2-b_cX_c\eta+a_c\eta^2/2$, with $a_c>0$
and energy-valued coefficients. Assemble before equilibrating $\eta$.
With $A=a_1+a_2$, $\vct b=(b_1,b_2)^{\mathsf T}$,
\begin{equation}
 \eta_{\rm rel}=\frac{b_1X_1+b_2X_2}{A},\qquad
 \mathsf H_{\rm rel}=\diag(k_1,k_2)-\frac{\vct b\vct b^{\mathsf T}}{A}.
 \label{eq:s3patch-shared-state-assembly}
\end{equation}
Choose coefficients for which this matrix is positive definite.
The cross curvature is $-b_1b_2/A$. Equilibrating two independent
copies of $\eta$ instead gives diagonal corrections $-b_c^2/a_c$
and loses that cross response: it represents two physical populations.
The calculation identifies an effective port coupling mediated by a
shared state, even without a direct spring between those ports.
\end{example}

Figure~\ref{fig:s34patch-shared-population-cells} connects the ownership rule to a seven-particle assembly and the cross curvature generated by its shared surface state.

\begin{figure}[H]
\centering\begingroup
\setlength{\unitlength}{1pt}
\scalebox{0.86}{%
\input{vlt-figures/figure-288.tex}%
}\endgroup
\caption{A shared state transmits a cross response between cell ports. Panel (a) gives an oblique view of two tetrahedral cell domains containing seven VLPs: the yellow $\beta$ body, its orientation and its one surface population belong to both cells. Pale hulls and straight edges identify cell ownership, rather than added physical springs. Panel (b) evaluates Example~\ref{ex:s3patch-shared-cell-state} with $k_1=k_2=3E_\star$, $a_1=a_2=b_1=b_2=E_\star$; matrix entries are divided by $E_\star$. Relaxing the shared population creates the displayed off-diagonal curvature. Assigning one independent population to each cell instead changes the physical model and gives the stated diagonal matrix. The two-port coefficients are assigned synthetic data, rather than inferred from the sketched geometry.}
\label{fig:s34patch-shared-population-cells}
\end{figure}

For two disjoint cells, a newly retained pair between $i$ and $j$ adds
the explicit block
\begin{equation}
 \Delta\mathsf K_{ij}
 =\begin{pmatrix}
 \mathsf S_i^{\mathsf T}\mathsf C_{ij}\mathsf S_i&
 -\mathsf S_i^{\mathsf T}\mathsf C_{ij}\mathsf S_j\\
 -\mathsf S_j^{\mathsf T}\mathsf C_{ij}\mathsf S_i&
 \mathsf S_j^{\mathsf T}\mathsf C_{ij}\mathsf S_j
 \end{pmatrix},
 \label{eq:s3-cross-cell-block-update}
\end{equation}
where $\mathsf S_i$ and $\mathsf S_j$ extract the endpoint displacements
from their respective cell columns. The pair changes both diagonal
blocks and both cross blocks. Its off-diagonal blocks carry the minus
sign of a relative-coordinate energy. Adding only an off-diagonal
``connection'' would generally spoil the pair translation identity and
the virtual-work account. This update concerns the declared reference;
if the cells relax after joining, their new positions and all affected
pair derivatives are recomputed before normal modes are assigned.

Equation~\eqref{eq:s3-conforming-cell-assembly} gives a precise meaning
to summing cell matrices: each local matrix is first embedded into the
same global coordinate space. A rotated copy of a reference cell also
rotates every directional curvature block by
$\mathsf C\mapsto\mathsf O\mathsf C\mathsf O^{\mathsf T}$, together
with the reference geometry. Reusing geometry alone while leaving its
force tensors unchanged would generally represent a different model.

\paragraph{Interface variables connect the blocks.}
For computation, choose disjoint physical cell ownership sets. Retain
as \emph{ports} every shared physical label, every site touched by a
cross-cell interaction and every site in the applied or measured
channels of interest. All other sites can be
treated as cell interiors. A port is an actual retained coordinate
through which a cell exchanges force and motion with its surroundings.
It is an algebraic interface rather than a new membrane or wall.
The same construction extends to any additional retained mechanical
coordinates whose inertia and force couplings have been declared.

\begin{proposition}[Exact hierarchical assembly through ports]
\label{prop:s3-cell-port-schur}
Partition the full dynamic stiffness
$\mathsf A(z)=\mathsf K-z^2\mathsf M$ into global port coordinates
$\Gamma$ and cell interiors $I_c$. Suppose every coupling between cells
touches ports, so $\mathsf A_{II}$ is block diagonal in the interiors.
At frequencies where the displayed inverses exist, the exact port
dynamic stiffness and driving force are
\begin{equation}
 \begin{aligned}
 \mathsf S_\Gamma(z)
   &=\mathsf A_{\Gamma\Gamma}(z)
     -\sum_c\mathsf A_{\Gamma I_c}(z)
        \mathsf A_{I_cI_c}(z)^{-1}\mathsf A_{I_c\Gamma}(z),\\
 \vct f_{\Gamma,\mathrm{eff}}(z)
   &=\vct f_\Gamma(z)
     -\sum_c\mathsf A_{\Gamma I_c}(z)
        \mathsf A_{I_cI_c}(z)^{-1}\vct f_{I_c}(z),\\
 [\mathsf A(z)^{-1}]_{\Gamma\Gamma}
   &=\mathsf S_\Gamma(z)^{-1}.
 \end{aligned}
 \label{eq:s3-cell-port-schur}
\end{equation}
Thus independent exact elimination of the cell interiors followed by
port assembly gives the same driven response as the full matrix.
\end{proposition}
\begin{proof}
Solve each interior row for
$\vct u_{I_c}=\mathsf A_{I_cI_c}^{-1}
(\vct f_{I_c}-\mathsf A_{I_c\Gamma}\vct u_\Gamma)$ and substitute into
the port rows. The block diagonal interior partition makes the
corrections a sum of independent cell contributions. The inverse-block
identity follows by applying port-only forcing.
\end{proof}

At $z=0$ this is static condensation. At finite frequency the interior
particles store inertia and produce frequency-dependent interface
response, as in~\eqref{eq:structure-dynamic-local-schur}. An interior
excitation appears in the effective force as well as in the operator;
elimination therefore retains the route by which a cell excites another
cell. Zero-initial-motion response uses the retarded prescription
already stated for that equation. Other initial preparations add their
corresponding source terms. If a retained electrostatic or reduced-medium
coupling directly connects two proposed interiors, those sites must
enter the ports, the groups must be enlarged, or the full coupled
interior block must be eliminated together.

This offers a scalable computation when the physically justified graph
has small interfaces. The interior solves can be cached for unchanged
cells, and a changed interface can be assembled from its actual local
pair updates. A fully retained long-range interaction can make the port
set large; locality and compression accuracy must then be established
from that interaction and the requested observable. A choice of cell
boundaries by itself creates no sparsity in the force law.

Figure~\ref{fig:s23-cell-port-reduction} identifies physical interface sites and the operator produced by eliminating the cell interiors.

\begin{figure}[H]
\centering\begingroup
\definecolor{vltTeal}{HTML}{176B64}
\definecolor{vltBlue}{HTML}{315D80}
\definecolor{vltGold}{HTML}{A87520}
\definecolor{vltPlum}{HTML}{725877}
\definecolor{vltRed}{HTML}{762B38}
\definecolor{vltYellow}{HTML}{D5AD3C}
\definecolor{vltNeutral}{HTML}{777D83}
\setlength{\unitlength}{1pt}
\scalebox{0.90}{%
\input{vlt-figures/figure-069.tex}%
}\endgroup
\caption{Gold sites carry the ports, including every cross-cell endpoint. Teal and blue sites form the two interiors; colors here identify computational roles. Force at site 2 and readout at site 6 use separate gains. The auxiliary static test has unit longitudinal edge curvature, unit masses and support curvature $0.2\Id$. Ordering $(\Gamma,I_1,I_2)$ exposes the zero interior-to-interior blocks. Eliminating the interiors leaves twelve port coordinates; back-substitution recovers the other twelve physical coordinates. The shaded blocks identify the retained system. Equation~\eqref{eq:s3-cell-port-schur} supplies the frequency-dependent operator and effective force for the same assembly.}
\label{fig:s23-cell-port-reduction}
\end{figure}

\paragraph{Reduce the complete interface law and its sources.}
Surface relaxation also enters the interface reduction. In the scaled
coordinates of~\eqref{eq:s2patch-generalized-inertia}, collect permitted
dimensionless surface and material perturbations in $\vct\chi$.
Let their quadratic blocks be $\mathsf B,\mathsf A$, with
$\mathsf A\succ0$, mobility $\mathsf L\succ0$, and direct mechanical
resistance $\mathsf C_0\succeq0$. Assume the joint stiffness
$\left(\begin{smallmatrix}\mathsf H_X&\mathsf B\\
\mathsf B^{\mathsf T}&\mathsf A\end{smallmatrix}\right)\succeq0$
on the declared admissible space. At a resting equilibrium, one
reciprocal linear realization is
\begin{equation}
 \begin{aligned}
 \mathsf G_X\Dtt{\vct X}+\mathsf C_0\Dt{\vct X}
       +\mathsf H_X\vct X+\mathsf B\vct\chi&=\vct f,\\
 \Dt{\vct\chi}&=-\mathsf L
        (\mathsf A\vct\chi+\mathsf B^{\mathsf T}\vct X).
 \end{aligned}
 \label{eq:s3patch-coupled-body-state-dynamics}
\end{equation}
The nonnegative joint quadratic energy, including mechanical kinetic
energy, obeys
$\dot E=\dot{\vct X}^{\mathsf T}\vct f
-\dot{\vct X}^{\mathsf T}\mathsf C_0\dot{\vct X}
-\vct g_\chi^{\mathsf T}\mathsf L\vct g_\chi$,
where $\vct g_\chi=\mathsf A\vct\chi+\mathsf B^{\mathsf T}\vct X$.
Mechanical work and hidden-state relaxation therefore use the same
couplings. A null mobility direction is retained as a frozen state,
rather than inverted in the following formula.

\begin{proposition}[Matched surface and cell elimination]
\label{prop:s3patch-matched-port-reduction}
For $e^{-i\omega t}$ and the invertible blocks in question, the
augmented dynamic stiffness and mechanical Schur operator are
\begin{equation}
 \begin{aligned}
 \mathcal A(\omega)&=
  \begin{pmatrix}
    \mathsf H_X-\omega^2\mathsf G_X-i\omega\mathsf C_0&\mathsf B\\
    \mathsf B^{\mathsf T}&\mathsf A-i\omega\mathsf L^{-1}
  \end{pmatrix},\\
 \mathsf Z_X(\omega)&=\mathsf H_X-\omega^2\mathsf G_X
    -i\omega\mathsf C_0
    -\mathsf B(\mathsf A-i\omega\mathsf L^{-1})^{-1}\mathsf B^{\mathsf T}.
 \end{aligned}
 \label{eq:s3patch-augmented-interface-pencil}
\end{equation}
Eliminating any interior block $I$ gives
$\mathsf S_\Gamma=\mathcal A_{\Gamma\Gamma}
-\mathcal A_{\Gamma I}\mathcal A_{II}^{-1}\mathcal A_{I\Gamma}$.
The same row map
$\mathsf T_{\rm out}=[\Id,-\mathcal A_{\Gamma I}\mathcal A_{II}^{-1}]$
acts on applied forces, preparation sources and random forcing.
Sequential elimination agrees with joint elimination wherever the
required intermediate inverses exist.
\end{proposition}
\begin{proof}
Multiply the surface equation by $\mathsf L^{-1}$ and solve the
eliminated rows. Substitution gives the Schur operator and its source
map. Both orders solve the same retained block equations, proving the
elimination identity.
\end{proof}

On a positive constrained equilibrium sector, the two augmented thermal force
spectra are $2k_{\rm B}T\mathsf C_0$ and
$2k_{\rm B}T\mathsf L^{-1}$, with zero cross spectrum in this
declared realization. The reduced spectrum is consequently
\begin{equation}
 \mathsf S_{\xi_\Gamma}
   =\mathsf T_{\rm out}\mathsf S_\xi\mathsf T_{\rm out}^{\dagger}
   =\frac{2k_{\rm B}T}{\omega}
               [-\operatorname{Im}_{\rm H}\mathsf S_\Gamma],
 \qquad \omega>0 .
 \label{eq:s3patch-port-noise-matching}
\end{equation}
The Schur inverse identity proves the last equality from
$\mathsf S_\xi=(2k_{\rm B}T/\omega)
[-\operatorname{Im}_{\rm H}\mathcal A]$; the Hermitian imaginary
part has the convention in~\eqref{eq:s2patch-charge-noise}.
Finite-time preparation is propagated by the same linear equations;
a stationary spectrum does not replace its initial state. Any discarded
output also needs its reconstruction map. Independent cell solves
require the full interior block to separate, including surface,
material and retained fluid couplings. This extends exact port
assembly to regulated patches without duplicating their relaxation or
losing its fluctuation law.

\paragraph{A controlled cell-mode approximation.}
When exact frequency-dependent interior elimination is too expensive,
the ports can be retained together with selected fixed-port normal
modes. For a cell with coordinates ordered as $(\Gamma_c,I_c)$ and
positive interior stiffness and mass,
$\mathsf K_{I_cI_c}\succ0$ and $\mathsf M_{I_cI_c}\succ0$, define
\begin{equation}
 \mathsf T_c=
 \begin{pmatrix}
 \Id&0\\
 -\mathsf K_{I_cI_c}^{-1}\mathsf K_{I_c\Gamma_c}&\mathsf V_c
 \end{pmatrix},\qquad
 \mathsf K_{I_cI_c}\vct v_a
     =\nu_a^2\mathsf M_{I_cI_c}\vct v_a,
 \label{eq:s3-cell-component-modes}
\end{equation}
where $\mathsf V_c$ contains selected independent interior modes. The
first columns permit interior accommodation of an imposed port motion;
the remaining columns retain internal resonances. This is the
component-mode construction of Craig and Bampton, adapted to the
declared particle coordinates~\cite{CraigBampton1968,BatheDong2014}.
After enforcing port compatibility, let the resulting full-column-rank
global trial map be $\mathsf T$. Then
\begin{equation}
 \mathsf M_r=\mathsf T^{\mathsf T}\mathsf M\mathsf T,\qquad
 \mathsf K_r=\mathsf T^{\mathsf T}\mathsf K\mathsf T,\qquad
 \vct f_r=\mathsf T^{\mathsf T}\vct f,
 \label{eq:s3-cell-reduced-mechanics}
\end{equation}
retain kinetic energy, potential energy and virtual work exactly on
the chosen trial subspace. In particular, $\mathsf M_r\succ0$ and
$\mathsf K_r\succeq0$ whenever the corresponding full matrices have
these properties. Every retained physical zero mode must also lie in
that trial subspace if its motion is to be reproduced.

The generalized min--max principle gives a useful one-sided check:
the ordered Ritz squared frequencies are no smaller than the
corresponding full squared frequencies, including zero modes in both
counts. Enlarging a nested trial space can only lower these estimates
toward the full values~\cite{HornJohnson2013}. The direction of this
bound does not by itself give its size. For an approximate physical
mode $\vct\phi\ne0$ with squared-frequency estimate $\lambda$, define
\begin{equation}
 \eta=
 \frac{\|\mathsf M^{-1/2}
          (\mathsf K\vct\phi-\lambda\mathsf M\vct\phi)\|}
      {\|\mathsf M^{1/2}\vct\phi\|},\qquad
 \operatorname{dist}\bigl(\lambda,\sigma(\mathsf D)\bigr)\le\eta.
 \label{eq:s3-cell-mode-residual}
\end{equation}
To verify the enclosure, expand the normalized column
$\mathsf M^{1/2}\vct\phi/\|\mathsf M^{1/2}\vct\phi\|$ in the exact
orthonormal eigenbasis of $\mathsf D$. Its residual norm squared is a
weighted sum of $(\omega_s^2-\lambda)^2$ and is at least the squared
distance to the spectrum. Both $\eta$ and $\lambda$ have units
$\mathrm{s^{-2}}$. A small residual identifies proximity to some
eigenvalue; assigning a particular mode or resolving a cluster of
near-degenerate modes additionally uses the spectral separation and
the relevant subspace.

A cell's compatible force--motion map, enriched as needed by internal
modes, is the computational building block. Four-particle modules can
therefore enter a thousand-particle calculation through their interfaces.
The assembled modes, transfer channels and error tests determine which
interior coordinates can be omitted for the intended prediction.

\subsection{Local restoring response and frequency operators}
\label{sec:local-response-frequency}

A local restoring measurement specifies which neighboring coordinates
are held or released. With the neighbors held, a unit polarization
$\vct p$ at site $i$ has curvature
$\vct p^{\mathsf T}\mathsf K_{ii}\vct p$. With all retained sites free
to respond on a fully pinned positive space, a force $F\vct\ell$ gives
$\vct u=F\mathsf K^{-1}\vct\ell$ and
$Y/F=\vct\ell^{\mathsf T}\mathsf K^{-1}\vct\ell$.
The following elimination expresses this collective accommodation.

\begin{proposition}[Clamped and statically relaxed local response]
\label{prop:s3-local-static-schur}
Partition a real symmetric positive-definite stiffness into a held
coordinate block $a$ and an unforced block $b$. Then the relaxed
curvature and the local compliance satisfy
\begin{equation}
 \boxed{\mathsf K_{a,\mathrm{rel}}
   =\mathsf K_{aa}-\mathsf K_{ab}\mathsf K_{bb}^{-1}\mathsf K_{ba},
 \qquad (\mathsf K^{-1})_{aa}=\mathsf K_{a,\mathrm{rel}}^{-1}.}
 \label{eq:structure-local-schur}
\end{equation}
The relaxed curvature is positive definite and no greater than the
clamped curvature in any trial direction.
\end{proposition}
\begin{proof}
Unforced equilibrium gives
$\vct u_b=-\mathsf K_{bb}^{-1}\mathsf K_{ba}\vct u_a$.
Substitution into the energy gives its minimum at fixed $\vct u_a$;
completion of the square proves the stated positivity and ordering.
Solving $\mathsf K\vct u=(\vct f_a,\vct0)^{\mathsf T}$ gives
the inverse-block identity.
\end{proof}

Unforced neighbors follow the imposed displacement and lower its stored
energy. A measured slope therefore constrains a specified loading
channel and support condition. Whole-particle motion, capsid indentation
and internal relative deformation use different mechanical maps.
Additional loading directions or spatially resolved motion can separate
support compliance from interparticle coupling.

Figure~\ref{fig:structure-local-schur-elimination} gives a direct static comparison of the held and unforced neighboring coordinates.

\begin{figure}[H]
\centering\begingroup
\definecolor{vltTeal}{HTML}{176B64}
\definecolor{vltBlue}{HTML}{315D80}
\definecolor{vltGold}{HTML}{A87520}
\definecolor{vltPlum}{HTML}{725877}
\definecolor{vltRed}{HTML}{762B38}
\definecolor{vltYellow}{HTML}{D5AD3C}
\definecolor{vltNeutral}{HTML}{777D83}
\setlength{\unitlength}{1pt}
\scalebox{0.90}{%
\input{vlt-figures/figure-070.tex}%
}\endgroup
\caption{The supported pair has $k_{\rm pin}=k_{\rm link}=k_\star$. Both rows share reference centres $0,a$ and a displacement gain of $12\,\mathrm{pt}/\ell_\star$. Site 1 moves by $\ell_\star$; site 2 is held or follows its unforced balance $u_2=u_1/2$. The outer supports stay fixed and coils follow their moved endpoints. Force slopes are $2k_\star$ and $3k_\star/2$; the shaded wedge is the quasistatic work saved by relaxation. The corresponding compliances are $1/(2k_\star)$ and $2/(3k_\star)$. This changes the imposed hold within the same conservative force law. The boxed result illustrated here is~\eqref{eq:structure-local-schur}.}
\label{fig:structure-local-schur-elimination}
\end{figure}

Neighboring ideal virions also carry inertia, so their accommodation depends
on the frequency of loading.

\begin{proposition}[Dynamic elimination of neighboring coordinates]
\label{prop:s3-local-dynamic-schur}
Let $\mathsf K\succ0$ and
$\mathsf M=\operatorname{diag}(\mathsf M_a,\mathsf M_b)\succ0$.
For a drive $e^{-\mathrm i zt}$ applied only to $a$, at a complex
frequency where the displayed inverses exist,
\begin{equation}
 \begin{split}
 \mathsf A_a(z)&=\mathsf K_{aa}-z^2\mathsf M_a
   -\mathsf K_{ab}(\mathsf K_{bb}-z^2\mathsf M_b)^{-1}
                              \mathsf K_{ba},\\
 [ (\mathsf K-z^2\mathsf M)^{-1}]_{aa}
                         &=\mathsf A_a(z)^{-1}.
 \end{split}
 \label{eq:structure-dynamic-local-schur}
\end{equation}
\end{proposition}
\begin{proof}
The unforced block gives
$\vct u_b=-(\mathsf K_{bb}-z^2\mathsf M_b)^{-1}
\mathsf K_{ba}\vct u_a$. Substitution into the driven block gives
$\mathsf A_a(z)\vct u_a=\vct f_a$ and the inverse-block identity.
\end{proof}
This is the retarded driven response with zero initial motion, first
defined for $\operatorname{Im}z>0$ and then approached as
$z=\omega+\mathrm i0^+$. Initial motion or forcing in $b$ supplies an
additional effective force on $a$. At $z=0$ the expression recovers
\eqref{eq:structure-local-schur}; at finite frequency the neighboring
centres follow with their own amplitudes and phases. The resulting
correction is frequency dependent, rather than a fixed replacement
spring constant. Poles of the eliminated block are not by themselves
normal-mode frequencies of the coupled lattice. At such real frequencies
the displayed elimination must be interpreted through the full
resolvent and its limiting response. In particular, the static ordering
between relaxed and clamped stiffness is not a universal ordering across
frequency. At large $|z|$ the leading compliance is
$-z^{-2}\mathsf M_a^{-1}$: the first response is set by the loaded
particle inertia. If constraints mix sites and make the mass matrix
nondiagonal in this partition, apply the Schur complement to the full
blocks of $\mathsf K-z^2\mathsf M$ instead.

Figure~\ref{fig:structure-dynamic-local-elimination} compares the eliminated dynamic stiffness with the corresponding full local compliance.

\begin{figure}[H]
\centering\begingroup
\definecolor{vltTeal}{HTML}{176B64}
\definecolor{vltBlue}{HTML}{315D80}
\definecolor{vltGold}{HTML}{A87520}
\definecolor{vltPlum}{HTML}{725877}
\definecolor{vltRed}{HTML}{762B38}
\definecolor{vltYellow}{HTML}{D5AD3C}
\definecolor{vltNeutral}{HTML}{777D83}
\setlength{\unitlength}{1pt}
\scalebox{0.80}{%
\input{vlt-figures/figure-071.tex}%
}\endgroup
\caption{Dynamic elimination \eqref{eq:structure-dynamic-local-schur} for equal masses and $\mathsf K/k_\star=\left(\begin{smallmatrix}2&-1\\-1&3\end{smallmatrix}\right)$, with drive and readout at site 1. Both maps use $x=\omega^2/\omega_\star^2$, $\omega_\star^2=k_\star/m$. The collective values $\lambda_\pm$ are zeros of the eliminated stiffness and poles of the full compliance. The eliminated coordinate instead has a pole at $x=3$, where the full compliance is zero. Real off-pole curves represent limits of the retarded response. Each curve is broken at its poles and clipped to the stated $[-4,4]$ ordinate range, so no line crosses a singularity. The comparison supplies no damping or linewidth.}
\label{fig:structure-dynamic-local-elimination}
\end{figure}

The static force slopes in Figure~\ref{fig:structure-local-schur-elimination}(b) use the same loading coordinate and declared supports.

\paragraph{Specify which body and surface coordinates can follow.}
A local force experiment can hold orientations fixed, release their
torques, or allow surface states to equilibrate. These give different
Schur responses. For the realization above, the relaxed mechanical
curvature and the leading added resistance are
\begin{equation}
 \begin{aligned}
 \mathsf H_{\rm rel}&=\mathsf H_X-\mathsf B\mathsf A^{-1}\mathsf B^{\mathsf T},\\
 \mathsf Z_X(\omega)&=\mathsf H_{\rm rel}-\omega^2\mathsf G_X
    -i\omega\left[\mathsf C_0+
       \mathsf B\mathsf A^{-1}\mathsf L^{-1}\mathsf A^{-1}\mathsf B^{\mathsf T}
               \right]+O(\omega^2).
 \end{aligned}
 \label{eq:s3patch-relaxed-stiffness-added-drag}
\end{equation}
The remainder includes the surface contribution at order $\omega^2$;
the displayed bare inertia is not the complete second-order expansion.
The added resistance is nonnegative. Thus finite-rate relaxation
produces phase and loss as well as static softening.

An explicit timescale test follows by setting
$\mathsf R=\mathsf A^{-1/2}\mathsf L^{-1}\mathsf A^{-1/2}$ and
$\tau_{\max}=\|\mathsf R\|_2$. For real $\omega$, the stiffness
correction $\Delta\mathsf H(\omega)=
-\mathsf B[(\mathsf A-i\omega\mathsf L^{-1})^{-1}
-\mathsf A^{-1}]\mathsf B^{\mathsf T}$ satisfies
\begin{equation}
 \|\Delta\mathsf H(\omega)\|_2
  \le \|\mathsf B\mathsf A^{-1/2}\|_2^2
                 \min(|\omega|\tau_{\max},1).
 \label{eq:s3patch-surface-adiabatic-error}
\end{equation}
Diagonalize the positive matrix $\mathsf R$ and use
$| (1-i\omega\tau)^{-1}-1 |
=|\omega|\tau/\sqrt{1+\omega^2\tau^2}$ to obtain the bound.
For a response comparison, multiply this correction by the relaxed
resolvent and apply the inverse-perturbation estimate
\eqref{eq:structure-internal-reduction-error-bound}. Near a soft mode
or a pole, a small coefficient error can still matter substantially.
The self-adjoint eigenproblem and positive square-root frequency
operator below belong to a specified conservative frozen or adiabatic
reference. The finite-rate operator~\eqref{eq:s3patch-augmented-interface-pencil}
instead has the poles of its augmented dynamics.

\begin{proposition}[Conservative frequency sensitivity]
\label{prop:s3-frequency-sensitivity}
At fixed masses and reference geometry, a positive-semidefinite increment
of an edge curvature adds a nonnegative matrix to $\mathsf K$, so the
ordered squared frequencies cannot decrease. For a differentiable real
symmetric $\mathsf D(\lambda)$ and a normalized simple eigenmode,
\begin{equation}
 \frac{\dd\omega_s^2}{\dd\lambda}
       =\bra{s}\frac{\dd\mathsf D}{\dd\lambda}\ket{s},\qquad
 \frac{\dd\omega_s}{\dd\lambda}
       =\frac1{2\omega_s}
           \bra{s}\frac{\dd\mathsf D}{\dd\lambda}\ket{s}
          \quad(\omega_s>0).
 \label{eq:mode-sensitivity}
\end{equation}

In particular, varying one unstressed spring constant $k_{ij}$ at
fixed geometry and masses gives
\begin{equation}
 \frac{\partial\omega_s^2}{\partial k_{ij}}
  =\left|\vct n_{ij}^{\mathsf T}
       (\vct\phi_{s,i}-\vct\phi_{s,j})\right|^2,
 \qquad \vct\phi_s=\mathsf M^{-1/2}\vct e_s.
 \label{eq:structure-edge-frequency-sensitivity}
\end{equation}
\end{proposition}
\begin{proof}
The min--max characterization gives monotonicity under a nonnegative
matrix increment~\cite{HornJohnson2013}. Differentiate the normalized
eigenvalue equation and multiply by its eigenbra; normalization cancels
the eigenvector derivatives and gives the first formula. The chain
rule gives the frequency derivative. A single spring increment has
quadratic form equal to its squared modal extension, giving the final
formula.
\end{proof}
Figure~\ref{fig:structure-bond-mode-sensitivity} compares a common pattern that leaves a link unchanged with a relative pattern whose frequency shifts when that link is stiffened.

\begin{figure}[H]
\centering\begingroup
\definecolor{vltTeal}{HTML}{176B64}
\definecolor{vltBlue}{HTML}{315D80}
\definecolor{vltGold}{HTML}{A87520}
\definecolor{vltPlum}{HTML}{725877}
\definecolor{vltRed}{HTML}{762B38}
\definecolor{vltYellow}{HTML}{D5AD3C}
\definecolor{vltNeutral}{HTML}{777D83}
\setlength{\unitlength}{1pt}
\scalebox{0.90}{%
\input{vlt-figures/figure-072.tex}%
}\endgroup
\caption{Common $(A,A)$ and relative $(A,-A)$ snapshots of the supported equal-mass pair share reference centres $0,a$, body size and a $12\,\mathrm{pt}/A$ gain. The common shift preserves the link length; the relative shift compresses it by $2A$. The exact squared-frequency lines vary the link curvature at fixed support, geometry and mass. Their slopes are determined by the squared mass-normalized link extensions. At zero link curvature, the perturbation selects these two directions within the degenerate eigenspace. Coil pitch follows the geometry; the printed parameters specify stiffness.}
\label{fig:structure-bond-mode-sensitivity}
\end{figure}

The derivative weights the bond by its modal extension. A link with zero
longitudinal relative motion has zero first-order sensitivity to this
unstressed stiffness increment. At degeneracy, diagonalizing the
perturbation within the complete eigenspace gives its frequency slopes.
If mass also varies, a physical normalized mode instead gives
$\dd\omega_s^2/\dd\lambda=
\vct\phi_s^{\mathsf T}
[\dd\mathsf K/\dd\lambda-\omega_s^2\dd\mathsf M/\dd\lambda]
\vct\phi_s$, with $\vct\phi_s^{\mathsf T}\mathsf M\vct\phi_s=1$.

Along equilibrated preparations, the derivative includes changing
reference positions and pair curvatures. The specified update determines
each mode's shift. Inference from that shift therefore records which
masses, supports and geometric inputs were fixed or independently measured.

An operator comparison is likewise made explicit. The dynamical matrix and
its frequency square root commute because they share a spectral
decomposition. For the following operator comparison, extend
$\mathsf W=\mathsf D^{1/2}$ to the full nonnegative finite space,
with value zero on $\ker\mathsf D$; its restriction to
$\mathcal H_+$ is the frequency operator used earlier.
\begin{definition}[Block-local restoring operator]
\label{def:s3-block-local-restoring}
Let $\mathsf\Pi_i$ be the orthogonal projector onto site $i$ in the
original mass-weighted coordinates. The \emph{block-local restoring
operator} and its commutator with the full dynamical matrix are
\begin{equation}
 \mathsf S=\sum_i\mathsf\Pi_i\mathsf D\mathsf\Pi_i,
 \qquad
 [\mathsf S,\mathsf D]_{ij}
       =\mathsf D_{ii}\mathsf D_{ij}
                       -\mathsf D_{ij}\mathsf D_{jj}.
 \label{eq:local-restoring-operator}
\end{equation}
\end{definition}
The block-diagonal operator $\mathsf S$ records mass-normalized local restoring
response and has units of inverse time squared. Its commutator is calculated
from the actual coupled blocks. A homogeneous cubic bulk response in the translational central
reference has
$\mathsf S=c\Id$, which commutes with $\mathsf W$. Boundaries, target coupling,
or heterogeneity may give a local response that does not commute with the
collective frequency operator. 
\begin{proposition}[Local-response and frequency commutation]
\label{prop:s3-frequency-commutation}
For finite nonnegative $\mathsf D$ and $\mathsf W=\mathsf D^{1/2}$,
$[\mathsf S,\mathsf W]=0$ if and only if
$[\mathsf S,\mathsf D]=0$.
\end{proposition}
\begin{proof}
Commutation with $\mathsf W$ implies commutation with its square.
Conversely, a map commuting with $\mathsf D$ preserves each of its
eigenspaces. Its square root is scalar on those same eigenspaces,
so it also commutes with that map.
\end{proof}

This definition uses the original mass-weighted site blocks. If fixed
constraints mix several sites, the admissible-coordinate map must
first be used to reconstruct the physical loading and readout; a
coordinate block of $\mathsf D_{\rm c}$ need not denote one virion.
If an operator comparison is made only on $\mathcal H_+$, use the
compression $\mathsf P_+\mathsf S\mathsf P_+$ there. In general
$\mathsf S$ itself need not preserve that subspace.

\begin{proposition}[Local spectral weights and frequency moments]
\label{prop:s3-local-spectral-moments}
Let $\mathsf D\succeq0$ have a complete orthonormal eigenbasis
$\vct e_s$, including zero modes, and let $\vct h$ be a unit
mass-weighted direction supported at site $i$ with unit polarization
$\vct p$. Its geometric spectral weights satisfy
\begin{equation}
 \begin{split}
 \rho_s^{(i,\vct p)}&=|\vct e_s^\dagger\vct h|^2,
       \qquad \sum_s\rho_s^{(i,\vct p)}=1,\\
 \sum_s\rho_s^{(i,\vct p)}\omega_s^2
       &=\vct p^\dagger\mathsf D_{ii}\vct p,\qquad
 \sum_s\rho_s^{(i,\vct p)}\omega_s
       =\vct p^\dagger\mathsf W_{ii}\vct p
       \le\sqrt{\vct p^\dagger\mathsf D_{ii}\vct p}.
 \end{split}
 \label{eq:structure-local-spectral-moments}
\end{equation}
Equality in the moment inequality holds exactly when all nonzero
weights share one frequency.
\end{proposition}
\begin{proof}
Completeness gives the unit sum of weights. The spectral
decompositions of $\mathsf D$ and $\mathsf W$ give the two moments.
Weighted Cauchy--Schwarz gives the final inequality, with equality
exactly for a frequency constant on the support of those weights.
\end{proof}
The clamped curvature divided by mass is thus a second frequency moment
of the collective modes sampled by that local direction~\cite{Teschl2009}.
Its square root is their root-mean-square frequency. At degeneracy, the
total weight at the shared frequency is basis independent. These
geometric weights determine how a localized initial displacement
decomposes into collective motion; occupation and thermal weights enter
through their separately specified preparations.

\begin{proposition}[Compression of the positive frequency operator]
\label{prop:s3-frequency-compression}
For $\mathsf D\succeq0$ and $\mathsf W=\mathsf D^{1/2}$ on the
full finite space, let $\bar i$ denote all sites other than $i$. Then
\begin{equation}
 \mathsf D_{ii}=\mathsf W_{ii}^{\,2}
                 +\mathsf W_{i\bar i}\mathsf W_{\bar i i},
 \qquad
 0\preceq\mathsf W_{ii}\preceq(\mathsf D_{ii})^{1/2}.
 \label{eq:structure-frequency-compression}
\end{equation}
Equality of the two block operators holds if and only if
$\mathsf W_{i\bar i}=0$, equivalently $\mathsf D_{i\bar i}=0$.
\end{proposition}
\begin{proof}
The diagonal block of $\mathsf W^2=\mathsf D$ gives the identity.
Its principal block $\mathsf W_{ii}$ and cross-block product are
nonnegative, so monotonicity of the positive
square root gives $\mathsf W_{ii}\preceq\mathsf D_{ii}^{1/2}$.
Equality makes the cross-block product zero and hence
$\mathsf W_{i\bar i}=0$. A block-diagonal matrix has a block-diagonal
positive square root, proving the equivalent condition for $\mathsf D$.
\end{proof}

The equality condition says that the site subspace is invariant under
the collective dynamics. This block statement needs neither simultaneous
diagonalization of the local matrices nor a positive spectral gap.
It makes precise why assigning one isolated ``site frequency'' from
the diagonal block discards the couplings responsible for relative
lattice motion.

Figure~\ref{fig:structure-local-collective-motion} illustrates the distinction dynamically: releasing a neighboring coordinate turns one clamped oscillation into a superposition of collective modes.

\begin{figure}[H]
\centering\begingroup
\definecolor{vltTeal}{HTML}{176B64}
\definecolor{vltBlue}{HTML}{315D80}
\definecolor{vltGold}{HTML}{A87520}
\definecolor{vltPlum}{HTML}{725877}
\definecolor{vltRed}{HTML}{762B38}
\definecolor{vltYellow}{HTML}{D5AD3C}
\definecolor{vltNeutral}{HTML}{777D83}
\setlength{\unitlength}{1pt}
\scalebox{0.90}{%
\input{vlt-figures/figure-073.tex}%
}\endgroup
\caption{The asymmetric supported pair has equal masses and the displayed $\mathsf K/k_\star$, with $\omega_\star^2=k_\star/m$. Reference centres are $0,a$ and the initial shifts are $(\ell_\star,0)$ at $15\,\mathrm{pt}/\ell_\star$, with zero velocities. Holding site 2 gives $u_1/\ell_\star=\cos(\sqrt2\tau)$; releasing it gives two collective cosines. Their site-1 weights $(1\pm1/\sqrt5)/2$ lie at $\lambda_\mp$. The dotted line locates their squared-frequency moment, two. All curves use the same interaction and support law; the hold selects the participating coordinates.}
\label{fig:structure-local-collective-motion}
\end{figure}

\begin{definition}[Operator spread of a mechanical direction]
\label{def:s3-directional-operator-spread}
For a normalized ket $\ket{\eta_{\rm dir}}$ and a self-adjoint operator
$A$ in the finite mechanical Hilbert space, its directional spread is
\[
 \Delta_{\eta_{\rm dir}} A=\sqrt{\bra{\eta_{\rm dir}}A^2\ket{\eta_{\rm dir}}
                       -\bigl(\bra{\eta_{\rm dir}}A\ket{\eta_{\rm dir}}\bigr)^2}.
\]
\end{definition}
\begin{proposition}[Joint directional-spread bound]
\label{prop:s3-directional-spread-bound}
For self-adjoint $\mathsf S,\mathsf W$ and a normalized mechanical
direction $\ket{\eta_{\rm dir}}$,
\begin{equation}
 \Delta_{\eta_{\rm dir}}\mathsf S\,\Delta_{\eta_{\rm dir}}\mathsf W
       \ge\frac12\left|\bra{\eta_{\rm dir}}
                   [\mathsf S,\mathsf W]\ket{\eta_{\rm dir}}\right|.
 \label{eq:conditional-resolution}
\end{equation}
\end{proposition}
\begin{proof}
Apply Cauchy--Schwarz to
$(\mathsf S-\langle\mathsf S\rangle)\ket{\eta_{\rm dir}}$ and
$(\mathsf W-\langle\mathsf W\rangle)\ket{\eta_{\rm dir}}$.
Their norms are the spreads. The magnitude of the imaginary part of
their inner product is half the absolute commutator expectation,
which proves the bound.
\end{proof}
The bound quantifies joint modal spread in a specified mechanical
direction~\cite{Shankar1994}. A direction concentrated in one
eigenspace has zero spread for that operator. The state and network
determine the bound, whose right-hand side can vanish. Connecting it
to measurement requires a physical probe and its error model.

Figure~\ref{fig:structure-operator-state-spread} evaluates this state dependence explicitly for the asymmetric supported pair and shows the additional information retained by the covariance term.

For real symmetric $\mathsf S,\mathsf W$ and a real mechanical
direction, the commutator is antisymmetric and its expectation
vanishes. Noncommutation alone therefore does not guarantee a
positive right-hand side in~\eqref{eq:conditional-resolution}.
\begin{proposition}[Covariance-resolved directional-spread bound]
\label{prop:s3-covariance-spread-bound}
For the same normalized direction, set
$\delta A=A-\langle A\rangle_{\eta}\Id$ and
$\operatorname{Cov}_{\eta}(\mathsf S,\mathsf W)
=\tfrac12\langle\{\delta\mathsf S,\delta\mathsf W\}\rangle_{\eta}$.
Then
\begin{equation}
 (\Delta_{\eta}\mathsf S)^2(\Delta_{\eta}\mathsf W)^2
 \ge \operatorname{Cov}_{\eta}(\mathsf S,\mathsf W)^2
      +\tfrac14|\langle[\mathsf S,\mathsf W]\rangle_{\eta}|^2.
 \label{eq:structure-covariance-resolution}
\end{equation}
Here $\eta$ abbreviates $\eta_{\rm dir}$ and
$\{A,B\}=AB+BA$ is the anticommutator.
\end{proposition}
\begin{proof}
Retain both parts of the centered inner product in the preceding
Cauchy--Schwarz argument. Its real part is the symmetrized covariance
and its imaginary part is the commutator expectation divided by $2i$. Their
squared magnitudes sum to its squared modulus.
\end{proof}
This finite-vector-space bound applies to classical energy directions
as well. Quantum preparations and detector errors receive their
physical definitions in Section~\ref{sec:operator-framework}.

\begin{figure}[H]
\centering\begingroup
\definecolor{vltTeal}{HTML}{176B64}
\definecolor{vltBlue}{HTML}{315D80}
\definecolor{vltGold}{HTML}{A87520}
\definecolor{vltPlum}{HTML}{725877}
\definecolor{vltRed}{HTML}{762B38}
\definecolor{vltYellow}{HTML}{D5AD3C}
\definecolor{vltNeutral}{HTML}{777D83}
\setlength{\unitlength}{1pt}
\scalebox{0.80}{%
\input{vlt-figures/figure-074.tex}%
}\endgroup
\caption{The covariance-resolved finite-vector bound \eqref{eq:structure-covariance-resolution} for the asymmetric pair. All displayed operators are normalized by their frequency units. The exact square root and commutator on the left act on $\eta_\theta$. On the right, the scaled spread product is $1/4+\sin^2\theta$ and the commutator contribution is $\sin^2\theta$. Their shaded difference is the covariance term $1/4$, saturating the full bound. At the real directions $\theta=0,\pi$, the expectation of the nonzero commutator vanishes. These are modal statistics of a defined finite-vector preparation; a detector-error model or quantum canonical postulate supplies its own observables.}
\label{fig:structure-operator-state-spread}
\end{figure}

\subsubsection{From shell geometry to deformation and loss of stability}
\label{sec:cmd-shell-mechanics}

The shell resolution in Section~\ref{sec:cmd-capsid-architecture}
connects an assembly load to a possible structural outcome. First find
the force-balanced reference under its actual supports, surface state
and loading. Then differentiate the complete energy on its admissible
coordinate space. \begin{proposition}[Shell tangent for length-dependent contacts]
\label{prop:cmd-shell-tangent}
For smooth contact-length energies in independent admissible coordinates,
the tangent stiffness is
\begin{equation}
 \boxed{\mathsf K_{\rm tan}
  =\mathsf J_\ell^{\mathsf T}\diag(\phi_e'')\mathsf J_\ell
    +\sum_e\phi_e'\nabla^2\ell_e
    +\nabla^2(\mathcal U_{\rm bend}+\mathcal U_{\rm other}).}
 \label{eq:cmd-shell-tangent}
\end{equation}
\end{proposition}
\begin{proof}
Differentiate each $\phi_e(\ell_e)$ twice. Its two contributions are
$\phi_e^{\prime\prime}\,\dd\ell_e\otimes\dd\ell_e$ and
$\phi_e^{\prime}\nabla^2\ell_e$. Sum the contacts and the remaining
energy Hessians on the same admissible coordinate chart.
\end{proof}

Figure~\ref{fig:cmd-shell-barriers} gives a graphical specialization of this boxed relation.

Here $\mathsf J_\ell$ differentiates the contact lengths with respect
to the chosen independent coordinates; primes hold the other retained
states fixed. This is the Hessian chain rule. Its second term is the
geometric stiffness from prestress, generalizing
\eqref{eq:edge-stiffness}. Retaining conserved or responsive surface
states then uses the constrained reductions of Section~4.

Partition this tangent into centre/orientation and shell blocks.
The existing Schur construction gives their relaxed coupling when
the eliminated block is positive and invertible. If $\vct b$ is a
calibrated force distribution and $Y=\vct d^{\mathsf T}\delta\vct X$
a displacement or shape readout, the local static response is
\begin{equation}
 \frac{\partial Y}{\partial F}
       =\vct d^{\mathsf T}\mathsf K_{\rm tan}^{-1}\vct b.
 \label{eq:cmd-shell-load-readout}
\end{equation}
Rigid zero modes are removed or supported before inversion. A bulk
force, wall contact and AFM tip generally have different $\vct b$.
They can therefore couple very differently to the same soft shell
direction. In an isotropic harmonic trap, uniform centre translation
does not by itself indent a freely moving shell. Contact, differential
traction or another specified coupling supplies that load path.

For an exactly icosahedral shell, the projectors in
\eqref{eq:icosahedral-mode-projectors} and the force selection in
\eqref{eq:icosahedral-invariant-force-selection} refine this design.
A uniform radial load samples invariant shell directions; a localized
contact carries additional symmetry components. A prospective soft
direction can be dark to a symmetric probe while visible to an
orientation-registered local load. Binding, support or a defect can
break that symmetry, so its complete tangent and detector map determine
the surviving selection rules.

Along a quasistatic force-controlled branch, a vanishing smallest
admissible eigenvalue of $\mathsf K_{\rm tan}$ signals local loss of
stability. A thermally activated transition can occur earlier, with
its rate depending on the barrier and loading time. Displacement
control, probe stiffness and support geometry change the combined
stability problem. Faceting of an elastic reference, reversible
indentation and persistent structural damage consequently receive
their own states and observations. The transport construction in
Section~\ref{sec:s7transport-delivery} continues those states after
unloading.

\begin{example}[Identical harmonic stiffness, different collapse barriers]
\label{ex:cmd-equal-stiffness-barriers}
Let $s=q/\ell_\star$, $f=F\ell_\star/E_\star$ and consider the
assigned reduced shell potentials
\begin{equation}
 \frac{\mathcal U_b(s;f)}{E_\star}
       =\frac{s^2}{2}-\frac{b s^3}{3}+\frac{s^4}{4}-fs,
       \qquad b\in\{2.2,2.6\}.
 \label{eq:cmd-shell-barrier-model}
\end{equation}
At zero load both have a local minimum at $s=0$ with stiffness
$E_\star/\ell_\star^2$. The positive barrier is at
$s_-=(b-\sqrt{b^2-4})/2$. The minimum reached from the origin loses
stability under increasing positive force at
\begin{equation}
 s_{\rm sp}=\frac{b-\sqrt{b^2-3}}{3},\qquad
 f_{\rm sp}=s_{\rm sp}-b s_{\rm sp}^2+s_{\rm sp}^3.
 \label{eq:cmd-shell-spinodal}
\end{equation}
These follow from $\partial_s\mathcal U=0$ and
$\partial_s^2\mathcal U=0$. The zero-load barriers are approximately
$0.0545E_\star$ and $0.0327E_\star$; the force-controlled spinodals
are approximately $0.129E_\star/\ell_\star$ and
$0.105E_\star/\ell_\star$, respectively.
\end{example}

The same initial capsomer geometry and harmonic Hessian can therefore
support different nonlinear failure predictions. Its low-amplitude
response does not determine the cubic and quartic coefficients.
The example supplies a controlled reason to reserve a load-dependent
shape, indentation or recovery measurement after harmonic calibration.
It assigns neither potential to a measured virus. For a physical shell,
the higher-order contact law and chemistry must constrain that landscape.
Figure~\ref{fig:cmd-shell-barriers} shows the matched local stiffness
and differing barriers, together with a possible shell deformation
projection.

\begin{figure}[H]
\centering\begingroup
\setlength{\unitlength}{1pt}
\scalebox{0.84}{%
\input{vlt-figures/figure-307.tex}%
}\endgroup
\caption{Geometry, tangent mechanics and a nonlinear structural test. Panel (a) preserves the 72-site graph while assigning a localized inward radial deformation $\vct X_a\mapsto[1-0.30\exp(-(1-\hat{\vct X}_a\cdot\vct e_z)/0.14)]\vct X_a$. It illustrates an admissible shape projection, rather than a force-balanced shell solution; the arrow locates an inward load. Panel (b) plots the two assigned zero-load potentials in \eqref{eq:cmd-shell-barrier-model}, with saddle points marked. Both have the same curvature at the origin, while their barriers and spinodals differ. The complete contact Hessian \eqref{eq:cmd-shell-tangent}, loading map and higher-order energy determine a physical prediction. The example explains why a calibrated harmonic response can leave a reserved indentation or recovery measurement informative.}
\label{fig:cmd-shell-barriers}
\end{figure}

\subsection{Bulk dispersion and a worked three-shell lattice}
\label{sec:worked-dispersion}

For an identical-particle bulk reference, insert a Bloch displacement into
\eqref{eq:bulk-difference-equation}. The result is the reciprocal-space
matrix~\eqref{eq:bulk-dispersion}, equivalently
\begin{equation}
 \mathsf D(\vct k)=\frac2m\sum_{\vct d\in\mathcal N_+}
       \mathsf C(\vct d)[1-\cos(\vct k\cdot\vct d)],\qquad
 \det[\mathsf D(\vct k)-\omega^2\Id_3]=0.
 \label{eq:paired-dispersion}
\end{equation}
Here $\mathcal N_+$ contains one representative of each opposite-offset pair.
Its three eigenvectors give the displacement polarizations at each wavevector.
The wavevector lies in the simple cubic first Brillouin zone
$[-\pi/a,\pi/a]^3$, with its periodically equivalent boundary points understood.

The Brillouin zone labels inequivalent spatial phase patterns: a
reciprocal-lattice shift preserves every sampled site phase.
Polarization specifies the displacement direction within a pattern.
Along suitable symmetry directions, longitudinal and transverse
polarizations are parallel and perpendicular to propagation. General
anisotropic directions require the actual eigenpolarizations.

\begin{example}[An unstressed three-shell reference]
\label{ex:three-shell}
Choose shell potentials
$V_\eta(r)=\tfrac12k_\eta(r-r_\eta)^2$ with
$k_\beta>0$, $k_\gamma>0$, and $k_\Omega\ge0$. Thus
$L_\eta=k_\eta$ and $T_\eta=0$. This is a specified effective spring model with
shell-dependent rest lengths; it provides an exactly tractable member of the
theory's model family.

Set $\theta_\mu=ak_\mu$, $c_\mu=\cos\theta_\mu$, and
$s_\mu=\sin\theta_\mu$. The reciprocal matrix entries are
\begin{equation}
 \begin{split}
 D_{xx}(\vct k)&=\frac2m\left[
 k_\beta(1-c_x)+k_\gamma(2-c_xc_y-c_xc_z)
                +\frac{4k_\Omega}{3}(1-c_xc_yc_z)\right],\\
 D_{xy}(\vct k)&=\frac2m\left[
 k_\gamma s_xs_y+\frac{4k_\Omega}{3}s_xs_yc_z\right].
 \end{split}
 \label{eq:three-shell-symbol}
\end{equation}
Permuting the coordinates gives the other entries. The sign-dependent
off-diagonal terms encode coupling between displacement components.

For propagation along $x$, $\vct k=(q,0,0)$, the polarizations separate into
one longitudinal and two transverse branches:
\begin{equation}
 \begin{split}
 \omega_L^2(q)&=\frac4m
       \left(k_\beta+2k_\gamma+\frac43k_\Omega\right)
                                  \sin^2\frac{qa}{2},\\
 \omega_{T_1}^2(q)=\omega_{T_2}^2(q)&=\frac4m
       \left(k_\gamma+\frac43k_\Omega\right)
                                  \sin^2\frac{qa}{2}.
 \end{split}
 \label{eq:three-shell-axis-dispersion}
\end{equation}
Axial nearest-neighbor springs alone have vanishing transverse frequencies
along this direction. Face and body diagonals add transverse restoring paths.
The long-wavelength speeds along $x$ are
$c_L^2=a^2(k_\beta+2k_\gamma+4k_\Omega/3)/m$ and
$c_T^2=a^2(k_\gamma+4k_\Omega/3)/m$.
\end{example}

\begin{proposition}[Stability of the unstressed three-shell bulk reference]
\label{prop:s3-three-shell-bulk-stability}
For Example~\ref{ex:three-shell}, with $k_\beta,k_\gamma>0$ and
$k_\Omega\ge0$, the bulk symbol is nonnegative and is positive definite
at every wavevector nonzero modulo the reciprocal lattice. For any
polarization $\vct p$,
\begin{equation}
 \vct p^\dagger\mathsf D(\vct k)\vct p
   =\frac1m\sum_{\vct d}
      k_{\eta(\vct d)}[1-\cos(\vct k\cdot\vct d)]
                          |\vct n_{\vct d}\cdot\vct p|^2\ge0.
 \label{eq:bulk-positive-form}
\end{equation}
\end{proposition}
\begin{proof}
If $\vct k$ is nonzero modulo the reciprocal lattice, at least one axial phase
is nonzero. Its spring forces the corresponding component of a zero-energy
polarization to vanish. Other nonzero axial phases remove their components in
the same way. Face-diagonal springs connecting to any remaining zero-phase
directions remove those components as well. Thus $k_\beta,k_\gamma>0$ give a
positive bulk matrix away from $\vct k=0$. Adding nonnegative body-diagonal
springs preserves this result.
\end{proof}

The actual assembled matrix supplies the corresponding boundary and
stability tests for the finite cylinder.

The axis branches also quantify the continuum approximation. For either
nonzero axis speed $c_{\mathrm{ax}}$ above, let $q>0$ and $0<qa\le\pi$.
Then
\begin{equation}
 \omega(q)=\frac{2c_{\mathrm{ax}}}{a}\sin\frac{qa}{2},\qquad
 \boxed{\frac{\omega(q)}{c_{\mathrm{ax}}q}
    =\frac{\sin(qa/2)}{qa/2}
    =1-\frac{(qa)^2}{24}+O((qa)^4).}
 \label{eq:structure-axis-continuum-error}
\end{equation}
The ratio compares the exact discrete frequency with its long-wave
prediction for the same mode. Its phase speed is $\omega/q$, whereas
the narrow-packet group speed is
$\dd\omega/\dd q=c_{\mathrm{ax}}\cos(qa/2)$.
Both approach $c_{\mathrm{ax}}$ at long wavelength; the group speed
vanishes at the zone edge even though its frequency is positive.
The exact bound $0\le1-\omega/(c_{\mathrm{ax}}q)\le(qa)^2/24$
gives a sufficient error criterion for this specific axis branch.
Figure~\ref{fig:structure-bulk-dispersions} compares the curves.

\begin{proposition}[Cubic moduli of the unstressed shell model]
\label{prop:s3-shell-moduli-isotropy}
The affine long-wavelength moduli of Example~\ref{ex:three-shell} are
\begin{equation}
 C_{11}=\frac{k_\beta+2k_\gamma+4k_\Omega/3}{a},\qquad
 C_{12}=C_{44}=\frac{k_\gamma+4k_\Omega/3}{a}.
 \label{eq:three-shell-moduli}
\end{equation}
They satisfy the cubic stability conditions
$C_{44}>0$, $C_{11}-C_{12}>0$ and $C_{11}+2C_{12}>0$.
This cubic elastic tensor is isotropic exactly when
$C_{11}=C_{12}+2C_{44}$, equivalently
$k_\beta=k_\gamma+8k_\Omega/3$.
\end{proposition}
\begin{proof}
Insert the three shell directions into
\eqref{eq:unstressed-elastic-tensor}. Their longitudinal fourth moments
give the displayed coefficients. The assumed spring signs make
$C_{44}$, $C_{11}-C_{12}$ and $C_{11}+2C_{12}$ positive.
Matching the cubic tensor to the isotropic Lam\'e form gives the stated
single additional relation.
\end{proof}

Cubic symmetry therefore fixes three independent moduli; the additional
relation selects its isotropic specialization.

Figure~\ref{fig:structure-cubic-isotropy} compares the isotropy condition with the actual directional acoustic speeds in one declared family of shell stiffnesses.

\begin{figure}[H]
\centering\begingroup
\definecolor{vltTeal}{HTML}{176B64}
\definecolor{vltBlue}{HTML}{315D80}
\definecolor{vltGold}{HTML}{A87520}
\definecolor{vltPlum}{HTML}{725877}
\definecolor{vltRed}{HTML}{762B38}
\definecolor{vltYellow}{HTML}{D5AD3C}
\definecolor{vltNeutral}{HTML}{777D83}
\setlength{\unitlength}{1pt}
\scalebox{0.80}{%
\input{vlt-figures/figure-075.tex}%
}\endgroup
\caption{The cubic family in \eqref{eq:three-shell-moduli}, with $k_\gamma=k_\star>0$, $k_\Omega=0$ and $k_\beta=rk_\star$. (a) Its Zener ratio equals one at $r=1$; the endpoint at $r=0$ is the continuous limit. (b) For propagation $(\cos\theta,\sin\theta,0)$, the in-plane eigenvalues of \eqref{eq:directional-acoustic-matrix} are constant at $1,3$ for $r=1$ and vary with direction for $r=2$. The unshown out-of-plane branch is one. Each ordinate is a squared speed in units $c_\star^2=a^2k_\star/m$. Thus the plotted family distinguishes cubic symmetry from its isotropic member using specified network curvatures.}
\label{fig:structure-cubic-isotropy}
\end{figure}

The equality $C_{12}=C_{44}$ is a further restriction of this
unstressed, affine central-force construction: the product
$n_pn_qn_rn_s$ in~\eqref{eq:unstressed-elastic-tensor} is unchanged
by any index permutation. If this specialization is also isotropic,
its Lam\'e moduli satisfy $\lambda_{\mathrm{el}}=\mu_{\mathrm{el}}$,
so that
\begin{equation}
 \nu_{\mathrm P}
   =\frac{\lambda_{\mathrm{el}}}
                  {2(\lambda_{\mathrm{el}}+\mu_{\mathrm{el}})}
   =\frac14,\qquad
 \frac{c_L}{c_T}=\sqrt3.
 \label{eq:central-force-isotropic-restriction}
\end{equation}
Here $\nu_{\mathrm P}$ is Poisson's ratio. These restrictions follow
from the unstressed affine central-force specialization. A wider range
of moduli can be modeled through internal relaxation, noncentral or
many-body interactions, or a consistent prestressed reference.
The medium retains its independently specified constitutive law.

\subsubsection{A repeated charged basis and its collective polarizations}
\label{sec:structure-charged-periodic-basis}

The one-site symbol applies when every primitive repeat has the same
particle and interaction environment. A repeated charge or mass pattern
is specified by the particles that repeat together.

\begin{definition}[Periodic charged ideal-virion basis]
\label{def:s3-periodic-charged-basis}
Let $\vct R_\ell$ be Bravais translations and let
$\vct d_\sigma$, $\sigma=1,\ldots,n_{\rm b}$, locate the retained
whole-particle centres in one primitive repeat. A periodic charged basis
assigns positions, masses and charges by
\begin{equation}
 \vct R_{\ell\sigma}=\vct R_\ell+\vct d_\sigma,\qquad
 m_{\ell\sigma}=m_\sigma>0,\qquad
 q_{\ell\sigma}=q_\sigma.
 \label{eq:structure-charged-basis-positions}
\end{equation}
All pair parameters and environmental terms entering the mechanical
operator repeat under these same translations.
\end{definition}
The index $\sigma$ labels a basis particle and is separate from the
interaction length $\sigma_{ij}$. A primitive basis counts distinct
particles repeated together; a coordination neighborhood classifies
neighbors relative to one centre. The complete masses, charges and
curvature tensors determine mechanical periodicity. Any smaller repeat
is verified on those coefficients.

Fix the radial laws and a stationary bulk reference before transforming
coordinates. Define the full Cartesian force constants
\begin{equation}
 \Phi_{0\sigma\mu,\ell\tau\nu}
   :=\left.
       \frac{\partial^2U}
       {\partial u_{0\sigma\mu}\partial u_{\ell\tau\nu}}
      \right|_{\vct u=0},\qquad
 \sum_{\ell,\tau}\Phi_{0\sigma\mu,\ell\tau\nu}=0
 \quad\hbox{for pair-only translation-invariant energy}.
 \label{eq:structure-basis-force-constants}
\end{equation}
The diagonal blocks contain the incident-pair sum, and the off-diagonal
blocks contain the negative pair tensors from~\eqref{eq:shell-curvatures}.
For the electrostatic term, the product $q_\sigma q_\tau$ enters the
same radial potential whose derivatives determine those tensors; charge
sign is not a separate label appended after diagonalization. The sum
rule expresses invariance under a common translation of all particles.
It does not remove the all-site force-balance condition or establish
positive curvature. A repeated unscreened Coulomb comparison additionally
requires its declared electrostatic summation and boundary prescription.
The formula below presumes that this prescription, screening or finite
range has made the force-constant transform well defined.

\begin{definition}[Multi-site mechanical Bloch matrix]
\label{def:s3-multisite-bloch-matrix}
For the periodic basis, use the cell-phase convention for the
mass-weighted displacement
$\vct x_{\ell\sigma}=\sqrt{m_\sigma}\vct u_{\ell\sigma}$.
With the well-defined force-constant transform specified above, the
mechanical Bloch matrix and polarization equation are
\begin{equation}
 \begin{aligned}
 \vct x_{\ell\sigma}(t)
  &=\operatorname{Re}\!\left[
      A_j\vct e_{\sigma j}(\vct k)
        e^{i(\vct k\cdot\vct R_\ell-\omega_jt)}\right],\\
 D_{\sigma\mu,\tau\nu}(\vct k)
  &=\frac1{\sqrt{m_\sigma m_\tau}}
       \sum_\ell\Phi_{0\sigma\mu,\ell\tau\nu}
                         e^{i\vct k\cdot\vct R_\ell},\qquad
 \mathsf D(\vct k)\vct e_j(\vct k)
       =\omega_j^2(\vct k)\vct e_j(\vct k).
 \end{aligned}
 \label{eq:structure-charged-basis-bloch}
\end{equation}
\end{definition}
Here $\sum_\sigma\vct e_{\sigma j}^\dagger\vct e_{\sigma j}=1$,
and $A_j$ carries units $\mathrm{kg^{1/2}\,m}$. Thus physical
displacement ratios are the components of
$\vct e_{\sigma j}/\sqrt{m_\sigma}$, not simply those of
$\vct e_{\sigma j}$. The matrix has $3n_{\rm b}$ rows; each eigenvector
specifies a relative motion of all basis particles at that wavevector.
This is the many-site form of the energy--inertia construction, with
the same Hessian signs and edge counting as the finite model
~\cite{Dove2011,Misra2012}.

\begin{proposition}[Bloch Hermiticity and phase-convention invariance]
\label{prop:s3-bloch-phase-invariance}
For real translation-invariant force constants derived from a symmetric
energy Hessian, assume the Bloch transform converges with index
rearrangements justified. Then
$\mathsf D(\vct k)^\dagger=\mathsf D(\vct k)$ and
$\mathsf D(-\vct k)=\mathsf D(\vct k)^*$.
The cell-phase matrix is reciprocal-lattice periodic. The corresponding
site-phase convention is related by
\begin{equation}
 \mathsf U(\vct k)=
  \operatorname{diag}_{\sigma}
       [e^{i\vct k\cdot\vct d_\sigma}\Id_3],\qquad
 \vct e_j^{\,\rm site}=\mathsf U^\dagger\vct e_j^{\,\rm cell},
 \qquad
 \mathsf D_{\rm site}=\mathsf U^\dagger\mathsf D_{\rm cell}\mathsf U.
 \label{eq:structure-bloch-phase-convention}
\end{equation}
The two conventions have identical spectra and reconstruct the same
physical displacement.
\end{proposition}
\begin{proof}
Hessian symmetry and translation give
$\Phi_{0\sigma\mu,\ell\tau\nu}
=\Phi_{0\tau\nu,-\ell\sigma\mu}$. Reversing the cell index in the
conjugate transpose proves Hermiticity; real coefficients give the
negative-wavevector identity. Reciprocal phases equal one on Bravais
translations. Factoring the site offsets out of each plane wave gives
the displayed unitary conjugation and preserves the reconstructed motion.
\end{proof}
Comparing entries or polarization phases requires matching the phase
convention. A reciprocal translation in the site-phase convention
can give a unitarily equivalent matrix, with the same periodic spectrum.

For an unpinned stable three-dimensional reference with exactly three
translation zero directions at $\vct k=0$, the normalized columns there
are proportional to
$\vct e_{\sigma a}=\sqrt{m_\sigma/M_{\rm cell}}\vct e_a$,
where $M_{\rm cell}=\sum_\sigma m_\sigma$ and
$a\in\{x,y,z\}$. Their physical displacements are common translations.
If the nearby spectrum is continuous and the remaining zero-wavevector
eigenvalues are strictly positive, three branches approach zero and the
other $3n_{\rm b}-3$ retain a gap. Additional mechanisms, loss of
stability or external pinning change this count. A gapped polarization
is a collective relative motion inside the repeat. Orthogonality to
the three translation columns gives
$\sum_\sigma\sqrt{m_\sigma}\vct e_{\sigma j}(\vct0)=\vct0$ for such
a mode, or zero physical centre-of-mass displacement of the basis at
zero wavevector. The gapped branch does not follow
from counting three neighbor shells or from assigning an internal
frequency to each particle. Its coupling to an electromagnetic field
would require an additional specified interaction.

Finite boundaries cut repeats, and disorder can break periodicity in
the interior. Equation~\eqref{eq:structure-finite-mode-preparation}
reconstructs the cylinder's actual motion from its finite assembly.
Bulk labels organize the periodic comparison.

\paragraph{A periodic basis of oriented VLPs.}
\begin{definition}[Translation--libration Bloch reference]
\label{def:s3patch-body-bloch}
For a resolved rigid-body repeat, periodicity includes orientations,
patch maps, frozen or equilibrated surface states and the environmental
coefficients. Reorder the scaled mechanical tangent into cell columns
$\vct X_\ell\in\mathbb R^{6n_{\rm b}}$. With positive cell inertia
$\mathsf G_{\rm b}$, let $\mathsf H_{0\ell}$ be its complete
conservative mechanical force-constant blocks. The well-defined Bloch
transform is
\begin{equation}
 \mathsf H_{\rm b}(\vct k)=\sum_\ell\mathsf H_{0\ell}
                   e^{i\vct k\cdot\vct R_\ell},\qquad
 \mathsf D_{\rm b}(\vct k)=\mathsf G_{\rm b}^{-1/2}
            \mathsf H_{\rm b}(\vct k)\mathsf G_{\rm b}^{-1/2}.
 \label{eq:s3patch-body-bloch}
\end{equation}
\end{definition}
Hessian symmetry and translation give
$\mathsf H_{0\ell}=\mathsf H_{0,-\ell}^{\mathsf T}$, so the
transformed matrix is Hermitian. The phase-convention result above
extends with the identity on each full six-component body block.
Its eigenvectors mix centre displacements and small angular oscillations,
or librations, according to their translational and rotational inertia
\cite{Dove2011}. Frozen and adiabatically relaxed surface references
give different conservative matrices; their actual finite-rate response
uses the augmented dynamics.

The common displacement $\vct u_{\ell\sigma}=\vct a$,
$\vct\theta_{\ell\sigma}=0$ remains a translation zero direction
for unpinned translation-invariant energy. If these are exactly its
three zero directions, the Bloch matrix is continuous and nonnegative
near zero, and its other zero-wavevector eigenvalues are positive,
three branches approach zero and the remaining
$6n_{\rm b}-3$ remain gapped. A global rotation of the finite body
assembly has both components in~\eqref{eq:s3patch-global-rigid-tangents};
its centre displacement grows with position and is not uniform spin
on a fixed periodic array. It supplies no automatic extra acoustic
branch. Individual orientations in an orientation-independent central
law instead have zero curvature and no restoring libration. Additional
gapless orientational propagation requires its own continuous symmetry
and spatial coupling; the icosahedral point group is discrete.
Translation invariance, rotational invariance and equilibrium therefore
remain distinct checks~\cite{LinPonceMarzari2022}.

\paragraph{The acoustic tensor after basis relaxation.}
The three translation columns identify the zero-wavevector subspace,
but their continuation does not in general move every basis particle
affinely. A slowly varying displacement can also induce small relative
shifts between the whole-particle centres inside a repeat. This is a
collective coordinate of several ideal virions, distinct from deformation
inside one rigid particle. Internal relaxation and the need to remove
translation zero directions before inversion are standard in crystal
elasticity~\cite{WuVanderbiltHamann2005}. Here the same distinction follows
directly from the declared mechanical Bloch matrix.

\begin{proposition}[Relaxed long-wave acoustic operator]
\label{prop:s3-relaxed-acoustic-operator}
In the site-phase convention, write $\vct k=s\widehat{\vct n}$ for a
fixed unit direction. Assume a twice continuously differentiable
Hermitian matrix near $s=0$, nonnegativity for both signs of $s$, and
exactly three translation zero directions at $s=0$. Choose orthonormal
translation columns $\mathsf E$ and complementary columns $\mathsf Z$.
Define
\begin{equation}
 \begin{aligned}
 \mathsf D(s\widehat{\vct n})
   &=\mathsf D_0+s\mathsf D_1+s^2\mathsf D_2+o(s^2),\\
 \mathsf H&=\mathsf Z^\dagger\mathsf D_0\mathsf Z\succ0,\\
 \mathsf A&=\mathsf E^\dagger\mathsf D_2\mathsf E,\qquad
 \mathsf B=\mathsf Z^\dagger\mathsf D_1\mathsf E .
 \end{aligned}
 \label{eq:structure-acoustic-relaxation-blocks}
\end{equation}
For an acoustic eigenvector
$\vct e=\mathsf E\vct a+\mathsf Z\vct b$, the leading relative
motion and ordered acoustic squared-frequency limits are
\begin{equation}
 \begin{aligned}
 \vct b&=-s\mathsf H^{-1}\mathsf B\vct a+o(s),\\
 \mathsf\Gamma(\widehat{\vct n})
   &=\mathsf A-\mathsf B^\dagger\mathsf H^{-1}\mathsf B
      \preceq\mathsf A,\\
 \lim_{s\to0}\frac{\omega_j^2(s\widehat{\vct n})}{s^2}
   &=\lambda_j[\mathsf\Gamma(\widehat{\vct n})],
       \qquad j=1,2,3 .
 \end{aligned}
 \label{eq:structure-relaxed-acoustic-tensor}
\end{equation}
\end{proposition}
\begin{proof}
Nonnegativity at both signs of $s$ forces
$\mathsf E^\dagger\mathsf D_1\mathsf E=0$.
The spectral gap of $\mathsf H$ separates the three small eigenvalues
from the complementary sector. The complementary equation can therefore
be inverted for these eigenvalues, giving
$\vct b=-s\mathsf H^{-1}\mathsf B\vct a+o(s)$.
Substitution into the translation equation gives
$s^2\mathsf\Gamma\vct a+o(s^2)=\omega^2\vct a$.
The min--max principle yields the ordered limits.
Finally $\mathsf B^\dagger\mathsf H^{-1}\mathsf B\succeq0$ proves
the softening inequality.
\end{proof}
The final eigenvalues have units of speed squared. Relative basis
motion softens the leading acoustic response by the nonnegative
correction $\mathsf B^\dagger\mathsf H^{-1}\mathsf B$, changing speeds
and potentially polarizations. The correction vanishes when
$\mathsf B=0$, or along a chosen polarization with $\mathsf B\vct a=0$.
A one-site translation-only basis has no complementary sector;
a one-body repeat with gapped librations has rotational
coordinates in that sector. A soft complementary mode
can enhance relaxation and narrow the useful elimination range; an
additional zero direction is retained explicitly.

Figure~\ref{fig:structure-acoustic-basis-relaxation} shows how relative basis motion lowers the leading acoustic restoring coefficient.

\begin{figure}[H]
\centering\begingroup
\definecolor{vltTeal}{HTML}{176B64}
\definecolor{vltBlue}{HTML}{315D80}
\definecolor{vltGold}{HTML}{A87520}
\definecolor{vltPlum}{HTML}{725877}
\definecolor{vltRed}{HTML}{762B38}
\definecolor{vltYellow}{HTML}{D5AD3C}
\definecolor{vltNeutral}{HTML}{777D83}
\setlength{\unitlength}{1pt}
\scalebox{0.90}{%
\input{vlt-figures/figure-076.tex}%
}\endgroup
\caption{(a) Pale frames identify repeats, dashed outlines reference centres, teal arrows the common shift and burgundy arrows its relative correction. The common shift is one display unit; the lower correction is $(-0.2,+0.2)$ at the same $16\,\mathrm{pt}$ gain. The body positions contain their sum. (b) A separate scalar quadratic has $A=2$, $B=1$, $H=2$ and $\beta=b/(sa)$ for nonzero leading amplitude $a$. Holding $\beta=0$ gives two; relaxation at $\beta=-1/2$ gives $3/2$. This depicts the nonnegative subtraction in~\eqref{eq:structure-relaxed-acoustic-tensor}, under its positive complementary-block and small-wavevector assumptions.}
\label{fig:structure-acoustic-basis-relaxation}
\end{figure}

These expansions require more than the existence of a Fourier transform.
A sufficient condition for two continuous derivatives is absolute
convergence of the force-constant sum weighted by
$1+(|\vct R_\ell|/\ell_{\rm ref})^2$, for any fixed reference length
$\ell_{\rm ref}>0$. Finite-range models and screened interactions can
satisfy it directly. An unscreened periodic Coulomb prescription requires
its own small-wavevector analysis; neutrality alone does not prove this
absolute moment or an analytic limit. Thus neither the relaxed tensor
above nor three ordinary linear acoustic dispersions follow solely from
the translation sum rule. The specified electric boundary prescription
and regularity remain part of that bulk comparison.

\subsubsection{Acoustic and optical branches}
\label{sec:internal-branch-hybridization}

The one-site translational example has three acoustic branches. A repeated
multi-site basis, or retained internal coordinates, supplies additional branches
with their own restoring mechanisms. For the measles example, a schematic
envelope coordinate $u_{\mathrm e}$ and internal-complex coordinate
$u_{\mathrm r}$ can be coupled by an internal spring $k_{\mathrm{int}}>0$.
For one polarization, their mass-weighted zero-wavevector block is
\begin{equation}
 \mathsf D_{\mathrm{int}}(0)=k_{\mathrm{int}}
 \begin{pmatrix}
 1/m_{\mathrm e}&-1/\sqrt{m_{\mathrm e}m_{\mathrm r}}\\
 -1/\sqrt{m_{\mathrm e}m_{\mathrm r}}&1/m_{\mathrm r}
 \end{pmatrix}.
 \label{eq:internal-optical-block}
\end{equation}
The common physical displacement is a zero mode; the relative coordinate has
$\omega_0^2=k_{\mathrm{int}}(1/m_{\mathrm e}+1/m_{\mathrm r})$.
The effective masses partition one virion's retained inertia,
$m_{\mathrm e}+m_{\mathrm r}=m$, while the two coordinates resolve
its common and relative motion.
Measles structure motivates these coordinate levels~\cite{Liljeroos2011,Ke2018}.
The coupling $k_{\mathrm{int}}$ is a separate mechanical input.

\begin{definition}[Common and relative internal coordinates]
\label{def:s3-common-relative-internal-coordinates}
For the two-coordinate unit with positive masses
$m_{\mathrm e},m_{\mathrm r}$ and
$m=m_{\mathrm e}+m_{\mathrm r}$, define the common displacement,
internal separation and reduced mass by
\begin{equation}
 u_{\rm cm}=\frac{m_{\mathrm e}u_{\mathrm e}
                         +m_{\mathrm r}u_{\mathrm r}}{m},\qquad
 w_{\rm int}=u_{\mathrm e}-u_{\mathrm r},\qquad
 \mu_{\rm int}=\frac{m_{\mathrm e}m_{\mathrm r}}{m}.
 \label{eq:structure-internal-coordinates}
\end{equation}
\end{definition}
For the isolated unit, its energy separates as
\begin{equation}
 H_{\rm unit}=\frac12m|\Dt{u_{\rm cm}}|^2
       +\frac12\mu_{\rm int}|\Dt{w_{\rm int}}|^2
       +\frac12k_{\mathrm{int}}w_{\rm int}^2,\qquad
 \omega_0^2=\frac{k_{\mathrm{int}}}{\mu_{\rm int}}.
 \label{eq:structure-internal-energy-split}
\end{equation}
The common coordinate carries the whole retained inertia, while the
relative coordinate stores deformation energy. The two-coordinate unit
selects one envelope--RNP motion from the richer internal structure.

To extend this unit to a repeated model, suppose that only the envelope
coordinates couple between cells. For one polarization of that envelope
network, let $\Lambda_s(\vct k)\ge0$ be its Fourier stiffness eigenvalue,
with units of stiffness and $\Lambda_s(\vct0)=0$. Identical isotropic
internal springs give the mass-weighted block
\begin{equation}
 \mathsf D_s(\vct k)=
 \begin{pmatrix}
 [\Lambda_s(\vct k)+k_{\mathrm{int}}]/m_{\mathrm e}
       &-k_{\mathrm{int}}/\sqrt{m_{\mathrm e}m_{\mathrm r}}\\
 -k_{\mathrm{int}}/\sqrt{m_{\mathrm e}m_{\mathrm r}}
       &k_{\mathrm{int}}/m_{\mathrm r}
 \end{pmatrix}.
 \label{eq:coupled-internal-bulk-block}
\end{equation}
Defining $A_s=(\Lambda_s+k_{\mathrm{int}})/m_{\mathrm e}$ and
$B=k_{\mathrm{int}}/m_{\mathrm r}$, its two branches are
\begin{equation}
 \omega_{\pm,s}^2(\vct k)
    =\frac{A_s+B\pm
       \sqrt{(A_s-B)^2+4k_{\mathrm{int}}^2/
                                      (m_{\mathrm e}m_{\mathrm r})}}{2}.
 \label{eq:internal-hybrid-branches}
\end{equation}
At zero wavevector they reduce to the common translation and the relative
mode above. For $\Lambda_s/k_{\mathrm{int}}\ll1$ the expansions are
\begin{equation}
 \begin{split}
 \omega_{-,s}^2&=\frac{\Lambda_s}{m}
               +O\!\left(\frac{\Lambda_s^2}{m k_{\mathrm{int}}}\right),\\
 \omega_{+,s}^2&=\omega_0^2+
          \frac{m_{\mathrm r}\Lambda_s}{m_{\mathrm e}m}
               +O\!\left(\frac{\Lambda_s^2}{m k_{\mathrm{int}}}\right).
 \end{split}
 \label{eq:internal-branch-low-k}
\end{equation}
The mass ratio is fixed in these expansions. At long wavelength the
acoustic mode carries the whole virion's inertia, although intercell
forces act on its envelope. Finite $\Lambda_s$ mixes common and
relative motion, and positive coupling keeps the two eigenvalues distinct.
Envelope and relative-displacement probes consequently weight the
branches differently. Anisotropic internal springs or nonidentical units
use the full coupled matrix. The term optical-like identifies the gapped
relative branch; coupling to an optical instrument additionally specifies
dipole, polarizability or another transduction mechanism.
Figure~\ref{fig:structure-bulk-dispersions} plots this declared mechanical
example with dimensionless parameters.

\begin{figure}[H]
\centering\begingroup
\definecolor{vltTeal}{HTML}{176B64}
\definecolor{vltBlue}{HTML}{315D80}
\definecolor{vltGold}{HTML}{A87520}
\definecolor{vltPlum}{HTML}{725877}
\definecolor{vltRed}{HTML}{762B38}
\definecolor{vltYellow}{HTML}{D5AD3C}
\definecolor{vltNeutral}{HTML}{777D83}
\setlength{\unitlength}{1pt}
\scalebox{0.80}{%
\input{vlt-figures/figure-077.tex}%
}\endgroup
\caption{Auxiliary homogeneous bulk dispersions and their limiting comparison. (a) A positive axis speed gives the exact branch $\omega a/c_{\rm ax}=2\sin(qa/2)$ in \eqref{eq:three-shell-axis-dispersion}; its dashed tangent is the long-wavelength law. (b) The envelope--internal example of \eqref{eq:internal-hybrid-branches} has $m_{\rm e}=m_{\rm r}=m_0$, $k_{\rm inter}=k_{\rm int}$, only envelopes coupled between cells, and $\omega_\star=\sqrt{k_{\rm int}/m_0}$. Its acoustic branch starts at zero and its internal branch at $\sqrt2\omega_\star$. (c) The boxed ratio \eqref{eq:structure-axis-continuum-error} is compared directly with its sufficient lower bound. Shading locates the gap between those two functions, including the continuous value at zero. The squared version is plotted in Figure~\ref{fig:structure-discrete-continuum}. These bulk examples keep their stated vertical scales; the finite cylinder retains discrete boundary-adapted modes.}
\label{fig:structure-bulk-dispersions}
\end{figure}

The displacement directions behind the branch labels are shown in Figure~\ref{fig:added-polarization-geometry}.

\begin{figure}[H]
\centering\begingroup
\definecolor{vltTeal}{HTML}{176B64}
\definecolor{vltBlue}{HTML}{315D80}
\definecolor{vltGold}{HTML}{A87520}
\definecolor{vltPlum}{HTML}{725877}
\definecolor{vltRed}{HTML}{762B38}
\definecolor{vltYellow}{HTML}{D5AD3C}
\definecolor{vltNeutral}{HTML}{777D83}
\setlength{\unitlength}{1pt}
\scalebox{0.90}{%
\input{vlt-figures/figure-078.tex}%
}\endgroup
\caption{Each polarization panel retains all twelve ideal virions of one $4\times3$ reference grid, with the same orthographic camera, body radius and displacement gain. Open points mark reference centres. Offset arrows are parallel to their actual shifts, $\vct u/A=\vct p\cos(qx)$ at $qa=\pi/2$, with $\vct p=\vct e_x,\vct e_y,\vct e_z$. Triads fix the viewing directions; separate wavevector arrows identify propagation. Pale grid lines locate reference coordinates. Colors distinguish polarizations in this figure. The two transverse branches are degenerate in an isotropic bulk model; an actual cubic or finite network uses its own spectrum.}
\label{fig:added-polarization-geometry}
\end{figure}

\subsubsection{What is lost when an internal coordinate is omitted?}
\label{sec:structure-internal-reduction}

The two-coordinate unit also tests the adequacy of a coarser model.
Choose one bulk polarization, suppress its index, and apply a force
$F_{\mathrm e}$ to its envelope coordinate. For the Fourier convention
$e^{-izt}$ with $\operatorname{Im}z>0$, the unforced internal equation gives
\begin{equation}
 u_{\mathrm r}(z)=\frac{k_{\mathrm{int}}}
                         {k_{\mathrm{int}}-m_{\mathrm r}z^2}
                         u_{\mathrm e}(z),\qquad
 \mathcal R_{\mathrm e}(z)
 =\frac{u_{\mathrm e}(z)}{F_{\mathrm e}(z)}
 =\frac{1}{\Lambda-z^2m_{\mathrm{eff}}(z)},
 \label{eq:structure-internal-effective-inertia}
\end{equation}
where
$m_{\mathrm{eff}}(z)=m_{\mathrm e}
+m_{\mathrm r}/(1-m_{\mathrm r}z^2/k_{\mathrm{int}})$.
This effective inertia expresses delayed conservative internal motion;
it is not the material mass of a newly defined particle. At
$|z|^2m_{\mathrm r}/k_{\mathrm{int}}\ll1$,
$m_{\mathrm{eff}}(z)=m+O(m_{\mathrm r}^2z^2/k_{\mathrm{int}})$.
The centre-only response $1/(\Lambda-mz^2)$ is thus the leading
low-frequency reduction, away from a pole where small denominator
errors can be amplified. Its validity is a statement about a
selected frequency band and readout.

The elimination formula has an apparent singularity at
$z^2=k_{\mathrm{int}}/m_{\mathrm r}$. The regular full response is
\begin{equation}
 \boxed{\mathcal R_{\mathrm e}(z)=
 \frac{k_{\mathrm{int}}-m_{\mathrm r}z^2}
 { [\Lambda+k_{\mathrm{int}}-m_{\mathrm e}z^2]
   [k_{\mathrm{int}}-m_{\mathrm r}z^2]-k_{\mathrm{int}}^2}.}
 \label{eq:structure-internal-antiresonance}
\end{equation}
For $\Lambda>0$ its positive-frequency poles are the two mixed modes in
\eqref{eq:internal-hybrid-branches}. At the intervening frequency
$\omega_{\mathrm{ar}}=\sqrt{k_{\mathrm{int}}/m_{\mathrm r}}$,
the numerator vanishes while the denominator equals
$-k_{\mathrm{int}}^2$: the envelope readout has an
\emph{antiresonance}, or cancellation of driven displacement.
The internal coordinate still moves, with
$u_{\mathrm r}=-F_{\mathrm e}/k_{\mathrm{int}}$ at that frequency.
These statements concern the driven harmonic component, obtained
as $z\to\omega+i0$. Arbitrary initial conditions can add free
oscillations that persist in the lossless model.
This antiresonance is interference between visible modal contributions
at a frequency between their poles. A dark eigenmode instead has zero
overlap with the selected channel at its own frequency.

Figure~\ref{fig:structure-internal-antiresonance} compares exact and
constant-inertia responses at the same parameters. The envelope signal
vanishes at the cancellation while an internal coordinate still responds.
A calibrated second coordinate can test that prediction. Damping,
heterogeneity and support motion enter through an expanded forward model
and can change the cancellation. Retaining the exact
$m_{\mathrm{eff}}(z)$ already reproduces the full envelope response;
the example specifically tests its constant-mass approximation.

\begin{figure}[H]
\centering\begingroup
\definecolor{vltTeal}{HTML}{176B64}
\definecolor{vltBlue}{HTML}{315D80}
\definecolor{vltGold}{HTML}{A87520}
\definecolor{vltPlum}{HTML}{725877}
\definecolor{vltRed}{HTML}{762B38}
\definecolor{vltYellow}{HTML}{D5AD3C}
\definecolor{vltNeutral}{HTML}{777D83}
\setlength{\unitlength}{1pt}
\scalebox{0.90}{%
\input{vlt-figures/figure-079.tex}%
}\endgroup
\caption{The lossless response \eqref{eq:structure-internal-antiresonance} with $m_{\rm e}=m_{\rm r}=m_0$ and fixed envelope-network curvature $\Lambda=k_{\rm int}$. The full envelope response is $(1-y^2)/(1-3y^2+y^4)$; the centre-only comparison is $1/(1-2y^2)$. Both are normalized by $k_{\rm int}$ and share the static compliance and leading inertial correction. Dotted guides mark their poles; every curve is interrupted at a pole and clipped outside $[-6,6]$. The open zero at $y=1$ is driven cancellation: the envelope stays still while the internal coordinate has $u_{\rm r}=-f_{\rm e}/k_{\rm int}$. Equal and opposite inset force arrows use one gain. The black internal dot is a schematic retained coordinate within one virion, not another particle or a normal-mode snapshot. No linewidth is assigned to this conservative example.}
\label{fig:structure-internal-antiresonance}
\end{figure}

\paragraph{A bandwise error test for constant inertia.}
\begin{proposition}[Bandwise constant-inertia response error]
\label{prop:s3-constant-inertia-response-error}
For the two-coordinate unit with the same envelope force and readout,
let $\mathcal R_0(z)=(\Lambda-mz^2)^{-1}$ and
$\delta m=m_{\mathrm{eff}}(z)-m$.
At a frequency with $\eta=|z|^2m_{\mathrm r}/k_{\mathrm{int}}<1$
and nonzero reduced denominator, define $\alpha(z)$ below.
If $\alpha(z)<1$, the inertia and relative response errors satisfy
\begin{equation}
 \begin{aligned}
 |\delta m|&\le\frac{m_{\mathrm r}\eta}{1-\eta},\qquad
 \alpha(z):=\frac{|z|^2m_{\mathrm r}\eta}
                {(1-\eta)|\Lambda-mz^2|}<1,\\
 \left|\frac{\mathcal R_{\mathrm e}(z)}{\mathcal R_0(z)}-1\right|
   &\le\frac{\alpha(z)}{1-\alpha(z)} .
 \end{aligned}
 \label{eq:structure-internal-reduction-error-bound}
\end{equation}
\end{proposition}
\begin{proof}
The exact difference is
$\delta m=m_{\mathrm{eff}}-m=m_{\mathrm r}^2z^2/
[k_{\mathrm{int}}(1-m_{\mathrm r}z^2/k_{\mathrm{int}})]$.
The inequality $|1-m_{\mathrm r}z^2/k_{\mathrm{int}}|\ge1-\eta$
bounds it as displayed. Writing
$\mathcal R_{\mathrm e}/\mathcal R_0=
[1-z^2\delta m/(\Lambda-mz^2)]^{-1}$ and using
$|z^2\delta m/(\Lambda-mz^2)|\le\alpha<1$ proves the second bound.
\end{proof}
The condition excludes a zero reduced denominator. A small ratio of
frequency squared to $k_{\mathrm{int}}/m_{\mathrm r}$ controls the inertia
expansion, whereas the additional denominator factor controls the
observable response. Close to a mechanical resonance, the latter can
fail even while the former is small. Requiring the two inequalities
uniformly on a declared band gives a sufficient response-error test for
that drive and readout; it is not an error bound for arbitrary internal
coordinates or all possible measurements. This tests a proposed
coarse description using its retained and omitted mechanics, rather
than identifying scale separation from particle shape alone.

\paragraph{Relaxational bands and a noncommuting long-wave limit.}
Fourier transforming the full surface/material realization gives the
wavevector-dependent version of
\eqref{eq:s3patch-augmented-interface-pencil}, with $\mathsf H_{\rm b}$
the clamped block at fixed $\vct\chi$. A relaxed block is rematched
before its eliminated states are restored. A conserved surface or
ionic population can have singular $\mathsf L(\vct k)$. For a Laplace
frequency $s$ with positive real part, eliminate its dynamic row without
inverting that mobility:
\begin{equation}
 \begin{aligned}
 \mathsf Z_{\rm b}(\vct k,s)
 &=s^2\mathsf G_{\rm b}+s\mathsf C_{\rm b}+\mathsf H_{\rm b}\\
 &\quad-\mathsf B_{\rm b}
  [s\Id+\mathsf L_{\rm b}\mathsf A_{\rm b}]^{-1}
       \mathsf L_{\rm b}\mathsf B_{\rm b}^{\dagger} .
 \end{aligned}
 \label{eq:s3patch-conserved-surface-bloch}
\end{equation}
All blocks on the right refer to the same $\vct k$. Preparation supplies
the additional source
$-\mathsf B_{\rm b}[s\Id+\mathsf L_{\rm b}\mathsf A_{\rm b}]^{-1}
\vct\chi(\vct k,0)$, together with mechanical initial data and
transformed noise. Conserved zero directions keep their prepared
inventory. Static relaxation is performed on the allowed subspace.

For a scalar relative or librational coordinate coupled by $b$ to a
diffusing population, take $A>0$ and $L(\vct k)=\ell|\vct k|^2$,
$\ell>0$. Its subtraction from the mechanical curvature is
\begin{equation}
 \Sigma(\vct k,s)=
   \frac{b^2\ell|\vct k|^2}{s+\ell A|\vct k|^2},\qquad
 \lim_{|\vct k|\to0}\lim_{s\to0^+}\Sigma=\frac{b^2}{A},\qquad
 \lim_{s\to0^+}\lim_{|\vct k|\to0}\Sigma=0 .
 \label{eq:s3patch-surface-limit-order}
\end{equation}
The same preparation can consequently look relaxed on one spatial and
temporal scale and frozen on another. The full augmented generator
has decay and oscillation poles $s_j(\vct k)$; surface relaxation does
not add particle mass or a positive-frequency phonon branch. The
conservative packet and occupation formulas below apply to their
specified inertial reference. A quasiparticle description of damped
motion additionally requires sufficiently resolved oscillatory poles.

\subsubsection{Packet propagation in the bulk approximation}
\label{sec:bulk-packet-transport}

For a differentiable nondegenerate positive bulk branch, a normalized
polarization $\vct e_s(\vct k)$ gives
\begin{equation}
 v_{g,s,\mu}=\partial_{k_\mu}\omega_s
       =\frac{\vct e_s^\dagger
              (\partial_{k_\mu}\mathsf D)\vct e_s}{2\omega_s}.
 \label{eq:group-velocity-mode-derivative}
\end{equation}
This derivative links the directional couplings to the motion of a narrow
packet. Expanding the frequency about its central wavevector $\vct k_0$
gives a leading envelope translated by $\vct v_{g,s}t$.
For spectral width $\Delta k$, neglect of dispersion requires
$t\|\nabla_{\vct k}^2\omega_s(\vct k_0)\|(\Delta k)^2\ll1$,
with polarization variation also small across the packet.
This bulk description applies before appreciable boundary reflection
and requires a packet scale compatible with the retained interactions.
At branch degeneracy a coupled polarization subspace replaces the
single-branch derivative.
The bandwidth criterion assumes that the Taylor remainder is controlled
throughout the packet's spectral neighborhood. If the quadratic dispersion
term vanishes, its first nonzero higher derivative sets the dispersive scale.
Phase speed describes the motion of a sinusoidal crest; group velocity
describes the leading envelope of a sufficiently narrow packet.
Neither determines that packet's amplitude or its observation lifetime.
Those also depend on its preparation, reflection and medium coupling.

\paragraph{An envelope is a field of collective centre motion.}
For the repeated basis of Section~\ref{sec:structure-charged-periodic-basis},
use its cell-phase polarizations, choose a smooth positive branch $j$
near $\vct k_0$, and write
$\omega_0=\omega_j(\vct k_0)>0$. A narrow superposition of its
energy amplitudes has a slowly varying scalar envelope
$\mathcal A(\vct R,t)$, with units $\mathrm{J}^{1/2}$ per repeat.
Freezing the polarization and reciprocal frequency at their central values
in the displacement reconstruction gives
\begin{equation}
 \vct u_{\ell\sigma}(t)\simeq
 \sqrt{\frac2{m_\sigma}}\operatorname{Re}\!\left[
   \frac{\vct e_{\sigma j}(\vct k_0)}{\omega_0}
   \mathcal A(\vct R_\ell,t)
   e^{i(\vct k_0\cdot\vct R_\ell-\omega_0t)}\right].
 \label{eq:structure-bulk-envelope-centre-motion}
\end{equation}
One envelope reconstructs correlated centre displacements throughout
each repeat. Masses and branch polarization fix their relative
amplitudes. With unitary lattice Fourier normalization, its leading
energy norm is $E_+\simeq\sum_\ell|\mathcal A(\vct R_\ell,t)|^2$.

Retaining the quadratic frequency expansion, rather than suppressing it,
gives the envelope equation
\begin{equation}
 \begin{aligned}
 i(\partial_t+\vct v_{g,j}\cdot\nabla)\mathcal A
  &=-\frac12\sum_{\mu,\nu}H^{(j)}_{\mu\nu}
                       \partial_\mu\partial_\nu\mathcal A,\\
 H^{(j)}_{\mu\nu}
       &:=\left.\partial_{k_\mu}\partial_{k_\nu}\omega_j\right|_{\vct k_0}.
 \end{aligned}
 \label{eq:structure-dispersive-envelope-equation}
\end{equation}
Each envelope Fourier component acquires the quadratic Taylor phase,
which directly proves the sign and factor in this equation.
The real symmetric curvature tensor has units $\mathrm{m^2\,s^{-1}}$;
it can produce anisotropic spreading or focusing without dissipation.
The approximation requires small variation of both polarization and
$1/\omega_j$, a controlled higher-order Taylor phase throughout the
occupied band, and no appreciable boundary interaction. In particular,
bounded third derivatives give the sufficient remainder criterion
$t\sup\|\nabla^3_{\vct k}\omega_j\|(\Delta k)^3\ll1$.
The envelope equation conserves its norm with periodic or decaying
envelope boundary conditions. Its complex form therefore describes
classical dispersive lattice mechanics as well as the corresponding
one-excitation evolution; it introduces no quantum preparation by itself.

The finite standing-mode expansion recovers vessel reflections and
interference. Superposition can move a packet or change its local
amplitude while every energy $E_s$ of the isolated diagonal harmonic
modes stays constant. This describes changing relative phase and spatial
energy distribution; transfer between modal energies requires the
appropriate forcing, coupling or nonlinear terms. Physical energy can
still flow between particles and regions, as quantified by the bond
currents in~\eqref{eq:edge-flux}. The next section adds causal medium
storage and loss. Section~\ref{sec:sup-mechanical} distinguishes global
modal orthogonality from interference in a local projection.

Figure~\ref{fig:final-packet-phase-group} turns this distinction into a
spatial readout. Here $A_0$ is the initial peak displacement amplitude and
$\omega_0=\omega(q_0)$; the reference scale and bandwidth are given in the
caption. A spatially resolved marker can follow a carrier phase, an
envelope arrival or a local amplitude. Comparing these quantities tests
the dispersion of the chosen contact network; they need not return the
same propagation speed. Assigning energy transfer additionally requires
the force--velocity current and its projection onto that network.

\begin{figure}[H]
\centering\begingroup
\definecolor{vltTeal}{HTML}{176B64}
\definecolor{vltBlue}{HTML}{315D80}
\definecolor{vltGold}{HTML}{A87520}
\definecolor{vltPlum}{HTML}{725877}
\definecolor{vltRed}{HTML}{762B38}
\definecolor{vltYellow}{HTML}{D5AD3C}
\definecolor{vltNeutral}{HTML}{777D83}
\setlength{\unitlength}{1pt}
\scalebox{0.80}{%
\input{vlt-figures/figure-080.tex}%
}\endgroup
\caption{Carrier phase and packet envelope for the unpinned monatomic branch $\omega(q)=2\omega_\star\sin(qa/2)$, with $\omega_\star=\sqrt{k/m}$. Both frames use $u/A_0=\exp[-(z-v_gt)^2/(4\sigma_z^2)]\cos(q_0z-\omega_0t)$, the first-order narrow-band approximation at $q_0a=\pi/2$ and $\sigma_z=5a$. Shading follows the upper and lower Gaussian envelopes. The burgundy guide moves with $v_g$; the blue dotted guide follows the same continuously tracked carrier-phase crest with $v_{\rm ph}$. At $\omega_\star t=30$ their positions are $21.21a$ and $27.01a$. The dot marks carrier phase inside the envelope; it need not be a local maximum of the product. Equal axes retain the speed comparison. Quadratic dispersion is omitted in these two schematic frames; its small parameter is $t|\omega^{\prime\prime}(q_0)|(\Delta q)^2\simeq0.106$ at the later time, with $\Delta q=1/(2\sigma_z)$.}
\label{fig:final-packet-phase-group}
\end{figure}

\subsubsection{Scattering, mode conversion and localized matrix updates}
\label{sec:s3-phonon-scattering-updates}

Spatial inhomogeneity gives a packet something to scatter from: a changed
mass, a different charge-dependent curvature, an interface, a boundary
or a newly established contact. Their common mathematical input is a
change of the assembled inertia and stiffness. This connects the
mechanical matrix formalism to lattice-wave scattering and Green-function
transport methods~\cite{WangWangLu2008,Tamura1983}. For a comparison on
the same full particle coordinate space, write
\begin{equation}
 \mathsf A_0(z)=\mathsf K_0-z^2\mathsf M_0,\qquad
 \mathsf W(z)=\Delta\mathsf K-z^2\Delta\mathsf M,\qquad
 \mathsf R_0(z)=\mathsf A_0(z)^{-1}.
 \label{eq:s3-scattering-dynamic-defect}
\end{equation}
All matrices use the same displacement units. The frequency-dependent
term in $\mathsf W$ is necessary for an inertia change; adding a
particle mass to a stiffness matrix would mix incompatible units.

\begin{definition}[Mechanical transition matrix]
\label{def:s3-mechanical-transition-matrix}
At a frequency where the inverses exist, define
\begin{equation}
 \mathsf T(z)
    :=\mathsf W(z)[\Id+\mathsf R_0(z)\mathsf W(z)]^{-1},\qquad
 \mathsf R(z)=[\mathsf A_0(z)+\mathsf W(z)]^{-1}.
 \label{eq:s3-mechanical-transition-matrix}
\end{equation}
The operator $\mathsf T$ sums repeated encounters with the changed
mechanical region in the chosen reference response.
\end{definition}

\begin{proposition}[Exact response update and its first approximation]
\label{prop:s3-transition-resolvent-update}
The perturbed response satisfies
\begin{equation}
 \mathsf R=\mathsf R_0-\mathsf R_0\mathsf T\mathsf R_0.
 \label{eq:s3-scattering-resolvent-update}
\end{equation}
If $\eta=\|\mathsf R_0\mathsf W\|_2<1$, the first Born approximation
and an operator-norm remainder bound are
\begin{equation}
 \mathsf R\simeq\mathsf R_0-\mathsf R_0\mathsf W\mathsf R_0,
 \qquad
 \|\mathsf R-\mathsf R_0+\mathsf R_0\mathsf W\mathsf R_0\|_2
       \le\frac{\|\mathsf R_0\|_2\eta^2}{1-\eta}.
 \label{eq:s3-scattering-born-remainder}
\end{equation}
\end{proposition}
\begin{proof}
Factor $\mathsf A_0+\mathsf W=
\mathsf A_0(\Id+\mathsf R_0\mathsf W)$ and use the inverse identity.
For $\eta<1$, expand $(\Id+\mathsf R_0\mathsf W)^{-1}$ in its
norm-convergent Neumann series and sum the tail from second order.
\end{proof}

The controlling parameter combines the perturbation with the reference
response. Near resonance, even a small stiffness change can require the
exact transition matrix. A local factorization
$\mathsf W=\mathsf L\mathsf C_d\mathsf L^{\mathsf T}$, with
dimensionless geometric map $\mathsf L$ and invertible
$p\times p$ matrix $\mathsf C_d$, gives
\begin{equation}
 \mathsf R=\mathsf R_0-
 \mathsf R_0\mathsf L
 [\mathsf C_d^{-1}+\mathsf L^{\mathsf T}\mathsf R_0\mathsf L]^{-1}
 \mathsf L^{\mathsf T}\mathsf R_0 .
 \label{eq:s3-scattering-local-port-update}
\end{equation}
Thus a cached factorization of the full reference matrix can support a
small $p\times p$ update. A changed central pair has at most three
independent relative-coordinate channels at fixed reference geometry;
a local three-dimensional mass change likewise has rank at most three.
Relaxing the reference can change additional blocks. A fitted charge change
updates every retained interaction affected by that charge, rather than
only one selected bond. New-particle comparisons can use a fixed
full-population space with inactive or weakly coupled members represented
explicitly, as in the recruitment construction.

Static conservative changes mix and reflect waves without consuming
their energy. In a translation-invariant open reference, incoming and
outgoing channels at the same frequency define elastic scattering;
bulk symmetry supplies the corresponding pseudomomentum selection rules.
The finite closed vessel instead has discrete standing modes and
recurrences. Its resolvent identity remains exact, while an asymptotic
incoming/outgoing scattering interpretation requires a declared host,
leads or observation regime. Irreversible absorption is assigned to
actual receiving channels, such as
\eqref{eq:s2-phonon-receiver-power}, rather than to mode conversion itself.

Figure~\ref{fig:s23-local-defect-scattering} evaluates a local inertia change through both particle motion and its reflected and transmitted fluxes.

\begin{figure}[H]
\centering\begingroup
\definecolor{vltTeal}{HTML}{176B64}
\definecolor{vltBlue}{HTML}{315D80}
\definecolor{vltGold}{HTML}{A87520}
\definecolor{vltPlum}{HTML}{725877}
\definecolor{vltRed}{HTML}{762B38}
\definecolor{vltYellow}{HTML}{D5AD3C}
\definecolor{vltNeutral}{HTML}{777D83}
\setlength{\unitlength}{1pt}
\scalebox{0.90}{%
\input{vlt-figures/figure-081.tex}%
}\endgroup
\caption{An infinite scalar chain has spacing $a$, stiffness $k$ and mass $m$, doubled at site 0. Nine displayed virions follow the exact scattering solution at one phase; the common ruler marks reference coordinates $x/a=j$. Dashed outlines fix the undeformed positions, and identical leads continue past both ends. Flux arrows have lengths $4:1:3$. Equal-sized bodies and five-turn coils follow the same coordinate scale. The frequency guide selects the snapshot; flux fractions sum to one. Displacement magnitudes $|r|,|t|$ give flux fractions through their squares in these identical leads. The local mass change supplies the rank-one matrix update; the auxiliary scatterer is lossless.}
\label{fig:s23-local-defect-scattering}
\end{figure}

Inelastic conversion requires terms beyond the fixed harmonic matrix.
For a smooth retained potential, let
$\Phi_{abc}^{(3)}=\partial_a\partial_b\partial_cU(\vct R)$ and
project its cubic term onto the real physical modes:
\begin{equation}
 \begin{aligned}
 U_3&=\frac1{3!}\sum_{s,t,u}\mathcal V_{stu}Q_sQ_tQ_u,\\
 \mathcal V_{stu}
   &:=\sum_{a,b,c}\Phi_{abc}^{(3)}
                 \phi_{s,a}\phi_{t,b}\phi_{u,c} .
 \end{aligned}
 \label{eq:s3-phonon-anharmonic-vertex}
\end{equation}
Replacing each $Q_s$ by its ladder quadrature gives matrix elements for
processes such as one mode quantum converting into two, with resonant
energy condition $\omega_s=\omega_t+\omega_u$. In a periodic bulk,
the projected vertex additionally selects
$\vct k_s=\vct k_t+\vct k_u+\vct G$; finite geometry replaces this
exact bulk rule by its actual spatial overlaps. The full conservative
Hamiltonian accounts for the transferred energy. A changing contact or
external control also needs its own stored energy and work, so an abrupt
replacement of one harmonic matrix by another cannot silently discard
the difference in energy or identify their mode occupations.

For an identifiable weakly damped phonon, an oscillatory response pole
has linewidth small compared with its frequency, and its reference
persists over several oscillation periods. Rapid rearrangement or strong
overdamping instead calls for the full time-dependent mechanical and
medium response. The matrix coordinates continue to describe correlated
motion in that regime, while a long-lived particle-like phonon need not
be a useful observational variable. This gives the excitation picture
an explicit computational domain alongside the assembly and recruitment
models.

\subsubsection{A documented STMV packing with a specified acoustic network}
\label{sec:stmv-crystal-network}

The reference cell \eqref{eq:stmv-reference-cell} permits an explicit
many-particle calculation. Start with one centre displacement $\vct u_i$
per whole virion and retain the body-centred orthorhombic translations.
This first network resolves acoustic displacement. Rotational and shell
coordinates can subsequently be retained through the nested construction;
their force constants are not inferred from the cell dimensions.

Choose one representative from each opposite-offset pair:
\begin{equation}
 \mathcal D_+=
 \{(a/2,\sigma b/2,\tau c/2)^{\mathsf T}:\sigma,\tau=\pm1\}
 \cup\{a\vct e_x,b\vct e_y,c\vct e_z\}.
 \label{eq:stmv-bond-offsets}
\end{equation}
Assign curvature $\kappa_D>0$ to each body diagonal and
$\kappa_A=\vartheta\kappa_D>0$ to each axial link, at its own rest length.
This is an unstressed effective network with fourteen neighbours per
site. The axial links are a positive finite-range completion; neither
their existence as molecular bonds nor their strength is read from the
space group. The quadratic energy is
\begin{equation}
 U_2=\frac12\sum_i\sum_{\vct d\in\mathcal D_+}
      \kappa_{\vct d}
      [\hat{\vct d}\cdot(\vct u_{i+\vct d}-\vct u_i)]^2.
 \label{eq:stmv-network-energy}
\end{equation}
Every undirected link occurs once. The reference is force balanced by
construction; a molecular potential at a different preparation instead
uses the full prestressed tangent \eqref{eq:edge-stiffness}.

\begin{proposition}[Bloch stiffness of the specified orthorhombic network]
\label{prop:stmv-bloch-stiffness}
For equal positive site mass $m_{\rm ac}$, the mass-normalized Bloch matrix is
\begin{equation}
 \boxed{\mathsf D_\vartheta(\vct k)=
 \frac{2}{m_{\rm ac}}\sum_{\vct d\in\mathcal D_+}
 \kappa_{\vct d}[1-\cos(\vct k\cdot\vct d)]
           \hat{\vct d}\hat{\vct d}^{\mathsf T}\succeq0.}
 \label{eq:stmv-bloch-stiffness}
\end{equation}
\end{proposition}
\begin{proof}
Insert $\vct u_i=\vct e\exp(i\vct k\cdot\vct r_i^0)$ in the
quadratic energy. Each opposite pair contributes
$2[1-\cos(\vct k\cdot\vct d)]$ times its nonnegative longitudinal
dyad. Their sum gives the displayed matrix.
\end{proof}

Write $\mathsf D_\vartheta=(\kappa_D/m_{\rm ac})\mathsf H_\vartheta$.
The dimensionless $\mathsf H_\vartheta$ contains geometry and the assigned
ratio $\vartheta$. The calculation below constrains an acoustic stiffness
per effective mass. Interpreting $m_{\rm ac}$ requires a decision about
entrained solvent; it is not automatically the dry capsid mass or the
buoyant mass. At the measured band this network is an effective
elastodynamic description of the hydrated assembly. A low-frequency
Stokes drag or incompressible Poiseuille law is not its acoustic bath.

In a chosen direction let $h_L$ be the eigenvalue of
$\mathsf H_\vartheta$ with the largest longitudinal overlap
$|\hat{\vct k}^{\mathsf T}\vct e_s|^2$. All branches and overlaps remain
available for the observation map. In the long-wavelength limit this
construction supplies a Christoffel tensor through
$D_{ik}=C_{ijkl}k_jk_l/\rho_{\rm ac}+O(|\vct k|^4)$.
One longitudinal line constrains only a projection of that tensor.
It cannot determine all orthorhombic elastic constants.

For a frequency-local passive acoustic reference, take
\begin{equation}
 \ddot{\vct u}_{\vct k}
 +\nu_L|\vct k|^2\dot{\vct u}_{\vct k}
 +\mathsf D_\vartheta(\vct k)\vct u_{\vct k}
 =\vct f_{\vct k}/m_{\rm ac},\qquad \nu_L>0.
 \label{eq:stmv-acoustic-dynamics}
\end{equation}
Each retained mode has a causal damped-oscillator response and
nonnegative loss. An equilibrium noise model uses the same resistance
$m_{\rm ac}\nu_L|\vct k|^2$. More detailed solvent or material states
use Section~4's causal reduction and compatible fluctuations. This
frequency-local reference does not imply frequency-independent loss
outside its stated band.

Internal and interparticle response can be joined without identifying
their springs. On a stable reference, partition the complete dynamic
stiffness $\mathsf A(s,\vct k)$ into retained rigid/acoustic coordinates
$r$ and internal coordinates $q$. Block elimination gives
\begin{equation}
 \mathsf A_{r,\rm eff}
  =\mathsf A_{rr}-\mathsf A_{rq}\mathsf A_{qq}^{-1}\mathsf A_{qr}.
 \label{eq:stmv-nested-dynamic-reduction}
\end{equation}
The inverse acts where the internal response is defined. At zero
frequency it is the usual constrained Schur reduction; at finite
frequency internal deformation carries dispersion, loss, preparation
and the corresponding reduced noise. Eliminating a coordinate also
changes a detector that observes it. Thus a crystal acoustic branch
and an internal shell vibration remain distinct possible contributions
to one measured spectrum. Figure~\ref{fig:stmv-cell-network} shows the
packing, link classes and a collective displacement.

\begin{figure}[H]
\centering\begingroup
\setlength{\unitlength}{1pt}
\scalebox{0.84}{%
\input{vlt-figures/figure-316.tex}%
}\endgroup
\caption{A specified effective network on the documented STMV reference packing. (a) Eight corner glyphs share boundaries and one glyph lies at the body centre, giving two whole particles per conventional cell. Blue lines are body diagonals; green lines show three representative axial links from the centre into neighbouring cells. The full network has both signs of every offset in \eqref{eq:stmv-bond-offsets}. These effective links have assigned curvatures, not recovered molecular bond strengths. (b) Eigenfrequencies of the positive Bloch tangent \eqref{eq:stmv-bloch-stiffness} along [100] for $\vartheta=1$, normalized at the longitudinal calibration wavevector. All three branches approach translation at zero wavevector. The curves are conservative stiffness information; damping and optical weights determine the spectral record. The lower blocks locate the internal reduction \eqref{eq:stmv-nested-dynamic-reduction}.}
\label{fig:stmv-cell-network}
\end{figure}

\subsection{Mode expansions and viral-phonon packet fields}
\label{sec:phonon-fields}

The displayed three-component packet fields in this subsection use the
translational baseline. For a fixed conservative body reference, the
same positive-frequency energy amplitudes act in the inertia-normalized
body space. Without mixed constraints, their physical reconstruction is
\begin{equation}
 \vct\zeta_i(t)=\sqrt2\,\mathsf S_i\mathsf G_{X,i}^{-1/2}
   \operatorname{Re}\sum_{s:\omega_s>0}\frac{a_s}{\omega_s}
                      \vct e_{s,i}e^{-i\omega_st}.
 \label{eq:s3patch-body-packet-reconstruction}
\end{equation}
Its two blocks give centre motion and small angular motion with their
own physical units. The corresponding admissible embedding is used
when constraints mix bodies or coordinate types. Surface and material
states with finite-rate dissipation remain in the augmented dynamics;
the autonomous conservative packet evolution is a specified reference.

\paragraph{Physical site blocks under fixed constraints.}
Fixed constraints can mix coordinates while the observations still track
individual particles. A mass-matched embedding connects these two descriptions.

\begin{definition}[Admissible mass-weighted site embedding]
\label{def:s3-admissible-site-embedding}
For the fixed linear admissible map in
\eqref{eq:structure-constrained-sector} and its reduced mass matrix,
define the embedding, embedded mode columns and admissible projector by
\begin{equation}
 \begin{aligned}
 \mathsf J_{\rm c}
   &:=\mathsf M^{1/2}\mathsf U_{\rm c}\mathsf M_{\rm c}^{-1/2},
 &\mathsf J_{\rm c}^{\mathsf T}\mathsf J_{\rm c}&=\Id,\\
 \vct e_s&:=\mathsf J_{\rm c}\vct e_{{\rm c},s}
          =\mathsf M^{1/2}\vct\phi_s,
 &\mathsf P_{\rm adm}&:=\mathsf J_{\rm c}\mathsf J_{\rm c}^{\mathsf T}.
 \end{aligned}
 \label{eq:structure-constrained-site-embedding}
\end{equation}
\end{definition}
Thus the embedded columns remain orthonormal in the full mass-weighted
site coordinates. Dividing their site blocks by $\sqrt{m_i}$ recovers
physical displacements, and the cell embedding below still preserves
their norm. Within the rest of this section, constrained calculations
interpret $\mathsf D$ and its spectral projectors on
$\operatorname{Ran}\mathsf P_{\rm adm}$. The admissible zero-mode
projector is $\mathsf P_0=\mathsf P_{\rm adm}-\mathsf P_+$;
forbidden displacements are not free drift modes.

A physical loading vector is first projected as
$\vct h=\mathsf P_{\rm adm}\mathsf M^{-1/2}\vct\ell$.
Its modal overlap is still
$\vct e_s^{\mathsf T}\vct h=\vct\phi_s^{\mathsf T}\vct\ell$,
so the same centre displacement, force conjugacy and response follow
from either representation. With no fixed constraints, one may choose
$\mathsf U_{\rm c}=\Id$, giving
$\mathsf J_{\rm c}=\mathsf P_{\rm adm}=\Id$. A reduced coordinate
block itself need not denote one ideal virion; the embedding supplies that
physical identification before spatial localization or detector weights
are interpreted.

The finite-network modes describe motion with the actual boundaries and
reference arrangement. All mode sums below run over positive frequencies.

\begin{definition}[Finite classical energy packet]
\label{def:s3-finite-energy-packet}
For the energy amplitude of~\eqref{eq:energy-amplitude}, set
$a_s=\braket{s}{\psi(0)}$. Its positive-sector mode superposition is
\begin{equation}
 \psi_A(t)=\sum_{s:\omega_s>0}a_s e_{s,A}e^{-i\omega_st},\qquad
 e_{s,A}=\braket{A}{s},\qquad
 \sum_A|\psi_A(t)|^2=\sum_s|a_s|^2=E_+.
 \label{eq:finite-energy-packet}
\end{equation}
\end{definition}
Orthonormality gives the conserved norm. Initial displacement and velocity
determine $a_s$; the boundary problem determines the mode columns and
frequencies. Together these reconstruct the prepared packet.

In particular, the displacement represented by these energy amplitudes
is, from~\eqref{eq:energy-ket-reconstruction},
\begin{equation}
 \vct u_{+,i}(t)=\sqrt{\frac2{m_i}}\,
       \operatorname{Re}\sum_s\frac{a_s}{\omega_s}
                   \vct e_{s,i}e^{-i\omega_st}.
 \label{eq:structure-packet-displacement-reconstruction}
\end{equation}
The site block $\vct\psi_i$ allocates the conserved representation
norm; $\vct u_i$ gives the length-valued displacement. The inverse
frequency weight makes equal-energy soft modes move farther than stiff
ones. Interference phases determine the time-dependent marker signal.

Figure~\ref{fig:structure-packet-representations} reconstructs both fields from one finite two-mode preparation, including their different frequency weights.

\begin{figure}[H]
\centering\begingroup
\definecolor{vltTeal}{HTML}{176B64}
\definecolor{vltBlue}{HTML}{315D80}
\definecolor{vltGold}{HTML}{A87520}
\definecolor{vltPlum}{HTML}{725877}
\definecolor{vltRed}{HTML}{762B38}
\definecolor{vltYellow}{HTML}{D5AD3C}
\definecolor{vltNeutral}{HTML}{777D83}
\setlength{\unitlength}{1pt}
\scalebox{0.90}{%
\input{vlt-figures/figure-082.tex}%
}\endgroup
\caption{One real preparation excites the common and opposed modes of the supported equal-mass pair with equal energy-ket amplitudes. Frequencies are $\omega_\star,\sqrt3\omega_\star$. The initial positions use $18\,\mathrm{pt}/\ell_E$ and fixed coil families. Site-norm fractions are $[1\pm\cos((\sqrt3-1)\tau)]/2$, while physical displacements are $[\cos\tau\pm\cos(\sqrt3\tau)/\sqrt3]/2$ in units $\ell_E$. At $\tau=0$ the energy ket lies entirely in site block 1 while both particles are displaced. The frequency weight in reconstruction produces this difference. Site norm, local mechanical energy and displacement are distinct specified readouts.}
\label{fig:structure-packet-representations}
\end{figure}

\begin{definition}[Cell field representation]
\label{def:s3-cell-field-representation}
Partition the vessel into cells $\mathcal V_i$ of volume $v_i>0$,
one per site. For real orthonormal mass-weighted mode columns, set
$\vct\varphi_s(\vct r)=\vct e_{s,i}/\sqrt{v_i}$ in cell
$\mathcal V_i$. The associated energy-amplitude field is
\begin{equation}
 \tilde{\vct\psi}(\vct r,t)=
       \sum_s a_s\vct\varphi_s(\vct r)e^{-i\omega_st},\qquad
 \int_{\mathcal V}|\tilde{\vct\psi}|^2\,\dd^3r=E_+
 \label{eq:energy-packet-field}
\end{equation}
\end{definition}
Integrating over each cell cancels its volume factor, so the field inner
products equal those of the site columns and the energy norm is
preserved exactly. Smooth continuum eigenfunctions provide the
corresponding long-wavelength representation.

This piecewise-constant field is an \emph{isometric embedding}: each
site amplitude is distributed over a chosen cell with precisely the
normalization needed to preserve inner products. Within that cell,
$|\tilde{\vct\psi}|^2=\|\vct\psi_i\|^2/v_i$, with units
$\mathrm{J\,m^{-3}}$, and integrating over the cell recovers
$\|\vct\psi_i\|^2$. Changing the cell convention changes this
pointwise density without changing the underlying mechanical state.
The chosen cells can include solvent and void between particles, so
$v_i$ is a representation volume. Particle-material volume is a
separate geometric quantity. A smooth field
requires an additional interpolation and a controlled spatial scale.
The physical local mechanical energy is assigned from the kinetic and bond energies in~\eqref{eq:edge-flux},
or their continuum counterparts; the square-root frequency operator generally
mixes site coordinates.

Figure~\ref{fig:structure-cell-isometry} makes the volume normalization explicit: different chosen cell sizes change field heights while preserving the norm assigned to each site block.

\begin{figure}[H]
\centering\begingroup
\definecolor{vltTeal}{HTML}{176B64}
\definecolor{vltBlue}{HTML}{315D80}
\definecolor{vltGold}{HTML}{A87520}
\definecolor{vltPlum}{HTML}{725877}
\definecolor{vltRed}{HTML}{762B38}
\definecolor{vltYellow}{HTML}{D5AD3C}
\definecolor{vltNeutral}{HTML}{777D83}
\setlength{\unitlength}{1pt}
\scalebox{0.80}{%
\input{vlt-figures/figure-083.tex}%
}\endgroup
\caption{The cell isometry for two representation cells with a common cross-section, lengths $\ell_\star,4\ell_\star$ and volumes $v_\star,4v_\star$. Both site coefficients equal $\sqrt{E_\star}$. Division by $\sqrt{v_i}$ gives field heights $1,1/2$ in the upper normalization and squared heights $1,1/4$ below. The lower shaded areas are therefore one each: $\int_{\Omega_i}|\Psi|^2\,dV=E_\star$ and $\|\Psi\|_{L^2}^2=2E_\star$. Line styles distinguish the cells in grayscale. A shared endpoint has no effect on the integrals. These are cells of an orthonormal representation, whose volumes set field normalization; a spatial mechanical-energy density requires a separately defined energy assignment.}
\label{fig:structure-cell-isometry}
\end{figure}

\paragraph{Spatial preparation and spectral bandwidth.}
\begin{proposition}[Band-limited spatial concentration]
\label{prop:s3-band-region-concentration}
For a nonempty retained band $I\subset(0,\infty)$, let
$\mathsf P_I=\sum_{\omega_s\in I}\ket{s}\bra{s}$ include complete
degenerate eigenspaces. For a set $S$ of particle sites, let
$\mathsf\Pi_S=\sum_{i\in S}\mathsf\Pi_i$. The maximal site-block
fraction among packets in that band is
\begin{equation}
 \max_{\substack{0\ne\psi\in\operatorname{Ran}\mathsf P_I}}
 \frac{\|\mathsf\Pi_S\psi\|^2}{\|\psi\|^2}
 =\lambda_{\max}\!\left(
       \left.\mathsf P_I\mathsf\Pi_S\mathsf P_I
                         \right|_{\operatorname{Ran}\mathsf P_I}\right)
 \in[0,1].
 \label{eq:structure-band-region-concentration}
\end{equation}
The maximum cannot increase when the band is restricted, and cannot
decrease when the site set is enlarged.
\end{proposition}
\begin{proof}
On $\operatorname{Ran}\mathsf P_I$, the fraction is the Rayleigh
quotient of the positive compression
$\mathsf P_I\mathsf\Pi_S\mathsf P_I\preceq\Id$.
Its largest eigenvalue is attained by an eigenvector.
Restriction reduces the set of admissible trial packets, whereas
enlarging $S$ increases each trial packet's site norm.
These variational facts give the two monotonicities~\cite{Teschl2009}.
\end{proof}

The maximizing eigenvector constructs a packet attaining the bound. A value below one proves that
no preparation using only that band is confined to those site blocks.
Free conservative evolution preserves the band, so the same upper bound
holds at every time, although the actual spatial fraction can vary.
These are preparation constraints imposed by the actual
finite mode shapes and boundaries, without assigning a wavevector to
each standing mode. The fraction measures concentration of the energy-ket representation
norm. Physical bond-and-kinetic energy in $S$ uses the specified local
energy assignment; a displacement signal uses its calibrated observable.

For the quantized description of Postulate~\ref{post:quantization}, the
one-phonon state remains
\[
 \ket{\Phi(t)}=\sum_sc_se^{-i\omega_st}\widehat b_s^\dagger\ket{0},
 \qquad \sum_s|c_s|^2=1.
\]
A normalized field representation is obtained by replacing
$a_s$ with $c_s$ in~\eqref{eq:energy-packet-field}; it represents the
one-excitation state in the selected orthonormal field basis.
Writing $\ket{\eta_{\rm dir}(t)}=\sum_sc_se^{-i\omega_st}\ket{s}$, a consistent
site-labelled measurement on $\mathcal H_+$ has effects
\begin{equation}
 \mathsf F_i=\left.\mathsf P_+\mathsf\Pi_i\mathsf P_+\right|_{\mathcal H_+},
 \qquad
 \sum_i\mathsf F_i=\Id_{\mathcal H_+},\qquad
 p_i(t)=\bra{\eta_{\rm dir}(t)}\mathsf F_i\ket{\eta_{\rm dir}(t)}
                                =\|\mathsf\Pi_i\eta_{\rm dir}(t)\|^2.
 \label{eq:site-packet-effects}
\end{equation}
The effects form a normalized positive-operator-valued measurement on
$\mathcal H_+$; compression can make them nonorthogonal.
A displacement detector uses the mass and frequency weights below.
The cell embedding remains a representation of the finite state;
a continuum field equation requires the separate controlled limit.

\begin{proposition}[One-phonon displacement statistics]
\label{prop:s3-one-phonon-displacement-statistics}
Under Postulate~\ref{post:quantization}, use real mass-weighted positive
mode columns and the normalized one-phonon preparation above.
Time-dependent operator averages are taken in the initial state
$\ket{\Phi(0)}$.
The displacement operator is
\begin{equation}
 \widehat{\vct u}_i(t)=
 \frac1{\sqrt{m_i}}\sum_s\sqrt{\frac{\hbar}{2\omega_s}}
 \vct e_{s,i}\left(\widehat b_se^{-i\omega_st}
                     +\widehat b_s^\dagger e^{i\omega_st}\right).
 \label{eq:quantum-displacement-field}
\end{equation}
It has zero mean in this preparation. Its vacuum-transition amplitude
and excess site mean-square displacement satisfy
\begin{equation}
 \vct d_i^{\rm tr}(t)=\bra{0}\widehat{\vct u}_i(t)\ket{\Phi(0)}
   =\frac1{\sqrt{m_i}}\sum_s\sqrt{\frac{\hbar}{2\omega_s}}
                           c_s\vct e_{s,i}e^{-i\omega_st},\qquad
 \Delta\langle|\widehat{\vct u}_i|^2\rangle=2|\vct d_i^{\rm tr}|^2.
 \label{eq:phonon-transition-amplitude}
\end{equation}
\end{proposition}
\begin{proof}
Insert the modal oscillator quadratures from
\eqref{eq:architecture-ladder-quadratures} into the physical mode
expansion. The displacement operator connects sectors differing by one
excitation, so its expectation within the one-excitation sector is zero.
The vacuum-transition matrix element retains the annihilation terms.
Evaluating the quadratic expectation with the canonical commutators
leaves the vacuum term plus twice the squared norm of that transition
amplitude.
\end{proof}
The operator and transition amplitude have length units; the excess
mean-square displacement is measured relative to the same site's vacuum
fluctuation. Coherent states instead have a classical mean-displacement
pattern, with $\alpha_s=a_s/\sqrt{\hbar\omega_s}$ and excitation energy
$\sum_s\hbar\omega_s|\alpha_s|^2=E_+$~\cite{Misra2012,Shankar1994}.

Figure~\ref{fig:structure-one-phonon-coordinate} illustrates why a number-state excitation can change the displacement fluctuations even when its mean motion vanishes.

\begin{figure}[H]
\centering\begingroup
\definecolor{vltTeal}{HTML}{176B64}
\definecolor{vltBlue}{HTML}{315D80}
\definecolor{vltGold}{HTML}{A87520}
\definecolor{vltPlum}{HTML}{725877}
\definecolor{vltRed}{HTML}{762B38}
\definecolor{vltYellow}{HTML}{D5AD3C}
\definecolor{vltNeutral}{HTML}{777D83}
\setlength{\unitlength}{1pt}
\scalebox{0.80}{%
\input{vlt-figures/figure-084.tex}%
}\endgroup
\caption{A quantized single normal coordinate under Postulate~\ref{post:quantization}. With $\ell_Q=\sqrt{\hbar/(2m\omega)}$, the normalized coordinate densities are $p_0(X)=e^{-X^2/2}/\sqrt{2\pi}$ and $p_1(X)=X^2p_0(X)$. Both have zero mean, while their second moments are one and three. On the right, the open arrows label the magnitudes $|\langle n+1|\widehat u|n\rangle|=\ell_Q\sqrt{n+1}$ and their Hermitian conjugates. Inserting the adjacent states $|0\rangle,|2\rangle$ gives the one-phonon variance $3\ell_Q^2$. These links specify a coordinate operator, rather than forces or preparation rates. The normal coordinate can represent collective virion motion through its mode shape; the plot adds no particle species or assumption of localized energy.}
\label{fig:structure-one-phonon-coordinate}
\end{figure}

These constructions distinguish a mechanical wave, a normalized excitation
state, and a displacement observable while preserving their shared mode
architecture.

\subsection{Density of states and spectral information}
\label{sec:density-states}

\begin{definition}[Finite positive-frequency density of states]
\label{def:s3-finite-dos}
For the stable finite reference, the positive-frequency counting measure
and its density per vessel volume are
\begin{equation}
 g_n(\omega)=\sum_{s:\omega_s>0}\delta(\omega-\omega_s),\qquad
 \int_0^\infty g_n(\omega)\,\dd\omega=r,\qquad
 d_n(\omega)=\frac{g_n(\omega)}{|\mathcal V|}.
 \label{eq:finite-dos-count}
\end{equation}
\end{definition}
Here $r=\dim\mathcal H_+$ is the positive-mode count used
in~\eqref{eq:occupation-space}. Each delta point mass counts a frequency
with its multiplicity. Replacing it by a normalized resolution kernel
produces a smooth presentation of the same finite spectrum. The chosen
resolution describes plotting or detection; damping, excitation and
channel sensitivity enter through their own physical inputs.

\paragraph{Counting moments constrain the assembled mechanics.}
Counting moments link the spectrum to the interactions that generated it.

\begin{proposition}[Spectral moments of the assembled lattice]
\label{prop:s3-dos-assembly-moments}
For a stable finite translational reference with scalar particle masses,
write $\mathsf K=\mathsf K_{\rm pair}+\mathsf K_{\rm sup}$ and retain
each unordered pair once. On the full coordinate space,
\begin{equation}
 \begin{aligned}
 \int_0^\infty\omega^2g_n(\omega)\,\dd\omega
  &=\tr\mathsf D\\
  &=\sum_{\{i,j\}}\left(\frac1{m_i}+\frac1{m_j}\right)
                         (L_{ij}+2T_{ij})
      +\tr(\mathsf M^{-1}\mathsf K_{\rm sup}),\\
 \int_0^\infty\omega^4g_n(\omega)\,\dd\omega
  &=\tr(\mathsf D^2)
   =\sum_{A,B}|D_{AB}|^2.
 \end{aligned}
 \label{eq:structure-dos-force-constant-moments}
\end{equation}
\end{proposition}
\begin{proof}
Spectral decomposition gives the traces as the second and fourth
frequency moments; zero directions contribute nothing.
Each pair contributes two diagonal Hessian blocks with
$\tr\mathsf C_{ij}=L_{ij}+2T_{ij}$ and the indicated inverse masses.
Finally $\tr(\mathsf D^2)=\sum_{A,B}|D_{AB}|^2$ because
$\mathsf D$ is Hermitian.
\end{proof}

For supports coupling no distinct sites, the last sum also equals
$\sum_i\tr(\mathsf D_{ii}^2)
+2\sum_{\{i,j\}}(L_{ij}^2+2T_{ij}^2)/(m_im_j)$.
Signed transverse curvature must be retained in these tests. Charge
patterns enter through the same pair derivatives as in the assembled
equations; counting identical shell distances cannot replace them.
The second and fourth moments have units $\mathrm{s^{-2}}$ and
$\mathrm{s^{-4}}$. They test a complete calculated spectrum against
its declared masses, interactions and supports. A finite detection band
or a resolution kernel with nonzero higher moments generally changes
these integrals. They do not reconstruct the force network uniquely:
isospectral matrices have the same moments and may have different
spatial eigenpatterns. Under mixed fixed constraints the trace identities
remain valid for the reduced dynamical matrix, while the displayed
unconstrained edge sum is replaced by that matrix's actual assembly.

For angular frequency, $g_n$ has units of time and $d_n$ units of
time per volume. If a plot instead uses ordinary frequency
$f=\omega/(2\pi)$, the count-preserving density is
$g_n^{(f)}(f)=2\pi g_n(2\pi f)$. The horizontal variable and its
Jacobian must be retained when comparing integrated peak weights.
A normalized resolution kernel on the full real frequency axis
preserves the count there; if its tails extend to negative frequency,
the positive half-axis integral need not remain $r$ unless a
positive-frequency boundary convention is explicitly imposed.

\paragraph{Finite boundaries and integrated spectral counts.}
A comparison with a bulk calculation first specifies compatible
coordinate spaces and the blocks that change.

\begin{proposition}[Finite-rank bound for integrated spectral counts]
\label{prop:s3-finite-rank-count-bound}
Compare two real symmetric $3n\times3n$ dynamical matrices on the same
sites, masses and positions. Suppose every changed block row and column
is supported on the sites $\mathcal I_\Delta$, which contain all endpoints
of changed edges and all sites with changed support blocks.
For $\Delta\mathsf D=\mathsf D_1-\mathsf D_0$ and full counting
functions $N_j(\xi)=\tr\,\mathbf1_{(-\infty,\xi]}(\mathsf D_j)$,
\begin{equation}
 \sup_{\xi\in\R}|N_1(\xi)-N_0(\xi)|
 \le r_{\Delta}:=\operatorname{rank}(\Delta\mathsf D)
 \le3|\mathcal I_{\Delta}|.
 \label{eq:structure-boundary-spectral-count}
\end{equation}
\end{proposition}
\begin{proof}
The range of $\Delta\mathsf D$ lies in the
$3|\mathcal I_\Delta|$ site coordinates, giving the rank bound.
The perturbation vanishes on a subspace of codimension $r_\Delta$;
the min--max characterization therefore interlaces the ordered
spectra with an index shift of at most $r_\Delta$~\cite{Teschl2009}.
\end{proof}
The argument needs neither positive individual edge tensors nor a
perturbation small in norm. It includes zero and negative eigenvalues.
For stable, force-balanced references, $\xi=\omega^2$ compares full
squared-frequency counts. Restricting to $(0,\omega^2]$ requires
subtracting the zero-threshold count, giving the safe bound
$2r_{\Delta}$ when the kernels differ.

For a compatible finite-range comparison, such as opening a periodic
patch with the same sites, the modified sites can lie in a boundary
layer. If $|\mathcal I_{\Delta}|/n\to0$, the normalized full counting functions
consequently agree in that limit. This controls integrated availability
of modes; individual boundary eigenvectors, gaps and local probe
weights can still differ. The comparison retains each model's stated boundary law. Removing particles changes the matrix dimension
and needs an explicit comparison of the retained spaces. Complete
long-range electrostatics can also change couplings throughout the
sample, so a geometrically thin boundary need not make
$\Delta\mathsf D$ low rank. Both the common-reference premise and the
retained interaction range must be checked before using this estimate.

\begin{definition}[Bulk density of states]
\label{def:s3-bulk-dos}
For a translation-invariant reference with well-defined branches
$\omega_\nu(\vct k)$, the density per unit occupied network volume is
the Brillouin-zone counting measure~\cite{Misra2012}
\begin{equation}
 d(\omega)=\frac1{(2\pi)^3}
    \sum_\nu\int_{\mathrm{BZ}}
       \delta[\omega-\omega_\nu(\vct k)]\,\dd^3k.
 \label{eq:bulk-dos-measure}
\end{equation}
\end{definition}
This normalization uses the occupied region $\mathcal V_{\mathrm{lat}}$,
whereas $d_n$ in~\eqref{eq:finite-dos-count} uses the whole vessel.
For a large uniformly occupied reference after suitable spectral coarse
graining, define $f_{\rm lat}=|\mathcal V_{\mathrm{lat}}|/|\mathcal V|$;
then $g_n/|\mathcal V_{\mathrm{lat}}|$ approaches $d$ and $d_n$ approaches
$f_{\rm lat}d$. This region fraction is not the particle-material volume
fraction: a reference cell also includes the space between particles.
\begin{proposition}[Regular isofrequency-surface counting]
\label{prop:s3-isofrequency-counting}
At a regular frequency, assume the contributing branches are
continuously differentiable and their isofrequency surfaces have
$\nabla_{\vct k}\omega_\nu\ne0$. The bulk measure has the density
\begin{equation}
 d(\omega)=\frac1{(2\pi)^3}\sum_\nu
    \int_{\omega_\nu(\vct k)=\omega}
       \frac{\dd S_{\vct k}}{|\nabla_{\vct k}\omega_\nu|}.
 \label{eq:dos-isofrequency-surfaces}
\end{equation}
\end{proposition}
\begin{proof}

The surface formula follows by choosing one coordinate normal to the
isofrequency surface: an increment $\dd k_\perp$ changes frequency
by $|\nabla_{\vct k}\omega_\nu|\dd k_\perp$.
\end{proof}
A nearly flat branch places many wavevector states in a narrow frequency
interval. Critical points with zero group velocity require a local
dispersion analysis~\cite{Misra2012}. Preparation and detection then
weight the available patterns. A net energy current additionally depends
on propagation direction and phase; opposite or standing preparations
can give zero transport at those same frequencies.

For a one-site translational lattice, its integral is $3/a^3$. A primitive
cell of volume $v_{\mathrm{cell}}$ with $n_{\mathrm b}$ virions and
$r_{\mathrm{int}}$ additional positive-inertia conservative internal
coordinates has a total branch count $3n_{\mathrm b}+r_{\mathrm{int}}$
and integral $(3n_{\mathrm b}+r_{\mathrm{int}})/v_{\mathrm{cell}}$.
Retaining rigid-body rotations changes that mechanical count to
$6n_{\mathrm b}+r_{\mathrm{int}}$. First-order chemical or material
states have their own relaxation spectra. The count assumes positive
branches apart from isolated symmetry zeros;
a flat zero-frequency branch belongs to the separately retained zero sector.

In an isotropic acoustic approximation,
$\omega_L\simeq c_L|\vct k|$ and
$\omega_{T_1}=\omega_{T_2}\simeq c_T|\vct k|$ at small wavenumber. Integrating
over spherical shells gives
\begin{equation}
 d_{\mathrm{ac}}(\omega)\simeq
       \frac{\omega^2}{2\pi^2}
             \left(\frac1{c_L^3}+\frac2{c_T^3}\right).
 \label{eq:acoustic-dos}
\end{equation}
For example,
$4\pi\int_0^\infty k^2\delta(\omega-ck)\,\dd k
=4\pi\omega^2/c^3$, which gives each branch's contribution and its units.
For anisotropic speeds $c_\nu(\vct{\hat k})$, the coefficient is instead
$(2\pi)^{-3}\sum_\nu\int_{\mathbb S^2}
c_\nu(\vct{\hat k})^{-3}\,\dd\Omega$.
The coefficient follows from the inverse cube of each directional speed,
so an isotropic average of the elastic constants does not generally give
the same density of states as an angular average of the actual branches.
The bulk limit precedes the limit $\omega\to0$. A finite cylinder
retains discrete levels and its smallest positive frequency.
In a finite-system comparison, a smooth bulk law requires many levels
inside the selected resolution window while that window remains small
on the scale of dispersion changes. Extending an acoustic law to all
frequencies would give an infinite count; any Debye cutoff must be
specified to reproduce the retained acoustic branch count.

An optical branch introduced by a specified basis or internal model has its
own spectral threshold. For one isolated isotropic minimum with
$\omega(\vct k)\simeq\omega_0+\alpha_{\mathrm o}|\vct k|^2$,
$\omega_0>0$, $\alpha_{\mathrm o}>0$, the coefficient $\alpha_{\mathrm o}$ has
units of length squared per time. The local near-threshold contribution is
\begin{equation}
 d_{\mathrm{op}}(\omega)\simeq
       \frac{\sqrt{\omega-\omega_0}}
                       {4\pi^2\alpha_{\mathrm o}^{3/2}}
                          \Theta(\omega-\omega_0).
 \label{eq:optical-threshold-dos}
\end{equation}
Here $\Theta$ is the step function. The relation follows from the same radial
integration with the quadratic dispersion. It describes a chosen local band
minimum; the complete Brillouin-zone dispersion supplies the rest of the band
and its upper cutoff. Optical modes therefore do not have one universal
density-of-states law independent of their restoring model.
The acoustic $\omega^2$ and optical square-root laws count three-dimensional
wavevectors near their stated dispersions; changing dimensionality,
confinement or branch curvature changes those powers. They are asymptotic
band descriptions, while the cylinder's actual spectrum remains discrete.
For an anisotropic quadratic minimum
$\omega-\omega_0\simeq\vct k^{\mathsf T}\mathsf A_{\mathrm o}\vct k$,
with $\mathsf A_{\mathrm o}$ positive definite, the corresponding
coefficient is $1/(4\pi^2\sqrt{\det\mathsf A_{\mathrm o}})$.
Separate equivalent minima and actual branch degeneracies contribute
their respective multiplicities. A flat branch contributes a delta
measure rather than this square-root edge.

Figure~\ref{fig:structure-dos-thresholds} contrasts these two local
three-dimensional counting laws before preparation or readout weights
are applied.

\begin{figure}[H]
\centering\begingroup
\definecolor{vltTeal}{HTML}{176B64}
\definecolor{vltBlue}{HTML}{315D80}
\definecolor{vltGold}{HTML}{A87520}
\definecolor{vltPlum}{HTML}{725877}
\definecolor{vltRed}{HTML}{762B38}
\definecolor{vltYellow}{HTML}{D5AD3C}
\definecolor{vltNeutral}{HTML}{777D83}
\setlength{\unitlength}{1pt}
\scalebox{0.80}{%
\input{vlt-figures/figure-085.tex}%
}\endgroup
\caption{Local three-dimensional density-of-states laws per occupied network volume, for one nondegenerate branch or minimum. Spherical $\vct k$-space counting gives $d_{\rm a}(\omega)=\omega^2/(2\pi^2c^3)$ for an acoustic branch and $d_{\rm o}(\omega)=\sqrt{\omega-\omega_0}\,\Theta(\omega-\omega_0)/(4\pi^2\alpha_{\rm o}^{3/2})$ near an isolated quadratic optical minimum. The separate vertical normalizations expose the powers $x^2$ and $\sqrt{x}$; they do not compare absolute mode counts. Equivalent branches or minima add their contributions. The gray region is below the optical threshold. These asymptotic curves use a locally bulk dispersion, rather than the discrete levels of a finite confined array; neither curve assigns occupation or a linewidth, and no ultraviolet cutoff is inferred from the plotted window.}
\label{fig:structure-dos-thresholds}
\end{figure}

Preparation specifies how the available modes are used.

\begin{definition}[Prepared spectral measures]
\label{def:s3-prepared-spectral-measures}
For classical energy amplitudes $a_s$ and normalized one-phonon
amplitudes $c_s$, define respectively
\begin{equation}
 \mathcal E(\omega)=\sum_s|a_s|^2\delta(\omega-\omega_s),\qquad
 \mathcal P_\Phi(\omega)=\sum_s|c_s|^2\delta(\omega-\omega_s).
 \label{eq:prepared-spectral-measures}
\end{equation}
\end{definition}
Their integrals are $E_+$ and one. The one-phonon excitation energy is
$\int_0^\infty\hbar\omega\mathcal P_\Phi(\omega)\,\dd\omega$.
These measures specify how a prepared state uses the spectrum; $g_n$ specifies
how many modes are available.

Figure~\ref{fig:structure-count-and-resolution} shows a finite counting measure at two declared resolutions; neither smoothing width is a mechanical damping rate.

\begin{figure}[H]
\centering\begingroup
\definecolor{vltTeal}{HTML}{176B64}
\definecolor{vltBlue}{HTML}{315D80}
\definecolor{vltGold}{HTML}{A87520}
\definecolor{vltPlum}{HTML}{725877}
\definecolor{vltRed}{HTML}{762B38}
\definecolor{vltYellow}{HTML}{D5AD3C}
\definecolor{vltNeutral}{HTML}{777D83}
\setlength{\unitlength}{1pt}
\scalebox{0.80}{%
\input{vlt-figures/figure-086.tex}%
}\endgroup
\caption{Counting and resolution for the same seven-site fixed-end scalar chain, with $\omega_s/\omega_\star=2\sin(s\pi/16)$ and $s=1,\ldots,7$. (a) The exact count is right-continuous: a filled upper point includes the level at each jump; an open lower point excludes it. Dotted vertical guides locate the jumps. (b) Each level is convolved with the normalized compact kernel $K_\delta(\Omega)=3[1-(\Omega/\delta)^2]/(4\delta)$ for $|\Omega|\le\delta$, and zero otherwise. Both resolutions integrate to seven over positive frequency, since every kernel support remains above zero. Black baseline ticks retain the exact levels beneath the smoothed curves. Changing the display resolution merges nearby peaks without changing the count. The kernel width specifies a counting resolution; damping, lifetime and preparation weights require their own models.}
\label{fig:structure-count-and-resolution}
\end{figure}

Comparison with a measured signal uses the prepared, probe-weighted
measure at the detector's resolution. A strong global DOS feature can
be dark in that channel, while one isolated, strongly weighted mode
can dominate it.

\subsection{Local spectral weight and displacement readouts}
\label{sec:local-spectral-readouts}

The global density counts modes. Spatially resolved spectral weights and
calibrated displacement channels specify where and how those modes
appear in an observation.

\begin{definition}[Positive-frequency matrix and site spectral measures]
\label{def:s3-matrix-site-spectral-measures}
For the stable conservative dynamical matrix, the
\emph{positive-frequency matrix spectral measure} assigns each positive
frequency its complete mass-weighted eigenspace projector. A
\emph{site spectral measure} is its trace against the projector
$\mathsf\Pi_i$ onto the retained coordinate block of site $i$.
\end{definition}

\begin{proposition}[Local spectral weights and operator moments]
\label{prop:s3-matrix-spectral-moments}
Let the site projectors partition the retained translational space, let
$\mathsf D\succeq0$, and choose a complete orthonormal mode basis. The
measures just defined and their first two even moments satisfy
\begin{equation}
 \begin{split}
 \mathsf A_+(\omega)&=\sum_{s:\omega_s>0}
                    \ket{s}\bra{s}\,\delta(\omega-\omega_s),\\
 g_i(\omega)&=\tr[\mathsf\Pi_i\mathsf A_+(\omega)]
       =\sum_s\|\vct e_{s,i}\|^2\delta(\omega-\omega_s),\\
 \sum_i g_i(\omega)&=g_n(\omega),\qquad
 \int_0^\infty\mathsf A_+(\omega)\,\dd\omega=\mathsf P_+,\qquad
 \int_0^\infty\omega^2\mathsf A_+(\omega)\,\dd\omega=\mathsf D.
 \end{split}
 \label{eq:matrix-local-dos}
\end{equation}
All modal sums in the positive-frequency measures run over
$\omega_s>0$.
\end{proposition}
\begin{proof}
Insert the finite spectral decomposition of $\mathsf D$.
Taking the site trace gives the squared block norm. Summing the site
projectors gives the total positive-mode density. Integrating the
atomic measure gives $\mathsf P_+$, and its squared-frequency moment
gives $\mathsf D$ because zero eigenvalues contribute zero.
\end{proof}
The site blocks partition the translational coordinates; additional
internal variables require a complete corresponding partition.
The integral $\int g_i\,\dd\omega=\tr(\mathsf\Pi_i\mathsf P_+)$
can be less than three when zero directions are retained. Complete
eigenspace projectors make the measure basis independent at degeneracy.
Its participation weight resolves where a mode lies in mass-weighted
coordinates. Physical displacement variance and detector intensity
use the calibrated mass, frequency and signal weights below.

\begin{definition}[A dark frequency eigenspace]
\label{def:s3-dark-eigenspace}
For a calibrated mass-weighted probe column $\vct h$, a positive-frequency
eigenspace is \emph{dark} to that channel when
$\mathsf P_{\omega_*}\vct h=\vct0$, where
$\mathsf P_{\omega_*}$ is its complete eigenspace projector.
\end{definition}

With
$\vct h=\mathsf P_{\rm adm}\mathsf M^{-1/2}\vct\ell$ for the calibrated probe
defined below, its site weight is
$\tr(\mathsf\Pi_i\mathsf P_{\omega_*})$ and its conjugate-probe
weight is $\|\mathsf P_{\omega_*}\vct h\|^2$.
The latter is zero only when the entire eigenspace is dark to that
channel. Rotating the basis within the eigenspace can redistribute
individual overlaps but cannot create or remove this total weight.
This matters when an apparently localized or invisible mode is
reported from one numerical eigenvector in a degenerate spectrum.

\paragraph{Centre, angular and marker participation.}
For the full conservative body space, let $\mathsf P_{\rm tr}$ and
$\mathsf P_{\rm rot}$ be the orthogonal projectors onto its
inertia-normalized translational and rotational blocks. Its positive
spectral measure then gives
\begin{equation}
 g_{\rm tr}(\omega)=\tr[\mathsf P_{\rm tr}\mathsf A_+(\omega)],\qquad
 g_{\rm rot}(\omega)=\tr[\mathsf P_{\rm rot}\mathsf A_+(\omega)],\qquad
 g_{\rm tr}+g_{\rm rot}=g_n .
 \label{eq:s3patch-translation-rotation-participation}
\end{equation}
These weights describe kinetic participation in mixed conservative
modes. Their integrals account for the actual zero sector; they need
not separately count $3n$ positive modes. If constraints mix these
blocks, pull the physical kinetic forms through the admissible map
before defining participation. A modal participation is distinct from
the intensity in a detector.

\begin{definition}[Registered surface-marker displacement]
\label{def:s3patch-marker-readout}
For a marker at known body arm $\vct a_{ia}$, its first-order
laboratory displacement is
\begin{equation}
 \delta\vct x_{ia}=\vct u_i+
       \vct\theta_i\times\vct b_{ia}
    =\vct u_i-[\vct b_{ia}]_\times\vct\theta_i,
 \qquad Y=\vct\ell_Y^{\mathsf T}\vct X .
 \label{eq:s3patch-marker-observation}
\end{equation}
The final form represents a specified scalar projection or calibrated
linear readout in scaled coordinates. Its coefficients include the
marker lever arm and the observation geometry.
\end{definition}
An orientation-sensitive channel requires a registered feature or
reporter with the stated symmetry and resolution. A centre track can
respond indirectly to a coupled libration, while a pure angular mode
can remain dark to it. On a positive conservative sector, a conjugate
source entering as $-F_Y Y$ gives the positive-sector response
\begin{equation}
 \mathcal R_{YY}^{+}(z)=\vct h_Y^{\mathsf T}
   (\mathsf D_X-z^2\Id)^{-1}\vct h_Y,\qquad
 \vct h_Y=\mathsf P_+\mathsf G_X^{-1/2}\vct\ell_Y .
 \label{eq:s3patch-mixed-channel-response}
\end{equation}
If $Y$ has output unit $[Y]$, then $F_Y$ has units
$\mathrm J/[Y]$ and the response has units $[Y]^2/\mathrm J$.
Any retained zero-sector response is accounted for separately.
Length-per-force compliance is the displacement specialization.
Surface reporters also require their explicit population response and
signal calibration. Retained finite-rate states use the full transfer
map of the augmented dynamics instead of the conservative spectral
projectors alone.

Figure~\ref{fig:s34patch-angular-marker-array} shows how a many-particle angular pattern reaches a registered surface-marker channel, even when its centre-displacement component vanishes.

\begin{figure}[H]
\centering\begingroup
\setlength{\unitlength}{1pt}
\scalebox{0.94}{%
\input{vlt-figures/figure-289.tex}%
}\endgroup
\caption{Angular participation and detector participation. Panel (a) uses twelve separated, stacked VLPs; arrows encode an assigned alternating small-angle pattern, and centre dots stay fixed. This is a kinematic illustration of a librational component, rather than a computed eigenvector of that array. Red markers remain attached to the same body arms throughout the motion; red/blue curved arrows distinguish the signs of angular motion. Panel (b) magnifies the green body and selects $\vct b=b\hat{\vct x}$, $\vct\theta=\theta_0\sin(\omega t)\hat{\vct z}$, with a registered $y$-marker readout. The solid curve is its first-order displacement; the centre signal is zero. The reconstruction follows~\eqref{eq:s3patch-marker-observation}. At this resolution, marker geometry and viewing direction determine which angular component appears in a record.}
\label{fig:s34patch-angular-marker-array}
\end{figure}

\subsubsection{A force and its conjugate displacement}
\label{sec:conservative-probe-response}

\begin{definition}[Calibrated conjugate displacement channel]
\label{def:s3-conjugate-displacement-channel}
Choose a real, dimensionless coordinate weight $\vct\ell$ and the
physical displacement $Y=\vct\ell^{\mathsf T}\vct u$.
For example, $\vct\ell$ can select a single particle polarization
or the difference between an envelope and an internal coordinate.
A weak preparation force $F(t)$ conjugate to this readout contributes
$-F(t)Y$ to the Hamiltonian and the force vector $F(t)\vct\ell$
to the equation of motion.
The applied force supplies power $F\Dt{Y}$; the unforced sealed
baseline is recovered after preparation.
\end{definition}
Choose a real orthonormal mode basis as in
Theorem~\ref{thm:s2-finite-normal-modes}, and define
$\vct h=\mathsf P_{\rm adm}\mathsf M^{-1/2}\vct\ell$ and
$h_s=\vct h^{\mathsf T}\ket{s}$.
\begin{proposition}[Conservative conjugate-channel response]
\label{prop:s3-conservative-channel-response}
For fixed conservative mechanics on the positive sector, with zero
directions constrained or $\mathsf P_0\vct h=\vct0$, the retarded
force-induced response is
\begin{equation}
 \mathcal R_Y(z)=\sum_s\frac{|h_s|^2}{\omega_s^2-z^2},
 \qquad \operatorname{Im}z>0,\qquad
 \delta Y(\omega)=\mathcal R_Y(\omega+i0)F(\omega).
 \label{eq:conservative-probe-susceptibility}
\end{equation}
\end{proposition}
\begin{proof}
Each mass-normalized coordinate satisfies
$\Dtt{Q_s}+\omega_s^2Q_s=h_sF(t)$ and the readout is
$Y=\sum_s h_sQ_s$. The retarded modal solution has denominator
$\omega_s^2-z^2$. Substitution into the readout gives the displayed
sum, with analyticity for $\operatorname{Im}z>0$.
\end{proof}
One overlap specifies how the force excites the pattern; the second
specifies how the same conjugate readout detects it, giving $|h_s|^2$.
Here $\delta Y$ is the displacement induced by the force; previously
prepared unforced motion contributes a separate homogeneous solution.
The positive-sector qualification can be implemented by constraining
the free coordinates or selecting $\mathsf P_0\vct h=0$.
Otherwise the full response includes
$-\|\mathsf P_0\vct h\|^2/z^2$, and a static force can cause
non-restoring drift.
The units of $\mathcal R_Y$ are length per force.
The $+i0$ prescription specifies retarded excitation; it does not
insert a phenomenological linewidth into the lossless system.

\begin{proposition}[Probe-weighted spectral measure]
\label{prop:s3-probe-spectral-measure}
Under Proposition~\ref{prop:s3-conservative-channel-response}, the
positive-frequency boundary value satisfies the distributional identity
\begin{equation}
 \frac{2\omega}{\pi}\operatorname{Im}\mathcal R_Y(\omega+i0)
     =\sum_s|h_s|^2\delta(\omega-\omega_s)
     =\vct h^{\mathsf T}\mathsf A_+(\omega)\vct h.
 \label{eq:probe-weighted-spectrum}
\end{equation}
\end{proposition}
\begin{proof}
For $\omega>0$, the imaginary part of each retarded modal denominator is
$\pi\delta(\omega-\omega_s)/(2\omega_s)$. Multiplying by
$2\omega/\pi$ gives its spectral weight. Contracting
$\mathsf A_+$ with the probe column gives the same modal sum.
\end{proof}
The overlap determines visibility: $h_s=0$ removes a mode from this
channel. Summing all mass-weighted coordinate channels recovers $g_n$;
a physical displacement channel retains its own weights. Cross response
uses $h_s^{(a)}h_s^{(b)}$ and records spatial correlation, including
its sign. This is the conservative limit of
\eqref{eq:damped-response-spectrum}. For an imaging or optical signal,
linearizing and calibrating both forcing and transduction supplies the
physical channel used in that comparison.

\paragraph{Several reciprocal channels.}
\begin{definition}[Reciprocal channel family]
\label{def:s3-reciprocal-channel-family}
For calibrated displacement channels, collect the dimensionless columns
$\vct\ell_a$ into $\mathsf L$, so that
$\vct Y=\mathsf L^{\mathsf T}\vct u$ and the conjugate loading is
$\mathsf L\vct F$. Retain the same positive-sector preparation as above
and put $\mathsf H=\mathsf P_+\mathsf M^{-1/2}\mathsf L$.
\end{definition}

\begin{proposition}[Reciprocity and positive spectral channel weights]
\label{prop:s3-reciprocal-spectral-weights}
For fixed real symmetric conservative mechanics and the conjugate channel
family above, write $\mathsf P_\lambda$ for the complete projector at
positive squared frequency $\lambda$. Then
\begin{equation}
 \begin{aligned}
 \mathsf R(z)&=\sum_{\lambda>0}
       \frac{\mathsf G_\lambda}{\lambda-z^2},
 &\mathsf G_\lambda&=\mathsf H^{\mathsf T}
                     \mathsf P_\lambda\mathsf H\succeq0,\\
 \mathsf R(z)^{\mathsf T}&=\mathsf R(z),
 &|(\mathsf G_\lambda)_{ab}|^2
      &\le(\mathsf G_\lambda)_{aa}(\mathsf G_\lambda)_{bb}.
 \end{aligned}
 \label{eq:structure-reciprocal-channel-matrix}
\end{equation}
\end{proposition}
\begin{proof}
Each $\mathsf G_\lambda$ is the Gram matrix of the real columns of
$\mathsf P_\lambda\mathsf H$, so it is symmetric nonnegative and its
entries obey Cauchy--Schwarz. Summing these symmetric matrices with
scalar modal denominators gives $\mathsf R(z)^{\mathsf T}=\mathsf R(z)$
~\cite{Teschl2009}.
\end{proof}

Every response
entry has units of length per force. The matrices $\mathsf G_\lambda$
are spectral weights, with units of inverse mass; residues at the poles
in $z$ carry an additional frequency factor. The diagonal entries are
nonnegative, whereas a cross-channel entry can have either sign. Its
magnitude obeys the displayed bound for each complete eigenspace,
including a degenerate one. A negative cross response therefore need not
indicate a negative stiffness. If one diagonal weight vanishes, all cross
weights involving that channel vanish at the same squared frequency.
For different, nonconjugate forcing and detection profiles the response
is rectangular and this Gram-matrix positivity does not apply directly.
For fixed linear constraints, the equivalent reduced channel map is
$\mathsf H_{\rm c}=\mathsf P_{{\rm c},+}\mathsf M_{\rm c}^{-1/2}
\mathsf U_{\rm c}^{\mathsf T}\mathsf L$;
its physical-site embedding is $\mathsf H=\mathsf J_{\rm c}\mathsf H_{\rm c}$.

The causal time-domain response makes the same point without a
frequency-space pole:
\begin{equation}
 \mathcal R_Y(t)=\Theta(t)\sum_s|h_s|^2
                  \frac{\sin(\omega_st)}{\omega_s},\qquad
 \delta Y(t)=\int_{-\infty}^{t}\mathcal R_Y(t-t')F(t')\,\dd t'.
 \label{eq:structure-retarded-impulse-response}
\end{equation}
It is obtained by integrating the modal equation across an impulse
and evolving the resulting velocity. The imaginary part in
\eqref{eq:probe-weighted-spectrum} consists of ideal spectral lines;
it does not assign an irreversible decay rate to these lossless modes.
A force exactly at a visible eigenfrequency can produce secular
growth, so a finite steady resonant amplitude requires the specified
loss or a finite observation history.

Figure~\ref{fig:structure-force-protocol-preparation} illustrates the causal response for a sudden force and distinguishes it from comparing separately equilibrated biased preparations.

\begin{figure}[H]
\centering\begingroup
\definecolor{vltTeal}{HTML}{176B64}
\definecolor{vltBlue}{HTML}{315D80}
\definecolor{vltGold}{HTML}{A87520}
\definecolor{vltPlum}{HTML}{725877}
\definecolor{vltRed}{HTML}{762B38}
\definecolor{vltYellow}{HTML}{D5AD3C}
\definecolor{vltNeutral}{HTML}{777D83}
\setlength{\unitlength}{1pt}
\scalebox{0.80}{%
\input{vlt-figures/figure-087.tex}%
}\endgroup
\caption{A force protocol and an equilibrium preparation for one classical harmonic coordinate, with $F_0/k=\ell_T$ and $\ell_T=\sqrt{k_{\rm B}T/k}$. (a) Switching on a constant force from an unbiased initial state of zero mean displacement and velocity gives $\langle u(t)\rangle/\ell_T=1-\cos\omega t$, as in \eqref{eq:structure-undamped-force-step}. The dashed line is the mean of a separately prepared Gibbs state under that bias. (b) The initial unbiased Gibbs density (black) and biased Gibbs density (burgundy) are unit-variance Gaussians centred at zero and one; the force shifts the mean while preserving the equilibrium covariance, as in \eqref{eq:structure-static-bias-fluctuations}. Conservative step-driven motion continues to oscillate about the biased mean. Equilibration would require the corresponding medium dynamics and noise, beyond this lossless protocol.}
\label{fig:structure-force-protocol-preparation}
\end{figure}

\begin{proposition}[Inertial and static response sum rules]
\label{prop:s3-response-sum-rules}
For the complete positive-sector response in
Proposition~\ref{prop:s3-conservative-channel-response},
\begin{equation}
 \begin{split}
 \frac2\pi\int_0^\infty\omega\,
       \operatorname{Im}\mathcal R_Y(\omega+i0)\,\dd\omega
    &=\vct h^{\mathsf T}\mathsf P_+\vct h,\\
 \frac2\pi\int_0^\infty
       \frac{\operatorname{Im}\mathcal R_Y(\omega+i0)}{\omega}
          \,\dd\omega&=\mathcal R_Y(0).
 \end{split}
 \label{eq:structure-response-moment-checks}
\end{equation}
\end{proposition}
\begin{proof}
Integrate the modal delta weights from
\eqref{eq:probe-weighted-spectrum}. The first integral sums
$|h_s|^2$; the second sums $|h_s|^2/\omega_s^2$. Completeness on the
positive sector and evaluation of the response at zero give the results.
\end{proof}

The first sum is the coefficient of $-1/z^2$ at large $|z|$; the second
is the static compliance. These are checks on a complete model spectrum. A detector covering only a finite band
need not exhaust them, and unresolved high-frequency coordinates
can change the inertial completion of a reduced model.

One experimental test changes loading and readout while holding the
arrangement fixed. Centre-of-mass and relative channels, each defined
by conjugate work, weight different mode patterns. For a free pair,
mass-proportional forces drive translation, while opposed forces drive
relative motion. Spatial readouts test those predicted overlaps in a
larger assembly. For supported particles, probe placement also records
any change in contact forces or support compliance.

\begin{example}[Clamped curvature and a calibrated two-site probe]
\label{ex:structure-pinned-probes}
Consider two equal translational masses $m$ in one polarization,
with a link spring $k_{\rm link}>0$ and smooth onsite pinning
$k_{\rm pin}>0$. This declared onsite potential supplies both positive
frequencies; an inactive hard wall would not supply it. The stiffness is
\begin{equation}
 \mathsf K_{\rm d}=
 \begin{pmatrix}k_{\rm pin}+k_{\rm link}&-k_{\rm link}\\
                -k_{\rm link}&k_{\rm pin}+k_{\rm link}\end{pmatrix},
 \quad
 \omega_{\rm com}^2=\frac{k_{\rm pin}}m,\qquad
 \omega_{\rm rel}^2=\frac{k_{\rm pin}+2k_{\rm link}}m.
 \label{eq:structure-pinned-probe-matrix}
\end{equation}
The mode shapes are $(1,1)^{\mathsf T}/\sqrt2$ and
$(1,-1)^{\mathsf T}/\sqrt2$ in mass-weighted coordinates.
For a force at site 1 and its displacement readout,
\begin{equation}
 \mathcal R_{u_1}(z)=\frac1{2m}
       \left[\frac1{\omega_{\rm com}^2-z^2}
             +\frac1{\omega_{\rm rel}^2-z^2}\right],\qquad
 \boxed{\mathcal R_{u_1}(0)
      =\frac{k_{\rm pin}+k_{\rm link}}
             {k_{\rm pin}(k_{\rm pin}+2k_{\rm link})}.}
 \label{eq:structure-pinned-probe-compliance}
\end{equation}
Holding site 2 fixed would instead give compliance
$1/(k_{\rm pin}+k_{\rm link})$. Allowing it to move increases
the site-1 compliance. Table~\ref{tab:structure-probe-weights} gives
three calibrated coordinates of this same system. Each response is
the sum of its listed weights divided by the corresponding
$\omega_s^2-z^2$.
\end{example}

\begin{table}[H]
\centering
\caption{Probe overlaps for the pinned two-site example.
The entries are $|h_s|^2$ with units of inverse mass. Each $Y$ has its
own conjugate force and stated normalization; the rows do not rank
instrument sensitivities. The two frequencies exist in every row's
mechanical spectrum, while a zero overlap removes that pole from the readout.}
\label{tab:structure-probe-weights}
\begin{tabular}{lll}
\toprule
Displacement readout $Y$ & Common-mode weight & Relative-mode weight\\
\midrule
$u_1$ & $1/(2m)$ & $1/(2m)$\\
$(u_1+u_2)/2$ & $1/(2m)$ & $0$\\
$u_1-u_2$ & $0$ & $2/m$\\
\bottomrule
\end{tabular}
\end{table}

For the mean-coordinate row, its conjugate force acts equally with
amplitude $F/2$ on the two sites; for the difference row it acts as
$(F,-F)$. A mode absent from one readout can therefore be visible in
another. The example concerns whole-particle translation; its link and
pinning coefficients are the declared translational model inputs.
Their relation to internal capsid deformation requires the corresponding
deformation coordinates.

Figure~\ref{fig:more-probe-dark-modes} gives a visual version of these
calibrated channels: the collective modes remain the same while their
visibility changes with the force and readout.

\begin{figure}[H]
\centering\begingroup
\definecolor{vltTeal}{HTML}{176B64}
\definecolor{vltBlue}{HTML}{315D80}
\definecolor{vltGold}{HTML}{A87520}
\definecolor{vltPlum}{HTML}{725877}
\definecolor{vltRed}{HTML}{762B38}
\definecolor{vltYellow}{HTML}{D5AD3C}
\definecolor{vltNeutral}{HTML}{777D83}
\setlength{\unitlength}{1pt}
\scalebox{0.90}{%
\input{vlt-figures/figure-088.tex}%
}\endgroup
\caption{Three conjugate channels use the same supported pair and $k_{\rm pin}=k_{\rm link}=k_\star$. Upper arrows are coefficients $\vct w$ in $Y=\vct w^{\mathsf T}\vct u$, at $22\,\mathrm{pt}$ per coefficient. Loading $F\vct w$ supplies work $F\,\dd Y$; these arrows are force profiles, not motion snapshots. Sticks give $m|h_s|^2$ at the two unchanged mode frequencies; open markers denote zero coefficients or weights. Dividing each weight by $(\omega_s/\omega_\star)^2$ gives the printed static compliance. Channel normalization belongs to calibration. Residues in the frequency variable additionally contain the pole-frequency factor. The boxed result illustrated here is~\eqref{eq:structure-pinned-probe-compliance}.}
\label{fig:more-probe-dark-modes}
\end{figure}

\subsubsection{Equilibrium fluctuations and soft-mode sensitivity}
\label{sec:conservative-displacement-fluctuations}

The following equilibrium formulas use the quadratic model on its full
retained linear coordinate space. For the finite-clearance cylinder,
they describe the local harmonic preparation when contact and
anharmonic regions carry negligible statistical weight. A preparation
explicitly conditioned on the hard configuration boundary uses the
corresponding normalized, truncated distribution.

\begin{proposition}[Thermal spectrum of a conjugate displacement]
\label{prop:s3-equilibrium-channel-spectrum}
Under Postulate~\ref{post:quantization}, take the Gibbs preparation at
$T>0$ on the positive harmonic sector and the calibrated channel of
Definition~\ref{def:s3-conjugate-displacement-channel}. Its symmetrized
displacement spectrum is
\begin{equation}
 \begin{split}
 S_Y^{\mathrm{sym}}(\omega)
  &:=\int_{-\infty}^{\infty}e^{i\omega t}
       \tfrac12\langle\{\hat Y(t),\hat Y(0)\}\rangle_T\,\dd t\\
  &=\sum_s\frac{\pi\hbar|h_s|^2}{2\omega_s}
       \coth\!\left(\frac{\hbar\omega_s}{2k_{\mathrm B}T}\right)
       [\delta(\omega-\omega_s)+\delta(\omega+\omega_s)].
 \end{split}
 \label{eq:equilibrium-displacement-spectrum}
\end{equation}
\end{proposition}
\begin{proof}
Independent thermal oscillator modes have the symmetrized coordinate
correlator
\[
 \frac{\hbar}{2\omega_s}
   \coth\!\left(\frac{\hbar\omega_s}{2k_{\mathrm B}T}\right)
   \cos(\omega_st).
\]
Projecting with $h_s$ and Fourier transforming the cosine gives the
two delta lines with the stated normalization.
\end{proof}
The equilibrium displacement mean is zero; for a nonzero mean the
connected correlator is used. The thermal oscillator calculation uses the displacement reconstruction
in~\eqref{eq:quantum-displacement-field}. With the stated Fourier and force conventions, it satisfies
$S_Y^{\mathrm{sym}}(\omega)=
\hbar\coth[\hbar\omega/(2k_{\mathrm B}T)]
\operatorname{Im}\mathcal R_Y(\omega+i0)$ for $\omega>0$,
the equilibrium fluctuation--dissipation relation~\cite{Clerk2010}.
\begin{proposition}[Equilibrium variance and static compliance]
\label{prop:s3-variance-static-compliance}
The preparation in Proposition~\ref{prop:s3-equilibrium-channel-spectrum}
has the variance below. Its classical harmonic Gibbs counterpart has
\begin{equation}
 \begin{split}
 \langle\hat Y^2\rangle_T
    &=\int_{-\infty}^{\infty}
                   S_Y^{\mathrm{sym}}(\omega)\frac{\dd\omega}{2\pi}
      =\sum_s\frac{\hbar|h_s|^2}{2\omega_s}
           \coth\!\left(\frac{\hbar\omega_s}{2k_{\mathrm B}T}\right),\\
 \langle Y^2\rangle_{\mathrm{cl}}
    &=k_{\mathrm B}T\sum_s\frac{|h_s|^2}{\omega_s^2}
      =k_{\mathrm B}T\,\mathcal R_Y(0).
 \end{split}
 \label{eq:probe-variance-compliance}
\end{equation}
\end{proposition}
\begin{proof}
Integration of the two spectral lines with the stated $1/(2\pi)$
convention gives the first sum. Classical equipartition gives
$\langle Q_s^2\rangle=k_{\mathrm B}T/\omega_s^2$; summing its probe
projections gives the static response in the second line. It is also
the limit of the quantum sum when every contributing
$\hbar\omega_s/(k_{\mathrm B}T)$ is small.
\end{proof}
The last equality concerns the positive-sector static compliance.
Unconstrained zero coordinates require their own finite-domain
preparation and cannot be assigned a harmonic Gibbs variance.
At $T\to0$ the quantum variance retains the zero-point term.
For $\hbar\omega_s\ll k_{\mathrm B}T$ the same spectral architecture
has the classical thermal variance.

Classical equipartition sets each positive mode's mean potential energy
to $k_{\mathrm B}T/2$, so softer modes have larger excursions.
Their probe projections connect equilibrium fluctuations to static
compliance. Thermal particle tracking in colloidal crystals provides
a precedent for this inference from fluctuations~\cite{Keim2004}.

\begin{proposition}[Classical harmonic coordinate covariance]
\label{prop:s3-classical-coordinate-covariance}
On a fully supported Cartesian translational space with
$\mathsf K\succ0$, the classical Gibbs density at $T>0$ is proportional to
$\exp[-\vct u^{\mathsf T}\mathsf K\vct u/(2k_{\mathrm B}T)]$.
For calibrated displacement weights $\vct\ell_a$, its covariance is
\begin{equation}
 \boxed{\mathsf C_u:=\langle\vct u\vct u^{\mathsf T}\rangle
         =k_{\mathrm B}T\mathsf K^{-1},\qquad
 \operatorname{Cov}(Y_a,Y_b)
         =k_{\mathrm B}T\vct\ell_a^{\mathsf T}
                          \mathsf K^{-1}\vct\ell_b.}
 \label{eq:structure-classical-coordinate-covariance}
\end{equation}
\end{proposition}
\begin{proof}
Diagonalize the real symmetric positive stiffness. Each independent
Gaussian coordinate has variance $k_{\mathrm B}T$ divided by its
stiffness eigenvalue. Transforming back gives $k_{\mathrm B}T\mathsf K^{-1}$;
linear projection gives the channel covariance.
\end{proof}
For a retained positive sector with unconstrained zero directions
excluded, the corresponding matrix is instead
$k_{\mathrm B}T\mathsf M^{-1/2}\mathsf D^+\mathsf M^{-1/2}$,
where $\mathsf D^+$ inverts positive eigenvalues and annihilates
the kernel. This specifies a mass-weighted sector preparation;
it is generally different from the Euclidean pseudoinverse of
$\mathsf K$. Fixed constraints use their own admissible map.

For the pinned pair of Example~\ref{ex:structure-pinned-probes},
\begin{equation}
 \mathsf C_{\rm pair}
  =\frac{k_{\mathrm B}T}
          {k_{\rm pin}(k_{\rm pin}+2k_{\rm link})}
       \begin{pmatrix}
       k_{\rm pin}+k_{\rm link}&k_{\rm link}\\
       k_{\rm link}&k_{\rm pin}+k_{\rm link}
       \end{pmatrix},\qquad
 \frac{(C_{\rm pair})_{12}}{(C_{\rm pair})_{11}}
       =\frac{k_{\rm link}}{k_{\rm pin}+k_{\rm link}}.
 \label{eq:structure-pinned-pair-covariance}
\end{equation}
For $k_{\rm pin}=k_{\rm link}=k_\star$, the diagonal covariance is
$2k_{\mathrm B}T/(3k_\star)$, the off-diagonal covariance is
$k_{\mathrm B}T/(3k_\star)$, and their correlation coefficient is $1/2$.
One set of interaction and support parameters predicts these
fluctuations, response poles and channel weights. The benchmark in
Section~\ref{sec:complete-benchmark} adds the frequency distribution
set by medium memory.

Figure~\ref{fig:structure-relaxed-covariance} shows the joint equilibrium contours and compares marginal fluctuations with a held-coordinate preparation, using that same pair and parameter set.

\begin{figure}[H]
\centering\begingroup
\definecolor{vltTeal}{HTML}{176B64}
\definecolor{vltBlue}{HTML}{315D80}
\definecolor{vltGold}{HTML}{A87520}
\definecolor{vltPlum}{HTML}{725877}
\definecolor{vltRed}{HTML}{762B38}
\definecolor{vltYellow}{HTML}{D5AD3C}
\definecolor{vltNeutral}{HTML}{777D83}
\setlength{\unitlength}{1pt}
\scalebox{0.80}{%
\input{vlt-figures/figure-089.tex}%
}\endgroup
\caption{Covariance and constraint in the same pinned pair, with $\mathsf K/k_\star=\left(\begin{smallmatrix}2&-1\\-1&2\end{smallmatrix}\right)$ and the classical Gibbs preparation. (a) Equal Cartesian scales show the density contours $\vct q^{\mathsf T}(\mathsf K/k_\star)\vct q=c$ at $c=1,4$. Their principal variances are one for the common coordinate and $1/3$ for the relative coordinate. The dotted lines locate those directions; the contours are density levels. (b) Integrating over $q_2$ gives variance $2/3$ at site 1, while conditioning on $q_2=0$ gives variance $1/2$. Each curve is normalized separately over the full real axis. Through \eqref{eq:structure-classical-coordinate-covariance}, the marginal measures the relaxed compliance $2/(3k_\star)$ and the conditional describes the held-coordinate preparation. Positive pinning makes this full harmonic position density normalizable.}
\label{fig:structure-relaxed-covariance}
\end{figure}

\paragraph{Which relative motions contribute to a compliance?}
The assembled matrix also resolves which relative particle motions
contribute to a static measurement.

\begin{proposition}[Bond-resolved compliance sensitivity]
\label{prop:s3-bond-compliance-sensitivity}
On a fully supported Cartesian centre-coordinate space with real
$\mathsf K\succ0$, define
$\vct v_a=\mathsf K^{-1}\vct\ell_a$, the displacement per unit force in
channel $a$. Let $\mathsf E_e$ extract the relative displacement of the
two ideal-virion centres on retained edge $e$. For an infinitesimal stiffness variation
at fixed reference geometry and fixed channel calibration,
\begin{equation}
 \begin{aligned}
 \delta R_{ab}(0)
   &=-\sum_e(\mathsf E_e\vct v_a)^{\mathsf T}
          \delta\mathsf C_e(\mathsf E_e\vct v_b)
     -\vct v_a^{\mathsf T}\delta\mathsf K_{\rm sup}\vct v_b,\\
 \operatorname{Cov}(\mathsf E_e\vct u)
   &=k_{\mathrm B}T\,\mathsf E_e\mathsf K^{-1}
                         \mathsf E_e^{\mathsf T}.
 \end{aligned}
 \label{eq:structure-bond-compliance-sensitivity}
\end{equation}
Here $\delta\mathsf K_{\rm sup}$ collects changes to non-pair quadratic
energy, and the covariance is for the classical Gibbs preparation at
fixed $T>0$.
\end{proposition}
\begin{proof}
Differentiating $\mathsf K^{-1}\mathsf K=\Id$ gives
$\delta\mathsf K^{-1}=-\mathsf K^{-1}(\delta\mathsf K)\mathsf K^{-1}$.
Insert the edge assembly and the fixed channel columns to obtain the
first identity. Projecting Proposition~\ref{prop:s3-classical-coordinate-covariance}
with $\mathsf E_e$ gives the second.
\end{proof}

The compliance variation has units of length per force, and the
relative-displacement covariance has units of length squared.
Both resolve motion between the retained particle centres.

At fixed edge direction,
$\delta\mathsf C_e=\delta L_e\vct n_e\vct n_e^{\mathsf T}
+\delta T_e(\Id_3-\vct n_e\vct n_e^{\mathsf T})$.
Both increments follow from the same varied pair energy.
Thus a conjugate compliance samples longitudinal and transverse
curvature changes through the relative motions generated by its own
unit loading. An edge whose endpoints move together under that loading
has no first-order contribution, even if it participates strongly in
another channel. A positive-semidefinite stiffness increment lowers a
conjugate static compliance; signed electrostatic curvature increments
do not give that ordering automatically. Changing charge, screening or
support may also move the equilibrium: a derivative along that family
must include the resulting geometry and direction changes. The formula
above isolates the fixed-reference mechanical derivative and is not an
assumption that the particles stay in force balance after an arbitrary
parameter change.

\begin{proposition}[Equilibrated bias and sudden conservative loading]
\label{prop:s3-static-bias-and-step}
Compare two fully supported classical harmonic Gibbs preparations at the
same temperature and stiffness, one with a constant bias $-F_0Y$. Then
\begin{equation}
 \Delta\langle Y\rangle=F_0\mathcal R_Y(0),\qquad
 \operatorname{Var}(Y)=k_{\mathrm B}T\mathcal R_Y(0).
 \label{eq:structure-static-bias-fluctuations}
\end{equation}
The bias shifts the mean without changing the harmonic covariance.
This is a comparison of equilibrated preparations within the same
quadratic basin. If the force is switched on suddenly in the lossless
model from zero initial mean displacement and velocity, the actual mean is
\begin{equation}
 \Delta\langle Y(t)\rangle=F_0\sum_s\frac{|h_s|^2}{\omega_s^2}
                         [1-\cos(\omega_st)],\qquad t\ge0.
 \label{eq:structure-undamped-force-step}
\end{equation}
\end{proposition}
\begin{proof}
Completing the biased quadratic energy shifts its mean by
$\mathsf K^{-1}\vct\ell F_0$ and leaves its covariance unchanged.
For the sudden force, the modal initial-value solution is
$Q_s(t)=F_0h_s[1-\cos(\omega_st)]/\omega_s^2$.
Projection onto $Y$ gives the second formula.
\end{proof}
It oscillates about the static shift instead of immediately equilibrating.
A selected preparation, slow protocol or specified medium relaxation
determines whether the static comparison describes an experiment.
An equal-time variance alone does not set a detector's force resolution
or its observation-time sensitivity.

At fixed nonzero overlap, softening increases a mode's compliance and
thermal excursions. Enhanced response therefore trades against the
clearance and harmonic-basin margins. The classical variance weights
contain $1/\omega_s^2$, information absent from a mode count.
These equilibrium predictions use a specified statistical preparation;
the following sections add finite loss, medium fluctuations and
nonequilibrium dynamics.

\subsubsection{Testing the conservative model and its observable variables}
\label{sec:structure-conservative-test}

Coordinates, interactions, inertia and supports now predict spatial
patterns, frequencies, impulse response and equilibrium covariance.
Testing one parameter set across these observations constrains the
mechanical model more strongly than fitting each separately.
Frequencies constrain stiffness relative to inertia; calibrated static
covariance constrains equilibrium curvature. Resolving an internal
deformation additionally requires a signal sensitive to that coordinate.

\begin{proposition}[Curvature recovered at the observed resolution]
\label{prop:s3-observed-covariance-curvature}
For the classical preparation in
Proposition~\ref{prop:s3-classical-coordinate-covariance}, known
temperature and complete calibrated coordinate covariance determine
$\mathsf K=k_{\mathrm B}T\mathsf C_u^{-1}$.
If only the block $\vct u_a$ is observed, its marginal covariance instead
determines
$k_{\mathrm B}T[(\mathsf C_u)_{aa}]^{-1}
=\mathsf K_{a,\mathrm{rel}}$, the relaxed curvature
in~\eqref{eq:structure-local-schur}.
\end{proposition}
\begin{proof}
The full statement inverts the positive covariance formula. Completing
the quadratic form in the unobserved coordinates and integrating them
out gives a Gaussian with Schur-complement stiffness
$\mathsf K_{a,\mathrm{rel}}$. Its covariance is
$k_{\mathrm B}T\mathsf K_{a,\mathrm{rel}}^{-1}$.
\end{proof}

Marginal curvature includes interactions mediated by omitted coordinates.
It is therefore an effective description at the observed resolution.
Recovering direct contacts additionally uses structural information and
controls for bandwidth, calibration, supports and nonequilibrium forcing.
Independent channel weights, cross-correlations and the cancellation
in~\eqref{eq:structure-internal-antiresonance} test whether the chosen
coordinate resolution is adequate.

The same covariance argument extends to a positive, locally flat
Gaussian tangent in position, angle and surface/material variables:
the inverse full covariance gives the joint Hessian, while a measured
subblock gives its Schur reduction. For finite-angle equilibrium the
physical rotational measure is Haar measure; it is generally not a
global Gaussian in three chosen angles. The tangent statement therefore
uses small angular excursions about the specified preparation.

The shared-state example~\eqref{eq:s3patch-shared-state-assembly}
shows the inverse consequence. A measured cross curvature can arise
from a surface coordinate shared by two ports. Static mechanical data
first identify that effective curvature. Frequency-dependent response,
angular motion and a separately calibrated surface channel can test
its origin. At finite rate, the accessible part is determined by
input--state--output dynamics of the augmented model. The symmetric
mechanical subspace theorem below applies to the declared conservative
reference; a general dissipative generator needs its corresponding
controllability and observability spaces. These comparisons make
variable selection a question about specified predictions and channels.

\paragraph{The part of the assembly determined by the channels.}
\begin{definition}[Channel-accessible mechanical subspace]
\label{def:s3-channel-accessible-subspace}
For the positive-sector channel matrix $\mathsf H$, the
\emph{channel-accessible mechanical subspace} is the smallest
$\mathsf D$-invariant subspace containing the range of $\mathsf H$.
\end{definition}

\begin{proposition}[Visible spectral mechanics and projected moments]
\label{prop:s3-visible-mechanical-reduction}
For fixed real symmetric stable conservative mechanics, set
$d_+=\operatorname{rank}\mathsf P_+$ and use the calibrated
positive-sector matrix $\mathsf H$. The accessible subspace and the
convergent large-frequency response expansion are
\begin{equation}
 \begin{aligned}
 \mathcal B&=\operatorname{span}
     \{\mathsf D^k\mathsf H\vct a:
           0\le k<d_+,\ \vct a\in\mathbb R^{p}\},\\
 \mathsf R(z)&=-\sum_{k=0}^{\infty}
       \frac{\mathsf H^{\mathsf T}\mathsf D^k\mathsf H}{z^{2k+2}},
       \qquad |z|^2>\|\mathsf D\|_2,
 \end{aligned}
 \label{eq:structure-visible-krylov-mechanics}
\end{equation}
where $p$ is the number of channels. Moreover,
$\mathcal B=\bigoplus_{\lambda>0}
\operatorname{range}(\mathsf P_\lambda\mathsf H)$; both
$\mathcal B$ and its positive-sector orthogonal complement are invariant.
The conjugate response depends only on the restriction to $\mathcal B$.
\end{proposition}
\begin{proof}
Cayley--Hamilton makes the finite power span invariant. Each complete
spectral projector is a polynomial in the restriction of $\mathsf D$,
giving the stated direct sum. Symmetry preserves the orthogonal
complement, which has zero overlap with $\mathsf H$ on both input and
output. Finally expand
$(\mathsf D-z^2\Id)^{-1}=-z^{-2}
\sum_{k\ge0}(\mathsf D/z^2)^k$ in its convergent norm range and project
with the channel matrix.
\end{proof}

The complement is neither excited by these forces nor detected in these
channels, although an independently prepared displacement can occupy it.
Within a degenerate eigenspace the accessible dimension can be smaller
than the degeneracy.

The convergent resolvent expansion shows that an exact complete
response fixes projected moments
$\mathsf H^{\mathsf T}\mathsf D^k\mathsf H$ and the visible spectral
weights. It leaves the dynamics on $\mathcal B^\perp$ unspecified.
This is the symmetric mechanical instance of the distinction between a
state-space model and its observable, controllable realization
\cite{Kalman1963}. Even a minimal visible realization identifies a
coordinate representation only up to compatible changes of basis;
it does not by itself assign unique particle contacts or charge products.
Known geometry, calibrated masses, support information and additional
independent channels supply extra physical constraints. This separates
an exact invisibility of part of the assembly from finite-band or noisy
data, which can limit reconstruction further.

\begin{example}[Recovering effective link and support stiffnesses]
\label{ex:s3-two-site-inverse-target}
In Example~\ref{ex:structure-pinned-probes}, suppose the common particle
mass $m$ is known and both positive frequencies are resolved. Then
$k_{\rm pin}=m\omega_{\rm com}^2$ and
$k_{\rm link}=m(\omega_{\rm rel}^2-\omega_{\rm com}^2)/2$.
The two coefficients are therefore determined within this specified
two-site model. If the masses are unspecified, these frequencies first
determine the corresponding stiffness-to-mass ratios.

The mean-coordinate channel $(u_1+u_2)/2$ alone has spectral weight
$1/(2m)$ at the common frequency and zero weight at the relative one.
Its full conservative response and classical equilibrium variance are
independent of $k_{\rm link}$. Every positive link stiffness gives the
same predictions in that channel at fixed $m$ and $k_{\rm pin}$.
Adding the relative channel supplies the missing link-dependent
frequency. This is an exact change in parameter information produced by
a change in observed motion, before introducing finite sampling or noise.

The recovered link coefficient is an effective tangent stiffness of the
declared energy at that reference. Separating its Lennard--Jones and
electrostatic contributions requires additional independently constrained
inputs. Identifying a spring coefficient and identifying its microscopic
origin are thus distinct inverse targets.
\end{example}

For a larger ideal-virion assembly, the same reasoning connects variable
integration to a specified prediction. Common, relative and spatially
resolved channels interrogate different collective displacements.
Retaining an internal coordinate is useful when its storage and coupling
change those predictions over the observation band. The explicit
response-error test in~\eqref{eq:structure-internal-reduction-error-bound}
quantifies one such decision: matching low-frequency inertia is
accompanied by control of the response denominator, especially near a
resonance. A richer retained state is assessed through predictions across
channels and preparations, with its additional inertia, energy and
initial data stated. These are concrete tasks within the broader open
problems of reconstructing a biophysical interaction network and
establishing variables adequate for its observed dynamics.

\subsection{Recruitment, breakup and transient lattice mechanics}
\label{sec:s23-transient-recruitment}

The fixed-reference matrix describes collective motion within one
persistent arrangement. A transient lattice is a succession of such
arrangements connected by encounter, accommodation and separation. The
full nonlinear interaction energy determines the connecting motion;
local Hessians describe the small excursions during residence in each
arrangement. This gives a single framework for a small nucleus, a growing
assembly and the fragments left by a breakup.

\subsubsection{From a distant interaction to a joining event}

For an unscreened cluster with total charge
$Q_{\mathcal C}=\sum_{j\in\mathcal C}q_j$, size $a_{\mathcal C}$,
and a test charge $q$ at distance $r\gg a_{\mathcal C}$, a homogeneous
dielectric with negligible enclosure corrections has the monopole limit
\begin{equation}
 W(r)=\frac{qQ_{\mathcal C}}{4\pi\varepsilon_m r}
       +O(r^{-2}),\qquad
 \vct F(r)=\frac{qQ_{\mathcal C}}{4\pi\varepsilon_m r^2}
                   \widehat{\vct r}+O(r^{-3}).
 \label{eq:s23-cluster-coulomb-tail}
\end{equation}
This inverse-square force is attractive when $qQ_{\mathcal C}<0$.
If the monopole vanishes, higher multipoles set the leading angular
dependence and distance power. Screening instead supplies the factor
$e^{-\kappa r}(1+\kappa r)$ of~\eqref{eq:screened-force}; the Coulomb
approximation requires negligible screening over the relevant separation.
For a screened cluster the bare-total-charge point approximation also
requires $\kappa a_{\mathcal C}\ll1$, in addition to
$a_{\mathcal C}/r\ll1$; otherwise its finite charge distribution supplies
the screened angular form factor. An inverse-square encounter
force therefore describes a particular interaction regime, rather than
the probability that a particle becomes bound.

\subsubsection{A conditional inverse-square absorption channel}
\label{sec:s23-wave-assisted-capture}

Electrodynamics supplies a useful way to organize the exchange question:
specify a source, the field carrying its energy, a receiving degree of
freedom, and the corresponding work account. Acoustic theory has its
own energy flux and source--receiver construction~\cite{Staelin2011}.
In the present setting, a vibrating cluster can couple to a propagating
mode of the surrounding material when that mode and its coupling are
retained. The carrier outside the cluster then belongs to the joint
particle--medium problem. The finite mechanical response and receiver
power in~\eqref{eq:s2-phonon-receiver-power} provide its matrix version.

\begin{definition}[Absorption-assisted capture channel]
\label{def:s23-wave-assisted-capture}
Consider a narrow-band source at angular frequency $\omega>0$ in a
homogeneous three-dimensional propagating-wave regime. Let
$P_{\rm rad}$ be its outward power at a reference sphere of radius
$r_0>0$ in the source's radiation far field, and let $a_{\rm att}\ge0$, in
$\mathrm{m^{-1}}$, be the amplitude-attenuation coefficient. For an
isotropic outgoing field with negligible reflected power in the
observation interval, define
\begin{equation}
 I(r,\omega)=\frac{P_{\rm rad}(\omega)}{4\pi r^2}
                  e^{-2a_{\rm att}(r-r_0)},\qquad r\ge r_0.
 \label{eq:s23-radiated-intensity}
\end{equation}
The receiving virion has absorption cross section
$\sigma_{\rm abs}(\omega)\ge0$, in $\mathrm{m^2}$, for the specified
incident carrier and preparation, so
$P_{\rm abs}=\sigma_{\rm abs}I$ in the weak-loading approximation.
This cross section is a response quantity, not the geometrical area
of the ideal virion~\cite{Silva2014AbsorbingParticle}.
\end{definition}

The inverse square in~\eqref{eq:s23-radiated-intensity} describes
geometrical spreading of power. The outgoing amplitude instead scales
as $r^{-1}$ before material attenuation~\cite{FitzpatrickSphericalWaves}.
Differentiating the power through the sphere gives
$\dd(4\pi r^2 I)/\dd r=-2a_{\rm att}(4\pi r^2 I)$, which verifies
the attenuation factor. For a directive source, an angular gain
$G(\widehat{\vct r},\omega)$ replaces isotropy, normalized by
$\int G\,\dd\Omega=4\pi$.

If the preparation resolves independent absorbed quanta of energy
$\hbar\omega$, let $p_{\rm cap}\in[0,1]$ be the conditional probability
that one such event leads to the declared persistent membership event
within the specified observation time. In a dilute, unsaturated channel
with negligible re-emission and a capture delay short compared with changes
in the source and receiver position, Poisson thinning gives
\begin{equation}
 \lambda_{\rm join}^{\rm wave}(r,\omega)
   =p_{\rm cap}\frac{\sigma_{\rm abs}(\omega)P_{\rm rad}(\omega)}
                        {4\pi\hbar\omega r^2}
                    e^{-2a_{\rm att}(r-r_0)}.
 \label{eq:s23-absorption-assisted-hazard}
\end{equation}
This rate has units $\mathrm{s^{-1}}$. The probability $p_{\rm cap}$
contains the subsequent geometry, binding and energy redistribution.
With those inputs fixed, weak attenuation gives an inverse-square
wave-assisted event rate. A delayed capture step instead uses a
completion-time kernel; under steady preparation the expression can
describe its stationary completion rate.

The event description applies, for example, to a receiving transition
initially in its lower state with a separately specified capture step.
At finite thermal occupation, upward and downward transitions must use
the occupation factors in~\eqref{eq:s2-phonon-exchange-weights}; net
classical absorbed power is not their separate counting rate. In the
classical regime, the same matrix model predicts deposited power and
the induced coordinate response directly. Capture is then computed
from that driven motion or from the mode backaction developed below.
Adding~\eqref{eq:s23-absorption-assisted-hazard} to a diffusive rate
requires genuinely distinct channels, so that the same trajectory is
not counted twice.

The cylinder's propagator includes reflections, guided modes, attenuation
and viscous response. A uniform traveling mode in a guide distributes
power across its cross section. Spherical spreading is the auxiliary
limit before substantial wall returns or in a large bulk-like region,
with a specified propagating carrier. Overdamped disturbances use the
actual causal response. Outward power depletes stored source energy
unless a maintained drive supplies work, so radiation, absorption and
capture share one energy account.

Figure~\ref{fig:s23-wave-spreading-capture} separates geometrical power spreading, material attenuation and the additional capture step.

\begin{figure}[H]
\centering\begingroup
\definecolor{vltTeal}{HTML}{176B64}
\definecolor{vltBlue}{HTML}{315D80}
\definecolor{vltGold}{HTML}{A87520}
\definecolor{vltPlum}{HTML}{725877}
\definecolor{vltRed}{HTML}{762B38}
\definecolor{vltYellow}{HTML}{D5AD3C}
\definecolor{vltNeutral}{HTML}{777D83}
\setlength{\unitlength}{1pt}
\scalebox{0.90}{%
\input{vlt-figures/figure-090.tex}%
}\endgroup
\caption{(a) A compact source and receiver B share the spherical construction; circles are sections of fronts with radii $r_0$, $2r_0$ and $3r_0$. B lies at $2.5r_0$. Icon sizes are schematic; the effective cross section $\sigma_{\rm abs}$ is a response quantity rather than a silhouette area. The guide below has a fixed cross section and is a separate ideal geometry. (b) All curves are normalized at $r_0$; dashed, solid, dash-dotted and dotted styles identify amplitude, spherical intensity, attenuated intensity and uniform guided intensity. The attenuation factor is $e^{-2a_{\rm att}(r-r_0)}$. (c) Units distinguish deposited power from event and joining rates; $p_{\rm cap}$ accounts for subsequent binding. The stated prompt independent-event approximation is auxiliary to the cylinder's full propagator.}
\label{fig:s23-wave-spreading-capture}
\end{figure}

\subsubsection{Diffusive encounters and energy accommodation}

A joining rate can be calculated from a transport and sticking model.
For an auxiliary dilute, unbounded, isotropic encounter with central
potential of mean force $W(r)$, $W(\infty)=0$, constant relative
diffusivity $D_{\rm rel}>0$ and perfect absorption at $r=a$, steady
Smoluchowski transport gives
\begin{equation}
 \partial_t c=\nabla\cdot
   D_{\rm rel}(\nabla c+\beta c\nabla W),\qquad
 k_{\rm diff}=
 \frac{4\pi D_{\rm rel}}
      {\displaystyle\int_a^\infty e^{\beta W(r)}r^{-2}\,\dd r},
 \qquad \beta=(k_{\rm B}T)^{-1}.
 \label{eq:s23-diffusive-capture}
\end{equation}
The coefficient has units $\mathrm{m^3s^{-1}}$; multiplying by the
distant number concentration gives an encounter hazard for one target.
The radial stationary flux is constant, so integrating
$4\pi r^2D_{\rm rel}e^{-\beta W}
\partial_r(e^{\beta W}c)$ from the absorbing surface to infinity
gives the denominator. For $W=-A/r$, $A>0$, this yields
\begin{equation}
 k_{\rm diff}=4\pi D_{\rm rel}
     \frac{\beta A}{1-e^{-\beta A/a}},\qquad
 \lim_{A\to0}k_{\rm diff}=4\pi D_{\rm rel}a.
 \label{eq:s23-coulomb-capture}
\end{equation}
Multiplying this concentration-dependent coefficient by the distant
concentration gives the target hazard. A finite surface reactivity,
with velocity units, represents imperfect sticking. The sealed cylinder
instead uses reflecting enclosure boundaries, depletion and its finite
particle pool; the unbounded coefficient supplies a local comparison
~\cite{SzaboSchultenSchulten1980}.

\paragraph{Encounter and capture inside the cylinder.}
The finite-domain problem can be stated on the actual retained coordinates.
For an isothermal overdamped reduction, let $\vct X$ contain the slow
Cartesian particle coordinates, let $W(\vct X)$ be their consistent potential of
mean force, and let $\mathsf L_{\rm mob}(\vct X)\succ0$ be a symmetric
mobility. The reversible It\^o diffusion has backward generator
\begin{equation}
 \mathcal L f=
  \left[-\mathsf L_{\rm mob}\nabla W
          +k_{\rm B}T\,\nabla\!\cdot\mathsf L_{\rm mob}\right]
               \cdot\nabla f
       +k_{\rm B}T\,\mathsf L_{\rm mob}:\nabla^2f .
 \label{eq:s23-confined-encounter-generator}
\end{equation}
The mobility and thermal drift belong to this specified Markov and
isothermal regime; a medium with resolved memory instead requires its
additional state or a corresponding nonlocal calculation.
Choose a separated return set $\mathcal A$ and a persistent bound set
$\mathcal B$ in configuration space. The capture committor $q$ and mean
first-passage time $\tau$ to $\mathcal B$ solve
\begin{equation}
 \begin{aligned}
 \mathcal L q&=0 \quad\hbox{outside }\mathcal A\cup\mathcal B,
       &q|_{\mathcal A}&=0, &q|_{\mathcal B}&=1,\\
 \mathcal L\tau&=-1 \quad\hbox{outside }\mathcal B,
       &&&\tau|_{\mathcal B}&=0 .
 \end{aligned}
 \label{eq:s23-confined-capture-boundary-problems}
\end{equation}
Each problem uses reflecting conormal conditions
$\vct n\cdot\mathsf L_{\rm mob}\nabla f=0$ on enclosure and excluded
collision boundaries. The normal is taken in the retained configuration
space. A residence-time criterion requires an age variable together with
its drift and reset rules in an extended generator, rather than a
position-only absorbing boundary.
These boundary problems make geometry, charge-dependent drift,
hydrodynamic mobility and the definition of joining explicit
\cite{SzaboSchultenSchulten1980,EVandenEijnden2006}.
Averaging $q$ over a deposition event's prepared states gives its
probability of reaching $\mathcal B$ before $\mathcal A$ under this
dynamics. A finite observation deadline uses the corresponding
time-dependent hitting-probability problem. Retained momentum or modal
preparation likewise requires its extended generator. A mean
first-passage time gives an inverse rate only when the preparation and
waiting-time approximation justify that reduction. The spatial problem
also resolves depletion and competing encounters that an unbounded
single-target coefficient omits.

\paragraph{Orientation-selected capture in the retained state.}
\begin{definition}[Resolved diffusion--reaction encounter]
\label{def:s3patch-resolved-encounter}
In a passive isothermal overdamped realization, let $X$ collect centres,
orientations and retained continuous material coordinates, and let
$\sigma$ label discrete surface occupations. Write
$\dd\nu=J(X)\dd X$ for physical configuration measure, including
rotational Haar measure. At fixed controls, choose the full free energy
$\mathcal F_\sigma(X)$ and
$\pi_\sigma=Z^{-1}e^{-\beta\mathcal F_\sigma}$ relative to this
measure. A reversible diffusion--jump generator is
\begin{equation}
 \begin{aligned}
 (\mathcal L f)_\sigma
  &=\frac{k_{\rm B}T}{J\pi_\sigma}
       \partial_a\!\left(J\pi_\sigma\mu_\sigma^{ab}
                                      \partial_bf_\sigma\right)\\
  &\quad+\sum_{\sigma'\ne\sigma}w_{\sigma'\sigma}(X)
                                   (f_{\sigma'}-f_\sigma),\\
 \pi_\sigma w_{\sigma'\sigma}
  &=\pi_{\sigma'}w_{\sigma\sigma'} .
 \end{aligned}
 \label{eq:s3patch-resolved-encounter-generator}
\end{equation}
Assume finite $Z$, smooth positive-definite mobility $\mu_\sigma$,
nonnegative finite off-diagonal jump rates and a well-posed reflected
process. The mobility maps conjugate forces to rates.
Occupation multiplicities and chemical work are included consistently
in $\mathcal F_\sigma$ or its stated measure. Allowed jumps and
boundary conditions preserve the specified particle and chemical
inventories. Continuous surface fractions instead use their declared
bounded kinetics and physical measure.
\end{definition}

The diffusion term is the backward form of
\eqref{eq:s2patch-reversible-configuration-diffusion}; integration by
parts and the displayed jump balance establish reversibility.
With prescribed reservoir pH, the free energy includes its chemical
potential; a closed proton pool retains the finite reservoir coordinates.
The local quadratic surface approximation is used within its occupancy
range, while a large encounter uses a physical global population law.

The boundary problems~\eqref{eq:s23-confined-capture-boundary-problems}
now act on $(X,\sigma)$. A target bound core specifies admissible
separations, compatible orientations and any required contact or
surface states; the return set specifies separation. A dwell-time
criterion still uses an age variable. Thus reaching a radial distance
and reaching a binding arrangement are distinct events. Rotational
diffusion, torque-driven alignment, residence and surface relaxation
compete during repeated encounters. Rotational--translational reaction
theory already establishes the importance of orientation-selective
reactive boundaries~\cite{ShoupLipariSzabo1981}; the present construction
carries that choice into the same finite VLP, surface and readout model.
A constant geometrical sticking multiplier is used only when derived
or calibrated for the stated regime.

Figure~\ref{fig:s34patch-orientation-capture-target} depicts an orientation-selected target within a larger assembly. The state-space view makes the binding and return sets explicit before their first-passage equations are solved.

\begin{figure}[H]
\centering\begingroup
\setlength{\unitlength}{1pt}
\scalebox{0.86}{%
\input{vlt-figures/figure-290.tex}%
}\endgroup
\caption{Encounter geometry and the target set. Panel (a) is an oblique nine-body view with an incoming ideal VLP beside an eight-member cluster; straight background lines identify the cluster topology. Translation, rotation and surface state all enter the encounter generator~\eqref{eq:s3patch-resolved-encounter-generator}. Panel (b) is a schematic two-coordinate slice: $\mathcal B$ requires $2.2\le r/b\le2.85$ and $|\phi|\le0.47$, whereas $\mathcal A$ is a separated return boundary at $r/b=4.9$. Arrows illustrate alignment followed by entry, or an encounter followed by return; they are not simulated trajectories or fitted rates. The marked states at $r/b=2.6$ differ in target membership. A full calculation uses the complete orientation manifold and surface/contact states, with an age variable when a minimum residence time is required. The boxed result illustrated here is~\eqref{eq:s4patch-stopping-task-closure}.}
\label{fig:s34patch-orientation-capture-target}
\end{figure}

\begin{proposition}[Energy accommodation in capture]
\label{prop:s23-pair-capture-energy}
An isolated two-body system retaining only relative translation in a
time-independent central potential with $V(\infty)=0$ cannot turn an
incoming state with nonnegative relative energy into a permanently bound
negative-energy state without energy transfer to additional degrees of
freedom.
\end{proposition}
\begin{proof}
The relative Hamiltonian
$E_{\rm rel}=p_r^2/(2\mu)+L^2/(2\mu r^2)+V(r)$ is conserved.
The asserted final state has a different value of this conserved energy.
\end{proof}

A many-particle assembly supplies energy-accommodation channels.
Binding energy can excite collective modes, move other particles or
enter the medium. In the unforced closed model, the complete energy is
conserved while the encounter coordinate loses energy. Redistribution
can prolong residence, with thermal return and recollisions still
possible. A phonon-emission description quantizes an identified transfer
under its mode and occupation assumptions. The nonlinear interaction
supplies the capture event; a driven protocol also includes applied work.

Figure~\ref{fig:s23-capture-energy-accommodation} follows the particle identities and energy account through an illustrative capture event.

\begin{figure}[H]
\centering\begingroup
\definecolor{vltTeal}{HTML}{176B64}
\definecolor{vltBlue}{HTML}{315D80}
\definecolor{vltGold}{HTML}{A87520}
\definecolor{vltPlum}{HTML}{725877}
\definecolor{vltRed}{HTML}{762B38}
\definecolor{vltYellow}{HTML}{D5AD3C}
\definecolor{vltNeutral}{HTML}{777D83}
\setlength{\unitlength}{1pt}
\scalebox{0.90}{%
\input{vlt-figures/figure-091.tex}%
}\endgroup
\caption{The same five labeled virions share one camera, body size and position scale before and after the illustrative capture. Gray coils retain the core; gold identifies the incoming member and two new interactions. The incident arrow is relative motion; opposite teal arrows prepare internal velocity with zero net momentum for equal masses. Their gains are independent of position. The energy bars share one axis and zero: the separated unexcited reference has zero potential energy, and $0.2=(-1.0)+1.2$ in both complete accounts. This illustrates accommodation followed by modal storage. The geometry supplies neither an integrated capture trajectory nor fitted equilibrium parameters.}
\label{fig:s23-capture-energy-accommodation}
\end{figure}

\subsubsection{Excitation changes the joining landscape through mode backaction}

Let $\xi$ be a slow joining coordinate, with units of length, and
construct stable fast modes about a conditional minimum at each fixed
$\xi$. This constrains the joining coordinate while relaxing the other
reference variables. The complete configuration need not be stationary
with respect to $\xi$. Frequencies $\omega_s(\xi)>0$ and their
derivatives then follow from the conditional mass and stiffness matrices.

\begin{definition}[Adiabatic mode-conditioned joining potential]
\label{def:s23-mode-conditioned-potential}
In a regime in which the joining motion is slow compared with the
retained mode periods and their actions remain approximately conserved,
the leading effective potential is
\begin{equation}
 W_{\rm ad}(\xi;\{I_s\})=
 W(\xi)+\sum_s I_s\omega_s(\xi),\qquad
 I_s=\frac{E_s}{\omega_s}.
 \label{eq:s23-adiabatic-mode-potential}
\end{equation}
Under the additional quantum occupation description,
$I_s=\hbar(N_s+1/2)$ for fixed occupations. The background $W$ excludes
the mode contribution explicitly added here.
\end{definition}

The action has units of energy times time. Its approximate constancy
requires small $|\dot\omega_s|/\omega_s^2$, weak intermode conversion
and an occupation-redistribution time longer than the joining motion.
Near a mode degeneracy, the coupled subspace and its dynamics replace an
unjustified independent-branch approximation. Inertia and the geometric
couplings of the moving coordinate basis also enter beyond this leading
adiabatic description.

\begin{proposition}[Matrix-computable excitation backaction]
\label{prop:s23-mode-backaction}
For differentiable isolated branches with
$\mathsf K_f\vct\phi_s=\omega_s^2\mathsf M_f\vct\phi_s$ and
$\vct\phi_s^{\mathsf T}\mathsf M_f\vct\phi_s=1$, the mode contribution
to the joining force is
\begin{equation}
 F_{\rm modes}=-\sum_s I_s\,\partial_\xi\omega_s,
 \label{eq:s23-mode-backaction}
\end{equation}
where the frequency slope is
\begin{equation}
 2\omega_s\partial_\xi\omega_s=
 \vct\phi_s^{\mathsf T}
   \bigl(\partial_\xi\mathsf K_f
         -\omega_s^2\partial_\xi\mathsf M_f\bigr)\vct\phi_s.
 \label{eq:s23-hf-reaction-slope}
\end{equation}
\end{proposition}
\begin{proof}
Differentiate the generalized eigenproblem and multiply on the left
by $\vct\phi_s^{\mathsf T}$. Symmetry and the eigenproblem cancel the
terms containing $\partial_\xi\vct\phi_s$; mass normalization gives
the frequency slope. Differentiating the mode energy at fixed actions
gives the force.
\end{proof}

An explicit two-mode construction shows how the sign is determined.
Take $\mathsf M_f=m\Id_2$ and
\begin{equation}
 \mathsf K_f=m
   \begin{pmatrix}\omega_0^2&-g(\xi)\\-g(\xi)&\omega_0^2\end{pmatrix},
 \quad0<g<\omega_0^2,\quad g'(\xi)<0,
 \qquad
 F_{\rm modes}=\frac{g'}2
       \left(\frac{I_-}{\omega_-}-\frac{I_+}{\omega_+}\right),
 \quad\omega_\pm^2=\omega_0^2\pm g.
 \label{eq:s23-softening-branch-recruitment}
\end{equation}
Increasing $\xi$ separates the objects. Approach increases $g$,
softens $\omega_-$ and hardens $\omega_+$. Preparation in the lower
branch therefore gives an inward contribution; preparation in the upper
branch gives an outward contribution. Capture uses the complete
landscape and energy accommodation. Maintaining excitation adds the
pump and its work to that account.

If fast equilibration with an approximately isothermal internal medium
occurs instead, fixed occupations are replaced by a thermal free energy:
$F_{\rm vib}=k_{\rm B}T\sum_s
\log[2\sinh(\beta\hbar\omega_s/2)]$ under the quantized harmonic
description. Its derivative gives the same force with the thermal mean
actions. An interaction already fitted as a potential of mean force
including those vibrations must not receive this contribution twice.
Thus matrix spectra supply a calculable mechanism by which a specified
mode preparation biases association or separation.

Figure~\ref{fig:s23-mode-conditioned-recruitment} connects the two fast displacement patterns to the sign of their force on the joining coordinate.

\begin{figure}[H]
\centering\begingroup
\definecolor{vltTeal}{HTML}{176B64}
\definecolor{vltBlue}{HTML}{315D80}
\definecolor{vltGold}{HTML}{A87520}
\definecolor{vltPlum}{HTML}{725877}
\definecolor{vltRed}{HTML}{762B38}
\definecolor{vltYellow}{HTML}{D5AD3C}
\definecolor{vltNeutral}{HTML}{777D83}
\setlength{\unitlength}{1pt}
\scalebox{0.90}{%
\input{vlt-figures/figure-092.tex}%
}\endgroup
\caption{The conditional two-mode model uses $\mathsf M_f=m\Id_2$ and $\mathsf K_f/m=\left(\begin{smallmatrix}\omega_0^2&-g\\-g&\omega_0^2\end{smallmatrix}\right)$. Pale frames identify the restoring environment; colored transverse arrows show fast displacement amplitudes. Black rulers define increasing separation, and lower arrows show the associated slow force. Dotted guides select $\xi/\ell=1$ in both plots. Tangents and marked values link each frequency slope to its force. Each curve prepares classical action $I>0$ in one branch only. A quantum excess-force reading subtracts its chosen vacuum background. Conditional supports and coupling are inputs; the displayed force excludes the background joining potential.}
\label{fig:s23-mode-conditioned-recruitment}
\end{figure}

\subsubsection{Breakup requires a landscape, a load and a time scale}

A breakup experiment specifies which fragments are sought, how force is
applied and how long it acts. For a declared set of separated endpoints
$\mathcal B$ and admissible configuration paths $\gamma$ beginning at
a bound reference $\vct R$, a mechanical barrier is
\begin{equation}
 \Delta U_{\mathcal C}^{\ddagger}
 =\inf_{\substack{\gamma(0)=\vct R\\\gamma(1)\in\mathcal B}}
    \max_{0\le s\le1}
     [U(\gamma(s))-U(\vct R)].
 \label{eq:s23-rupture-barrier}
\end{equation}
The admissible paths include particle exclusion and the actual supports.
A finite-temperature reduction uses the corresponding free-energy
landscape and dynamical mobility. A graph cut identifies candidate
fragments; it does not determine this barrier. A globally quadratic
spring law has no dissociation threshold, so its stiffness alone cannot
predict breakup into separated members.

\begin{example}[A force-controlled Lennard--Jones pair]
\label{ex:s23-lj-rupture}
For $V/\epsilon=x^{-12}-2x^{-6}$, $x=r/r_m$, a tensile force $F>0$
tilts the energy to $V(r)-Fr$. The stable and unstable stationary
branches merge when $F=V'(r)$ and $V''(r)=0$, giving
\begin{equation}
 x_c=(13/7)^{1/6},\qquad
 F_c=\frac{72}{7}\frac{\epsilon}{r_m}
                    (7/13)^{13/6}
       \simeq2.69\,\frac{\epsilon}{r_m}.
 \label{eq:s23-lj-tensile-threshold}
\end{equation}
This is a deterministic force-controlled fold of the stated isolated
pair law, within its admissible separation range. Electrostatics,
multibody relaxation, support and loading geometry change the result.
\end{example}

Below such a deterministic fold, thermal fluctuations can already cause
separation. For a smooth one-dimensional free energy
$W_F(\xi)=W(\xi)-F\xi$, with a high barrier between minimum $\xi_a$
and saddle $\xi_b$, the overdamped Markov approximation gives
\begin{equation}
 \begin{aligned}
 k_{\rm off}(F)&\simeq
  \frac{\sqrt{W_F''(\xi_a)|W_F''(\xi_b)|}}{2\pi\zeta_\xi}
      e^{-\beta[W_F(\xi_b)-W_F(\xi_a)]},\\
 S(t)&=\exp\!\left[-\int_0^t k_{\rm off}(F(s))\,\dd s\right].
 \end{aligned}
 \label{eq:s23-escape-survival}
\end{equation}
Here $\zeta_\xi$ is the coordinate friction in $\mathrm{N\,s\,m^{-1}}$,
and $S$ is the surviving bound fraction under a quasistatic loading
protocol. The formula assumes intrawell equilibration and a barrier
large compared with $k_{\rm B}T$; memory and rapid forcing require a
corresponding escape calculation\cite{Kramers1940,HanggiTalknerBorkovec1990}.
Force-dependent activation and loading-rate dependence are established
tools of molecular force spectroscopy\cite{DudkoHummerSzabo2006}.
They give a probability of breakup over time rather than a universal
force attached to a virion count.

Figure~\ref{fig:s23-force-controlled-breakup} connects a specified tensile loading geometry to the fold and escape barrier of the Lennard--Jones example.

\begin{figure}[H]
\centering\begingroup
\definecolor{vltTeal}{HTML}{176B64}
\definecolor{vltBlue}{HTML}{315D80}
\definecolor{vltGold}{HTML}{A87520}
\definecolor{vltPlum}{HTML}{725877}
\definecolor{vltRed}{HTML}{762B38}
\definecolor{vltYellow}{HTML}{D5AD3C}
\definecolor{vltNeutral}{HTML}{777D83}
\setlength{\unitlength}{1pt}
\scalebox{0.90}{%
\input{vlt-figures/figure-093.tex}%
}\endgroup
\caption{The held left virion fixes $z=0$; equal reference rulers use $z/r_m$ in both loading views. Body size, position scale, force-arrow gain and seven-turn coil family are shared. The right-hand tensile forces are $f=1.5,2.6$. Solid and dashed stationary branches have positive and negative curvature; filled $x_a$ and open $x_b$ locate the minimum and saddle for $f=1.5$. Their merger is the force-controlled fold. The barrier panel uses the same markers and $W_F=V-Fr$, shifted to zero at $x_a$. All curves use $V/\epsilon=x^{-12}-2x^{-6}$. The barrier sets the activated exponential under the stated regime; supports, multidimensional relaxation and dynamical prefactors enter separately.}
\label{fig:s23-force-controlled-breakup}
\end{figure}

\paragraph{Prepared survival and a multidimensional escape comparison.}
\begin{definition}[Survival of a prepared aggregate]
\label{def:s3patch-prepared-survival}
At fixed controls and separated target $\mathcal B$, set
$s_\sigma(X,t)=\Pr_{X,\sigma}\{\tau_{\mathcal B}>t\}$.
For a normalized preparation $\rho_{0,\sigma}$ outside $\mathcal B$,
\begin{equation}
 \begin{aligned}
 \partial_t s&=\mathcal Ls,\qquad
 s|_{\mathcal B}=0,\qquad s(X,0)=1\quad(X\notin\mathcal B),\\
 S_{\rho_0}(t)&=\sum_\sigma\int s_\sigma(X,t)
                  \rho_{0,\sigma}(X)\,\dd\nu,\qquad
 h_{\rho_0}(t)=-\frac{\dd}{\dd t}\log S_{\rho_0}(t).
 \end{aligned}
 \label{eq:s3patch-prepared-survival}
\end{equation}
The physical reflecting boundaries remain in the generator domain.
\end{definition}
The backward survival equation follows from the Markov property and
stopping at the first target hit. It permits nonexponential residence.
Opposite retained material preparations at identical particle positions
can give different survival and recovery, extending the local test in
Figure~\ref{fig:s2patch-prepared-pair-futures} to nonlinear separation.
Time-dependent controls use the corresponding two-time backward or
forward evolution with its declared reservoir/work account.

\begin{proposition}[An overdamped saddle-point escape limit]
\label{prop:s3patch-langer-escape}
For an auxiliary Euclidean realization with constant physical measure,
constant positive mobility $\mathsf\mu$, a locally equilibrated
nondegenerate minimum $a$ and one isolated index-one saddle $b$,
remove all zero directions. In the high-barrier small-noise regime,
the leading escape rate through that saddle is
\begin{equation}
 k_{a\to b}\simeq\frac{\lambda_+}{2\pi}
    \sqrt{\frac{\det\mathsf H_a}{|\det\mathsf H_b|}}
      e^{-\beta(\mathcal F_b-\mathcal F_a)},\qquad
 \lambda_+\in\sigma(-\mathsf\mu\mathsf H_b),\quad\lambda_+>0 .
 \label{eq:s3patch-langer-escape}
\end{equation}
Here both Hessians use the same coordinate convention and
$\lambda_+$ is the unique unstable deterministic growth rate.
\end{proposition}
This is the overdamped multidimensional saddle-point result of
Langer~\cite{Langer1969}. A Gaussian well population and stable
saddle directions give the determinant ratio; flux along the one
unstable direction supplies $\lambda_+/(2\pi)$. In one dimension,
$\lambda_+=|\mathcal F_b''|/\zeta$ and the formula reduces to
\eqref{eq:s23-escape-survival}. Multiple isolated exit saddles add
their fluxes when the same metastable approximation holds. Rotational
measure, variable mobility, unresolved slow states or unremoved rigid
modes require their proper generator and prefactor treatment. Memory
can change barrier transmission at an unchanged equilibrium landscape,
as established in reaction-rate theory~\cite{GroteHynes1980}.
The full survival problem remains available when local equilibration
does not justify a single rate.

Aggregation and recovery under changed ion availability have been
observed for T4 phage~\cite{SzermerOlearnik2017}. This establishes a
preparation-specific whole-virion comparison, with its own structure
and solution conditions. For the synthetic ideal VLP system, the
corresponding test fixes a separation landscape, orientation law,
ionic controls and material preparation, then predicts survival and
recovery with the same inputs. Rigid intact nodes describe interparticle
dissociation; predicting capsid damage additionally retains deformation
or chemical failure pathways.

\subsubsection{A population generator coupled to sparse mechanics}

During residence in a basin $g$, assemble its reference and tangent
operators from the same nonlinear model. When an encounter changes the
arrangement, continue the full positions and momenta, then recompute the
conditional reference and spectrum. The basis change re-expresses the
same physical state; it is not an unexplained energy source. If actual
internal contact states switch at fixed coordinates, their energy jump
must be supplied by work or exchanged with retained modes and medium.
An observed membership relabelling has no such physical energy jump.

For long-lived basin states, a coarse partition generator provides a
useful additional matrix. Let $p_g$ be state probabilities and let
$\mathsf Q_{gh}\ge0$ for $g\ne h$ be rates from $h$ to $g$, with
$\sum_g\mathsf Q_{gh}=0$. If $n_s(g)$ counts size-$s$ clusters, then
\begin{equation}
 \dot{\vct p}=\mathsf Q\vct p,\qquad
 c_s=\sum_g n_s(g)p_g,\qquad
 \sum_s s\,c_s=N_{\rm tot}.
 \label{eq:s23-partition-master}
\end{equation}
The last identity follows because every allowed partition contains the
same particle pool and the generator conserves probability. The rates
depend on the encounter, sticking and escape mechanisms; cluster size
alone need not determine them. Two clusters of the same size can have
different charge patterns, soft modes and escape routes. A Markov
description requires sufficient intrabasin relaxation relative to
waiting times, or an enlarged state containing the missing history.
Enumerating every partition becomes impractical for a large population;
a useful generator retains an experimentally justified family of basins
or a tested coarse description. The size averages $c_s$ need not obey a
closed equation by themselves. Charge arrangement, shape or mode
preparation can supply the additional variables needed for closure.
Transient-network mechanics provides a biophysical example in which
attachment kinetics and collective relaxation are distinct scales
\cite{Broedersz2010}.

\paragraph{Reversible populations, traps and a finite observation deadline.}
\begin{definition}[Equilibrium basin weight and first-arrival yield]
\label{def:s3patch-basin-yield}
For disjoint basin regions $\Omega_g$ partitioning the complete retained
state, their
equilibrium weights and a consistent reversible Markov reduction obey
\begin{equation}
 \pi_g=\sum_\sigma\int_{\Omega_{g,\sigma}}
                   \pi_\sigma(X)\,\dd\nu,\qquad
 \mathsf Q_{gh}\pi_h=\mathsf Q_{hg}\pi_g .
 \label{eq:s3patch-basin-detailed-balance}
\end{equation}
The convention remains $\mathsf Q_{gh}:h\to g$. For a chosen target
family $B$, let $\mathsf Q^{\rm abs}$ stop paths when they first
reach $B$. At a deadline $T$, define
\begin{equation}
 Y_B^{\rm first}(T)=\vct1_B^{\mathsf T}
                   e^{T\mathsf Q^{\rm abs}}\vct p_0,\qquad
 P_B^{\rm end}(T)=\vct1_B^{\mathsf T}
                   e^{T\mathsf Q}\vct p_0 .
 \label{eq:s3patch-first-arrival-yield}
\end{equation}
\end{definition}
Stopping paths is a counting convention; the physical process still
permits target breakup. First-arrival yield and endpoint population
therefore answer different questions. Basin entropy includes allowed
orientations, surface populations and fluctuations, in addition to
minimum energy. At equilibrium, integrating microscopic reversible
transition fluxes gives the basin balance above; constant Markov rates
additionally require adequate relaxation and a tested state partition.

A kinetic trap is an arrangement with slow escape or correction relative
to the specified deadline and target. Size alone can assign identical
labels to states with different misalignment, contact topology or
retained material strain. The test is whether those states have the
same conditional future records. Capsid-subunit simulations already
show mechanisms for which cluster size fails as a reaction coordinate
\cite{HaganElradJack2011}. Here whole-VLP trapping concerns competing
arrangements of intact bodies. A model of malformed individual HPV
shells instead changes the building blocks to capsomers and supplies
their directional and conformational interactions. The two resolutions
can share matrix methods without sharing uncalibrated parameters.

For implementation, construct a force--torque-balanced basin from the
chosen nonlinear patch law, assemble its mechanical and surface blocks,
and calculate the channel responses and state-resolved first-passage
records. Test a reduced state on reserved preparations or controls.
Adding orientational mismatch or a retained material coordinate is
justified when it resolves a stated predictive failure over the chosen
time and measurement band. This connects the mechanical matrix, the
population generator and the selection of biophysical variables through
calculable comparisons.

Local interaction changes also preserve computational structure. With
$\mathsf E_{ij}\vct u=\vct u_i-\vct u_j$, a changed central-pair
tangent at fixed reference gives
\begin{equation}
 \Delta\mathsf K=\mathsf E_{ij}^{\mathsf T}
                    \Delta\mathsf C_{ij}\mathsf E_{ij},
 \qquad \operatorname{rank}(\Delta\mathsf K)\le3.
 \label{eq:s23-edge-update}
\end{equation}
Sparse assembly and low-rank resolvent updates can therefore track
individual central contacts efficiently; an unstressed central spring
has a rank-one tangent. A resolved patch contact uses its full
translation--rotation blocks and feature Jacobian instead. A site-pair
term touches two body blocks, while elimination of a shared surface
state can mediate couplings across several cells. The sparsity of the
retained model and that of its reduced operator are assessed separately. Reference relaxation, charge changes or altered
inertia contribute their additional updates. Combining this mechanical
matrix with the partition generator makes recruitment experimentally
specific: membership histories constrain residence and transition rates,
while displacement spectra and spatial readouts constrain the evolving
force and storage network. The smallest successful state is the one
that predicts both classes of observation over the declared time range.

The next section adds causal viscoelastic storage, attenuation and
fluctuations using the same particle coordinates and calibrated channels.

\clearpage

\section{Viscoelastic Dynamics: Memory, Attenuation, and Fluctuations}
\label{sec:viscoelastic-dynamics}

Sections~2--3 construct finite ideal-virion assemblies, their collective
motions, and their recruitment and breakup dynamics. This section adds
the surrounding material: particle motion loads the medium, the medium
stores strain and transports momentum, and its reaction changes the next
particle force. Explicit material coordinates resolve the
frequency-dependent response introduced by the internal-coordinate
reduction in Subsection~\ref{sec:structure-internal-reduction}.

The cylinder contains the fixed pool of $N_{\rm tot}$ rigid ideal virions.
A selected cluster has $n_{\mathcal C}(t)$ members under
Definition~\ref{def:s23-cluster-membership}. Within one reference window,
the calculation retains $n$ centres: the complete pool, or a declared
subset coupled to the remaining particles explicitly or by reduction.
Compatible cell and interface maps assemble either a small nucleus or a
large population on the same physical particle labels. The centre-only
reference has $3N_{\rm tot}$ translational coordinates before constraints;
resolved rigid bodies also retain orientation and angular momentum.
Masses and particle identities persist through membership changes.
Fixed assigned charges define the baseline, while regulated surface
populations evolve with their specified chemical inventories.

The extensions below carry the Section~3 force--torque and surface-state
construction into material response. They connect angular drag, stored
interfacial strain, population relaxation and recovery after separation.
The length-valued $\vct u,\mathsf M,\mathsf K$ equations retain their
translational meaning. Coupled body calculations use the scaled
$\vct X,\mathsf G_X$ coordinates of
\eqref{eq:s2patch-generalized-inertia}, with their dual force/torque
and observation maps. A material coordinate records a physical
preparation, independently of the cell or mode basis used to compute it.

A displacement-and-hold experiment makes the material state tangible.
Immediately after loading, both the particle arrangement and the medium
can resist deformation. While the centres remain fixed, internal strain
can relax and the holding force can change. Releasing the particles at
different stages therefore prepares different motions, even from the
same visible geometry. The relaxing coordinates below retain the
information needed to distinguish those preparations.

An equivalent description can express the medium force through the
past particle motion. This memory representation trades explicit
material coordinates for a causal history dependence. At equilibrium,
the eliminated coupling also determines the compatible fluctuating force.
Storage, loss and noise then become linked predictions of one material
model. Their comparison provides a physical test of how much medium
resolution the chosen experiment requires.

Blood provides a concrete medium in which plasma, deformable cells and
ionic degrees of freedom contribute on different scales. The construction
also describes passive inclusions in a polymer solution or hydrated
network once the couplings and observation window are specified.
Reference force balance, conservative curvature and medium resistance
are independently specified inputs.

The first model describes a passive window with approximately fixed
coefficients. Subsequent constructions carry its stored strain through
recruitment and breakup. In the unforced, thermally isolated realization,
particle, medium and unresolved thermal energy share a conserved total
after preparation. An isothermal reduction instead includes exchange with
a fixed-temperature reservoir. These thermal conditions supplement the
vessel's material boundary condition.

The development proceeds from medium resolution to retained state,
causal memory and measured response. Holding a deformation, holding a
force and releasing a preparation probe different parts of that response.
Registered particle identities, branch strain and membership history then
extend the comparison to an evolving assembly.

\subsection{Biophysical mechanisms and the level of resolution}
\label{sec:viscoelastic-biophysical-mechanisms}

\subsubsection{Blood at a specified mechanical resolution}
\label{sec:blood-medium-resolution}

Whole blood is a cellular suspension in a protein-containing ionic liquid.
Red-cell deformation and aggregation contribute to its mechanical
response. Small-amplitude oscillatory measurements resolve storage and
loss over a stated linear frequency window~\cite{Tomaiuolo2016}.
The hematocrit $f_{\mathrm{RBC}}=V_{\mathrm{RBC}}/V_{\mathrm{blood}}$
measures red-cell volume fraction; virion number density and the
occupied-network fraction $f_{\rm lat}$ of Section~3 describe different
populations. A constitutive specification includes temperature, cellular
state, plasma composition, deformation history and observation band.
Appreciable aggregation or deformation during the observation calls for
an evolving reference.

Two spatial averages enter independently. Smoothing the particle network
requires $|\vct k|a\ll1$. Homogenizing blood requires an averaging region
representative of its cellular microstructure and small relative to the
resolved variation. A particle between cells can instead sample local
plasma and nearby interfaces. High-frequency blood microrheology probes
a response attributed primarily to plasma within its measurement
band~\cite{GuzmanSepulveda2022}; single- and two-point microrheology also
distinguish local tracer response from bulk response~\cite{Crocker2000}.
Figure~\ref{fig:blood-resolution-channels} locates these resolutions.
Assigning drag therefore requires matching the constitutive law to the
particle's scale and fluid boundary conditions.

\begin{figure}[H]
\centering\begingroup
\definecolor{vltTeal}{HTML}{176B64}
\definecolor{vltBlue}{HTML}{315D80}
\definecolor{vltGold}{HTML}{A87520}
\definecolor{vltPlum}{HTML}{725877}
\definecolor{vltRed}{HTML}{762B38}
\definecolor{vltYellow}{HTML}{D5AD3C}
\definecolor{vltNeutral}{HTML}{777D83}
\setlength{\unitlength}{1pt}
\scalebox{0.90}{%
\input{vlt-figures/figure-094.tex}%
}\endgroup
\caption{The same four retained virion centres at two medium resolutions. (a) Symbolic gray cells locate the cellular suspension and local plasma. (b) An effective medium averages those surroundings over a specified window. Rigid glyphs, reference outlines and displacement arrows use matching gains in both panels; the open arrow changes resolution. The lower boxes identify mechanical and electrical channels, whose connection requires calibrated work-conjugate couplings and boundary data. Cell sizes, concentration and flow are schematic.}
\label{fig:blood-resolution-channels}
\end{figure}

Table~\ref{tab:blood-medium-inputs} separates reference descriptors from
constitutive functions. A function of probing frequency is a response law,
not an instantaneous state coordinate. The internal variables introduced
below supply the additional state information.
\begin{table}[H]
\centering
\caption{Selected medium descriptors at a declared resolution. Their values
and observation window are inputs, not universal blood constants. Whole-blood
averages and local plasma quantities need not coincide.}
\label{tab:blood-medium-inputs}
\begin{tabular}{p{0.22\textwidth}p{0.56\textwidth}p{0.13\textwidth}}
\toprule
Quantity & Physical role & SI units\\
\midrule
$T$, $f_{\mathrm{RBC}}$ & Thermal state and cellular volume fraction & K, $1$\\
$\rho_{\rm b}$ & Mass density of a separately resolved blood medium & $\mathrm{kg\,m^{-3}}$\\
$G_{\rm b}^*(\omega)$ & Stress response to a specified shear strain & Pa\\
$\eta_{\rm b}^*(\omega)$ & Stress response to the associated strain rate & $\mathrm{Pa\,s}$\\
$\sigma_{\rm dc,b}$ & Explicit ionic conduction in the electrical convention below & $\mathrm{S\,m^{-1}}$\\
$\varepsilon_{\rm pol,b}(\omega)$ & Absolute polarization permittivity, with that conduction excluded & $\mathrm{F\,m^{-1}}$\\
\bottomrule
\end{tabular}
\end{table}

\subsubsection{Mechanical and electrical response have different couplings}
\label{sec:blood-electrical-response}

Blood's electrical response also depends on frequency, temperature and
hematocrit. Ionic conduction and polarization at cell interfaces are
distinct contributions~\cite{Wolf2011}. In a chosen homogeneous, isotropic,
linear electrical approximation, write
\begin{equation}
 \vct D_{\rm b}(\omega)=\varepsilon_{\rm pol,b}(\omega)\vct E(\omega),
 \qquad
 \vct j_{\rm tot}(\omega)
       =[\sigma_{\rm dc,b}-i\omega\varepsilon_{\rm pol,b}(\omega)]
                                                \vct E(\omega).
 \label{eq:blood-electrical-response}
\end{equation}
Here $\vct E$ is electric field in $\mathrm{V\,m^{-1}}$,
$\vct D_{\rm b}$ electric displacement in $\mathrm{C\,m^{-2}}$,
and $\vct j_{\rm tot}$ total current density in $\mathrm{A\,m^{-2}}$.
The explicit DC conductivity is excluded from
$\varepsilon_{\rm pol,b}$ to prevent counting conduction twice.
Equivalently it can be incorporated once as
$\varepsilon_{\rm tot,b}=\varepsilon_{\rm pol,b}
+i\sigma_{\rm dc,b}/\omega$ for $\omega>0$.
For $e^{-i\omega t}$, the mean electrical power per volume is
\begin{equation}
 \overline p_{\rm el}
 =\tfrac12\operatorname{Re}(\vct E^*\cdot\vct j_{\rm tot})
 =\tfrac12[\sigma_{\rm dc,b}
       +\omega\operatorname{Im}\varepsilon_{\rm pol,b}]|\vct E|^2.
 \label{eq:blood-electrical-power}
\end{equation}
A passive total response makes the bracket nonnegative.
The dielectric loss has a positive imaginary sign in this convention,
whereas the mechanical stiffness below has a negative imaginary loss.
The signs differ because current is related to the time derivative
of electric displacement, while mechanical force is related to displacement.

A mechanical electrical coupling specifies the retained charge,
polarization or electrokinetic coordinate, its conjugate work and its
reaction. Cell-membrane Maxwell--Wagner polarization and the mechanical
Maxwell branch below describe distinct mechanisms with independently
specified times. An applied field supplies power; a prepared electrical
imbalance supplies stored energy. Section~\ref{sec:phonon-ionic-extension}
gives an explicit ionic coupling. Mechanical, ionic and electrical wall
conditions complete the corresponding boundary data.

Figure~\ref{fig:mechanical-electrical-loss-signs} reads each complex response through its corresponding real power, making the sign convention explicit.

\begin{figure}[H]
\centering\begingroup
\definecolor{vltTeal}{HTML}{176B64}
\definecolor{vltBlue}{HTML}{315D80}
\definecolor{vltGold}{HTML}{A87520}
\definecolor{vltPlum}{HTML}{725877}
\definecolor{vltRed}{HTML}{762B38}
\definecolor{vltYellow}{HTML}{D5AD3C}
\definecolor{vltNeutral}{HTML}{777D83}
\setlength{\unitlength}{1pt}
\scalebox{0.80}{%
\input{vlt-figures/figure-095.tex}%
}\endgroup
\caption{Power fixes the passive half-plane for the convention $e^{-i\omega t}$. (a) Holding stiffness $k^*=k^\prime-ik^{\prime\prime}$ has mean material power $\omega k^{\prime\prime}|u|^2/2$, giving the shaded lower half-plane. Inertia is separate. (b) Electrical admittance $Y_{\rm el}=\sigma_{\rm dc,b}-i\omega\varepsilon_{\rm pol,b}$ has power density $\operatorname{Re}Y_{\rm el}|E|^2/2$, giving the shaded right half-plane. Each plane has equal component scales. Positive reference units are $k_\star$ in $\mathrm{N\,m^{-1}}$ and $Y_\star$ in $\mathrm{S\,m^{-1}}$. Vectors illustrate sign conventions without assigning a mechanical--electrical coupling.}
\label{fig:mechanical-electrical-loss-signs}
\end{figure}

\subsubsection{A local fluid benchmark and its limitations}
\label{sec:local-fluid-benchmark}

Conservative interaction curvature and resistance to motion enter different
operators. Whole-virion translation can couple to solvent flow and to
neighboring surfaces; envelope--RNP motion can couple to retained internal
deformation and friction. A binding or ionic state can also relax as the
geometry changes, as the explicit free-energy example of
Section~\ref{sec:phonon-ionic-extension} illustrates. These mechanisms
supply different response times and coupling patterns. Structural context
motivates coordinates, while their constitutive coefficients remain
measured, calculated, or declared model inputs.

At the chosen resolution, the medium can retain solvent momentum,
deformable surface layers or internal particle coordinates.
Viscoelasticity combines recoverable deformation with internal
rearrangement over time. The virion centre remains a displacement
coordinate; capsid, envelope and internal structure enter through
calibrated couplings or additional coordinates. Assigning a fitted branch
to a particular process requires evidence for that coupling as well as
agreement of relaxation times.

A tractable benchmark specializes the local fluid to a Newtonian law
within the chosen deformation and frequency range. Plasma can have a
Newtonian shear approximation alongside viscoelastic extensional
response~\cite{Brust2013}, so this is a specified local-fluid limit.
Let $b_v$ be the radius of an isolated rigid no-slip sphere in an
unbounded incompressible Newtonian fluid with viscosity $\eta_f$,
density $\rho_f$ and kinematic viscosity $\nu_f=\eta_f/\rho_f$.
Linear unsteady flow gives~\cite{Grimm2012}
\begin{equation}
 F_h(\omega)=-Z_h(\omega)V(\omega),\qquad
 Z_h(\omega)=6\pi\eta_f b_v
 \left[1+b_v\sqrt{\frac{-i\omega}{\nu_f}}
                       -\frac{i\omega b_v^2}{9\nu_f}\right].
 \label{eq:unsteady-sphere-resistance}
\end{equation}
Choose the causal square-root branch with nonnegative real part.
For $\omega>0$, $\sqrt{-i\omega/\nu_f}=(1-i)\sqrt{\omega/(2\nu_f)}$.
The terms give static Stokes drag, vorticity-diffusion memory, and added
mass $m_a=2\pi\rho_f b_v^3/3$. For bare particle mass $m_p$ and a
specified spring $k_y$,
\begin{equation}
 \begin{aligned}
 \mathcal R_h(\omega)
  &=[k_y-m_p\omega^2-i\omega Z_h(\omega)]^{-1}\\
  &=[k_y-(m_p+m_a)\omega^2
      -i\omega6\pi\eta_f b_v(1+b_v\sqrt{-i\omega/\nu_f})]^{-1}.
 \end{aligned}
 \label{eq:hydrodynamic-added-mass-matching}
\end{equation}
The two representations allocate added mass to either inertia or
resistance, once. The benchmark assumes small particle Reynolds number
and continuum incompressible flow in the stated band. A quasisteady drag
approximation additionally requires $\omega b_v^2/\nu_f\ll1$; small
convective inertia alone leaves unsteady fluid inertia in the balance.

Figure~\ref{fig:viscoelastic-hydrodynamic-crossover} separates the dissipative and reactive terms of this same benchmark.

\begin{figure}[H]
\centering\begingroup
\definecolor{vltTeal}{HTML}{176B64}
\definecolor{vltBlue}{HTML}{315D80}
\definecolor{vltGold}{HTML}{A87520}
\definecolor{vltPlum}{HTML}{725877}
\definecolor{vltRed}{HTML}{762B38}
\definecolor{vltYellow}{HTML}{D5AD3C}
\definecolor{vltNeutral}{HTML}{777D83}
\setlength{\unitlength}{1pt}
\scalebox{0.80}{%
\input{vlt-figures/figure-096.tex}%
}\endgroup
\caption{Components of the unsteady-sphere resistance~\eqref{eq:unsteady-sphere-resistance}, in Stokes units with $q\geq0$. (a) Dissipative resistance $1+\sqrt{q/2}$, compared with the quasisteady value one. (b) Reactive resistance $\sqrt{q/2}+q/9$, separated into dashed vorticity and dotted added-mass terms. The shaded gap is $q/9$, with $m_a=2\pi\rho_fb_v^3/3$. Panels share axis scales. These exact curves use the unbounded spherical continuum benchmark; a confined faceted array requires its own fluid boundary calculation.}
\label{fig:viscoelastic-hydrodynamic-crossover}
\end{figure}

Particle motion disturbs the surrounding fluid, whose momentum takes
time to diffuse. At angular frequency $\omega$, the penetration length
$\delta_f=\sqrt{2\nu_f/\omega}$ gives a correction proportional to
$(1-i)b_v/\delta_f$. A Newtonian solvent can thus produce dynamical
memory through its motion. The reactive resistance describes delayed
momentum exchange and its nonnegative real part describes loss;
the conservative interaction curvature is a separate input.

A defined solvent series offers a conditional way to distinguish these
contributions. If the retained particle geometry and static curvature are
unchanged, independently measured $\eta_f$ and $\rho_f$ predict how the
Stokes resistance and the crossover $\omega b_v^2/\nu_f$ change. Comparing
that prediction with the phase and decay of the same displacement channel
tests the fluid contribution; paired quasistatic compliance checks the
curvature. High-bandwidth Brownian measurements demonstrate this strategy
for calibrated trapped spheres~\cite{Grimm2012}. A comparison involving virions additionally requires matched structural
and buffer states, because a solvent change can also alter surface
interactions.

Hydrodynamic memory arises from solvent momentum transport and can occur
in a Newtonian liquid, as resolved experimentally for trapped
spheres~\cite{Franosch2011}. It differs from material stress relaxation;
its square-root response is not exactly a finite Maxwell sum. For an array in a
cylinder, the resistance becomes a geometry-dependent matrix: fluid flow
can couple particles outside the short-range conservative graph. Surface
clearances, shape, permeability, and electrical or mechanical boundary
conditions can change that matrix. Independent Stokes drags require
adequate separation and their own frequency assumptions. The particle's
specular collision law, the material's power-free displacement boundary,
and a fluid no-slip condition specify different dynamical surfaces.

The internal-variable model below is a tractable passive constitutive
choice. Its branches can resolve selected relaxation mechanisms or fit
a specified observation band. An explicit solvent field or another
microscopic medium realization can instead be retained when its inertia,
geometry, or nonlocal response is essential.

The relation between viscosity and a particle resistance also needs
geometry. For a selected engineering shear strain phasor $e_{\rm sh}$, define
stress $t_{\rm sh}=G_{\rm b}^*e_{\rm sh}
=\eta_{\rm b}^*(-i\omega e_{\rm sh})$, so that
\begin{equation}
 G_{\rm b}^*=-i\omega\eta_{\rm b}^*,\qquad
 \eta_{\rm b}^*=\frac{G_{\rm b}''}{\omega}
                          +i\frac{G_{\rm b}'}{\omega},
 \qquad G_{\rm b}^*=G_{\rm b}'-iG_{\rm b}''.
 \label{eq:blood-modulus-viscosity-convention}
\end{equation}
For an isolated rigid sphere in an effectively unbounded, homogeneous,
isotropic, incompressible linear medium, with no slip and negligible
unsteady fluid inertia, the quasisteady correspondence gives
$Z\simeq6\pi b_v\eta_{\rm b}^*$ and
$-i\omega Z\simeq6\pi b_vG_{\rm b}^*$.
The factors have units of drag resistance and stiffness, respectively.
This correspondence underlies passive microrheological inference under its
continuum assumptions~\cite{MasonWeitz1995}. Nearby cells, interfaces and
many-particle flow can change the matching law. The relaxed particle-network stiffness remains distinct from a
medium shear modulus; a persistent network stiffness does not assert
that fluid blood has a permanent equilibrium shear modulus.

\subsubsection{Matching the fluid force to the finite array}
\label{sec:s4-bio-fluid-array-matching}

A fluid resistance can be calculated from the same particle geometry that
defines the lattice. Let $B_i$ be the rigid body occupied by ideal virion $i$
at its reference centre and orientation, and let
$\mathcal F_0=\mathcal V\setminus\bigcup_i\overline{B_i}$ be the
connected fluid region. For small displacements compared with all surface
clearances, linearize the fluid problem on this reference domain.
The Newtonian benchmark becomes
\begin{equation}
 \begin{aligned}
 \rho_f\partial_t\vct v_f&=-\nabla p_f+\eta_f\nabla^2\vct v_f,
       &\nabla\!\cdot\vct v_f&=0\quad\hbox{in }\mathcal F_0,\\
 \vct v_f|_{\partial B_i}&=\Dt{\vct u_i},
       &\vct v_f|_{\partial\mathcal V}&=0.
 \end{aligned}
 \label{eq:s4-bio-reference-fluid-problem}
\end{equation}
Here $p_f$ is pressure, $\vct v_f$ fluid velocity, and the outer cylinder
is stationary and no slip. The particle boundary condition retains
translation with fixed orientation. This is the usual surface-velocity
matching for a rigid suspension~\cite{GuazzelliMorris2012}.
It applies to the faceted body itself; replacing it by a sphere is a
separate shape approximation. A fitted hydrodynamic radius belongs to
this resistance law, whereas an exclusion radius belongs to the
geometric nonoverlap condition.

Solving this boundary-value problem for harmonic velocities
$\vct V=-i\omega\vct U$ gives the stacked fluid force
\begin{equation}
 \vct F_h=-\mathsf Z_h(\omega)\vct V,
 \qquad
 (\vct F_h)_i=-\sum_{j=1}^{n}\mathsf Z_{h,ij}(\omega)\vct V_j.
 \label{eq:s4-bio-array-resistance}
\end{equation}
Each block is $3\times3$ and has units $\mathrm{kg\,s^{-1}}$.
A moving body sets fluid into motion around other bodies, so a block
$\mathsf Z_{h,ij}$ can be nonzero when the conservative pair block
$\mathsf K_{ij}$ is zero. The cylinder enters through the fluid boundary
problem as well as through any separately declared mechanical support.
Harmonic resistance describes the response after flow transients;
a prepared fluid motion additionally supplies initial-state data.

The sign and admissible matrix structure follow from work. With
$\mathsf E_f=[\nabla\vct v_f+(\nabla\vct v_f)^{\mathsf T}]/2$
now denoting the strain-rate phasor, integration of the fluid momentum
equation against $\vct v_f^*$ gives
\begin{equation}
 -\tfrac12\operatorname{Re}(\vct V^\dagger\vct F_h)
 =\tfrac12\vct V^\dagger\mathsf Z_{h,\mathrm H}\vct V
 =\eta_f\int_{\mathcal F_0}\mathsf E_f^*:\mathsf E_f\,\dd V\geq0,
 \qquad
 \mathsf Z_{h,\mathrm H}=\tfrac12(\mathsf Z_h+\mathsf Z_h^\dagger).
 \label{eq:s4-bio-fluid-resistance-power}
\end{equation}
The mean fluid kinetic-energy derivative vanishes over a cycle, leaving
the viscous loss on the right. Interchanging two solutions at the same
frequency also gives $\mathsf Z_h^{\mathsf T}=\mathsf Z_h$ for this
reciprocal fluid and these fixed boundaries; the quasisteady reciprocity
and dissipation arguments are standard~\cite{GuazzelliMorris2012}. Thus a negative
off-diagonal resistance entry can be compatible with passivity: the
nonnegative quantity is the complete power quadratic form.

Figure~\ref{fig:viscoelastic-collective-medium-drag} resolves that power into common and relative particle motions in a two-coordinate resistance example.

\begin{figure}[H]
\centering\begingroup
\definecolor{vltTeal}{HTML}{176B64}
\definecolor{vltBlue}{HTML}{315D80}
\definecolor{vltGold}{HTML}{A87520}
\definecolor{vltPlum}{HTML}{725877}
\definecolor{vltRed}{HTML}{762B38}
\definecolor{vltYellow}{HTML}{D5AD3C}
\definecolor{vltNeutral}{HTML}{777D83}
\setlength{\unitlength}{1pt}
\scalebox{0.90}{%
\input{vlt-figures/figure-097.tex}%
}\endgroup
\caption{Eighteen rigid ideal virions illustrate a collective resistance: two nine-site layers have reference centres $\vct R/a\in\{-1,0,1\}^2\times\{-1/2,1/2\}$. Both views share an orthographic camera, body radius $0.13a$ and position scale $49\,\mathrm{pt}/a$. Teal and blue identify layers. Offset black arrows show $x$ velocities at $15\,\mathrm{pt}/v_0$; the geometry is a fixed reference, not a displaced snapshot. The assigned reciprocal resistance acts on the eighteen $x$ components, with $c>0$ in $\mathrm{kg\,s^{-1}}$. Its common and opposed-layer eigenvectors give per-particle forces $-cv_i/2$ and $-3cv_i/2$. Bars show instantaneous loss, normalized by $18cv_0^2$; all resistance eigenvalues are positive. This auxiliary matrix demonstrates shared-medium coupling for a larger population. A resistance calculated for this geometry would additionally require the fluid boundary problem in~\eqref{eq:s4-bio-array-resistance}.}
\label{fig:viscoelastic-collective-medium-drag}
\end{figure}

The separation of length scales makes this matching testable. Besides the
single-body scale $b_v^2/\nu_f$, a particle separation $d$ introduces
$d^2/\nu_f$, and the distance $L_f$ over which a flow reaches the wall
introduces $L_f^2/\nu_f$. A quasisteady many-body approximation requires
small $\omega\ell^2/\nu_f$ on the fluid lengths $\ell$ it resolves.
Consequently $\omega b_v^2/\nu_f\ll1$ alone does not make an entire
confined array quasisteady. Even at zero frequency, small surface gaps
alter the resistance through constrained flow. Changes in decay or
phase can therefore report confinement or shared-fluid transport while
the matched static interaction curvature is unchanged. The two-probe
example in Section~\ref{sec:phonon-two-probe-application} develops
one such distinction through lagged displacement correlations.

\paragraph{Angular boundary motion and fluid torque.}
The same fluid calculation extends to the whole rigid-body motion.
At a resting reference, replace the particle boundary velocity by
\begin{equation}
 \vct v_f(\vct x)|_{\partial B_i}
   =\vct V_i+\vct\Omega_i\times(\vct x-\vct R_i),\qquad
 \vct\nu=(\vct V_1,\ldots,\vct V_n,
                    \vct\Omega_1,\ldots,\vct\Omega_n).
 \label{eq:s4patch-rigid-fluid-boundary}
\end{equation}
Fluid traction supplies both centre force and intrinsic torque. Stack
them in the work-dual column $\vct g_h=(\vct F_h,\vct T_h)$.
For the reciprocal reference fluid problem, the resulting resistance
has translation, rotation and cross blocks~\cite{KimKarrila1991}:
\begin{equation}
 \boxed{\begin{aligned}
 \vct g_h&=-\mathsf Z_{\rm rb}(\omega)\vct\nu,
       &\mathsf Z_{\rm rb}^{\mathsf T}&=\mathsf Z_{\rm rb},\\
 -\tfrac12\operatorname{Re}(\vct\nu^\dagger\vct g_h)
  &=\tfrac12\vct\nu^\dagger
        \operatorname{Re}_{\rm H}\mathsf Z_{\rm rb}\vct\nu
    =\eta_f\int_{\mathcal F_0}\mathsf E_f^*:\mathsf E_f\,\dd V
    \geq0 .
 \end{aligned}}
 \label{eq:s4patch-rigid-fluid-power}
\end{equation}
The same integration by parts as in
\eqref{eq:s4-bio-fluid-resistance-power} now includes
$\sum_i\vct\Omega_i\cdot\vct T_{h,i}$ in the boundary power.
The raw translation--translation, cross and rotation--rotation blocks
have units $\mathrm{N\,s\,m^{-1}}$, $\mathrm{N\,s}$ and
$\mathrm{N\,m\,s}$, respectively. For
$(\vct u,\vct\theta)=\mathsf S\vct X$, the congruence
$\mathsf Z_X=\mathsf S^{\mathsf T}\mathsf Z_{\rm rb}\mathsf S$
gives energy-times-time units in every entry. The quasisteady isolated
no-slip spherical reference has translational resistance
$6\pi\eta_f b_v$ and rotational resistance $8\pi\eta_f b_v^3$;
the faceted, confined array uses its actual boundary solution.
Prepared fluid motion also supplies a force/torque source. Angular
relaxation and a translation--rotation cross response can therefore
test a mode whose centre displacement alone is small. The fixed-domain
calculation remains local in its displacement and angular excursions.

Figure~\ref{fig:s34patch-fluid-rigid-ports} pairs an oblique array with its axial cross-section and identifies the force--torque ports used in the fluid power balance.

\begin{figure}[H]
\centering\begingroup
\setlength{\unitlength}{1pt}
\scalebox{0.94}{%
\input{vlt-figures/figure-291.tex}%
}\endgroup
\caption{Fluid coupling acts on full rigid-body motion. The oblique view shows twelve VLPs at three axial stations, with particles above, below and behind one another; panel (b) looks down the cylinder at the middle station. Circumradius is $0.53\ell_\star$, cylinder radius $2.7\ell_\star$, and centres are $(x,\pm1.2,\pm1.2)\ell_\star$ for $x=-2.8,0,2.8$, leaving positive wall and interparticle clearances. Velocity arrows are boundary inputs, not computed fluid streamlines or force vectors. The lower map includes translation--rotation cross blocks and pairs the output force and torque with their conjugate velocities. Thus the dissipated power in~\eqref{eq:s4patch-rigid-fluid-power} includes angular work as well as centre work. Body colours identify selected particles consistently across the two views.}
\label{fig:s34patch-fluid-rigid-ports}
\end{figure}

\subsubsection{Retaining a moving material component}
\label{sec:s4-bio-two-component-medium}

A hydrated polymer network supplies a concrete extension when the
surroundings themselves carry elastic displacement. Denote that material
displacement by $\vct w_m(\vct r,t)$ and the interpenetrating solvent
velocity by $\vct v_f$. The field $\vct w_m$ describes the surrounding
material, separately from the ideal-virion lattice displacement $\vct u$.
An isotropic passive coupling with coefficient $\Gamma_m>0$ assigns
equal and opposite force densities
\begin{equation}
 \vct f_{m\leftarrow f}=-\Gamma_m(\partial_t\vct w_m-\vct v_f),
 \qquad
 \vct f_{f\leftarrow m}=-\vct f_{m\leftarrow f},
 \qquad
 d_{\rm rel}=\Gamma_m|\partial_t\vct w_m-\vct v_f|^2.
 \label{eq:s4-bio-relative-medium-drag}
\end{equation}
Here $\Gamma_m$ has units $\mathrm{kg\,m^{-3}\,s^{-1}}$ and
$d_{\rm rel}$ is power lost per volume. The two force powers sum to
$-d_{\rm rel}$: common motion costs no intercomponent drag, while
relative motion dissipates. This is the force balance used in a
two-fluid description of a network permeated by solvent~\cite{LevineLubensky2001}.
Particle--solvent and particle--network surface matching then specify
which combination a virion motion excites. A no-slip solvent surface does
not by itself specify attachment to the material network.

One useful consequence follows without assigning a molecular process to
a fitted time. Take a dilute-volume-fraction, homogeneous elastic
material network with Lam\'e coefficients $\lambda_m,\mu_m$, neglect
its inertia, and use an incompressible solvent. For a bulk longitudinal
Fourier component at nonzero $\vct k$, incompressibility sets the
longitudinal solvent velocity to zero in this dilute approximation.
With $M_m=\lambda_m+2\mu_m>0$, longitudinal force balance gives
\begin{equation}
 \Gamma_m\partial_t w_{m,L}(\vct k,t)
       =-M_m|\vct k|^2w_{m,L}(\vct k,t),
 \qquad
 \tau_{\rm drain}(\vct k)=\frac{\Gamma_m}{M_m|\vct k|^2}.
 \label{eq:s4-bio-drainage-scale}
\end{equation}
The relaxation is elastic compression relieved by relative solvent
transport. A length of order $1/|\vct k|$ therefore produces a time
proportional to its square; boundary conditions select the permitted
spatial patterns in a finite specimen. This predicts a geometry-dependent
relaxation even when the microscopic friction and elastic moduli stay
fixed. It provides a physical distinction between drainage and a local
stress branch whose declared relaxation time is independent of wave
number. The two-fluid sphere calculation likewise finds a frequency
window for reduction to the generalized Stokes--Einstein relation,
set by material and probe properties~\cite{LevineLubensky2001}.

For passive biophysical interpretation, these constructions establish
variables by the work they carry: particle centres locate the retained
inclusions; fluid velocity carries momentum; a material displacement
carries elastic strain; relative velocity determines dissipative exchange.
Bulk moduli and resistance matrices are response laws connecting these
variables to forces. Eliminating the fields gives the history force
developed below, while retaining them gives a coupled initial-boundary
value problem. A finite set of internal variables is a further reduction
that should reproduce the force channels and band selected for the
particle readout.

\subsection{From the finite lattice to a viscoelastic equation of motion}
\label{sec:viscoelastic-force-balance}

Fix the reference arrangement and $\lambda$, and let $\mathsf K$ be its
relaxed conservative Hessian over the permitted internal directions at
fixed conserved inventory. The positive-rate branch models use independent
relaxing states; the conserved-sector construction below retains frozen
inventories explicitly. Any static medium contribution already included in
the effective potential belongs to this same $\mathsf K$. The new
branches supply the delayed response relative to that reference.

A branch coordinate retains the part of material preparation needed to
predict force from the present state. Add branches when their response
can be constrained within the observation band. Their number describes
the retained relaxation structure, independently of the numbers of
virions or blood constituents.

A useful finite model assigns a real internal variable $\vct z_a$ to each
relaxation branch $a=1,\ldots,r_{\rm R}$. The branch count
$r_{\rm R}$ is independent of the positive mechanical mode count $r$. A dimensionless real matrix $\mathsf G_a$
extracts the relevant displacement combinations. For an edge branch it can
extract $\vct u_i-\vct u_j$; more general maps describe selected collective or
internal deformations. The equations are
\begin{equation}
 \begin{split}
 \mathsf M\Dtt{\vct u}+\mathsf C_0\Dt{\vct u}
      +\mathsf K\vct u+\sum_a\mathsf G_a^{\mathsf T}
                                      \mathsf H_a\vct z_a&=\vct f,\\
 \Dt{\vct z_a}+\tau_a^{-1}\vct z_a
                                      &=\mathsf G_a\Dt{\vct u}.
 \end{split}
 \label{eq:finite-maxwell-system}
\end{equation}
Here $\tau_a>0$ is a relaxation time,
$\mathsf H_a=\mathsf H_a^{\mathsf T}\geq0$ is a branch stiffness, and
$\mathsf C_0=\mathsf C_0^{\mathsf T}\geq0$ is an instantaneous drag matrix.
The variables $\vct u$ and $\vct z_a$ have units of length,
$\mathsf H_a$ has units of force per length, and $\mathsf C_0$ has units of
mass per time. Unlike the force-valued scalar $z$ in
\eqref{eq:architecture-scalar-memory-realization}, $\vct z_a$ here is a
displacement-like branch extension; the force is
$\mathsf G_a^{\mathsf T}\mathsf H_a\vct z_a$. This is a generalized Maxwell construction: each branch stores
elastic deformation while its stress relaxes. A relaxed spring in parallel
with one such branch gives the standard linear solid~\cite{Roylance2001}.
Edge-local branches retain the shell organization and sparse energy
assembly. Long-range conservative or shared-medium couplings retain
their actual cross-cell blocks.

For one scalar deformation, put
$z=u-y$, where $y$ is the branch's relaxed displacement. Its spring and
dashpot are in series: both carry the branch force $k_1z$, while the
dashpot changes $y$. With dashpot resistance $\eta_1=k_1\tau$, the equations
become, with $m>0$, $c_0\geq0$, $k_0\geq0$, and $k_1>0$,
\begin{equation}
 \boxed{\begin{aligned}
    m\frac{\dd^2u}{\dd t^2}+c_0\frac{\dd u}{\dd t}
                     +k_0u+k_1(u-y)&=f,\\
    \eta_1\frac{\dd y}{\dd t}=k_1(u-y),\qquad
                         \eta_1&=k_1\tau.
 \end{aligned}}
 \label{eq:scalar-maxwell-physical-branch}
\end{equation}
The dashpot resistance $\eta_1$ has units of force times time per length;
the fluid viscosity $\eta_f$ has different units. The internal coordinate
$y$ describes relaxation of omitted material. At fixed $u$, it approaches
$u$, leaving the relaxed force $k_0u$. During rapid motion it changes
little, so the branch supplies additional elastic resistance.
In the array, $\mathsf G_a$ selects the deformation and $\mathsf H_a$
sets its branch force and stored energy.

Figure~\ref{fig:scalar-maxwell-ledger} draws the branch in~\eqref{eq:scalar-maxwell-physical-branch} and identifies which elements store energy and which transfer it to the omitted medium.

\begin{figure}[H]
\centering\begingroup
\definecolor{vltTeal}{HTML}{176B64}
\definecolor{vltBlue}{HTML}{315D80}
\definecolor{vltGold}{HTML}{A87520}
\definecolor{vltPlum}{HTML}{725877}
\definecolor{vltRed}{HTML}{762B38}
\definecolor{vltYellow}{HTML}{D5AD3C}
\definecolor{vltNeutral}{HTML}{777D83}
\setlength{\unitlength}{1pt}
\scalebox{0.90}{%
\input{vlt-figures/figure-098.tex}%
}\endgroup
\caption{One scalar standard linear solid, shown as a topology and a prepared state. (a) The whole-virion coordinate is $u$; the internal dot is $y$, with $z=u-y$ and series-branch resistance $\eta_1=k_1\tau$. The three parallel paths identify constitutive elements; their drawn lengths carry no displacement scale. The ledger assigns storage and loss to their elements. (b) Holding $u=u_0>0$ gives $y/u_0=1-e^{-t/\tau}$ and $z/u_0=e^{-t/\tau}$. Equal body sizes, natural spring span and six-turn coils share one displacement gain; excess coil span follows $z$. The anchor is a declared support and coefficients stay fixed. The boxed result illustrated here is~\eqref{eq:scalar-maxwell-physical-branch}.}
\label{fig:scalar-maxwell-ledger}
\end{figure}

The finite state is
$(\vct u,\Dt{\vct u},\vct z_1,\ldots,\vct z_{r_{\rm R}})$,
with prescribed forcing and fixed coefficients. Integrating a
concentration, charge or gating variable requires its own coordinate map,
units and conjugate work. More generally, passive internal relaxation
can follow a declared energy or free-energy gradient with nonnegative
mobility. The spring--dashpot law is one explicit realization.

\paragraph{Deriving branches from relaxing biophysical variables.}
An internal variable can be introduced by specifying the quantity that
relaxes, its energetic coupling to displacement and its kinetic law.
This gives a concrete route from a physical description to the Maxwell
coefficients. Let $\vct\chi$ contain small, dimensionless deviations of
selected medium variables, each divided by a declared physical scale.
Conservation constraints are imposed before choosing these independent
coordinates. A component can represent a polarization coordinate or a
population deviation of a structural state; the fixed particle count $n$
and assigned baseline charges $q_i$ retain their definitions.
At a stationary isothermal reference, take the quadratic free energy
\begin{equation}
 \begin{split}
 \mathcal F^{(2)}(\vct u,\vct\chi)
 &=\tfrac12\vct u^{\mathsf T}\mathsf K_{\rm cl}\vct u
    -\vct u^{\mathsf T}\mathsf B_\chi\vct\chi
    +\tfrac12\vct\chi^{\mathsf T}\mathsf A_\chi\vct\chi,\\
 \mathsf K&=\mathsf K_{\rm cl}
           -\mathsf B_\chi\mathsf A_\chi^{-1}\mathsf B_\chi^{\mathsf T}
           \geq0,\qquad \mathsf A_\chi>0.
 \end{split}
 \label{eq:s4-root-internal-free-energy}
\end{equation}
Here $\mathsf A_\chi$ has units of energy, $\mathsf B_\chi$ units of
force and $\mathsf K_{\rm cl}$ units of stiffness. The mixed derivative
$-\mathsf B_\chi$ states how changing an internal variable changes
mechanical force. Minimizing over $\vct\chi$ gives the relaxed
$\mathsf K$ already used in the lattice; holding $\vct\chi$ fixed gives
the clamped curvature $\mathsf K_{\rm cl}$. This is the static Schur
reduction of the same joint free energy, rather than a second static
contribution appended to the pair Hessian.

For time-even internal variables with reciprocal dissipative kinetics,
choose the linear Onsager law~\cite{Onsager1931,Doi2011}
\begin{equation}
 \dot{\vct\chi}=-\mathsf L_\chi\vct g_\chi,\qquad
 \vct g_\chi=\mathsf A_\chi\vct\chi
                         -\mathsf B_\chi^{\mathsf T}\vct u,
 \qquad \mathsf L_\chi=\mathsf L_\chi^{\mathsf T}>0.
 \label{eq:s4-root-internal-kinetics}
\end{equation}
The mobility $\mathsf L_\chi$ has units $(\mathrm{J\,s})^{-1}$.
Its symmetry is an explicit reciprocal-model assumption; positivity
ensures downhill free-energy relaxation. With
$\vct f_{\rm hold}=\partial_{\vct u}\mathcal F^{(2)}$, differentiation gives
$\dot{\mathcal F}^{(2)}=\dot{\vct u}^{\mathsf T}\vct f_{\rm hold}
-\vct g_\chi^{\mathsf T}\mathsf L_\chi\vct g_\chi$.
This is an isothermal available-work balance. The thermodynamic
distinction from a thermal internal-energy ledger remains the one used
in Proposition~\ref{prop:maxwell-energy} below.

For harmonic motion after preparation transients, eliminating
$\vct\chi$ produces the mechanical stiffness
\begin{equation}
 \mathsf K_\chi^*(\omega)=\mathsf K_{\rm cl}
 -\mathsf B_\chi
   (\mathsf A_\chi-i\omega\mathsf L_\chi^{-1})^{-1}
                          \mathsf B_\chi^{\mathsf T}.
 \label{eq:s4-root-internal-dynamic-stiffness}
\end{equation}
The relaxation times and branch weights can now be derived. Diagonalize
$\mathsf A_\chi^{1/2}\mathsf L_\chi\mathsf A_\chi^{1/2}
=\mathsf O\,\diag(\tau_j^{-1})\mathsf O^{\mathsf T}$,
where $\mathsf O$ is orthogonal, and let $\vct o_j$ denote its columns.
Putting $\vct b_j=\mathsf B_\chi\mathsf A_\chi^{-1/2}\vct o_j$ gives
\begin{equation}
 \mathsf K_\chi^*(\omega)=\mathsf K+
       \sum_j\mathsf K_j
           \frac{-i\omega\tau_j}{1-i\omega\tau_j},
 \qquad \mathsf K_j=\vct b_j\vct b_j^{\mathsf T}\geq0.
 \label{eq:s4-root-derived-relaxation-spectrum}
\end{equation}
The result follows by factoring the inverse in
\eqref{eq:s4-root-internal-dynamic-stiffness} through
$\mathsf A_\chi^{-1/2}$ and using the stated eigendecomposition.
It has exactly the positive Maxwell form, without requiring the
mobility and free-energy curvature to commute. The frozen limit is
$\mathsf K_{\rm cl}=\mathsf K+\sum_j\mathsf K_j$.
Each $\vct b_j$ has units of the square root of stiffness.
To express its branch in the dimensionless-map convention above, choose
a reference stiffness $h_*>0$ and set
$\mathsf G_j=\vct b_j^{\mathsf T}/\sqrt{h_*}$ and $\mathsf H_j=h_*$;
then $\mathsf G_j^{\mathsf T}\mathsf H_j\mathsf G_j=\mathsf K_j$.
For repeated relaxation times, the sum of the corresponding
$\mathsf K_j$ is invariant under rotations of the relaxation eigenbasis;
individual rank-one terms depend on that basis choice.

The construction separates internal susceptibility
$\mathsf A_\chi^{-1}$, kinetics $\mathsf L_\chi$ and mechanical coupling
$\mathsf B_\chi$. A relaxation with $\vct b_j=0$ evolves without loading
the mechanical coordinates. A visible branch can combine several
internal coordinates through $\vct o_j$; identifying its physical
origin therefore benefits from a matched internal-state readout.
Section~\ref{sec:phonon-ionic-extension} uses this coupled-variable
principle. The calculation above derives the contribution of its linear
overdamped sector to particle mechanics.

\paragraph{Reciprocal coupling of mechanical and internal relaxation rates.}
\begin{definition}[Joint resistance at a resolved-body reference]
\label{def:s4patch-joint-resistance}
Use dimensionless mechanical coordinates $\vct X$ and independent
dimensionless, time-even surface/material coordinates $\vct\chi$.
At a resting reference, prescribe
\begin{equation}
 \begin{aligned}
 \mathcal F_2&=\tfrac12\vct X^{\mathsf T}\mathsf H\vct X
       +\vct X^{\mathsf T}\mathsf B\vct\chi
       +\tfrac12\vct\chi^{\mathsf T}\mathsf A\vct\chi,\\
 \mathsf H-\mathsf B\mathsf A^{-1}\mathsf B^{\mathsf T}&\succeq0,
      \qquad \mathsf A\succ0,\qquad
 \mathsf R_{\rm d}=\begin{pmatrix}
       \mathsf C&\mathsf R\\\mathsf R^{\mathsf T}&\mathsf D
                     \end{pmatrix}\succeq0,\quad \mathsf D\succ0 .
 \end{aligned}
 \label{eq:s4patch-joint-storage-resistance}
\end{equation}
The resistance acts on $(\dot{\vct X},\dot{\vct\chi})$; all its
entries have units $\mathrm{J\,s}$. The energy blocks have units
$\mathrm J$. The plus cross-energy convention here corresponds to
$\mathsf B=-\mathsf S^{\mathsf T}\mathsf B_\chi$ for the scaled
translational version of~\eqref{eq:s4-root-internal-free-energy}.
\end{definition}

The reciprocal constitutive equations are
\begin{equation}
 \begin{aligned}
 \mathsf G_X\ddot{\vct X}+\partial_X\mathcal F_2
       +\mathsf C\dot{\vct X}+\mathsf R\dot{\vct\chi}
       &=\vct f_X,\\
 \partial_\chi\mathcal F_2+\mathsf R^{\mathsf T}\dot{\vct X}
                         +\mathsf D\dot{\vct\chi}&=0 .
 \end{aligned}
 \label{eq:s4patch-cross-rayleigh-dynamics}
\end{equation}
Differentiating kinetic-plus-free-energy storage gives supplied power
$\dot{\vct X}^{\mathsf T}\vct f_X$ minus
$(\dot{\vct X},\dot{\vct\chi})^{\mathsf T}
\mathsf R_{\rm d}(\dot{\vct X},\dot{\vct\chi})$.
This is a reciprocal Rayleigh/Onsager construction for the stated
time-even coordinates~\cite{Onsager1931,Doi2011}. The cross block
permits motion and interfacial rearrangement to share a dissipative
channel. It is a constitutive input, not a consequence of net charge.

\begin{proposition}[Cross resistance preserves a matched thermal reduction]
\label{prop:s4patch-cross-resistance-reduction}
In this fixed linear window, harmonic elimination gives
\begin{equation}
 \begin{aligned}
 \mathsf Z_{\rm m}(\omega)
   &=\mathsf H-\omega^2\mathsf G_X-i\omega\mathsf C\\
   &\quad-(\mathsf B-i\omega\mathsf R)
        (\mathsf A-i\omega\mathsf D)^{-1}
        (\mathsf B^{\mathsf T}-i\omega\mathsf R^{\mathsf T}).
 \end{aligned}
 \label{eq:s4patch-cross-dissipative-schur}
\end{equation}
At common-temperature classical equilibrium, a matched jointly Gaussian
white-noise realization has primitive generalized-force spectrum
\begin{equation}
 \boxed{\mathsf S_\xi=2k_{\rm B}T\mathsf R_{\rm d},\qquad
 \mathsf S_{\xi,\rm eff}
    =\frac{2k_{\rm B}T}{\omega}
                   [-\operatorname{Im}_{\rm H}\mathsf Z_{\rm m}],
 \quad\omega>0 .}
 \label{eq:s4patch-cross-resistance-noise}
\end{equation}
The second identity uses the full correlated force map of
Proposition~\ref{prop:s4-cell-port-fdt}. A normalizable equilibrium
sector and its matched preparation are assumed.
\end{proposition}
\begin{proof}
Eliminate the internal row of~\eqref{eq:s4patch-cross-rayleigh-dynamics}.
The augmented pencil has negative Hermitian imaginary part
$\omega\mathsf R_{\rm d}$. Applying the same Schur row map to its
primitive noise gives the displayed loss/noise identity.
\end{proof}

A constant work-preserving change of internal coordinates makes the
relation to the previous Maxwell realization explicit. Set
$\mathsf S_\chi=\mathsf D^{-1}\mathsf R^{\mathsf T}$ and
$\vct\chi'=\vct\chi+\mathsf S_\chi\vct X$. Then
\begin{equation}
 \begin{aligned}
 \mathsf C_{\rm eff}&=\mathsf C-\mathsf R\mathsf D^{-1}\mathsf R^{\mathsf T}
                         \succeq0,\\
 \mathsf B'&=\mathsf B-\mathsf S_\chi^{\mathsf T}\mathsf A,\\
 \mathsf H'&=\mathsf H-\mathsf B\mathsf S_\chi
       -\mathsf S_\chi^{\mathsf T}\mathsf B^{\mathsf T}
       +\mathsf S_\chi^{\mathsf T}\mathsf A\mathsf S_\chi,\\
 \mathsf H'-\mathsf B'\mathsf A^{-1}\mathsf B'^{\mathsf T}
       &=\mathsf H-\mathsf B\mathsf A^{-1}\mathsf B^{\mathsf T}.
 \end{aligned}
 \label{eq:s4patch-resistance-whitening}
\end{equation}
Substitution completes the resistance square and gives these energy
congruences. The transformed body force is
$\vct\xi'_X=\vct\xi_X-\mathsf R\mathsf D^{-1}\vct\xi_\chi$;
its intensity is $2k_{\rm B}T\mathsf C_{\rm eff}$ and its cross
covariance with $\vct\xi_\chi$ is zero. Independent primitive noises
would instead change the original model. The formal rapid-drive limit
locks $\vct\chi'$, so $\vct\chi\simeq-\mathsf S_\chi\vct X$ and
the frozen storage is $\mathsf H'$. This limit is used within a band
where neglected microscopic inertia and kinetics remain negligible.
At slow drive the added resistance is
$\mathsf B'\mathsf A^{-1}\mathsf D\mathsf A^{-1}\mathsf B'^{\mathsf T}$.
Mechanical records alone can admit the two equivalent representations;
a calibrated surface channel also transforms through
$\vct\chi=\vct\chi'-\mathsf S_\chi\vct X$.

Figure~\ref{fig:s34patch-cross-resistance-noise} evaluates the response--noise identity with a nonzero reciprocal cross resistance. The same reduction must act on the correlated primitive force sources.

\begin{figure}[H]
\centering\begingroup
\setlength{\unitlength}{1pt}
\scalebox{0.86}{%
\input{vlt-figures/figure-292.tex}%
}\endgroup
\caption{A reciprocal cross resistance carries correlated thermal forcing. In units $E_\star$, $\tau_\star$, the scalar construction uses $H=3$, $G=0.2$, $A=2$, $B=-1$, $C=D=1$ and $R=0.6$; the full storage and resistance matrices are positive definite. Panel (b) evaluates the reduced force spectrum with the row map $[1,t_\chi]$. The green curve is the matched spectrum; open circles independently evaluate $-\operatorname{Im}Z_{\rm m}/\omega$. The burgundy dashed comparison holds the response and marginal primitive noises fixed but sets their cross spectrum to zero. Its discrepancy shows the contribution of the shared primitive noise. The plotted force is in scaled mechanical coordinates, and the spectrum uses the classical two-sided convention of~\eqref{eq:s4patch-cross-resistance-noise}. These assigned coefficients demonstrate a reduction identity, not measured virion noise.}
\label{fig:s34patch-cross-resistance-noise}
\end{figure}

\paragraph{Conserved populations select the accessible relaxed curvature.}
For zero cross resistance, allow a semidefinite hidden mobility
$\mathsf L\succeq0$ in
$\dot{\vct\chi}=-\mathsf L(\mathsf A\vct\chi+
\mathsf B^{\mathsf T}\vct X)$. Define
\begin{equation}
 \mathsf Q_\chi=\mathsf A^{1/2}\mathsf L\mathsf A^{1/2},\qquad
 \mathsf V_\chi=\mathsf B\mathsf A^{-1/2},\qquad
 \mathsf P_0=\mathbf1_{\{0\}}(\mathsf Q_\chi),\quad
 \mathsf P_R=\Id-\mathsf P_0 .
 \label{eq:s4patch-conserved-relaxation-projectors}
\end{equation}
The inventory $\vct c_0=\mathsf P_0\mathsf A^{1/2}\vct\chi$ is
constant. Minimizing only the accessible directions at fixed inventory
gives
\begin{equation}
 \boxed{\begin{aligned}
 \mathsf H_{\rm acc}&=\mathsf H-
       \mathsf V_\chi\mathsf P_R\mathsf V_\chi^{\mathsf T},\\
 \mathsf H_{\rm acc}-\mathsf H_{\rm rel}
    &=\mathsf V_\chi\mathsf P_0\mathsf V_\chi^{\mathsf T}\succeq0,
 \qquad \mathsf H_{\rm rel}=\mathsf H-
       \mathsf B\mathsf A^{-1}\mathsf B^{\mathsf T}.
 \end{aligned}}
 \label{eq:s4patch-constrained-relaxed-curvature}
\end{equation}
Indeed completion of the energy square leaves
$\mathcal F_{\rm acc}=\vct X^{\mathsf T}\mathsf H_{\rm rel}\vct X/2
+\|\vct c_0+\mathsf P_0\mathsf V_\chi^{\mathsf T}\vct X\|^2/2$.
Its linear holding-force term is $\mathsf V_\chi\vct c_0$; the
mechanical bias force has the opposite sign. Its Hessian is
$\mathsf H_{\rm acc}$. The decaying kernel is
$\mathsf\Gamma_R(t)=\mathsf V_\chi e^{-\mathsf Q_\chi t}
\mathsf P_R\mathsf V_\chi^{\mathsf T}$, containing only positive-rate
sectors. A frozen inventory supplies its prepared bias and curvature.
For a supported positive Gaussian sector its conditional mechanical
covariance is $k_{\rm B}T\mathsf H_{\rm acc}^{-1}$.
An unconstrained Schur minimum would instead allow the inventory to
exchange. Conservation constraints are imposed before using
$\mathsf D^{-1}$ in the preceding realization; mixing coordinates
transforms those constraints too. This gives a finite-dimensional version of
the order-of-limits issue in~\eqref{eq:s3patch-surface-limit-order}.

Figure~\ref{fig:s34patch-conserved-curvature} shows how a conserved population restricts the relaxation path, changes the measured curvature and retains a preparation-dependent holding-force bias.

\begin{figure}[H]
\centering\begingroup
\setlength{\unitlength}{1pt}
\scalebox{0.86}{%
\input{vlt-figures/figure-293.tex}%
}\endgroup
\caption{Accessible relaxation in a two-population example. All coordinates are dimensionless and $\mathcal F/E_\star=2X^2+X(\chi_1+\chi_2)+(\chi_1^2+\chi_2^2)/2$, with mobility proportional to $\diag(1,0)$. At $X=1$, contours are centred on the unrestricted internal minimum $(-1,-1)$ (open green marker). The fixed inventory $\chi_2=c_0=0.6$ instead selects $(-1,0.6)$ (filled burgundy marker). Panel (b) gives mechanical curvatures in units $E_\star$: $4$ with both populations held, $3$ when only $\chi_1$ relaxes, and $2$ for the unrestricted comparison. The fixed inventory also adds the holding-force bias $c_0E_\star$. This realizes~\eqref{eq:s4patch-constrained-relaxed-curvature} on a positive Gaussian sector; the population coordinates denote local perturbations, not unrestricted global occupation fractions.}
\label{fig:s34patch-conserved-curvature}
\end{figure}

The relaxed and frozen references can be seen directly by defining
$\vct y_a=\mathsf G_a\vct u-\vct z_a$. The branch law becomes
$\tau_a\Dt{\vct y_a}=\mathsf G_a\vct u-\vct y_a$, with quadratic storage
\begin{equation}
 U_{\rm R}(\vct u,\{\vct y_a\})=
   \frac12\vct u^{\mathsf T}\mathsf K\vct u+
   \frac12\sum_a(\mathsf G_a\vct u-\vct y_a)^{\mathsf T}
               \mathsf H_a(\mathsf G_a\vct u-\vct y_a).
 \label{eq:relaxed-frozen-branch-reference}
\end{equation}
Static minimization over $\vct y_a$ leaves $\mathsf K$; holding the
branch coordinates fixed gives the instantaneous curvature
$\mathsf K_\infty=\mathsf K+\sum_a\mathsf K_a$, where
$\mathsf K_a=\mathsf G_a^{\mathsf T}\mathsf H_a\mathsf G_a$.
Thus a statically relaxed contribution is not appended a second time as
a new spring. An energetic mechanical realization interprets $U_{\rm R}$
as recoverable internal mechanical energy. If $\mathsf K$ or a branch
is instead obtained from a potential of mean force at fixed $T$, the
same quadratic algebra describes available free energy; its thermal
interpretation must then be supplied separately.

\paragraph{Which motion does the medium resist?}
For whole ideal virions, a constitutive map specifies an actual combination
of centre motions. Reuse the relative-displacement extractor
$\mathsf E_{ij}$ of~\eqref{eq:global-edge-assembly}. A longitudinal branch
on an unordered pair $e=\{i,j\}$ has
\begin{equation}
 \mathsf G_e=\vct n_{ij}^{\mathsf T}\mathsf E_{ij},\qquad
 z_e\in\mathbb R,\quad H_e=h_e\geq0,\qquad
 \mathsf K_e=h_e\mathsf E_{ij}^{\mathsf T}
              \vct n_{ij}\vct n_{ij}^{\mathsf T}\mathsf E_{ij}.
 \label{eq:s4-memory-edge-branch}
\end{equation}
It senses the first-order change of separation along the reference bond.
The branch force is assembled with equal and opposite node contributions,
just as the conservative pair force. Its rate of dissipation is
$h_ez_e^2/\tau_e$. Geometry therefore determines which displacement
patterns load the branch; $h_e$ and $\tau_e$ determine its stored-energy
and relaxation scales. The conservative tensor $\mathsf C_{ij}$ still
contains both the longitudinal curvature and transverse prestress of the
chosen Lennard--Jones and charge interaction. A positive Maxwell weight
is an additional constitutive input, rather than an identification with
a possibly signed transverse conservative curvature.

Let $\vct t_{\vct c}=(\vct c,\ldots,\vct c)$ be a common translation.
For every relative-motion branch,
\begin{equation}
 \mathsf G_a\vct t_{\vct c}=\vct0,\qquad
 \mathsf K_a\vct t_{\vct c}=\vct0,\qquad
 \vct t_{\vct c}^{\mathsf T}\mathsf G_a^{\mathsf T}
                      \mathsf H_a\vct z_a=0.
 \label{eq:s4-memory-translation-identity}
\end{equation}
Thus internal pair relaxation cannot damp a common translation by itself.
This is the dissipative counterpart of the conservative sum rule
\eqref{eq:vlp-finite-acoustic-sum-rule}~\cite{Misra2012}.
The identity holds for unequal masses and mixed assigned charges.
Longitudinal maps also annihilate an infinitesimal rigid rotation about
the reference. More general vector-valued branches can represent
orientation-sensitive coupling to retained surrounding material; their
torque exchange and rotational invariance require the corresponding
medium coordinates. Equal and opposite forces alone establish zero net
force, not zero net torque.

Drag on each particle relative to a stationary surrounding medium senses
a different motion. If medium velocity $\vct v_{\rm b}$ is retained,
let a dimensionless map $\mathsf S$ interpolate it to the particle centres
and set $\vct v=\Dt{\vct u}$. A symmetric resistance $\mathsf C\geq0$
gives the matched forces and total power
\begin{equation}
 \begin{aligned}
 \vct f_{\rm p}&=-\mathsf C(\vct v-\mathsf S\vct v_{\rm b}),\\
 \vct f_{\rm b}&=\mathsf S^{\mathsf T}
                      \mathsf C(\vct v-\mathsf S\vct v_{\rm b}),\\
 \vct v^{\mathsf T}\vct f_{\rm p}
      +\vct v_{\rm b}^{\mathsf T}\vct f_{\rm b}
   &=-(\vct v-\mathsf S\vct v_{\rm b})^{\mathsf T}
       \mathsf C(\vct v-\mathsf S\vct v_{\rm b})\leq0.
 \end{aligned}
 \label{eq:s4-memory-retained-medium-drag}
\end{equation}
The medium's own momentum equation receives $\vct f_{\rm b}$.
Setting its velocity to zero recovers body drag with an assigned reaction
partner. Allowing the medium to move makes shared motion and relative
motion distinct measurable channels. A geometry-dependent dense
$\mathsf C$ can couple distant particles through the medium even when
the retained conservative graph is sparse. Assigned charge affects the
reference geometry and conservative modes; an additional charge-dependent
resistance requires its own medium coupling law.

The force $\vct f$ has units of force. It can describe a preparation pulse,
a separately retained medium coordinate, or an explicitly supplied drive.
After a preparation pulse the closed reference evolution has no external
drive. If $\vct f$ describes an internal interaction, its reaction and energy
exchange belong to the retained partner subsystem. Its mechanical power is
$\Dt{\vct u}^{\mathsf T}\vct f$. A work term is consequently a scalar energy
or power, rather than the vector appearing in the force balance.

\subsubsection{Compatible cells carry both restoring forces and material memory}
\label{sec:s4-cell-memory-assembly}

Assembly cells from Subsection~\ref{sec:s2-assembly-cells} can contribute
inertia, relaxed curvature, resistance and branch strain. Shared sites
have one physical displacement, so calibrated pieces combine into a
larger viscoelastic array. The following construction fixes a reference
window; the transient extension carries its preparation through joining
and breakup.

\begin{definition}[Viscoelastic cell data and ownership]
\label{def:s4-cell-memory-data}
Use the restrictions $\vct u^{(c)}=\mathsf R_c\vct u$ and physical-label
ownership of Definition~\ref{def:s3-cell-restriction}. Assign each retained
energy contribution and each physical dissipation channel once. A cell
$c$ carries $\mathsf M_c,\mathsf K_c,\mathsf C_{0c}$ and branches
$(\mathsf G_{ca},\mathsf H_{ca},\tau_{ca},\vct z_{ca})$, where its
branch map acts on $\vct u^{(c)}$. Shared-particle inertia uses the mass
fractions already defined in Section~3. Then
\begin{equation}
 \begin{aligned}
 \mathsf M&=\sum_c\mathsf R_c^{\mathsf T}\mathsf M_c\mathsf R_c,
 &\mathsf K&=\sum_c\mathsf R_c^{\mathsf T}\mathsf K_c\mathsf R_c,\\
 \mathsf C_0&=\sum_c\mathsf R_c^{\mathsf T}\mathsf C_{0c}\mathsf R_c,
 &\mathsf G_{ca}^{\rm glob}&=\mathsf G_{ca}\mathsf R_c,\\
 \mathsf K_{ca}^{\rm glob}
   &=\mathsf R_c^{\mathsf T}\mathsf G_{ca}^{\mathsf T}
       \mathsf H_{ca}\mathsf G_{ca}\mathsf R_c.
 \end{aligned}
 \label{eq:s4-cell-memory-congruence}
\end{equation}
The local coordinate list includes all endpoints needed by an owned
cross-cell channel. Its local branch extension appears once in the
retained state.
\end{definition}

\begin{proposition}[Passive compatible assembly]
\label{prop:s4-cell-passive-assembly}
Suppose the reference restrictions are fixed, the assembled physical
mass is positive definite, and the assembled relaxed stiffness is
nonnegative on the declared admissible sector. If
$\mathsf C_{0c}\succeq0$, $\mathsf H_{ca}\succeq0$ and $\tau_{ca}>0$,
the assembled equations have the storage and power balance of
Proposition~\ref{prop:maxwell-energy}, with loss
\begin{equation}
 \mathcal D_{\rm asm}
  =\sum_c\dot{\vct u}^{(c)\mathsf T}
                 \mathsf C_{0c}\dot{\vct u}^{(c)}
   +\sum_{c,a}\tau_{ca}^{-1}
                 \vct z_{ca}^{\mathsf T}\mathsf H_{ca}\vct z_{ca}
   \geq0.
 \label{eq:s4-cell-assembled-dissipation}
\end{equation}
Consequently compatible assembly preserves passivity with respect to
the global force--velocity supply.
\end{proposition}
\begin{proof}
Substitute $\dot{\vct u}^{(c)}=\mathsf R_c\dot{\vct u}$ into each
kinetic, potential and resistance form. The congruences assemble the
same global storage and virtual work. Each branch force contributes
$\dot{\vct u}^{\mathsf T}\mathsf R_c^{\mathsf T}
\mathsf G_{ca}^{\mathsf T}\mathsf H_{ca}\vct z_{ca}$, which cancels
the branch-storage exchange term. The remaining terms are
$\dot E_{\rm res}=\dot{\vct u}^{\mathsf T}\vct f-\mathcal D_{\rm asm}$.
The stated positive forms make the global storage nonnegative.
\end{proof}

The global storage--supply statement is the criterion of dissipative
systems theory~\cite{Willems1972,Willems1972Linear}. A conservative prestress contribution
assigned to one cell can be signed; stability is tested on the assembled
reference. Calling every such local Hessian a separately passive material
would require additional positivity conditions. The proposition uses the
global stability condition and positive physical resistance channels.

The fixed $3N_{\rm tot}$ coordinate space follows particle identities as
the selected cluster changes. Cross-cell branches retain their pair
forces and storage across computational interfaces. Ownership weights
count each physical mass, charge and branch once, including sites shared
by several cell descriptions.

Cells disturbing a shared solvent have coupled resistance blocks and
potentially correlated fluctuating forces. Retain the fluid operator
\eqref{eq:s4-bio-array-resistance} once, either as a shared component or
as justified physical channels with all endpoints. At a common
equilibrium temperature, independent physical channels can carry
independent noise assembled by the same coordinate maps; a shared
channel carries its shared covariance.

For example, a pair of neighboring cell groups may have a sparse set of
short-range conservative links but a broader hydrodynamic resistance.
The maps in~\eqref{eq:s4-cell-memory-congruence} distinguish those
patterns. A measured loss associated with relative motion can be fitted
to an interface branch, while collective translation can load the
shared solvent. Charge changes the prepared geometry and conservative
curvatures through the declared interactions; a change in a dissipative
branch additionally requires its constitutive coupling. This separates
what the cell arrangement determines from the medium inputs that must
be calibrated.

Figure~\ref{fig:s4-compatible-cell-memory} relates the local cell maps to a five-virion assembly with shared physical sites.

\begin{figure}[H]
\centering\begingroup
\definecolor{vltTeal}{HTML}{176B64}
\definecolor{vltBlue}{HTML}{315D80}
\definecolor{vltGold}{HTML}{A87520}
\definecolor{vltPlum}{HTML}{725877}
\definecolor{vltRed}{HTML}{762B38}
\definecolor{vltYellow}{HTML}{D5AD3C}
\definecolor{vltNeutral}{HTML}{777D83}
\setlength{\unitlength}{1pt}
\scalebox{0.90}{%
\input{vlt-figures/figure-099.tex}%
}\endgroup
\caption{Two tetrahedral cells share face $(1,2,3)$. Eight local site occurrences assemble into five physical virions and nine distinct pair links. Gold identifies shared sites; the assembled orthographic view fixes their spatial relationship. For equal masses $m$, half ownership of shared masses gives common-speed kinetic energy $5mV^2/2$. The same ownership rule counts each charge, branch and dissipation channel once. Springs identify the selected deformation channels; additional conservative and fluid couplings enter their declared blocks. The congruences in~\eqref{eq:s4-cell-memory-congruence} preserve shared coordinates and virtual work.}
\label{fig:s4-compatible-cell-memory}
\end{figure}

\subsubsection{Nonlinear motion, recruitment and prepared material strain}
\label{sec:s4-transient-nonlinear-memory}

During the nonlinear approach of
Subsection~\ref{sec:s23-transient-recruitment}, a virion disturbs the medium
before satisfying a membership criterion. The resulting strain can
persist after attachment or release. Following the fixed particle pool
and its material preparation therefore connects successive local
harmonic references through one particle--medium trajectory.

\begin{definition}[Nonlinear prepared particle--medium state]
\label{def:s4-transient-nonlinear-state}
Let $\vct r$ collect the $3N_{\rm tot}$ particle-centre coordinates before
fixed constraints, and let the constant $\mathsf M$ contain their positive material
masses. On a smooth collision-free configuration domain, branch $a$ has
a length-valued deformation map $\vct\psi_a(\vct r)$, relaxed coordinate
$\vct y_a$, and extension
\begin{equation}
 \vct z_a=\vct\psi_a(\vct r)-\vct y_a,\qquad
 \mathsf G_a(\vct r)=D_{\vct r}\vct\psi_a(\vct r).
 \label{eq:s4-transient-deformation-state}
\end{equation}
The map $\mathsf G_a$ is dimensionless. The prepared state contains
$(\vct r,\vct v,\{\vct y_a\})$, or equivalently
$(\vct r,\vct v,\{\vct z_a\})$, together with the declared interaction
laws, supports and constitutive parameters.
\end{definition}

For a declared branch family between physical contact-state events,
take the nonlinear conservative energy $U(\vct r)$ of the same particle
model, constant $\mathsf H_a\succeq0$ and $\tau_a>0$, and symmetric
instantaneous resistance $\mathsf C(\vct r)\succeq0$. One passive
extension is
\begin{equation}
 \begin{aligned}
 \mathsf M\dot{\vct v}
   &=-\nabla U(\vct r)-\mathsf C(\vct r)\vct v
      -\sum_a\mathsf G_a(\vct r)^{\mathsf T}
                    \mathsf H_a\vct z_a+\vct f,\\
 \dot{\vct r}&=\vct v,\qquad
 \dot{\vct z}_a=\mathsf G_a(\vct r)\vct v-\tau_a^{-1}\vct z_a.
 \end{aligned}
 \label{eq:s4-transient-nonlinear-system}
\end{equation}
Equivalently $\dot{\vct y}_a=\tau_a^{-1}\vct z_a$. The force uses the
transpose of the same deformation map as the branch evolution, preserving
their conjugate work. A distance-sensing branch can use
$\psi_a=|\vct r_i-\vct r_j|-\ell_a$, with declared reference length
$\ell_a$. The local linearization at an unstrained reference gives the
relative-motion map of~\eqref{eq:s4-memory-edge-branch}. More general
maps can describe a resolved surface deformation or relative motion
against a retained material component, with its reaction included.

\begin{proposition}[Passive nonlinear energy account]
\label{prop:s4-transient-nonlinear-energy}
For smooth solutions of~\eqref{eq:s4-transient-nonlinear-system}, define
\begin{equation}
 \begin{aligned}
 E&=\tfrac12\vct v^{\mathsf T}\mathsf M\vct v+U(\vct r)
       +\tfrac12\sum_a\vct z_a^{\mathsf T}\mathsf H_a\vct z_a,\\
 \dot E&=\vct v^{\mathsf T}\vct f
      -\vct v^{\mathsf T}\mathsf C(\vct r)\vct v
      -\sum_a\tau_a^{-1}\vct z_a^{\mathsf T}\mathsf H_a\vct z_a.
 \end{aligned}
 \label{eq:s4-transient-nonlinear-energy}
\end{equation}
Power-free ideal reactions add no term to this identity. If $U$ has a
lower bound $U_{\min}$, $E-U_{\min}$ is a nonnegative storage function.
\end{proposition}
\begin{proof}
Differentiate kinetic, conservative and branch energies. The chain rule
gives $\dot{\vct\psi}_a=\mathsf G_a\vct v$, cancelling the two branch
exchange terms. Symmetry and positivity give the remaining losses.
\end{proof}

The displayed realization holds its constitutive parameters fixed
between events. A continuous physical control protocol adds its explicit
parameter-work term $\partial_\lambda E\,\dot\lambda$ to the balance.

Binding can lower $U$ while raising kinetic or branch energy. This is the
energy-accommodation mechanism of Section~3 with material memory retained.
The total vessel population and material mass stay fixed as the selected
cluster changes. In an energetic mechanical realization the omitted
thermal account receives the lost mechanical energy. An isothermal
potential of mean force instead requires the free-energy and heat
distinction already made in Proposition~\ref{prop:maxwell-energy}; thermal
noise additionally contributes its exchanged stochastic work.

Eliminating a branch along the actual trajectory also gives a concrete
history law. Between contact-state events, its force contribution is
\begin{equation}
 \begin{aligned}
 \vct f_{\rm br}(t)
   &=\vct h_{\rm prep}(t)
       +\int_{t_0}^t\mathsf\Gamma_{\rm traj}(t,s)\vct v(s)\,\dd s,\\
 \mathsf\Gamma_{\rm traj}(t,s)
   &=\sum_a\mathsf G_a(\vct r(t))^{\mathsf T}\mathsf H_a
               e^{-(t-s)/\tau_a}\mathsf G_a(\vct r(s)),\\
 \vct h_{\rm prep}(t)
   &=\sum_a\mathsf G_a(\vct r(t))^{\mathsf T}\mathsf H_a
                      e^{-(t-t_0)/\tau_a}\vct z_a(t_0).
 \end{aligned}
 \label{eq:s4-transient-trajectory-memory}
\end{equation}
Here $\vct f_{\rm br}$ is the resisting term on the left of the force
balance. Solving the branch equation retains both the earlier loading
geometry and the current geometry transmitting its force. Rearrangement
therefore gives a two-time kernel even for constant relaxation times.
Fixed maps recover convolution. Retaining branch states computes this
same history without storing the trajectory. Thermal statistics follow
from the separately specified stochastic dynamics and preparation.

Figure~\ref{fig:s4-trajectory-memory-directions} shows explicitly how the old loading direction and current force direction give a two-time kernel.

\begin{figure}[H]
\centering\begingroup
\definecolor{vltTeal}{HTML}{176B64}
\definecolor{vltBlue}{HTML}{315D80}
\definecolor{vltGold}{HTML}{A87520}
\definecolor{vltPlum}{HTML}{725877}
\definecolor{vltRed}{HTML}{762B38}
\definecolor{vltYellow}{HTML}{D5AD3C}
\definecolor{vltNeutral}{HTML}{777D83}
\setlength{\unitlength}{1pt}
\scalebox{0.90}{%
\input{vlt-figures/figure-100.tex}%
}\endgroup
\caption{The same relaxation time can produce a two-time memory law during rearrangement. (a) A prescribed relative trajectory has $r(t)/\ell_*=2+0.3\sin[\pi t/(3\tau)]$ and direction $\theta(t)=\pi t/(6\tau)$; the same two ideal virions are shown at three times. The straight connector marks their distance sensor; the material extension also depends on its internal relaxed coordinate. (b) On relative increments $\vct q=\vct r_{\rm right}-\vct r_{\rm left}$, the displayed component is $\Gamma_{yx}(t,s)/h=e^{-(t-s)/\tau}\sin\theta(t)\cos\theta(s)$ for $s\le t$ and zero for $s>t$. Darker shading means a larger component. (c) At the fixed lag $t-s=\tau$, this component still varies with the earlier time. The dashed comparison freezes the direction at $45^\circ$. This is the exact history kernel along the declared trajectory; it is not a state-independent response or an unforced rotational solution.}
\label{fig:s4-trajectory-memory-directions}
\end{figure}

\paragraph{Prepared strain changes the tangent as well as the force.}
Linearization about a strained branch retains a geometric term. With
$\vct y_a$ held fixed during the instantaneous variation,
\begin{equation}
 \nabla_{\vct r}^{2}
      \left(\tfrac12\vct z_a^{\mathsf T}\mathsf H_a\vct z_a\right)
  =\mathsf G_a^{\mathsf T}\mathsf H_a\mathsf G_a
    +\sum_\alpha(\mathsf H_a\vct z_a)_\alpha
                        \nabla_{\vct r}^{2}\psi_{a,\alpha}.
 \label{eq:s4-transient-geometric-tangent}
\end{equation}
For the scalar distance branch with $H_a=h_a\geq0$, its relative-coordinate
tensor is
\begin{equation}
 \mathsf C_a^{\rm br}
    =h_a\vct n\vct n^{\mathsf T}
       +\frac{h_az_a}{r_{ij}}
                       (\Id_3-\vct n\vct n^{\mathsf T}),\qquad
 \vct n=\frac{\vct r_i-\vct r_j}{r_{ij}}.
 \label{eq:s4-transient-distance-tangent}
\end{equation}
The transverse term has the same prestress structure as the conservative
pair tensor in~\eqref{eq:edge-stiffness}. Compression and extension can
therefore alter incremental stability differently even with the same
nonnegative branch stiffness. The full constrained tangent determines
stability; the power identity determines passivity. At $z_a=0$ the
geometric term vanishes, recovering the positive frozen branch increment
used in the fixed-reference Maxwell model.

Figure~\ref{fig:s4-prepared-geometric-tangent} compares the force and transverse tangent of two prepared branches at the same virion separation.

\begin{figure}[H]
\centering\begingroup
\definecolor{vltTeal}{HTML}{176B64}
\definecolor{vltBlue}{HTML}{315D80}
\definecolor{vltGold}{HTML}{A87520}
\definecolor{vltPlum}{HTML}{725877}
\definecolor{vltRed}{HTML}{762B38}
\definecolor{vltYellow}{HTML}{D5AD3C}
\definecolor{vltNeutral}{HTML}{777D83}
\setlength{\unitlength}{1pt}
\scalebox{0.84}{%
\input{vlt-figures/figure-101.tex}%
}\endgroup
\caption{Equal particle geometry with opposite prepared branch strains. The distance branch has $r=2\ell_*$, $\ell_a=1.8\ell_*$ and $H=h>0$. Relaxed coordinates $y=0,0.4\ell_*$ give $z=+0.2\ell_*,-0.2\ell_*$. Burgundy arrows show force on the right virion. Internal dots move on the same length scale as strain; both rows share body size, natural spring span and six coil turns. At fixed $y$, the transverse holding-force slope is $hz/r$ and the longitudinal slope is $h$. Equal branch energies thus coexist with opposite transverse contributions; stability uses the complete constrained tangent.}
\label{fig:s4-prepared-geometric-tangent}
\end{figure}

\subsubsection{Nonlinear body motion and persistent surface--material state}
\label{sec:s4patch-nonlinear-body-memory}

The nonlinear translational construction has a direct rigid-body
extension. Keep $Q=(\vct r_i,\mathsf O_i)_{i=1}^{N_{\rm tot}}$ on
the positive-clearance body configuration domain and let
$\vct\chi=(\boldsymbol\alpha,\vct y)$ contain the permitted surface
and material coordinates in their declared frames and scales. A useful
attachment realization is
\begin{equation}
 \mathcal F(Q,\boldsymbol\alpha,\vct y;\lambda)
   =\mathcal F_0(Q,\boldsymbol\alpha;\lambda)
       +\tfrac12\vct e^{\mathsf T}\mathsf A_h\vct e,
 \qquad \vct e=\vct h(Q,\boldsymbol\alpha;\lambda)-\vct y,
 \quad \mathsf A_h\succ0 .
 \label{eq:s4patch-body-attachment-energy}
\end{equation}
The fixed matrix $\mathsf A_h$ acts on a fixed-dimensional material
space. Section~2's persistent material coordinate is $\vct y$ here;
it is distinct from the branch extension $\vct e$ or $\vct z_a$.
Choose both $\vct h$ and its mechanical derivative to vanish in the
separated region for a contact-localized coupling. A permanent
distance-sensing connector, $\psi=r_{ij}-\ell_a$, has a different
global meaning. After localized contact separates, $\vct y$ continues
its declared evolution with positive self-energy while its mechanical
force vanishes; null mobility directions remain frozen.
The Gaussian material integral leaves $\mathcal F_0$ plus a constant
under the hypotheses of Proposition~\ref{prop:s2patch-material-marginal}.

Mechanical force and torque are the work-dual derivatives $-D_Q\mathcal F$.
Surface and material affinities include the reciprocal attachment terms:
\begin{equation}
 \vct g_\alpha=\partial_\alpha\mathcal F_0
       +(D_\alpha\vct h)^{\mathsf T}\mathsf A_h\vct e,
 \qquad \vct g_y=-\mathsf A_h\vct e .
 \label{eq:s4patch-body-surface-affinities}
\end{equation}
Thus a charge-induced change of $\vct h$ changes both the force and
its chemical affinity. Retaining only its force would discard the
reciprocal exchange.

\begin{proposition}[Nonlinear rigid-body storage and supply]
\label{prop:s4patch-rigid-nonlinear-passivity}
Let the lab angular momentum be $\vct\ell_i$ and let
$\mathsf J_i(\mathsf O_i)=\mathsf O_i\mathsf J_i^0\mathsf O_i^{\mathsf T}$,
with $\mathsf J_i^0\succ0$. Set
$\vct V_i=\vct p_i/m_i$,
$\vct\Omega_i=\mathsf J_i(\mathsf O_i)^{-1}\vct\ell_i$ and retain
\begin{equation}
 \begin{aligned}
 \dot{\vct r}_i&=\vct V_i,\qquad
 \dot{\mathsf O}_i=[\vct\Omega_i]_\times\mathsf O_i,\\
 (\dot{\vct p},\dot{\vct\ell})
    &=-D_Q\mathcal F-\mathsf C_{\rm rb}(Q,\vct\chi)\vct\nu
                               +(\vct f,\vct t),\\
 \dot{\vct\chi}&=-\mathsf L_\chi(Q,\vct\chi)\vct g_\chi,
      \qquad \vct g_\chi=\partial_\chi\mathcal F .
 \end{aligned}
 \label{eq:s4patch-rigid-nonlinear-state}
\end{equation}
For symmetric nonnegative resistance and mobility, smooth admissible
solutions have storage
$\mathcal E=\sum_i[|\vct p_i|^2/(2m_i)
+\vct\ell_i^{\mathsf T}\mathsf J_i^{-1}\vct\ell_i/2]+\mathcal F$ and
\begin{equation}
 \boxed{\dot{\mathcal E}
    =\sum_i(\vct V_i\cdot\vct f_i+\vct\Omega_i\cdot\vct t_i)
       +\partial_\lambda\mathcal F\,\dot\lambda
       -\vct\nu^{\mathsf T}\mathsf C_{\rm rb}\vct\nu
       -\vct g_\chi^{\mathsf T}\mathsf L_\chi\vct g_\chi .}
 \label{eq:s4patch-rigid-storage-supply}
\end{equation}
\end{proposition}
\begin{proof}
Rigid-body kinematics gives rotational kinetic power
$\vct\Omega_i\cdot\dot{\vct\ell}_i$; equivalently the gyroscopic
term in body-frame Euler dynamics does no work. The virtual-work
identity for $D_Q\mathcal F$ cancels reversible translational and
angular exchange. Substitution of the internal gradient law gives
the remaining nonnegative losses.
\end{proof}

Figure~\ref{fig:cmd-mixed-storage-certificate} gives a graphical specialization of this boxed relation.

A lower-bounded $\mathcal F$ supplies nonnegative shifted storage.
For isothermal free energy this is an available-work identity; a full
internal-energy account retains entropy, heat and reservoir fluxes.
In particular a grand-potential parameter derivative is not the entire
chemical energy ledger. Smooth collision-free evolution and actual
contact/chemical events use their stated boundary and transition laws.
Large rotations evolve on $\operatorname{SO}(3)$; they do not use a
constant angular-coordinate mass matrix globally.

\paragraph{Earlier rotation and charge changes load later force.}
For constant $\mathsf A_h,\mathsf L_h\succ0$ and a block-diagonal
material mobility, set $\mathsf R_h=\mathsf L_h\mathsf A_h$,
$\mathsf J_Q=D_Q\vct h$ as a map on the body twist, and
$\mathsf J_\alpha=D_\alpha\vct h$. Then
\begin{equation}
 \begin{aligned}
 \dot{\vct e}&=\mathsf J_Q\vct\nu
       +\mathsf J_\alpha\dot{\boldsymbol\alpha}
       +\partial_\lambda\vct h\,\dot\lambda-\mathsf R_h\vct e,\\
 \mathsf\Gamma_{Qb}(t,s)&=\mathsf J_Q(t)^{\mathsf T}\mathsf A_h
        e^{-\mathsf R_h(t-s)}\mathsf J_b(s),\quad b=Q,\alpha .
 \end{aligned}
 \label{eq:s4patch-angular-surface-memory}
\end{equation}
Solving this equation gives the propagated initial strain plus its
mechanical, surface and control loading integrals. Multiplying by
$\mathsf J_Q(t)^{\mathsf T}\mathsf A_h$ gives the resisting force
and torque. These histories are mutually coupled when the surface
state also evolves; they are not independent imposed sources.
State-dependent noncommuting relaxation uses the ordered propagator
solving $\partial_t\mathsf U_h(t,s)=-\mathsf R_h(t)\mathsf U_h(t,s)$.
Scalar patch distances and relative-orientation invariants can supply
objective attachment features. Vector material coordinates in moving
frames retain their transport terms. This is a nonlinear prepared-state
reduction of the type motivated by shifted-bath constructions
\cite{Zwanzig1973}. Thermal statistics on a self-consistent stochastic
trajectory are determined by its joint augmented model, rather than
by treating the geometry factors as independent of the noise.

\begin{example}[Load, separate, wait and return]
\label{ex:s4patch-load-wait-return}
Hold the surface state and controls fixed. Start separated with
$\vct h=0$ and $\vct y=0$; hold a configuration $Q_b$, with
$\vct h_b=\vct h(Q_b)$, for $T_L$; separate to $\vct h=0$ and
wait $T_W$; then return to the same $Q_b$. In the prescribed rapid-ramp
limit relative to material relaxation,
\begin{equation}
 \boxed{\begin{aligned}
 \vct y_{\rm ret}
   &=e^{-\mathsf R_hT_W}(\Id-e^{-\mathsf R_hT_L})\vct h_b,\\
 \Delta\vct f_h(Q_b)
   &=\mathsf J_Q(Q_b)^{\mathsf T}\mathsf A_h\vct y_{\rm ret} .
 \end{aligned}}
 \label{eq:s4patch-recontact-recovery}
\end{equation}
The second line is the reversible generalized force difference from
a fresh contact at the same geometry. Its angular components are
torque differences. The material equation
$\dot{\vct y}=\mathsf R_h(\vct h-\vct y)$ proves both formulas
by propagation through the two holds. Finite ramps use its exact
Duhamel integral. Matched additive material noise preserves the same
conditional mean formulas for a prescribed path.
\end{example}
The force difference requires overlap with the chosen loading and
readout directions. The same $\mathsf R_h$ sets force relaxation
and waiting-time recovery, giving two records to compare under one
input set. This is a synthetic controlled protocol, not an unforced
position jump or a measured HPV recovery law. At fixed chemical
controls, the integrated storage identity gives
$W_{\rm mech}=\Delta T_{\rm rb}+\Delta\mathcal F+
\int\mathcal D_{\rm rb}\,\dd t$, where
$\mathcal D_{\rm rb}=\vct\nu^{\mathsf T}\mathsf C_{\rm rb}\vct\nu
+\vct g_\chi^{\mathsf T}\mathsf L_\chi\vct g_\chi$.
Endpoints at rest remove the kinetic-energy difference. Returning
particle geometry alone need not return material storage.
Full-state periodicity is needed before cycle work equals
dissipation alone. Evolving material reference states have established
transient-network precedents~\cite{MengPritchardTerentjev2016}.

Figure~\ref{fig:s34patch-material-return-protocol} turns the recovery calculation into a controlled multi-particle protocol. Returning to the same geometry restores the coupling to a material coordinate that has continued to evolve during separation.

\begin{figure}[H]
\centering\begingroup
\setlength{\unitlength}{1pt}
\scalebox{0.94}{%
\input{vlt-figures/figure-294.tex}%
}\endgroup
\caption{Separation removes a localized mechanical coupling while its material state persists. The upper sequence keeps the same eight labelled VLPs in two four-body groups; prescribed rapid ramps bring the groups to $Q_b$, separate them, and return them to that geometry. Views are oblique and use a common particle and position scale. Gauges encode the selected scalar material coordinate $y/h_b$, for $T_L=T_W=\tau$, rather than contact force. In the separated region $h=D_Qh=0$ and the selected material force vanishes while $y$ decays. The lower curves evaluate~\eqref{eq:s4patch-recontact-recovery} for one positive-rate scalar branch, $\tau=(L_hA_h)^{-1}$, and a readout with $J_QA_hh_b\ne0$. The open marker matches the upper protocol. Curves report a difference from a fresh contact at identical geometry; finite ramps use the stated history integral. The construction is synthetic, with no fitted biological recovery time.}
\label{fig:s34patch-material-return-protocol}
\end{figure}

\subsubsection{A growing cluster and a fixed-domain continuum window}
\label{sec:s4-growing-continuum-account}

The particle description also specifies what a growing cluster means for
a continuum field. Let $h_i^{\mathcal C}(t)$ equal one when physical
particle $i$ belongs to the selected cluster and zero otherwise. Its
mass measure, mass flux and membership source are
\begin{equation}
 \begin{aligned}
 \rho_{\mathcal C}(\vct x,t)
   &=\sum_{i=1}^{N_{\rm tot}}h_i^{\mathcal C}(t)m_i
                    \delta(\vct x-\vct r_i(t)),\\
 \vct j_{\mathcal C}(\vct x,t)
   &=\sum_{i=1}^{N_{\rm tot}}h_i^{\mathcal C}(t)m_i\vct v_i(t)
                    \delta(\vct x-\vct r_i(t)),\\
 s_{\mathcal C}(\vct x,t)
   &=\sum_{i=1}^{N_{\rm tot}}\dot h_i^{\mathcal C}(t)m_i
                    \delta(\vct x-\vct r_i(t)),\\
 \partial_t\rho_{\mathcal C}
             +\nabla\cdot\vct j_{\mathcal C}&=s_{\mathcal C}.
 \end{aligned}
 \label{eq:s4-growing-cluster-mass-balance}
\end{equation}
The last equality holds distributionally. Differentiating each moving
delta contributes $-\vct v_i\cdot\nabla\delta$, which cancels its
flux divergence; differentiating the membership indicator leaves the
source. At a discrete joining or leaving event, $\dot h_i^{\mathcal C}$
contains the corresponding time impulse. The measures have density,
mass-flux and mass-source units. Consistent spatial smoothing gives
coarse fields, with the enclosure handled in the smoothing convention.
The displayed density is in current coordinates; the constant density
used below is the inertial density of its fixed reference patch.

For the complete particle pool every $h_i=1$ and the source vanishes.
The complement of $\mathcal C$ has the opposite membership source, so
recruitment transfers assigned mass between two descriptions while
the vessel retains $\sum_i m_i$. A mode quantum transfers excitation
energy without adding a material term to this mass account. This also
distinguishes the growing assembly from a fixed material patch whose
particles merely deform.

The continuum equations below describe a fixed reference patch during
its declared window. Their constant density, material domain and
constitutive tensors belong to that reference. Extending them across
association requires the corresponding mass and momentum transport,
interface power and evolving memory variables. The discrete full-pool
model provides those accounts directly. Increasing $n_{\mathcal C}$
therefore calls for a registered assembly update or a transport model,
rather than a density change inserted into an otherwise unchanged
fixed-domain equation.

The long-wavelength construction of Section~\ref{sec:continuum-bridge} gives
the continuum counterpart on the occupied material region
$\mathcal V_{\mathrm{lat}}$ of Section~3. This region need not fill
the vessel; its particle mass is approximately
$\rho|\mathcal V_{\mathrm{lat}}|\simeq mn$ for the filled bulk reference,
with finite boundary counting corrections. Write
$F_{p\mu}=\partial_\mu u_p$ for the displacement gradient and introduce real
branch fields $Z_{a,p\mu}$ with the same dimensionless units. We use the
contraction $\mathsf F:\mathcal A:\mathsf F
=F_{p\mu}\mathcal A_{p\mu q\nu}F_{q\nu}$. The incremental stress and branch laws are
\begin{equation}
 \begin{split}
 T_{p\mu}&=\mathcal A_{p\mu q\nu}\partial_\nu u_q
     +\mathcal D^{\mathrm v}_{p\mu q\nu}\partial_\nu\Pdt{u_q}
     +\sum_a\mathcal A^{(a)}_{p\mu q\nu}Z_{a,q\nu},\\
 \Pdt{Z_{a,p\mu}}+\tau_a^{-1}Z_{a,p\mu}
                                      &=\partial_\mu\Pdt{u_p}.
 \end{split}
 \label{eq:continuum-maxwell-stress}
\end{equation}
The relaxed gradient tensor $\mathcal A$ is the tensor of
\eqref{eq:continuum-tensor}; $\mathcal A^{(a)}$ are nonnegative branch tensors
with the major symmetry under $(p,\mu)\leftrightarrow(q,\nu)$, and
$\mathcal D^{\mathrm v}$ is a nonnegative viscosity tensor. Their units are pressure,
pressure, and pressure times time, respectively. In an unstressed elastic
model the corresponding laws may be expressed through symmetric strain and
ordinary symmetric Cauchy stress. In the full-gradient formulation,
$T_{p\mu}$ denotes the incremental nominal stress work-conjugate to
$\partial_\mu u_p$ about the declared reference. This interpretation
retains prestress information carried by the discrete Hessian.
For an unstressed, frame-indifferent material branch, the strain map
must annihilate a rigid infinitesimal rotation; nonnegativity of a
gradient tensor alone is not that objectivity condition. A medium
law may therefore use the symmetric strain explicitly. An effective
network coupled to supports can retain the declared full-gradient map.

\begin{figure}[H]
\centering\begingroup
\definecolor{vltTeal}{HTML}{176B64}
\definecolor{vltBlue}{HTML}{315D80}
\definecolor{vltGold}{HTML}{A87520}
\definecolor{vltPlum}{HTML}{725877}
\definecolor{vltRed}{HTML}{762B38}
\definecolor{vltYellow}{HTML}{D5AD3C}
\definecolor{vltNeutral}{HTML}{777D83}
\setlength{\unitlength}{1pt}
\scalebox{0.90}{%
\input{vlt-figures/figure-102.tex}%
}\endgroup
\caption{Force, power and displacement-gradient readings of \eqref{eq:causal-continuum-pde}. (a) A longitudinal slice of area $S_\perp$ has outward face forces $-S_\perp T(x)$ and $+S_\perp T(x+\Delta x)$. The example $T(x)=T_\star>0$, $T(x+\Delta x)=3T_\star/2$ gives the calibrated arrow-length ratio $2:3$ and a positive stress-gradient force. (b) Total traction enters boundary power through $\vct v\cdot\vct t$. Storage $w$ and nonnegative loss use \eqref{eq:continuum-branch-storage-density}, with any declared local pinning included in storage. The two power arrows identify channels; their lengths assign no power values. (c) Three rigid virions sample $u(x)=\epsilon x$ at equally spaced reference centres, with $\epsilon=0.12$ exaggerated for display and one $90\,\mathrm{pt}/a$ spatial gain. Matched coils acquire the same increment $\epsilon a$. This strip visualizes the gradient; the independently specified traction example above fixes its own stress values. The energy account belongs to occupied lattice material.}
\label{fig:continuum-force-power}
\end{figure}

The units and force sign can be checked through continuum storage.
In the constant-coefficient energetic realization, let
\begin{equation}
 w_{\rm R}=\tfrac12\rho|\Pdt{\vct u}|^2
       +\tfrac12\mathsf F:\mathcal A:\mathsf F
       +\tfrac12\sum_a\mathsf Z_a:\mathcal A^{(a)}:\mathsf Z_a,
 \qquad
 d_{\rm R}=\partial_t\mathsf F:\mathcal D^{\rm v}:
                  \partial_t\mathsf F
       +\sum_a\tau_a^{-1}\mathsf Z_a:\mathcal A^{(a)}:\mathsf Z_a.
 \label{eq:continuum-branch-storage-density}
\end{equation}
Here $w_{\rm R}$ is energy per volume and $d_{\rm R}$ power per volume;
the constant reference density is the retained lattice inertia.
Differentiating the branch storage gives
\begin{equation}
 \Pdt{\left(\tfrac12\sum_a\mathsf Z_a:
                       \mathcal A^{(a)}:\mathsf Z_a\right)}
 =\sum_a(\mathcal A^{(a)}:\mathsf Z_a):\partial_t\mathsf F
       -\sum_a\tau_a^{-1}\mathsf Z_a:
                                      \mathcal A^{(a)}:\mathsf Z_a.
 \label{eq:continuum-branch-storage-rate}
\end{equation}
The first term is exchange with the resolved deformation; the second
is nonnegative loss with a minus sign. Combining it with momentum
balance gives boundary power plus body-force work minus $d_{\rm R}$
and any declared body-drag loss. Local pinning adds its own
$\tfrac12\vct u^{\mathsf T}\mathsf B_\lambda\vct u$ to storage.
This is the continuum form of the finite cancellation derived below,
not an identification of lattice mass or energy density with all blood
inside the cylinder.

Balance of momentum gives the real initial-boundary value problem
\begin{equation}
 \boxed{\rho\Pdtt{u_p}+b_{pq}\Pdt{u_q}
       +B_{\lambda,pq}u_q-\partial_\mu T_{p\mu}=f_p.}
 \label{eq:causal-continuum-pde}
\end{equation}
The tensor $\mathsf b=\mathsf b^{\mathsf T}\geq0$ describes optional drag
relative to eliminated medium motion. The symmetric $\mathsf B_\lambda$
describes an optional stable local wall or target restoring term; a nonlocal
conservative target requires its own symmetric operator. The pair-network
part remains in $\mathcal A$. With $\rho=m/a^3$ for the identical-particle
bulk specialization, every term in~\eqref{eq:causal-continuum-pde} has units of force per
volume. Specifically, $\mathsf b$ has units of mass per volume per time and
$\mathsf B_\lambda$ has units of pressure per length squared. The density
refers to the retained inertial coordinates. A model with an independently
moving medium adds its density and displacement field; it does not simply
replace the particle density by a fluid density.

The construction assumes $|\vct k|a\ll1$ at the specified spacing and a linear
response about the chosen reference. No infinite density or additional
virions are introduced by the interpolation. The initial data are
$\vct u(\vct r,0)$, $\Pdt{\vct u}(\vct r,0)$, and each
$\mathsf Z_a(\vct r,0)$; the last specify the medium's prepared memory.

Partition the material boundary into an engaged stationary slip portion
$\Gamma_{\mathrm w}$ and a remaining free portion
$\Gamma_{\mathrm f}$. The chosen power-free conditions use the
\emph{total} traction:
\begin{equation}
 \vct u\cdot\vct n=0,\qquad
 (\Id_3-\vct n\vct n^{\mathsf T})\vct t=0,
 \quad\text{on }\Gamma_{\mathrm w},\qquad t_p=n_\mu T_{p\mu},
 \qquad \vct t=0\quad\text{on }\Gamma_{\mathrm f}.
 \label{eq:viscoelastic-slip-wall}
\end{equation}
Thus $\Pdt{\vct u}\cdot\vct t=0$. The wall may exchange normal momentum
with the material, while its mechanical power is zero. A lattice filling
the cylinder with an engaged slip boundary is the specialization
$\mathcal V_{\mathrm{lat}}=\mathcal V$ and
$\Gamma_{\mathrm w}=\partial\mathcal V_{\mathrm{lat}}$.
A collision-free array away from the wall instead has a free mechanical
surface unless additional interface forces are prescribed. Nonzero
traction from a retained medium or target adds its boundary power and
partner energy account. These continuum conditions are additional
boundary idealizations; they do not follow from inactive hard-wall
particle reflection. A phonon
packet is encoded in the initial displacement, velocity, or quantum state;
its energy or probability amplitude is not an additional force density.

Figure~\ref{fig:continuum-force-power} shows how opposing face tractions produce the stress-divergence term in~\eqref{eq:causal-continuum-pde}, and how the same forces enter the power balance.

\paragraph{Rotational continuum power uses couple traction.}
For a continuum comparison retaining the independent body rotation
of~\eqref{eq:s3patch-micropolar-strains}, material branches can act on
the scaled strain column
$\vct\varepsilon_{\rm rb}=(\operatorname{vec}\mathsf E,
\ell_*\operatorname{vec}\mathsf K_\theta)$, with fixed length
$\ell_*>0$. Both blocks are dimensionless. Match its relaxed,
viscous and branch quadratic forms to the resolved cell construction.
For energy density $w$, their derivatives give force stress
$\mathsf\sigma=\partial w/\partial\mathsf E$ and couple stress
$\mathsf m=\ell_*\,\partial w/\partial(\ell_*\mathsf K_\theta)$,
after undoing vectorization. Angular inertia density has units
$\mathrm{kg\,m^{-1}}$. The combined boundary power is
\begin{equation}
 \mathcal P_{\partial\mathcal V_{\rm lat}}
    =\int_{\partial\mathcal V_{\rm lat}}
       [\dot{\vct u}\cdot\mathsf\sigma\vct n
        +\dot{\vct\theta}\cdot\mathsf m\vct n]\,\dd S .
 \label{eq:s4patch-micropolar-boundary-power}
\end{equation}
The same branch cancellation gives storage change equal to this power
plus body-force/couple supply minus nonnegative loss. Boundary data
specify both traction and couple channels; zero force traction alone
does not close the angular power account. The displacement-only PDE
above retains its stated translational specialization.

\subsection{Causal memory, medium energy, and complex response}
\label{sec:causal-complex-response}

The finite internal variables can be eliminated explicitly:
\begin{equation}
 \vct z_a(t)=e^{-t/\tau_a}\vct z_a(0)
       +\int_0^t e^{-(t-s)/\tau_a}\mathsf G_a\frac{\dd\vct u(s)}{\dd s}\,\dd s.
 \label{eq:maxwell-history}
\end{equation}

The first term propagates prepared branch strain; the integral
accumulates earlier displacement increments with weight
$e^{-(t-s)/\tau_a}$. Equal present displacement and velocity can thus
coexist with different forces. The retained $\vct z_a$ records exactly
the additional history needed by this closure.

Figure~\ref{fig:viscoelastic-causal-history-map} separates the propagated preparation from the accumulated causal history.

\begin{figure}[H]
\centering\begingroup
\definecolor{vltTeal}{HTML}{176B64}
\definecolor{vltBlue}{HTML}{315D80}
\definecolor{vltGold}{HTML}{A87520}
\definecolor{vltPlum}{HTML}{725877}
\definecolor{vltRed}{HTML}{762B38}
\definecolor{vltYellow}{HTML}{D5AD3C}
\definecolor{vltNeutral}{HTML}{777D83}
\setlength{\unitlength}{1pt}
\scalebox{0.80}{%
\input{vlt-figures/figure-103.tex}%
}\endgroup
\caption{Causal accumulation at observation time $t=3\tau$, with branch index suppressed. An increment at $s$ retains weight $e^{-(t-s)/\tau}$; shading selects $0\leq s\leq t$. The ordinate is a memory weight. Propagated initial strain and accumulated displacement increments add to $\vct z(t)$, in the same displacement-like units. Open arrows identify algebraic maps to branch strain and force. Setting the initial term to zero specifies a relaxed preparation.}
\label{fig:viscoelastic-causal-history-map}
\end{figure}

At a held displacement $u=U_0$, the branch force is
$k_1z(0)e^{-t/\tau}$. Figure~\ref{fig:prepared-memory-state} compares an
unrelaxed preparation $z(0)=U_0$ with the relaxed preparation $z(0)=0$.
Their positions and velocities agree, while their holding forces and
stored branch energies differ.

\begin{figure}[H]
\centering\begingroup
\definecolor{vltTeal}{HTML}{176B64}
\definecolor{vltBlue}{HTML}{315D80}
\definecolor{vltGold}{HTML}{A87520}
\definecolor{vltPlum}{HTML}{725877}
\definecolor{vltRed}{HTML}{762B38}
\definecolor{vltYellow}{HTML}{D5AD3C}
\definecolor{vltNeutral}{HTML}{777D83}
\setlength{\unitlength}{1pt}
\scalebox{0.90}{%
\input{vlt-figures/figure-104.tex}%
}\endgroup
\caption{Two preparations at the same whole-virion displacement $u_0$ and zero velocity. A follows rapid placement from a relaxed origin; B follows a long hold. The placement impulse precedes the comparison. Internal dots differ while rigid bodies, natural span and six-turn coils match. During the subsequent hold, A has $z=u_0e^{-t/\tau}$ and B has $z=0$. Burgundy marks the branch force on A, opposite its holding contribution. The common holding term is $k_0u_0$. The clamp performs zero work during this hold while A releases branch storage. Curves describe held displacement after $0^+$, with fixed coefficients. The boxed result illustrated here is~\eqref{eq:maxwell-memory-equation}.}
\label{fig:prepared-memory-state}
\end{figure}

Consequently, the memory equation has the form
\begin{equation}
 \boxed{\begin{aligned}
 \mathsf M\Dtt{\vct u}+\mathsf K\vct u
      +\mathsf C_0\Dt{\vct u}
      +\int_0^t\mathsf\Gamma_{\mathrm R}(t-s)
                                      \frac{\dd\vct u(s)}{\dd s}\,\dd s
      &=\vct f(t)-\vct h_0(t),\\
 \mathsf\Gamma_{\mathrm R}(t)
      &=\sum_a\mathsf K_a e^{-t/\tau_a},\qquad
          \mathsf K_a=\mathsf G_a^{\mathsf T}\mathsf H_a\mathsf G_a,\\
 \vct h_0(t)&=\sum_a\mathsf G_a^{\mathsf T}\mathsf H_a
                                        e^{-t/\tau_a}\vct z_a(0).
 \end{aligned}}
 \label{eq:maxwell-memory-equation}
\end{equation}
The kernel has units of stiffness and depends only on past motion.
The initial-memory term is essential when the branches have already been
strained. Its sign follows from moving their initial stresses to the
right-hand side.

The scalar kernel $k_1e^{-t/\tau}$ weights delayed force per displacement
increment. For velocity varying slowly over $\tau$, its integral
approaches the resistance $k_1\tau\Dt u$. This low-frequency reduction
applies after the initial-memory transient; an initially strained branch
retains that additional contribution until it relaxes.

\paragraph{Superposition with a prepared material state.}
\label{sec:sup-prepared-response}
Linear memory preserves superposition when its kernel, coordinates and
protocol are fixed. Retaining the material states gives a linear
augmented equation, possibly with prescribed time-dependent coefficients.
Write its deterministic response or conditional mean as
$X[\vct f;X_0]$, with the same complete preparation $X_0$.
Its force dependence is affine:
\begin{equation}
 X[\vct f_1+\vct f_2;X_0]
 =X[\vct f_1;X_0]+X[\vct f_2;X_0]-X[0;X_0].
 \label{eq:sup-prepared-affine-response}
\end{equation}
Subtracting the unforced prepared trajectory prevents counting the
initial strain and motion twice. Equivalently, response increments about
that trajectory add linearly. A common prescribed sequence of linear
event maps preserves this property. If the combined forcing changes
the actual contact sequence, charge population, material coefficients
or basin, those solutions use their updated dynamical laws. The sum of
two separately small motions must itself satisfy the harmonic and
clearance conditions.

The stochastic model also keeps one physical thermal bath. Adding two
independently sampled responses from the same thermal model doubles
their thermal covariance. The combined-force experiment uses the matched
noise law once. For correlated random drives, output spectra retain
their cross-spectra. The driven mean, ensemble covariance and an
individual noise realization therefore require their respective
superposition or averaging rules.

\paragraph{Memory acts on a basis of collective particle motions.}
The conservative modes remain an exact coordinate basis for the finite
problem. Express all matrices first in the justified admissible coordinate
space, and let $\mathsf\Phi=(\vct\phi_1,\ldots,\vct\phi_d)$ contain
a complete real mass-normalized conservative basis in that space:
\begin{equation}
 \mathsf\Phi^{\mathsf T}\mathsf M\mathsf\Phi=\Id_d,\qquad
 \mathsf\Phi^{\mathsf T}\mathsf K\mathsf\Phi
       =\mathsf\Omega^2=\diag(\omega_1^2,\ldots,\omega_d^2),\qquad
 \vct u=\mathsf\Phi\vct Q.
 \label{eq:s4-memory-complete-modal-basis}
\end{equation}
Here $d$ counts all admissible displacement coordinates, including any
unconstrained zero modes. Keeping only selected modes is a further
reduction. Multiplying~\eqref{eq:maxwell-memory-equation} by
$\mathsf\Phi^{\mathsf T}$ gives, with no modal approximation,
\begin{equation}
 \begin{aligned}
 \Dtt{\vct Q}+\mathsf\Omega^2\vct Q
  +\mathsf C_{\rm m}\Dt{\vct Q}
  +\int_0^t\mathsf\Gamma_{\rm m}(t-s)\Dt{\vct Q}(s)\,\dd s
       &=\mathsf\Phi^{\mathsf T}(\vct f-\vct h_0),\\
 \mathsf C_{\rm m}&=\mathsf\Phi^{\mathsf T}\mathsf C_0\mathsf\Phi,\\
 (\mathsf\Gamma_{\rm m}(t))_{ss'}
       &=\sum_a (\mathsf G_a\vct\phi_s)^{\mathsf T}
             \mathsf H_a(\mathsf G_a\vct\phi_{s'})e^{-t/\tau_a}.
 \end{aligned}
 \label{eq:s4-memory-modal-kernel}
\end{equation}
An off-diagonal entry couples two conservative patterns through a shared
relaxing coordinate. A prepared pattern can therefore generate force in
another pattern within linear dynamics. Diagonal entries measure
self-coupling. The memory kernel has units of inverse time squared and
$\mathsf C_{\rm m}$ inverse time. Conservative modes remain a coordinate
basis while the medium determines their coupled evolution.

Independent scalar memory equations result when $\mathsf C_{\rm m}$ and
all distinct-time branch sums are diagonal in one conservative basis.
A sufficient test is pairwise commutation of the real symmetric
mass-weighted matrices $\mathsf D$,
$\mathsf M^{-1/2}\mathsf C_0\mathsf M^{-1/2}$ and
$\mathsf M^{-1/2}(\sum_{a:\tau_a=\tau}\mathsf K_a)
 \mathsf M^{-1/2}$ for each distinct $\tau$.
Within a degenerate conservative eigenspace the same common
diagonalization must be checked. Statistical independence of thermal
channels additionally refers to the matched Gaussian equilibrium
preparation, rather than following from diagonal deterministic equations
alone. This basis-level distinction is familiar in projected
memory equations~\cite{Mori1965}.

For example, two equal scalar masses $m$ with exchange-symmetric
conservative supports have common and relative columns
$\vct\phi_\pm=(1,\pm1)^{\mathsf T}/\sqrt{2m}$.
A single relative branch $\mathsf G=(1,-1)$, $\mathsf H=h$ contributes
\begin{equation}
 \mathsf\Phi^{\mathsf T}\mathsf K_a\mathsf\Phi
      =\begin{pmatrix}0&0\\0&2h/m\end{pmatrix},\qquad
 \mathsf\Phi^{\mathsf T}\diag(\zeta_1,\zeta_2)\mathsf\Phi
      =\frac1{2m}
       \begin{pmatrix}\zeta_1+\zeta_2&\zeta_1-\zeta_2\\
                       \zeta_1-\zeta_2&\zeta_1+\zeta_2\end{pmatrix}.
 \label{eq:s4-memory-two-particle-channels}
\end{equation}
The relative branch senses only separation change. Unequal local medium
drags mix common and relative motion through their off-diagonal term.
Recording the centre average and separation therefore distinguishes
these couplings more directly than assigning one damping time to each
particle. The coordinates describe whole ideal virion translations; the
branch variable describes the medium deformation they share.

Figure~\ref{fig:viscoelastic-memory-pattern-mixing} follows a prepared common pattern as one relaxing medium channel generates relative motion.

\begin{figure}[H]
\centering\begingroup
\definecolor{vltTeal}{HTML}{176B64}
\definecolor{vltBlue}{HTML}{315D80}
\definecolor{vltGold}{HTML}{A87520}
\definecolor{vltPlum}{HTML}{725877}
\definecolor{vltRed}{HTML}{762B38}
\definecolor{vltYellow}{HTML}{D5AD3C}
\definecolor{vltNeutral}{HTML}{777D83}
\setlength{\unitlength}{1pt}
\scalebox{0.90}{%
\input{vlt-figures/figure-105.tex}%
}\endgroup
\caption{One five-state evolution generates all snapshots and traces. The pinned equal-mass pair has $\mathsf K/k_\star=\left(\begin{smallmatrix}2&-1\\-1&2\end{smallmatrix}\right)$, $\omega_\star^2=k_\star/m$, $\mathsf C_0=0.15m\omega_\star\Id$, $h=0.8k_\star$, $\tau\omega_\star=1$ and site-1 branch coupling. Displacements reconstruct as $\vct u=\mathsf\Phi\vct Q$, with $\vct\phi_\pm=(1,\pm1)/\sqrt{2m}$ and gain $14\,\mathrm{pt}$ per $A/\sqrt m$. Shared rulers locate reference particles 1 and 2 at $x/a=0,1$, where $a$ is their reference spacing. Fixed supports and matched coils follow each displacement. A common preparation develops a relative component through the displayed off-diagonal branch matrix.}
\label{fig:viscoelastic-memory-pattern-mixing}
\end{figure}

\begin{proposition}[Storage and transfer within the sealed system]
\label{prop:maxwell-energy}
For constant coefficients in~\eqref{eq:finite-maxwell-system}, define
\begin{equation}
 \begin{split}
 E_{\mathrm{res}}&=\tfrac12\Dt{\vct u}^{\mathsf T}\mathsf M\Dt{\vct u}
       +\tfrac12\vct u^{\mathsf T}\mathsf K\vct u
       +\tfrac12\sum_a\vct z_a^{\mathsf T}\mathsf H_a\vct z_a,\\
 \mathcal D_{\mathrm{res}}&=
       \Dt{\vct u}^{\mathsf T}\mathsf C_0\Dt{\vct u}
       +\sum_a\tau_a^{-1}\vct z_a^{\mathsf T}\mathsf H_a\vct z_a.
 \end{split}
 \label{eq:maxwell-stored-energy}
\end{equation}
Then $\Dt{E_{\mathrm{res}}}=
\Dt{\vct u}^{\mathsf T}\vct f-\mathcal D_{\mathrm{res}}$.
If $\mathsf K\geq0$, the stored energy and dissipation rate are nonnegative.
For the energetic mechanical realization, an unresolved internal-energy
account with $\Dt{E_{\mathrm{th}}}=\mathcal D_{\mathrm{res}}$ gives
$\Dt{E_{\mathrm{res}}}+\Dt{E_{\mathrm{th}}}=0$ during the unforced evolution.
\end{proposition}
\begin{proof}
Multiply the first equation by $\Dt{\vct u}^{\mathsf T}$ and use
\[
 \vct z_a^{\mathsf T}\mathsf H_a\Dt{\vct z_a}
 =\vct z_a^{\mathsf T}\mathsf H_a\mathsf G_a\Dt{\vct u}
  -\tau_a^{-1}\vct z_a^{\mathsf T}\mathsf H_a\vct z_a.
\]
The branch exchange terms cancel. The remaining terms are the stated
power and nonnegative losses.
\end{proof}

The resolved energy combines particle inertia, relaxed configuration
storage and unrelaxed branch deformation. Force times velocity gives
their power balance. Passivity bounds mechanical work by preparation
and supplied work, allowing a prepared branch to return its stored
energy to the particles.

The proposition follows particle and branch storage together; either
part can temporarily gain energy from the other. Its thermal ledger
applies to the declared energetic spring--dashpot realization.
Free-energy storage additionally needs entropy and first-law terms,
as in Section~\ref{sec:phonon-ionic-extension}. Appreciable temperature
change requires compatible thermal evolution and constitutive laws.
With fluctuating forces, the medium account also includes exchanged
random work, the chosen stochastic convention and any drift correction.
That exchange replaces deterministic monotonicity along individual
thermal trajectories.

\begin{example}[Relaxation and creep probe different times]
\label{ex:relaxation-creep-times}
An ideal prescribed displacement step $\vct u=\vct U$ from an initially
relaxed state gives, for $t>0$, the material restoring force
\begin{equation}
 \vct f_{\rm hold}(t)=\mathsf R_{\rm step}(t)\vct U,
 \qquad\mathsf R_{\rm step}(t)=\mathsf K+
                               \sum_a\mathsf K_a e^{-t/\tau_a}.
 \label{eq:step-relaxation-matrix}
\end{equation}
The step's instantaneous impulse and controller work are outside this
$t>0$ constitutive expression. For one scalar branch, denote its relaxed
spring by $k_0>0$ and branch spring by $k_1>0$. In a separately justified
quasistatic creep experiment, negligible resolved inertia and body drag
give $F_0=k_0u+k_1z$ with $\Dt{z}=\Dt{u}-z/\tau$ after a force step.
Solving these equations yields
\begin{equation}
 \frac{u(t)}{F_0}=\frac1{k_0}
    \left[1-\frac{k_1}{k_0+k_1}e^{-t/\tau_c}\right],
 \qquad\tau_c=\tau\frac{k_0+k_1}{k_0}.
 \label{eq:scalar-creep-retardation-time}
\end{equation}
The initial and final compliances are $1/(k_0+k_1)$ and $1/k_0$.
The force-relaxation time $\tau$ and creep time $\tau_c$ differ even
in this one-branch material. A measured decay time must therefore be
associated with its forcing protocol and readout; neither test is the
unforced inertial ringdown of a viral-phonon mode.
\end{example}

Figure~\ref{fig:viscoelastic-relaxation-creep} puts the two protocols on
the same dimensionless time axis. A displacement clamp measures a decreasing
holding force; a force clamp measures an increasing displacement. In a
finite array, the selected displacement and force patterns also matter:
a local deformation and a whole-array displacement need not excite the
same branches.

\begin{figure}[H]
\centering\begingroup
\definecolor{vltTeal}{HTML}{176B64}
\definecolor{vltBlue}{HTML}{315D80}
\definecolor{vltGold}{HTML}{A87520}
\definecolor{vltPlum}{HTML}{725877}
\definecolor{vltRed}{HTML}{762B38}
\definecolor{vltYellow}{HTML}{D5AD3C}
\definecolor{vltNeutral}{HTML}{777D83}
\setlength{\unitlength}{1pt}
\scalebox{0.80}{%
\input{vlt-figures/figure-106.tex}%
}\endgroup
\caption{Displacement hold and quasistatic force hold for one standard linear solid, $k_1=k_0>0$, initially relaxed. Their normalized curves are $1+e^{-t/\tau}$ and $1-\tfrac12e^{-t/(2\tau)}$. Dashed lines give relaxed limits; dots mark one exponential time in each protocol. Curves start at $0^+$ after the ideal step. Lower sketches identify controlled and measured quantities of a whole-virion coordinate. The creep reduction neglects inertia and body drag. Clamp work and branch storage follow~\eqref{eq:maxwell-stored-energy}.}
\label{fig:viscoelastic-relaxation-creep}
\end{figure}

Positive branch sums give a completely monotone relaxation family:
for fixed $\vct U$, the transient quadratic form has nonnegative
$(-1)^j$-weighted time derivatives for every $j\geq0$.
Continuous nonnegative spectra extend this family~\cite{Hanyga2013}.
The oscillator bath below and the hydrodynamic benchmark illustrate
other causal passive memory structures.

There is also a conservative mechanical realization of reduced memory.
Introduce medium oscillator coordinates $Q_\ell$, momenta $P_\ell$, masses
$m_\ell^{\mathrm b}>0$, frequencies $\nu_\ell>0$, and real coupling vectors
$\vct c_\ell$. A quadratic oscillator-bath realization, consistent with generalized
Langevin reduction~\cite{Mori1965,Zwanzig1973}, is
\begin{equation}
 \begin{split}
 H_{\mathrm{full}}={}&\tfrac12\vct p^{\mathsf T}\mathsf M^{-1}\vct p
                 +\tfrac12\vct u^{\mathsf T}\mathsf K\vct u\\
 &+\sum_\ell\left[
       \frac{P_\ell^2}{2m_\ell^{\mathrm b}}
       +\frac{m_\ell^{\mathrm b}\nu_\ell^2}{2}
        \left(Q_\ell-
             \frac{\vct c_\ell^{\mathsf T}\vct u}
                  {m_\ell^{\mathrm b}\nu_\ell^2}\right)^2\right].
 \end{split}
 \label{eq:closed-oscillator-bath}
\end{equation}
The square includes a counterterm: after static minimization over $Q_\ell$,
the relaxed lattice Hessian remains $\mathsf K$. Hamilton's equations conserve
$H_{\mathrm{full}}$. Solving the bath equations and integrating the driven
solution by parts gives the exact reduced equation~\cite{Zwanzig1973}
\begin{equation}
 \begin{split}
 \mathsf M\Dtt{\vct u}+\mathsf K\vct u
   +\int_0^t\mathsf\Gamma_{\mathrm H}(t-s)
                                  \frac{\dd\vct u(s)}{\dd s}\,\dd s&=\vct F_0(t),\\
 \mathsf\Gamma_{\mathrm H}(t)&=
   \sum_\ell\frac{\vct c_\ell\vct c_\ell^{\mathsf T}}
                         {m_\ell^{\mathrm b}\nu_\ell^2}\cos(\nu_\ell t),\\
 \vct F_0(t)&=\sum_\ell\vct c_\ell
      \left[\delta Q_\ell(0)\cos(\nu_\ell t)
       +\frac{P_\ell(0)}{m_\ell^{\mathrm b}\nu_\ell}\sin(\nu_\ell t)\right],\\
 \delta Q_\ell(0)&=Q_\ell(0)
       -\frac{\vct c_\ell^{\mathsf T}\vct u(0)}
                           {m_\ell^{\mathrm b}\nu_\ell^2}.
 \end{split}
 \label{eq:exact-bath-memory}
\end{equation}
No random dynamics is required for this identity; $\vct F_0$ is determined
by the medium's initial coordinates. A finite oscillator bath permits
recurrences. Smooth irreversible relaxation describes a dense-spectrum,
coarse-grained regime over times for which recurrence is unresolved.
The Maxwell closure and the exact oscillator construction are two descriptions
of medium response at different resolutions.

The shifted oscillator square also makes the role of the Hamiltonian
explicit. Its kinetic terms track momenta, and its potential terms track
lattice and medium storage together. Eliminating the medium changes the
equation used for the observed lattice, not the full energy conservation
law. For the same nonsingular quadratic kinetic energy, a Lagrangian with
these coordinates is equally suitable for deriving the motion; the
Hamiltonian is especially convenient here for displaying the energy ledger,
canonical quantization, and the subsystem being traced out.

For a conditional canonical preparation of the classical bath at temperature
$T$, centered on the prescribed $\vct u(0)$, its fluctuating force satisfies
\begin{equation}
 \langle\vct F_0(t)\rangle=0,\qquad
 \langle\vct F_0(t)\vct F_0(t')^{\mathsf T}\rangle
                 =k_{\mathrm B}T\mathsf\Gamma_{\mathrm H}(t-t').
 \label{eq:classical-bath-fdt}
\end{equation}
This follows directly from
$\langle\delta Q_\ell^2\rangle=k_{\mathrm B}T/
(m_\ell^{\mathrm b}\nu_\ell^2)$ and
$\langle P_\ell^2\rangle=m_\ell^{\mathrm b}k_{\mathrm B}T$.
The preparation specifies the ensemble; each realization still conserves its
complete energy. A bath initially centered elsewhere supplies a corresponding
initial transient. Equilibrium noise and dissipation must therefore be
specified together when thermal fluctuations are retained.

For the Maxwell closure itself, a stationary classical thermal reduction
at approximately constant $T$ requires a fluctuating force $\vct\Xi$
matched to the chosen loss:
\begin{align}
 \langle\vct\Xi(t)\vct\Xi(t')^{\mathsf T}\rangle
  &=2k_BT\mathsf C_0\delta(t-t')+
      k_BT\sum_a\mathsf K_a e^{-|t-t'|/\tau_a},\nonumber\\
 \mathsf S_\Xi(\omega)
  &=2k_BT\left[\mathsf C_0+
        \sum_a\frac{\tau_a\mathsf K_a}{1+\omega^2\tau_a^2}\right].
 \label{eq:maxwell-stationary-thermal-force}
\end{align}
The exponential part can be realized by independent stationary
Ornstein--Uhlenbeck force processes. Initial branch fluctuations and
their preparation determine whether this stationary covariance applies;
an arbitrary strained deterministic branch does not have it. White body
drag and colored branch noise are thus specified together. Equilibrium
fluctuations balance mean dissipative loss rather than obeying a purely
decaying displacement. This extends the Markov thermal construction of
Section~\ref{sec:operator-thermal-stochastic} to resolved memory.

The covariance specifies what ``fluctuation'' means in this reduction:
deviations about the prepared ensemble mean, with correlations between two
times. The Dirac delta describes an ideal white force with negligible
correlation time on the observation scale, whereas the exponential terms
describe colored forces whose correlations persist for $\tau_a$.
An Ornstein--Uhlenbeck realization is a linear randomly driven relaxation
process with precisely this exponential stationary covariance.

\paragraph{A complete thermal preparation of the finite branch model.}
The colored-force formula can be generated by explicit finite state
equations. Auxiliary-variable realizations and their covariance identities
provide a useful route from memory to such stochastic
states~\cite{CeriottiBussiParrinello2010}. For the present mechanical
branches, use $\mathsf K>0$ on a declared supported or constrained
coordinate space and discard inactive null directions of each $\mathsf H_a$, leaving
$\mathsf H_a>0$ on its $d_a$ retained branch coordinates. At fixed
$T$, one matched additive-noise realization is
\begin{equation}
 \begin{aligned}
 \dd\vct u&=\mathsf M^{-1}\vct p\,\dd t,\\
 \dd\vct p&=\left[-\mathsf K\vct u
       -\mathsf C_0\mathsf M^{-1}\vct p
       -\sum_a\mathsf G_a^{\mathsf T}\mathsf H_a\vct z_a\right]\dd t
       +(2k_{\rm B}T\mathsf C_0)^{1/2}\dd\vct W_0,\\
 \dd\vct z_a&=(\mathsf G_a\mathsf M^{-1}\vct p
                    -\tau_a^{-1}\vct z_a)\dd t
       +(2k_{\rm B}T\tau_a^{-1}\mathsf H_a^{-1})^{1/2}
                                                    \dd\vct W_a.
 \end{aligned}
 \label{eq:s4-memory-thermal-state-embedding}
\end{equation}
The $\vct W_a$ are independent standard Wiener processes; positive
matrix square roots set the force and branch noise covariances.
Because the coefficients are constant, the It\^o and Stratonovich forms
have the same state drift. The conservative exchange between $\vct p$
and $\vct z_a$ is antisymmetric in the energy balance. Each branch's
relaxation and noise form a matched Ornstein--Uhlenbeck pair.

Writing $E_{\rm res}$ of~\eqref{eq:maxwell-stored-energy} in
$(\vct u,\vct p,\{\vct z_a\})$ gives the invariant density and its
nonzero covariance blocks
\begin{equation}
 \begin{aligned}
 \rho_{\rm eq}&=Z^{-1}e^{-E_{\rm res}/(k_{\rm B}T)},\\
 \langle\vct u\vct u^{\mathsf T}\rangle_{\rm eq}
      &=k_{\rm B}T\mathsf K^{-1},\\
 \langle\vct p\vct p^{\mathsf T}\rangle_{\rm eq}
      &=k_{\rm B}T\mathsf M,\\
 \langle\vct z_a\vct z_b^{\mathsf T}\rangle_{\rm eq}
      &=\delta_{ab}k_{\rm B}T\mathsf H_a^{-1}.
 \end{aligned}
 \label{eq:s4-memory-thermal-state-covariance}
\end{equation}
All cross blocks between $\vct u$, $\vct p$ and the branches vanish in
these variables. Eliminating the branch equations with this preparation
recovers~\eqref{eq:maxwell-stationary-thermal-force}: the freely relaxing
branch fluctuation, including its propagated initial fluctuation and
subsequent branch noise, has covariance
$k_{\rm B}T\mathsf H_a^{-1}e^{-|t-t'|/\tau_a}$, and multiplying by
$-\mathsf G_a^{\mathsf T}\mathsf H_a$ gives its force covariance
$k_{\rm B}T\mathsf K_a e^{-|t-t'|/\tau_a}$.
A prescribed nonzero mean branch strain adds the decaying preparation
force already identified in~\eqref{eq:maxwell-memory-equation}.
Specifying $\vct z_a(0)=\vct0$ exactly instead removes its equilibrium
variance and produces a different initial ensemble.

\paragraph{Finite preparation gives an explicit covariance correction.}
In the fixed linear realization~\eqref{eq:s4patch-resistance-whitening},
let $\vct r_\chi=\vct\chi'+\mathsf A^{-1}\mathsf B'^{\mathsf T}\vct X$,
$\mathsf E_\chi(t)=e^{-\mathsf D^{-1}\mathsf A t}$ and
$\mathsf\Gamma_\chi(t)=\mathsf B'\mathsf E_\chi(t)
\mathsf A^{-1}\mathsf B'^{\mathsf T}$. Suppose the prepared
$\vct r_\chi(0)$ has mean $\vct m_0$ and covariance
$\mathsf\Sigma_0$, independent of subsequent Wiener increments.
The freely propagated hidden force on the mechanical right-hand side
has mean $-\mathsf B'\mathsf E_\chi(t)\vct m_0$ and centred covariance
\begin{equation}
 \begin{aligned}
 \operatorname{Cov}[\vct\xi_h(t),\vct\xi_h(t')]
  &=k_{\rm B}T\mathsf\Gamma_\chi(|t-t'|)\\
  &\quad+\mathsf B'\mathsf E_\chi(t)
     [\mathsf\Sigma_0-k_{\rm B}T\mathsf A^{-1}]
            \mathsf E_\chi(t')^{\mathsf T}\mathsf B'^{\mathsf T}.
 \end{aligned}
 \label{eq:s4patch-prepared-force-covariance}
\end{equation}
Solving its linear Ornstein--Uhlenbeck equation propagates the initial
covariance and integrates the matched noise covariance. Subtracting
the invariant covariance $k_{\rm B}T\mathsf A^{-1}$ gives the
displayed correction. The independent whitened body force adds
$2k_{\rm B}T\mathsf C_{\rm eff}\delta(t-t')$.
A nonzero prepared mean can coexist with stationary centred force
covariance when $\mathsf\Sigma_0=k_{\rm B}T\mathsf A^{-1}$.
Fixing the initial strain exactly instead sets $\mathsf\Sigma_0=0$.
Neither choice alone makes particle records stationary, and any
mechanical--hidden initial cross covariance is propagated too.
Finite-time spectral differences can therefore reflect preparation
even with an unchanged passive bath. Gaussian population fluctuations
apply locally within their physical occupancy range; a bounded or
jump model supplies the corresponding global statistics.

Figure~\ref{fig:s34patch-finite-preparation-covariance} separates preparation age from lag. A matched memory kernel fixes the stationary reference, while the initial internal covariance determines the transient correction.

\begin{figure}[H]
\centering\begingroup
\setlength{\unitlength}{1pt}
\scalebox{0.86}{%
\input{vlt-figures/figure-295.tex}%
}\endgroup
\caption{Preparation controls the early-time force covariance. A scalar retained state relaxes with time $\tau$, the mechanical coordinate is held fixed, and both preparations have zero mean. The thermal initial state has $\Sigma_0=k_BT/A$; the prepared state has $r_0=0$ and the same subsequent matched thermal forcing. With $K=k_BT(B^{\prime})^2/A$, their coloured-force covariances are respectively $K e^{-|t-t^{\prime}|/\tau}$ and $K[e^{-|t-t^{\prime}|/\tau}-e^{-(t+t^{\prime})/\tau}]$. The left plot evaluates equal times; the right fixes the lag at $\tau$ and varies preparation age. These are exact scalar evaluations of~\eqref{eq:s4patch-prepared-force-covariance}. The white instantaneous-drag contribution is excluded from both plotted quantities. The internal coordinate is a local Gaussian perturbation; the upper curves symbolize its initial probability law rather than a particle displacement trajectory.}
\label{fig:s34patch-finite-preparation-covariance}
\end{figure}

The ensemble energy identity makes the thermal exchange explicit:
\begin{equation}
 \frac{\dd\langle E_{\rm res}\rangle}{\dd t}
 =-\langle\mathcal D_{\rm res}\rangle
   +k_{\rm B}T\operatorname{tr}(\mathsf M^{-1}\mathsf C_0)
   +k_{\rm B}T\sum_a\frac{d_a}{\tau_a}.
 \label{eq:s4-memory-thermal-energy-balance}
\end{equation}
The positive terms are the It\^o noise contributions. At the invariant
state they exactly balance the mean dissipative loss. For a
displacement readout $Y=\vct w^{\mathsf T}\vct u$, the equal-time
variance remains $k_{\rm B}T\vct w^{\mathsf T}\mathsf K^{-1}\vct w$,
while $\mathsf G_a$, $\mathsf H_a$ and $\tau_a$ change the two-time
correlations and spectrum. Thus equal static fluctuations can coexist
with different histories and decay patterns.

Figure~\ref{fig:viscoelastic-static-dynamic-fluctuations} compares equal static variance with distinct lagged correlations in two matched thermal Maxwell media.

\begin{figure}[H]
\centering\begingroup
\definecolor{vltTeal}{HTML}{176B64}
\definecolor{vltBlue}{HTML}{315D80}
\definecolor{vltGold}{HTML}{A87520}
\definecolor{vltPlum}{HTML}{725877}
\definecolor{vltRed}{HTML}{762B38}
\definecolor{vltYellow}{HTML}{D5AD3C}
\definecolor{vltNeutral}{HTML}{777D83}
\setlength{\unitlength}{1pt}
\scalebox{0.80}{%
\input{vlt-figures/figure-107.tex}%
}\endgroup
\caption{Changing relaxation time at fixed mass, relaxed stiffness, drag and branch stiffness. The supported coordinate has $G=1$ and $\omega_\star^2=k_\star/m$. Each medium carries its matched equilibrium noise at the same temperature. Exact three-state propagation gives $C_u(t)=\langle u(t)u(0)\rangle$ with common variance $k_{\rm B}T/k_\star$ and different time dependence. The identical virions locate the same reference; symmetric arrows denote fluctuations. Static variance constrains relaxed stiffness, while the correlations additionally probe memory.}
\label{fig:viscoelastic-static-dynamic-fluctuations}
\end{figure}
 An invariant density is
distinct from convergence toward it from every preparation; the
undamped-subspace test in~\eqref{eq:finite-damping-detectability} controls
that issue for the finite linear model. The constant-$T$ realization
represents an equilibrium reservoir over the observation window. A finite
thermally isolated medium instead needs its evolving thermal state in
the complete energy account.

In the blood example, a passive fluctuation model therefore needs
more than a rheological fit: it needs a sufficiently stationary
reference and a justified equilibrium preparation, or a locally
equilibrated reservoir approximated at a common temperature over the
coupled observation window. Maintained or evolving cellular or
chemical activity and imposed flow can require additional fluctuating
forces whose covariance is not fixed by the equilibrium formula.
Stationary force statistics also do not guarantee stationary position
statistics in an unconstrained zero mode. A finite positive or properly
constrained restoring sector is required for the harmonic equilibrium
displacement covariance used in Section~3.

The factor $k_BT$ sets the thermal energy scale; the resistance determines
how that fluctuating forcing is coupled to each displacement combination
\cite{Cotterill2002}. The same channel that removes a prepared deformation
also exchanges energy with thermally moving medium coordinates.

Now define the temporal transform by
\begin{equation}
 \widehat{\vct u}(\omega)=\int_{-\infty}^{\infty}
                  \vct u(t)e^{i\omega t}\,\dd t,
 \qquad \vct u(t)=\frac1{2\pi}\int_{-\infty}^{\infty}
             \widehat{\vct u}(\omega)e^{-i\omega t}\,\dd\omega.
 \label{eq:viscoelastic-fourier-convention}
\end{equation}
A complex phasor denotes a real motion through its real part. For a stationary
harmonic response after preparation transients, the Maxwell dynamic stiffness is
\begin{equation}
 \begin{split}
 \mathsf K^*(\omega)&=\mathsf K-i\omega\mathsf C_0
          +\sum_a\mathsf K_a\frac{-i\omega\tau_a}{1-i\omega\tau_a}
                       =\mathsf K'(\omega)-i\mathsf K''(\omega),\\
 \mathsf K'(\omega)&=\mathsf K+
             \sum_a\mathsf K_a\frac{(\omega\tau_a)^2}
                                      {1+(\omega\tau_a)^2},\\
 \mathsf K''(\omega)&=\omega\mathsf C_0+
             \sum_a\mathsf K_a\frac{\omega\tau_a}
                                      {1+(\omega\tau_a)^2},\qquad\omega>0.
 \end{split}
 \label{eq:maxwell-dynamic-stiffness}
\end{equation}

\begin{figure}[H]
\centering\begingroup
\definecolor{vltTeal}{HTML}{176B64}
\definecolor{vltBlue}{HTML}{315D80}
\definecolor{vltGold}{HTML}{A87520}
\definecolor{vltPlum}{HTML}{725877}
\definecolor{vltRed}{HTML}{762B38}
\definecolor{vltYellow}{HTML}{D5AD3C}
\definecolor{vltNeutral}{HTML}{777D83}
\setlength{\unitlength}{1pt}
\scalebox{0.90}{%
\input{vlt-figures/figure-108.tex}%
}\endgroup
\caption{The same one-branch material in frequency and strain coordinates, using \eqref{eq:maxwell-dynamic-stiffness}. Storage increases from the relaxed $k_0$ toward $2k_0$; positive loss peaks at $\omega\tau=1$ with value $k_0/2$. The $e^{-i\omega t}$ convention makes $k^{\prime\prime}=-\operatorname{Im}k^*$ and mean material power $\omega k^{\prime\prime}|U|^2/2\ge0$. The lower snapshots sample the periodic steady state at $\omega t=0$, where $u=U>0$, $y/U=1/(1+x_\omega^2)$ and $z/U=x_\omega^2/(1+x_\omega^2)$. Body displacement, six coil turns and natural span are shared; one $14\,\mathrm{pt}$ gain moves both particle and material node. The excess coil span is $14z/U$ in points. These instantaneous strains complement the full-cycle constitutive coefficients; the internal dot represents the medium coordinate.}
\label{fig:viscoelastic-storage-loss}
\end{figure}

The branch sum also allows more than one relaxation scale, as the dimensionless example in Figure~\ref{fig:added-two-relaxation-scales} illustrates.

\begin{figure}[H]
\centering\begingroup
\definecolor{vltTeal}{HTML}{176B64}
\definecolor{vltBlue}{HTML}{315D80}
\definecolor{vltGold}{HTML}{A87520}
\definecolor{vltPlum}{HTML}{725877}
\definecolor{vltRed}{HTML}{762B38}
\definecolor{vltYellow}{HTML}{D5AD3C}
\definecolor{vltNeutral}{HTML}{777D83}
\setlength{\unitlength}{1pt}
\scalebox{0.80}{%
\input{vlt-figures/figure-109.tex}%
}\endgroup
\caption{Two positive branch weights sum to one in $K^*/k_0=1+\sum_a\rho_a[-i\omega\tau_a/(1-i\omega\tau_a)]$. Storage rises from one to two. Branch centres occur at $\omega\tau_1=1,50$; the right panel separates their nonnegative losses beneath the black sum. Individual contributions peak at their centres, while maxima of the sum can shift. Distinct dash patterns and direct labels preserve identification in grayscale. Both branches load the same coordinate.}
\label{fig:added-two-relaxation-scales}
\end{figure}

The storage part $\mathsf K'$ is the force component in phase with
displacement; the loss part $\mathsf K''$ is the component in phase with
velocity for this convention. For a scalar deformation
$u=A\cos(\omega t)$, the constitutive holding force, excluding the inertial
term, is $f_{\rm hold}=A[k'\cos(\omega t)-k''\sin(\omega t)]$.
Its storage component returns the work done over a complete cycle;
its velocity component contributes positive net work to the material.
For one branch with no instantaneous drag, $x_\omega=\omega\tau\ll1$
allows relaxation during each cycle, while $x_\omega\gg1$ nearly freezes
the internal coordinate. Figure~\ref{fig:viscoelastic-storage-loss}
shows the crossover between these limits. Its branch loss maximum is
a constitutive relaxation feature, not by itself a mechanical resonance.

For later fluctuation formulas, define the \emph{total} Maxwell resistance,
including the instantaneous drag,
\begin{equation}
 \mathsf Z_{\rm eff}(z)=\mathsf C_0+
       \sum_a\frac{\tau_a\mathsf K_a}{1-iz\tau_a},\qquad
 \mathsf K^*(z)=\mathsf K-iz\mathsf Z_{\rm eff}(z).
 \label{eq:total-maxwell-resistance}
\end{equation}
Thus $\mathsf K''(\omega)=\omega\operatorname{Re}_{\rm H}
\mathsf Z_{\rm eff}(\omega)$ for real positive $\omega$.
Here $\operatorname{Re}_{\rm H}\mathsf X=(\mathsf X+\mathsf X^\dagger)/2$;
for these real symmetric branch matrices it is the elementwise real part.

\begin{samepage}
\noindent The corresponding continuum tensor is
\begin{equation}
 \mathcal A^*(\omega)=\mathcal A-i\omega\mathcal D^{\mathrm v}
       +\sum_a\mathcal A^{(a)}
                              \frac{-i\omega\tau_a}{1-i\omega\tau_a}.
 \label{eq:maxwell-dynamic-tensor}
\end{equation}
\end{samepage}
For this $e^{-i\omega t}$ convention the loss part has a \emph{negative}
imaginary sign. The cycle-averaged mechanical power delivered to the finite
passive material by a displacement phasor $\vct U$ is
\begin{equation}
 \overline{\mathcal P}_{\mathrm{loss}}
          =\frac{\omega}{2}\vct U^\dagger
                              \mathsf K''(\omega)\vct U\geq0.
 \label{eq:average-loss-power}
\end{equation}
It follows by averaging the product of the real restoring force and velocity.
The continuum strain or gradient contraction gives the analogous power per
volume. The material can have real time-domain equations and a complex
frequency response simultaneously.

For the scalar test, integrating the power over one period gives
\begin{equation}
 \boxed{W_{\rm cycle}=\oint f_{\rm hold}\,\dd u
                =\pi A^2k''(\omega),\qquad
         \overline{\mathcal P}_{\rm loss}
                =\frac{\omega}{2\pi}W_{\rm cycle}.}
 \label{eq:scalar-cycle-loss-work}
\end{equation}
The enclosed force--displacement loop area is the work dissipated by the
selected material response in one cycle. Frequency-dependent phase and loop area provide
two directly interpretable ways of reporting the same constitutive loss.

Figure~\ref{fig:maxwell-cycle-work} shows the loop for three rates in
one fixed branch model. The loop is narrow when the branch nearly
relaxes within a cycle and also when it is nearly frozen. Its largest
normalized area occurs in the intervening range $\omega\tau=1$.
A narrow high-rate loop does not imply a small holding force: the
frozen branch increases storage stiffness. Likewise, the net work
per cycle and the mean power differ by the cycle duration, so their
frequency dependence need not have the same maximum.

\begin{figure}[H]
\centering\begingroup
\definecolor{vltTeal}{HTML}{176B64}
\definecolor{vltBlue}{HTML}{315D80}
\definecolor{vltGold}{HTML}{A87520}
\definecolor{vltPlum}{HTML}{725877}
\definecolor{vltRed}{HTML}{762B38}
\definecolor{vltYellow}{HTML}{D5AD3C}
\definecolor{vltNeutral}{HTML}{777D83}
\setlength{\unitlength}{1pt}
\scalebox{0.90}{%
\input{vlt-figures/figure-110.tex}%
}\endgroup
\caption{Periodic holding force for $k_1=k_0>0$, $c_0=0$ and $u=U\cos\theta$. Each loop uses $f/(k_0U)=[1+x_\omega^2/(1+x_\omega^2)]\cos\theta-[x_\omega/(1+x_\omega^2)]\sin\theta$ on shared axes. Open arrows advance phase; dashed in-phase lines enclose zero work. Shaded areas give~\eqref{eq:scalar-cycle-loss-work}. The lower row samples the middle loop: rulers mark $u/U=-1,0,1$ at $11\,\mathrm{pt}/U$, and burgundy holding-force arrows use $12\,\mathrm{pt}/(k_0U)$. The two zero-displacement phases have opposite forces. Resolved inertia is treated separately.}
\label{fig:maxwell-cycle-work}
\end{figure}

Causality imposes more than a sign at one frequency. For an integrable real
memory kernel, its retarded transform
$\mathsf Z(z)=\int_0^\infty\mathsf\Gamma(t)e^{izt}\,\dd t$
is analytic for $\operatorname{Im}z>0$, and obeys the reality symmetry
$\mathsf Z(-\omega)=\mathsf Z(\omega)^*$. The Maxwell poles lie at
$z=-i/\tau_a$. The storage and loss cannot be chosen independently over the
whole frequency axis. For a response component $h(z)$ that decays sufficiently
at infinity, the Kramers--Kronig identities are~\cite{Tokmakoff2009}
\begin{equation}
 \begin{split}
 \operatorname{Re}h(\omega)&=\frac1\pi\operatorname{PV}
        \int_{-\infty}^{\infty}
                  \frac{\operatorname{Im}h(\omega')}{\omega'-\omega}
                                                        \,\dd\omega',\\
 \operatorname{Im}h(\omega)&=-\frac1\pi\operatorname{PV}
        \int_{-\infty}^{\infty}
                  \frac{\operatorname{Re}h(\omega')}{\omega'-\omega}
                                                        \,\dd\omega'.
 \end{split}
 \label{eq:retarded-kramers-kronig}
\end{equation}
Here $\operatorname{PV}$ denotes the Cauchy principal value: the
symmetric limiting integral obtained by excluding an increasingly
small interval around $\omega'=\omega$. These identities link the real and imaginary response quadratures across
frequency as a consequence of causality. Passivity imposes the additional
work constraint; it does not follow from analyticity alone. For
$\mathsf K^*$, subtract the instantaneous polynomial
$\mathsf K+\sum_a\mathsf K_a-i\omega\mathsf C_0$ before applying this
decaying-response form componentwise. An arbitrary complex coefficient
depending on time supplies neither this analyticity nor a memory law.

Removing the loss terms recovers conservative motion with the appropriate
retained storage coordinates. Keeping a real viscous term still dissipates
resolved energy and gives a non-self-adjoint evolution generator. Reality of
the coefficients and self-adjointness of conservative mechanical operators
are separate properties.

A complex susceptibility consequently describes the phase and magnitude
of a real displacement readout. It is not a claim that a biological
measurement has an imaginary value. Self-adjointness pertains to the
complete conservative Hamiltonian and its real observable spectrum;
the reduced damped generator accounts for transfer into omitted coordinates.

\subsubsection{Reducing cell interiors preserves their loss and thermal forcing}
\label{sec:s4-cell-retarded-ports}

The conservative port construction in~\eqref{eq:s3-cell-port-schur}
extends to the full causal viscoelastic operator. It allows a cell to
be replaced, for a specified calculation, by the force--motion relation
at its retained interface and observation sites. Eliminated interior
particles continue to store inertia, dissipate energy and transmit
their prepared or thermal forcing. This is a finite-matrix instance of
the memory and fluctuating-force pairing in mechanical reduction
\cite{Mori1965,Zwanzig1973}.

\begin{definition}[Retarded cell port operator]
\label{def:s4-cell-retarded-port}
At a fixed reference and real $\omega>0$, put
$\mathsf A=\mathsf K^*(\omega)-\omega^2\mathsf M$ and order the
displacements as $(\Gamma,I)$, retaining the physical ports of
Section~3. Where $\mathsf A_{II}$ is invertible, define
\begin{equation}
 \begin{aligned}
 \mathsf S_\Gamma
   &=\mathsf A_{\Gamma\Gamma}
       -\mathsf A_{\Gamma I}\mathsf A_{II}^{-1}\mathsf A_{I\Gamma},\\
 \mathsf L_\Gamma
   &=[\Id,\,-\mathsf A_{\Gamma I}\mathsf A_{II}^{-1}],
 &\mathsf P_\Gamma
   &=\begin{pmatrix}\Id\\-\mathsf A_{II}^{-1}\mathsf A_{I\Gamma}\end{pmatrix}.
 \end{aligned}
 \label{eq:s4-cell-retarded-port-maps}
\end{equation}
Here $\mathsf L_\Gamma$ maps a full force to its effective port force,
and $\mathsf P_\Gamma$ reconstructs the displacement under port-only
forcing. Both maps are dimensionless. The operator $\mathsf S_\Gamma$
has stiffness units. With deterministic force $\vct F$ and fluctuating
force $\vct\Xi$,
\begin{equation}
 \mathsf S_\Gamma\vct U_\Gamma
       =\mathsf L_\Gamma\vct F+\vct\Xi_\Gamma,
 \qquad \vct\Xi_\Gamma=\mathsf L_\Gamma\vct\Xi.
 \label{eq:s4-cell-effective-port-forcing}
\end{equation}
Prepared interior motion or branch strain supplies its corresponding
effective source. No independent relaxation of that preparation is
inserted merely because a coordinate is eliminated.
\end{definition}

\begin{proposition}[Loss and equilibrium noise survive exact port reduction]
\label{prop:s4-cell-port-fdt}
Write $\operatorname{Im}_{\rm H}\mathsf X
=(\mathsf X-\mathsf X^\dagger)/(2i)$ and define
$\mathsf S_\Gamma''=-\operatorname{Im}_{\rm H}\mathsf S_\Gamma$.
For the passive Maxwell realization,
$-\operatorname{Im}_{\rm H}\mathsf A=\mathsf K''\succeq0$.
For the coherent port-only response, with no independent interior drive,
the loss and mean supplied power obey
\begin{equation}
 \mathsf S_\Gamma''
  =\mathsf P_\Gamma^\dagger\mathsf K''\mathsf P_\Gamma
  =\mathsf L_\Gamma\mathsf K''\mathsf L_\Gamma^\dagger\succeq0,
 \qquad
 \overline{\mathcal P}_\Gamma
   =\frac\omega2\vct U_\Gamma^\dagger
                  \mathsf S_\Gamma''\vct U_\Gamma\geq0.
 \label{eq:s4-cell-port-loss}
\end{equation}
For the stationary classical equilibrium preparation at common
temperature $T$, with the two-sided force spectrum already specified,
the reduced force spectrum is
\begin{equation}
 \mathsf S_{\Xi_\Gamma}(\omega)
   =\mathsf L_\Gamma\mathsf S_\Xi(\omega)\mathsf L_\Gamma^\dagger
   =\frac{2k_{\rm B}T}{\omega}\mathsf S_\Gamma''(\omega).
 \label{eq:s4-cell-port-thermal-spectrum}
\end{equation}
The reduced displacement response equals the full port block:
$\mathsf S_\Gamma^{-1}=[\mathsf A^{-1}]_{\Gamma\Gamma}$ wherever
these inverses exist.
\end{proposition}
\begin{proof}
Block elimination gives
$\mathsf A\mathsf P_\Gamma=(\mathsf S_\Gamma,0)^{\mathsf T}$ and
$\mathsf L_\Gamma\mathsf A=(\mathsf S_\Gamma,0)$.
Hence $\mathsf P_\Gamma^\dagger\mathsf A\mathsf P_\Gamma
=\mathsf S_\Gamma
=\mathsf L_\Gamma\mathsf A\mathsf L_\Gamma^\dagger$.
Taking negative Hermitian imaginary parts proves the loss identities.
The power follows from $\vct F_\Gamma=\mathsf S_\Gamma\vct U_\Gamma$
and $\vct V_\Gamma=-i\omega\vct U_\Gamma$. Finally,
$\mathsf S_\Xi=2k_{\rm B}T\operatorname{Re}_{\rm H}\mathsf Z_{\rm eff}
=(2k_{\rm B}T/\omega)\mathsf K''$; applying the force map proves
the noise identity. The inverse-block relation is the usual Schur
identity.
\end{proof}

Even with no deterministic force applied to an interior particle, that
particle's coupled bath contributes to $\vct\Xi_\Gamma$. Keeping the
reduced compliance while replacing its noise by the original port-only
noise would generally give the wrong fluctuation spectrum. Interior--port
noise correlations must also remain in the congruence. A driven or
otherwise nonequilibrium source uses its actual covariance instead of
the equilibrium equality. Exact mechanical elimination preserves a
matched equilibrium preparation; it does not turn an arbitrary prepared
state into equilibrium.

The resulting interface response can contain hidden-particle resonances
as well as material relaxation. It need not remain a finite positive
Maxwell sum after inertial particle coordinates are removed. Its full
retarded operator retains those poles and the associated effective
forcing. This supplies a practical criterion for retaining an interior
mode explicitly: it is needed when its resonance or preparation changes
the selected force/displacement or fluctuation channel over the required
band. A fit judged only by the static port stiffness can miss that
effect.

Independent cell elimination is exact when the retained interface set
makes $\mathsf A_{II}$ block diagonal, including the medium couplings.
Then the Schur corrections can be computed and cached cell by cell.
Long-range electrostatic or shared-fluid couplings can instead connect
two proposed interiors; those coordinates are promoted to ports or the
coupled interior block is eliminated jointly. The same tests apply to a
four-particle module and to a thousand-particle assembly. Computational
economy follows from the actual interface structure and the retained
observation channels, while particle count and energy accounting remain
global.

Figure~\ref{fig:s4-retarded-cell-port-reduction} compares a full cell with its exact port model, including the interior source and its matched fluctuation spectrum.

\begin{figure}[H]
\centering\begingroup
\definecolor{vltTeal}{HTML}{176B64}
\definecolor{vltBlue}{HTML}{315D80}
\definecolor{vltGold}{HTML}{A87520}
\definecolor{vltPlum}{HTML}{725877}
\definecolor{vltRed}{HTML}{762B38}
\definecolor{vltYellow}{HTML}{D5AD3C}
\definecolor{vltNeutral}{HTML}{777D83}
\setlength{\unitlength}{1pt}
\scalebox{0.90}{%
\input{vlt-figures/figure-111.tex}%
}\endgroup
\caption{Exact interior elimination in a supported three-virion scalar cell. Physical sites are $1,2,3$; ports $\Gamma_1,\Gamma_2$ are sites 1 and 3, and $I$ is site 2. The auxiliary matrices are $M/m_*=I$, $K/(m_*\omega_*^2)=0.7I+L_{\rm path}$ and $C/(m_*\omega_*)=\operatorname{diag}(0.12,0.6,0.12)$. The reduced port response retains interior force and noise through $L_\Gamma$. Panel (c) gives effective force at either port for unit interior forcing; (d) gives diagonal and cross spectra, compared with the original port-only level $0.12$. Here $c_*=m_*\omega_*$ and the classical equilibrium noise spectrum is two-sided. Reduction preserves the interior influence and its thermal channel.}
\label{fig:s4-retarded-cell-port-reduction}
\end{figure}

\subsection{Damped viral-phonon modes and prepared states}
\label{sec:damped-phonon-states}

The fixed array of ideal virions can move together, move against one another,
or exchange stored mechanical energy with the surrounding medium. The
reference Hessian determines the relaxed restoring forces, while the
retained medium variables determine how those forces depend on the history
of motion. A preparation specifies the disturbance whose evolution is to
be predicted.

\begin{definition}[Collective mechanical excitation]
\label{def:s4-collective-excitation}
For a declared reference and coordinate resolution, a \emph{collective
mechanical excitation} is a prepared departure of the retained mechanical
state from that reference. In the finite classical Maxwell realization,
this state consists of particle displacements $\vct u$, momenta
$\vct p=\mathsf M\dot{\vct u}$, and retained branch extensions
$\{\vct z_a\}$. Its subsequent motion can be oscillatory, traveling, or
relaxational. The particle count remains $n$.
\end{definition}

For example, the same initial particle positions and velocities can lead
to different later accelerations when the medium carries different initial
branch stresses. Conversely, common translation and a change of
interparticle separation prepare different collective motions even if a
single detector assigns them the same characteristic frequency. The
state variables specify what was prepared; the response equations specify
how it evolves.

A normal mode identifies a pattern of motion; an excitation specifies how
that pattern is populated or prepared. A finite classical network has a
finite list of mode frequencies, and cavity boundaries can discretize
allowed wavelengths. The amplitudes and energies of these classical modes
remain continuously variable. Under the additional canonical quantization postulate, a positive
harmonic mode has occupation energies
$E_{N_s}=\hbar\omega_s(N_s+\tfrac12)$, $N_s=0,1,\ldots$; one phonon is
one quantum of that mode~\cite{Misra2012,Simon2011}. An excitation can also
start from a thermally occupied state. General quantum Hamiltonians may
have discrete bound levels, continuous spectra, or both~\cite{Teschl2009}.
An excitation thus specifies a prepared disturbance within the chosen
model. Resolving quantum occupation experimentally additionally requires
measurements that distinguish its energy increments and state statistics.

\subsubsection{Response poles and the lifetime of an oscillation}

The frequency-domain finite equation is
\begin{equation}
 [\mathsf K^*(\omega)-\omega^2\mathsf M]\widehat{\vct u}(\omega)
                                      =\widehat{\vct f}_{\mathrm{eff}}.
 \label{eq:finite-retarded-pencil}
\end{equation}
The effective forcing includes the specified preparation and initial-memory
terms. This equation relates temporal Fourier transforms. For harmonic
forcing, the same pencil relates a displacement phasor $\vct U$ to a force
phasor $\vct F$, with units of length and force respectively.
\begin{definition}[Retarded mechanical response and damped mode]
\label{def:s4-retarded-mode}
Where $\mathsf T(z)=\mathsf K^*(z)-z^2\mathsf M$ is invertible, its
retarded inverse $\mathsf R_{\rm mech}(z)=\mathsf T(z)^{-1}$ is the
\emph{mechanical compliance}: it maps a force to a displacement and has
units $\mathrm{m\,N^{-1}}$. An isolated pole of its causal continuation
defines a \emph{damped mode of the displacement response}. A pole
$\widetilde\omega_s=\overline\omega_s-i\Gamma_s$, with
$\Gamma_s>0$, contributes motion with temporal factor
$e^{-i\overline\omega_s t}e^{-\Gamma_s t}$.
\end{definition}

This definition uses the full particle displacement response; a chosen
readout can still suppress that pole. The augmented
state also retains any medium relaxation whose particle displacement is
zero. Thus the state spectrum and the poles actually present in a measured
compliance can contain different information. With several relaxation
branches, additional nonoscillatory transients occur. General damping
mixes conservative modes, so the moving pattern follows from the coupled
pencil rather than from an independently assigned decay factor for each
conservative eigenvector.
These complex poles belong to the reduced causal response. In an exactly
finite stable quadratic lattice--bath system, the self-adjoint energy
observable and its coupled normal modes have real energy and frequency
spectra, and energy can return to the lattice. Irreversible decay uses
the reservoir or coarse-grained observation limit stated in Section~4.3.

\begin{definition}[Resolved phonon resonance]
\label{def:s4-resolved-resonance}
A \emph{resolved phonon resonance} is a positive-frequency oscillatory
mechanical pole whose decay rate satisfies
$0<\Gamma_s\ll\overline\omega_s$, whose linewidth is
small compared with neighboring mode separations, and whose preparation,
readout and regular response weights vary slowly across that line. In this
\emph{quasiparticle approximation}, the collective disturbance can be
tracked as a carrier with a frequency and lifetime.
\end{definition}

When a single pole dominates a channel and its residue is effectively
constant across the line, the squared magnitude of that contribution has
the Lorentzian denominator
$(\omega-\overline\omega_s)^2+\Gamma_s^2$ and full width
$2\Gamma_s$. This gives a concrete meaning to the linewidth condition.
A stable broad or purely decaying response is still a mechanical
excitation; its displacement and branch coordinates provide the useful
description. The energy of a canonically quantized conservative mode
continues to use its real reference frequency and self-adjoint Hamiltonian.

For example, if $\mathsf M^{-1/2}\mathsf C_0\mathsf M^{-1/2}$ is diagonal
in the conservative mode basis and the relaxation branches are absent, a
mass-weighted modal coordinate satisfies
\begin{equation}
 \Dtt{Q_s}+2\gamma_s\Dt{Q_s}+\omega_s^2Q_s=0,
 \qquad \gamma_s\geq0.
 \label{eq:proportional-modal-damping}
\end{equation}
In the underdamped regime $\gamma_s<\omega_s$, let
$\Omega_s=\sqrt{\omega_s^2-\gamma_s^2}$. The exact real motion is
\begin{equation}
 Q_s(t)=e^{-\gamma_s t}
       \left[Q_s(0)\cos(\Omega_s t)
       +\frac{\Dt{Q_s}(0)+\gamma_s Q_s(0)}{\Omega_s}
                                               \sin(\Omega_s t)\right].
 \label{eq:exact-modal-ringdown}
\end{equation}
Its poles are $\pm\Omega_s-i\gamma_s$. Critical and overdamped modes have
their corresponding repeated or purely decaying roots; they do not have an
oscillatory packet of this form.

Figure~\ref{fig:more-damping-thresholds} compares a release from rest with
a harmonic-force measurement at the same conservative frequency. An
oscillatory decay and a nonzero-frequency displacement-response maximum
therefore need separate criteria.

\begin{figure}[H]
\centering\begingroup
\definecolor{vltTeal}{HTML}{176B64}
\definecolor{vltBlue}{HTML}{315D80}
\definecolor{vltGold}{HTML}{A87520}
\definecolor{vltPlum}{HTML}{725877}
\definecolor{vltRed}{HTML}{762B38}
\definecolor{vltYellow}{HTML}{D5AD3C}
\definecolor{vltNeutral}{HTML}{777D83}
\setlength{\unitlength}{1pt}
\scalebox{0.90}{%
\input{vlt-figures/figure-112.tex}%
}\endgroup
\caption{An oscillatory release and a displacement-amplitude resonance have distinct thresholds in the memory-free scalar model. The curves use $d=\gamma_s/\omega_s$ and $Q(0)=Q_\star$, $\Pi(0)=0$. (a) Exact release trajectories. (b) Exact conjugate-force compliance; the vertical dashed line marks the conservative frequency. Curves for $d=0.8$ are dashed and those for $d=1$ dotted in both panels. The $d=0.8$ trajectory has oscillatory poles although its amplitude response peaks at zero frequency. The lower pair identifies the common particle pattern used by both protocols. Ghost bodies mark equilibrium: its interior separation remains fixed, while equal common displacement stretches one end connector and shortens the other. All corresponding connectors retain their coil count, amplitude and end leads. The displayed damping ratios are supplied coefficients, independent of spring rendering.}
\label{fig:more-damping-thresholds}
\end{figure}

The conservative energy coordinate
$\psi_s=(\omega_s Q_s+i\Dt{Q_s})/\sqrt2$ now obeys
\begin{equation}
 \Dt{\psi_s}=-i\omega_s\psi_s-\gamma_s\psi_s
                                      +\gamma_s\psi_s^*.
 \label{eq:damped-energy-ket-equation}
\end{equation}
The coupling to its conjugate shows why multiplying the original energy ket
by $e^{-\gamma_s t}$ is not an exact replacement. An exact exponential
coordinate for the underdamped oscillator can instead be defined by
$w_s=[\Omega_s Q_s+i(\Dt{Q_s}+\gamma_sQ_s)]/\sqrt2$;
it satisfies $\Dt{w_s}=(-\gamma_s-i\Omega_s)w_s$, while
$|w_s|^2$ is not the harmonic energy previously assigned to $|\psi_s|^2$.

The distinction between $\psi_s$ and $w_s$ also answers part of the
effective-variable question. A convenient decay coordinate need not have
the energy or measurement interpretation of the coordinate from which it
was constructed. \begin{definition}[Conservative modal action]
\label{def:s4-modal-action}
For the isolated positive-frequency conservative reference, put
$\Pi_s=\Dt{Q_s}$. Its \emph{modal action} is
\begin{equation}
 \mathcal I_s:=\frac{E_s}{\omega_s}
       =\frac{\Pi_s^2+\omega_s^2Q_s^2}{2\omega_s}
       =\frac{|\psi_s|^2}{\omega_s}.
 \label{eq:excitation-modal-action}
\end{equation}
\end{definition}
Here $Q_s$ has units $\mathrm{kg}^{1/2}\mathrm m$, $\Pi_s$ has units
$\mathrm{kg}^{1/2}\mathrm m\,\mathrm s^{-1}$, and $\mathcal I_s$ has units
$\mathrm{J\,s}$. The action is conserved for this time-independent,
unforced harmonic reference. It is a useful measure of amplitude relative
to frequency; damping, forcing, and changes of the reference generally
change it. The pair $(Q_s,\Pi_s)$ and complex $\psi_s$ retain the same two
real pieces of information. Energy or action alone discards the oscillation
phase and cannot reconstruct a trajectory, interference, or instantaneous
force. Figure~\ref{fig:excitation-state-phase} shows this information loss
using the reconstruction in \eqref{eq:architecture-modal-energy-phase}.

\begin{figure}[H]
\centering\begingroup
\definecolor{vltTeal}{HTML}{176B64}
\definecolor{vltBlue}{HTML}{315D80}
\definecolor{vltGold}{HTML}{A87520}
\definecolor{vltPlum}{HTML}{725877}
\definecolor{vltRed}{HTML}{762B38}
\definecolor{vltYellow}{HTML}{D5AD3C}
\definecolor{vltNeutral}{HTML}{777D83}
\setlength{\unitlength}{1pt}
\scalebox{0.90}{%
\input{vlt-figures/figure-113.tex}%
}\endgroup
\caption{Classical phase in $\psi=(\omega_sQ+i\Pi)/\sqrt2$. A and B share one energy circle but give displacement histories $\cos(\omega_st)$ and $-\sin(\omega_st)$. The supported pair reconstructs their initial common motion: green arrows show displacement, burgundy arrows velocity. Both rulers locate reference centres at $x/a=0,1$, with equal gains. A preserves its interior span while changing the end spans oppositely; B retains reference spans. The invertible state-to-$\psi$ map preserves phase, whereas energy and action discard it. Quantum occupation uses the separate quantization postulate.}
\label{fig:excitation-state-phase}
\end{figure}

For weak memory corrections, keep the conservative normalized eigenvector
$\vct e_s$ and frequency $\omega_s>0$. Define
$\mathsf R(\omega)=\mathsf M^{-1/2}
[\mathsf K'(\omega)-\mathsf K]\mathsf M^{-1/2}$ and
$\mathsf L(\omega)=\mathsf M^{-1/2}\mathsf K''(\omega)\mathsf M^{-1/2}$.
First-order perturbation of the pencil gives
\begin{equation}
 \delta\omega_s=\frac{\vct e_s^\dagger\mathsf R(\omega_s)\vct e_s}
                                    {2\omega_s},\qquad
 \Gamma_s=\frac{\vct e_s^\dagger\mathsf L(\omega_s)\vct e_s}
                                    {2\omega_s}\geq0.
 \label{eq:weak-memory-frequency-shift}
\end{equation}
The corrections must be small on the oscillation and mode-separation scales,
and the response must vary smoothly across the linewidth. Degenerate or
strongly mixed modes require the corresponding coupled pencil. The Maxwell
example stiffens relative to its relaxed reference; other medium models can
also produce inertial frequency shifts. There is no universal shift obtained
by subtracting an independently chosen imaginary damping constant.

\subsubsection{Residence windows and rearrangement-induced spectral change}
\label{sec:s4-exchange-residence-spectrum}

Mechanical ringdown and structural residence supply two lifetimes.
A pattern can decay within a persistent arrangement, or the arrangement
can change first. Cluster membership
(Definition~\ref{def:s23-cluster-membership}) and a response pole
(Definition~\ref{def:s4-retarded-mode}) quantify these events.
Joint membership and motion records can distinguish them.

\begin{definition}[Residence-conditioned resonance window]
\label{def:s4-exchange-residence-window}
Let $g$ identify one persistent arrangement, with approximately constant
conditional mechanical and medium coefficients during its residence.
A residence-conditioned resonance record uses data from that arrangement,
with its preparation, branch state and input/readout maps specified.
In the auxiliary continuous-time Markov description with constant exit
rate $\Lambda_g=-\mathsf Q_{gg}\geq0$, the probability of no exit during
a window of length $T$ is
\begin{equation}
 \begin{aligned}
 S_g(T)&=e^{-\Lambda_gT},\\
 T&\leq\frac{-\log(1-\epsilon_{\rm exit})}{\Lambda_g}
 \quad\Longrightarrow\quad
 \Pr\{\hbox{an exit during }T\}\leq\epsilon_{\rm exit}.
 \end{aligned}
 \label{eq:s4-exchange-residence-survival}
\end{equation}
for $0<\epsilon_{\rm exit}<1$. The upper bound is infinite when
$\Lambda_g=0$.
\end{definition}

A direct single-window Fourier estimate of the intrinsic linewidth of a pole
$\overline\omega_s-i\Gamma_s$ uses several carrier periods
and a frequency resolution substantially finer than that linewidth:
\begin{equation}
 \frac{2\pi}{\overline\omega_s}\ll T,\qquad
 \frac{2\pi}{T}\ll\Gamma_s,
 \qquad
 \Lambda_gT\leq-\log(1-\epsilon_{\rm exit}).
 \label{eq:s4-exchange-resonance-window}
\end{equation}
The inequalities specify a resolved single-residence Fourier protocol,
alongside mode separation and preparation conditions. Reference drift
must remain small over $T$. When no common window exists, use registered
transient records or conditional responses of the changing arrangement.
Strain-dependent escape also makes residence selection a selection of
mechanical states, which enters the observation model. Branch memory
can persist across residences through its state; its time need not be
shorter than each residence.

A small exactly solvable example shows how rearrangement can change a
line without being identified with a viscous loss coefficient.
Random-frequency modulation is an established spectral construction
\cite{Anderson1954}. Let a phase carrier obey
$\dot a=-i[\omega_0+\Delta\sigma(t)]a$, where
$\sigma=\pm1$ switches at rate $\kappa$ in each direction and begins
in its equal stationary distribution. Take $\kappa>0$ and
$\omega_0>|\Delta|$ for positive carrier frequencies. Here $\Delta$ and $\kappa$ have
units $\mathrm{s^{-1}}$. Each realization has constant $|a|$, while
its ensemble correlation contains
\begin{equation}
 \begin{aligned}
 C_\phi(t)&=
 \left\langle e^{-i\Delta\int_0^t\sigma(s)\,\dd s}\right\rangle
 =e^{-\kappa t}
   \left[\cosh(dt)+\frac{\kappa}{d}\sinh(dt)\right],\\
 d&=\sqrt{\kappa^2-\Delta^2},\\
 \gamma_\phi&=\kappa-\sqrt{\kappa^2-\Delta^2}
       \quad(\kappa>|\Delta|),\\
 \gamma_\phi&\simeq\frac{\Delta^2}{2\kappa}
       \quad(\kappa\gg|\Delta|).
 \end{aligned}
 \label{eq:s4-exchange-telegraph-phase}
\end{equation}
The formula has its continuous limit at $d=0$ and trigonometric
continuation for $|\Delta|>\kappa$. To derive it, propagate the two
conditional phase weights with
$\left(\begin{smallmatrix}-\kappa-i\Delta&\kappa\\
\kappa&-\kappa+i\Delta\end{smallmatrix}\right)$ and sum them;
the sum solves $\ddot C_\phi+2\kappa\dot C_\phi+\Delta^2C_\phi=0$,
with $C_\phi(0)=1$ and $\dot C_\phi(0)=0$.
In the fast-switching regime, the long-time correlation has an additional
decay scale $\gamma_\phi$. This is a kinematic phase-modulation example,
not a complete variable-stiffness mechanical model. A physical change
of lattice energy retains its work and state-transition account.
The example demonstrates why a linewidth alone cannot assign the
observed change uniquely to medium dissipation or structural kinetics.

Figure~\ref{fig:new-s4-spectra-residence} separates the residence window from the phase decorrelation caused by structural exchange.

\begin{figure}[H]
\centering\begingroup
\definecolor{vltTeal}{HTML}{176B64}
\definecolor{vltBlue}{HTML}{315D80}
\definecolor{vltGold}{HTML}{A87520}
\definecolor{vltPlum}{HTML}{725877}
\definecolor{vltRed}{HTML}{762B38}
\definecolor{vltYellow}{HTML}{D5AD3C}
\definecolor{vltNeutral}{HTML}{777D83}
\setlength{\unitlength}{1pt}
\scalebox{0.90}{%
\input{vlt-figures/figure-114.tex}%
}\endgroup
\caption{Residence and mechanical decay constrain different parts of a spectral protocol. (a) The same four particle identities illustrate two registered arrangements; weak exterior couplings remain drawn. The labels associate these arrangements with the two frequencies of the auxiliary telegraph phase model, rather than deriving a frequency from the sketch. (b) Exact no-exit survival. The pale interval simultaneously gives $2\pi/T\le\Gamma/2$ and exit probability at most $0.1$ for the stated illustrative ratios. Its Fourier condition is an explicitly chosen finite-resolution threshold. (c) Exact $C_\phi$ for two switching rates at fixed $\Delta$, with dashed $\kappa/\Delta=0.25$ and solid $\kappa/\Delta=2$ curves; the grey dashed unit line denotes the constant amplitude magnitude of every realization. Exchange changes ensemble phase correlation even with this unchanged realization magnitude. The curves follow~\eqref{eq:s4-exchange-telegraph-phase}; they describe phase modulation, while any physical stiffness transition retains its separate work and state account.}
\label{fig:new-s4-spectra-residence}
\end{figure}

\subsubsection{The moving pattern and its measured response weight}
\label{sec:s4-modes-pole-shapes}

A damped mode specifies the relative amplitudes and phases of ideal-virion
centre motion. The relaxed Hessian retains the reference Lennard--Jones,
Coulomb and support curvatures; the medium supplies delayed reaction
forces. Their coupled response determines both the moving pattern and
its decay rate, even when the conservative network stays fixed.

Write the analytically continued physical-coordinate pencil as
$\mathsf T(z)=\mathsf K^*(z)-z^2\mathsf M$ on $d$ selected mechanical
coordinates. Assume analyticity near $z_s$, a determinant not identically
zero there, and $\operatorname{rank}\mathsf T(z_s)=d-1$.
Choose nonzero right and left null vectors satisfying
\begin{equation}
 \mathsf T(z_s)\vct r_s=0,\qquad
 \vct\ell_s^\dagger\mathsf T(z_s)=0,\qquad
 N_s^{\rm res}:=\vct\ell_s^\dagger
                \partial_z\mathsf T(z_s)\vct r_s\ne0.
 \label{eq:s4-modes-left-right-pole}
\end{equation}
The rank condition and $N_s^{\rm res}\ne0$ specify a simple resonance
away from a constitutive singularity. Here $N_s^{\rm res}$ is its derivative
normalization. The simple-pole expansion is
\begin{equation}
 \mathsf T(z)^{-1}=
 \frac{\vct r_s\vct\ell_s^\dagger}
      {(z-z_s)N_s^{\rm res}}+\mathsf R_s^{\rm reg}(z),
 \label{eq:s4-modes-resolvent-residue}
\end{equation}
where $\mathsf R_s^{\rm reg}$ is analytic locally. This is the
simple-eigenvalue form of the nonlinear resolvent expansion
\cite{Beyn2012,GuettelTisseur2017}. To see its normalization, let
$\mathsf A_{-1}$ be the residue of $\mathsf T^{-1}$. The singular
terms of both $\mathsf T\mathsf T^{-1}=\Id$ and
$\mathsf T^{-1}\mathsf T=\Id$ place its columns and rows in the right
and left nullspaces, so
$\mathsf A_{-1}=c\vct r_s\vct\ell_s^\dagger$. Multiplying the
constant term by the left null row gives $cN_s^{\rm res}=1$.
Rescaling either null vector therefore cancels between numerator and
denominator: a convenient plotted normalization of a shape leaves its
physical response weight unchanged.

For the Maxwell model the required derivative is explicit:
\begin{equation}
 \partial_z\mathsf T(z)=-2z\mathsf M-i\mathsf C_0
       -i\sum_a\frac{\tau_a\mathsf K_a}{(1-iz\tau_a)^2}.
 \label{eq:s4-modes-pencil-derivative}
\end{equation}
The branch term accounts for the change of medium deformation as the
frequency changes. Replacing this denominator by a conservative modal
mass generally changes the residue. In a reciprocal model
$\mathsf T^{\mathsf T}=\mathsf T$, the left row can be chosen
$\vct\ell_s^\dagger=\vct r_s^{\mathsf T}$; complex symmetry involves
the transpose, rather than the Hermitian adjoint of $\vct r_s$.
The positive stored energy remains the real augmented energy of
Proposition~\ref{prop:maxwell-energy}. The complex residue factor is a
response normalization, not an additional stored energy.

Let a dimensionless real vector $\vct a$ specify a force pattern
$\vct F=\vct a F_0$, and let the displacement readout be
$X=\vct d^{\mathsf T}\vct U$ with dimensionless real $\vct d$.
Its compliance is
\begin{equation}
 \frac{X}{F_0}=\vct d^{\mathsf T}\mathsf T(z)^{-1}\vct a
 =\frac{(\vct d^{\mathsf T}\vct r_s)
              (\vct\ell_s^\dagger\vct a)}
            {(z-z_s)N_s^{\rm res}}+
       \vct d^{\mathsf T}\mathsf R_s^{\rm reg}(z)\vct a.
 \label{eq:s4-modes-channel-residue}
\end{equation}
\begin{definition}[Preparation and readout visibility]
\label{def:s4-channel-visibility}
For the force pattern $\vct a$ and displacement readout $\vct d$, a simple
mode is \emph{visible in that channel} when the residue of
$\vct d^{\mathsf T}\mathsf R_{\rm mech}(z)\vct a$ at its pole is nonzero.
The factors $\vct\ell_s^\dagger\vct a$ and
$\vct d^{\mathsf T}\vct r_s$ specify, respectively, its preparation
and detection overlaps.
\end{definition}

\begin{proposition}[A mode can disappear from one measured channel]
\label{prop:s4-simple-pole-visibility}
Under the analytic and simple-resonance hypotheses of
\eqref{eq:s4-modes-left-right-pole}, the channel has a pole at $z_s$
exactly when both overlaps in Definition~\ref{def:s4-channel-visibility}
are nonzero. If either overlap vanishes, that channel is analytic at
$z_s$, although the full mechanical compliance has a pole there.
\end{proposition}
\begin{proof}
Equation~\eqref{eq:s4-modes-channel-residue} has a single possible singular
term at $z_s$. Its coefficient is the product of the two overlaps divided
by the nonzero $N_s^{\rm res}$. The regular remainder is analytic, and
the full residue in \eqref{eq:s4-modes-resolvent-residue} is nonzero.
\end{proof}

For a concrete centre-motion example, take two identical ideal virions whose
reference, supports and medium couplings are invariant under exchanging
the particles. In their axial subspace the normalized common and relative
patterns are $\vct a_+=(1,1)^{\mathsf T}/\sqrt2$ and
$\vct a_-=(1,-1)^{\mathsf T}/\sqrt2$. The readouts
$X_+=(u_1+u_2)/\sqrt2$ and $X_-=(u_1-u_2)/\sqrt2$ measure common
motion and changing separation. Exchange symmetry makes these sectors
invariant. A common force has zero relative response, while opposite
forces excite the relative sector; the two measurements examine
different motions of the same particles. A relative medium branch can
therefore be absent from the common-motion data while strongly affecting
the separation response. Fixed supports can give the common motion a
positive restoring curvature; without them its translation must be
treated as the rigid sector specified in Section~3.

This is an exact channel-selection statement within the symmetric model.
A small asymmetry gives a previously absent resonance a small weight,
rather than creating a new particle degree of freedom. Measurements of
phase, linewidth and spatial pattern then supply complementary constraints
on interaction curvature and medium coupling. A response maximum can also
shift through interference with neighboring poles or the regular
background. These geometric and spectral effects explain why a single
peak position is only one part of the inverse mechanical problem.

The same denominator gives a sensitivity formula. If a real reference
parameter $p$ changes smoothly while the resonance remains simple, then
\begin{equation}
 \frac{\dd z_s}{\dd p}=-
 \frac{\vct\ell_s^\dagger(\partial_p\mathsf T)(z_s,p)\vct r_s}
      {\vct\ell_s^\dagger(\partial_z\mathsf T)(z_s,p)\vct r_s}.
 \label{eq:s4-modes-pole-sensitivity}
\end{equation}
This follows by differentiating $\mathsf T(z_s,p)\vct r_s=0$;
the left null vector removes the unknown shape derivative
\cite{GuettelTisseur2017}. A declared charge or separation changes
the conservative curvature through $\partial_p\mathsf K$; a relaxation
time changes the branch factors; a mass changes the inertial term.
If the equilibrium positions also change, their contribution belongs in
the total parameter derivative of the assembled reference matrices.
One frequency shift cannot uniquely identify all of these changes.
Approaching a repeated pole also invalidates the simple-pole formulas;
the augmented state or the corresponding coupled spectral subspace then
provides the appropriate description.

For an isolated conservative mode the leading shape change is especially
transparent. With $\mathsf\Delta_s=\mathsf R(\omega_s)
-i\mathsf L(\omega_s)$ and the normalization choice
$\vct e_s^\dagger\delta\vct e_s=0$,
\begin{equation}
 \delta\vct e_s=
 \sum_{t\ne s}\vct e_t
 \frac{\vct e_t^\dagger\mathsf\Delta_s\vct e_s}
                 {\omega_s^2-\omega_t^2}.
 \label{eq:s4-modes-first-shape-change}
\end{equation}
The sum runs over the complete selected conservative coordinate space,
with any rigid motions retained or constrained as in Section~3.
This first-order expansion supplements
\eqref{eq:weak-memory-frequency-shift}: the diagonal correction shifts
the pole, whereas the off-diagonal corrections mix displacement patterns.
It requires small corrections, smooth frequency dependence and separated
eigenvalues. Nearly degenerate patterns must be retained together, even
when the damping is weak relative to each individual frequency.

Figure~\ref{fig:viscoelastic-damped-pattern-readout} reconstructs a complex particle pattern and compares the response weights seen by two readouts of the same coupled system.

\begin{figure}[H]
\centering\begingroup
\definecolor{vltTeal}{HTML}{176B64}
\definecolor{vltBlue}{HTML}{315D80}
\definecolor{vltGold}{HTML}{A87520}
\definecolor{vltPlum}{HTML}{725877}
\definecolor{vltRed}{HTML}{762B38}
\definecolor{vltYellow}{HTML}{D5AD3C}
\definecolor{vltNeutral}{HTML}{777D83}
\setlength{\unitlength}{1pt}
\scalebox{0.90}{%
\input{vlt-figures/figure-115.tex}%
}\endgroup
\caption{A complex shape and two readouts for the pair of Figure~\ref{fig:viscoelastic-memory-pattern-mixing}. The lower positive-frequency pole comes from its exact five-state generator. Its largest displacement component is real and positive; snapshots use $14\,\mathrm{pt}$ per unit component of $\operatorname{Re}(\vct r_se^{-i\theta})$ after dividing out decay. Equal rulers locate reference sites at $x/a=0,1$; matched coils follow their moved surfaces. A common drive $\vct a=(1,1)/\sqrt2$ is read through $\vct d_\pm=(1,\pm1)/\sqrt2$. Full-pencil responses share the same poles; dotted guides locate their real parts, while channel residues set visible weights.}
\label{fig:viscoelastic-damped-pattern-readout}
\end{figure}

\subsubsection{Occupation, coherence, and transfer to the medium}

The quantum viral-phonon postulate quantizes the mechanical patterns of
the declared reference. The particle count and the occupation of one of
these patterns are separate observables.

\begin{definition}[Reference-mode occupation]
\label{def:s4-reference-occupation}
For the positive reference mode its lowering operator is
$\hat b_s=(\omega_s\hat Q_s+i\hat\Pi_s)/\sqrt{2\hbar\omega_s}$,
and $\hat N_s=\hat b_s^\dagger\hat b_s$ is dimensionless. The mode energy
is $\hbar\omega_s(\hat N_s+\tfrac12\Id)$; occupation counts energy quanta,
not virions. Under coupling to the medium, this bare-mode occupation is a
specified reference observable.
\end{definition}

Exact occupation energies of a finite
coupled quadratic Hamiltonian instead belong to its coupled normal modes.
Independent bare-mode loss is an additional weak-coupling approximation.

Occupation and coherence describe complementary features of a quantum
preparation. A number state can have positive
$\langle\hat N_s\rangle$ and zero mean displacement; a coherent
preparation with the same mean occupation can have an oscillating mean.
The energy can reside in mean motion, in centered fluctuations, or in both.
Section~\ref{sec:complex-field-observables} unfolds this distinction into
coordinate moments and the energy split
\eqref{eq:mean-motion-fluctuation-energy-split}. Thus a displacement wave,
a state amplitude, and a mode population require different interpretations
even when they refer to the same mechanical pattern.

At the complete oscillator-bath level, the coupled Hamiltonian has unitary
evolution and conserves total energy. After medium coordinates are traced out,
the lattice is generally described by a density operator, rather than a
normalized state vector evolving under a complex frequency alone.
For weak independent mode losses into a large medium near its ground state,
with a short bath correlation time and well-separated retained mode
frequencies, a Markov and secular approximation gives a loss model on
the vacuum-plus-one-phonon space
$\operatorname{span}\{\ket0,\ket{1_s}\}$. Define the bounded lowering maps
$\hat L_s=\ket0\bra{1_s}$. Then
\begin{equation}
 \begin{split}
 \Dt{\hat\varrho}&=-\frac{i}{\hbar}
                 [\hat H_{\mathrm{ph}},\hat\varrho]
                   +\sum_s\kappa_s\mathcal D[\hat L_s]\hat\varrho,\\
 \hat H_{\mathrm{ph}}&=\sum_s\hbar\overline\omega_s\ket{1_s}\bra{1_s},
 \qquad
 \mathcal D[\hat L]\hat\varrho=
     \hat L\hat\varrho\hat L^\dagger
                  -\tfrac12\{\hat L^\dagger\hat L,\hat\varrho\}.
 \end{split}
 \label{eq:phonon-loss-master-equation}
\end{equation}
Here $\hat\varrho$ is positive with unit trace, the braces denote an
anticommutator, and $\kappa_s\geq0$ is an occupation decay rate.
This Lindblad form preserves probability while permitting excitation transfer
to the medium~\cite{Lindblad1976}. The amplitude rate is
$\Gamma_s=\kappa_s/2$ in this approximation.

For a normalized initial packet, $\sum_s|c_s|^2=1$, the explicit reduced state is
\begin{equation}
 \begin{split}
 \ket{\widetilde\Phi(t)}&=\sum_s c_s
           e^{-i\overline\omega_s t-\kappa_s t/2}\ket{1_s},\\
 P_1(t)&=\sum_s|c_s|^2e^{-\kappa_s t},\qquad
 \hat\varrho(t)=\ket{\widetilde\Phi(t)}\bra{\widetilde\Phi(t)}
                           +[1-P_1(t)]\ket0\bra0.
 \end{split}
 \label{eq:one-phonon-survival-state}
\end{equation}
These $c_s$ are occupation-state amplitudes, not classical displacements.
A state confined to the one-phonon sector has zero displacement first
moment, while its second moments can retain intermode coherence.
The decaying ket has norm squared $P_1$, the probability that the excitation
remains in the retained modes. Its normalized conditional version describes
the surviving one-phonon sector. The unconditional state also contains the
vacuum sector, so its trace remains one. Its excitation energy is
$\sum_s\hbar\overline\omega_s|c_s|^2e^{-\kappa_s t}$; the transferred
energy belongs to the enclosed medium. Virion number remains $n$.

Figure~\ref{fig:viscoelastic-phonon-loss-account} distinguishes loss of retained occupation from normalization and from preservation of the full oscillator algebra.

\begin{figure}[H]
\centering\begingroup
\definecolor{vltTeal}{HTML}{176B64}
\definecolor{vltBlue}{HTML}{315D80}
\definecolor{vltGold}{HTML}{A87520}
\definecolor{vltPlum}{HTML}{725877}
\definecolor{vltRed}{HTML}{762B38}
\definecolor{vltYellow}{HTML}{D5AD3C}
\definecolor{vltNeutral}{HTML}{777D83}
\setlength{\unitlength}{1pt}
\scalebox{0.80}{%
\input{vlt-figures/figure-116.tex}%
}\endgroup
\caption{Two distinct unit accounts in specified Markov reservoir models. (a) Starting with one retained phonon, \eqref{eq:one-phonon-survival-state} gives $P_1=e^{-\kappa t}$ and $P_0=1-e^{-\kappa t}$; their dotted sum is one. (b) The separate full-oscillator input model \eqref{eq:markov-phonon-input-noise} preserves $[b(t),b^\dagger(t)]=1$. Each stack combines the initial-operator contribution $e^{-\kappa t}$ with the input-noise contribution $1-e^{-\kappa t}$. Probability normalization and operator algebra use different state spaces and meanings, despite their matching exponential factors. The lower reference diagrams retain the same whole-particle count. Both number states have zero mean modal displacement; the glyphs identify the quantized coordinates rather than a classical trajectory. The accounts apply within their stated reservoir reductions.}
\label{fig:viscoelastic-phonon-loss-account}
\end{figure}

The vacuum in this state is the \emph{zero-phonon occupation} of the retained
mechanical modes, not an empty cylinder or the removal of a virion.
Unit trace counts all alternatives available to that subsystem, including
the loss of its excitation to the medium. A change in particle count,
loss of infectivity, or chemical transformation is a different event and needs a
corresponding state and rate model.

The finite-temperature Markov extension requires the full oscillator
occupation space, since absorption can create a second phonon. With
$\bar N_s^{(T)}=[e^{\hbar\overline\omega_s/(k_BT)}-1]^{-1}$,
the independently damped weak-coupling model is
\begin{equation}
 \Dt{\varrho}=-\frac{i}{\hbar}[H_{\rm ph},\varrho]
   +\sum_s\kappa_s\left[(\bar N_s^{(T)}+1)\mathcal D[b_s]\varrho
                       +\bar N_s^{(T)}\mathcal D[b_s^\dagger]\varrho\right].
 \label{eq:thermal-phonon-damping-generator}
\end{equation}
Here $N_s=b_s^\dagger b_s$ and
$H_{\rm ph}=\sum_s\hbar\overline\omega_s N_s$;
oscillator operators are used on states with the required finite
moments. This is not the bounded vacuum-plus-one-phonon generator.
The downward and upward factors out of level $N$ are respectively
$(\bar N_s^{(T)}+1)N$ and $\bar N_s^{(T)}(N+1)$. Paired adjacent
transitions instead obey
\begin{equation}
 \frac{w_{N\to N+1}}{w_{N+1\to N}}=
       \frac{\bar N_s^{(T)}}{\bar N_s^{(T)}+1}
       =e^{-\hbar\overline\omega_s/(k_BT)},
 \label{eq:thermal-adjacent-detailed-balance}
\end{equation}
which is thermal detailed balance~\cite{Clerk2010,Potts2026}.
In particular,
\begin{equation}
 \frac{\dd}{\dd t}\langle N_s\rangle
       =-\kappa_s(\langle N_s\rangle-\bar N_s^{(T)}),\qquad
 \langle N_s(t)\rangle=\bar N_s^{(T)}+
   [\langle N_s(0)\rangle-\bar N_s^{(T)}]e^{-\kappa_st}.
 \label{eq:thermal-occupation-relaxation}
\end{equation}
A medium can therefore raise the energy of a mode initially below its
thermal population. At a fixed reservoir temperature, the deviation
$\delta N_s=\langle N_s\rangle-\bar N_s^{(T)}$ is a useful relaxation
variable: it can be positive or negative, whereas occupation itself is
nonnegative. It describes excess or deficient population relative to the
chosen thermal reference, not an independently created degree of freedom.
A nearly constant-temperature reservoir within a
sealed vessel is an approximation over the chosen observation window;
an exactly finite isolated medium must retain its changing energy and
state. Strong coupling or memory requires the fuller coupled description.

The dimensionless ratio $\hbar\overline\omega_s/(k_BT)$ compares a
mode's energy increment with thermal energy. When this ratio is small,
$\bar N_s^{(T)}\simeq k_BT/(\hbar\overline\omega_s)$ is large and the
harmonic equilibrium motion admits its classical high-occupation limit.
When it is large, thermal occupation is suppressed but quantum vacuum
fluctuations remain. Room or body temperature alone does not select a
regime: the frequency, linewidth, and retained coordinate must also be
specified. Temperature can change stiffness, viscosity, relaxation times,
and the prepared biological structure as well as occupations. The
temperature comparison in Section~\ref{sec:phonon-temperature-comparison}
keeps these effects separate.

The same Markov approximation also illustrates why operator noise
cannot be omitted. For one full oscillator, a narrow-band stationary
input field obeys $[b_{\rm in}(t),b_{\rm in}^\dagger(t')]=\delta(t-t')$,
and
\begin{equation}
 b(t)=e^{-(i\overline\omega+\kappa/2)t}b(0)
       +\sqrt\kappa\int_0^t
           e^{-(i\overline\omega+\kappa/2)(t-s)}b_{\rm in}(s)\,\dd s.
 \label{eq:markov-phonon-input-noise}
\end{equation}
For independent initial system and input operators, the equal-time
commutator is
\[
 e^{-\kappa t}+\kappa\int_0^t e^{-\kappa(t-s)}\dd s=1.
\]
Omitting the noise would instead give
$e^{-\kappa t}$. Thermal inputs have
$\langle b_{\rm in}^\dagger(t)b_{\rm in}(t')\rangle
=\bar N^{(T)}\delta(t-t')$ and the reverse order has
$(\bar N^{(T)}+1)\delta(t-t')$~\cite{Clerk2010}.
The delta-correlated annihilation input is a carrier-band Markov model;
it is not the exact broadband mechanical force of the finite bath.

For the oscillator realization, define the matrix bath
spectral density
\begin{equation}
 \mathsf J(\nu)=\frac\pi2\sum_\ell
       \frac{\vct c_\ell\vct c_\ell^{\mathsf T}}
                                      {m_\ell^{\mathrm b}\nu_\ell}
                          \delta(\nu-\nu_\ell),\qquad \nu>0.
 \label{eq:matrix-bath-spectral-density}
\end{equation}
It is nonnegative and determines
$\mathsf\Gamma_{\mathrm H}(t)=
(2/\pi)\int_0^\infty\mathsf J(\nu)\cos(\nu t)\,\dd\nu/\nu$.
Its retarded resistance is
$\mathsf Z_{\mathrm H}(z)=\int_0^\infty
\mathsf\Gamma_{\mathrm H}(t)e^{izt}\,\dd t$, with the boundary value
understood distributionally for a finite oscillator bath. This bath
resistance is distinct from $\mathsf Z_{\rm eff}$ of the Maxwell model.
For a stationary equilibrium quantum bath, let $\hat{\vct F}_{\mathrm{eq}}$
denote its reduced fluctuating force. Its symmetrized spectrum satisfies
the fluctuation--dissipation relation~\cite{Ford1988}
\begin{equation}
 \begin{split}
 \mathsf S_F^{\mathrm{sym}}(\omega)
 &=\int_{-\infty}^{\infty}e^{i\omega t}
       \tfrac12\langle
          \{\hat{\vct F}_{\mathrm{eq}}(t),
                          \hat{\vct F}_{\mathrm{eq}}(0)^{\mathsf T}\}
                                               \rangle\,\dd t\\
 &=\hbar|\omega|\coth\!\left(
                     \frac{\hbar|\omega|}{2k_{\mathrm B}T}\right)
                  \operatorname{Re}\mathsf Z_{\mathrm H}(\omega+i0),
 \qquad
 \operatorname{Re}\mathsf Z_{\mathrm H}(\omega+i0)
                              =\frac{\mathsf J(|\omega|)}{|\omega|}.
 \end{split}
 \label{eq:quantum-bath-fdt}
\end{equation}
The symmetrization is componentwise. This equilibrium statement requires the
corresponding reservoir preparation or stationary regime; arbitrary coupled
quantum initial data need not give the same force covariance. The classical limit is
$2k_{\mathrm B}T\operatorname{Re}\mathsf Z_{\mathrm H}$; the ground-state limit retains
the quantum fluctuation spectrum. A complex resonance frequency alone does
not specify those fluctuations or the full quantum state.

Figure~\ref{fig:viscoelastic-quantum-bath-crossover} visualizes the thermal crossover and the paired transition-rate ratio without identifying either quantity with a displacement trajectory.

\begin{figure}[H]
\centering\begingroup
\definecolor{vltTeal}{HTML}{176B64}
\definecolor{vltBlue}{HTML}{315D80}
\definecolor{vltGold}{HTML}{A87520}
\definecolor{vltPlum}{HTML}{725877}
\definecolor{vltRed}{HTML}{762B38}
\definecolor{vltYellow}{HTML}{D5AD3C}
\definecolor{vltNeutral}{HTML}{777D83}
\setlength{\unitlength}{1pt}
\scalebox{0.80}{%
\input{vlt-figures/figure-117.tex}%
}\endgroup
\caption{A shared thermal scale with two specified readings. (a) The equilibrium bath relation \eqref{eq:quantum-bath-fdt}, normalized at $T>0$ and $\operatorname{Re}Z_H>0$, gives $(x_T/2)\coth(x_T/2)$, continued by one at zero. Dashed and dotted lines mark its classical limit and large-$x_T$ vacuum asymptote. (b) A separately justified thermal Markov oscillator has paired transitions between $N$ and $N+1$. The common multiplicity cancels, giving $W_\uparrow/W_\downarrow=e^{-x_T}$. The level diagram labels their individual rates and the required Bose occupation. These rates require the declared reservoir and oscillator reduction. At $T=0$ the unnormalized force spectrum tends to $\hbar\omega\operatorname{Re}Z_H$; the finite-temperature normalization is then replaced by that limit. Frequency and reservoir response, together with temperature, determine the applicable regime.}
\label{fig:viscoelastic-quantum-bath-crossover}
\end{figure}

The same equilibrium construction also specifies the reduced coordinate
variance. For a scalar coordinate $Q$ with retarded susceptibility
$\mathcal R^{\rm ret}_{QQ}$ to its conjugate force, and convergent
equilibrium spectra, the coordinate fluctuation--dissipation relation gives
\begin{equation}
 \langle(\delta\widehat Q)^2\rangle
 =\frac{\hbar}{\pi}\int_0^\infty
    \coth\!\left(\frac{\hbar\Omega}{2k_{\rm B}T}\right)
    \operatorname{Im}\mathcal R^{\rm ret}_{QQ}(\Omega+i0)\,\dd\Omega.
 \label{eq:qfa-dressed-equilibrium-variance}
\end{equation}
This is the integral of the symmetrized coordinate spectrum, using its
evenness and the stated Fourier convention~\cite{Ford1988,Clerk2010}.
A narrow weakly coupled positive pole recovers the isolated oscillator
variance. With substantial bath coupling, the whole spectral measure
determines the reduced equilibrium state; it need not be the Gibbs
state of the bare oscillator Hamiltonian. A physically specified bath
cutoff and the measured response band are therefore part of a
quantitative prediction. Quantum Brownian motion remains meaningful
for overdamped coordinates, while a long-lived occupation of one
bare mode requires a further controlled reduction.

\subsection{Plane-wave dispersion and a worked relaxation model}
\label{sec:viscoelastic-dispersion}

The discrete periodic version starts from the compatible cells before
taking a continuum limit. Under the periodic geometry and electrostatic
conventions of Subsection~\ref{sec:structure-charged-periodic-basis},
repeat the medium parameters and their coupling maps with the same
particle basis. Let $\mathsf M_{\rm b}$ and $\mathsf K_{\rm b}(\vct k)$
be its mass and relaxed Bloch stiffness, and let
$\mathsf G_{a,\rm b}(\vct k)$ be the dimensionless Bloch map of an
owned relaxation channel. For real $\vct k$, elimination of its branch
gives
\begin{equation}
 \begin{aligned}
 \mathsf K_{a,\rm b}(\vct k)
   &=\mathsf G_{a,\rm b}(\vct k)^\dagger
                 \mathsf H_a\mathsf G_{a,\rm b}(\vct k)\succeq0,\\
 \mathsf A_{\rm b}(\vct k,\omega)
   &=\mathsf K_{\rm b}(\vct k)-\omega^2\mathsf M_{\rm b}
                       -i\omega\mathsf C_{0,\rm b}(\vct k)\\
   &\quad+\sum_a\frac{-i\omega\tau_a}{1-i\omega\tau_a}
                             \mathsf K_{a,\rm b}(\vct k).
 \end{aligned}
 \label{eq:s4-cell-bloch-memory-pencil}
\end{equation}
For passive repeated channels and real $\omega>0$,
\begin{equation}
 -\operatorname{Im}_{\rm H}\mathsf A_{\rm b}
  =\omega\mathsf C_{0,\rm b}
    +\sum_a\frac{\omega\tau_a}{1+\omega^2\tau_a^2}
                       \mathsf K_{a,\rm b}\succeq0.
 \label{eq:s4-cell-bloch-memory-loss}
\end{equation}
The finite-assembly work identity gives the congruence, and its scalar
imaginary parts give the loss. A basis with differently charged virions
can have collective polarizations that load the branches differently.
The Bloch construction retains this cell-scale motion before any
homogeneous acoustic approximation.

Zeros of $\det\mathsf A_{\rm b}$ away from branch denominators identify
the corresponding mechanical dispersion under retarded continuation.
The complete augmented cell generator also retains mechanically dark
internal relaxations and handles those denominators. A finite growing
cluster uses its assembled retarded operator during each reference
window; a bulk band description additionally requires the stated
translation symmetry or a separately controlled local approximation.

Figure~\ref{fig:new-s4-spectra-bloch} reconstructs a repeated basis polarization and its matched passive response and power loss.

\begin{figure}[H]
\centering\begingroup
\definecolor{vltTeal}{HTML}{176B64}
\definecolor{vltBlue}{HTML}{315D80}
\definecolor{vltGold}{HTML}{A87520}
\definecolor{vltPlum}{HTML}{725877}
\definecolor{vltRed}{HTML}{762B38}
\definecolor{vltYellow}{HTML}{D5AD3C}
\definecolor{vltNeutral}{HTML}{777D83}
\setlength{\unitlength}{1pt}
\scalebox{0.90}{%
\input{vlt-figures/figure-118.tex}%
}\endgroup
\caption{A repeated two-virion basis links physical motion to the passive Bloch pencil. (a) Two phases of the relaxed lower polarization at $ka=\pi/2$, with identical particle identities and displacement gain. Green and burgundy identify the two basis sites; black links are intracell and blue links intercell. Ghosts mark the fixed reference. (b,c) Exact forced response (solid $\alpha$, dashed $\beta$) and cycle-averaged power for the same scalar axial basis, with a Maxwell channel on each intercell relative displacement. In units $k_\star=m\omega_\star^2$, the intracell stiffness is $k_\star$, intercell stiffness $0.5k_\star$, support $0.2k_\star$, branch stiffness $0.8k_\star$, $\tau\omega_\star=1$ and $c_0/(m\omega_\star)=0.12$. The force $F_\star$ acts on the $\alpha$ site, and $P_\star=F_\star^2\omega_\star/k_\star$. The black power curve sums the two nonnegative contributions from~\eqref{eq:s4-cell-bloch-memory-loss}. The motion sketch uses the relaxed polarization; the frequency-dependent driven response is computed from the full pencil and need not have that same polarization. All spring spans are reconstructed from the displaced body surfaces with fixed coil geometry. The periodic repeat is an auxiliary translation-symmetric reference.}
\label{fig:new-s4-spectra-bloch}
\end{figure}

In a homogeneous auxiliary bulk material, take
$\vct u=\operatorname{Re}[\vct U
e^{i\vct k\cdot\vct r-i\omega t}]$ and a force-density phasor
$\vct{\mathcal F}(\vct k,\omega)$ with the same space-time factor.
Define the dynamic acoustic tensor
$Q^*_{pq}(\vct k,\omega)
=\mathcal A^*_{p\mu q\nu}(\omega)k_\mu k_\nu$.
Equation~\eqref{eq:causal-continuum-pde} becomes
\begin{equation}
 [\mathsf Q^*(\vct k,\omega)+\mathsf B_\lambda
                    -\rho\omega^2\Id_3-i\omega\mathsf b]\vct U
                                          =\vct{\mathcal F}(\vct k,\omega).
 \label{eq:bulk-viscoelastic-pencil}
\end{equation}
Setting the determinant to zero gives unforced dispersion; the forcing
weights its observable response. Frequency-dependent tensors make this
a nonlinear spectral problem. The augmented formulation also retains
internal relaxation states and handles constitutive denominators.

For an isotropic, unstressed material without body drag or target pinning,
write the complex bulk modulus as $\mathcal B^*(\omega)$ and shear modulus
as $\mu^*(\omega)$. Longitudinal and transverse waves satisfy
\begin{equation}
 \begin{aligned}
 \rho\omega^2&=\mathcal L^*(\omega)k^2,
 &\qquad \mathcal L^*&=\mathcal B^*+\tfrac43\mu^*,\\
 \rho\omega^2&=\mu^*(\omega)k^2,
 \end{aligned}
 \label{eq:isotropic-viscoelastic-branches}
\end{equation}
respectively. The two transverse polarizations are degenerate in this case.

Longitudinal motion compresses and expands material along propagation;
transverse motion probes shear across that direction. The specified
interactions and medium determine whether a restoring shear branch
exists. Away from the cubic model's isotropy condition, the full
directional acoustic tensor gives the polarizations.

\subsubsection{A finite lattice calculation behind the bulk equation}
\label{sec:s4-modes-finite-to-bulk}

A one-dimensional auxiliary calculation makes the connection to the
finite particle coordinates explicit. Take $n$ equal-mass ideal virion
centres with reference spacing $a$, retain their axial translations
$u_j$, and attach both ends to declared fixed supports:
$u_0=u_{n+1}=0$. Each edge, including its support attachments, has the
same relaxed longitudinal tangent stiffness $\kappa_0>0$ and Maxwell branch stiffness
$\kappa_1>0$ with time $\tau$. The conservative coefficient is the
longitudinal interaction curvature at the force-balanced reference;
the branch represents delayed medium response to the same relative
motion. This scalar reduction keeps the particles rigid and the count
fixed. Applying the nearest-neighbor displacement equation
\cite{Misra2012} to these fixed ends gives the finite sine patterns
\begin{equation}
 u_j=U_\ell\sin(q_\ell ja)e^{st},\qquad
 q_\ell=\frac{\ell\pi}{(n+1)a},\qquad
 \lambda_\ell=4\sin^2(q_\ell a/2),\quad \ell=1,\ldots,n.
 \label{eq:s4-modes-finite-chain-shapes}
\end{equation}
All relaxed and branch edge operators here are multiples of the same
finite difference matrix. They preserve these shapes exactly. For each
$\ell$, eliminating the branch extension gives
\begin{equation}
 m\tau s^3+ms^2+
 \tau(\kappa_0+\kappa_1)\lambda_\ell s
                  +\kappa_0\lambda_\ell=0.
 \label{eq:s4-modes-finite-chain-cubic}
\end{equation}
The cubic supplies the temporal roots associated with one collective
particle pattern; it does not supply three independent particle patterns.
Retained branch preparations fix the weights of its transients.
If all $n+1$ edges retain independent branch extensions, one further
equal-tension branch state has $u_j=0$ and $s=-1/\tau$.
Its edge forces balance at every particle; this mode lies in the nullspace
of the edge-to-particle force map and has zero displacement residue.
When $q_\ell a\ll1$, $\lambda_\ell=(q_\ell a)^2+
O((q_\ell a)^4)$. With a declared averaging cross-section $A_c$,
$\rho=m/(A_ca)$, $\mathcal L_0=\kappa_0a/A_c$ and
$\Delta\mathcal L=\kappa_1a/A_c$, division by $A_ca$ yields
\eqref{eq:sls-wave-cubic} to this order. Thus a continuum wavelength
comes from a slowly varying finite displacement pattern, and the
constitutive relaxation remains tied to relative particle motion.
For the actual cylinder, its geometry, transverse couplings and chosen
surface conditions determine the finite shapes. Spatially varying
branch coefficients generally remove the simultaneous sine
diagonalization. Fixed supports in this calculation specify a mechanical
constraint; a specular particle-wall collision law supplies a different
boundary problem.

Figure~\ref{fig:viscoelastic-finite-chain-patterns} reconstructs three of these finite sine patterns using the particle coordinates and their changing separations.

\begin{figure}[H]
\centering\begingroup
\definecolor{vltTeal}{HTML}{176B64}
\definecolor{vltBlue}{HTML}{315D80}
\definecolor{vltGold}{HTML}{A87520}
\definecolor{vltPlum}{HTML}{725877}
\definecolor{vltRed}{HTML}{762B38}
\definecolor{vltYellow}{HTML}{D5AD3C}
\definecolor{vltNeutral}{HTML}{777D83}
\setlength{\unitlength}{1pt}
\scalebox{0.90}{%
\input{vlt-figures/figure-119.tex}%
}\endgroup
\caption{Three exact finite sine patterns of seven rigid ideal virions between fixed endpoints. Each row uses $u_j=U\sin(\ell\pi j/8)$, the same $U/a=1/7$ and identical body size. Ghosts mark the equally spaced reference centres; arrows show axial displacement. The small signed stems give each edge extension $\Delta\ell_j/U=(u_{j+1}-u_j)/U$, including the end attachments. Burgundy stems denote longer spans and green stems shorter spans relative to this reference; the ticks index edges $0$ through $7$. All spring symbols retain the same turn count, coil amplitude and end leads; their free spans follow the actual body surfaces. Uniform relaxed and Maxwell edge operators preserve these shapes, while the displayed cubic supplies their temporal roots. It supplies three temporal components per pattern, rather than three additional particle patterns. The continuum approximation further requires $q_\ell a\ll1$.}
\label{fig:viscoelastic-finite-chain-patterns}
\end{figure}

\subsubsection{Longitudinal relaxation and temporal root character}

\begin{example}[One relaxation time in a longitudinal branch]
\label{ex:one-relaxation-wave}
Let
\begin{equation}
 \mathcal L^*(\omega)=\mathcal L_0
        +\Delta\mathcal L\frac{-i\omega\tau}{1-i\omega\tau},
 \qquad \mathcal L_0>0,\quad\Delta\mathcal L>0,\quad\tau>0.
 \label{eq:sls-longitudinal-modulus}
\end{equation}
The relaxed speed is $c_0=\sqrt{\mathcal L_0/\rho}$ and the instantaneous
speed is $c_\infty=\sqrt{(\mathcal L_0+\Delta\mathcal L)/\rho}$.
With temporal growth variable $s=-i\omega$, the characteristic polynomial is
\begin{equation}
 \rho\tau s^3+\rho s^2
       +\tau k^2(\mathcal L_0+\Delta\mathcal L)s
       +k^2\mathcal L_0=0.
 \label{eq:sls-wave-cubic}
\end{equation}
For $k>0$ all roots have negative real part: the coefficients are positive
and the cubic Routh--Hurwitz inequality reduces to
$\rho\tau k^2\Delta\mathcal L>0$.
At $k=0$, the translational sector retains its zero roots.
For $c_0k\tau\ll1$, expansion gives
\begin{equation}
 \omega(k)=c_0k
             -i\frac{\Delta\mathcal L\,\tau}{2\rho}k^2+O(k^3).
 \label{eq:sls-low-frequency-wave}
\end{equation}
In the high-frequency regime $c_\infty k\tau\gg1$, provided the spatial
continuum condition still holds, the positive-frequency branch approaches
\begin{equation}
 \omega(k)=c_\infty k
          -i\frac{\Delta\mathcal L}
                        {2(\mathcal L_0+\Delta\mathcal L)\tau}+O(k^{-1}).
 \label{eq:sls-high-frequency-wave}
\end{equation}
These are asymptotic regimes of the specified constitutive model. They show
how the same relaxation mechanism changes propagation speed and attenuation.
The local viscous-stress limit instead has
$\mathcal L^*=\mathcal L_0-i\omega\eta_{\mathrm L}$,
with longitudinal viscosity $\eta_{\mathrm L}\geq0$.
It gives $\rho\Pdtt{u}=\mathcal L_0\partial_x^2u
+\eta_{\mathrm L}\partial_x^2\Pdt{u}$.
Such Kelvin--Voigt gradient damping has a different short-scale structure
and does not inherit a finite wavefront speed merely from the elastic part.
\end{example}

A passive constitutive law can organize different wavelengths into
oscillatory and purely relaxational responses. For this same standard
linear solid, put
$R_{\rm L}=(\mathcal L_0+\Delta\mathcal L)/\mathcal L_0$,
$x=(c_0k\tau)^2$, and $y=s\tau$. The cubic is
$y^3+y^2+R_{\rm L}xy+x=0$, with discriminant
\begin{equation}
 \Delta_{\rm cub}=x\left[-4+
       (R_{\rm L}^2+18R_{\rm L}-27)x-4R_{\rm L}^3x^2\right].
 \label{eq:sls-wave-discriminant}
\end{equation}
When $R_{\rm L}>9$, define $A_{\rm L}=R_{\rm L}^2+18R_{\rm L}-27$ and
\begin{equation}
 x_\pm=\frac{A_{\rm L}\pm
          \sqrt{(R_{\rm L}-1)(R_{\rm L}-9)^3}}{8R_{\rm L}^3}.
 \label{eq:sls-intermediate-overdamped-band}
\end{equation}
\begin{proposition}[Oscillation and relaxation in the passive branch]
\label{prop:s4-sls-root-character}
For $R_{\rm L}>1$ and $x>0$, all temporal roots of this longitudinal
standard linear solid are stable. When $1<R_{\rm L}<9$, one negative
real root accompanies an oscillatory conjugate pair at every $x>0$.
When $R_{\rm L}>9$, the interval $x_-<x<x_+$ has three distinct negative
real roots; its endpoints have repeated roots, and the oscillatory pair
is present outside the closed interval. At $R_{\rm L}=9$, the exceptional
point $x=1/27$ has the triple root $y=-1/3$; elsewhere the pair is
oscillatory.
\end{proposition}
\begin{proof}
The Routh--Hurwitz condition for $y^3+y^2+R_{\rm L}xy+x$ is
$R_{\rm L}x>x$, which holds. The discriminant in
\eqref{eq:sls-wave-discriminant} has the sign of the downward-opening
quadratic $-4+A_{\rm L}x-4R_{\rm L}^3x^2$. Its own discriminant is
$A_{\rm L}^2-64R_{\rm L}^3=(R_{\rm L}-1)(R_{\rm L}-9)^3$.
It is negative for $1<R_{\rm L}<9$, and for $R_{\rm L}>9$ the two
positive zeros are exactly \eqref{eq:sls-intermediate-overdamped-band}.
The sign criterion for the roots of a real cubic gives the stated
classification; stability supplies their negative real parts. At the
exceptional point the cubic equals $(y+1/3)^3$.
\end{proof}
For $R_{\rm L}=16$, the boundaries are approximately
$x_-=0.013589$ and $x_+=0.017967$. Within the continuum range $ka\ll1$,
this law changes from oscillatory to relaxational response and back as
wavelength varies. The augmented state predicts both characters with
the same coordinates: an oscillatory pair describes vibration and real
roots describe relaxation. In the nonoscillatory band, mechanical and
branch coordinates provide a direct description of the disturbance.
High-frequency recovery applies within the justified constitutive and
continuum ranges.

Figure~\ref{fig:viscoelastic-sls-root-characters} shows the change of root character in the worked constitutive model.

\begin{figure}[H]
\centering\begingroup
\definecolor{vltTeal}{HTML}{176B64}
\definecolor{vltBlue}{HTML}{315D80}
\definecolor{vltGold}{HTML}{A87520}
\definecolor{vltPlum}{HTML}{725877}
\definecolor{vltRed}{HTML}{762B38}
\definecolor{vltYellow}{HTML}{D5AD3C}
\definecolor{vltNeutral}{HTML}{777D83}
\setlength{\unitlength}{1pt}
\scalebox{0.90}{%
\input{vlt-figures/figure-120.tex}%
}\endgroup
\caption{Temporal roots of \eqref{eq:sls-wave-cubic} at $R_{\rm L}=16$, with $y=s\tau$ and real $q=c_0k\tau\ge0$. (a) The three decay rates are ordered by size and plotted with multiplicity: smallest solid green, middle dashed burgundy, largest dotted black. Coincident real parts overlap outside the shaded band. (b) The oscillatory pair has zero frequency for $q_-<q<q_+$, where all three roots are negative real; endpoints have repeated roots. (c--e) Equal-component-scale complex planes display the three roots at selected wavenumbers, including the three-real-root case. All lie in the stable half-plane. At $q=0$ two translational zero roots remain. The band boundaries follow \eqref{eq:sls-intermediate-overdamped-band}. This temporal classification requires no added particle coordinates; using the continuum model also requires $ka\ll1$ and a valid constitutive window.}
\label{fig:viscoelastic-sls-root-characters}
\end{figure}

Temporal damping and spatial attenuation answer different experiments.
For real $\vct k$, a pole with $\operatorname{Im}\omega<0$ describes decay
in time. For a real drive frequency and a wave traveling in the positive
$x$ direction, the causal passive branch of $k(\omega)$ has
$\operatorname{Im}k\geq0$, since $e^{ikx}$ attenuates with $x$.
The finite reflecting cylinder has a boundary-dependent standing response.
Its spatial modes remain those of the selected boundary problem only when
the constitutive operators preserve that mode basis. General heterogeneous
or nonproportional response changes the shapes as well as the frequencies.

The real-frequency spatial root can also be evaluated without a small-loss
approximation. For the longitudinal model, write
$\mathcal L^*=\mathcal L'-i\mathcal L''$ and
$L_{\rm abs}=|\mathcal L^*|$, where $\mathcal L'>0$ and
$\mathcal L''\geq0$ at $\omega>0$. The outgoing root is
\begin{equation}
 \begin{split}
 k_{\rm r}&=\omega
   \sqrt{\frac{\rho(L_{\rm abs}+\mathcal L')}{2L_{\rm abs}^{\,2}}},\\
 k_{\rm i}&=\omega
   \sqrt{\frac{\rho(L_{\rm abs}-\mathcal L')}{2L_{\rm abs}^{\,2}}},
 \qquad k=k_{\rm r}+ik_{\rm i}.
 \end{split}
 \label{eq:s4-modes-exact-spatial-root}
\end{equation}
Squaring this expression gives $k^2=\rho\omega^2/\mathcal L^*$.
It selects decay toward positive $x$ and positive mean longitudinal
energy flux, since
$\overline J(x)=(\omega/2)|U|^2e^{-2k_{\rm i}x}
\operatorname{Re}(\mathcal L^*k)
=(\rho\omega^3/2)|U|^2e^{-2k_{\rm i}x}
\operatorname{Re}(1/k)>0$.
Causal viscoelastic dispersion treats attenuation and phase together
\cite{Hanyga2013}. This driven spatial solution exists at real
frequency even when a corresponding real-wavenumber temporal problem
has three decaying roots. The two root constructions specify different
preparations. Comparing transmission phase and amplitude with a release
measurement therefore tests one constitutive law in two ways, provided
the boundaries and the retained coordinates are matched.

\begin{definition}[Temporal lifetime and spatial attenuation length]
\label{def:s4-decay-scales}
For a simple temporal pole $\omega=\omega_{\rm r}-i\Gamma$ with
$\Gamma>0$, the \emph{amplitude decay time} is $1/\Gamma$; its
squared-amplitude envelope decays on $1/(2\Gamma)$. For a real-frequency
traveling solution with $k=k_{\rm r}+ik_{\rm i}$ and $k_{\rm i}>0$,
the \emph{amplitude attenuation length} is $1/k_{\rm i}$, while its
squared amplitude attenuates over $1/(2k_{\rm i})$.
\end{definition}

These scales specify both the preparation and the measured quantity.
They allow release and transmission data to constrain the same medium
law. Reflection and interference in the cylinder are described by its
full standing response, with the declared boundary conditions retained
in that comparison.

For an isolated weakly attenuated traveling branch with smooth real
dispersion and positive group speed $v_{\rm g}=\dd\omega_{\rm r}/\dd k$,
analytic continuation to a real drive frequency gives, to leading order,
$k_{\rm i}\simeq\Gamma/v_{\rm g}$. The amplitude decay length is then the
distance traveled in its amplitude decay time. This approximation requires
small loss and controlled dispersion derivatives; it breaks down near a
vanishing group speed or strongly mixed or overdamped branches.

Figure~\ref{fig:viscoelastic-temporal-spatial-attenuation} keeps the preparation, measured quantity and decay scale separate.

\begin{figure}[H]
\centering\begingroup
\definecolor{vltTeal}{HTML}{176B64}
\definecolor{vltBlue}{HTML}{315D80}
\definecolor{vltGold}{HTML}{A87520}
\definecolor{vltPlum}{HTML}{725877}
\definecolor{vltRed}{HTML}{762B38}
\definecolor{vltYellow}{HTML}{D5AD3C}
\definecolor{vltNeutral}{HTML}{777D83}
\setlength{\unitlength}{1pt}
\scalebox{0.80}{%
\input{vlt-figures/figure-121.tex}%
}\endgroup
\caption{Free temporal decay and driven spatial attenuation, each with amplitude $e^{-\xi}$ and squared amplitude $e^{-2\xi}$. Guides locate $\xi=1$. The two upper particle rows sample the same phase at $\Gamma t=0,1$ with $\omega_{\rm r}/\Gamma=2\pi$. The lower row samples a driven field at fixed time with $k_ix_j=0.4j$ and $k_{\rm r}x_j=j\pi/2$. Rulers mark reference site index $j=0,\ldots,6$ in every row. Identical glyphs and one displacement gain reconstruct the samples. Only for an isolated weak-loss branch with positive group speed does $k_i\simeq\Gamma/v_g$ connect these two experiments.}
\label{fig:viscoelastic-temporal-spatial-attenuation}
\end{figure}

\subsection{Functional formulation and well-posedness}
\label{sec:viscoelastic-wellposedness}

The prepared particle--medium state determines the evolution operator.
Particle coordinates, cylinder constraints and $\mathsf K$ specify the
admissible mechanics. Medium coordinates and their coupling maps specify
which motions load stored strain. The operator analysis below tests this
joint state and its energy metric.

\begin{definition}[Prepared mechanical state]
\label{def:s4-prepared-mechanical-state}
A prepared mechanical state is the admissible tuple of particle
displacements, particle velocities, and retained medium variables at one
time, sufficient to determine their subsequent deterministic evolution
for a specified applied force. For the Maxwell realization it is
$X=(\vct u,\vct v,\vct z_1,\ldots,\vct z_{r_{\rm R}})$, with the
declared supports, constraints, reference geometry and constitutive
coefficients held fixed. A boundary value or a conservation constraint is
part of this specification; it is not an extra independent coordinate.
\end{definition}

Moving a virion and then waiting can change medium strain at the same
particle position. The prepared state records that stored work and its
relaxation, including information absent from a position image.

The finite system provides a direct starting point for operator analysis.
With $\vct v=\Dt{\vct u}$ and state
$X=(\vct u,\vct v,\vct z_1,\ldots,\vct z_{r_{\rm R}})$, write
\begin{equation}
 \begin{aligned}
 \Dt{X}&=\mathsf A_{\mathrm R}X+
                  (0,\mathsf M^{-1}\vct f,0,\ldots,0),\\[2pt]
 \mathsf A_{\mathrm R}X&=
 \begin{pmatrix}
   \vct v\\
   -\mathsf M^{-1}(\mathsf K\vct u+\mathsf C_0\vct v
                 +\sum_a\mathsf G_a^{\mathsf T}\mathsf H_a\vct z_a)\\
   \mathsf G_1\vct v-\tau_1^{-1}\vct z_1\\
   \vdots\\
   \mathsf G_{r_{\rm R}}\vct v-\tau_{r_{\rm R}}^{-1}\vct z_{r_{\rm R}}
 \end{pmatrix}.
 \end{aligned}
 \label{eq:augmented-relaxation-generator}
\end{equation}
For any initial state and locally integrable force this finite linear system
has a unique global solution, given by the matrix exponential and Duhamel's
integral. This statement does not require all modes to decay.
When the storage quadratic forms are positive on the selected state space,
the energy inner product has weight
$\diag(\mathsf K,\mathsf M,\mathsf H_1,\ldots,\mathsf H_{r_{\rm R}})$ and satisfies
\begin{equation}
 \operatorname{Re}\langle X,\mathsf A_{\mathrm R}X\rangle_E
             =-\vct v^\dagger\mathsf C_0\vct v
                -\sum_a\tau_a^{-1}\vct z_a^\dagger\mathsf H_a\vct z_a.
 \label{eq:generator-dissipativity}
\end{equation}
\begin{proposition}[Contraction in the physical energy metric]
\label{prop:s4-energy-contraction}
On a finite retained space with $\mathsf K>0$, $\mathsf M>0$ and positive
branch storage on its energetic range, define
$\mathsf W_E=\diag(\mathsf K,\mathsf M,\mathsf H_1,\ldots,
\mathsf H_{r_{\rm R}})$ and
$\|X\|_E^2=X^\dagger\mathsf W_EX=2E_{\mathrm{res}}$.
The unforced propagation is a contraction in this norm. Two solutions
with the same coefficients and force difference $\Delta\vct f$ obey
\begin{equation}
 \begin{aligned}
 \mathsf W_E\mathsf A_{\mathrm R}
      +\mathsf A_{\mathrm R}^{\dagger}\mathsf W_E
   &=-2\diag\bigl(0,\mathsf C_0,
                  \tau_1^{-1}\mathsf H_1,\ldots,
                  \tau_{r_{\rm R}}^{-1}\mathsf H_{r_{\rm R}}\bigr),\\
 \|\Delta X(t)\|_E
   &\leq\|\Delta X(0)\|_E+
       \int_0^t\|\mathsf M^{-1/2}\Delta\vct f(s)\|\,\dd s.
 \end{aligned}
 \label{eq:s4-state-energy-contraction}
\end{equation}
\end{proposition}
\begin{proof}
The off-diagonal elastic and branch terms cancel in the displayed
matrix sum. With zero force its quadratic form is nonpositive, so
differentiating $\|X\|_E^2$ proves contraction. Duhamel's formula then
gives the second inequality: the energy norm of an instantaneous forcing
vector $(0,\mathsf M^{-1}\Delta\vct f,0,\ldots)$ is exactly
$\|\mathsf M^{-1/2}\Delta\vct f\|$.
\end{proof}

This is a storage--supply statement~\cite{Willems1972}.
Each contribution to $\|X\|_E^2$ has energy units; the force integral
has units $\mathrm J^{1/2}$. An individual coordinate can grow as it
receives prepared energy while total resolved energy contracts after
forcing ends. Semidefinite storage requires treating zero-energy
coordinates separately; the finite ODE still has existence and
uniqueness even when its storage form is not a norm.

Figure~\ref{fig:viscoelastic-augmented-state-release} gives a free-release example of this augmented-state propagation and its passive energy contraction.

\begin{figure}[H]
\centering\begingroup
\definecolor{vltTeal}{HTML}{176B64}
\definecolor{vltBlue}{HTML}{315D80}
\definecolor{vltGold}{HTML}{A87520}
\definecolor{vltPlum}{HTML}{725877}
\definecolor{vltRed}{HTML}{762B38}
\definecolor{vltYellow}{HTML}{D5AD3C}
\definecolor{vltNeutral}{HTML}{777D83}
\setlength{\unitlength}{1pt}
\scalebox{0.90}{%
\input{vlt-figures/figure-122.tex}%
}\endgroup
\caption{Two freely released preparations of the same scalar particle--Maxwell model: $k_1=k_0$, $c_0=0$ and $\tau\sqrt{k_0/m}=1$. The traces are exact dimensionless solutions for $X\prime=V$, $V\prime=-X-Z$ and $Z\prime=V-Z$ with the initial states in the key; the energy includes particle inertia, the relaxed spring and branch storage. The same initial displacement and velocity give different accelerations because B retains branch strain. The lower rows show the branch only; the relaxed parallel spring is omitted. The small black point is an internal medium coordinate, and dashed body outlines mark the particle reference. Both rigid virion glyphs have the same current position. One gain, $20\,\mathrm{pt}$ per unit normalized displacement, moves the particle and the medium node. Identical turns, coil amplitude and leads give excess free spring span $20Z\,\mathrm{pt}$ in each row. All energy traces decrease although their displacement histories differ; no external work or noise is supplied.}
\label{fig:viscoelastic-augmented-state-release}
\end{figure}

Two preparations can have identical $\vct u$ and $\vct v$ but different
$\vct z_a$. Their branch contributions to acceleration differ by
$-\mathsf M^{-1}\sum_a\mathsf G_a^{\mathsf T}\mathsf H_a
\Delta\vct z_a$. The question of which biophysical variables to retain
can therefore be posed as a precise question about prediction.

\begin{definition}[Exact deterministic state closure]
\label{def:s4-state-exact-closure}
For the specified finite model $\dot X=\mathsf A_{\mathrm R}X+
\mathsf B_f\vct f$, where
$\mathsf B_f\vct f=(0,\mathsf M^{-1}\vct f,0,\ldots)$, let
$\vct y_c=\mathsf P_cX$ be a fixed surjective linear choice of retained
coordinates. These coordinates have an exact deterministic state closure
if one matrix $\mathsf A_c$ gives
$\dot{\vct y}_c=\mathsf A_c\vct y_c+
\mathsf P_c\mathsf B_f\vct f$ for every admissible initial state and
every common applied force in this model.
\end{definition}

\begin{proposition}[Closure criterion for retained variables]
\label{prop:s4-state-exact-closure}
The closure in Definition~\ref{def:s4-state-exact-closure} exists if and
only if
\begin{equation}
 \mathsf P_c\mathsf A_{\mathrm R}=\mathsf A_c\mathsf P_c
 \quad\Longleftrightarrow\quad
 \mathsf A_{\mathrm R}(\ker\mathsf P_c)
                      \subseteq\ker\mathsf P_c.
 \label{eq:s4-state-exact-closure}
\end{equation}
Equivalently, any two full preparations with the same retained
coordinates must have the same retained-coordinate derivative.
\end{proposition}
\begin{proof}
If a closed equation holds for every initial state, compare its
derivative at time zero with the projected full equation to obtain the
matrix identity. That identity maps every vector in
$\ker\mathsf P_c$ back into the kernel. Conversely, if the kernel is
invariant, define
$\mathsf A_c(\mathsf P_cX)=\mathsf P_c\mathsf A_{\mathrm R}X$.
Two representatives of the same retained state differ by a kernel
vector, so this definition is independent of the representative.
Surjectivity makes it a linear operator on the entire retained space.
\end{proof}

Discarding a branch that exerts force makes
$\vct y_c=(\vct u,\vct v)$ fail the closure test. Discarding velocity
usually makes position alone fail even for a conservative oscillator.
One can retain the missing state or eliminate it with its derived
history and preparation terms. A complex viscosity characterizes the
frequency response; branch strain supplies instantaneous state data.

The test applies to every admissible preparation of the declared model.
Equilibrated hidden variables or a slow observation band can justify a
narrower reduction. A small set of outputs can likewise close an
input--output description while leaving other coordinates unresolved;
Section~\ref{sec:viscoelastic-resonance} develops this distinction.
Variable integration then identifies the omitted strain, polarization
or structural coordinate that changes future force at fixed retained
state. Assigned charges remain parameters in the fixed-charge specialization.
Resolved surface populations instead use the coupled affinities,
kinetics and chemical inventories specified above.

Figure~\ref{fig:s4-state-closure-projection} places the invariant-kernel test beside two mechanically identical particle preparations with different branch forces.

\begin{figure}[H]
\centering\begingroup
\definecolor{vltTeal}{HTML}{176B64}
\definecolor{vltBlue}{HTML}{315D80}
\definecolor{vltGold}{HTML}{A87520}
\definecolor{vltPlum}{HTML}{725877}
\definecolor{vltRed}{HTML}{762B38}
\definecolor{vltYellow}{HTML}{D5AD3C}
\definecolor{vltNeutral}{HTML}{777D83}
\setlength{\unitlength}{1pt}
\scalebox{0.90}{%
\input{vlt-figures/figure-123.tex}%
}\endgroup
\caption{The invariant-kernel criterion \eqref{eq:s4-state-exact-closure}. (a) Two full states differing by $\Delta X\in\ker\mathsf P_c$ have the same retained value. An exact closed evolution requires their projected derivatives to agree, giving the displayed test. Open arrows denote projection maps. (b) In the scalar realization $m=k=h=\tau=1$, retaining only $(U,V)$ merges two distinct branch preparations. Both whole-virion coordinates have $U=V=0$, while $Z=0,1$ gives accelerations $0,-1$. The internal node is $Y=U-Z$; the drawn branch has one natural span, six turns and an $18\,\mathrm{pt}$ displacement gain. Its excess free span is $18Z$ in points. The relaxed parallel spring is unstrained and omitted. The acceleration arrow belongs to a prepared state comparison, while the particle population remains the same.}
\label{fig:s4-state-closure-projection}
\end{figure}

\subsubsection{Continuing the state through a change of arrangement}
\label{sec:s4-transient-reference-continuation}

The fixed-coefficient results apply during residence near one reference.
The nonlinear system~\eqref{eq:s4-transient-nonlinear-system} connects
residences; a new local basis then represents its continuing particle
motion and stored medium strain.

\begin{definition}[Compatible reference chart]
\label{def:s4-transient-reference-chart}
On the same full or fixed-linearly constrained particle space with constant
$\mathsf M$, let
$\vct R_g$ belong to the same admissible affine configuration space
and let $\mathsf\Phi_g$ be a complete admissible basis
with $\mathsf\Phi_g^{\mathsf T}\mathsf M\mathsf\Phi_g=\Id$.
It includes any retained rigid directions. Define
$\vct q_g=\mathsf\Phi_g^{\mathsf T}\mathsf M(\vct r-\vct R_g)$ and
$\vct\nu_g=\mathsf\Phi_g^{\mathsf T}\mathsf M\vct v$.
The medium variables accompany this chart; they are not inferred anew
from the current particle positions.
\end{definition}

\begin{proposition}[Reference changes preserve the physical preparation]
\label{prop:s4-transient-chart-change}
Changing from chart $g$ to $h$ for the same physical state gives
\begin{equation}
 \begin{aligned}
 \vct q_h&=\mathsf O_{hg}\vct q_g
       +\mathsf\Phi_h^{\mathsf T}\mathsf M(\vct R_g-\vct R_h),\\
 \vct\nu_h&=\mathsf O_{hg}\vct\nu_g,\qquad
 \mathsf O_{hg}=\mathsf\Phi_h^{\mathsf T}\mathsf M\mathsf\Phi_g,
 \qquad \mathsf O_{hg}^{\mathsf T}\mathsf O_{hg}=\Id.
 \end{aligned}
 \label{eq:s4-transient-chart-change}
\end{equation}
Positions, velocities, kinetic energy and physical branch extensions
remain unchanged.
\end{proposition}
\begin{proof}
Substitute $\vct r=\vct R_g+\mathsf\Phi_g\vct q_g$ and
$\vct v=\mathsf\Phi_g\vct\nu_g$. Completeness on the same admissible
space makes the overlap matrix orthogonal. The branch state follows from
the unchanged $\vct r$ and $\vct y_a$.
\end{proof}

For resolved bodies, keep the actual orientation continuous when changing
the reference. In a common laboratory angular convention, local rotation
charts obey
\begin{equation}
 \exp([\vct\theta_i^{+}]_\times)
   =\exp([\vct\theta_i^{-}]_\times)
       \mathsf O_i^{0,-}(\mathsf O_i^{0,+})^{\mathsf T} .
 \label{eq:s4patch-orientation-reference-change}
\end{equation}
Take a consistent local logarithm where that chart is regular. The
chart differential transports velocities, while covectors transform
dually. The affine formula~\eqref{eq:s4-transient-chart-change} is
exact for its stated translational affine space; it is not a finite
rotation-logarithm rule. A membership label changes neither physical
orientation nor surface population, material preparation or fluid
initial data. A shared physical state remains owned once across cells.
The stress-free branch-birth rule below is a specified physical creation
law; it is not applied to every computational joining label. Actual
chemical/contact transitions retain their energy, entropy and kinetic
accounts.

A continuously updated reference also retains its frame-motion terms.
For a differentiable complete basis in the same fixed admissible space,
let $\mathsf\Omega_g=\mathsf\Phi_g^{\mathsf T}
\mathsf M\dot{\mathsf\Phi}_g$. Differentiation gives
\begin{equation}
 \begin{aligned}
 \mathsf\Omega_g^{\mathsf T}&=-\mathsf\Omega_g,\\
 \dot{\vct q}_g
   &=\vct\nu_g-\mathsf\Phi_g^{\mathsf T}\mathsf M\dot{\vct R}_g
                        -\mathsf\Omega_g\vct q_g,\\
 \dot{\vct\nu}_g
   &=\mathsf\Phi_g^{\mathsf T}\mathsf M\dot{\vct v}
                        -\mathsf\Omega_g\vct\nu_g.
 \end{aligned}
 \label{eq:s4-transient-moving-frame}
\end{equation}
The skew connection has units of inverse time and redistributes modal
components without changing $\vct\nu_g^{\mathsf T}\vct\nu_g/2$ through
that term. Motion relative to a moving reference is thus distinct from
laboratory velocity. At a degeneracy a continuous subspace frame replaces
the derivative of an arbitrary eigenvector label. These terms preserve
the original mechanics when instantaneous modes are used computationally.

If branch coordinates also change by an invertible map
$\vct z'_a=\mathsf T_a\vct z_a$, use
$\mathsf H'_a=\mathsf T_a^{-\mathsf T}\mathsf H_a\mathsf T_a^{-1}$
and transform their conjugate force maps consistently. Then
$\vct z_a^{\prime\mathsf T}\mathsf H'_a\vct z'_a
=\vct z_a^{\mathsf T}\mathsf H_a\vct z_a$.
If $\mathsf T_a$ varies with time, its transformed evolution also
contains $\dot{\mathsf T}_a\mathsf T_a^{-1}\vct z'_a$, and
$\mathsf H'_a$ follows the same time-dependent congruence.
Resetting a prestrained branch to zero would instead remove its stored
energy and alter the future force. A truncated modal space also loses
components unless their discarded preparation and response are accounted
for, as in the controlled cell reduction of
Subsection~\ref{sec:s3-compatible-cell-assembly}.

The conservative Taylor expansion makes the energy baseline explicit:
\begin{equation}
 \begin{aligned}
 U(\vct R_g+\mathsf\Phi_g\vct q_g)
   ={}&U(\vct R_g)+\nabla U(\vct R_g)^{\mathsf T}
                       \mathsf\Phi_g\vct q_g\\
     &+\tfrac12\vct q_g^{\mathsf T}
          \mathsf\Phi_g^{\mathsf T}\mathsf K_g\mathsf\Phi_g\vct q_g
        +\mathcal R_g(\vct q_g).
 \end{aligned}
 \label{eq:s4-transient-taylor-ledger}
\end{equation}
Reference energies preserve binding-depth differences, and the linear
term preserves force at a nonstationary reference. The local Hessians
approximate one nonlinear landscape with their stated remainders.
Changing the chart supplies no physical work.

A physical attachment has a different role from an observational edge
label. Its birth law specifies its relaxed coordinate and initial strain.
For a newly formed stress-free branch at $t_*$, setting
$\vct y_a(t_*^+)=\vct\psi_a(\vct r(t_*))$ sets its new strain to zero
while existing branches retain their preparation. Removing a strained
branch transfers its stored energy to declared mechanical, thermal or
chemical channels. Contact-age and evolving reference states are explicit
variables in transient-network mechanics~\cite{MengPritchardTerentjev2016};
the present construction assigns its own ideal-virion forces and couplings.

For an actual constitutive-state or control jump, the event account is
\begin{equation}
 E^+-E^-=W_{\rm event}+Q_{\rm in}-\Delta E_{\rm other}.
 \label{eq:s4-transient-event-first-law}
\end{equation}
Here $W_{\rm event}$ and $Q_{\rm in}$ are work and heat entering the
chosen resolved-plus-other account from outside that partition;
$E_{\rm other}$ is separately retained source, medium or chemical energy.
An internally retained source is counted through $\Delta E_{\rm other}$,
rather than again as external work. A mere membership
relabel has zero physical event energy. A free-energy constitutive model
adds its entropy terms rather than identifying this heat with a free-energy
drop~\cite{Seifert2012}.

With $U,\vct\psi_a$ twice continuously differentiable,
$\mathsf C$ locally Lipschitz and a prescribed locally integrable force,
the nonlinear state has a unique local solution away from force
singularities and excluded collisions. It extends
while the state remains in a compact subset of that smooth domain. For
finitely many isolated events with specified single-valued admissible
reset maps, continuation is the composition of these flow segments and
resets. The event law must provide unique transitions and exclude infinitely
many switches in finite time. Hard-wall impacts retain their separate
impact and energy law. Thus the existing linear well-posedness theorem
describes each frozen episode, while this explicit state and event law
describe how episodes are connected.

Figure~\ref{fig:s4-chart-change-and-physical-event} distinguishes a new description of the same prepared state from a physical change of its constitutive channels.

\begin{figure}[H]
\centering\begingroup
\definecolor{vltTeal}{HTML}{176B64}
\definecolor{vltBlue}{HTML}{315D80}
\definecolor{vltGold}{HTML}{A87520}
\definecolor{vltPlum}{HTML}{725877}
\definecolor{vltRed}{HTML}{762B38}
\definecolor{vltYellow}{HTML}{D5AD3C}
\definecolor{vltNeutral}{HTML}{777D83}
\setlength{\unitlength}{1pt}
\scalebox{0.90}{%
\input{vlt-figures/figure-124.tex}%
}\endgroup
\caption{Changing a complete reference chart preserves the physical preparation. Panels (a,b) show the same three rigid ideal virions and the same prepared branch; crosses mark alternative reference centres of particle 2, and the small axes show an in-plane basis rotation. The unshown normal directions complete the admissible charts. Blue arrows are projected displacement differences; the red velocity arrow has the same gain and direction in both panels. The identical coil geometry and endpoints of the existing branch make its unchanged physical extension explicit. Panel (c) keeps the same centres, scale and velocity while introducing a specified material branch at the displayed configuration. Its relaxed coordinate is chosen stress free at birth, while old branch strains continue. Only the selected material channels are drawn; other conservative interactions are omitted. The lower ledger applies to a physical event, whereas a chart change alone supplies no work \eqref{eq:s4-transient-chart-change}--\eqref{eq:s4-transient-event-first-law}. Geometry is schematic.}
\label{fig:s4-chart-change-and-physical-event}
\end{figure}

For the continuum, take the real displacement space
\begin{equation}
 V=\{\vct w\in H^1(\mathcal V_{\mathrm{lat}};\R^3):
                 \vct w\cdot\vct n=0\text{ on }\Gamma_{\mathrm w}\},
 \qquad H=L^2(\mathcal V_{\mathrm{lat}};\R^3).
 \label{eq:slip-energy-spaces}
\end{equation}
Assume a bounded Lipschitz material region and admissible boundary
portions. The normal constraint is a trace condition on
$\Gamma_{\mathrm w}$. Tangential traction there and full zero
traction on $\Gamma_{\mathrm f}$ are natural boundary conditions. The vector space $V$ is not
the clamped space $H_0^1$, since tangential displacement is allowed.

Here $L^2$ requires finite integrated squared amplitude, and $H^1$
additionally requires finite integrated squared first spatial derivatives.
These conditions admit fields with finite kinetic and elastic energy without
demanding smoothness at every point. A trace is the field's boundary value
in this energy-space sense. The weak equation tests momentum balance against
admissible displacement patterns and integrates over the material, allowing
the same force balance to apply at this lower regularity.
Define the relaxed bilinear form and mass form by
\begin{equation}
 \begin{split}
 a(\vct u,\vct w)&=\int_{\mathcal V_{\mathrm{lat}}}
       [\partial_\mu w_p\mathcal A_{p\mu q\nu}\partial_\nu u_q
                    +\vct w\cdot\mathsf B_\lambda\vct u]\,\dd^3r,\\
 m(\vct v,\vct w)&=\int_{\mathcal V_{\mathrm{lat}}}\rho\vct v\cdot\vct w\,\dd^3r.
 \end{split}
 \label{eq:continuum-energy-forms}
\end{equation}
Assume $0<\rho_-\leq\rho\leq\rho_+<\infty$, bounded symmetric material
tensors, and $0<\tau_-\leq\tau_a\leq\tau_+<\infty$.
Require a stable relaxed form: after any rigid-motion kernel is treated,
$a(\vct w,\vct w)\geq c\|\vct w\|_{H^1}^2$ for some $c>0$.
Rigid coordinates may be retained as a separate finite-dimensional sector
when the constitutive operators preserve that sector, or removed by a
compatible pinning or restriction. This assumption concerns the assembled
tensor and boundary problem. It does not follow just from positive particle
masses or a positive scalar viscosity.

Coercivity controls the $H^1$ size of an admissible deformation by its
restoring energy after rigid motions are handled. This lower bound
allows energy estimates for finite approximations to control the
continuum limit.

For clarity, consider $\mathcal D^{\mathrm v}=0$ in
\eqref{eq:continuum-maxwell-stress}, while retaining bounded body drag
and the Maxwell branches. The weak equation, for every $\vct w\in V$, is
\begin{equation}
 m(\Pdtt{\vct u},\vct w)+a(\vct u,\vct w)
   +\int_{\mathcal V_{\mathrm{lat}}}\vct w\cdot\mathsf b\Pdt{\vct u}\,\dd^3r
   +\sum_a\int_{\mathcal V_{\mathrm{lat}}}
        \partial_\mu w_p\mathcal A^{(a)}_{p\mu q\nu}Z_{a,q\nu}\,\dd^3r
                  =\int_{\mathcal V_{\mathrm{lat}}}\vct w\cdot\vct f\,\dd^3r.
 \label{eq:maxwell-weak-equation}
\end{equation}
The branch laws have their integrated history interpretation when
$\nabla\Pdt{\vct u}$ is not a classical field. Writing
$\mathsf Y_a=\nabla\vct u-\mathsf Z_a$ gives the equivalent regular law
$\Pdt{\mathsf Y_a}=(\nabla\vct u-\mathsf Y_a)/\tau_a$.
This formulation uses displacement in $V$, velocity in $H$, and branch
fields in their stored-energy $L^2$ spaces. For semidefinite branch
tensors these spaces are compatible weighted completions or quotients
by the zero-storage directions. Equivalently, the stored variables
$(\mathcal A^{(a)})^{1/2}\mathsf Z_a$ lie in their $L^2$ ranges;
branch storage does not control an arbitrary unweighted $L^2$ norm
of $\mathsf Z_a$.

Under the stated coercivity assumptions, initial data of finite stored
energy and $\vct f\in L^1(0,T;H)$ yield a unique finite-energy weak solution
on each finite time interval. One construction projects onto finitely many
admissible displacement functions, solves the resulting augmented linear
system, and uses the energy estimate to pass to the limit. The branch history
law passes to the limit with the displacement gradient. Applying the same
estimate to the difference of two solutions proves uniqueness and continuous
dependence. The estimate is displayed in the next subsection. Adding a
nonnegative gradient viscosity supplies further dissipation and requires
the corresponding velocity regularity in the weak formulation.

Existence and uniqueness, passivity, and exponential decay answer
different questions. A rigid or loss-free motion can persist in a
well-posed passive model. A stable finite system decays exponentially
when its augmented generator has no imaginary-axis eigenvalues.
A continuum additionally needs uniform control over its infinitely many
modes before a physical or numerical cutoff.

\begin{theorem}[Finite-space criterion for complete relaxation]
\label{thm:s4-finite-relaxation}
Consider the unforced finite system with $\mathsf M>0$,
$\mathsf K>0$, $\mathsf C_0\geq0$, $0<\tau_a<\infty$, and
$\mathsf H_a>0$ on each retained energetic branch space. The following
statements are equivalent: every initial state decays exponentially;
all eigenvalues of $\mathsf A_{\mathrm R}$ have strictly negative real
part; and
\begin{equation}
 \ker(\mathsf K-\omega^2\mathsf M)\cap\ker\mathsf C_0
       \cap\bigcap_a\ker(\mathsf H_a^{1/2}\mathsf G_a)=\{0\}
       \quad\text{for every real }\omega\ne0.
 \label{eq:finite-damping-detectability}
\end{equation}
\end{theorem}
\begin{proof}
Proposition~\ref{prop:s4-energy-contraction} excludes eigenvalues with
positive real part. If $\mathsf A_{\mathrm R}X=i\omega X$ for real
$\omega\ne0$, its energy identity makes
$\mathsf C_0^{1/2}\vct v=0$ and every $\vct z_a=0$. The branch
equations and $\vct v=i\omega\vct u$ then imply
$\mathsf H_a^{1/2}\mathsf G_a\vct u=0$, and momentum balance gives
$(\mathsf K-\omega^2\mathsf M)\vct u=0$.
Conversely, a nonzero vector in this intersection produces the
eigenstate $(\vct u,i\omega\vct u,0,\ldots)$.
At $\omega=0$ the state equations give $\vct v=0$,
$\vct z_a=0$ and $\mathsf K\vct u=0$, hence no nonzero eigenstate.
The criterion therefore excludes the entire imaginary axis. In finite
dimension the remaining eigenvalues lie strictly in the left half-plane;
their Jordan factors can be absorbed into an exponential bound.
Exponential decay conversely excludes an imaginary-axis eigenstate.
\end{proof}

The conclusion means that some constants $C<\infty$ and $\alpha>0$
give $\|X(t)\|_E\leq C e^{-\alpha t}\|X(0)\|_E$ for this one fixed
system. The criterion tests whether a common loss-free displacement
pattern exists; it does not specify the numerical relaxation time.
Near a loss-free pattern the rate can be arbitrarily small even though
the test still passes. The intersection must be checked on each entire
degenerate eigenspace, where linear combinations can cancel the branch
extensions. Mechanical connectivity alone therefore does not decide
relaxation. An unrestricted continuum family additionally requires
uniform control over its infinitely many modes to obtain one common
$C$ and $\alpha$.

For example, take the pinned two-site system of
Example~\ref{ex:structure-pinned-probes}, but let every lossy branch sense only
$u_1-u_2$ and set $\mathsf C_0=0$. Its common mode has $u_1=u_2$, so none of
those branches is strained; that mode remains undamped even though the pair
spring connects the sites. A branch coupled to common displacement or a
positive body drag changes this conclusion. The loss maps must therefore
be specified along with the conservative interaction graph. Darkness to
a loss channel and darkness to a detector are different properties: an
undamped mode may be easily measured, while a damped mode may cancel in
one displacement readout. Neither invisibility establishes absence of
stored excitation energy.

Figure~\ref{fig:viscoelastic-loss-dark-mode} illustrates the finite-system loss-kernel criterion with the pinned two-site example.

\begin{figure}[H]
\centering\begingroup
\definecolor{vltTeal}{HTML}{176B64}
\definecolor{vltBlue}{HTML}{315D80}
\definecolor{vltGold}{HTML}{A87520}
\definecolor{vltPlum}{HTML}{725877}
\definecolor{vltRed}{HTML}{762B38}
\definecolor{vltYellow}{HTML}{D5AD3C}
\definecolor{vltNeutral}{HTML}{777D83}
\setlength{\unitlength}{1pt}
\scalebox{0.90}{%
\input{vlt-figures/figure-125.tex}%
}\endgroup
\caption{A supported pair with one specified loss channel in \eqref{eq:finite-damping-detectability}. Equal masses have $\mathsf K/k_\star=\left(\begin{smallmatrix}2&-1\\-1&2\end{smallmatrix}\right)$ and conservative mass-weighted patterns $\vct e_\pm=(1,\pm1)/\sqrt2$. (a) Common and relative snapshots use displacements $(a,a)$ and $(a,-a)$ with $a=9\,\mathrm{pt}$. Matched coils attach to the actual rigid-body surfaces and fixed end supports. Common motion preserves the interior span; the displayed relative phase compresses it by $2a$. (b) Each normalized modal release begins at one with zero velocity and relaxed branch. Exact augmented evolution preserves common-mode energy and dissipates relative-mode energy. The overlap with $G$ explains the contrast even though both patterns belong to the same connected mechanics. Darkness to this loss channel is distinct from a detector overlap. The full criterion includes complete degenerate eigenspaces.}
\label{fig:viscoelastic-loss-dark-mode}
\end{figure}

\subsection{Mechanical energy transport and internal dissipation}
\label{sec:viscoelastic-energy-transport}

A displacement can decay as energy relaxes locally, propagates out of
the observed region or enters recoverable medium storage. The local
power balance assigns these processes to dissipation, boundary transport
and stored energy, respectively.

\begin{definition}[Power-conjugate mechanical boundary variables]
\label{def:s4-boundary-power-variables}
On an oriented material boundary, traction $\vct t=\mathsf T\vct n$
and boundary velocity $\vct v=\partial_t\vct u$ are power-conjugate
variables: their product $\vct v\cdot\vct t$ is the rate of mechanical
work supplied to the enclosed material per unit area. The boundary is
power-free when the integral of this product vanishes for every
admissible motion under its stated boundary law.
\end{definition}

The declared supported-slip wall condition sets normal velocity to zero. A tangentially
slipping, traction-free material boundary also removes tangential work,
whereas frictional contact with a separately moving medium exchanges
power through the relative motion. A stationary vessel thus fixes a
geometric boundary; its mechanical, thermal and electrical exchange laws
still supply different physical data. The following balance uses the
total incremental traction, including the elastic, viscous and branch
contributions already specified.

The continuum stored energy density for the specified branch realization is
\begin{equation}
 e_{\mathrm{res}}=\tfrac12\rho|\Pdt{\vct u}|^2
       +\tfrac12\mathsf F:\mathcal A:\mathsf F
       +\tfrac12\vct u\cdot\mathsf B_\lambda\vct u
       +\tfrac12\sum_a\mathsf Z_a:\mathcal A^{(a)}:\mathsf Z_a.
 \label{eq:continuum-stored-energy-density}
\end{equation}
For a prestressed reference, the local elastic quadratic density can
be signed; nonnegative total storage follows from the stable assembled
form, rather than from a pointwise interpretation as a probability.
Kinetic and branch storage retain their specified nonnegative forms.

With constant coefficients in time, multiply momentum balance by velocity
and differentiate the storage terms. The identity
$\Pdt{u_p}\partial_\mu T_{p\mu}
=\partial_\mu(\Pdt{u_p}T_{p\mu})
-T_{p\mu}\partial_\mu\Pdt{u_p}$ cancels the stress work against storage
and relaxation. The resulting local balance is
\begin{equation}
 \begin{split}
 \partial_t e_{\mathrm{res}}+\partial_\mu S_\mu
                         &=\Pdt{\vct u}\cdot\vct f-\mathfrak d,\\
 S_\mu&=-\Pdt{u_p}T_{p\mu},\\
 \mathfrak d&=\Pdt{\vct u}\cdot\mathsf b\Pdt{\vct u}
        +\nabla\Pdt{\vct u}:\mathcal D^{\mathrm v}:\nabla\Pdt{\vct u}
        +\sum_a\tau_a^{-1}\mathsf Z_a:\mathcal A^{(a)}:\mathsf Z_a
                                                              \geq0.
 \end{split}
 \label{eq:viscoelastic-local-flux}
\end{equation}
The mechanical flux $\vct S$ has units of power per area. Its outward normal
component is $\vct S\cdot\vct n=-\Pdt{\vct u}\cdot\vct t$.
Thus the chosen slip/free conditions eliminate material boundary power,
while $\mathfrak d$ records the resolved storage loss.
For the energetic mechanical closure, the unresolved internal-energy
balance receives $+\mathfrak d$. A thermally sealed vessel has zero
outward heat flux, and the complete energy account is conserved. If
the constitutive storage represents free energy, additional entropy,
thermoelastic, chemical, or interface terms belong in that thermal
balance. Internal material boundaries can exchange energy with the
enclosed medium even though the outer vessel is sealed.

In particular, $E_{\mathrm{res}}=\int_{\mathcal V_{\mathrm{lat}}}e_{\mathrm{res}}\,\dd^3r$
obeys
\begin{equation}
 \begin{split}
 \Dt{E_{\mathrm{res}}}+\int_{\mathcal V_{\mathrm{lat}}}\mathfrak d\,\dd^3r
                              &=\int_{\mathcal V_{\mathrm{lat}}}\Pdt{\vct u}\cdot\vct f\,\dd^3r,\\
 \sqrt{E_{\mathrm{res}}(t)}&\leq\sqrt{E_{\mathrm{res}}(0)}
        +\frac1{\sqrt2}\int_0^t
                            \|\vct f(s)/\sqrt\rho\|_{L^2}\,\dd s.
 \end{split}
 \label{eq:continuum-passive-energy-estimate}
\end{equation}
The bound follows from Cauchy--Schwarz and
$\int\rho|\Pdt{\vct u}|^2\leq2E_{\mathrm{res}}$; at zero energy it is
understood by regularization. It controls the Galerkin solutions uniformly
over finite time intervals and makes the dependence on preparation explicit.

For a harmonic field, the cycle-averaged flux is
$\overline S_\mu=(\omega/2)
\operatorname{Im}(U_p^*T_{p\mu})$.
In a lossless scalar traveling wave this equals the usual positive energy
flux in the direction of propagation. In a viscoelastic medium the same
stress-velocity expression applies, with the constitutive loss included in
the local balance. It complements the discrete edge currents of
\eqref{eq:edge-flux}.

For a fixed material subregion $\Omega\subseteq\mathcal V_{\mathrm{lat}}$
with outward normal $\vct n_\Omega$, the same identity gives
\begin{equation}
 \begin{aligned}
 \frac{\dd E_\Omega}{\dd t}+D_\Omega
    &=P_\Omega^{\rm body}+P_\Omega^{\partial},\\
 E_\Omega&=\int_\Omega e_{\mathrm{res}}\,\dd^3r,
 &D_\Omega&=\int_\Omega\mathfrak d\,\dd^3r,\\
 P_\Omega^{\rm body}&=\int_\Omega
                    \partial_t\vct u\cdot\vct f\,\dd^3r,
 &P_\Omega^{\partial}&=\int_{\partial\Omega}
      \partial_t\vct u\cdot(\mathsf T\vct n_\Omega)\,\dd A.
 \end{aligned}
 \label{eq:s4-subregion-power-balance}
\end{equation}
Each power has units of watts. Velocity continuity and opposite
tractions cancel the powers of an ideal bonded interface. A compliant
interface adds its own storage and relative-motion law. This work
compatibility permits passive subsystems to be assembled~\cite{Willems1972}.
At prestress, $E_\Omega$ can be a signed local quadratic contribution;
the global energy estimate uses positivity of the assembled form.

\begin{proposition}[Resolved loss from a complete mechanical work account]
\label{prop:s4-work-loss-inference}
For an interval $[t_0,t_1]$ of duration $T_o>0$, define the supplied work
$W_\Omega=\int_{t_0}^{t_1}
(P_\Omega^{\rm body}+P_\Omega^{\partial})\,\dd t$ and the stored-energy
change $\Delta E_\Omega=E_\Omega(t_1)-E_\Omega(t_0)$. Then
\begin{equation}
 \frac1{T_o}\int_{t_0}^{t_1}D_\Omega\,\dd t
       =\frac{W_\Omega-\Delta E_\Omega}{T_o}\geq0.
 \label{eq:s4-work-loss-window}
\end{equation}
For a periodic steady state observed over whole periods,
$\Delta E_\Omega=0$ and the supplied average power equals the average
resolved dissipation.
\end{proposition}
\begin{proof}
Integrate~\eqref{eq:s4-subregion-power-balance}; the inequality follows
from $\mathfrak d\geq0$, and periodicity cancels the endpoint storage.
\end{proof}

For the ideal virion array this gives a useful inverse constraint. A force
and its conjugate particle velocity determine supplied power; a
calibrated displacement and velocity field, together with the stated
medium state, determines the model's storage change. Their difference
tests the total resolved loss. A position trace in a small part of the
cylinder does not by itself separate local dissipation from energy
transport across that region. The boundary term states exactly which
additional mechanical information that separation requires.

Figure~\ref{fig:s4-local-work-inference} shows how the boundary-work term distinguishes two identical local storage decreases.

\begin{figure}[H]
\centering\begingroup
\definecolor{vltTeal}{HTML}{176B64}
\definecolor{vltBlue}{HTML}{315D80}
\definecolor{vltGold}{HTML}{A87520}
\definecolor{vltPlum}{HTML}{725877}
\definecolor{vltRed}{HTML}{762B38}
\definecolor{vltYellow}{HTML}{D5AD3C}
\definecolor{vltNeutral}{HTML}{777D83}
\setlength{\unitlength}{1pt}
\scalebox{0.90}{%
\input{vlt-figures/figure-126.tex}%
}\endgroup
\caption{Work channels for a fixed material subregion, with whole-particle coordinates shown explicitly. The upper auxiliary instantaneous ledger has boundary input $3\,\mathrm W$, output $1\,\mathrm W$, internal resolved loss $1\,\mathrm W$ and storage growth $1\,\mathrm W$; body-force power is zero. The three power-arrow lengths share the scale $30\,\mathrm{pt\,W^{-1}}$ and are not velocity vectors. The springs show a selected interaction topology across the fixed boundary, with no inferred strain or force magnitude. The two lower rows are independent one-second accounts: equal storage decreases can represent internal relaxation or exported work. All powers and energies are illustrative ledger data, not estimates from a displacement movie. The inference in \eqref{eq:s4-work-loss-window} needs boundary work and the complete assigned storage; its nonnegative loss does not require a positive pointwise prestress allocation.}
\label{fig:s4-local-work-inference}
\end{figure}

Even a complete average-power measurement determines the sum of the
loss-channel contributions in this balance. Separating local drag,
relative-particle branches and a retained network relaxation requires
their different spatial couplings, times or additional medium readouts.
The same force pattern can load several channels at once, while a
channel can be invisible to a particular pattern. This connects the
state-closure question to the inverse question: a candidate biophysical
variable earns a mechanical role by a declared stored-work term,
coupling and evolution law whose consequences can be compared with the
observed work and motion. If the constitutive storage is an isothermal
free energy, this resolved mechanical loss is the decrease of available
work; identifying a temperature rise or a chemical heat release needs
the additional thermal and chemical account. The following
susceptibility formulation organizes what the chosen force and readout
channels can distinguish.

Figure~\ref{fig:viscoelastic-wave-flux-loss} relates a traveling-wave attenuation envelope to its local mechanical flux and dissipation.

\begin{figure}[H]
\centering\begingroup
\definecolor{vltTeal}{HTML}{176B64}
\definecolor{vltBlue}{HTML}{315D80}
\definecolor{vltGold}{HTML}{A87520}
\definecolor{vltPlum}{HTML}{725877}
\definecolor{vltRed}{HTML}{762B38}
\definecolor{vltYellow}{HTML}{D5AD3C}
\definecolor{vltNeutral}{HTML}{777D83}
\setlength{\unitlength}{1pt}
\scalebox{0.90}{%
\input{vlt-figures/figure-127.tex}%
}\endgroup
\caption{Traveling mechanical flux and local dissipation in the unforced homogeneous scalar bulk example. At real drive frequency, $\overline S=S_0e^{-2k_ix}$ and $\partial_x\overline S=-\overline{\mathfrak d}$. (a) Remaining flux and cumulative transfer sum to the input. (b) The local loss has the stated power-per-volume normalization. (c) An internal region with $k_iL=0.5$ has input and output mean-power arrows in one $58\,\mathrm{pt}/(AS_0)$ gain. Its reference spatial gain is $600\,\mathrm{pt}$ per unit $\xi$, with $k_iA_{\rm disp}=0.02$. Rigid virion shifts therefore use $12\,\mathrm{pt}$ per displacement amplitude and sample $e^{-\xi}\cos(k_rx-\omega t)$ at the labelled locations, with successive phases $0,\pi/2,\pi$. Energy transfer is accompanied by oscillatory centre motion within this internal region. The outer reflecting cylinder requires its full boundary solution; temperature evolution requires a heat account.}
\label{fig:viscoelastic-wave-flux-loss}
\end{figure}

These currents transport mechanical energy through the selected coordinates;
they do not transport virions across the reflecting wall. A disturbance can
redistribute energy while the particles remain in their reference basin.
In the quantized harmonic description a change of mode occupation transfers
energy $\hbar\omega_s$ per quantum without changing particle count. The
local balance above is an energy balance, not a general conservation law
for phonon number. A local phonon-population current requires an additional
resolved carrier or kinetic approximation. Broadband, strongly damped,
and purely relaxational disturbances are described directly by the
mechanical storage, flux, and medium-transfer terms.

\subsubsection{Energy accounting for a growing selected cluster}
\label{sec:s4-transient-cluster-energy}

The fixed material-region balance and a growing-cluster balance describe
different observations. A fixed region can lose energy through its boundary;
a selected cluster can additionally gain an entire member's assigned energy
when its membership changes. This contribution is essential when comparing
spectral energy before and after recruitment.

Choose once a particle-resolved allocation $e_i$ of the full resolved
mechanical storage, with $\sum_i e_i=E$. For a pairwise conservative
potential, a possible convention assigns half of each pair energy to each
endpoint. Branch and other coupling energies likewise have declared weights
that sum to one over the fixed ideal-virion pool. A separately retained
medium subsystem can instead keep a named storage $E_{\rm med}$:
then $E=\sum_i e_i+E_{\rm med}$, and the same cluster allocation is
applied only to the particle-assigned terms while $E_{\rm med}$ keeps
its own balance.
Local allocations can be signed even when the complete storage is bounded
below. The allocation and common energy zero are held fixed throughout
the comparison.

\begin{definition}[Selected-cluster storage]
\label{def:s4-transient-selected-energy}
Let $h_i^{\mathcal C}(t)\in\{0,1\}$ record membership in the selected
cluster, as in Definition~\ref{def:s23-cluster-membership}. Its assigned
resolved storage is
\begin{equation}
 E_{\mathcal C}(t)=\sum_i h_i^{\mathcal C}(t)e_i(t).
 \label{eq:s4-transient-cluster-storage}
\end{equation}
This is a partition of the declared full storage, including its chosen
allocation of cross-cluster interactions.
\end{definition}

For smooth mechanical evolution and isolated observational membership
updates at times $t_k$, with continuous physical energies $e_i$, the
distributional balance is
\begin{equation}
 \frac{\dd E_{\mathcal C}}{\dd t}
   =\sum_i h_i^{\mathcal C}\dot e_i
      +\sum_k\left[\sum_i
           \Delta h_i^{\mathcal C}(t_k)e_i(t_k)\right]\delta(t-t_k).
 \label{eq:s4-transient-membership-energy-transfer}
\end{equation}
It follows directly by differentiating the weighted sum between events
and taking its jump at each event. The first term contains the actual
mechanical transport, work and dissipation under the chosen allocation;
the second records membership transfer. Complementary membership cancels
the latter exactly, recovering the full-vessel account. Actual contact-state
energy jumps additionally obey~\eqref{eq:s4-transient-event-first-law}.
If a physical energy jump and membership update coincide, their exact
increment is instead evaluated with declared one-sided values:
\begin{equation}
 \Delta E_{\mathcal C}
   =\sum_i\bigl(h_i^{\mathcal C,+}e_i^+
                       -h_i^{\mathcal C,-}e_i^-\bigr)
   =\sum_i h_i^{\mathcal C,-}\Delta e_i
                       +\sum_i\Delta h_i^{\mathcal C}e_i^+.
 \label{eq:s4-transient-simultaneous-event}
\end{equation}
This accounts for the entering member's post-event assigned energy
without an ambiguous product of discontinuous quantities.

This gives a practical route for a growing ideal-virion assembly: retain
particle identities and strain history, resolve the work and energy
exchange during an encounter, then update the conditional cluster spectrum.
A rise in its assigned energy can contain imported member energy as well
as newly deposited excitation. Comparing the same registered particles,
or recording the membership term explicitly, separates these contributions.
The capture and escape mechanisms of Section~3 and the viscoelastic power
balance then constrain one trajectory and one energy ledger.

Figure~\ref{fig:s4-selected-cluster-event-ledger} evaluates this one-sided update and its complementary account for a four-particle event.

\begin{figure}[H]
\centering\begingroup
\definecolor{vltTeal}{HTML}{176B64}
\definecolor{vltBlue}{HTML}{315D80}
\definecolor{vltGold}{HTML}{A87520}
\definecolor{vltPlum}{HTML}{725877}
\definecolor{vltRed}{HTML}{762B38}
\definecolor{vltYellow}{HTML}{D5AD3C}
\definecolor{vltNeutral}{HTML}{777D83}
\setlength{\unitlength}{1pt}
\scalebox{0.90}{%
\input{vlt-figures/figure-128.tex}%
}\endgroup
\caption{Simultaneous physical energy change and recruitment, using the one-sided rule \eqref{eq:s4-transient-simultaneous-event}. The four rigid ideal virions keep their identities and displayed positions; the pale boundary selects the observed cluster and particle 4 changes membership. The connecting coils have identical endpoints and geometry in the two frames; membership colour does not change a force law. The table stipulates assigned energies in an arbitrary positive unit $E_\star$, with one fixed allocation and energy zero. It is an algebraic event ledger, not an integrated trajectory: the physical energy update can include declared changes of momentum, material storage or interaction state. The entering particle contributes its post-event assigned energy $1.4E_\star$. That membership term cancels in the complementary cluster, leaving the full-pool change $-0.6E_\star$. The latter still needs the physical event work/heat/source account; it is not generated by selecting more particles.}
\label{fig:s4-selected-cluster-event-ledger}
\end{figure}

\subsection{Resonance, susceptibility, and controlled target coupling}
\label{sec:viscoelastic-resonance}

The susceptibility to a specified force is
\begin{equation}
 \mathsf R_{\rm mech}(\omega)
       =[\mathsf K^*(\omega)-\omega^2\mathsf M]^{-1}.
 \label{eq:mechanical-susceptibility}
\end{equation}
The force pattern and readout select the observed spectrum from this
operator. Pole frequency, displacement peak and absorbed-power peak
are different features. For one viscously damped mode in
\eqref{eq:proportional-modal-damping}, driven by mass-weighted force
phasor $F_s$,
\begin{equation}
 |Q_s(\omega)|^2=
       \frac{|F_s|^2}{(\omega_s^2-\omega^2)^2+4\gamma_s^2\omega^2}.
 \label{eq:forced-modal-lineshape}
\end{equation}
For frequency-independent $F_s$ its nonzero displacement peak is at
\begin{equation}
 \omega_{\mathrm{peak}}=\sqrt{\omega_s^2-2\gamma_s^2},
 \qquad 0<\gamma_s<\omega_s/\sqrt2.
 \label{eq:displacement-resonance-peak}
\end{equation}

The strict positive-loss condition gives a finite steady response.
At zero loss, a force exactly at the conservative frequency instead
produces secular growth, and the resonant phasor inverse is singular.
This differs from the ring-down frequency
$\Omega_s=\sqrt{\omega_s^2-\gamma_s^2}$.
The average supplied power is
$\overline P_s=\gamma_s\omega^2|Q_s(\omega)|^2$.
For constant nonzero force amplitude and $\gamma_s>0$, the absorbed-power
peak is instead exactly
$\omega_s$. Its positive half-power frequencies are
\begin{equation}
 \omega_\pm=\sqrt{\omega_s^2+\gamma_s^2}\pm\gamma_s,
 \qquad \omega_+-\omega_-=2\gamma_s.
 \label{eq:absorbed-power-half-widths}
\end{equation}
Thus a power peak can persist even when the unforced mode is overdamped;
it alone does not establish oscillatory ringdown. For weak damping the
positive-frequency squared-displacement line is approximately
Lorentzian, with full width $2\gamma_s$ in angular frequency, and
$\mathcal Q_s\simeq\omega_s/(2\gamma_s)$.
Memory response, nearby modes, or a frequency-dependent drive require the
full susceptibility rather than these single-mode formulas.

\subsubsection{Finite preparation, spectral selection, and measured work}
\label{sec:finite-pulse-excitation}

An excitation requires a coupling that supplies or redistributes energy.
For a real normalized conservative mode $\vct e_s$, the physical force
pattern $\vct f(t)$ produces the modal force
$F_s(t)=\vct e_s^{\mathsf T}\mathsf M^{-1/2}\vct f(t)$.
This spatial overlap can vanish even at resonance. Its units are
$\mathrm{kg}^{1/2}\mathrm m\,\mathrm s^{-2}$, and $F_s\,\dd Q_s$ is work.
The analogous quantum requirement is a nonzero matrix element of the
interaction operator between the proposed states. Frequency matching
cannot replace the required coupling~\cite{Harrow2016}.

Consider a transparent reference calculation: one positive conservative
mode starts at rest; a prescribed force acts for $0<t<T_p$; loss is
negligible over this preparation. Its driven Hamiltonian and equations are
\begin{equation}
 H_s(t)=\tfrac12(\Pi_s^2+\omega_s^2Q_s^2)-F_s(t)Q_s,
 \qquad \Dt{\psi_s}=-i\omega_s\psi_s+\frac{iF_s(t)}{\sqrt2}.
 \label{eq:excitation-pulse-hamiltonian}
\end{equation}
Integrating the latter equation gives
\begin{equation}
 \begin{split}
 \psi_s(T_p)&=\frac{i e^{-i\omega_sT_p}}{\sqrt2}
              \int_0^{T_p}F_s(t)e^{i\omega_st}\,\dd t,\\
 E_s^{\rm after}&=\tfrac12
             \left|\int_0^{T_p}F_s(t)e^{i\omega_st}\,\dd t\right|^2
          =\int_0^{T_p}F_s(t)\Dt{Q_s(t)}\,\dd t.
 \end{split}
 \label{eq:excitation-finite-pulse-energy}
\end{equation}
The pulse integral has units $\mathrm J^{1/2}$; its squared modulus is
the energy left after switch-off. The work equality follows from zero
initial energy. A prescribed drive supplies this work; an isolated
description retains the source and its changing energy explicitly.

For $F_s(t)=F_0\cos(\omega_dt)$ during the pulse, define
$\operatorname{sinc}x=\sin x/x$ with $\operatorname{sinc}0=1$. The exact
Fourier integral is
\begin{equation}
 \int_0^{T_p}F_s(t)e^{i\omega_st}\,\dd t
 =\frac{F_0T_p}{2}\left[
 e^{i(\omega_s-\omega_d)T_p/2}
       \operatorname{sinc}\!\frac{(\omega_s-\omega_d)T_p}{2}
 +e^{i(\omega_s+\omega_d)T_p/2}
       \operatorname{sinc}\!\frac{(\omega_s+\omega_d)T_p}{2}\right].
 \label{eq:excitation-pulse-frequency-window}
\end{equation}
For $\omega_dT_p\gg1$ and small relative detuning, dropping the second
component gives $E_s^{\rm after}\simeq F_0^2T_p^2
\operatorname{sinc}^2[(\omega_s-\omega_d)T_p/2]/8$.
Figure~\ref{fig:excitation-finite-pulse} compares it with the exact
real-force result. Pulse duration sets a frequency window of order
$1/T_p$.

\begin{figure}[H]
\centering\begingroup
\definecolor{vltTeal}{HTML}{176B64}
\definecolor{vltBlue}{HTML}{315D80}
\definecolor{vltGold}{HTML}{A87520}
\definecolor{vltPlum}{HTML}{725877}
\definecolor{vltRed}{HTML}{762B38}
\definecolor{vltYellow}{HTML}{D5AD3C}
\definecolor{vltNeutral}{HTML}{777D83}
\setlength{\unitlength}{1pt}
\scalebox{0.80}{%
\input{vlt-figures/figure-129.tex}%
}\endgroup
\caption{Energy after a finite five-cycle force pulse. (a) $F_s=F_0\cos\omega_dt$ acts for $0<t<T_p$, with $\omega_dT_p=10\pi$; open endpoint markers indicate the specified pulse interval. (b) The exact real-force integral and its near-resonant sinc window are plotted against $\delta=(\omega_s-\omega_d)T_p/(2\pi)$, using the energy normalization on the ordinate. The initially resting conservative mode retains this energy after switch-off. The approximation drops the counterrotating contribution; it agrees near resonance under the stated long-carrier condition.}
\label{fig:excitation-finite-pulse}
\end{figure}

Under canonical quantization, the same ideal linear pulse acting on the
mode vacuum prepares a coherent state, up to an overall phase. With the
rotating-frame amplitude
\begin{equation}
 \beta_s=\frac{i}{\sqrt{2\hbar\omega_s}}
               \int_0^{T_p}F_s(t)e^{i\omega_st}\,\dd t,
 \qquad
 \langle\hat N_s\rangle=|\beta_s|^2,
 \qquad
 \Delta\langle\hat H_s^{(0)}\rangle
                =\hbar\omega_s|\beta_s|^2=E_s^{\rm after}.
 \label{eq:excitation-pulse-quantum-correspondence}
\end{equation}
Here $\hat H_s^{(0)}=\tfrac12(\hat\Pi_s^2+\omega_s^2\hat Q_s^2)$
is the unforced reference energy; $\Delta$ compares its expectations before
and after preparation, excluding the drive coupling during the pulse.
The laboratory-frame coherent amplitude at switch-off is
$e^{-i\omega_sT_p}\beta_s$. Individual occupation measurements are
integer, but their mean and the mean added energy vary continuously with
the pulse. Linear coordinate coupling connects neighboring oscillator
levels in first-order perturbation theory; finite drive can populate
many levels. The full coherent occupation statistics and their
single-phonon preparation limit are developed in
\eqref{eq:phonon-driven-coherent-amplitude}--\eqref{eq:phonon-linear-drive-one-quantum-limit}.
Thus a classical force pulse is not a deterministic one-phonon source.

A long weak monochromatic perturbation selects allowed level differences
near its frequency; a finite pulse broadens that selection, and medium
coupling can add a resonance linewidth~\cite{Harrow2016}.
Both source and medium enter energy conservation. Prepared-state energy
variance is a separate quantity. When damping is appreciable during
preparation, causal susceptibility or the augmented state replaces the
isolated calculation and includes energy transferred to the medium.

\subsubsection{Passivity, probe noise, and static compliance}
\label{sec:damped-probe-fluctuations}

Let $\operatorname{Im}_{\rm H}\mathsf R_{\rm mech}
=(\mathsf R_{\rm mech}-\mathsf R_{\rm mech}^\dagger)/(2i)$ denote the Hermitian
absorptive part. On real positive frequencies where the inverse exists,
the pencil is Hermitian storage minus $i\mathsf K''$, and the inverse
identity gives
\begin{equation}
 \begin{aligned}
 \operatorname{Im}_{\rm H}\mathsf R_{\rm mech}
     &=\mathsf R_{\rm mech}^\dagger\mathsf K''\mathsf R_{\rm mech}\\
     &=\mathsf R_{\rm mech}\mathsf K''\mathsf R_{\rm mech}^\dagger\ge0,\\
 \overline P&=\frac\omega2\vct F^\dagger
                  \operatorname{Im}_{\rm H}\mathsf R_{\rm mech}\vct F.
 \end{aligned}
 \label{eq:susceptibility-passive-inverse-identity}
\end{equation}
No simultaneous diagonalization of mass, storage, and loss is needed.
In the present reciprocal model this matrix equals the componentwise
imaginary part; a cross-channel entry by itself need not be positive.

For the physical readout $Y=\vct\ell^{\mathsf T}\vct u$ with real
$\vct\ell$, driven through the conjugate interaction $-F_Y Y$,
$\mathcal R_Y=\vct\ell^{\mathsf T}\mathsf R_{\rm mech}\vct\ell$.
Its connected stationary equilibrium spectrum obeys
\begin{align}
 S_Y^{\rm sym}(\omega)
   &=\hbar\coth\!\left(\frac{\hbar\omega}{2k_BT}\right)
                   \operatorname{Im}\mathcal R_Y(\omega+i0),&&\omega>0,\nonumber\\
 \operatorname{Var}_T(Y)
   &=\frac\hbar\pi\int_0^\infty
       \coth\!\left(\frac{\hbar\omega}{2k_BT}\right)
                   \operatorname{Im}\mathcal R_Y(\omega+i0)\,\dd\omega.
 \label{eq:damped-probe-equilibrium-fdt}
\end{align}
This is the finite-loss counterpart of Section~3's displacement
FDT~\cite{CallenWelton1951,Clerk2010}. In a full finite equilibrium Hamiltonian the
formula is understood through its discrete coupled spectral measure.
For a regular stationary reservoir response with decaying preparation
transients, it can also be obtained by filtering the matched force noise,
$\mathsf S_u=\mathsf R_{\rm mech}\mathsf S_F\mathsf R_{\rm mech}^\dagger$, using
$\mathsf K''=\omega\operatorname{Re}_{\rm H}\mathsf Z_{\rm eff}$ for
the Maxwell realization, whose classical force spectrum is
$\mathsf S_\Xi=2k_BT\operatorname{Re}_{\rm H}\mathsf Z_{\rm eff}$.
That filtering step does not justify discarding nondecaying initial
terms or correlations in an arbitrary finite closed bath.

On a stable sector with finite static compliance, the classical limit
and the causal static sum rule give
\begin{equation}
 \operatorname{Var}_{\rm cl}(Y)
   =\frac{2k_BT}{\pi}\int_0^\infty
                   \frac{\operatorname{Im}\mathcal R_Y(\omega)}\omega\,\dd\omega
   =k_BT\mathcal R_Y(0).
 \label{eq:damped-probe-static-compliance}
\end{equation}
Changing passive damping redistributes the fluctuation spectrum without
necessarily reducing its integrated classical variance when the relaxed
stiffness is held fixed. For the completed-square classical bath,
integration of its shifted Gaussian coordinates gives
$\mathcal R_Y(0)=\vct\ell^{\mathsf T}\mathsf K^{-1}\vct\ell$ on that sector.
At appreciable quantum coupling, the reduced equilibrium covariance
instead follows the full coupled spectral integral; it is generally
not the Gibbs covariance of the uncoupled bare oscillator
\cite{TalknerHanggi2020}. The independent-mode equilibrium formulas
remain the selected harmonic reference or a justified weak-coupling limit.

For dimensionless $\vct\ell$, the readout $Y$ has length units,
$\mathcal R_Y$ compliance units and the two-sided $S_Y$ units of length
squared times time. The connected correlation subtracts the ensemble
mean; its spectrum integrates to variance with factor $1/(2\pi)$.
At fixed relaxed curvature and temperature, damping redistributes that
variance across frequencies. Temperature comparisons additionally
control viscosity, relaxation times and reference curvature, as in
Section~\ref{sec:phonon-temperature-comparison}.

An ideal classical measurement with unit gain in the positive-frequency band
$[\omega_{\rm lo},\omega_{\rm hi}]$ and its negative-frequency mirror, zero
gain elsewhere, and $0\leq\omega_{\rm lo}<\omega_{\rm hi}$ has
\begin{equation}
 \operatorname{Var}_{\rm band}(Y)=\frac1\pi
     \int_{\omega_{\rm lo}}^{\omega_{\rm hi}}
                             S_Y^{\rm cl}(\omega)\,\dd\omega.
 \label{eq:classical-detector-band-variance}
\end{equation}
Here $S_Y^{\rm cl}$ is the connected two-sided classical displacement spectrum.
Moving spectral weight outside that band can reduce the recorded jitter
without changing $k_BT\mathcal R_Y(0)$. An actual instrument's frequency
transfer and its own noise must be included before interpreting that change
as altered lattice stiffness.

Figure~\ref{fig:opening-five-equilibrium-band} makes the distinction between total equilibrium variance and a finite recording band explicit. At fixed mass and relaxed curvature, matched thermal damping changes the spectral distribution and the fraction a detector records, while the full classical displacement variance remains fixed.

\begin{figure}[H]
\centering\begingroup
\definecolor{vltTeal}{HTML}{176B64}
\definecolor{vltBlue}{HTML}{315D80}
\definecolor{vltGold}{HTML}{A87520}
\definecolor{vltPlum}{HTML}{725877}
\definecolor{vltRed}{HTML}{762B38}
\definecolor{vltYellow}{HTML}{D5AD3C}
\definecolor{vltNeutral}{HTML}{777D83}
\setlength{\unitlength}{1pt}
\scalebox{0.80}{%
\input{vlt-figures/figure-130.tex}%
}\endgroup
\caption{Stationary classical fluctuations at fixed $m,k,T$, with $\omega_0=\sqrt{k/m}$ and $d=\gamma/\omega_0$. The Markov model $m\ddot x+2m\gamma\dot x+kx=\xi$ uses matched noise $\langle\xi(t)\xi(t^\prime)\rangle=4m\gamma k_{\rm B}T\delta(t-t^\prime)$. (a) Its two-sided displacement spectrum is $S_x(\omega)=[k_{\rm B}T/(m\omega_0^3)]f_d(r)$, where $f_d(r)=4d/[(1-r^2)^2+4d^2r^2]$. (b) Integrating over the symmetric band $|\omega|\le\Omega$ gives $B_d(H)=\pi^{-1}\int_0^Hf_d(r)\,\dd r$. Damping redistributes the spectral weight while each full integral gives the same variance. All displayed finite-band fractions remain below one. Solid, dashed and dotted curves identify the same damping assignments in both panels. The comparison concerns matched equilibrium preparations, with fixed restoring curvature.}
\label{fig:opening-five-equilibrium-band}
\end{figure}

For a stable mass-weighted finite system, a useful displacement response
spectrum in mass-weighted coordinates is
\begin{equation}
 \begin{split}
 \mathsf G_{\mathrm R}(\omega)
  &=\bigl[\mathsf M^{-1/2}\mathsf K^*(\omega)\mathsf M^{-1/2}
                                             -\omega^2\Id\bigr]^{-1},\\
 g_{\mathrm{resp}}(\omega)
  &=\frac{2\omega}{\pi}
                           \operatorname{Im}\tr\mathsf G_{\mathrm R}(\omega),
                                                \qquad\omega>0.
 \end{split}
 \label{eq:damped-response-spectrum}
\end{equation}
For the passive Maxwell model, the imaginary part of this inverse is a
nonnegative quadratic form. In the lossless limit with the retarded
$\omega+i0$ prescription, each conservative mode contributes
$\delta(\omega-\omega_s)$, recovering the density of states of Section~3.
At finite loss this object is a coordinate response spectrum, with broadened
resonances and preparation-independent response weights. Counting complex
poles does not give the finite mode-counting measure defined earlier.
Mode preparation still supplies its own energy or occupation weights.
If zero-frequency coordinates couple to the measured positive-frequency
sector, project the full inverse onto that sector rather than deleting
those couplings before inversion.

For a strictly positive finite reference and this causal Maxwell response,
the inertial high-frequency behavior $\mathsf G_{\rm R}(z)\sim-\Id/z^2$
also fixes its spectral weight:
\begin{equation}
 \int_0^\infty\frac{2\omega}{\pi}
                  \operatorname{Im}_{\rm H}\mathsf G_{\rm R}(\omega+i0)
                   \,\dd\omega=\Id,
 \qquad\int_0^\infty g_{\rm resp}(\omega)\,\dd\omega=p.
 \label{eq:damped-response-inertial-sum-rule}
\end{equation}
Here $p$ is the number of retained coordinates in that positive reference;
any lossless lines are included as spectral measures. This normalization
constrains a coordinate response and does not count all bath modes.
Unconstrained zero modes require their separate singular spectral treatment.
Higher response moments also depend on the short-time closure: an ideal
instantaneous drag gives a $1/\omega^2$ response-weight tail in its damped
directions and divergent higher frequency moments. A measured linewidth
is consequently not the state frequency variance used in an uncertainty
relation. Physical bandwidth and ultraviolet modeling matter when such
moments are inferred.

\begin{figure}[H]
\centering\begingroup
\definecolor{vltTeal}{HTML}{176B64}
\definecolor{vltBlue}{HTML}{315D80}
\definecolor{vltGold}{HTML}{A87520}
\definecolor{vltPlum}{HTML}{725877}
\definecolor{vltRed}{HTML}{762B38}
\definecolor{vltYellow}{HTML}{D5AD3C}
\definecolor{vltNeutral}{HTML}{777D83}
\setlength{\unitlength}{1pt}
\scalebox{0.84}{%
\input{vlt-figures/figure-131.tex}%
}\endgroup
\caption{Input, retained state and readout form a calibrated channel. Open arrows are mathematical maps; the transfer $\mathsf H(s)$ uses zero initial state. Preparation adds $\mathsf L_{\rm out}e^{t\mathsf A_{\rm R}}X_0$. A simple-pole residue factors into input and output overlaps. Below, arrows show normalized vector components on the same equal-mass exchange-symmetric pair: a common channel reads its centre, and a relative channel reads separation. The bodies and spring spans match because these are channel profiles, rather than motion snapshots.}
\label{fig:excitation-input-state-readout}
\end{figure}

The tunable target construction now enters through its conservative Hessian
and, if selected, additional passive interaction branches. Comparing fixed
values of $\lambda$ compares different prepared mechanical systems. A
conservative target may change both mode shapes and restoring frequencies;
additional target-coordinate dynamics can change the memory and linewidth.
If $\lambda$ varies during evolution, the control power is
$\Dt{\lambda}\,\partial_\lambda H_{\mathrm{full}}$, as in Section~2.
This supplies a precise way to study responsiveness without assigning an
unspecified source of energy to the virions.

In the measles example, whole-virion translation and envelope--RNP
deformation require their respective coordinates and couplings. A peak
then identifies a feature of the chosen mechanical susceptibility.
Predicting entry or structural failure additionally requires the
constitutive process coupling that motion to the biological event.

A practical comparison can pair a calibrated small-force response with
spontaneous fluctuations of the same displacement combination at the same
reference state. Their phase, spectral weight, and static compliance test
one jointly specified passive model. An instrument may instead report a
surface displacement, an internal deformation, or an optical signal; its
transduction and forcing pattern must then be mapped to those coordinates.
Equilibrium fluctuation--dissipation is a test of a stationary thermal
description, not an assumption that every biological preparation is at
equilibrium. Sustained chemical or electrical work belongs in the explicit
energy and state extension of Section~\ref{sec:phonon-biochemical-extensions}.

\subsubsection{Inverse identification from collective motion}
\label{sec:s4-inverse-mechanics}

The forward problem maps positions, masses, assigned charges, interaction
curvatures and medium laws to motion. The inverse problem asks which
inputs specified observations distinguish. In the finite cylinder,
common motion, changing separation and material strain provide concrete
channels that can carry different information despite sharing a peak.
Input and readout maps specify the comparison.

\begin{definition}[Calibrated mechanical channel]
\label{def:s4-calibrated-channel}
On the selected admissible displacement space, let a known matrix
$\mathsf P_{\rm in}$ map applied force amplitudes $\vct F_{\rm in}$
to particle forces, and let $\vct Y=\mathsf W_{\rm out}^{\mathsf T}\vct U$
be the reported displacement amplitudes. For calibrated force and length
coordinates these maps are dimensionless. The channel compliance is
\begin{equation}
 \mathsf H_{\rm ch}(\omega;\vct\theta)
 =\mathsf W_{\rm out}^{\mathsf T}
       \mathsf R_{\rm mech}(\omega;\vct\theta)\mathsf P_{\rm in},
 \qquad \vct Y=\mathsf H_{\rm ch}\vct F_{\rm in}.
 \label{eq:s4-inverse-calibrated-channel}
\end{equation}
Here $\vct\theta$ denotes the unknown constitutive or reference parameters
within a declared family. The observation specification also includes its
frequency band, preparation, detector transfer and noise model.
\end{definition}

A tracked centre supplies a local row of $\mathsf W_{\rm out}^{\mathsf T}$;
a difference of two rows reports changing separation. Optical or
electrical signals carry their own calibrated transduction maps and units.
Detector filtering is applied to the predicted channel before comparison
with the record. A frequency rejected by the detector supplies no direct
constraint through that record. An uncertain gain or an uncertain initial
branch strain belongs among the inferred quantities if it has not been
fixed independently.

\begin{definition}[Response equivalence and identifiability]
\label{def:s4-response-equivalence}
For a declared model family and observation specification, two parameter
sets are \emph{response equivalent} when their ideal predicted records agree
for every input and preparation admitted in that specification. A
parameter, or a function of parameters, is \emph{structurally identifiable}
when this equality forces its value to agree within the admissible family.
\emph{Practical identifiability} concerns how accurately the available
finite, noisy records distinguish nearby admissible values.
\end{definition}

Response identification can recover an effective law or distinguish
physical mechanisms within a specified parameter family. Structural and
practical identifiability formalize these tasks~\cite{Raue2009}.
Here lattice geometry, energy and medium couplings define that family.

\begin{proposition}[Complete response and projected response]
\label{prop:s4-inverse-response-recovery}
At a real frequency where the mechanical pencil is invertible, a known
mass matrix and the complete calibrated displacement compliance determine
the dynamic stiffness uniquely at that frequency:
\begin{equation}
 \mathsf K^*(\omega)=\mathsf R_{\rm mech}(\omega)^{-1}
                                      +\omega^2\mathsf M.
 \label{eq:s4-inverse-full-stiffness}
\end{equation}
For fixed input and readout maps, the first variation of a projected
channel instead satisfies
\begin{equation}
 \delta\mathsf H_{\rm ch}
 =\mathsf W_{\rm out}^{\mathsf T}\mathsf R_{\rm mech}
        (\omega^2\delta\mathsf M-\delta\mathsf K^*)
             \mathsf R_{\rm mech}\mathsf P_{\rm in}.
 \label{eq:s4-inverse-channel-sensitivity}
\end{equation}
\end{proposition}
\begin{proof}
The first statement rearranges the inverse defining
$\mathsf R_{\rm mech}$. For the second, differentiate
$\mathsf T\mathsf T^{-1}=\Id$ with
$\mathsf T=\mathsf K^*-\omega^2\mathsf M$, obtaining
$\delta\mathsf T^{-1}=-\mathsf T^{-1}(\delta\mathsf T)\mathsf T^{-1}$,
then apply the fixed channel maps.
\end{proof}

Complete response means independently resolved forces and displacements
spanning the admissible space; one spectral peak provides a smaller
observation. A projected matrix variation can vanish between the two
channel maps while changing other particle motions. Even the recovered
full $\mathsf K^*$ is an assembled law. Separating its Lennard--Jones,
Coulomb, support and relaxed-medium contributions uses the declared
parameterization and independent constraints on those contributions.
When a parameter changes the force-balanced reference positions, that
change is included in the total derivative of the assembled matrices.

Figure~\ref{fig:s4-inverse-common-relative} gives an exact pair example: a common channel cannot determine the pair curvature, while a relative channel can distinguish it.

\begin{figure}[H]
\centering\begingroup
\definecolor{vltTeal}{HTML}{176B64}
\definecolor{vltBlue}{HTML}{315D80}
\definecolor{vltGold}{HTML}{A87520}
\definecolor{vltPlum}{HTML}{725877}
\definecolor{vltRed}{HTML}{762B38}
\definecolor{vltYellow}{HTML}{D5AD3C}
\definecolor{vltNeutral}{HTML}{777D83}
\setlength{\unitlength}{1pt}
\scalebox{0.90}{%
\input{vlt-figures/figure-132.tex}%
}\endgroup
\caption{Common and relative channels use conjugate normalized vectors $\vct e_+=(1,1)/\sqrt2$ and $\vct e_-=(-1,1)/\sqrt2$. Shared rulers locate reference sites at $x/a=0,1$ in both snapshots; displacements use an $8\,\mathrm{pt}$ gain and matched coils. Their common and relative shifts preserve and increase separation, respectively. The responses compare two pair curvatures at fixed masses, supports and drag. Common-channel curves coincide for both laws; relative peaks distinguish $k_c$. Snapshot arrows illustrate the channel patterns independently of harmonic-response amplitudes.}
\label{fig:s4-inverse-common-relative}
\end{figure}

Assigned charge gives a concrete exact ambiguity already present in this
system. Equation~\eqref{eq:vlp-global-charge-reversal} preserves every
Coulomb pair product under simultaneous reversal of all $q_i$. With the
other reference inputs and charge-even medium laws fixed, the same
mechanical response follows. This observation class is sensitive to
charge pair products while remaining invariant under a common sign reversal.
A separately calibrated charge-sensitive observation can supply the
missing distinction. More generally, the hidden-parameter family in
\eqref{eq:phonon-hidden-parameter-equivalence} preserves a relaxation
response while changing internal susceptibility and kinetic inputs.
It exhibits physical parameter ambiguity even with a fixed definition
and unit for the hidden variable.

Local sensitivity makes the available distinctions calculable. Assume a
continuously differentiable data map near an interior admissible reference.
Choose independent dimensionless parameter coordinates and form a real data
vector $\vct d(\vct\theta)$ from the real and imaginary channel values
at the measured frequencies, together with any separately specified
preparation or fluctuation observations. Its Jacobian and a weighted
sensitivity matrix are
\begin{equation}
 \mathsf J_{\rm d}=\frac{\partial\vct d}{\partial\vct\theta},
 \qquad
 \mathsf I_{\rm loc}=\mathsf J_{\rm d}^{\mathsf T}
                    \mathsf\Sigma_{\rm d}^{-1}\mathsf J_{\rm d},
 \qquad \mathsf\Sigma_{\rm d}>0.
 \label{eq:s4-inverse-local-sensitivity}
\end{equation}
The declared scales make the relative sizes of parameter variations
meaningful. Full column rank of $\mathsf J_{\rm d}$ is a sufficient local
injectivity condition at a regular reference: a square nonsingular
submatrix selects data components to which the inverse function theorem
applies. This local statement leaves possible distant equivalent
assignments, such as a common charge reversal. A null direction gives a
first-order invisible perturbation.
An exact symmetry or an explicit equivalent family establishes exact
nonidentifiability; a zero derivative at an isolated singular point alone
does not establish it. Small singular values of
$\mathsf\Sigma_{\rm d}^{-1/2}\mathsf J_{\rm d}$ describe weak local
constraints at the chosen noise and parameter scales. For a Gaussian
data model with known parameter-independent covariance
$\mathsf\Sigma_{\rm d}$, the displayed matrix is the local Fisher
information. Profile likelihood can assess the wider nonlinear parameter
range when that local quadratic description is inadequate~\cite{Raue2009}.

Figure~\ref{fig:s4-inverse-sensitivity-geometry} visualizes the noise-weighted Jacobian and the new distinction supplied by a complementary channel.

\begin{figure}[H]
\centering\begingroup
\definecolor{vlfGreen}{HTML}{176B64}
\definecolor{vlfDarkRed}{HTML}{762B38}
\definecolor{vlfRed}{HTML}{762B38}
\definecolor{vlfBlue}{HTML}{315D80}
\definecolor{vlfCopper}{HTML}{762B38}
\setlength{\unitlength}{1pt}
\scalebox{0.80}{%
\input{vlt-figures/figure-133.tex}%
}\endgroup
\caption{Static-channel sensitivity at $(\theta_1,\theta_2)=(1,1)$, with $\theta_1=k_0/k_\star$, $\theta_2=k_c/k_\star$, data $d_+=1/\theta_1$, $d_-=1/(\theta_1+2\theta_2)$ and independent noise standard deviation $0.05$. The unit parameter circle maps to a line for the whitened common channel and an ellipse when the relative channel is added. Both data panels use the same $22$-unit scale. Ellipse singular values are approximately $20.13,4.42$. These derivative images measure local first-order constraints; Figure~\ref{fig:s4-inverse-common-relative} establishes the exact common-channel ambiguity.}
\label{fig:s4-inverse-sensitivity-geometry}
\end{figure}

Complementary channels add data: separation can reveal a motion
suppressed by a centre average, and an internal-state readout can resolve
constituents of an effective coefficient. At known temperature, complete
conjugate equilibrium response fixes its displacement spectrum through
\eqref{eq:damped-probe-equilibrium-fdt}. Independently measured
fluctuations test this closure and can constrain inputs unresolved by
partial response. Their independent value depends on what was used in
calibration, as in Section~\ref{sec:complete-benchmark}.

\subsubsection{Joint membership and motion as an inference problem}
\label{sec:s4-identification-switching-state}

Motion within a residence constrains conditional stiffness and medium
response; arrivals, departures and waiting times constrain transitions.
Equal-sized clusters can differ in charge pattern, shape and prepared
strain. The retained state records the distinctions that change the next
force or event probability.

\begin{example}[The same cluster size with two mechanical responses]
\label{ex:s4-identification-response-mixture}
Suppose two persistent arrangements have the same retained mass $m$,
the same scalar resistance $c>0$, and different conditional restoring
coefficients $k_1,k_2>0$. Their calibrated scalar compliances are
$H_g(\omega)=(k_g-m\omega^2-i\omega c)^{-1}$. In separately prepared
records with negligible switching during each measurement, let
$\pi_1+\pi_2=1$ be the arrangement weights. With a common coherent
force phasor $F$, the mean phasor and mean squared response satisfy
\begin{equation}
 \begin{aligned}
 \langle U\rangle&=(\pi_1H_1+\pi_2H_2)F,\\
 \langle|U|^2\rangle-|\langle U\rangle|^2
      &=\pi_1\pi_2|H_1-H_2|^2|F|^2\geq0.
 \end{aligned}
 \label{eq:s4-identification-mixture-response}
\end{equation}
Expanding the two squares proves the second identity. The pooled record
contains arrangement variability even though the within-arrangement
resistance is unchanged. Conditioning on the arrangement separates
these compliances; conditioning only on a common particle count need
not do so. A mean squared displacement response is also distinct from
the squared mean phasor. For this specified common resistance, mean
dissipated power is $\omega^2c\langle|U|^2\rangle/2$.
This example describes an ensemble of frozen arrangements, while the
switching model below retains transitions within a record.
\end{example}

\begin{definition}[Common-space switching mechanical state]
\label{def:s4-identification-switching-state}
For a declared finite family of arrangements, let
$g(t)\in\{1,\ldots,G\}$ denote the arrangement and let $X$ contain a
common list of retained physical particle and medium coordinates.
The complete vessel still has $N_{\rm tot}$ ideal virions. A useful auxiliary
linear ensemble model is
\begin{equation}
 \dd X=(\mathsf A_gX+\vct b_g)\dd t+\mathsf B_g\dd\vct W,
 \qquad
 \Pr\{g(t+\dd t)=h\mid g(t)=g,X(t)\}
       =\mathsf Q_{hg}\dd t+o(\dd t),\quad h\ne g.
 \label{eq:s4-identification-switching-model}
\end{equation}
The rates are state independent in this auxiliary closure,
$\mathsf Q_{hg}\geq0$ and $\sum_h\mathsf Q_{hg}=0$.
The matrices, affine drifts and rates are constant for each arrangement
in this auxiliary model.
The Wiener drivers and switching clock are independent.
The coordinate $X$ is continuous at a switch in this representation;
physical resets, changes of state basis and their energy terms require
their declared maps instead. The affine term includes specified forcing
and reference offsets; $\mathsf D_g=\mathsf B_g\mathsf B_g^{\mathsf T}$
is the within-arrangement diffusion covariance.
\end{definition}

The dimensions and physical meanings of the components of $X$ remain
those of the retained state. Declared scales can put them into
dimensionless numerical coordinates. Conditional operators are assembled
from compatible cells and their declared medium couplings. Their
transition account includes the energy of a switch as well as the
nonnegative within-residence loss.
An independent clock can represent a prescribed switching protocol or
an approximate ensemble model. Autonomous thermal transitions between
configuration-dependent energies generally couple their rate ratios to
the energy change through local detailed balance~\cite{Seifert2012}.
Such a physical rate law can depend on $X$ and therefore require the
more complete closure discussed below.

\begin{proposition}[Exact moment propagation for the auxiliary closure]
\label{prop:s4-identification-switching-moments}
With finite second moments, define unnormalized conditional quantities
$p_g=\Pr\{g(t)=g\}$,
$\vct m_g=\langle X\mathbf1_{g(t)=g}\rangle$ and
$\mathsf V_g=\langle XX^{\mathsf T}\mathbf1_{g(t)=g}\rangle$.
They obey the finite coupled linear equations
\begin{equation}
 \begin{aligned}
 \dot p_g&=\sum_h\mathsf Q_{gh}p_h,\\
 \dot{\vct m}_g&=\mathsf A_g\vct m_g+\vct b_gp_g
                     +\sum_h\mathsf Q_{gh}\vct m_h,\\
 \dot{\mathsf V}_g&=\mathsf A_g\mathsf V_g
                    +\mathsf V_g\mathsf A_g^{\mathsf T}
                    +\vct b_g\vct m_g^{\mathsf T}
                    +\vct m_g\vct b_g^{\mathsf T}
                    +\mathsf D_gp_g
                    +\sum_h\mathsf Q_{gh}\mathsf V_h.
 \end{aligned}
 \label{eq:s4-identification-switching-moments}
\end{equation}
\end{proposition}
\begin{proof}
Apply the joint generator to $\mathbf1_g$, $X\mathbf1_g$ and
$XX^{\mathsf T}\mathbf1_g$. The drift supplies the affine first- and
second-moment terms; It\^o's product rule supplies $\mathsf D_gp_g$.
An identity transport of $X$ at a switch transfers each moment with the
same probability generator. These are precisely the displayed terms.
\end{proof}

Consequently, a matrix exponential can propagate the probabilities,
means and second moments together after vectorizing the independent
entries of $\mathsf V_g$. A time-dependent prescribed force gives the
corresponding nonautonomous moment equations and forcing terms.
For an arrangement-dependent calibrated linear
readout $\vct Y=\mathsf L_gX$, the observable mean and uncentered second
moment are
\begin{equation}
 \langle\vct Y\rangle=\sum_g\mathsf L_g\vct m_g,\qquad
 \langle\vct Y\vct Y^{\mathsf T}\rangle
       =\sum_g\mathsf L_g\mathsf V_g\mathsf L_g^{\mathsf T}.
 \label{eq:s4-identification-switching-readouts}
\end{equation}
Subtracting the outer product of the mean gives covariance. The
conditional moments are weighted by occupancy: $\mathsf V_g$ is not
the covariance conditioned on $g$. Even with Gaussian preparation and
linear dynamics between events, averaging switching histories can
produce a non-Gaussian output law. The closed moments therefore predict
these statistics without establishing that they specify the full state
distribution.
For state dimension $d_X$, the closure has $G$ probabilities,
$Gd_X$ first moments and $Gd_X(d_X+1)/2$ independent second-moment
entries. A useful calculation therefore selects a justified family of
arrangements and exploits its matrix structure. Reducing further to
selected measurement channels requires its own closure or controlled
response approximation; a small number of outputs does not automatically
close their moment dynamics.

Actual capture rates can depend on separation, force, branch strain,
excitation or residence age
(Subsection~\ref{sec:s23-transient-recruitment}). Replacing
$\mathsf Q_{gh}$ by $\mathsf Q_{gh}(X)$ introduces terms such as
$\langle\mathsf Q_{gh}(X)X\mathbf1_h\rangle$ and their second-moment
counterparts. Closing them generally requires additional state,
age structure or a controlled distributional approximation. The moment
equations thus locate a specific variable-integration problem.

For fully observed state-independent switching, the jump part of the
path likelihood separates counts from exposure: if $N_{gh}$ counts
$h\to g$ events and $T_h$ is total time in $h$, its rate-dependent log
likelihood is
\begin{equation}
 \log\mathcal L_Q
 =\sum_{g\ne h}\left[N_{gh}\log\mathsf Q_{gh}
                           -\mathsf Q_{gh}T_h\right]+\mathrm{const}.
 \label{eq:s4-identification-membership-likelihood}
\end{equation}
The ratio $N_{gh}/T_h$ estimates an allowed rate when $T_h>0$.
The sum uses allowed positive-rate edges; structural zeros are omitted.
It uses a specified observation time unit inside the logarithm, with
parameter-independent terms absorbed in the constant. Unobserved
events, uncertain cluster assignments and state-dependent hazards need
their observation and exposure models. Within-residence force and
displacement data then constrain the conditional viscoelastic operators.
The combined fit tests whether a long correlation belongs to stored
medium strain, long residence, or both; neither record alone universally
identifies the microscopic binding mechanism.

Figure~\ref{fig:s4-joint-membership-motion} separates the residence information that constrains switching from the motion information that constrains the conditional dynamics.

\begin{figure}[H]
\centering\begingroup
\definecolor{vltTeal}{HTML}{176B64}
\definecolor{vltBlue}{HTML}{315D80}
\definecolor{vltGold}{HTML}{A87520}
\definecolor{vltPlum}{HTML}{725877}
\definecolor{vltRed}{HTML}{762B38}
\definecolor{vltYellow}{HTML}{D5AD3C}
\definecolor{vltNeutral}{HTML}{777D83}
\setlength{\unitlength}{1pt}
\scalebox{0.90}{%
\input{vlt-figures/figure-134.tex}%
}\endgroup
\caption{Joint membership and motion inference in a two-arrangement auxiliary closure. (a) Four labelled ideal virions remain in the vessel; the selected set changes between $A$ and $B$. Thin links indicate selected arrangement structure, not an abrupt force reset. The continuous scaled coordinate obeys $\dot X=-\gamma_gX$, with state-independent rates $A\to B:a=0.18$, $B\to A:b=0.08$, and $(\gamma_A,\gamma_B)=(0.25,0.65)$. (b,c) Exact matrix exponentials propagate $\dot{\vct p}=\mathsf Q\vct p$ and $\dot{\vct m}=[\mathsf Q-\operatorname{diag}(\gamma_A,\gamma_B)]\vct m$, starting in $A$ with $X=1$. The plotted $m_g$ are occupancy-weighted first moments, rather than conditional means or probabilities. Observed transition counts and residence exposure constrain $a,b$; registered motion constrains the within-arrangement operators. This scalar illustration isolates the closure structure of \eqref{eq:s4-identification-switching-moments}; state-dependent capture rates require their additional variables and do not generally retain this closed moment system.}
\label{fig:s4-joint-membership-motion}
\end{figure}

\subsubsection{Response along an evolving assembly}
\label{sec:s4-nonstationary-response}

A pulse before joining can leave strain that acts after joining; the
same pulse applied later loads a different prepared assembly. A two-time
response retains this dependence on preparation and forcing times.

\begin{definition}[Path-conditioned mechanical response]
\label{def:s4-path-conditioned-response}
Choose a full, fixed-dimension particle--medium state and a registered
reference evolution. Discrete-event times are held fixed in this
conditional calculation. Consider deterministic augmented dynamics,
or additive noise held fixed across the comparison. Its linear perturbations, with zero initial
perturbation, obey
\begin{equation}
 \delta\dot X=\mathsf A(t)\delta X
                    +\mathsf B_{\rm in}(t)\delta\vct f_{\rm in}(t),
 \qquad \delta\vct Y=\mathsf L_{\rm out}(t)\delta X.
 \label{eq:s4-path-tangent-dynamics}
\end{equation}
The coefficients are derivatives of the declared evolution and
calibrated input/readout maps along that reference. Let
$\mathsf T(t,s)$ solve
$\partial_t\mathsf T(t,s)=\mathsf A(t)\mathsf T(t,s)$ and
$\mathsf T(s,s)=\Id$ between registered events. The conditional causal
response kernel is
\begin{equation}
 \begin{aligned}
 \mathsf H_{\rm ch}(t,s)
   &=\mathsf L_{\rm out}(t)\mathsf T(t,s)\mathsf B_{\rm in}(s),
                  &&t\geq s,\\
 \delta\vct Y(t)
   &=\int_{t_0}^{t}\mathsf H_{\rm ch}(t,s)
                          \delta\vct f_{\rm in}(s)\,\dd s.
 \end{aligned}
 \label{eq:s4-two-time-response}
\end{equation}
For a displacement output and force input, this time-domain kernel
has units $\mathrm{m\,N^{-1}s^{-1}}$.
\end{definition}

Variation of constants proves the integral formula. It retains the
initially prepared memory in the reference evolution and its induced
perturbations in $\delta X$. An independent preparation perturbation
adds $\mathsf L_{\rm out}(t)\mathsf T(t,t_0)\delta X(t_0)$ to the output.
State-dependent diffusion requires its stochastic tangent terms rather
than this ordinary variational equation alone. For two constant-coefficient episodes
separated at a registered time $t_*$, a declared linear perturbation
map $\mathsf J$ gives
\begin{equation}
 \mathsf T(t,s)=e^{\mathsf A_+(t-t_*)}\mathsf J
                      e^{\mathsf A_-(t_*-s)},\qquad s<t_*<t.
 \label{eq:s4-two-episode-propagator}
\end{equation}
In common physical coordinates a mere membership relabelling has
$\mathsf J=\Id$. A coordinate-chart change uses its compatible map,
as specified in the transient-state construction. Replacing the two
generators by their average generally changes this product because
their actions need not commute.

The response conditions on the registered event sequence and times.
If forcing changes a capture probability or event time, the full
response also includes that change in transition law. A finite
perturbation selecting another basin likewise requires its new trajectory
rather than the local tangent alone.

For a persistent fixed reference and fixed channels, the kernel depends
only on $t-s$. Its retarded frequency transform reduces to
\begin{equation}
 \mathsf H_{\rm ch}(\omega)
  =\mathsf L_{\rm out}
       [-i\omega\Id-\mathsf A_{\rm R}]^{-1}\mathsf B_{\rm in},
 \label{eq:s4-stationary-channel-limit}
\end{equation}
where the inverse exists or has its declared retarded interpretation.
Eliminating the branch coordinates recovers the mechanical compliance
in~\eqref{eq:s4-inverse-calibrated-channel}. A stationary frequency
fit thus has a precise role within the evolving model: it represents
one persistent episode or a separately justified effective limit.
Equilibrium fluctuation identities keep their stated ensemble and
stationarity assumptions; changing the reference does not supply a
new equilibrium preparation.

Figure~\ref{fig:s4-ordered-two-time-response} shows how a force pulse is propagated across two registered mechanical episodes.

\begin{figure}[H]
\centering\begingroup
\definecolor{vltTeal}{HTML}{176B64}
\definecolor{vltBlue}{HTML}{315D80}
\definecolor{vltGold}{HTML}{A87520}
\definecolor{vltPlum}{HTML}{725877}
\definecolor{vltRed}{HTML}{762B38}
\definecolor{vltYellow}{HTML}{D5AD3C}
\definecolor{vltNeutral}{HTML}{777D83}
\setlength{\unitlength}{1pt}
\scalebox{0.90}{%
\input{vlt-figures/figure-135.tex}%
}\endgroup
\caption{Ordered two-time response of a prepared particle--medium state. (a) A unit momentum impulse $J$ acts on virion~1 at $t=0$; virion~2 is read out. The scaled auxiliary model has $\mathsf M=\Id$, $\mathsf K=0.6\Id+k_{\rm rel}\vct b\vct b^{\mathsf T}$, $\mathsf C_0=0.12\Id$, $\vct b=(-1,1)^{\mathsf T}$ and one branch $h=0.7$, $\tau=1.2$, initially unstrained. (b) Exact matrix exponentials compare $k_{\rm rel}:0.25\to1.3$ (solid) with $1.3\to0.25$ (dashed) at the same registered time $t_*=2$. The full state, including the inherited branch strain, remains continuous. The product in \eqref{eq:s4-two-episode-propagator} retains the episode order. This is a prescribed stiffness change, whose work is $\tfrac12\Delta k_{\rm rel}(u_2-u_1)^2$; mere membership relabelling would leave the physical operator unchanged. The matched spring glyphs indicate equal instantaneous pair geometry, rather than encoding the different stiffnesses through coil pitch. Coefficients are illustrative scaled parameters, not fitted virion data.}
\label{fig:s4-ordered-two-time-response}
\end{figure}

\paragraph{Registering the particles in a growing readout.}
A detector may report the centre of the selected assembly rather than
the positions of fixed particle identities. For a nonempty cluster,
define its selected material mass and mass-weighted centre by
\begin{equation}
 m_{\mathcal C}=\sum_i h_i^{\mathcal C}m_i,\qquad
 \overline{\vct r}_{\mathcal C}
    =\frac{\sum_i h_i^{\mathcal C}m_i\vct r_i}{m_{\mathcal C}}.
 \label{eq:s4-membership-centre-readout}
\end{equation}
If one particle $j$ is included at a registration event while all
physical positions and velocities remain continuous, direct
substitution gives
\begin{equation}
 \overline{\vct r}_{\mathcal C}^{\,+}
       -\overline{\vct r}_{\mathcal C}^{\,-}
  =\frac{m_j}{m_{\mathcal C}^{-}+m_j}
       (\vct r_j-\overline{\vct r}_{\mathcal C}^{\,-}).
 \label{eq:s4-registration-centre-shift}
\end{equation}
The reported centre changes because its membership weights changed.
For example, two equal masses centred at zero and a newly included
equal mass at $2a$ give a reported shift $2a/3$ even at a fixed
instantaneous geometry. The full particle mass and physical motion
still have their original accounts.

Tracking a persistent core, retaining particle labels, or recording
the membership mask alongside the centre allows this change of
readout to be predicted. Registered relative displacements can then
be compared across episodes without assigning a registration shift
to an impulsive mechanical force. This makes variable selection
concrete: a population-dependent readout needs its population state,
while a fixed-particle displacement channel can retain a constant
physical meaning through the same recruitment event.

Figure~\ref{fig:s4-membership-centre-registration} separates a change of selected population from a change of particle position.

\begin{figure}[H]
\centering\begingroup
\definecolor{vltTeal}{HTML}{176B64}
\definecolor{vltBlue}{HTML}{315D80}
\definecolor{vltGold}{HTML}{A87520}
\definecolor{vltPlum}{HTML}{725877}
\definecolor{vltRed}{HTML}{762B38}
\definecolor{vltYellow}{HTML}{D5AD3C}
\definecolor{vltNeutral}{HTML}{777D83}
\setlength{\unitlength}{1pt}
\scalebox{0.90}{%
\input{vlt-figures/figure-136.tex}%
}\endgroup
\caption{The selected centre changes while physical positions stay fixed. Equal masses occupy $r_1=-a$, $r_2=a$, $r_3=2a$ in both frames. Brackets select $\{1,2\}$ before registration and $\{1,2,3\}$ afterward. Dashed burgundy lines locate their mass-weighted centres. Gray traces in (c) track the same three positions; the burgundy readout jumps by $2a/3$. The event changes selection weights and supplies no impulse. Registered membership or a persistent-core readout resolves this contribution.}
\label{fig:s4-membership-centre-registration}
\end{figure}

\subsubsection{Which excitation variables are sufficient?}
\label{sec:excitation-variable-closure}

Inverse identification asks which inputs observations distinguish.
State sufficiency asks which retained description predicts those
observations across preparations and applied forces.

\begin{definition}[Observation-sufficient retained description]
\label{def:s4-observation-sufficient-state}
For specified dynamics, input and preparation classes, readouts and
prediction horizon, a retained description is \emph{observation sufficient}
when any two admissible initial states assigned the same retained values
give the same future readouts under every common allowed input. For a
stochastic model the requirement concerns the conditional output laws,
with the noise and preparation ensemble specified.
\end{definition}

The full prepared state of Definition~\ref{def:s4-prepared-mechanical-state}
is sufficient for the deterministic finite branch model. Conservative
modal coordinates reorganize its mechanical part; branch strains retain
the medium preparation. Energy and action summarize how much excitation
was prepared, while phase specifies where that preparation lies in its
motion. Additional shape, polarization, ionic or configurational
coordinates become useful when their coupling changes the predictions
resolved by the selected readouts. The following table organizes these
roles within the same fixed ideal-virion population.

\begin{table}[H]
\centering\small
\begin{tabular}{@{}p{0.28\textwidth}p{0.67\textwidth}@{}}
\toprule
Retained description & Predictive content and limitation \\
\midrule
$X=(\vct u,\vct v,\vct z_a)$
 & Complete instantaneous state of the specified deterministic branch model;
   the prepared memory variables affect future forces. \\[4pt]
$(Q_s,\Pi_s)$ or $\psi_s$
 & Equivalent mechanical coordinates with amplitude and phase; a full medium
   model also retains its branch or bath coordinates. \\[4pt]
$E_s$ or $\mathcal I_s$
 & Energy or action summary of the conservative reference; drops phase and
   generally does not close coupled, driven, or damped dynamics. \\[4pt]
$\langle\hat N_s\rangle$ or mode populations
 & Reference quantum excitation content; omits coherences and intermode
   correlations. A density operator specifies the general reduced state. \\[4pt]
Means and covariances
 & Determine a Gaussian statistical state with a declared noise/preparation
   model; general non-Gaussian states need additional information. \\[4pt]
Stiffness, loss, $\mathsf K^*(\omega)$
 & Constitutive parameters and response functions; these describe how a
   selected state evolves, rather than its current preparation. \\
\bottomrule
\end{tabular}
\caption{Excitation variables retain different amounts of information.
Sufficiency concerns a specified dynamics, preparation, observation window,
and readout; it is not a universal list of biological state variables.
Figure~\ref{fig:excitation-state-phase} shows the simplest information-losing
projection, and \eqref{eq:augmented-relaxation-generator} shows why memory
coordinates restore closure for the branch model.}
\label{tab:excitation-variable-hierarchy}
\end{table}

Visibility has an equally explicit state-space meaning. Let
$\mathsf B_{\rm in}$ map a calibrated input to the augmented equation and
$\mathsf L_{\rm out}$ map its state to the reported signal, with no direct
input term in this example. For a complex Laplace variable $s$ in
the resolvent set of $\mathsf A_{\rm R}$, the transfer is
\begin{equation}
 \mathsf T_{\rm out,in}(s)=\mathsf L_{\rm out}
          (s\Id-\mathsf A_{\mathrm R})^{-1}\mathsf B_{\rm in}.
 \label{eq:excitation-input-readout-transfer}
\end{equation}
The resolvent set consists of those $s$ for which the displayed inverse
exists. This transfer describes the driven response with zero initial
state; a prepared state adds its separately propagated contribution.
For a simple eigenvalue $s_j$ with right and left eigenvectors
$\vct r_j,\vct l_j$ normalized by
$\vct l_j^\dagger\vct r_j=1$, its residue is
$(\mathsf L_{\rm out}\vct r_j)
(\vct l_j^\dagger\mathsf B_{\rm in})$.
Either factor can vanish. A mode may therefore store energy yet be dark
to the chosen drive or readout. This extends the conservative overlap
test in \eqref{eq:architecture-modal-readout-residues} to memory and
relaxation excitations. It also distinguishes detector visibility from
coupling to the loss channels in \eqref{eq:finite-damping-detectability}.

Figure~\ref{fig:excitation-input-state-readout} separates the prepared state from the input and detector maps that determine which excitation is observed.

For a virion in the specified blood-like environment, a response feature
may belong to centre motion, retained particle deformation, or resolved
medium relaxation. Bulk rheology does not identify which of these a
detector measures; the coordinate, coupling, and transduction map do.
Likewise, a calibrated displacement line area estimates the weighted
coordinate variance within the instrument's band. It is not by itself a
phonon count. Even the bare reference-oscillator relation
$\langle\hat N_s\rangle=[\langle\hat\Pi_s^2\rangle+
\omega_s^2\langle\hat Q_s^2\rangle]/(2\hbar\omega_s)-\tfrac12$
requires the momentum moment, mass normalization, frequency, and state
convention, as well as physically finite moments. Strong medium coupling
does not justify substituting a bare Gibbs occupation.

For a known finite augmented generator, the relevant state distinctions
can be stated directly in terms of future observations. In the same
applied input, two initial states differ at the output by
$\mathsf L_{\rm out}e^{t\mathsf A_{\rm R}}\Delta X(0)$.
If $d_X$ is the full augmented dimension, define
\begin{equation}
 \begin{aligned}
 \mathcal N_{\rm obs}
   &=\bigcap_{k=0}^{d_X-1}
        \ker(\mathsf L_{\rm out}\mathsf A_{\rm R}^{k}),\\
 \Delta X(0)\in\mathcal N_{\rm obs}
 &\quad\Longleftrightarrow\quad
 \mathsf L_{\rm out}e^{t\mathsf A_{\rm R}}\Delta X(0)=0
 \quad(t\geq0).
 \end{aligned}
 \label{eq:s4-inverse-unobservable-state}
\end{equation}
The equivalence follows by differentiating at zero and using the
Cayley--Hamilton theorem to reduce every higher power of the generator.
The resulting kernel is invariant under $\mathsf A_{\rm R}$; its quotient
therefore retains the distinctions needed for these outputs. For
zero-state forced response, only the reachable space
$\mathcal V_{\rm reach}=\operatorname{span}_{0\leq k<d_X}
\operatorname{ran}(\mathsf A_{\rm R}^{k}\mathsf B_{\rm in})$
is prepared. Its observable quotient has dimension
$\dim\mathcal V_{\rm reach}
-\dim(\mathcal V_{\rm reach}\cap\mathcal N_{\rm obs})$, the standard
minimal-realization count~\cite{Kalman1963}. Additional initial-state
preparations can reveal observable directions outside that reachable
space.

Figure~\ref{fig:s4-observable-state-quotient} compares these dimensions with two exact release records of the same particle pair.

\begin{figure}[H]
\centering\begingroup
\definecolor{vlfGreen}{HTML}{176B64}
\definecolor{vlfDarkRed}{HTML}{762B38}
\definecolor{vlfRed}{HTML}{762B38}
\definecolor{vlfBlue}{HTML}{315D80}
\definecolor{vlfCopper}{HTML}{762B38}
\setlength{\unitlength}{1pt}
\scalebox{0.80}{%
\input{vlt-figures/figure-137.tex}%
}\endgroup
\caption{A channel-dependent state quotient, illustrating \eqref{eq:s4-inverse-unobservable-state}. For the pair in Figure~\ref{fig:s4-inverse-common-relative}, take $k_c/k_0=1$ and $c/\sqrt{mk_0}=0.2$. The normalized common and relative sectors have reference frequencies $1$ and $\sqrt3$ in units $\sqrt{k_0/m}$. Prepare A with $(q_+,v_+,q_-,v_-)=(1,0,0,0)$ and B with $(1,0,0.7,0)$ in the stated length and time scales, then release without forcing. (a) Both common-coordinate traces coincide exactly. (b) The relative-coordinate detector distinguishes the additional prepared motion. Curves are exact damped-oscillator solutions, not an arbitrary schematic. With only the common readout, the two-dimensional relative sector is the invariant unobservable space. Adding the relative readout makes the four-state mechanical model observable, yet a common force starting from rest still reaches only the common two-dimensional sector. The zero-state driven realization can therefore remain two-dimensional while arbitrary prepared-state prediction requires four. No medium branches are included in this illustrative pair.}
\label{fig:s4-observable-state-quotient}
\end{figure}

This dimension belongs to the specified channels. A state-basis change
preserves predictions when the generator, input, readout and energy
metric transform together. Physical identification also assigns units,
constraints, storage and calibrated couplings. A minimal response model
can therefore determine the number of dynamical combinations before
their molecular assignment is known. Appending unknown parameters with
$\dot{\vct\theta}=0$ gives a generally nonlinear state--parameter model
requiring its own identifiability analysis~\cite{Villaverde2016}.
Changing membership additionally carries its transition and energy
account within the fixed $N_{\rm tot}$ pool.

\paragraph{A state adequate for response may miss a stopping event.}
\begin{proposition}[Sufficient closure for a first-passage task]
\label{prop:s4patch-stopped-process-closure}
Let $x$ be the complete Markov state of a specified stochastic
particle--surface--material realization, with generator $\mathcal L$.
For a retained map $P$, suppose the projected martingale problem is
well posed and, on an appropriate core with compatible boundaries and
event maps,
\begin{equation}
 \boxed{\mathcal L(\varphi\circ P)
       =(\overline{\mathcal L}\varphi)\circ P,
       \qquad \mathcal B=P^{-1}(\overline{\mathcal B}).}
 \label{eq:s4patch-stopping-task-closure}
\end{equation}
Then the reduced process with generator $\overline{\mathcal L}$
preserves the first-hitting laws of the target $\mathcal B$.
\end{proposition}
\begin{proof}
The generator identity projects the full martingale equations onto the
retained variables. Uniqueness identifies their law with the specified
reduced process. Since the target is a union of complete projection
fibres, stopping at its first hit commutes with that projection.
\end{proof}
This is a sufficient criterion for the declared task and boundary
problem. The linear observable quotient above concerns its selected
linear records; it need not preserve recruitment or rupture targets.
For the persistent-material model, identical positions can have
different drifts and survival laws when $\vct y$ differs. A dwell-time
definition also retains the required age variable. A diagnostic for a
specified preparation and deadline is
$\operatorname{Var}[s_T(x)\mid P(x)]>0$, where $s_T$ is the full
conditional survival probability, rather than a binary outcome.
It detects information lost for that prediction. Vanishing variation
for one deadline does not establish universal closure. Comparing the
holding-force, recovery and first-passage records therefore tests which
surface/material variables are needed at the stated resolution.

The angular target in Figure~\ref{fig:s34patch-orientation-capture-target}
gives a geometric test of the second condition: projection onto separation
alone identifies its marked states at the same radius, although their
target membership differs. Retaining the relevant orientation resolves
that distinction; its projected dynamics must satisfy the generator
condition as well.

\paragraph{Integrating a variable while controlling the approximation.}
A useful extension starts with a physical quantity that can change and
states how it acts on the retained particles. A medium strain carries
stress; a polarization coordinate changes electrical work; a structural
population changes a conditional force or free energy. Conservation and
normalization constraints select their independent components before
couplings are assembled. The joint free energy and reciprocal kinetics
in~\eqref{eq:s4-root-internal-free-energy}--\eqref{eq:s4-root-derived-relaxation-spectrum}
then show how an added variable produces both a relaxed curvature and
memory. At equilibrium its elimination also carries the matching noise
and preparation terms. These three outputs make an integrated variable
testable through storage, phase and fluctuations.

The exact closure test of Proposition~\ref{prop:s4-state-exact-closure}
concerns every admissible full preparation. A finite observation band
allows a useful, explicitly approximate reduction. For one Maxwell
branch, write its stiffness factor as $f(x)=-ix/(1-ix)$ with $x=\omega\tau_a$.
The exact remainder identities are
\begin{equation}
 \begin{aligned}
 f(x)+ix&=\frac{x^2}{1-ix},&
 f(x)-1&=-\frac1{1-ix},\\
 |f(x)+ix|&=\frac{x^2}{\sqrt{1+x^2}},&
 |f(x)-1|&=\frac1{\sqrt{1+x^2}}.
 \end{aligned}
 \label{eq:s4-inverse-branch-reduction-errors}
\end{equation}
For $|\omega|\tau_a\ll1$, this channel is approximated by the drag
$\tau_a\mathsf K_a$ with stiffness error at most
$\|\mathsf K_a\|(|\omega|\tau_a)^2$. For
$|\omega|\tau_a\gg1$, it contributes the nearly frozen elastic
$\mathsf K_a$, with error at most
$\|\mathsf K_a\|/(|\omega|\tau_a)$. Sampling the crossover supplies
sensitivity to the branch time and weight when that branch contributes
distinguishably to the calibrated channels. Identifiability still follows
the declared parameter family and observation maps. A band confined to
one asymptotic regime chiefly constrains the corresponding combined
coefficient. This states why the observation window helps decide whether
to retain a relaxation coordinate.

\begin{proposition}[Response error from a reduced mechanical pencil]
\label{prop:s4-inverse-reduction-error}
At a selected frequency let
$\widetilde{\mathsf T}=\mathsf T+\delta\mathsf T$, with
$\mathsf R=\mathsf T^{-1}$. In a fixed subordinate matrix norm,
if $\|\mathsf R\|\,\|\delta\mathsf T\|<1$, then
\begin{equation}
 \|\widetilde{\mathsf T}^{-1}-\mathsf R\|
 \leq\frac{\|\mathsf R\|^2\|\delta\mathsf T\|}
                  {1-\|\mathsf R\|\,\|\delta\mathsf T\|}.
 \label{eq:s4-inverse-resolvent-error}
\end{equation}
\end{proposition}
\begin{proof}
Factor $\widetilde{\mathsf T}=\mathsf T(\Id+\mathsf R\delta\mathsf T)$.
The Neumann series bounds the latter inverse by
$(1-\|\mathsf R\|\|\delta\mathsf T\|)^{-1}$. Apply this bound to
$\widetilde{\mathsf T}^{-1}-\mathsf R
=-(\Id+\mathsf R\delta\mathsf T)^{-1}
                    \mathsf R\delta\mathsf T\mathsf R$.
\end{proof}

Multiplication by $\|\mathsf W_{\rm out}^{\mathsf T}\|$ and
$\|\mathsf P_{\rm in}\|$ gives a corresponding channel-error bound.
The norms use the declared physical coordinate scales, so the product
in the denominator is dimensionless and the response bound has compliance
units. Near a strong resonance, a small constitutive error can produce a
large displacement error. An effective drag or frozen stiffness is
therefore assessed through the readout it must predict. A transient
reduction also checks the discarded branch's initial strain, while a
fluctuation prediction checks the reduced noise covariance. Agreement in
harmonic response alone supplies a narrower test.

Figure~\ref{fig:s4-branch-reduction-accuracy} compares the two branch reductions against their exact errors across the observation band.

\begin{figure}[H]
\centering\begingroup
\definecolor{vlfGreen}{HTML}{176B64}
\definecolor{vlfDarkRed}{HTML}{762B38}
\definecolor{vlfRed}{HTML}{762B38}
\definecolor{vlfBlue}{HTML}{315D80}
\definecolor{vlfCopper}{HTML}{762B38}
\setlength{\unitlength}{1pt}
\scalebox{0.80}{%
\input{vlt-figures/figure-138.tex}%
}\endgroup
\caption{Accuracy of replacing one Maxwell branch. For $x=\omega\tau_a>0$, panel (a) gives storage $x^2/(1+x^2)$ and loss $x/(1+x^2)$; panel (b) gives exact errors of the drag replacement $-ix$ and frozen replacement $1$. Logarithmic coordinates are labeled explicitly. Multiplication by the branch stiffness norm bounds its matrix error. The lower topology sketches identify the limiting elements without assigning a prepared displacement. Equation~\eqref{eq:s4-inverse-resolvent-error} transfers stiffness error to response error; transient and fluctuation reductions also retain preparation and matched noise.}
\label{fig:s4-branch-reduction-accuracy}
\end{figure}

Retain a variable when its specified coupling and evolution explain a
reproducible difference between matched preparations, or when its
elimination exceeds prediction tolerance. Combine or eliminate it when
the derived memory, preparation and noise preserve the selected
observations to that accuracy. Inverse constraints distinguish physical
inputs; closure and error bounds test the resulting predictions.

A predictive reduction states its reference, work account and retained
state, then uses one parameter set for response and matched fluctuations.
A different preparation or readout supplies a further test without
refitting hidden state into material constants.
Section~\ref{sec:complete-benchmark} implements this comparison.
Disagreement directs attention to the interaction law, retained memory
and observation model.
\subsection{Self-excitation and the boundary of the passive model}
\label{sec:self-excitation-passive-boundary}

Self-excitation adds a converter that draws on a resource and feeds work
into retained motion. Feedback can select an oscillation amplitude while
replacing dissipated energy. The passive balance
\eqref{eq:continuum-passive-energy-estimate} instead bounds motion by
prepared and supplied energy. Prepared strain can return energy and
lossless motion can preserve it; sustained attracting motion requires
the additional work pathway specified below.

\begin{definition}[Self-excitation]
\label{def:s4-self-excitation}
\emph{Self-excitation} is conversion of an available resource into
mechanical motion through feedback, without an imposed periodic force.
\end{definition}
The Van der Pol oscillator exemplifies gain at small excursions and
saturation at larger ones~\cite{Stewart2014}. Its biophysical assignment
specifies a moving coordinate, a force-generating converter and a resource
paying for that force. Thermal fluctuations and externally forced
resonance provide comparison descriptions for an oscillatory record.

\subsubsection{Gain, saturation, and selection of an oscillation amplitude}

For one positive reference mode with $\Pi_s=\Dt{Q_s}$, consider
\begin{equation}
 \Dtt{Q_s}+2\left[\gamma_{\rm p}-\gamma_{\rm a}
             +\gamma_{\rm n}\left(\frac{Q_s}{Q_c}\right)^2\right]
                      \Dt{Q_s}+\omega_s^2Q_s=0.
 \label{eq:self-excitation-modal-model}
\end{equation}

The passive rate $\gamma_{\rm p}\geq0$, active gain rate
$\gamma_{\rm a}\geq0$, and nonlinear damping rate $\gamma_{\rm n}>0$
have units $\mathrm s^{-1}$. The positive scale $Q_c$ has the mass-weighted
displacement units $\mathrm{kg}^{1/2}\mathrm m$; it is a constitutive
amplitude scale, not generally the final oscillation amplitude. In this
first model the rates are fixed over the observation window. An isolated
mode and a sufficiently fast reduced feedback are assumed; slow gating,
medium memory, or competing modes require their own retained coordinates.
The nonlinear damping is an added constitutive law, while the restoring
energy remains the selected local harmonic reference.

Multiplying by $\Pi_s$ makes the work account explicit:
\begin{equation}
 \begin{split}
 E_s&=\tfrac12(\Pi_s^2+\omega_s^2Q_s^2),\qquad
 \boxed{\Dt{E_s}=P_{\rm a}-P_{\rm d}},\\
 P_{\rm a}&=2\gamma_{\rm a}\Pi_s^2,\qquad
 P_{\rm d}=2\left[\gamma_{\rm p}
              +\gamma_{\rm n}(Q_s/Q_c)^2\right]\Pi_s^2.
 \end{split}
 \label{eq:self-excitation-power-balance}
\end{equation}
Both powers are nonnegative and measured in watts. The active force
supplies $P_{\rm a}$, and passive and nonlinear loss remove $P_{\rm d}$.
Net negative damping means supplied work exceeds loss. Rest remains an
exact deterministic solution, with small preparation errors or specified
noise able to seed instability above onset. On a periodic cycle,
supplied work equals loss per period even though $E_s$ varies within it.
Figure~\ref{fig:self-excitation-phase-portrait} compares the resulting
trajectories.

Put $\delta_s=\gamma_{\rm a}-\gamma_{\rm p}$. Substituting a small
displacement proportional to $e^{\lambda_{\rm g}t}$ into the linearized
equation gives the growth exponents
$\lambda_{\rm g}=\delta_s\pm\sqrt{\delta_s^2-\omega_s^2}$.
A positive real part means amplification; near onset it crosses zero
at $\delta_s=0$. For small net damping and saturation
on the oscillation scale, write
$Q_s\simeq A_s(t)\cos[\omega_st+\varphi_s(t)]$ with slowly changing
amplitude. Cycle averages give
$\langle\Pi_s^2\rangle\simeq\omega_s^2A_s^2/2$ and
$\langle Q_s^2\Pi_s^2\rangle\simeq\omega_s^2A_s^4/8$. Thus
\begin{equation}
 \Dt{A_s}\simeq\delta_sA_s
              -\frac{\gamma_{\rm n}}{4Q_c^2}A_s^3,
 \qquad
 A_s^*\simeq2Q_c\sqrt{\frac{\delta_s}{\gamma_{\rm n}}}
                 \quad(\delta_s>0).
 \label{eq:self-excitation-averaged-onset}
\end{equation}
The approximation requires $|\delta_s|/\omega_s\ll1$ and
$\gamma_{\rm n}A_s^2/(Q_c^2\omega_s)\ll1$. Radial disturbances about
the nonzero averaged amplitude decay at rate $2\delta_s$. This is a
supercritical Hopf bifurcation: rest loses stability and a small attracting
periodic orbit emerges continuously as the gain crosses the loss rate,
with $\gamma_{\rm n}>0$ held fixed.
At threshold the linear damping vanishes but nonlinear damping remains;
the full threshold model is not a family of conservative harmonic orbits.
The predicted amplitude must also remain inside the reference basin and
the domain where the constitutive truncation is valid.

\begin{figure}[H]
\centering\begingroup
\definecolor{vlfGreen}{HTML}{176B64}
\definecolor{vlfDarkRed}{HTML}{762B38}
\definecolor{vlfRed}{HTML}{762B38}
\definecolor{vlfBlue}{HTML}{315D80}
\definecolor{vlfCopper}{HTML}{762B38}
\setlength{\unitlength}{1pt}
\scalebox{0.90}{%
\input{vlt-figures/figure-139.tex}%
}\endgroup
\caption{Passive, conservative and active motion in the same scaled phase plane. Curves solve $x^{\prime\prime}+2(-d+a_nx^2)x^\prime+x=0$, with $x=Q_s/Q_c$, $v=x^\prime$, $\tau=\omega_st$, $d=(\gamma_{\rm a}-\gamma_{\rm p})/\omega_s$ and $a_n=\gamma_{\rm n}/\omega_s$. (a) $(-0.2,0)$ gives decay. (b) $(0,0)$ gives a family of preparation-dependent closed orbits. (c) $(0.2,0.2)$ gives the numerically computed attracting cycle (black); gray releases from $(x,v)=(0.35,0),(2.75,0)$ approach it. Dots mark initial states and arrows advance time. (d) Three phases of one conservative common pattern reconstruct rigid-virion shifts $u_1=u_2=15\,\mathrm{pt}\cos\varphi$. Reference marks fix the common displacement gain. Matched pair coils preserve their span; the fixed environment supplies restoring curvature. At zero displacement the motion arrows vanish while velocity is nonzero. The active portrait assumes maintained gain over its observation window. The boxed result illustrated here is~\eqref{eq:self-excitation-power-balance}.}
\label{fig:self-excitation-phase-portrait}
\end{figure}

Time-shifted trajectories lie on the same autonomous cycle, so amplitude
selection leaves a phase freedom~\cite{Tedrake2023}. Amplitude and phase
reorganize the retained motion under the averaging assumptions;
synchronization needs an additional coupling. Figure~\ref{fig:self-excitation-onset-variables}
connects onset to the energy resource supplying gain.

The amplitude--phase distinction also has an exact stability measure.
For a maintained periodic orbit of period $T_c$, compare nearby trajectories
after one cycle. Their linear return factors are called Floquet multipliers:
the tangent factor is $1$ because shifting the time origin gives another
solution, while the transverse factor for this two-coordinate model is
\begin{equation}
 m_\perp=\exp\!\left\{\int_0^{T_c}
       2\left[\delta_s-\gamma_{\rm n}Q_s^2(t)/Q_c^2\right]\dd t\right\}.
 \label{eq:self-excitation-orbital-return}
\end{equation}
The exponent is the integral of the phase-space vector field's divergence,
which measures local expansion or contraction of nearby state-space area.
For $0<m_\perp<1$, a displacement across the orbit contracts on each
return; the unchanged tangent factor still preserves a phase offset.
Using the weak-rate amplitude above gives
$m_\perp\simeq\exp(-2\delta_sT_c)$, recovering the envelope relaxation
rate. Figure~\ref{fig:self-excitation-phase-portrait} illustrates this
contraction. These stationary-cycle factors apply to maintained gain;
a trajectory with a depleting resource is not a periodic orbit of its
complete state.

\begin{figure}[H]
\centering\begingroup
\definecolor{vlfGreen}{HTML}{176B64}
\definecolor{vlfDarkRed}{HTML}{762B38}
\definecolor{vlfRed}{HTML}{762B38}
\definecolor{vlfBlue}{HTML}{315D80}
\definecolor{vlfCopper}{HTML}{762B38}
\setlength{\unitlength}{1pt}
\scalebox{0.90}{%
\input{vlt-figures/figure-140.tex}%
}\endgroup
\caption{Maintained-gain onset and a finite-resource energy ledger. (a) With $\gamma_{\rm p}/\omega_s=\gamma_{\rm n}/\omega_s=0.05$, the rest state is stable below $g=\gamma_{\rm a}/\omega_s=0.05$. Above threshold, the dashed weak-rate prediction is $A/Q_c\simeq2\sqrt{(\gamma_{\rm a}-\gamma_{\rm p})/\gamma_{\rm n}}$; dots are late positive maxima from direct integration. At threshold, nonlinear relaxation is algebraic. (b) Positive power $P_{\rm a}=2\gamma_{\rm a}\dot Q_s^2$ transfers source energy to the mode, while $P_{\rm d}=2[\gamma_{\rm p}+\gamma_{\rm n}(Q_s/Q_c)^2]\dot Q_s^2$ transfers it to the receiver. With no external exchange, the three stored energies sum to a constant. A finite-source realization therefore retains its resource dependence; the maintained-gain curve applies over its separately specified window. The lower inset identifies the force and velocity whose calibrated modal projection gives the work balance in \eqref{eq:self-excitation-power-balance}. Its arrows locate the coupling without prescribing amplitudes.}
\label{fig:self-excitation-onset-variables}
\end{figure}

\subsubsection{The resource variable in a finite closed system}

Fixed gain freezes or maintains the resource supplying $P_{\rm a}$.
To keep a finite closed energy account explicit, let $\mathcal E_{\rm a}$
be nonnegative stored energy available to the idealized feedback, and let
$\mathcal E_{\rm h}$ be the unresolved internal energy receiving
$P_{\rm d}$. Replace the gain by $\gamma_{\rm a}(\mathcal E_{\rm a})$ and
take the minimal energetic closure
\begin{equation}
 \Dt{\mathcal E_{\rm a}}=-P_{\rm a},\qquad
 \Dt{\mathcal E_{\rm h}}=P_{\rm d},\qquad
 \boxed{\Dt{(E_s+\mathcal E_{\rm a}+\mathcal E_{\rm h})}=0}.
 \label{eq:self-excitation-finite-energy-ledger}
\end{equation}

Figure~\ref{fig:self-excitation-resource-transient} gives a graphical specialization of this boxed relation.

Work leaves the source, enters the mode and reaches unresolved internal
energy. Conversion losses debit the source and credit the receiver.
A chemical realization specifies populations, affinities and entropy
production; a thermal realization supplies temperature and its feedback
on force. Section~\ref{sec:phonon-biochemical-extensions} develops these
conversion relations. The quantities above are energies; a free-energy
budget additionally specifies temperature and exchange conditions.

For example, a specified gain law with $\mathcal E_{\rm ref}>0$ and $\gamma_{{\rm a}0}\geq0$ is
\begin{equation}
 \gamma_{\rm a}(\mathcal E_{\rm a})
    =\gamma_{{\rm a}0}\frac{\mathcal E_{\rm a}}{\mathcal E_{\rm ref}},
 \qquad
 \mathcal E_{\rm a}(t)=\mathcal E_{\rm a}(0)
     \exp\!\left[-\frac{2\gamma_{{\rm a}0}}{\mathcal E_{\rm ref}}
                         \int_0^t\Pi_s^2(t')\,\dd t'\right].
 \label{eq:self-excitation-resource-positivity}
\end{equation}
The resource stays nonnegative. Its state changes the available gain,
and the exact finite-budget bound is
$\int_0^tP_{\rm d}\,\dd t'\leq E_s(0)+\mathcal E_{\rm a}(0)$.
A nonzero cycle dissipating positive work each period cannot persist
indefinitely from this finite budget. A constant-gain attracting cycle
can approximate a replenished source or a slowly changing resource over
a declared window; it is not a perpetual cycle of the complete isolated
system. To resolve the transient rather than just its budget, put
$c_{\rm a}=\gamma_{{\rm a}0}/\mathcal E_{\rm ref}$ and
$c_{\rm n}=\gamma_{\rm n}/(\omega_s^2Q_c^2)$, both in
$\mathrm J^{-1}\mathrm s^{-1}$. The leading cycle energy is
$\overline E_s=\omega_s^2A_s^2/2$. Averaging the same work and resource
equations gives the coupled slow description
\begin{equation}
 \begin{split}
 \Dt{\overline E_s}&=2(c_{\rm a}\mathcal E_{\rm a}-\gamma_{\rm p})
                 \overline E_s-c_{\rm n}\overline E_s^2,\\
 \Dt{\mathcal E_{\rm a}}&=-2c_{\rm a}\mathcal E_{\rm a}\overline E_s.
 \end{split}
 \label{eq:self-excitation-resource-slow-flow}
\end{equation}
These equations define the leading averaged model. Resource work feeds
the mode, linear loss removes it and the quadratic term saturates growth.
For $c_{\rm a}>0$ and $\overline E_s>0$, the growth boundary is
\begin{equation}
 \mathcal E_{\rm crit}=\frac{\gamma_{\rm p}}{c_{\rm a}},\qquad
 \mathcal E_{\rm a}=\mathcal E_{\rm crit}
              +\frac{c_{\rm n}}{2c_{\rm a}}\overline E_s.
 \label{eq:self-excitation-resource-nullcline}
\end{equation}
This line is an energy-growth \emph{nullcline}: the envelope derivative
vanishes there. Zero envelope energy is a separate nullcline for every
resource value. Above the nonzero-growth line the envelope grows;
below it the envelope decays.
At a nonzero envelope maximum, the resource is still decreasing, so this
point is not an equilibrium. The maximum generally precedes crossing the
small-amplitude threshold $\mathcal E_{\rm crit}$ because nonlinear
losses already compete with gain.

The resource must change slowly within an oscillation for this averaging
to be appropriate. In addition to the earlier weak-rate conditions,
its relative drift requires $2c_{\rm a}\overline E_s/\omega_s\ll1$.
Following the \emph{instantaneous attracting amplitude} is a stronger
assumption. With $\delta_s(t)=c_{\rm a}\mathcal E_{\rm a}(t)-\gamma_{\rm p}>0$,
its fractional drift must be small compared with the relaxation rate:
$|\Dt{\delta_s}|\ll4\delta_s^2$. As onset is approached, the restoring
rate for amplitude disturbances tends to zero. A resource can therefore
vary slowly per period while the envelope fails to track its frozen-gain
cycle. Figure~\ref{fig:self-excitation-resource-transient} plots the
evolving system instead of assigning that reference cycle to every instant.
The two slow energy variables predict the envelope; they omit phase and
cannot reconstruct the full displacement waveform by themselves.

\begin{proposition}[Eventual rest with a residual resource]
\label{prop:self-excitation-finite-resource-rest}
Use the gain law in~\eqref{eq:self-excitation-resource-positivity}, with
$\gamma_{\rm p}>0$, $c_{\rm a}>0$, positive reference frequency and finite
nonnegative initial energies. Suppose the deterministic constitutive law
remains valid along the trajectory. Then
\begin{equation}
 \begin{split}
 (Q_s,\Pi_s)&\longrightarrow(0,0),\qquad
 \mathcal E_{\rm a}(t)\longrightarrow\mathcal E_{{\rm a},\infty},\\
 \mathcal E_{{\rm a},\infty}&\geq\mathcal E_{\rm a}(0)
  \exp\!\left[-\frac{c_{\rm a}[E_s(0)+\mathcal E_{\rm a}(0)]}
                    {\gamma_{\rm p}}\right]>0
       \quad\bigl(\mathcal E_{\rm a}(0)>0\bigr).
 \end{split}
 \label{eq:self-excitation-residual-resource}
\end{equation}
For a nonzero initial mechanical preparation,
$\mathcal E_{{\rm a},\infty}\leq\mathcal E_{\rm crit}$.
\end{proposition}

\begin{proof}
The nonnegative quantity $W=E_s+\mathcal E_{\rm a}$ satisfies
$\Dt{W}=-P_{\rm d}\leq-2\gamma_{\rm p}\Pi_s^2$. Its sublevel set on the reduced state space
$(Q_s,\Pi_s,\mathcal E_{\rm a}\geq0)$ bounds all three coordinates
and is compact and forward invariant.
Within $\Pi_s=0$, a nonzero $Q_s$ generates restoring acceleration, so
the largest invariant subset has $Q_s=\Pi_s=0$. LaSalle's principle
therefore gives mechanical rest~\cite{Tedrake2023,Khalil2002}.
The resource decreases to a limit. Integrating the inequality gives
$\int_0^\infty\Pi_s^2\,\dd t\leq W(0)/(2\gamma_{\rm p})$;
substitution in~\eqref{eq:self-excitation-resource-positivity} gives the
stated positive lower bound. If a nonzero mechanical preparation ended
with $\mathcal E_{{\rm a},\infty}>\mathcal E_{\rm crit}$, then at
sufficiently late times the positive small-amplitude gain would dominate
the vanishing amplitude-dependent loss. Uniqueness of the finite-dimensional state equation keeps $E_s(t)>0$
at every finite time for a nonzero initial mechanical preparation.
That energy would then be nondecreasing at late times, contradicting
its convergence to zero.
\end{proof}

Exact rest consumes no resource, even above threshold. With nonzero
preparation, motion ceases when conversion can no longer overcome loss;
stored source energy can remain. The proposition assumes noiseless
dynamics and positive passive loss. A thermal or active noise background
has its own law, and the residual-resource bound excludes
$\gamma_{\rm p}=0$.

\begin{figure}[H]
\centering\begingroup
\definecolor{vlfGreen}{HTML}{176B64}
\definecolor{vlfDarkRed}{HTML}{762B38}
\definecolor{vlfRed}{HTML}{762B38}
\definecolor{vlfBlue}{HTML}{315D80}
\definecolor{vlfCopper}{HTML}{762B38}
\setlength{\unitlength}{1pt}
\scalebox{0.90}{%
\input{vlt-figures/figure-141.tex}%
}\endgroup
\caption{Finite-resource self-excitation in a classical auxiliary model. Use scaled time $\tau=\omega_st$, displacement $x=Q_s/Q_c$, velocity $v=\mathrm dx/\mathrm d\tau$, and energies $(e,r,h)=(E_s,\mathcal E_{\rm a},\mathcal E_{\rm h})/(\omega_s^2Q_c^2)$, with instantaneous $e=(x^2+v^2)/2$. (a) Resource $r$ and receiver $h$ follow the full equations; the modal curve is the centered one-period mean $\bar e$. The constant line is the exact unaveraged total. (b) The leading cycle-averaged trajectory (black) is compared with the same one-period means of the full solution (gray circles). The energy maximum lies on $r=r_{\rm crit}+n\bar e/(2k)$, above $r_{\rm crit}=p/k$; the mode can decline while unused source energy remains. For $\gamma_{\rm a}=c_{\rm a}\mathcal E_{\rm a}$, the illustrative parameters are $p=\gamma_{\rm p}/\omega_s=0.006$, $n=\gamma_{\rm n}/\omega_s=0.016$, $k=c_{\rm a}\omega_sQ_c^2=0.02$, with initial $(x,v,r,h)=(0.1,0,0.8,0)$. The lower diagram gives the storage and positive power-transfer channels of \eqref{eq:self-excitation-finite-energy-ledger}. The source energy constrains conversion duration; a retained converter force would additionally determine its response time and phase. No measured virion or blood coefficients are assigned.}
\label{fig:self-excitation-resource-transient}
\end{figure}

\paragraph{Recruitment, absorption and the available work.}
\label{sec:s4-exchange-recruitment-work}

The source--mode--receiver construction of
Subsection~\ref{sec:s2-phonon-source-absorber} makes excitation-assisted
growth an energy-transfer question. In the passive response,
\eqref{eq:s2-phonon-receiver-power} and
\eqref{eq:average-loss-power} measure power deposited in a specified
loss channel. That deposited power can enter local heating or a retained
structural process according to its declared coupling. A persistent
joining event additionally requires encounter, accommodation and the
membership criterion. The conditional rate
\eqref{eq:s23-absorption-assisted-hazard} supplies one explicit
independent-event model of that additional step. At finite temperature,
net dissipated power is distinguished from the separate upward and
downward quantum transition counts in
\eqref{eq:s2-phonon-exchange-weights}.

An increase of energy reported for a selected cluster can have several
accounted causes: work supplied by a source, energy carried in by an
arriving virion, and binding energy redistributed into collective motion.
The full $N_{\rm tot}$ particle--medium system includes those arrivals
before they become cluster members. Its energy account therefore
distinguishes redistribution and selection from maintained active gain.
Preparing an occupied mode can bias a slow joining coordinate through
\eqref{eq:s23-mode-backaction}; preserving that preparation against loss
requires its source and converter state. These links make the final
self-excitation model predictive for an evolving assembly: a measured
growth or oscillation record is compared with membership, source work
and medium loss over the same interval. The fixed vessel population,
the cluster membership and the excitation occupations remain separate
variables throughout.

Figure~\ref{fig:s4-source-absorption-recruitment} identifies the work, absorption and association measurements needed to connect excitation to recruitment.

\begin{figure}[H]
\centering\begingroup
\definecolor{vltTeal}{HTML}{176B64}
\definecolor{vltBlue}{HTML}{315D80}
\definecolor{vltGold}{HTML}{A87520}
\definecolor{vltPlum}{HTML}{725877}
\definecolor{vltRed}{HTML}{762B38}
\definecolor{vltYellow}{HTML}{D5AD3C}
\definecolor{vltNeutral}{HTML}{777D83}
\setlength{\unitlength}{1pt}
\scalebox{0.90}{%
\input{vlt-figures/figure-142.tex}%
}\endgroup
\caption{A source, four-member assembly and receiver $j$ occupy one five-virion vessel. The orthographic view preserves particle identities and attaches six reference links to the three-dimensional body surfaces. Blue identifies applied force; separate power arrows locate receiver deposition and medium loss, without magnitude or travel-time scales. Lower boxes connect source work, deposited energy and the additional encounter/accommodation variables entering~\eqref{eq:s23-absorption-assisted-hazard}. Open arrows connect calculations. The full-pool energy account includes the receiver before membership changes; the geometry specifies neither fitted couplings nor capture probabilities.}
\label{fig:s4-source-absorption-recruitment}
\end{figure}

\subsubsection{Retaining the variables that generate force}
\label{sec:self-excitation-feedback-closure}

\begin{definition}[Resource and converter coordinates]
\label{def:self-excitation-resource-converter}
A \emph{resource coordinate} records the state of the source that enters
the force-generating law; here $\mathcal E_{\rm a}$ is its available
stored energy, in joules. A \emph{converter coordinate} records the
retained force-generating state, such as the active force
$F_{{\rm a},s}$ below. Its relaxation law specifies how the source and
mechanical motion generate that force. Each coordinate has both a physical
meaning and a specified dynamical role.
\end{definition}

For the instantaneous resource-dependent law,
$(Q_s,\Pi_s,\mathcal E_{\rm a})$ predicts mode and resource motion.
The internal-energy coordinate $\mathcal E_{\rm h}$ completes the energy
account; it becomes a feedback variable when temperature or another
constitutive dependency is coupled to it. Two preparations with the same
$Q_s,\Pi_s$ and different resource energies have acceleration difference
$2[\gamma_{\rm a}(\mathcal E_{{\rm a},1})-
\gamma_{\rm a}(\mathcal E_{{\rm a},2})]\Pi_s$.
For $\Pi_s\ne0$, this is a direct test of the resource's dynamical role.
A fixed gain is a useful reduction while the resource changes negligibly
over the prediction window. A systematic change of subsequent growth at
the same mechanical state calls for retaining the source state or the
preparation history that determines it.

Converter response time controls when the instantaneous-force reduction
applies. For a maintained resource, retain active modal force
$F_{{\rm a},s}$ with response time $\tau_{\rm a}>0$:
\begin{equation}
 \begin{split}
 \Dtt{Q_s}+2\left[\gamma_{\rm p}
       +\gamma_{\rm n}(Q_s/Q_c)^2\right]\Dt{Q_s}
                  +\omega_s^2Q_s&=F_{{\rm a},s},\\
 \tau_{\rm a}\Dt{F_{{\rm a},s}}+F_{{\rm a},s}
                  &=2\gamma_{\rm a}\Dt{Q_s}.
 \end{split}
 \label{eq:self-excitation-retained-feedback}
\end{equation}
This relaxation variable describes how an active force builds and
decays, with units $\mathrm{kg}^{1/2}\mathrm m\,\mathrm s^{-2}$.
A prepared force persists for a time of order $\tau_{\rm a}$ even when
the mechanical coordinates have been matched between preparations.
After that initial transient, eliminating the converter recovers the
instantaneous gain in~\eqref{eq:self-excitation-modal-model} over a band
with $|\Omega|\tau_{\rm a}\ll1$. This condition includes the carrier,
the envelope and the force variations used in the prediction. An initial
force transient must either be resolved or have decayed within the stated
accuracy. At finite response time, the maintained-source state is
$(Q_s,\Pi_s,F_{{\rm a},s})$: equal positions and velocities with different
active forces have different accelerations immediately. A physical motor
or other converter supplies its own force-generation law when a single
relaxation time is insufficient.

For a sinusoidal velocity phasor $\Pi_s(\Omega)$ using the paper's
$e^{-i\Omega t}$ convention,
\begin{equation}
 F_{{\rm a},s}(\Omega)=
       \frac{2\gamma_{\rm a}}{1-i\Omega\tau_{\rm a}}\Pi_s(\Omega),
 \qquad
 \langle P_{{\rm a},s}\rangle_{\rm cycle}
       =\frac{\gamma_{\rm a}|\Pi_s(\Omega)|^2}
                    {1+\Omega^2\tau_{\rm a}^2}.
 \label{eq:self-excitation-feedback-work}
\end{equation}
Taking real parts of the phasors fixes the cycle-average factor. Force
lags velocity: its in-phase component supplies mean work and its
quadrature component shifts frequency. For nonzero gain in this law,
their ratio is $\Omega\tau_{\rm a}$. Measuring phase across a band tests
the single converter time separately from source energy. A resolved
departure can require further converter coordinates at the same
mechanical state and resource.

Substituting a small perturbation proportional to
$e^{\lambda_{\rm g}t}$ into the linearization at rest gives
\begin{equation}
 (1+\tau_{\rm a}\lambda_{\rm g})
   (\lambda_{\rm g}^2+2\gamma_{\rm p}\lambda_{\rm g}+\omega_s^2)
                 -2\gamma_{\rm a}\lambda_{\rm g}=0.
 \label{eq:self-excitation-feedback-characteristic}
\end{equation}
Here $\operatorname{Re}\lambda_{\rm g}$ is the perturbation growth
or decay rate; $|\operatorname{Im}\lambda_{\rm g}|$ gives its angular
oscillation frequency. The cubic stability criterion locates the boundary where a
conjugate pair of these growth rates reaches the imaginary axis. For
$\gamma_{\rm p}>0$, its exact threshold and neutral frequency are
\begin{equation}
 \gamma_{{\rm a},H}=\gamma_{\rm p}
   \left[1+\frac{\omega_s^2\tau_{\rm a}^2}
                   {1+2\gamma_{\rm p}\tau_{\rm a}}\right],
 \qquad
 \Omega_H=\frac{\omega_s}{\sqrt{1+2\gamma_{\rm p}\tau_{\rm a}}}.
 \label{eq:self-excitation-feedback-onset}
\end{equation}
Rest is linearly stable below this gain and unstable above it; the remaining
growth rate at threshold is $-(1/\tau_{\rm a}+2\gamma_{\rm p})$.
Figure~\ref{fig:self-excitation-feedback-state} shows how retaining
the force state raises the required gain and lowers the onset frequency.
This is a linear oscillatory-instability result. The stability and amplitude
of a nonlinear periodic orbit need their own calculation; the earlier
instantaneous-model saturation formula cannot simply be transferred.

The sign of instantaneous work $F_{{\rm a},s}\Pi_s$ records the
direction of energy exchange between converter and mode. Lag permits
some work to return during a cycle even when the mean supplied work is
positive. A finite-source realization therefore specifies where returned
work is stored or dissipated, alongside source depletion and conversion
losses. The force coordinate and its response equation describe the
converter's dynamics; an actuator storage law and thermodynamic resource
constraints complete its energy account. These are distinct requirements:
the first finite-resource example tracks how much source energy remains,
and the lagged-force example tracks how quickly that source can generate
the force seen by the mode.

\begin{figure}[H]
\centering\begingroup
\definecolor{vltTeal}{HTML}{176B64}
\definecolor{vltBlue}{HTML}{315D80}
\definecolor{vltGold}{HTML}{A87520}
\definecolor{vltPlum}{HTML}{725877}
\definecolor{vltRed}{HTML}{762B38}
\definecolor{vltYellow}{HTML}{D5AD3C}
\definecolor{vltNeutral}{HTML}{777D83}
\setlength{\unitlength}{1pt}
\scalebox{0.90}{%
\input{vlt-figures/figure-143.tex}%
}\endgroup
\caption{Retained force dynamics and a collective load. (a,b) Exact maintained-source onset and neutral-frequency curves from \eqref{eq:self-excitation-feedback-onset}, with $p=0.1$. (c) At $\Omega\tau_{\rm a}=1$, unit sinusoidal velocity produces normalized force $(\cos\Omega t+\sin\Omega t)/2$: its amplitude is $1/\sqrt2$ and its maximum occurs $\pi/4$ later. The two normalizations are written in the legend. (d) A possible load realization has four equal masses, equal positive onsite restoring curvature and the normalized common column shown below the oblique assembly. Reference links connect the rigid virion surfaces; common translation preserves their spans. Burgundy arrows indicate common force components, while the open arrow maps the converter state into that pattern. The geometry and component arrows are a coupling sketch; the onset curves apply to the calibrated modal equation. Source storage and nonlinear amplitude selection retain their separate accounts.}
\label{fig:self-excitation-feedback-state}
\end{figure}

\subsubsection{Biophysical coupling and experimental discrimination}

A concrete biophysical assignment places a mechanical probe in contact
with a force-generating soft structure. Its centre displacement is the
load coordinate; the surrounding structure contains the converter and
its energy source. Collective molecular-motor theory supplies a physical
precedent: chemically driven internal states can act through an elastic
load to produce spontaneous oscillation~\cite{JulicherProst1997}.
This example explains why converter time scales and attachment geometry
enter beside the load's stiffness and inertia. Assigning the role to an
ideal virion requires a specified mechanical coupling to such an active
structure. The passive cavity model continues to describe flow-driven
and thermal particle motion with its declared medium law.

For a real translational pattern $\vct e_s$ that is dimensionless and
unit-normalized in mass-weighted coordinates, a physical active-force
vector $\vct f_{\rm a}$ has the same modal projection used for finite
pulse excitation in Section~\ref{sec:finite-pulse-excitation} and the
associated power:
\begin{equation}
 F_{{\rm a},s}=\vct e_s^{\mathsf T}\mathsf M^{-1/2}\vct f_{\rm a},
 \qquad P_{{\rm a},s}=F_{{\rm a},s}\Pi_s.
 \label{eq:self-excitation-active-modal-projection}
\end{equation}
The mass normalization makes this the physical power exchanged through
that pattern. A converter can be active while its force has zero overlap
with the selected displacement pattern. Likewise, a measured surface
deformation is not automatically a centre-displacement mode. The force
projection, readout, and structural resolution identify which excitation
receives and reveals the work~\cite{Leake2016}.
In the measles-motivated envelope/RNP example of
Section~\ref{sec:measles-phase-response}, relative deformation and whole-particle
translation consequently require different projections of the applied force.

Equal stored availability can coexist with different chemical
concentrations, transition rates or populations of force-bearing converter
configurations. Those preparations can exert different forces on the same
load. A resource energy is therefore sufficient for the stipulated gain
law, while a chemical realization can require the relevant composition
or converter population as an additional state. Its physical definition
and calibrated force coupling determine what that variable predicts.

The thermodynamic account specifies controlled temperature, volume or
pressure, chemical work and exchange conditions before introducing a
free-energy budget~\cite{Cotterill2002,Seifert2012}. Availability then
constrains useful work, and the internal-energy account records storage,
work and heat consistently. Maintained concentrations correspond to
replenishment or a sufficiently short observation window. The common
chemical and electrical construction in
Section~\ref{sec:phonon-biochemical-extensions} supplies those conversion
relations. Here the distinction establishes a practical criterion: retain
the composition or population when equal-resource preparations generate
different subsequent forces or readouts.

A matched comparison first constrains passive stiffness, inertia and
loss, then uses the same force/readout map to test growth, saturation
and recovery. Within the resolved-envelope and fast-feedback regime,
the small-amplitude slope of $\log A_s$ tests gain minus passive loss;
larger amplitudes test the saturation term
in~\eqref{eq:self-excitation-averaged-onset}. Calibrated force and velocity
give the work exchanged during a cycle.

Resource depletion predicts preparation-dependent envelopes and gain
decline in~\eqref{eq:self-excitation-resource-slow-flow}. A calibrated
source readout tests its work account. Converter delay predicts a
force--velocity lag, the onset shift
\eqref{eq:self-excitation-feedback-onset} and an initial-force transient.
Band-resolved phase and repeated preparation at matched mechanical
coordinates distinguish these effects. Attachment, stiffness and
detector drift are constrained over the same window.

A resource predicts the work budget and duration of conversion;
a converter predicts force timing and phase; a configuration population
can predict occupancy of a force-bearing state. Establishing such a
variable means assigning its physical meaning and evolution, then
testing a reproducible readout change across preparations with the other
inputs fixed or independently constrained.

The strongest signature joins autonomous growth or amplitude recovery
to a positive supplied-work account and the specified source dynamics.
Thermal motion filtered by passive resonance, an imposed periodic drive
and temporary energy return provide concrete comparison descriptions for
the same spectral peak. Thermal and converter noise enter through a
nonequilibrium stochastic law matched to that realization. The exact
approach-to-rest proposition describes the loss of coherent deterministic
motion while a separately modeled thermal background can persist.
The active extension thus links source storage, force generation and
mechanical readout through a defined physical pathway.

The next section develops complex state representations. Particle
identities, transition laws and source--medium energy accounts continue
across reference windows.

\clearpage

\section{The Complex-Valued Displacement Field}
\label{sec:complex-displacement}

Complex amplitudes record the pattern and timing of the real motion
developed in Sections~3--4. The labelled ideal virions retain their masses,
body orientations, specified surface preparation and particle--medium
couplings. Fixed assigned charges give one reference; regulated surfaces
retain their population dynamics. Within a persistent
reference, a phasor describes steady oscillation. Across recruitment or
breakup, the continued physical state and registered particle identities
connect successive descriptions.

Relative phase has a direct mechanical role. Equal-amplitude particles
can move together, oppose one another or reach their extrema a quarter
cycle apart. Common motion preserves their separation, while opposing
motion cyclically stretches and compresses their interaction. Complex
notation retains these timing relations and simplifies superposition,
mode expansion and frequency-dependent response.

\begin{figure}[H]
\centering\begingroup
\definecolor{vlfGreen}{HTML}{176B64}
\definecolor{vlfDarkRed}{HTML}{762B38}
\definecolor{vlfRed}{HTML}{762B38}
\definecolor{vlfBlue}{HTML}{315D80}
\definecolor{vlfCopper}{HTML}{762B38}
\setlength{\unitlength}{1pt}
\scalebox{0.90}{%
\input{vlt-figures/figure-144.tex}%
}\endgroup
\caption{The real harmonic $u/A=\cos(\theta-\pi/3)$, with $\theta=\omega t$, has coefficient $U/A=e^{i\pi/3}$. The complex-plane point rotates clockwise; its horizontal projection gives the displacement plotted at right. Filled and open points mark $\theta=0,\pi/2$. Separate line styles resolve the cosine and sine quadratures. Lower snapshots translate the same fixed-orientation ideal virion with gain $22\,\mathrm{pt}/A$. Equal rulers mark $u/A=-1,0,1$; dashed lines locate the reference and arrows show displacement. The upper complex axes are signal quadratures, whereas the lower rulers locate physical motion.}
\label{fig:complex-phasor-quadratures}
\end{figure}

The retained mechanical coordinates describe real centre motion, angular
motion and any specified deformation. Surface populations and material
coordinates carry their own coupled kinetics and storage. Two real
quadratures encode its amplitude and timing relative to a clock.
The calibrated force and readout maps give those coefficients their
physical meaning. For example, a spatial average can suppress an
opposing pattern even when individual particles move appreciably.

A phase-resolved record contains more information than an oscillation
amplitude alone. Referencing the displacement to a known drive reveals
when each coordinate reaches its maximum and how the material delays
the response. Combining the two timing components into one complex
coefficient keeps this information available when patterns are added,
projected or compared across particles.

The weighting used in that comparison is equally important. A measured
length, a mass-weighted pattern and an energy-normalized amplitude
answer different questions about the same motion. Their corresponding
inner products make those questions precise. This section builds the
representations from the real displacement and its work account, so that
each later complex formula can be translated back into a physical record.

A resolved fluid adds velocity, pressure and material stress. About a
stationary reference, differentiating an oscillatory material
displacement gives velocity. Sustained transport instead uses the
velocity field and boundary flux. The pressure--flow application below
connects these fluid variables to particle motion and supplied work.

A Hilbert space compares entire displacement patterns through an inner
product and is complete in its associated norm. Spatial positions index
a field; directions in this vector space distinguish patterns. This
geometry supports orthogonal projection and approximation. The energy
law, preparation and detector map supply the physical interpretation;
the phonon postulate separately specifies quantized occupations.

Table~\ref{tab:complex-representation-guide} fixes the units and roles
used throughout. A displacement phasor records a steady oscillation;
the mass metric compares spatial patterns. The energy amplitude in
Section~\ref{sec:conservative-complex-generator} instead combines
instantaneous displacement and velocity with an energy norm.
The occupation space carries the separately postulated quantum state.

\begin{table}[H]
\centering\small
\begin{tabular}{@{}p{3.0cm}p{2.9cm}p{7.1cm}@{}}
\toprule
\raggedright Representation & \raggedright Units or normalization
& \raggedright Physical content \tabularnewline
\midrule
\raggedright Displacement phasor $\vct U$ or signal $\vct z_{\rm disp}$
& \raggedright Length
& \raggedright Real displacement amplitude and timing. Mechanical energy
requires inertial, restoring, and material-storage weights. \tabularnewline
\raggedright Classical energy ket $\psi$
& \raggedright $\sqrt{\text{energy}}$ for a finite lattice;
$\|\psi\|^2=E_{\rm harm}$
& \raggedright Invertible encoding of displacement and velocity on the
positive conservative sector; not a probability distribution. \tabularnewline
\raggedright Quantum occupation ket $\ket\Phi$ or density operator $\varrho$
& \raggedright Dimensionless occupation-basis coefficients; unit norm or
unit trace
& \raggedright Probabilities and correlations of specified quantum
observables in an independently declared quantum state. \tabularnewline
\bottomrule
\end{tabular}
\caption{The same mechanical reference supports different representations
with different physical jobs. A complex displacement does not become a
quantum state through notation alone. The continuum energy field uses the
mass-weighted norm specified in
Section~\ref{sec:conservative-complex-generator}.}
\label{tab:complex-representation-guide}
\end{table}

\subsection{Real motion and its complex amplitude}
\label{sec:displacement-phasor}

\begin{definition}[Displacement phasor]
\label{def:s5-displacement-phasor}
At a declared real frequency $\omega>0$, coordinate basis and time
origin, a \emph{displacement phasor} is the complex column or field
$\vct U$ whose real-part reconstruction is
$\vct u(t)=\operatorname{Re}[\vct Ue^{-i\omega t}]$.
Its two real components are the cosine and sine displacement
coefficients. A translational phasor has length units; it specifies a
harmonic history relative to the chosen clock.
\end{definition}

Begin with a single scalar harmonic readout,
$u(t)=u_{\rm R}\cos\omega t+u_{\rm I}\sin\omega t$, at a real frequency
$\omega>0$. Both coefficients have length units. Euler's identity
$e^{-i\omega t}=\cos\omega t-i\sin\omega t$ lets us combine them as
$U=u_{\rm R}+iu_{\rm I}$ and recover the real motion by taking the real
part. Applying the same construction componentwise gives
\begin{equation}
 \vct u(\vct r,t)
   =\operatorname{Re}[\vct U(\vct r;\omega)e^{-i\omega t}],
 \qquad \vct U=\vct U_{\mathrm R}+i\vct U_{\mathrm I},
 \label{eq:physical-phasor-displacement}
\end{equation}
and, on multiplying out,
\begin{equation}
 \vct u(\vct r,t)=\vct U_{\mathrm R}(\vct r)\cos\omega t
                      +\vct U_{\mathrm I}(\vct r)\sin\omega t.
 \label{eq:displacement-quadratures}
\end{equation}
The two real fields are \emph{quadratures}: coefficients of oscillations
separated by a quarter period. For one coordinate, $U=Ae^{i\phi}$ means
$u(t)=A\cos(\omega t-\phi)$, where $A=|U|$ and
$\phi=\operatorname{atan2}(u_{\rm I},u_{\rm R})$ when $A>0$.
Multiplying $U$ by $i$ delays this real harmonic trajectory by a quarter
period under the stated convention. No phase is defined for zero amplitude.
Figure~\ref{fig:complex-phasor-quadratures} illustrates this correspondence.

Time differentiation becomes multiplication by $-i\omega$, turning a
constant-coefficient linear equation into algebraic force balance while
retaining both real quadratures~\cite{Georgi1992}. In viscoelastic
response, the displacement's phase relative to force then resolves
timing information discarded by an amplitude-only measurement.

The reference clock matters. Resetting it to $t'=t-t_0$ gives the same
trajectory with $U'=Ue^{-i\omega t_0}$. Real and imaginary coefficients
therefore change together; relative phase between two coordinates at the
same frequency, or between force and response, survives a common reset.
At the chosen origin a purely harmonic field satisfies
$\vct u(\vct r,0)=\vct U_{\rm R}$ and
$\Pdt{\vct u}(\vct r,0)=\omega\vct U_{\rm I}$.
The imaginary coefficient therefore contributes to the real velocity;
its interpretation here is kinematic.

For the discrete ideal-virion array this reconstruction reads
\begin{equation}
 \begin{aligned}
 \vct r_i(t)&=\vct R_i+
        \operatorname{Re}[\vct U_i e^{-i\omega t}],\\
 \dot{\vct r}_i(t)&=
        \operatorname{Re}[-i\omega\vct U_i e^{-i\omega t}],
        \qquad i=1,\ldots,n.
 \end{aligned}
 \label{eq:s5-phasor-node-reconstruction}
\end{equation}
Each component follows the same labelled particle through the cycle.
Its rigid facets fix body geometry. The phasor specifies motion about
the reference; assigned charges enter the force law and the reference
itself.

\paragraph{Reconstructing the oriented body and its registered marker.}
\begin{definition}[Rigid-body motion phasor]
\label{def:s5patch-body-phasor}
Near a resting body reference, let $\vct U_i$ be a length-valued centre
phasor and $\vct\Theta_i$ a dimensionless laboratory angular-tangent
phasor. Their real reconstruction is
\begin{equation}
 \begin{aligned}
 \vct r_i(t)&=\vct R_i+\operatorname{Re}(\vct U_i e^{-i\omega t}),\qquad
 \vct\theta_i(t)=\operatorname{Re}(\vct\Theta_i e^{-i\omega t}),\\
 \mathsf O_i(t)&=\exp([\vct\theta_i(t)]_\times)\mathsf O_i^0 .
 \end{aligned}
 \label{eq:s5patch-body-phasor-reconstruction}
\end{equation}
The angular velocity has phasor $-i\omega\vct\Theta_i$ to first order.
Finite rotations use the exponential-map differential and the full
rigid-body equations of Section~4.
\end{definition}

Figure~\ref{fig:s5patch-registered-marker-motion-source} distinguishes translation and rotation in controlled marker motions, and gives their conjugate force and torque.

\begin{figure}[H]
\centering\begingroup
\definecolor{vltTeal}{HTML}{176B64}
\definecolor{vltBlue}{HTML}{315D80}
\definecolor{vltGold}{HTML}{A87520}
\definecolor{vltPlum}{HTML}{725877}
\definecolor{vltRed}{HTML}{762B38}
\definecolor{vltYellow}{HTML}{D5AD3C}
\definecolor{vltNeutral}{HTML}{777D83}
\setlength{\unitlength}{1pt}
\scalebox{0.84}{%
\input{vlt-figures/figure-296.tex}%
}\endgroup
\caption{A registered surface marker responds to both centre and angular motion. The same projected icosahedron, body-frame attachment and length gain are used in (a,b). Open and filled gold marks identify reference and current marker positions; dashed silhouettes register the original body. (a) Translation moves the centre and marker by the same vector. (b) A $0.26$-rad rotation about the laboratory $z$ axis leaves the centre fixed; the blue marker arc uses the exact rotation. (c) The differential readout and virtual-work map are evaluated at the reference: a force applied at the marker supplies a centre force and a lever-arm torque. The panels isolate controlled motions; inferring both from an unknown preparation requires sufficient independent readouts. Body dimensions and attachment are illustrative.}
\label{fig:s5patch-registered-marker-motion-source}
\end{figure}

The orientation is a real rotation matrix along this path. The same
centre history can accompany different surface-site histories. A
registered marker at reference arm $\vct b=\mathsf O_i^0\vct a$ makes
this distinction visible and fixes its lever arm in the detector map.

\begin{proposition}[Geometric harmonics of a rigid marker]
\label{prop:s5patch-marker-harmonics}
For the prescribed body motion above and small angular amplitude, the
marker displacement $\delta\vct x=\vct r_i-\vct R_i+
(\exp([\vct\theta_i]_\times)-\Id)\vct b$ has the expansion
\begin{equation}
 \boxed{\begin{aligned}
 \delta\vct x(t)&=\vct d_0+
   \operatorname{Re}[\vct d_1e^{-i\omega t}
                        +\vct d_2e^{-2i\omega t}]
           +O(|\vct b|\|\vct\Theta_i\|^3),\\
 \vct d_1&=\vct U_i+\vct\Theta_i\times\vct b,\\
 \vct d_0&=\tfrac14\operatorname{Re}
       [\vct\Theta_i\times(\vct\Theta_i^*\times\vct b)],\qquad
 \vct d_2=\tfrac14\vct\Theta_i\times(\vct\Theta_i\times\vct b).
 \end{aligned}}
 \label{eq:s5patch-marker-harmonics}
\end{equation}
\end{proposition}
\begin{proof}
Expand the rotation as $\Id+[\vct\theta_i]_\times+
[\vct\theta_i]_\times^2/2+O(\|\vct\theta_i\|^3)$.
Writing each real angular component as the sum of its two conjugate
frequency parts gives the constant and doubled-frequency terms.
\end{proof}

Figure~\ref{fig:cmd-marker-harmonics} gives a graphical specialization of this boxed relation.

\begin{figure}[H]
\centering\begingroup
\setlength{\unitlength}{1pt}
\scalebox{0.84}{%
\input{vlt-figures/figure-314.tex}%
}\endgroup
\caption{A geometric harmonic generated by one registered surface marker. The body rotates about $z$ through $\theta=\epsilon\cos\tau$, with $\epsilon=0.2$ and marker reference $\vct b=\ell\vct e_x$. All four frames use the same size, centre and marker attachment; open gold marks and dashed wireframes register the reference. The exact planar marker coordinates are $x=\ell\cos(\epsilon\cos\tau)$ and $y=\ell\sin(\epsilon\cos\tau)$. Solid traces show their normalized displacements; dashed traces give $-(1+\cos2\tau)/4$ and $\cos\tau$. The offset and second harmonic in $x$ arise from the second-order rotational geometry of \eqref{eq:s5patch-marker-harmonics}, with no second harmonic assigned to the centre motion. The upper views are a geometry illustration rather than a force-driven solution.}
\label{fig:cmd-marker-harmonics}
\end{figure}

Thus marker geometry alone supplies a mean shift and a second harmonic
when the prescribed angular tangent is monochromatic. A mechanical
nonlinearity can produce additional harmonics through the dynamics;
the registered geometric map is applied to that prediction as well.
Marker position, viewing direction, exposure and angular excursion set
the measured amplitudes. A sphere-like centre track and a resolved
surface reporter can therefore answer different questions about one VLP.

Figure~\ref{fig:added-traveling-phase-snapshots} applies the same real-part reconstruction to a spatial phase pattern.

\begin{figure}[H]
\centering\begingroup
\definecolor{vlfGreen}{HTML}{176B64}
\definecolor{vlfDarkRed}{HTML}{762B38}
\definecolor{vlfRed}{HTML}{762B38}
\definecolor{vlfBlue}{HTML}{315D80}
\definecolor{vlfCopper}{HTML}{762B38}
\setlength{\unitlength}{1pt}
\scalebox{0.84}{%
\input{vlt-figures/figure-145.tex}%
}\endgroup
\caption{Eight fixed identities sample the scalar-ring pattern $U_j=Ae^{iqj}$, $q=\pi/4$, with $u_j/A=\cos(qj-\theta)$. Each row uses one body size and gain $8\,\mathrm{pt}/A$; dashed marks and the shared site-index ruler register the reference centres. Numbers under bodies give $u_j/A$, rounded where necessary. Identical coils follow the exact changes $u_{j+1}-u_j$ between displayed neighbours. The unwrapped ring omits its periodic link $7\to0$. Equal masses $m$ and scalar springs $k_b>0$ give $\omega=2\sqrt{k_b/m}\sin(\pi/8)$. Phase advances two site indices per quarter cycle while each particle oscillates about its own reference. The reflecting cylinder uses its own boundary modes.}
\label{fig:added-traveling-phase-snapshots}
\end{figure}

\begin{example}[Complex polarization without mechanical loss]
\label{ex:lossless-complex-polarization}
Two degenerate conservative Cartesian modes admit
\begin{equation}
 \vct U=(A,iB,0)^{\mathsf T},\qquad
 \vct u(t)=(A\cos\omega t,B\sin\omega t,0)^{\mathsf T},
 \qquad A,B\in\R.
 \label{eq:elliptic-displacement-example}
\end{equation}
The physical coordinate traces an ellipse about its reference position.
With isotropic stiffness $m\omega^2\Id_3$, the kinetic and potential
contributions exchange during each period, while their sum is the constant
$E_{\rm harm}=m\omega^2(A^2+B^2)/2$. The imaginary quadrature records
timing and polarization even though this example has no dissipation.
\end{example}

\emph{Polarization} describes the direction followed by the real
displacement. Equal-frequency Cartesian components can make a line,
ellipse, or circle, depending on their relative phase and amplitude.
Different-frequency components generally do not maintain this fixed
ellipse. These are trajectories of the retained coordinate; an internal
deformation pattern requires its own displacement and inertia maps.

\begin{proposition}[Clock-independent harmonic orbit]
\label{prop:s5-harmonic-orbit-geometry}
For a finite length-valued monochromatic pattern, let
$\mathsf C_U=(\vct U_{\rm R},\vct U_{\rm I})$ in the declared physical
coordinate axes. Its real trajectory is the image of the unit circle
under $\mathsf C_U$. It lies in a real subspace of dimension at most
two. Its nonzero squared semiaxes are the eigenvalues of
\begin{equation}
 \mathsf C_U^{\mathsf T}\mathsf C_U=
 \begin{pmatrix}
  |\vct U_{\rm R}|^2&\vct U_{\rm R}\cdot\vct U_{\rm I}\\
  \vct U_{\rm R}\cdot\vct U_{\rm I}&|\vct U_{\rm I}|^2
 \end{pmatrix}.
 \label{eq:s5-phasor-orbit-gram}
\end{equation}
A common shift of the time origin leaves these semiaxes unchanged.
\end{proposition}
\begin{proof}
Equation~\eqref{eq:displacement-quadratures} is
$\vct u(t)=\mathsf C_U(\cos\omega t,\sin\omega t)^{\mathsf T}$.
The singular values of a real matrix are the semiaxes of its image of
a unit circle. A clock shift rotates the two columns by a real
orthogonal $2\times2$ matrix, which preserves these singular values.
\end{proof}
For one particle the image is a spatial orbit; for the full array it is
an ellipse in configuration space. Its particle projections can be
different ellipses or lines. Rank one gives line motion and rank zero
rest. The orbit is independent of the clock phase used to sample it.
These semiaxes use the listed coordinates' Euclidean metric; the mass
metric below supplies inertial weighting.

For $\vct U=(A,Be^{i\phi},0)^{\mathsf T}$ in the isotropic point-coordinate
example, a direct substitution gives
\begin{equation}
 \boxed{L_z=m\left(u_x\frac{\dd u_y}{\dd t}
                    -u_y\frac{\dd u_x}{\dd t}\right)
          =m\omega AB\sin\phi,\qquad
                |L_z|\leq\frac{E_{\rm harm}}\omega.}
 \label{eq:polarization-orbital-moment}
\end{equation}
The bound follows from $2|AB\sin\phi|\leq A^2+B^2$. Circular motion with
equal component amplitudes and a quarter-period phase difference saturates
it; linear motion has zero orbital moment. The isotropic restoring force
has zero torque about the reference point, so this moment is conserved in
the example. Figure~\ref{fig:complex-polarization-orbits} compares these
preparations at equal energy. This displacement-orbit moment differs from
a virion's intrinsic rotation about its own centre. Intrinsic orientation and boundary torques require their own retained
coordinates and force laws.

\begin{figure}[H]
\centering\begingroup
\definecolor{vlfGreen}{HTML}{176B64}
\definecolor{vlfDarkRed}{HTML}{762B38}
\definecolor{vlfRed}{HTML}{762B38}
\definecolor{vlfBlue}{HTML}{315D80}
\definecolor{vlfCopper}{HTML}{762B38}
\setlength{\unitlength}{1pt}
\scalebox{0.80}{%
\input{vlt-figures/figure-146.tex}%
}\endgroup
\caption{Three conservative displacement orbits at equal energy $E_{\rm harm}=m\omega^2u_\star^2$, with isotropic stiffness $m\omega^2\Id_2$. In (a,b), $A=\sqrt{8/5}\,u_\star$ and $B=\sqrt{2/5}\,u_\star$: in-phase components give a line, while $U_y=iB$ gives the counterclockwise ellipse. Equal amplitudes $A=B=u_\star$ with the same quarter-cycle offset give the circle in (c). Filled points and fixed-orientation glyphs mark $t=0$; arrows advance time. Shared physical axis scales give normalized orbital moments $0,4/5,1$. Reversing relative phase reverses the moment. Shading marks orbit area, independently of material loss or intrinsic virion rotation. The boxed result illustrated here is~\eqref{eq:polarization-orbital-moment}.}
\label{fig:complex-polarization-orbits}
\end{figure}

\paragraph{From phase to interparticle deformation.}
The pair network makes the physical content of phase explicit. Let
$\Delta\vct U_{ij}=\vct U_i-\vct U_j$ and use the reference direction
$\vct n_{ij}^{0}$ from \eqref{eq:edge-stiffness}.
\begin{definition}[Phase-resolved pair extension]
\label{def:s5-pair-extension-phasor}
The \emph{longitudinal pair-extension phasor} is
$D_{ij}=(\vct n_{ij}^{0})^{\mathsf T}\Delta\vct U_{ij}$.
The linear change of centre separation is
$\delta r_{ij}(t)=\operatorname{Re}[D_{ij}e^{-i\omega t}]$.
The transverse relative displacement is the complementary projection
of $\Delta\vct U_{ij}$.
\end{definition}
Here $\delta r_{ij}$ denotes the first-order separation change; the
geometric second-order effect of transverse motion is already retained
in the prestress part of the pair Hessian.

\begin{proposition}[Phase-dependent pair loading]
\label{prop:s5-phase-pair-loading}
For a real symmetric central pair curvature $\mathsf C_{ij}$, the
cycle-averaged quadratic pair contribution is
\begin{equation}
 \begin{aligned}
 \overline{V_{ij}^{(2)}}
 &=\frac14\Delta\vct U_{ij}^{\dagger}
                        \mathsf C_{ij}\Delta\vct U_{ij}\\
 &=\frac14 V''_{ij}(R_{ij})|D_{ij}|^2
   +\frac{V'_{ij}(R_{ij})}{4R_{ij}}
       \left(|\Delta\vct U_{ij}|^2-|D_{ij}|^2\right).
 \end{aligned}
 \label{eq:s5-phase-pair-loading}
\end{equation}
For collinear amplitudes
$\vct U_i=A_i e^{i\phi_i}\vct n_{ij}^{0}$ and
$\vct U_j=A_j e^{i\phi_j}\vct n_{ij}^{0}$, with $A_i,A_j\geq0$,
\begin{equation}
 |D_{ij}|^2=A_i^2+A_j^2
               -2A_iA_j\cos(\phi_i-\phi_j),\qquad
 \overline{\delta r_{ij}^{\,2}}=\frac12|D_{ij}|^2.
 \label{eq:s5-phase-relative-extension}
\end{equation}
\end{proposition}
\begin{proof}
Over a complete period, the cosine and sine squared averages are
$1/2$ and their product average is zero. Applying this to
$\tfrac12\Delta\vct u_{ij}^{\mathsf T}
\mathsf C_{ij}\Delta\vct u_{ij}$ gives the first identity.
Substitution of \eqref{eq:edge-stiffness} gives the longitudinal and
transverse terms. Expanding the modulus of
$A_i e^{i\phi_i}-A_j e^{i\phi_j}$ proves the final formula.
\end{proof}
Prestress can give a signed pair contribution; stability uses the
assembled Hessian with supports and medium terms. Lennard--Jones and
Coulomb curvatures enter this same loading formula through their
coefficients. The phasors determine the deformation. A common
translation has $\Delta\vct U_{ij}=0$ on each pair while potentially
loading supports or branches sensing absolute displacement.

\paragraph{Relative phase also loads oriented patches.}
For the frozen site interaction of Proposition~\ref{prop:s3patch-site-second-variation},
the first-order relative site phasor is
\begin{equation}
 \widehat{\vct a}_e=\vct U_i-\vct U_j+
   \vct\Theta_i\times\vct b_{ia}
       -\vct\Theta_j\times\vct b_{jb} .
 \label{eq:s5patch-site-relative-phasor}
\end{equation}
Cycle averaging its complete tangent second variation gives
\begin{equation}
 \begin{aligned}
 \overline{\mathcal F_e^{(2)}}
  &=\tfrac14\widehat{\vct a}_e^\dagger
                       \mathsf C_e\widehat{\vct a}_e\\
  &\quad+\frac{v'_e}{4}\vct n_e\cdot\operatorname{Re}
   [\vct\Theta_i\times(\vct\Theta_i^*\times\vct b_{ia})
    -\vct\Theta_j\times(\vct\Theta_j^*\times\vct b_{jb})] .
 \end{aligned}
 \label{eq:s5patch-phase-patch-loading}
\end{equation}
The identity follows from the harmonic average of the Cartesian
quadratic and the angular path-curvature term in
\eqref{eq:s3patch-site-hessian-geometric}. It retains site prestress.
Even with equal centre phasors, different angular phasors can load the
same patch channel. A common rigid motion instead rotates centres and
body orientations together, as in~\eqref{eq:s3patch-global-rigid-tangents}.
The assembled energy includes every owned interaction, support and
surface/material contribution. The central pair formula above is its
translation-only specialization.

For example, take two equal axial masses with equal support curvature
$k_0>0$ and a relaxed pair curvature $k_c>0$, as in
Figure~\ref{fig:s4-inverse-common-relative}. Compare harmonic patterns
$U_1=A$ and $U_2=Ae^{i\phi}$ at the same drive frequency. Their mean
kinetic energy is $m\omega^2A^2/2$ for every $\phi$, while their pair
storage is $k_cA^2(1-\cos\phi)/2$. Equal phase leaves that pair
unextended; opposite phase maximizes its extension at fixed $A$.
The corresponding free common and relative directions have frequencies
$\sqrt{k_0/m}$ and $\sqrt{(k_0+2k_c)/m}$.
An arbitrary intermediate phase at one prescribed frequency is a forced
pattern whose force follows from the full response pencil. Thus two
equal-amplitude particle records can probe different parts of the
interaction network, and amplitude alone cannot reconstruct its loading.

In active microrheology, calibrated force and tracked motion probe the
response; passive microrheology uses equilibrium fluctuation relations
under its stated assumptions~\cite{SquiresMason2010Review}.
Centre-average and separation phasors provide complementary channels
in this array. A material coordinate earns predictive value when its
coupling explains additional force or phase relations across those
channels and preparations. Section~4's inverse and closure tests then
assess which physical variables the phase-resolved record distinguishes
and requires.

Figure~\ref{fig:s5-phase-pair-loading} reconstructs the phase-dependent loading as common translation, pair compression and pair extension.

\begin{figure}[H]
\centering\begingroup
\definecolor{vlfGreen}{HTML}{176B64}
\definecolor{vlfDarkRed}{HTML}{762B38}
\definecolor{vlfRed}{HTML}{762B38}
\definecolor{vlfBlue}{HTML}{315D80}
\definecolor{vlfCopper}{HTML}{762B38}
\setlength{\unitlength}{1pt}
\scalebox{0.84}{%
\input{vlt-figures/figure-147.tex}%
}\endgroup
\caption{Phase-resolved deformation for $U_1=A$, $U_2=Ae^{i\phi}$ and $\theta=\omega t$. The common reference spacing is $a$; each ruler locates sites at $x/a=0,1$. One $11\,\mathrm{pt}/A$ displacement gain gives common translation at $\phi=0$, compression by $2A$ at $(\phi,\theta)=(\pi,0)$ and extension by $2A$ at $(\pi,\pi)$. Coils retain their turns and transverse amplitude. The cycle-average curve is $\overline V_{12}=k_cA^2(1-\cos\phi)/2$ for relaxed curvature $k_c>0$; both opposed-phase snapshots lie at its right endpoint. Declared equal supports are omitted. Intermediate one-frequency patterns use the force given by the full response pencil.}
\label{fig:s5-phase-pair-loading}
\end{figure}

A pulse or packet contains multiple frequencies.

\begin{definition}[Positive-frequency displacement signal]
\label{def:s5-positive-frequency-displacement-signal}
For a real time history with both the history and its Fourier transform
integrable, and with a static component treated separately, the
\emph{positive-frequency displacement signal} uses the convention
in~\eqref{eq:viscoelastic-fourier-convention}:
\begin{equation}
 \vct z_{\mathrm{disp}}(\vct r,t)
       =\frac1\pi\int_0^\infty
          \widehat{\vct u}(\vct r,\omega)e^{-i\omega t}\,\dd\omega,
 \qquad \vct u=\operatorname{Re}\vct z_{\mathrm{disp}}.
 \label{eq:positive-frequency-displacement}
\end{equation}
\end{definition}

Conjugate positive- and negative-frequency parts reconstruct a real
history. Doubling the positive part preserves that motion, with any
static component included once separately. Finite harmonic sums use
the same construction. The result is length-valued; energy requires
the mechanical weights.

The normalization distinguishes a phasor from a Fourier amplitude.
For an exactly persistent harmonic, with Fourier frequency denoted by
$\nu$ to separate it from the fixed carrier $\omega$, the distributional
transform is
\begin{equation}
 \widehat{\vct u}(\nu)=\pi\left[
    \vct U\,\delta(\nu-\omega)
        +\vct U^*\,\delta(\nu+\omega)\right].
 \label{eq:s5-harmonic-phasor-fourier-pair}
\end{equation}
The delta distribution has time units when its argument is an angular
frequency. Accordingly $\vct U$ has length units, while
$\widehat{\vct u}$ has length times time units. The positive-frequency
construction recovers $\vct Ue^{-i\omega t}$ with the same displacement
normalization. A finite pulse has a frequency distribution in place of
these delta lines; it requires its full amplitudes and phases to recover
the time history~\cite{Wunsch2005,FeynmanLectures1963}.

For real $L^2$ time histories, the same construction uses the temporal
Hilbert transform,
\begin{equation}
 (\mathcal H_t\vct u)(t)=\frac1\pi\operatorname{PV}
       \int_{-\infty}^{\infty}\frac{\vct u(s)}{t-s}\,\dd s,
 \qquad \vct z_{\mathrm{disp}}=\vct u-i\mathcal H_t\vct u.
 \label{eq:displacement-hilbert-transform}
\end{equation}
The integral is understood in the principal-value/$L^2$ sense. In our
Fourier convention its multiplier is $i\operatorname{sgn}\omega$, so
$\widehat{\vct z}_{\rm disp}=2\Theta(\omega)\widehat{\vct u}$ away from
zero. The harmonic illustration $\mathcal H_t\cos\omega t=\sin\omega t$
accordingly gives $z_{\rm disp}=e^{-i\omega t}$; this persistent harmonic
is a generalized Fourier example rather than an $L^2(\R)$ history.
The sign follows the stated convention~\cite{Wunsch2005}.

\begin{proposition}[Norm of the temporal quadrature representation]
\label{prop:s5-positive-frequency-displacement-norm}
For a real $L^2(\R)$ displacement history and the signal
$\vct z_{\rm disp}=\vct u-i\mathcal H_t\vct u$,
\begin{equation}
 \int_{-\infty}^{\infty}|\vct z_{\mathrm{disp}}(t)|^2\,\dd t
       =2\int_{-\infty}^{\infty}|\vct u(t)|^2\,\dd t.
 \label{eq:analytic-displacement-parseval}
\end{equation}
\end{proposition}
\begin{proof}
The Hilbert-transform Fourier multiplier has unit modulus away from
zero, so Plancherel gives $\|\mathcal H_t\vct u\|_2=\|\vct u\|_2$.
Both fields are real. Their squared norms therefore add as the real
and imaginary quadratures of $\vct z_{\rm disp}$, giving the factor two.
\end{proof}

The norm measures time-integrated squared displacement. The Hilbert
transform constructs a temporal quadrature; the Hilbert-space inner
product compares vectors. A finite record additionally specifies its
window and extension beyond the observed interval. This two-sided
signal reconstruction serves a different purpose from the causal force
law and instantaneous state.

A phase-sensitive measurement can instead estimate the quadratures against
a known reference. For a purely harmonic real readout
$y(t)=y_{\rm R}\cos\omega t+y_{\rm I}\sin\omega t$ observed over complete
cycles from the chosen phase origin,
\begin{equation}
 \begin{split}
 T_{\rm obs}&=2\pi N_{\rm cyc}/\omega,
                  \qquad N_{\rm cyc}\in\{1,2,\ldots\},\\
 y_{\rm R}&=\frac2{T_{\rm obs}}\int_0^{T_{\rm obs}}
                           y(t)\cos\omega t\,\dd t,\qquad
 y_{\rm I}=\frac2{T_{\rm obs}}\int_0^{T_{\rm obs}}
                           y(t)\sin\omega t\,\dd t.
 \end{split}
 \label{eq:phase-sensitive-quadrature-readout}
\end{equation}
These equalities follow from the sine/cosine orthogonality over complete
periods. A real calibrated linear readout
$y=\vct\ell^{\mathsf T}\vct u$ yields the same projection of both
displacement quadratures. The instrument and the biological coordinate
being measured determine this map~\cite{Leake2016}.
Noise, incomplete cycles, nearby frequencies,
and transients alter the estimates; Section~\ref{sec:phase-window-coherence}
addresses that observation-window problem.

Complex notation also represents the \emph{linear} real augmented model
of Section~\ref{sec:viscoelastic-wellposedness}: its complexification
$X_{\rm C}=X_{\rm R}+iX_{\rm I}$ evolves as two independent real linear
solutions, with the physical one selected by the appropriate real data.
Real data and force preserve reality. This statement does not extend by
substitution to the nonlinear damping in
Section~\ref{sec:self-excitation-passive-boundary}; nonlinear products
would couple the formal real and imaginary parts.

A Fourier series describes maintained periodic motion; a time-dependent
spectrum describes a finite-resource burst. Slow amplitude and phase
can be used when justified, while growth, saturation and harmonics
require more than one fixed phasor. Prediction still uses the retained
mechanical, memory, converter and resource state. The energy encoding
in Section~\ref{sec:conservative-complex-generator} preserves that
state without assuming complex-linear active dynamics.

\subsection{Displacement Hilbert spaces and the boundary operator}
\label{sec:complex-field-hilbert-space}

A spatial pattern lists displacement at every retained node or material
point. Complex linear combinations organize superposition within the
linear reference. The physical preparation must also stay within its
small-excursion basin; nonlinear evolution retains its own law.

An inner product defines the size and overlap of displacement patterns.

\begin{definition}[Finite displacement Hilbert space]
\label{def:s5-finite-displacement-space}
The complex displacement space of $n$ unconstrained whole-virion
translations is $\C^{3n}$ with the positive-mass inner product
\begin{equation}
 \langle\vct U,\vct V\rangle_M
     =\vct U^\dagger\mathsf M\vct V
     =\sum_{i=1}^n m_i\vct U_i^*\cdot\vct V_i.
 \label{eq:mass-displacement-inner-product}
\end{equation}
\end{definition}

The sum is written for unconstrained whole-virion translations; the
restricted positive mass matrix supplies the corresponding inner product
after linear constraints, and can be nondiagonal. Rotational or other
generalized coordinates require their specified inertia metric and units.
The dagger means conjugate transpose.
Conjugation makes $\langle\vct U,\vct U\rangle_M$ real and positive for
a nonzero pattern; $\langle\vct U,\vct V\rangle_M
=\langle\vct V,\vct U\rangle_M^*$. The product is conjugate linear in
its first argument and linear in its second. Its norm
$\|\vct U\|_M=\sqrt{\langle\vct U,\vct U\rangle_M}$ measures size in
this inertial metric. The corresponding displacement bra is
$\bra U_M=\vct U^\dagger\mathsf M$. Bra--ket notation records this
mass-metric overlap compactly; its physical meaning remains a comparison
of the declared displacement patterns.

Mass weighting has a mechanical purpose. The same physical velocity
costs more kinetic energy at a heavier node, so unweighted geometric
overlap need not be the appropriate comparison. For length-valued
patterns, $\|\vct U\|_M^2$ has units $\mathrm{kg\,m^2}$, not joules.
A harmonic phasor has mean kinetic energy
$\langle T\rangle=\omega^2\|\vct U\|_M^2/4$. Restoring and medium
storage must also be included before calling a quantity total energy.
Multiplying by $\mathsf M^{1/2}$ converts this inner product into the
Euclidean one. Figure~\ref{fig:complex-mass-metric-projection} shows how two
patterns that appear oblique in physical-coordinate axes become
perpendicular after that change of coordinates.

\begin{figure}[H]
\centering\begingroup
\definecolor{vlfGreen}{HTML}{176B64}
\definecolor{vlfDarkRed}{HTML}{762B38}
\definecolor{vlfRed}{HTML}{762B38}
\definecolor{vlfBlue}{HTML}{315D80}
\definecolor{vlfCopper}{HTML}{762B38}
\setlength{\unitlength}{1pt}
\scalebox{0.80}{%
\input{vlt-figures/figure-148.tex}%
}\endgroup
\caption{Mass-orthogonal projection in a two-coordinate slice, with $\vct q=\vct U/u_\star$ and $\widehat{\mathsf M}=\mathsf M/m_\star=\operatorname{diag}(1,4)$. Projecting $\vct q=(1,0.6)^{\mathsf T}$ onto $(1,1)^{\mathsf T}$ gives $\vct q_\parallel=(0.68,0.68)^{\mathsf T}$ and residual $(0.32,-0.08)^{\mathsf T}$. The shaded mass-norm ellipse becomes a circle under $\vct y=\widehat{\mathsf M}^{1/2}\vct q$; the same orthogonal decomposition becomes a Euclidean right angle. Squared norms carry the factor $m_\star u_\star^2$. Lower equal-size virions identify the unequal masses and displacement components; upper arrows are pattern vectors.}
\label{fig:complex-mass-metric-projection}
\end{figure}

The finite column space is complete: every Cauchy sequence converges
within that space. The continuum analogue requires this property explicitly, so
that converging approximations to a field do not leave the chosen
mathematical setting~\cite{Teschl2009}. The selected norm fixes which
approximations converge. Mechanical stability and measurement coverage
are established through the restoring law and the calibrated readouts.

\paragraph{The inertia metric for translation and libration.}
Use the dimensionless admissible body tangent $\vct X$ and inertia
$\mathsf G_X\succ0$ of~\eqref{eq:s2patch-generalized-inertia}.
For its complex phasors the corresponding geometry is
\begin{equation}
 \langle\widehat{\vct X},\widehat{\vct V}\rangle_{G_X}
    =\widehat{\vct X}^{\dagger}\mathsf G_X\widehat{\vct V},
 \qquad
 \overline T=\tfrac14\omega^2
             \widehat{\vct X}^{\dagger}\mathsf G_X\widehat{\vct X} .
 \label{eq:s5patch-body-inertia-metric}
\end{equation}
The squared norm has units $\mathrm{J\,s^2}$, and the kinetic average
has energy units. The adjoint, projection and compatible-copy arguments
below apply to this metric with full body restrictions. Every occurrence
of a shared VLP copies the same six mechanical components in a common
tangent frame. Its surface/material state is shared too, with its own
storage and first-order kinetics. Chemical populations receive their
specified free-energy weights; the inertial metric acts on the retained
mechanical coordinates.

For a locally pure libration about unit axis $\vct k$, let
$J_k=\vct k^{\mathsf T}\mathsf J_i\vct k>0$ and let a length-valued
marker channel project along unit direction $\vct e$. Its sensitivity
and effective readout inertia, for angular displacement $\theta$ and
phasor $\Theta$, are
\begin{equation}
 h_\theta=\vct e\cdot(\vct k\times\vct b),\qquad
 y=h_\theta\theta,\quad Y=h_\theta\Theta,\qquad
 m_{\rm eff}=\frac{J_k}{|h_\theta|^2}\quad(h_\theta\ne0).
 \label{eq:s5patch-libration-marker-inertia}
\end{equation}
Indeed $J_k\dot\theta^2/2=m_{\rm eff}\dot y^2/2$.
The result has mass units because the lever arm has length units.
A centre-only channel has zero direct sensitivity to this pure angular
component. Two registered markers with different lever arms can test
its angular assignment while sampling the same mechanics.

A continuum comparison retaining independent body rotation likewise
uses the sum of displacement and spin-inertia pairings,
$\int[\rho\vct U^*\cdot\vct V+
\vct\Theta^\dagger\mathsf j_\rho\vct\Phi]\,\dd V$,
where $\mathsf j_\rho\succ0$ is rotational inertia per volume, with
units $\mathrm{kg\,m^{-1}}$. Its elastic form uses the strain and
rotation-gradient variables of~\eqref{eq:s3patch-micropolar-strains}
and the traction/couple-traction work of
\eqref{eq:s4patch-micropolar-boundary-power}. The displacement-only
continuum form below is the specified reduced reference.

\begin{proposition}[Adjoint in the displacement mass metric]
\label{prop:s5-displacement-mass-adjoint}
For $\mathsf M\succ0$ and any finite linear operator $\mathsf T$, the
adjoint in the mass inner product is
\begin{equation}
 \mathsf T^{\dagger_M}=\mathsf M^{-1}\mathsf T^\dagger\mathsf M.
 \label{eq:mass-weighted-adjoint}
\end{equation}
In particular, $\mathsf M^{-1}\mathsf K$ is self-adjoint in this
metric when $\mathsf K$ is real symmetric.
\end{proposition}
\begin{proof}
The defining identity for the adjoint equates
$\vct U^\dagger\mathsf M\mathsf T\vct V$ with
$(\mathsf T^{\dagger_M}\vct U)^\dagger\mathsf M\vct V$.
Solving it gives the displayed expression. Substituting
$\mathsf T=\mathsf M^{-1}\mathsf K$ and $\mathsf K^\dagger=\mathsf K$
proves the final assertion.
\end{proof}

Thus $(\mathsf M^{-1}\mathsf K)^{\dagger_M}
=\mathsf M^{-1}\mathsf K$, while its mass-weighted representation is
the real symmetric matrix $\mathsf D$ of Section~3. Self-adjointness
depends on the metric as well as the operator. Unequal masses can make
the ordinary coordinate matrix look nonsymmetric without violating
conservative mechanics.

Let $\vct\phi_s=\mathsf M^{-1/2}\vct e_s$ be a conservative displacement
basis, with $\vct e_s$ the real Euclidean-orthonormal mass-weighted
patterns already defined. Then
$\langle\vct\phi_s,\vct\phi_t\rangle_M=\delta_{st}$.
The normalized shapes have units $\mathrm{kg}^{-1/2}$; their complex
phasor coefficients $\mathcal Q_s$ have units
$\mathrm{kg}^{1/2}\mathrm m$.

\begin{proposition}[Mass-orthogonal mode restriction]
\label{prop:s5-mass-orthogonal-mode-restriction}
For a mass-orthonormal conservative displacement basis and a selected
mode set $\mathcal S$, the orthogonal projection of $\vct U$ obeys
\begin{equation}
 \begin{split}
 \mathcal Q_s&=\langle\vct\phi_s,\vct U\rangle_M,\qquad
 \vct U_{\mathcal S}=\sum_{s\in\mathcal S}\mathcal Q_s\vct\phi_s,\\
 \langle\vct\phi_s,\vct U-\vct U_{\mathcal S}\rangle_M&=0
                      \quad(s\in\mathcal S),\\
 \|\vct U\|_M^2&=\|\vct U_{\mathcal S}\|_M^2
                       +\|\vct U-\vct U_{\mathcal S}\|_M^2.
 \end{split}
 \label{eq:complex-displacement-mode-projection}
\end{equation}
It uniquely minimizes the mass-norm error over that selected span.
\end{proposition}
\begin{proof}
Mass orthonormality makes the residual orthogonal to each selected
column. Expanding its inner product with the projected component gives
the Pythagorean identity. Any other vector in the selected span adds
the squared norm of its difference from the projection to the residual
error, proving the minimum and uniqueness.
\end{proof}

The residual measures approximation error in the mass norm. Orthogonal
patterns may involve the same virions: their weighted overlap vanishes.
A degenerate frequency specifies a subspace within which an orthonormal
basis can be chosen.

An experimental readout adds a second, separately calibrated projection.
For a real dimensionless finite weight $\vct\ell$,
$Y=\vct\ell^{\mathsf T}\vct U$ is the displacement readout phasor. Its
mass-metric representative and a useful bound are
\begin{equation}
 \begin{split}
 \vct g&=\mathsf M^{-1}\vct\ell,\qquad
 Y=\langle\vct g,\vct U\rangle_M
   =\sum_s(\vct\ell^{\mathsf T}\vct\phi_s)\mathcal Q_s,\\
 |Y|^2&\leq(\vct\ell^{\mathsf T}\mathsf M^{-1}\vct\ell)
                                      \|\vct U\|_M^2.
 \end{split}
 \label{eq:complex-displacement-readout-bound}
\end{equation}
The last inequality is Cauchy--Schwarz: overlap cannot exceed the product
of the two norms. A mode with $\vct\ell^{\mathsf T}\vct\phi_s=0$ is invisible
to this readout even when excited. A force has its own modal overlap;
ability to excite a pattern and ability to detect it are distinct. For
instance, equal opposite displacements can cancel in a two-particle mean
while remaining visible in their separation. Identifying centre,
surface, or internal deformation readouts is therefore part of applying
this geometry to a virion system~\cite{Leake2016}.

\begin{proposition}[A readout-sensitive mode-truncation bound]
\label{prop:s5-readout-truncation}
On the positive finite conservative sector, let $\vct U_{\mathcal S}$ be
the mass-orthogonal projection in
\eqref{eq:complex-displacement-mode-projection}, and put
$Y_{\mathcal S}=\vct\ell^{\mathsf T}\vct U_{\mathcal S}$. Then
\begin{equation}
 |Y-Y_{\mathcal S}|^2\leq
 \left(\sum_{\substack{s\notin\mathcal S\\\omega_s>0}}
       \frac{|\vct\ell^{\mathsf T}\vct\phi_s|^2}{\omega_s^2}\right)
 \left(\sum_{\substack{s\notin\mathcal S\\\omega_s>0}}
                         \omega_s^2|\mathcal Q_s|^2\right).
 \label{eq:s5-readout-truncation-bound}
\end{equation}
The second factor is the omitted elastic quadratic form
$(\vct U-\vct U_{\mathcal S})^\dagger
\mathsf K(\vct U-\vct U_{\mathcal S})$.
\end{proposition}
\begin{proof}
Expand the readout error over the omitted positive modes, write each term
as $(\vct\ell^{\mathsf T}\vct\phi_s/\omega_s)
(\omega_s\mathcal Q_s)$, and apply Cauchy--Schwarz. The generalized
eigenvalue equation and mass orthonormality diagonalize the elastic
quadratic form, giving the stated second factor.
\end{proof}

The bound weights omitted patterns by detector overlap and softness.
A soft mode can carry little elastic energy yet contribute appreciable
displacement. A detector-dark mode contributes zero direct error to
that readout. The phasor elastic quadratic is four times its
cycle-averaged potential energy; medium storage is accounted for
separately. Retain zero modes whenever the preparation or detector uses
them. A readout-error tolerance can justify a compact prediction,
while Section~4's closure test asks whether it predicts every
admissible preparation.

\paragraph{Compatible complex patterns on assembled cells.}
\label{sec:s5-cell-compatible-patterns}

Shared cells must assign the same complex displacement to each shared
virion. Their copies therefore satisfy compatibility at a common frequency,
in common laboratory axes and with a common clock
(Definition~\ref{def:s5-displacement-phasor}).

\begin{definition}[Compatible cell-copy space]
\label{def:s5-cell-copy-space}
On laboratory translational coordinates before constraints, for a
finite assembly with $n=n_{\rm asm}$ physical particle labels,
stack the restrictions of Definition~\ref{def:s3-cell-restriction} into
$\mathsf J=(\mathsf R_1^{\mathsf T},\ldots,
\mathsf R_{N_{\rm cell}}^{\mathsf T})^{\mathsf T}$.
The map $\mathsf J$ copies a physical displacement into its cell lists.
Choose strictly positive mass fractions on each occurrence, summing
to one for each physical site, and put
\begin{equation}
 \mathsf M_{\rm cp}=\operatorname{diag}_c\mathsf M_c,
 \qquad \mathsf J^\dagger\mathsf M_{\rm cp}\mathsf J=\mathsf M,
 \qquad \mathcal H_{\rm comp}=\operatorname{ran}\mathsf J.
 \label{eq:s5-cell-copy-metric}
\end{equation}
The direct sum of the complex cell columns has the mass inner product
defined by $\mathsf M_{\rm cp}>0$. Its compatible subspace
$\mathcal H_{\rm comp}$ consists of copies that assign the same
displacement to every occurrence of a shared physical label.
\end{definition}

The positive fractions are a convenient geometry for the copy space;
they do not divide a particle physically. The more general zero-weight
ownership convention in Section~3 instead requires its appropriate
zero-norm quotient if used as an inner-product space. For two four-virion
tetrahedra sharing one face, there are eight local occurrences but five
physical particles. The copy column has 24 complex entries and its
compatible subspace has dimension 15. With equal physical masses and a
common displacement $\vct b$, the mass norm is $5m|\vct b|^2$;
counting a full mass in all eight occurrences would change that norm.
Each shared site likewise retains one physical orientation and one surface
population history; a fixed assigned charge is its frozen specialization.
Linear kinematic constraints can subsequently be imposed on the
physical patterns and their copies together; they do not create
additional shared-site coordinates.

\begin{proposition}[Compatible reconstruction and its discrepancy]
\label{prop:s5-cell-compatible-reconstruction}
For the data above, define
\begin{equation}
 \mathsf A_{\rm cp}
       =\mathsf M^{-1}\mathsf J^\dagger\mathsf M_{\rm cp},\qquad
 \mathsf\Pi_{\rm comp}=\mathsf J\mathsf A_{\rm cp}.
 \label{eq:s5-cell-copy-reconstruction}
\end{equation}
Then $\mathsf A_{\rm cp}\mathsf J=\Id$ and
$\mathsf\Pi_{\rm comp}$ is the mass-orthogonal projector onto
$\mathcal H_{\rm comp}$. For every complex copy column $\vct V$,
\begin{equation}
 \|\vct V\|_{M_{\rm cp}}^2
   =\|\mathsf A_{\rm cp}\vct V\|_M^2
      +\|(\Id-\mathsf\Pi_{\rm comp})\vct V\|_{M_{\rm cp}}^2.
 \label{eq:s5-cell-copy-discrepancy}
\end{equation}
Thus reconstruction is a contraction between these two mass spaces.
\end{proposition}
\begin{proof}
Equation~\eqref{eq:s5-cell-copy-metric} gives the left inverse.
It implies $\mathsf\Pi_{\rm comp}^2=\mathsf\Pi_{\rm comp}$ and
$\mathsf\Pi_{\rm comp}^\dagger\mathsf M_{\rm cp}
=\mathsf M_{\rm cp}\mathsf\Pi_{\rm comp}$. The projection and residual
are orthogonal, while the norm of $\mathsf J\mathsf A_{\rm cp}\vct V$
equals that of $\mathsf A_{\rm cp}\vct V$. Pythagoras proves the result.
\end{proof}

At a shared site, $\mathsf A_{\rm cp}$ averages copies with the declared
mass fractions. The residual measures amplitude and phase disagreement
between overlapping cell calculations or records. Physical assembly
requires both compatibility and force balance. The projection
discrepancy measures copy-norm error; experimental uncertainty can
instead define a calibrated statistical fitting metric.

Figure~\ref{fig:s5-compatible-cell-copy-geometry} connects the compatible copy space to a five-virion assembly and a concrete mass-weighted projection.

\begin{figure}[H]
\centering\begingroup
\definecolor{vltTeal}{HTML}{176B64}
\definecolor{vltBlue}{HTML}{315D80}
\definecolor{vltGold}{HTML}{A87520}
\definecolor{vltPlum}{HTML}{725877}
\definecolor{vltRed}{HTML}{762B38}
\definecolor{vltYellow}{HTML}{D5AD3C}
\definecolor{vltNeutral}{HTML}{777D83}
\setlength{\unitlength}{1pt}
\scalebox{0.90}{%
\input{vlt-figures/figure-149.tex}%
}\endgroup
\caption{Two tetrahedra share sites 1--3: five physical virions have eight cell-list occurrences. Each shared occurrence owns $m/2$; copies refer to one physical mass, orientation and continued surface state per shared site. At shared site 1 in one common direction, proposed copies $(1+i)\ell_*$ and $(1-i)\ell_*$ reconstruct $U_1=\ell_*$; all other components vanish in this norm example. Their weighted squared norm splits into reconstructed norm and orthogonal discrepancy, as the lower ledger shows. Solid and dashed vectors identify copies in grayscale. The discrepancy tests compatibility; force balance separately determines the assembled motion.}
\label{fig:s5-compatible-cell-copy-geometry}
\end{figure}

\begin{proposition}[Assembly of retained cell patterns]
\label{prop:s5-cell-mode-compatible}
Let $\mathsf Z=\operatorname{diag}_c\mathsf V_c$ contain linearly
independent retained displacement patterns within each cell. Choose a
full-column-rank matrix $\mathsf B$ whose columns span
$\ker[(\Id-\mathsf\Pi_{\rm comp})\mathsf Z]$.
The assembled compatible basis
$\mathsf Y=\mathsf A_{\rm cp}\mathsf Z\mathsf B$ then satisfies
\begin{equation}
 \mathsf J\mathsf Y=\mathsf Z\mathsf B,\qquad
 \mathsf Y^\dagger\mathsf M\mathsf Y
   =\mathsf B^\dagger\mathsf Z^\dagger\mathsf M_{\rm cp}
                    \mathsf Z\mathsf B>0.
 \label{eq:s5-cell-mode-compatible-metric}
\end{equation}
Here positivity refers to a nonempty retained coefficient space.
\end{proposition}
\begin{proof}
The defining kernel gives
$\mathsf\Pi_{\rm comp}\mathsf Z\mathsf B=\mathsf Z\mathsf B$.
Substitute $\mathsf J\mathsf A_{\rm cp}=\mathsf\Pi_{\rm comp}$ and
then use~\eqref{eq:s5-cell-copy-metric}. Independent columns of
$\mathsf Z\mathsf B$ give the positive Gram matrix.
\end{proof}

Cell coefficients combine into one trajectory per shared virion. Their
assembled Gram matrix supplies inertia and can differ from the identity
even for locally normalized bases. Interior and interface patterns can
be retained together as in Section~3. Compatibility tests shared motion;
the force residual and response denominator
in~\eqref{eq:s5-quadrature-residual-bound} additionally test the
resulting displacement prediction.

\paragraph{Membership selects coordinates within the same particle pool.}
On the full unconstrained laboratory space $\C^{3N_{\rm tot}}$ with
the site-diagonal material mass matrix $\mathsf M_{\rm tot}$, let
$\mathsf S_{\mathcal C}$ select the three coordinates of each member
of the current cluster. With
$\mathsf M_{\mathcal C}=\mathsf S_{\mathcal C}\mathsf M_{\rm tot}
\mathsf S_{\mathcal C}^{\mathsf T}$, its zero extension and mask are
\begin{equation}
 \mathsf E_{\mathcal C}=\mathsf S_{\mathcal C}^{\mathsf T},\qquad
 \mathsf P_{\mathcal C}=\mathsf E_{\mathcal C}\mathsf S_{\mathcal C},
 \qquad
 \|\vct U\|_{M_{\rm tot}}^2
   =\|\mathsf S_{\mathcal C}\vct U\|_{M_{\mathcal C}}^2
      +\|\mathsf S_{\mathcal C^c}\vct U\|_{M_{\mathcal C^c}}^2.
 \label{eq:s5-pool-membership-mask}
\end{equation}
Partitioning the site sum proves the identity. Small and large clusters
are selections from the same fixed pool. With constraints coupling
selected and unselected sites, the ambient mask still defines an
observation; a work-compatible constrained restriction supplies the
admissible mechanics and inertia.

Masking a pattern does not eliminate forces exerted by the complement.
An effective cluster response retains their reduced impedance, forcing
and preparation according to Section~\ref{sec:s4-cell-retarded-ports}.
The membership mask, displacement pattern and excitation state are
separate inputs to that prediction.

\paragraph{Assemble the mechanics before forming its energy amplitude.}
Compatible displacements, velocities and branch states carry the
physical assembly. Their energy contributions use the ownership rules
of~\eqref{eq:s4-cell-memory-congruence}. The global energy amplitude
in Section~\ref{sec:conservative-complex-generator} is then constructed
from the assembled $\mathsf M$ and $\mathsf K$.
Its square-root operator generally couples different cells.
This particle amplitude represents conservative kinetic and elastic
energy; material branch strains retain their separate storage.
For an equal-mass pair with support curvature $k_0>0$ and link curvature
$k>0$, the common and relative square-root eigenvalues are
$\sqrt{k_0/m}$ and $\sqrt{(k_0+2k)/m}$. Their difference produces an
off-diagonal square-root entry in the particle basis. Independent
uncoupled cell amplitudes would omit that interaction.

A signed cell prestress need not admit a positive square root, even
when the assembled reference is stable. The mass-copy geometry remains
valid, while the energy amplitude uses the positive global sector.
Component-mode and residual tests in
Subsection~\ref{sec:s3-compatible-cell-assembly} control omitted
interior motion. Compatible mechanical states and interface responses
thus provide the building blocks; their complex representations retain
the cross-cell exchange of motion and work.

Figure~\ref{fig:s5-global-square-root-assembly} shows explicitly how an interaction enters the global energy amplitude.

\begin{figure}[H]
\centering\begingroup
\definecolor{vltTeal}{HTML}{176B64}
\definecolor{vltBlue}{HTML}{315D80}
\definecolor{vltGold}{HTML}{A87520}
\definecolor{vltPlum}{HTML}{725877}
\definecolor{vltRed}{HTML}{762B38}
\definecolor{vltYellow}{HTML}{D5AD3C}
\definecolor{vltNeutral}{HTML}{777D83}
\setlength{\unitlength}{1pt}
\scalebox{0.90}{%
\input{vlt-figures/figure-150.tex}%
}\endgroup
\caption{The supported equal-mass pair has $\mathsf K=\left(\begin{smallmatrix}2&-1\\-1&2\end{smallmatrix}\right)$ and $\mathsf M=\Id$ in declared scaled units. The snapshot $\vct u=(1,0)$, $\vct v=0$ uses $18\,\mathrm{pt}$ per displacement unit. A reference ruler registers sites 1 and 2 at $x/a=0,1$; matched coils show the extended left support, compressed pair and unchanged right support. The global square root has eigenvalues $1,\sqrt3$; its printed entries are rounded. The second energy-amplitude component is nonzero although site 2 is undisplaced. Support and pair storage each contribute $1/2$ to the unit total. The component squares encode the global energy metric; spatial allocation and material-branch storage have their own definitions.}
\label{fig:s5-global-square-root-assembly}
\end{figure}

\begin{definition}[Continuum displacement mass space]
\label{def:s5-continuum-displacement-mass-space}
For the occupied material region and its declared positive density,
the continuum displacement space is
\begin{equation}
 \begin{aligned}
 H_\rho&=L^2(\mathcal V_{\mathrm{lat}},\rho\,\dd^3r;\C^3),\\
 \langle\vct U,\vct V\rangle_\rho
          &=\int_{\mathcal V_{\mathrm{lat}}}\rho\vct U^*\cdot\vct V\,\dd^3r.
 \end{aligned}
 \label{eq:continuum-complex-mass-space}
\end{equation}
\end{definition}

The space $L^2$ contains fields with finite squared norm, identifying
fields equal except on a zero-volume set. Convergence controls
integrated squared amplitude. The retained density is bounded above
and away from zero as in Section~4. The occupied material region and
the field's physical resolution are specified before this space is
constructed.

A finite-resolution volume readout with real kernel $\vct w$ is
\begin{equation}
 \begin{split}
 Y&=\int_{\mathcal V_{\rm lat}}\vct w\cdot\vct U\,\dd^3r
   =\langle\vct w/\rho,\vct U\rangle_\rho,\\
 |Y|^2&\leq\left(\int\frac{|\vct w|^2}{\rho}\,\dd^3r\right)
                                      \|\vct U\|_\rho^2.
 \end{split}
 \label{eq:complex-continuum-detector-bound}
\end{equation}
All integrals are over the occupied region. If $\vct w$ has units of
inverse volume, $Y$ has length units; boundedness requires
$\vct w/\sqrt\rho\in L^2$. This condition says that small errors in
field norm produce bounded errors in the readout. An ideal point
evaluation is generally not a bounded functional on $L^2$, and not on
general $H^1$ fields in three dimensions. It requires stronger regularity
or a resolved spatial averaging model. A discrete particle-site readout,
in contrast, is bounded in the finite column space. Instrument resolution
is consequently a mathematical input, as well as a biophysical one.

Figure~\ref{fig:complex-finite-resolution-readout} illustrates how finite
spatial averaging controls readout sensitivity to field-norm errors.

\begin{figure}[H]
\centering\begingroup
\definecolor{vlfGreen}{HTML}{176B64}
\definecolor{vlfDarkRed}{HTML}{762B38}
\definecolor{vlfRed}{HTML}{762B38}
\definecolor{vlfBlue}{HTML}{315D80}
\definecolor{vlfCopper}{HTML}{762B38}
\setlength{\unitlength}{1pt}
\scalebox{0.80}{%
\input{vlt-figures/figure-151.tex}%
}\endgroup
\caption{A bounded volume readout on the dimensionless unit ball in three dimensions. The profiles $f_a(R)=Ca^{-3/2}(1-R/a)_+$, $C=\sqrt{15/(2\pi)}$, $0<a\leq1$, all have unit $L^2$ norm. Narrowing them increases $f_a(0)/C=a^{-3/2}$ while decreasing the full-ball average $\overline f_a/C=a^{3/2}/4$. Panel (b) uses a logarithmic ordinate. Cauchy--Schwarz bounds that average by $1/\sqrt{4\pi/3}$ for every unit-norm field. This auxiliary family tests readout continuity in~\eqref{eq:complex-continuum-detector-bound}; the actual material domain and detector kernel are specified independently.}
\label{fig:complex-finite-resolution-readout}
\end{figure}

Elastic deformation also requires control of spatial gradients.

\begin{definition}[Complex elastic energy form]
\label{def:s5-complex-elastic-energy-form}
Let $V_{\rm C}$ be the complexification of
\eqref{eq:slip-energy-spaces}: its fields have finite squared first
derivatives and satisfy the normal constraint on the engaged boundary
$\Gamma_{\rm w}$. The conservative elastic pairing is
\begin{equation}
 a_{\mathrm C}(\vct U,\vct V)=\int_{\mathcal V_{\mathrm{lat}}}
       [\partial_\mu U_p^*\mathcal A_{p\mu q\nu}\partial_\nu V_q
                      +\vct U^*\cdot\mathsf B_\lambda\vct V]\,\dd^3r.
 \label{eq:complex-elastic-form}
\end{equation}
\end{definition}

Tangential motion remains allowed on $\Gamma_{\rm w}$; the remaining
material surface $\Gamma_{\rm f}$ is free in this idealization.

This \emph{sesquilinear} form is conjugate linear in the test field and
linear in the second field. Its diagonal value measures twice the
conservative elastic energy of a real displacement. Under Section~4's
density, symmetry, stable-form and boundary assumptions it is densely
defined, closed, and nonnegative. Closedness means completeness in a
form norm that controls both displacement and elastic deformation;
one may use the squared form norm
$\|\vct U\|_\rho^2+\omega_\star^{-2}a_{\rm C}(\vct U,\vct U)$
with any fixed reference frequency $\omega_\star>0$ to keep the units alike.

\begin{proposition}[Elastic operator associated with the energy form]
\label{prop:s5-elastic-form-operator}
For the densely defined closed nonnegative form $a_{\rm C}$ on
$H_\rho$, there is a unique nonnegative self-adjoint elastic operator
with domain and action
\begin{equation}
 \begin{split}
 D(\mathcal L_{\mathrm{el}})
    =\{\vct V\in V_{\mathrm C}:{}&\text{ there is }\vct g\in H_\rho
       \text{ with }a_{\mathrm C}(\vct U,\vct V)
           =\langle\vct U,\vct g\rangle_\rho
                              \text{ for every }\vct U\in V_{\mathrm C}\},\\
 \mathcal L_{\mathrm{el}}\vct V&=\vct g.
 \end{split}
 \label{eq:elastic-operator-domain}
\end{equation}
\end{proposition}
\begin{proof}
The representation theorem for densely defined closed nonnegative
forms gives the operator characterized by the displayed test identity
~\cite{Teschl2009}. Density makes $\vct g$ unique, while the form
representation gives self-adjointness and nonnegativity.
\end{proof}

The acceleration $-\mathcal L_{\rm el}\vct V$ expresses restoring
force against every admissible test pattern. Finite storage uses the
form domain; evaluating acceleration in $H_\rho$ uses the stronger
operator domain. For smooth fields it is
$-\rho^{-1}[-\partial_\mu(\mathcal A_{p\mu q\nu}\partial_\nu V_q)
+B_{\lambda,pq}V_q]$. The conservative boundary law combines zero
tangential traction on $\Gamma_{\rm w}$, zero full traction on
$\Gamma_{\rm f}$ and the prescribed normal-displacement constraint.
The full memory response instead uses Section~4's total traction.

\begin{proposition}[Finite-domain spectrum and square-root energy domain]
\label{prop:s5-elastic-square-root-domain}
Under the form assumptions above, suppose $V_{\rm C}$ is compactly
embedded in $H_\rho$. The elastic operator has compact resolvent and
a complete mass-orthonormal eigenbasis. With
$\mathcal W=\mathcal L_{\rm el}^{1/2}$,
\begin{equation}
 D(\mathcal W)=V_{\mathrm C},\qquad
 a_{\mathrm C}(\vct U,\vct V)
       =\langle\mathcal W\vct U,\mathcal W\vct V\rangle_\rho.
 \label{eq:elastic-square-root-form}
\end{equation}
\end{proposition}
\begin{proof}
The shifted resolvent maps $H_\rho$ boundedly into the form domain;
the compact embedding makes its composition into $H_\rho$ compact.
The compact self-adjoint spectral theorem gives the eigenbasis.
The closed-form representation identifies its form domain with the
square-root domain and its pairing with the displayed inner product.
\end{proof}

The spectral square root multiplies a mode by $\omega_s$, while the
elastic operator multiplies it by $\omega_s^2$. Completeness means
that finite mode sums approximate every displacement in the norm.
The calibrated detector determines which of those modes are visible.

The zero-frequency sector is handled separately. On the positive sector
$H_{\rho,+}$, the stated coercivity supplies a gap $\omega_{\min}>0$
and a bounded $\mathcal W^{-1}$. These properties belong to the assembled
material and boundary problem; a periodic bulk symbol has different
assumptions and cannot replace its finite-domain spectrum.

Figure~\ref{fig:complex-hilbert-domain-ladder} separates displacement, elastic-energy and force regularity in an exactly solvable interval example.

\begin{figure}[H]
\centering\begingroup
\definecolor{vlfGreen}{HTML}{176B64}
\definecolor{vlfDarkRed}{HTML}{762B38}
\definecolor{vlfRed}{HTML}{762B38}
\definecolor{vlfBlue}{HTML}{315D80}
\definecolor{vlfCopper}{HTML}{762B38}
\setlength{\unitlength}{1pt}
\scalebox{0.80}{%
\input{vlt-figures/figure-152.tex}%
}\endgroup
\caption{An explicit interval model separates displacement, strain and force regularity. With $L=-\mathrm d^2/\mathrm dX^2$ on $(0,1)$ and Dirichlet endpoints, $g_n=\sqrt2\sin(n\pi X)$ gives $\|g_n\|=1$, $\|g_n^\prime\|^2=(n\pi)^2$ and $\|Lg_n\|^2=(n\pi)^4$. The nested domains are $H^2\cap H_0^1\subset H_0^1\subset L^2$, corresponding to $D(L)$, $D(L^{1/2})$ and displacement space. Arrows denote continuous inclusion. The cylinder retains its material tensor, inertia and mixed boundary problem; its operator regularity follows those assumptions.}
\label{fig:complex-hilbert-domain-ladder}
\end{figure}

\begin{definition}[Positive-sector conservative energy state]
\label{def:s5-positive-conservative-energy-state}
With the positive-sector gap specified above, the energy state retains
both displacement and velocity in
$\mathcal E_+=D(\mathcal W|_{H_{\rho,+}})\oplus H_{\rho,+}$.
Its conservative wave generator and domain are
\begin{equation}
 \mathcal A_0(\vct u,\vct v)
       =(\vct v,-\mathcal L_{\mathrm{el}}\vct u),\qquad
 D(\mathcal A_0)=D(\mathcal L_{\mathrm{el}}|_{H_{\rho,+}})
                      \oplus D(\mathcal W|_{H_{\rho,+}}).
 \label{eq:continuum-conservative-wave-domain}
\end{equation}
\end{definition}

Its skew-adjointness uses the energy inner product defined explicitly in
the next Hamiltonian discussion. Displacement space and energy state space
are different because a second-order equation needs velocity as well.
If a bulk or softening limit closes the gap, excluding exact zero modes
does not alone bound $\mathcal W^{-1}$. Reconstructing an $L^2$
displacement then requires the real part of an energy amplitude in
$D(\mathcal W^{-1})$, or a separately declared homogeneous energy
completion. Free zero-mode motion is
$\vct u_0(t)=\vct u_0(0)+t\vct v_0(0)$ outside that inverse.

\paragraph{Restricting patterns while preserving mechanical work.}
A trial pattern specifies how the retained virion centres or material
points participate in one displacement coordinate. Common motion and
changing separation are distinct patterns on the same particles.
Adding patterns refines the motion description at fixed population.
The following restriction preserves their mechanical work.

\begin{definition}[Work-compatible pattern restriction]
\label{def:s5-work-compatible-restriction}
In the finite translational model let the columns of a real
full-column-rank matrix $\mathsf S_r$ be dimensionless admissible
patterns, and approximate $\vct u=\mathsf S_r\vct q$. A
\emph{work-compatible restriction} uses the same map for trial
displacements and virtual-work testing. The reduced force is
$\vct f_r=\mathsf S_r^{\mathsf T}\vct f$, so
$\vct f^{\mathsf T}\dot{\vct u}
=\vct f_r^{\mathsf T}\dot{\vct q}$. The coordinates $\vct q$ retain
length units. In the continuum its conforming analogue chooses trial
patterns in $V_{\rm C}$ and restricts both arguments of the mass and
elastic forms to their span.
\end{definition}

\begin{proposition}[Storage, inertia and passive loss under restriction]
\label{prop:s5-restriction-preserves-work}
At a real frequency $\omega>0$, with the same reference, supports
and material coefficients, write
$\mathsf P(\omega)=\mathsf K^*(\omega)-\omega^2\mathsf M$ and
$\mathsf L_{\rm loss}(\omega)=\mathsf K''(\omega)\geq0$. The finite
restriction has
\begin{equation}
 \begin{aligned}
 \mathsf M_r&=\mathsf S_r^{\mathsf T}\mathsf M\mathsf S_r>0,
 &\mathsf K_r&=\mathsf S_r^{\mathsf T}\mathsf K\mathsf S_r,\\
 \mathsf P_r(\omega)&=\mathsf S_r^{\mathsf T}
                       \mathsf P(\omega)\mathsf S_r,
 &\mathsf L_r(\omega)&=\mathsf S_r^{\mathsf T}
                  \mathsf L_{\rm loss}(\omega)\mathsf S_r\geq0.
 \end{aligned}
 \label{eq:s5-work-compatible-reduced-matrices}
\end{equation}
The restricted kinetic and elastic energies equal the full model's
energies evaluated on $\mathsf S_r\vct q$, and the cycle work of a
restricted phasor is
$\pi\vct q^\dagger\mathsf L_r\vct q$.
\end{proposition}
\begin{proof}
Substitute $\vct u=\mathsf S_r\vct q$ into the quadratic energies and
the virtual-work equation. Positive definiteness of $\mathsf M$ and
injectivity of $\mathsf S_r$ give $\mathsf M_r>0$. For every complex
$\vct q$, the loss quadratic form is
$\vct q^\dagger\mathsf L_r\vct q
=(\mathsf S_r\vct q)^\dagger\mathsf L_{\rm loss}
(\mathsf S_r\vct q)\geq0$. The same substitution in the full cycle
work gives the last statement. Restricting the continuum forms gives
the identical argument for conforming trial fields.
\end{proof}

This is a conforming Galerkin approximation. Charges, interaction
curvatures and medium terms enter the assembled operators before
restriction. Projecting displacement, inertia and force together
preserves the work exchanged by a retained collective motion.
Independent coefficient averages need their own compatibility check.

For the equal supported pair in
Figure~\ref{fig:s4-inverse-common-relative}, choose
$\mathsf S_1=(1,1)^{\mathsf T}/\sqrt2$. Then
$M_1=m$, $K_1=k_0$, and $f_1=(f_1^{\rm site}+f_2^{\rm site})/\sqrt2$:
the pair curvature $k_c$ contributes no work along a common displacement.
The relative pattern supplies the missing change of separation and has
curvature $k_0+2k_c$. In this exchange-symmetric example the common
subspace is invariant, so its reduced conservative motion is exact for
preparations in that subspace. With unequal masses or heterogeneous
couplings, the same common-pattern restriction still preserves work,
but generally approximates motion that can transfer into omitted
patterns. Exact state closure additionally requires the invariant-kernel
condition of Proposition~\ref{prop:s4-state-exact-closure}; a retained
medium branch must be projected with its own force-bearing variables.

Figure~\ref{fig:s5-work-compatible-projection} connects the common and relative particle patterns to the dual force map that preserves work.

\begin{figure}[H]
\centering\begingroup
\definecolor{vltTeal}{HTML}{176B64}
\definecolor{vltBlue}{HTML}{315D80}
\definecolor{vltGold}{HTML}{A87520}
\definecolor{vltPlum}{HTML}{725877}
\definecolor{vltRed}{HTML}{762B38}
\definecolor{vltYellow}{HTML}{D5AD3C}
\definecolor{vltNeutral}{HTML}{777D83}
\setlength{\unitlength}{1pt}
\scalebox{0.90}{%
\input{vlt-figures/figure-153.tex}%
}\endgroup
\caption{Work-compatible restriction on an auxiliary $3\times3\times2$ array of eighteen equal-mass ideal virions. Centres lie at $\vct R/a\in\{-1,0,1\}^2\times\{-1/2,1/2\}$. Thirty-three nearest-axis central links are unstressed with longitudinal curvature $k_z>0$; uniform onsite curvature $k_0\Id_{54}>0$ supplies the declared supports. The two columns of $\mathsf S$ are $\vct1_{18}\otimes\vct e_z/\sqrt{18}$ and $(\vct1_9,-\vct1_9)\otimes\vct e_z/\sqrt{18}$. Common motion preserves every link length; opposed layers compress all nine vertical links by $2A$, with horizontal spans unchanged. Both views share a camera, radius $0.13a$ and position gain $52\,\mathrm{pt}/a$. Dashed bodies locate references; offset black displacement arrows use $130\,\mathrm{pt}/a$. Coils share radius, turns and leads and attach to the actual three-dimensional facets. These patterns are invariant for the stated symmetric reference. The reduced inertia and curvatures follow from the same map as virtual work; the array illustrates the restriction rather than a fitted viral interaction law.}
\label{fig:s5-work-compatible-projection}
\end{figure}

Conservative frequencies have a further variational check. The Rayleigh
quotient is
$\mathcal R_{\rm el}(\vct U)
=a_{\rm C}(\vct U,\vct U)/\|\vct U\|_\rho^2$ in the continuum,
and $\vct U^\dagger\mathsf K\vct U/
(\vct U^\dagger\mathsf M\vct U)$ in the finite lattice. Its
minimum--maximum characterization supplies the following bound.

\begin{proposition}[Conforming positive-sector Ritz bounds]
\label{prop:s5-conforming-ritz-bounds}
Order the full positive conservative eigenvalues as
$\lambda_j=\omega_j^2$. For finite-dimensional conforming nested
positive-sector trial spaces $V_r\subseteq V_{r+1}$ of the same
conservative form and boundary problem,
\begin{equation}
 \lambda_j\leq\lambda_j^{(r+1)}\leq\lambda_j^{(r)},
 \qquad 1\leq j\leq\dim V_r.
 \label{eq:s5-conforming-ritz-bounds}
\end{equation}
\end{proposition}
\begin{proof}
The minimum--maximum principle minimizes the maximal Rayleigh quotient
over all $j$-dimensional form-domain subspaces. Restricting to $V_r$
reduces that choice; enlarging to $V_{r+1}$ adds choices and can only
lower the restricted minimum~\cite{Teschl2009,HornJohnson2013}.
\end{proof}

These upper bounds refer to
the same conservative reference and boundary problem. They provide a
way to test the resolution of a collective description before fitting
its physical inputs. Damped poles and frequency-dependent hidden-variable
eliminations require their own response analysis, as in Section~4.

\subsection{Coupled real quadratures of the frequency response}
\label{sec:real-quadrature-system}

The displacement space supplies patterns and projections. The mechanical
law now determines which pattern responds to a specified force. For the reciprocal passive Maxwell model of Section~4, storage and loss
enter one frequency-dependent force--displacement law.

\begin{definition}[Mechanical response pencil]
\label{def:s5-mechanical-response-pencil}
At a real frequency $\omega>0$, the mechanical response pencil is the
operator family in $\mathsf P(\omega)\vct U=\vct F$, with
\begin{equation}
 \mathsf P(\omega)=\mathsf A(\omega)-i\mathsf L_{\mathrm{loss}}(\omega),
 \qquad
 \mathsf A=\mathsf K'-\omega^2\mathsf M,
 \quad\mathsf L_{\mathrm{loss}}=\mathsf K''.
 \label{eq:real-imaginary-mechanical-pencil}
\end{equation}
\end{definition}
Its real part combines restoring storage and
inertial opposition; its imaginary part represents the specified
constitutive loss with the paper's $e^{-i\omega t}$ sign convention.
Both matrices are real symmetric here, and
$\mathsf L_{\rm loss}\geq0$ for the passive realization. The phasors
$\vct U$ and $\vct F$ have displacement and force units respectively;
a continuous Fourier transform would carry an additional time unit.

\begin{proposition}[Real quadrature force balance]
\label{prop:s5-quadrature-force-balance}
For real $\mathsf A$ and $\mathsf L_{\rm loss}$, the complex equation
is equivalent to the doubled real system below. Multiplication gives
$\mathsf P\vct U=(\mathsf A\vct U_{\rm R}
+\mathsf L_{\rm loss}\vct U_{\rm I})
+i(\mathsf A\vct U_{\rm I}-\mathsf L_{\rm loss}\vct U_{\rm R})$.
Equating real and imaginary parts to those of the force yields
\begin{equation}
 \boxed{\begin{pmatrix}
     \mathsf A&\mathsf L_{\mathrm{loss}}\\
     -\mathsf L_{\mathrm{loss}}&\mathsf A
 \end{pmatrix}
 \begin{pmatrix}\vct U_{\mathrm R}\\\vct U_{\mathrm I}\end{pmatrix}
       =\begin{pmatrix}\vct F_{\mathrm R}\\\vct F_{\mathrm I}\end{pmatrix}.}
 \label{eq:physical-quadrature-block-system}
\end{equation}
\end{proposition}
\begin{proof}
Insert $\vct U=\vct U_{\rm R}+i\vct U_{\rm I}$ and equate its two real
coefficients to those of the force. Conversely, combining the two block
rows reconstructs the original complex equation.
\end{proof}
The doubled column gives the cosine and sine coefficients of one
physical history. Loss couples their timing even for a cosine-only
input. The continuum counterpart uses the frequency-domain operator
with total-traction boundary conditions.

Figure~\ref{fig:complex-realification-particle-map} connects one translated particle history to the row and column roles of the real quadrature system.

\begin{figure}[H]
\centering\begingroup
\definecolor{vlfGreen}{HTML}{176B64}
\definecolor{vlfDarkRed}{HTML}{762B38}
\definecolor{vlfRed}{HTML}{762B38}
\definecolor{vlfBlue}{HTML}{315D80}
\definecolor{vlfCopper}{HTML}{762B38}
\setlength{\unitlength}{1pt}
\scalebox{0.84}{%
\input{vlt-figures/figure-154.tex}%
}\endgroup
\caption{Four times of one real trajectory, $u/u_\star=\cos\theta+\tfrac12\sin\theta$, $\theta=\omega t$. The same rigid virion translates with gain $36\,\mathrm{pt}/u_\star$ and one fixed reference; the common lower ruler calibrates every row. The block system maps displacement quadratures to their force quadratures. Signs follow $e^{-i\omega t}$, with $\mathsf A=\mathsf K^\prime-\omega^2\mathsf M$ and passive $\mathsf L_{\rm loss}$. Shading selects loss blocks. Storage and loss can have different eigendirections; panel (a) supplies a coefficient example rather than a fitted modal response. The boxed result illustrated here is~\eqref{eq:physical-quadrature-block-system}.}
\label{fig:complex-realification-particle-map}
\end{figure}

Reciprocity gives $\mathsf P^{\mathsf T}=\mathsf P$ in this model;
with loss, it does not give $\mathsf P^\dagger=\mathsf P$.
Consequently this complex symmetric pencil is a response operator, not
a self-adjoint observable. Storage and loss also need not have common
normal modes. For example,
$[\mathsf P,\mathsf P^\dagger]=2i[\mathsf A,\mathsf L_{\rm loss}]$.
Here $[\mathsf A,\mathsf L]=\mathsf A\mathsf L-\mathsf L\mathsf A$
measures dependence on the order of multiplication. A nonzero commutator makes the pencil nonnormal and rules out its
unitary diagonalization into independent scalar responses at that
frequency. It does not create an active energy source or invalidate the
passive work test below.

\begin{definition}[Realification]
\label{def:s5-realification}
The \emph{realification} of a complex linear map is its real matrix on
the concatenated real and imaginary coordinate columns. It represents
the same map while recording its two quadratures separately.
\end{definition}

\begin{proposition}[Realification and invertibility]
\label{prop:s5-realification-invertibility}
For any complex square matrix,
\begin{equation}
 \mathfrak R(\mathsf P)=
   \begin{pmatrix}\operatorname{Re}\mathsf P&-\operatorname{Im}\mathsf P\\
                   \operatorname{Im}\mathsf P&\operatorname{Re}\mathsf P
   \end{pmatrix},\qquad
 \det\mathfrak R(\mathsf P)=|\det\mathsf P|^2.
 \label{eq:realification-determinant}
\end{equation}
\end{proposition}
\begin{proof}
Complexify the real block space and change variables to $\vct U_{\rm R}+i\vct U_{\rm I}$ and
$\vct U_{\rm R}-i\vct U_{\rm I}$. The operator becomes
$\diag(\mathsf P,\mathsf P^*)$, so its determinant is the product of
the two conjugate determinants.
\end{proof}

Realification preserves invertibility for arbitrary storage and loss
matrices. At real frequency the absolute-value determinant tests the
same singular response. Temporal poles instead use analytic
continuation of the causal pencil or its augmented time generator.

\begin{proposition}[Real quadratures preserve response conditioning]
\label{prop:s5-realification-singular-values}
After a fixed choice of physical coordinate and force scales, every
singular value of a complex finite pencil $\mathsf P$ occurs twice in
its real representation $\mathfrak R(\mathsf P)$. In particular,
\begin{equation}
 \begin{aligned}
 \|\mathfrak R(\mathsf P)\|_2&=\|\mathsf P\|_2,\qquad
 \sigma_{\min}(\mathfrak R(\mathsf P))=\sigma_{\min}(\mathsf P),\\
 \kappa_2(\mathfrak R(\mathsf P))&=\kappa_2(\mathsf P)
 \end{aligned}
 \label{eq:s5-realification-singular-values}
\end{equation}
when the last expression is defined for an invertible pencil.
\end{proposition}
\begin{proof}
The real representation preserves products and converts the conjugate
transpose into the real transpose. A complex unitary matrix therefore
becomes a real orthogonal matrix. Applying this construction to a
singular-value decomposition $\mathsf P=\mathsf U\mathsf\Sigma
\mathsf V^\dagger$ gives real orthogonal factors and the central block
$\diag(\mathsf\Sigma,\mathsf\Sigma)$. Its diagonal entries are the
singular values of the real representation, proving the statements
\cite{HornJohnson2013}.
\end{proof}

The real system has the same small response denominators as the complex
one. Declared scales for displacements, forces and detector channels
fix the meaning of numerical perturbations. For computed phasor
$\vct U_h$ and force residual $\vct r_F=\vct F-\mathsf P\vct U_h$,
\begin{equation}
 \|\vct U-\vct U_h\|_2
       \leq\frac{\|\vct r_F\|_2}{\sigma_{\min}(\mathsf P)}.
 \label{eq:s5-quadrature-residual-bound}
\end{equation}
Applying $\mathsf P^{-1}$ to the residual proves this estimate. A small
force-balance residual can therefore coexist with appreciable
displacement error near a compliant collective response. Calibrated
readouts multiply this bound by their operator norm. Such an error
estimate helps decide whether adding a spatial pattern or a medium
variable improves a prediction at the instrument's resolution.

\paragraph{Assembling and eliminating cells in real quadratures.}
The complex cell calculation can be performed directly or through its
real quadrature representation. Let the owned local causal pencils be
$\mathsf P_c(\omega)$ in a common frame, with
$\mathsf P=\sum_c\mathsf R_c^{\mathsf T}\mathsf P_c\mathsf R_c$.
Because the coordinate restrictions are real, the realification already
defined above gives
\begin{equation}
 \mathfrak R(\mathsf P)
  =\sum_c\widetilde{\mathsf R}_c^{\mathsf T}
          \mathfrak R(\mathsf P_c)\widetilde{\mathsf R}_c,
 \qquad
 \widetilde{\mathsf R}_c=\operatorname{diag}
                       (\mathsf R_c,\mathsf R_c).
 \label{eq:s5-cell-quadrature-assembly}
\end{equation}
This follows from preservation of addition and multiplication under
realification. Each physical site has a cosine and sine coefficient;
the two copies use identical physical ownership.

Exact interior elimination commutes with the same construction.
Order both quadratures of the retained ports first, followed by both
quadratures of the interiors. If $\mathsf S_\Gamma$ is the complex
Schur complement in~\eqref{eq:s4-cell-retarded-port-maps}, the Schur
complement of this reordered real system is
$\mathfrak R(\mathsf S_\Gamma)$. Multiplicativity and
$\mathfrak R(\mathsf P_{II}^{-1})=
\mathfrak R(\mathsf P_{II})^{-1}$ prove the assertion wherever the
interior inverse exists. The permutation matters: placing all real
coordinates first is a different storage order from grouping both
quadratures by physical port.

For the Hermitian loss form
$\mathsf L_{\rm loss}=-\operatorname{Im}_{\rm H}\mathsf P\succeq0$
and $\vct u_q=(\vct U_{\rm R},\vct U_{\rm I})$, its cycle work at
$\omega>0$ is
\begin{equation}
 W_{\rm cycle}
  =\pi\vct U^\dagger\mathsf L_{\rm loss}\vct U
  =\pi\vct u_q^{\mathsf T}
             \mathfrak R(\mathsf L_{\rm loss})\vct u_q.
 \label{eq:s5-cell-quadrature-work}
\end{equation}
The two representations therefore retain the same loss. Eliminated
particle and medium coordinates also retain the effective forcing and
noise congruence of~\eqref{eq:s4-cell-port-thermal-spectrum}. Shared
fluid fluctuations remain correlated in the doubled solver; resolving
two quadratures does not introduce two independent physical baths.
These identities allow cell assembly, interface solves and their error
checks to use the most convenient algebra while predicting the same
registered particle motion and supplied work.

Figure~\ref{fig:s5-cell-realification-port-square} shows the two exact routes from a complex cell model to its real port equations.

\begin{figure}[H]
\centering\begingroup
\definecolor{vltTeal}{HTML}{176B64}
\definecolor{vltBlue}{HTML}{315D80}
\definecolor{vltGold}{HTML}{A87520}
\definecolor{vltPlum}{HTML}{725877}
\definecolor{vltRed}{HTML}{762B38}
\definecolor{vltYellow}{HTML}{D5AD3C}
\definecolor{vltNeutral}{HTML}{777D83}
\setlength{\unitlength}{1pt}
\scalebox{0.90}{%
\input{vlt-figures/figure-155.tex}%
}\endgroup
\caption{Two routes give identical port motion. Three complex virion phasors become six real quadratures; interior elimination leaves two complex ports or four real port quadratures. The real route orders both port quadratures before both interior quadratures and transforms the effective force accordingly. The auxiliary law uses $M/m_*=I$, $K/k_*=0.7I+L_{\rm path}$, $C/(m_*\omega_*)=\operatorname{diag}(0.12,0.6,0.12)$, $k_*=m_*\omega_*^2$, $\omega=0.9\omega_*$ and $\vct F=k_*\ell_*(1,0,0)^{\mathsf T}$. Declared onsite support supplies $0.7I$. The listed port responses and equal positive cycle work test both routes. Upper springs show reference topology.}
\label{fig:s5-cell-realification-port-square}
\end{figure}

\paragraph{A surface reporter survives the mechanical reduction.}
\begin{definition}[Joint mechanical and surface response]
\label{def:s5patch-joint-response}
In a fixed linear window, use the dimensionless mechanical tangent
$\vct X$, independent relaxing surface/material coordinates $\vct\chi$
and the real storage/resistance blocks of
Definition~\ref{def:s4patch-joint-resistance}. Let
$\widehat{\vct X},\vct\Xi$ be their phasors, with generalized
sources $\vct F_X,\vct a_\chi$ in energy units. Write
\begin{equation}
 \begin{aligned}
 \mathcal P(\omega)
  \binom{\widehat{\vct X}}{\vct\Xi}
   &=\binom{\vct F_X}{\vct a_\chi},\\
 \mathcal P(\omega)&=
  \begin{pmatrix}
   \mathsf H-\omega^2\mathsf G_X-i\omega\mathsf C&\mathsf B_\omega\\
   \mathsf B_\omega^{\mathsf T}&\mathsf A_\omega
  \end{pmatrix},\qquad
 \mathsf B_\omega=\mathsf B-i\omega\mathsf R,
 \quad\mathsf A_\omega=\mathsf A-i\omega\mathsf D .
 \end{aligned}
 \label{eq:s5patch-joint-response-pencil}
\end{equation}
Here $\mathsf D\succ0$ acts on the permitted relaxing directions.
Conserved inventories retain the restricted curvature and bias of
\eqref{eq:s4patch-constrained-relaxed-curvature}. A prescribed chemical
potential perturbation can supply $\vct a_\chi$ through its specified
stoichiometric coupling; a closed finite reservoir is retained in the
state instead.
\end{definition}

\begin{proposition}[Reduced mechanics with a mixed physical readout]
\label{prop:s5patch-mixed-readout}
Suppose $\mathsf A_\omega$ and the mechanical Schur operator
$\mathsf Z_X$ are invertible. For calibrated channels
$\vct Y=\mathsf L_X\widehat{\vct X}+\mathsf L_\chi\vct\Xi$, define
\begin{equation}
 \boxed{\begin{aligned}
 \mathsf Z_X&=\mathsf H-\omega^2\mathsf G_X-i\omega\mathsf C
        -\mathsf B_\omega\mathsf A_\omega^{-1}
                                    \mathsf B_\omega^{\mathsf T},\\
 \mathsf L_{\rm e}&=\mathsf L_X-
       \mathsf L_\chi\mathsf A_\omega^{-1}\mathsf B_\omega^{\mathsf T},\\
 \mathsf U_\chi&=(\mathsf L_\chi-
                \mathsf L_{\rm e}\mathsf Z_X^{-1}\mathsf B_\omega)
                                               \mathsf A_\omega^{-1},\\
 \vct Y&=\mathsf L_{\rm e}\mathsf Z_X^{-1}\vct F_X
                                      +\mathsf U_\chi\vct a_\chi .
 \end{aligned}}
 \label{eq:s5patch-mixed-response-readout}
\end{equation}
\end{proposition}
\begin{proof}
Solve the second row for
$\vct\Xi=\mathsf A_\omega^{-1}
(\vct a_\chi-\mathsf B_\omega^{\mathsf T}\widehat{\vct X})$.
Substitute first in the mechanical row, then in the output. Collecting
the coefficients of the two sources gives the stated maps.
\end{proof}

Figure~\ref{fig:cmd-complementary-ports} gives a graphical specialization of this boxed relation.

The transpose in this reciprocal pencil is ordinary transpose. Hermitian
conjugation enters power and covariance. A surface reporter has a direct
response to its source and a mechanically mediated response. Replacing
only the stiffness while keeping its old detector row discards part of
that prediction. A calibrated centre/marker channel and a surface-state
channel can therefore distinguish reductions with the same centre
response. Such a surface channel requires an independently justified
signal--population calibration.

For an asymptotically stable supported common-temperature equilibrium
sector of Section~4, the
same map acts on the correlated primitive random forces:
\begin{equation}
 \mathsf T_Y=[\mathsf L_{\rm e}\mathsf Z_X^{-1},\mathsf U_\chi],
 \qquad
 \mathsf S_Y=\mathsf T_Y(2k_{\rm B}T\mathsf R_{\rm d})
                  \mathsf T_Y^\dagger .
 \label{eq:s5patch-mixed-channel-noise}
\end{equation}
This is the fluctuation-preserving output counterpart of
\eqref{eq:s4patch-cross-resistance-noise}. The direct and mediated
surface-noise paths share a primitive source. Their cross terms are
part of the model spectrum. Independent detector noise is added only
with its separately calibrated statistics. Finite preparation supplies
the initial-state response, treated explicitly below.

\begin{proposition}[A loss-kernel test for a real-frequency response]
\label{prop:s5-real-frequency-loss-kernel}
For $\mathsf P=\mathsf A-i\mathsf L_{\rm loss}$ with real symmetric
$\mathsf A$ and $\mathsf L_{\rm loss}\geq0$,
\begin{equation}
 \ker\mathsf P=\ker\mathsf A\cap\ker\mathsf L_{\rm loss}.
 \label{eq:s5-real-frequency-loss-kernel}
\end{equation}
If $\mathsf L_{\rm loss}\geq\ell_{\min}\Id$ with
$\ell_{\min}>0$ in the stated coordinates, then
$\sigma_{\min}(\mathsf P)\geq\ell_{\min}$ and
$\|\mathsf P^{-1}\|_2\leq\ell_{\min}^{-1}$.
\end{proposition}
\begin{proof}
For $\mathsf P\vct U=0$, the imaginary part of
$\vct U^\dagger\mathsf P\vct U$ is
$-\vct U^\dagger\mathsf L_{\rm loss}\vct U=0$. Positivity implies
$\mathsf L_{\rm loss}\vct U=0$, and the original equation then gives
$\mathsf A\vct U=0$. The converse is immediate. Under strict loss,
Cauchy--Schwarz gives
$\|\vct U\|_2\|\mathsf P\vct U\|_2
\geq|\vct U^\dagger\mathsf P\vct U|
\geq\vct U^\dagger\mathsf L_{\rm loss}\vct U
\geq\ell_{\min}\|\vct U\|_2^2$, yielding the stated bounds.
\end{proof}

A real-frequency singularity requires conservative force balance in a
loss-free pattern. Semidefinite loss can still give a unique phasor
when its kernel contains no null vector of $\mathsf A$. This assembled
test allows distinct storage and loss eigendirections. The strict-loss
bound has compliance units because $\ell_{\min}$ is a stiffness.
Prepared medium strain contributes its separate transient alongside
the steady response.

\begin{proposition}[Mechanical cycle work]
\label{prop:s5-mechanical-cycle-work}
For a bounded steady harmonic response of the passive pencil, the
velocity phasor is $-i\omega\vct U$. Its mean supplied power and
work per period are
\begin{equation}
 \begin{split}
 \langle P_{\rm in}\rangle_{\rm cycle}
   &=\tfrac12\operatorname{Re}
                   [\vct F^\dagger(-i\omega\vct U)]
     =\frac\omega2\vct U^\dagger\mathsf L_{\rm loss}\vct U\\
   &=\frac\omega2(\vct U_{\rm R}^{\mathsf T}\mathsf L_{\rm loss}\vct U_{\rm R}
                  +\vct U_{\rm I}^{\mathsf T}\mathsf L_{\rm loss}\vct U_{\rm I})
       \geq0,\\
 W_{\rm cycle}&=\pi\vct U^\dagger\mathsf L_{\rm loss}\vct U.
 \end{split}
 \label{eq:complex-response-cycle-work}
\end{equation}
\end{proposition}
\begin{proof}
The product of two real harmonic histories has cycle average one half
the real part of the corresponding phasor product. Substitute
$\vct F=(\mathsf A-i\mathsf L_{\rm loss})\vct U$; the real symmetric
storage term contributes no net work. Multiplication by $2\pi/\omega$
gives the work per cycle.
\end{proof}
The work $W_{\rm cycle}$ has energy units. Conservative work cancels
over the closed cycle, leaving the force-conjugate loss term. Thus
phase indicates dissipation through its relation to applied force
and the loss form. A prepared branch can separately return work
during a transient, as in Section~4.

\begin{definition}[Conjugate and cross mechanical channels]
\label{def:s5-conjugate-cross-channels}
For a real dimensionless force pattern $\vct g$, the input
$\vct F=\vct g F_g$ is \emph{work conjugate} to the readout
$Y_g=\vct g^{\mathsf T}\vct U$: their real histories obey
$\vct f\cdot\dot{\vct u}=f_g\dot y_g$. A \emph{cross channel}
instead applies force through one pattern and observes displacement
through a different one. Its readout describes transmitted motion,
while the conjugate velocity at the input determines supplied work.
\end{definition}

\begin{proposition}[Positive conjugate compliance and signed cross phase]
\label{prop:s5-compliance-loss-identity}
At an invertible real-frequency pencil define
$\mathsf R=\mathsf P^{-1}$. Its Hermitian imaginary part satisfies
\begin{equation}
 \operatorname{Im}_{\rm H}\mathsf R
 :=\frac{\mathsf R-\mathsf R^\dagger}{2i}
   =\mathsf R^\dagger\mathsf L_{\rm loss}\mathsf R\geq0.
 \label{eq:s5-compliance-loss-identity}
\end{equation}
Consequently $h_{gg}=\vct g^{\mathsf T}\mathsf R\vct g$ has
$\operatorname{Im}h_{gg}\geq0$, and a conjugate sinusoidal drive gives
\begin{equation}
 \langle P_{\rm in}\rangle_{\rm cycle}
    =\frac{\omega}{2}|F_g|^2\operatorname{Im}h_{gg},
 \qquad
 W_{\rm cycle}=\pi|F_g|^2\operatorname{Im}h_{gg}.
 \label{eq:s5-conjugate-compliance-work}
\end{equation}
No analogous sign condition applies to the imaginary part of a general
cross entry $\vct w^{\mathsf T}\mathsf R\vct g$.
\end{proposition}
\begin{proof}
Use
$\mathsf R-\mathsf R^\dagger
=\mathsf R^\dagger(\mathsf P^\dagger-\mathsf P)\mathsf R$
and $\mathsf P^\dagger-\mathsf P=2i\mathsf L_{\rm loss}$.
Sandwiching this identity between $\vct g^{\mathsf T}$ and $\vct g$
proves positivity of the scalar imaginary part. Substituting
$Y_g=h_{gg}F_g$ into the conjugate force--velocity cycle average
gives the two work formulas. Positivity controls diagonal quadratic
forms, rather than the sign of each off-diagonal entry.
\end{proof}

\begin{example}[Conjugate and cross response of a passive particle pair]
\label{ex:s5-passive-pair-cross-response}
Let
two identical ideal virions have equal end-support curvature $k_\star$,
pair curvature $3k_\star/2$, mass $m_\star$, and independent drag
$c=0.2\sqrt{m_\star k_\star}$. With
$r=\omega\sqrt{m_\star/k_\star}$, put
\begin{equation}
 \begin{aligned}
 \frac{\mathsf K}{k_\star}
    &=\begin{pmatrix}5/2&-3/2\\-3/2&5/2\end{pmatrix},
 &a_+(r)&=\frac1{1-r^2-0.2ir},\\
 a_-(r)&=\frac1{4-r^2-0.2ir},
 &k_\star R_{11}&=\frac{a_++a_-}{2},\qquad
   k_\star R_{21}=\frac{a_+-a_-}{2}.
 \end{aligned}
 \label{eq:s5-passive-pair-cross-phase}
\end{equation}
Driving particle 1 and observing that same centre gives $R_{11}$;
observing particle 2 gives $R_{21}$. The latter combines the common
and relative motions with opposite signs. Its imaginary part is
positive near $r=1$ and negative near $r=2$, although both the drag
matrix and $\operatorname{Im}_{\rm H}\mathsf R$ remain positive.
The total supplied work follows $R_{11}$. This phase reversal records
how a relative pattern transmits motion between two particles.
\end{example}

A second tracked centre can distinguish collective coupling hidden in
a local trace; work pairs the applied force with its conjugate
velocity. Probe microrheology likewise specifies force and displacement
directions when inferring material properties~\cite{SquiresMason2010}.
An added coupling or material coordinate can be tested against a
cross-phase change, while the conjugate power balance determines
the work stored, supplied or dissipated.

Figure~\ref{fig:s5-conjugate-cross-response} places the signed cross response beside the force-conjugate response of the same passive pair.

\begin{figure}[H]
\centering\begingroup
\definecolor{vlfGreen}{HTML}{176B64}
\definecolor{vlfDarkRed}{HTML}{762B38}
\definecolor{vlfRed}{HTML}{762B38}
\definecolor{vlfBlue}{HTML}{315D80}
\definecolor{vlfCopper}{HTML}{762B38}
\setlength{\unitlength}{1pt}
\scalebox{0.80}{%
\input{vlt-figures/figure-156.tex}%
}\endgroup
\caption{Conjugate and transmitted response of a supported equal-mass pair: $m=m_\star$, supports $k_\star$, pair curvature $3k_\star/2$ and local drag $0.2\sqrt{m_\star k_\star}$. A force acts at reference site 1. For $r=\omega\sqrt{m_\star/k_\star}$, define $a_+=(1-r^2-0.2ir)^{-1}$ and $a_-=(4-r^2-0.2ir)^{-1}$. The curves give $k_\star\operatorname{Im}R_{11}=\operatorname{Im}(a_++a_-)/2$ and $k_\star\operatorname{Im}R_{21}=\operatorname{Im}(a_+-a_-)/2$. The force-site response fixes positive cycle work $\pi|F_1|^2\operatorname{Im}R_{11}$; opposite modal weights give the signed transmitted response. Reference coils denote curvature; drag is omitted from the sketch.}
\label{fig:s5-conjugate-cross-response}
\end{figure}

\begin{example}[Quadratures of one damped mechanical mode]
\label{ex:s5-single-damped-quadratures}
For an independently retained mass-weighted damped mode, take a real
force phasor $F_s=\vct e_s^{\mathsf T}\mathsf M^{-1/2}\vct F$ and
$a=\omega_s^2-\omega^2$, $\ell=2\gamma_s\omega$ with
$\gamma_s>0$. The scalar equation is
$(a-i\ell)Q_s(\omega)=F_s$. Multiplying numerator and denominator by
$a+i\ell$ gives
\begin{equation}
 Q_{s,\mathrm R}=\frac{aF_s}{a^2+\ell^2},\qquad
 Q_{s,\mathrm I}=\frac{\ell F_s}{a^2+\ell^2}.
 \label{eq:single-mode-response-quadratures}
\end{equation}
The phasor $Q_s(\omega)$ has mass-weighted displacement units, whereas
its modal force has units $\mathrm{kg}^{1/2}\mathrm m\,\mathrm s^{-2}$.
The susceptibility $\mathcal R_s=(a-i\ell)^{-1}$ has units of time
squared. Its phase lag is
$\theta_{{\rm lag},s}=\arg\mathcal R_s
=\operatorname{atan2}(\ell,a)$; the dimensional net gain $\delta_s$ of
Section~\ref{sec:self-excitation-passive-boundary} retains its earlier meaning. Below the conservative reference
frequency the cosine coefficient is positive and the lag lies between
$0$ and $\pi/2$; at $\omega=\omega_s$ the response is in the sine
quadrature. Above it the cosine coefficient is negative and the lag lies
between $\pi/2$ and $\pi$.
The amplitude maximum need not occur exactly at that reference frequency.
\end{example}

Figure~\ref{fig:complex-response-quadratures-work} plots these coefficients
and the associated positive force--displacement loop work. A phase-resolved
readout can thus test both response magnitude and the work-relevant phase
using one parameter set. Its biological interpretation still depends on
the measured coordinate and how the force is transmitted. A displacement
image or spectral peak alone need not determine $\gamma_s$, and the scalar
approximation requires an isolated or independently specified mode.

\begin{figure}[H]
\centering\begingroup
\definecolor{vlfGreen}{HTML}{176B64}
\definecolor{vlfDarkRed}{HTML}{762B38}
\definecolor{vlfRed}{HTML}{762B38}
\definecolor{vlfBlue}{HTML}{315D80}
\definecolor{vlfCopper}{HTML}{762B38}
\setlength{\unitlength}{1pt}
\scalebox{0.80}{%
\input{vlt-figures/figure-157.tex}%
}\endgroup
\caption{One forced mass-weighted mode with $\gamma_s/\omega_s=0.12$ and real $F_s>0$. For $r=\omega/\omega_s$ and $q=\omega_s^2Q_s/F_s$, the plotted coefficients are $q_{\rm R}=(1-r^2)/D$ and $q_{\rm I}=0.24r/D$, where $D=(1-r^2)^2+(0.24r)^2$. Distinct styles identify cosine and sine response. At $r=1$, displacement lags by $\pi/2$ and the normalized force--displacement loop is circular. Its clockwise-in-time parameterization in the plotted force-versus-displacement axes gives positive $\oint f_s\,\dd Q_s=\pi F_sQ_\star$. The loop is a work plot with conjugate axes.}
\label{fig:complex-response-quadratures-work}
\end{figure}

The nonlinear active extension has a different status. Growth about rest
or small perturbations about a maintained cycle can be analyzed by an
appropriate linearization. The periodic trajectory generally contains
harmonics, and linearization along that trajectory has periodic
coefficients even when the original equations are autonomous. Its force-generation
response and resource must be retained as in
Section~\ref{sec:self-excitation-feedback-closure}. A fitted phase alone
does not identify an active mechanism; effective gain requires the
specified resource and force-generation closure.

\subsubsection{Pressure-driven flow and complex viscosity in a cylindrical medium}
\label{sec:s5-cylindrical-fluid-context}

A pressure-driven capillary connects complex response to fluid work.
In a hypodermic needle, the lumen's hydraulic load contributes to
syringe-plunger force alongside device friction and other loads
~\cite{Verwulgen2018,Thueer2018}. This example relates pressure, flow,
material memory and particle motion through the same work account.

Pressure ports extend the architecture by exchanging work with the
fluid. Closing them and restoring the boundary preparation recovers
the sealed reference. The benchmark uses a straight rigid circular
lumen of radius $R_h$ and length $L_h$, a stationary no-slip wall and
fully developed axial flow. These dimensions refer to the selected
capillary segment. First solve its homogeneous carrier medium;
the populated cylinder then uses its actual particle surfaces and
fluid domain in the coupled problem.

\begin{definition}[Hydraulic channel and material viscosity]
\label{def:s5-hydraulic-channel}
Let $\Delta p=p_{\rm in}-p_{\rm out}$ and let $Q$ be volume flow toward
the outlet. Their real histories supply power $\Delta p(t)Q(t)$.
At $\omega>0$, define hydraulic admittance $\mathcal Y_h$ and impedance
$Z_h^{\rm tube}$ by
\begin{equation}
 \begin{aligned}
 \widehat Q&=\mathcal Y_h(\omega)\widehat{\Delta p},
 &Z_h^{\rm tube}&=\mathcal Y_h^{-1},\\
 \widehat\tau_{rz}&=\eta_f^*(\omega)\partial_r\widehat v_z,
 &G_f^*&=-i\omega\eta_f^*.
 \end{aligned}
 \label{eq:s5-hydraulic-material-channel}
\end{equation}
Here $\eta_f^*$ is the homogeneous isotropic medium's causal linear
shear viscosity, with the $e^{-i\omega t}$ convention.
\end{definition}

The material has complex viscosity; the geometry converts it into a
hydraulic impedance. The units of $\eta_f^*$, $Z_h^{\rm tube}$ and
$\mathcal Y_h$ are respectively $\mathrm{Pa\,s}$,
$\mathrm{Pa\,s\,m^{-3}}$ and $\mathrm{m^3\,Pa^{-1}\,s^{-1}}$.
This distinction also separates a hydraulic resistance from the
particle drag matrix in Section~\ref{sec:viscoelastic-dynamics}.
For an ideal rigid, no-leak piston of area $A_{\rm pl}$ referenced to
the outlet pressure, the fluid load is
$F_{\rm pl,fluid}=A_{\rm pl}\Delta p$ and
$Q=A_{\rm pl}\dot x_{\rm pl}$. Their power product is again
$\Delta p\,Q$. Friction, device compliance and other fluid segments
supply their additional loads before a measured plunger force is
interpreted as a material viscosity~\cite{Thueer2018}.

Consistently with~\eqref{eq:blood-modulus-viscosity-convention},
$G_f^*=G_f'-iG_f''$ gives
$\eta_f^*=G_f''/\omega+iG_f'/\omega$: its real part supplies shear
loss, while its imaginary part describes the constitutive reactive
response. One fluid-like passive realization is
\begin{equation}
 \eta_f^*(\omega)=\eta_\infty+
       \sum_{a=1}^{N_f}\frac{\eta_a}{1-i\omega\tau_a},
 \qquad \eta_\infty\ge0,\quad\eta_a>0,\quad\tau_a>0,
 \qquad \eta_0=\eta_\infty+\sum_a\eta_a.
 \label{eq:s5-fluid-maxwell-viscosity}
\end{equation}
The finite positive zero-frequency viscosity $\eta_0>0$ permits steady flow. A
material with a persistent equilibrium shear modulus instead has a
different zero-frequency limit. Oscillatory flow of Maxwell and
Oldroyd-B media provides a well-developed setting for this constitutive
distinction~\cite{CasanellasOrtin2011}.
For real histories $\eta_f^*(-\omega)=\eta_f^*(\omega)^*$, with the
storage and loss parts linked by the causal constitutive law.

Figure~\ref{fig:new-s5-flow-material-geometry} connects material viscosity and capillary geometry through the piston and hydraulic work pairing.

\begin{figure}[H]
\centering\begingroup
\definecolor{vltTeal}{HTML}{176B64}
\definecolor{vltBlue}{HTML}{315D80}
\definecolor{vltGold}{HTML}{A87520}
\definecolor{vltPlum}{HTML}{725877}
\definecolor{vltRed}{HTML}{762B38}
\definecolor{vltYellow}{HTML}{D5AD3C}
\definecolor{vltNeutral}{HTML}{777D83}
\setlength{\unitlength}{1pt}
\scalebox{0.90}{%
\input{vlt-figures/figure-158.tex}%
}\endgroup
\caption{An ideal piston translates the pressure--flow port into a mechanical force--velocity port. (a) The rigid no-leak piston is referenced to the outlet pressure; the capillary is an unobstructed homogeneous carrier benchmark with stationary no-slip walls. The same supplied power is $F_{\rm pl}\dot x_{\rm pl}=\Delta p\,Q$. (b) A one-time passive solvent--Maxwell viscosity separates dissipative real and reactive imaginary parts under the $e^{-i\omega t}$ convention. (c) Its quasistatic hydraulic impedance at $\omega\tau=1$ varies as $R_h^{-4}$ at fixed length and material law. The geometric curve uses $Z_0$ defined in the panel. The Poiseuille correspondence assumes $\rho_f\omega R_h^2/|\eta_f^*|\ll1$ throughout the compared geometries. Device friction, compliance and a populated particle geometry have their own matched loads; the schematic supplies an ideal work identity and no empirical injection protocol.}
\label{fig:new-s5-flow-material-geometry}
\end{figure}

For an incompressible fluid of density $\rho_f>0$, linear momentum
balance in the fully developed segment is
\begin{equation}
 \begin{aligned}
 -i\omega\rho_f\widehat v_z(r)
  &=\frac{\widehat{\Delta p}}{L_h}
     +\eta_f^*(\omega)\frac1r\partial_r
                            [r\partial_r\widehat v_z(r)],
 \\
 \widehat v_z(R_h)&=0,\qquad
 \partial_r\widehat v_z(0)=0.
 \end{aligned}
 \label{eq:s5-linear-pipe-momentum}
\end{equation}
Taking real parts reconstructs fluid velocity. A small oscillatory
material excursion about a stationary reference satisfies
$\widehat v_z=-i\omega\widehat w_{f,z}$. Sustained transport instead
follows the advected material through velocity and mass balance.

\begin{proposition}[Complex-viscosity Poiseuille limit]
\label{prop:s5-complex-poiseuille}
For the homogeneous no-slip segment, neglect unsteady fluid inertia
while retaining the frequency dependence of $\eta_f^*\ne0$. The unique
regular axial solution and its hydraulic impedance are
\begin{equation}
 \begin{aligned}
 \widehat v_{\rm P}(r)&=
  \frac{\widehat{\Delta p}}{4L_h\eta_f^*}(R_h^2-r^2),\\
 \widehat Q_{\rm P}&=2\pi\int_0^{R_h}\widehat v_{\rm P}(r)r\,\dd r
       =\frac{\pi R_h^4}{8L_h\eta_f^*}\widehat{\Delta p},\\
 Z_{\rm P}(\omega)&=\frac{8L_h\eta_f^*(\omega)}{\pi R_h^4}.
 \end{aligned}
 \label{eq:s5-complex-poiseuille}
\end{equation}
The steady Newtonian case uses the real viscosity $\eta_0$.
\end{proposition}
\begin{proof}
Integrate $\partial_r(r\partial_r\widehat v_z)
=-r\widehat{\Delta p}/(L_h\eta_f^*)$. Regularity removes the
logarithmic homogeneous solution; no slip fixes the remaining constant.
Cross-sectional integration gives $Q$ and the impedance.
\end{proof}

Momentum balance, circular geometry and no slip give the parabolic
Hagen--Poiseuille profile and its linear viscoelastic correspondence
~\cite{Womersley1955}. Memory changes its amplitude and phase.
The approximation requires
$\epsilon_h=\rho_f\omega R_h^2/|\eta_f^*(\omega)|\ll1$ in the
observation band, in addition to small convective inertia.
The Deborah number $\omega\tau_a$ measures material memory; amplitude
must also lie within the linear-rheology range. Entrance effects,
slip, wall deformation and heterogeneity enter their corresponding
boundary or constitutive descriptions.

\subsubsection{Unsteady flow, pressure phase and supplied work}
\label{sec:s5-unsteady-hydraulic-response}

\begin{proposition}[Unsteady cylindrical response]
\label{prop:s5-complex-womersley}
For~\eqref{eq:s5-linear-pipe-momentum} with
$\operatorname{Re}\eta_f^*(\omega)>0$, put
$\chi^2=i\omega\rho_fR_h^2/\eta_f^*(\omega)$.
With $J_\nu$ the Bessel function of the first kind, the response is
\begin{equation}
 \begin{aligned}
 \widehat v_z(r)&=
  \frac{\widehat{\Delta p}}{-i\omega\rho_fL_h}
       \left[1-\frac{J_0(\chi r/R_h)}{J_0(\chi)}\right],\\
 \mathcal Y_h(\omega)&=
  \frac{\pi R_h^2}{-i\omega\rho_fL_h}
       \left[1-\frac{2J_1(\chi)}{\chi J_0(\chi)}\right],\\
 \mathcal Y_h(\omega)&=
  \frac{\pi R_h^4}{8L_h\eta_f^*}
       [1+\chi^2/6+O(|\chi|^4)]\quad(\chi\to0).
 \end{aligned}
 \label{eq:s5-complex-womersley}
\end{equation}
The ratios are independent of the sign chosen for $\chi$.
\end{proposition}
\begin{proof}
The regular homogeneous radial solution is $J_0(\chi r/R_h)$; a
constant particular solution is
$\widehat{\Delta p}/(-i\omega\rho_fL_h)$. No slip gives the first
line. Integrate using
$\int_0^{R_h}rJ_0(\chi r/R_h)\,\dd r=R_h^2J_1(\chi)/\chi$.
The Bessel series gives the final limit. Strict shear loss makes the
homogeneous real-frequency problem unique by its energy identity.
\end{proof}

For real Newtonian $\eta_f^*=\eta_0$,
$\alpha=R_h\sqrt{\rho_f\omega/\eta_0}$ is the Womersley number and
$\chi^2=i\alpha^2$ in this paper's time convention. The result resolves
the radial profile and pressure--flow phase beyond the parabolic limit
~\cite{Womersley1955,CasanellasOrtin2011}. In particular,
\begin{equation}
 Z_h^{\rm tube}(\omega)=
       \frac{8\eta_0L_h}{\pi R_h^4}
       -i\omega\frac{4\rho_fL_h}{3\pi R_h^2}
       +O(\omega^2)\qquad(\omega\to0).
 \label{eq:s5-newtonian-hydraulic-inertance}
\end{equation}
Even a Newtonian material therefore has a reactive hydraulic response.
The second term is fluid inertance, with the coefficient set by the
nonuniform velocity profile. A measured phase is interpreted through
both momentum transport and the material law.

\begin{proposition}[Hydraulic cycle power]
\label{prop:s5-hydraulic-cycle-power}
For a bounded steady-periodic solution of the uniform segment above,
with no other boundary work, let $\mathcal A_h$ denote its cross-section.
The cycle-averaged supplied power is
\begin{equation}
 \begin{aligned}
 \overline P_h
  &=\frac12\operatorname{Re}
        [\widehat{\Delta p}\,\widehat Q^{\,*}]
   =\frac12\operatorname{Re}Z_h^{\rm tube}|\widehat Q|^2\\
  &=\frac{L_h}{2}\operatorname{Re}\eta_f^*
          \int_{\mathcal A_h}|\partial_r\widehat v_z|^2\,\dd A
       \ge0.
 \end{aligned}
 \label{eq:s5-hydraulic-cycle-power}
\end{equation}
Equivalently, when $\widehat Q=\mathcal Y_h\widehat{\Delta p}$,
$\overline P_h=\tfrac12\operatorname{Re}\mathcal Y_h
|\widehat{\Delta p}|^2$.
\end{proposition}
\begin{proof}
Multiply the momentum equation by $\widehat v_z^*$, integrate over the
cross-section and integrate the radial derivative by parts. The wall
term vanishes by no slip and the axis term by regularity. The inertial
term is purely imaginary. Taking real parts and multiplying by $L_h/2$
gives the identity and its sign.
\end{proof}

Pressure is conjugate to transported volume, whose rate is $Q$.
Hydraulic loss therefore uses the real part of pressure-to-flow
impedance; displacement compliance uses an imaginary part and a factor
$\omega$. Both express the same real work pairing. Fluid inertia and
material stress exchange storage within the cycle, with zero net
periodic storage change. Prepared stress supplies an additional
transient contribution.

Figure~\ref{fig:new-s5-flow-womersley} shows the radial profile and reactive hydraulic response supplied by the same unsteady momentum balance.

\begin{figure}[H]
\centering\begingroup
\definecolor{vltTeal}{HTML}{176B64}
\definecolor{vltBlue}{HTML}{315D80}
\definecolor{vltGold}{HTML}{A87520}
\definecolor{vltPlum}{HTML}{725877}
\definecolor{vltRed}{HTML}{762B38}
\definecolor{vltYellow}{HTML}{D5AD3C}
\definecolor{vltNeutral}{HTML}{777D83}
\setlength{\unitlength}{1pt}
\scalebox{0.90}{%
\input{vlt-figures/figure-159.tex}%
}\endgroup
\caption{Unsteady inertia changes an unobstructed tube response even when the Newtonian viscosity is the constant real $\eta_0$. (a,b) Exact no-slip profiles from~\eqref{eq:s5-complex-womersley}, normalized by the complex cross-sectional mean $\hat{\bar v}=\hat Q/(\pi R_h^2)$. The real ratio is the instantaneous profile at that case's maximum mean flow, in its own mean-flow amplitude units. The arrow lengths in (a) and the horizontal profile coordinate in (b) use this same normalization; $y$ is a signed diametral coordinate with $r=|y|$. (c) Exact hydraulic resistance and negative reactance, divided by $Z_{\rm P}(0)=8\eta_0L_h/(\pi R_h^4)$. The dotted low-frequency inertial approximation is $-\operatorname{Im}[Z_h/Z_{\rm P}(0)]\simeq\alpha^2/6$. All profiles are regular at the axis, vanish at the wall, and satisfy the same pressure-work/viscous-loss identity. The fully developed, rigid, circular, linear-flow assumptions define this carrier calibration; resolved virion obstructions require the populated boundary-value problem.}
\label{fig:new-s5-flow-womersley}
\end{figure}

\subsubsection{Coupling the fluid to ideal-virion motion and inverse variables}
\label{sec:s5-fluid-particle-inference}

An array of ideal virions modifies the fluid domain. The relevant calculation
resolves flow around the icosahedral surfaces, matches each surface
velocity to its retained rigid-body translation and rotation as in
\eqref{eq:s4patch-rigid-fluid-boundary}, and includes the stated
wall and pressure-port conditions. This is the hydraulic extension of
the array resistance problem in Section~\ref{sec:viscoelastic-dynamics}.
The unobstructed tube formula calibrates a carrier-medium benchmark;
the populated geometry supplies its own pressure--flow and
particle--fluid coupling.
If a measured suspension viscosity already averages over the same
particles, those effects are matched before adding explicit particle
obstructions, so their mechanical contribution is counted once.

\begin{definition}[Reciprocal particle--hydraulic resistance]
\label{def:s5-particle-hydraulic-resistance}
For the translational specialization, linearize about a quiescent reference.
Let $\vct V$ collect particle centre velocities, let $\vct F_h$ be fluid
force on those particles, and
choose pressure ports whose boundary work is $\Delta p\,Q$.
The fluid resistance map is
\begin{equation}
 \begin{pmatrix}-\vct F_h\\\widehat{\Delta p}\end{pmatrix}
  =\begin{pmatrix}
      \mathsf Z_{pp}&\vct Z_{pQ}\\
      \vct Z_{Qp}&Z_{QQ}
    \end{pmatrix}
    \begin{pmatrix}\vct V\\\widehat Q\end{pmatrix}
   =:\mathsf Z_{\rm port}
    \begin{pmatrix}\vct V\\\widehat Q\end{pmatrix}.
 \label{eq:s5-particle-hydraulic-port-map}
\end{equation}
The first component is force delivered by the particles to the fluid.
For reciprocal passive linear fluid mechanics,
$\mathsf Z_{\rm port}^{\mathsf T}=\mathsf Z_{\rm port}$ and
$\operatorname{Re}_{\rm H}\mathsf Z_{\rm port}\succeq0$ in these
work-conjugate variables.
\end{definition}

The map follows by solving the boundary-value problem with independent
particle velocities and port flux. Lorentz reciprocity gives its
transpose symmetry; the fluid power identity gives its nonnegative
Hermitian real part~\cite{GuazzelliMorris2012}. Here
$\operatorname{Re}_{\rm H}\mathsf Z=(\mathsf Z+\mathsf Z^\dagger)/2$.
Its blocks have the units required by their different input variables;
numerical condition numbers require declared reference scales. Initial
fluid or material excitations add their prepared forcing, just as in
the reduced medium equations.

\begin{proposition}[Pressure-driven particle response]
\label{prop:s5-pressure-driven-particles}
For the port map above with $Z_{QQ}\ne0$, fixed particle mass and
conservative stiffness, and $\vct V=-i\omega\vct U$, eliminating the
flow variable at prescribed pressure gives
\begin{equation}
 \begin{aligned}
 \mathsf Z_{\rm eff}
   &=\mathsf Z_{pp}-\vct Z_{pQ}Z_{QQ}^{-1}\vct Z_{Qp},\\
 [\mathsf K-\omega^2\mathsf M-i\omega\mathsf Z_{\rm eff}]\vct U
   &=\vct F_{\rm ext}
       -\vct Z_{pQ}Z_{QQ}^{-1}\widehat{\Delta p},\\
 \widehat Q&=Z_{QQ}^{-1}
           [\widehat{\Delta p}-\vct Z_{Qp}\vct V].
 \end{aligned}
 \label{eq:s5-pressure-driven-particle-response}
\end{equation}
At zero pressure difference the reduced resistance remains passive.
\end{proposition}
\begin{proof}
Solve the second port equation for $Q$ and substitute in the first.
Insert the resulting fluid force into particle momentum balance. For
zero pressure difference the port power equals the particle-to-fluid
power alone; its nonnegativity gives
$\operatorname{Re}_{\rm H}\mathsf Z_{\rm eff}\succeq0$.
\end{proof}

Zero pressure difference is a pressure-equilibrated termination that
can allow flow generated by moving particles. Sealed ends instead
impose $Q=0$, with pressure providing the required reaction; the
particle resistance is then $\mathsf Z_{pp}$. These are distinct
boundary preparations of the same coupled system.

The blocks predict pressure-driven particle motion and its feedback
on flow. The conservative $\mathsf K$ retains interaction and support
curvatures; the fluid response supplies momentum transfer, storage and
loss at the actual geometry. Match inertia to the chosen impedance as
in~\eqref{eq:hydrodynamic-added-mass-matching}, counting each
contribution once. Electrokinetic forcing additionally specifies its
fields, charge transport and boundary work.

\paragraph{Pressure couples to angular and surface motion through one port law.}
The complete linear construction uses
$\vct y=(\vct X,\vct\chi)$, with joint storage Hessian
$\mathsf H_y$ from~\eqref{eq:s4patch-joint-storage-resistance} and
$\mathsf G_y=\diag(\mathsf G_X,0)$. The zero block denotes the
first-order surface/material kinetics. Choose a fixed reference volume
$V_\star>0$ and set $q_h=Q/V_\star$, $p_h=V_\star\Delta p$;
then $p_hq_h=\Delta p\,Q$. All rates in
$(\dot{\vct y},q_h)$ have inverse-time units, and the work-dual
sources have energy units. In phasor form a reciprocal causal
resistance $\mathsf Z_h$ gives
\begin{equation}
 \begin{aligned}
 (\mathsf H_y-\omega^2\mathsf G_y)\widehat{\vct y}
     -i\omega\mathsf Z_{yy}\widehat{\vct y}
       +\vct Z_{yq}\widehat q_h&=\vct F_y,\\
 \widehat p_h&=-i\omega\vct Z_{qy}\widehat{\vct y}
                                       +Z_{qq}\widehat q_h,\\
 \mathsf Z_h&=
 \begin{pmatrix}\mathsf Z_{yy}&\vct Z_{yq}\\
                 \vct Z_{qy}&Z_{qq}\end{pmatrix},\qquad
 \mathsf Z_h^{\mathsf T}=\mathsf Z_h,\quad
 \operatorname{Re}_{\rm H}\mathsf Z_h\succeq0 .
 \end{aligned}
 \label{eq:s5patch-body-surface-hydraulic-port}
\end{equation}
The entries have units $\mathrm{J\,s}$. Here
$\vct F_y=(\vct F_X,\vct a_\chi)$, and fluid, surface and material
losses are assembled once at the stated resolution. A surface--flow
cross coefficient is a specified constitutive coupling. Net charge
alone does not determine it. Rigid-body resistance and its geometric
translation--rotation blocks follow the actual fluid boundary problem
\cite{KimKarrila1991}.

For $Z_{qq}\ne0$, prescribed-pressure elimination gives
\begin{equation}
 \begin{aligned}
 \mathsf Z_y^{(p)}&=\mathsf Z_{yy}
               -\vct Z_{yq}Z_{qq}^{-1}\vct Z_{qy},\\
 [\mathsf H_y-\omega^2\mathsf G_y-i\omega\mathsf Z_y^{(p)}]
           \widehat{\vct y}
       &=\vct F_y-\vct Z_{yq}Z_{qq}^{-1}\widehat p_h .
 \end{aligned}
 \label{eq:s5patch-pressure-body-surface-reduction}
\end{equation}
At zero imposed pressure, the full port power reduces to the retained
power, proving $\operatorname{Re}_{\rm H}\mathsf Z_y^{(p)}\succeq0$.
The subsequent surface Schur solve uses the mixed readout and source
maps above. Either elimination order solves the same retained equations
where the intermediate inverses exist. Prepared fluid forcing and
equilibrium random forces undergo those same row transformations.
For a quasisteady real resistance, multiplying the time-domain equations
by $(\dot{\vct y},q_h)$ gives
$\dot E=\vct F_y\cdot\dot{\vct y}+p_hq_h-
(\dot{\vct y},q_h)^{\mathsf T}\mathsf Z_h
(\dot{\vct y},q_h)$, with the declared joint storage.
Fluid inertia or memory carries its own storage in the resolved realization.

Holding a body orientation sets its angular velocity to zero.
Zero external torque instead leaves angular motion dynamical.
After terminating the other ports, partition the full mechanical
pencil into translational and angular sectors. The torque-free
translational response then uses
\begin{equation}
 \begin{aligned}
 \mathcal P_{t}^{\rm free\ torque}
  &=\mathcal P_{tt}-\mathcal P_{t\theta}
                \mathcal P_{\theta\theta}^{-1}\mathcal P_{\theta t},\\
 \vct F_{t,\rm eff}&=\vct F_t-\mathcal P_{t\theta}
                  \mathcal P_{\theta\theta}^{-1}\vct F_\theta .
 \end{aligned}
 \label{eq:s5patch-torque-free-response}
\end{equation}
Here the source blocks include the continued hydraulic, chemical and
preparation forcing; zero externally applied torque can coexist with a
hydraulically induced torque. The pencil includes angular stiffness,
inertia and relaxation. A resistance-only
reduction belongs to its quasisteady restoring-torque-free limit.
Symmetry can suppress a translation--rotation block for an isolated
ideal icosahedron; neighbours, walls and specified surface boundary asymmetry
can change that conclusion. The block is calculated for the actual
preparation rather than inferred from shape alone.

\paragraph{Electrical termination changes a hydraulic measurement.}
\begin{example}[A reciprocal electrohydraulic comparison]
\label{ex:s5patch-electrohydraulic-termination}
For a near-equilibrium isothermal electrolyte channel, include terminal
voltage difference $\Delta V$ and current $I$. In the real DC linear
limit without time-reversal-breaking control, use
\begin{equation}
 \begin{aligned}
 \binom{Q}{I}&=
 \begin{pmatrix}L_{pp}&L_{pE}\\L_{Ep}&L_{EE}\end{pmatrix}
 \binom{\Delta p}{\Delta V},\qquad P_{\rm in}=\Delta p\,Q+\Delta V I,\\
 L_{pE}&=L_{Ep},\qquad\mathsf L\succeq0,\qquad L_{EE}>0 .
 \end{aligned}
 \label{eq:s5patch-electrohydraulic-admittance}
\end{equation}
Setting $\Delta V=0$ or solving $I=0$ gives respectively
\begin{equation}
 \boxed{\begin{aligned}
 \mathcal Y_h^{\rm short}&=L_{pp},\qquad
 \mathcal Y_h^{\rm open}=L_{pp}-L_{pE}L_{EE}^{-1}L_{Ep},
 \\
 0&\leq\mathcal Y_h^{\rm open}\leq\mathcal Y_h^{\rm short}.
 \end{aligned}}
 \label{eq:s5patch-electrical-hydraulic-termination}
\end{equation}
The inequality is the positive Schur-complement property. Streaming
potential feedback can therefore increase measured hydraulic resistance
at fixed carrier viscosity. This is an established electrokinetic
mechanism~\cite{MortensenOlesenBruus2006}.
\end{example}

Figure~\ref{fig:cmd-complementary-ports} gives a graphical specialization of this boxed relation.

\begin{figure}[H]
\centering\begingroup
\setlength{\unitlength}{1pt}
\scalebox{0.84}{%
\input{vlt-figures/figure-313.tex}%
}\endgroup
\caption{Two complementary-channel calculations. (a) The displayed dimensionless positive reciprocal conductance has hydraulic conductance $2$ at $\Delta V=0$ and $2-0.8^2/1.5$ at $I=0$, the two electrical terminations in \eqref{eq:s5patch-electrical-hydraulic-termination}. (b) Assign $\ddot x+0.4\dot x+2x-0.6\chi=f$, $\dot\chi+2\chi-0.6x=a_\chi$ and $Y=x+0.4\chi$. An impulse conjugate to $Y$ gives $f=\delta(t)$ and $a_\chi=0.4\delta(t)$, hence $(x,\dot x,\chi)(0^+)=(0,1,0.4)$. The centre-only comparison applies and reads only the mechanical port. The mixed response follows \eqref{eq:s5patch-mixed-response-readout} and has the prompt value and derivative of \eqref{eq:s6patch-mixed-impulse-asymptotics}. This scalar affine contrast also illustrates the additive structural contribution in \eqref{eq:s6patch-linear-body-contrast}; a scattering experiment supplies its actual wavevector-dependent coefficients. The two panels are separate positive linear realizations, not fitted parameters of one sample.}
\label{fig:cmd-complementary-ports}
\end{figure}

For the VLP application, the coefficients require a specified electrolyte,
surface law and populated-domain electrochemical boundary problem.
Comparing electrical terminations, pressure/flow and angular readouts
can test that feedback with the same registered arrangement. Fixed
surface charge, fixed surface potential and kinetic charge regulation
give different boundary specifications. At finite frequency the terminal
current includes the required charging contributions; ionic conservation,
electrodes and driven chemical reservoirs enter the corresponding state
and work account. Those dynamics determine $\mathsf L(\omega)$.
The real DC ordering above is the stated limit of this construction.

The inverse questions now have explicit observables. For the
unobstructed carrier calibration segment in the parabolic limit,
calibrated geometry and nonzero pressure--flow phasors determine
\begin{equation}
 \eta_f^*(\omega)=\frac{\pi R_h^4}{8L_h}
        \frac{\widehat{\Delta p}}{\widehat Q}.
 \label{eq:s5-hydraulic-viscosity-inference}
\end{equation}
An unknown radius instead leaves the combination
$\eta_f^*L_h/R_h^4$ in this relation. Even small geometric uncertainty
therefore matters: to first order
$\delta Z_{\rm P}/Z_{\rm P}=
\delta\eta_f^*/\eta_f^*+\delta L_h/L_h-4\delta R_h/R_h$.
Outside the small-$\epsilon_h$ band, the unsteady profile must be fitted
with the same density, geometry and boundary assumptions. Pressure--flow
phase then contains both material memory and fluid inertia.

Tracked motion $\vct U$ and calibrated readout
$Y=\vct w^{\mathsf T}\vct U$ add spatial information to hydraulic data.
The fluid boundary problem relates bulk viscosity to local resistance
at their respective resolutions. Fluid momentum and branch stresses
are state variables; viscosity, times and dimensions are parameters;
pressure, flow and displacement are specified inputs or outputs.
These roles make the biophysical variable assignment explicit.

A small oscillatory pressure protocol can interrogate a held or locally
stationary array while its original particle labels remain in the
observation region. Sustained transport requires advection, changing
reference geometry and particle flux at the ports over the corresponding
time scale. A fixed-population local-mode calculation uses the interval
on which its preparation remains valid. The conservative complex
coordinates developed next then describe the retained closed reference;
the hydraulic extension supplies an explicit externally driven channel
and its work account.

\subsubsection{Hydraulic response across recruitment and rearrangement}
\label{sec:s5-flow-evolving-reference}

A recruited virion changes both the associated particle group and the fluid
geometry around that group. The constitutive memory of the carrier and
the geometrical resistance of the populated lumen then enter the same
measured pressure--flow record. A registered reference records the particle
positions, wall and pressure-port conditions, and prepared fluid and branch
states used to predict that record. The compatible interface construction
of Subsection~\ref{sec:s4-cell-retarded-ports} can reuse unchanged cell
interiors, while retaining shared-fluid couplings and induced forcing.
Its interface maps assemble a single reciprocal force--flow operator,
with the shared medium and each physical energy channel retained once.

\paragraph{Association and passage through an observation segment.}
The fixed population $N_{\rm tot}$ refers to the complete declared system.
A pressure-driven experiment can retain that particle pool with specified
particle-retaining port conditions, whose supports and hydraulic load are
included in the boundary model. A local segment of a larger apparatus can
instead exchange virions with the rest of that apparatus. Its observed number
then has an advective contribution even if the global pool is fixed.
Let $w_i(t)$ indicate that particle $i$ lies in the observed segment and
let $h_i^{\mathcal C}$ indicate association with a selected cluster. Then
\begin{equation}
 n_{\rm obs}=\sum_i w_i,\qquad
 n_{\mathcal C\cap\rm obs}=\sum_i h_i^{\mathcal C}w_i.
 \label{eq:s5-flow-observation-membership}
\end{equation}
Tracking $w_i$ and $h_i^{\mathcal C}$ distinguishes passage across the
segment boundary from joining or leaving the associated group. The
mass-transport account in~\eqref{eq:s4-growing-cluster-mass-balance}
supplies the corresponding distinction between material transport and
membership assignment. For separate inlet and outlet fluid fluxes, the
moving fluid volume $V_f$ of the incompressible carrier instead obeys
\begin{equation}
 \begin{aligned}
 \dot V_f&=Q_{\rm in}-Q_{\rm out},\qquad
 Q_m=\tfrac12(Q_{\rm in}+Q_{\rm out}),\quad
 \bar p=\tfrac12(p_{\rm in}+p_{\rm out}),\\
 P_p&=p_{\rm in}Q_{\rm in}-p_{\rm out}Q_{\rm out}
      =\Delta p\,Q_m+\bar p\,\dot V_f .
 \end{aligned}
 \label{eq:s5patch-open-segment-fluid-volume-power}
\end{equation}
Thus the one-flux port applies when the two fluid fluxes agree. Rigid
VLPs wholly inside a rigid cylinder preserve its total excluded volume
during rearrangement. Passage across observation planes, wall compliance
or changing excluded volume uses the additional volume/transport account;
work on transported particles and reservoir exchange are included too.
The reciprocal passive port is referenced to quiescent linear response.
A frozen-geometry creeping Newtonian problem retains Stokes reciprocity;
a strongly driven, deforming or nonlinear constitutive reference uses
its own incremental law and pump-work account.

\begin{definition}[Registered finite-window hydraulic channel]
\label{def:s5-flow-window-channel}
Use a consistently resolved finite particle--fluid--memory state, and
linearize along a registered reference without a physical reset inside
the observation window. In its common coordinates let
$\delta\dot X=\mathsf A(t)\delta X+
\mathsf B_p(t)\delta p(t)$ and
$\delta Q=\mathsf L_Q(t)\delta X$, where $p=\Delta p$ and
$\delta X(0)=0$. This realization retains the finite-inertia fluid
variables needed for an output without algebraic pressure feedthrough.
For a complexified calculation driven by
$\delta p(t)=P_\nu e^{-i\nu t}$, define
\begin{equation}
 \begin{aligned}
 \delta\widehat Q_T(\omega)
   &=\frac1T\int_0^T\delta Q(t)e^{i\omega t}\,\dd t
      =\mathcal Y_T(\omega,\nu)P_\nu,\\
 \mathcal Y_T(\omega,\nu)
   &=\frac1T\int_0^T\int_0^t
       \mathsf L_Q(t)\mathsf T(t,s)\mathsf B_p(s)
                  e^{i\omega t-i\nu s}\,\dd s\,\dd t.
 \end{aligned}
 \label{eq:s5-flow-two-frequency-channel}
\end{equation}
Here $\mathsf T(t,s)$ is the registered propagator of
Definition~\ref{def:s4-path-conditioned-response}. The quantity
$\mathcal Y_T$ has hydraulic-admittance units. The physical drive is
the real part; a real Fourier record includes its conjugate contribution.
\end{definition}

Variation of constants proves the formula. Independent changes of initial
material strain add their preparation response, while registered events
use the compatible continuation of Section~4. A measured response at
$\omega\ne\nu$ can contain ordinary finite-window leakage and preparation
transients even for fixed coefficients. A controlled comparison therefore
uses the same window, source onset and readout for its fixed-reference
prediction. The remaining discrepancy can test changing geometry, material
state or their coupling. The single-frequency viscosity inference in
\eqref{eq:s5-hydraulic-viscosity-inference} belongs to its persistent
unobstructed reference. The two-frequency channel instead retains the
ordering of input and response during an evolving populated geometry.
If pressure changes event probabilities or event times, their response
is included in the transition model rather than in a fixed-event average.

\begin{proposition}[Controlled frozen-reference hydraulic approximation]
\label{prop:s5-flow-frozen-reference-bound}
Express state, pressure and flow in fixed dimensionless coordinates using
declared reference scales. Apply the estimate to
$\dot X=\mathsf A(t)X+\mathsf B_p(t)p$; any affine offsets are included
as known inputs or a constant homogeneous coordinate.
Suppose $\|\mathsf T(t,s)\|\le
\mathcal M e^{\alpha(t-s)}$ on $0\le s\le t\le T$. Let $X_0$ solve
the frozen model with constant $\mathsf A_0,\mathsf B_0,\mathsf L_0$,
the same initial state and the same input $p$. Assume
\begin{equation}
 \begin{aligned}
 \|\mathsf A(t)-\mathsf A_0\|&\le\epsilon_A,&
 \|\mathsf B_p(t)-\mathsf B_0\|&\le\epsilon_B,&
 \|\mathsf L_Q(t)-\mathsf L_0\|&\le\epsilon_L,\\
 \|X_0(t)\|&\le x_{\max},&
 |p(t)|&\le p_{\max},&
 \|\mathsf L_Q(t)\|&\le L_{\max}.
 \end{aligned}
 \label{eq:s5-flow-frozen-error-inputs}
\end{equation}
Then, with $\phi_\alpha(T)=(e^{\alpha T}-1)/\alpha$ and
$\phi_0(T)=T$,
\begin{equation}
 \begin{aligned}
 \sup_{0\le t\le T}|Q(t)-Q_0(t)|
 &\le L_{\max}\mathcal M\phi_\alpha(T)
       (\epsilon_Ax_{\max}+\epsilon_Bp_{\max})\\
 &\quad+\epsilon_Lx_{\max}.
 \end{aligned}
 \label{eq:s5-flow-frozen-error-bound}
\end{equation}
The same bound controls the difference of normalized finite-window
Fourier coefficients. Multiplying by the reference flow scale restores
physical volume-flow units.
\end{proposition}
\begin{proof}
For $E=X-X_0$,
$\dot E=\mathsf A(t)E+
(\mathsf A(t)-\mathsf A_0)X_0+
(\mathsf B_p(t)-\mathsf B_0)p$ and $E(0)=0$.
Variation of constants bounds its norm by
$\mathcal M\phi_\alpha(T)(\epsilon_Ax_{\max}+
\epsilon_Bp_{\max})$. Substituting
$Q-Q_0=\mathsf L_Q E+(\mathsf L_Q-\mathsf L_0)X_0$ gives the result;
the normalized Fourier integral is bounded by the same supremum.
\end{proof}

If a Stokes or differential--algebraic reduction instead gives
$\delta Q=\mathsf L_Q(t)\delta X+\mathsf D_p(t)\delta p$, its direct
pressure channel adds
\begin{equation}
 \frac1T\int_0^T\mathsf D_p(t)e^{i(\omega-\nu)t}\,\dd t
 \label{eq:s5-flow-direct-pressure-channel}
\end{equation}
to $\mathcal Y_T(\omega,\nu)$. With a frozen direct coefficient
$\mathsf D_0$ and $\sup_t\|\mathsf D_p(t)-\mathsf D_0\|\le\epsilon_D$,
the output and Fourier bounds gain $\epsilon_Dp_{\max}$. Thus the
same comparison also covers a calibrated quasisteady flow port.

The constants bound drift of the coefficients, source and readout,
including amplification over the window. Bounded finite matrices
permit a finite growth estimate in fixed scales; a common energy
estimate can sharpen it. Separate passive metrics need not give a
common contraction. Assess geometric updates against output accuracy,
retain long-lived strain in $X(0)$, and calculate the populated-domain
resistance once with its shared particle interfaces.

Figure~\ref{fig:new-s5-flow-evolving-channel} compares a continued material preparation and its registered finite-window pressure--flow channel.

\begin{figure}[H]
\centering\begingroup
\definecolor{vltTeal}{HTML}{176B64}
\definecolor{vltBlue}{HTML}{315D80}
\definecolor{vltGold}{HTML}{A87520}
\definecolor{vltPlum}{HTML}{725877}
\definecolor{vltRed}{HTML}{762B38}
\definecolor{vltYellow}{HTML}{D5AD3C}
\definecolor{vltNeutral}{HTML}{777D83}
\setlength{\unitlength}{1pt}
\scalebox{0.90}{%
\input{vlt-figures/figure-160.tex}%
}\endgroup
\caption{A registered hydraulic record carries prepared memory through a change in coupling. (a) The same four ideal virions illustrate two reference geometries, with their labels retained by the declared holding/boundary preparation. The reduced reciprocal channel printed below uses $R_\infty/R_\star=0.7$, $R_1/R_\star=1.1$ and the indicated $g$; these coefficients are declared, rather than inferred from the sketch. Its stored branch energy is $\tau z^2/(2R_1)$ and its pressure power equals its storage rate plus $R_\infty Q^2+z^2/R_1$. (b) For $\Delta p/p_\star=\cos(t/\tau)$, $p_\star=R_\star Q_\star$ and $z(0)=0.3p_\star$, the black trace preserves $z$; the dashed comparison resets it at $t_*$ and therefore changes the preparation. The dotted trace isolates the additive initial-state contribution. (c) Exact $R_\star|\mathcal Y_T(\omega,\nu)|$ at $T/\tau=8$, with zero initial perturbation and source onset at zero. Both matrices use identical windows and input/output frequencies. The algebraic direct term is $D_p=1/R_\infty$ and is retained in every entry. Even the fixed reference has off-diagonal finite-window leakage. This quasisteady auxiliary channel permits a flow jump while its stored material state remains continuous; an actual geometry transition retains its particle, transport and work account.}
\label{fig:new-s5-flow-evolving-channel}
\end{figure}

\paragraph{Joint hydraulic and particle records identify different inputs.}
\label{sec:s5-flow-joint-identification}

\begin{example}[An exact mechanical-equivalence family of fluid couplings]
\label{ex:s5-flow-equivalent-couplings}
Take one calibrated translational coordinate and one hydraulic channel
in a quasisteady auxiliary band, with real reciprocal resistance
\begin{equation}
 \mathsf Z_{\rm port}=
  \begin{pmatrix}\zeta&\lambda_h\\\lambda_h&r_h\end{pmatrix},
 \qquad r_h>0,\quad
 \zeta=c_{\rm eff}+\lambda_h^2/r_h,\quad c_{\rm eff}>0.
 \label{eq:s5-flow-equivalent-family}
\end{equation}
The blocks retain their work-conjugate units from
\eqref{eq:s5-particle-hydraulic-port-map}; in particular
$[\lambda_h]=\mathrm{N\,s\,m^{-3}}$ and
$[r_h]=\mathrm{Pa\,s\,m^{-3}}$.
The determinant is $c_{\rm eff}r_h>0$, so this whole family is passive.
At zero imposed pressure difference, eliminating flow gives the same
$c_{\rm eff}$ for every allowed $r_h,\lambda_h$. With fixed mass $m$
and restoring coefficient $k$, the mechanical compliance is always
$(k-m\omega^2-i\omega c_{\rm eff})^{-1}$.

Two additional calibrated protocols can separate the three resistance
coefficients within this family. Hold the particle coordinate fixed to
measure $r_h=\widehat{\Delta p}/\widehat Q$, then drive the particle
at zero pressure difference to measure
\begin{equation}
 \frac{\widehat Q}{V}=-\frac{\lambda_h}{r_h},\qquad
 \lambda_h=-r_h\frac{\widehat Q}{V},\qquad
 \zeta=c_{\rm eff}+\frac{\lambda_h^2}{r_h}.
 \label{eq:s5-flow-joint-reconstruction}
\end{equation}
All protocols use the same geometry, port boundary model and linear
band. They identify this stipulated coupled resistance model, while
assigning its coefficients to a viscosity or microscopic mechanism
still uses the populated fluid boundary-value problem.
\end{example}

Figure~\ref{fig:new-s5-flow-reciprocal-ports} compares hydraulic terminations and the extra measurements that separate an exact mechanical-equivalence family.

\begin{figure}[H]
\centering\begingroup
\definecolor{vltTeal}{HTML}{176B64}
\definecolor{vltBlue}{HTML}{315D80}
\definecolor{vltGold}{HTML}{A87520}
\definecolor{vltPlum}{HTML}{725877}
\definecolor{vltRed}{HTML}{762B38}
\definecolor{vltYellow}{HTML}{D5AD3C}
\definecolor{vltNeutral}{HTML}{777D83}
\setlength{\unitlength}{1pt}
\scalebox{0.90}{%
\input{vlt-figures/figure-161.tex}%
}\endgroup
\caption{A particle--hydraulic port viewed inside the populated cylinder. (a) One selected virion translates axially while two registered surfaces are held. The oblique reference arrangement locates the conjugate velocity and flow ports; its geometry does not assign their coefficients. The reciprocal matrix pairs $(-F_{\rm h},\Delta p)$ with $(V,Q)$ in the stated power account. (b) Pressure-equilibrated ends give $c_{\rm eff}=\zeta-\lambda_h^2/r_h$; sealed ends give $\zeta$. Responses use $m=k=1$ and the stated loads. (c) The three positive matrices retain the exact mechanical-equivalence family in \eqref{eq:s5-flow-equivalent-family}; displayed entries are rounded. A held-particle pressure--flow record determines $r_h$, and the zero-pressure driven ratio determines $\lambda_h$ and $\zeta$ through the reconstruction below. All protocols use fixed work-conjugate scales and the same populated boundary-value problem. The spatial arrows locate ports without prescribing force or velocity amplitudes.}
\label{fig:new-s5-flow-reciprocal-ports}
\end{figure}

Registered positions and membership constrain geometry, pressure and
flow constrain hydraulic work, and relative motion constrains the
mechanical load. Comparing them across episodes tests changes in
carrier memory, coupling or readout. The example supplies an exact
equivalence family and the extra measurements that distinguish its
members within the declared model.

\subsection{Conservative evolution and the meaning of a Hamiltonian}
\label{sec:conservative-complex-generator}

A phasor records harmonic motion relative to a clock. The instantaneous
conservative state instead retains displacement and velocity. We now
encode that pair in a complex amplitude whose squared norm is harmonic
energy, exposing modal energy and its evolution generator.
Prepared medium coordinates remain part of the viscoelastic state.

Start with the real canonical harmonic Hamiltonian already established,
\begin{equation}
 \begin{split}
 H_{\rm harm}(\vct u,\vct p)
   &=\tfrac12\vct p^{\mathsf T}\mathsf M^{-1}\vct p
                      +\tfrac12\vct u^{\mathsf T}\mathsf K\vct u\\
   &=\tfrac12(\vct v^{\mathsf T}\vct v
                         +\vct x^{\mathsf T}\mathsf D\vct x),\\
 \vct x&=\mathsf M^{1/2}\vct u,\qquad
 \vct v=\mathsf M^{-1/2}\vct p=\Dt{\vct x}.
 \end{split}
 \label{eq:complex-harmonic-hamiltonian-start}
\end{equation}
Momentum $\vct p=\mathsf M\Dt{\vct u}$ is conjugate to the physical
displacement. The mass-weighted pair $(\vct x,\vct v)$ is canonical
too; here $\vct v$ is the mass-weighted momentum as well as the
mass-weighted velocity. Hamilton's equations are
$\Dt{\vct x}=\vct v$ and $\Dt{\vct v}=-\mathsf D\vct x$.
The function $H_{\rm harm}$ measures kinetic plus restoring energy;
its derivatives generate the real trajectory. It is the particle
reference energy, not automatically the total energy of an enclosed
medium or an externally driven preparation.

\begin{figure}[H]
\centering\begingroup
\definecolor{vlfGreen}{HTML}{176B64}
\definecolor{vlfDarkRed}{HTML}{762B38}
\definecolor{vlfRed}{HTML}{762B38}
\definecolor{vlfBlue}{HTML}{315D80}
\definecolor{vlfCopper}{HTML}{762B38}
\setlength{\unitlength}{1pt}
\scalebox{0.90}{%
\input{vlt-figures/figure-162.tex}%
}\endgroup
\caption{Equal-unit coordinates $(\omega_sQ_s,\Pi_s)$ map to $\psi_s=(\omega_sQ_s+i\Pi_s)/\sqrt2$, with $E_s=E_\star>0$. Their unnormalized radii are $\sqrt{2E_\star}$ and $\sqrt{E_\star}$; plotted axes divide by $\sqrt{E_\star}$ and have equal component scales. Colored points match phases of one clockwise orbit. Lower snapshots reconstruct a supported virion with $u/a=\cos\theta$ and $\dot u/(\omega_sa)=-\sin\theta$. Green displacement arrows use $10\,\mathrm{pt}$ per normalized unit; burgundy velocity arrows use $12\,\mathrm{pt}$. Matched coils follow the fixed-support extension and compression; ghosts locate the reference. The boxed result illustrated here is~\eqref{eq:complex-energy-coordinate-equivalence}.}
\label{fig:classical-energy-amplitude-map}
\end{figure}

\begin{proposition}[Conservative generator in the energy metric]
\label{prop:s5-energy-metric-generator}
On a fixed strictly positive conservative sector, let
$\mathsf D=\mathsf D^{\mathsf T}>0$. The first-order generator of the
real displacement--velocity pair and its energy metric satisfy
\begin{equation}
 \mathsf A_0=
       \begin{pmatrix}0&\Id\\-\mathsf D&0\end{pmatrix},\qquad
 \mathsf Q_E=\diag(\mathsf D,\Id),\qquad
 \mathsf A_0^\dagger\mathsf Q_E+\mathsf Q_E\mathsf A_0=0.
 \label{eq:conservative-energy-skew-adjointness}
\end{equation}
\end{proposition}
\begin{proof}
Block multiplication gives the displayed metric-skew identity.
For $X=(\vct x,\vct v)$ obeying $\dot X=\mathsf A_0X$, differentiating
$X^{\mathsf T}\mathsf Q_EX/2$ gives zero by that identity.
\end{proof}

Skew-adjointness holds in the energy metric, whose weights give
displacement and velocity a common physical comparison.

Figure~\ref{fig:s5-zero-and-positive-motion} contrasts free common motion with the relative oscillator of the same fixed-particle pair.

\begin{figure}[H]
\centering\begingroup
\definecolor{vlfGreen}{HTML}{176B64}
\definecolor{vlfDarkRed}{HTML}{762B38}
\definecolor{vlfRed}{HTML}{762B38}
\definecolor{vlfBlue}{HTML}{315D80}
\definecolor{vlfCopper}{HTML}{762B38}
\setlength{\unitlength}{1pt}
\scalebox{0.84}{%
\input{vlt-figures/figure-163.tex}%
}\endgroup
\caption{The same unpinned equal-mass pair has a free common coordinate and positive relative frequency $\omega_r=\sqrt{2k_c/m}$. Left rows advance by $9\,\mathrm{pt}$ per interval $t_\star=\pi/(2\omega_r)$, preserving separation. Right rows sample a relative preparation of component amplitude $10\,\mathrm{pt}$ at $\omega_rt=0,\pi/2,\pi$, giving compressed, reference and extended spans. Body sizes and coil families match. Shared rulers locate reference centres at $x/a=0,1$ throughout each column. The preparations can have different energies; each stays within its declared clearance window. The zero coordinate retains its initial position and momentum separately from the positive energy amplitude.}
\label{fig:s5-zero-and-positive-motion}
\end{figure}

Let $\mathsf W=\mathsf D^{1/2}$, and compare equal-unit coordinates
$Y=(\mathsf W\vct x,\vct v)$:
\begin{equation}
 \Dt{Y}=\mathsf A_Y Y,\qquad
 \mathsf A_Y=\begin{pmatrix}0&\mathsf W\\-\mathsf W&0\end{pmatrix},
 \qquad \mathsf A_Y^\dagger=-\mathsf A_Y,
 \qquad E_{\mathrm{harm}}=\tfrac12\|Y\|^2.
 \label{eq:energy-isometric-generator}
\end{equation}
Both entries of $Y$ have units $\sqrt{\mathrm J}$. The transformation
is an isometry for the declared energy metric: it expresses physical
energy as a squared Euclidean size.

\begin{proposition}[Invertible conservative energy-coordinate representation]
\label{prop:s5-conservative-energy-coordinates}
For fixed real $\mathsf D>0$ and $\mathsf W=\mathsf D^{1/2}$, the real
harmonic state is equivalently represented by
\begin{equation}
 \boxed{\begin{aligned}
 \ket\psi&=\frac1{\sqrt2}(\mathsf W\vct x+i\vct v),\\
 i\partial_t\ket\psi&=\mathsf W\ket\psi,\\
 \braket\psi\psi&=E_{\mathrm{harm}}.
 \end{aligned}}
 \label{eq:complex-energy-coordinate-equivalence}
\end{equation}
\end{proposition}
\begin{proof}
The real and imaginary parts recover
$\vct x=\sqrt2\mathsf W^{-1}\operatorname{Re}\psi$ and
$\vct v=\sqrt2\operatorname{Im}\psi$.
Substituting $\dot{\vct x}=\vct v$ and
$\dot{\vct v}=-\mathsf W^2\vct x$ gives the first-order equation.
Its squared norm is the sum of the kinetic and restoring quadratic
energies because both quadratures are real.
\end{proof}
For real data, the inverse is
$\vct x=\sqrt2\mathsf W^{-1}\operatorname{Re}\psi$ and
$\vct v=\sqrt2\operatorname{Im}\psi$. One complex column therefore
retains the original displacement and momentum. Its imaginary part has
energy-amplitude units and encodes velocity; a phasor's imaginary part
is a length-valued sine coefficient.

For a single mode,
$E_s=(\omega_s^2Q_s^2+\Pi_s^2)/2$ and
$\psi_s=(\omega_sQ_s+i\Pi_s)/\sqrt2$.
The trajectory $Q_s=A\cos\omega_st$,
$\Pi_s=-\omega_sA\sin\omega_st$ gives
$\psi_s=\omega_sA e^{-i\omega_st}/\sqrt2$.
Figure~\ref{fig:classical-energy-amplitude-map} shows the resulting clockwise
rotation and radius $\sqrt{E_s}$. At a fixed clock a phase rotation of
$\psi_s$ changes the real displacement and momentum preparation.
This invertible encoding preserves the number of real mechanical
degrees of freedom and retains the absolute energy scale.

The Hamiltonian and the frequency generator now have distinct, precise
jobs. In a real normal-mode basis, the inherited canonical Poisson
bracket gives
\begin{equation}
 \begin{split}
 H_{\rm harm}&=\sum_s|\psi_s|^2,\qquad
 \{\psi_s,\psi_t^*\}=-i\omega_s\delta_{st},\\
 \Dt{\psi_s}&=\{\psi_s,H_{\rm harm}\}=-i\omega_s\psi_s.
 \end{split}
 \label{eq:complex-classical-hamiltonian-bracket}
\end{equation}
The bracket follows by differentiating the real-coordinate definition of
$\psi_s$ using $\{Q_s,\Pi_t\}=\delta_{st}$.
$H_{\rm harm}$ is an energy function on classical phase space;
$\mathsf W$ in the first-order equation is a self-adjoint operator with
frequency units. The bracket connects these two roles. This first-order form expresses classical Hamiltonian evolution in
energy coordinates. The separate quantization postulate supplies the
occupation space and its measurement probabilities.

\begin{definition}[Classical action amplitude]
\label{def:s5-hamiltonian-action-amplitude}
For the fixed positive conservative reference, the \emph{action amplitude}
is the classical column
\begin{equation}
 \begin{aligned}
 \vct\alpha_{\rm cl}
   &=\mathsf W^{-1/2}\psi
     =\frac{\mathsf W^{1/2}\vct x
                    +i\mathsf W^{-1/2}\vct v}{\sqrt2},\\
 \mathcal I_{\rm tot}
   &:=\vct\alpha_{\rm cl}^{\dagger}\vct\alpha_{\rm cl}
     =\sum_s\frac{E_s}{\omega_s}.
 \end{aligned}
 \label{eq:s5-hamiltonian-action-amplitude}
\end{equation}
Its components have units $\sqrt{\mathrm{J\,s}}$, and
$\mathcal I_{\rm tot}$ is the sum of the conservative modal actions.
\end{definition}

This normalization compares excitation relative to frequency. The energy
column $\psi$ gives equal squared radius to equal energies; the action
column gives equal squared radius to equal values of $E_s/\omega_s$.
A softer collective motion therefore carries more action at the same
energy. Both columns encode the same real particle positions and momenta
on this positive sector.

\begin{proposition}[Canonical action normalization]
\label{prop:s5-hamiltonian-action-normalization}
In any fixed real orthonormal basis of the mass-weighted coordinates,
the inherited canonical brackets and the harmonic energy satisfy
\begin{equation}
 \begin{aligned}
 \{\alpha_{{\rm cl},j},\alpha_{{\rm cl},k}^*\}
       &=-i\delta_{jk},
 &\{\alpha_{{\rm cl},j},\alpha_{{\rm cl},k}\}&=0,\\
 H_{\rm harm}&=\vct\alpha_{\rm cl}^{\dagger}
                         \mathsf W\vct\alpha_{\rm cl},
 &i\Dt{\vct\alpha_{\rm cl}}&=\mathsf W\vct\alpha_{\rm cl}.
 \end{aligned}
 \label{eq:s5-hamiltonian-action-bracket}
\end{equation}
Consequently both $H_{\rm harm}$ and $\mathcal I_{\rm tot}$ are conserved
under this unforced, time-independent harmonic evolution.
\end{proposition}
\begin{proof}
The real canonical brackets are $\{x_j,v_k\}=\delta_{jk}$.
Differentiating the displayed linear transformation gives
$\{\psi_j,\psi_k^*\}=-iW_{jk}$ and
$\{\psi_j,\psi_k\}=0$. Multiplication by the real symmetric
$\mathsf W^{-1/2}$ on both indices gives the stated action brackets.
Substitution of $\psi=\mathsf W^{1/2}\vct\alpha_{\rm cl}$ into
$H_{\rm harm}=\psi^\dagger\psi$ gives its quadratic expression;
Hamilton's equation or the fixed transformation of $\dot\psi$ gives
the evolution. Self-adjointness of $\mathsf W$ then preserves both
quadratic forms.
\end{proof}

Transforming the Hamiltonian and Poisson structure together preserves
classical dynamics~\cite[Sections~2.2 and~2.6--2.8]{Deriglazov2010}.
Energy and action amplitudes are two coordinates for that same state.
Converting either to measured centre motion uses the masses, shapes
and detector calibration. The quantum correspondence below specifies
its occupation-space normalization separately.

The doubled real state also has a precise complex structure.

\begin{definition}[Complex structure of equal-unit phase space]
\label{def:s5-energy-complex-structure}
On the doubled equal-unit energy coordinates, define the quarter-turn
and its complexified projections by
\begin{equation}
 \mathsf J=\begin{pmatrix}0&-\Id\\\Id&0\end{pmatrix},\qquad
 \mathsf J^2=-\Id,\qquad
 \mathsf\Pi_\pm=\tfrac12(\Id\mp i\mathsf J).
 \label{eq:energy-phase-space-complex-structure}
\end{equation}
\end{definition}
\begin{proposition}[Reality-paired conservative frequencies]
\label{prop:s5-reality-paired-frequencies}
For the fixed positive conservative generator,
$\mathsf\Pi_\pm$ project onto the positive and negative spectral parts
of $i\mathsf A_Y$. The unitary map
$Y\mapsto(Y_1+iY_2,Y_1-iY_2)/\sqrt2$ sends this operator to
$\diag(\mathsf W,-\mathsf W)$. For real $Y$, the new coordinates are
$(\psi,\psi^*)$.
\end{proposition}
\begin{proof}
The real skew-adjoint matrix $\mathsf J$ has complexified eigenvalues
$\pm i$, giving the stated orthogonal projections.
Since $\mathsf A_Y=-\mathsf J\diag(\mathsf W,\mathsf W)$, its two
frequency blocks are $\mathsf W$ and $-\mathsf W$. Substituting real
$Y_1,Y_2$ gives conjugate coordinate columns.
\end{proof}

The negative-frequency copy is fixed by the reality of the mechanical
state. It carries the same particle data in the conjugate representation.

Figure~\ref{fig:complex-conjugate-frequency-pair} shows the paired frequencies and the reality constraint in the single-mode case.

\begin{figure}[H]
\centering\begingroup
\definecolor{vlfGreen}{HTML}{176B64}
\definecolor{vlfDarkRed}{HTML}{762B38}
\definecolor{vlfRed}{HTML}{762B38}
\definecolor{vlfBlue}{HTML}{315D80}
\definecolor{vlfCopper}{HTML}{762B38}
\setlength{\unitlength}{1pt}
\scalebox{0.80}{%
\input{vlt-figures/figure-164.tex}%
}\endgroup
\caption{Positive and negative frequency copies can encode one real conservative oscillator. For one positive mode with $E_\star>0$, take $\psi(t)=\sqrt{E_\star}e^{-i\theta}$ and $\theta=\omega_st$. (a) The energy amplitude rotates clockwise; (b) its conjugate rotates counterclockwise. Both panels have equal axis scales. Filled dots mark $\theta=0$ and open dots mark $\theta=\pi/2$. The paired first-order equations are $i\partial_t(\psi,\psi^*)^{\mathsf T}=\diag(\omega_s,-\omega_s)(\psi,\psi^*)^{\mathsf T}$. The negative frequency is the conjugate needed to reconstruct a real history, not a negative mechanical energy or an independently prepared second oscillator. The unitary change from $Y=(\omega_sQ_s,\Pi_s)$ to $(\psi,\psi^*)$ preserves the doubled norm $\|Y\|^2=|\psi|^2+|\psi^*|^2=2E_\star$; the single-copy energy remains $|\psi|^2=E_\star$. This illustrates \eqref{eq:energy-phase-space-complex-structure} without imposing quantum-state normalization. The lower reconstruction joins both constrained copies to a single ideal-virion displacement coordinate. Connecting arrows denote reconstruction, not force or energy transfer.}
\label{fig:complex-conjugate-frequency-pair}
\end{figure}

For free conservative finite-mode motion, the positive-frequency
displacement signal has the exact relation
\begin{equation}
 \vct z_{\mathrm{disp}}(t)
       =\sqrt2\mathsf M^{-1/2}\mathsf W^{-1}\ket{\psi(t)},\qquad
 E_+=\tfrac12\|\mathsf W\mathsf M^{1/2}
                           \vct z_{\mathrm{disp}}\|^2.
 \label{eq:conservative-signal-energy-map}
\end{equation}
This uses the real conservative mode basis and its positive sector.
The mass and frequency factors explain why equal displacement amplitude
need not mean equal energy. For an isolated mode read as $y=h_sQ_s$,
$h_s\ne0$, its effective inertia is $m_s^{\rm eff}=|h_s|^{-2}$;
a displacement amplitude $A_s$ then has
$E_s=m_s^{\rm eff}\omega_s^2|A_s|^2/2$.
The inertia is specific to the mode and readout. A mode with $h_s=0$
is invisible and has no unit-readout normalization. A softer mode can
give a larger excursion at the same energy, possibly leaving the
harmonic basin. Driven motion or material memory does not in general
obey this instantaneous signal-to-energy identity.

The energy representation also makes a useful inverse statement possible.
When only part of the motion is measured, one can calculate the least
harmonic energy compatible with those calibrated coordinates. This is
a state constraint within a specified mechanical model.

\begin{proposition}[Least energy compatible with a mechanical readout]
\label{prop:s5-hamiltonian-readout-energy}
Let $\mathsf K>0$ and $\mathsf M>0$ on the selected finite displacement
space. Let the real dimensionless matrix $\mathsf P_{\rm obs}$ have independent
columns, and suppose the simultaneous displacement and velocity readouts
are $\vct y=\mathsf P_{\rm obs}^{\mathsf T}\vct u$ and
$\vct w=\mathsf P_{\rm obs}^{\mathsf T}\dot{\vct u}$. Define
\begin{equation}
 \mathsf C_K=\mathsf P_{\rm obs}^{\mathsf T}\mathsf K^{-1}\mathsf P_{\rm obs},
 \qquad
 \mathsf C_M=\mathsf P_{\rm obs}^{\mathsf T}\mathsf M^{-1}\mathsf P_{\rm obs}.
 \label{eq:s5-hamiltonian-readout-gramians}
\end{equation}
The harmonic energy at that instant obeys the sharp bound
\begin{equation}
 \begin{aligned}
 E_{\rm harm}&\geq E_{\rm min}(\vct y,\vct w)
       :=\tfrac12\vct y^{\mathsf T}\mathsf C_K^{-1}\vct y
          +\tfrac12\vct w^{\mathsf T}\mathsf C_M^{-1}\vct w,\\
 \vct u_{\rm min}&=\mathsf K^{-1}\mathsf P_{\rm obs}\mathsf C_K^{-1}\vct y,
 &\vct p_{\rm min}&=\mathsf P_{\rm obs}\mathsf C_M^{-1}\vct w.
 \end{aligned}
 \label{eq:s5-hamiltonian-readout-energy-bound}
\end{equation}
The displayed displacement and momentum attain equality.
\end{proposition}
\begin{proof}
Both Gram matrices are positive definite by the independent-column
assumption. Write $\vct u=\vct u_{\rm min}+\vct r_u$ and
$\vct p=\vct p_{\rm min}+\vct r_p$. The measured constraints imply
$\mathsf P_{\rm obs}^{\mathsf T}\vct r_u=0$ and
$\mathsf P_{\rm obs}^{\mathsf T}\mathsf M^{-1}\vct r_p=0$.
The elastic and kinetic cross terms therefore vanish, leaving
\begin{equation}
 E_{\rm harm}=E_{\rm min}
       +\tfrac12\vct r_u^{\mathsf T}\mathsf K\vct r_u
       +\tfrac12\vct r_p^{\mathsf T}\mathsf M^{-1}\vct r_p.
 \label{eq:s5-hamiltonian-hidden-energy-residual}
\end{equation}
The residual terms are nonnegative and vanish at the stated state.
\end{proof}

$\mathsf C_K$ is a calibrated static compliance matrix, with units
$\mathrm{m\,N^{-1}}$, while $\mathsf C_M$ has inverse-mass units.
Each term of the bound is consequently an energy. For one scalar
readout, its two terms are $y^2/(2C_K)$ and $w^2/(2C_M)$.
For the equal-mass supported pair of
Figure~\ref{fig:s4-inverse-common-relative}, a normalized common
coordinate has $C_K=1/k_0$ and $C_M=1/m$; a normalized relative
coordinate has $C_K=1/(k_0+2k_c)$ and the same $C_M$.
Thus the same measured excursion requires different minimum restoring
energy depending on whether the particles move together or change
their separation. The assigned charges and pair curvatures enter this
comparison through the assembled $k_c$.

For tracked ideal-virion centres, the bound quantifies the mechanical
storage certified by the resolved displacement and velocity. Extra
motion in the two residual subspaces adds hidden energy while leaving
those instantaneous readouts unchanged. A position-only record supplies
the elastic term of the bound; reconstructing a velocity requires the
time resolution and noise treatment of the observation model.
Unmeasured medium strain can add further storage in the stated passive
branch realization. Recovering total particle--medium energy therefore
uses the independently calibrated mechanical law and the state variables
that carry that storage. An upper energy estimate additionally needs
information about unresolved motion or admissible preparation sizes.
This relates the inverse problem to variable selection: a readout can
certify part of the energy while a different preparation or a new
detector is needed to resolve the remaining state.

The continuum energy metric on the positive sector is
\begin{equation}
 \begin{split}
 \langle(\vct u,\vct v),(\vct w,\vct z)\rangle_E
  &=\langle\mathcal W\vct u,\mathcal W\vct w\rangle_\rho
                      +\langle\vct v,\vct z\rangle_\rho,\\
 \psi_\rho&=\frac{\mathcal W\vct u+i\Pdt{\vct u}}{\sqrt2}.
 \end{split}
 \label{eq:complex-continuum-energy-metric}
\end{equation}
The state norm squared is twice harmonic energy, whereas
$\|\psi_\rho\|_\rho^2$ equals energy. The field $\psi_\rho$ has velocity
units; density and volume supply its energy norm. Multiplication by
$\sqrt\rho$ gives the equivalent unweighted $L^2$ field. Spatial
storage still uses the energy allocation of Section~4, since the
spectral square root can couple distant locations.

Self-adjoint $\mathcal W$ generates the unitary group
$e^{-it\mathcal W}$~\cite{Teschl2009}. Unitary means inner products,
and hence the energy norm, are preserved. The equation holds strongly
for energy amplitudes in $D(\mathcal W)$ and as a mild unitary evolution
for general finite-energy data. Strong evolution permits the time
derivative in the space; mild evolution specifies the solution by the
propagator without demanding that derivative for every initial field.
The doubled equal-unit generator has domain
$D(\mathcal W)\oplus D(\mathcal W)$. The bounded-domain positive gap
makes displacement reconstruction continuous; that control deteriorates
if the gap closes. Zero-frequency translations retain their separate
free-coordinate description.

The positive-sector restriction has an explicit mechanical meaning.
In the finite reference let $\mathsf P_0$ be the orthogonal projector
onto $\ker\mathsf D$, with $\mathsf P_+=\Id-\mathsf P_0$.
The unforced zero sector and a positive sector with
$\mathsf W|_+\geq\omega_{\rm gap}\Id>0$ obey
\begin{equation}
 \begin{aligned}
 \vct x_0(t)&=\vct x_0(0)+t\vct v_0(0),
 &E_0&=\tfrac12\|\vct v_0(0)\|^2,\\
 \|\vct x_+(t)\|&\leq\frac{\sqrt{2E_+}}{\omega_{\rm gap}},
 &\|\vct v_+(t)\|&\leq\sqrt{2E_+}.
 \end{aligned}
 \label{eq:s5-hamiltonian-zero-and-gap-reconstruction}
\end{equation}
These follow respectively by projecting $\ddot{\vct x}=-\mathsf D\vct x$
and by the spectral lower bound in the positive energy. A permitted
rigid translation is a free coordinate: its initial position and
velocity remain part of the mechanical preparation. The energy
amplitude on that zero sector would keep velocity and lose its initial
position, which is why the positive amplitude and the free coordinate
are retained separately. In the cylinder the actual constraint and
support laws decide whether such a translation is admissible.

The gap quantifies control of displacement by energy. Softening a
support or approaching a mechanism permits larger excursions at the
same energy. Reconstructed centres and pair clearances therefore test
the harmonic basin. Without a positive gap, displacement reconstruction
uses its specified spectral domain or additional low-frequency data.

For a quantum phonon state, a different norm is prescribed. A normalized
occupation ket $\ket\Phi=\sum_s c_s\ket{1_s}$ has
$\sum_s|c_s|^2=1$, while its energy above the vacuum is
$\sum_s\hbar\omega_s|c_s|^2$. The norm expresses probability, not that
energy. A common phase multiplying this quantum ket leaves its measurement
statistics unchanged. Only after the quantization postulate and this occupation-space
identification does the one-phonon excitation Hamiltonian become
$\hbar\mathsf W$. The full oscillator Hamiltonian includes all integer
occupations and the stated zero-point terms. Its relation to classical
first moments and displacement statistics belongs to the quantum
construction, not to a choice of complex notation~\cite{Shankar1994}.

\paragraph{A changing stiffness and its work account.}
The preceding unitary energy-amplitude law uses a fixed conservative
reference. A controlled change of a support or an effective interaction
curvature changes the energy stored in a given displacement. Its work
can be retained exactly in the same coordinates.

\begin{proposition}[Energy amplitudes for a controlled positive stiffness]
\label{prop:s5-hamiltonian-parametric-work}
Keep the coordinate reference and $\mathsf M$ fixed, let
$\mathsf K(t)>0$ be continuously differentiable on the chosen finite
space, and define $\mathsf D(t)=\mathsf M^{-1/2}\mathsf K(t)\mathsf M^{-1/2}$
and $\mathsf W(t)=\mathsf D(t)^{1/2}$. For the real driven dynamics
$\ddot{\vct x}+\mathsf D(t)\vct x=\vct f_x(t)$, the instantaneous
energy amplitude $\psi=(\mathsf W(t)\vct x+i\vct v)/\sqrt2$ obeys
\begin{equation}
 \begin{aligned}
 \Dt{\psi}
  &=-i\mathsf W\psi
     +\tfrac12\dot{\mathsf W}\mathsf W^{-1}(\psi+\psi^*)
     +\frac{i}{\sqrt2}\vct f_x,\\
 \Dt{E_{\rm harm}}
  &=\vct v^{\mathsf T}\vct f_x
       +\tfrac12\vct x^{\mathsf T}\dot{\mathsf D}\vct x,
 \qquad E_{\rm harm}=\psi^\dagger\psi.
 \end{aligned}
 \label{eq:s5-hamiltonian-parametric-amplitude-work}
\end{equation}
The derivative of the square root is specified uniquely by
\begin{equation}
 \mathsf W\dot{\mathsf W}+\dot{\mathsf W}\mathsf W
       =\dot{\mathsf D}.
 \label{eq:s5-hamiltonian-square-root-derivative}
\end{equation}
No commutation between the stiffness and its rate is assumed.
\end{proposition}
\begin{proof}
Differentiate the definition of $\psi$, substitute
$\dot{\vct v}=-\mathsf D\vct x+\vct f_x$ and
$\vct x=\mathsf W^{-1}(\psi+\psi^*)/\sqrt2$.
Differentiating the real kinetic and elastic quadratic forms gives
the work identity. Finally differentiate $\mathsf W^2=\mathsf D$.
In an instantaneous eigenbasis of $\mathsf W$, the last equation gives
$(\dot W)_{st}=(\dot D)_{st}/(\omega_s+\omega_t)$;
positive eigenfrequencies make this solution unique.
\end{proof}

Applied force supplies the first power term. Changing the restoring
law at fixed displacement supplies the second, parametric term.
Both are measured in watts. For one mode,
$\dot\psi_s=-i\omega_s\psi_s+
(\dot\omega_s/2\omega_s)(\psi_s+\psi_s^*)+if_{x,s}/\sqrt2$.
The conjugate term retains the changing normalization and stiffness
work. The matrix equation also handles changing eigendirections.
The transformed generator must accompany a time-dependent coordinate
map~\cite[Sections~2.6 and~3.2]{Deriglazov2010}.

A curvature-changing variable needs an evolution and energy law when
its future is part of the prediction. A prescribed stiffness instead
enters as a control history with its work. This choice specifies
the physical inputs independently of the complex representation.

\paragraph{Registering an instantaneous energy amplitude.}
A recruitment record can select a new cluster, or a new reference can
describe the same moving population. In either case the physical
positions, momenta and retained material strains provide the starting
state. Section~\ref{sec:s4-transient-reference-continuation} gives its
real-coordinate registration. The corresponding energy amplitude has
an additional dependence on the restoring operator used to measure
displacement energy.

\begin{proposition}[Finite registration between positive energy references]
\label{prop:s5-reference-energy-registration}
Use the same full-pool mass matrix, laboratory axes and fixed positive
coordinate sector in two harmonic descriptions $g$ and $h$. Let
$\vct R_g,\vct R_h$ be their reference centres,
$\vct x_g=\mathsf M^{1/2}(\vct r-\vct R_g)$,
$\vct v=\mathsf M^{1/2}\dot{\vct r}$ and
$\mathsf W_g=\mathsf D_g^{1/2}>0$, with the analogous definitions for
$h$. Define
\begin{equation}
 \begin{aligned}
 \vct b_{hg}&=\mathsf M^{1/2}(\vct R_g-\vct R_h),&
 \mathsf T_{hg}&=\mathsf W_h\mathsf W_g^{-1},\\
 \mathsf A_{hg}&=\tfrac12(\mathsf T_{hg}+\Id),&
 \mathsf B_{hg}&=\tfrac12(\mathsf T_{hg}-\Id),\qquad
 \vct c_{hg}=\mathsf W_h\vct b_{hg}/\sqrt2.
 \end{aligned}
 \label{eq:s5-reference-energy-registration-data}
\end{equation}
For continuous physical position and velocity, their exact registration is
\begin{equation}
 \boxed{\displaystyle
 \psi_h=\mathsf A_{hg}\psi_g+\mathsf B_{hg}\psi_g^*
                                      +\vct c_{hg}.}
 \label{eq:s5-reference-energy-registration}
\end{equation}
It is an invertible affine map of the real displacement--velocity
state. Its linear part couples $\psi_g$ to its complex conjugate
whenever the restoring normalization changes.
\end{proposition}
\begin{proof}
Use $\vct x_h=\vct x_g+\vct b_{hg}$ together with
$\mathsf W_g\vct x_g=(\psi_g+\psi_g^*)/\sqrt2$ and
$i\vct v=(\psi_g-\psi_g^*)/\sqrt2$ in
$\psi_h=(\mathsf W_h\vct x_h+i\vct v)/\sqrt2$.
In real and imaginary quadratures the linear map is
$\diag(\mathsf T_{hg},\Id)$, which is invertible on the stated
positive sector. No commutation of $\mathsf W_g$ and $\mathsf W_h$
is needed.
\end{proof}

This registration encodes the instantaneous physical state with its
restoring metric. Displacement frames have their own mass-overlap map
below. Near a soft mode, retain displacement and momentum directly
and compare calibrated readouts. Zero and newly retained directions
remain explicit; truncated directions carry a residual estimate.
Medium strains continue under Section~4's event law.

The norm of a registered amplitude must be interpreted using the
declared event. Re-expanding a single nonlinear potential about a new
reference changes its local quadratic approximation. The reference
energy, linear term and Taylor remainder in
\eqref{eq:s4-transient-taylor-ledger} complete that account. A genuine
externally controlled jump $\mathsf K_g\to\mathsf K_h$ at fixed
reference, continuous displacement and continuous momentum instead
supplies the work
\begin{equation}
 \Delta W_{\rm ctrl}
       =\tfrac12\vct u^{\mathsf T}(\mathsf K_h-\mathsf K_g)\vct u.
 \label{eq:s5-reference-stiffness-jump-work}
\end{equation}
For a single mass-weighted oscillator with
$\eta=\omega_h/\omega_g$ and
$\psi_g=\sqrt{E_g}\,e^{-i\theta}$, this gives
\begin{equation}
 \frac{E_h}{E_g}=\eta^2\cos^2\theta+\sin^2\theta,\qquad
 \frac{\Delta W_{\rm ctrl}}{E_g}
                         =(\eta^2-1)\cos^2\theta.
 \label{eq:s5-reference-phase-dependent-jump}
\end{equation}
Stiffening at a displacement extremum supplies the greatest work;
stiffening at zero displacement supplies none at that instant. Thus
phase can affect energy accommodation during a controlled mechanical
change. A spontaneous joining or breakup event requires its actual
interaction landscape, material evolution and energy balance from
\eqref{eq:s4-transient-event-first-law}; the controlled jump is a
worked limiting case. The conjugate term in the classical registration
does not by itself assign quantum occupation or a phonon-emission rate.

Figure~\ref{fig:s5-energy-reference-registration} reconstructs the affine registration and contrasts it with a controlled change that supplies work.

\begin{figure}[H]
\centering\begingroup
\definecolor{vltTeal}{HTML}{176B64}
\definecolor{vltBlue}{HTML}{315D80}
\definecolor{vltGold}{HTML}{A87520}
\definecolor{vltPlum}{HTML}{725877}
\definecolor{vltRed}{HTML}{762B38}
\definecolor{vltYellow}{HTML}{D5AD3C}
\definecolor{vltNeutral}{HTML}{777D83}
\setlength{\unitlength}{1pt}
\scalebox{0.90}{%
\input{vlt-figures/figure-165.tex}%
}\endgroup
\caption{Instantaneous registration of a classical energy amplitude, illustrating the boxed map \eqref{eq:s5-reference-energy-registration}. (a,b) In scaled scalar coordinates, the same physical states have $\omega_g=1$, $\omega_h=2$ and reference offset $b_{hg}=0.3$. The unit circle in $\psi_g$ maps to a displaced ellipse in $\psi_h$; matching points identify identical position--velocity preparations. Both panels use equal Cartesian axis scales. The real part is rescaled and shifted while the imaginary part is retained, producing both the conjugate and affine terms. These are instantaneous encodings, not trajectories under two simultaneous dynamics. (c) A separately declared physical stiffness change has $b=0$, $k_h/k_g=4$ and $\Delta W_{\rm ctrl}/E_g=3\cos^2\theta$. Displacement arrows are teal; the burgundy velocity at the zero crossing $\theta=\pi/2$ is negative. A change of reference alone supplies no work; the controlled change uses \eqref{eq:s5-reference-stiffness-jump-work}, and a spontaneous association needs its actual event ledger. All plotted parameters are illustrative scaled values.}
\label{fig:s5-energy-reference-registration}
\end{figure}

Passive reduction changes the evolution while leaving the mathematical
space available. The augmented generator is dissipative in its full
storage metric; under the assumptions of Section~4 its evolution is a
contraction there. The linear real model has the complexified mild law
\begin{equation}
 X_{\mathrm C}(t)=T(t)X_{\mathrm C}(0)
       +\int_0^tT(t-s)F_{\mathrm C}(s)\,\dd s,\qquad
 T(t)(X_{\mathrm R}+iX_{\mathrm I})
                         =T(t)X_{\mathrm R}+iT(t)X_{\mathrm I}.
 \label{eq:complexified-mild-evolution}
\end{equation}
Here $F_{\rm C}$ is the augmented input column. Complexification sums
the storage forms of two real copies, and damping acts on both. A
classical amplitude need not preserve its norm under passive loss;
energy transfers to retained or unresolved receiver coordinates. This
is different from probability normalization of a quantum density
operator, which a consistent reduced quantum model must preserve.

The particle energy amplitude can still track exchange but generally
does not close by itself. On a positive sector invariant under the
constitutive maps, set
$\mathsf C_x=\mathsf M^{-1/2}\mathsf C_0\mathsf M^{-1/2}$,
$\mathsf R_a=\mathsf M^{-1/2}\mathsf G_a^{\mathsf T}\mathsf H_a$,
and $\vct f_x=\mathsf M^{-1/2}\vct f$.

\begin{proposition}[Exact energy amplitudes with retained Maxwell states]
\label{prop:s5-exact-maxwell-energy-amplitudes}
For the real linear Maxwell realization with fixed conservative
$\mathsf W>0$ on this invariant positive sector, the particle energy
amplitude and prepared branch coordinates obey
\begin{align}
 \Dt{\psi}&=-i\mathsf W\psi
       -\tfrac12\mathsf C_x(\psi-\psi^*)
       +\frac{i}{\sqrt2}\left(\vct f_x-\sum_a\mathsf R_a\vct z_a\right),
       \label{eq:matrix-damped-energy-amplitude}\\
 \Dt{\vct z_a}&=-\frac{i}{\sqrt2}\mathsf G_a\mathsf M^{-1/2}
                         (\psi-\psi^*)-\frac{\vct z_a}{\tau_a}.
       \label{eq:branch-energy-amplitude-coupling}
\end{align}
\end{proposition}
\begin{proof}
Differentiate $\psi=(\mathsf W\vct x+i\vct v)/\sqrt2$ and use the
particle equation including drag and branch force.
The identity $\vct v=(\psi-\psi^*)/(i\sqrt2)$ gives the drag term
and the corresponding substitution in each branch equation.
\end{proof}

The difference $\psi-\psi^*$ reconstructs velocity, so exact mechanics
retains the conjugate and prepared branch states. Equal particle
amplitudes can evolve differently at different branch strains.
Full augmented storage contracts, while particle energy can receive
returned branch energy. Zero-sector couplings retain their coordinates.
The construction allows noncommuting damping and frequency matrices.

Dropping conjugate couplings requires a justified slow-amplitude
approximation. In an interaction picture they oscillate at sums
$\omega_s+\omega_t$, so loss rates must be small relative to those
sums. For memory, reactive shifts and prepared forcing must also be
weak and narrow-band or included in a larger conservative reference;
a slowly relaxing branch can supply small loss but large storage.
Independent-mode reduction further requires separation of frequency
differences and must retain degenerate blocks. These classical reductions
do not specify quantum reservoir noise. Nor does the augmented memory
generator automatically share the conservative elastic operator's
compact resolvent or complete orthogonal mode basis.

The exact damped example in Figure~\ref{fig:complex-damped-amplitude-conjugation} shows why the conjugate amplitude cannot generally be omitted when passing from real mechanics to complex coordinates.

\begin{figure}[H]
\centering\begingroup
\definecolor{vlfGreen}{HTML}{176B64}
\definecolor{vlfDarkRed}{HTML}{762B38}
\definecolor{vlfRed}{HTML}{762B38}
\definecolor{vlfBlue}{HTML}{315D80}
\definecolor{vlfCopper}{HTML}{762B38}
\setlength{\unitlength}{1pt}
\scalebox{0.80}{%
\input{vlt-figures/figure-166.tex}%
}\endgroup
\caption{The damped oscillator has $d=\gamma/\omega_0=0.2$, $\tau=\omega_0t$, $b=\sqrt{1-d^2}$ and $z=(\omega_0Q+i\Pi)/\sqrt{2E_\star}$. Its exact propagator is $T_\tau z_0=A z_0+Bz_0^*$, with $A=e^{-d\tau}[\cos(b\tau)-i\sin(b\tau)/b]$ and $B=e^{-d\tau}d\sin(b\tau)/b$. Equal-energy initial values $1,i$ give different spirals on shared scales. The defect at right is $2|B|$. Lower supported-virion sketches reconstruct positive displacement at rest and reference displacement with positive velocity. Green marks displacement, burgundy velocity; the drag retained in the curves is omitted from these sketches.}
\label{fig:complex-damped-amplitude-conjugation}
\end{figure}

\paragraph{Energy amplitudes with coupled surface and material storage.}
The body/surface realization also admits an exact energy representation.
Use the resistance-whitened $\vct\chi'$, $\mathsf H'$, $\mathsf B'$
and $\mathsf C_{\rm e}=\mathsf C-\mathsf R\mathsf D^{-1}\mathsf R^{\mathsf T}$
of~\eqref{eq:s4patch-resistance-whitening}, with no independent
chemical drive in this calculation. On a fixed positive sector define
\begin{equation}
 \begin{aligned}
 \mathsf H_{\rm r}&=\mathsf H'-\mathsf B'\mathsf A^{-1}\mathsf B'^{\mathsf T}
           \succ0,\qquad
 \vct r=\vct\chi'+\mathsf A^{-1}\mathsf B'^{\mathsf T}\vct X,\\
 \vct x&=\mathsf G_X^{1/2}\vct X,\quad \vct v=\dot{\vct x},\quad
 \mathsf W=(\mathsf G_X^{-1/2}\mathsf H_{\rm r}\mathsf G_X^{-1/2})^{1/2},
 \qquad \psi=(\mathsf W\vct x+i\vct v)/\sqrt2 .
 \end{aligned}
 \label{eq:s5patch-coupled-energy-reference}
\end{equation}
The completed square assigns a real mechanical curvature and a residual
internal displacement $\vct r$. Conserved inventories and their holding
bias are restricted and shifted first, as in Section~4; inverses here
act on the independent relaxing coordinates. A frozen reference can
likewise be chosen explicitly. A frequency-dependent complex reduced
stiffness is used in response, while this $\mathsf W$ uses the declared
real positive conservative curvature.

\begin{proposition}[Exact body--surface energy amplitudes]
\label{prop:s5patch-coupled-energy-amplitudes}
Put $\mathsf C_x=\mathsf G_X^{-1/2}\mathsf C_{\rm e}\mathsf G_X^{-1/2}$,
$\mathsf B_x=\mathsf G_X^{-1/2}\mathsf B'$ and
$\vct f_x=\mathsf G_X^{-1/2}\vct f_X$. The deterministic reciprocal
linear realization is equivalently
\begin{equation}
 \begin{aligned}
 \dot\psi&=-i\mathsf W\psi-\tfrac12\mathsf C_x(\psi-\psi^*)
                         +\frac{i}{\sqrt2}(\vct f_x-\mathsf B_x\vct r),\\
 \dot{\vct r}&=-\mathsf D^{-1}\mathsf A\vct r
       -\frac{i}{\sqrt2}\mathsf A^{-1}\mathsf B'^{\mathsf T}
                      \mathsf G_X^{-1/2}(\psi-\psi^*) .
 \end{aligned}
 \label{eq:s5patch-coupled-amplitude-dynamics}
\end{equation}
Its storage and supply are
\begin{equation}
 \boxed{\begin{aligned}
 E&=\psi^\dagger\psi+\tfrac12\vct r^{\mathsf T}\mathsf A\vct r,\\
 \dot E&=\dot{\vct X}^{\mathsf T}\vct f_X
   -\dot{\vct X}^{\mathsf T}\mathsf C_{\rm e}\dot{\vct X}
   -\vct r^{\mathsf T}\mathsf A\mathsf D^{-1}\mathsf A\vct r .
 \end{aligned}}
 \label{eq:s5patch-coupled-amplitude-storage}
\end{equation}
\end{proposition}
\begin{proof}
Completing the energy square gives the first line. The whitened state
equations become
$\mathsf G_X\ddot{\vct X}+\mathsf C_{\rm e}\dot{\vct X}
+\mathsf H_{\rm r}\vct X+\mathsf B'\vct r=\vct f_X$ and
$\dot{\vct r}=-\mathsf D^{-1}\mathsf A\vct r+
\mathsf A^{-1}\mathsf B'^{\mathsf T}\dot{\vct X}$.
Substitute $\vct v=(\psi-\psi^*)/(i\sqrt2)$ to obtain the amplitude
equations. The reciprocal exchange terms cancel in $\dot E$.
\end{proof}

Figure~\ref{fig:cmd-mixed-storage-certificate} gives a graphical specialization of this boxed relation.

This represents translational and angular mechanics together with
first-order surface/material relaxation. An internal preparation can
return energy to particle motion while total storage decreases. Its
fluctuation realization uses the whitened matched noise and the prepared
covariance of Section~4. Physical surface reporters are transformed
with $\vct\chi=\vct\chi'-\mathsf D^{-1}\mathsf R^{\mathsf T}\vct X$.
The state and readout change together under this coordinate choice.

Self-excitation gives a direct nonlinear test of the same distinction.
For the \emph{instantaneous} finite-resource model in Section~\ref{sec:self-excitation-passive-boundary}, define
the same $\psi_s$ and put
$\delta_s(\mathcal E_{\rm a})=\gamma_{\rm a}(\mathcal E_{\rm a})
-\gamma_{\rm p}$. Its exact classical equations become
\begin{equation}
 \begin{split}
 \Dt{\psi_s}&=-i\omega_s\psi_s+
    \left[\delta_s(\mathcal E_{\rm a})
      -\frac{\gamma_{\rm n}(\psi_s+\psi_s^*)^2}
                         {2\omega_s^2Q_c^2}\right](\psi_s-\psi_s^*),\\
 \Dt{\mathcal E_{\rm a}}&=-\gamma_{\rm a}(\mathcal E_{\rm a})
                                      |\psi_s-\psi_s^*|^2.
 \end{split}
 \label{eq:complex-active-energy-amplitude}
\end{equation}
Because $\psi_s+\psi_s^*=\sqrt2\omega_sQ_s$ and
$\psi_s-\psi_s^*=i\sqrt2\Pi_s$, this is an invertible rewriting of
the real model. It introduces no quantum self-excitation postulate.
$|\psi_s|^2$ remains the instantaneous harmonic mode energy, but its
derivative is the supplied work minus loss of
\eqref{eq:self-excitation-power-balance}. Gain, saturation, and resource
depletion therefore require $\psi_s^*$ and the resource state, rather
than a closed complex-linear unitary equation. For the separately
retained-force model, $F_{{\rm a},s}$ remains an additional state instead;
its delayed gain cannot be counted again as instantaneous gain.

A biophysical active realization specifies the host converter,
attachment and resource pathway supplying work. The complex amplitude
organizes the selected motion while these added states predict its
energy exchange. A maintained cycle can exchange energy throughout a
period; a finite-resource episode carries depletion in its full state.
The representation change preserves the variables established in
Section~\ref{sec:self-excitation-passive-boundary}.

Finally, real linear time-domain laws pair their temporal poles as
$\mathsf P(-z^*)=\mathsf P(z)^*$. A pole $z=\Omega-i\Gamma$ is paired
with $-z^*=-\Omega-i\Gamma$, with the same decay rate; their combination
reconstructs real damped motion. Conjugation without frequency reversal
would give the wrong causal partner. With the state, units, and evolution
now specified, the next subsection can ask which phase patterns redistribute
energy and which of them a chosen mechanical observable reveals.

\subsection{Phase, mechanical observables, and energy localization}
\label{sec:complex-field-observables}

An observable selects a particle centre, surface marker, internal
projection or spatially weighted signal. The readout map from
Section~\ref{sec:complex-field-hilbert-space} determines how each
pattern appears there. We now follow its mean, fluctuations, energy
and relative phase across specified preparations.

For a harmonic translational phasor, the reconstructed quantities are
$\vct r_i(t)=\vct R_i+\operatorname{Re}[\vct U_i e^{-i\omega t}]$
and $\vct v_i(t)=\operatorname{Re}[-i\omega\vct U_i e^{-i\omega t}]$.
A calibrated real linear detector has
$y_a(t)=\vct\ell_a^{\mathsf T}\vct u(t)$ and phasor
$Y_a=\vct\ell_a^{\mathsf T}\vct U$. Its coefficients specify which
material motion and projection are observed; a surface signal is a centre
coordinate only when the assumed rotation and deformation contributions
are negligible or separately resolved~\cite{Leake2016}.
Stress requires the constitutive and prepared memory state, as in
Section~4. A displacement trace alone need not determine it.
For the classical energy amplitude, the inverse map in
\eqref{eq:complex-energy-coordinate-equivalence} supplies the same real
coordinate and momentum. These reconstructions give the representation a
physical meaning before any statistical interpretation is assigned.

\paragraph{Controlled interference in a calibrated channel.}
\label{sec:sup-driven-phase-control}
Two force components at the same frequency give a concrete use of
superposition. Let their calibrated force columns be $\vct f_1$ and
$\vct f_2$, let $\mathcal R(\omega)$ be the full mechanical response,
and let $\vct\ell$ define a linear displacement readout. Set
$H_j=\vct\ell^{\mathsf T}\mathcal R(\omega)\vct f_j$.
Changing only their relative phase gives
\begin{equation}
 \begin{aligned}
 Y(\theta)&=H_1+e^{i\theta}H_2,\\
 \min_\theta|Y|^2&=(|H_1|-|H_2|)^2,\qquad
 \max_\theta|Y|^2=(|H_1|+|H_2|)^2.
 \end{aligned}
 \label{eq:sup-driven-interference-extrema}
\end{equation}
For nonzero $H_1,H_2$, the minimum occurs at
$\theta=\pi+\arg H_1-\arg H_2$ modulo $2\pi$;
an exact zero additionally requires equal magnitudes. These statements
follow by expanding $|Y|^2$. A second readout generally has different
overlaps and can remain nonzero at that phase. Sweeping a known relative
phase thus tests the response map and exposes cancellation hidden by
a single amplitude measurement. The same applied forces supply the
associated work; a quiet channel is not an energy balance.

Complementary projections make phase cancellation explicit in Figure~\ref{fig:added-phase-interference}: a quiet readout can coexist with a nonzero displacement pattern.

\begin{figure}[H]
\centering\begingroup
\definecolor{vltTeal}{HTML}{176B64}
\definecolor{vltBlue}{HTML}{315D80}
\definecolor{vltGold}{HTML}{A87520}
\definecolor{vltPlum}{HTML}{725877}
\definecolor{vltRed}{HTML}{762B38}
\definecolor{vltYellow}{HTML}{D5AD3C}
\definecolor{vltNeutral}{HTML}{777D83}
\setlength{\unitlength}{1pt}
\scalebox{0.90}{%
\input{vlt-figures/figure-167.tex}%
}\endgroup
\caption{Relative phase redistributes a fixed squared displacement-phasor norm between two readouts. At $t=0$, $\vct U=A(1,e^{i\delta})^{\mathsf T}$ gives the displayed real shifts. Reference ticks, rigid bodies, five-turn coils and one arrow gain register all frames. At $\delta=\pi/2$, complex addition combines real and imaginary quadratures. Squared common and opposing readouts sum to the fixed norm; their styles remain distinct in grayscale. This is a prescribed one-frequency pattern family whose required force follows the mechanical pencil.}
\label{fig:added-phase-interference}
\end{figure}

\subsubsection{Mean motion, fluctuations, and their energy}

A fluctuation is the difference from a specified mean. Let angle brackets
first denote an ensemble of classical preparations at the same time, with
finite second moments, and set
$\overline{\vct u}=\langle\vct u\rangle$,
$\delta\vct u=\vct u-\overline{\vct u}$, and likewise for momentum.
The covariances
$\mathsf C_{uu}=\langle\delta\vct u\,\delta\vct u^{\mathsf T}\rangle$
and $\mathsf C_{pp}=\langle\delta\vct p\,\delta\vct p^{\mathsf T}\rangle$
describe the spread of the retained coordinates. Their entries have units
of length squared and momentum squared. They do not describe uncertainty
in a fitted stiffness or in a detector calibration.
\begin{proposition}[Classical mean and fluctuation energy]
\label{prop:s5-classical-mean-fluctuation-energy}
For a fixed positive harmonic reference, a classical preparation with
finite second moments and calibrated real readout matrix $\mathsf L_p$
satisfies
\begin{equation}
 \boxed{\begin{aligned}
 \langle H_{\rm harm}\rangle
 &=H_{\rm harm}(\overline{\vct u},\overline{\vct p})
   +\tfrac12\tr\!\left(\mathsf M^{-1}\mathsf C_{pp}
                                      +\mathsf K\mathsf C_{uu}\right),\\
 \mathsf C_{Y}&=\mathsf L_p\mathsf C_{uu}\mathsf L_p^{\mathsf T}.
 \end{aligned}}
 \label{eq:complex-classical-mean-covariance-energy}
\end{equation}
\end{proposition}
\begin{proof}
Expand displacement and momentum into their means and centered
fluctuations. Their linear cross terms average to zero, while each
quadratic fluctuation average is the trace of its stiffness or
inverse-mass weight against the corresponding covariance. Linear
projection gives the readout covariance.
\end{proof}
Rows of $\mathsf L_p$ are detector vectors, and $\mathsf C_Y$ is their
ideal displacement covariance. The identity uses first and second
moments and applies to correlated non-Gaussian preparations as well
as equilibrium ones. Detector noise enters its separate measurement
model. A resolved Maxwell state adds branch storage and covariance.
Below, $E_{\rm mean}$ names the first term and $E_{\rm fl}^{\rm cl}$
the trace term; this classical split uses the full classical energy.

For one positive mass-weighted mode, every trajectory
$Q_s=A\cos(\omega_st-\varphi)$,
$\Pi_s=-\omega_sA\sin(\omega_st-\varphi)$ has energy
$E_s=\omega_s^2A^2/2$. Repeating one phase gives the same nonzero mean
motion. Uniformly randomizing the phase gives zero means but
$\operatorname{Var}Q_s=A^2/2$ and
$\operatorname{Var}\Pi_s=\omega_s^2A^2/2$, with exactly the same mean
energy. Figure~\ref{fig:classical-equal-energy-ensembles} shows this
distinction without a quantization assumption. This fixed-energy random
phase ensemble is not a thermal ensemble: a classical thermal harmonic
mode has an energy distribution as well as a phase distribution.
A zero ensemble mean is therefore compatible with mechanical excitation.
Replacing an ensemble by a long time record requires a stationary
preparation and an appropriate ergodic or repeatability assumption.

\begin{figure}[H]
\centering\begingroup
\definecolor{vltTeal}{HTML}{176B64}
\definecolor{vltBlue}{HTML}{315D80}
\definecolor{vltGold}{HTML}{A87520}
\definecolor{vltPlum}{HTML}{725877}
\definecolor{vltRed}{HTML}{762B38}
\definecolor{vltYellow}{HTML}{D5AD3C}
\definecolor{vltNeutral}{HTML}{777D83}
\setlength{\unitlength}{1pt}
\scalebox{0.90}{%
\input{vlt-figures/figure-168.tex}%
}\endgroup
\caption{A fixed classical mode energy with two ensemble preparations. On $X=\omega_sQ_s/\sqrt{2E}$, $V=\Pi_s/\sqrt{2E}$, one repeated phase gives deterministic mean motion; uniform phases give zero means with variances $E/\omega_s^2,E$. The origin marks the latter mean and perimeter dots sample its continuous phase law. The physical relative pattern is $\vct\phi_s=(1,-1)^{\mathsf T}/\sqrt{2m}$, with the centre unexcited. Reference ticks, fixed body size and matched six-turn coils register all snapshots. Burgundy arrows are velocity. The three right-hand frames sample the continuous fixed-energy ensemble; a thermal ensemble additionally distributes energy. The boxed result illustrated here is~\eqref{eq:complex-classical-mean-covariance-energy}.}
\label{fig:classical-equal-energy-ensembles}
\end{figure}

\begin{definition}[Second-order mechanical preparation]
\label{def:s55-second-order-preparation}
For a declared finite real state column $X$ with finite second moments,
its \emph{second-order preparation} is the pair
$(\vct m_X,\mathsf C_X)$, where $\vct m_X=\langle X\rangle$ and
$\mathsf C_X=\langle(X-\vct m_X)(X-\vct m_X)^{\mathsf T}\rangle$.
The full covariance includes cross blocks between displacement,
momentum and retained medium variables. For the branch model one may
use $X=(\vct u,\vct p,\vct z_1,\ldots)$, with the momentum version
of the generator in \eqref{eq:augmented-relaxation-generator}.
\end{definition}

\begin{proposition}[Prediction from a second-order preparation]
\label{prop:s55-second-order-prediction}
For fixed finite linear dynamics
$\dot X=\mathsf A_X X+\vct F_X(t)$ with the same deterministic input
in every preparation, put $\mathsf T_X(t)=e^{t\mathsf A_X}$ and let
$\vct d_X(t)$ be the zero-initial-state forced solution. Then
\begin{equation}
 \begin{aligned}
 \vct m_X(t)&=\mathsf T_X(t)\vct m_X(0)+\vct d_X(t),\\
 \mathsf C_X(t)&=\mathsf T_X(t)\mathsf C_X(0)
                                  \mathsf T_X(t)^{\mathsf T}.
 \end{aligned}
 \label{eq:s55-second-order-propagation}
\end{equation}
Two ensembles with the same full second-order preparation consequently
predict the same future means and covariances of every fixed linear
readout, and the same expectations of quadratic mechanical observables.
This assertion does not require Gaussian initial data.
\end{proposition}
\begin{proof}
The affine solution is $X(t)=\mathsf T_X(t)X(0)+\vct d_X(t)$.
Taking its expectation gives the first formula. Subtracting that mean
gives $\delta X(t)=\mathsf T_X(t)\delta X(0)$, which proves the second.
Every quadratic expectation is determined by these two moments, as in
\eqref{eq:complex-classical-mean-covariance-energy}.
\end{proof}

For a Gaussian ensemble these moments specify its complete probability
law under this linear evolution. General ensembles can share them and
still have different tails or higher-order correlations. Independently
specified additive noise contributes its propagated covariance as well;
the corresponding thermal model is developed in
Section~\ref{sec:operator-thermal-stochastic}. Random forcing correlated with
the preparation requires the corresponding cross terms. These choices
state exactly which statistical information a retained description
predicts.

The displacement--momentum cross block has a direct mechanical role.
For a centered conservative mode, let
$\operatorname{Var}Q_s=\sigma_Q^2$,
$\operatorname{Var}\Pi_s=\sigma_\Pi^2$ and
$\operatorname{Cov}(Q_s,\Pi_s)=c$, with
$|c|\leq\sigma_Q\sigma_\Pi$. Exact oscillator evolution gives
\begin{equation}
 \operatorname{Var}Q_s(t)
 =\sigma_Q^2\cos^2\omega_st
   +\frac{\sigma_\Pi^2}{\omega_s^2}\sin^2\omega_st
   +\frac{c}{\omega_s}\sin(2\omega_st).
 \label{eq:s55-cross-covariance-variance-evolution}
\end{equation}
Reversing $c$ preserves the initial variances and mean energy but
reverses the initial variance-growth rate $2c$. Equal position images
and energy estimates can thus lead to different later spreads.
Cross blocks with medium strain play the same predictive role when
the readout resolves their effect.

\paragraph{Carrying complex fluctuations through an event.}
The full real covariance remains sufficient for the linear prediction
just established. When the state is stored as an instantaneous energy
amplitude, the same information uses two complex second moments.
This distinction becomes useful when an arrangement change mixes an
amplitude with its conjugate.

\begin{definition}[Complete second moments of an energy amplitude]
\label{def:s5-reference-complex-second-moments}
For $\mu_\psi=\langle\psi\rangle$ and
$\delta\psi=\psi-\mu_\psi$, define its Hermitian covariance and
symmetric complementary covariance by
\begin{equation}
 \mathsf C_\psi=\langle\delta\psi\,\delta\psi^\dagger\rangle,
 \qquad
 \mathsf P_\psi=\langle\delta\psi\,\delta\psi^{\mathsf T}\rangle,
 \qquad
 \mathcal C_\psi=
 \begin{pmatrix}
  \mathsf C_\psi&\mathsf P_\psi\\
  \mathsf P_\psi^*&\mathsf C_\psi^*
 \end{pmatrix}\succeq0.
 \label{eq:s5-reference-augmented-covariance}
\end{equation}
Their entries have units of energy. Together they encode the covariance
of both real quadratures. Setting $\mathsf P_\psi=0$ is an additional
second-order statistical assumption, conventionally called propriety.
\end{definition}
The need for both moments is standard in the statistical description
of complex data~\cite{Picinbono1996,SchreierScharf2003}. Here their
values follow from the prepared mechanical state and its specified
energy metric.

\begin{proposition}[Second-moment registration with conjugate mixing]
\label{prop:s5-reference-second-moment-registration}
For the same deterministic affine registration
$\psi_h=\mathsf A\psi_g+\mathsf B\psi_g^*+\vct c$ in every member
of an ensemble, let
$\mathcal T=\left(\begin{smallmatrix}\mathsf A&\mathsf B\\
\mathsf B^*&\mathsf A^*\end{smallmatrix}\right)$. Then
\begin{equation}
 \begin{aligned}
 \mu_h&=\mathsf A\mu_g+\mathsf B\mu_g^*+\vct c,\\
 \mathcal C_h&=\mathcal T\mathcal C_g\mathcal T^\dagger,\\
 \langle E_{{\rm harm},h}\rangle
                    &=\|\mu_h\|^2+\tr\mathsf C_h.
 \end{aligned}
 \label{eq:s5-reference-second-moment-registration}
\end{equation}
The congruence preserves positive semidefiniteness and all second-order
quadrature information on the retained positive sector.
\end{proposition}
\begin{proof}
Subtract the transformed mean, stack the resulting amplitude and its
conjugate, and take their Hermitian outer-product expectation. Taking
the expected squared amplitude gives the last identity.
\end{proof}

For the scalar frequency change above, a centered preparation with
$C_g=C>0$ and $P_g=0$ yields
\begin{equation}
 C_h=\tfrac12(\eta^2+1)C,\qquad
 P_h=\tfrac12(\eta^2-1)C.
 \label{eq:s5-reference-scalar-complementary-covariance}
\end{equation}
The complementary moment $P_h$ retains unequal quadrature variances.
Discarding it preserves mean energy but changes the predicted later
displacement variance. Uncertain joining times, reference positions
or event amplitudes additionally require conditional moments and
cross terms; a random offset $\vct c$ contributes to covariance.
Apply this deterministic map within each registration, then combine
event-conditioned statistics. Medium strain and detector noise retain
their calibrated blocks.

Figure~\ref{fig:s5-complementary-covariance-prediction} shows the displacement information lost when complementary covariance is discarded.

\begin{figure}[H]
\centering\begingroup
\definecolor{vltTeal}{HTML}{176B64}
\definecolor{vltBlue}{HTML}{315D80}
\definecolor{vltGold}{HTML}{A87520}
\definecolor{vltPlum}{HTML}{725877}
\definecolor{vltRed}{HTML}{762B38}
\definecolor{vltYellow}{HTML}{D5AD3C}
\definecolor{vltNeutral}{HTML}{777D83}
\setlength{\unitlength}{1pt}
\scalebox{0.80}{%
\input{vlt-figures/figure-169.tex}%
}\endgroup
\caption{Complete complex second moments after controlled stiffening. Starting from centered Gaussian $C_g=1$, $P_g=0$, $\omega_g=1$, a change to $\omega_h=2$ at zero reference offset gives $C_h=2.5$, $P_h=1.5$. The contour in (a) has real/imaginary variances $2,1/2$. Setting $P_h=0$ gives the circular contour in (b) at the same mean energy. Axes and Mahalanobis radius match. Free evolution gives $\operatorname{Var}Q_h=(2.5+1.5\cos2\theta)/4$, compared with the circular prediction $2.5/4$. Contours describe distributions; the lower curves describe their evolving displacement variance in scaled classical coordinates.}
\label{fig:s5-complementary-covariance-prediction}
\end{figure}

\begin{proposition}[Storage certified by mixed retained-state readouts]
\label{prop:s5patch-mixed-storage-certificate}
In the positive linear realization above set
$\vct Z=(\vct X,\dot{\vct X},\vct r)$ and
$\mathsf S_E=\diag(\mathsf H_{\rm r},\mathsf G_X,\mathsf A)\succ0$.
For calibrated real instantaneous channels $\vct y=\mathsf L\vct Z$
with independent rows, the sharp storage bound is
\begin{equation}
 \boxed{\begin{aligned}
 E&\geq\tfrac12\vct y^{\mathsf T}
       (\mathsf L\mathsf S_E^{-1}\mathsf L^{\mathsf T})^{-1}\vct y,\\
 \vct Z_{\min}&=\mathsf S_E^{-1}\mathsf L^{\mathsf T}
       (\mathsf L\mathsf S_E^{-1}\mathsf L^{\mathsf T})^{-1}\vct y .
 \end{aligned}}
 \label{eq:s5patch-mixed-storage-certificate}
\end{equation}
\end{proposition}
\begin{proof}
The displayed state satisfies the measured constraint. Writing
$\vct Z=\vct Z_{\min}+\vct r_Z$ gives
$\mathsf L\vct r_Z=0$, so the storage cross term vanishes.
The remaining $\vct r_Z^{\mathsf T}\mathsf S_E\vct r_Z/2$ is nonnegative.
\end{proof}

Figure~\ref{fig:cmd-mixed-storage-certificate} gives a graphical specialization of this boxed relation.

This extends the mechanical certificate of
Proposition~\ref{prop:s5-hamiltonian-readout-energy} to calibrated
orientation and surface/material observations. Additional compatible
channels shrink the feasible state set and can raise the certified
storage. Units and noise calibration belong to $\mathsf L$.
A derivative of a mixed chemical/marker signal contains the declared
internal kinetics; its row is derived from that state equation before
being used as a velocity channel. The certificate concerns deterministic
resolved states. Ensemble energy also includes the covariance contribution
already specified above.

\begin{figure}[H]
\centering\begingroup
\setlength{\unitlength}{1pt}
\scalebox{0.84}{%
\input{vlt-figures/figure-311.tex}%
}\endgroup
\caption{A scalar positive-storage specialization of the mixed mechanical/material account. Assign $m=1$, $\mathcal U=x^2-xz+z^2$, $v=\dot x$ and $\eta\dot z=x-2z$. Completing the square gives $r=z-x/2$ and $E=v^2/2+3x^2/4+r^2$. In (a), contours use $v=0$; the measured line $Y=x=1$ has its minimum at $r=0$, while additional residual strain raises storage without changing $Y$. This is the sharp certificate \eqref{eq:s5patch-mixed-storage-certificate}. Panel (b) writes the same energy in the amplitude variables of \eqref{eq:s5patch-coupled-amplitude-storage}; its power balance is a scalar fixed-control case of \eqref{eq:s4patch-rigid-storage-supply}. The internal dissipation follows directly from the assigned evolution. Additional measured channels can constrain the residual energy; they do not alter the stored-energy law.}
\label{fig:cmd-mixed-storage-certificate}
\end{figure}

\subsubsection{The separate quantum first-moment correspondence}

Under the viral-phonon postulate, self-adjoint displacement operators
and a specified state determine measurement statistics. A number
state has zero mean displacement and nonzero variance. Establishing
its quantum character uses the canonical operator structure and
measurement predictions; centered classical ensembles also have
fluctuations.

A quadratic Hamiltonian links classical and quantum first-moment
dynamics through the same mechanical coefficients. The conservative
realization includes the retained particle and internal coordinates and,
when needed, the explicit oscillator-bath coordinates of Section~4.
Effective relaxing Maxwell strains instead retain the storage and loss
law of their passive realization.

\begin{proposition}[First moments of a quadratic canonical realization]
\label{prop:s5-quadratic-first-moment-correspondence}
Let $\hat{\vct q},\hat{\vct p}$ satisfy the canonical commutators on a
common invariant domain, with real constant
$\mathsf M_{\mathrm{all}}>0$, symmetric $\mathsf K_{\mathrm{all}}$ and
real prescribed force $\vct F(t)$. For the Hamiltonian
\[
 \hat H=\hat{\vct p}^{\mathsf T}\mathsf M_{\mathrm{all}}^{-1}
             \hat{\vct p}/2
       +\hat{\vct q}^{\mathsf T}\mathsf K_{\mathrm{all}}\hat{\vct q}/2
       -\vct F(t)^{\mathsf T}\hat{\vct q}.
\]
the Heisenberg first moments, whenever the commutators and required
expectations are defined, satisfy
\begin{equation}
 \frac{\dd}{\dd t}\langle\hat{\vct q}\rangle
             =\mathsf M_{\mathrm{all}}^{-1}\langle\hat{\vct p}\rangle,
 \qquad
 \frac{\dd}{\dd t}\langle\hat{\vct p}\rangle
             =-\mathsf K_{\mathrm{all}}\langle\hat{\vct q}\rangle+\vct F(t).
 \label{eq:quadratic-ehrenfest-coordinate-dynamics}
\end{equation}
\end{proposition}
\begin{proof}
The canonical commutators give
$[\hat H,\hat{\vct q}]=-i\hbar\mathsf M_{\mathrm{all}}^{-1}\hat{\vct p}$
and
$[\hat H,\hat{\vct p}]=i\hbar
(\mathsf K_{\mathrm{all}}\hat{\vct q}-\vct F)$.
The Heisenberg equation and linearity of expectation then give the two
classical first-moment equations.
\end{proof}
Their first moments obey the classical linear equations exactly for every
state with the required finite moments~\cite{Shankar1994}.
Coherent preparations provide nonzero means following those equations,
while a centered occupation state carries its excitation in second moments.
The correspondence is exact for the specified quadratic Hamiltonian;
nonlinear forces generally require higher moments and do not close this way.
The limitation already exists in a classical ensemble: the nonlinear
damping of Section~4.9 contains $\langle Q_s^2\Pi_s\rangle$, which is
not fixed by $\langle Q_s\rangle$ and $\langle\Pi_s\rangle$ alone.
A quantum counterpart additionally requires its operator ordering,
converter and reservoir model. Section~\ref{sec:phonon-quantum-active}
develops a separately specified pump-and-loss construction; it is not
obtained by changing the notation in this classical converter law.

For an isolated positive normal mode, write its mass-weighted coordinate
and momentum as $\hat Q_s$ and $\hat\Pi_s$. Its ladder operator is
$\hat b_s=(\omega_s\hat Q_s+i\hat\Pi_s)/\sqrt{2\hbar\omega_s}$.
With $\alpha_s=\langle\hat b_s\rangle$, the energy amplitude of the mean
motion is therefore
\begin{equation}
 \psi_s^{\rm mean}
       =\frac{\omega_s\langle\hat Q_s\rangle
                            +i\langle\hat\Pi_s\rangle}{\sqrt2}
       =\sqrt{\hbar\omega_s}\,\alpha_s.
 \label{eq:coherent-mean-energy-amplitude}
\end{equation}
It evolves with $e^{-i\omega_st}$ in the unforced harmonic reference.
For a multimode coherent preparation this reconstructs the classical
trajectory, with excitation energy $\sum_s\hbar\omega_s|\alpha_s|^2$.
The ladder normalization and coherent-state construction are standard
\cite{Glauber1963,Shankar1994,Zwiebach2013}; their application here uses
the viral-phonon postulate and the chosen mechanical coordinates.

\begin{definition}[Coherent oscillator preparation]
\label{def:s5-coherent-oscillator-preparation}
A coherent state is a normalized eigenstate
$\hat b_s\ket{\alpha_s}=\alpha_s\ket{\alpha_s}$.
For the finite retained modes, a product coherent preparation specifies
one such state in each mode.
\end{definition}

It displaces the oscillator's quantum means while retaining its reference
vacuum variances. A number state instead has a sharp occupation and a centred
displacement distribution. These definitions explain why a reproducible
mean trajectory and a known occupation are different preparation data.

Here the mode sums run over a finite retained positive-frequency sector;
the Fock space still allows arbitrary occupations of those modes. The
reference oscillator Hamiltonian is
$\hat H_{\rm ref}=\sum_s\hbar\omega_s
(\hat b_s^\dagger\hat b_s+\tfrac12\Id)$, including its vacuum energy
$E_0=\tfrac12\sum_s\hbar\omega_s$. This is the selected unforced harmonic
reference, not the driven full Hamiltonian or the total enclosed energy.
An untruncated continuum vacuum sum need not be finite; that limit
requires a declared cutoff or a normal-ordered excitation construction.
\begin{proposition}[Mean motion and harmonic excitation energy]
\label{prop:s5-quantum-mean-fluctuation-energy}
For the finite retained positive reference, let the preparation have
finite mean energy, put $\alpha_s=\langle\hat b_s\rangle$ and
$\delta\hat b_s=\hat b_s-\alpha_s$. Then
\begin{equation}
 \langle\hat H_{\rm ref}\rangle-E_0
    =\underbrace{\sum_s\hbar\omega_s|\alpha_s|^2}_{E_{\rm mean}}
       +\sum_s\hbar\omega_s
                    \langle\delta\hat b_s^\dagger\delta\hat b_s\rangle
    \geq E_{\rm mean}.
 \label{eq:mean-motion-fluctuation-energy-split}
\end{equation}
Equality holds for the product coherent preparation on these modes.
\end{proposition}
\begin{proof}
Expand $\hat b_s=\alpha_s+\delta\hat b_s$ in the number expectation.
The centered linear terms vanish. Each
$\delta\hat b_s^\dagger\delta\hat b_s$ is nonnegative, proving the
bound; the joint coherent preparation makes every centered
annihilation term zero.
\end{proof}
A number state or centered thermal state can have $E_{\rm mean}=0$
while its excitation energy is positive. A displacement trace therefore
does not reconstruct the full phonon state or all its energy; fluctuations
and correlations supply additional information.

This split can be unfolded into ordinary coordinate moments. Define the
excitation energy $E_{\rm exc}=\langle\hat H_{\rm ref}\rangle-E_0$ and
the centered contribution $E_{\rm fl}=E_{\rm exc}-E_{\rm mean}$. Then
\begin{equation}
 \begin{split}
 E_{\rm mean}&=\frac12\sum_s
       \left[\langle\hat\Pi_s\rangle^2+
                      \omega_s^2\langle\hat Q_s\rangle^2\right],\\
 E_{\rm fl}&=\frac12\sum_s
       \left[\operatorname{Var}(\hat\Pi_s)+
           \omega_s^2\operatorname{Var}(\hat Q_s)-\hbar\omega_s\right]
           \geq0.
 \end{split}
 \label{eq:coordinate-mean-fluctuation-energy}
\end{equation}
The vacuum variance is subtracted because $E_{\rm exc}$ is measured above
the reference vacuum energy.

For each mode, the canonical uncertainty relation gives
$\operatorname{Var}(\hat Q_s)\operatorname{Var}(\hat\Pi_s)\geq\hbar^2/4$,
which implies
$\operatorname{Var}(\hat\Pi_s)+\omega_s^2\operatorname{Var}(\hat Q_s)
\geq\hbar\omega_s$. This proves the nonnegative centered contribution
even when one variance is smaller than its vacuum value. In a coherent state the centered variances
remain at those vacuum values; a number-state excitation can instead reside
entirely in centered fluctuations. Table~\ref{tab:equal-energy-phonon-readouts}
compares two preparations with the same one-mode excitation energy.

\begin{table}[H]
\centering\small
\begin{tabular}{@{}llll@{}}
\toprule
Preparation & $\langle\hat Q_s(t)\rangle$
 & $\operatorname{Var}(\hat Q_s)$ & $E_{\rm exc}$ \\
\midrule
$\ket{\alpha_s(t)}$, $\alpha_s(t)=e^{-i\omega_st}$
 & $\sqrt{2\hbar/\omega_s}\cos\omega_st$
 & $\hbar/(2\omega_s)$ & $\hbar\omega_s$ \\
$\ket{1_s}$ & $0$ & $3\hbar/(2\omega_s)$ & $\hbar\omega_s$ \\
\bottomrule
\end{tabular}
\caption{Equal excitation energy with different displacement statistics in
one positive reference mode. $Q_s$ is mass weighted and the remaining modes
are in their reference vacuum. The coherent occupation mean is one, but its
occupation is not sharp; the one-phonon number state has sharp occupation one.
These are theoretical quantum preparations under the viral-phonon postulate.}
\label{tab:equal-energy-phonon-readouts}
\end{table}

Figure~\ref{fig:more-equal-energy-readouts} depicts the displacement
densities and subsequent mean motion for these equal-energy preparations.
The distribution and its mean provide complementary readouts of the
energy split in \eqref{eq:mean-motion-fluctuation-energy-split}.

\begin{figure}[H]
\centering\begingroup
\definecolor{vltTeal}{HTML}{176B64}
\definecolor{vltBlue}{HTML}{315D80}
\definecolor{vltGold}{HTML}{A87520}
\definecolor{vltPlum}{HTML}{725877}
\definecolor{vltRed}{HTML}{762B38}
\definecolor{vltYellow}{HTML}{D5AD3C}
\definecolor{vltNeutral}{HTML}{777D83}
\setlength{\unitlength}{1pt}
\scalebox{0.80}{%
\input{vlt-figures/figure-170.tex}%
}\endgroup
\caption{Equal excitation energy and different mode-coordinate readouts, for the preparations in Table~\ref{tab:equal-energy-phonon-readouts}. (a) Densities with respect to $\dd x$ are $\pi^{-1/2}e^{-(x-\sqrt2)^2}$ for $|\alpha=1\rangle$ and $2\pi^{-1/2}x^2e^{-x^2}$ for $|1\rangle$, with the displayed means and variances. (b) Harmonic evolution rotates the coherent mean; the number-state density and zero mean remain stationary. Both preparations have $E_{\rm exc}=\hbar\omega_s$ above the reference vacuum, partitioned between mean motion and centered fluctuations as in \eqref{eq:mean-motion-fluctuation-energy-split}. Identical color and line-style assignments join the panels. This is the separately postulated quantum mode coordinate; a virion-position readout also requires its physical mode-shape weights and preparation mechanism.}
\label{fig:more-equal-energy-readouts}
\end{figure}

The phase distinction established in Section~\ref{sec:conservative-complex-generator} applies here too:
rotating a coherent parameter $\alpha_s$ changes its mechanical means,
whereas multiplying the entire quantum ket by a common phase leaves its
measurement statistics unchanged. Classical ensemble phase alignment,
cross-spectral coherence, and a quantum coherent state are three different
preparation or statistical statements.

The constant-inertia quadratic quantum correspondence just discussed
also applies to angular librations in a fixed small-angle chart, with
their rotational inertia and canonical momentum. Finite body rotation
uses the configuration manifold $\operatorname{SO}(3)$ and its rotor
kinetics. The local harmonic correspondence retains that stated scale
and does not replace the global rigid-body dynamics.

\subsubsection{Recoverable storage and spatial concentration}

For a driven steady oscillation, recoverable storage includes the
resolved medium as well as particle motion. In the Maxwell realization
the branch phasors are
$\vct Z_a=[-i\omega\tau_a/(1-i\omega\tau_a)]\mathsf G_a\vct U$.
\begin{proposition}[Cycle-averaged storage in the Maxwell realization]
\label{prop:s5-maxwell-cycle-storage}
For a steady harmonic solution at $\omega>0$ with the specified real
particle and branch-storage quadratic forms, the cycle average is
\begin{equation}
 \begin{split}
 \overline E_{\mathrm{res}}
  &=\tfrac14\omega^2\vct U^\dagger\mathsf M\vct U
       +\tfrac14\vct U^\dagger\mathsf K\vct U
       +\tfrac14\sum_a\vct Z_a^\dagger\mathsf H_a\vct Z_a\\
  &=\tfrac14\vct U^\dagger
                  [\omega^2\mathsf M+\mathsf K'(\omega)]\vct U.
 \end{split}
 \label{eq:phasor-average-stored-energy}
\end{equation}
\end{proposition}
\begin{proof}
The cycle average of each real quadratic storage term is one quarter
of its phasor quadratic form. Insert the branch phasor and use
$|{-i\omega\tau_a}/(1-i\omega\tau_a)|^2
=(\omega\tau_a)^2/[1+(\omega\tau_a)^2]$ together with
$\mathsf K_a=\mathsf G_a^{\mathsf T}\mathsf H_a\mathsf G_a$.
The resulting branch sum is the storage part $\mathsf K'(\omega)$ of
this realization.
\end{proof}

These storage terms belong to the specified Maxwell realization.
An isothermal free-energy branch retains its thermodynamic
interpretation as in Section~4. Storage and loss
\eqref{eq:average-loss-power} both use their mechanical weights.

\begin{proposition}[Cycle storage with coupled relaxing coordinates]
\label{prop:s5patch-coupled-cycle-storage}
For the fixed whitened realization of
Proposition~\ref{prop:s5patch-coupled-energy-amplitudes}, with zero
independent internal source, put
$\mathsf V_\omega=(\mathsf A-i\omega\mathsf D)^{-1}$ and
$\mathsf K_\chi=\mathsf H'-\mathsf B'\mathsf V_\omega\mathsf B'^{\mathsf T}$.
The steady residual phasor and mean storage obey
\begin{equation}
 \begin{aligned}
 \widehat{\vct r}&=(\mathsf A^{-1}-\mathsf V_\omega)
                  \mathsf B'^{\mathsf T}\widehat{\vct X},\\
 \overline E&=\tfrac14\omega^2\widehat{\vct X}^\dagger\mathsf G_X\widehat{\vct X}
      +\tfrac14\widehat{\vct X}^\dagger\mathsf H_{\rm r}\widehat{\vct X}
      +\tfrac14\widehat{\vct r}^{\dagger}\mathsf A\widehat{\vct r}\\
   &=\tfrac14\widehat{\vct X}^\dagger
             [\omega^2\mathsf G_X+\operatorname{Re}_{\rm H}\mathsf K_\chi]
                                              \widehat{\vct X} .
 \end{aligned}
 \label{eq:s5patch-coupled-cycle-storage}
\end{equation}
\end{proposition}
\begin{proof}
Solve the internal row for its phasor and use
$\operatorname{Re}_{\rm H}\mathsf V_\omega
=\mathsf V_\omega^\dagger\mathsf A\mathsf V_\omega$.
It follows that
\[
(\mathsf A^{-1}-\mathsf V_\omega)^\dagger\mathsf A
(\mathsf A^{-1}-\mathsf V_\omega)
=\mathsf A^{-1}-\operatorname{Re}_{\rm H}\mathsf V_\omega .
\]
Substitution gives the last line without requiring $\mathsf A$ and
$\mathsf D$ to commute.
\end{proof}
The compact storage expression therefore extends to this specified
coupled realization. A fitted complex stiffness alone need not identify
the corresponding internal storage or preparation. A mixed readout of
that storage uses its full reconstruction in
\eqref{eq:s5patch-mixed-response-readout}.

Resolved storage and extracted work are connected by the declared
mechanical ports. A port specifies where a force acts and which
velocity is conjugate to it. This makes the operational meaning of
recovering energy precise within the passive model.

\begin{definition}[Available mechanical work through specified ports]
\label{def:s55-available-mechanical-work}
In the deterministic passive closure, let
$\vct f=\mathsf B_{\rm p}\vct f_{\rm p}$ be the applied particle
force and
$\vct v_{\rm p}=\mathsf B_{\rm p}^{\mathsf T}\mathsf M^{-1}\vct p$
its conjugate port velocity, with a fixed real dimensionless
$\mathsf B_{\rm p}$. For a prepared state $X_0$ and horizon
$T_{\rm w}>0$, choose an admissible class $\mathcal U_{T_{\rm w}}$
of finite-work inputs containing the zero input. The \emph{available
mechanical work} is
\begin{equation}
 \mathcal W_{\rm av}(X_0;\mathsf B_{\rm p},T_{\rm w})
 =\sup_{\vct f_{\rm p}\in\mathcal U_{T_{\rm w}}}
        \left[-\int_0^{T_{\rm w}}
                 \vct f_{\rm p}^{\mathsf T}\vct v_{\rm p}\,\dd t\right].
 \label{eq:s55-available-port-work}
\end{equation}
All mechanical supply enters through these ports, and the passive
storage law and boundaries are fixed during this comparison.
\end{definition}

This is the available-storage idea of dissipative systems theory
applied to calibrated mechanical force and velocity~\cite{Willems1972}.
The definition specifies both access to the system and the allowed
observation or work-extraction interval. The same prepared storage can
give different available work through different ports.
For the pressure-driven cylindrical application, pressure difference
and volume flow rate form the conjugate port, with mechanical supply
rate $\Delta p\,Q$. Applying the storage bound to that experiment uses
the complete fluid--particle realization and includes this pressure
work among its inputs.

\begin{proposition}[Storage bounds available work]
\label{prop:s55-available-work-bound}
For nonnegative resolved storage satisfying
$\dot E_{\rm res}=\vct f_{\rm p}^{\mathsf T}\vct v_{\rm p}
-D_{\rm res}$ with $D_{\rm res}\geq0$, every admissible input obeys
\begin{equation}
 \begin{aligned}
 \mathcal W_{\rm ext}
  &=E_{\rm res}(0)-E_{\rm res}(T_{\rm w})
           -\int_0^{T_{\rm w}}D_{\rm res}(t)\,\dd t,\\
 0&\leq\mathcal W_{\rm av}\leq E_{\rm res}(0).
 \end{aligned}
 \label{eq:s55-storage-available-work-bound}
\end{equation}
Here $\mathcal W_{\rm ext}$ is the bracketed work for that input.
\end{proposition}
\begin{proof}
Integrate the storage balance and change the sign of supplied work.
Terminal storage and integrated dissipation are nonnegative, giving
the upper bound for every input and hence for the supremum. The zero
input extracts zero work and supplies the lower bound.
\end{proof}

For example, the conservative exchange-symmetric supported pair separates
into independently evolving common and relative sectors. Prepare energy only in the relative
sector, and allow a common force port. The initially resting common
sector receives all port work; its own passive balance makes net
extracted work nonpositive for every such input. The available work
through that port is therefore zero, even while the opposing particle
motion carries positive energy. A relative port accesses a different
coordinate. This is a consequence of the specified coupling and
preparation, rather than an absence of stored mechanical energy.

The storage law locates retained work; probe attachment and force maps
determine access to it. Available work also depends on extraction
time and intervening relaxation. Thermal or active work enters as an
additional supply in the same account.

The distinction between DC and a slowly varying harmonic preparation
also matters. On a fixed strictly positive finite sector, set
$\mathsf C_{\rm eff}=\mathsf C_0+\sum_a\tau_a\mathsf K_a$. Then
\begin{equation}
 \mathsf R_{\rm mech}(\omega)=\mathsf K^{-1}
       +i\omega\mathsf K^{-1}\mathsf C_{\rm eff}\mathsf K^{-1}
       +O(\omega^2),\qquad \omega\to0.
 \label{eq:low-frequency-phasor-compliance}
\end{equation}
The out-of-phase response vanishes continuously for a real force.
However, the cycle-averaged storage of a cosine force with fixed amplitude
approaches $\vct F^{\mathsf T}\mathsf K^{-1}\vct F/4$; a genuinely
constant force stores $\vct F^{\mathsf T}\mathsf K^{-1}\vct F/2$
after relaxation. A full-period average and setting the frequency to zero
are different preparations. Static displacement is therefore carried
separately rather than substituted into a positive-frequency average.
This expansion is not uniform when a restoring gap closes. Its slow
harmonic limit also requires observation over full, increasingly long
cycles.

For the continuum branch realization, the corresponding density is
\begin{equation}
 \overline e_{\mathrm{res}}=\tfrac14\left[
     \rho\omega^2|\vct U|^2
       +\nabla\vct U^*:\mathcal A'(\omega):\nabla\vct U
       +\vct U^*\cdot\mathsf B_\lambda\vct U\right].
 \label{eq:phasor-continuum-storage-density}
\end{equation}
The same constitutive realization determines the local stress-velocity flux
and dissipated power. In an unstressed positive-spring realization with
nonnegative local branch storage and pinning, this density gives a
nonnegative energy assignment. For other prestressed coarse-grained models,
the local allocation must be defined consistently with their mechanical
energy functional; global stability alone does not fix a unique positive
local allocation.

\begin{definition}[Regional energy fraction and enrichment]
\label{def:s5-regional-energy-enrichment}
For a fixed nonnegative storage allocation, a subregion
$\mathcal D\subset\mathcal V_{\mathrm{lat}}$ of nonzero volume and
positive total stored energy, define
\begin{equation}
 \eta_{\mathcal D}=
       \frac{\int_{\mathcal D}\overline e_{\mathrm{res}}\,\dd^3r}
            {\int_{\mathcal V_{\mathrm{lat}}}\overline e_{\mathrm{res}}\,\dd^3r},
 \qquad
 \mathcal L_{\mathcal D}=
                  \frac{\eta_{\mathcal D}}{|\mathcal D|/|\mathcal V_{\mathrm{lat}}|}.
 \label{eq:energy-localization-ratio}
\end{equation}
\end{definition}
Positive denominator energy makes the share well defined. Uniform
storage gives $\mathcal L_{\mathcal D}=1$; enrichment above one means
a larger energy share than volume share. An unforced pulse uses its
instantaneous storage allocation in place of the cycle average.

Relative phases determine interference between mode contributions.
Selective attenuation changes their relative weights, so a surviving
pattern can become concentrated in a region even while total resolved
energy decreases. The energy functional and flux quantify this change.
Its localization ratio need not vary monotonically, and an increase can
arise because energy outside the region decays faster, without an increase
of the region's own stored energy.

Comparisons use the same allocation: shifting an assigned bond share
between endpoints changes a local ratio while preserving mechanics.
For $f_{\mathcal D}=|\mathcal D|/|\mathcal V_{\rm lat}|$,
nonnegative storage gives $0\leq\eta_{\mathcal D}\leq1$ and
$0\leq\mathcal L_{\mathcal D}\leq1/f_{\mathcal D}$.
The ratio depends on both preparation and chosen region.

The instantaneous finite-state counterpart makes the geometry of this
allocation explicit.
Use the real prepared column $X=(\vct u,\vct p,\vct z_1,\ldots)$
on a positive storage sector, with the physical coordinate scales held
fixed, $E_{\rm res}=X^{\mathsf T}\mathsf S_E X/2$ and
$\mathsf S_E>0$. For the Maxwell realization,
$\mathsf S_E=\diag(\mathsf K,\mathsf M^{-1},\mathsf H_1,\ldots)$
on its energetic ranges. An assigned regional energy
$E_{\mathcal D}=X^{\mathsf T}\mathsf S_{\mathcal D}X/2$ is nonnegative
and no larger than the total when
$0\leq\mathsf S_{\mathcal D}\leq\mathsf S_E$.

\begin{proposition}[Regional storage in the energy metric]
\label{prop:s55-regional-storage-geometry}
For such a fixed allocation and $X\ne0$, define
$\mathsf B_{\mathcal D}
=\mathsf S_E^{-1/2}\mathsf S_{\mathcal D}\mathsf S_E^{-1/2}$.
Then
\begin{equation}
 \begin{aligned}
 \lambda_{\min}(\mathsf B_{\mathcal D})
 &\leq\eta_{\mathcal D}(X)
 \leq\lambda_{\max}(\mathsf B_{\mathcal D}),\\
 \eta_{\mathcal D}(X)
 &=\frac{X^{\mathsf T}\mathsf S_{\mathcal D}X}
             {X^{\mathsf T}\mathsf S_E X},
 \qquad 0\leq\mathsf B_{\mathcal D}\leq\Id.
 \end{aligned}
 \label{eq:s55-regional-energy-rayleigh-bounds}
\end{equation}
For an ensemble with mean $\vct m_X$, covariance $\mathsf C_X$ and
positive mean energy, the ratio of mean regional and total energies is
\begin{equation}
 \eta_{\mathcal D}^{\rm ens}
 :=\frac{\langle E_{\mathcal D}\rangle}{\langle E_{\rm res}\rangle}
 =\frac{\vct m_X^{\mathsf T}\mathsf S_{\mathcal D}\vct m_X
                         +\tr(\mathsf S_{\mathcal D}\mathsf C_X)}
        {\vct m_X^{\mathsf T}\mathsf S_E\vct m_X
                         +\tr(\mathsf S_E\mathsf C_X)}.
 \label{eq:s55-ensemble-regional-energy-ratio}
\end{equation}
\end{proposition}
\begin{proof}
Set $\vct\xi=\mathsf S_E^{1/2}X$. The fraction becomes the Rayleigh
quotient of the real symmetric $\mathsf B_{\mathcal D}$; its spectral
bounds and $0\leq\mathsf B_{\mathcal D}\leq\Id$ follow from the
allocation inequalities. Expanding each quadratic form about the
ensemble mean gives the second identity.
\end{proof}

The eigenvalue bounds refer to the chosen energy allocation, while
actual preparation by a particular force port has its own dynamical
reachability. They can be sharper than the interval $[0,1]$ without
introducing a new material coefficient. The ensemble ratio weights
each realization by its total energy; it generally differs from an
unweighted average of the individual $\eta_{\mathcal D}$. The choice
must therefore accompany an ensemble localization report.

Correlations are essential to the assigned interaction energy. For a
centered scalar pair with positive axial curvature $k_c$,
\begin{equation}
 \langle E_{ij}^{\rm pair}\rangle
   =\frac{k_c}{2}
      \bigl(C_{ii}+C_{jj}-2C_{ij}\bigr),
 \qquad E_{ij}^{\rm pair}=\frac{k_c}{2}(u_i-u_j)^2.
 \label{eq:s55-pair-covariance-storage}
\end{equation}
Equal marginal variances can accompany different common-motion and
separation fluctuations, hence different pair storage. Tracking two
centres or structural projections provides the missing correlation.
Retain it when it changes interaction work or a future readout.
The matrix allocation assumes nonnegative storage; prestressed signed
allocations use their signed local balance.

For an instantaneous unforced comparison, write $E_{\mathcal D}(t)$ for
the assigned energy inside the region and $E_{\rm res}(t)>0$ for the total
resolved energy. At times when $E_{\mathcal D}>0$,
\begin{equation}
 \frac{\dd}{\dd t}\log\mathcal L_{\mathcal D}
    =\frac{1}{E_{\mathcal D}}\Dt{E_{\mathcal D}}
       -\frac{1}{E_{\rm res}}\Dt{E_{\rm res}}.
 \label{eq:complex-localization-rate}
\end{equation}
The ratio rises whenever the region loses a smaller fraction of its energy
per unit time than the whole system. This need not involve inward energy
transport. The local balance of Section~4 distinguishes boundary flux,
recoverable storage, passive loss, and applied or active work; that balance,
rather than the ratio alone, identifies the mechanism.
Figure~\ref{fig:local-energy-versus-enrichment} illustrates the point
with two decreasing positive energy accounts and no transfer between them.

\begin{figure}[H]
\centering\begingroup
\definecolor{vltTeal}{HTML}{176B64}
\definecolor{vltBlue}{HTML}{315D80}
\definecolor{vltGold}{HTML}{A87520}
\definecolor{vltPlum}{HTML}{725877}
\definecolor{vltRed}{HTML}{762B38}
\definecolor{vltYellow}{HTML}{D5AD3C}
\definecolor{vltNeutral}{HTML}{777D83}
\setlength{\unitlength}{1pt}
\scalebox{0.90}{%
\input{vlt-figures/figure-171.tex}%
}\endgroup
\caption{Absolute energy, regional share and volume-normalized enrichment. (a) An oblique reference view retains four labelled virions and the fixed full-cross-section slab $\mathcal D$, occupying one quarter of the cylinder length and volume. The dashed section marks its boundary; shading identifies the region rather than particle concentration. (b) The stipulated energies $E_{\mathcal D}/E_\star=e^{-2\tau}$ and $E_{\rm out}/E_\star=3e^{-6\tau}$ both decrease, along with their sum. (c) Their share $\eta_{\mathcal D}=1/(1+3e^{-4\tau})$ and enrichment $\mathcal L_{\mathcal D}=4\eta_{\mathcal D}$ increase because the outside storage decays faster. Each curve has an explicit line style and normalization. Geometry specifies the allocation region; these auxiliary energy histories do not specify a flow or decay mechanism. Boundary work, storage and loss records complete that physical account.}
\label{fig:local-energy-versus-enrichment}
\end{figure}

A biophysical region identifies the resolved material: particle nodes,
a shell patch or an internal component with assigned storage.
Centre motion and internal strain can differ substantially.
Connecting their energy concentration to a conformational event
requires the event coordinate and its mechanical coupling.

Self-excitation sharpens the need to keep these distinctions. A maintained
oscillator can have a reproducible amplitude but an unspecified phase
relative to repeated preparations. Its ensemble mean can then vanish.
Its energy supply is still determined by force--velocity work and the
resource or force-generation state in Section~\ref{sec:self-excitation-passive-boundary}. Neither a nonzero
variance, a surviving spatial pattern, nor an increasing energy fraction
alone establishes that supply.

\subsubsection{Registering patterns through recruitment and breakup}
\label{sec:s5-register-patterns}

Mode coefficients can change sign while laboratory motion continues.
Recruitment can also change the useful pattern subspace. Comparisons
therefore retain particle identities and match the pattern spaces.
Starting from~\eqref{eq:s4-transient-chart-change}, the following
overlap quantifies motion retained in a selected displacement band;
the event law still propagates the physical state.

For the resolved body tangent, registration also carries the orientation
and the force-dual momentum. On overlapping angular charts with
$\vct X_h=f_{hg}(\vct X_g)$, let $\mathsf J_{hg}=Df_{hg}$ at the
same physical state. Then
$\delta\vct X_h=\mathsf J_{hg}\delta\vct X_g$ and
$\vct P_h=\mathsf J_{hg}^{-\mathsf T}\vct P_g$ preserve canonical
virtual work. Inertia transforms by the corresponding congruence.
The exact rotation identity is
\eqref{eq:s4patch-orientation-reference-change}; an angular offset is
used only in its local linear chart. A shared surface/material history
continues in its declared frame through this registration. A physical
contact reset or changed constitutive reference is specified separately,
with its work and preparation. The exact affine energy-reference maps
above retain their stated translational scope.

\begin{definition}[Registered displacement band]
\label{def:s5-register-band}
Use persistent identities for all $N_{\rm tot}$ ideal virions in the cylindrical
pool and the same positive, constant material mass matrix $\mathsf M$.
For arrangement $g$, let $\mathsf\Phi_g$ have $r_g$ mass-orthonormal
columns on that common particle space:
$\mathsf\Phi_g^\dagger\mathsf M\mathsf\Phi_g=\Id$.
They span a chosen conservative displacement-mode band, or another declared
displacement subspace. Define
\begin{equation}
 \mathsf\Pi_g=\mathsf\Phi_g\mathsf\Phi_g^\dagger\mathsf M,
 \qquad
 \vct c_g=\mathsf\Phi_g^\dagger\mathsf M\vct U,
 \qquad
 \mathsf O_{hg}=\mathsf\Phi_h^\dagger\mathsf M\mathsf\Phi_g.
 \label{eq:s5-register-overlap}
\end{equation}
Here $\vct U$ is a length-valued analytic displacement or phasor with a
common clock and registered reference positions. The coefficients
$\vct c_g$ have units $\mathrm{kg}^{1/2}\mathrm m$; they are displacement
coordinates, rather than the energy amplitudes $\vct\psi$.
\end{definition}

A cluster-only basis is embedded in the full pool before taking overlaps;
its neglected displacement tails remain a reduction error. Reference-centre
changes retain the affine term in \eqref{eq:s4-transient-chart-change}.
Mass orthogonality applies to these displacement frames and is not an
assumption that right eigenvectors of a damped, nonnormal pencil are
orthogonal.

\begin{proposition}[Pattern retention and detector error]
\label{prop:s5-register-retention}
For the same physical pattern $\vct U=\mathsf\Phi_g\vct c_g$, its projection
onto band $h$ obeys
\begin{equation}
 \begin{aligned}
 \vct c_h&=\mathsf O_{hg}\vct c_g,\\
 \|\vct c_g\|^2-\|\mathsf O_{hg}\vct c_g\|^2
  &=\|(\Id-\mathsf\Pi_h)\vct U\|_{\mathsf M}^2\geq0,\\
 |\ell^\dagger\vct U-
       \ell^\dagger\mathsf\Pi_h\vct U|
  &\leq\|\mathsf M^{-1/2}\ell\|\,
                 \|(\Id-\mathsf\Pi_h)\vct U\|_{\mathsf M}.
 \end{aligned}
 \label{eq:s5-register-retention}
\end{equation}
The last line bounds the error in a fixed calibrated displacement readout.
\end{proposition}
\begin{proof}
The projector is orthogonal in the mass inner product. Decompose
$\vct U=\mathsf\Pi_h\vct U+(\Id-\mathsf\Pi_h)\vct U$ and apply
Pythagoras. Cauchy--Schwarz after multiplication by
$\mathsf M^{1/2}$ gives the readout bound.
\end{proof}

The singular values of rectangular $\mathsf O_{hg}$ are cosines of
principal angles between the bands. Their deficit measures motion
omitted by the new band, including when its dimension changes.
Dissipation instead follows force, velocity and storage balance;
the restoring metric converts displacement into energy.

This residual also controls whether a reported detector phase is stable.
Put $Y=\ell^\dagger\vct U$, $\widehat Y=\ell^\dagger\mathsf\Pi_h\vct U$
and let $\epsilon$ be the last bound in
\eqref{eq:s5-register-retention}. If $|Y|>\epsilon$, elementary geometry
of the disk $|\widehat Y-Y|\leq\epsilon$ gives
\begin{equation}
 |\arg\widehat Y-\arg Y|_{\mathrm{circ}}
       \leq\arcsin(\epsilon/|Y|),
 \label{eq:s5-register-phase-error}
\end{equation}
where the left side is the smaller angular separation. The tangent from
the origin to that disk gives the maximal angle. Near a readout zero the
condition fails, correctly leaving phase uncontrolled. Thus a small
discarded displacement can be harmless for a bright channel but decisive
for a nearly cancelled channel. The bound supplies a channel-specific
criterion for retaining extra modes when comparing recruitment episodes.

\begin{proposition}[A matched frame for equal-rank bands]
\label{prop:s5-register-polar}
For $r_g=r_h=r$, write
$\mathsf O_{hg}=\mathsf A\mathsf\Sigma\mathsf B^\dagger$ by a singular-value
decomposition. The unitary matrix
$\mathsf Q_{hg}=\mathsf A\mathsf B^\dagger$ minimizes
\begin{equation}
 \|\mathsf\Phi_h\mathsf Q-\mathsf\Phi_g\|_{\mathsf M,F}^2,
 \qquad
 \min_{\mathsf Q^\dagger\mathsf Q=\Id}
 \|\mathsf\Phi_h\mathsf Q-\mathsf\Phi_g\|_{\mathsf M,F}^2
      =2r-2\operatorname{tr}\mathsf\Sigma.
 \label{eq:s5-register-polar}
\end{equation}
The weighted Frobenius norm sums the mass norms of the columns.
\end{proposition}
\begin{proof}
Expanding the square gives
$2r-2\operatorname{Re}\operatorname{tr}(\mathsf Q^\dagger\mathsf O_{hg})$.
The trace is maximized by $\mathsf Q=\mathsf A\mathsf B^\dagger$, since
the diagonal entries of a unitary matrix have real parts at most one.
\end{proof}

This is the unitary polar-factor alignment~\cite{Higham1986}.
If the two spaces coincide it removes an arbitrary basis rotation exactly.
For example, replacing one real mode by its negative changes its
coefficient phase by $\pi$ while preserving every particle displacement.
More generally $\mathsf\Phi\mapsto\mathsf\Phi\mathsf V$ requires
$\vct c\mapsto\mathsf V^\dagger\vct c$ and
$\ell^\dagger\mathsf\Phi\mapsto\ell^\dagger\mathsf\Phi\mathsf V$.
The detector amplitude is invariant. A rotation within a nearly degenerate
band requires retaining its full projected operator; individual diagonal
mode frequencies do not remain a diagonal representation under an
arbitrary rotation.

Only matched equal-rank bands receive a square unitary alignment.
New, discarded and zero-frequency directions remain explicit. Singular
overlap leaves a nonunique alignment in unmatched directions; it supplies
no reproducible phase there. A closing eigenvalue gap calls for tracking
the enlarged isolated subspace rather than relabelling individual
eigenvectors~\cite{DavisKahan1970}. At zero frequency the real zero-sector
coordinates of Section~\ref{sec:conservative-complex-generator} remain
necessary. These choices allow a small overlap matrix to track patterns in
a large virion pool without manufacturing a phase match at a breakup event.

Figure~\ref{fig:s5-retained-band-readout-error} places rectangular pattern retention beside the stability of a registered detector phase.

\begin{figure}[H]
\centering\begingroup
\definecolor{vltTeal}{HTML}{176B64}
\definecolor{vltBlue}{HTML}{315D80}
\definecolor{vltGold}{HTML}{A87520}
\definecolor{vltPlum}{HTML}{725877}
\definecolor{vltRed}{HTML}{762B38}
\definecolor{vltYellow}{HTML}{D5AD3C}
\definecolor{vltNeutral}{HTML}{777D83}
\setlength{\unitlength}{1pt}
\scalebox{0.90}{%
\input{vlt-figures/figure-172.tex}%
}\endgroup
\caption{A three-coordinate pattern projected onto $(1,1,0)/\sqrt2$ and $(0,0,1)$. The same equal-mass virions reconstruct the full and projected real phase with gain $16\,\mathrm{pt}/\ell_\star$; opposing pair motion is omitted by the new band. Reference ghosts and fixed coil parameters register the change. The rectangular overlap records the lost mass norm. The middle strip transforms coefficient and detector together under a basis sign change. Lower complex planes give separate readout-error tests with equal scales and radius $\epsilon=Y_\star/4$: $Y=Y_\star$ admits the tangent phase bound, while $Y=Y_\star/10$ leaves zero inside the error disk.}
\label{fig:s5-retained-band-readout-error}
\end{figure}

\subsection{Finite observation windows and measured phase coherence}
\label{sec:phase-window-coherence}

A finite record selects and weights part of a trajectory. Two modes
accumulating little relative phase over that interval can produce
similar signals in a shared detector. Preparation, window and spatial
readout therefore enter the distinguishability calculation together.

Record duration and sampling interval constrain different parts of a measurement. Figure~\ref{fig:added-sampling-alias} isolates the ambiguity introduced by discrete sampling.

\begin{figure}[H]
\centering\begingroup
\definecolor{vltTeal}{HTML}{176B64}
\definecolor{vltBlue}{HTML}{315D80}
\definecolor{vltGold}{HTML}{A87520}
\definecolor{vltPlum}{HTML}{725877}
\definecolor{vltRed}{HTML}{762B38}
\definecolor{vltYellow}{HTML}{D5AD3C}
\definecolor{vltNeutral}{HTML}{777D83}
\setlength{\unitlength}{1pt}
\scalebox{0.90}{%
\input{vlt-figures/figure-173.tex}%
}\endgroup
\caption{Two histories share all integer-time samples: $u_a/A=\cos(\pi s/3)$ and $u_b/A=\cos(7\pi s/3)$, $s=t/\Delta t$. Black dots identify seven samples; solid and dashed curves give the intervening motion. At unsampled $s=1/2$, the same virion lies on opposite sides of its reference in the two alternatives. Equal rulers and gain $28\,\mathrm{pt}/A$ calibrate those shifts; arrows denote displacement. Their frequency difference $2\pi/\Delta t$ preserves the alias for longer records at the same sampling interval. Exposure, filtering and noise enter their own measurement maps.}
\label{fig:added-sampling-alias}
\end{figure}

\subsubsection{Window averaging, beats, and spectral resolution}

A phase comparison depends on the duration and position of that interval.

\begin{definition}[Normalized rectangular observation window]
\label{def:s5-rectangular-observation-window}
For a window of positive duration $T_{\rm obs}$ centered on $t_c$, its
normalized phase factor is
\begin{equation}
 \begin{aligned}
 B_{T_{\rm obs}}(\Delta\omega)
    &=\frac1{T_{\rm obs}}\int_{t_c-T_{\rm obs}/2}^{t_c+T_{\rm obs}/2}
                           e^{-i\Delta\omega t}\,\dd t\\
    &=e^{-i\Delta\omega t_c}
       \operatorname{sinc}\!\left(\frac{\Delta\omega T_{\rm obs}}2\right),\\
 \operatorname{sinc}x&=\frac{\sin x}{x},
 \end{aligned}
 \label{eq:finite-window-phase-factor}
\end{equation}
with $\operatorname{sinc}0=1$.
\end{definition}
Direct integration gives the sinc form. Its complex phase locates the
window relative to the chosen preparation time.

\begin{proposition}[Finite-window quadratic displacement signal]
\label{prop:s5-finite-window-quadratic-signal}
For a finite real probe signal
$y(t)=\operatorname{Re}\sum_s A_s e^{-i\omega_st}$, the normalized
rectangular-window average is
\begin{equation}
 \overline{y^2}_{T_{\rm obs}}=\frac12\operatorname{Re}\sum_{s,t}
   \left[A_sA_t^*B_{T_{\rm obs}}(\omega_s-\omega_t)
        +A_sA_tB_{T_{\rm obs}}(\omega_s+\omega_t)\right].
 \label{eq:finite-window-probe-intensity}
\end{equation}
\end{proposition}
\begin{proof}
Write the real signal as half the sum of its complex expansion and its
conjugate. Squaring gives difference-frequency and sum-frequency
terms. Integrating each exponential over the same normalized window
gives its $B_{T_{\rm obs}}$ factor; conjugate pairs combine into the
displayed real part.
\end{proof}
Difference-frequency cross terms average away when
$|\Delta\omega|T_{\rm obs}\gg1$; summed-frequency terms additionally
require many carrier cycles. Exact degeneracy preserves interference,
and near degeneracy can preserve beats throughout the observation.
Local quadratic storage has the same temporal factors, weighted by
its mass, spring and branch-energy allocation.

A simple beat makes this window effect tangible. Two equal real amplitudes
give
\begin{equation}
 A\cos\omega_1t+A\cos\omega_2t
   =2A\cos\!\left(\frac{\omega_1-\omega_2}{2}t\right)
        \cos\!\left(\frac{\omega_1+\omega_2}{2}t\right).
 \label{eq:two-frequency-displacement-beat}
\end{equation}
The slowly varying interference envelope can appear constant in a short
record. For the rectangular window, the first zero of its difference-frequency
factor occurs at $|\Delta\omega|T_{\rm obs}=2\pi$. This gives the window's
frequency-separation scale; practical estimation also depends on noise,
sampling and preparation. It does not replace an uncertainty relation for
specified quantum observables.

Figure~\ref{fig:final-window-doublet} isolates duration broadening for a
phase-averaged two-line preparation. For an experimental comparison, retain
the same stationary particle state and analysis convention while varying
the recording duration, with sampling and exposure specified separately.
A changing width or line separation can then be compared with the known
window kernel before fitting a mechanical decay law. A reproducible phase
between prepared modes instead retains the cross terms in
\eqref{eq:finite-window-probe-intensity}; a power-sum plot is then a
different preparation average.

\begin{figure}[H]
\centering\begingroup
\definecolor{vltTeal}{HTML}{176B64}
\definecolor{vltBlue}{HTML}{315D80}
\definecolor{vltGold}{HTML}{A87520}
\definecolor{vltPlum}{HTML}{725877}
\definecolor{vltRed}{HTML}{762B38}
\definecolor{vltYellow}{HTML}{D5AD3C}
\definecolor{vltNeutral}{HTML}{777D83}
\setlength{\unitlength}{1pt}
\scalebox{0.80}{%
\input{vlt-figures/figure-174.tex}%
}\endgroup
\caption{The same undamped doublet observed for two durations. Analytic amplitudes $A$ have independent uniform phases, frequencies $\omega_c\mp\Delta\omega/2$ and $\omega_c=100\Delta\omega$. Phase averaging gives $\langle|\bar A_T|^2\rangle/A^2=\sum_j\operatorname{sinc}^2[(\omega-\omega_j)T/2]$. Both upper panels share axes; guides locate the fixed line frequencies. The full ordinate area in the displayed frequency coordinate is $4\pi/(\Delta\omega T)$. Lower unit-integral rectangular kernels change height reciprocally with duration. Exposure, intrinsic decay and real-signal sum-frequency terms are specified separately.}
\label{fig:final-window-doublet}
\end{figure}

Camera exposure averages motion and introduces blur alongside
localization noise~\cite{Berglund2010}. A midpoint-reported rectangular
exposure multiplies a harmonic amplitude by
$\operatorname{sinc}(\omega T_{\rm exp}/2)$.
Uniform sampling at $\Delta t$ identifies temporal frequencies modulo
$2\pi/\Delta t$ unless the band is resolved by additional information.
Exposure, sampling interval and record length are distinct measurement
inputs. Apply their calibrated maps to the predicted signal before
inferring a mechanical lifetime.

\subsubsection{Which modal variables can the record distinguish?}

The window factor also provides a direct test of the observation problem.
Work in a declared finite set of positive conservative modes with known
frequencies and known detector weights. Suppose that the positive-frequency
band has been isolated without losing its quadratures. \begin{definition}[Analytic modal displacement readout]
\label{def:s5-analytic-modal-readout}
For the ideal analytic displacement readout,
\begin{equation}
 y_{\rm an}(t)=\sum_s d_s\psi_s(0)e^{-i\omega_st},\qquad
 d_s=\frac{\sqrt2}{\omega_s}\vct\ell^{\mathsf T}\vct\phi_s,
 \label{eq:complex-modal-observation-map}
\end{equation}
where $\vct\phi_s=\mathsf M^{-1/2}\vct e_s$ is the existing
mass-normalized displacement pattern.
\end{definition}
The energy amplitude $\psi_s$ has
units $\sqrt{\mathrm J}$, and $d_s$ has units
$\mathrm m/\sqrt{\mathrm J}$. A coefficient $d_s=0$ describes a pattern
invisible to this detector. The analytic readout is a representation of
the measured real signal; the following calculation assumes its carrier
band and both quadratures are available. Equation~\eqref{eq:finite-window-probe-intensity}
remains the real-signal calculation when they are not.

\begin{definition}[Finite-time observation Gramian]
\label{def:s5-finite-time-observation-gramian}
For the analytic modal readout, the observation Gramian is the Hermitian
matrix whose quadratic form is the normalized squared signal size:
\begin{equation}
 \begin{split}
 \frac1{T_{\rm obs}}\int_{\rm window}|y_{\rm an}(t)|^2\,\dd t
       &=\psi(0)^\dagger\mathsf G_{\rm obs}\psi(0),\\
 (G_{\rm obs})_{st}
       &=d_s^*d_t\,e^{i(\omega_s-\omega_t)t_c}
          \operatorname{sinc}\!\left[
                   \frac{(\omega_s-\omega_t)T_{\rm obs}}2\right].
 \end{split}
 \label{eq:complex-window-observation-gram}
\end{equation}
\end{definition}
The defining squared signal proves nonnegativity for every coefficient
column. Its entries follow by integrating the modal exponential products.
This finite-time observability Gramian has entries equal to the overlaps
of the signals produced by unit modal coefficients. It specializes the
linear-system distinction between a retained state and the part recoverable
from an input/output record~\cite{Kalman1963}. Its diagonal records how strongly
each mode is seen; off-diagonal terms record how similar their finite
records are. \begin{proposition}[Finite-time modal distinguishability]
\label{prop:s5-finite-time-modal-distinguishability}
For two modes with $|d_1|=|d_2|=d>0$, the observation Gramian has
eigenvalues
\begin{equation}
 \boxed{\lambda_\pm=d^2\left[1\pm
       \left|\operatorname{sinc}\!\left(
                \frac{\Delta\omega T_{\rm obs}}2\right)\right|\right].}
 \label{eq:complex-two-mode-observability}
\end{equation}
More generally, distinct known frequencies with nonzero detector weights
give a positive-definite Gramian on every interval of positive duration
in the stipulated finite analytic model.
\end{proposition}
\begin{proof}
The two-mode matrix has diagonal $d^2$ and off-diagonal magnitude
$d^2|\operatorname{sinc}(\Delta\omega T_{\rm obs}/2)|$, giving the
two eigenvalues. For distinct frequencies, a zero signal norm makes
the finite exponential sum vanish throughout the interval. Its
successive derivatives at an interior point form a nonsingular
Vandermonde system, so every modal coefficient vanishes.
\end{proof}
Figure~\ref{fig:finite-window-modal-observability} shows their meaning.
At exact degeneracy, one detector sees only one combination of the two
coefficients. With equal real weights, $(\psi_1,\psi_2)\propto(1,-1)$
is invisible, whereas $(1,1)$ is visible. Changing basis inside that
degenerate subspace does not create a second measured quantity.

This ideal uniqueness still leaves a quantitative conditioning question:
how much can the inferred coefficients change when the recorded motion
changes slightly? With
$\delta=|\Delta\omega|T_{\rm obs}\ll1$,
$\lambda_-\simeq d^2\delta^2/24$ and
$\lambda_+/\lambda_-\simeq48/\delta^2$.
Two almost indistinguishable records can then correspond to substantially
different modal coefficients. If two trajectories of this stipulated
finite model produce analytic records differing by a root-mean-square
amount $\epsilon_y$ over the same window, their coefficient difference
satisfies the bound
$\|\delta\psi(0)\|\leq\epsilon_y/\sqrt{\lambda_-}$.
Model truncation, uncertain frequencies and detector calibration add
separate errors. A sinc zero is consequently a useful orthogonality scale
for this window, rather than a universal threshold below which inference
is impossible.

\begin{figure}[H]
\centering\begingroup
\definecolor{vltTeal}{HTML}{176B64}
\definecolor{vltBlue}{HTML}{315D80}
\definecolor{vltGold}{HTML}{A87520}
\definecolor{vltPlum}{HTML}{725877}
\definecolor{vltRed}{HTML}{762B38}
\definecolor{vltYellow}{HTML}{D5AD3C}
\definecolor{vltNeutral}{HTML}{777D83}
\setlength{\unitlength}{1pt}
\scalebox{0.80}{%
\input{vlt-figures/figure-175.tex}%
}\endgroup
\caption{The observation Gramian measures separation of temporal signatures, illustrating the boxed result \eqref{eq:complex-two-mode-observability}. (a) Two equal nonzero detector weights give $\lambda_\pm/|d|^2=1\pm|\operatorname{sinc}(\delta/2)|$, where $\delta=|\Delta\omega|T$. The small eigenvalue is $\delta^2/24+O(\delta^4)$ near degeneracy. (b) At exact degeneracy the coefficient combinations $(1,1)/\sqrt2$ and $(1,-1)/\sqrt2$ are bright and dark for the common analytic readout. The axes show a real coefficient slice of the complex modal space. (c) For the initially dark coefficients $\psi_*(1,-1)/\sqrt2$ referred to the window centre, the exact envelope is $\sqrt2|\sin(\delta s/2)|$. Finite frequency separation makes the record nonzero while its conditioning can remain weak. The centered unit-duration window, known frequencies, retained quadratures and calibrated detector define this auxiliary example; noise, exposure and unknown frequencies require their additional inference model.}
\label{fig:finite-window-modal-observability}
\end{figure}

Calibrated channels add their Gram matrices. Independent spatial
weights resolve a degenerate band when their restricted detector
matrix has full column rank. Proportional channels add no rank.
These tests distinguish a state sufficient for prediction from its
recovery in one record. Memory uses its augmented propagator;
nonlinear active motion uses the trajectory and retained source state.

\paragraph{The calibrated sampled record.}
Tracked centres provide discrete exposures and spatial projections.
Their measurement map gives a finite-record counterpart to the
continuous Gramian, testing which displacement and momentum
combinations survive observation.

\begin{definition}[Finite-record measurement design]
\label{def:s56-sampled-design}
For the same conservative release on $r$ retained positive modes, let
sample $j$ use detector row $a_j$ and a known real exposure kernel
$w_j(t)$ with $\int w_j(t)\,\dd t=1$. Define
\begin{equation}
 \begin{aligned}
 h_{js}&=d_{a_js}\int w_j(t)e^{-i\omega_st}\,\dd t,\\
 \vct\beta&=\begin{pmatrix}\operatorname{Re}\psi(0)\\
                         \operatorname{Im}\psi(0)\end{pmatrix},\\
 \mathsf V&=(\operatorname{Re}\mathsf h,
                                  -\operatorname{Im}\mathsf h),
 &\vct y&=\mathsf V\vct\beta+\vct n.
 \end{aligned}
 \label{eq:s56-sampled-real-design}
\end{equation}
Here $d_{as}$ is the analytic displacement weight in
\eqref{eq:complex-modal-observation-map}, $\vct n$ is additive
measurement noise, and $\mathsf V$ has one real row per recorded sample.
For centred rectangular exposure, the integral is
$e^{-i\omega_st_j}\operatorname{sinc}(\omega_sT_{{\rm exp},j}/2)$.
\end{definition}

This construction acts directly on the real data. It retains both
quadratures and their finite-window coupling without presuming that a
finite record has already supplied an exact analytic signal. Each
$h_{js}$ has units $\mathrm m/\sqrt{\mathrm J}$, while the components
of $\vct\beta$ have units $\sqrt{\mathrm J}$.

\begin{proposition}[Recoverability and covariance of a sampled record]
\label{prop:s56-sampled-recovery}
Treat the prepared amplitude as a fixed unknown. With frequencies,
exposure kernels and detector maps fixed, two
preparations give the same noiseless record exactly when their
$\vct\beta$ difference lies in $\ker\mathsf V$. If $\mathsf V$ has
full column rank and $\vct n$ has zero mean and known covariance
$\mathsf\Sigma_{\rm rec}>0$, weighted least squares gives
\begin{equation}
 \begin{aligned}
 \mathsf I_\beta&=\mathsf V^{\mathsf T}
                         \mathsf\Sigma_{\rm rec}^{-1}\mathsf V,\\
 \widehat{\vct\beta}&=\mathsf I_\beta^{-1}
        \mathsf V^{\mathsf T}\mathsf\Sigma_{\rm rec}^{-1}\vct y,
 &\operatorname{Cov}(\widehat{\vct\beta})&=\mathsf I_\beta^{-1}.
 \end{aligned}
 \label{eq:s56-sampled-estimator-covariance}
\end{equation}
The estimator is unbiased. For Gaussian noise with fixed covariance,
$\mathsf I_\beta$ is also the Fisher information for these real
amplitude coordinates.
\end{proposition}
\begin{proof}
Subtract the two noiseless measurement equations to obtain the kernel
statement. Full column rank makes $\mathsf I_\beta$ positive definite.
Substitution of $\vct y=\mathsf V\vct\beta+\vct n$ gives
$\widehat{\vct\beta}=\vct\beta+
\mathsf I_\beta^{-1}\mathsf V^{\mathsf T}
\mathsf\Sigma_{\rm rec}^{-1}\vct n$; taking its mean and covariance
proves the claims. Differentiating the Gaussian log likelihood gives
the same information matrix.
\end{proof}

The matrix formulation is standard linear inverse estimation
\cite{Wunsch2005BasicMachinery}; here its columns are explicitly
the measured histories of particle-motion patterns. The estimator
covariance quantifies measurement uncertainty at a fixed preparation;
spread between repeated preparations is the separate ensemble covariance
of Section~\ref{sec:complex-field-observables}. It distinguishes
loss of information from large statistical uncertainty. A zero exposure
factor can remove a mode altogether. With uniform times
$t_j=t_0+j\Delta t$ and fixed exposure, frequencies differing by
$2\pi k/\Delta t$, $k\in\mathbb Z$, have proportional temporal
columns in one detector whenever their exposure factors are nonzero.
Their unknown amplitudes can then compensate exactly, regardless of
record duration. Real sampling also identifies the corresponding
conjugate-frequency aliases through the same cosine--sine column space.
Independent spatial channels can separate coincident temporal columns
when the full stacked columns become independent. This makes exposure,
temporal sampling and particle-pattern visibility parts of one concrete
inference problem.

Known-frequency amplitude recovery is a different question from fitting
the frequencies or medium coefficients themselves. The distinction can
also be calculated rather than assigned a universal duration threshold.
Let a differentiable real predicted record be
$\vct m(\vct\beta,\vct\theta)$, where $\vct\theta$ are declared
dimensionless physical parameter coordinates. At an interior reference
write $\mathsf V=\partial\vct m/\partial\vct\beta$ and
$\mathsf J_\theta=\partial\vct m/\partial\vct\theta$.

\begin{proposition}[Parameter information after allowing amplitude changes]
\label{prop:s56-profiled-parameter-information}
For full-column-rank $\mathsf V$ and the same fixed noise covariance,
put $\mathsf V_w=\mathsf\Sigma_{\rm rec}^{-1/2}\mathsf V$ and
$\mathsf J_w=\mathsf\Sigma_{\rm rec}^{-1/2}\mathsf J_\theta$.
After arbitrary first-order changes of the unknown amplitudes, the
remaining squared change of the whitened predicted record is
\begin{equation}
 \begin{aligned}
 \min_{\delta\vct\beta}
 \|\mathsf V_w\delta\vct\beta+
                       \mathsf J_w\delta\vct\theta\|^2
    &=\delta\vct\theta^{\mathsf T}
                     \mathsf I_{\theta\mid\beta}\delta\vct\theta,\\
 \mathsf I_{\theta\mid\beta}
    &=\mathsf J_w^{\mathsf T}(\Id-\mathsf\Pi_V)\mathsf J_w,
 &\mathsf\Pi_V&=\mathsf V_w
        (\mathsf V_w^{\mathsf T}\mathsf V_w)^{-1}\mathsf V_w^{\mathsf T}.
 \end{aligned}
 \label{eq:s56-profiled-parameter-information}
\end{equation}
For fixed-covariance Gaussian noise this is the local information for
$\vct\theta$ after profiling the amplitude coordinates.
\end{proposition}
\begin{proof}
Least squares cancels the component of
$\mathsf J_w\delta\vct\theta$ in $\operatorname{ran}\mathsf V_w$.
The residual is its orthogonal projection through $\Id-\mathsf\Pi_V$.
This projector is symmetric and idempotent, so its squared norm gives
the displayed form. Equivalently it is the Schur complement of the
amplitude block in the joint Gaussian information matrix.
\end{proof}

A parameter perturbation in the amplitude span is invisible to first
order after preparation is refitted. Exact ambiguity requires a
separate argument. Another spatial channel, preparation or resolved
transient can add a component outside that span and distinguish
initial excitation from force-law changes.

\subsubsection{Cross-spectra, calibration, and sources of coherence}

\begin{definition}[Stationary channel cross-spectrum and coherence]
\label{def:s5-stationary-channel-coherence}
For centered stationary classical displacement fluctuations on frequencies
with regular spectral densities, choose real probe vectors $\vct\ell_a$ and $Y_a=\vct\ell_a^{\mathsf T}\vct u$.
Their cross-spectrum and normalized coherence are
\begin{align}
 S_{ab}(\omega)&=\int_{-\infty}^{\infty}e^{i\omega t}
       \langle\delta Y_a(t)\delta Y_b(0)\rangle\,\dd t,
       \label{eq:phase-resolved-probe-cross-spectrum}\\
 \gamma_{ab}(\omega)&=\frac{S_{ab}(\omega)}
          {\sqrt{S_{aa}(\omega)S_{bb}(\omega)}},\qquad
                 |\gamma_{ab}(\omega)|\leq1,
       \label{eq:classical-probe-coherence}
\end{align}
where the two auto-spectra are positive.
\end{definition}
If both readouts have length
units, $S_{ab}$ has units $\mathrm m^2\,\mathrm s$ in this angular-frequency
convention, whereas $\gamma_{ab}$ is dimensionless. Some analyses report
the squared magnitude $|\gamma_{ab}|^2$ instead of the magnitude; the
chosen convention must be stated. The bound follows from positivity
of their $2\times2$ spectral matrix. Exact finite conservative spectra
can instead be measures with discrete lines; their coherence uses specified
frequency-bin spectral matrices rather than a pointwise ratio of delta
functions. Its phase is defined only when
$S_{ab}\ne0$ and relative timing is calibrated.

\begin{proposition}[Equilibrium cross-phase for time-even channels]
\label{prop:s5patch-equilibrium-cross-phase}
Consider real instantaneous classical laboratory readouts $Y_a,Y_b$ that are even
under time reversal, in a stationary reciprocal thermal equilibrium
without a time-reversal-breaking control. When the covariance spectra
exist, their equilibrium relation is
\begin{equation}
 C_{ab}(t)=C_{ba}(t)=C_{ab}(-t),\qquad
 S_{ab}(\omega)=S_{ba}(\omega)\in\R .
 \label{eq:s5patch-equilibrium-cross-phase}
\end{equation}
Thus the cross-phase is $0$ or $\pi$ wherever the cross-spectrum is nonzero.
\end{proposition}
\begin{proof}
Microscopic reversibility exchanges the two time-even observables in
the equilibrium correlation. Stationarity and commutativity of the
classical readouts give $C_{ba}(t)=C_{ab}(-t)$. The real correlation
is therefore even, and its Fourier transform is real and symmetric.
\end{proof}
This is an equilibrium consistency test based on time-reversal symmetry
\cite{Onsager1931,OhgaItoKolchinsky2023}. Centre coordinates, registered
small angular coordinates and surface populations are time-even;
velocity or current channels have different parity. A driven transfer
function can still have a material lag in this same passive model.
The proposition concerns unforced stationary fluctuations, with relative
detector timing and filters accounted for. Initial preparation, event
conditioning and maintained drive require their own joint statistics.
A real cross-spectrum is necessary under the stated equilibrium
hypotheses; it alone does not identify equilibrium or a microscopic
mechanism. The following detector map makes the timing requirement explicit.

\begin{proposition}[Pure detector delay and cross-phase]
\label{prop:s5-detector-delay-cross-phase}
For fixed detector delay $t_a^{\rm det}$ in each stationary channel,
the recorded signal $Y_a^{\rm rec}(t)=Y_a(t-t_a^{\rm det})$ has
\begin{equation}
 S_{ab}^{\rm rec}(\omega)=
       e^{i\omega(t_a^{\rm det}-t_b^{\rm det})}S_{ab}(\omega).
 \label{eq:detector-delay-cross-phase}
\end{equation}
Its pointwise coherence magnitude is unchanged where defined.
\end{proposition}
\begin{proof}
A delayed signal has Fourier multiplier $e^{i\omega t_a^{\rm det}}$
under the stated convention. Multiplying one channel by the conjugate
of the other gives the displayed cross-phase. Each multiplier has unit
modulus, so the auto-spectra and coherence magnitude are unchanged.
\end{proof}
Such fixed pure delays rotate the cross-phase while leaving the pointwise
coherence magnitude unchanged. A differential channel delay can therefore
appear as a mechanical phase lag until it is calibrated. Timing jitter,
frequency-dependent detector transfer and added noise require their own
measurement models. In the stationary linear
reservoir setting of Section~4, collecting probe rows into $\mathsf L_p$
gives
\begin{equation}
 \mathsf S_Y=\mathsf L_p\mathsf R_{\rm mech}\mathsf S_F
                       \mathsf R_{\rm mech}^\dagger\mathsf L_p^{\mathsf T}.
 \label{eq:probe-coherence-response-filter}
\end{equation}
A single common random drive can give unit coherence for two nonzero
readouts even with unequal amplitudes and a phase lag.

For example, two linear readouts, even from mechanically uncoupled
particles, of one centered stationary force $\xi$ have
Fourier amplitudes $\widehat Y_a=h_a\widehat\xi$. Wherever the common
force spectrum and both transfers are nonzero,
\begin{equation}
 S_{ab}=h_ah_b^*S_\xi,\qquad
 \gamma_{ab}=\frac{h_ah_b^*}{|h_a||h_b|},\qquad
                                  |\gamma_{ab}|=1.
 \label{eq:common-force-readout-coherence}
\end{equation}
Coherence measures the selected readouts' frequency-resolved phase
relationship, independently of their absolute displacement or energy.
Estimation needs declared averaging or smoothing; averaged windowed
periodograms provide one method~\cite{Welch1967}. A single Fourier-vector
outer product gives unit apparent coherence wherever both components
are nonzero~\cite{Wunsch2005}. This statistic describes classical
channel correlation; quantum state preparation requires its separately
specified tests.

More generally, calibrated linear channel responses $D_a(\omega)$ and
additive detector noises $n_a$ give
$\widehat Y_a^{\rm rec}=D_a\widehat Y_a+\widehat n_a$.
When the noises are uncorrelated with the mechanical signal and with one
another,
\begin{equation}
 S_{ab}^{\rm rec}=D_aD_b^*S_{ab}\quad(a\ne b),\qquad
 S_{aa}^{\rm rec}=|D_a|^2S_{aa}+N_a,
 \label{eq:complex-detector-noise-coherence}
\end{equation}
where $N_a\geq0$ is the noise auto-spectrum. Nonzero deterministic filters
alone preserve the coherence magnitude at a frequency; independent added
noise generally reduces it. Shared detector noise instead contributes to
the cross-spectrum and can create apparent agreement between channels.
Pure timing delays are the special case already displayed above.

A calibrated pressure modulation provides one possible common
mechanical input in a driven capillary comparison. It can produce
correlated particle-displacement channels through the same surrounding
flow, even when those channels do not identify a direct pair contact.
Recording the pressure input and the spatial response tests that declared
force--readout law. Mutual coherence alone does not separate a shared
fluid drive, a direct interparticle coupling and a shared detector
artifact; the calibrated input and complementary channels supply that
additional physical information.
Coherence should therefore be reported with its averaging convention,
instrument response and noise assumptions~\cite{Wunsch2005}.

\paragraph{Spectral averaging also averages phase relations.}
The coherence of a resolved frequency and the coherence of a finite
spectral estimate are related through the chosen window. The relation
can be retained at the level of the whole spectral matrix.

\begin{proposition}[Expected windowed spectral matrix]
\label{prop:s56-windowed-spectral-matrix}
Let the centred stationary readout vector have finite second moments
and integrable spectral density matrix $\mathsf S_Y(\nu)$. For a
nonzero real compactly supported window $w\in L^2$, define
$\widetilde{\vct Y}_w(\omega)=
\int w(t)\delta\vct Y(t)e^{i\omega t}\,\dd t$ and
$\mathcal N_w=\int |w(t)|^2\,\dd t$. Then
\begin{equation}
 \begin{aligned}
 \mathsf S_Y^{(w)}(\omega)
  &:=\frac{\langle\widetilde{\vct Y}_w(\omega)
             \widetilde{\vct Y}_w(\omega)^\dagger\rangle}{\mathcal N_w}
    =\int K_w(\omega-\nu)\mathsf S_Y(\nu)\,\dd\nu,\\
 K_w(\zeta)&=\frac{|\widehat w(\zeta)|^2}{2\pi\mathcal N_w},
 &K_w&\geq0,\qquad\int K_w(\zeta)\,\dd\zeta=1.
 \end{aligned}
 \label{eq:s56-windowed-spectral-matrix}
\end{equation}
Consequently $\mathsf S_Y^{(w)}$ is nonnegative Hermitian, and its
normalized two-channel coherence has magnitude at most one wherever
the diagonal entries are positive.
\end{proposition}
\begin{proof}
Insert the inverse Fourier representation of the stationary covariance
into the double integral for the numerator. Integrating the two window
factors gives $|\widehat w(\omega-\nu)|^2$; the Fourier convention
supplies $1/(2\pi)$. Parseval gives the unit integral of $K_w$.
Nonnegative averaging preserves Hermitian positivity, and the
$2\times2$ principal-minor inequality gives the coherence bound.
\end{proof}

For a rectangular record, $K_w(\zeta)=
T_{\rm obs}\operatorname{sinc}^2(\zeta T_{\rm obs}/2)/(2\pi)$.
This is the duration kernel underlying the line broadening in
Figure~\ref{fig:final-window-doublet} and the modified-periodogram
construction~\cite{Welch1967,Wunsch2005}. The ratio formed from
$\mathsf S_Y^{(w)}$ is a population coherence of the windowed matrix;
it is not the expectation of the nonlinear sample-coherence estimator.

Distinct phase relations inside the same spectral window can cancel
even when each resolved contribution has unit coherence. For example,
equal weighted spectral contributions $P>0$ from common and relative
two-particle patterns give
\begin{equation}
 \begin{aligned}
 \mathsf S^{(+)}&=P\begin{pmatrix}1&1\\1&1\end{pmatrix},
 &\mathsf S^{(-)}&=P\begin{pmatrix}1&-1\\-1&1\end{pmatrix},\\
 \mathsf S^{(+)}+\mathsf S^{(-)}&=2P\Id_2.
 \end{aligned}
 \label{eq:s56-windowed-phase-cancellation}
\end{equation}
Here $P$ carries the units of the smoothed spectral matrix. Each
contribution is rank one, but their sum has zero cross-spectrum.
For the ideal-virion pair the two centre readouts see the common pattern
with equal signs and the relative pattern with opposite signs. A
window pooling their equally weighted contributions therefore reports
zero coherence even while each resolved pattern is fully correlated
between the two readouts. This is spectral pooling, in addition to the
independent-noise mechanism already described above. Comparing the same
preparation across resolved bands or windows helps decide whether an
observed coherence change calls for an added dynamical variable or is
already predicted by the known particle patterns and observation map.

\paragraph{Finite preparation enters the measured spectral matrix.}
\begin{proposition}[Windowed covariance after a prepared internal state]
\label{prop:s5patch-prepared-window-spectrum}
In a fixed stable linear episode, let a real augmented state obey
$\dd\vct z=\mathsf A_0\vct z\,\dd t+\mathsf B_0\dd\vct W$,
with invariant covariance $\mathsf\Sigma_{\rm eq}$ and prepared
covariance $\mathsf\Sigma_0$ at time $t_0$, independent of later
Wiener increments. Use fixed calibrated readouts
$\vct Y=\mathsf L_0\vct z$, subtract their time-dependent ensemble
mean, and choose a real window supported at $t\geq t_0$. Define
\begin{equation}
 \mathsf H_w(\omega)=\int w(t)\mathsf L_0
       e^{\mathsf A_0(t-t_0)}e^{i\omega t}\,\dd t,
 \qquad \mathcal N_w=\int|w(t)|^2\,\dd t>0 .
 \label{eq:s5patch-window-preparation-map}
\end{equation}
The windowed spectral matrix satisfies
\begin{equation}
 \boxed{\mathsf S_{Y,\rm prep}^{(w)}(\omega)
    =\mathsf S_{Y,\rm eq}^{(w)}(\omega)
      +\frac{\mathsf H_w(\omega)
             (\mathsf\Sigma_0-\mathsf\Sigma_{\rm eq})
                     \mathsf H_w(\omega)^\dagger}{\mathcal N_w}.}
 \label{eq:s5patch-prepared-window-spectrum}
\end{equation}
\end{proposition}
\begin{proof}
Variation of constants gives the equilibrium two-time covariance plus
$e^{\mathsf A_0(t-t_0)}(\mathsf\Sigma_0-\mathsf\Sigma_{\rm eq})
e^{\mathsf A_0^{\mathsf T}(s-t_0)}$.
Insert this expression in the double window integral. The correction
factors into the two $\mathsf H_w$ maps, while the equilibrium term is
the stationary window average of Proposition~\ref{prop:s56-windowed-spectral-matrix}.
\end{proof}

Figure~\ref{fig:cmd-prepared-projected-records} gives a graphical specialization of this boxed relation.

\begin{figure}[H]
\centering\begingroup
\setlength{\unitlength}{1pt}
\scalebox{0.84}{%
\input{vlt-figures/figure-312.tex}%
}\endgroup
\caption{Two assigned Gaussian examples give distinct tests of retained information. (a) Let $Y$ be the normalized sum of independent stationary Ornstein--Uhlenbeck coordinates with rates $1$ and $4$. Then $K(t)=(e^{-t}+e^{-4t})/2$, $K(0)=1$ and the boxed conditional covariance \eqref{eq:s6patch-gaussian-record-defect} gives $\Delta(t,t)=(e^{-t}-e^{-4t})^2/4$. (b) A separate scalar process has drift $-1$, equilibrium variance one and a rectangular window of duration two. Its excess periodogram expectation is $\Delta\Sigma|1-e^{(-1+i\omega)2}|^2/[2(1+\omega^2)]$, illustrating \eqref{eq:s5patch-prepared-window-spectrum}. Both initial variances $1+\Delta\Sigma$ are positive; a negative excess does not make the total spectrum negative. The examples use dimensionless time and compare different records, rather than identifying a unique hidden biological mechanism.}
\label{fig:cmd-prepared-projected-records}
\end{figure}

The preparation correction can have either sign; the complete covariance
matrix remains nonnegative. If the time-dependent mean is left in the
record, its Fourier outer product contributes as well. This pushes the
internal-force correction in~\eqref{eq:s4patch-prepared-force-covariance}
through the actual readout and window. Waiting time, observation duration
and detector calibration can consequently separate a changing preparation
from a changed stationary bath law. A prescribed time-dependent protocol
uses its propagator. Conditioning on a state-dependent observed
recruitment or survival history instead uses the corresponding joint
conditional law, as in the next subsection.

\begin{example}[A transient cross-phase from prepared correlations]
\label{ex:s5patch-prepared-two-rate-phase}
Let two dimensionless time-even coordinates obey independent passive
Ornstein--Uhlenbeck equations
$\dd z_j=-\lambda_jz_j\dd t+\sqrt{2\lambda_j}\dd W_j$,
with invariant covariance $\Id_2$. Prepare a centred Gaussian with
$\mathsf\Sigma_0=\left(\begin{smallmatrix}1&c\\c&1\end{smallmatrix}\right)$,
$0<c<1$, at $t_0=0$. Their marginal laws remain stationary, whereas
$C_{12}(t,s)=c e^{-\lambda_1t-\lambda_2s}$ relaxes. A rectangular
window $[t_a,t_a+T]$ gives
\begin{equation}
 S_{12}^{(w)}(\omega)=\frac cT h_1(\omega)h_2(\omega)^*,\qquad
 h_j(\omega)=e^{(-\lambda_j+i\omega)t_a}
        \frac{1-e^{-(\lambda_j-i\omega)T}}{\lambda_j-i\omega} .
 \label{eq:s5patch-two-rate-window-cross-spectrum}
\end{equation}
For $\lambda_1=1/\tau_\star$, $\lambda_2=3/\tau_\star$,
$T=\tau_\star$ and $\omega=\pi/\tau_\star$, its cross-phase is
$\tan^{-1}\pi-\tan^{-1}(\pi/3)\simeq26.0^\circ$.
Its magnitude decays as $e^{-4t_a/\tau_\star}$, with unchanged marginal
auto-spectra. This synthetic Fourier-bin phase comes from a prepared
joint correlation relaxing under fixed passive dynamics. Comparing
preparation ages tests that account; stationary equilibrium requires
the joint preparation, not just its two marginal variances.
\end{example}

Stationary spectra describe ensemble correlations or a justified
stationary record; a driven phasor describes a particular response.
Thermal resonance, imposed drive, beats and maintained self-excitation
can each give narrow features. Their preparation and work balances
distinguish them. A finite-resource burst retains its evolving state
or uses an explicitly justified local-stationary approximation.

Figure~\ref{fig:complex-common-source-coherence} distinguishes a common source from a direct mechanical coupling and shows how independent detector noise changes coherence.

\begin{figure}[H]
\centering\begingroup
\definecolor{vltTeal}{HTML}{176B64}
\definecolor{vltBlue}{HTML}{315D80}
\definecolor{vltGold}{HTML}{A87520}
\definecolor{vltPlum}{HTML}{725877}
\definecolor{vltRed}{HTML}{762B38}
\definecolor{vltYellow}{HTML}{D5AD3C}
\definecolor{vltNeutral}{HTML}{777D83}
\setlength{\unitlength}{1pt}
\scalebox{0.80}{%
\input{vlt-figures/figure-176.tex}%
}\endgroup
\caption{Two mechanically uncoupled coordinates respond to the same scalar input through $R_j$. Open arrows are signal dependencies; summing nodes add independent detector noises, also independent of $f$. For equal nonzero response magnitudes and noise spectra $N$, squared coherence is $(1+q)^{-2}$ with $q=N/(|R_1|^2S_f)$, $S_f>0$. The zero-noise endpoint is one. This calculation applies where the signal spectra are nonzero and tests the common-source measurement model; transferred work requires its force-conjugate account.}
\label{fig:complex-common-source-coherence}
\end{figure}

\subsubsection{Phase records conditioned on an assembly history}
\label{sec:s5-phase-assembly-record}

\begin{definition}[Reproducible assembly phase record]
\label{def:s5-phase-record}
A registered record specifies persistent virion identities, calibrated detector
channels and their delays, the common drive or detector clock, the
analytic-signal or demodulation filter, and the exposure window. It records
the membership history and the reference and mode-frame maps used in
reporting collective coefficients. Event-aligned comparisons use the same
declared time $\tau=t-t_*$ relative to a registered recruitment or breakup
event, together with any retained phase of the external drive.
\end{definition}

The experimental question is whether the same virion pair retains a
reproducible relative phase after a neighbour joins. Calibrated particle
channels preserve identity; a cluster-centre channel additionally
uses membership weights and~\eqref{eq:s4-registration-centre-shift}.
Inter-episode comparisons retain the drive phase and clock registration.
Within an episode, same-frequency relative phase stays invariant under
a common clock shift.

\begin{proposition}[Registered moments and arrangement pooling]
\label{prop:s5-phase-registered-moments}
For a complex displacement coefficient with conditional mean $\vct\mu_c$
and Hermitian covariance $\mathsf C_c$, a unitary registration
$\vct c'=\mathsf Q^\dagger\vct c$ and detector row matrix
$\mathsf D'=\mathsf D\mathsf Q$ give
\begin{equation}
 \vct\mu'_c=\mathsf Q^\dagger\vct\mu_c,
 \qquad
 \mathsf C'_c=\mathsf Q^\dagger\mathsf C_c\mathsf Q,
 \qquad
 \mathsf D'\mathsf C'_c\mathsf D'^\dagger
       =\mathsf D\mathsf C_c\mathsf D^\dagger.
 \label{eq:s5-phase-gauge-covariance}
\end{equation}
For the same registered analytic detector vector $\vct Z(\tau)$, let
$p_g$ be the probability of arrangement $g$ in the declared event-aligned
ensemble, and let $\vct\mu_g$ and $\mathsf C_g$ be its conditional moments.
Then
\begin{equation}
 \begin{aligned}
 \vct\mu&=\sum_g p_g\vct\mu_g,\\
 \mathsf C&=\sum_g p_g\mathsf C_g+
   \sum_g p_g(\vct\mu_g-\vct\mu)(\vct\mu_g-\vct\mu)^\dagger.
 \end{aligned}
 \label{eq:s5-phase-arrangement-covariance}
\end{equation}
Both contributions to $\mathsf C$ are positive semidefinite.
\end{proposition}
\begin{proof}
Linear substitution gives the registration identities. For pooling, write
$\vct Z-\vct\mu=(\vct Z-\vct\mu_g)+(\vct\mu_g-\vct\mu)$ conditional
on $g$. The first term has zero conditional mean, so the cross terms
vanish after averaging. Each remaining outer product is positive
semidefinite.
\end{proof}

The first covariance term describes variability within arrangements;
the second measures variability between their conditional means. A
large pooled cross-covariance can therefore arise because two channels
both change their means when a virion joins. It does not establish a
persistent phase relation within either arrangement. Conversely, pooling
episodes with different reproducible phases can reduce the complex mean.
The stated Hermitian moments describe normal complex correlations;
complete real-quadrature statistics also retain the symmetric moment
$\mathbb E[(\vct Z-\vct\mu)(\vct Z-\vct\mu)^{\mathsf T}]$ when it is
nonzero.

An overlap congruence $\mathsf O\mathsf C_c\mathsf O^\dagger$ describes
projection of the same random displacement pattern. It does not determine
the physical post-recruitment preparation: that prediction also uses the
event dynamics, newly resolved coordinates and prepared medium memory.
For nonstationary events the two-time response
\eqref{eq:s4-two-time-response} propagates a specified source to a
registered receiver. Residence-dependent phase evolution in
\eqref{eq:s4-exchange-telegraph-phase} and the joint membership/readout
description in \eqref{eq:s4-identification-switching-readouts} then
distinguish phase dispersion between episodes from local mechanical
relaxation. Conditioning on survival requires that same joint law when
escape depends on strain or excitation; surviving records can select a
different preparation distribution.

Equal event weights describe typical recruitment events; pooling every
sample weights arrangements by residence time. These measures can
differ when lifetime depends on preparation. Report conditional
statistics with event counts or residence durations. If episodes are
shorter than demodulation windows, the filtered record mixes histories;
its registered trace can still be compared with the two-time prediction.

In a cylindrical suspension this yields a practical comparison between a
persistent core, a newly recruited particle and the surrounding fluid.
Record core and newcomer motion, membership and calibrated forcing in the
same episodes; compare their conditional response and residual fluctuations
before pooling. A claim of source--receiver energy transfer additionally
uses force-conjugate velocities, source work and the relevant stored and
dissipated energies. Phase lag and high coherence alone do not set the
direction or amount of mechanical power.

Figure~\ref{fig:s5-assembly-conditioned-statistics} separates conditional fluctuations, changing arrangement means and the weighting of recorded episodes.

\begin{figure}[H]
\centering\begingroup
\definecolor{vltTeal}{HTML}{176B64}
\definecolor{vltBlue}{HTML}{315D80}
\definecolor{vltGold}{HTML}{A87520}
\definecolor{vltPlum}{HTML}{725877}
\definecolor{vltRed}{HTML}{762B38}
\definecolor{vltYellow}{HTML}{D5AD3C}
\definecolor{vltNeutral}{HTML}{777D83}
\setlength{\unitlength}{1pt}
\scalebox{0.90}{%
\input{vlt-figures/figure-177.tex}%
}\endgroup
\caption{Conditional statistics on one four-virion pool. Reference centres and identities match in the cylinder views; selected membership changes from $\{1,2\}$ to $\{1,2,3\}$. Core-channel means are $\pm\ell_\star(1,i)^{\mathsf T}$ with independent centered noises $\ell_\star\{1,-1,i,-i\}/2$. Panel (c) displays all sixteen noise combinations per arrangement, using dot area for multiplicity and crosses for means. Equal event weights give the displayed within/between covariance split; all pooled cross-covariance comes from changing means. The duration strip changes the weights when pooling by observation time. These are registered statistical preparations with a specified geometry.}
\label{fig:s5-assembly-conditioned-statistics}
\end{figure}

\subsection{A phase-resolved envelope--RNP illustration}
\label{sec:measles-phase-response}

Envelope and internal motion supply two readouts of one structural
model. Measles membrane-associated components and the RNA--protein
ribonucleoprotein assembly motivate distinct coordinates
~\cite{Liljeroos2011,Ke2018}. The Section~3 surrogate assigns inertia
and coupling to their projections. One applied force then produces
different readouts at fixed mechanical parameters, including small
envelope motion alongside substantial internal motion.

\subsubsection{Structural coordinates and work-conjugate forces}

Specialize to one retained displacement direction. Let $u_{\rm e}(t)$
and $u_{\rm r}(t)$ be real envelope-associated and internal RNP-associated
coordinates, with positive effective inertias $m_{\rm e}$ and $m_{\rm r}$.
They describe the chosen projections of motion. They need not be the
geometric centres of two rigid biological pieces, and a deformation
coordinate requires its declared mode shape and inertial normalization.
For this translational two-mass surrogate, define

\begin{equation}
 U_{\mathrm c}=
       \frac{m_{\mathrm e}u_{\mathrm e}+m_{\mathrm r}u_{\mathrm r}}{m},
 \qquad w=u_{\mathrm e}-u_{\mathrm r},\qquad
 m=m_{\mathrm e}+m_{\mathrm r},\qquad
 \mu=\frac{m_{\mathrm e}m_{\mathrm r}}m.
 \label{eq:center-relative-biophysical-coordinates}
\end{equation}
Here $U_{\mathrm c}$ is a real center displacement and $w$ is a real relative
coordinate; their phasors will be denoted $\mathcal U_{\mathrm c}$ and
$\mathcal W_{\mathrm r}$. The kinetic energy is
$m\Dt{U_{\mathrm c}}^2/2+\mu\Dt{w}^2/2$.
The two effective masses belong to one virion, so this refinement preserves
the fixed virion count.

\begin{proposition}[Work-conjugate centre and relative forces]
\label{prop:s5-centre-relative-conjugate-forces}
For the constant positive masses and coordinate transformation above,
preparation forces $f_{\mathrm e}$ and $f_{\mathrm r}$ give the
generalized forces and power identity
\begin{equation}
 F_{\mathrm c}=f_{\mathrm e}+f_{\mathrm r},\qquad
 F_{\mathrm r}=
                \frac{m_{\mathrm r}f_{\mathrm e}-m_{\mathrm e}f_{\mathrm r}}m,
 \qquad
 f_{\mathrm e}\Dt{u_{\mathrm e}}+f_{\mathrm r}\Dt{u_{\mathrm r}}
                    =F_{\mathrm c}\Dt{U_{\mathrm c}}+F_{\mathrm r}\Dt{w}.
 \label{eq:internal-coordinate-force-power}
\end{equation}
\end{proposition}
\begin{proof}
The inverse map is $u_{\rm e}=U_{\rm c}+\beta w$ and
$u_{\rm r}=U_{\rm c}-(1-\beta)w$, with $\beta=m_{\rm r}/m$.
Substitute its velocities into the applied power and collect the
coefficients of $\dot U_{\rm c}$ and $\dot w$.
\end{proof}
The forces are work conjugate to these coordinates: multiplying each by
its corresponding velocity gives power in watts. Changing coordinates
therefore transforms the force as well as the displacement. A force on
the surface need not act as a pure centre force in the refined model.
Figure~\ref{fig:envelope-rnp-coordinate-cancellation} shows the coordinate
and force geometry used below.

\begin{figure}[H]
\centering\begingroup
\definecolor{vltTeal}{HTML}{176B64}
\definecolor{vltBlue}{HTML}{315D80}
\definecolor{vltGold}{HTML}{A87520}
\definecolor{vltPlum}{HTML}{725877}
\definecolor{vltRed}{HTML}{762B38}
\definecolor{vltYellow}{HTML}{D5AD3C}
\definecolor{vltNeutral}{HTML}{777D83}
\setlength{\unitlength}{1pt}
\scalebox{0.90}{%
\input{vlt-figures/figure-178.tex}%
}\endgroup
\caption{Two structural projections refine one whole-particle identity. The single faceted glyph maps to the envelope-associated and RNP-associated coordinate ports; the boxes are effective inertias, not extra virions or a literal anatomical drawing. The upper spring/dashpot symbols specify the constitutive topology $A=\Lambda_s^*-i\omega c_{\rm e}$ and $B$ without encoding a deformed snapshot. The left anchor supplies the specified surrounding-network reference. For equal participating masses, $B=k_{\rm int}$, $f_{\rm r}=0$ and $\omega^2=k_{\rm int}/m_{\rm r}$, choose $U_\star=\mathcal F_{\rm e}/k_{\rm int}>0$. The four signed phasors use one horizontal scale: the two envelope contributions cancel, while the internal readout remains $-U_\star$ \eqref{eq:antiresonance-phase-cancellation}. Their arrows are phasor contributions along the real axis. The cycle-averaged internal kinetic-plus-spring energy is still $|\mathcal F_{\rm e}|^2/(2k_{\rm int})$. This exact surrogate assigns no measured viral parameters.}
\label{fig:envelope-rnp-coordinate-cancellation}
\end{figure}

Additional patterns can resolve extended RNP motion, shell bending,
matrix contacts or a marked glycoprotein. The retained resolution
follows the measured motion; numerical curvatures require independent
mechanical inputs beyond structural images.

\subsubsection{Coupled response and an observable cancellation}

The shell-only interaction model of Section~3 makes the coupling between
these coordinates explicit. At a selected bulk wave vector and direction,
let $A(\vct k,\omega)$ be the passive shell dynamic stiffness, including
any specified envelope resistance, and let $B(\omega)=k_{\rm int}^*(\omega)$
be the passive internal stiffness. Define $\beta=m_{\mathrm r}/m$.
Their physical envelope/RNP pencil and transformed center/relative pencil are
\begin{align}
 \mathsf P_{\rm er}&=
  \begin{pmatrix}A+B-m_{\mathrm e}\omega^2&-B\\
                 -B&B-m_{\mathrm r}\omega^2\end{pmatrix},
       \label{eq:coupled-envelope-rnp-pencil}\\
 \mathsf P_{\rm cr}&=
  \begin{pmatrix}A-m\omega^2&\beta A\\
                 \beta A&B+\beta^2 A-\mu\omega^2\end{pmatrix}.
       \label{eq:coupled-center-relative-pencil}
\end{align}
Here $A$ and $B$ have stiffness units; their negative imaginary parts
represent the specified passive loss at $\omega>0$. For example,
$A=\Lambda_s^*(\vct k,\omega)-i\omega c_{\mathrm e}$ with
$c_{\mathrm e}\geq0$, and
$B=k_{\rm int}+\Delta k_{\rm int}[-i\omega\tau/(1-i\omega\tau)]$
with $k_{\rm int}>0$ and $\Delta k_{\rm int}\geq0$.
\begin{proposition}[Work-preserving change of response coordinates]
\label{prop:s5-work-preserving-response-congruence}
For the stated two-coordinate model, the real coordinate map and
transformed dynamical pencil satisfy
\begin{equation}
 \begin{aligned}
 \begin{pmatrix}u_{\rm e}\\u_{\rm r}\end{pmatrix}
   &=\mathsf T_{\rm cr}\begin{pmatrix}U_{\rm c}\\w\end{pmatrix},
 &\quad\mathsf T_{\rm cr}&=\begin{pmatrix}1&\beta\\1&-(1-\beta)\end{pmatrix},\\[2pt]
 \mathsf P_{\rm cr}&=\mathsf T_{\rm cr}^{\mathsf T}
                                  \mathsf P_{\rm er}\mathsf T_{\rm cr}.
 \end{aligned}
 \label{eq:complex-work-preserving-congruence}
\end{equation}
The force column transforms by $\mathsf T_{\rm cr}^{\mathsf T}$.
Both virtual work and the passive loss quadratic form are preserved.
\end{proposition}
\begin{proof}
Substitute the real coordinate map into the physical-coordinate
equation and multiply by $\mathsf T_{\rm cr}^{\mathsf T}$.
The same substitution in force times virtual displacement proves force
conjugacy. Because the map is real, taking the negative imaginary part
commutes with this congruence, preserving the loss form.
\end{proof}
The corresponding force column is multiplied by
$\mathsf T_{\rm cr}^{\mathsf T}$, exactly as virtual work requires.
This congruence preserves virtual work and the loss quadratic form.
A mass-normalized generator instead transforms by similarity.
The internal relaxed reference frequency in an isolated cell is
$\omega_0=\sqrt{k_{\rm int}/\mu}$. For the same independent Maxwell
storage realization, the selected cell's cycle-averaged energy and loss are
\begin{align}
 \overline E_{\rm cell}&=\tfrac14\left[
       \omega^2(m|\mathcal U_{\mathrm c}|^2+\mu|\mathcal W_{\mathrm r}|^2)
       +\operatorname{Re}A|\mathcal U_{\mathrm c}
                                  +\beta\mathcal W_{\mathrm r}|^2
       +\operatorname{Re}B|\mathcal W_{\mathrm r}|^2\right],
       \label{eq:coupled-cell-storage}\\
 \overline{\mathcal P}_{\rm cell}
    &=\frac\omega2\left[
       (-\operatorname{Im}A)|\mathcal U_{\mathrm c}
                                  +\beta\mathcal W_{\mathrm r}|^2
       +(-\operatorname{Im}B)|\mathcal W_{\mathrm r}|^2\right]\geq0.
       \label{eq:coupled-cell-loss}
\end{align}
The cross term records phase-dependent envelope storage. The allocation
is per selected normalized bulk component; finite real-space arrays
use their bond allocation. Under the specified Maxwell and viscous
laws, shell coupling $\beta A$ mixes translation and relative
deformation. The finite boundary problem uses its assembled pencil.

In particular, envelope-only viscous drag becomes
$c_{\mathrm e}\vct a\vct a^{\mathsf T}$ in center/relative coordinates,
where $\vct a=(1,\beta)^{\mathsf T}$. Within Section~4's stationary
classical Markov reservoir approximation, its force covariance is
$2k_BT c_{\mathrm e}\vct a\vct a^{\mathsf T}\delta(t-t')$.
Both transformed forces arise from the same fluctuating envelope force;
they are correlated. Internal drag has a different coupling pattern.
Replacing this rank-one drag by two independent scalar losses would
change the mechanical and thermal model.

For comparison, a separately specified diagonal center/relative model
would give the earlier uncoupled reference,
\begin{equation}
 \mathcal U_{\mathrm c}=
         \frac{\mathcal F_{\mathrm c}}
                         {k_{\mathrm c}^*(\vct k,\omega)-m\omega^2},
 \qquad
 \mathcal W_{\mathrm r}=
         \frac{\mathcal F_{\mathrm r}}
                                  {k_{\mathrm r}^*(\omega)-\mu\omega^2}.
 \label{eq:center-relative-phasor-response}
\end{equation}
The diagonal reference sets centre/relative cross coupling to zero.
For shell-only neighbours the coupled response instead satisfies
$\mathsf P_{\rm cr}(\mathcal U_{\mathrm c},\mathcal W_{\mathrm r})^{\mathsf T}
=(\mathcal F_{\mathrm c},\mathcal F_{\mathrm r})^{\mathsf T}$.

The envelope and RNP phasors are reconstructed by
\begin{equation}
 \mathcal U_{\mathrm e}=\mathcal U_{\mathrm c}
                              +\frac{m_{\mathrm r}}m\mathcal W_{\mathrm r},
 \qquad
 \mathcal U_{\mathrm r}=\mathcal U_{\mathrm c}
                              -\frac{m_{\mathrm e}}m\mathcal W_{\mathrm r}.
 \label{eq:internal-biophysical-phasor-reconstruction}
\end{equation}
A force applied only to the envelope has
$F_{\mathrm c}=f_{\mathrm e}$ and $F_{\mathrm r}=\beta f_{\mathrm e}$.
Writing
$\Delta=(A+B-m_{\mathrm e}\omega^2)(B-m_{\mathrm r}\omega^2)-B^2$,
its displacement readouts, away from poles, are
\begin{equation}
 \frac{\mathcal U_{\mathrm e}}{\mathcal F_{\mathrm e}}
       =\frac{B-m_{\mathrm r}\omega^2}{\Delta},\qquad
 \frac{\mathcal U_{\mathrm r}}{\mathcal F_{\mathrm e}}
       =\frac{B}{\Delta},\qquad
 \frac{\mathcal W_{\mathrm r}}{\mathcal F_{\mathrm e}}
       =-\frac{m_{\mathrm r}\omega^2}{\Delta}.
 \label{eq:envelope-driven-transfer-readouts}
\end{equation}
The reciprocal cross-transfer is $B/\Delta$ in both drive directions;
it is not in general real or Hermitian at a dissipative frequency. The
force/readout pairing determines which combinations are visible. Its
absorbed power is obtained from the positive loss quadratic form of
Section~4.

\begin{example}[An envelope readout zero with internal motion]
\label{ex:envelope-internal-antiresonance}
With a lossless internal spring $B=k_{\rm int}$ and no other force on the
RNP, the envelope-only transfer has an exact zero at
\begin{equation}
 \omega_{\rm ar}^2=\frac{k_{\rm int}}{m_{\mathrm r}},\qquad
 \mathcal U_{\mathrm e}=0,\quad
 \mathcal U_{\mathrm r}=-\frac{\mathcal F_{\mathrm e}}{k_{\rm int}},\quad
 \mathcal W_{\mathrm r}=\frac{\mathcal F_{\mathrm e}}{k_{\rm int}}.
 \label{eq:envelope-antiresonance-motion}
\end{equation}
At this frequency $\Delta=-k_{\rm int}^2\ne0$: the zero is a readout
antiresonance, not a missing normal mode. It survives envelope-only drag
because the envelope velocity vanishes. The cycle-averaged internal
kinetic-plus-spring energy is still
$|\mathcal F_{\mathrm e}|^2/(2k_{\rm int})$ in this model. Internal loss
with $\operatorname{Im}B(\omega_{\rm ar})\ne0$ removes this real-frequency
numerator zero and generally rounds the dip. Additional RNP drag or
cross-couplings also change the transfer. These are defined theoretical
readouts, not measured measles frequencies. A missing envelope feature
therefore need not imply absence of an internal excitation.
\end{example}

The cancellation is visible in the centre/relative coordinates:
\begin{equation}
 \boxed{\mathcal U_{\rm c}=-\beta\frac{\mathcal F_{\rm e}}{k_{\rm int}},
 \qquad \mathcal W_{\rm r}=\frac{\mathcal F_{\rm e}}{k_{\rm int}},
 \qquad \mathcal U_{\rm e}=\mathcal U_{\rm c}
                                +\beta\mathcal W_{\rm r}=0.}
 \label{eq:antiresonance-phase-cancellation}
\end{equation}
The two contributions to the envelope coordinate have opposite phase.
The envelope phase itself is undefined at the exact zero. Also,
$\omega_{\rm ar}/\omega_0=\sqrt{m_{\rm e}/m}<1$: this transfer zero differs
from the isolated relative-motion frequency and from the coupled system's
normal-mode frequencies. Figure~\ref{fig:complex-envelope-antiresonance}
shows both component readouts under one specified envelope force.
In the ideal envelope-loss-only steady response, the mean supplied power
vanishes at this zero even though internal storage is positive; preparation
transients supply that storage. Internal friction changes this balance and
rounds the dip as stated above.

\begin{figure}[H]
\centering\begingroup
\definecolor{vltTeal}{HTML}{176B64}
\definecolor{vltBlue}{HTML}{315D80}
\definecolor{vltGold}{HTML}{A87520}
\definecolor{vltPlum}{HTML}{725877}
\definecolor{vltRed}{HTML}{762B38}
\definecolor{vltYellow}{HTML}{D5AD3C}
\definecolor{vltNeutral}{HTML}{777D83}
\setlength{\unitlength}{1pt}
\scalebox{0.90}{%
\input{vlt-figures/figure-179.tex}%
}\endgroup
\caption{Exact two-coordinate passive response to one envelope force. Set $m_{\rm e}=m_{\rm r}=m_0$, $k_{\rm sh}=k_{\rm int}=k_\star$, $A=k_\star-i\omega c_{\rm e}$, $B=k_\star$ and $c_{\rm e}/(m_0\omega_{\rm ar})=1/2$, with $\omega_{\rm ar}=\sqrt{k_\star/m_0}$ and no RNP drag. The magnitude curves share one force/stiffness normalization. The signed ratio is exactly $1-r^2$; its sign gives the relative phase away from the envelope zero. At $r=1$ the envelope phase is undefined and the RNP readout is finite, as shown by the projection ports below. The single faceted glyph retains one virion identity. The zero is an observable cancellation, distinct from a pole and from the isolated relative frequency $\sqrt2\,\omega_{\rm ar}$. \eqref{eq:envelope-driven-transfer-readouts} supplies both plotted readouts; no empirical frequency is assigned. The boxed result illustrated here is~\eqref{eq:antiresonance-phase-cancellation}.}
\label{fig:complex-envelope-antiresonance}
\end{figure}

A common-acceleration force pattern
$f_{\mathrm e}/m_{\mathrm e}=f_{\mathrm r}/m_{\mathrm r}$ has
$F_{\mathrm r}=0$ by virtual work. It excites no relative coordinate in
the diagonal reference. In the shell-coupled model, however, a nonzero
center response drives relative motion through $\beta A$ even with
$F_{\mathrm r}=0$. For a pure relative preparation with $U_{\mathrm c}=0$,
$m_{\mathrm e}u_{\mathrm e}+m_{\mathrm r}u_{\mathrm r}=0$;
the component displacements have opposite signs and obey
$u_{\mathrm r}=-(m_{\mathrm e}/m_{\mathrm r})u_{\mathrm e}$.
That preparation is an eigenmode only when the full pencil supports it. Phase and amplitude
thus distinguish external drive selection, mode mixing and the selected
observation coordinate.

\subsubsection{Hidden motion, reduction, and a second readout}

Eliminating the undriven RNP coordinate generalizes
\eqref{eq:structure-internal-effective-inertia} to the specified passive
internal response. Away from $B-m_{\rm r}\omega^2=0$,
\begin{equation}
 \left[A-\omega^2\left(m_{\rm e}
              +\frac{m_{\rm r}B}{B-m_{\rm r}\omega^2}\right)\right]
                  \mathcal U_{\rm e}=\mathcal F_{\rm e}.
 \label{eq:complex-internal-schur-response}
\end{equation}
Solving the internal row and substituting gives the Schur response.
Its frequency-dependent inertia is an effective coefficient at the
envelope readout. At the lossless antiresonance this coefficient
diverges while the full pencil remains regular; that full system
evaluates response and storage there. Transient elimination also
carries internal displacement and momentum as prepared forcing or
memory. These states can change future envelope motion at identical
present envelope displacement and velocity.

\paragraph{Eliminating the internal coordinate with its thermal force.}
An exact one-coordinate response also specifies what happens to the
internal fluctuation source. This can be calculated in the same model,
while allowing a passive internal loss in $B$. Write
$A=A'-iA''$ and $B=B'-iB''$ at a real $\omega>0$, with
$A'',B''\geq0$. Suppose these two constitutive channels have independent
stationary classical equilibrium force sources $\eta_{\rm e}$ and
$\eta_{\rm i}$ at the same temperature, with spectra
$S_{\eta_{\rm e}}=2k_{\rm B}TA''/\omega$ and
$S_{\eta_{\rm i}}=2k_{\rm B}TB''/\omega$.
The internal source is conjugate to $w=u_{\rm e}-u_{\rm r}$, so its
physical force column is $(\eta_{\rm i},-\eta_{\rm i})^{\mathsf T}$.
The full thermal force column is consequently
$(\eta_{\rm e}+\eta_{\rm i},-\eta_{\rm i})^{\mathsf T}$.
This equal-and-opposite pair follows from virtual work; it is the
stochastic counterpart of the internal coupling.

\begin{definition}[Stochastic envelope reduction]
\label{def:s57-stochastic-envelope-reduction}
At a frequency where $D_{\rm r}:=B-m_{\rm r}\omega^2\ne0$, define
the reduced envelope pencil
$\mathcal P_{\rm e}^{\rm red}
=A-m_{\rm e}\omega^2-m_{\rm r}\omega^2 B/D_{\rm r}$.
With a specified envelope input and no imposed internal input, its
\emph{stochastic envelope reduction} retains the effective force
\begin{equation}
 \begin{aligned}
 \mathcal P_{\rm e}^{\rm red}\widehat u_{\rm e}
    &=\widehat f_{\rm e}+\widehat\eta_{\rm eff},\\
 \widehat\eta_{\rm eff}
    &=\widehat\eta_{\rm e}
       -\frac{m_{\rm r}\omega^2}{D_{\rm r}}
                          \widehat\eta_{\rm i}.
 \end{aligned}
 \label{eq:s57-stochastic-schur-force}
\end{equation}
Hats here denote temporal Fourier amplitudes or the corresponding
stationary spectral representation. They are distinct from a single
deterministic harmonic phasor. Prepared internal initial data supply
their separate transient term.
\end{definition}

\begin{proposition}[Passive noise matching under internal elimination]
\label{prop:s57-passive-noise-schur}
Under the preceding equilibrium and independent-channel assumptions,
the reduced pencil and effective force obey
\begin{equation}
 \begin{aligned}
 -\operatorname{Im}\mathcal P_{\rm e}^{\rm red}
   &=A''+\frac{m_{\rm r}^{\,2}\omega^4}{|D_{\rm r}|^2}B''\geq0,\\
 S_{\eta_{\rm eff}}
   &=\frac{2k_{\rm B}T}{\omega}
                   [-\operatorname{Im}\mathcal P_{\rm e}^{\rm red}],\\
 S_{\rm ee}
   &=|R_{\rm ee}|^2S_{\eta_{\rm eff}}
     =\frac{2k_{\rm B}T}{\omega}\operatorname{Im}R_{\rm ee},
 \qquad R_{\rm ee}=\frac1{\mathcal P_{\rm e}^{\rm red}}.
 \end{aligned}
 \label{eq:s57-reduced-passivity-fdt}
\end{equation}
The last line concerns the fluctuation about any specified deterministic
mean, at frequencies where the displayed inverses exist.
\end{proposition}
\begin{proof}
The second physical-coordinate equation gives
$\widehat u_{\rm r}=(B\widehat u_{\rm e}-\widehat\eta_{\rm i})/D_{\rm r}$.
Substitution into the first equation proves the reduced force.
For real $a=m_{\rm r}\omega^2$, use
$B/(B-a)=1+a/(B-a)$ and
$\operatorname{Im}(B-a)^{-1}=B''/|B-a|^2$ to obtain the loss identity.
Independence of the two sources removes their cross spectrum and gives
$S_{\eta_{\rm eff}}
=S_{\eta_{\rm e}}+a^2S_{\eta_{\rm i}}/|D_{\rm r}|^2$.
Finally $\operatorname{Im}(1/\mathcal P)
=[-\operatorname{Im}\mathcal P]/|\mathcal P|^2$ proves the scalar
response--fluctuation identity.
\end{proof}

The effective force spectrum has units of force squared times time;
the compliance factor converts it to displacement squared times time.

Thus a passive internal coordinate can appear in an envelope-only
description as both frequency-dependent response and colored force.
Their weights are fixed together. In the Markov example below,
$B''=0$ and $A''=\omega c_{\rm e}$, so the surviving force spectrum
is exactly $2k_{\rm B}Tc_{\rm e}$. If internal loss is retained, its
equal-and-opposite source is filtered through the same internal
denominator as the response. This is a concrete instance of the
memory--fluctuation connection in statistical reduction
\cite{Zwanzig1973,Kubo1957}; the displayed identities follow directly
from this two-coordinate law.

The same constitutive law fixes mean response and effective noise
weights. Transients additionally retain internal displacement,
momentum, branch strain and their preparation statistics. At
$D_{\rm r}=0$, use the regular full pencil. Keeping an internal variable
or its exact memory, prepared force and matched noise gives equivalent
predictions at this readout. Constant inertia and white noise instead
require a band-specific approximation check.

A second displacement projection can expose some of what the first hides.
\begin{proposition}[Conditional two-readout constitutive reconstruction]
\label{prop:s5-two-readout-constitutive-reconstruction}
For the stipulated two-coordinate law, an envelope-only harmonic force,
known $m_{\rm e},m_{\rm r}$ and phase-calibrated component amplitudes,
the constitutive coefficients are recovered, where both denominators
are nonzero, by
\begin{equation}
 \begin{aligned}
 B(\omega)&=\frac{m_{\rm r}\omega^2\mathcal U_{\rm r}}
                         {\mathcal U_{\rm r}-\mathcal U_{\rm e}},\\
 A(\omega)&=m_{\rm e}\omega^2+
  \frac{\mathcal F_{\rm e}-B(\mathcal U_{\rm e}-\mathcal U_{\rm r})}
       {\mathcal U_{\rm e}}.
 \end{aligned}
 \label{eq:complex-two-readout-constitutive-inference}
\end{equation}
\end{proposition}
\begin{proof}
The undriven internal row gives
$B(\mathcal U_{\rm r}-\mathcal U_{\rm e})
=m_{\rm r}\omega^2\mathcal U_{\rm r}$.
Solving it gives $B$; inserting that value into the driven envelope row
and solving for $A$ gives the second identity.
\end{proof}
Finite precision can limit practical identifiability even when this
conditional inversion is unique~\cite{Raue2009}. The effective masses and applied force are assumed
known independently. At the exact envelope zero, the full equations fix
$B=m_{\rm r}\omega^2$, but the zero itself cannot determine $A$.
Near either small denominator, measurement errors are strongly amplified.
If inertia is also unknown, the data constrain different parameter
combinations; an algebraic inversion under known masses is not a proof
that all microscopic properties are identifiable. The result concerns
the two declared mechanical projections. Imaging contrast or an
unresolved marker motion must first be related to those projections by
its own readout map.

\paragraph{A transfer zero and visibility over time.}
The harmonic inverse calculation concerns unknown coefficients in a
chosen response law. A different question starts with those coefficients
known and asks whether a time record distinguishes different prepared
states. The supported passive surrogate gives an exact answer.

\begin{proposition}[Envelope state observability and passive relaxation]
\label{prop:s57-envelope-state-observability}
Take $A=k_{\rm sh}-i\omega c_{\rm e}$ and $B=k_{\rm int}$ with
$m_{\rm e},m_{\rm r},k_{\rm sh},k_{\rm int},c_{\rm e}>0$.
For a known continuously differentiable deterministic envelope force,
the four-state mechanical model is observable from its ideal continuous
envelope-position record. Its internal displacement and velocity are
reconstructed by
\begin{equation}
 \begin{aligned}
 u_{\rm r}
  &=\frac{m_{\rm e}\ddot u_{\rm e}
       +c_{\rm e}\dot u_{\rm e}
       +(k_{\rm sh}+k_{\rm int})u_{\rm e}-f_{\rm e}}
          {k_{\rm int}},\\
 \dot u_{\rm r}
  &=\frac{m_{\rm e}u_{\rm e}^{(3)}
       +c_{\rm e}\ddot u_{\rm e}
       +(k_{\rm sh}+k_{\rm int})\dot u_{\rm e}-\dot f_{\rm e}}
          {k_{\rm int}}.
 \end{aligned}
 \label{eq:s57-envelope-time-state-reconstruction}
\end{equation}
Its unforced deterministic motion also decays exponentially. With no
imposed deterministic force and the matched envelope Markov noise at
temperature $T>0$, it has a unique
stationary Gaussian state with the Gibbs covariance of this positive
harmonic reference.
\end{proposition}
\begin{proof}
The first identity is the real envelope force balance solved for
$u_{\rm r}$; differentiating gives the second. Thus equal envelope
histories under the same known input fix both component positions and
velocities. This is the initial-state meaning of observability
\cite{Kalman1963}.

The unforced positive energy satisfies
$\dot E=-c_{\rm e}\dot u_{\rm e}^{\,2}$, so every growth rate has
nonpositive real part. A zero growth rate is excluded by the positive
stiffness. At a nonzero imaginary growth rate, the vanishing loss
forces the envelope component to vanish; its force-balance equation
then gives $k_{\rm int}u_{\rm r}=0$. There is consequently no nonzero
undamped eigenvector, and all four growth rates have negative real
part. Write the state as $(\vct u,\vct v)$, its generator as
$\mathsf A_{\rm cell}$, and
$\mathsf M_{\rm er}=\diag(m_{\rm e},m_{\rm r})$,
$\mathsf C_{\rm er}=\diag(c_{\rm e},0)$.
The positive stiffness $\mathsf K_{\rm er}$ is the static part of
\eqref{eq:coupled-envelope-rnp-pencil} with $A=k_{\rm sh}$ and
$B=k_{\rm int}$. Direct multiplication gives
\begin{equation}
 \begin{aligned}
 \mathsf\Sigma_G
   &=k_{\rm B}T\diag(\mathsf K_{\rm er}^{-1},
                                      \mathsf M_{\rm er}^{-1}),\\
 \mathsf A_{\rm cell}\mathsf\Sigma_G
  +\mathsf\Sigma_G\mathsf A_{\rm cell}^{\mathsf T}
   &=-2k_{\rm B}T\diag\left(0,
       \mathsf M_{\rm er}^{-1}\mathsf C_{\rm er}
                                  \mathsf M_{\rm er}^{-1}\right).
 \end{aligned}
 \label{eq:s57-rank-one-gibbs-relaxation}
\end{equation}
The negative right side is the opposite of the matched Markov diffusion
covariance. Exponential stability makes its stationary covariance
unique; the resulting Ornstein--Uhlenbeck stationary state is Gaussian.
\end{proof}

A transfer zero concerns a particular forcing frequency; state
observability concerns distinctions over time. At antiresonance,
known force and coefficients still determine internal motion.
For unknown preparation, the transient envelope record supplies
state information. The derivative identities establish ideal
uniqueness, while exposure, bandwidth and noise set practical accuracy
through Section~\ref{sec:phase-window-coherence}. Parameter inference
retains its separate calibration assumptions.

The internal spring transmits envelope forcing to the other coordinate,
so one random force can equilibrate both in this model. Instantaneous
noise rank and stationary covariance rank therefore differ. Resolving
the internal coordinate preserves that shared physical source.

\paragraph{Carrying the structural readouts into an evolving lattice.}
Envelope and RNP channels retain their declared structural meaning for a
persistent virion while its neighbours change. Their local two-coordinate
law is then a conditional component of the larger particle--medium model.
A stationary transfer zero belongs to a persistent episode with specified
surrounding couplings and input channel. During recruitment, a weak pooled
envelope record can instead reflect changing transfer functions or phase
cancellation across episodes. Resolve the two structural channels and the
assembly history before attributing that suppression to internal loss.
For a pulse spanning an event, the registered two-time response replaces
a single stationary compliance; its subsequent envelope and internal
motions remain predictions of the same continued mechanical preparation.

\subsubsection{The same passive parameters predict fluctuations}

There is a complementary equilibrium prediction with no new fit
parameters. Choose the explicit positive finite-cell surrogate
$A=k_{\rm sh}-i\omega c_{\rm e}$ and $B=k_{\rm int}$, with
$k_{\rm sh},k_{\rm int},c_{\rm e}>0$. In this comparison $k_{\rm sh}$
is a fixed supporting stiffness, not a frequency-independent surrogate
for an arbitrary medium. The masses are the same as above. Within
Section~4's stationary classical Markov-reservoir approximation to an
equilibrated medium at temperature $T>0$, envelope drag is accompanied by
an envelope random force with two-sided angular-frequency spectrum
$S_f=2k_{\rm B}Tc_{\rm e}$. There is no independent RNP force or drag
in this stipulated example. The positive internal connection makes each
normal mode move the damped envelope coordinate, so both modes thermalize
in this Markov model despite the rank-one forcing. Writing
$R_{\rm ee}=\mathcal U_{\rm e}/\mathcal F_{\rm e}$ and
$R_{\rm re}=\mathcal U_{\rm r}/\mathcal F_{\rm e}$, the response filter
in \eqref{eq:probe-coherence-response-filter} gives the following linked
prediction.

\begin{proposition}[Matched envelope and internal thermal readouts]
\label{prop:s5-matched-internal-thermal-readouts}
For this positive surrogate and its stated single-source classical
equilibrium Markov preparation, the two displacement spectra and their
cross-spectrum are
\begin{equation}
 \begin{split}
 S_{\rm ee}&=2k_{\rm B}Tc_{\rm e}|R_{\rm ee}|^2,
 \qquad S_{\rm rr}=2k_{\rm B}Tc_{\rm e}|R_{\rm re}|^2,\\
 S_{\rm er}&=2k_{\rm B}Tc_{\rm e}R_{\rm ee}R_{\rm re}^*,\\
 S_{\rm ee}(\omega_{\rm ar})&=0,
 \qquad S_{\rm rr}(\omega_{\rm ar})
                     =\frac{2k_{\rm B}Tc_{\rm e}}{k_{\rm int}^2}>0.
 \end{split}
 \label{eq:complex-antiresonance-matched-fluctuations}
\end{equation}
\end{proposition}
\begin{proof}
Filter the common force of spectrum $2k_{\rm B}Tc_{\rm e}$ through the
two component compliances. Their product gives the cross-spectrum.
At the internal antiresonance the full response gives
$R_{\rm ee}=0$ and $R_{\rm re}=-1/k_{\rm int}$, yielding the last line.
\end{proof}

The units are displacement squared times time. The same cancellation
that suppresses the forced envelope readout suppresses its ideal
equilibrium spectrum at that frequency, while the internal spectrum is
nonzero. Away from a transfer zero, this single-source model has unit
coherence magnitude. At the exact zero the normalized coherence is
undefined because an auto-spectrum vanishes. Additional internal
friction would require its matched fluctuating force, and detector noise
would fill the recorded spectral dip. Neither can be inserted into only
one of the response and fluctuation calculations.

Figure~\ref{fig:complex-internal-matched-spectra} applies this same response model to both spectral readouts and their integrated equilibrium variances.

The equal-time statistics give a further check. The configurational energy
is $[k_{\rm sh}u_{\rm e}^2+k_{\rm int}(u_{\rm e}-u_{\rm r})^2]/2$.
Its positive Gibbs covariance is $k_{\rm B}T\mathsf K^{-1}$, yielding
\begin{equation}
 \operatorname{Var}u_{\rm e}=\frac{k_{\rm B}T}{k_{\rm sh}},\qquad
 \operatorname{Var}w=\frac{k_{\rm B}T}{k_{\rm int}},\qquad
 \operatorname{Cov}(u_{\rm e},w)=0.
 \label{eq:complex-internal-equilibrium-covariance}
\end{equation}
The integral of $S_{\rm ee}$ over $\dd\omega/(2\pi)$ stays positive
despite the isolated zero. Also,
$\operatorname{Cov}(u_{\rm e},u_{\rm r})=k_{\rm B}T/k_{\rm sh}$:
uncorrelated envelope/relative coordinates become correlated
envelope/RNP coordinates. Frequency-resolved phase adds information
beyond equal-time covariance. These are conditional predictions
of the stated equilibrium surrogate.

\begin{figure}[H]
\centering
\begingroup
\definecolor{vlfGreen}{HTML}{176B64}
\definecolor{vlfDarkRed}{HTML}{762B38}
\definecolor{vlfRed}{HTML}{762B38}
\definecolor{vlfBlue}{HTML}{315D80}
\definecolor{vlfCopper}{HTML}{762B38}

\setlength{\unitlength}{1pt}
\scalebox{0.80}{%
\input{vlt-figures/figure-180.tex}%
}%
\endgroup
\caption{The same passive parameters predict two equilibrium spectra. Use the surrogate in~\eqref{eq:complex-antiresonance-matched-fluctuations}, with $m_{\rm e}=m_{\rm r}=m_0$, $k_{\rm sh}=k_{\rm int}=k_\star>0$ and $c_{\rm e}/(m_0\omega_{\rm ar})=1/2$, where $\omega_{\rm ar}=\sqrt{k_\star/m_0}$. The Markov force spectrum is $S_f=2k_{\rm B}Tc_{\rm e}$ at $T>0$. Define $\widehat S_{jj}=k_\star^2S_{jj}/(2k_{\rm B}Tc_{\rm e})$. (a) $\widehat S_{\rm ee}=|k_\star R_{\rm ee}|^2$ has an exact isolated zero at $\omega=\omega_{\rm ar}$, whereas $\widehat S_{\rm rr}=|k_\star R_{\rm re}|^2=1$ there. (b) The cumulative positive-and-negative-frequency variances are normalized by $k_{\rm B}T/k_\star$; they approach $1$ and $2$, respectively. The dip therefore does not remove the envelope's positive integrated fluctuation variance. These are stationary classical model spectra, with no fitted viral parameters; additional material friction must bring its matched noise.}
\label{fig:complex-internal-matched-spectra}
\end{figure}

The covariance geometry in Figure~\ref{fig:complex-internal-covariance-map} shows how the same equilibrium state appears in structural and detector coordinates.

\begin{figure}[H]
\centering
\begingroup
\definecolor{vlfGreen}{HTML}{176B64}
\definecolor{vlfDarkRed}{HTML}{762B38}
\definecolor{vlfRed}{HTML}{762B38}
\definecolor{vlfBlue}{HTML}{315D80}
\definecolor{vlfCopper}{HTML}{762B38}

\setlength{\unitlength}{1pt}
\scalebox{0.80}{%
\input{vlt-figures/figure-181.tex}%
}%
\endgroup
\caption{The same equilibrium Gaussian in structural and detector coordinates. With $k_{\rm sh}=k_{\rm int}=k_\star$, let $\ell_T=\sqrt{k_{\rm B}T/k_\star}$ and $w=u_{\rm e}-u_{\rm r}$. (a) Coordinates $(u_{\rm e},w)/\ell_T$ have covariance $\Id_2$; the unit-Mahalanobis contour is a circle. (b) The readout map $u_{\rm r}=u_{\rm e}-w$ gives $2X^2-2XY+Y^2=1$ for $(X,Y)=(u_{\rm e},u_{\rm r})/\ell_T$, with covariance $\left(\begin{smallmatrix}1&1\\1&2\end{smallmatrix}\right)$ and correlation $1/\sqrt2$. Equal Cartesian scales are used. The map has unit absolute determinant, and both contours enclose the same Gaussian probability. These are covariance contours rather than trajectories; frequency-dependent phase requires the cross-spectrum.}
\label{fig:complex-internal-covariance-map}
\end{figure}

The internal coordinate produces the transfer zero, hidden storage
and second displacement projection. Its exact memory and prepared
force can reproduce envelope response in a reduced description.
A constant-inertia, single-spring reduction loses those effects.
The comparison identifies the approximation and what a second readout
can recover. Section~7 extends the shared-parameter test to collective
response and fluctuations.

An active converter would require an additional force-generation and
resource law as in Section~\ref{sec:self-excitation-passive-boundary}. Passive interference can make a surface
signal small or return stored energy during a transient without supplying
sustained work. A spectral feature, cross-phase, or fitted negative
resistance cannot replace that active closure. A quantum extension would
also require its separate reservoir and occupation-state construction;
the classical two-coordinate calculation does not specify it.

Real-motion reconstruction, mechanical propagation and calibrated
measurement connect state to observation. Through changing assembly,
particle identities, continued material strain and matched pattern
frames preserve that connection. Modal coefficients, energy shares
and coherence quantify different aspects of the same experiment.
Their value for selecting biophysical variables lies in predicting
and distinguishing motion across preparations, readouts and residence
windows. The next section develops operators on these specified
spaces with the same domains, units and measurement meanings.

\clearpage

\section{Operators and Observables of the Viral Lattice}
\label{sec:operator-framework}

An operator specifies how a retained state is transformed or read out.
It connects the mechanical state and measured signals of
Section~\ref{sec:complex-displacement} through a defined action, domain
and physical units. The same labelled ideal virions can participate in
changing associated groups while their laboratory trajectories and
prepared medium state continue.

Different maps extract different aspects of that motion. A surface
readout can miss internal displacement; a changing membership mask can
shift a cluster-centre signal. Persistent episodes admit spectral
operators, and registered event maps connect their predictions across
recruitment, rearrangement and breakup.

Consider observing an assembly through a few tracked markers. The full
mechanical state contains motions that those markers combine or suppress.
Changing the applied force or adding a separation measurement can expose
a pattern absent from the first record. Operators express these choices
as definite maps between the physical state, the applied perturbation
and the measured channel.

Oriented patch interactions make the observation geometry especially
consequential. A registered surface marker moves when its centre
translates or its body rotates; a responsive surface can change contrast
while that marker remains stationary. These contributions have different
conjugate perturbations and short-time response. The constructions below
carry the body tangent, relaxing surface state and prepared medium through
the same detector map. Marker-rank tests, mixed-response normalization
and calibrated lag statistics then state which combinations a proposed
experiment can recover. The final benchmark adds a coupled angular
coordinate to demonstrate the response--fluctuation comparison with
one consistent set of mechanical and thermal inputs.

This also clarifies what a change of representation accomplishes.
Particle coordinates and modal coefficients can describe the same full
state, with all force and detector maps transformed consistently.
Selecting only a subset of either description discards information and
raises a further predictive question: does the retained part determine
the target record over the required time window? The spectral and
closure results below give calculable tests of that question.

For example, the stiffness map sends a displacement to its linear
restoring-force coefficient, while the susceptibility sends an applied
force to a displacement response:
\begin{equation}
 \vct U\longmapsto\mathsf K\vct U,
 \qquad
 \vct F\longmapsto\mathsf R_{\rm mech}(\omega)\vct F.
 \label{eq:operator-physical-map-examples}
\end{equation}
The conservative restoring force is $-\mathsf K\vct U$. For
length-valued coordinates, the two displayed outputs have force and
length units, respectively; other coordinates use their conjugate units.
A detector selects a component or weighted sum of the resulting motion.
Matrix size alone therefore leaves the physical role unspecified:
curvature, susceptibility and quantum displacement act on their declared
spaces with their own units.

The construction starts with the conservative reference and then admits
the passive and statistical extensions developed later in this section.
It uses the operator foundations of mechanics and quantum theory
\cite{FeynmanLectures1963,Shankar1994,Teschl2009}; the later algebraic
transformations use the stated Lie-algebra framework
\cite{NeumaierWestra2011,Sternberg2004}. The viral-phonon postulate is
the additional step that assigns quantum occupation to mechanical modes.
Spectral decomposition itself applies to the classical reference.

A biophysical readout specifies its structural coordinate and instrument
map. Tracked centres, surface markers and envelope--RNP differences
select different motions, with the force and detector maps of
Section~\ref{sec:complex-displacement}. Structural observations motivate
that resolution~\cite{Leake2016,Ke2018}; the interaction law, boundary
and preparation supply its mechanical inputs.
Table~\ref{tab:operator-biophysical-readouts} organizes the resulting
observation tasks.

\begin{table}[H]
\centering\small
\begin{tabular}{@{}p{3.1cm}p{4.1cm}p{5.8cm}@{}}
\toprule
\raggedright Physical question & \raggedright Defined quantity
& \raggedright Information needed \tabularnewline
\midrule
\raggedright Which coordinate moves?
& \raggedright $Y=\vct\ell^{\mathsf T}\vct u$, a displacement
& \raggedright Structural projection, spatial resolution and calibration
\tabularnewline\addlinespace[3pt]
\raggedright Which force excites it?
& \raggedright $\mathcal R^{\rm ret}_{ab}(\Omega)$, displacement per force
& \raggedright Force channel, readout channel, boundary and constitutive law
\tabularnewline\addlinespace[3pt]
\raggedright Which motions fluctuate together?
& \raggedright $C_{ab}$, displacement squared
& \raggedright Preparation, observation interval and detector covariance
\tabularnewline\addlinespace[3pt]
\raggedright How many virions occupy a region?
& \raggedright $n[w]$ for an indicator weight $w$
& \raggedright Particle positions and the selected spatial region
\tabularnewline\addlinespace[3pt]
\raggedright Where does mechanical energy flow?
& \raggedright $J_{i\to j}$, energy per time
& \raggedright Interaction forces, velocities and local energy allocation
\tabularnewline\addlinespace[3pt]
\raggedright Which phonon state is prepared?
& \raggedright Occupations, canonical moments and $\varrho$
& \raggedright Quantization postulate and the corresponding measurement protocol
\tabularnewline
\bottomrule
\end{tabular}
\caption{Biophysical questions specify different readouts of one retained
mechanical system. A structural image helps define coordinates and contrast;
force response, fluctuation statistics and phonon-state measurements each
require their own preparation and calibration.}
\label{tab:operator-biophysical-readouts}
\end{table}

\paragraph{An observable as an inference question.}
The observation map is also part of the inverse problem. Clermont and
Zenker emphasize the roles of incomplete observations, uncertain inputs
and prediction beyond the data used to fit a biological model
\cite{ClermontZenker2015}. Here those questions are posed within the
declared lattice mechanics: its coordinates, force law, boundary,
preparation and detector determine a forward prediction before any
parameter is inferred.

\begin{definition}[Identifiable prediction for an observation task]
\label{def:s6-identifiable-prediction}
Fix an admissible mechanical model family and an observation task
$\mathcal E$, including its force channels, preparations, readouts and
recording windows. Let $\zeta$ collect the unknown mechanical parameters,
preparation data and calibration variables. Write
$\mathfrak D_{\mathcal E}(\zeta)$ for the specified ideal observations:
these may be deterministic records, declared statistical moments or
output probability laws, with that choice fixed in advance. A target
$\mathfrak Q(\zeta)$ is an \emph{identifiable prediction} for this task
when
\begin{equation}
 \mathfrak D_{\mathcal E}(\zeta_1)
       =\mathfrak D_{\mathcal E}(\zeta_2)
 \quad\Longrightarrow\quad
 \mathfrak Q(\zeta_1)=\mathfrak Q(\zeta_2)
 \label{eq:s6-prediction-on-data-fibres}
\end{equation}
for all admissible assignments. Thus the target has one value on each
family of observationally equivalent assignments.
\end{definition}

A target can be a later displacement, an energy, another channel's
response or a parameter combination. Its forward law uses the declared
deterministic preparation, classical ensemble or quantum density
operator. Inference from finite noisy data additionally specifies an
error law and uncertainty or likelihood construction. Compatible model
assignments describe uncertainty in that prediction; an ensemble of
particle preparations describes physical variability.

At a fixed mechanical reference the distinction between an identifiable
state and an identifiable target has a particularly simple test.

\begin{proposition}[Target recovery from partial linear observations]
\label{prop:s6-observable-target-recovery}
Use a finite real prepared-state column $\vct\beta$, with the known
forced contribution subtracted from the record, and fixed matrices
$\vct d=\mathsf V\vct\beta$ and
$\vct q=\mathsf C\vct\beta$. Over the full prepared-state vector
space, the linear target is determined by the record if and only if
\begin{equation}
 \ker\mathsf V\subseteq\ker\mathsf C
 \quad\Longleftrightarrow\quad
 \mathsf C=\mathsf L\mathsf V
 \quad\hbox{for some }\mathsf L.
 \label{eq:s6-linear-target-observation-criterion}
\end{equation}
For a nonnegative quadratic storage
$E(\vct\beta)=\vct\beta^{\mathsf T}\mathsf H\vct\beta/2$,
$\mathsf H\geq0$, the corresponding condition is
\begin{equation}
 E\hbox{ is determined by }\mathsf V\vct\beta
       \quad\Longleftrightarrow\quad
                  \ker\mathsf V\subseteq\ker\mathsf H.
 \label{eq:s6-storage-observation-criterion}
\end{equation}
In particular, positive-definite total storage requires an injective
record map for exact recovery on this unrestricted state space.
\end{proposition}
\begin{proof}
Two preparations have the same record precisely when their difference
lies in $\ker\mathsf V$. Their linear targets agree precisely when
that difference also lies in $\ker\mathsf C$. Under this inclusion,
define $\mathsf L(\mathsf V\vct\beta)=\mathsf C\vct\beta$ on
$\operatorname{ran}\mathsf V$; the definition is independent of the
representative and extends linearly to the record space.
For storage, compare any $\vct v\in\ker\mathsf V$ with the zero
preparation. Equality of their energies gives
$\vct v^{\mathsf T}\mathsf H\vct v=0$, hence
$\mathsf H\vct v=0$ by nonnegativity. Conversely, if
$\mathsf H\vct v=0$, adding $\vct v$ changes neither the quadratic
term nor its cross term with any prepared state. This proves both
directions of the storage criterion.
\end{proof}

Let $\mathsf V^{+}$ denote the Moore--Penrose pseudoinverse. For the
linear target one may choose $\mathsf L=\mathsf C\mathsf V^{+}$.
With fixed physical coordinate and record scales, the stable recovery
statement is
\begin{equation}
 \vct q=\mathsf C\mathsf V^{+}\vct d,
 \qquad
 \|\delta\vct q\|\leq
       \|\mathsf C\mathsf V^{+}\|\,\|\delta\vct d\|.
 \label{eq:s6-target-reconstruction-error}
\end{equation}
The singular-value geometry is the one used in linear inverse estimation
\cite{Wunsch2005BasicMachinery}. Physical target and record scales fix
the amplification of data errors. Restricting admissible preparations
can sharpen recovery; the proposition gives the unrestricted
vector-space criterion.

For example, in the conservative exchange-symmetric supported pair, a
record of common motion can predict its later common displacement even
when the relative sector is unresolved. Positive energy in that relative
sector remains invisible to the common record. The sharp lower-energy
bound of Proposition~\ref{prop:s5-hamiltonian-readout-energy} then
quantifies storage compatible with an incomplete instantaneous observation,
while the criterion above specifies when all of that storage is uniquely
determined. Another detector or a resolved transient can distinguish
additional preparations without changing the force law.

Figure~\ref{fig:s6-target-fibre-and-storage} compares a determined collective target with storage that remains unresolved on the same observation fibre.

\begin{figure}[H]
\centering\begingroup
\definecolor{vltTeal}{HTML}{176B64}
\definecolor{vltBlue}{HTML}{315D80}
\definecolor{vltGold}{HTML}{A87520}
\definecolor{vltPlum}{HTML}{725877}
\definecolor{vltRed}{HTML}{762B38}
\definecolor{vltYellow}{HTML}{D5AD3C}
\definecolor{vltNeutral}{HTML}{777D83}
\setlength{\unitlength}{1pt}
\scalebox{0.90}{%
\input{vlt-figures/figure-182.tex}%
}\endgroup
\caption{Two initial preparations of one supported pair with $m=k_0=k_b=1$ in reference units and zero velocities. Set $c=(u_1+u_2)/\sqrt2$, $r=(u_1-u_2)/\sqrt2$, displacement unit $u_\star$, $E_\star=k_0u_\star^2$ and $\omega_\star=\sqrt{k_0/m}$. Both have $c=1/2$, while $r=0,0.9$ gives the printed different energies. Rigid bodies use $16\,\mathrm{pt}/u_\star$ and one coil family; rulers locate the common reference spacing $a=95u_\star/16$. Declared supports are omitted. The vertical observation fibre retains all relative preparations at fixed common record. Its energy varies quadratically, while future common motion is determined. Body colors register alternatives A and B throughout the figure.}
\label{fig:s6-target-fibre-and-storage}
\end{figure}

The fixed matrices in the proposition describe state inference within
one known model. Unknown curvatures, masses, charge assignments or
detector gains also change the forward map and require the wider
assignment comparison in Definition~\ref{def:s6-identifiable-prediction}.
Prediction in a new channel tests that entire specification. An operator
representation makes these questions explicit while preserving the
mechanical constraints from which its matrices were derived.

Geometry, assigned charge and short-range interactions enter through
the force law and Hessian; a readout selects the resulting motion.
Two tests guide variable integration: whether observations determine a
target, and whether retained variables determine their own evolution.
Propositions~\ref{prop:s62-quadratic-statistic-identification}
and~\ref{prop:s6-canonical-readout-closure} make these observation
and closure tests explicit for the lattice.

\subsection{One lattice, several representations}
\label{sec:operator-representations}

\subsubsection{Patterns, energy amplitudes, and occupation states}

Within a persistent registered episode, fix the reference geometry and
coupling parameter $\lambda$ of the same ideal-virion pool. Write
$n=N_{\rm tot}$ for its physical particle count. Let $p$ count the
retained real inertial mechanical coordinates, including specified
internal projections. Pure unconstrained translations give $p=3n$;
independent rigid-body centre and angular tangents give $p=6n$ before
constraints. Relaxing surface populations and material coordinates retain
their first-order kinetics and are counted separately. The positive inertia matrix $\mathsf M$ and conservative
Hessian $\mathsf K$ describe the constrained reference. Assume that
reference is stable, so $\mathsf K\geq0$, and define
\[
 \mathsf D=\mathsf M^{-1/2}\mathsf K\mathsf M^{-1/2},
 \qquad \mathsf W=\mathsf D^{1/2}.
\]
Here $\mathsf D$ has units $\mathrm s^{-2}$ and $\mathsf W$ has units
$\mathrm s^{-1}$. Work on the positive-frequency sector
$\mathcal H_+=\operatorname{ran}\mathsf P_+$, of dimension $r$.
Separated zero modes keep their free-coordinate dynamics. Choose real
orthonormal patterns
$\mathsf E=(\ket{1},\ldots,\ket{r})$, with
$\mathsf E\mathsf E^{\mathsf T}=\mathsf P_+$ and
$\mathsf E^{\mathsf T}\mathsf D\mathsf E
 =\diag(\omega_1^2,\ldots,\omega_r^2)$.
One pattern may move many virions or several components of one virion.
Thus $r$ counts retained positive modes, rather than particles.

The displacement column $\vct U\in\C^p$ records length and phase for
translations, with the mass metric of Section~\ref{sec:complex-displacement}.
The energy amplitude $\ket\psi\in\mathcal H_+$ combines instantaneous
mass-weighted displacement and momentum, satisfying
$\braket\psi\psi=E_+$ and $i\partial_t\ket\psi=\mathsf W\ket\psi$
in the conservative reference. For $E_+>0$, the direction
$\ket\eta=\ket\psi/\sqrt{E_+}$ gives modal energy fractions.
Displacement and work predictions also retain the scale $E_+$.

Under the quantization postulate, the oscillator occupation state belongs
to a different space,
\begin{equation}
 \begin{aligned}
 \mathcal F_+&=\bigoplus_{N=0}^{\infty}\operatorname{Sym}^N\mathcal H_+,\\
 \ket{\vct N}&=\ket{N_1,\ldots,N_r},\qquad N=\sum_sN_s.
 \end{aligned}
 \label{eq:operator-fock-space}
\end{equation}
The direct sum collects sectors with different total excitation number.
The symmetric tensor power encodes the bosonic occupation postulate for
these normal-mode excitations; it does not specify the exchange statistics
of whole virions. Each $N_s$ is a nonnegative integer. The vacuum sector
is one dimensional, the one-phonon sector is isomorphic to $\mathcal H_+$,
and the $N$-phonon sector has dimension
$\binom{N+r-1}{N}$ for $r\geq1$.
For example, two modes have the one-phonon states $\ket{1,0}$ and
$\ket{0,1}$, and the two-phonon states $\ket{2,0}$, $\ket{1,1}$ and
$\ket{0,2}$. Arbitrarily large occupations make $\mathcal F_+$ infinite
dimensional even though $r$ is finite.

Figure~\ref{fig:mode-directions-versus-occupation-counts} distinguishes the mode
labels from the occupation coordinates. A one-phonon ket
$\ket\Phi=\sum_sc_s\ket{1_s}$ has $\sum_s|c_s|^2=1$, where
$\ket{1_s}=\hat b_s^\dagger\ket0$. An identical coefficient column can represent $\eta$, but the
normalization has a different physical interpretation:
$|c_s|^2$ is now a probability for a specified occupation measurement.
An arbitrary occupation state need not have one excitation, a fixed
excitation number, or a nonzero displacement mean. The fixed virion count
is a separate assumption of the mechanical model.

\begin{figure}[H]
\centering\begingroup
\definecolor{vltTeal}{HTML}{176B64}
\definecolor{vltBlue}{HTML}{315D80}
\definecolor{vltGold}{HTML}{A87520}
\definecolor{vltPlum}{HTML}{725877}
\definecolor{vltRed}{HTML}{762B38}
\definecolor{vltYellow}{HTML}{D5AD3C}
\definecolor{vltNeutral}{HTML}{777D83}
\setlength{\unitlength}{1pt}
\scalebox{0.80}{%
\input{vlt-figures/figure-183.tex}%
}\endgroup
\caption{A real two-mode coefficient slice and its occupation labels. The unit vector $(\sqrt3/2,1/2)^{\mathsf T}$ can encode a classical energy direction or a one-phonon probability amplitude under the respective normalization. Integer points label $\ket{N_1,N_2}$; diagonals have constant total occupation. The raising operation $\hat b_1^\dagger\ket{2,1}=\sqrt3\ket{3,1}$ supplies the marked coefficient. Both sets of axes are abstract. The occupation grid is cropped at three for display and has no formal upper bound.}
\label{fig:mode-directions-versus-occupation-counts}
\end{figure}

\subsubsection{What an operator and its adjoint specify}

\begin{definition}[Operator and observable]
A linear operator $A$ maps a specified linear domain $D(A)$ in an input
Hilbert space to an output Hilbert space. Its action and domain together
define the operator. For densely defined $A$, the adjoint maps in the
reverse direction and is specified by
$\langle f,Ag\rangle=\langle A^\dagger f,g\rangle$ for every
$g\in D(A)$ and every $f$ in its adjoint domain. On a common input/output
space, a spectral observable is represented by a self-adjoint operator:
$A=A^\dagger$, including
equality of domains. A response map or evolution generator need not be
self-adjoint; its physical output follows from the declared force,
displacement, or energy map.
\end{definition}

Linearity means $A(af+bg)=aAf+bAg$ on the domain. A finite matrix
represents that action; its ordinary adjoint is the conjugate transpose.
With the mass metric,
$A^{\dagger_{\mathsf M}}=\mathsf M^{-1}A^\dagger\mathsf M$.
Mass weighting converts this adjoint into the Euclidean one. The metric
therefore remains part of the operator specification.

For self-adjoint $A$ and $f\in D(A)$,
$\langle f,Af\rangle^*=\langle Af,f\rangle=\langle f,Af\rangle$,
so the quadratic value is real. Distinct eigenspaces are orthogonal.
Its units follow from $A$: frequency statistics have inverse-time units,
whereas a Hamiltonian has energy units. A biophysical interpretation
additionally specifies the coordinate and measurement; dissipative
response uses its own operator properties.

An unbounded operator requires more than a differential expression. Its
domain says which vectors have a square-integrable image and which
boundary behavior is permitted. A symmetric operator satisfies
$A\subseteq A^\dagger$; this inclusion alone does not give equal domains
or a unique self-adjoint extension~\cite{Teschl2009}.
For a simple one-coordinate illustration, integration by parts gives
\begin{equation}
 \langle f,-i\hbar\partial_q g\rangle
       -\langle-i\hbar\partial_q f,g\rangle
            =-i\hbar[f(q)^*g(q)]_{0}^{L_q}.
 \label{eq:operator-domain-boundary-form}
\end{equation}
Setting both functions to zero at the endpoints makes this momentum
expression symmetric on that restricted domain. Its adjoint has a larger
domain, so it is not self-adjoint there. In contrast, the momentum expression
$-i\hbar\partial_q$
with a fixed quasi-periodic condition
$g(L_q)=e^{i\vartheta}g(0)$ on its $H^1$ domain is self-adjoint for real
$\vartheta$. These are different operators with different boundary data.
The example illustrates a domain issue; it does not prescribe the physical
cylinder's wall law. The elastic continuum operator in Section~\ref{sec:complex-displacement} already
uses its own closed form and mechanical boundary condition.

Figure~\ref{fig:operator-domain-boundary-comparison} makes the endpoint and domain distinction explicit for this auxiliary momentum example.

\begin{figure}[H]
\centering
\begingroup
\definecolor{vlfGreen}{HTML}{176B64}
\definecolor{vlfDarkRed}{HTML}{762B38}
\definecolor{vlfRed}{HTML}{762B38}
\definecolor{vlfBlue}{HTML}{315D80}
\definecolor{vlfCopper}{HTML}{762B38}

\setlength{\unitlength}{1pt}
\scalebox{0.80}{%
\input{vlt-figures/figure-184.tex}%
}%
\endgroup
\caption{The momentum expression $-i\hbar\partial_q$ with two domains on an auxiliary interval. Dirichlet data give the symmetric $H^1_0$ restriction, whose adjoint has domain $H^1$; the left sine illustrates domain membership. The right quasi-periodic $H^1$ domain, $g(L_q)=e^{i\vartheta}g(0)$, is self-adjoint. At $\vartheta=\pi/2$, the curves show the real and imaginary parts of $L_q^{-1/2}e^{i\vartheta q/L_q}$ on equal scales. Its eigenvalue is $\hbar\vartheta/L_q$; the full spectrum adds $2\pi\hbar j/L_q$, $j\in\mathbb Z$. Endpoint phases distinguish the domains. The coordinate belongs to this operator example; the cylinder retains its mechanical wall law.}
\label{fig:operator-domain-boundary-comparison}
\end{figure}

A core is dense in the graph norm, which controls both a vector and its
operator image. Agreement on a common core identifies the closures of
the operators. Hilbert-norm approximation alone need not control the
image, a distinction needed when finite calculations approximate a
continuum field.

\subsubsection{Which kind of expectation is being computed?}

For a classical energy direction, define
$\langle A\rangle_\eta=\bra\eta A\ket\eta$.
This is a modal quadratic statistic. A classical detector displacement
is instead the linear functional $\vct\ell^{\mathsf T}\vct u$.
For a quantum density operator $\varrho\geq0$ with
$\tr\varrho=1$, the quantum expectation is $\tr(\varrho\hat A)$ when
the required moment is finite. Their shared linear algebra retains distinct normalizations and
experimental interpretations.

\paragraph{The mechanical information in a complex second moment.}
The same distinction applies to statistical preparations of the classical
lattice. Work in the fixed real normal-mode basis of $\mathcal H_+$, and
write $\vct Q$ and $\vct\Pi$ for its mass-weighted displacement and
momentum columns. With
$\mathsf\Omega=\diag(\omega_1,\ldots,\omega_r)$, the component column is
$\vct\psi=(\mathsf\Omega\vct Q+i\vct\Pi)/\sqrt2$.
Both quadratures carry the square root of energy. Their mean and
fluctuations describe a physical ensemble of prepared motions at the
chosen time; a family of uncertain parameter assignments has a different
interpretation.

\begin{definition}[Complex second-order mechanical preparation]
\label{def:s61-complex-second-order-preparation}
For a classical preparation with finite second moments, define
\begin{equation}
 \begin{aligned}
 \vct\mu_\psi&=\langle\vct\psi\rangle,
 &\delta\vct\psi&=\vct\psi-\vct\mu_\psi,\\
 \mathsf G_\psi&=\langle\delta\vct\psi\,
                    \delta\vct\psi^\dagger\rangle,
 &\mathsf P_\psi&=\langle\delta\vct\psi\,
                    \delta\vct\psi^{\mathsf T}\rangle.
 \end{aligned}
 \label{eq:s61-complex-second-order-preparation}
\end{equation}
Here $\mathsf G_\psi$ is Hermitian and positive semidefinite, whereas
$\mathsf P_\psi$ is complex symmetric. Both have energy units. The second
matrix is the relation matrix of a complex random vector; retaining it
supplies the second-order information absent from a general complex
covariance alone~\cite{Picinbono1996}.
\end{definition}

\begin{proposition}[Recovering real mechanical covariance]
\label{prop:s61-complex-real-covariance}
For the real energy-quadrature column
$\vct Y_E=((\mathsf\Omega\vct Q)^{\mathsf T},
\vct\Pi^{\mathsf T})^{\mathsf T}$, its centered covariance is
\begin{equation}
 \mathsf C_{Y_E}=
 \begin{pmatrix}
  \operatorname{Re}(\mathsf G_\psi+\mathsf P_\psi)
   &\operatorname{Im}(\mathsf P_\psi-\mathsf G_\psi)\\
  \operatorname{Im}(\mathsf P_\psi+\mathsf G_\psi)
   &\operatorname{Re}(\mathsf G_\psi-\mathsf P_\psi)
 \end{pmatrix}.
 \label{eq:s61-complex-real-covariance}
\end{equation}
Consequently, $\vct\mu_\psi$, $\mathsf G_\psi$ and $\mathsf P_\psi$
determine the first and second moments of the selected mechanical pair.
Conversely, a Hermitian $\mathsf G_\psi$ and symmetric
$\mathsf P_\psi$ are jointly admissible as centered second moments
exactly when
\begin{equation}
 \begin{pmatrix}
  \mathsf G_\psi&\mathsf P_\psi\\
  \overline{\mathsf P_\psi}&\overline{\mathsf G_\psi}
 \end{pmatrix}\geq0.
 \label{eq:s61-complex-moment-admissibility}
\end{equation}
\end{proposition}

\begin{proof}
Write $\delta\vct\psi=\vct a+i\vct b$, with real centered columns.
Expanding the products gives
$\operatorname{Re}\mathsf G_\psi=
\langle\vct a\vct a^{\mathsf T}+\vct b\vct b^{\mathsf T}\rangle$,
$\operatorname{Re}\mathsf P_\psi=
\langle\vct a\vct a^{\mathsf T}-\vct b\vct b^{\mathsf T}\rangle$,
$\operatorname{Im}\mathsf G_\psi=
\langle\vct b\vct a^{\mathsf T}-\vct a\vct b^{\mathsf T}\rangle$
and
$\operatorname{Im}\mathsf P_\psi=
\langle\vct a\vct b^{\mathsf T}+\vct b\vct a^{\mathsf T}\rangle$.
Since $\delta\vct Y_E=\sqrt2(\vct a^{\mathsf T},
\vct b^{\mathsf T})^{\mathsf T}$, their sums and differences give the
four blocks above. Inverting $\mathsf\Omega$ on the positive sector then
recovers the displacement and momentum moments.
For admissibility, the block matrix is the covariance of the augmented
column $(\delta\vct\psi^{\mathsf T},
\overline{\delta\vct\psi}^{\mathsf T})^{\mathsf T}$. Its invertible
linear relation to the real column $\delta\vct Y_E$ makes positivity
equivalent to $\mathsf C_{Y_E}\geq0$. Every real nonnegative covariance
is realizable by a centered real random vector, as its eigenvector
decomposition shows. This also proves sufficiency.
\end{proof}

The reconstruction tests whether displacement, velocity and correlation
channels describe one admissible preparation. Separate fits of the two
complex matrices can violate joint positivity even when
$\mathsf G_\psi\geq0$. A consistent fit uses one real covariance and
the declared observation map.

For a concrete one-mode example, compare an equal mixture of
$\psi=\pm\sqrt E$ with an equal mixture of
$\psi=\pm i\sqrt E$, where $E>0$. Both preparations have zero mean,
$G_\psi=E$ and identical Hermitian quadratic modal statistics. In the
first, $P_\psi=E$, $\operatorname{Var}Q=2E/\omega^2$ and
$\operatorname{Var}\Pi=0$; in the second, $P_\psi=-E$,
$\operatorname{Var}Q=0$ and $\operatorname{Var}\Pi=2E$.
They prepare different particle motions at the observation time despite
having the same modal energy. For a declared particle-centre or structural
projection, this difference changes the displacement fluctuations recorded
by a linear detector. Thus the normalized energy-weighted matrix used
below is sufficient for its declared Hermitian quadratic modal statistics;
general displacement--momentum covariances also require the relation
matrix. These are first- and second-moment descriptions of the retained
mechanical coordinates. Higher moments or additional medium variables
require their corresponding preparation data.

\paragraph{A detector calibration can hide a mechanical scale.}
An observable also requires a conversion from physical motion to a
recorded signal. Consider a stipulated passive displacement coordinate
$u$, with positive effective inertia $m$, drag $c$ and stiffness $k$,
driven by its independently calibrated conjugate force $f$.
Let $y=gu$ be the detector output, with an unknown real positive,
frequency-independent gain $g$. Using the $e^{-i\omega t}$ convention,
\begin{equation}
 \begin{aligned}
 m\ddot u+c\dot u+ku&=f,\qquad y=gu,\\
 H_{\mathrm{rec}}(\omega)
   :=\frac{Y(\omega)}{F(\omega)}
   &=\frac{g}{k-m\omega^2-ic\omega}.
 \end{aligned}
 \label{eq:s61-gain-calibrated-response}
\end{equation}
The transfer function has detector-output units per force. This example
may describe a declared collective or structural projection when its
single-coordinate constitutive law is justified. Treat all four
coefficients $(m,c,k,g)$ as unknown here; an independently known effective
inertia already supplies an absolute scale.

\begin{proposition}[A response-scale ambiguity and its thermal test]
\label{prop:s61-gain-mechanical-scale}
For every dimensionless $a>0$, replacing
$(m,c,k,g)$ by $(am,ac,ak,ag)$ leaves
$H_{\mathrm{rec}}(\omega)$ unchanged at every frequency.
Suppose additionally that the same coordinate is in stationary classical
thermal equilibrium at known temperature $T>0$, with Markov drag,
two-sided thermal force spectrum $2k_{\mathrm B}Tc$ and no additional
detector noise. Then
\begin{equation}
 \begin{aligned}
 S_{yy}(\omega)
   &=\frac{2k_{\mathrm B}Tc\,g^2}
      {|k-m\omega^2-ic\omega|^2},\\
 g&=\frac{\omega S_{yy}(\omega)}
          {2k_{\mathrm B}T\operatorname{Im}
                         H_{\mathrm{rec}}(\omega)},
       \qquad \omega>0.
 \end{aligned}
 \label{eq:s61-gain-thermal-calibration}
\end{equation}
Along the scaled family, $S_{yy}$ changes by the factor $a$.
\end{proposition}

\begin{proof}
The scaling multiplies the numerator and denominator of
$H_{\mathrm{rec}}$ by the same $a$. With
$Z=k-m\omega^2-ic\omega$, the thermal output is the force spectrum
times $|g/Z|^2$. Also
$\operatorname{Im}H_{\mathrm{rec}}=gc\omega/|Z|^2>0$ for
$\omega>0$. Substitution yields the gain formula. Scaling multiplies the
thermal force spectrum by $a$ while leaving $|g/Z|^2$ unchanged, so the
output spectrum scales by $a$.
\end{proof}

Driven transfer contains one power of gain; the fluctuation spectrum
contains two. Their comparison resolves this scale ambiguity under the
stated thermal and calibration assumptions
\cite{CallenWelton1951,SquiresMason2010Review}. Force calibration,
temperature, reservoir response, filtering and detector noise determine
whether that comparison is available. At fixed physical units the scaled
family represents distinct mechanics and gains. A coordinate change,
by comparison, transports the same preparation and its maps together.
Coupled or viscoelastic models require their corresponding fluctuation
and response laws.

Figure~\ref{fig:s6-gain-response-fluctuation-calibration} exhibits the exact response-scale ambiguity and its thermal comparison.

\begin{figure}[H]
\centering\begingroup
\definecolor{vltTeal}{HTML}{176B64}
\definecolor{vltBlue}{HTML}{315D80}
\definecolor{vltGold}{HTML}{A87520}
\definecolor{vltPlum}{HTML}{725877}
\definecolor{vltRed}{HTML}{762B38}
\definecolor{vltYellow}{HTML}{D5AD3C}
\definecolor{vltNeutral}{HTML}{777D83}
\setlength{\unitlength}{1pt}
\scalebox{0.90}{%
\input{vlt-figures/figure-185.tex}%
}\endgroup
\caption{Driven response and thermal fluctuations constrain different powers of detector gain in \eqref{eq:s61-gain-calibrated-response}--\eqref{eq:s61-gain-thermal-calibration}. Use baseline $m/m_\star=1$, $k/(m_\star\omega_\star^2)=2$, $c/(m_\star\omega_\star)=0.6$ and $g/g_\star=1$. Scaling all four coefficients by $a=3$ preserves the complete driven transfer, so its two curves coincide in (b). At the same known temperature, the equilibrium two-sided output spectrum in (c) is multiplied by three. Units are $H_\star=g_\star/(m_\star\omega_\star^2)$ and $S_\star=k_{\rm B}Tg_\star^2/(m_\star\omega_\star^3)$. The incoming filled arrow is a calibrated conjugate force; open arrows are signal maps. This auxiliary single-coordinate model assumes Markov drag and no added detector noise.}
\label{fig:s6-gain-response-fluctuation-calibration}
\end{figure}

Self-excited motion uses the force-generation and resource state of
Section~\ref{sec:self-excitation-passive-boundary} to predict growth,
saturation and depletion. Its displacement patterns can use the same
Hilbert-space geometry; the conservative operator below supplies the
reference for extending their dynamics.

\subsection{Spectral observables and joint modal spread}
\label{sec:spectral-observables}

\subsubsection{Spectral decomposition as a question about a preparation}

An eigenvector is a pattern that an operator scales without changing its
direction. If $A\ket a=a\ket a$, its component in a superposition is
multiplied by $a$. For a finite self-adjoint matrix this gives
$A\ket\eta=\sum_a a\mathsf P_a\ket\eta$, where the orthogonal
projectors keep the full eigenspaces, including degeneracy. The spectral
theorem extends this construction to a projection-valued measure
$\mathsf P_A$:
\begin{equation}
 A=\int_{\R}a\,\mathsf P_A(\dd a),\qquad
 f(A)=\int_{\R}f(a)\,\mathsf P_A(\dd a).
 \label{eq:operator-spectral-measure}
\end{equation}
Functional calculus means applying the scalar function to each spectral
value while keeping its spectral subspace. Thus $\mathsf W=\sqrt{\mathsf D}$
is a spectral square root, not an entrywise matrix square root.
For a measurable set $B$, $\mathsf P_A(B)$ is self-adjoint and
idempotent; applying it twice gives the same selected component.
Disjoint spectral sets have orthogonal projections, and their full
resolution gives the identity~\cite{Teschl2009}.

Every normalized $\eta$ defines a nonnegative unit-mass measure
$\mu_\eta(B)=\bra\eta\mathsf P_A(B)\ket\eta$.
For unbounded $A$, $\eta\in D(A)$ precisely when
$\int a^2\,\dd\mu_\eta(a)<\infty$. This is the condition for a finite
operator-image norm and second moment, rather than merely normalization
of the state. An orthonormal eigenbasis is available for the finite
lattice and the compact-resolvent conservative problem of Section~\ref{sec:complex-displacement};
general self-adjoint operators can also have continuous spectra.

On the finite positive sector,
\begin{equation}
 \mathsf W=\sum_{s=1}^r\omega_s\ket s\bra s,\qquad
 \mathsf D=\sum_{s=1}^r\omega_s^2\ket s\bra s,\qquad
 \mathsf P_B=\sum_{\omega_s\in B}\ket s\bra s.
 \label{eq:frequency-spectral-projectors}
\end{equation}
For a classical energy preparation, the band weight
$w_B=\bra\eta\mathsf P_B\ket\eta$ gives $E_B=E_+w_B$.
When $w_B>0$, the restricted direction is
$\mathsf P_B\eta/\sqrt{w_B}$. This spectral selection specifies a
component of the state. Its physical preparation needs a force protocol;
time filtering of a detector acts on the recorded signal.

A detector has coefficients
$d_s=\sqrt2\,\vct\ell^{\mathsf T}\vct\phi_s/\omega_s$ and analytic
signal $y_{\rm an}(t)=\sum_sd_s\psi_s(0)e^{-i\omega_st}$.
Its transfer $D_{\rm det}$ multiplies each line by
$D_{\rm det}(\omega_s)$. Detector overlap, interference within
degenerate subspaces and the observation window thus determine measured
line weights and peak shapes. Modes with $d_s=0$ are dark to this channel.

For a normalized one-phonon ket $\Phi$, the same matrix projector instead
gives a frequency probability. Its energy fraction includes a frequency
weight:
\begin{equation}
 p_B=\sum_{\omega_s\in B}|c_s|^2,\qquad
 \frac{E_B^{(1)}}{E^{(1)}}=
 \frac{\sum_{\omega_s\in B}\hbar\omega_s|c_s|^2}
      {\sum_s\hbar\omega_s|c_s|^2}.
 \label{eq:operator-band-probability-energy}
\end{equation}
Figure~\ref{fig:spectral-weights-versus-energy-fractions} illustrates the difference.
With two frequencies $\omega_\star$ and $2\omega_\star$ and equal
coefficient magnitudes, a classical energy direction assigns one half of
its energy to each mode, while the one-phonon state assigns one half of
its frequency probability but one third and two thirds of its excitation
energy. The two columns have been normalized for different physical jobs.
For general occupation states, band occupation and energy use
$\sum_{\omega_s\in B}\hat N_s$ and
$\sum_{\omega_s\in B}\hbar\omega_s\hat N_s$ on the occupation space;
a mode-space projector alone is not their full many-excitation operator.

\begin{figure}[H]
\centering
\begingroup
\definecolor{vlfGreen}{HTML}{176B64}
\definecolor{vlfDarkRed}{HTML}{762B38}
\definecolor{vlfRed}{HTML}{762B38}
\definecolor{vlfBlue}{HTML}{315D80}
\definecolor{vlfCopper}{HTML}{762B38}

\setlength{\unitlength}{1pt}
\scalebox{0.80}{%
\begin{picture}(470,332)%
{\color{black}\put(119,316){\makebox(0,0)[c]{\fontfamily{lmss}\fontsize{10.500}{12.500}\selectfont \bfseries (a) Spectral weights}}}%
{\color{black}\put(356,316){\makebox(0,0)[c]{\fontfamily{lmss}\fontsize{10.500}{12.500}\selectfont \bfseries (b) Energy fractions}}}%
{\color{black}\put(235,289){\makebox(0,0)[c]{\fontfamily{lmss}\fontsize{10.500}{12.500}\selectfont $\eta=c=(1,1)^{\mathsf T}/\sqrt2,\qquad \omega_2=2\omega_1=2\omega_\star$}}}%
{\color{black}\vlB{%
\vSc{38.000000}{111.000000}{0.000000}{3.000000}%
}}%
{\color{black}\put(34,111){\makebox(0,0)[r]{\fontfamily{lmss}\fontsize{10.500}{12.500}\selectfont $0$}}}%
{\color{black}\vlB{%
\vSc{38.000000}{181.000000}{0.000000}{3.000000}%
}}%
{\color{black}\put(34,181){\makebox(0,0)[r]{\fontfamily{lmss}\fontsize{10.500}{12.500}\selectfont $1/2$}}}%
{\color{black!20}\vlB{%
\vSan{41}{181}{0}{168}%
}}%
{\color{black}\vlB{%
\vSc{38.000000}{251.000000}{0.000000}{3.000000}%
}}%
{\color{black}\put(34,251){\makebox(0,0)[r]{\fontfamily{lmss}\fontsize{10.500}{12.500}\selectfont $1$}}}%
{\color{black!20}\vlB{%
\vSan{41}{251}{0}{168}%
}}%
{\color{black}\vlB{%
\vSc{41.000000}{111.000000}{0.000000}{168.000000}%
\vSc{40.500000}{111.000000}{90.000000}{140.000000}%
}}%
{\color{black}\put(125,80){\makebox(0,0)[c]{\fontfamily{lmss}\fontsize{10.500}{12.500}\selectfont }}}%
{\color{black}\put(41,267){\makebox(0,0)[l]{\fontfamily{lmss}\fontsize{10.500}{12.500}\selectfont $p_s$}}}%
{\color{black}\vlB{%
\vSat{83}{111}{90}{70}%
}}%
{\color{black}\put(97,193){\makebox(0,0)[c]{\fontfamily{lmss}\fontsize{10.500}{12.500}\selectfont $1/2$}}}%
{\color{black}\put(97,91){\makebox(0,0)[c]{\fontfamily{lmss}\fontsize{10.500}{12.500}\selectfont $\omega_1$}}}%
{\color{black}\vlB{%
\vSat{153}{111}{90}{70}%
}}%
{\color{black}\put(167,193){\makebox(0,0)[c]{\fontfamily{lmss}\fontsize{10.500}{12.500}\selectfont $1/2$}}}%
{\color{black}\put(167,91){\makebox(0,0)[c]{\fontfamily{lmss}\fontsize{10.500}{12.500}\selectfont $\omega_2$}}}%
{\color{black}\vlB{%
\vSc{279.000000}{111.000000}{0.000000}{3.000000}%
}}%
{\color{black}\put(275,111){\makebox(0,0)[r]{\fontfamily{lmss}\fontsize{10.500}{12.500}\selectfont $0$}}}%
{\color{black}\vlB{%
\vSc{279.000000}{181.000000}{0.000000}{3.000000}%
}}%
{\color{black}\put(275,181){\makebox(0,0)[r]{\fontfamily{lmss}\fontsize{10.500}{12.500}\selectfont $1/2$}}}%
{\color{black!20}\vlB{%
\vSan{282}{181}{0}{169}%
}}%
{\color{black}\vlB{%
\vSc{279.000000}{251.000000}{0.000000}{3.000000}%
}}%
{\color{black}\put(275,251){\makebox(0,0)[r]{\fontfamily{lmss}\fontsize{10.500}{12.500}\selectfont $1$}}}%
{\color{black!20}\vlB{%
\vSan{282}{251}{0}{169}%
}}%
{\color{black}\vlB{%
\vSc{282.000000}{111.000000}{0.000000}{169.000000}%
\vSc{281.500000}{111.000000}{90.000000}{140.000000}%
}}%
{\color{black}\put(366.5,80){\makebox(0,0)[c]{\fontfamily{lmss}\fontsize{10.500}{12.500}\selectfont }}}%
{\color{black}\put(282,267){\makebox(0,0)[l]{\fontfamily{lmss}\fontsize{10.500}{12.500}\selectfont Energy share}}}%
{\color{vlfGreen}\vlB{%
\vSau{304.8425}{111}{90}{70}%
}}%
{\color{black}\put(312.8425,193){\makebox(0,0)[c]{\fontfamily{lmss}\fontsize{10.500}{12.500}\selectfont $1/2$}}}%
{\color{black}\put(312.8425,99){\makebox(0,0)[c]{\fontfamily{lmss}\fontsize{10.500}{12.500}\selectfont C}}}%
{\color{vlfDarkRed!75}\vlB{%
\vSau{327.6575}{111}{90}{46.666667}%
}}%
{\color{black}\put(335.6575,169.666667){\makebox(0,0)[c]{\fontfamily{lmss}\fontsize{10.500}{12.500}\selectfont $1/3$}}}%
{\color{black}\put(335.6575,99){\makebox(0,0)[c]{\fontfamily{lmss}\fontsize{10.500}{12.500}\selectfont Q}}}%
{\color{black}\put(324.25,81){\makebox(0,0)[c]{\fontfamily{lmss}\fontsize{10.500}{12.500}\selectfont $\omega_1$}}}%
{\color{vlfGreen}\vlB{%
\vSau{389.3425}{111}{90}{70}%
}}%
{\color{black}\put(397.3425,193){\makebox(0,0)[c]{\fontfamily{lmss}\fontsize{10.500}{12.500}\selectfont $1/2$}}}%
{\color{black}\put(397.3425,99){\makebox(0,0)[c]{\fontfamily{lmss}\fontsize{10.500}{12.500}\selectfont C}}}%
{\color{vlfDarkRed!75}\vlB{%
\vSau{412.1575}{111}{90}{93.333333}%
}}%
{\color{black}\put(420.1575,216.333333){\makebox(0,0)[c]{\fontfamily{lmss}\fontsize{10.500}{12.500}\selectfont $2/3$}}}%
{\color{black}\put(420.1575,99){\makebox(0,0)[c]{\fontfamily{lmss}\fontsize{10.500}{12.500}\selectfont Q}}}%
{\color{black}\put(408.75,81){\makebox(0,0)[c]{\fontfamily{lmss}\fontsize{10.500}{12.500}\selectfont $\omega_2$}}}%
{\color{black}\put(119,65){\makebox(0,0)[c]{\fontfamily{lmss}\fontsize{10.500}{12.500}\selectfont $p_1+p_2=1$}}}%
{\color{black}\put(356,58){\makebox(0,0)[c]{\fontfamily{lmss}\fontsize{10.500}{12.500}\selectfont C: classical energy direction}}}%
{\color{black}\put(356,40){\makebox(0,0)[c]{\fontfamily{lmss}\fontsize{10.500}{12.500}\selectfont Q: one-phonon excitation energy}}}%
{\color{black}\put(235,16){\makebox(0,0)[c]{\fontfamily{lmss}\fontsize{10.500}{12.500}\selectfont A projector weight needs a stated state normalization and interpretation.}}}%
\end{picture}%
%
}%
\endgroup
\caption{Equal projector weights with two normalizations. For frequencies $\omega_\star,2\omega_\star$ and columns $\eta=c=(1,1)^{\mathsf T}/\sqrt2$, both weights are $1/2$. They are classical modal energy fractions for $\eta$ and frequency probabilities for the one-phonon state $c$. The latter excitation-energy fractions include the frequency factor and become $1/3,2/3$, with mean excitation energy $3\hbar\omega_\star/2$. Bars exclude vacuum energy. All plotted shares are dimensionless; no equality of total energies between the two preparations is assumed.}
\label{fig:spectral-weights-versus-energy-fractions}
\end{figure}

\paragraph{Repeated preparations and energy-weighted spectral statistics.}
A modal direction describes one prepared motion. Repeated preparations
can distribute energy among different directions, including directions
with correlated or randomized relative phases. Their spectral statistics
have a finite operator representation of their own.

\begin{definition}[Classical energy-preparation matrix]
\label{def:s62-energy-preparation-matrix}
Fix the conservative positive sector and all its mechanical coefficients.
For an ensemble of classical energy columns $\psi$ with finite positive
mean energy $\overline E_+=\langle\psi^\dagger\psi\rangle>0$, define
\begin{equation}
 \mathsf R_E=\frac{\langle\psi\psi^\dagger\rangle}{\overline E_+},
 \qquad \mathsf R_E\geq0,\qquad\tr\mathsf R_E=1.
 \label{eq:s62-energy-preparation-matrix}
\end{equation}
Its entries record normalized Hermitian correlations of the energy
amplitudes. The band share of the ensemble's mean energy is
$\tr(\mathsf R_E\mathsf P_B)$. For a finite self-adjoint $A$, set
$\langle A\rangle_{\mathsf R_E}=\tr(\mathsf R_EA)$ and
$(\Delta_{\mathsf R_E}A)^2
=\tr[\mathsf R_E(A-\langle A\rangle_{\mathsf R_E}\Id)^2]$.
\end{definition}

This dimensionless matrix records Hermitian modal second moments; the
absolute energy is $\overline E_+$. Full real displacement--momentum
covariance also needs the mean and complementary anomalous moment from
Section~\ref{sec:operator-representations}. Randomizing phase can make
$\langle\psi\rangle=0$ while preserving excitation and Hermitian
correlations in $\mathsf R_E$.

\begin{proposition}[Intrinsic and between-preparation modal spread]
\label{prop:s62-energy-mixture-variance}
For a finite ensemble with probabilities $p_\alpha$, nonzero energies
$E_\alpha$ and directions $\eta_\alpha=\psi_\alpha/\sqrt{E_\alpha}$,
put $\pi_\alpha=p_\alpha E_\alpha/\overline E_+$ and
$a_\alpha=\langle A\rangle_{\eta_\alpha}$. Then
\begin{equation}
 \begin{aligned}
 \mathsf R_E&=\sum_\alpha\pi_\alpha
                        \ket{\eta_\alpha}\bra{\eta_\alpha},
 &\langle A\rangle_{\mathsf R_E}&=\sum_\alpha\pi_\alpha a_\alpha,\\
 (\Delta_{\mathsf R_E}A)^2
   &=\sum_\alpha\pi_\alpha(\Delta_{\eta_\alpha}A)^2
     +\sum_\alpha\pi_\alpha
          (a_\alpha-\langle A\rangle_{\mathsf R_E})^2.
 \end{aligned}
 \label{eq:s62-energy-mixture-variance}
\end{equation}
Zero-energy preparations contribute no weight.
\end{proposition}
\begin{proof}
Substitute $\psi_\alpha=\sqrt{E_\alpha}\eta_\alpha$ into the
definition of $\mathsf R_E$ and take traces. For each component,
$\langle A^2\rangle_{\eta_\alpha}
=(\Delta_{\eta_\alpha}A)^2+a_\alpha^2$.
The weighted sum minus the square of the weighted mean gives the
displayed decomposition.
\end{proof}

Repeated preparations contribute both their intrinsic spectral spreads
and the variation of their spectral means. The weights are proportional
to energy; averaging unit directions with probabilities $p_\alpha$
alone gives another statistic when energies differ. This construction
uses one mechanical law. Uncertainty in that law instead defines a family
of conditional models.

The full covariance inequality in \eqref{eq:full-modal-uncertainty}
also holds with these trace expectations. Apply Cauchy--Schwarz to
$\delta A\mathsf R_E^{1/2}$ and $\delta B\mathsf R_E^{1/2}$ in the
Hilbert--Schmidt inner product $\tr(X^\dagger Y)$; its real and
imaginary parts give the same covariance and commutator terms
\cite{HornJohnson2013}. This extends the spectral-spread calculation
to repeated classical preparations without converting that spread
into a parameter-error bar. Under free motion,
$\mathsf R_E(t)=e^{-it\mathsf W}\mathsf R_E(0)e^{it\mathsf W}$,
so the mean band shares remain fixed while phase-sensitive quadratic
statistics can change.

\subsubsection{Local curvature and collective frequency need not align}

The local restoring operator of \eqref{eq:local-restoring-operator} is
compressed to the positive sector as
\begin{equation}
 \mathsf S_+=\left.\mathsf P_+\mathsf S\mathsf P_+\right|_{\mathcal H_+},
 \qquad \mathsf S=\sum_i\mathsf\Pi_i\mathsf D\mathsf\Pi_i.
 \label{eq:compressed-local-response}
\end{equation}
The blocks $\mathsf\Pi_i$ retain each node's specified coordinates.
Their diagonal dynamical blocks give the curvature with other
coordinates held fixed in the assembled network. For $\mathsf D\geq0$
this operator is nonnegative, with units $\mathrm s^{-2}$. Compression
retains its action on the positive sector when local operations also
reach zero modes. For normalized $\eta\in\mathcal H_+$,
\begin{equation}
 (\Delta_\eta\mathsf S)^2
    =(\Delta_\eta\mathsf S_+)^2
      +\|(\Id-\mathsf P_+)\mathsf S\eta\|^2.
 \label{eq:operator-compression-variance}
\end{equation}
The first moment is unchanged by compression, but its variance excludes
the nonnegative contribution carried outside the selected sector.
In a fully positive lattice $\mathsf S_+=\mathsf S$.

The operators $\mathsf W$ and $\mathsf D$ commute because the former
is a spectral function of the latter. The separately constructed
$\mathsf S_+$ can mix frequency subspaces. Its commutator
$[A,B]=AB-BA$ compares the two orders of action and diagnoses that
mixing. Quantum disturbance and dissipative linewidths require their
respective state and dynamical laws.

\begin{example}[A mechanically derived noncommuting pair]
\label{ex:operator-mechanical-pair}
Take two equal retained masses $m>0$ with supporting springs
$k_\star$, $2k_\star$ and a coupling spring $k_\star$, where
$k_\star=m\omega_\star^2>0$. Their conservative energy and operators are
\begin{equation}
 \begin{split}
 V&=\frac{k_\star}{2}\left[u_1^2+2u_2^2+(u_1-u_2)^2\right],\\
 \mathsf D&=\omega_\star^2\begin{pmatrix}2&-1\\-1&3\end{pmatrix},
 \qquad \mathsf S=\omega_\star^2\begin{pmatrix}2&0\\0&3\end{pmatrix},\\
 \mathsf W&=\frac{\omega_\star}{g}
    \begin{pmatrix}2+\sqrt5&-1\\-1&3+\sqrt5\end{pmatrix},
 \qquad g=\sqrt{5+2\sqrt5}.
 \end{split}
 \label{eq:operator-mechanical-pair}
\end{equation}
Both modes are positive, with
$\omega_\pm=\omega_\star\sqrt{(5\pm\sqrt5)/2}$, and direct multiplication
gives
\begin{equation}
 [\mathsf S,\mathsf W]
    =\frac{\omega_\star^3}{g}
           \begin{pmatrix}0&1\\-1&0\end{pmatrix}\ne0.
 \label{eq:operator-mechanical-commutator}
\end{equation}
The mismatch follows from the specified couplings and unequal local
supports. These supporting curvatures are declared conservative
constraints of the example; a reflecting cylinder alone would not
generate them. No additional internal degree of freedom is required for this
two-coordinate example. The parameters are theoretical reference values,
not a measured viral stiffness.
\end{example}

In an application, unequal local curvature could represent independently
constrained attachments or local restoring environments at the selected
resolution. The comparison is then between the specified local response
and the coupled mode patterns, rather than between a biological label and
an assumed material constant. A commutator diagnoses that decomposition;
it does not identify a unique molecular origin for the curvatures.

\begin{figure}[H]
\centering\begingroup
\definecolor{vltTeal}{HTML}{176B64}
\definecolor{vltBlue}{HTML}{315D80}
\definecolor{vltGold}{HTML}{A87520}
\definecolor{vltPlum}{HTML}{725877}
\definecolor{vltRed}{HTML}{762B38}
\definecolor{vltYellow}{HTML}{D5AD3C}
\definecolor{vltNeutral}{HTML}{777D83}
\setlength{\unitlength}{1pt}
\scalebox{0.90}{%
\input{vlt-figures/figure-187.tex}%
}\endgroup
\caption{Local and collective operators for a supported ideal-virion pair. The same mass $m$ and reference spring spans are used in (a); the right support has curvature $2k_\star$, while the other two have $k_\star=m\omega_\star^2$. Coil pitch depicts reference geometry and does not encode curvature. (b) For $\eta(0)=(1,i)^{\mathsf T}/\sqrt2$ in the mass-weighted coordinate basis and $\mathsf W=\sqrt{\mathsf D}$, the local-restoring statistic is $5/2+\sin\tau/\sqrt5$. Here $\omega_\pm=\omega_\star\sqrt{(5\pm\sqrt5)/2}$. The total classical energy and its equal normal-mode shares remain fixed. The ordinate is a quadratic restoring statistic of the normalized preparation, with units of frequency squared before scaling.}
\label{fig:mechanical-local-collective-noncommutation}
\end{figure}

The generally nonlocal map $\mathsf W=\sqrt{\mathsf D}$ mixes site
coordinates. Thus $\langle\mathsf S\rangle_\eta$ is a restoring
statistic in the energy representation. Anatomical storage and energy
transport use the local allocation and current of
Sections~\ref{sec:viscoelastic-dynamics} and~\ref{sec:complex-displacement}.

\subsubsection{Unfolding the covariance relation}

Let $A,B$ be self-adjoint finite matrices and $\eta$ normalized. Define
$\delta A=A-\langle A\rangle_\eta\Id$ and
$\Delta_\eta A=\|\delta A\eta\|$, with the corresponding definitions
for $B$. These spreads describe the distribution of spectral values in
that prepared direction. Applying Cauchy--Schwarz to
$\delta A\eta$ and $\delta B\eta$ gives
$\Delta_\eta A\,\Delta_\eta B
 \geq|\langle\delta A\eta,\delta B\eta\rangle|$.
Its real and imaginary parts are respectively the symmetric covariance
and half the commutator expectation. Thus
\begin{equation}
 (\Delta_\eta A)^2(\Delta_\eta B)^2
 \geq
 \left|\frac{\langle[A,B]\rangle_\eta}{2i}\right|^2
 +\left|\frac{\langle\{\delta A,\delta B\}\rangle_\eta}{2}\right|^2.
 \label{eq:full-modal-uncertainty}
\end{equation}
The braces denote $\{X,Y\}=XY+YX$. The full covariance bound retains both terms~\cite{Robertson1934};
discarding covariance can weaken it even when the commutator is known.
For unbounded observables, the centered-vector argument is meaningful
when the corresponding second moments are finite. Writing $AB$ and $BA$
as operator products additionally requires their product domains.

For the mechanical pair and $\eta_0=(1,i)^{\mathsf T}/\sqrt2$ in the
original mass-weighted site basis,
\begin{equation}
 \begin{aligned}
 \Delta_{\eta_0}\mathsf S&=\omega_\star^2/2,
 &\Delta_{\eta_0}\mathsf W&=\sqrt5\,\omega_\star/(2g),\\
 \left|\frac{\langle[\mathsf S,\mathsf W]\rangle_{\eta_0}}{2i}\right|
       &=\omega_\star^3/(2g),
 &\frac{\langle\{\delta\mathsf S,\delta\mathsf W\}\rangle_{\eta_0}}2
       &=\omega_\star^3/(4g).
 \end{aligned}
 \label{eq:operator-mechanical-spread-values}
\end{equation}
The squared product equals the sum of the two squared terms. The full
bound is attained, while the commutator-only bound is strictly smaller.
For a frequency eigenvector, however, $\Delta\mathsf W=0$ and the
commutator expectation vanishes. A nonzero commutator therefore does not
impose one positive lower product on all preparations.

Figure~\ref{fig:more-modal-spread-bound} varies the preparation for this
same spring model. It uses the ordered real frequency basis $(-,+)$,
oriented so the off-diagonal entry of $\mathsf S/\omega_\star^2$ is
$1/\sqrt5$, and the family
$(\cos\theta,e^{i\pi/4}\sin\theta)^{\mathsf T}$.
Its operator parameters stay fixed while the spreads change. In the
classical interpretation these are modal spectral spreads, not errors in
estimated spring constants. Under the quantum measurement postulate the
corresponding self-adjoint spectral observables have state variances.
Thermal fluctuations, finite-window spectral resolution and instrumental
estimation error are separate quantities with their own definitions.

\begin{figure}[H]
\centering\begingroup
\definecolor{vltTeal}{HTML}{176B64}
\definecolor{vltBlue}{HTML}{315D80}
\definecolor{vltGold}{HTML}{A87520}
\definecolor{vltPlum}{HTML}{725877}
\definecolor{vltRed}{HTML}{762B38}
\definecolor{vltYellow}{HTML}{D5AD3C}
\definecolor{vltNeutral}{HTML}{777D83}
\setlength{\unitlength}{1pt}
\scalebox{0.80}{%
\input{vlt-figures/figure-188.tex}%
}\endgroup
\caption{Modal spreads for the supported pair of Figure~\ref{fig:mechanical-local-collective-noncommutation}. In its ordered $(-,+)$ frequency basis use $S/\omega_\star^2=\left(\begin{smallmatrix}5/2-1/(2\sqrt5)&1/\sqrt5\\1/\sqrt5&5/2+1/(2\sqrt5)\end{smallmatrix}\right)$ and $W/\omega_\star=\operatorname{diag}(\sqrt{(5-\sqrt5)/2},\sqrt{(5+\sqrt5)/2})$. The displayed family varies preparation at fixed mechanics. The product and the magnitudes $C=|\langle[S,W]\rangle/(2i)|$, $V=|\langle\{\delta S,\delta W\}\rangle/2|$ are divided by $\omega_\star^3$. Open circles give $\sqrt{C^2+V^2}$, equal to the product throughout this pure two-dimensional family. Endpoint frequency spreads vanish. Line styles distinguish contributions in grayscale; the spreads refer to modal content under the stated classical or quantum interpretation.}
\label{fig:more-modal-spread-bound}
\end{figure}

\paragraph{Which measured spectral statistics determine another?}
The inverse question can be made precise for a known reference operator.
Frequency-band shares constrain diagonal parts of a preparation matrix;
a statistic sensitive to correlations between those bands asks for
additional information.

\begin{definition}[Calibrated quadratic observation family]
\label{def:s62-quadratic-observation-family}
After fixed unit scales have been declared, let
$\mathsf A_1,\ldots,\mathsf A_m$ be dimensionless Hermitian matrices
on the $r$-dimensional positive sector. A \emph{quadratic observation
family} reports $a_j=\tr(\mathsf R_E\mathsf A_j)$, with its
specified reconstruction and calibration. The identity's statistic
$\tr\mathsf R_E=1$ is already known. Each such quadratic value needs
the coordinate, momentum and mechanical weights appropriate to its
operator.
\end{definition}

\begin{proposition}[Identification of a quadratic target statistic]
\label{prop:s62-quadratic-statistic-identification}
Let $\mathcal V_A=\operatorname{span}_{\R}
\{\Id,\mathsf A_1,\ldots,\mathsf A_m\}$.
For a dimensionless Hermitian target $\mathsf B$, its value
$\tr(\mathsf R_E\mathsf B)$ is determined by these data for every
nonnegative unit-trace preparation matrix precisely when
\begin{equation}
 \mathsf B\in\mathcal V_A.
 \label{eq:s62-quadratic-statistic-criterion}
\end{equation}
If the condition fails, two positive preparation matrices can have
identical reported statistics and distinct target values.
\end{proposition}
\begin{proof}
Membership gives $\mathsf B=c_0\Id+\sum_jc_j\mathsf A_j$, so
the target is $c_0+\sum_jc_ja_j$. Conversely, use the real
Hilbert--Schmidt inner product $\tr(XY)$ on Hermitian matrices.
If $\mathsf B\notin\mathcal V_A$, its nonzero orthogonal residual
$\mathsf Z$ obeys $\tr\mathsf Z=\tr(\mathsf Z\mathsf A_j)=0$
and $\tr(\mathsf Z\mathsf B)=\tr(\mathsf Z^2)>0$.
For sufficiently small $\epsilon>0$,
$\mathsf R_\pm=\Id_r/r\pm\epsilon\mathsf Z$ are positive and
unit trace. Their measured statistics agree, while their target
values differ by $2\epsilon\tr(\mathsf Z^2)$.
\end{proof}

Every such matrix is realizable as an ensemble of classical energy
directions by its spectral decomposition, with an arbitrary common
positive energy scale. The scale can be chosen within the stated
harmonic basin. The proposition concerns all preparations in that
class; additional preparation restrictions define a smaller inference
problem. Determining one target statistic can require fewer independent
quadratic data than reconstructing the full preparation matrix.

The existing supported pair gives a direct example. In its ordered real
frequency basis, write $\mathsf B=\mathsf S_+/\omega_\star^2$, whose
off-diagonal entry is $1/\sqrt5$. The two directions
$\eta_\pm=(1,\pm1)^{\mathsf T}/\sqrt2$ give
\begin{equation}
 \begin{aligned}
 \tr(\mathsf R_\pm\mathsf P_-)
 &=\tr(\mathsf R_\pm\mathsf P_+)=\tfrac12,
 \qquad \mathsf R_\pm=\ket{\eta_\pm}\bra{\eta_\pm},\\
 \tr(\mathsf R_\pm\mathsf B)&=\frac52\pm\frac1{\sqrt5}.
 \end{aligned}
 \label{eq:s62-equal-bands-distinct-statistic}
\end{equation}
Their frequency-band shares coincide, while their phase-sensitive local
curvature statistics differ. The masses, supports, pair interactions
and ideal-virion count are the same. A calibrated additional quadratic
projection can distinguish this correlation without changing the
mechanical law. With degeneracy, a frequency projector reports only
the total within its eigenspace; internal contrasts in that eigenspace
also require an independently specified observation. The operator
family therefore states which preparation distinctions the data retain.

\subsubsection{The same commutator controls conservative evolution}

For a time-independent mode operator $A$ and fixed positive energy,
$i\partial_t\eta=\mathsf W\eta$ gives
\begin{equation}
 \frac{\dd}{\dd t}\langle A\rangle_\eta
       =i\langle[\mathsf W,A]\rangle_\eta.
 \label{eq:modal-heisenberg-statistic}
\end{equation}
This follows by differentiating the bra and ket. It is already an exact
classical energy-amplitude identity; no factor of $\hbar$ is needed
because $\mathsf W$ is a frequency generator. If $A$ commutes with
$\mathsf W$, its statistic is conserved. Otherwise relative phases can
change it while the total energy remains fixed.

In the spring example with the original-site preparation $\eta_0$,
\begin{equation}
 \frac{\langle\mathsf S\rangle_{\eta(t)}}{\omega_\star^2}
   =\frac52+\frac1{\sqrt5}
                  \sin[(\omega_+-\omega_-)t].
 \label{eq:operator-mechanical-statistic-evolution}
\end{equation}
Figure~\ref{fig:mechanical-local-collective-noncommutation} shows this variation
beside its mechanical origin. The two frequency-band energy fractions
are constant at one half; a statistic tied to another operator still
oscillates. Such variation is interference within a specified conservative
state, rather than evidence of dissipative energy gain or self-excitation.
An active model would add its force and resource evolution before this
conservative identity could be extended.

\begin{proposition}[Finite-time change and separated-band mixing]
\label{prop:s62-commutator-controlled-mixing}
For finite self-adjoint $\mathsf W$ and $A$, the energy-space operator
norm obeys
\begin{equation}
 \|e^{it\mathsf W}Ae^{-it\mathsf W}-A\|_2
       \leq |t|\,\|[\mathsf W,A]\|_2.
 \label{eq:s62-commutator-time-bound}
\end{equation}
Let $\mathsf P$ and $\mathsf Q$ be spectral projectors onto two
nonempty bands with
$\max\sigma(\mathsf W|_{\operatorname{ran}\mathsf P})\leq\Omega_0$
and
$\min\sigma(\mathsf W|_{\operatorname{ran}\mathsf Q})\geq\Omega_0+g$
for $g>0$. Then
\begin{equation}
 \|\mathsf Q A\mathsf P\|_2
       \leq\frac{\|\mathsf Q[\mathsf W,A]\mathsf P\|_2}{g}.
 \label{eq:s62-separated-band-mixing}
\end{equation}
\end{proposition}
\begin{proof}
Integrate the derivative
$i e^{it\mathsf W}[\mathsf W,A]e^{-it\mathsf W}$ and use unitary
norm invariance for the first estimate. For the second, set
$X=\mathsf Q A\mathsf P$ and
$C=\mathsf Q[\mathsf W,A]\mathsf P$. On the two spectral ranges,
$\mathsf W_QX-X\mathsf W_P=C$. Its solution is
$X=\int_0^\infty
e^{-s(\mathsf W_Q-\Omega_0\Id)}C
e^{s(\mathsf W_P-\Omega_0\Id)}\,\dd s$.
The integrand norm is at most $e^{-sg}\|C\|_2$.
Integrating proves the bound; differentiating the integrand verifies
the Sylvester equation and its solution.
\end{proof}

The first estimate also bounds the change of the $A$ statistic in a
normalized direction or $\mathsf R_E$. It has units of $A$, since the
commutator has units of $A$ per time. The second controls mixing between
the specified frequency bands, whose separation is $g$.

These estimates quantify approximate conservation and band preservation.
Close bands can require their correlations to be retained together even
for a small commutator. The bounds control the chosen statistic;
closure and observation tests assess the other predictions required of
the reduced variables. Parameter-estimation uncertainty uses its own
data-error model.

\FloatBarrier
\subsection{Canonical displacement, momentum, and the phonon Hamiltonian}
\label{sec:canonical-phonon-operators}

\subsubsection{From collective coordinates to a configuration representation}

The spectral matrices organize patterns. A canonical displacement asks a
different question: which value of a mechanical coordinate is obtained in
the declared quantum preparation? Start with the real mass-weighted pair
$\vct x=\mathsf M^{1/2}\vct u$ and
$\vct p_x=\mathsf M^{-1/2}\vct p=\Dt{\vct x}$, and its positive-mode
coordinates
$\vct Q=\mathsf E^{\mathsf T}\vct x$,
$\vct\Pi=\mathsf E^{\mathsf T}\vct p_x$.
Classically these are functions on phase space with
$\{Q_s,\Pi_t\}_{\rm PB}=\delta_{st}$.
Their harmonic energy is
$\sum_s(\Pi_s^2+\omega_s^2Q_s^2)/2$.
For translational normalization, $Q_s$ has units
$\mathrm{kg}^{1/2}\mathrm m$ and $\Pi_s$ has units
$\mathrm{kg}^{1/2}\mathrm m\,\mathrm s^{-1}$.

\paragraph{Canonical normalization of the resolved body tangent.}
The dimensionless body coordinates of
\eqref{eq:s2patch-generalized-inertia} use their own positive inertia
$\mathsf G_X$ and real conservative curvature $\mathsf H_X$.
Choose the frozen, relaxed or inventory-constrained reference explicitly;
retained relaxing states carry their additional storage as in
\eqref{eq:s5patch-coupled-amplitude-storage}.
The canonical mechanical momentum is
$\vct P_X=\mathsf G_X\dot{\vct X}$, with action units. Let
$\mathsf E_X$ have real orthonormal eigenvectors spanning the retained
positive sector of
$\mathsf D_X=\mathsf G_X^{-1/2}\mathsf H_X\mathsf G_X^{-1/2}$.
Under the same oscillator postulate, its reconstruction is
\begin{equation}
 \begin{aligned}
 \hat{\vct X}_+&=\mathsf G_X^{-1/2}\mathsf E_X\hat{\vct Q}_X,
 \qquad\hat{\vct P}_{X,+}=\mathsf G_X^{1/2}\mathsf E_X\hat{\vct\Pi}_X,\\
 [\hat X_{+,a},\hat P_{X,+,b}]
   &=i\hbar(\mathsf G_X^{-1/2}\mathsf P_{+,X}
                              \mathsf G_X^{1/2})_{ab}\Id,\\
 \mathsf P_{+,X}&=\mathsf E_X\mathsf E_X^{\mathsf T}.
 \end{aligned}
 \label{eq:s6patch-body-canonical-reconstruction}
\end{equation}
Here $\vct Q_X$ and $\vct\Pi_X$ have units
$\sqrt{\mathrm J}\,\mathrm s$ and $\sqrt{\mathrm J}$.
The identity follows by inserting the modal canonical commutators.
A calibrated row $\mathsf L_a$ has modal overlap
$g_{as}=\mathsf L_a\mathsf G_X^{-1/2}\vct e_{X,s}$.
Length readouts give inverse-mass normalization; angular readouts give
inverse-moment-of-inertia normalization. First-order surface populations
and material strains retain their free-energy weights and kinetics.
The inertial oscillator count includes only the declared mechanical
projections. The $\mathsf M$ formulas are the translational specialization.

Under quantization, the configuration representation is
$L^2(\R^r,\dd\vct Q)$ with $\int|\Psi|^2\dd\vct Q=1$.
Each point $\vct Q$ specifies all retained collective amplitudes;
the wavefunction's units compensate this configuration-volume element.
Products of normalized oscillator eigenfunctions identify it unitarily
with the occupation basis of $\mathcal F_+$. Continuous coordinates and
discrete occupations thus represent the same quantum oscillator model.

Figure~\ref{fig:physical-and-quantum-configuration} separates the physical mode pattern from a probability distribution over its collective coordinate.

\begin{figure}[H]
\centering\begingroup
\definecolor{vltTeal}{HTML}{176B64}
\definecolor{vltBlue}{HTML}{315D80}
\definecolor{vltGold}{HTML}{A87520}
\definecolor{vltPlum}{HTML}{725877}
\definecolor{vltRed}{HTML}{762B38}
\definecolor{vltYellow}{HTML}{D5AD3C}
\definecolor{vltNeutral}{HTML}{777D83}
\setlength{\unitlength}{1pt}
\scalebox{0.90}{%
\input{vlt-figures/figure-189.tex}%
}\endgroup
\caption{Three configurations of the same pair, $u_1=u_2=Q/\sqrt{2m}$. With $q_c=\sqrt{\omega/\hbar}\,Q$, common physical shifts are $q_ca_\star$, $a_\star=\sqrt{\hbar/(2m\omega)}$. The gain is $12\,\mathrm{pt}/a_\star$; gray outlines and the shared ruler register reference spacing $a=97a_\star/12$. All link spans remain unchanged and supports are omitted. The right ground-state density is normalized on the collective coordinate, with variance $1/2$. Its shaded area describes a preparation over whole pair configurations. The physical harmonic basin determines the applicability of this whole-line oscillator law.}
\label{fig:physical-and-quantum-configuration}
\end{figure}

The coordinate and momentum actions are
\begin{equation}
 \begin{aligned}
 (\hat Q_s\Psi)(\vct Q)&=Q_s\Psi(\vct Q),\\
 (\hat\Pi_s\Psi)(\vct Q)&=-i\hbar\,\partial_{Q_s}\Psi(\vct Q).
 \end{aligned}
 \label{eq:mode-canonical-representation}
\end{equation}
Multiplication weights the amplitude according to its coordinate value;
the derivative measures variation along that configuration coordinate.
The factor $\hbar$ gives the latter momentum units. The Schwartz space
$\mathcal S(\R^r)$ of smooth rapidly decreasing functions is a common
invariant core. The self-adjoint closures have maximal domains specified
by $Q_s\Psi\in L^2$ and by the square-integrable weak derivative
$\partial_{Q_s}\Psi\in L^2$, respectively
\cite{Shankar1994,Teschl2009}.

The whole-line configuration extends the harmonic reference. Its use
requires preparations concentrated within the collision-free basin and
wall clearance of the physical assembly. High occupations can exceed
that range. An exact hard-wall model instead specifies its boundary
Hamiltonian and momentum domain.

On the common core, the product rule gives
$Q_s(-i\hbar\partial_{Q_t}\Psi)
 -[-i\hbar\partial_{Q_t}(Q_s\Psi)]=i\hbar\delta_{st}\Psi$.
Thus
\begin{equation}
 [\hat Q_s,\hat\Pi_t]=i\hbar\delta_{st}\Id,\qquad
 [\hat Q_s,\hat Q_t]=[\hat\Pi_s,\hat\Pi_t]=0.
 \label{eq:mode-canonical-commutators}
\end{equation}
This canonical commutator compares coordinate multiplication and
translation in quantum configuration space. The mechanical commutator
$[\mathsf S_+,\mathsf W]$ compares finite maps derived from the Hessian.
Their spaces, units and postulates fix the meaning of the same algebraic
operation.

Finite body rotation also has a precise operator description. On
$L^2(\operatorname{SO}(3),\dd\mu_{\rm H})$, orientation acts by
multiplication and spatial angular momentum generates laboratory
rotations. On the smooth rotor core,
\begin{equation}
 [\hat\ell_a,\hat\ell_b]=i\hbar\epsilon_{abc}\hat\ell_c,\qquad
 [\hat O_{ij},\hat\ell_a]
       =i\hbar([\vct e_a]_\times\hat{\mathsf O})_{ij}.
 \label{eq:s6patch-spatial-rotation-algebra}
\end{equation}
The spatial and body-fixed momentum algebras have opposite signs
\cite{StoperaMorales2022}. Classically,
$\mathsf O=\exp([\vct\theta]_\times)\mathsf O^0$ gives
$\vct\Omega=\mathsf J_{\rm l}(\vct\theta)\dot{\vct\theta}$ and
$\vct P_\theta=\mathsf J_{\rm l}(\vct\theta)^{\mathsf T}\vct\ell$:
the cotangent relation preserves angular power. Local librations use
canonical chart momentum, equal to spatial angular momentum to leading
order at the reference. Finite rotations retain the exponential-map
differential. Exact quantum chart momenta additionally require their
measure, ordering and domains. The flat harmonic configuration
representation above is the localized tangent model; a finite physical
rotation uses the rotor algebra rather than a Weyl shift of three
unrestricted angular coordinates.

\subsubsection{Ladder operations, energy levels, and finite moments}

For $\omega_s>0$, define the dimensionless ladder operators
\begin{equation}
 \begin{gathered}
 \hat b_s=\sqrt{\frac{\omega_s}{2\hbar}}\hat Q_s
             +\frac{i}{\sqrt{2\hbar\omega_s}}\hat\Pi_s,\\
 [\hat b_s,\hat b_t^\dagger]=\delta_{st}\Id,\qquad
 \hat N_s=\hat b_s^\dagger\hat b_s.
 \end{gathered}
 \label{eq:canonical-ladder-operators}
\end{equation}
These are combinations of coordinate and momentum, with coefficients
chosen to remove their units and normalize the commutator.
Their actions
$\hat b_s\ket{\vct N}=\sqrt{N_s}\ket{\vct N-\vct e_s}$ and
$\hat b_s^\dagger\ket{\vct N}
 =\sqrt{N_s+1}\ket{\vct N+\vct e_s}$
change one occupation. The vacuum $\ket0$ satisfies $\hat b_s\ket0=0$ and contains no excitation
of that mode; it is not a classical state with perfectly zero coordinate
and momentum. Its coordinate and momentum variances remain
$\hbar/(2\omega_s)$ and $\hbar\omega_s/2$.
The finite linear span of occupation vectors, $\mathcal D_{\rm fin}$,
is a common invariant core for the polynomial identities used here.

The harmonic Hamiltonian and its excitation part are
\begin{equation}
 \begin{split}
 \hat H_0&=\frac12\sum_s(\hat\Pi_s^2+\omega_s^2\hat Q_s^2)
       =\sum_s\hbar\omega_s\left(\hat N_s+\frac12\Id\right),\\
 \hat H_{\rm ph}&=\hat H_0-E_{\rm vac}\Id,\qquad
 E_{\rm vac}=\frac12\sum_s\hbar\omega_s.
 \end{split}
 \label{eq:canonical-phonon-hamiltonian}
\end{equation}
Raising mode $s$ adds $\hbar\omega_s$ at fixed virion count.
The isolated harmonic Hamiltonian conserves total occupation. A force
coupling $-F_s(t)\hat Q_s$ can change it through supplied work;
autonomous active motion additionally retains its force-generation
mechanism and resource.

This is the energy of the selected conservative oscillator coordinates.
Enclosed medium storage, passive reduction, and externally changed
parameters keep the energy accounts already specified in Section~\ref{sec:viscoelastic-dynamics}.
Likewise, $\hbar\mathsf W$ is the excitation Hamiltonian only on the
identified one-phonon sector, whereas the full Fock Hamiltonian includes
all integer occupations.

For $\Psi=\sum_{\vct N}c_{\vct N}\ket{\vct N}$, write
$E_{\vct N}=\sum_s\hbar\omega_s(N_s+\tfrac12)$. Then
\begin{equation}
 \begin{aligned}
 \Psi\in D(\hat H_0)&\ \Longleftrightarrow\ 
       \sum_{\vct N}E_{\vct N}^2|c_{\vct N}|^2<\infty,\\
 \Psi\in D(\hat H_0^{1/2})&\ \Longleftrightarrow\ 
       \sum_{\vct N}E_{\vct N}|c_{\vct N}|^2<\infty.
 \end{aligned}
 \label{eq:operator-hamiltonian-moment-domains}
\end{equation}
The first is the operator domain, needed for the Hamiltonian image to
lie in Hilbert space. The second is its form domain, appropriate to
finite mean energy. A normalized state need not satisfy either moment
condition. Finite mode count avoids the continuum ultraviolet vacuum
sum but does not bound the occupations or automatically make every
operator bounded. For a number state,
$\langle\hat Q_s^2\rangle=\hbar(N_s+1/2)/\omega_s$;
mapping that excursion back through the physical mode shape tests the
harmonic resolution, rather than the mere existence of the Fock vector.

\subsubsection{Unitary evolution and the meaning of conserved probability}

\begin{proposition}[Conservative occupation evolution]
The self-adjoint harmonic Hamiltonian generates the strongly continuous
unitary group $\hat{\mathcal U}(t)=e^{-it\hat H_0/\hbar}$ on
$\mathcal F_+$. Every initial vector has a norm-preserving evolution;
vectors in $D(\hat H_0)$ additionally satisfy
$i\hbar\partial_t\Psi=\hat H_0\Psi$ as a strong equation.
\end{proposition}
\begin{proof}
In the occupation basis, $c_{\vct N}(t)=
e^{-itE_{\vct N}/\hbar}c_{\vct N}(0)$. Each coefficient has a
unit-modulus phase. Dominated convergence gives continuity in Hilbert
norm, and the energy-squared summability above permits differentiation
in that norm. This is the finite-mode realization of the self-adjoint
unitary-group construction~\cite{Teschl2009}.
\end{proof}

The conserved norm is total probability over the retained oscillator
configurations. For a density operator,
$\sum_{\vct N}\bra{\vct N}\varrho(t)\ket{\vct N}=1$.
This does not require the probability of one coordinate range to remain
fixed. For sufficiently regular configuration wavefunctions, the same
fact has a local form:
\begin{equation}
 \partial_t|\Psi|^2+\sum_s\partial_{Q_s}\mathcal J_s=0,
 \qquad \mathcal J_s=\hbar\operatorname{Im}
                              (\Psi^*\partial_{Q_s}\Psi).
 \label{eq:operator-configuration-probability-current}
\end{equation}
Multiplying the Schr\"odinger equation by $\Psi^*$ and subtracting its
conjugate cancels the real potential and yields this divergence. Decay
at configuration infinity removes the total boundary flux, preserving
the integral of $|\Psi|^2$. The current is in the collective $Q$ space;
it is not the bond-energy current or a material flow inside the cylinder.
Probability can move between sets of configurations while its total
remains one. The virion count is fixed by the modeled system, rather than
by this normalization identity. A reduced quantum state can also exchange
energy or lose coherence while remaining trace normalized.

For an observable $\hat A$, Heisenberg evolution is
$\hat A(t)=\hat{\mathcal U}(t)^\dagger\hat A\hat{\mathcal U}(t)$,
with transported domain
$D(\hat A(t))=\hat{\mathcal U}(t)^\dagger D(\hat A)$.
It preserves self-adjointness. On a common differentiable core,
$\Dt{\hat A}(t)=(i/\hbar)[\hat H_0,\hat A(t)]$.
In particular, $\hat b_s(t)=e^{-i\omega_st}\hat b_s$,
$\Dt{\hat Q_s}=\hat\Pi_s$ and
$\Dt{\hat\Pi_s}=-\omega_s^2\hat Q_s$.
These equations connect the occupation construction back to the mechanical
coordinate motion, under its separate quantum assumptions.

\subsubsection{A canonical pair for a calibrated displacement readout}

A collective canonical coordinate need not be one normal mode. For a
real dimensionless detector vector $\vct\ell$, let
$\vct a=\mathsf E^{\mathsf T}\mathsf M^{-1/2}\vct\ell$ and
$\hat Y=\vct\ell^{\mathsf T}\hat{\vct u}_+
 =\vct a^{\mathsf T}\hat{\vct Q}$ on the retained positive sector.
For $\vct a\ne0$, define
\begin{equation}
 \begin{aligned}
 \mathcal M_Y^{-1}&=\vct a^{\mathsf T}\vct a,
 &\hat P_Y&=\mathcal M_Y\vct a^{\mathsf T}\hat{\vct\Pi},\\
 [\hat Y,\hat P_Y]&=i\hbar\Id,
 &\Delta Y\,\Delta P_Y&\geq\hbar/2.
 \end{aligned}
 \label{eq:operator-canonical-readout-pair}
\end{equation}
Here $Y$ has length units, $\mathcal M_Y$ mass units, and $P_Y$ momentum
units. The commutator follows by summing the mode commutators with their
weights, and the variance bound uses states with the necessary finite
moments. The same projected inverse inertia appears in the response
normalization of Section~\ref{sec:operator-probe-response}.

The mass also characterizes classical kinetic cost: minimizing
$\|\vct\Pi\|^2/2$ at $\vct a^{\mathsf T}\vct\Pi=v_Y$ gives
$T_{\min}=\mathcal M_Yv_Y^2/2$, attained parallel to $\vct a$.
Under conservative evolution, $\hat P_Y=\mathcal M_Y\Dt{\hat Y}$.
Other velocity directions can store further energy, and the restoring
law can couple $Y$ to other coordinates. A closed oscillator therefore
also needs an invariant subspace. For $\vct a=0$ the readout is dark on
this sector. Instrument precision is assessed separately through its
calibration and noise map.

\subsubsection{Several readouts and a closed canonical subsystem}

Several tracked displacements usually overlap the same particles and modes.
Their effective inertias must therefore be assembled together. Suppose that
$k\leq r$ independent calibrated displacement readouts have the form
$\hat{\vct Y}=\mathsf A_{\rm o}\hat{\vct Q}$, where
$\mathsf A_{\rm o}\in\R^{k\times r}$ has full row rank. Its rows are
the vectors $\vct a^{\mathsf T}$ constructed above. All components of
$\vct Y$ have length units; the entries of $\mathsf A_{\rm o}$ have
units $\mathrm{kg}^{-1/2}$.

\begin{definition}[Canonical readout family]
\label{def:s6-canonical-readout-family}
The projected inverse-inertia matrix, effective mass matrix, and conjugate
momentum of these independent readouts are
\begin{equation}
 \begin{gathered}
 \mathsf G_{\rm o}=\mathsf A_{\rm o}\mathsf A_{\rm o}^{\mathsf T}>0,
 \qquad \mathsf M_{\rm o}=\mathsf G_{\rm o}^{-1},\qquad
 \hat{\vct P}_{\rm o}=\mathsf M_{\rm o}\mathsf A_{\rm o}\hat{\vct\Pi},\\
 [\hat Y_a,\hat P_{{\rm o},b}]=i\hbar\delta_{ab}\Id,\qquad
 [\hat Y_a,\hat Y_b]=[\hat P_{{\rm o},a},\hat P_{{\rm o},b}]=0.
 \end{gathered}
 \label{eq:s6-canonical-readout-family}
\end{equation}
The quantum identities are read on the common canonical core. Replacing
the commutators divided by $i\hbar$ with Poisson brackets gives the
corresponding classical construction.
\end{definition}

The same mass matrix is obtained from the least kinetic energy consistent
with the readout velocity in
Proposition~\ref{prop:s5-hamiltonian-readout-energy}:
$T\geq\Dt{\vct Y}^{\mathsf T}\mathsf M_{\rm o}\Dt{\vct Y}/2$.
Here $\hat{\vct P}_{\rm o}=\mathsf M_{\rm o}\Dt{\hat{\vct Y}}$
under the conservative dynamics. Off-diagonal entries of
$\mathsf G_{\rm o}$ record the mechanical overlap between readouts.
They can occur for independent detector errors because the same ideal
virions contribute to both displacement coordinates. Two named readouts
therefore need not describe two dynamically independent motions.

\begin{proposition}[Exact conservative closure of canonical readouts]
\label{prop:s6-canonical-readout-closure}
Use the normal-mode representation
$\mathsf W^2=\diag(\omega_1^2,\ldots,\omega_r^2)>0$ and the
fixed conservative harmonic Hamiltonian. The family
$(\hat{\vct Y},\hat{\vct P}_{\rm o})$ has an autonomous linear canonical
evolution for every retained initial state if and only if
\begin{equation}
 \mathsf A_{\rm o}\mathsf W^2=\mathsf D_{\rm o}\mathsf A_{\rm o},
 \qquad
 \mathsf D_{\rm o}=
 \mathsf A_{\rm o}\mathsf W^2\mathsf A_{\rm o}^{\mathsf T}
                                     \mathsf G_{\rm o}^{-1}.
 \label{eq:s6-canonical-readout-invariance}
\end{equation}
Equivalently, $\ker\mathsf A_{\rm o}$, or its orthogonal complement,
is invariant under $\mathsf W^2$. In this case the visible subsystem has
\begin{equation}
 \begin{gathered}
 \hat H_{\rm o}=\tfrac12\hat{\vct P}_{\rm o}^{\mathsf T}
                       \mathsf G_{\rm o}\hat{\vct P}_{\rm o}
       +\tfrac12\hat{\vct Y}^{\mathsf T}
                       \mathsf M_{\rm o}\mathsf D_{\rm o}\hat{\vct Y},\\
 \Dt{\hat{\vct Y}}=\mathsf G_{\rm o}\hat{\vct P}_{\rm o},\qquad
 \Dt{\hat{\vct P}_{\rm o}}=
                       -\mathsf M_{\rm o}\mathsf D_{\rm o}\hat{\vct Y}.
 \end{gathered}
 \label{eq:s6-closed-canonical-readout-hamiltonian}
\end{equation}
The stiffness matrix $\mathsf M_{\rm o}\mathsf D_{\rm o}$ is symmetric
positive definite. $H_{\rm o}$ is the energy of this invariant subsystem;
unobserved coordinates may carry additional energy.
\end{proposition}
\begin{proof}
The canonical identities follow from
$\mathsf A_{\rm o}\mathsf A_{\rm o}^{\mathsf T}
\mathsf M_{\rm o}=\Id_k$. The projected restoring acceleration is
$\ddot{\vct Y}=-\mathsf A_{\rm o}\mathsf W^2\vct Q$.
For it to depend only on $\vct Y$, this expression must agree for
$\vct Q$ and $\vct Q+\vct z$ whenever
$\vct z\in\ker\mathsf A_{\rm o}$. Hence the kernel must be invariant.
Factoring the restoring map through $\mathsf A_{\rm o}$ and multiplying
by its right inverse gives~\eqref{eq:s6-canonical-readout-invariance}.
Conversely that identity gives the two closed equations directly.
Finally
$\mathsf M_{\rm o}\mathsf D_{\rm o}=
 \mathsf G_{\rm o}^{-1}\mathsf A_{\rm o}\mathsf W^2
 \mathsf A_{\rm o}^{\mathsf T}\mathsf G_{\rm o}^{-1}>0$,
which verifies both the energy and its canonical equations. The same
calculation applies to the classical coordinates and to the quantum
Heisenberg equations on their common core.
\end{proof}

The invariance test determines whether specified centre and relative
displacements form a closed canonical state. If they fail it, initially
hidden motion can later drive the readouts. One then retains the coupled
coordinates or derives their memory and prepared forcing as in
Section~\ref{sec:viscoelastic-dynamics}. This supplies a mechanical
criterion for integrating measured displacement variables.

\paragraph{Shared cell lists retain one canonical particle coordinate.}
The assembly cells of Sections~2--3 provide local lists of the same
physical virions. Their compatible displacement copies in
Section~\ref{sec:s5-cell-compatible-patterns} can also organize
operators. An operator acting on a shared particle keeps that particle's
single canonical coordinate, however many cells contain its label.
This is particularly useful when a small cell library is reused to
describe a much larger particle pool.

\begin{proposition}[Canonical commutators of shared cell copies]
\label{prop:s6-cell-copy-canonical}
Before kinematic constraints, retain the laboratory translational
coordinates of all $N_{\rm tot}$ labelled particles and their canonical
momenta in the configuration representation. Put $d=3N_{\rm tot}$,
so $[\hat u_j,\hat p_k]=i\hbar\delta_{jk}\Id$ on the common
Schwartz core. For binary cell restrictions $\mathsf R_c$, define
$\hat{\vct u}_c=\mathsf R_c\hat{\vct u}$ and
$\hat{\vct p}_c=\mathsf R_c\hat{\vct p}$. Then
\begin{equation}
 [\hat u_{c,a},\hat p_{e,b}]
     =i\hbar(\mathsf R_c\mathsf R_e^{\mathsf T})_{ab}\Id.
 \label{eq:s6-cell-copy-commutator}
\end{equation}
If $\mathsf J$ stacks a cell cover of all physical coordinates,
the copied position--momentum commutator matrix has rank $2d$,
independently of the number of cell occurrences:
\begin{equation}
 \mathsf J_{\rm cp}^{\rm can}
  =\begin{pmatrix}0&\mathsf J\mathsf J^{\mathsf T}\\
       -\mathsf J\mathsf J^{\mathsf T}&0\end{pmatrix},\qquad
 \operatorname{rank}\mathsf J_{\rm cp}^{\rm can}=2d.
 \label{eq:s6-cell-copy-canonical-rank}
\end{equation}
\end{proposition}
\begin{proof}
Insert the real restrictions into the canonical commutator and sum over
the physical coordinate index. A complete cell cover makes $\mathsf J$
injective, so $\operatorname{rank}(\mathsf J\mathsf J^{\mathsf T})=d$.
The off-diagonal block form has twice that rank.
\end{proof}

For two tetrahedra sharing a face, eight listed virion occurrences
represent five particles. Their 24 copied position entries have only
15 independent physical coordinates. The momentum of a shared vertex
has commutator $i\hbar\Id$ with that vertex's position in either
cell. Declaring the two cell copies independent would change this
algebra. After constraints, one first chooses independent canonical
coordinates and applies their actual restriction maps. Free directions
keep their coordinate--momentum description; positive modes supply the
oscillator occupation representation.

Figure~\ref{fig:s6-shared-cell-canonical-algebra} displays the shared-face example and its cross-cell commutator entries.

\begin{figure}[H]
\centering\begingroup
\definecolor{vltTeal}{HTML}{176B64}
\definecolor{vltBlue}{HTML}{315D80}
\definecolor{vltGold}{HTML}{A87520}
\definecolor{vltPlum}{HTML}{725877}
\definecolor{vltRed}{HTML}{762B38}
\definecolor{vltYellow}{HTML}{D5AD3C}
\definecolor{vltNeutral}{HTML}{777D83}
\setlength{\unitlength}{1pt}
\scalebox{0.90}{%
\input{vlt-figures/figure-190.tex}%
}\endgroup
\caption{Two regular tetrahedra share face $(1,2,3)$: five physical virions have eight cell occurrences and nine unique reference links. The links specify topology without a strain or curvature scale. For one Cartesian component, the displayed list orders rows and columns of the $8\times8$ copy commutator. Each shaded unit entry joins occurrences of the same particle, including cross-cell copies; the matrix has rank five. Three components give 24 copied positions, 15 independent positions and a position--momentum commutator rank of 30. The cover assembles one canonical coordinate, mass and charge per physical particle.}
\label{fig:s6-shared-cell-canonical-algebra}
\end{figure}

A detector assembled from cell rows pulls back to the physical operator
by the same rule as a force:
\begin{equation}
 \hat Y=\sum_c\vct\ell_c^{\mathsf T}\hat{\vct u}_c
       =\left(\sum_c\mathsf R_c^{\mathsf T}\vct\ell_c\right)^{\mathsf T}
           \hat{\vct u}.
 \label{eq:s6-cell-observable-pullback}
\end{equation}
Ownership weights assemble each shared-site count, charge or detector
contribution once. Independent physical coordinates or orthogonal mode
subspaces can define canonical tensor partitions; their Hamiltonian may
still couple the parts. Overlapping cell lists organize assembly while
preserving those physical subsystem choices.

\subsubsection{Why an occupation cutoff changes the algebra}

A finite number of mode labels does not permit finite canonical position
and momentum matrices. In dimension $d<\infty$, cyclicity of trace gives
$\tr[Q,\Pi]=0$, whereas exact $[Q,\Pi]=i\hbar\Id$ would require
$0=i\hbar d$. Numerical truncation must therefore be distinguished from
the exact finite mechanical matrices $\mathsf D$, $\mathsf W$ and
$\mathsf S_+$.

For one selected oscillator, project onto $\ket0,\ldots,\ket J$ with
$\mathsf P_{\leq J}$ and set
$\hat b_{(J)}=\mathsf P_{\leq J}\hat b\mathsf P_{\leq J}$ restricted
to this $(J+1)$-dimensional space. The parentheses mark a cutoff, not a
mode label. Applying the two ordered ladder products to each basis vector
gives
\begin{equation}
 \boxed{[\hat b_{(J)},\hat b_{(J)}^\dagger]
           =\Id_{J+1}-(J+1)\ket J\bra J.}
 \label{eq:operator-cutoff-commutator}
\end{equation}
Below the top level their difference is one; at the top it is $-J$
because the upward transition has been removed. The trace is zero as
required. The compressed $\hat Q_{(J)}$ and $\hat\Pi_{(J)}$ give the
same defect multiplied by $i\hbar$.
Figure~\ref{fig:occupation-cutoff-boundary-commutator} displays the exact boundary
term for $J=3$.

\begin{figure}[H]
\centering
\begingroup
\definecolor{vlfGreen}{HTML}{176B64}
\definecolor{vlfDarkRed}{HTML}{762B38}
\definecolor{vlfRed}{HTML}{762B38}
\definecolor{vlfBlue}{HTML}{315D80}
\definecolor{vlfCopper}{HTML}{762B38}

\setlength{\unitlength}{1pt}
\scalebox{0.80}{%
\begin{picture}(470,357)%
{\color{black}\put(119,339){\makebox(0,0)[c]{\fontfamily{lmss}\fontsize{10.500}{12.500}\selectfont \bfseries (a) Projected raising ladder}}}%
{\color{black}\put(356,339){\makebox(0,0)[c]{\fontfamily{lmss}\fontsize{10.500}{12.500}\selectfont \bfseries (b) Commutator diagonal}}}%
{\color{black}\put(235,314){\makebox(0,0)[c]{\fontfamily{lmss}\fontsize{10.500}{12.500}\selectfont $J=3,\qquad b_{(3)}=P_{\leq3}bP_{\leq3}$}}}%
{\color{black}\vlB{%
\vSc{54.000000}{92.000000}{0.000000}{84.000000}%
}}%
{\color{black}\put(153,92){\makebox(0,0)[c]{\fontfamily{lmss}\fontsize{10.500}{12.500}\selectfont $\ket{0}$}}}%
{\color{black}\vlB{%
\vSar{95.328}{96}{90}{41}%
\vSar{95.376064}{137}{-68.198591}{5.385165}%
\vSar{93.376064}{137}{-111.801409}{5.385165}%
}}%
{\color{black}\put(82,117){\makebox(0,0)[r]{\fontfamily{lmss}\fontsize{10.500}{12.500}\selectfont $1$}}}%
{\color{black}\vlB{%
\vSc{54.000000}{142.000000}{0.000000}{84.000000}%
}}%
{\color{black}\put(153,142){\makebox(0,0)[c]{\fontfamily{lmss}\fontsize{10.500}{12.500}\selectfont $\ket{1}$}}}%
{\color{black}\vlB{%
\vSar{95.328}{146}{90}{41}%
\vSar{95.376064}{187}{-68.198591}{5.385165}%
\vSar{93.376064}{187}{-111.801409}{5.385165}%
}}%
{\color{black}\put(82,167){\makebox(0,0)[r]{\fontfamily{lmss}\fontsize{10.500}{12.500}\selectfont $\sqrt2$}}}%
{\color{black}\vlB{%
\vSc{54.000000}{192.000000}{0.000000}{84.000000}%
}}%
{\color{black}\put(153,192){\makebox(0,0)[c]{\fontfamily{lmss}\fontsize{10.500}{12.500}\selectfont $\ket{2}$}}}%
{\color{black}\vlB{%
\vSar{95.328}{196}{90}{41}%
\vSar{95.376064}{237}{-68.198591}{5.385165}%
\vSar{93.376064}{237}{-111.801409}{5.385165}%
}}%
{\color{black}\put(82,217){\makebox(0,0)[r]{\fontfamily{lmss}\fontsize{10.500}{12.500}\selectfont $\sqrt3$}}}%
{\color{black}\vlB{%
\vSc{54.000000}{242.000000}{0.000000}{84.000000}%
}}%
{\color{black}\put(153,242){\makebox(0,0)[c]{\fontfamily{lmss}\fontsize{10.500}{12.500}\selectfont $\ket{3}$}}}%
{\color{vlfDarkRed}\vlB{%
\vSs{95.425000}{247.000000}{90.000000}{28.000000}%
\vSs{95.466126}{275.000000}{-68.198591}{4.308132}%
\vSs{93.866126}{275.000000}{-111.801409}{4.308132}%
}}%
{\color{black}\vlB{%
\vSas{87.664}{264}{36.869898}{20}%
\vSas{87.664}{276}{-36.869898}{20}%
}}%
{\color{black}\put(119,292){\makebox(0,0)[c]{\fontfamily{lmss}\fontsize{10.500}{12.500}\selectfont $b_{(3)}^\dagger\ket3=0$}}}%
{\color{black}\put(119,68){\makebox(0,0)[c]{\fontfamily{lmss}\fontsize{10.500}{12.500}\selectfont Only the top transition is blocked.}}}%
{\color{black}\vlB{%
\vSc{306.500000}{92.000000}{-90.000000}{3.000000}%
}}%
{\color{black}\put(307,79){\makebox(0,0)[c]{\fontfamily{lmss}\fontsize{10.500}{12.500}\selectfont $0$}}}%
{\color{black}\vlB{%
\vSc{343.166667}{92.000000}{-90.000000}{3.000000}%
}}%
{\color{black}\put(343.666667,79){\makebox(0,0)[c]{\fontfamily{lmss}\fontsize{10.500}{12.500}\selectfont $1$}}}%
{\color{black}\vlB{%
\vSc{379.833333}{92.000000}{-90.000000}{3.000000}%
}}%
{\color{black}\put(380.333333,79){\makebox(0,0)[c]{\fontfamily{lmss}\fontsize{10.500}{12.500}\selectfont $2$}}}%
{\color{black}\vlB{%
\vSc{416.500000}{92.000000}{-90.000000}{3.000000}%
}}%
{\color{black}\put(417,79){\makebox(0,0)[c]{\fontfamily{lmss}\fontsize{10.500}{12.500}\selectfont $3$}}}%
{\color{black}\vlB{%
\vSc{282.000000}{106.500000}{0.000000}{3.000000}%
}}%
{\color{black}\put(278,106.5){\makebox(0,0)[r]{\fontfamily{lmss}\fontsize{10.500}{12.500}\selectfont $-3$}}}%
{\color{black!20}\vlB{%
\vSan{285}{106.5}{0}{154}%
}}%
{\color{black}\vlB{%
\vSc{282.000000}{142.750000}{0.000000}{3.000000}%
}}%
{\color{black}\put(278,142.75){\makebox(0,0)[r]{\fontfamily{lmss}\fontsize{10.500}{12.500}\selectfont $-2$}}}%
{\color{black!20}\vlB{%
\vSan{285}{142.75}{0}{154}%
}}%
{\color{black}\vlB{%
\vSc{282.000000}{179.000000}{0.000000}{3.000000}%
}}%
{\color{black}\put(278,179){\makebox(0,0)[r]{\fontfamily{lmss}\fontsize{10.500}{12.500}\selectfont $-1$}}}%
{\color{black!20}\vlB{%
\vSan{285}{179}{0}{154}%
}}%
{\color{black}\vlB{%
\vSc{282.000000}{215.250000}{0.000000}{3.000000}%
}}%
{\color{black}\put(278,215.25){\makebox(0,0)[r]{\fontfamily{lmss}\fontsize{10.500}{12.500}\selectfont $0$}}}%
{\color{black!20}\vlB{%
\vSan{285}{215.25}{0}{154}%
}}%
{\color{black}\vlB{%
\vSc{282.000000}{251.500000}{0.000000}{3.000000}%
}}%
{\color{black}\put(278,251.5){\makebox(0,0)[r]{\fontfamily{lmss}\fontsize{10.500}{12.500}\selectfont $1$}}}%
{\color{black!20}\vlB{%
\vSan{285}{251.5}{0}{154}%
}}%
{\color{black}\vlB{%
\vSc{285.000000}{92.000000}{0.000000}{154.000000}%
\vSc{284.500000}{92.000000}{90.000000}{174.000000}%
}}%
{\color{black}\put(362,61){\makebox(0,0)[c]{\fontfamily{lmss}\fontsize{10.500}{12.500}\selectfont $N$}}}%
{\color{black}\put(285,282){\makebox(0,0)[l]{\fontfamily{lmss}\fontsize{10.500}{12.500}\selectfont $\bra N[b_{(3)},b_{(3)}^\dagger]\ket N$}}}%
{\color{black}\vlB{%
\vSc{285.000000}{215.250000}{0.000000}{154.000000}%
\vSav{297.5}{215.25}{90}{36.25}%
}}%
{\color{black}\put(307,262.5){\makebox(0,0)[c]{\fontfamily{lmss}\fontsize{10.500}{12.500}\selectfont $1$}}}%
{\color{black}\vlB{%
\vSav{334.166667}{215.25}{90}{36.25}%
}}%
{\color{black}\put(343.666667,262.5){\makebox(0,0)[c]{\fontfamily{lmss}\fontsize{10.500}{12.500}\selectfont $1$}}}%
{\color{black}\vlB{%
\vSav{370.833333}{215.25}{90}{36.25}%
}}%
{\color{black}\put(380.333333,262.5){\makebox(0,0)[c]{\fontfamily{lmss}\fontsize{10.500}{12.500}\selectfont $1$}}}%
{\color{vlfDarkRed}\vlB{%
\vSav{407.5}{215.25}{-90}{108.75}%
}}%
{\color{black}\put(433,109.5){\makebox(0,0)[l]{\fontfamily{lmss}\fontsize{10.500}{12.500}\selectfont $-3$}}}%
{\color{black}\put(235,38){\makebox(0,0)[c]{\fontfamily{lmss}\fontsize{10.500}{12.500}\selectfont $[b_{(3)},b_{(3)}^\dagger]=I-4\ket3\bra3,\qquad \operatorname{tr}[b_{(3)},b_{(3)}^\dagger]=0$}}}%
{\color{black}\put(235,15){\makebox(0,0)[c]{\fontfamily{lmss}\fontsize{10.500}{12.500}\selectfont The upper-boundary defect comes from truncation, not new mechanics.}}}%
\end{picture}%
%
}%
\endgroup
\caption{One oscillator cut off at $J=3$. Raising coefficients $1,\sqrt2,\sqrt3$ survive; the upward transition from the top level is removed. Consequently $[b_{(3)},b_{(3)}^\dagger]$ has diagonal $(1,1,1,-3)$ and trace zero. The same diagonal gives $[Q_{(3)},\Pi_{(3)}]/(i\hbar)$. Its expectation defect is $4\Pr(N=3)$. This is a numerical boundary diagnostic; errors of particular observables and driven evolutions require their own convergence bounds. The vertical occupation coordinate labels excitations of one retained mode. The boxed result illustrated here is~\eqref{eq:operator-cutoff-commutator}.}
\label{fig:occupation-cutoff-boundary-commutator}
\end{figure}

Projection also fails to commute with forming products. If
$\hat H_{0,(J)}=\mathsf P_{\leq J}\hat H_0\mathsf P_{\leq J}$ on this
space, then
\begin{equation}
 \tfrac12(\hat\Pi_{(J)}^2+\omega^2\hat Q_{(J)}^2)
       =\hat H_{0,(J)}
             -\tfrac12\hbar\omega(J+1)\ket J\bra J.
 \label{eq:operator-cutoff-hamiltonian-defect}
\end{equation}
Squaring already truncated coordinate matrices therefore does not give
the projected harmonic Hamiltonian at its upper boundary. For a normalized
cutoff state, the commutator expectation differs from $i\hbar$ by
$-i\hbar(J+1)\Pr(N=J)$. Small boundary occupation is a useful algebraic
diagnostic, but it is not by itself a full error bound for every observable
or driven trajectory. Raising the cutoff and checking the chosen
predictions remain numerical convergence requirements.

\subsubsection{What a cutoff can certify about a prediction}

The boundary commutator tests an algebraic identity. A prediction needs an
error bound for its own observable. On the finite set of retained positive
modes define the total-occupation projection
$\hat{\mathcal P}_J=\mathbf1_{[0,J]}(\hat N_{\rm ph})$, where
$\hat N_{\rm ph}=\sum_s\hat N_s$ and $J$ is a nonnegative integer.
This is a finite-dimensional numerical space, of dimension
$\binom{J+r}{r}$, inside the infinite occupation space.

\begin{proposition}[Cutoff control of bounded observations]
\label{prop:s6-cutoff-bounded-prediction}
Let $\Psi$ be a normalized pure preparation with finite mean total
occupation $\bar N=\langle\hat N_{\rm ph}\rangle$. Set
$\varepsilon_J=\|(\Id-\hat{\mathcal P}_J)\Psi\|^2$ and, when
$\varepsilon_J<1$, let
$\Psi_J=\hat{\mathcal P}_J\Psi/\sqrt{1-\varepsilon_J}$.
For $\omega_{\rm gap}=\min_s\omega_s>0$,
\begin{equation}
 \begin{gathered}
 \varepsilon_J\leq\frac{\bar N}{J+1}
       \leq\frac{\langle\hat H_{\rm ph}\rangle}
                        {\hbar\omega_{\rm gap}(J+1)},\\
 \big|\langle\hat F\rangle_\Psi-
                    \langle\hat F\rangle_{\Psi_J}\big|
                  \leq\sqrt{\varepsilon_J}
            \quad(0\leq\hat F\leq\Id),\\
 \big|\langle\hat A\rangle_\Psi-
                    \langle\hat A\rangle_{\Psi_J}\big|
                  \leq2\|\hat A\|\sqrt{\varepsilon_J}
            \quad(\hat A\hbox{ bounded}).
 \end{gathered}
 \label{eq:s6-cutoff-bounded-observation-error}
\end{equation}
Here an effect $\hat F$ is the operator for one measurement outcome or
one set of outcomes, and its expectation is their probability. The
bound concerns the same full-space measurement and its compression to the
cutoff space.
\end{proposition}
\begin{proof}
Every omitted occupation vector has total occupation at least $J+1$;
also $\hat H_{\rm ph}\geq\hbar\omega_{\rm gap}\hat N_{\rm ph}$.
These two spectral inequalities give the first line. If
$0<\varepsilon_J<1$, decompose $\Psi$ into the orthogonal retained and
omitted normalized vectors. In their two-dimensional span the difference
$\ket\Psi\bra\Psi-\ket{\Psi_J}\bra{\Psi_J}$ has eigenvalues
$\pm\sqrt{\varepsilon_J}$ and vanishes on the orthogonal complement.
Its trace norm is therefore $2\sqrt{\varepsilon_J}$. A bounded operator
gives the last bound by the trace inequality. An effect weights each of
the positive and negative one-dimensional parts between zero and one,
giving the sharper probability bound. The case $\varepsilon_J=0$ is
immediate.
\end{proof}

The same bounds apply to a density operator $\varrho$ with finite mean
occupation. In that case
$\varepsilon_J=\tr[\varrho(\Id-\hat{\mathcal P}_J)]$ and the
normalized retained state is
$\varrho_J=\hat{\mathcal P}_J\varrho\hat{\mathcal P}_J/
(1-\varepsilon_J)$. To see the probability bound, purify $\varrho$
and project its oscillator factor with $\hat{\mathcal P}_J$.
The pure-state argument above applies to every effect
$\hat F\otimes\Id$ and hence to the original system after the auxiliary
factor is traced out. The bounded-operator bound and the spectral tail
inequalities follow in the same way. Thus the criterion also covers
thermal mixtures and other statistically prepared quantum states.

A displacement-interval projection is a bounded effect even though the
coordinate operator is unbounded. Its threshold probability can therefore
have a certified cutoff error. The coordinate-to-particle map fixes the
physical length and observed virion group; contact geometry determines
whether that threshold lies within the harmonic basin.

Mean energy requires a weighted tail. For one mode and integers
$L>\max(J,2)$, the normalized states
$\Psi^{(L)}=\sqrt{1-L^{-1}}\ket0+L^{-1/2}\ket L$ have
$\varepsilon_J=L^{-1}\to0$, yet
$\langle\hat H_{\rm ph}\rangle_{\Psi^{(L)}}=\hbar\omega$ and their
normalized cutoff state is the vacuum. Each state has finite energy;
the disappearing probability at an increasingly energetic level carries
a fixed omitted mean energy. Small tail probability alone thus gives no
uniform energy-error guarantee, even in a family with bounded mean energy.
For an individual state of finite mean energy the energy-weighted tail
does tend to zero. Quantitatively, if a moment of order $1+\delta$ is
known for some $\delta>0$, then
\begin{equation}
 \begin{aligned}
 E_{{\rm tail},J}
   &:=\langle\hat H_{\rm ph}(\Id-\hat{\mathcal P}_J)\rangle
     \leq\frac{\langle\hat H_{\rm ph}^{\,1+\delta}\rangle}
                   {[\hbar\omega_{\rm gap}(J+1)]^\delta},\\
 \langle\hat H_{\rm ph}\rangle-
                  \langle\hat H_{\rm ph}\rangle_{\Psi_J}
   &=\frac{E_{{\rm tail},J}-
                   \varepsilon_J\langle\hat H_{\rm ph}\rangle}
                  {1-\varepsilon_J}.
 \end{aligned}
 \label{eq:s6-cutoff-energy-tail}
\end{equation}
These energy identities also hold for $\varrho$ and $\varrho_J$.
The inequality follows by comparing the powers of energy on the omitted
spectral set; the equality also accounts for normalization. Higher
unbounded moments need their corresponding weighted tails. Under the
fixed free harmonic evolution, $\hat{\mathcal P}_J$ commutes with
$\hat H_0$, so the probability bound holds at every time with the same
$\varepsilon_J$. A driven displacement or squeeze can move probability
beyond the retained set, and needs a bound over the preparation interval.
Convergence inside the harmonic model and validity of that model are
distinct checks.

Mechanical matrices act on patterns, whereas canonical operators act on
oscillator configurations or occupations. Their actions, normalizations
and domains determine the physical meaning of a commutator. The following
subsection uses these distinctions to describe collective transformations.

\FloatBarrier

\subsection{The Heisenberg--Weyl algebra and collective transformations}
\label{sec:lattice-lie-algebra}

Controlled preparations act on the canonical coordinates of
Section~\ref{sec:canonical-phonon-operators}. A force shifts a mode's
mean; curvature control changes its quadrature fluctuations; coupling
can transfer excitation between modes. The algebra organizes these
operations and their composition, while a physical implementation
specifies the force and energy source.

\subsubsection{Canonical translations and their mechanical preparation}

The real span of $\hat Q_s,\hat\Pi_s,\Id$, with bracket
$[A,B]/(i\hbar)$, is the Heisenberg Lie algebra. Its identity element is
central: it commutes with every generator. Closure means that taking the
bracket of two generators stays within the same span. There are $2r+1$
independent generators for $r$ retained modes, although their oscillator
representation is infinite dimensional. Classical canonical coordinates
realize the corresponding Poisson algebra. For the linear and quadratic
expressions here, their canonical bracket structure agrees with the
quantum commutator structure; the quantum state and measurement postulate
remain additional physical assumptions.

The bounded unitary Weyl operators implement finite translations
\cite{NeumaierWestra2011,Shankar1994}:
\begin{equation}
 \begin{split}
 \hat{\mathcal D}(\vct\xi)
   &=\exp\!\left[\sum_s(\xi_s\hat b_s^\dagger-\xi_s^*\hat b_s)\right],\\
 \hat{\mathcal D}(\vct\xi)\hat{\mathcal D}(\vct\xi')
   &=\exp\!\left[\frac{\vct\xi\cdot\vct\xi'^*
                           -\vct\xi^*\cdot\vct\xi'}2\right]
                          \hat{\mathcal D}(\vct\xi+\vct\xi').
 \end{split}
 \label{eq:weyl-displacements}
\end{equation}
Here $\xi_s$ is dimensionless and indexes a preparation, rather than a
new spatial coordinate. For real desired shifts $q_{0s},p_{0s}$, set
$\xi_s=\sqrt{\omega_s/(2\hbar)}q_{0s}
       +ip_{0s}/\sqrt{2\hbar\omega_s}$. The same exponent is
$i\sum_s(p_{0s}\hat Q_s-q_{0s}\hat\Pi_s)/\hbar$.
Thus $q_{0s}$ has the mass-weighted displacement units of $Q_s$, and
$p_{0s}$ has the units of $\Pi_s$. Conjugation gives
\begin{equation}
 \delta\langle\hat Q_s\rangle
      =\sqrt{\frac{2\hbar}{\omega_s}}\operatorname{Re}\xi_s=q_{0s},
 \qquad
 \delta\langle\hat\Pi_s\rangle
      =\sqrt{2\hbar\omega_s}\operatorname{Im}\xi_s=p_{0s}.
 \label{eq:weyl-mechanical-mean-shifts}
\end{equation}
The state changes as $\varrho\mapsto\hat{\mathcal D}\varrho
\hat{\mathcal D}^\dagger$. Its centered covariance is unchanged: a
translation moves the mean without narrowing or broadening the state.
Bounded Weyl operators are defined on the whole Hilbert space; the mean
and energy statements additionally require the corresponding finite moments.

Two translations add their shifts but also acquire the phase in
\eqref{eq:weyl-displacements}. In physical shifts this phase is
$\exp[i\sum_s(p_{0s}q'_{0s}-q_{0s}p'_{0s})/(2\hbar)]$.
It records the oriented canonical area between the two operations.
A closed translation loop can return the coordinate and momentum means
while leaving a phase. A lone global phase changes no displacement
probability; detecting a relative loop phase requires an interference
reference. This canonical geometry is distinct from the virion's shape
or the cylinder's rotational symmetry.

For the fixed harmonic Hamiltonian, expansion after the shift gives
\begin{equation}
 \Delta E_{\mathcal D}
  =\sum_s\left[p_{0s}\langle\hat\Pi_s\rangle
      +\omega_s^2q_{0s}\langle\hat Q_s\rangle
      +\frac{p_{0s}^2+\omega_s^2q_{0s}^2}{2}\right].
 \label{eq:operator-displacement-preparation-work}
\end{equation}
For a centered state the work is $\sum_s\hbar\omega_s|\xi_s|^2$.
For an already moving state, the cross terms can permit extraction.
This is the mechanical energy change under the translation; the full
preparation account also includes actuator losses.

A specified force makes this map calculable. Write the drive on a mode as
$-F_s(t)\hat Q_s$, with $F_s$ conjugate to the mass-weighted coordinate.
In translational normalization,
$F_s=\vct e_s^{\mathsf T}\mathsf M^{-1/2}\vct f_{\rm ext}$
has units $\mathrm{kg}^{1/2}\mathrm m\,\mathrm s^{-2}$, where
$\vct f_{\rm ext}$ is the applied force pattern in newtons.
For a pulse supported on $0\leq t\leq t_p$, the displacement of its
annihilation mean relative to free evolution is exactly
\begin{equation}
 \xi_s(t_p)=\frac{i}{\sqrt{2\hbar\omega_s}}
       \int_0^{t_p}F_s(t')e^{-i\omega_s(t_p-t')}\,\dd t'.
 \label{eq:operator-force-pulse-displacement}
\end{equation}
It follows by integrating
$\Dt{\langle\hat b_s\rangle}=-i\omega_s\langle\hat b_s\rangle
 +iF_s/\sqrt{2\hbar\omega_s}$.
The frequency overlap and spatial force projection select the shifted
modes. Harmonic forcing changes means while the centered covariance
follows its unforced law. Surface contacts, distributed electric forces
and optical actuators each need a calibrated projection onto $F_s$.
Autonomous growth instead uses the converter and resource state of
Section~\ref{sec:self-excitation-passive-boundary}.

\subsubsection{Mixing modes, preserving number, and accounting for work}

The bilinears $\hat E_{st}=\hat b_s^\dagger\hat b_t$ obey
\begin{equation}
 [\hat E_{st},\hat E_{uv}]
       =\delta_{tu}\hat E_{sv}-\delta_{vs}\hat E_{ut}.
 \label{eq:bilinear-mode-algebra}
\end{equation}
One lowering and one raising operation move an excitation between mode
labels without changing total occupation. More generally, an $r\times r$
matrix $B$ has the number-conserving lift
\begin{equation}
 \dd\Gamma(B)=\sum_{s,t}B_{st}\hat b_s^\dagger\hat b_t,
 \qquad
 [\dd\Gamma(B),\dd\Gamma(C)]=\dd\Gamma([B,C])
                 \quad\hbox{on }\mathcal D_{\rm fin}.
 \label{eq:second-quantized-mode-algebra}
\end{equation}
The notation $\dd\Gamma$ is a lift, not a time derivative. It turns a
one-excitation-space operator into the sum of its actions on the
excitation factors in each fixed-total-occupation sector; it does not
add new mechanical coordinates. Diagonalizing
a finite Hermitian $B$ gives its self-adjoint occupation-sum realization.
The skew-Hermitian matrices organize $\mathfrak u(r)$, the algebra of
unitary mode mixing. In the normal-mode basis,
$\hat H_{\rm ph}=\dd\Gamma(\hbar\mathsf W)$.
The lift $\dd\Gamma(\mathsf S_+)$ instead retains stiffness-per-inertia
units and is an extensive local-response statistic; it is not an elastic
energy solely because it has an occupation representation.

For dimensionless Hermitian $B$ and real $\theta$, let
$\hat T_B(\theta)=\exp[-i\theta\dd\Gamma(B)]$ and
$\mathsf T_B=\exp(-i\theta B)$. Then
\begin{equation}
 \begin{gathered}
 \hat T_B^\dagger\hat{\vct b}\hat T_B=\mathsf T_B\hat{\vct b},
 \qquad\hat T_B^\dagger\hat N_{\rm ph}\hat T_B=\hat N_{\rm ph},\\
 \Delta E_B=\hbar\left\langle
     \hat{\vct b}^\dagger
       (\mathsf T_B^\dagger\mathsf W\mathsf T_B-\mathsf W)
                                      \hat{\vct b}\right\rangle .
 \end{gathered}
 \label{eq:operator-mode-mixing-work}
\end{equation}
Transfer between unequal frequencies can change energy at fixed total
occupation. The whole mixing family preserves the Hamiltonian when
$[B,\mathsf W]=0$, as for mixing within a degenerate frequency space.
A passive basis change transports state and operators together and
preserves every physical prediction. Actual mode conversion has the
work account above.

\subsubsection{Squeezing and changing the harmonic reference}

Quadratic generators containing $\hat b_s\hat b_t$ and their adjoints
also change occupation. A finite-mode Bogoliubov map
$\hat{\vct b}'=\mathsf U_{\rm B}\hat{\vct b}
                 +\mathsf V_{\rm B}\hat{\vct b}^{\dagger}$
is canonical precisely when
\begin{equation}
 \begin{aligned}
 \mathsf U_{\rm B}\mathsf U_{\rm B}^\dagger
       -\mathsf V_{\rm B}\mathsf V_{\rm B}^\dagger&=\Id_r,\\
 \mathsf U_{\rm B}\mathsf V_{\rm B}^{\mathsf T}
       &=\mathsf V_{\rm B}\mathsf U_{\rm B}^{\mathsf T}.
 \end{aligned}
 \label{eq:bogoliubov-conditions}
\end{equation}
Both lowering and raising arrays are columns. Substituting this map into
$[\hat b'_s,\hat b_t'{}^\dagger]$ and $[\hat b'_s,\hat b'_t]$
gives the two conditions. They preserve the canonical algebra, rather
than asserting a symmetry of the energy. For finitely many modes, these homogeneous linear canonical maps have
unitary oscillator implementations, determined up to phase. Polynomial identities are first read on a suitable common
core; finite-moment expectations require their stated domains.

For one mode define dimensionless quadratures
$q_{\rm c}=\sqrt{\omega/\hbar}\,Q$ and
$p_{\rm c}=\Pi/\sqrt{\hbar\omega}$, so
$[\hat q_{\rm c},\hat p_{\rm c}]=i\Id$.
A displacement moves the covariance ellipse's centre. An ideal squeeze
$\mathsf S_{\rm sq}=\diag(e^{-r_{\rm sq}},e^{r_{\rm sq}})$
instead sends a vacuum covariance $\Id_2/2$ to
$\diag(e^{-2r_{\rm sq}},e^{2r_{\rm sq}})/2$.
Its determinant remains $1/4$, while its excitation energy becomes
$\hbar\omega\sinh^2 r_{\rm sq}$. Reduced coordinate variance is
accompanied by increased conjugate variance. Aligned squeezed Gaussians,
as well as coherent states, can attain the canonical variance product.
Figure~\ref{fig:canonical-displacement-and-squeezing} shows these
distinct operations. The contours describe Gaussian covariance, not
particle trajectories or a universal positive probability density for
all quantum states.

\begin{figure}[H]
\centering
\begingroup
\definecolor{vlfGreen}{HTML}{176B64}
\definecolor{vlfDarkRed}{HTML}{762B38}
\definecolor{vlfRed}{HTML}{762B38}
\definecolor{vlfBlue}{HTML}{315D80}
\definecolor{vlfCopper}{HTML}{762B38}

\setlength{\unitlength}{1pt}
\scalebox{0.80}{%
\input{vlt-figures/figure-192.tex}%
}%
\endgroup
\caption{Gaussian preparations on $q_c=\sqrt{\omega/\hbar}\,Q$, $p_c=\Pi/\sqrt{\hbar\omega}$, with $[q_c,p_c]=i$. Each curve is a unit-Mahalanobis covariance contour. A displacement $\xi=(0.85+0.65i)/\sqrt2$ moves the vacuum mean to $(0.85,0.65)$ at covariance $I/2$. A centered squeeze $r_{\rm sq}=0.7$ gives $V=\operatorname{diag}(e^{-2r_{\rm sq}},e^{2r_{\rm sq}})/2$. Both contours have area $\pi/2$; their excitation energies are $\hbar\omega|\xi|^2$ and $\hbar\omega\sinh^2r_{\rm sq}$. Dashed curves register the vacuum. Contours encode state covariance; physical preparation supplies the drive and work.}
\label{fig:canonical-displacement-and-squeezing}
\end{figure}

\begin{example}[Stiffness control and the meaning of phonon creation]
Keep the physical pair $(\hat Q,\hat\Pi)$ fixed and set
$\hat H_0(t)=[\hat\Pi^2+\omega(t)^2\hat Q^2]/2$, with
$\omega(t)>0$. The instantaneous oscillator definition gives
\begin{equation}
 \frac{\dd\hat b(t)}{\dd t}
   =-i\omega(t)\hat b(t)
       +\frac{\Dt{\omega}(t)}{2\omega(t)}\hat b^\dagger(t),
 \qquad
 \frac{\dd\langle\hat H_0\rangle}{\dd t}
       =\frac12\frac{\dd\omega^2}{\dd t}\langle\hat Q^2\rangle.
 \label{eq:phonon-stiffness-control}
\end{equation}
The derivative contains Heisenberg evolution and the explicit change
of the annihilation operator's definition. The ratio
$|\Dt{\omega}|/\omega^2$ measures the local size of the mixing term
relative to the oscillation term. A smooth, controlled slow ramp can be
adiabatic; a small instantaneous ratio alone does not guarantee an
arbitrarily long protocol has negligible accumulated conversion.

For an ideal sudden change $\omega_i\to\omega_f$, $Q$ and $\Pi$
are continuous across the switch, and
\begin{equation}
 \hat b_f=u_q\hat b_i+v_q\hat b_i^\dagger,
 \quad
 u_q=\frac12\left(\sqrt{\frac{\omega_f}{\omega_i}}
                         +\sqrt{\frac{\omega_i}{\omega_f}}\right),
 \quad
 v_q=\frac12\left(\sqrt{\frac{\omega_f}{\omega_i}}
                         -\sqrt{\frac{\omega_i}{\omega_f}}\right).
 \label{eq:phonon-sudden-quench-map}
\end{equation}
Thus $u_q^2-v_q^2=1$. An initial vacuum has
$\langle\hat N_f\rangle=v_q^2$, while the controller's mean work is
\begin{equation}
 \langle W\rangle
       =\frac{\hbar(\omega_f^2-\omega_i^2)}{4\omega_i}.
 \label{eq:phonon-sudden-quench-work}
\end{equation}
More generally an initial thermal occupation $\bar N_i$ gives
\begin{equation}
 \langle\hat N_f\rangle=(u_q^2+v_q^2)\bar N_i+v_q^2,
 \qquad
 \langle W\rangle
       =\frac{\hbar(2\bar N_i+1)(\omega_f^2-\omega_i^2)}{4\omega_i}.
 \label{eq:operator-thermal-quench-occupation-work}
\end{equation}
The work is the change in the full quadratic energy at the switch,
not simply $\hbar\omega_f$ times an excitation count. Softening can
extract work while leaving positive occupation relative to the final
vacuum, because the vacuum reference also changes. A changing contact,
confinement or internal restoring environment could motivate such a
protocol, but its curvature law and driving timescale must be supplied
independently. A coordinate relabeling alone performs no quench.
The zero-frequency limit needs free-coordinate mechanics, and both the
initial and final excursions must remain in the specified harmonic basin.

Figure~\ref{fig:quench-occupation-work-reference} displays this distinction between created occupation and the controller's complete energy account.

\begin{figure}[H]
\centering
\begingroup
\definecolor{vlfGreen}{HTML}{176B64}
\definecolor{vlfDarkRed}{HTML}{762B38}
\definecolor{vlfRed}{HTML}{762B38}
\definecolor{vlfBlue}{HTML}{315D80}
\definecolor{vlfCopper}{HTML}{762B38}

\setlength{\unitlength}{1pt}
\scalebox{0.80}{%
\input{vlt-figures/figure-193.tex}%
}%
\endgroup
\caption{Changing stiffness changes both the oscillator and its vacuum reference. Take the sudden protocol of \eqref{eq:phonon-sudden-quench-map}, an initial vacuum, fixed mass-normalized $(Q,\Pi)$, and $r_q=\omega_f/\omega_i>0$. (a) The final occupation is $\langle N_f\rangle=(r_q+r_q^{-1}-2)/4\geq0$, vanishing only without a quench. (b) In units $\hbar\omega_i$, the controller work is $(r_q^2-1)/4$, the final excitation energy is $r_q\langle N_f\rangle=(r_q-1)^2/4$, and the vacuum-energy shift is $(r_q-1)/2$. The latter two sum exactly to the work. Softening ($r_q<1$) extracts work while still leaving positive occupation relative to the new vacuum. The switch is idealized as faster than both oscillation periods, with $Q$ and $\Pi$ continuous, and both preparations must remain within the harmonic basin. These dimensionless examples assign no measured viral stiffness or timescale.}
\label{fig:quench-occupation-work-reference}
\end{figure}

\end{example}

\paragraph{A coupled reference change in canonical oscillator variables.}
Recruitment can change several restoring directions at once. The scalar
quench above extends to noncommuting restoring matrices without matching
old and new eigenvectors individually. Work on one common finite positive
sector in fixed real mass-weighted laboratory axes. Let
$\mathsf D_g,\mathsf D_h>0$ be its two restoring matrices and
$\mathsf W_j=\mathsf D_j^{1/2}$. Its reference centres satisfy
$\hat{\vct x}_h=\hat{\vct x}_g+\vct b_{hg}$ at unchanged laboratory
momentum $\hat{\vct p}_x$. Define canonical ladder arrays by
\begin{equation}
 \hat{\vct b}_j
   =\frac{\mathsf W_j^{1/2}\hat{\vct x}_j
                 +i\mathsf W_j^{-1/2}\hat{\vct p}_x}{\sqrt{2\hbar}},
 \qquad j\in\{g,h\}.
 \label{eq:s6-reference-canonical-array}
\end{equation}
Each array is dimensionless and has canonical bosonic commutators.
An orthogonal diagonalization of its $\mathsf W_j$ gives the usual
normal-mode ladders. The half powers here enforce canonical normalization;
the classical energy amplitude of Section~5 instead uses
$(\mathsf W_j\vct x_j+i\vct p_x)/\sqrt2$.

\begin{proposition}[Canonical registration of two positive references]
\label{prop:s6-reference-matrix-bogoliubov}
With the data above put
\begin{equation}
 \begin{aligned}
 \mathsf X_{hg}&=\mathsf W_h^{1/2}\mathsf W_g^{-1/2},&
 \mathsf Y_{hg}&=\mathsf W_h^{-1/2}\mathsf W_g^{1/2},\\
 \mathsf U_{hg}&=\tfrac12(\mathsf X_{hg}+\mathsf Y_{hg}),&
 \mathsf V_{hg}&=\tfrac12(\mathsf X_{hg}-\mathsf Y_{hg}),\\
 \vct d_{hg}&=\mathsf W_h^{1/2}\vct b_{hg}/\sqrt{2\hbar}.
 \end{aligned}
 \label{eq:s6-reference-matrix-bogoliubov-data}
\end{equation}
The same physical coordinate--momentum pair obeys
\begin{equation}
 \hat{\vct b}_h=\mathsf U_{hg}\hat{\vct b}_g
             +\mathsf V_{hg}\hat{\vct b}_g^\dagger+\vct d_{hg},
 \label{eq:s6-reference-matrix-bogoliubov}
\end{equation}
where the raising array is a column. The real matrices satisfy both
Bogoliubov identities in~\eqref{eq:bogoliubov-conditions} even when
$[\mathsf W_g,\mathsf W_h]\ne0$.
\end{proposition}
\begin{proof}
Solve \eqref{eq:s6-reference-canonical-array} at $g$ for its real
coordinate and momentum, then substitute them at $h$.
Moreover $\mathsf X_{hg}\mathsf Y_{hg}^{\mathsf T}
=\mathsf Y_{hg}\mathsf X_{hg}^{\mathsf T}=\Id$.
Expanding $\mathsf U\mathsf U^{\mathsf T}-
\mathsf V\mathsf V^{\mathsf T}$ and
$\mathsf U\mathsf V^{\mathsf T}-\mathsf V\mathsf U^{\mathsf T}$
therefore gives $\Id$ and zero, respectively.
\end{proof}

This is an explicit affine canonical registration within the finite-mode
framework~\cite{Weedbrook2012}. Its powers use the assembled restoring
operator, including cross-cell interactions. A passive re-expression
transforms observables and the state together. In particular, reconstructing
$\hat{\vct x}_h-\vct b_{hg}$ recovers the same laboratory displacement
and all of its calibrated moments. Changing the definition of occupation
does not prepare a new state. For the old vacuum, the new number operator
has the expectation
\begin{equation}
 \left\langle0_g\left|
       \hat{\vct b}_h^\dagger\!\cdot\hat{\vct b}_h
                       \right|0_g\right\rangle
       =\tr(\mathsf V_{hg}\mathsf V_{hg}^{\mathsf T})
                             +\vct d_{hg}^{\mathsf T}\vct d_{hg}.
 \label{eq:s6-reference-vacuum-number}
\end{equation}
It measures occupation relative to the new reference. If a controller
actually switches the quadratic Hamiltonian at fixed origin, continuous
coordinate and momentum instead give, for the old vacuum,
\begin{equation}
 \langle W_{g\to h}\rangle
   =\frac{\hbar}{4}\tr\!\left[
       (\mathsf D_h-\mathsf D_g)\mathsf W_g^{-1}\right].
 \label{eq:s6-reference-vacuum-quench-work}
\end{equation}
This follows from its coordinate covariance
$\hbar\mathsf W_g^{-1}/2$ and the changed potential energy.
An autonomous contact change uses the interaction and medium account of
Section~4 instead of an unassigned controller.

Figure~\ref{fig:s6-coupled-canonical-reference} compares the same physical preparation in two coupled oscillator references.

\begin{figure}[H]
\centering\begingroup
\definecolor{vltTeal}{HTML}{176B64}
\definecolor{vltBlue}{HTML}{315D80}
\definecolor{vltGold}{HTML}{A87520}
\definecolor{vltPlum}{HTML}{725877}
\definecolor{vltRed}{HTML}{762B38}
\definecolor{vltYellow}{HTML}{D5AD3C}
\definecolor{vltNeutral}{HTML}{777D83}
\setlength{\unitlength}{1pt}
\scalebox{0.90}{%
\input{vlt-figures/figure-194.tex}%
}\endgroup
\caption{Two positive restoring references at zero centre shift, with $\omega_\star=\hbar=1$ in the calculation and $x_\star=\sqrt{\hbar/\omega_\star}$. The unit-Mahalanobis position contours use common laboratory axes. Re-expression preserves the solid old-vacuum law; the dashed contour belongs to the distinct new vacuum. Bars give old-vacuum occupations in the ordered new normal modes, obtained by rotating $\mathsf V\mathsf V^{\mathsf T}$ into that basis. Their sum is occupation relative to the new reference. The printed work instead uses an actual fixed-origin Hamiltonian switch and the full energy law in~\eqref{eq:s6-reference-vacuum-quench-work}.}
\label{fig:s6-coupled-canonical-reference}
\end{figure}

The fixed virion pool supplies a common physical coordinate space across
these descriptions. A shorter retained band is a projection of that
space, with the residual and detector tests of
Section~\ref{sec:s5-register-patterns}. It is generally not an invertible
canonical transformation: discarded conjugate directions and their
preparation must be represented by an explicit reduction. At a zero-mode
crossing the inverse powers above cease to apply; retain the corresponding
free coordinates and momenta until a positive oscillator reference is
available. These rules allow a joining event to change the useful
occupation basis while preserving physical identity, continued motion
and the source of any actual energy transfer.

\subsubsection{Canonical preparations, visible moments, and physical work}

Translations, mode mixing and squeezing can be compared in the same
physical energy account. For this purpose introduce the dimensionless
column $\hat{\vct\zeta}=(\hat q_1,\ldots,\hat q_r,
\hat p_1,\ldots,\hat p_r)^{\mathsf T}$, where
$\hat q_s=\sqrt{\omega_s/\hbar}\hat Q_s$ and
$\hat p_s=\hat\Pi_s/\sqrt{\hbar\omega_s}$. Write
$\mathsf J_{\rm c}=\left(\begin{smallmatrix}0&\Id_r\\-\Id_r&0
\end{smallmatrix}\right)$, so that
$[\hat\zeta_j,\hat\zeta_k]=i(\mathsf J_{\rm c})_{jk}\Id$.
These dimensionless scalings use the fixed positive reference frequencies;
they do not change the mechanical coordinates or their physical units.

\begin{definition}[Affine canonical preparation]
\label{def:s6-affine-canonical-preparation}
In the finite-mode quantum model, an affine canonical preparation is a
unitary $\hat{\mathcal T}$ with
$\hat{\mathcal T}^\dagger\hat{\vct\zeta}\hat{\mathcal T}
=\mathsf S_{\rm c}\hat{\vct\zeta}+\vct d_{\rm c}$,
where $\mathsf S_{\rm c}$ and $\vct d_{\rm c}$ are real and
$\mathsf S_{\rm c}\mathsf J_{\rm c}\mathsf S_{\rm c}^{\mathsf T}
=\mathsf J_{\rm c}$. With mean $\vct m=\langle\hat{\vct\zeta}\rangle$
and symmetrized centered covariance
$\Sigma_{jk}=\langle\{\delta\hat\zeta_j,
\delta\hat\zeta_k\}\rangle/2$, finite second moments give
\begin{equation}
 \vct m'=\mathsf S_{\rm c}\vct m+\vct d_{\rm c},\qquad
 \mathsf\Sigma'=\mathsf S_{\rm c}\mathsf\Sigma
                                     \mathsf S_{\rm c}^{\mathsf T}.
 \label{eq:s6-affine-canonical-moments}
\end{equation}
\end{definition}
This standard affine symplectic action~\cite{Weedbrook2012} applies to
the first two moments of any state for which they exist. A Gaussian state
is completely specified by those moments; a general state is not.
The canonical covariance admissibility condition is given in
\eqref{eq:phonon-full-covariance-condition}. It is preserved by the
symplectic congruence, rather than becoming a new independent constraint
for each detector channel. A measured covariance must first be related to
this state covariance through the calibration and noise map of
Section~\ref{sec:operator-representations}.

\begin{proposition}[Work at a fixed harmonic reference]
\label{prop:s6-affine-canonical-work}
Let $\mathsf K_\omega=\diag(\mathsf W,\mathsf W)$, so
$\hat H_0=\hbar\hat{\vct\zeta}^{\mathsf T}
\mathsf K_\omega\hat{\vct\zeta}/2$. Define the real symmetric matrix
$\mathsf A_{\rm w}=\mathsf S_{\rm c}^{\mathsf T}
\mathsf K_\omega\mathsf S_{\rm c}-\mathsf K_\omega$ and the real
vector $\vct b_{\rm w}=\mathsf S_{\rm c}^{\mathsf T}
\mathsf K_\omega\vct d_{\rm c}$. For any finite-second-moment initial
state the preparation changes the full fixed-reference harmonic energy by
\begin{equation}
 \begin{aligned}
 \Delta E=\frac{\hbar}{2}\big[&
   \vct m^{\mathsf T}\mathsf A_{\rm w}\vct m
       +\tr(\mathsf A_{\rm w}\mathsf\Sigma)\\
       &{}
       +2\vct b_{\rm w}^{\mathsf T}\vct m
       +\vct d_{\rm c}^{\mathsf T}\mathsf K_\omega\vct d_{\rm c}
                              \big].
 \end{aligned}
 \label{eq:s6-affine-canonical-work}
\end{equation}
\end{proposition}
\begin{proof}
The expectation of a real symmetric quadratic form is its value at the
mean plus its trace against the symmetrized covariance. Insert
\eqref{eq:s6-affine-canonical-moments} in
$\langle\hat H_0\rangle=
\hbar[\vct m^{\mathsf T}\mathsf K_\omega\vct m+
\tr(\mathsf K_\omega\mathsf\Sigma)]/2$ and subtract the initial value.
The antisymmetric commutator part drops out of the symmetric quadratic
form.
\end{proof}

The formula recovers translation, mode-mixing and squeezing work from
first and second moments. The same mean-plus-covariance calculation
applies to a classical canonical distribution. The energy metric here
is fixed; physical curvature control uses the initial and final metrics,
as in the quench example.

There is also a precise test of which measured moments suffice for this
work prediction.

\begin{proposition}[Visibility of canonical preparation work]
\label{prop:s6-work-visible-moments}
Let a real full-row-rank matrix $\mathsf L_{\rm c}$
specify calibrated channels with state means
$\vct y=\mathsf L_{\rm c}\vct m$ and symmetrized state covariance
$\mathsf C_{\rm y}=\mathsf L_{\rm c}\mathsf\Sigma
\mathsf L_{\rm c}^{\mathsf T}$. For fixed known preparation and energy
metric, these two measured moments determine~\eqref{eq:s6-affine-canonical-work}
for every finite-second-moment state of the retained oscillator model
if and only if there exist a real symmetric $\mathsf B_{\rm y}$ and a
real $\vct c_{\rm y}$ such that
\begin{equation}
 \mathsf A_{\rm w}=\mathsf L_{\rm c}^{\mathsf T}
                         \mathsf B_{\rm y}\mathsf L_{\rm c},\qquad
 \vct b_{\rm w}=\mathsf L_{\rm c}^{\mathsf T}\vct c_{\rm y}.
 \label{eq:s6-work-visible-moment-criterion}
\end{equation}
\end{proposition}
\begin{proof}
Sufficiency follows by replacing the three state-dependent terms with
$\vct y^{\mathsf T}\mathsf B_{\rm y}\vct y$,
$\tr(\mathsf B_{\rm y}\mathsf C_{\rm y})$, and
$2\vct c_{\rm y}^{\mathsf T}\vct y$. For necessity, take coherent
preparations with fixed covariance $\Id_{2r}/2$ and arbitrary means.
Means $\vct m$ and $\vct m+t\vct k$, with
$\vct k\in\ker\mathsf L_{\rm c}$, give identical recorded moments.
Their work polynomial is equal for all $\vct m,t$ precisely when
$\mathsf A_{\rm w}\vct k=0$ and
$\vct b_{\rm w}^{\mathsf T}\vct k=0$. Factoring through the full-row-rank
map gives~\eqref{eq:s6-work-visible-moment-criterion}.
\end{proof}
Restrictions on the admissible preparations can supply additional
information; that information must be stated rather than inferred from
the channel count.

For a single-mode squeeze, for example,
$\mathsf A_{\rm w}=\omega\diag(e^{-2r_{\rm sq}}-1,
e^{2r_{\rm sq}}-1)$. When $r_{\rm sq}\ne0$, a displacement-only
channel $\mathsf L_{\rm c}=(1,0)$ fails the criterion: the momentum
variance also contributes to work. A narrower distribution of tracked
ideal-virion displacements thus has a direct kinematic meaning, while its
energetic interpretation also needs the relevant velocity statistics or
an independently justified preparation law. If channel quadratures do
not commute, their covariance entries require an appropriate measurement
protocol on matched preparations; they are not automatically the joint
outcomes of one noiseless measurement. This connects the canonical
description to the actual observable without treating detector noise,
physical state variability and missing coordinates as the same object.

\FloatBarrier
\subsection{Local displacement, force, and energy-current observables}
\label{sec:local-mechanical-operators}

A local measurement selects a particle, contact or structural component
of a collective pattern. Reconstructing that coordinate connects mode
coefficients to the probe. Amplitude, restoring force and energy current
then follow their respective mechanical definitions.

\subsubsection{Reconstructing a displacement at the selected resolution}

For whole-virion translations with mass blocks $m_i\Id_3$, the positive-mode
coordinate and velocity at node $i$ are
\begin{equation}
 \hat{\vct u}_i=\sum_s\vct d_{is}(\hat b_s+\hat b_s^\dagger),
 \quad
 \vct d_{is}=\sqrt{\frac{\hbar}{2m_i\omega_s}}\,\vct e_{s,i},
 \quad
 \hat{\vct v}_i=-i\sum_s\omega_s\vct d_{is}
                                      (\hat b_s-\hat b_s^\dagger).
 \label{eq:local-canonical-displacement}
\end{equation}
The real dimensionless block $\vct e_{s,i}$ describes that site's portion
of a mass-weighted mode. The coefficient $\vct d_{is}$ has length units;
it combines mode shape, inertia and the frequency-dependent quantum
displacement scale. It is not the displacement amplitude of every
preparation. That amplitude or fluctuation follows from the state of the
ladder operators. The corresponding classical reconstruction uses the
real mode coordinates $Q_s$ rather than occupation operators.

For a general positive inertia matrix the reconstruction is
$\hat{\vct u}=\mathsf M^{-1/2}\mathsf E\hat{\vct Q}$,
$\hat{\vct v}=\mathsf M^{-1/2}\mathsf E\hat{\vct\Pi}$ and
$\hat{\vct p}=\mathsf M^{1/2}\mathsf E\hat{\vct\Pi}$.
Retaining fewer than all coordinates gives
\begin{equation}
 [\hat u_a,\hat p_b]
       =i\hbar(\mathsf M^{-1/2}\mathsf P_+
                            \mathsf M^{1/2})_{ab}\Id.
 \label{eq:projected-site-commutator}
\end{equation}
This follows by inserting $[\hat Q_s,\hat\Pi_t]=i\hbar\delta_{st}\Id$
and $\mathsf E\mathsf E^{\mathsf T}=\mathsf P_+$.
The projected site coordinates are therefore linear combinations of
the retained canonical pairs. Assigning a new independent oscillator to
every projected component would double count motion. Including the
separated zero-mode canonical coordinates restores the full-coordinate
commutator, but their free dynamics still differ from positive-frequency
oscillator motion.

Structural resolution fixes the reconstruction. A centre readout tracks
translation; a capsid contact or membrane marker can also follow rigid
rotation and shape deformation, as
in~\eqref{eq:structured-virion-positions}. Envelope and RNP markers
select different rows of one augmented map. Their calibrated difference
compares those mechanical coordinates, with particle geometry retained
in the force and readout laws.

\subsubsection{An amplitude statistic is not a density or an energy}

Define the site-valued annihilation amplitude
$\hat{\vct A}_i=\sum_s\vct d_{is}\hat b_s$.
Here \emph{field} means an operator-valued amplitude assigned to the
retained nodes; no independent molecular field has been introduced.
Its adjoint creates a mode excitation, not a virion. The measurable
displacement is the self-adjoint sum
$\hat{\vct u}_i=\hat{\vct A}_i+\hat{\vct A}_i^\dagger$.
The normally ordered amplitude statistic is
\begin{equation}
 \hat I_i=\hat{\vct A}_i^\dagger\cdot\hat{\vct A}_i
       =\dd\Gamma(B_i),
 \qquad
 (B_i)_{st}=\vct d_{is}\cdot\vct d_{it}.
 \label{eq:local-amplitude-intensity}
\end{equation}
The nonnegative Gram matrix $B_i$ defines a self-adjoint nonnegative
occupation-sum realization. With units of length squared, $I_i$ measures
the stated amplitude correlation on states with finite quadratic
moments. Particle counts use positions; mechanical energy uses kinetic
and interaction terms; optical intensity uses its transduction law.

For a scalar amplitude $A=X+iY$ with self-adjoint quadratures,
$A^\dagger A=X^2+Y^2+i[X,Y]$. The commutator term cannot be
discarded. In a one-phonon state
$\ket\Phi=\sum_s c_s\ket{1_s}$,
$\langle\hat I_i\rangle_\Phi=|\sum_s\vct d_{is}c_s|^2$.
Relative mode phase can therefore change a local amplitude statistic
without changing the total excitation number. More generally,
\begin{equation}
 \langle\hat{\vct u}_i\cdot\hat{\vct u}_i\rangle
  =2\langle\hat I_i\rangle+\sum_s|\vct d_{is}|^2
       +2\operatorname{Re}
                   \langle\hat{\vct A}_i\cdot\hat{\vct A}_i\rangle.
 \label{eq:displacement-amplitude-moments}
\end{equation}
The middle term is the vacuum contribution. The final, anomalous
correlation vanishes in a fixed-total-occupation state but need not vanish
in a displaced or squeezed preparation. It is a quadrature correlation,
not automatically an instability.

For one scalar contribution $u=d(b+b^\dagger)$, a number state has
$\langle u\rangle=0$ and $\langle u^2\rangle=(2N+1)d^2$.
A coherent state instead has $\langle u\rangle=2d\operatorname{Re}\xi$
and centered variance $d^2$ at that phase. Thus a nonzero occupation need
not give a nonzero mean position, and the same mean does not fix its
fluctuation width. A biological record must distinguish estimates of mean motion,
fluctuation spectra and amplitude correlations before comparison with
the corresponding predictions.

\subsubsection{Forces and a local energy balance}

For the conservative pair network, the force is the negative derivative
of its declared potential:
\begin{equation}
 \hat{\vct F}_i=-(\mathsf K\hat{\vct u})_i
       =-\sum_{j\sim i}\mathsf C_{ij}
                      (\hat{\vct u}_i-\hat{\vct u}_j)
                   -(\mathsf K_{\rm wall,target}\hat{\vct u})_i.
 \label{eq:local-restoring-force}
\end{equation}
The pair expression uses the symmetric bond curvatures and coordinate
resolution assembled in $\mathsf K$; general internal couplings use
$-\mathsf K\hat{\vct u}$ directly. Real force components are
self-adjoint linear canonical observables. External drive, drag and
random force enter through their work-conjugate couplings and the
specified medium model.

For translational mass blocks, assign half of every bond potential to
each endpoint. In an undriven pair network with local conservative
time-independent quadratic pinning energy $V_i$, use
\begin{equation}
 \hat e_i=\frac{m_i}{2}\hat{\vct v}_i^2+V_i(\hat{\vct u}_i)
       +\frac14\sum_{j\sim i}
          (\hat{\vct u}_i-\hat{\vct u}_j)^{\mathsf T}
           \mathsf C_{ij}(\hat{\vct u}_i-\hat{\vct u}_j).
 \label{eq:operator-half-bond-local-energy}
\end{equation}
The two quarter-bond terms sum to the bond's full one-half quadratic
potential. Differentiating on the common harmonic core and inserting
the force cancels the local pinning power. The remaining pair term gives
\begin{equation}
 \hat J_{i\to j}
   =\frac14\sum_{\alpha=1}^3
      \left\{\hat v_{i,\alpha}+\hat v_{j,\alpha},
         [\mathsf C_{ij}
              (\hat{\vct u}_i-\hat{\vct u}_j)]_\alpha\right\},
 \qquad \hat J_{j\to i}=-\hat J_{i\to j}.
 \label{eq:quantized-bond-current}
\end{equation}
Consequently $\Dt{\hat e_i}=-\sum_j\hat J_{i\to j}$.
The anticommutator makes the quantum force--velocity product symmetric;
for a classical product it reduces to
$J_{i\to j}=(\vct v_i+\vct v_j)\cdot
\mathsf C_{ij}(\vct u_i-\vct u_j)/2$.
In the finite-mode canonical representation the real symmetrized quadratic
current is essentially self-adjoint on the common Schwartz core; its
closure defines the current observable. The polynomial balance identities
are first read on that core, with the required finite moments for
expectations.
This is the local allocation behind~\eqref{eq:edge-flux} and the usual
harmonic-network energy continuity~\cite{Dhar2008}.
It has energy-per-time units and a sign fixed by the chosen direction.
Displacement times momentum alone lacks both the restoring interaction
and the required power units.

\begin{proposition}[Pair power for resolved rigid bodies]
\label{prop:s6patch-pair-wrench-current}
Consider independent labelled bodies with separable material kinetic
energy $\sum_iT_i$, coordinate-valued local potentials and pair
interactions $V_{ij}$. Hold surface parameters fixed. Define the
endpoint wrench
$\vct g_i^{ij}=(\vct F_i^{ij},\vct T_i^{ij})=-D_iV_{ij}$ and
$\vct\nu_i=(\vct v_i,\vct\Omega_i)$ by virtual work.
Half interaction storage at each endpoint gives the outward current
\begin{equation}
 \begin{aligned}
 J_{i\to j}
   &=\tfrac12(\vct\nu_j\cdot\vct g_j^{ij}
                         -\vct\nu_i\cdot\vct g_i^{ij}),\\
 \hat J_{i\to j}
   &=\frac{i}{2\hbar}[\hat T_i-\hat T_j,\hat V_{ij}]
                       =-\hat J_{j\to i}.
 \end{aligned}
 \label{eq:s6patch-pair-wrench-current}
\end{equation}
The quantum identity uses a common invariant core on which the indicated
products and commutators exist.
\end{proposition}
\begin{proof}
The classical chain rule gives
$\dot V_{ij}=-\vct\nu_i\cdot\vct g_i^{ij}
            -\vct\nu_j\cdot\vct g_j^{ij}$.
Combine half this derivative with each endpoint's kinetic and local
potential balance. Quantum coordinate potentials commute; the
Heisenberg derivative of half-pair storage gives the displayed
commutator.
\end{proof}
In a constant-inertia body tangent the quantum current is one quarter
of the difference of the endpoint wrench--velocity anticommutators.
This recovers \eqref{eq:quantized-bond-current}. Endpoint torques need
not be opposite. These formulas use independent body variables before
noncanonical reduction; constraints and coupled effective inertia
retain their reaction and cross-kinetic work. Relaxing surface/material
coordinates keep their separately assigned storage and loss.

A signed prestressed bond need not carry a positive local quadratic
allocation, even when the assembled reference is stable. The balance law
still follows from that allocation. Assigning bond energy differently can
redistribute instantaneous local storage and current terms; the assigned
balance and complete conserved energy must change together. Additional
target, wall or many-coordinate interactions require their corresponding
storage and power. A time-dependent force adds its symmetrized work rate;
a reduced medium adds the already specified exchange with unresolved
coordinates. The continuum analogue is the symmetrized stress--velocity
flux in~\eqref{eq:viscoelastic-local-flux}, interpreted with spatial
smearing or a retained finite mode set in the quantum construction.

An explicit pinned pair illustrates transfer without autonomous gain.
Let $k_0=k_b=m\omega_0^2>0$ and
$V=k_0(u_1^2+u_2^2)/2+k_b(u_1-u_2)^2/2$.
Common and relative frequencies are $\omega_0$ and $\sqrt3\omega_0$.
The classical solution with small displacement scale $a_{\rm osc}>0$,
$u_1=a_{\rm osc}[\cos(\omega_0t)+\cos(\sqrt3\omega_0t)]$,
$u_2=a_{\rm osc}[\cos(\omega_0t)-\cos(\sqrt3\omega_0t)]$,
has
\begin{equation}
 J_{1\to2}=-2k_0a_{\rm osc}^2\omega_0
                    \cos(\sqrt3\omega_0t)\sin(\omega_0t),
 \qquad e_1+e_2=4k_0a_{\rm osc}^2.
 \label{eq:operator-pinned-pair-energy-transfer}
\end{equation}
The current reverses while total energy stays constant, as
Figure~\ref{fig:bond-energy-current-continuity} shows. In either pure
common or pure relative standing pattern the current is zero under this
allocation. Motion and mechanical transfer are therefore different
questions. In stationary thermal or number-state preparations of the
time-reversal-invariant reference, the mean current is zero; its
fluctuations need not vanish. None of these finite-pair statements by
itself establishes a bulk thermal conductivity or sustained active work.

\begin{figure}[H]
\centering\begingroup
\definecolor{vltTeal}{HTML}{176B64}
\definecolor{vltBlue}{HTML}{315D80}
\definecolor{vltGold}{HTML}{A87520}
\definecolor{vltPlum}{HTML}{725877}
\definecolor{vltRed}{HTML}{762B38}
\definecolor{vltYellow}{HTML}{D5AD3C}
\definecolor{vltNeutral}{HTML}{777D83}
\setlength{\unitlength}{1pt}
\scalebox{0.90}{%
\input{vlt-figures/figure-195.tex}%
}\endgroup
\caption{Particle motion and a signed local energy ledger for the pinned conservative pair $k_b=k_0=m\omega_0^2$. Set $\tau=\omega_0t$, $E_\star=k_0a_{\rm osc}^2$, $J_\star=E_\star\omega_0$ and $u_{1,2}/a_{\rm osc}=\cos\tau\pm\cos(\sqrt3\tau)$. (a) Oblique snapshots at $\tau=0.5,1.6$ use the same camera, rigid-body orientation and world displacement gain, equivalent to $10\,\mathrm{pt}$ per amplitude before projection. Support planes supply the declared pinning. All three helices attach to the current surfaces, with fixed coil geometry within each family. Teal arrows show displacement; burgundy current arrows reverse and use $18\,\mathrm{pt}$ per $J_\star$ along the projected bond. (b,c) Half-bond allocation gives $e_i=mv_i^2/2+k_0u_i^2/2+k_b(u_1-u_2)^2/4$ and $J=k_b(u_1-u_2)(v_1+v_2)/2$. Their exact trajectories redistribute a constant $4E_\star$ without applied drive. The faint cylinder locates the reference environment; the supports and clearance determine this auxiliary excursion window.}
\label{fig:bond-energy-current-continuity}
\end{figure}

\subsubsection{Transfer between particle groups and the storage convention}

A group of ideal virions gives a concrete spatial boundary for the energy
account. Summing its node balances cancels every internal bond current;
only contacts crossing the group's boundary remain. The resulting power
must be compared with the storage assigned to that same group.

\begin{definition}[Bond allocation and cut storage]
\label{def:s65-bond-allocation}
For the pair model above, write the complete bond potential as
$\hat B_{ij}=(\hat{\vct u}_i-\hat{\vct u}_j)^{\mathsf T}
\mathsf C_{ij}(\hat{\vct u}_i-\hat{\vct u}_j)/2$.
A time-independent allocation assigns fraction $\alpha_{ij}$ to node
$i$, with $\alpha_{ij}+\alpha_{ji}=1$. Relative to half-bond storage,
define
\begin{equation}
 \hat\chi_{ij}=(\alpha_{ij}-\tfrac12)\hat B_{ij}
       =-\hat\chi_{ji},\qquad
 \hat e_i^{(\alpha)}=\hat e_i+\sum_j\hat\chi_{ij},\qquad
 \hat\chi_S=\sum_{i\in S,\,j\notin S}\hat\chi_{ij}.
 \label{eq:s65-bond-allocation-storage}
\end{equation}
Here $S$ is a fixed set of particle labels, rather than a region whose
membership changes as particles move. The total Hamiltonian is unchanged.
For positive bond energies, $0\leq\alpha_{ij}\leq1$ gives a literal
partition; the balance identity also holds for signed bond curvatures.
\end{definition}

\begin{proposition}[Cut current under a change of energy allocation]
\label{prop:s65-cut-current-allocation}
Use the undriven time-independent pair model and preparations with the
finite moments needed for its quadratic balances. On the common core,
$\hat J_{i\to j}^{(\alpha)}=
\hat J_{i\to j}-\Dt{\hat\chi}_{ij}$ gives the balance for
$\hat e_i^{(\alpha)}$. If
$\hat J_S^{(\alpha)}=\sum_{i\in S,j\notin S}
\hat J_{i\to j}^{(\alpha)}$ is outward cut power, then
\begin{equation}
 \int_{t_0}^{t_1}\!(
     \langle\hat J_S^{(\alpha)}\rangle
                 -\langle\hat J_S\rangle)\dd t
       =-\langle\hat\chi_S(t_1)-\hat\chi_S(t_0)\rangle.
 \label{eq:s65-cut-current-storage-identity}
\end{equation}
The same identity holds for classical trajectories and their ensemble
means. If $|\langle\hat\chi_S(t)\rangle|$ is uniformly bounded,
the difference of the two time-averaged cut powers tends to zero as
$t_1-t_0\to\infty$.
\end{proposition}
\begin{proof}
Differentiate the reallocated energy and substitute
$\Dt{\hat e_i}=-\sum_j\hat J_{i\to j}$. Both currents and
$\hat\chi_{ij}$ are antisymmetric under exchange of endpoints, so
internal terms cancel when node balances are summed over $S$.
Integrating the remaining derivative proves the identity. Dividing by
the interval length and bounding the two endpoint expectations proves
the final statement.
\end{proof}

The identity expresses the storage convention in microscopic current
constructions~\cite{Ercole2016}. Over finite intervals, allocated
transfer changes by the boundary-storage term, while actuator work and
total energy retain their definitions. Comparing transfer between virion
groups therefore specifies their partition, storage allocation and
observation interval. A bulk conductivity requires additional transport
assumptions.

The data needed for current also differ from a displacement width.
For one bond put $\Delta\vct u=\vct u_i-\vct u_j$ and
$\Sigma\vct v=\vct v_i+\vct v_j$, and define
$(\mathsf C_{uv}^{ij})_{ab}=
\langle\{\delta\hat{\Delta u}_a,
\delta\hat{\Sigma v}_b\}\rangle/2$. Then half-bond current has mean
\begin{equation}
 \langle\hat J_{i\to j}\rangle
   =\tfrac12\overline{\Sigma\vct v}^{\mathsf T}
                      \mathsf C_{ij}\overline{\Delta\vct u}
      +\tfrac12\tr(\mathsf C_{ij}\mathsf C_{uv}^{ij}).
 \label{eq:s65-current-mean-cross-covariance}
\end{equation}
Expanding the anticommutator in
\eqref{eq:quantized-bond-current} gives this formula directly; the
classical version uses ordinary centered products. A mean transfer can
thus be carried by displacement--velocity correlations even when the
mean positions and velocities vanish. Reconstructing that term requires
the relevant time-resolved, calibrated channels and the specified bond
curvature. Equal-time position covariance alone supplies different
information. This makes the distinction between correlated motion and
mechanical transfer precise in the variables of the lattice.

\subsection{Probe response, commutators, and inertial normalization}
\label{sec:operator-probe-response}

A force experiment probes the response selected by its drive and
detector. Inertia, interactions and boundary determine propagation;
preparation determines fluctuations. In the harmonic model the
commutator recovers the mechanical response, while the noise also
retains the preparation's statistical information.

\subsubsection{From a work-conjugate force to a causal kernel}

Choose real dimensionless calibration vectors $\vct\ell_a$ for
length-valued coordinates and define
$\hat Y_a=\vct\ell_a^{\mathsf T}\hat{\vct u}$ and
$g_{as}=\vct\ell_a^{\mathsf T}\mathsf M^{-1/2}\vct e_s$.
Then $Y_a=\sum_s g_{as}Q_s$ has length units and $g_{as}$ has
units $\mathrm{kg}^{-1/2}$ in translational normalization.
A force $F_b$ in newtons couples as $\hat H_d=-F_b\hat Y_b$,
so $F_b\,\dd Y_b$ is work. Other generalized coordinates require
their own conjugate units. With the perturbation absent before $t_0$,
the retarded kernel gives
\begin{equation}
 \begin{split}
 \delta\langle\hat Y_a(t)\rangle
   &=\int_{t_0}^t\mathcal R^{\rm ret}_{ab}(t-t')F_b(t')\,\dd t',\\
 \mathcal R^{\rm ret}_{ab}(t)
   &=\frac{i}{\hbar}\Theta(t)
                   \langle[\hat Y_a(t),\hat Y_b(0)]\rangle.
 \end{split}
 \label{eq:phonon-kubo-probe}
\end{equation}
The step function $\Theta$ enforces causality: a later force cannot change
an earlier coordinate. The positive $i/\hbar$ follows from the negative
force coupling. This is the retarded linear-response construction~\cite{Kubo1957};
the opposite interaction sign gives the opposite commutator
prefactor~\cite{Clerk2010}.
For general systems this is linear response about a stated stationary
preparation; a nonstationary reference needs a two-time kernel.
For the harmonic reference driven linearly in position, the change in
the mean is exactly linear in the applied force as long as the harmonic
model remains physically valid.

Indeed the unforced solution is
$\hat Q_s(t)=\hat Q_s(0)\cos\omega_st
      +\hat\Pi_s(0)\sin\omega_st/\omega_s$.
Inserting it into the commutator gives
\begin{align}
 \mathcal K_{ab}(t)
   &=\sum_s g_{as}g_{bs}\frac{\sin\omega_st}{\omega_s}
      =\vct\ell_a^{\mathsf T}\mathsf M^{-1/2}\mathsf E
       \diag\!\left(\frac{\sin\omega_st}{\omega_s}\right)
               \mathsf E^{\mathsf T}\mathsf M^{-1/2}\vct\ell_b,
       \nonumber\\
 [\hat Y_a(t),\hat Y_b(0)]
   &=-i\hbar\mathcal K_{ab}(t)\Id,
 \qquad \mathcal R^{\rm ret}_{ab}(t)=\Theta(t)\mathcal K_{ab}(t).
 \label{eq:phonon-probe-commutator-kernel}
\end{align}
The commutator is a scalar multiple of the identity, so the harmonic
response is independent of initial occupation, temperature and squeezing.
The classical equation $\Dtt{Q_s}+\omega_s^2Q_s=g_{bs}F_b$ gives
the same kernel. A response peak therefore identifies a visible
mechanical mode; testing quantized occupation requires further data.

With the Fourier convention $e^{i\Omega t}$ and complex frequency
$z$ in the upper half-plane, integration gives
\begin{equation}
 \boxed{\begin{aligned}
 \mathcal R^{\rm ret}_{ab}(z)
    &=\sum_s\frac{g_{as}g_{bs}}{\omega_s^2-z^2}\\
    &=\vct\ell_a^{\mathsf T}\mathsf M^{-1/2}\mathsf E
       (\diag(\omega_s^2)-z^2\Id_r)^{-1}
          \mathsf E^{\mathsf T}\mathsf M^{-1/2}\vct\ell_b.
 \end{aligned}}
 \label{eq:operator-probe-modal-resolvent}
\end{equation}

Figure~\ref{fig:operator-matched-response-thermal-lines} gives a graphical specialization of this boxed relation.

The boundary value $z=\Omega+i0$ is retarded. Frequency-domain
susceptibility has units metres per newton; its time-domain kernel has
units seconds per kilogram because convolution includes $\dd t'$.
The poles are the mode frequencies, while the products $g_{as}g_{bs}$
specify excitation and detection. A mode may be present and still have
zero residue in this channel. For a diagonal channel the spectral weights
are nonnegative; cross-channel residues can have either sign and encode
the relative pattern of the two readouts. The impulse diagram below
represents this same resolvent and its time-domain kernel.

\subsubsection{The detector response through an assembly event}
\label{sec:s6-readout-event-response}

A pulse can begin before recruitment and be detected afterward. The
two-time kernel of Definition~\ref{def:s4-path-conditioned-response}
continues the same full-pool state, including prepared carrier strain.
Recruitment can change the reference, loading and illuminated particles.
Compatible cells assemble those mechanical couplings, and the detector
acts on the resulting continued history.

\begin{definition}[Registered causal detector response]
\label{def:s6-readout-causal-detector}
On a specified event history, use the fixed-dimension tangent state
$\delta\dot X=\mathsf A(t)\delta X+\mathsf B(t)\delta f(t)$.
Let the calibrated instantaneous receiver be
$\delta Y(r)=\mathsf L(r)\delta X(r)$ and let a causal exposure or
electronic filter give
$\delta y(t)=\int_{t_0}^{t}h(t,r)\delta Y(r)\,\dd r$.
Here $h$ has inverse-time units when its integral is one. With zero
initial perturbation, the recorded response is
\begin{equation}
 \begin{aligned}
 \delta y(t)&=\int_{t_0}^{t}\mathsf H_{\rm det}(t,s)
                                  \delta f(s)\,\dd s,\\
 \mathsf H_{\rm det}(t,s)
   &=\int_s^t h(t,r)\mathsf L(r)\mathsf T(r,s)
                                  \mathsf B(s)\,\dd r.
 \end{aligned}
 \label{eq:s6-readout-nonstationary-detector}
\end{equation}
The propagator includes the declared compatible event maps. Instantaneous
detection is the corresponding delta-kernel limit.
\end{definition}

Substituting the mechanical response and interchanging the causal
integrals proves the formula for integrable coefficients and inputs.
An independent preparation perturbation adds the filtered history of
$\mathsf L(r)\mathsf T(r,t_0)\delta X(t_0)$. An algebraic input-to-output
term adds its separately calibrated direct channel. The memory strain
already present in the reference is preserved even when this initial
\emph{perturbation} vanishes. For a density receiver, a row of
$\mathsf L(r)$ is the derivative of its actual spatial weight at the
reference positions. A tracked marker, changing illumination and changing
membership can therefore give different rows on the same mechanical
trajectory. Camera exposure and localization noise have distinct effects
even for simple diffusion~\cite{Berglund2010}; here the two-time observation
map retains exposure across a recruitment event instead of assigning
the entire frame to its final arrangement.

\begin{proposition}[Event selection contributes to an ensemble response]
\label{prop:s6-readout-selection-response}
Let a dimensionless pulse scale $\epsilon$ change the probability law
$P_\epsilon$ of registered histories $g$. Suppose these laws have
differentiable densities on a common history space, with score
$S(g)=\partial_\epsilon\log P_\epsilon(g)|_0$, and conditional mean
records $m_\epsilon(t,g)$. Under differentiation and integration
conditions sufficient for these derivatives,
\begin{equation}
 \left.\partial_\epsilon\mathbb E_\epsilon[y(t)]\right|_0
 =\mathbb E_0[\partial_\epsilon m_\epsilon(t,g)|_0]
       +\operatorname{Cov}_0(m_0(t,g),S(g)).
 \label{eq:s6-readout-selection-score}
\end{equation}
\end{proposition}
\begin{proof}
Differentiate $\int m_\epsilon(t,g)P_\epsilon(\dd g)$ by the product
rule. Normalization gives $\mathbb E_0S=0$, so the probability-law
derivative is the stated covariance.
\end{proof}

For two recorded outcomes, let $p_\epsilon=p_0+\epsilon\rho+o(\epsilon)$
be the recruitment probability. The derivative is
$p_0m'_1+(1-p_0)m'_0+\rho(m_1-m_0)$.
Two equal-mass core virions centred at zero and a newcomer at $2a$ have
registered centres $0$ and $2a/3$. At an observation instant with
$m'_0=m'_1=0$, their ensemble centre still has derivative $2a\rho/3$.
The selection term quantifies a change in the observed population.
The conditional derivatives also include any change of the preparation
distribution selected by that history; holding an event sequence fixed
computes only the corresponding controlled contribution. The complete
prediction uses the pulse-dependent transition law when the force changes
capture or escape probabilities. The score formula assumes regular
history densities; deterministic threshold events require the appropriate
event-time variation instead.

Figure~\ref{fig:s6-event-spanning-detector-exposure} compares a continued exposure with a frozen-arrangement approximation.

\begin{figure}[H]
\centering\begingroup
\definecolor{vltTeal}{HTML}{176B64}
\definecolor{vltBlue}{HTML}{315D80}
\definecolor{vltGold}{HTML}{A87520}
\definecolor{vltPlum}{HTML}{725877}
\definecolor{vltRed}{HTML}{762B38}
\definecolor{vltYellow}{HTML}{D5AD3C}
\definecolor{vltNeutral}{HTML}{777D83}
\setlength{\unitlength}{1pt}
\scalebox{0.90}{%
\input{vlt-figures/figure-196.tex}%
}\endgroup
\caption{A causal exposure across a controlled link change. The equal-mass pair has supports $k_0$, local drag $0.3m\omega_\star$ and $\omega_\star=\sqrt{k_0/m}$. At $t=0$, an impulse sets $(\vct u,\dot{\vct u})=(0,0,u_\star\omega_\star,0)$. At $\omega_\star t_e=1.5$, link curvature changes from $0.2k_0$ to $1.2k_0$ with continuous position and velocity and accounted stiffness work. Upper glyphs show identical reference topology. The solid site-2 record averages the continued trajectory over a causal window of width $0.8/\omega_\star$. Dashed curves use the final generator for the entire history. Shading marks straddling windows; early windows include zero pre-impulse motion. Dotted curves give instantaneous displacement.}
\label{fig:s6-event-spanning-detector-exposure}
\end{figure}

\subsubsection{Spectral weight, impulse slope, and projected inertia}

For a single probe $Y$, write $g_s=g_{Ys}$.
The positive-frequency spectral measure is
\begin{equation}
 \operatorname{Im}\mathcal R^{\rm ret}_{YY}(\Omega+i0)
     =\frac\pi2\sum_s\frac{g_s^2}{\omega_s}
                      \delta(\Omega-\omega_s),\qquad \Omega>0.
 \label{eq:operator-probe-positive-spectral-measure}
\end{equation}
Delta functions specify integrated line weights. Observation windows and
declared damping mechanisms determine their measured broadening and
finite peak heights.

The double commutator removes the restoring potential because coordinate
functions commute with $Y$:
\begin{equation}
 [\hat Y,[\hat H_0,\hat Y]]
       =\hbar^2\mathcal M_Y^{-1}\Id,
 \qquad
 \mathcal M_Y^{-1}=\sum_sg_s^2
       =\vct\ell^{\mathsf T}\mathsf M^{-1/2}\mathsf P_+
                            \mathsf M^{-1/2}\vct\ell.
 \label{eq:phonon-probe-double-commutator}
\end{equation}
It is the same effective mass obtained from the canonical readout pair
in~\eqref{eq:operator-canonical-readout-pair}, rather than a new fitted
mass for each resonance. Combining the line weights gives
\begin{equation}
 \int_0^\infty\frac{2\Omega}{\pi}
        \operatorname{Im}\mathcal R^{\rm ret}_{YY}(\Omega+i0)\,\dd\Omega
       =\mathcal M_Y^{-1}.
 \label{eq:phonon-probe-inertial-sum}
\end{equation}
Equivalently $\mathcal R_{YY}^{\rm ret}(z)
 =-\mathcal M_Y^{-1}/z^2+O(|z|^{-4})$ for the finite reference.
Static compliance instead weights each mode by its softness:
$\mathcal R^{\rm ret}_{YY}(0)=\sum_s g_s^2/\omega_s^2$.
Inertia controls the initial acceleration; restoring curvature controls
the shifted force-balanced equilibrium under a static force. An undamped
step response need not settle at that equilibrium.

An impulse $I_F=\int F(t)\,\dd t$ adds
$\delta\Pi_s=g_s I_F$. It leaves coordinates continuous and gives
\begin{equation}
 \delta\langle\hat Y(t)\rangle
       =I_F\mathcal R^{\rm ret}_{YY}(t)
       =I_F\mathcal M_Y^{-1}t+O(t^3),\qquad t\downarrow0.
 \label{eq:probe-short-time-inertia}
\end{equation}
Thus a calibrated impulse slope and a complete weighted response spectrum
test the same projected inverse inertia. Figure~\ref{fig:opening-five-impulse-inertia}
shows both for the equal-pinning/equal-link pair, also used for the local
current. A probe invisible to all retained modes has $g_s=0$ and neither
signal. Free coordinates have their own zero-frequency dynamics and are
excluded from this positive-mode sum.

\begin{figure}[H]
\centering
\begingroup
\definecolor{vlfGreen}{HTML}{176B64}
\definecolor{vlfDarkRed}{HTML}{762B38}
\definecolor{vlfRed}{HTML}{762B38}
\definecolor{vlfBlue}{HTML}{315D80}
\definecolor{vlfCopper}{HTML}{762B38}

\setlength{\unitlength}{1pt}
\scalebox{0.80}{%
\input{vlt-figures/figure-197.tex}%
}%
\endgroup
\caption{Integrated spectral weight and short-time response for $Y=u_1$ in the supported pair $k_b=k_0=m\omega_0^2$. The measure $(2\Omega/\pi)\operatorname{Im}\mathcal R^{\rm ret}(\Omega+i0)\,\dd\Omega$ has masses $1/(2m)$ at $\omega_0,\sqrt3\omega_0$. Multiplying by $m$ gives unit total. At right, $m\omega_0\mathcal R^{\rm ret}(t)=[\sin\tau+\sin(\sqrt3\tau)/\sqrt3]/2$, $\tau=\omega_0t>0$. Its tangent at zero gives the same inverse inertia $1/m$. Static compliance instead equals $2/(3m\omega_0^2)$. Both classical and quantized harmonic mechanics predict these diagnostics.}
\label{fig:opening-five-impulse-inertia}
\end{figure}

A finite-excursion receiver also changes the inertial normalization.
For a smooth real coordinate observable $Y(\vct X)$ and Hamiltonian
$\hat H=\hat{\vct P}_X^{\mathsf T}\mathsf G_X^{-1}\hat{\vct P}_X/2+V(\vct X)$
with constant $\mathsf G_X$ and coordinate potential,
the common-core product rule gives
\begin{equation}
 [\hat Y,[\hat H,\hat Y]]
    =\hbar^2(\nabla_XY)^{\mathsf T}\mathsf G_X^{-1}\nabla_XY .
 \label{eq:s6patch-geometric-probe-inertia}
\end{equation}
For an unconstrained full rigid body, the marker
$Y=\vct e\cdot(\vct r+\mathsf O\vct a)$, $|\vct e|=1$, gives instead
the exact kinetic-metric weight
$1/m+(\vct b\times\vct e)^{\mathsf T}
\mathsf J(\mathsf O)^{-1}(\vct b\times\vct e)$, with
$\vct b=\mathsf O\vct a$.
A conjugate point force therefore excites translation and torque.
For stationary preparations with convergent spectral moments, the
weighted response integral equals the expectation of this weight.
Its local projected linearization recovers the preceding modal sum.

Finite bandwidth can miss part of this nonnegative sum. Interpretation
also includes force calibration, instrument compliance and frequency
response. Eliminating hidden coordinates can produce frequency-dependent
apparent inertia even with a fixed full mass matrix. The sum rule tests
the specified coordinate and spectral coverage; several hidden coupling
networks may satisfy the same integral.

\begin{proposition}[Instantaneous and inertial parts of a mixed response]
\label{prop:s6patch-mixed-impulse-normalization}
Use Section~4's fixed resistance-whitened passive realization, with
$\mathsf G_X,\mathsf A,\mathsf D\succ0$, positive joint storage and an
asymptotically stable generator. For real calibrated channels define
\begin{equation}
 \begin{aligned}
 \vct Y&=\mathsf L_X\vct X+\mathsf L_\chi\vct\chi
          =\mathsf L_{\rm m}\vct X+\mathsf L_\chi\vct\chi',\\
 \mathsf S_\chi&=\mathsf D^{-1}\mathsf R^{\mathsf T},\qquad
 \vct\chi'=\vct\chi+\mathsf S_\chi\vct X,\qquad
 \mathsf L_{\rm m}=\mathsf L_X-\mathsf L_\chi\mathsf S_\chi .
 \end{aligned}
 \label{eq:s6patch-mixed-conjugate-row}
\end{equation}
The perturbation $-\vct f^{\mathsf T}\vct Y$ supplies the conjugate
sources $(\mathsf L_{\rm m}^{\mathsf T}\vct f,
\mathsf L_\chi^{\mathsf T}\vct f)$. Thus
\begin{equation}
 \begin{aligned}
 \mathsf h_Y(\omega)&=\mathsf L'\mathcal P'(\omega)^{-1}
                                  \mathsf L'^{\mathsf T},
 \qquad \mathsf L'=[\mathsf L_{\rm m},\mathsf L_\chi],\\
 \mathcal P'(\omega)&=
 \begin{pmatrix}
 \mathsf H'-\omega^2\mathsf G_X-i\omega\mathsf C_{\rm e}&\mathsf B'\\
 \mathsf B'^{\mathsf T}&\mathsf A-i\omega\mathsf D
 \end{pmatrix}.
 \end{aligned}
 \label{eq:s6patch-mixed-conjugate-response}
\end{equation}
Put
$\mathsf K_\chi=\mathsf L_\chi\mathsf D^{-1}\mathsf L_\chi^{\mathsf T}$,
$\mathsf K_{\rm m}=\mathsf L_{\rm m}\mathsf G_X^{-1}
\mathsf L_{\rm m}^{\mathsf T}$ and
$\mathsf K_{\chi,1}=\mathsf L_\chi\mathsf D^{-1}\mathsf A
\mathsf D^{-1}\mathsf L_\chi^{\mathsf T}$.
Let $\mathsf R_Y$ be the causal kernel whose Fourier transform is
$\mathsf h_Y$. Its initial values and formal large-frequency expansion satisfy
\begin{equation}
 \boxed{\begin{aligned}
 \mathsf R_Y(0^+)&=\mathsf K_\chi,\\
 \dot{\mathsf R}_Y(0^+)&=\mathsf K_{\rm m}-\mathsf K_{\chi,1},\\
 \mathsf h_Y(\omega)&=\frac{i}{\omega}\mathsf K_\chi
       +\frac{\mathsf K_{\chi,1}-\mathsf K_{\rm m}}{\omega^2}
                                  +O(\omega^{-3}).
 \end{aligned}}
 \label{eq:s6patch-mixed-impulse-asymptotics}
\end{equation}
\end{proposition}
\begin{proof}
For a conjugate impulse $\vct f=\vct I\delta(t)$ and zero perturbation
beforehand, $\vct X$ is continuous, while
$\Delta\dot{\vct X}=\mathsf G_X^{-1}\mathsf L_{\rm m}^{\mathsf T}\vct I$
and $\Delta\vct\chi'=\mathsf D^{-1}\mathsf L_\chi^{\mathsf T}\vct I$.
The unforced right-hand derivatives give the first two identities.
Expanding the block inverse gives the last; $\mathsf B'$ first enters
higher orders.
\end{proof}

Figure~\ref{fig:cmd-complementary-ports} gives a graphical specialization of this boxed relation.

Since $\mathsf D\succ0$,
$\ker\mathsf K_\chi=\ker\mathsf L_\chi^{\mathsf T}$.
With its work-conjugate contrast input, a calibrated contrast
$\mathsf W\vct Y$ has no instantaneous jump
precisely when $\mathsf W\mathsf L_\chi=0$. Then
$\mathsf W\mathsf L_{\rm m}=\mathsf W\mathsf L_X$; its leading inertial
coefficient is
$\mathsf W\mathsf L_X\mathsf G_X^{-1}
\mathsf L_X^{\mathsf T}\mathsf W^{\mathsf T}$.
Internal relaxation can still affect its later response.

The mixed first-order channel has a relaxational jump, rather than the
pure mechanical impulse slope alone. Extending its $i/\omega$ tail to
infinite frequency makes the preceding inertial integral diverge.
The expansion characterizes this coarse-grained realization; physical
use retains its calibrated frequency band and the omitted microscopic
completion. A detector-only surface signal and a centre-force actuator
instead define a cross response with their actual separate source maps.

\subsubsection{What unresolved spectral weight can still constrain}

An incomplete spectrum can give a quantitative consistency test when
the same probe's static compliance and projected inverse inertia are
known independently. The test uses positive diagonal response weights,
so it retains information even when individual omitted contacts or
mode patterns cannot be reconstructed.

\begin{proposition}[Missing-weight compliance bounds]
\label{prop:s66-missing-spectral-weight}
Fix the finite conservative positive sector, the calibrated conjugate
probe $Y$, and the reference parameters. Let $B$ be a specified set of
resolved modes whose frequencies and corrected weights $g_s^2$ are
known. Define the residual inverse inertia and static compliance by
\begin{equation}
 I_{\rm miss}=\mathcal M_Y^{-1}-\sum_{s\in B}g_s^2,
 \qquad
 C_{\rm miss}=\mathcal R_{YY}^{\rm ret}(0)
                         -\sum_{s\in B}\frac{g_s^2}{\omega_s^2}.
 \label{eq:s66-missing-inertia-compliance}
\end{equation}
Suppose every omitted mode with nonzero probe weight has frequency in
the declared interval $[\omega_L,\omega_H]$, with
$0<\omega_L\leq\omega_H<\infty$. Then
\begin{equation}
 I_{\rm miss}\geq0,\qquad
 \frac{I_{\rm miss}}{\omega_H^2}
       \leq C_{\rm miss}\leq
                   \frac{I_{\rm miss}}{\omega_L^2}.
 \label{eq:s66-missing-weight-compliance-bounds}
\end{equation}
These bounds are sharp with only that support and residual-weight
information. If $I_{\rm miss}>0$, then $C_{\rm miss}>0$ and
$\omega_{\rm eff,miss}^2:=I_{\rm miss}/C_{\rm miss}$ lies between
$\omega_L^2$ and $\omega_H^2$.
\end{proposition}
\begin{proof}
The complete sums give
$I_{\rm miss}=\sum_{s\notin B}g_s^2$ and
$C_{\rm miss}=\sum_{s\notin B}g_s^2/\omega_s^2$.
Multiply $1/\omega_H^2\leq1/\omega_s^2\leq1/\omega_L^2$
by each nonnegative weight and sum. Concentrating the residual weight
at an endpoint attains the corresponding bound; distributing it among
the two endpoints realizes the intermediate ratios. The positivity
and ratio statements follow from the same sums.
\end{proof}

Here $I_{\rm miss}$ has inverse-mass units and $C_{\rm miss}$ has
compliance units. The effective squared frequency is a weighted harmonic
mean of omitted squared frequencies, rather than an additional
eigenfrequency. If the only unresolved weight lies above $\omega_L$,
its contribution to the static compliance is at most
$I_{\rm miss}/\omega_L^2$. A low-frequency unresolved component can
instead contribute appreciable compliance with little inverse-inertia
weight. This separates two physically different reasons for a broad
structural fluctuation or a large static displacement under force.

Projection, inverse inertia and spectral sums refer to the same positive
sector. Before assigning resolved modes to $B$, account for window
mixing, filtering and degeneracy. Signed cross-channel residues require
other bounds. Inconsistent residual sums reject the assumed hidden
support; consistent sums constrain it without specifying a unique
network.

Figure~\ref{fig:s6-unresolved-spectral-compliance} locates a hidden spectral population within its sharp compliance bounds.

\begin{figure}[H]
\centering\begingroup
\definecolor{vltTeal}{HTML}{176B64}
\definecolor{vltBlue}{HTML}{315D80}
\definecolor{vltGold}{HTML}{A87520}
\definecolor{vltPlum}{HTML}{725877}
\definecolor{vltRed}{HTML}{762B38}
\definecolor{vltYellow}{HTML}{D5AD3C}
\definecolor{vltNeutral}{HTML}{777D83}
\setlength{\unitlength}{1pt}
\scalebox{0.80}{%
\input{vlt-figures/figure-198.tex}%
}\endgroup
\caption{Resolved response weight and sharp bounds on hidden compliance, using \eqref{eq:s66-missing-weight-compliance-bounds}. The auxiliary conservative spectrum has frequencies $(1,2,4)\omega_\star$ and inverse-inertia weights $(0.5,0.2,0.3)/m_\star$, with $k_\star=m_\star\omega_\star^2$. (a) Only the first line is resolved; the hidden support is stipulated to lie in $[2,4]\omega_\star$. (b) Its residual weight $I_{\rm miss}=0.5/m_\star$ and compliance $C_{\rm miss}=0.06875/k_\star$ locate the marked point inside the sharp wedge $I_{\rm miss}/\omega_H^2\le C_{\rm miss}\le I_{\rm miss}/\omega_L^2$. The ratio frequency $\sqrt{I_{\rm miss}/C_{\rm miss}}$ summarizes those hidden weights. It adds no spectral line. Sticks denote integrated response masses before detector filtering or broadening.}
\label{fig:s6-unresolved-spectral-compliance}
\end{figure}

\subsubsection{Response and fluctuations from the same parameter set}

The response fixes how an imposed force propagates. Fluctuations also
need a preparation. For independent harmonic modes in a Gibbs state at
$\beta=(k_{\rm B}T)^{-1}$, let
$\bar N_s=(e^{\beta\hbar\omega_s}-1)^{-1}$.
The ordered coordinate spectrum is
\begin{equation}
 \begin{aligned}
 S^{>}_{YY}(\Omega)
   &:=\int_{-\infty}^{\infty}e^{i\Omega t}
                   \langle\delta\hat Y(t)\delta\hat Y(0)\rangle_\beta\,\dd t
       \\
   &=2\pi\sum_s\frac{\hbar g_s^2}{2\omega_s}
       [(\bar N_s+1)\delta(\Omega-\omega_s)
                       +\bar N_s\delta(\Omega+\omega_s)].
 \end{aligned}
 \label{eq:operator-probe-equilibrium-ordered-noise}
\end{equation}
The extra one in the positive-frequency weight is the canonical vacuum
contribution. The symmetrized spectrum
$S^{\rm sym}_{YY}(\Omega)=[S^{>}_{YY}(\Omega)+S^{>}_{YY}(-\Omega)]/2$
obeys, on the nonzero spectral lines,
\begin{equation}
 S^{\rm sym}_{YY}(\Omega)
      =\hbar\coth\!\left(\frac{\beta\hbar\Omega}{2}\right)
                    \operatorname{Im}\mathcal R^{\rm ret}_{YY}(\Omega+i0).
 \label{eq:operator-probe-symmetrized-fdt}
\end{equation}
Both sides are spectral measures for the ideal isolated harmonic reference.
At high occupation this reduces to
$S_{YY}(\Omega)=(2k_{\rm B}T/\Omega)
\operatorname{Im}\mathcal R_{YY}^{\rm ret}$,
and integration yields
$\operatorname{Var}Y=k_{\rm B}T\mathcal R_{YY}^{\rm ret}(0)$.
The units are displacement squared times time for the spectra.
The same response frequencies and $g_s$ predict both forced motion and
equilibrium fluctuations. Temperature and the Gibbs preparation supply
the remaining statistical input, as required by the quantum
fluctuation--dissipation relation~\cite{CallenWelton1951,Kubo1957,Clerk2010}.

Figure~\ref{fig:operator-matched-response-thermal-lines} compares the integrated response and ordered-noise lines without assigning a linewidth to the isolated reference.

\begin{figure}[H]
\centering\begingroup
\definecolor{vltTeal}{HTML}{176B64}
\definecolor{vltBlue}{HTML}{315D80}
\definecolor{vltGold}{HTML}{A87520}
\definecolor{vltPlum}{HTML}{725877}
\definecolor{vltRed}{HTML}{762B38}
\definecolor{vltYellow}{HTML}{D5AD3C}
\definecolor{vltNeutral}{HTML}{777D83}
\setlength{\unitlength}{1pt}
\scalebox{0.80}{%
\input{vlt-figures/figure-199.tex}%
}\endgroup
\caption{Integrated response and ordered-noise lines for independent modes $r_s=\omega_s/\omega_\star=1,2$, $g_s^2=1/m_\star$. Response masses are $(m_\star\omega_\star/\pi)\int_{\rm line}\operatorname{Im}\mathcal R^{\rm ret}_{YY}\,\dd\Omega=1/(2r_s)$; negative partners are omitted in (a). Noise masses are $(m_\star\omega_\star/\hbar)\int_{\rm line}S^>_{YY}\,\dd\Omega/(2\pi)$. Filled positive-frequency and open negative-frequency endpoints distinguish the lines. Their ratio is $e^{\beta\hbar\omega_s}$ and difference $1/(2r_s)$. Common ordinate scales compare dimensionless integrated weights with no assigned linewidth; a detector supplies its ordering and transduction law. The boxed result illustrated here is~\eqref{eq:operator-probe-modal-resolvent}.}
\label{fig:operator-matched-response-thermal-lines}
\end{figure}

Passive medium reduction replaces the conservative susceptibility and
equilibrium noise together, following
Section~\ref{sec:viscoelastic-dynamics}. Their matched parameters
preserve the response--fluctuation test. Active and self-excited
preparations instead supply their nonequilibrium correlations.

For a sinusoidal diagonal force, the regular dissipative boundary value
has cycle-averaged input power
$\overline P=\Omega|F_0|^2\operatorname{Im}\mathcal R_{YY}^{\rm ret}/2$.
Positive absorption is the passive sign convention. In the exact finite
conservative limit there is no off-resonant irreversible absorption;
the resonant spectral lines describe the ideal limiting response and
resonant energy accumulation, not a steady finite-width dissipation law.

\subsubsection{What the response bound does and does not measure}

The same commutator gives a two-time state-variance inequality:
\begin{equation}
 \Delta Y_a(t)\,\Delta Y_b(0)
       \geq\frac\hbar2|\mathcal K_{ab}(t)|.
 \label{eq:phonon-two-time-probe-bound}
\end{equation}
This is Robertson's inequality applied to two Heisenberg observables in
one state with finite second moments~\cite{Robertson1929}. Equal-time positions commute; dynamics
can make their unequal-time commutator nonzero. The bound is therefore
tied to the same inertial coupling and mode phases as the response.
It does not specify a sequential measurement history or a sensor's error.

For virion contact, virtual work fixes the force channel, including
contact direction, attachment and sensor compliance
\cite{Michel2006,Cotterill2002}. Quantum detection also requires coupling,
back-action force and output noise~\cite{Clerk2010}; classical marker
tracking requires localization, exposure and calibration errors. These
observation laws distinguish mechanical susceptibility, prepared-state
variance and instrumental precision. The next subsection resolves their
spatial observation map.

\FloatBarrier
\subsection{Virion density, spatial readouts, and correlations}
\label{sec:spatial-readouts-correlations}

Spatial readouts assign weights to the assembly's motion. Particle counts,
image contrast and mode-pattern weights answer different questions.
For a layered virion, a membrane label, capsid contact and centre record
can select different parts of the same motion. The following maps
specify those weights for the finite cylinder before using correlations
to infer mechanics.

\subsubsection{Resolving a mode pattern versus counting particles}

Let $\mathsf\Pi_i$ project onto node $i$'s three translational components
in the full mass-weighted coordinate space. Its positive-sector compression
is
\begin{equation}
 \mathsf F_i=\left.\mathsf P_+\mathsf\Pi_i
                       \mathsf P_+\right|_{\mathcal H_+},
 \qquad 0\leq\mathsf F_i\leq\Id_+,
 \qquad \sum_i\mathsf F_i=\Id_+.
 \label{eq:site-mode-effects}
\end{equation}
The sum assumes the complete translational coordinate partition; retained
internal blocks complete it in an augmented model. Positivity follows from
$\langle\eta|\mathsf F_i|\eta\rangle
=\|\mathsf\Pi_i\eta\|^2$. Compression generally destroys mutual orthogonality. These positive
effects form a positive-operator-valued measure on the retained sector,
but need not be projections.
For a classical energy direction $\eta$, the expectation gives its
coordinate-block weight. Anatomical storage uses the local energy
allocation because $\sqrt{\mathsf D}$ generally mixes sites. In a
one-phonon preparation the same effects give probabilities for the
specified site-effect measurement. Virion counts follow the position
readout below.

Particle count follows the moving centres themselves. For a bounded real
spatial weight $w$, define
\begin{equation}
 n[w](t)=\sum_{i=1}^n w(\vct R_i^{(0)}+\vct u_i(t)),
 \qquad
 \hat n[w]=\sum_{i=1}^n w(\vct R_i^{(0)}+\hat{\vct u}_i).
 \label{eq:virion-density-readout}
\end{equation}
An indicator weight counts centres in a selected region; a smooth
point-spread weight gives a spatially averaged count. Equal-time
displacement components commute, even after positive-mode projection,
so joint functional calculus defines the quantum expression. It is bounded
and self-adjoint with $\|\hat n[w]\|\leq n\|w\|_\infty$.
For $0\leq w\leq1$ its value lies between zero and $n$, and
$\hat n[1]=n\Id$ exactly. A normalized smoothing kernel $W_h$ with
units inverse volume instead defines the number density
$n_h(\vct r)=\sum_i W_h(\vct r-\vct R_i)$, with units
particles per volume and integral $n$.

The distributional classical density is
$n(\vct r,t)=\sum_i\delta(\vct r-\vct R_i(t))$.
Its weak continuity equation follows by differentiating a test weight:
\begin{equation}
 \begin{aligned}
 \frac{\dd n[w]}{\dd t}
       &=\sum_i\vct v_i\cdot\nabla w(\vct R_i),\\
 \frac{\dd\hat n[w]}{\dd t}
       &=\frac12\sum_{i,\alpha}
          \{\hat v_{i,\alpha},
                  (\partial_\alpha w)(\vct R_i^{(0)}+\hat{\vct u}_i)\}.
 \end{aligned}
 \label{eq:operator-smeared-count-continuity}
\end{equation}
For smooth bounded test weights with bounded derivatives, the quantum
identity is first read on the common Schwartz core; taking its expectation
also requires the appropriate velocity moments. These expressions give
$\partial_t n+\nabla\cdot\vct j_n=0$ with the corresponding
classical or symmetrized particle flux. This flux transports particle
count; the bond current of Section~\ref{sec:local-mechanical-operators} transports mechanical energy.
For $w=1$ the derivative vanishes. The fixed mass $\sum_i m_i$ and the
fixed virion count therefore coexist with an excitation number
$\hat N_{\rm ph}=\sum_s\hat N_s$ that can change under a drive.
The harmonic readout remains an approximation about its declared basin;
it does not replace exact hard-wall or excluded-volume particle dynamics.

\subsubsection{Counting a transient cluster within the complete pool}
\label{sec:s6-readout-selected-density}

The same positions support a full-pool count and a cluster-selected
record. Write $N_{\rm tot}$ for the fixed vessel population and
$h_i^{\mathcal C}(t)$ for the registered association indicator. A smooth
spatial weight may also depend on the detector time. These distinctions
can be retained in one weak balance.

\begin{definition}[Membership- and contrast-weighted readout]
\label{def:s6-readout-selected-weight}
For classical centre trajectories continuous at isolated registration
events, define
\begin{equation}
 A_{\mathcal C}[w](t)
   =\sum_{i=1}^{N_{\rm tot}}h_i^{\mathcal C}(t)a_i(t)
                                  w(t,\vct R_i(t)).
 \label{eq:s6-readout-selected-weight}
\end{equation}
The coefficients $a_i$ are specified channel weights, smooth between
events. Number uses $a_i=1$; assigned particle charge uses $a_i=q_i$;
a coherent scattering amplitude uses its calibrated, generally complex
contrast. These choices retain their respective units and meanings.
\end{definition}

Between registration events, the product and chain rules give
\begin{equation}
 \begin{aligned}
 \frac{\dd A_{\mathcal C}[w]}{\dd t}
  =\sum_i h_i^{\mathcal C}\bigl[
       a_i(\partial_t w+\vct v_i\cdot\nabla w)
                       +\dot a_i w\bigr]_{(t,\vct R_i)}.
 \end{aligned}
 \label{eq:s6-readout-selected-smooth-balance}
\end{equation}
At an event $t_*$, the exact jump is
\begin{equation}
 [A_{\mathcal C}[w]]_*
  =\sum_i [h_i^{\mathcal C}a_i]_*w(t_*,\vct R_i(t_*)).
 \label{eq:s6-readout-selected-jump}
\end{equation}
Here the weight $w$ is continuous at the event; a detector-weight jump
uses the full product jump instead. For continuous $a_i$, the displayed
increment is simply $\sum_i a_i\Delta h_i^{\mathcal C}w$.
Integrating the smooth equation and adding these jumps gives the
finite-interval balance exactly. It is the measured-weight version of
the membership-source account in
\eqref{eq:s4-growing-cluster-mass-balance}.

With $a_i=1$ and time-independent $w$, the smooth term is physical
particle transport across the weighted region. The jump records its
assignment to a selected group. Summing the cluster and its complement
cancels that assignment term. Consequently the complete-pool count at
$w=1$ stays $N_{\rm tot}$ while the cluster count changes. An indicator
region is read through smooth approximation or distributional flux, with
actual boundary passages retained. The force model keeps its declared
long-range tails on both sides of the association criterion.

For fixed assigned $q_i$, the same cancellation holds for positive,
negative and mixed-charge pools. A changing charge assignment instead
has the explicit $\dot q_i$ or charge-jump term. A physical charging
mechanism then includes the transferred charge and its reservoir or
ionic current in the enlarged account. A changing optical contrast has
its own transduction term. It can change a scattering record even when
centre positions are instantaneously unchanged. Thus a selected
zero-wavevector amplitude can change through membership or contrast,
whereas the full-pool number amplitude retains its exact constant value.

Paired full-pool and membership-selected images at matched exposure
separate these effects. At zero wavevector, the selected-number
correlation is $\operatorname{Cov}(N_{\mathcal C}(t),N_{\mathcal C}(s))$,
which probes association histories. At finite wavevector the products
$h_i^{\mathcal C}(t)a_i(t)w(t,\vct R_i(t))$ retain the joint effects
of motion, contrast and membership. Event-aligned positions and counts
then test whether a brightness-relaxation time reflects residence or
mechanical damping.

Conditional quantum readouts can use the same externally registered
$h_i^{\mathcal C}$ and channel weights, multiplying the position
operators already defined above. A membership decision inferred from a
quantum measurement record additionally needs that measurement's joint
state-update law. It is not supplied by the equal-time functional
calculus for position alone.

Figure~\ref{fig:s6-membership-number-charge-ledger} separates membership jumps from material transport and checks complementary pool balances.

\begin{figure}[H]
\centering\begingroup
\definecolor{vltTeal}{HTML}{176B64}
\definecolor{vltBlue}{HTML}{315D80}
\definecolor{vltGold}{HTML}{A87520}
\definecolor{vltPlum}{HTML}{725877}
\definecolor{vltRed}{HTML}{762B38}
\definecolor{vltYellow}{HTML}{D5AD3C}
\definecolor{vltNeutral}{HTML}{777D83}
\setlength{\unitlength}{1pt}
\scalebox{0.90}{%
\input{vlt-figures/figure-200.tex}%
}\endgroup
\caption{A changing selector on eighteen stationary ideal virions. Gray bodies show the complement. Both orthographic views use the same centres $\vct R/a\in\{-1,0,1\}^2\times\{-1/2,1/2\}$, radius $0.13a$, camera and gain $54\,\mathrm{pt}/a$; small triads show laboratory directions. The thirty-three nearest-axis links, including nine vertical links, show unchanged reference topology. At $t=0$, the nine lower-layer members acquire the three upper-layer members at $y=-a$. Teal selects membership; gray retains the complete complement in both scenes. Signs denote fixed charges $q_i=q(-1)^{x_i/a+y_i/a+\ell_i}$, with $\ell_i=0,1$ for lower and upper layers and $q>0$. Their mixed-charge pool is neutral. The equal-mass selected centre changes as printed although no particle moves. Selected/complement counts change from $(9,9)$ to $(12,6)$, and charges from $(q,-q)$ to $(0,0)$. The lower ledgers show their conserved full-pool sums; post-event zero-charge curves coincide. Open and filled endpoints distinguish pre- and post-event values. This registered selection illustrates the weak readout balance; physical capture additionally needs its force and event law.}
\label{fig:s6-membership-number-charge-ledger}
\end{figure}

\subsubsection{Finite spatial visibility and a density spectrum}

A Fourier weight asks which motions change a particular spatial pattern.
For a specified inverse-length wavevector $\vct q$, set
$\hat n_{\vct q}=\sum_i\exp[-i\vct q\cdot
(\vct R_i^{(0)}+\hat{\vct u}_i)]$.
This bounded complex readout satisfies
$\hat n_{-\vct q}=\hat n_{\vct q}^\dagger$; its real and imaginary
parts are self-adjoint observables. Here $\vct q$ labels the probe's spatial
Fourier component, not an assumed Bloch wavevector of a finite lattice.
If a scattering experiment uses this variable, the associated physical
momentum transfer is $\hbar\vct q$.

Centering the chosen preparation and expanding to first order gives
\begin{equation}
 \delta\hat n_{\vct q}
       =-i\sum_s\mathcal G_s(\vct q)
                          (\delta\hat b_s+\delta\hat b_s^\dagger),
 \qquad
 \mathcal G_s(\vct q)=\sum_i e^{-i\vct q\cdot\vct R_i^{(0)}}
                                     \vct q\cdot\vct d_{is}.
 \label{eq:phonon-density-mode-overlap}
\end{equation}
The dimensionless overlap combines spatial phase, displacement direction
and mode amplitude. It obeys
$\mathcal G_s(-\vct q)=-\mathcal G_s(\vct q)^*$.
The expansion requires both the mean shift and the rms fluctuation of
$\vct q\cdot\hat{\vct u}_i$ to be small. Small mechanical displacement
relative to a contact length alone is insufficient at arbitrarily large
probe wavevector.

Density time correlations provide the standard connection to
energy-resolved scattering in the Born approximation~\cite{VanHove1954}.
For independent thermal modes, the leading ordered connected spectrum is
\begin{align}
 S_n(\vct q,\Omega)
   &:=\int_{-\infty}^{\infty}e^{i\Omega t}
       \langle\delta\hat n_{\vct q}(t)
                          \delta\hat n_{-\vct q}(0)\rangle_\beta\,\dd t
        \nonumber\\
   &=2\pi\sum_s|\mathcal G_s(\vct q)|^2
      [(\bar N_s+1)\delta(\Omega-\omega_s)
                              +\bar N_s\delta(\Omega+\omega_s)].
 \label{eq:phonon-density-ordered-spectrum}
\end{align}
With this angular-frequency convention $S_n$ has units of time. Detailed
balance reads
$S_n(-\vct q,-\Omega)=e^{-\beta\hbar\Omega}S_n(\vct q,\Omega)$
as a spectral-measure identity. A zero overlap hides a mode from this
channel without deleting that mode or its occupation. At $\vct q=0$,
the exact count is $n\Id$, so its connected spectrum vanishes to every
order. The full exponential at nonzero wavevector also contains elastic
and multiphonon structure beyond the first-order expression.

A sudden stiffness change makes the distinction operational. At the
instant of the ideal quench in Section~\ref{sec:lattice-lie-algebra}, the physical $Q,\Pi$ and
their covariances are continuous. The occupation definition and the
reconstruction coefficient proportional to $\omega_s^{-1/2}$ change
together. A new phonon count therefore does not require an instantaneous
jump in centre displacement or number density. Subsequent motion follows
the new Hamiltonian. This is one reason to keep the physical coordinate
and its representation separate when choosing effective variables.

\paragraph{A spatial surface channel beyond total charge.}
For acidic surface populations $\alpha_a$ with site capacities $N_a$,
write $Q_a=Q_a^0-eN_a\alpha_a$. At fixed body geometry, linearizing the declared static or quasistatic
ionic boundary problem of Section~2 gives, at a specified probe,
\begin{equation}
 \delta Q_{\Sigma}=-e\sum_aN_a\delta\alpha_a,
 \qquad
 \delta\psi_p=-e\sum_aN_aG_{pa}\delta\alpha_a .
 \label{eq:s6patch-spatial-surface-readout}
\end{equation}
Here $G_{pa}$ is the linear potential response to charge at site $a$,
with the stated cavity, ionic and dielectric boundaries. If
$\sum_aN_a\delta\alpha_a=0$, total charge is unchanged; the spatially
weighted signal can still vary when the $G_{pa}$ differ. This is a
concrete complementary readout for surface redistributions hidden from
a net-charge measurement. HPV16 VLP studies provide experimental context
for changing surface electrostatic distributions~\cite{Vega2016}; the
coefficients and rates here remain preparation-specific inputs. Sensor
transduction, screening and backaction require calibration. Independently
resolved ionic deviations add their own observation rows; ionic or
reservoir variables complete the charge ledger.

\subsubsection{Scattering contrast and a mode that becomes dark}

In a specified weak single-scattering channel, let the particle's
calibrated coherent amplitude be $f_i(\vct q)$, held fixed during the
interval. Then
\begin{equation}
 \begin{split}
 \hat{\mathcal A}_{\vct q}
   &=\sum_i f_i(\vct q)e^{-i\vct q\cdot
                    (\vct R_i^{(0)}+\hat{\vct u}_i)},\\
 \mathcal G_s^{(f)}(\vct q)
   &=\sum_i f_i(\vct q)e^{-i\vct q\cdot\vct R_i^{(0)}}
                                \vct q\cdot\vct d_{is}.
 \end{split}
 \label{eq:contrast-weighted-mechanical-scattering}
\end{equation}
The leading connected amplitude spectrum replaces $|\mathcal G_s|^2$
by $|\mathcal G_s^{(f)}|^2$ in~\eqref{eq:phonon-density-ordered-spectrum}.
Its units carry $|f|^2$, and its definition uses
$\langle\delta\hat{\mathcal A}_{\vct q}(t)
\delta\hat{\mathcal A}_{\vct q}^\dagger(0)\rangle$.
If all $f_i=f$, then $S_{\mathcal A}=|f|^2S_n$.
Shape, orientation, polarization and the scattering-density distribution
determine these form factors~\cite{Misra2012,Cotterill2002}.
They are properties of the selected channel, rather than universal
mechanical charges.

For the equal-mass pinned pair at reference positions $-a/2,+a/2$ on
one axis, choose common and relative patterns $(1,1)/\sqrt2$ and
$(1,-1)/\sqrt2$. With $k_b=k_0=m\omega_0^2$, their labels mean
$\omega_+=\omega_0$ for common motion and
$\omega_-=\sqrt3\omega_0$ for relative motion. Here the signs label
patterns, rather than the ordered frequencies of the different pair
in Section~\ref{sec:spectral-observables}. Equal contrasts give
\begin{equation}
 |\mathcal G_+^{(f)}(q)|^2
       =\frac{|f|^2\hbar q^2}{m\omega_+}\cos^2(qa/2),
 \qquad
 |\mathcal G_-^{(f)}(q)|^2
       =\frac{|f|^2\hbar q^2}{m\omega_-}\sin^2(qa/2).
 \label{eq:operator-dimer-density-selection}
\end{equation}
At $qa=\pi$ the common mode is dark and the relative mode is visible;
near zero wavevector the relative weight starts at order $q^4$, while
the common weight starts at order $q^2$. This suppression follows from
cancellation of the two displacement phases, without a periodic crystal.
Changing coherent contrast can interchange or soften these selection
rules, as Figure~\ref{fig:density-overlap-and-coherent-contrast} shows.
The plotted envelopes remove the stated frequency and $q^2$ prefactors;
they are not complete intensity or temperature spectra.

\begin{figure}[H]
\centering\begingroup
\definecolor{vltTeal}{HTML}{176B64}
\definecolor{vltBlue}{HTML}{315D80}
\definecolor{vltGold}{HTML}{A87520}
\definecolor{vltPlum}{HTML}{725877}
\definecolor{vltRed}{HTML}{762B38}
\definecolor{vltYellow}{HTML}{D5AD3C}
\definecolor{vltNeutral}{HTML}{777D83}
\setlength{\unitlength}{1pt}
\scalebox{0.90}{%
\input{vlt-figures/figure-201.tex}%
}\endgroup
\caption{One supported pair, two displacement patterns and two spatial channels. With $k_b=k_0=m\omega_0^2$, common and relative frequencies are $\omega_0,\sqrt3\omega_0$. At gain $12\,\mathrm{pt}/a_{\rm osc}$, common motion preserves the link and relative motion compresses it by $2a_{\rm osc}$. Rulers register the same reference spacing $a$ in both rows; supports remain in the equations. For $d_s=\sqrt{\hbar/(2m\omega_s)}$, dividing the count overlaps by $q^2d_s^2$ gives $1\pm\cos(qa)$. Coherent weights $(1,-1)$ interchange those envelopes. Their continuous plotted limits at $q=0$ coexist with a vanishing full density overlap there. Occupation, form factors and instrument response supply the additional measured-spectrum weights.}
\label{fig:density-overlap-and-coherent-contrast}
\end{figure}

Internal deformation or rotation that changes a form factor adds
$\sum_i e^{-i\vct q\cdot\vct R_i^{(0)}}\delta f_i$ at first order.
A capsid-shape or membrane-contrast channel can therefore see an internal
motion that a centre-count channel misses. The geometry and coordinate
extension must be specified before attributing such a signal to a mode.
An isotropic orientation average, if used, must average the calculated
channel rather than assume a single oriented pair describes the sample.

For a Gaussian equal-time coordinate preparation with mean $\bar{\vct u}_i$
and site covariance $\mathsf C_{ii}$, the characteristic function gives
\begin{equation}
 \langle\hat{\mathcal A}_{\vct q}\rangle
   =\sum_i f_i(\vct q)
      e^{-i\vct q\cdot(\vct R_i^{(0)}+\bar{\vct u}_i)}
      e^{-\vct q^{\mathsf T}\mathsf C_{ii}\vct q/2}.
 \label{eq:operator-gaussian-mean-scattering-amplitude}
\end{equation}
The final factor attenuates the coherent mean through Gaussian
displacement fluctuations. Full amplitude correlations also contain
fluctuations about that mean, and an undamped system can have additional
elastic contributions. Intensity, heterodyne and photon-count records
apply their optical and noise maps. Those maps also specify whether
ordered quantum, symmetrized or classical high-occupation spectra are
being compared~\cite{Clerk2010}.

\paragraph{Oriented form factors and responsive surface contrast.}
The fixed $f_i$ approximation can be extended using the same body sites
that generate the interaction. In a declared weak scalar single-scattering
channel, let $\rho_i(\vct a;\vct\chi_i)$ be the body-frame contrast density,
including the chosen hydration and solvent reference. Define
\begin{equation}
 \begin{aligned}
 F_i(\vct k,\vct\chi_i)&=\int\rho_i(\vct a;\vct\chi_i)
                         e^{-i\vct k\cdot\vct a}\,\dd\vct a,\\
 \mathcal A_{\vct q}&=\sum_i e^{-i\vct q\cdot\vct r_i}
                  F_i(\mathsf O_i^{\mathsf T}\vct q,\vct\chi_i).
 \end{aligned}
 \label{eq:s6patch-body-form-factor}
\end{equation}
Coordinate-based macromolecular scattering provides a practical route
to this input~\cite{SvergunBarberatoKoch1995}. The contrast density belongs
to the specified probe; it need not equal the electrostatic site-charge
distribution. Polarization-sensitive scattering requires its corresponding
tensor amplitude and instrument contraction.

\begin{proposition}[Linear body and surface contrast]
\label{prop:s6patch-body-contrast}
At a resting registered reference set
$\vct k_i=(\mathsf O_i^0)^{\mathsf T}\vct q$ and
$f_i=F_i(\vct k_i,\vct\chi_i^0)$. For differentiable form factors,
the first variation is
\begin{equation}
 \boxed{\begin{aligned}
 \delta\mathcal A_{\vct q}
   &=\sum_i e^{-i\vct q\cdot\vct R_i^0}\delta a_i,\\
 \delta a_i&=-if_i\vct q\cdot\vct u_i
       +\vct g_i\cdot\vct\theta_i
       +(\partial_{\vct\chi_i}F_i)\delta\vct\chi_i .
 \end{aligned}}
 \label{eq:s6patch-linear-body-contrast}
\end{equation}
where $\vct g_i=(\mathsf O_i^0\nabla_{\vct k}F_i)\times\vct q$,
and the derivatives are evaluated at the reference.
\end{proposition}
\begin{proof}
Differentiate the centre phase and
$\mathsf O_i^{\mathsf T}\vct q$ using
$\delta\mathsf O_i=[\vct\theta_i]_\times\mathsf O_i^0$.
The scalar triple product gives the displayed angular coefficient.
\end{proof}

Figure~\ref{fig:cmd-complementary-ports} gives a graphical specialization of this boxed relation.

A centre-count channel has $F_i=1$ and no angular term. A radial scalar
form factor likewise has $\vct g_i=0$, whereas an oriented, patterned
capsid can contribute angular contrast. At $\vct q=0$ a rigid rotation
preserves the integrated scalar contrast. Surface redistribution can
change other channels even when every centre remains fixed. This gives
distinct detector rows for the coupled state of Section~5.

An intensity measurement applies a further map. With a coherent reference
$\mathcal A_0$ and a small amplitude increment,
\begin{equation}
 I-|\mathcal A_0|^2
   =2\operatorname{Re}(\mathcal A_0^*\delta\mathcal A)
                         +|\delta\mathcal A|^2 .
 \label{eq:s6patch-amplitude-intensity-map}
\end{equation}
The linear intensity channel selects one real amplitude quadrature.
At a dark reference $\mathcal A_0=0$, intensity starts at second order;
its covariance can require fourth-order amplitude statistics. A calibrated
phase reference can select additional quadratures. Thus amplitude spectra
are passed through the actual optical map before comparison with counts
or intensity fluctuations.

\subsubsection{Other spatial observables and calibrated covariance}

The position observable for a specified particle is
$\hat{\vct R}_i=\vct R_i^{(0)}\Id+\hat{\vct u}_i$.
Its expectation gives a mean history. A joint quantum measurement history
additionally requires a sequential measurement law; equal-time position
statistics do not define that law. Classical trajectories follow directly
from the chosen mechanical initial state and evolution.

For the conservative continuum, stress is
$T_{p\mu}=\mathcal A_{p\mu q\nu}\partial_\nu u_q$ and the
small continuum rotation is $\vct\vartheta=\nabla\times\vct u/2$.
Smearing stress with a specified real probe gives a linear functional
on the elastic form domain; finite-mode quantization gives a real linear
combination of $\hat Q_s$. A finite contact therefore measures a weighted
stress, rather than an unregularized point field. The curl map acts on
$H(\operatorname{curl})$, and its adjoint depends on the boundary term
involving $\vct n\cdot(\vct U^*\times\vct V)$.
The cylinder's mechanical boundary conditions do not automatically make
curl self-adjoint on all vector fields. Torsional mode content is still
calculable without adding an unproved curl eigenbasis. A continuum
rotation gradient is also distinct from a virion's independently retained
rigid orientation. Kinematic helicity alone establishes no topological
protection.

For a classical ensemble, covariance means
$C_{ab}=\mathbb E[(u_a-\bar u_a)(u_b-\bar u_b)]$.
For quantum displacements with finite second moments use
\begin{equation}
 C_{ab}=\frac12\langle
              \{\delta\hat u_a,\delta\hat u_b\}\rangle.
 \label{eq:mechanical-covariance}
\end{equation}
Equal-time displacement components commute, so both constructions give
$\vct\ell^{\mathsf T}\mathsf C\vct\ell\geq0$, the variance of a
real displacement readout. Symmetry alone would not imply positivity;
the variance construction does. Covariance eigenvectors identify
correlated patterns in that preparation. They coincide with mechanical
normal modes only under additional conditions, such as a compatible
mass metric and equilibrium harmonic law.

\paragraph{Which rigid-body coordinates can markers recover?}
A reporter resolves body motion only through its registered geometry.
For a marker arm $\vct b_a=\mathsf O^0\vct a_a$, its local displacement is
\begin{equation}
 \delta\vct x_a=\vct u-[\vct b_a]_\times\vct\theta,
 \qquad \mathsf L_a=(\Id_3,-[\vct b_a]_\times).
 \label{eq:s6patch-marker-observation-matrix}
\end{equation}
\begin{proposition}[Rigid-marker observation rank]
\label{prop:s6patch-marker-rank}
On the six-dimensional unconstrained body tangent, full three-dimensional
positions of one marker, two distinct markers and three noncollinear
markers give stacked observation ranks $3$, $5$ and $6$, respectively.
If the centre is independently known, two nonparallel nonzero marker
arms suffice to recover all three angular components.
\end{proposition}
\begin{proof}
A null motion satisfies $\vct u=\vct b_a\times\vct\theta$ for every
marker. Subtracting two such equations leaves
$(\vct b_a-\vct b_b)\times\vct\theta=0$.
One separation permits rotation about that line, with compensating centre
translation; two nonparallel separations remove it. With a known centre,
the null angular direction must be parallel to every nonzero arm.
\end{proof}
For stacked independent calibrated marker noise, subtract its covariance
from the record covariance. A full-column-rank $\mathsf L$ then permits
$\mathsf C_{\rm body}=\mathsf L^+
(\mathsf C_{\rm rec}-\mathsf C_\epsilon)(\mathsf L^+)^{\mathsf T}$
within this linear instantaneous model. Otherwise only observed
combinations are determined. A projected image uses its actual viewing
matrix and can have lower rank. Single-emitter position/orientation
microscopy demonstrates why these calibrations matter~\cite{Backlund2012};
recovering a complete VLP body additionally requires a known attachment
geometry and any reporter wobble model.

The covariance and Hessian use the same local chart and physical scales.
To linear order at a stationary reference, a nonsingular reparameterization with Jacobian
$\mathsf J_c$ gives $\mathsf C'=\mathsf J_c\mathsf C\mathsf J_c^{\mathsf T}$
and $\mathsf H'=\mathsf J_c^{-\mathsf T}\mathsf H\mathsf J_c^{-1}$.
Consequently raw covariance eigenvectors depend on the coordinate metric.
Broad orientation distributions and discrete surface populations use
their actual probability laws; the inverse-Hessian Gaussian law describes
the localized stable tangent preparation.

Figure~\ref{fig:s56-marker-reconstruction-rank} makes the observation nullspace geometric: two fixed reporters can hide a rigid-body motion that also shifts the centre.

\begin{figure}[H]
\centering\begingroup
\definecolor{vltTeal}{HTML}{176B64}
\definecolor{vltBlue}{HTML}{315D80}
\definecolor{vltGold}{HTML}{A87520}
\definecolor{vltPlum}{HTML}{725877}
\definecolor{vltRed}{HTML}{762B38}
\definecolor{vltYellow}{HTML}{D5AD3C}
\definecolor{vltNeutral}{HTML}{777D83}
\setlength{\unitlength}{1pt}
\scalebox{0.90}{%
\input{vlt-figures/figure-297.tex}%
}\endgroup
\caption{How registered markers constrain a rigid ideal VLP. Coloured points are distinct reporters; facets register body orientation. Each panel uses the same reference body and marker geometry. With full three-dimensional marker positions, the stacked tangent observation matrix has ranks $3$, $5$ and $6$ for one, two distinct and three noncollinear markers. Filled bar segments count observed body directions. In (b), a finite rotation of $0.38$ rad about the burgundy dashed line through markers 1 and 2 leaves both reporters exactly fixed while the centre moves from the open to the filled point; the grey wireframe is the reference. This exaggerated finite motion illustrates the one-dimensional infinitesimal nullspace, rather than an applied force protocol. Projection, localization error and reporter attachment enter the calibrated observation map before using these ranks. See Proposition~\ref{prop:s6patch-marker-rank}.}
\label{fig:s56-marker-reconstruction-rank}
\end{figure}

For instantaneous calibrated classical records
$\vct y_{\rm det}=\mathsf L_{\rm det}\vct u+\vct\epsilon_{\rm det}$,
independent zero-mean detector noise gives
\begin{equation}
 \mathsf C_{\rm rec}
    =\mathsf L_{\rm det}\mathsf C_u\mathsf L_{\rm det}^{\mathsf T}
                                               +\mathsf C_\epsilon.
 \label{eq:calibrated-record-covariance}
\end{equation}
The rows of $\mathsf L_{\rm det}$ select structural components and convert
them to calibrated displacement channels. For a common real exposure
kernel $h$ with integral one, a stationary record instead obeys
\begin{equation}
 \begin{aligned}
 \mathsf C_{\rm rec}
  &=\int_{-\infty}^{\infty}\frac{\dd\Omega}{2\pi}
      |H_{\rm det}(\Omega)|^2\mathsf L_{\rm det}
         \mathsf S_u(\Omega)\mathsf L_{\rm det}^{\mathsf T}
        +\mathsf C_\epsilon,\\
 H_{\rm det}(\Omega)&=\int h(s)e^{i\Omega s}\,\dd s.
 \end{aligned}
 \label{eq:operator-exposure-filtered-covariance}
\end{equation}
A uniform exposure of duration $\tau_{\rm exp}$ has
$|H_{\rm det}|^2=\operatorname{sinc}^2(\Omega\tau_{\rm exp}/2)$,
where $\operatorname{sinc}x=\sin x/x$.
Exposure suppresses fast motion, whereas independent localization noise
adds variance~\cite{Berglund2010}. Treating either contribution as equilibrium mechanical
variance can bias a curvature inference in opposite directions. Different
channel filters or correlated noise require their full matrix observation
law, following Sections~\ref{sec:damped-probe-fluctuations} and
\ref{sec:phase-window-coherence}.

Figure~\ref{fig:exposure-filtered-operator-covariance} gives an exact two-mode example of this observation law, including an independently declared localization variance.

\begin{figure}[H]
\centering
\begingroup
\definecolor{vlfGreen}{HTML}{176B64}
\definecolor{vlfDarkRed}{HTML}{762B38}
\definecolor{vlfRed}{HTML}{762B38}
\definecolor{vlfBlue}{HTML}{315D80}
\definecolor{vlfCopper}{HTML}{762B38}

\setlength{\unitlength}{1pt}
\scalebox{0.80}{%
\input{vlt-figures/figure-202.tex}%
}%
\endgroup
\caption{Exposure filtering and independent localization noise for an equal-mass pinned pair, $k_b=k_0=m\omega_0^2>0$, in undamped classical Gibbs equilibrium. (a) The observation chain specifies uniform exposure averaging followed by independent zero-mean localization noise. (b) Set $a=\omega_0\tau_{\rm exp}$ and $s_\pm=\operatorname{sinc}(a/2),\operatorname{sinc}(\sqrt3a/2)$. In units $k_{\rm B}T/(m\omega_0^2)$, the mechanical site variance is $s_+^2/2+s_-^2/6$ and the cross-covariance is $s_+^2/2-s_-^2/6$, with zero-exposure limits $2/3$ and $1/3$. An illustrative calibrated record-noise variance $0.05$ raises the diagonal alone. Its exposure independence is stipulated for this example; measured detector noise supplies its own law. Curves are ensemble covariances, so reduced recorded width can arise from averaging at unchanged stiffness.}
\label{fig:exposure-filtered-operator-covariance}
\end{figure}

\subsubsection{Relaxed, clamped, and dynamically hidden coordinates}

For a full-line classical harmonic Gibbs reference at known $T>0$,
assume $\mathsf K>0$, so $\mathsf C_u=k_{\rm B}T\mathsf K^{-1}$.
Split coordinates into observed $o$ and unobserved $h$. Completing the
square in $u_h$ and integrating the hidden Gaussian gives
\begin{equation}
 \boxed{\begin{aligned}
 \mathsf C_{oo}&=k_{\rm B}T\mathsf K_{\rm eff}^{-1},\\
 \mathsf K_{\rm eff}&=\mathsf K_{oo}
        -\mathsf K_{oh}\mathsf K_{hh}^{-1}\mathsf K_{ho}.
 \end{aligned}}
 \label{eq:observed-covariance-effective-curvature}
\end{equation}
The Schur complement $\mathsf K_{\rm eff}$ is the relaxed free-energy
curvature seen by the observed variables. The subtracted term is
nonnegative: hidden relaxation cannot stiffen this positive quadratic
reference relative to fixing those hidden coordinates. At fixed $u_h$,
the conditional mean shifts, but the conditional covariance is
$k_{\rm B}T\mathsf K_{oo}^{-1}$.
This distinction is the statistical counterpart of the energy reduction
in~\eqref{eq:architecture-completed-square}.

An internal coordinate that relaxes during loading changes an envelope
marker's effective stiffness; clamping it defines a different protocol.
Figure~\ref{fig:final-hidden-coordinate-curvature} compares their
calibrated widths. The fluctuation--compliance inference follows
the same mechanics as~\eqref{eq:probe-variance-compliance}
\cite{CallenWelton1951,Cotterill2002}. Contact-network models of protein
fluctuations offer another biophysical example~\cite{Bahar1997}, at their
own residue resolution and interaction scale. For either resolution,
multiple hidden networks can share one relaxed Schur complement.

\begin{figure}[H]
\centering
\begingroup
\definecolor{vlfGreen}{HTML}{176B64}
\definecolor{vlfDarkRed}{HTML}{762B38}
\definecolor{vlfRed}{HTML}{762B38}
\definecolor{vlfBlue}{HTML}{315D80}
\definecolor{vlfCopper}{HTML}{762B38}

\setlength{\unitlength}{1pt}
\scalebox{0.80}{%
\input{vlt-figures/figure-203.tex}%
}%
\endgroup
\caption{Hidden-coordinate relaxation in the quadratic Gibbs model $U=k_0(x^2-2\rho xy+y^2)/2$, $\rho=4/5$, with $\ell_T=\sqrt{k_{\rm B}T/k_0}$. (a) Contours $U/(k_{\rm B}T)=1/2,2$ surround the minimizing line $y=\rho x$. (b) The relaxed marginal has $\operatorname{Var}x=25\ell_T^2/9$ and effective curvature $9k_0/25$; clamping $y=0$ gives variance $\ell_T^2$ and curvature $k_0$. Both displayed densities, multiplied by $\ell_T$, are normalized on the full line. Color and line style distinguish the two protocols. Their difference is the hidden relaxation included by the observation, as quantified in \eqref{eq:observed-covariance-effective-curvature}; the underlying quadratic coefficients are identical.}
\label{fig:final-hidden-coordinate-curvature}
\end{figure}

Static elimination is not generally a closed dynamics. For block-diagonal
inertia, zero hidden forcing and conservative harmonic response,
eliminating $u_h$ at complex frequency $z$ gives
\begin{equation}
 \mathsf A_{\rm eff}(z)=\mathsf K_{oo}-z^2\mathsf M_{oo}
       -\mathsf K_{oh}(\mathsf K_{hh}-z^2\mathsf M_{hh})^{-1}
                                                    \mathsf K_{ho}.
 \label{eq:operator-observed-dynamic-curvature}
\end{equation}
Where this inverse exists, the observed susceptibility is
$\mathsf A_{\rm eff}(z)^{-1}$. Its zero-frequency stiffness is
$\mathsf K_{\rm eff}$, but its frequency dependence retains the hidden
inertia and couplings. With inertia cross blocks, one must take the
Schur complement of the complete pencil $\mathsf K-z^2\mathsf M$.
Initial hidden motion also supplies a prepared force/history term in a
time-domain reduction. Thus a useful low-dimensional coordinate and its
static curvature do not by themselves form a sufficient instantaneous
mechanical state. Discarded motion that affects the predictions must reappear as explicit
coordinates, memory or preparation-dependent forcing and fluctuations
\cite{Mori1965}.

These inverse-curvature identities require the full-line Gaussian basin.
Engaged hard-wall or excluded-volume restrictions truncate its law; free
modes, nonlinear excursions and nonequilibrium drift need their own
treatment. The fluctuation scale must fit the admissible harmonic basin,
as assessed in~\eqref{eq:harmonic-basin-excursion-bound}. A state with
sustained active oscillation can have phase-dependent means and
covariances; replacing its variance by $k_{\rm B}T/K$ would not define
its restoring law without an equilibrium justification.

\subsubsection{What a complete pair and a heterogeneous sample can identify}

\begin{example}[Separating pinning from a contact coupling]
Retain two complete scalar translations with mass $m>0$ and
\begin{equation}
 V=\frac{k_0}{2}(u_1^2+u_2^2)
       +\frac{k_b}{2}(u_1-u_2)^2,\qquad k_0>0,\quad k_b\geq0.
 \label{eq:classical-pinned-pair-readouts}
\end{equation}
Here pinning is the declared quadratic confinement and $k_b$ the contact
curvature, both in newtons per metre. The normalized readouts
$Y_\pm=(u_1\pm u_2)/\sqrt2$ have frequencies
$\sqrt{k_0/m}$ and $\sqrt{(k_0+2k_b)/m}$.
At classical Gaussian equilibrium,
$\operatorname{Var}(Y_+)=k_{\rm B}T/k_0$ and
$\operatorname{Var}(Y_-)=k_{\rm B}T/(k_0+2k_b)$, so
\begin{equation}
 k_0=\frac{k_{\rm B}T}{\operatorname{Var}(Y_+)},\qquad
 k_b=\frac{k_{\rm B}T}{2}
     \left[\frac1{\operatorname{Var}(Y_-)}
                       -\frac1{\operatorname{Var}(Y_+)}\right].
 \label{eq:pinned-pair-covariance-inference}
\end{equation}
The nonnegative-link model predicts
$\operatorname{Var}(Y_-)\leq\operatorname{Var}(Y_+)$.
With independently known $m$, the inferred curvatures must also predict
the response frequencies and static compliances. A violation challenges
the declared model, equilibrium or calibration; it does not identify one
specific missing biological mechanism. A single site's variance mixes
the two patterns and cannot make this separation by itself.
Figure~\ref{fig:operator-probe-covariance} varies the readout direction
for $k_b=k_0$, the same pinned reference used above. Partial observation
of a larger assembly instead has the effective-curvature interpretation.
\end{example}

\begin{figure}[H]
\centering
\begingroup
\definecolor{vlfGreen}{HTML}{176B64}
\definecolor{vlfDarkRed}{HTML}{762B38}
\definecolor{vlfRed}{HTML}{762B38}
\definecolor{vlfBlue}{HTML}{315D80}
\definecolor{vlfCopper}{HTML}{762B38}

\setlength{\unitlength}{1pt}
\scalebox{0.80}{%
\input{vlt-figures/figure-204.tex}%
}%
\endgroup
\caption{A calibrated probe rotates across the same complete pinned pair, with $\mathsf K=k_0\left(\begin{smallmatrix}2&-1\\-1&2\end{smallmatrix}\right)$ and classical Gibbs covariance $\mathsf C_u=(k_{\rm B}T/3k_0)\left(\begin{smallmatrix}2&1\\1&2\end{smallmatrix}\right)$. For $Y_\theta=\cos\theta\,u_1+\sin\theta\,u_2$, (a) common and relative weights are $w_\pm=(1\pm\sin2\theta)/2$, and (b) the normalized variance is $k_0\operatorname{Var}Y_\theta/(k_{\rm B}T)=w_++w_-/3$. Marked probes at $-\pi/4,0,\pi/4$ select relative motion, site 1 and common motion, giving $1/3,2/3,1$. Solid teal and dashed burgundy distinguish the weights; the variance curve summarizes their combination. Both positive modes remain in the preparation. Detector noise and time averaging enter the measured record separately.}
\label{fig:operator-probe-covariance}
\end{figure}

\subsubsection{How accurately covariance determines a relaxed curvature}

The pair example determines two curvatures because both collective
coordinates are observed and the equilibrium law is known. For a larger
assembly, inverse covariance has a similarly precise but resolution-
dependent meaning. For the observed block above, let
$\mathsf C=\mathsf C_{oo}>0$ and
$\mathsf K_{\rm stat}=k_{\rm B}T\mathsf C^{-1}
=\mathsf K_{\rm eff}$ at known $T>0$. It is the curvature of the
observed marginal free energy. With every retained coordinate observed,
it is the complete $\mathsf K$; with hidden coordinates, their relaxed
contribution remains in $\mathsf K_{\rm eff}$.

\begin{proposition}[Stability of covariance-derived curvature]
\label{prop:s67-covariance-curvature-stability}
Use this fixed full-line classical harmonic Gibbs reference, with
calibrated observed coordinates and known temperature. Let
$\widetilde{\mathsf C}=\mathsf C+\Delta\mathsf C$ be a real symmetric
covariance estimate and suppose
\begin{equation}
 \varepsilon=\|\mathsf C^{-1/2}\Delta\mathsf C
                                  \mathsf C^{-1/2}\|_2<1.
 \label{eq:s67-relative-covariance-error}
\end{equation}
Then $\widetilde{\mathsf C}>0$, and the inferred curvature
$\widetilde{\mathsf K}_{\rm stat}
=k_{\rm B}T\widetilde{\mathsf C}^{-1}$ satisfies
\begin{equation}
 \begin{gathered}
 \frac{\mathsf K_{\rm stat}}{1+\varepsilon}
       \leq\widetilde{\mathsf K}_{\rm stat}
       \leq\frac{\mathsf K_{\rm stat}}{1-\varepsilon},\\
 \|\mathsf K_{\rm stat}^{-1/2}
       (\widetilde{\mathsf K}_{\rm stat}-\mathsf K_{\rm stat})
                      \mathsf K_{\rm stat}^{-1/2}\|_2
          \leq\frac{\varepsilon}{1-\varepsilon}.
 \end{gathered}
 \label{eq:s67-curvature-inference-error-bound}
\end{equation}
The matrix inequalities are in the nonnegative quadratic-form order.
\end{proposition}
\begin{proof}
Put $\mathsf E_C=\mathsf C^{-1/2}\Delta\mathsf C
\mathsf C^{-1/2}$. Its eigenvalues lie in
$[-\varepsilon,\varepsilon]$, so $\Id+\mathsf E_C>0$ and
$(1+\varepsilon)^{-1}\Id\leq(\Id+\mathsf E_C)^{-1}
\leq(1-\varepsilon)^{-1}\Id$. Congruence with
$\sqrt{k_{\rm B}T}\mathsf C^{-1/2}$ gives the first line.
The normalized curvature difference in the second line equals
$(\Id+\mathsf E_C)^{-1}-\Id$. Its eigenvalues are
$-e/(1+e)$ for $|e|\leq\varepsilon$, giving the stated norm bound.
This is the positive-matrix inverse perturbation calculation in the
physical covariance metric~\cite{HornJohnson2013}.
\end{proof}

The error scale is the covariance along each resolved direction, not
an absolute position tolerance common to all patterns. A stiff direction
can have a small thermal width, making a fixed localization error large
in this relative metric. Noise subtraction and exposure correction in
\eqref{eq:calibrated-record-covariance}--
\eqref{eq:operator-exposure-filtered-covariance} therefore precede
the curvature interpretation. The bound is deterministic: a sampling or
calibration analysis must supply $\varepsilon$ for the actual correlated
record. It does not convert the number of frames into independent
equilibrium samples.

For variable integration, these bounds quantify how covariance errors
affect the relaxed cost of specified collective displacements and their
compliances. Frequency, force and history information additionally test
hidden contacts and dynamical closure through
\eqref{eq:operator-observed-dynamic-curvature}. Measurement resolution,
static restoring cost and predictive state are thus related parts of
the mechanical specification.

Figure~\ref{fig:added-directional-covariance} shows a minimal reconstruction of a classical covariance from three directional projections.

\begin{figure}[H]
\centering\begingroup
\definecolor{vltTeal}{HTML}{176B64}
\definecolor{vltBlue}{HTML}{315D80}
\definecolor{vltGold}{HTML}{A87520}
\definecolor{vltPlum}{HTML}{725877}
\definecolor{vltRed}{HTML}{762B38}
\definecolor{vltYellow}{HTML}{D5AD3C}
\definecolor{vltNeutral}{HTML}{777D83}
\setlength{\unitlength}{1pt}
\scalebox{0.80}{%
\input{vlt-figures/figure-205.tex}%
}\endgroup
\caption{Directional reconstruction of $\mathsf C/\ell_\star^2=\left(\begin{smallmatrix}2&1\\1&1\end{smallmatrix}\right)$. Equal spatial scales show $\vct u^{\mathsf T}\mathsf C^{-1}\vct u=1$ and dotted principal directions. Colored arrows use one gain for unit readout directions. The scan $v_\alpha=\operatorname{Var}(\vct n_\alpha\cdot\vct u)$ links those directions to three measured widths; the displayed polarization identity recovers $C_{xy}$. Extrema are $(3\pm\sqrt5)\ell_\star^2/2$. These are second moments of one preparation with detector contributions modeled or removed. A confidence probability for the ellipse additionally needs an ensemble law.}
\label{fig:added-directional-covariance}
\end{figure}

Structural variation adds a different source of covariance. For registered
readouts conditioned on a structural or preparation class $c$,
\begin{equation}
 \mathsf C_{\rm ensemble}
      =\mathbb E_c[\operatorname{Cov}(\vct u\mid c)]
          +\operatorname{Cov}_c(\mathbb E[\vct u\mid c]).
 \label{eq:structural-class-total-covariance}
\end{equation}
The second term measures variation between class means. Cryo-ET
resolves measles structural layers and particle heterogeneity
\cite{Ke2018}; M coverage and RNP organization can define classes.
Time correlations require dynamic records on appropriately matched
preparations, with registration, exposure and detector errors retained.
Conditioning on class separates within-basin motion from differences
between specimens before assigning a pooled width to thermal mechanics.

\subsubsection{Full canonical covariance and phase-sensitive coherence}

Position covariance alone does not specify a quantum oscillator state.
Choose $\omega_{\rm ref}>0$ and form
$\hat{\vct R}_{\rm can}=(\sqrt{\omega_{\rm ref}}\hat{\vct Q},
\hat{\vct\Pi}/\sqrt{\omega_{\rm ref}})^{\mathsf T}$.
Every component has units of square root of action. For finite second
moments define
$V_{jk}=\langle\{\delta\hat R_{{\rm can},j},
\delta\hat R_{{\rm can},k}\}\rangle/2$. Positivity of
$\langle(\vct z\cdot\delta\hat{\vct R}_{\rm can})^\dagger
(\vct z\cdot\delta\hat{\vct R}_{\rm can})\rangle$ for every
complex $\vct z$ gives
\begin{equation}
 \mathsf V+\frac{i\hbar}{2}\boldsymbol\Omega_{\rm can}\geq0,
 \qquad
 \boldsymbol\Omega_{\rm can}=
          \begin{pmatrix}0&\Id_r\\-\Id_r&0\end{pmatrix}.
 \label{eq:phonon-full-covariance-condition}
\end{equation}
For one canonical pair this contains $\det\mathsf V\geq\hbar^2/4$.
The covariance ellipse in Figure~\ref{fig:canonical-displacement-and-squeezing}
is its dimensionless vacuum/squeezed specialization. The condition is
necessary for every quantum state with those moments; for Gaussian states
it is the covariance admissibility condition, but moments do not generally
determine a non-Gaussian state. A positive classical covariance need not
satisfy this stronger canonical inequality. Projected readouts
$\mathsf L\hat{\vct R}_{\rm can}$ carry the commutator matrix
$\mathsf L\boldsymbol\Omega_{\rm can}\mathsf L^{\mathsf T}$,
rather than acquiring new independent canonical pairs.

Phase-sensitive mode correlations can be specified by
\begin{equation}
 \hat X_{st}=\frac{\hat b_s^\dagger\hat b_t
                              +\hat b_t^\dagger\hat b_s}{2},
 \qquad
 \hat Y_{st}=\frac{\hat b_s^\dagger\hat b_t
                              -\hat b_t^\dagger\hat b_s}{2i}.
 \label{eq:mode-coherence-quadratures}
\end{equation}
Their expectations recover the real and imaginary parts of
$\langle\hat b_s^\dagger\hat b_t\rangle$.
Together with occupations and anomalous moments
$\langle\delta\hat b_s\delta\hat b_t\rangle$, they separate
normal-mode coherence from quadrature squeezing. Classical complex mode
products describe the corresponding wave phases without supplying a
quantum state. Correlated envelope and RNP displacements, correlated
whole-virion translations, and a quantum correlation witness are therefore
different claims. Shared forcing can correlate uncoupled coordinates;
an active oscillation can have a reproducible phase without establishing
quantum squeezing or entanglement.

Inference now has a concrete sequence: specify the structural coordinate
and force, predict response and fluctuations from the same mechanics,
apply the spatial and temporal readout, and compare calibrated records
across preparations. Contact and defect sensitivity ask which changes
those records can resolve.

\subsection{Target susceptibility, defects, and spectral sensitivity}
\label{sec:operator-target-defects}

Sensitivity predicts how response and covariance change with a contact,
support, inertia or internal coupling. A spectral shift constrains those
parameters through the declared force and detector maps. Its molecular
interpretation depends on the proposed perturbation model.

\subsubsection{A control changes an energy, not just a mode label}

At fixed physical coordinates and masses, let the dimensionless control
$\lambda$ select $\mathsf K(\lambda)=\mathsf K_0+\lambda\mathsf K_{\rm T}$.
The extra potential is
$V_{\rm T}=\lambda\vct u^{\mathsf T}\mathsf K_{\rm T}\vct u/2$.
The symbol \emph{target} identifies this supplied mechanical interaction;
it is not an infectivity, binding-affinity or biochemical-rate variable.
Its work-conjugate observable and driven energy balance are
\begin{equation}
 \hat{\mathcal V}_{\rm T}
       =\frac{\partial\hat H}{\partial\lambda}
       =\frac12\hat{\vct u}^{\mathsf T}\mathsf K_{\rm T}\hat{\vct u},
 \qquad
 \frac{\dd}{\dd t}\langle\hat H(\lambda(t))\rangle
       =\Dt{\lambda}\,\langle\hat{\mathcal V}_{\rm T}\rangle.
 \label{eq:target-generalized-work}
\end{equation}
The controller's generalized force has the opposite sign to
$\partial_\lambda H$. The balance assumes unitary driven evolution,
the necessary moments/regularity, and no other work channel. Its classical
counterpart follows by differentiating the same energy along Hamilton's
equations. A fixed $\lambda$ selects a conservative reference; varying it
supplies or extracts explicitly accounted work.

The derivative holds the physical coordinate representation fixed.
Reexpressing the state in instantaneous modes does not remove controller
work. In particular the physical quadratic Hamiltonian includes its
$\lambda$-dependent vacuum offset. Using
$\hat H_{\rm ph}=\hat H_0-E_{\rm vac}(\lambda)\Id$ instead requires
restoring $\Dt{\lambda}\,\partial_\lambda E_{\rm vac}$ in the work
account. If the control is an internal dynamical resource rather than a
prescribed protocol, its own coordinates and energy must be added, as in
the active closure of Section~\ref{sec:viscoelastic-dynamics}.

\subsubsection{Sensitivity along an equilibrated reference family}

A change in assigned charge, support or contact law can move the
force-balanced arrangement as well as change its curvature. The
appropriate derivative is then taken along that family of arrangements.
This is the total derivative used in the earlier lattice sensitivity
formulas, now expressed in the physical coordinates from which the
Hessian is assembled.

\begin{proposition}[Relaxation of the reference under a parameter change]
\label{prop:s68-equilibrium-family-sensitivity}
Let $V(\vct q,\lambda)$ be a $C^3$ potential in a fixed smooth
admissible coordinate chart, with dimensionless parameter $\lambda$.
Suppose an equilibrium $\vct q_\star$ is isolated in that chart and
its constrained Hessian $\mathsf K_\star=\nabla_q^2V>0$.
Locally there is a unique differentiable equilibrium branch, with
\begin{equation}
 \begin{aligned}
 \frac{\dd\vct q_\star}{\dd\lambda}
    &=-\mathsf K_\star^{-1}
               \partial_\lambda\nabla_q V(\vct q_\star,\lambda),\\
 \frac{\dd\mathsf K_\star}{\dd\lambda}
    &=\partial_\lambda\nabla_q^2 V(\vct q_\star,\lambda)
       +\nabla_q^3V(\vct q_\star,\lambda)
                     \left[\frac{\dd\vct q_\star}{\dd\lambda}\right].
 \end{aligned}
 \label{eq:s68-equilibrium-and-curvature-derivatives}
\end{equation}
The last term is the third-derivative tensor contracted with the
reference displacement per unit parameter.
\end{proposition}
\begin{proof}
The implicit-function theorem applies to
$\nabla_qV(\vct q_\star,\lambda)=0$ because its coordinate
Jacobian is the invertible constrained Hessian. Differentiating that
identity gives the first line. The chain rule applied to
$\nabla_q^2V(\vct q_\star(\lambda),\lambda)$ gives the second.
All derivatives are evaluated in the same admissible chart.
\end{proof}

The curvature derivative separates changed interactions at fixed
positions from changes in separation, direction and prestress after
relaxation. Both enter $\delta\mathsf K$ in
\eqref{eq:mass-stiffness-mode-perturbation}; inertia uses its total
derivative as well. These are clamped and equilibrated comparisons of
the same physical law. Work within one admissible chart, separating
free rigid directions before inversion. Engaging a new constraint needs
a new chart. A soft restoring direction can amplify the geometry change
under a small constitutive perturbation.

These are derivatives between equilibrated references. Along the branch,
$\dd V(\vct q_\star(\lambda),\lambda)/\dd\lambda
=\partial_\lambda V$, because the equilibrium gradient vanishes.
The controller work in~\eqref{eq:target-generalized-work} therefore
continues to use the partial derivative at fixed physical coordinates;
reference relaxation is not an additional independent work term.

\subsubsection{Frequency shifts and the directions that carry them}

For a simple positive eigenfrequency, normalize
$\mathsf K\vct U_s=\omega_s^2\mathsf M\vct U_s$ by
$\vct U_s^\dagger\mathsf M\vct U_s=1$.
Differentiating the generalized eigenproblem and canceling the two
eigenvector-derivative terms gives
\begin{equation}
 \delta\omega_s
       =\frac{\vct U_s^\dagger
           (\delta\mathsf K-\omega_s^2\delta\mathsf M)\vct U_s}
                          {2\omega_s}+O(\epsilon^2),
 \label{eq:mass-stiffness-mode-perturbation}
\end{equation}
where the smooth perturbations are $O(\epsilon)$ and the eigenvalue
remains separated. Added curvature and added inertia enter differently:
the first raises a mode's restoring cost when positive along its pattern,
while positive added inertia lowers its frequency at fixed curvature.
Geometry, attachment or hydration can change several inputs together;
the formula does not assign their microscopic origins from the shift alone.

For fixed mass and an orthogonal phase convention, a simple
mass-normalized pattern changes by
\begin{equation}
 \delta\ket s=\sum_{t\ne s}
       \frac{\bra t\delta\mathsf D\ket s}{\omega_s^2-\omega_t^2}
                    \ket t+O(\epsilon^2).
 \label{eq:mode-vector-perturbation}
\end{equation}
The sum uses the full conservative basis, including separated zero modes
if present. Small frequency gaps can strongly rotate a pattern and hence
its visibility in a detector. At degeneracy, diagonalize the perturbation
inside the degenerate sector; near a crossing, retain the coupled subspace.
A rapidly changing eigenvector is not automatically a mechanical
instability. Stability concerns the restoring spectrum and the permitted
coordinate domain.

The positive frequency \emph{operator} can be more regular than a chosen
eigenbasis. On a fixed strictly positive coordinate sector, differentiation
of $\mathsf W^2=\mathsf D$ gives the Sylvester equation
\begin{equation}
 \mathsf W\,\delta\mathsf W+\delta\mathsf W\,\mathsf W=\delta\mathsf D,
 \qquad
 \langle s|\delta\mathsf W|t\rangle
       =\frac{\langle s|\delta\mathsf D|t\rangle}{\omega_s+\omega_t}.
 \label{eq:operator-frequency-frechet-derivative}
\end{equation}
It is not entrywise differentiation of a square root. The sum of positive
frequencies in the denominator stays nonzero at a positive degeneracy,
unlike the eigenvector gap. Equivalently
$\delta\mathsf W=\int_0^\infty e^{-t\mathsf W}\delta\mathsf D
e^{-t\mathsf W}\,\dd t$, so
$\|\delta\mathsf W\|_2\leq\|\delta\mathsf D\|_2/(2\omega_{\min})$.
Approaching a soft zero mode removes that uniform control and requires
the zero-mode treatment already separated from oscillator occupation.

\subsubsection{Contact loading, detection, and an exact response update}

Let $y_{ij}=\vct\ell_{ij}^{\mathsf T}\vct u$ be the extension of one
declared longitudinal contact, with site blocks $\vct n_{ij}$ and
$-\vct n_{ij}$. Adding $k_{ij}y_{ij}^2/2$ changes the Hessian by
$k_{ij}\vct\ell_{ij}\vct\ell_{ij}^{\mathsf T}$. At fixed reference,
direction and inertia,
\begin{equation}
 \boxed{\frac{\partial\omega_s}{\partial k_{ij}}
       =\frac{|\vct\ell_{ij}^{\mathsf T}\vct U_s|^2}{2\omega_s}\geq0.}
 \label{eq:contact-extension-frequency-sensitivity}
\end{equation}
Relative extension selects sensitivity. In the pinned equal-mass pair,
common motion does not extend the linking spring, so
$\partial_{k_b}\omega_{\rm com}=0$; relative motion gives
$\partial_{k_b}\omega_{\rm rel}=1/(m\omega_{\rm rel})$.
A common-only force/readout can therefore miss the very branch that
identifies the link curvature. Figure~\ref{fig:contact-loading-and-frequency-sensitivity}
separates loading from visibility. This monotonicity is for the specified
positive spring addition. Changing prestress or a biological interaction
can also alter directions and force balance, requiring the full Hessian
perturbation rather than this one block.

\begin{figure}[H]
\centering\begingroup
\definecolor{vltTeal}{HTML}{176B64}
\definecolor{vltBlue}{HTML}{315D80}
\definecolor{vltGold}{HTML}{A87520}
\definecolor{vltPlum}{HTML}{725877}
\definecolor{vltRed}{HTML}{762B38}
\definecolor{vltYellow}{HTML}{D5AD3C}
\definecolor{vltNeutral}{HTML}{777D83}
\setlength{\unitlength}{1pt}
\scalebox{0.90}{%
\input{vlt-figures/figure-206.tex}%
}\endgroup
\caption{A contact loads only the relative pattern of the supported equal-mass pair. Gray outlines and rulers locate the common reference; rigid bodies and arrows use $12\,\mathrm{pt}/a_{\rm osc}$, with reference spacing $a$. Common motion leaves the link unchanged, while the opposed shifts compress it by $2a_{\rm osc}$. Omitted supports have curvature $k_0>0$. At $\omega_0=\sqrt{k_0/m}$ the exact branches are $1$ and $\sqrt{1+2k_b/k_0}$ after normalization. At $k_b=0$, the displayed patterns diagonalize the link perturbation in the degenerate sector. Curves describe positive longitudinal curvature added at fixed geometry and inertia. The boxed result illustrated here is~\eqref{eq:contact-extension-frequency-sensitivity}.}
\label{fig:contact-loading-and-frequency-sensitivity}
\end{figure}

The same parameter change propagates into the measured response. For
$\mathsf R(z)=(\mathsf K-z^2\mathsf M)^{-1}$ away from poles,
differentiating the inverse gives
\begin{equation}
 \delta\mathsf R(z)
       =-\mathsf R(z)(\delta\mathsf K-z^2\delta\mathsf M)\mathsf R(z)
           +O(\epsilon^2).
 \label{eq:operator-response-parameter-derivative}
\end{equation}
Fixed drive and detector maps select the response derivative's matrix
element. Changed attachments or contrast add their own derivatives,
allowing a measured amplitude change to be assigned to mechanics,
loading or visibility.

For the rank-one spring update, the stronger finite identity is
\begin{equation}
 \mathsf R_k(z)=\mathsf R_0(z)
      -\frac{k\,\mathsf R_0(z)\vct\ell\vct\ell^{\mathsf T}\mathsf R_0(z)}
                     {1+k\vct\ell^{\mathsf T}\mathsf R_0(z)\vct\ell}.
 \label{eq:operator-contact-resolvent-update}
\end{equation}
It follows by multiplying by
$\mathsf K_0+k\vct\ell\vct\ell^{\mathsf T}-z^2\mathsf M$;
the displayed inverse and denominator must exist. For static response,
$\mathsf K_0>0$ and $k\geq0$ make the denominator positive.
The correction to any diagonal calibrated compliance is then nonpositive,
and vanishes if that probe has zero cross-compliance with the contact.
At finite frequency, interference and shifted poles prevent the same
simple ordering of response magnitudes.

\subsubsection{Separating changing mechanics from changing channels}
\label{sec:s6-readout-transient-sensitivity}

The static resolvent update has a temporal counterpart that separates
where a perturbation enters a registered pulse experiment. Work in
common full-pool physical coordinates, holding registered event times
and continuation maps fixed. For bounded differentiable coefficient
families, write $\delta$ for their first parameter derivative. Then
\begin{equation}
 \begin{aligned}
 \delta\mathsf H(t,s)
  ={}&\delta\mathsf L(t)\mathsf T(t,s)\mathsf B(s)
       +\mathsf L(t)\mathsf T(t,s)\delta\mathsf B(s)\\
    &+\mathsf L(t)\int_s^t\mathsf T(t,r)\delta\mathsf A(r)
                   \mathsf T(r,s)\,\dd r\,\mathsf B(s).
 \end{aligned}
 \label{eq:s6-readout-transient-sensitivity}
\end{equation}
Differentiating the propagator equation and applying variation of
constants proves the integral term. If a physical continuation map
$\mathsf J_*$ also changes, add
$\mathsf L(t)\mathsf T(t,t_*^+)\delta\mathsf J_*
\mathsf T(t_*^-,s)\mathsf B(s)$ for each crossed event.
The detector filter acts on these terms as in
\eqref{eq:s6-readout-nonstationary-detector}. A changed contact enters
$\delta\mathsf A$ through its assembled Hessian and medium couplings;
changed force attachment enters $\delta\mathsf B$; changed illuminated
particles or contrast enters $\delta\mathsf L$. Their different insertion
times allow several controlled pulses to constrain different causes.

An exact ambiguity shows why crossing the event can add information.
In episode $g$, an unknown scalar receiver gain and source calibration
give $\mathsf H_{gg}=\mathsf L_g\mathsf T_g\mathsf B_g$.
For any nonzero real $c_g$, the replacements
$\mathsf L_g\mapsto c_g\mathsf L_g$ and
$\mathsf B_g\mapsto\mathsf B_g/c_g$ preserve every within-episode
response. For a pulse in $g$ received after recruitment into $h$,
\begin{equation}
 \mathsf H_{hg}(t,s)
   =\mathsf L_h\mathsf T_h(t,t_*)\mathsf J_{hg}
                         \mathsf T_g(t_*,s)\mathsf B_g
   \ \mapsto\ \frac{c_h}{c_g}\mathsf H_{hg}(t,s).
 \label{eq:s6-readout-cross-event-gain}
\end{equation}
A nonzero cross-event response therefore constrains the relative gains
when the physical continuation and dynamics are independently specified.
A common absolute scale still requires calibration. Fitting an arbitrary
continuation map could hide the same ambiguity again; pure membership
registration uses the identity in common physical coordinates. In
practice, persistent marker channels, force calibration and the recorded
newcomer positions make this comparison tangible. The pulse probes one
continuing particle--medium system, including its inherited memory,
instead of two independently prepared arrays.

Envelope contacts, internal coupling and whole-particle connections
occupy distinct Hessian blocks. Structural observations motivate the
envelope--RNP resolution~\cite{Liljeroos2011,Ke2018}; an interaction
law supplies its curvatures~\cite{Cotterill2002}. Joint response and
fluctuations test that law. Rupture, uncoating and rearrangement also
need a nonlinear landscape or failure criterion beyond local harmonic
sensitivity.

\FloatBarrier
\subsection{Symmetries, filters, and changes of representation}
\label{sec:operator-symmetries-transforms}

A symmetry preserves the physical problem, a representation change
transports its coordinates, and a filter selects part of a state or
signal. Their domains, metrics and action distinguish these operations
even when the matrices look similar.

\subsubsection{Symmetry belongs to the reference and its boundary}

A real invertible coordinate map $\mathsf R$ represents a conservative
symmetry when it preserves the admissible coordinate/boundary space and
\begin{equation}
 \mathsf R^{\mathsf T}\mathsf M\mathsf R=\mathsf M,
 \qquad \mathsf R^{\mathsf T}\mathsf K\mathsf R=\mathsf K.
 \label{eq:mechanical-symmetry-conditions}
\end{equation}
Physical rotations and permutations supply familiar orthogonal examples.
In mass-normalized coordinates
$\mathsf U=\mathsf M^{1/2}\mathsf R\mathsf M^{-1/2}$ is unitary
and preserves $\mathsf D$; hence it commutes with $\mathsf D$ and
$\mathsf W$. A continuous family has infinitesimal Lie-algebra generators
only when those conditions hold along the family. The group is a property
of the mass distribution, interactions and boundary together, rather than
of the container's outline alone.

An axisymmetric isotropic continuum cylinder can admit continuous axial
rotations. A finite particle arrangement, anisotropic constitutive law or
localized contact can reduce or remove that symmetry. For the pinned pair,
exchange of identical sites leaves $\mathsf M,\mathsf K$ unchanged.
Common and relative patterns are its even and odd sectors. An exchange-even
force has zero overlap with the odd sector; an exchange-even detector
also misses it. This is the symmetry origin of the contact-blind common
channel in Section~\ref{sec:operator-target-defects}. Unequal supports break exchange symmetry and can
mix those patterns. The mode and readout must then be recalculated.

For a polar continuum displacement, a spatial rotation acts on position
and vector components:
$(\mathcal U_R\vct U)(\vct r)=R\vct U(R^{-1}\vct r)$.
Inversion gives $-\vct U(-\vct r)$. Time reversal instead sends
real conservative data $(\vct u,\vct v)$ to $(\vct u,-\vct v)$;
in the real-coordinate quantum representation it conjugates amplitudes
antiunitarily and reverses canonical momentum. Chirality compares a
structure to its mirror under proper rotations. These are distinct
operations. Dissipative reduction also differs from microscopic
time-reversal symmetry: reversing the resolved velocities alone does
not reverse an unrecorded medium history. Periodic translation and Bloch
labels belong to the auxiliary bulk reference, whereas the finite
cylinder retains its own boundary-adapted modes.

\subsubsection{A small symmetry breaking can enter a signal at second order}

The order at which a parameter enters a readout matters for inference.
Symmetry can make a real mechanical change invisible to the first
derivative of one channel, while another channel responds linearly.

\begin{example}[Registered channels distinguish the sign of a support contrast]
\label{ex:s69-support-contrast-channels}
Keep equal scalar masses $m>0$, link curvature $k_b\geq0$ and
fixed reference centres. Let the two support curvatures be
$k_0+\varepsilon$ and $k_0-\varepsilon$, with
$k_0>|\varepsilon|$. The calibrated common and relative displacements
are $Y_c=(u_1+u_2)/\sqrt2$ and $Y_r=(u_1-u_2)/\sqrt2$.
In that basis their Hessian and the response to a force conjugate to
$Y_c$ are
\begin{equation}
 \begin{gathered}
 \mathsf K_{cr}=
       \begin{pmatrix}k_0&\varepsilon\\
                       \varepsilon&k_0+2k_b\end{pmatrix},
 \qquad a(z)=k_0-mz^2,\quad b(z)=k_0+2k_b-mz^2,\\
 \mathcal R_{cc}(z)=\frac{b(z)}{a(z)b(z)-\varepsilon^2},
 \qquad
 \mathcal R_{rc}(z)=-\frac{\varepsilon}
                              {a(z)b(z)-\varepsilon^2}.
 \end{gathered}
 \label{eq:s69-support-contrast-response}
\end{equation}
These expressions follow by inverting the two-by-two mechanical pencil
where its determinant is nonzero. A common-force/common-readout record
is identical for $\varepsilon$ and $-\varepsilon$ at every such
frequency. It cannot identify which labelled support is stiffer.
At $\varepsilon=0$ its first derivative vanishes, and away from the
reference poles
$\mathcal R_{cc}=a^{-1}+\varepsilon^2/(a^2b)+O(\varepsilon^4)$.
The relative readout instead has
$\partial_\varepsilon\mathcal R_{rc}|_0=-1/(ab)$, so its
signed, phase-calibrated response distinguishes that contrast to first
order. The labels and the sign of the relative detector are registered
physical conventions; exchanging both particle labels also transports
those conventions.
\end{example}

The contrast changes the force coupling between common and relative
motion. The even channel sees a round trip through that coupling and
therefore acquires a quadratic correction; the cross channel sees one
conversion and acquires a linear correction. This provides an exact
mechanical instance of Definition~\ref{def:s6-identifiable-prediction}:
the magnitude of a contrast and its spatial assignment can be different
inference targets. In a multi-particle or layered biophysical reference,
the same symmetry test asks which force/readout combinations resolve an
asymmetric motion. The small-contrast expansion is made at fixed frequency
away from poles; the exact response is used when nearby resonances move.

\subsubsection{A phase can change motion without changing modal weights}

A spectral filter acts as $f(\mathsf W)$. An indicator selects a frequency
subspace; $|f|\leq1$ gives an energy-norm contraction. For the classical
energy amplitude,
\begin{equation}
 E_f=\|f(\mathsf W)\psi\|^2
       =\sum_s|f(\omega_s)|^2|\psi_s|^2\leq E_+.
 \label{eq:operator-filter-retained-energy}
\end{equation}
As a calculation this reports a selected component. Preparing that
filtered mechanical state requires a mechanism and an account of the
discarded energy; normalizing the filtered direction hides that energy
scale. A detector's finite-time filter instead acts on the recorded
signal, as in Section~\ref{sec:complex-displacement}, and need not prepare the selected modal state.

The phase map
$\mathsf U_\theta=\sum_s e^{i\theta_s}\ket s\bra s$ is unitary.
It preserves energy and frequency weights, but generally changes real
coordinate motion. For a common phase $\theta$ and the canonical
mass-normalized energy amplitude
$\psi=(\mathsf W\vct x+i\vct p_x)/\sqrt2$, reconstruction gives
\begin{equation}
 \vct x'=\vct x\cos\theta-\mathsf W^{-1}\vct p_x\sin\theta,
 \qquad
 \vct p_x'=\mathsf W\vct x\sin\theta+\vct p_x\cos\theta
 \quad\hbox{on the positive sector}.
 \label{eq:operator-classical-common-phase-motion}
\end{equation}
Thus a common \emph{classical energy-ket} phase leaves quadratic modal
statistics unchanged while rotating displacement and momentum
quadratures. It is not the unobservable overall phase of a quantum
state vector. A physical clock shift instead applies
$e^{-i\omega_st_0}$ to each mode, which is a common phase only for a
single frequency or a specially matched set of phases. In the full
occupation construction, $e^{i\theta\hat N_{\rm ph}}$ rotates oscillator
means and is not a global phase across different total-occupation
sectors. A constant multiple of the entire quantum state vector remains
physically unobservable. The representation and the prepared object
must therefore be stated when discussing phase.

\subsubsection{Transforming the metric preserves physical predictions}

For a bounded invertible map $\mathsf T$, define
$X'=\mathsf T X$ and $A'=\mathsf T A\mathsf T^{-1}$.
If the original state norm is $X^\dagger\mathsf QX$, then
\begin{equation}
 \mathsf Q'=\mathsf T^{-\dagger}\mathsf Q\mathsf T^{-1},
 \qquad X'{}^\dagger\mathsf Q'X'=X^\dagger\mathsf QX.
 \label{eq:operator-representation-metric}
\end{equation}
Force and output maps transform as well. Similarity preserves spectrum
and, in finite dimensions, generalized eigenspaces and Jordan structure.
For an unbounded operator its domain becomes $D(A')=\mathsf T D(A)$.
A coordinate rescaling can change Euclidean condition numbers while
leaving every physical response and energy estimate unchanged in the
transformed norm. It cannot create a spectral gap, an instability or
damping in an unchanged model.

For spatial multiplication, let
$(\mathcal G_{g_{\rm rep}}U)(\vct r)=e^{g_{\rm rep}(\vct r)}U(\vct r)$,
with dimensionless complex $g_{\rm rep}$. Bounded real part above and
below makes this a bounded invertible map on $L^2$; vanishing real part
makes it unitary. When sufficiently smooth, transforming a differential
operator also transforms its derivative terms and boundary domain.
For example
$\mathcal G_{g_{\rm rep}}(-i\nabla)\mathcal G_{g_{\rm rep}}^{-1}
=-i\nabla+i\nabla g_{\rm rep}$ on the transported domain.
This representation term does not supply an electromagnetic field or a
charge. Those need their own physical coupling and energy. Similarly,
rewriting a fixed pattern in a virion's body frame differs from actually
reorienting that body against anisotropic contacts, which can change its
Hessian. Figure~\ref{fig:symmetry-representation-and-filter} illustrates
the distinction between preservation, rescaling and selection.

\begin{figure}[H]
\centering
\begingroup
\definecolor{vlfGreen}{HTML}{176B64}
\definecolor{vlfDarkRed}{HTML}{762B38}
\definecolor{vlfRed}{HTML}{762B38}
\definecolor{vlfBlue}{HTML}{315D80}
\definecolor{vlfCopper}{HTML}{762B38}

\setlength{\unitlength}{1pt}
\scalebox{0.80}{%
\input{vlt-figures/figure-207.tex}%
}%
\endgroup
\caption{Symmetry, coordinate representation and spectral selection on a real slice of the classical energy ket. Exchange symmetry of the equal-pin pair sends $(\psi_c,\psi_r)$ to $(\psi_c,-\psi_r)$ and preserves the circle. Rescaling $\vct y=\operatorname{diag}(2,1/2)\vct\psi$ gives an ellipse with metric $\operatorname{diag}(1/4,4)$. Projecting $(\sqrt3/2,1/2)$ onto the common component instead retains squared norm $3/4$. The first two preserve the energy prediction with the appropriate transported maps; the last reports a selected component. Preparing it requires a physical selection and energy account. Panel colors distinguish operations.}
\label{fig:symmetry-representation-and-filter}
\end{figure}

\paragraph{A variable set that remains closed through an arrangement change.}
The compatible pattern spaces of Section~5 supply several useful ways
to record one physical assembly. Predictive reduction adds a dynamical
requirement: the retained variables must determine their subsequent
motion, including their continuation when a new cell joins. This is a
test on operators and event maps together. The distinction between
continuous motion and discrete switching is standard in hybrid
mechanics and control~\cite{Liberzon2003}; here the continuous state
retains the full ideal-virion pool and its material memory.

\begin{definition}[Closed variables along a registered assembly history]
\label{def:s6-history-closed-variables}
On each persistent episode $g$, take a finite real linear state with
$\dot X_g=\mathsf A_gX_g+\mathsf F_g\vct f$ and readout
$Y_g=\mathsf L_gX_g$. A registered event continues it by
$X_h^+=\mathsf J_{hg}X_g^-$. Known affine offsets have been removed
or retained as known inputs. Full-row-rank maps $\mathsf B_g$ define
variables $Z_g=\mathsf B_gX_g$. They are \emph{closed along the
registered history} when their continuous evolution, event continuation
and readouts are determined by $Z_g$, the declared forcing and that
history. This is a pointwise requirement for every state in each
episode's full linear space, with the history supplied independently
for this test. Conditioning on observed event times can restrict that
space and defines a separate inference problem.
\end{definition}

\begin{proposition}[Continuous and event closure]
\label{prop:s6-history-closed-variables}
The variables just defined are closed if and only if there exist
matrices satisfying, on every allowed episode and event,
\begin{equation}
 \begin{aligned}
 \mathsf B_g\mathsf A_g&=\overline{\mathsf A}_g\mathsf B_g,\\
 \mathsf B_h\mathsf J_{hg}&=
                  \overline{\mathsf J}_{hg}\mathsf B_g,\\
 \mathsf L_g&=\overline{\mathsf L}_g\mathsf B_g.
 \end{aligned}
 \label{eq:s6-history-closure-factorization}
\end{equation}
Equivalently, $\mathsf A_g\ker\mathsf B_g\subseteq
\ker\mathsf B_g$, while
$\ker\mathsf B_g\subseteq\ker(\mathsf B_h\mathsf J_{hg})$
and $\ker\mathsf B_g\subseteq\ker\mathsf L_g$.
Their reduced laws are then
\begin{equation}
 \dot Z_g=\overline{\mathsf A}_gZ_g+
                           \mathsf B_g\mathsf F_g\vct f,
 \qquad Z_h^+=\overline{\mathsf J}_{hg}Z_g^-,
 \qquad Y_g=\overline{\mathsf L}_gZ_g.
 \label{eq:s6-history-reduced-evolution}
\end{equation}
\end{proposition}
\begin{proof}
Preparations with the same $Z_g$ differ by a vector in
$\ker\mathsf B_g$. Their instantaneous reduced derivatives,
continued variables and readouts agree precisely when the three
kernel conditions hold. Each resulting map on equivalence classes
factors through $\mathsf B_g$, giving
\eqref{eq:s6-history-closure-factorization}. Conversely those
factorizations give the displayed reduced differential equation and
event map. Uniqueness on each finite linear episode and induction over
the registered events prove sufficiency for the entire history.
\end{proof}

The retained dimension may change between episodes. Closure requires
each new entry to be determined by the old variables or supplied as
an event input. For a cluster selector, $\mathsf J_{hg}=\Id$ in
laboratory coordinates: a newcomer's displacement, velocity and prepared
environment already exist in the full-pool state. Tracking or a
conditional preparation law supplies them to an expanded cluster record.

As a mechanical test, take two equally supported particles of mass $m$.
Initially their scalar displacements are independent. At $t=0$ a
controlled connection adds $k(u_1-u_2)^2/2$ to the energy, with $k>0$.
The old variables $(u_1,\dot u_1)$ close before connection. Two
preparations with $u_1=\dot u_1=\dot u_2=0$ and $u_2=\pm a$
have the same old variables but give
\begin{equation}
 \ddot u_1(0^+)=\pm ka/m,
 \qquad \Delta W_{\rm ctrl}=ka^2/2.
 \label{eq:s6-history-new-neighbour-input}
\end{equation}
The new restoring force resolves the omitted neighbour coordinate;
both preparations have the same connection work. Adding the neighbour
state, its exact reduced history, or a specified incoming-state input
restores a predictive description. The example stipulates the connection
time; a spontaneous joining mechanism additionally supplies its capture
and transition law.

Figure~\ref{fig:s6-new-neighbour-closure-test} shows two incoming preparations that agree on the old variables but produce opposite post-connection motion.

\begin{figure}[H]
\centering\begingroup
\definecolor{vltTeal}{HTML}{176B64}
\definecolor{vltBlue}{HTML}{315D80}
\definecolor{vltGold}{HTML}{A87520}
\definecolor{vltPlum}{HTML}{725877}
\definecolor{vltRed}{HTML}{762B38}
\definecolor{vltYellow}{HTML}{D5AD3C}
\definecolor{vltNeutral}{HTML}{777D83}
\setlength{\unitlength}{1pt}
\scalebox{0.90}{%
\input{vlt-figures/figure-208.tex}%
}\endgroup
\caption{A controlled connection reveals an omitted neighbour coordinate. Both initial preparations have $u_1=\dot u_1=\dot u_2=0$, with $u_2=\pm a$. Equal supports and the new link have curvature $k$; $\omega_\star=\sqrt{k/m}$. Rulers locate reference spacing $d=25a/3$ at gain $12\,\mathrm{pt}/a$. Matched coils are longer for the positive neighbour shift. Upper arrows give particle-1 acceleration, with one gain for $\pm ka/m$; supports are omitted. Exact subsequent records are $u_1/a=\pm[\cos(\omega_\star t)-\cos(\sqrt3\omega_\star t)]/2$, at equal connection work $ka^2/2$. Future prediction therefore uses the incoming neighbour state or its exact reduction in addition to the old variables.}
\label{fig:s6-new-neighbour-closure-test}
\end{figure}

Observation and closure test complementary properties. An accurate
cluster record can still omit a neighbour or strain that controls its
future. Conversely, several partial records can identify a closed
collective state. Equation~\eqref{eq:s6-linear-target-observation-criterion}
tests recovery; the closure factorizations test prediction. If the
arrangement history is predicted as well, event rates must satisfy
$\lambda_{hg}(X_g)=\overline\lambda_{hg}(\mathsf B_gX_g)$,
with a corresponding condition on deterministic event boundaries.
Approximate closure instead specifies propagated readout errors over
the residence window. This makes variable selection a prediction test.

\subsubsection{Decomposing a calibrated response into gain and direction}

Let $\mathsf H(\omega)=\mathsf R_{\rm out}\mathsf R_{\rm mech}(\omega)
\mathsf F_{\rm in}$ map specified force channels to displacement channels.
With fixed input/output units and norms, its polar decomposition away from
poles is
\begin{equation}
 \mathsf H=\mathsf V_{\rm pol}|\mathsf H|,
 \qquad |\mathsf H|=(\mathsf H^\dagger\mathsf H)^{1/2}.
 \label{eq:response-polar-decomposition}
\end{equation}
The nonnegative factor measures gain along input directions. The partial
isometry maps the complement of the kernel to the output range; it is
unitary only for the appropriate invertible square map. In a singular
value decomposition, the largest singular value maximizes measured
displacement per unit input norm, and its right singular vector identifies
the corresponding force combination. The left vector specifies the
observed motion. This is not a universal mechanical efficiency: changing
the force-cost norm, detector calibration or observation bandwidth changes
the question. Noise-weighted readouts require their stated noise metric.

Such a decomposition can compare an envelope-only channel with a joint
envelope--RNP observation without confusing an invisible motion with an
absent mode. It provides an amplitude/direction analysis of an actual
response, rather than a general self-adjoint oscillator phase operator.
The next subsection keeps that physical norm and those channels explicit
when the resolved dynamics is dissipative and its mode vectors need not
be orthogonal.

\FloatBarrier
\subsection{Passive generators, resolvents, and nonorthogonal modes}
\label{sec:passive-operator-spectrum}

Viscoelastic evolution acts on the augmented state carrying recoverable
energy. Identical displacement and velocity can accompany different
branch strains and future forces. Their generator therefore includes
that material preparation. Its spectrum describes mechanical evolution
with loss; phonon-energy measurements use the separately defined
self-adjoint occupation Hamiltonian.

\subsubsection{Closure in an augmented energy space}

For the finite relaxation model of Section~\ref{sec:viscoelastic-dynamics}, use
$\vct X=(\vct u,\vct v,\vct z_1,\ldots,\vct z_{r_R})^{\mathsf T}$,
where $\vct v=\Dt{\vct u}$. The branch equation
$\Dt{\vct z_a}=\mathsf G_a\vct v-\vct z_a/\tau_a$
retains the deformation whose restoring force is
$\mathsf G_a^{\mathsf T}\mathsf H_a\vct z_a$.
Two preparations with identical $(\vct u,\vct v)$ but different
$\vct z_a$ are different mechanical states. Eliminating the latter is
permitted, but returns a memory kernel and its prepared-history term,
as in~\eqref{eq:maxwell-memory-equation}.

For one scalar branch, the closed equations are
$m\Dtt{u}+c_0\Dt{u}+k_0u+k_1z=0$ and
$\Dt{z}=\Dt{u}-z/\tau$. Starting from $u=\Dt{u}=0$, a prepared
$z=Z$ gives initial acceleration $-k_1Z/m$; the relaxed preparation
$z=0$ gives zero. The former has stored branch energy $k_1Z^2/2$.
Motion released from that preparation is therefore possible without
an active force source. A position-only state would miss both the
initial force and the energy that supplies its subsequent motion.
This is why enlarging the state is a predictive requirement rather
than additional notation.

On the positive energy space, set
$\mathsf Q=\diag(\mathsf K,\mathsf M,\mathsf H_1,\ldots,
\mathsf H_{r_R})$ and $E_{\rm res}=\vct X^\dagger\mathsf Q\vct X/2$.
Null displacement or branch directions are treated by a compatible
energy-space quotient, or an invariant restriction when available.
They are not discarded if they still carry a force or a required history.
With $\mathsf Q>0$ on the retained space, the real unforced matrix
$\mathsf A_{\rm r}$ obeys
\begin{equation}
 \Dt{\vct X}=\mathsf A_{\rm r}\vct X,\qquad
 \mathsf A_{\rm r}^\dagger\mathsf Q+
             \mathsf Q\mathsf A_{\rm r}=-2\mathsf R,\qquad
 \mathsf R=\diag\!\left(0,\mathsf C_0,
          \frac{\mathsf H_1}{\tau_1},\ldots,
          \frac{\mathsf H_{r_R}}{\tau_{r_R}}\right)\geq0.
 \label{eq:passive-generator-metric}
\end{equation}
Multiplication by $\vct X^\dagger$ and $\vct X$ gives
$\Dt{E_{\rm res}}=-\vct X^\dagger\mathsf R\vct X$.
Thus $\|\vct X\|_{\mathsf Q}^2=\vct X^\dagger\mathsf Q\vct X$
is twice stored energy, and
$\|e^{t\mathsf A_{\rm r}}\vct X\|_{\mathsf Q}
\leq\|\vct X\|_{\mathsf Q}$ for $t\geq0$.
The loss is transferred to unresolved medium energy in the complete
ledger; it is not destruction of energy in the sealed enlarged system.

The blocks of $\mathsf Q$ convert displacement, velocity and branch
strain into the same energy units. An unweighted Euclidean norm of these
different physical quantities has no such meaning. In the coordinates
$\mathsf Q^{1/2}\vct X$, the generator
$\mathsf Q^{1/2}\mathsf A_{\rm r}\mathsf Q^{-1/2}$ has inverse-time
units and a nonpositive Hermitian part. This normalization expresses the existing energy and dissipation in
equal-unit coordinates. For a continuum generator the
analogous statement needs the admissible domain and a maximality or range
condition; a formally dissipative differential expression alone does not
establish a contraction evolution.

A marker $Y=\vct\ell^{\mathsf T}\vct u$ may nevertheless grow as
kinetic or branch energy becomes displacement energy. For a positive
retained stiffness and a readout supported on it,
\begin{equation}
 |Y(t)|^2\leq
    2E_{\rm res}(0)\,
        \vct\ell^{\mathsf T}\mathsf K^{-1}\vct\ell.
 \label{eq:passive-marker-energy-bound}
\end{equation}
This is Cauchy--Schwarz applied to
$\mathsf K^{1/2}\vct u$; the remaining stored terms are nonnegative.
On a compatible positive subspace the inverse is taken there.
Softness can make a marker's allowed excursion large, while the complete
energy still decreases. The bound does not apply to an unconstrained
free translation whose displacement costs no restoring energy.

\subsubsection{How much energy is lost within an observation interval}

The loss-kernel condition also has a quantitative finite-time form. A
particular initial displacement may have zero instantaneous dissipation
because its velocity and branch strain vanish. Subsequent mechanical
motion can still carry that preparation into dissipative directions.

\begin{definition}[Accumulated-loss matrix]
\label{def:s610-accumulated-loss}
For the fixed unforced finite generator in
\eqref{eq:passive-generator-metric}, with $\mathsf Q>0$, define at
an observation horizon $T>0$
\begin{equation}
 \mathsf D_T=\int_0^T e^{t\mathsf A_{\rm r}^\dagger}
                    \mathsf R e^{t\mathsf A_{\rm r}}\,\dd t,
 \qquad
 \eta_T=\lambda_{\min}
       (\mathsf Q^{-1/2}\mathsf D_T\mathsf Q^{-1/2}).
 \label{eq:s610-accumulated-loss-matrix}
\end{equation}
For an initial state $\vct X_0$, the quantity
$\vct X_0^\dagger\mathsf D_T\vct X_0$ is the stored energy lost
by time $T$. Thus $\eta_T$ is dimensionless and measures the least
loss per squared energy norm among all nonzero initial directions.
\end{definition}

\begin{proposition}[Loss visibility and a uniform decay bound]
\label{prop:s610-finite-loss-contraction}
Let $d_X$ be the retained real state dimension. Then
\begin{equation}
 \begin{aligned}
 e^{T\mathsf A_{\rm r}^\dagger}\mathsf Q
                e^{T\mathsf A_{\rm r}}
       &=\mathsf Q-2\mathsf D_T,\\
 \|e^{T\mathsf A_{\rm r}}\|_{\mathsf Q}^2
       &=1-2\eta_T,\qquad 0\leq\eta_T<\tfrac12.
 \end{aligned}
 \label{eq:s610-exact-finite-loss-contraction}
\end{equation}
Moreover $\eta_T>0$, for any and hence every $T>0$, exactly when
$\bigcap_{j=0}^{d_X-1}
\ker(\mathsf R^{1/2}\mathsf A_{\rm r}^j)=\{0\}$.
In that case the energy obeys
$E_{\rm res}(t)\leq(1-2\eta_T)^{\lfloor t/T\rfloor}
E_{\rm res}(0)$ for every $t\geq0$.
\end{proposition}
\begin{proof}
Differentiate
$e^{t\mathsf A_{\rm r}^\dagger}\mathsf Q
 e^{t\mathsf A_{\rm r}}$ and integrate the metric identity.
Congruence by $\mathsf Q^{-1/2}$ makes its largest eigenvalue the
squared induced energy norm and gives the second line. Finite-time
matrix evolution is invertible, so this largest eigenvalue is positive.
An initial vector has zero accumulated loss if and only if
$\mathsf R^{1/2}e^{t\mathsf A_{\rm r}}\vct X_0=0$ throughout
the interval. Analyticity of the matrix exponential makes this equivalent
to vanishing of all its Taylor coefficients; Cayley--Hamilton reduces
them to the displayed finite kernel intersection. Finally iterate the
norm bound over complete intervals of length $T$ and use contraction
over the remaining interval.
\end{proof}

The observability Gramian $\mathsf D_T$ measures the loss output
$\mathsf R^{1/2}\vct X$ in the energy ledger's units; a camera Gramian
instead measures recorded motion~\cite{Kalman1963}. Small positive
$\eta_T$ identifies a preparation retaining most energy over this
window even if every mode eventually decays. Its minimizing direction
need not be invariant or reconstructible from a partial marker record.
Closure and observation tests assess those further properties.

\paragraph{Accumulated loss across registered episodes.}
Recruitment changes the relevant tangent and may redirect motion into
different dissipative directions. The accumulated-loss operator can
follow that sequence while retaining the material preparation. Each
episode contributes a positive loss form, pulled back through the
earlier evolution to the same initial state. This also exposes the
energy contribution of a declared event map.

\begin{proposition}[Composition of registered quadratic energy accounts]
\label{prop:s6-history-energy-composition}
Consider $m+1$ finite unforced linear episodes $j=0,\ldots,m$, with
durations $\Delta_j>0$, total horizon $T=\sum_j\Delta_j$ and
$\mathsf A_j^\dagger\mathsf Q_j+\mathsf Q_j\mathsf A_j
=-2\mathsf R_j$, where $\mathsf Q_j>0$ and $\mathsf R_j\geq0$.
At the intervening events use the specified homogeneous linear
continuations $X_{j+1}^+=\mathsf J_jX_j^-$, including the retained
material coordinates. Define
\begin{equation}
 \begin{aligned}
 \mathsf S_j&=e^{\Delta_j\mathsf A_j},&
 \mathsf D_j&=\int_0^{\Delta_j}
       e^{t\mathsf A_j^\dagger}\mathsf R_j e^{t\mathsf A_j}\,\dd t,\\
 \mathsf F_0&=\Id,&
 \mathsf F_{j+1}&=\mathsf J_j\mathsf S_j\mathsf F_j,\\
 \mathsf\Lambda_j&=\tfrac12
  (\mathsf Q_j-\mathsf J_j^\dagger\mathsf Q_{j+1}\mathsf J_j),&
 \mathsf\Phi_T&=\mathsf S_m\mathsf F_m.
 \end{aligned}
 \label{eq:s6-history-energy-data}
\end{equation}
Then the total quadratic-storage deficit is
\begin{equation}
 \begin{aligned}
 \mathsf D_{\rm hist}
  &=\sum_{j=0}^{m}\mathsf F_j^\dagger\mathsf D_j\mathsf F_j
    +\sum_{j=0}^{m-1}\mathsf F_j^\dagger\mathsf S_j^\dagger
                          \mathsf\Lambda_j\mathsf S_j\mathsf F_j,\\
 \mathsf\Phi_T^\dagger\mathsf Q_m\mathsf\Phi_T
  &=\mathsf Q_0-2\mathsf D_{\rm hist}.
 \end{aligned}
 \label{eq:s6-history-energy-deficit}
\end{equation}
If all $\mathsf\Lambda_j\geq0$, this gives a passive composition and
\begin{equation}
 \|\mathsf\Phi_T\|_{\mathsf Q_0\to\mathsf Q_m}^2
 =1-2\eta_{\rm hist},\qquad
 \eta_{\rm hist}=
  \lambda_{\min}(\mathsf Q_0^{-1/2}\mathsf D_{\rm hist}
                                  \mathsf Q_0^{-1/2})\geq0.
 \label{eq:s6-history-energy-contraction}
\end{equation}
The induced norm compares final and initial stored energy using their
respective metrics.
\end{proposition}
\begin{proof}
Integrating each episode's metric identity gives
$\mathsf S_j^\dagger\mathsf Q_j\mathsf S_j
=\mathsf Q_j-2\mathsf D_j$. Its event changes the quadratic
storage by $-(X_j^-)^\dagger\mathsf\Lambda_jX_j^-$.
Substitute $X_j^-=\mathsf S_j\mathsf F_jX_0$ and sum; intermediate
energies telescope. Congruence by $\mathsf Q_0^{-1/2}$ and taking
the largest eigenvalue prove the induced-norm identity. Each term in
$\mathsf D_{\rm hist}$ is nonnegative under the stated assumption.
\end{proof}

For a pure coordinate registration,
$\mathsf Q_j=\mathsf J_j^\dagger\mathsf Q_{j+1}\mathsf J_j$,
so $\mathsf\Lambda_j=0$: the event introduces no storage change.
Selecting a different cluster readout changes its observation operator,
while the complete physical state continues without a membership reset.
Constitutive loss is the first sum in
\eqref{eq:s6-history-energy-deficit}. A positive event deficit requires
its own destination in the full energy account. If it results from
discarding resolved coordinates, it represents omitted storage until
an actual transfer into the medium is specified.

A controlled stiffness change illustrates the other sign. In laboratory
coordinates with continuous state, $\mathsf J_j=\Id$, and increasing
the stiffness at nonzero displacement can make
$\mathsf\Lambda_j$ negative in that direction. The energy increase
is the work in \eqref{eq:s5-reference-stiffness-jump-work}, even
though every frozen episode has a passive generator. General contact
energies, reference offsets and nonlinear remainders use the complete
event balance \eqref{eq:s4-transient-event-first-law}; the proposition
states the homogeneous quadratic calculation explicitly.

Figure~\ref{fig:s6-registered-event-energy-ledger} separates the accumulated material loss from a controlled event contribution.

\begin{figure}[H]
\centering\begingroup
\definecolor{vltTeal}{HTML}{176B64}
\definecolor{vltBlue}{HTML}{315D80}
\definecolor{vltGold}{HTML}{A87520}
\definecolor{vltPlum}{HTML}{725877}
\definecolor{vltRed}{HTML}{762B38}
\definecolor{vltYellow}{HTML}{D5AD3C}
\definecolor{vltNeutral}{HTML}{777D83}
\setlength{\unitlength}{1pt}
\scalebox{0.90}{%
\input{vlt-figures/figure-209.tex}%
}\endgroup
\caption{Loss and work across a controlled stiffness change. The scalar coordinate obeys $m\ddot u+c\dot u+k_ju=0$, with $\omega_\star=\sqrt{k_\star/m}$, $c/(m\omega_\star)=0.4$, $u(0)=u_\star$ and $\dot u(0)=0$. At $\omega_\star t_e=1.2$, stiffness changes from $k_\star$ to $3k_\star$ with continuous state. Both upper glyphs show that event displacement on equal rulers and gain $18\,\mathrm{pt}/u_\star$; coil geometry stays fixed. Matrix-exponential trajectories give storage $E$, integrated loss $L=\int c\dot u^2\,\dd t$ and supplied work $W$, in units $E_\star=k_\star u_\star^2$. Their balance includes the positive controller-work jump. Each frozen oscillator remains passive.}
\label{fig:s6-registered-event-energy-ledger}
\end{figure}

Cell-assembled generators and registration maps compute this account for
different episode orderings. Intervening lossless motion can rotate a
protected pattern into a lossy direction, changing the longest-lived
preparation. Its persistence through rearrangements can therefore differ
from any frozen relaxation time. Pulling back detector forms instead of
loss forms computes the corresponding camera Gramian with the same
propagation machinery.

\subsubsection{Decay rates, nonorthogonal modes, and resolvents}

Writing $i\Dt{\vct X}=\mathcal G\vct X$, with
$\mathcal G=i\mathsf A_{\rm r}$, gives the energy-metric decomposition
\begin{equation}
 \begin{aligned}
 \mathcal G&=\mathcal H_{\rm g}-i\mathcal L_{\rm g},\\
 \mathcal H_{\rm g}
   &=\frac{\mathcal G+\mathcal G^{\dagger_{\mathsf Q}}}{2},\\
 \mathcal L_{\rm g}
   &=\frac{\mathcal G^{\dagger_{\mathsf Q}}-\mathcal G}{2i}
       =\mathsf Q^{-1}\mathsf R\geq_{\mathsf Q}0.
 \end{aligned}
 \label{eq:generator-storage-loss}
\end{equation}
Here $A^{\dagger_{\mathsf Q}}=\mathsf Q^{-1}A^\dagger\mathsf Q$.
Both displayed parts are self-adjoint in that metric; the complete
$\mathcal G$ generally is not. It is a first-order mechanical frequency
generator, rather than the quantum occupation Hamiltonian $\hat H_0$.
If $\mathsf A_{\rm r}\vct X=\lambda\vct X$, its energy identity yields
$\operatorname{Re}\lambda
=-\vct X^\dagger\mathsf R\vct X/
(\vct X^\dagger\mathsf Q\vct X)\leq0$.
The corresponding frequency $z=i\lambda$ lies in the lower half-plane,
consistent with $e^{-izt}$. Strict decay requires that no nonzero
trajectory remain entirely in the loss kernel. Positive loss in only
some coordinates is not itself that condition.

The resolvent $\mathsf R_{\mathsf A}(z)=(z\Id-\mathsf A_{\rm r})^{-1}$
describes the response of the augmented state to forcing at a specified
complex growth rate. Its argument here has inverse-time units and is
not the phonon occupation energy. Dissipativity gives a useful bound:
\begin{equation}
 \boxed{\|\mathsf R_{\mathsf A}(z)\|_{\mathsf Q}
              \leq\frac1{\operatorname{Re}z},
              \qquad\operatorname{Re}z>0.}
 \label{eq:passive-energy-resolvent-bound}
\end{equation}

Figure~\ref{fig:passive-critical-damping-and-resolvent} gives a graphical specialization of this boxed relation.

Indeed the real part of
$\langle\vct X,(z\Id-\mathsf A_{\rm r})\vct X\rangle_{\mathsf Q}$
is at least $(\operatorname{Re}z)\|\vct X\|_{\mathsf Q}^2$;
Cauchy--Schwarz then bounds the inverse. Finite dimensionality and the
left-half-plane spectrum ensure its existence in this region.
In a continuum this bound follows after the generator and its resolvent
have been established, not from the formal expression alone.

\begin{figure}[H]
\centering\begingroup
\definecolor{vltTeal}{HTML}{176B64}
\definecolor{vltBlue}{HTML}{315D80}
\definecolor{vltGold}{HTML}{A87520}
\definecolor{vltPlum}{HTML}{725877}
\definecolor{vltRed}{HTML}{762B38}
\definecolor{vltYellow}{HTML}{D5AD3C}
\definecolor{vltNeutral}{HTML}{777D83}
\setlength{\unitlength}{1pt}
\scalebox{0.80}{%
\input{vlt-figures/figure-210.tex}%
}\endgroup
\caption{A passive defective generator and the energy resolvent bound \eqref{eq:passive-energy-resolvent-bound}. For the critically damped oscillator $m\ddot u+2m\omega_0\dot u+m\omega_0^2u=0$, use $\tau=\omega_0t$, $x=u/u_\star$ and $v=\dot u/(\omega_0u_\star)$. Its displayed generator has energy metric $Q=I$ and repeated eigenvalue $-1$. (a) From $(x,v)=(0,1)$, displacement $x=\tau e^{-\tau}$ grows initially while $E/E_0=x^2+v^2$ decreases with derivative $-4v^2$. (b) The solid imaginary-axis singular gain exceeds the dashed normal-matrix pole-distance comparison $1/\sqrt{1+\nu^2}$. (c) In the right half-plane, direct singular values at $z=\sigma+i$ remain below $1/\sigma$; dimensional rates multiply $z$ by $\omega_0$. Thus the nonnormal excess and the passive energy bound coexist. These state-space resolvents precede a particular choice of force and receiver.}
\label{fig:passive-critical-damping-and-resolvent}
\end{figure}

A nonnormal matrix need not have orthogonal eigenvectors in the chosen
metric. Its resolvent can therefore be much larger than inverse distance
to the spectrum, subject to the energy bound where applicable
\cite{Trefethen1997}. A measured force-to-marker gain additionally uses
the input and output maps of Section~\ref{sec:operator-symmetries-transforms}. With the Fourier-transform convention
$e^{i\Omega t}$ it contains
$( -i\Omega\Id-\mathsf A_{\rm r})^{-1}$, whereas eliminating branch
variables gives
$\mathsf R_{\rm mech}(\Omega)=
[\mathsf K^*(\Omega)-\Omega^2\mathsf M]^{-1}$.
These are compatible representations with different state and channel
spaces. A component's transient growth, or a large singular response,
does not contradict contraction of the complete energy norm. A diagonal
complex filter is normal; complex entries alone do not cause nonnormality.

For a contour $\mathcal C_\lambda$ enclosing one isolated eigenvalue,
the finite-dimensional generalized eigenspace is selected by
\begin{equation}
 \mathsf P_\lambda=\frac1{2\pi i}
     \oint_{\mathcal C_\lambda}\mathsf R_{\mathsf A}(z)\,\dd z,
 \qquad
 \mathsf N_\lambda=(\mathsf A_{\rm r}-\lambda\Id)\mathsf P_\lambda.
 \label{eq:riesz-jordan-projection}
\end{equation}
The generally oblique $\mathsf P_\lambda$ is a Riesz projection, not a
quantum probability effect. If $\mathsf N_\lambda^{m_\lambda}=0$,
\begin{equation}
 e^{t\mathsf A_{\rm r}}\mathsf P_\lambda
   =e^{\lambda t}\left[\mathsf P_\lambda+
      \sum_{j=1}^{m_\lambda-1}\frac{t^j}{j!}
                                      \mathsf N_\lambda^j\right].
 \label{eq:jordan-evolution}
\end{equation}
These polynomial factors are required at a defective eigenvalue, where
ordinary eigenvectors do not span its generalized eigenspace
\cite{HornJohnson2013}. In a contraction evolution, imaginary-axis
eigenvalues are semisimple: a nonzero Jordan factor there would grow
polynomially and violate the norm bound. Defective decaying modes are
allowed. For a continuum, isolated spectrum is needed for this contour,
and a sum of such projections need not be complete.

For example, the passive critical oscillator
$m\Dtt{u}+2m\omega_0\Dt{u}+m\omega_0^2u=0$ has a repeated decay rate.
With $x=u/u_\star$, $w=\Dt{u}/(\omega_0u_\star)$ and
$\tau=\omega_0t$, its dimensionless matrix is
\begin{equation}
 \mathsf A_c=\begin{pmatrix}0&1\\-1&-2\end{pmatrix},\qquad
 (\mathsf A_c+\Id)^2=0,\qquad
 x(\tau)=\tau e^{-\tau},\quad
 w(\tau)=(1-\tau)e^{-\tau}
 \label{eq:passive-critical-damping-example}
\end{equation}
for initial $(x,w)=(0,1)$. Here $u_\star$ is a prescribed displacement
scale. The stored energy divided by
$m\omega_0^2u_\star^2$ is $(x^2+w^2)/2$, with derivative
$-2w^2$ in $\tau$. Displacement first grows, but energy never does.
The physical susceptibility is $[m(\omega_0-iz)^2]^{-1}$ for the
complex Fourier frequency $z$; its double pole at $z=-i\omega_0$
corresponds to the polynomial $t e^{-\omega_0t}$ in time.
Figure~\ref{fig:passive-critical-damping-and-resolvent} shows the
same example's state history and resolvent gain. A biological marker can
likewise become more displaced through passive redistribution; autonomous
self-excitation would require an additional source and its resource ledger.

For the critical example, the dimensionless imaginary-axis argument
$i\nu$ gives
\begin{equation}
 \| (i\nu\Id-\mathsf A_c)^{-1}\|_2
    =\frac{\sqrt{\nu^2+2}+1}{1+\nu^2},\qquad
 \operatorname{dist}(i\nu,\operatorname{spec}\mathsf A_c)^{-1}
    =\frac1{\sqrt{1+\nu^2}}.
 \label{eq:critical-state-resolvent-gain}
\end{equation}
The first exceeds the pole-distance expression because the matrix is
nonnormal, although its energy metric is already the identity. At zero
argument the two gains are $1+\sqrt2$ and $1$.
The state norm allows both input and output components. A force confined
to acceleration with displacement alone recorded instead has normalized
transfer $[(1-i\nu)^2]^{-1}$ and magnitude $1/(1+\nu^2)$.
Different channels give different gains under the same passive dynamics.
The real generator has equal singular resolvent norms at $\pm i\nu$;
the Fourier convention fixes the response phase.

\subsubsection{Passivity in calibrated force and velocity channels}

A measured gain can also be tested directly against the work supplied
to the retained assembly. The relevant output is the velocity conjugate
to the applied force. It is fixed by the energy balance, rather than
chosen by the size of a displacement signal.

\begin{definition}[Work-conjugate port]
\label{def:s610-work-conjugate-port}
For a real input matrix $\mathsf B_f$ and force-channel column
$\vct f$, let
$\Dt{\vct X}=\mathsf A_{\rm r}\vct X+\mathsf B_f\vct f$.
Its work-conjugate output is
$\vct y_v=\mathsf B_f^{\mathsf T}\mathsf Q\vct X$, so that
\begin{equation}
 \Dt{E_{\rm res}}=-\vct X^{\mathsf T}\mathsf R\vct X
                         +\vct f^{\mathsf T}\vct y_v,
 \qquad
 \mathsf H_v(s)=\mathsf B_f^{\mathsf T}\mathsf Q
                   (s\Id-\mathsf A_{\rm r})^{-1}\mathsf B_f.
 \label{eq:s610-power-port-transfer}
\end{equation}
For translational forces applied in patterns $\mathsf P_f$,
$\mathsf B_f=\left(\begin{smallmatrix}0\\
\mathsf M^{-1}\mathsf P_f\\0\\\vdots\end{smallmatrix}\right)$
and $\vct y_v=\mathsf P_f^{\mathsf T}\vct v$. The force amplitudes
have force units, their outputs have velocity units and their product
is power. The growth variable $s$ has inverse-time units.
\end{definition}

\begin{proposition}[Positive-real mechanical admittance]
\label{prop:s610-positive-real-admittance}
Let $\mathsf Z_s=(s\Id-\mathsf A_{\rm r})^{-1}$ and
$\operatorname{Re}_{\rm H}A=(A+A^\dagger)/2$. For
$\operatorname{Re}s>0$, the port transfer is analytic and satisfies
\begin{equation}
 \operatorname{Re}_{\rm H}\mathsf H_v(s)
   =\mathsf B_f^\dagger\mathsf Z_s^\dagger
       [(\operatorname{Re}s)\mathsf Q+\mathsf R]
                         \mathsf Z_s\mathsf B_f\geq0.
 \label{eq:s610-positive-real-port-identity}
\end{equation}
The nonnegative boundary identity also holds at $s=-i\Omega$
where the inverse exists.
\end{proposition}
\begin{proof}
Insert $\mathsf Z_s^{-1}=s\Id-\mathsf A_{\rm r}$ on the two
sides of the sum $\mathsf Q\mathsf Z_s+
\mathsf Z_s^\dagger\mathsf Q$. Its middle factor becomes
$(s+s^*)\mathsf Q-
(\mathsf A_{\rm r}^\dagger\mathsf Q+
\mathsf Q\mathsf A_{\rm r})$.
The metric identity reduces this to
$2[(\operatorname{Re}s)\mathsf Q+\mathsf R]$.
Multiplication by the two input maps and division by two prove the
formula. The passive resolvent bound ensures analyticity in the open
right half-plane; taking a finite regular boundary value proves the
last statement.
\end{proof}

For a harmonic force with complex amplitude $\vct f_\Omega$, the
mean supplied power in a steady response is
$\vct f_\Omega^\dagger
\operatorname{Re}_{\rm H}\mathsf H_v(-i\Omega)
\vct f_\Omega/2\geq0$.
This is the explicit force--velocity form of passive storage and supply
\cite{Willems1972,Willems1972Linear}. If the displacement response uses
the same force patterns for loading and readout, its mobility is
$\mathsf H_v(-i\Omega)=-i\Omega\mathsf H_u(\Omega)$, and
the test becomes
$\Omega\operatorname{Im}_{\rm H}\mathsf H_u(\Omega)\geq0$.
A different structural marker is a valid observable but is generally not
the work-conjugate output, so its cross-transfer need not satisfy this
positivity by itself. In micromechanical or biophysical inference, this
distinction supplies an energy-consistency check on a calibrated passive
fit while retaining the separate detector map. It also shows why gain,
normalization and a list of decaying poles are incomplete descriptions
of the physical response.

\subsubsection{Controlled coefficients and normalized quantum operations}

Time-dependent constitutive controls give a two-parameter evolution
$\mathsf T(t,s)$ with
$\mathsf T(t,s)\mathsf T(s,a)=\mathsf T(t,a)$.
For $\Dt{\vct X}=\mathsf A(t)\vct X+\vct f_X$ and differentiable
positive $\mathsf Q(t)$,
\begin{equation}
 \Dt{E}=\frac12\vct X^\dagger
       [\Dt{\mathsf Q}+\mathsf A^\dagger\mathsf Q+
                                \mathsf Q\mathsf A]\vct X
       +\operatorname{Re}(\vct X^\dagger\mathsf Q\vct f_X).
 \label{eq:time-dependent-operator-energy}
\end{equation}
The metric derivative includes parameter work. Periodic curvature
control can supply energy even when every frozen oscillator has stable
modes; stability of the controlled evolution needs its time-dependent
law. An autonomous extension retains the variables generating modulation,
gain and saturation, as in
Section~\ref{sec:self-excitation-passive-boundary}.

Nonunitary quantum conditioning is another distinct operation. A bounded measurement operator $\hat T$ with
$\hat T^\dagger\hat T\leq\Id$ has
outcome probability $p_T=\tr(\hat T\varrho\hat T^\dagger)$ and, for
$p_T>0$, conditional state
\begin{equation}
 \varrho_T=\frac{\hat T\varrho\hat T^\dagger}{p_T}.
 \label{eq:conditioned-phonon-state}
\end{equation}
An unconditioned measurement or reduced channel includes all its outcomes,
with $\sum_a\hat T_a^\dagger\hat T_a=\Id$, so total probability remains
one. The similarity $\hat T\varrho\hat T^{-1}$ generally preserves
neither positivity nor the stated physical probabilities. It is a
representation formula, not a substitute for this channel law.

On the explicitly restricted span $\{\ket0,\ket1\}$, a zero-temperature
loss channel with survival $p$ has
$K_0=\ket0\bra0+\sqrt p\ket1\bra1$ and
$K_1=\sqrt{1-p}\ket0\bra1$.
Their $K_0^\dagger K_0+K_1^\dagger K_1$ is the identity on that span.
The excitation population becomes $p\varrho_{11}$, while the vacuum gains
$(1-p)\varrho_{11}$ and the coherence is multiplied by $\sqrt p$.
Discarding the loss outcome would instead give an unnormalized branch.
This elementary version of the reduced construction in Section~\ref{sec:viscoelastic-dynamics} retains
the vacuum and changes no virion count~\cite{Lindblad1976}; higher
occupations need the corresponding full oscillator channel.
Its trace-preserving quantum probability law and the classical
energy contraction above answer different questions about a declared
system, even when both describe reduced loss.

\FloatBarrier
\subsection{Thermal fluctuations and stochastic evolution}
\label{sec:operator-thermal-stochastic}

Thermal dynamics pairs attenuation with the medium's fluctuating return
force. The same inertia, curvatures and dissipation predict both
relaxation and fluctuations. Here a classical probability law evolves
on mechanical phase space; the optional quantum Gibbs construction
below uses oscillator occupation space.

\subsubsection{A thermal force with declared units and preparation}

On $r$ positive mechanical coordinates, let $\vct x$ be mass-weighted
displacement and $\vct v=\Dt{\vct x}$, in an orthonormal coordinate
basis. The real matrices satisfy $\mathsf D>0$ and
$\mathsf C=\mathsf C^{\mathsf T}\geq0$.
Here $\mathsf D$ has inverse-time-squared units and the mass-normalized
drag $\mathsf C$ has inverse-time units. A Markov approximation at
temperature $T>0$ is
\begin{equation}
 \dd\vct x=\vct v\,\dd t,\qquad
 \dd\vct v=(-\mathsf D\vct x-\mathsf C\vct v)\,\dd t
                    +\mathsf B\,\dd\vct{\mathcal B}_t,\qquad
 \mathsf B\mathsf B^{\mathsf T}=2k_{\rm B}T\,\mathsf C.
 \label{eq:thermal-lattice-langevin}
\end{equation}
The increments of a standard real Wiener process have mean zero and
covariance $\mathbb E[\dd\mathcal B_a\dd\mathcal B_b]
=\delta_{ab}\dd t$. They scale as the square root of time, rather
than as a smooth force history. In translational normalization,
$x$ has units $\mathrm{kg}^{1/2}\mathrm m$, $v$ has units
$\mathrm{kg}^{1/2}\mathrm m\,\mathrm s^{-1}$, and
$B$ has units $\mathrm{kg}^{1/2}\mathrm m\,\mathrm s^{-3/2}$.
Thus both sides of its covariance prescription have energy-per-time
units. The noise factor $B$ is not unique; its covariance $BB^{\mathsf T}$
specifies the same Gaussian forcing law.

The reservoir remains effectively at fixed $T$ over the observation
interval, an equilibrium or large-reservoir approximation for the finite
sealed system. Exchanged energy remains in the enlarged ledger. White
noise also assumes short unresolved correlation times; Maxwell memory
needs matched colored noise or an augmented stochastic realization.
These reservoir assumptions fix the force covariance used in an
equilibrium test~\cite{Cotterill2002,Seifert2012}.

For a smooth classical function $f(\vct x,\vct v)$, It\^o's rule gives
the backward generator
\begin{equation}
 \mathcal Lf=\vct v\cdot\nabla_{\vct x}f
       -(\mathsf D\vct x+\mathsf C\vct v)\cdot\nabla_{\vct v}f
       +k_{\rm B}T\sum_{a,b}C_{ab}
                          \partial_{v_a}\partial_{v_b}f.
 \label{eq:thermal-backward-generator}
\end{equation}
This backward generator evolves expected functions. Its forward adjoint
evolves the probability law, $\partial_tg=\mathcal L^*g$ when a
density exists. Singular preparations and dark sectors can use weak
measure-valued laws. Linear drift and constant noise give a unique
finite-dimensional stochastic solution for each initial state and
finite time.

\subsubsection{Finite-time covariance and the conditions for thermalization}

Set $\vct X_{\rm th}=(\vct x,\vct v)^{\mathsf T}$,
$\mathsf A_{\rm th}=\left(\begin{smallmatrix}0&\Id\\
-\mathsf D&-\mathsf C\end{smallmatrix}\right)$, and
$\mathsf B_{\rm th}=\left(\begin{smallmatrix}0\\\mathsf B
\end{smallmatrix}\right)$.
For an initial distribution independent of future Wiener increments,
its mean and covariance with finite second moments obey
\begin{equation}
 \begin{aligned}
 \bar{\vct X}_{\rm th}(t)&=e^{t\mathsf A_{\rm th}}
                                      \bar{\vct X}_{\rm th}(0),\\
 \mathsf V_{\rm th}(t)
   &=e^{t\mathsf A_{\rm th}}\mathsf V_{\rm th}(0)
                         e^{t\mathsf A_{\rm th}^{\mathsf T}}
    +\int_0^t e^{s\mathsf A_{\rm th}}
        \mathsf B_{\rm th}\mathsf B_{\rm th}^{\mathsf T}
                        e^{s\mathsf A_{\rm th}^{\mathsf T}}\,\dd s.
 \end{aligned}
 \label{eq:thermal-finite-time-covariance}
\end{equation}
The first term propagates the initial covariance; the integral
accumulates reservoir forcing. The stochastic convolution is Gaussian,
while the initial law may be non-Gaussian. Mean and covariance specify
the full law when its Gaussian character is established.

The Gibbs law $g_{\rm eq}\propto e^{-E_{\rm mech}/(k_{\rm B}T)}$,
with $E_{\rm mech}=(\vct v^{\mathsf T}\vct v+
\vct x^{\mathsf T}\mathsf D\vct x)/2$, is invariant. Its covariance
satisfies the stationary Lyapunov identity
\begin{equation}
 \mathsf V_{\rm th,eq}=k_{\rm B}T\diag(\mathsf D^{-1},\Id),\qquad
 \mathsf A_{\rm th}\mathsf V_{\rm th,eq}
 +\mathsf V_{\rm th,eq}\mathsf A_{\rm th}^{\mathsf T}
 +\mathsf B_{\rm th}\mathsf B_{\rm th}^{\mathsf T}=0.
 \label{eq:thermal-stationary-lyapunov}
\end{equation}
Direct multiplication verifies this even when $\mathsf C$ and
$\mathsf D$ do not commute. Consequently the finite-time covariance
also equals
$\mathsf V_{\rm th,eq}+e^{t\mathsf A_{\rm th}}
[\mathsf V_{\rm th}(0)-\mathsf V_{\rm th,eq}]
e^{t\mathsf A_{\rm th}^{\mathsf T}}$.
For a stationary preparation its two-time covariance is
$\mathbb E[\delta\vct X_{\rm th}(t)
\delta\vct X_{\rm th}(0)^{\mathsf T}]
=e^{t\mathsf A_{\rm th}}\mathsf V_{\rm th,eq}$ for $t\geq0$.
The same drift therefore predicts the relaxation of a perturbation and
the time dependence of equilibrium correlations.

Invariance alone does not ensure attraction. At $T>0$ the exact
reachability condition in this symmetric finite-dimensional model is
\begin{equation}
 \ker(\mathsf D-\omega^2\Id_r)\cap\ker\mathsf C=\{0\}
                      \quad\hbox{for every eigenfrequency}.
 \label{eq:phonon-thermal-reachability}
\end{equation}
Equivalently,
$\operatorname{span}\{\mathsf D^j\operatorname{ran}\mathsf C:
0\leq j<r\}=\R^r$.
No mechanical eigenspace may contain a nonzero direction invisible to
the reservoir. This is the corresponding Ornstein--Uhlenbeck Kalman
condition~\cite{Ottobre2015}; it also makes the drift strictly stable
and the invariant Gibbs law unique. Positive definite drag suffices,
but is not necessary.

For example, $\mathsf C=\gamma\diag(1,0)$, $\gamma>0$, directly
forces only the first coordinate. For
$\mathsf D=\left(\begin{smallmatrix}a&b\\b&d\end{smallmatrix}\right)>0$,
$\vct e_1$ and $\mathsf D\vct e_1$ span both coordinates exactly
when $b\ne0$. Mechanical coupling then thermalizes an internal
coordinate through an envelope-coupled bath without inventing a second
independent noise source. Figure~\ref{fig:thermal-reachability-one-reservoir}
shows that startup from zero uncertainty need not be stationary.
At $b=0$ the second coordinate retains its undamped preparation.
Degeneracy requires particular care: if $\mathsf D=\omega_0^2\Id_2$
and $\mathsf C=\gamma\vct h\vct h^{\mathsf T}$ with
$\vct h=(1,1)^{\mathsf T}/\sqrt2$, each displayed coordinate has a
nonzero diagonal drag, yet $(1,-1)^{\mathsf T}$ is a dark eigenvector.
Checking diagonal damping numbers alone would miss it.

When this reachability condition holds, the stationary displacement
spectrum follows directly by propagating the white force:
\begin{equation}
 \mathsf S_x(\Omega)
   =2k_{\rm B}T\,\mathsf R_x(\Omega)\mathsf C
                                \mathsf R_x(\Omega)^\dagger,
 \qquad
 \mathsf R_x(\Omega)=(\mathsf D-\Omega^2\Id-i\Omega\mathsf C)^{-1}.
 \label{eq:thermal-markov-displacement-spectrum}
\end{equation}
Here $\mathsf S_x$ uses the two-sided angular-frequency convention of
Section~\ref{sec:operator-probe-response}. The identity
$\operatorname{Im}_{\rm H}\mathsf R_x
=\Omega\mathsf R_x\mathsf C\mathsf R_x^\dagger$, where
$\operatorname{Im}_{\rm H}A=(A-A^\dagger)/(2i)$, recovers the classical
fluctuation--response relation without requiring simultaneous
diagonalization of drag and stiffness. The stochastic startup in the
figure is not stationary and must instead use its finite-time covariance.
An undamped dark sector retains its initial fluctuations; one cannot
infer those from the noise-driven spectrum alone.

\begin{figure}[H]
\centering\begingroup
\definecolor{vltTeal}{HTML}{176B64}
\definecolor{vltBlue}{HTML}{315D80}
\definecolor{vltGold}{HTML}{A87520}
\definecolor{vltPlum}{HTML}{725877}
\definecolor{vltRed}{HTML}{762B38}
\definecolor{vltYellow}{HTML}{D5AD3C}
\definecolor{vltNeutral}{HTML}{777D83}
\setlength{\unitlength}{1pt}
\scalebox{0.90}{%
\input{vlt-figures/figure-211.tex}%
}\endgroup
\caption{A single matched thermal channel on a supported pair with $D/\omega_0^2=\left(\begin{smallmatrix}2&-1\\-1&2\end{smallmatrix}\right)$ and $C/\omega_0=\operatorname{diag}(1,0)$. The open bath arrow identifies its stochastic input; coils show reference topology. Starting from $(\vct x,\vct v)=0$, covariance evolves as $V_{\rm eq}-e^{At}V_{\rm eq}e^{A^{\mathsf T}t}$. Coupling brings both position variances to $2k_{\rm B}T/(3\omega_0^2)$. The dotted comparison instead uses $D/\omega_0^2=I$ with the same single bath, leaving site 2 unforced and at zero variance. Curves describe ensemble startup toward equilibrium.}
\label{fig:thermal-reachability-one-reservoir}
\end{figure}

\subsubsection{Which thermal time correlations identify a drift?}

Equilibrium widths and relaxation times supply different mechanical
information. At known temperature and in the complete mass-weighted
coordinate basis, the equilibrium position covariance already determines
$\mathsf D=k_{\rm B}T(\mathsf V_{{\rm th,eq},xx})^{-1}$.
It contains no independent value of $\mathsf C$. Time correlations test
how that same preparation moves and loses memory.

\begin{definition}[Complete-state lag covariance]
\label{def:s611-complete-lag-covariance}
For the stationary classical preparation above, define
$\mathsf K_{\rm lag}(t)=
\mathbb E[\delta\vct X_{\rm th}(t)
\delta\vct X_{\rm th}(0)^{\mathsf T}]$, $t\geq0$.
A \emph{complete-state observation} here resolves all retained
displacements and velocities, with the mass normalization, coordinate
units, temporal response and detector-noise covariance calibrated.
The covariance used in this definition is that of the mechanical state,
after applying the declared observation-law corrections.
\end{definition}

\begin{proposition}[Drift identification from a complete lag family]
\label{prop:s611-lag-identifies-drift}
Under the stationary thermal model and reachability condition above,
$\mathsf V=\mathsf K_{\rm lag}(0)>0$. The complete continuous lag
family determines the drift through
\begin{equation}
 \begin{aligned}
 \mathsf K_{\rm lag}(t)\mathsf V^{-1}
     &=e^{t\mathsf A_{\rm th}},\\
 \mathsf A_{\rm th}
     &=\mathsf K_{\rm lag}'(0^+)\mathsf V^{-1},\\
 \mathsf B_{\rm th}\mathsf B_{\rm th}^{\mathsf T}
     &=-\mathsf A_{\rm th}\mathsf V
                          -\mathsf V\mathsf A_{\rm th}^{\mathsf T}.
 \end{aligned}
 \label{eq:s611-lag-drift-identification}
\end{equation}
Consequently the lower drift blocks identify $-\mathsf D$ and
$-\mathsf C$, and the last line tests their compatibility with the
thermal force covariance $2k_{\rm B}T\mathsf C$ in the velocity block.
\end{proposition}
\begin{proof}
The stochastic solution is
$\vct X_{\rm th}(t)=e^{t\mathsf A_{\rm th}}\vct X_{\rm th}(0)$
plus a zero-mean stochastic convolution independent of the initial
state. Multiplication by the centered initial state and averaging gives
$\mathsf K_{\rm lag}(t)=e^{t\mathsf A_{\rm th}}\mathsf V$.
Invertibility and the right derivative at zero give the first two
identities. Stationarity in the covariance equation gives the third.
\end{proof}

This identity applies to ideal complete data. Finite-record estimation
instead fits the mechanical constraints and observation law. One sampled
lag gives $e^{\Delta t\mathsf A_{\rm th}}$; selecting its logarithm
uses additional frequency or structural information. Gibbs covariance
and mechanical block structure can supply those constraints. Partial
channels give $\mathsf L e^{t\mathsf A_{\rm th}}\mathsf V\mathsf L^{\mathsf T}$,
whose zero-lag inverse generally leaves part of the propagator unresolved.

For an ideal-virion assembly, this distinguishes a change in contact
curvature from a change in medium drag: a classical equilibrium width
constrains the former, whereas its time dependence also constrains the
latter. A centre-only movie may omit an internal motion or an unresolved
velocity, and an exposure average modifies its lag law. The appropriate
predictive variables are therefore the ones required by the measured
time correlations, with unresolved effects carried by the previously
derived memory and preparation terms.

\paragraph{Testing Gaussian closure from lagged records.}
A partial record can reveal whether its present value retains the information needed for its own future. This test is distinct from identifying the physical origin of a missing coordinate.

\begin{proposition}[Gaussian record closure]
\label{prop:s6patch-gaussian-record-closure}
Let $\vct Y(t)$ be a centred stationary real Gaussian vector process, continuous in mean square, with
$\mathsf K(t)=\mathbb E[\vct Y(t)\vct Y(0)^{\mathsf T}]$ and
$\mathsf K_0=\mathsf K(0)\succ0$. For $s,t\geq0$,
\begin{equation}
 \boxed{\begin{aligned}
 \mathsf\Delta(t,s)
   &:=\mathsf K(t+s)-\mathsf K(t)\mathsf K_0^{-1}\mathsf K(s)\\
   &=\operatorname{Cov}[\vct Y(t+s),\vct Y(0)\mid\vct Y(s)] .
 \end{aligned}}
 \label{eq:s6patch-gaussian-record-defect}
\end{equation}
The process is Markov with respect to its natural filtration if and only if
$\mathsf\Delta(t,s)=0$ for every $s,t\geq0$.
\end{proposition}
\begin{proof}
Gaussian conditioning gives the displayed Schur complement. Its vanishing makes each future residual independent of every earlier finite collection, conditional on the present; Gaussianity converts zero conditional cross-covariance into independence. These finite collections generate the past filtration. The converse follows from the Markov property. Equivalently,
$\mathsf K(t)\mathsf K_0^{-1}$ is a continuous finite-dimensional semigroup, hence equals $e^{t\mathsf A_Y}$.
\end{proof}

Figure~\ref{fig:cmd-prepared-projected-records} gives a graphical specialization of this boxed relation.

This is the Gaussian covariance characterization of Markov evolution
\cite{BrownErnstSheppWolpert2014}. A singular $\mathsf K_0$ first requires restriction to its supported record space. Outside the Gaussian class, this covariance test alone does not establish closure.

\begin{example}[A hidden contrast preserves the same equilibrium width]
\label{ex:s6patch-two-rate-hidden-contrast}
Let independent stationary dimensionless coordinates satisfy
$\dd z_j=-\lambda_jz_j\dd t+\sqrt{2\lambda_j}\dd W_j$, with
$\lambda_j>0$, and observe $Y=(z_1+z_2)/\sqrt2$.
Its variance is one, while, for $t\geq0$,
\begin{equation}
 K(t)=\frac{e^{-\lambda_1t}+e^{-\lambda_2t}}2,\qquad
 \Delta(t,s)=\frac14
 (e^{-\lambda_1t}-e^{-\lambda_2t})
 (e^{-\lambda_1s}-e^{-\lambda_2s}).
 \label{eq:s6patch-two-rate-closure-defect}
\end{equation}
For distinct rates and positive lags the defect is nonzero. Retaining
$\eta=(z_1-z_2)/\sqrt2$ restores a two-dimensional Markov state. Its assignment to surface chemistry, material relaxation or another motion requires independent physical constraints.
\end{example}
A position-only inertial record can also fail closure because velocity is unresolved. Exposure and detector noise must likewise enter the forward comparison before a residual is attributed to an additional material variable.

\subsubsection{Exact frame dynamics with motion during exposure}
\label{sec:operator-exact-camera-frames}

A camera integrates motion while the same thermal force continues to
change the assembly. Consequently, the random motion contributing to
one image also contributes to the next mechanical state. An exact frame
model makes this connection calculable from the retained drift and noise
matrices. Linear stochastic discretization supplies the state transition
\cite{SarkkaSolin2019}; motion-blur filtering treats the corresponding
correlation in particle tracking~\cite{Berglund2010,Calderon2016}.

\begin{definition}[Exposure-resolved frame model]
\label{def:operator-exact-camera-frames}
Within one persistent linear episode, subtract the deterministic mean and
write $\dd\vct X=\mathsf A\vct X\,\dd t+\mathsf B\,\dd\vct W_t$,
with constant real matrices and a standard Wiener process independent
of the initial state. Here $\mathsf A=\mathsf A_{\rm th}$ and
$\mathsf B=\mathsf B_{\rm th}$, or their specified material-state
augmentation. For $t_k=k\Delta$ and $0<\tau_e\leq\Delta$, define
$\vct X_k=\vct X(t_k)$ and the calibrated frame
\begin{equation}
 \vct Y_k=\frac1{\tau_e}\int_0^{\tau_e}
             \mathsf L\vct X(t_k+s)\,\dd s+\vct\epsilon_k.
 \label{eq:operator-camera-frame-definition}
\end{equation}
The localization errors $\vct\epsilon_k$ are independent, zero mean,
have covariance $\mathsf R_{\rm loc}$, and are independent of the
mechanical process. The frame is assigned to its exposure interval
$[t_k,t_k+\tau_e]$.
\end{definition}

\begin{proposition}[Joint frame and state increments]
\label{prop:operator-camera-frame-increments}
The exact discrete representation is
\begin{equation}
 \begin{aligned}
 \vct X_{k+1}&=\mathsf F_\Delta\vct X_k+\vct\eta_k,
 &\mathsf F_\Delta&=e^{\Delta\mathsf A},\\
 \vct Y_k&=\mathsf H_e\vct X_k+\vct\zeta_k+\vct\epsilon_k,
 &\mathsf H_e&=\frac1{\tau_e}\int_0^{\tau_e}
                        \mathsf L e^{s\mathsf A}\,\dd s.
 \end{aligned}
 \label{eq:operator-camera-exact-increments}
\end{equation}
Set $\mathsf E(r)=e^{(\Delta-r)\mathsf A}\mathsf B$ and
$\mathsf J_e(r)=\tau_e^{-1}\int_r^{\tau_e}
\mathsf L e^{(s-r)\mathsf A}\mathsf B\,\dd s$ for
$0\leq r\leq\tau_e$. The increment pairs
$(\vct\eta_k,\vct\zeta_k)$ are centered Gaussian, independent
between intervals and of $\vct X_k$, with
\begin{equation}
 \begin{aligned}
 \mathsf Q_\Delta&=\mathbb E[\vct\eta_k\vct\eta_k^{\mathsf T}]
       =\int_0^\Delta\mathsf E(r)\mathsf E(r)^{\mathsf T}\,\dd r,\\
 \mathsf Q_e&=\mathbb E[\vct\zeta_k\vct\zeta_k^{\mathsf T}]
       =\int_0^{\tau_e}\mathsf J_e(r)\mathsf J_e(r)^{\mathsf T}\,\dd r,\\
 \mathsf S_{\Delta e}
       &=\mathbb E[\vct\eta_k\vct\zeta_k^{\mathsf T}]
       =\int_0^{\tau_e}\mathsf E(r)\mathsf J_e(r)^{\mathsf T}\,\dd r.
 \end{aligned}
 \label{eq:operator-camera-joint-noise}
\end{equation}
In particular, the two increments within one interval generally share
noise even when localization error is independent.
\end{proposition}

\begin{proof}
Variation of constants gives the state increment as the stochastic
integral of $\mathsf E(r)$ over the interval. Substitution in the
exposure integral and interchange of the deterministic and stochastic
integrations gives $\vct\zeta_k=\int_0^{\tau_e}\mathsf J_e(r)
\,\dd\vct W_{t_k+r}$. These finite-dimensional deterministic kernels
are square integrable. It\^o isometry gives all three covariances;
independent Wiener increments give the interval and initial-state
independence.
\end{proof}

For a stationary covariance $\mathsf V$, the continuous Lyapunov law
implies $\mathsf V=\mathsf F_\Delta\mathsf V
\mathsf F_\Delta^{\mathsf T}+\mathsf Q_\Delta$. The frame statistics are
\begin{equation}
 \boxed{\begin{aligned}
 \operatorname{Cov}(\vct Y_k)
    &=\mathsf H_e\mathsf V\mathsf H_e^{\mathsf T}
                            +\mathsf Q_e+\mathsf R_{\rm loc},\\
 \operatorname{Cov}(\vct Y_{k+1},\vct Y_k)
    &=\mathsf H_e\bigl(\mathsf F_\Delta\mathsf V
                     \mathsf H_e^{\mathsf T}+\mathsf S_{\Delta e}\bigr).
 \end{aligned}}
 \label{eq:operator-camera-frame-covariances}
\end{equation}
Thus an exposure-averaged initial state with an independently added
frame error omits a definite contribution to the lag covariance.
The joint increment covariance supplies the observation model for a
Gaussian likelihood when the initial state and localization errors
are also Gaussian. Its parameters still require the identifiability
and calibration checks above.

The scalar overdamped coordinate of
\eqref{eq:opening-thermal-reduction} gives a closed comparison.
Set $\tau_y=c_{\rm loc}/k_y$, $\sigma_y^2=k_{\rm B}T/k_y$,
$a=\tau_e/\tau_y$ and $d=\Delta/\tau_y\geq a$. With zero
localization error, stationary integration gives
\begin{equation}
 \frac{\operatorname{Var}Y_k}{\sigma_y^2}
      =\frac{2(a-1+e^{-a})}{a^2},\qquad
 \frac{\operatorname{Cov}(Y_{k+1},Y_k)}{\sigma_y^2}
      =e^{-d}\frac{e^a+e^{-a}-2}{a^2}.
 \label{eq:operator-camera-scalar-check}
\end{equation}
Both follow by integrating
$\langle y(t)y(s)\rangle=\sigma_y^2e^{-|t-s|/\tau_y}$ over
the two exposure windows. As $a\to0$, they recover the instantaneous
variance and lag covariance. Figure~\ref{fig:operator-exact-camera-frames}
separates the conditional mean from motion accumulated during exposure.
For virion tracking, these terms help distinguish a changed material
relaxation time from a changed acquisition time. The same matrix
construction includes retained material memory; an interval spanning
a rearrangement uses the registered event and exposure maps of
Section~\ref{sec:s6-readout-event-response}.

\begin{figure}[H]
\centering\begingroup
\definecolor{vltTeal}{HTML}{176B64}
\definecolor{vltBlue}{HTML}{315D80}
\definecolor{vltGold}{HTML}{A87520}
\definecolor{vltPlum}{HTML}{725877}
\definecolor{vltRed}{HTML}{762B38}
\definecolor{vltYellow}{HTML}{D5AD3C}
\definecolor{vltNeutral}{HTML}{777D83}
\setlength{\unitlength}{1pt}
\scalebox{0.90}{%
\input{vlt-figures/figure-212.tex}%
}\endgroup
\caption{Motion continues while a virion coordinate is imaged. (a) An illustrative stationary Ornstein--Uhlenbeck realization, in rms and relaxation-time units, is integrated over two pale windows; thick teal segments show their averages. The spring is at reference geometry and the arrow names the tracked direction. (b) The exact exposure variance includes both the random initial state and motion during the window. (c) At fixed $\Delta/\tau_y=3$, the shared increment raises the next-frame covariance above the initial-state contribution alone. Black curves follow \eqref{eq:operator-camera-scalar-check}; dashed blue curves retain only $H_e^2\sigma_y^2$ and $H_e^2\sigma_y^2e^{-3}$, respectively. Shading isolates the omitted positive contributions. Localization variance is zero. The point-exposure limit is included at $a=0$; the dots mark $a=1$. The boxed result illustrated here is~\eqref{eq:operator-camera-frame-covariances}.}
\label{fig:operator-exact-camera-frames}
\end{figure}

\paragraph{Acquisition can generate apparent record memory.}
The scalar camera check gives an exact counterexample to interpreting a nonzero observed closure defect as material memory. Take its stationary scalar Gaussian Ornstein--Uhlenbeck coordinate, with $\sigma_y^2>0$, rectangular exposure and no localization error. Put
$a=\tau_e/\tau_y>0$, $d=\Delta/\tau_y\geq a$, and
$c_n=\operatorname{Cov}(Y_{k+n},Y_k)$. Integration of the exponential covariance gives
\begin{equation}
 \begin{aligned}
 v_e(a)&=\frac{2(a-1+e^{-a})}{a^2},&
 j_e(a)&=\frac{e^a+e^{-a}-2}{a^2},\\
 c_0&=\sigma_y^2v_e(a),&
 c_n&=\sigma_y^2j_e(a)e^{-nd}\quad(n\geq1).
 \end{aligned}
 \label{eq:s6patch-exposure-lag-family}
\end{equation}
The three-frame conditional covariance is consequently
\begin{equation}
 c_2-\frac{c_1^2}{c_0}
 =\sigma_y^2j_e(a)e^{-2d}
       \left[1-\frac{j_e(a)}{v_e(a)}\right]\ne0,
 \qquad
 j_e(a)-v_e(a)=\frac{2(\sinh a-a)}{a^2}>0.
 \label{eq:s6patch-exposure-markov-defect}
\end{equation}
The latent coordinate is Markov, but its Gaussian exposure-averaged frame sequence is not first-order Markov. The shared increment in
\eqref{eq:operator-camera-joint-noise} predicts this effect. Localization error adds its own variance and changes the comparison again. A variable-selection test therefore compares the observed lag family with the complete calibrated acquisition model; remaining discrepancies can then test the retained mechanical, surface and material state.

\paragraph{Information in a correlated thermal record.}
Preparation, relaxation and acquisition parameters can change the covariance even when the predicted mean remains zero. Stack a finite real record with a justified Gaussian law
$\vct y\sim\mathcal N(\vct m(\vct\theta),\mathsf\Sigma(\vct\theta))$,
where $\mathsf\Sigma\succ0$ and both functions are differentiable.
The covariance includes shared process/frame noise and propagated initial uncertainty.

\begin{proposition}[Gaussian record information]
\label{prop:s6patch-gaussian-record-information}
For the stated finite record, the Fisher information is
\begin{equation}
 \begin{aligned}
 I_{ij}
  &=(\partial_i\vct m)^{\mathsf T}\mathsf\Sigma^{-1}
                                     (\partial_j\vct m)\\
  &\quad+\frac12\tr\!\left(
       \mathsf\Sigma^{-1}(\partial_i\mathsf\Sigma)
       \mathsf\Sigma^{-1}(\partial_j\mathsf\Sigma)\right).
 \end{aligned}
 \label{eq:s6patch-gaussian-record-information}
\end{equation}
\end{proposition}
\begin{proof}
Differentiate the multivariate Gaussian log density. Its centred third moments vanish, and its fourth-moment pairings give the covariance term~\cite{TegmarkTaylorHeavens1997}.
\end{proof}
For a parameter variation, the resulting quadratic form is
$\|\mathsf\Sigma^{-1/2}\delta\vct m\|^2+
\|\mathsf\Sigma^{-1/2}\delta\mathsf\Sigma
\mathsf\Sigma^{-1/2}\|_{\rm F}^2/2$.
It vanishes precisely when both record variations vanish. Preparation, gain and timing uncertainties enter as nuisance parameters; their regular information block is removed by the same Schur complement as in
\eqref{eq:s56-profiled-parameter-information}.
This quantifies local distinguishability within the stated Gaussian model. Global target identifiability still uses
Definition~\ref{def:s6-identifiable-prediction}, and nonlinear orientation or contact-switching records require their own probability law.

\subsubsection{Energy exchange, probability relaxation, and a slow limit}

Applying It\^o's rule to the same mechanical energy yields
\begin{equation}
 \dd E_{\rm mech}
  =[-\vct v^{\mathsf T}\mathsf C\vct v
                    +k_{\rm B}T\tr\mathsf C]\,\dd t
       +\vct v^{\mathsf T}\mathsf B\,\dd\vct{\mathcal B}_t.
 \label{eq:phonon-thermal-ito-energy}
\end{equation}
The second drift term comes from the quadratic Wiener increment; it
would be lost by treating the noise as an ordinary differentiable force.
At Gibbs equilibrium, $\mathbb E[\vct v\vct v^{\mathsf T}]
=k_{\rm B}T\Id$ makes mean injection equal mean dissipation.
A thermal path can gain resolved energy temporarily. That random return
from the reservoir is not evidence of negative damping or an autonomous
limit cycle~\cite{Seifert2012}.

For sufficiently regular densities with decaying boundary terms, the
relative entropy $\mathcal D(g\Vert g_{\rm eq})
=\int g\ln(g/g_{\rm eq})\,\dd\vct x\,\dd\vct v$ obeys
\begin{equation}
 \frac{\dd\mathcal D}{\dd t}
   =-k_{\rm B}T\int g\,
       (\nabla_{\vct v}\ln(g/g_{\rm eq}))^{\mathsf T}
         \mathsf C\nabla_{\vct v}\ln(g/g_{\rm eq})
                                      \,\dd\vct x\,\dd\vct v\leq0.
 \label{eq:thermal-relative-entropy-dissipation}
\end{equation}
Hamiltonian streaming cancels in this integral and carries no entropy
production of this kind; friction and matched diffusion produce the
displayed term. The full backward generator is nevertheless not
self-adjoint in the Gibbs inner product, because streaming is
antisymmetric there. Probability conservation and Gibbs invariance do
not require quantum unitary evolution. Zero instantaneous entropy
dissipation need not mean equilibrium: mechanical transport can bring
initial position structure into dissipative velocity directions later.
Long-time relaxation still needs the reachability condition above.
Equilibrium Hamiltonian streaming can circulate probability in phase
space while its irreversible friction--diffusion current vanishes.
The underdamped time-reversal comparison includes changing the sign of
velocity; it does not require the entire streaming generator to be
self-adjoint in the unchanged variables.

A separate overdamped approximation removes resolved velocity only
after an appropriate time-scale separation. With constant positive
mobility $\boldsymbol\mu$, it gives
$\dd\vct x=-\boldsymbol\mu\mathsf D\vct x\,\dd t
+\sqrt{2k_{\rm B}T\boldsymbol\mu}\,\dd\vct{\mathcal B}_t$ and
\begin{equation}
 \partial_tg=\nabla_{\vct x}\cdot
   [\boldsymbol\mu(\mathsf D\vct x\,g+
                                      k_{\rm B}T\nabla_{\vct x}g)].
 \label{eq:overdamped-lattice-fokker-planck}
\end{equation}
If all velocity directions relax rapidly, $\boldsymbol\mu$ can be
$\mathsf C^{-1}$ in these mass-weighted coordinates. A singular drag
matrix cannot simply be inverted, even when coupling eventually
thermalizes every mode. The overdamped model cannot predict the inertial
oscillations it has removed. For $\mathsf C=\gamma\Id$, a concrete
separation is $\max_s\omega_s/\gamma\ll1$, with observation after the
velocity time $\gamma^{-1}$ and slow position rates
$\omega_s^2/\gamma$. At a reflecting configuration boundary its
normal probability current vanishes; this statistical boundary law is
different from a continuum zero-power slip condition. A restrictive
configuration boundary also truncates the full-line Gibbs law.

\paragraph{Rotational Brownian motion and a registered surface marker.}
The configuration-dependent mobility law of
Proposition~\ref{prop:s2patch-passive-configuration-generator} includes the rotational measure and thermal drift. On one labelled body its laboratory rotational derivatives are
\begin{equation}
 (\mathcal D_a f)(\mathsf O)
 =\left.\frac{\dd}{\dd\epsilon}
 f(e^{\epsilon[\vct e_a]_\times}\mathsf O)\right|_{\epsilon=0},
 \qquad \mathcal D_a^*=-\mathcal D_a
 \quad\hbox{in }L^2(\dd\mathsf O),
 \label{eq:s6patch-haar-rotational-derivatives}
\end{equation}
where $\dd\mathsf O$ is Haar measure. Their adjoints follow from invariance under rotations. Translation--rotation coupling and configuration-dependent mobility use the full generator, rather than independent Cartesian angular kicks~\cite{DelongUsabiagaDonev2015}.

For a held centre with unrestricted orientation, no torque and constant isotropic rotational diffusion $D_{\rm rot}>0$, take
$\mathcal L_{\rm rot}=D_{\rm rot}\sum_a\mathcal D_a^2$.
A registered marker $\vct b(t)=\mathsf O(t)\vct a$, with
$a_0=\|\vct a\|>0$, has Haar-equilibrium covariance
\begin{equation}
 \operatorname{Cov}[\vct b(t),\vct b(0)]
   =\frac{a_0^2}{3}e^{-2D_{\rm rot}|t|}\Id_3 .
 \label{eq:s6patch-free-marker-covariance}
\end{equation}
Indeed $\sum_a\mathcal D_a^2\vct b=-2\vct b$, and isotropic equilibrium gives
$\langle\vct b\vct b^{\mathsf T}\rangle=a_0^2\Id_3/3$.

For a unit viewing direction $\vct e$, the instantaneous
$Y=\vct e\cdot\vct b$ is uniform on $[-a_0,a_0]$, so
$\langle Y^4\rangle=a_0^4/5$. A Gaussian of the same variance would give
$a_0^4/3$. This bounded coordinate is itself Markov, with generator
\begin{equation}
 (\mathcal L_Yf)(y)
 =D_{\rm rot}\big[(a_0^2-y^2)f''(y)-2yf'(y)\big].
 \label{eq:s6patch-projected-marker-generator}
\end{equation}
Thus exponential covariance alone does not identify a Gaussian oscillator law. Rectangular exposures nevertheless obey the second-moment formulas
\eqref{eq:operator-camera-scalar-check}, with
$\sigma_y^2=a_0^2/3$ and $\tau_y=(2D_{\rm rot})^{-1}$.
The Gaussian frame-increment law does not follow globally; the fourth-moment comparison above concerns the instantaneous marker. Walls, patch torques and anisotropic mobility instead use their specified configuration generator.

\subsubsection{Thermal motion with resolved contact states}
\label{sec:s6-thermal-contact-state}

Recruitment can change a contact's mechanical state while the same
particles and prepared medium continue moving. An equilibrium model of
that process specifies both the conditional forces and the probabilities
of changing contact state. A reference Hessian alone supplies neither
the relative thermodynamic weights nor the transition kinetics.

\begin{definition}[Joint contact-state preparation]
\label{def:s6-thermal-contact-preparation}
Let $g$ belong to a finite set of retained physical contact or
conformational states. Use the same full-pool mechanical coordinates
$X$ in every state, including momentum and each retained material strain.
Write $f_g(X,t)$ for their joint probability densities, with
$\sum_g\int f_g\,\dd X=1$. For identity-continuation events, the
physical coordinates remain continuous while $g$ changes, and
\begin{equation}
 \partial_t f_g=\mathcal L_g^*f_g+
       \sum_{h\ne g}[k_{hg}(X)f_h-k_{gh}(X)f_g].
 \label{eq:s6-thermal-contact-forward-law}
\end{equation}
Here $k_{gh}$ is the rate from $g$ to $h$; $\mathcal L_g^*$ includes
the forces and consistently matched thermal fluctuations at fixed $g$.
Other physical event maps require their transported probability kernels.
Identity continuation uses common physical memory coordinates; a strain
defined relative to a state-dependent rest configuration instead uses
its actual registration map.
\end{definition}

A resolved contact state changes force, rest configuration or stored
energy at the same positions. A geometrical label $g=g(\vct r)$
instead partitions the existing configuration law, with probabilities
set by its current across the membership boundary. Independent contact
weights are appropriate when that additional physical state is retained.
Both descriptions use the same fixed particle pool.

\begin{proposition}[An equilibrium law for continued contact states]
\label{prop:s6-thermal-contact-equilibrium}
At a common fixed temperature, let
$\mathcal E_g(X)$ be the specified Gibbs-weight function, including
the relative contact-state offsets. Assume
$\mathcal L_g^*e^{-\beta\mathcal E_g}=0$ on the common state domain,
with the stated no-flux boundary laws and finite partition integrals.
If the jump rates satisfy
\begin{equation}
 k_{gh}(X)e^{-\beta\mathcal E_g(X)}
   =k_{hg}(X)e^{-\beta\mathcal E_h(X)},\qquad
 \pi_g(X)=\frac{e^{-\beta\mathcal E_g(X)}}
                    {\sum_h\int e^{-\beta\mathcal E_h}\,\dd X},
 \label{eq:s6-thermal-contact-detailed-balance}
\end{equation}
then the joint law $\pi_g$ is stationary under
\eqref{eq:s6-thermal-contact-forward-law}.
\end{proposition}
\begin{proof}
Each continuous generator annihilates its Gibbs weight. The displayed
pairwise identity cancels every incoming jump flux with its reverse
outgoing flux at the same physical state.
\end{proof}

The identity is the equilibrium detailed-balance requirement for the
contact transitions. It leaves their common kinetic scale undetermined.
Hamiltonian streaming retains the momentum-reversal qualification
already stated above; stationarity alone also does not prove attraction.
Connectivity, bath reachability and confinement still matter.
The isothermal reservoir assumption and local thermodynamic balance
are standard in stochastic descriptions of biomolecular transitions
\cite{Seifert2012}. Here retained strains continue through a jump;
they are not replaced by newly thermalized draws. Their conditional
storage belongs in $\mathcal E_g$ and in its force law.

When fast molecular variables have been integrated out,
$\mathcal E_g$ can contain their conditional free energy, including
intrinsic-entropy offsets. The associated rate ratio uses that free
energy; heat and work still require the corresponding internal-energy
and entropy account~\cite[Section~9.2]{Seifert2012}. Chemical driving
or a finite changing reservoir instead needs its supplied affinity or
resource state and the event ledger of
\eqref{eq:s4-transient-event-first-law}. Arbitrarily switching stable
harmonic models and resetting their noise preparation does not establish
the equilibrium law in the proposition.

For transport and delivery, the same hybrid construction can use
structural-integrity classes rather than only contact classes.
Section~\ref{sec:s7transport-delivery} specifies their probability kets,
state-dependent loading and calibrated class readout. Centre and marker
records then test within-basin motion, while a separate structural
measurement tests the proposed transition and any delayed loss.

\paragraph{A local curvature and a population fluctuation measure different things.}
For a common set of $d$ length-valued harmonic coordinates, take
$V_g(\vct q)=\epsilon_g+
(\vct q-\vct a_g)^{\mathsf T}\mathsf K_g
(\vct q-\vct a_g)/2$ with $\mathsf K_g>0$ and common inertia.
Here $V_g$ includes any state-dependent conditional free energy from
variables already integrated out; all remaining partition factors are
assumed common across $g$. Full-line Gaussian integration gives
\begin{equation}
 \begin{aligned}
 p_g&=\frac{e^{-\beta\epsilon_g}(\det\mathsf K_g)^{-1/2}}
              {\sum_h e^{-\beta\epsilon_h}(\det\mathsf K_h)^{-1/2}},\\
 \mathsf C_q&=\sum_g p_g k_{\rm B}T\mathsf K_g^{-1}
       +\sum_g p_g(\vct a_g-\bar{\vct a})
                      (\vct a_g-\bar{\vct a})^{\mathsf T},\\
 \bar{\vct a}&=\sum_g p_g\vct a_g.
 \end{aligned}
 \label{eq:s6-thermal-contact-mixture-covariance}
\end{equation}
Common momentum factors cancel in $p_g$. The determinant factor
records the available displacement volume within each harmonic state;
the offset $\epsilon_g$ supplies information absent from its Hessian.
These formulas require the declared unconstrained harmonic basins;
finite geometric boundaries replace their Gaussian partition factors.

\begin{example}[Equal contact curvatures and a broad equilibrium record]
\label{ex:s6-thermal-two-contact-basins}
Let one tracked relative displacement $q$ have two physical contact
states $V_\pm(q)=k(q\mp a)^2/2$, with $k>0$ and equal offsets.
At joint equilibrium each state has probability $1/2$ and conditional
variance $k_{\rm B}T/k$, but
\begin{equation}
 \operatorname{Var}q=\frac{k_{\rm B}T}{k}+a^2,\qquad
 k_{\rm width}:=\frac{k_{\rm B}T}{\operatorname{Var}q}
                   =\frac{k}{1+ka^2/(k_{\rm B}T)}.
 \label{eq:s6-thermal-two-contact-width}
\end{equation}
Thus the pooled width can be large with neither conditional spring
softening. Under a static force conjugate to $q$, the potential becomes
$V_g-fq$. Differentiating the normalized equilibrium mean gives
$\partial_f\langle q\rangle|_{f=0}=\beta\operatorname{Var}q$.
Thus $k_{\rm width}$ is the inverse fully equilibrated static compliance
of this model, while $k$ is its within-contact curvature. Both have a
mechanical meaning. A calibrated displacement record conditioned on
contact state, together with a force response and residence statistics,
can distinguish them experimentally.
An explicit continuous-state transition law is
$k_{gh}(q)=\kappa/[1+e^{\beta(V_h(q)-V_g(q))}]$, $\kappa>0$;
its forward/reverse ratio satisfies
\eqref{eq:s6-thermal-contact-detailed-balance}. Changing $\kappa$
changes residence and time correlations while preserving this same
equilibrium position law.
\end{example}

Eliminating the contact label also changes the marginal force law.
Define $F_{\rm marg}(\vct q)=-k_{\rm B}T\ln
\sum_g e^{-\beta V_g(\vct q)}$ and let conditional brackets at
$\vct q$ use weights proportional to these exponentials.
Differentiation gives
\begin{equation}
 \nabla^2 F_{\rm marg}
  =\langle\nabla^2 V_g\rangle_{g\mid\vct q}
       -\beta\operatorname{Cov}_{g\mid\vct q}(\nabla V_g).
 \label{eq:s6-thermal-marginal-curvature}
\end{equation}
Indeed the derivative of a conditional weight is
$-\beta w_g(\nabla V_g-\langle\nabla V\rangle)$; substitution
proves the identity. In the example,
$F_{\rm marg}=k(q^2+a^2)/2-k_{\rm B}T\ln[2\cosh(\beta kaq)]$
and $F_{\rm marg}''(0)=k-\beta k^2a^2$. A nonquadratic or locally
negative marginal curvature can therefore coexist with positive
conditional curvatures. This is a statement about resolving contact
states, not a negative local Lennard--Jones spring assignment.

Figure~\ref{fig:s6-contact-resolved-and-pooled-fluctuations} distinguishes conditional stiffness, pooled static width and marginal local curvature.

\begin{figure}[H]
\centering\begingroup
\definecolor{vltTeal}{HTML}{176B64}
\definecolor{vltBlue}{HTML}{315D80}
\definecolor{vltGold}{HTML}{A87520}
\definecolor{vltPlum}{HTML}{725877}
\definecolor{vltRed}{HTML}{762B38}
\definecolor{vltYellow}{HTML}{D5AD3C}
\definecolor{vltNeutral}{HTML}{777D83}
\setlength{\unitlength}{1pt}
\scalebox{0.90}{%
\input{vlt-figures/figure-213.tex}%
}\endgroup
\caption{Two physical contact states with $V_\pm(q)=k(q\mp a)^2/2$, equal offsets and $a/\ell_T=1.2$, $\ell_T=\sqrt{k_{\rm B}T/k}$. At fixed centre, $q=(u_2-u_1)/\sqrt2$. Upper rulers locate a common $q=0$ reference of spacing $d=25\ell_T/3$; bodies use $12\,\mathrm{pt}/\ell_T$. Each state is at its own relaxed separation, so different spans encode different rest configurations. Lower colored joint densities each have mass $1/2$ and their black sum has mass one. The three curvature tests give $k$, $k/2.44$ and $-0.44k$: conditional stiffness, equilibrated inverse compliance and marginal local curvature at zero. A change of detailed-balanced transition rates changes residence correlations while preserving these equilibrium widths.}
\label{fig:s6-contact-resolved-and-pooled-fluctuations}
\end{figure}

The marginal is an exact static result. Closed Markov dynamics for it
additionally needs fast conditional contact relaxation or an explicit
memory law. Contact histories and lag correlations test that reduction.
For a virion suspension, conditioning separates within-arrangement
fluctuations from transitions while continuing particle motion and
material storage.

\subsubsection{Quantum Gibbs fluctuations and the admissible harmonic basin}

For independently quantized positive modes at fixed parameters,
\begin{equation}
 \varrho_\beta=\frac{e^{-\beta\hat H_{\rm ph}}}{Z},\qquad
 Z=\prod_s(1-e^{-\beta\hbar\omega_s})^{-1},\qquad
 \bar N_s=\frac1{e^{\beta\hbar\omega_s}-1},\quad
 \beta=(k_{\rm B}T)^{-1}.
 \label{eq:thermal-phonon-gibbs}
\end{equation}
The geometric occupation sums converge for finite mode count,
$\omega_s>0$ and $T>0$. A free coordinate is not included in those sums.
Restoring vacuum energy multiplies $Z$ by $e^{-\beta E_{\rm vac}}$
without changing normalized occupation probabilities. Parameter work
must still use the full energy account of Section~\ref{sec:operator-target-defects}.
The canonical covariances are
\begin{equation}
 \begin{aligned}
 \langle\hat Q_s^2\rangle_\beta
    &=\frac{\hbar}{2\omega_s}
             \coth\!\left(\frac{\beta\hbar\omega_s}{2}\right),\\
 \langle\hat\Pi_s^2\rangle_\beta
    &=\frac{\hbar\omega_s}{2}
             \coth\!\left(\frac{\beta\hbar\omega_s}{2}\right).
 \end{aligned}
 \label{eq:thermal-phonon-covariance}
\end{equation}
At $k_{\rm B}T\gg\hbar\omega_s$ they approach
$k_{\rm B}T/\omega_s^2$ and $k_{\rm B}T$, reproducing classical
equipartition. At low temperature the stipulated quantum oscillator
retains vacuum variance. Temperature alone therefore does not identify
the regime: it must be compared with each retained $\hbar\omega_s$.
These are optional quantized predictions, not evidence that a biological
preparation realizes the required quantum state
\cite{NeumaierWestra2011,Shankar1994}.

At fixed mode parameters the resolved equilibrium heat capacity is
\begin{equation}
 C_V=k_{\rm B}\sum_s\frac{x_s^2e^{x_s}}{(e^{x_s}-1)^2},\qquad
 x_s=\frac{\hbar\omega_s}{k_{\rm B}T}.
 \label{eq:phonon-heat-capacity}
\end{equation}
It is a temperature derivative of mean energy, not a state-independent
linear observable. Its high-temperature limit is $rk_{\rm B}$ for these
$r$ oscillators, not the entire virion's heat capacity. Temperature-dependent
effective curvatures or additional molecular degrees of freedom need
their own contribution. Classical softness remains directly visible in
the covariance even after each oscillator's heat capacity approaches
the same limit.

Finally the fluctuation scale must fit the geometric resolution. Suppose
the translational ball $\|\vct u\|_2<\ell_{\rm safe}$ lies inside the
collision-free harmonic basin about the declared reference. For finite
equal-time covariance and mean, Markov's inequality gives
\begin{equation}
 \boxed{\Pr(\|\vct u\|_2\geq\ell_{\rm safe})
       \leq\frac{\tr\mathsf C_u+\|\bar{\vct u}\|_2^2}
                         {\ell_{\rm safe}^2}.}
 \label{eq:harmonic-basin-excursion-bound}
\end{equation}
The numerator is the mean squared \emph{physical} translational
excursion, after mapping mass-weighted mode covariance back to $\vct u$.
The bound can be capped at one; it assumes no Gaussian law. The commuting
equal-time quantum displacements have a joint distribution to which the
same argument applies. Figure~\ref{fig:thermal-variance-and-harmonic-basin}
compares a centered one-dimensional Gaussian tail with this general bound,
beside the thermal quantum/classical variance crossover.

\begin{figure}[H]
\centering
\begingroup
\definecolor{vlfGreen}{HTML}{176B64}
\definecolor{vlfDarkRed}{HTML}{762B38}
\definecolor{vlfRed}{HTML}{762B38}
\definecolor{vlfBlue}{HTML}{315D80}
\definecolor{vlfCopper}{HTML}{762B38}

\setlength{\unitlength}{1pt}
\scalebox{0.80}{%
\input{vlt-figures/figure-214.tex}%
}%
\endgroup
\caption{Thermal variance and a geometric excursion test. (a) The quantized mass-normalized oscillator has $\omega\langle Q^2\rangle/\hbar=\tfrac12\coth[1/(2\theta)]$, $\theta=k_{\rm B}T/(\hbar\omega)$, approaching $1/2$ at zero temperature and the classical line $\theta$ at high occupation. (b) For a separate centered Gaussian displacement, $s=\sigma/\ell_{\rm safe}$ gives exact one-time tail $\operatorname{erfc}[1/(\sqrt2s)]$. The second-moment bound in \eqref{eq:harmonic-basin-excursion-bound} gives $\min(1,s^2)$. Solid teal curves are the specified laws; dashed blue curves are their classical comparison or distribution-independent bound. Geometry and harmonic validity supply $\ell_{\rm safe}$. The tail concerns one time, while a first-exit probability requires the motion over the observation interval.}
\label{fig:thermal-variance-and-harmonic-basin}
\end{figure}

A small bound supports validity at the stated time. First exit over an
interval needs the trajectory test below; quantum measurement sequences
need their update law. Geometry fixes $\ell_{\rm safe}$ through contact
clearance, walls and the harmonic expansion's range. Significant
truncation of the preparation calls for a constrained or nonlinear law.
The validity interval is thus part of the selected biophysical resolution.

\paragraph{A classical bound over an observation interval.}
The same energy ledger also gives a conservative first-exit estimate.
It uses the continuous classical Langevin trajectory, including
excursions between recorded frames.

\begin{proposition}[Stopped-energy control of a translational exit]
\label{prop:s611-harmonic-first-exit}
Let $\vct u=\mathsf S_u\vct x$ be the retained physical translational
reconstruction and let $\ell_{\rm safe}>0$ be a declared limit on its
excursion.
For the classical Langevin model above, assume finite initial mean
energy $\overline E_0$ and independence from future noise increments.
Put $\kappa_u=\|\mathsf S_u\mathsf D^{-1/2}\|_2^2>0$ and
$\tau_{\rm safe}=\inf\{t\geq0:\|\vct u(t)\|_2
\geq\ell_{\rm safe}\}$. For every finite $t_{\rm obs}\geq0$,
\begin{equation}
 \Pr(\tau_{\rm safe}\leq t_{\rm obs})
 \leq\min\!\left\{1,
 \frac{2\kappa_u[\overline E_0+
                   k_{\rm B}T\tr(\mathsf C)t_{\rm obs}]}
      {\ell_{\rm safe}^2}\right\}.
 \label{eq:s611-harmonic-first-exit}
\end{equation}
If $\mathsf S_u=0$, the selected translational excursion is identically
zero and no such exit occurs.
\end{proposition}
\begin{proof}
At every time,
$\|\vct u\|_2^2\leq\kappa_u\vct x^{\mathsf T}\mathsf D\vct x
\leq2\kappa_u E_{\rm mech}$.
An exit from the ball therefore requires energy to reach
$E_*=\ell_{\rm safe}^2/(2\kappa_u)$.
Stop the energy process at its first crossing of $E_*$.
Before that stopping time the velocity is energy bounded, so the stopped
stochastic integral in~\eqref{eq:phonon-thermal-ito-energy} has zero
expectation on every finite interval. Its dissipative drift is
nonpositive and its thermal injection is
$k_{\rm B}T\tr\mathsf C$. Consequently the expected stopped energy
at $t_{\rm obs}$ is at most
$\overline E_0+k_{\rm B}T\tr(\mathsf C)t_{\rm obs}$.
It is at least $E_*$ times the probability of reaching that threshold.
Continuity of the classical paths then bounds the displacement-exit
event by this energy-crossing event, proving the result.
\end{proof}

The factor $\kappa_u$ has units of length squared per energy and carries
the masses, mode patterns and soft restoring directions into the
estimate. The bound can be loose and grows with the observation time;
its value is that it needs no independent-snapshot assumption. It applies
to the linear stochastic model stopped at the declared translational limit.
It does not predict the nonlinear motion after an exit. A quantum
sequence of position measurements instead needs its measurement and
back-action law; the equal-time quantum bound above is not being turned
into a trajectory statement. Together the one-time and first-exit tests
make the validity interval part of the chosen biophysical resolution.
Restrictions on internal deformation, orientation or other retained
coordinates require their own stopping regions. In particular, a
noninjective $\mathsf S_u$ leaves directions untested by this translational
bound, so satisfying it alone does not establish validity of the complete
internal harmonic expansion.

\FloatBarrier
\subsection{Phonon coherence and a defined entanglement witness}
\label{sec:phonon-correlations-witness}

A quantum correlation test specifies an occupation state, a subsystem
partition and a measurement protocol. Classical phase-resolved motion
and covariance remain described by the preceding mechanical readouts.
This subsection defines population and coherence statistics, then
constructs a witness for the additional quantum claim.

\subsubsection{What one-body coherence retains and discards}

For finite mean occupation, define
$\gamma_{st}=\tr(\varrho\hat b_t^\dagger\hat b_s)$.
This nonnegative one-body matrix has
$\tr\gamma=\langle\hat N_{\rm ph}\rangle$; it is not generally a
trace-one quantum state. A one-phonon ket
$\ket\Phi=\sum_s c_s\ket{1_s}$ gives $\gamma=\vct c\vct c^\dagger$.
An incoherent mixture with the same populations keeps the diagonal while
removing those off-diagonal terms. Yet a product of coherent oscillator
states also gives a rank-one $\gamma=\vct\alpha\vct\alpha^\dagger$
and can have the same normalized matrix. One-body information does not
specify the full occupation distribution or its entanglement.

For Hermitian $B$, the mean lifted readout is
$\langle\dd\Gamma(B)\rangle=\tr(B\gamma)$, whereas
\begin{equation}
 \begin{split}
 \operatorname{Var}[\dd\Gamma(B)]
    ={}&\tr(B^2\gamma)-[\tr(B\gamma)]^2\\
      &+\sum_{s,t,u,v}B_{st}B_{uv}
            \langle\hat b_s^\dagger\hat b_u^\dagger\hat b_t\hat b_v\rangle.
 \end{split}
 \label{eq:phonon-many-body-readout-variance}
\end{equation}
The first term comes from commuting one lowering operator past one
raising operator; the remaining normally ordered term needs two-body
occupation correlations. Finite second occupation moments suffice in
finitely many modes. Finite mean harmonic energy ensures canonical
coordinate second moments, but does not ensure occupation or energy
second moments. The stronger requirements therefore cannot be inferred
from normalization or a finite mean alone.

For example, $\ket{1,1}$ in two unequal-frequency normal modes has
exact energy $\hbar(\omega_1+\omega_2)$ and zero excitation-energy
variance. Its normalized $\gamma/2$ nevertheless assigns half its
one-body weight to each frequency. The resulting one-body frequency
spread is not the variance of the full system's energy. Centering also
matters: with $\alpha_s=\langle\hat b_s\rangle$,
$\gamma_{st}=\alpha_s\alpha_t^*
+\langle(\hat b_t-\alpha_t)^\dagger(\hat b_s-\alpha_s)\rangle$.
The first term is coherent mean motion; the second describes centered
normal correlations. Anomalous moments and canonical covariance carry
additional quadrature information, as in Section~\ref{sec:spatial-readouts-correlations}.

\begin{example}[Quantum superposition and a population mixture]
\leavevmode\label{ex:sup-quantum-versus-mixture}
Under the quantization postulate, take two orthogonal retained modes
$a,b$ with the same positive frequency $\omega$. In their one-excitation
sector compare
\begin{equation}
 \begin{aligned}
 \ket{\Psi_\phi}&=\frac{\ket{1,0}+e^{i\phi}\ket{0,1}}{\sqrt2},\\
 \varrho_{\rm mix}&=\tfrac12\ket{1,0}\bra{1,0}
                         +\tfrac12\ket{0,1}\bra{0,1},\\
 \varrho_{\rm sup}-\varrho_{\rm mix}
 &=\tfrac12\left(e^{-i\phi}\ket{1,0}\bra{0,1}
                   +e^{i\phi}\ket{0,1}\bra{1,0}\right),
 \quad\varrho_{\rm sup}=\ket{\Psi_\phi}\bra{\Psi_\phi}.
 \end{aligned}
 \label{eq:sup-quantum-density-comparison}
\end{equation}
Both preparations have $\langle N_a\rangle=\langle N_b\rangle=1/2$,
excitation energy $\hbar\omega$ above the two-mode vacuum, and zero
displacement first moments. Their off-diagonal matrix elements differ.
Coherent oscillator states form the particular superposition family of
Definition~\ref{def:s5-coherent-oscillator-preparation}; the present
comparison holds the total occupation at one.
For the projection onto $\ket{\Psi_\theta}$ defined by the same formula,
and for the canonical mode coordinates, the contrast is
\begin{equation}
 \begin{aligned}
 \langle\Psi_\theta|\varrho_{\rm sup}|\Psi_\theta\rangle
    &=\frac{1+\cos(\phi-\theta)}2,
 &\langle\Psi_\theta|\varrho_{\rm mix}|\Psi_\theta\rangle&=\tfrac12,\\
 \langle\widehat Q_a\widehat Q_b\rangle_{\rm sup}
    &=\frac{\hbar}{2\omega}\cos\phi,
 &\langle\widehat Q_a\widehat Q_b\rangle_{\rm mix}&=0.
 \end{aligned}
 \label{eq:sup-quantum-phase-sensitive-records}
\end{equation}
Expansion in the number basis proves both identities. The phase-sensitive
projector changes a probability while leaving the two occupation
populations fixed; the covariance supplies a distinct canonical readout.
Transducing these quantities to virion positions uses the physical mode
weights and detector map. A classical covariance can reproduce particular
second moments, so a quantum interpretation requires the specified state
preparation and measurement, not the covariance alone. The calibrated
witness below addresses its own stronger correlation claim
\cite{Shankar1994,Glauber1963}.
\end{example}

\subsubsection{Population fluctuations and a missing phase}

For $0<\langle\hat N_s\rangle$ and finite second occupation moment,
define
\begin{equation}
 g_s^{(2)}=\frac{\langle\hat N_s(\hat N_s-1)\rangle}
                              {\langle\hat N_s\rangle^2}.
 \label{eq:phonon-occupation-second-coherence}
\end{equation}
Normal ordering removes the same-quantum contribution from the pair
count, following the standard oscillator coherence
construction~\cite{Glauber1963}. Coherent, thermal and number states have respectively
$g^{(2)}=1$, $2$ and $1-1/N$ for $N\geq1$.
Their Fano factor is
$F_s^{(N)}=\operatorname{Var}(\hat N_s)/\langle\hat N_s\rangle
=1+\langle\hat N_s\rangle[g_s^{(2)}-1]$.
Figure~\ref{fig:operator-occupation-statistics} compares complete
distributions at the same mean excitation. A number-dephased coherent
state retains its Poisson populations and both diagnostics while losing
its mean oscillation. Occupation and phase readouts therefore answer
different questions. An optical intensity correlation measures its
optical detection operator; identifying it with this phonon statistic
requires an independently specified transduction coupling and protocol.

\begin{figure}[H]
\centering\begingroup
\definecolor{vltTeal}{HTML}{176B64}
\definecolor{vltBlue}{HTML}{315D80}
\definecolor{vltGold}{HTML}{A87520}
\definecolor{vltPlum}{HTML}{725877}
\definecolor{vltRed}{HTML}{762B38}
\definecolor{vltYellow}{HTML}{D5AD3C}
\definecolor{vltNeutral}{HTML}{777D83}
\setlength{\unitlength}{1pt}
\scalebox{0.80}{%
\input{vlt-figures/figure-215.tex}%
}\endgroup
\caption{Different occupation laws at the same mean excitation energy. The coherent, thermal and number-state laws are $P_N=e^{-1}/N!$, $2^{-N-1}$ and $\delta_{N1}$, respectively. Their mean is one; the Fano factors and normally ordered second-order correlations printed below the bars use the full laws. Only $N=0,\ldots,6$ are displayed, without renormalization. Omitted masses are $1-e^{-1}\sum_{N=0}^{6}1/N!$, $1/128$ and zero. Axes and displayed truncation are shared. Complete number-basis dephasing preserves the coherent counting law, $F^{(N)}$ and $g^{(2)}$, while taking $\langle b\rangle$ from one to zero; that change needs a phase-sensitive readout. Identifying an experimental counting channel with these phonon occupations requires its transduction model.}
\label{fig:operator-occupation-statistics}
\end{figure}

\subsubsection{From a phonon statistic to detected counts}
\label{sec:qfa-counting}

A detector tests the counting law only through a physical conversion.
For a mode-preserving readout, model the detected optical mode by
$\widehat a_d=\sqrt\eta\,\widehat b+
\sqrt{1-\eta}\,\widehat v$, with $0<\eta\leq1$ and an independent
vacuum input $\widehat v$. The total efficiency includes conversion,
collection and detection. With no added counts, normal ordering gives
$\langle N_d\rangle=\eta\langle N\rangle$ and
$\langle N_d(N_d-1)\rangle=\eta^2\langle N(N-1)\rangle$.
Consequently ideal loss preserves $g^{(2)}$, while reducing the
available coincidences. Such a map requires a calibrated linear
quantum transducer, not just a proportionality between average powers.

\begin{proposition}[Independent background in a converted counting channel]
\label{prop:qfa-count-background}
Let the signal undergo the preceding vacuum-loss channel and let an
independent, noninterfering Poisson count background have mean $b_d$ in
the same acquisition window. Write $\bar n=\langle N\rangle>0$ and
$f_2=\langle N(N-1)\rangle$. The total counts obey
\begin{equation}
 \begin{aligned}
 \langle N_d\rangle&=\eta\bar n+b_d,\\
 g_d^{(2)}&=
 \frac{\eta^2 f_2+2\eta\bar n b_d+b_d^2}
      {(\eta\bar n+b_d)^2}.
 \end{aligned}
 \label{eq:qfa-detected-count-statistic}
\end{equation}
\end{proposition}
\begin{proof}
Independent thinning scales each normally ordered signal moment by its
corresponding power of $\eta$. For independent counts $X,Y$,
$(X+Y)(X+Y-1)=X(X-1)+Y(Y-1)+2XY$.
The Poisson factorial moment is $b_d^2$; taking expectations proves
the result.
\end{proof}

For an initially ideal single phonon, $\eta=0.2$ and $b_d=0.01$ give
$g_d^{(2)}\simeq0.093$. These are assigned detector parameters.
The formula describes additive distinguishable background counts;
a coherent optical field interfering with the signal, thermal added
noise or a detector with memory requires its own channel model.

The distinguishing statistic has a precise classical comparison.
A positive Glauber--Sudarshan mixture of coherent states has
$g^{(2)}=\mathbb E[I^2]/\mathbb E[I]^2\geq1$ for nonzero mean
intensity $I$~\cite{Glauber1963}. A calibrated, normally ordered
$g^{(2)}<1$ therefore excludes that family, whereas a thermal or
coherent counting law alone does not. Heralded single-phonon
preparation followed by optical conversion and a Hanbury Brown--Twiss
measurement has been demonstrated in an engineered mechanical
resonator~\cite{Hong2017}. Transferring the protocol to a virion mode
requires its actual coupling, efficiency, bath and damage constraints.

Operationally, specify the herald, the two output count channels,
the storage delay and the coincidence window before acquiring the test
record. Calibrate dead time, afterpulsing, background, pulse leakage and
the response to a coherent reference: these effects can bias a
correlation estimate. A reported violation uses uncertainty on the
correctly conditioned statistic, with its upper confidence bound below
one under the stated nuisance model. Shared heralds and repeated
preparations retain their sampling dependencies. The uncertainty
construction in Section~\ref{sec:hpvu-budget} then applies to detector
parameters as well as to the mechanical inputs. The finite thermal
window derived in Section~\ref{sec:qfa-lifetimes} sets a separate
constraint on the delay and readout duration.

\subsubsection{A bounded witness on the full occupation space}

First declare two orthogonal one-mode subspaces, with
$\mathcal H_{AB}=\mathcal H_A\oplus\mathcal H_B\subseteq\mathcal H_+$ and
$\mathcal F(\mathcal H_{AB})\simeq\mathcal F(\mathcal H_A)\otimes
\mathcal F(\mathcal H_B)$. The two-channel state $\varrho_{AB}$ is obtained
by tracing out the complementary modes, if present.
Compressed site effects from~\eqref{eq:site-mode-effects} do not
automatically supply this tensor product. A separable state is a
trace-norm limit of convex mixtures of product states across the declared
division. A witness is a one-sided test: negative expectation excludes
every such mixture; a nonnegative result can remain inconclusive.

For two selected oscillator channels, choose
\begin{equation}
 \ket{\Phi_q}=\sqrt q\,\ket1_A\ket0_B
       +e^{i\phi}\sqrt{1-q}\,\ket0_A\ket1_B,
       \qquad 0<q<1.
 \label{eq:two-channel-entangled-phonon}
\end{equation}
Here the scalar $q$ is a population weight, distinct from the spatial
probe wavevector of Section~\ref{sec:spatial-readouts-correlations}. The reduced state on $A$ has eigenvalues
$q,1-q$ and entropy $-q\ln q-(1-q)\ln(1-q)$.
Both nonzero weights make the pure state entangled across this partition.
Let $\lambda_{\max}=\max(q,1-q)$ and define
\begin{equation}
 \hat{\mathcal W}_q=\lambda_{\max}\Id_{AB}
                                     -\ket{\Phi_q}\bra{\Phi_q}.
 \label{eq:defined-entanglement-witness}
\end{equation}
It is bounded and self-adjoint on the full two-channel Fock space.

The Schmidt-overlap construction gives the separable bound
\cite{Guff2022}; its extension beyond a truncated occupation sector can
be proved directly here. Only the coefficients at $0$ and $1$ in each
factor of a normalized product vector overlap the target. Their truncated
norms are at most one, and the target's two-by-two Schmidt matrix has
operator norm $\sqrt{\lambda_{\max}}$. The squared product overlap is
therefore at most $\lambda_{\max}$ even if higher occupations are present.
Convexity proves the bound for mixtures, and boundedness extends it
continuously to their trace-norm limits. Hence
$\tr(\hat{\mathcal W}_q\varrho_{\rm sep})\geq0$, whereas the
target gives $\lambda_{\max}-1<0$. For $q=1/2$ this is $-1/2$;
the equal incoherent mixture of $\ket{1,0},\ket{0,1}$ gives zero.

The required data are exact sector populations and coherence in the same
unconditioned reduced preparation. Define
$P_{10}=\langle1,0|\varrho_{AB}|1,0\rangle$,
$P_{01}=\langle0,1|\varrho_{AB}|0,1\rangle$ and
$R_{AB}=\langle1,0|\varrho_{AB}|0,1\rangle$. Then
\begin{equation}
 \begin{aligned}
 F_q&:=\langle\Phi_q|\varrho_{AB}|\Phi_q\rangle\\
    &=qP_{10}+(1-q)P_{01}
          +2\sqrt{q(1-q)}\operatorname{Re}(e^{i\phi}R_{AB}),\\
 \langle\hat{\mathcal W}_q\rangle&=\lambda_{\max}-F_q.
 \end{aligned}
 \label{eq:phonon-witness-sector-fidelity}
\end{equation}
With $\Pi_1=\ket{1,0}\bra{1,0}+\ket{0,1}\bra{0,1}$,
$R_{AB}=\tr(\varrho_{AB}\Pi_1\hat b_B^\dagger\hat b_A\Pi_1)$.
Unprojected $\langle\hat b_B^\dagger\hat b_A\rangle$ also contains
higher-occupation contributions and cannot replace this matrix element.
For the separable coherent product with amplitudes $\alpha_A=\alpha_B=1$,
raw coherence equals one, but the phase-matched balanced target fidelity
is only $2e^{-2}<1/2$. A negative witness cannot be inferred from the
raw coherence alone.

\subsubsection{A calibrated margin for a correlation claim}

The separable bound is useful only when the measured fidelity clears it
by more than the uncertainty of its reconstruction. The bounded sector
observables in~\eqref{eq:phonon-witness-sector-fidelity} allow that
comparison without an occupation cutoff.

\begin{proposition}[Robust sector-fidelity certificate]
\label{prop:s612-robust-witness-margin}
Fix $q,\phi$ and the two-channel partition. Suppose estimates
$\widehat P_{10},\widehat P_{01},\widehat R_{AB}$ obey simultaneous
absolute error bounds $\epsilon_{10},\epsilon_{01},\epsilon_R$ for
the corresponding unconditioned sector quantities. Form
$\widehat F_q$ by their substitution in
\eqref{eq:phonon-witness-sector-fidelity} and put
\begin{equation}
 \begin{gathered}
 \epsilon_q=q\epsilon_{10}+(1-q)\epsilon_{01}
                       +2\sqrt{q(1-q)}\epsilon_R,\\
 \widehat F_q-\epsilon_q>\lambda_{\max}
       \ \Longrightarrow\ \varrho_{AB}\ \hbox{is entangled}.
 \end{gathered}
 \label{eq:s612-robust-witness-margin}
\end{equation}
Alternatively, let $\widetilde{\mathsf P}_q$ be a bounded self-adjoint
calibrated fidelity observable satisfying
$\|\widetilde{\mathsf P}_q-\ket{\Phi_q}\bra{\Phi_q}\|
\leq\epsilon_{\rm cal}$. If a direct estimate
$\widehat F_q^{\,\rm cal}$ has error at most $\epsilon_{\rm stat}$
relative to $\tr(\varrho_{AB}\widetilde{\mathsf P}_q)$, then
$\widehat F_q^{\,\rm cal}-\epsilon_{\rm stat}-\epsilon_{\rm cal}
>\lambda_{\max}$ certifies the same conclusion.
\end{proposition}
\begin{proof}
The triangle inequality in the sector formula gives
$|\widehat F_q-F_q|\leq\epsilon_q$.
For the calibration contribution, positivity and unit trace give
$|\tr[\varrho_{AB}(\widetilde{\mathsf P}_q-
\ket{\Phi_q}\bra{\Phi_q})]|\leq\epsilon_{\rm cal}$.
Every separable state has $F_q\leq\lambda_{\max}$ by the full-space
witness proof above. A lower fidelity bound strictly above that value
therefore excludes separability.
\end{proof}

Simultaneous error bounds with coverage $1-\alpha$ give the certificate
that coverage under the calibrated measurement model. Statistics,
transduction uncertainty and preparation variability each contribute
their justified bounds. These data are bounded sector matrix elements.
Replacing them by raw $\hat b_B^\dagger\hat b_A$ coherence requires
occupation-weighted tail control, as in the cutoff analysis.

A coherent projection exactly equal to $\Pi_1$ also gives the useful
identity $F_q=p_1F_{q\mid1}$, where $p_1=\tr(\varrho_{AB}\Pi_1)$
and $F_{q\mid1}$ is the target fidelity of its normalized conditional
state. Thus simultaneous lower bounds $p_-,f_-\in[0,1]$ certify the original state with
this witness only when $p_-f_->\lambda_{\max}$. The success probability
is part of the unconditioned test. This use of sector data assumes the
stated projection and its preserved coherence; a different selection or
feed-forward operation requires its own transported observable.

For a prospective biophysical correlation measurement, this separates
the chosen quantum prediction from the instrument needed to test it.
The mechanical coordinate map supplies the mode channels. The state
preparation, quantum transduction and calibrated uncertainty determine
whether their sector fidelity is accessible. Ordinary correlated
particle motion remains described by its classical covariance and
response, without requiring this additional measurement postulate.

\subsubsection{Conditioning changes which state the test certifies}

The normalization of sector data is consequential. A coherent global
projection onto total occupation one sends a product preparation to
\begin{equation}
 \Pi_1\ket{\alpha_A}_A\ket{\alpha_B}_B
       \ \propto\ \alpha_A\ket{1,0}+\alpha_B\ket{0,1},
 \qquad
 p_1=e^{-(|\alpha_A|^2+|\alpha_B|^2)}(|\alpha_A|^2+|\alpha_B|^2).
 \label{eq:operator-global-sector-postselection}
\end{equation}
For nonzero $\alpha_A,\alpha_B$, the selected state is entangled although the
input is a product. The operation is global and retains coherence between
the two alternatives. Measuring both channel occupations separately and
then keeping records with total one generally reveals which channel was
occupied and removes that coherence. These are different protocols.
A negative conditional witness certifies the selected state; it does not
prove the original unconditioned input was entangled. The success
probability and the physical selection operation must be reported, as in
the normalized conditioning law of Section~\ref{sec:passive-operator-spectrum}.

The experiment therefore specifies orthogonal channels, sector
populations, phase-sensitive sector coherence and all selection outcomes.
A phonon transduction law connects those quantities to an optical or
other detector. Together they test the declared quantum preparation.

\subsubsection{Loss: survival, witness sensitivity, and actual separability}

For the balanced state under equal zero-temperature losses, let
$P=P_1(t)$ be its survival probability. With the evolved relative phase
included in the target,
$\varrho(t)=P\ket{\Phi(t)}\bra{\Phi(t)}
+(1-P)\ket{0,0}\bra{0,0}$ has witness expectation $1/2-P$.
The test detects entanglement for $P>1/2$; its failure below that threshold
does not prove separability. The vacuum outcome loses an excitation,
not a virion.

Partial transpose exchanges bra and ket indices in channel $B$. A separable state's partial transpose remains
positive~\cite{Peres1996}. For this finite-support
mixture, its $\ket{0,0},\ket{1,1}$ block is
$\left(\begin{smallmatrix}1-P&P/2\\P/2&0\end{smallmatrix}\right)$
after a local phase choice, with smaller eigenvalue
\begin{equation}
 \lambda_{\rm neg}(P)
       =\frac{1-P-\sqrt{(1-P)^2+P^2}}2<0,
       \qquad 0<P\leq1.
 \label{eq:loss-mixture-partial-transpose-eigenvalue}
\end{equation}
It proves entanglement at every positive survival probability for this
ideal state, even where the chosen witness becomes inconclusive.
Figure~\ref{fig:loss-witness-and-partial-transpose} shows the difference.

\begin{figure}[H]
\centering
\begingroup
\definecolor{vlfGreen}{HTML}{176B64}
\definecolor{vlfDarkRed}{HTML}{762B38}
\definecolor{vlfRed}{HTML}{762B38}
\definecolor{vlfBlue}{HTML}{315D80}
\definecolor{vlfCopper}{HTML}{762B38}

\setlength{\unitlength}{1pt}
\scalebox{0.80}{%
\input{vlt-figures/figure-216.tex}%
}%
\endgroup
\caption{Witness sensitivity within a specified excitation-loss family. For $\ket\Phi=(\ket{1,0}+\ket{0,1})/\sqrt2$, take $\varrho=P\ket\Phi\bra\Phi+(1-P)\ket{0,0}\bra{0,0}$ under a fixed two-mode partition and phase-matched readout. The burgundy witness $\langle\mathcal W\rangle=1/2-P$ detects entanglement for $P>1/2$. The dashed teal partial-transpose eigenvalue $\lambda_{\rm neg}=[1-P-\sqrt{(1-P)^2+P^2}]/2$ remains negative for every $P>0$, establishing entanglement for this family even where that witness is inconclusive. At $P=0$ the preparation is the separable vacuum. Here $P$ tracks one excitation; the virion population stays fixed.}
\label{fig:loss-witness-and-partial-transpose}
\end{figure}

The dependence on preparation can be made more general. Independent
zero-temperature survival probabilities $\eta_A,\eta_B$ and a dephasing
factor $d$ with $|d|\leq1$ give populations
$a=q\eta_A$, $b=(1-q)\eta_B$, vacuum weight $v=1-a-b$, and
coherence $c=e^{-i\phi}\sqrt{q(1-q)\eta_A\eta_B}\,d$.
For this state supported on $\ket{0,0},\ket{1,0},\ket{0,1}$, the
partially transposed block has smaller eigenvalue
\begin{equation}
 \lambda_- =\frac{v-\sqrt{v^2+4|c|^2}}2,
 \qquad F_{1/2}=\frac{a+b}{2}+|c|
                   \quad\hbox{for a phase-matched balanced target}.
 \label{eq:operator-loss-dephasing-witness}
\end{equation}
Nonzero coherence proves entanglement in this particular sparse state;
loss and dephasing reduce the witness's sensitivity differently. Heating
or population outside that support invalidates the sparse-state argument
and requires the complete state or a valid measured witness. Independent
normal-mode Gibbs states are products across those modes and do not
produce this target entanglement without a specified preparation.

\paragraph{Arrangement-conditioned quantum claims.}
A classical recruitment record can accompany an optional quantum
preparation. Compare its conditional states on one common full-pool
canonical space, retaining newly resolved and omitted coordinates as
specified in the canonical construction. The arrangement record is a
classical flag, represented by orthogonal labels $\ket g$; these labels
count neither virions nor phonons. Transport the chosen physical subsystem
partition and its measurement operators into the same representation
before combining records.

\begin{proposition}[A witness with a recorded classical arrangement]
\label{prop:s6-thermal-flagged-witness}
For the same physical division $A|B$, let the normalized conditional
states be $\widetilde\varrho_g$ and their recorded probabilities $p_g$.
If each bounded $\hat{\mathcal W}_g$ is a valid witness for that
division, define
\begin{equation}
 \begin{aligned}
 \varrho_{\rm flag}
    &=\sum_g p_g\ket g\bra g\otimes\widetilde\varrho_g,\\
 \hat{\mathcal W}_{\rm flag}
    &=\sum_g\ket g\bra g\otimes\hat{\mathcal W}_g,\\
 \tr(\varrho_{\rm flag}\hat{\mathcal W}_{\rm flag})
    &=\sum_g p_g\tr(\widetilde\varrho_g\hat{\mathcal W}_g).
 \end{aligned}
 \label{eq:s6-thermal-flagged-witness}
\end{equation}
If every conditional state is separable across the stated division,
this expectation is nonnegative. A negative value certifies that at
least one contributing conditional preparation is entangled there.
\end{proposition}
\begin{proof}
Orthogonality of the classical labels gives the trace identity. Each
conditional witness is nonnegative on separable states, so their
nonnegative weighted sum has the same property.
\end{proof}

A fixed witness is one choice across arrangements. Transporting it and
the subsystem algebras with a mode-basis change preserves the claim.
A cluster/complement division that changes with recruitment instead
defines episode-specific partitions; persistence needs a common
division or transported test family. Excitation-dependent selection
also retains its probabilities and state-update operation. The contact
record thus specifies which state and correlation the calibrated
experiment certifies.

Figure~\ref{fig:s6-recorded-arrangement-witness} gives a phase-resolved example of the flagged witness and its unconditioned marginal.

\begin{figure}[H]
\centering\begingroup
\definecolor{vltTeal}{HTML}{176B64}
\definecolor{vltBlue}{HTML}{315D80}
\definecolor{vltGold}{HTML}{A87520}
\definecolor{vltPlum}{HTML}{725877}
\definecolor{vltRed}{HTML}{762B38}
\definecolor{vltYellow}{HTML}{D5AD3C}
\definecolor{vltNeutral}{HTML}{777D83}
\setlength{\unitlength}{1pt}
\scalebox{0.90}{%
\input{vlt-figures/figure-217.tex}%
}\endgroup
\caption{Two equally probable recorded arrangements with one fixed channel partition $A|B$. The declared states $\ket{\Phi_\pm}=(\ket{10}\pm\ket{01})/\sqrt2$ occupy only the displayed single-excitation blocks. Phase-matched bounded witnesses $\mathcal W_g=I/2-\ket{\Phi_g}\bra{\Phi_g}$ each give $-1/2$. Retaining the classical flag certifies those conditional states; discarding it cancels coherences and gives a separable marginal. Open arrows carry classical records. Occupation labels refer to the two quantum modes, while the common subsystem partition and flag probabilities are specified inputs.}
\label{fig:s6-recorded-arrangement-witness}
\end{figure}

\subsubsection{A thermal state can depend on the subsystem partition}

\begin{example}[Coupled equilibrium and the subsystem partition]
Retain two complete pinned mass-normalized site pairs
$[\hat x_j,\hat p_k]=i\hbar\delta_{jk}\Id$ with
\begin{equation}
 \hat H=\frac12(\hat p_1^2+\hat p_2^2)
       +\frac12\omega_0^2(\hat x_1^2+\hat x_2^2)
       +\frac12\kappa_c(\hat x_1-\hat x_2)^2,
 \quad\omega_0>0,\quad\kappa_c\geq0.
 \label{eq:phonon-pinned-two-site-hamiltonian}
\end{equation}
The coupling $\kappa_c$ has inverse-time-squared units. Common and
relative frequencies are $\omega_+=\omega_0$ and
$\omega_-=\sqrt{\omega_0^2+2\kappa_c}$. The Gibbs state is a
product across these normal modes. The physical site partition uses
different canonical subalgebras and can instead be entangled.

Write $c_\pm=\coth(\beta\hbar\omega_\pm/2)$.
Normal-coordinate and momentum variances are
$X_\pm=\hbar c_\pm/(2\omega_\pm)$ and
$P_\pm=\hbar\omega_\pm c_\pm/2$.
The site covariance follows by the orthogonal common/relative transform.
In scaled local quadratures
$(\sqrt{\omega_0}x_1,p_1/\sqrt{\omega_0},
\sqrt{\omega_0}x_2,p_2/\sqrt{\omega_0})$, it has action units.
Partial transpose reflects $p_2$, exchanging the common/relative momentum
variances in that transformed test. Its smallest symplectic eigenvalue is
\begin{equation}
 \widetilde\nu_- =\sqrt{X_-P_+}
       =\frac\hbar2\sqrt{\frac{\omega_+}{\omega_-}c_+c_-}.
 \label{eq:phonon-two-site-thermal-entanglement}
\end{equation}
For one Gaussian oscillator per party, the partial-transpose criterion
is necessary and sufficient: entanglement occurs exactly when
$\widetilde\nu_-<\hbar/2$~\cite{Simon2000}.
At zero temperature any $\kappa_c>0$ satisfies it; increasing thermal
occupation eventually removes it. The threshold, when coupling is
positive, solves $c_+c_-=\omega_-/\omega_+$. Pinning makes the Gibbs
state normalizable and retains complete site canonical pairs. Compressed
site effects or a bath with arbitrary noise would not establish this
same calculation.
\end{example}

Symplectic eigenvalues are the positive magnitudes of the eigenvalues
of $i\boldsymbol\Omega\mathsf V$ for the chosen canonical ordering;
they measure the covariance's canonical uncertainty scales. Partial
transpose can make a tested scale fall below $\hbar/2$, violating
quantum covariance admissibility and thus revealing entanglement in the
original state. Changing the subsystem partition changes the question;
a mere passive change of coordinates with the partition transported
does not create entanglement. This distinction, the required quantum
preparation, and the thermal frequency scales must remain explicit when
connecting the calculation to a possible biophysical readout.

Figure~\ref{fig:thermal-state-subsystem-partition} follows the same Gibbs preparation across its two possible subsystem partitions and displays the site-entanglement threshold.

\begin{figure}[H]
\centering\begingroup
\definecolor{vltTeal}{HTML}{176B64}
\definecolor{vltBlue}{HTML}{315D80}
\definecolor{vltGold}{HTML}{A87520}
\definecolor{vltPlum}{HTML}{725877}
\definecolor{vltRed}{HTML}{762B38}
\definecolor{vltYellow}{HTML}{D5AD3C}
\definecolor{vltNeutral}{HTML}{777D83}
\setlength{\unitlength}{1pt}
\scalebox{0.90}{%
\input{vlt-figures/figure-218.tex}%
}\endgroup
\caption{The same coupled Gibbs state across normal-mode and site partitions. Equal pins and an axial intersite link define the ideal-virion reference at left; common and relative canonical pairs factor its normal-mode Gibbs state. For $r=\sqrt{1+2\kappa_c/\omega_0^2}$, the site test is $2\widetilde\nu_-/\hbar=\sqrt{\coth[1/(2\theta)]\coth[r/(2\theta)]/r}$, tending to $r^{-1/2}$ at zero temperature. Values below one certify site entanglement; at or above one the two-mode Gaussian site state is separable. Open circles mark thresholds for positive coupling. The test uses complete canonical pairs and the stipulated quantized thermal preparation.}
\label{fig:thermal-state-subsystem-partition}
\end{figure}

\FloatBarrier
\subsection{Geometry and additional couplings as specified extensions}
\label{sec:operator-geometric-extensions}

Extending the geometry means specifying centres, orientations,
interactions and constraints, then assembling equilibrium and the
Hessian. The same matrix procedure can describe a capsid's internal
network or an assembly of whole virions, with each model retaining its
own material coordinates and force law.

\subsubsection{Adding shape without counting rigid motion twice}

For a structured particle, specify body-frame surface/subunit positions
$\vct a_{i\alpha}$ and orientation $\mathsf O_i\in SO(3)$.
A retained material point has position
\begin{equation}
 \vct r_{i\alpha}=\vct R_i+\mathsf O_i\vct a_{i\alpha}
                                             +\vct w_{i\alpha}.
 \label{eq:structured-virion-positions}
\end{equation}
The list can represent protein contacts, membrane patches or selected
RNP coordinates. Translation changes $\vct R_i$, rotation changes
$\mathsf O_i$, and $\vct w_{i\alpha}$ describes retained deformation.
An interaction can now depend on centre separation, relative orientation
and these internal amplitudes. Facets or patches acquire mechanical
meaning through those interactions and their derivatives, rather than
through a geometrical label alone.

For a complete positive-mass material discretization, impose
$\sum_\alpha m_{i\alpha}=m_i$ and
$\sum_\alpha m_{i\alpha}\vct a_{i\alpha}=0$.
Let $\vct b_{i\alpha}=\mathsf O_i^{(0)}\vct a_{i\alpha}$ at the
fixed reference. A small lab-frame rotation $\delta\vct\theta_i$
gives $\delta\vct r_{i\alpha}=\delta\vct R_i
+\delta\vct\theta_i\times\vct b_{i\alpha}+\vct w_{i\alpha}$.
Choose the internal subspace by
\begin{equation}
 \sum_\alpha m_{i\alpha}\vct w_{i\alpha}=0,
 \qquad
 \sum_\alpha m_{i\alpha}\vct b_{i\alpha}\times\vct w_{i\alpha}=0.
 \label{eq:structured-body-deformation-orthogonality}
\end{equation}
These are mass-orthogonality conditions to rigid translation and small
rotation. They prevent a displacement from being represented twice.
For nondegenerate rotational inertia on the retained rotation space,
the resulting quadratic kinetic energy is
\begin{equation}
 \begin{aligned}
 T_i&=\frac{m_i}{2}|\Dt{\vct R_i}|^2
       +\frac12\Dt{\vct\theta_i}^{\mathsf T}\mathsf J_i\Dt{\vct\theta_i}
       +\frac12\sum_\alpha m_{i\alpha}|\Dt{\vct w_{i\alpha}}|^2,\\
 \mathsf J_i&=\sum_\alpha m_{i\alpha}
          (|\vct b_{i\alpha}|^2\Id_3
                  -\vct b_{i\alpha}\vct b_{i\alpha}^{\mathsf T}).
 \end{aligned}
 \label{eq:operator-structured-kinetic-split}
\end{equation}
The cross terms vanish by the centre-of-mass condition and
\eqref{eq:structured-body-deformation-orthogonality}. Rotational angles
are dimensionless and $\mathsf J_i$ has mass-times-length-squared units;
translation uses mass instead. Expanding the specified potential gives
translation--rotation--deformation Hessian blocks, with normalization by
the corresponding positive inertia matrix. Selected marker positions
alone define an observation map; this complete material realization also
needs a mass distribution. Large rotations/deformations require nonlinear
coordinates beyond the local split.

Figure~\ref{fig:operator-structured-coordinate-split} illustrates the three coordinate families and their mass-orthogonality conditions on an ideal faceted particle.

\begin{figure}[H]
\centering\begingroup
\definecolor{vltTeal}{HTML}{176B64}
\definecolor{vltBlue}{HTML}{315D80}
\definecolor{vltGold}{HTML}{A87520}
\definecolor{vltPlum}{HTML}{725877}
\definecolor{vltRed}{HTML}{762B38}
\definecolor{vltYellow}{HTML}{D5AD3C}
\definecolor{vltNeutral}{HTML}{777D83}
\setlength{\unitlength}{1pt}
\scalebox{0.90}{%
\input{vlt-figures/figure-219.tex}%
}\endgroup
\caption{Three trial changes of one faceted virion. Gray edges register a common regular icosahedron. Translation preserves orientation; a $24^\circ$ display rotation preserves the centre; the shape trial uses $\vct w_\alpha=0.15\operatorname{diag}(1,-1,0)\vct b_\alpha$. Equal masses at twelve vertices make both mass-orthogonality sums vanish. Arrows distinguish coordinate changes and carry no common force scale. These exaggerated views illustrate the local translation--rotation--deformation split; a material realization additionally specifies the vertex masses and constitutive interactions.}
\label{fig:operator-structured-coordinate-split}
\end{figure}

\subsubsection{What icosahedral symmetry predicts}

An exact isolated icosahedral internal reference has the proper rotation
group $I$ of order 60. Let $\mathsf U_g$ act on mass-normalized
displacement patterns by rotating vectors and permuting equivalent nodes.
If the chosen masses, constraints and interactions obey
$[\mathsf D,\mathsf U_g]=0$ for every $g\in I$, the mode problem
decouples into symmetry sectors. For an irreducible representation
$\mathsf U^{(\alpha)}(g)$ of dimension $d_\alpha$, the projector is
\begin{equation}
 \mathsf P_\alpha=\frac{d_\alpha}{60}\sum_{g\in I}
       \tr[\mathsf U^{(\alpha)}(g)]^*\mathsf U_g,
 \qquad d_\alpha\in\{1,3,4,5\}.
 \label{eq:icosahedral-mode-projectors}
\end{equation}
The five inequivalent irreducible representations include two distinct
three-dimensional ones. The projector keeps all copies of the selected
representation, an \emph{isotypic} sector, rather than necessarily one
mode. Within a sector with multiplicity $\nu_\alpha$, a suitable basis
has $\mathsf D=\Id_{d_\alpha}\otimes\mathsf A_\alpha$.
The Hermitian multiplicity matrix can mix equivalent copies. Each of its
frequencies consequently has a degeneracy of at least $d_\alpha$, subject
to the stated exact symmetry. These are mode multiplets, not counts of
proteins or virions~\cite{PeetersTaormina2009}.

A mass-normalized force $\vct h_{\rm sym}$ invariant under every
$\mathsf U_g$ satisfies
\begin{equation}
 \mathsf P_{\rm triv}\vct h_{\rm sym}=\vct h_{\rm sym},
 \qquad
 \mathsf P_\alpha\vct h_{\rm sym}=0
                    \quad\hbox{for nontrivial sectors }\alpha.
 \label{eq:icosahedral-invariant-force-selection}
\end{equation}
Uniform radial loading of an exactly symmetric shell is an example;
a localized contact has other symmetry content. Even in the trivial
sector, several modes can exist and the actual drive requires their
force overlaps. Group averaging a local force extracts its invariant
component; it does not assert that every remaining force component is
an internal vibration. Rigid components must still be handled by the
chosen coordinate split. Figure~\ref{fig:icosahedral-loading-and-gyroscopic-response}
illustrates the forcing patterns for an ideal reference.

Contacts, binding, maturation and anisotropic confinement can split the
isolated shell's multiplets. The complete particle--medium system uses
the symmetry of all its retained layers and constraints. Chiral protein
structure also requires inversion symmetry to be checked separately.

An icosahedral particle can occupy a site in a periodic array of centres;
the array need not share its internal fivefold symmetry. A claim of
icosahedral quasicrystalline order concerns the order of the full
arrangement. Mackay-type constructions supply separately specified
multi-shell positions with their own packing and strain constraints
\cite{Canestrari2025}. Neither internal capsid symmetry nor visual
roundness establishes that a cluster of whole virions has such an
equilibrium. Its force balance and Hessian must still be supplied.

\subsubsection{Testing a spherical approximation dynamically}

A triangulation number describes capsid organization; it does not provide
a universal error bound for replacing its mechanics by a sphere.
For two stable conservative references in the same retained coordinate
space with matched inertia, let
$\mathsf D_{\rm shape}=\mathsf D_{\rm sph}+\delta\mathsf D$.
Hermitian eigenvalue perturbation gives
\begin{equation}
 |\omega_{s,\rm shape}^2-\omega_{s,\rm sph}^2|
                                   \leq\|\delta\mathsf D\|_2
 \label{eq:geometry-spectral-error-bound}
\end{equation}
for ordered squared frequencies. The bound concerns actual curvature
changes. A small perturbation relative to positive eigenvalues controls
frequency shifts; individual patterns additionally require separated
spectral gaps. Near a response pole even a small curvature change can
strongly alter a signal, as the resolvent identity in Section~\ref{sec:operator-target-defects} shows.
A useful spherical approximation must therefore be assessed for its
selected observables and frequency band. If the spherical description
has fewer coordinates, first specify a projection or reduction; comparing
matrices of unequal physical resolution is not this same-space bound.

\subsubsection{Testing retained shape variables by their dynamical residual}

A spherical or low-dimensional shape description can be tested directly
against the motion it predicts. This comparison starts from one common
material coordinate space and its inertia, so the chosen shape amplitudes
have a specified meaning in the structured-particle reconstruction.

\begin{definition}[Shape trial subspace and coupling residual]
\label{def:s613-shape-trial-residual}
On a strictly positive conservative mass-normalized coordinate space,
let $\mathsf V_{\rm tr}$ have real orthonormal columns spanning the
chosen trial patterns. Rigid coordinates have already been separated by
\eqref{eq:structured-body-deformation-orthogonality}. Define
\begin{equation}
 \mathsf D_{\rm tr}=\mathsf V_{\rm tr}^{\mathsf T}
                    \mathsf D\mathsf V_{\rm tr}>0,
 \qquad
 \mathsf R_{\rm tr}=
 (\Id-\mathsf V_{\rm tr}\mathsf V_{\rm tr}^{\mathsf T})
                         \mathsf D\mathsf V_{\rm tr}.
 \label{eq:s613-shape-trial-residual}
\end{equation}
For retained amplitudes $\vct a$, the omitted restoring acceleration
is $-\mathsf R_{\rm tr}\vct a$. Vanishing of the residual is exact
invariance of the trial subspace under $\mathsf D$.
\end{definition}

\begin{proposition}[Residual control of a shape-readout error]
\label{prop:s613-shape-readout-error}
Let $\Dtt{\vct x}+\mathsf D\vct x=0$ be the full conservative
motion and let
$\Dtt{\vct a}+\mathsf D_{\rm tr}\vct a=0$ be its declared trial
model after preparation. Put
$\vct e=\vct x-\mathsf V_{\rm tr}\vct a$ and
$\eta_e(t)=(\|\Dt{\vct e}\|_2^2+
\vct e^{\mathsf T}\mathsf D\vct e)^{1/2}$.
For a calibrated linear structural readout
$Y=\vct h^{\mathsf T}\vct x$ and
$Y_{\rm tr}=\vct h^{\mathsf T}\mathsf V_{\rm tr}\vct a$,
\begin{equation}
 \begin{aligned}
 \eta_e(t)&\leq\varepsilon_{\rm tr}(t):=\eta_e(0)+
              \int_0^t\|\mathsf R_{\rm tr}\vct a(s)\|_2\,\dd s,\\
 |Y(t)-Y_{\rm tr}(t)|&\leq
   \sqrt{\vct h^{\mathsf T}\mathsf D^{-1}\vct h}\,
                         \varepsilon_{\rm tr}(t).
 \end{aligned}
 \label{eq:s613-shape-readout-error}
\end{equation}
\end{proposition}
\begin{proof}
Subtracting the embedded trial equation gives
$\Dtt{\vct e}+\mathsf D\vct e=
-\mathsf R_{\rm tr}\vct a$.
The unforced full propagator is an isometry in the norm $\eta_e$.
Variation of constants therefore bounds its forced norm by the initial
norm plus the time integral of the residual force, giving the first
line. Cauchy--Schwarz applied to
$\mathsf D^{-1/2}\vct h$ and $\mathsf D^{1/2}\vct e$ gives the
second. No eigenvector-gap assumption is required for this finite-time
bound.
\end{proof}

Here $\eta_e$ has square-root-of-energy units. The readout factor
converts that energy error to the physical units of the selected marker,
contact displacement or other linear structural projection. An exact
symmetry sector can supply an invariant trial space, while a radial
ansatz in a faceted or contacted particle generally has a nonzero
residual. One can evaluate that residual along the reduced motion and
ask whether the predicted detector error remains below its declared
tolerance over the observation interval.

Add shape variables when omitted restoring acceleration or initial
excitation changes the chosen readout beyond tolerance. The residual
tests the marker trajectory directly. This result uses the unforced
harmonic reference; damping, branch memory, unmatched histories and
forcing outside the trial space add their corresponding residual terms.

\subsubsection{Prescribed flow and a nondissipative velocity coupling}

A geometric motion is a dynamical preparation. For example,
\begin{equation}
 \Dt{\vct R_i}=\vct b(\vct R_i,t),
 \qquad
 \vct b(\vct r,t)=\Omega_{\rm rot}(\vct r,t)
                 \vct e_z\times(\vct r-\vct r_{\rm c}).
 \label{eq:specified-rotational-flow}
\end{equation}
The angular rate $\Omega_{\rm rot}$ has inverse-time units. This flow
is tangent to an axial cylinder centred on $\vct r_{\rm c}$; a radial
clustering flow needs its own boundary compatibility, interactions or
external drive. Integrating angular rates produces angles, and successive
geometric maps compose. Their coordinate images are not added as linear
occupation operators. Large rearrangements can leave the original
harmonic basin and require a new reference.

A different extension changes the reversible equations themselves.
For constant real $\mathsf G_{\rm gyro}^{\mathsf T}=-\mathsf G_{\rm gyro}$,
\begin{equation}
 \mathsf M\Dtt{\vct u}+\mathsf G_{\rm gyro}\Dt{\vct u}
             +\mathsf K\vct u=\vct f,
 \qquad \Dt{\vct u}^{\mathsf T}\mathsf G_{\rm gyro}\Dt{\vct u}=0.
 \label{eq:operator-gyroscopic-mechanical-balance}
\end{equation}
The skew term changes polarization but supplies no power. For a planar
isotropic oscillator, take $\mathsf M=m\Id_2$, $\mathsf K=k\Id_2$,
$\mathsf G_{\rm gyro}=g\left(\begin{smallmatrix}0&-1\\1&0\end{smallmatrix}\right)$
 and $g\geq0$. For length coordinates, both $\mathsf G_{\rm gyro}$ and
 $g$ have units of mass per time, as does viscous drag, but their skew
 symmetry makes their mechanical effect different. Its two positive
 circular frequencies are
\begin{equation}
 \omega_{\pm,\rm gyro}=\frac{\sqrt{g^2+4mk}\pm g}{2m},
 \qquad \omega_{+,\rm gyro}\omega_{-,\rm gyro}=k/m.
 \label{eq:operator-gyroscopic-frequency-pair}
\end{equation}
Frequency splitting here does not imply attenuation or self-excitation.
Zero power also does not imply time-reversal invariance at a fixed
background: reversing time changes the sign of velocity, so the equation
recovers its form only when $\mathsf G_{\rm gyro}$ is reversed as well,
as for reversal of a magnetic field or frame angular velocity.
The frequency branches and zero-power property appear
beside the symmetry example in Figure~\ref{fig:icosahedral-loading-and-gyroscopic-response}.

Canonical momentum also changes. With constant $\mathsf G_{\rm gyro}$,
the Lagrangian and Hamiltonian are
\begin{equation}
 \begin{aligned}
 L&=\frac12\vct v^{\mathsf T}\mathsf M\vct v
       +\frac12\vct v^{\mathsf T}\mathsf G_{\rm gyro}\vct u
       -\frac12\vct u^{\mathsf T}\mathsf K\vct u,\\
 \vct p&=\mathsf M\vct v+\tfrac12\mathsf G_{\rm gyro}\vct u,\\
 H&=\frac12(\vct p-\tfrac12\mathsf G_{\rm gyro}\vct u)^{\mathsf T}
      \mathsf M^{-1}(\vct p-\tfrac12\mathsf G_{\rm gyro}\vct u)
      +\frac12\vct u^{\mathsf T}\mathsf K\vct u.
 \end{aligned}
 \label{eq:operator-gyroscopic-canonical-energy}
\end{equation}
Euler--Lagrange differentiation gives the skew velocity term above;
the canonical momentum is the velocity derivative of this Lagrangian,
rather than a prescribed mass-times-velocity rule~\cite{Deriglazov2010}.
Canonical quantization would use $\vct p$, while the mechanical momentum
is $\boldsymbol\pi=\mathsf M\vct v
=\vct p-\mathsf G_{\rm gyro}\vct u/2$.
Its components obey
$[\hat\pi_a,\hat\pi_b]=-i\hbar(G_{\rm gyro})_{ab}\Id$.
One cannot retain the old canonical identification with mass times
velocity and merely replace the oscillator frequencies. A Lorentz
realization additionally needs defined charged coordinates and a magnetic
field; a rotating-frame realization has its associated inertial terms.
A time-dependent field or control also needs its work law. None is
generated by choosing a complex phase for displacement.

\begin{figure}[H]
\centering
\begingroup
\definecolor{vlfGreen}{HTML}{176B64}
\definecolor{vlfDarkRed}{HTML}{762B38}
\definecolor{vlfRed}{HTML}{762B38}
\definecolor{vlfBlue}{HTML}{315D80}
\definecolor{vlfCopper}{HTML}{762B38}

\setlength{\unitlength}{1pt}
\scalebox{0.90}{%
\input{vlt-figures/figure-220.tex}%
}%
\endgroup
\caption{Symmetry-selected loading and an independent gyroscopic extension. (a) The regular icosahedral reference has twelve vertices and thirty edges. Averaging a unit radial force at one vertex over its sixty proper rotations gives radial amplitude $1/12$ at every vertex and invariant squared-norm fraction $1/12$. A localized force also contains rigid translation; its remaining component therefore includes more than internal vibration. The two force drawings use separate arrow gains. The projection requires masses, interactions and constraints that preserve the group. (b) For $M=mI$, $K=kI$ and $G=g\left(\begin{smallmatrix}0&-1\\1&0\end{smallmatrix}\right)$, let $\omega_0=\sqrt{k/m}$ and $\gamma_g=g/(m\omega_0)$. The two circular branches obey $\omega_{\rm hi,lo}/\omega_0=\sqrt{1+\gamma_g^2/4}\pm\gamma_g/2$ and have product $\omega_0^2$. Their splitting occurs at zero gyroscopic power, $\vct v^{\mathsf T}G\vct v=0$. This coupling is an additional constitutive choice, independent of the symmetry example.}
\label{fig:icosahedral-loading-and-gyroscopic-response}
\end{figure}

\subsubsection{Closing the operator construction}

Additional molecular, electronic or electromagnetic coordinates require
their own energies, domains, interactions and boundary conditions. A
specified self-adjoint total Hamiltonian supplies their conservative
coupling; eliminating a subsystem can then create memory, fluctuations
and an effective non-self-adjoint generator. The envelope--RNP example
already gives the simpler mechanical version. An extension is justified when it changes a defined prediction or
resolves a demonstrated failure of a coarser description.

The operator framework connects specified mechanics to observations.
Hessian and inertia determine conservative patterns; an augmented state
carries material memory; preparation determines fluctuations. Force and
detector maps then select response, transport and spatial correlations.
Canonical quantization supplies the additional occupation and quantum
measurement laws.

Variable selection is a prediction task with stated force channels,
detector outputs and observation time. Target identifiability
(Definition~\ref{def:s6-identifiable-prediction}) tests what a record
determines. Closure or the shape-readout error bound
(Proposition~\ref{prop:s613-shape-readout-error}) tests predictive
sufficiency. Registered histories additionally use the continuation
criterion of Proposition~\ref{prop:s6-history-closed-variables}.
Accumulated-loss operators measure energy retention within an episode
and through a sequence
(Propositions~\ref{prop:s610-finite-loss-contraction}
and~\ref{prop:s6-history-energy-composition}).

Centre displacement, relaxed shape and stored medium strain can
contribute to one virion signal while serving different predictive roles.
Their integration follows the force law, preparation, calibrated
observations and error tolerance. Other coupled biophysical particles
or structural domains admit the same tests once those inputs are
specified. The following benchmark combines these ingredients in one
passive pair, linking independent calibration to response, fluctuations
and a controlled link change.

\subsection{A complete collective response--fluctuation benchmark}
\label{sec:complete-benchmark}

A joint test fixes one mechanical reference, one parameter set and
its comparison models before prediction. Combining the pinned-pair
covariance developed above with the passive branch of Section~4 gives
a minimal collective benchmark. Common motion is a control; relative
motion tests a resolved material relaxation. Two-particle phase and
response measurements provide a colloidal precedent~\cite{Hough2002},
and viscoelastic Brownian models supply matched white and colored
fluctuations~\cite{Paul2021}. The synthetic inputs below connect
independent calibration to complementary predictions.

\paragraph{Mechanical reference and biological resolution.}
Two identical particles retain only their axial centre displacements
$\vct u=(u_1,u_2)^{\mathsf T}$ and equal resolved mass $m$.
For a concrete geometric reference, take nominal particle radius
$100\,\mathrm{nm}$, centre separation $1\,\mu\mathrm m$, and
centres at $z=\pm0.5\,\mu\mathrm m$ in a cylindrical cavity of radius
$2\,\mu\mathrm m$ and length $4\,\mu\mathrm m$, centred at $z=0$.
The radius is a clearance scale for this mechanical surrogate, rather
than a statement about capsid symmetry. Ideal quadratic supports pin the
centres, and an elastic link responds to their separation. Their positive
stiffnesses are prescribed independently of the reflecting boundary:
\begin{equation}
 U_0(\vct u)=\frac{k_p}{2}(u_1^2+u_2^2)
              +\frac{k_l}{2}(u_1-u_2)^2,\qquad
 \mathsf K_0=\begin{pmatrix}k_p+k_l&-k_l\\-k_l&k_p+k_l\end{pmatrix}.
 \label{eq:benchmark-reference}
\end{equation}
The support stiffness is $k_p>0$ and the relaxed link curvature is
$k_l\geq0$. Force balance has removed constant terms. The retained
axial motion is local to a persistent contact network, with wall
collisions, rotation and deformation omitted at this resolution.
The exact Gaussian calculation extends $\vct u$ to $\R^2$, so
the specified cavity is a clearance check. A biological realization
establishes the same coordinates, interactions and independent supports.

\begin{definition}[Pinned-pair Maxwell state]
\label{def:s7-benchmark-maxwell-state}
Let $\vct d=(1,-1)^{\mathsf T}$. A single Maxwell branch adds an internal
relaxed displacement $y$, a stiffness $\Delta k>0$ and relaxation time
$\tau>0$, with total stored energy
\begin{equation}
 E_b=\frac{m}{2}\vct v^{\mathsf T}\vct v+U_0(\vct u)
       +\frac{\Delta k}{2}(\vct d^{\mathsf T}\vct u-y)^2,
 \qquad \vct v=\Dt{\vct u}.
 \label{eq:benchmark-storage}
\end{equation}
\end{definition}
The branch senses separation through extension
$\vct d^{\mathsf T}\vct u-y$. Its material coordinate $y$ relaxes
through a dashpot of coefficient $\Delta k\tau$, as in Section~4.
Minimization over $y$ returns $U_0$: the link has relaxed stiffness
$k_l$ and frozen stiffness $k_l+\Delta k$.

\begin{figure}[H]
\centering\begingroup
\definecolor{vltTeal}{HTML}{176B64}
\definecolor{vltBlue}{HTML}{315D80}
\definecolor{vltGold}{HTML}{A87520}
\definecolor{vltPlum}{HTML}{725877}
\definecolor{vltRed}{HTML}{762B38}
\definecolor{vltYellow}{HTML}{D5AD3C}
\definecolor{vltNeutral}{HTML}{777D83}
\setlength{\unitlength}{1pt}
\scalebox{0.90}{%
\input{vlt-figures/figure-221.tex}%
}\endgroup
\caption{Two benchmark displacements and one material relaxation. (a) Supports $k_p$ and relaxed link $k_l$ act with a Maxwell branch of stiffness $\Delta k$ and dashpot resistance $\Delta k\tau$. The internal black point labels $y$. This topology view specifies connectivity, without prescribing strain. (b,c) Patterns $(a,a)$ and $(a,-a)$ use $8\,\mathrm{pt}$ per amplitude $a$, common reference rulers and matched coils. Each ruler uses separation $d$ and origin at its left reference centre. Common motion preserves span; relative motion compresses it by $2a$. At frozen $y=0$, $z=u_1-u_2>0$. Omitted lower supports remain part of the model. At $k_l=k_p$, all three comparison models share the printed covariance and differ dynamically.}
\label{fig:benchmark-mechanical-reference}
\end{figure}

\paragraph{One complete thermal dynamics.}
Specify an instantaneous resistance $c_0\Id$, with $c_0>0$, and an
isothermal medium at temperature $T>0$. Independent standard Wiener
increments $\dd\vct B$ and $\dd B_y$ give the augmented Langevin model
\begin{equation}
 \begin{aligned}
  \dd\vct u&=\vct v\,\dd t,\\
  m\,\dd\vct v&=\bigl[-\mathsf K_0\vct u
       -\Delta k\vct d(\vct d^{\mathsf T}\vct u-y)
       -c_0\vct v+\vct f\bigr]\dd t
       +\sqrt{2k_{\rm B}Tc_0}\,\dd\vct B,\\
  \dd y&=\frac{\vct d^{\mathsf T}\vct u-y}{\tau}\,\dd t
       +\sqrt{\frac{2k_{\rm B}T}{\Delta k\tau}}\,\dd B_y .
 \end{aligned}
 \label{eq:benchmark-thermal-dynamics}
\end{equation}
Each noise strength is fixed by the resistance or internal mobility at
the same temperature. No separate fluctuation amplitude is fitted.
\begin{proposition}[Benchmark passivity and thermal stationarity]
\label{prop:s7-benchmark-passive-thermal}
With noise omitted, differentiation of the storage gives
\begin{equation}
 \Dt{E_b}=\vct f^{\mathsf T}\vct v
       -c_0\vct v^{\mathsf T}\vct v
       -\frac{\Delta k}{\tau}(\vct d^{\mathsf T}\vct u-y)^2.
 \label{eq:benchmark-passivity}
\end{equation}
Thus the branch is passive: supplied work can be stored or dissipated,
but the internal relaxation produces no mechanical energy. At zero drive,
the stationary density is proportional to $\exp[-E_b/(k_{\rm B}T)]$.
\end{proposition}
\begin{proof}
Differentiate $E_b$ along the deterministic equations. The coupling
terms cancel between the particle and internal-coordinate powers,
leaving the displayed losses and supplied force power. At zero drive,
each friction--diffusion pair cancels on the stated Gibbs density,
while Hamiltonian streaming preserves it. Positive support, mass and
branch coefficients make that density normalizable.
\end{proof}
This is a reduced thermal description over a medium-equilibration window,
as specified in Section~4 and the thermal-state construction above;
an arbitrary finite sealed preparation
need not maintain a constant bath temperature. The diagonal resistance is
also a constitutive assumption, not a hydrodynamic consequence of the
cylinder. Measured fluid-mediated cross-resistance or wall corrections
would require an enlarged resistance model with its matched noise.

\paragraph{Inputs fixed before the joint prediction.}
Table~\ref{tab:benchmark-inputs} specifies all dimensional parameters.
Their values are chosen model inputs, and the final column states how an
empirical counterpart could constrain them without using the held-out
fluctuation record. Calibration and held-out tests may use the same instrument, but must
use distinct observations.

\begin{table}[H]
\centering\small
\begin{tabular}{@{}p{1.1cm}p{3.8cm}p{9.8cm}@{}}
\toprule
Input & \raggedright Prescribed synthetic value & Prospective independent constraint\\
\midrule
$m$ & $4\times10^{-18}\,\mathrm{kg}$
& Resolved inertia from mass characterization and a declared added-mass approximation.\\
$k_p$ & $10^{-4}\,\mathrm{N\,m^{-1}}$
& Quasistatic common-coordinate force gradient.\\
$k_l$ & $10^{-4}\,\mathrm{N\,m^{-1}}$
& Difference of relaxed relative and common force gradients, divided by two.\\
$\Delta k$ & $2\times10^{-4}\,\mathrm{N\,m^{-1}}$
& Excess force after a prescribed relative displacement, with particle positions held.\\
$\tau$ & $2\times10^{-5}\,\mathrm s$
& Decay time of that independently calibrated excess force.\\
$c_0$ & $2\times10^{-9}\,\mathrm{kg\,s^{-1}}$
& Common-motion drag calibration within the instantaneous-resistance band.\\
$T$ & $300\,\mathrm K$
& Thermometry during calibration and the held-out stationary record.\\
\bottomrule
\end{tabular}
\caption{One specified benchmark and its calibration requirements.
These are prescribed inputs, not measured virion properties. Detector
gain, exposure, transfer function and noise covariance must additionally
be constrained using the observation laws developed above before
comparing measured readouts.}
\label{tab:benchmark-inputs}
\end{table}

For calibration, define the normalized relative coordinate
$q_-=(u_1-u_2)/\sqrt2$, its conjugate force
$f_-=(f_1-f_2)/\sqrt2$, and branch strength $\kappa_b=2\Delta k$.
Start with the branch relaxed and hold a small relative displacement
$q_-=q_0$ after a rapid displacement compared with $\tau$.
For $t>0$ in this ideal-step limit, its mean holding force is
$f_-(t)=(k_p+2k_l)q_0+2\Delta kq_0\exp(-t/\tau)$.
The excess amplitude and decay constrain $\Delta k$ and $\tau$
without using the spontaneous displacement variance. For a linear rise
of duration $t_r$, the excess force at $t\geq t_r$ is instead
$\kappa_bq_0[\tau(1-e^{-t_r/\tau})/t_r]e^{-(t-t_r)/\tau}$;
its finite-rise attenuation must be included in the calibration.
The remaining held-out
predictions are the relative complex response over a frequency band,
the common and relative fluctuation spectra, and their particle readout
covariance. Inferring stiffness from that same covariance would make it
a calibration quantity rather than an independent test. Parameter
uncertainties should be propagated into the prediction, rather than
removed by refitting the held-out observations.

Figure~\ref{fig:benchmark-finite-rise-calibration} makes the calibration
window explicit. If the rise takes as long as the branch relaxes, the
ideal-step excess force is attenuated by $1-e^{-1}\simeq0.632$.
For $t_r\gg\tau$, the branch relaxes during loading and the excess is
only of relative order $\tau/t_r$. A missed early transient can therefore
make a real branch appear weaker; fitting its noise independently would
hide this calibration error. Displacement-control bandwidth, time stamps
and the measured ramp must be included before the relaxation amplitude
is interpreted as a stiffness. These are measurement requirements of the
benchmark, not additional fitted constitutive parameters.

\begin{figure}[H]
\centering\begingroup
\definecolor{vlfGreen}{HTML}{176B64}
\definecolor{vlfDarkRed}{HTML}{762B38}
\definecolor{vlfRed}{HTML}{762B38}
\definecolor{vlfBlue}{HTML}{315D80}
\definecolor{vlfCopper}{HTML}{762B38}
\setlength{\unitlength}{1pt}
\scalebox{0.80}{%
\input{vlt-figures/figure-222.tex}%
}\endgroup
\caption{Finite loading rate attenuates a relaxation-calibration amplitude. After a linear displacement ramp of duration $t_r$, hold $q_-=q_0$ from a previously relaxed branch. For $s=(t-t_r)/\tau\ge0$ and $r_r=t_r/\tau$, the excess holding force is $\Delta f_-/(\kappa_bq_0)=A(r_r)e^{-s}$, $A(r_r)=(1-e^{-r_r})/r_r$. (a) Curves at $r_r=0,1/4,1,3$ share one decay time; zero denotes the ideal-step limit. (b) The amplitude tends to one for rapid loading and scales as $1/r_r$ for slow loading. Ignoring it underestimates branch strength. The time origin is the end of the ramp. This force calibration fixes the same branch strength and timescale used in the held-out response and fluctuation predictions.}
\label{fig:benchmark-finite-rise-calibration}
\end{figure}

\paragraph{Common and relative response from the same parameters.}
The orthonormal columns
$\vct e_+=(1,1)^{\mathsf T}/\sqrt2$ and
$\vct e_-=(1,-1)^{\mathsf T}/\sqrt2$ define
$q_\pm=\vct e_\pm^{\mathsf T}\vct u$ and conjugate forces
$f_\pm=\vct e_\pm^{\mathsf T}\vct f$. In particular,
$\vct f^{\mathsf T}\vct u=f_+q_++f_-q_-$; this normalization
fixes the response units without an extra factor of two.
Their relaxed curvatures are $k_+=k_p$ and $k_-=k_p+2k_l$,
and the relative branch strength is $\kappa_b=2\Delta k$.
Eliminating $y$ with the $e^{-i\omega t}$ convention gives
\begin{equation}
 \begin{aligned}
 \mathcal R_+(\omega)&=\frac{1}{k_+-m\omega^2-i\omega c_0},\\
 \mathcal R_-(\omega)&=
 \left[k_--m\omega^2-i\omega c_0
       +\kappa_b\frac{-i\omega\tau}{1-i\omega\tau}\right]^{-1}.
 \end{aligned}
 \label{eq:benchmark-responses}
\end{equation}
The common coordinate leaves the separation unchanged, so the internal
branch cannot act on it. In relative motion, the branch follows slow
displacements but resists fast ones. Its real stiffness increment grows
from zero to $\kappa_b$; its positive loss stiffness is
$\kappa_b\omega\tau/[1+(\omega\tau)^2]$. Both follow from the
same relaxation law. A particle force and readout are reconstructed by
$\mathcal R_{11}=(\mathcal R_++\mathcal R_-)/2$ and
$\mathcal R_{12}=(\mathcal R_+-\mathcal R_-)/2$, with
$\mathcal R_{22}=\mathcal R_{11}$ and
$\mathcal R_{21}=\mathcal R_{12}$.

The inertial time $m/c_0=2\,\mathrm{ns}$ is much shorter than
$\tau=20\,\mu\mathrm s$, placing the branch feature in a
predominantly relaxational regime. For dimensional comparison, the
unbounded no-slip spherical Stokes formula gives
$c_0/(6\pi a)\simeq1.06\,\mathrm{mPa\,s}$ at nominal
$a=100\,\mathrm{nm}$~\cite{SquiresMason2010}. An actual cavity
experiment additionally constrains wall corrections, surface conditions
and cross-resistance.

\paragraph{Fluctuations and static covariance without an extra fit.}
Use the two-sided angular-frequency convention
$S_\pm(\omega)=\int_{-\infty}^{\infty}e^{i\omega t}
\langle q_\pm(t)q_\pm(0)\rangle\,\dd t$ at zero drive.
The same elimination that produces memory also produces colored force
noise. Its prediction is
\begin{equation}
 \begin{aligned}
  S_+(\omega)&=2k_{\rm B}Tc_0|\mathcal R_+(\omega)|^2,\\
  S_-(\omega)&=2k_{\rm B}T
    \left[c_0+\frac{\kappa_b\tau}{1+(\omega\tau)^2}\right]
    |\mathcal R_-(\omega)|^2
    =\frac{2k_{\rm B}T}{\omega}\operatorname{Im}\mathcal R_-(\omega).
 \end{aligned}
 \label{eq:benchmark-spectra}
\end{equation}
The last identity applies for $\omega\ne0$, with its continuous limit
at zero. Replacing the colored term by an independently fitted white
noise would break this response--fluctuation link. Marginalizing the
Gibbs density over $y$ instead gives the relaxed potential $U_0$, so
\begin{equation}
 \boxed{\langle q_\pm^2\rangle=\frac{k_{\rm B}T}{k_\pm},\qquad
 \mathsf C_u=\frac{k_{\rm B}T}{2}
 \begin{pmatrix}
 k_+^{-1}+k_-^{-1}&k_+^{-1}-k_-^{-1}\\
 k_+^{-1}-k_-^{-1}&k_+^{-1}+k_-^{-1}
 \end{pmatrix}.}
 \label{eq:benchmark-covariance}
\end{equation}

Figure~\ref{fig:benchmark-covariance-bases} displays this covariance in the particle and normal-coordinate bases with equal geometric scales.

\begin{figure}[H]
\centering\begingroup
\definecolor{vlfGreen}{HTML}{176B64}
\definecolor{vlfDarkRed}{HTML}{762B38}
\definecolor{vlfRed}{HTML}{762B38}
\definecolor{vlfBlue}{HTML}{315D80}
\definecolor{vlfCopper}{HTML}{762B38}
\setlength{\unitlength}{1pt}
\scalebox{0.80}{%
\input{vlt-figures/figure-223.tex}%
}\endgroup
\caption{One relaxed Gibbs covariance in two orthonormal bases. With $k_+=k_p$, $k_-=3k_p$ and $\ell_T^2=k_BT/k_p$, (a) the equal-scale unit-Mahalanobis contour has axes $\ell_T,\ell_T/\sqrt3$ along $(1,\pm1)/\sqrt2$. It encloses probability $1-e^{-1/2}\simeq0.393$ for this two-dimensional Gaussian. (b) Modal variances are $1,1/3$ in units $\ell_T^2$ and modal cross-covariance is zero. Rotation gives particle correlation $1/2$. The full, L and E models share these static statistics. The boxed result illustrated here is~\eqref{eq:benchmark-covariance}.}
\label{fig:benchmark-covariance-bases}
\end{figure}

The mode cross-covariance vanishes. The particle spectra are correspondingly
$S_{11}=S_{22}=(S_++S_-)/2$ and $S_{12}=S_{21}=(S_+-S_-)/2$.
Each modal spectrum integrates to its boxed variance:
$\langle q_\pm^2\rangle=(2\pi)^{-1}\int_{-\infty}^{\infty}
S_\pm(\omega)\,\dd\omega$.
Relaxation changes the distribution of fluctuation power over frequency,
while the total area remains set by the relaxed restoring force.

The observation window must resolve equilibration. In the augmented
state $(\vct u,\vct v,y)$, the stable drift evolves covariance by
the Lyapunov law developed above. Stationary spectra apply after transients
relax or for a correctly prepared joint Gibbs state. Detector exposure
and windowing then filter the same process used for the driven response;
the variance identities refer to its unfiltered full-band spectrum.

\paragraph{Two explicitly matched simpler models.}
An elastic-network reference retains $m,\mathsf K_0,c_0$ and its
compatible white thermal noise, but omits the Maxwell branch. Its
relative susceptibility is
$\mathcal R_-^{\rm E}=[k_--m\omega^2-i\omega c_0]^{-1}$ and
$S_-^{\rm E}=2k_{\rm B}Tc_0|\mathcal R_-^{\rm E}|^2$.
A stronger memory-free Langevin comparison matches the low-frequency
resistance $c_0+\kappa_b\tau$ of the full relative coordinate:
\begin{equation}
 \mathcal R_-^{\rm L}=\frac{1}{k_--m\omega^2
                      -i\omega(c_0+\kappa_b\tau)},\qquad
 S_-^{\rm L}=2k_{\rm B}T(c_0+\kappa_b\tau)
                  |\mathcal R_-^{\rm L}|^2.
 \label{eq:benchmark-memory-free}
\end{equation}
In particle coordinates its resistance is
$c_0\Id+\Delta k\tau\vct d\vct d^{\mathsf T}$, with the
corresponding correlated white force noise. Both comparisons have the
same common response and the same equilibrium covariance as the full
model. The Langevin comparison also agrees to leading low-frequency
order; it is an appropriate reduction when $|\omega\tau|\ll1$.
The full model is itself Brownian and is Markovian on $(\vct u,\vct v,y)$.
Its advantage is additional resolved constitutive memory, rather than a
replacement of Brownian mechanics. A genuinely lossless elastic network
can also have a Gibbs ensemble, but has discrete delta-function spectra
and no dissipative equilibration. It is not the finite-linewidth reference
used here. Freezing the branch instead would change the relative static
variance to $k_{\rm B}T/(k_-+\kappa_b)$ and therefore compare a different
equilibrium energy.

\paragraph{Calculated discrimination at the relaxation crossover.}
Define $r=\omega\tau$, the thermal length
$\ell_T=(k_{\rm B}T/k_p)^{1/2}$, and
$\mu=m/(k_p\tau^2)$. Table~\ref{tab:benchmark-inputs} gives
$k_-/k_p=3$, $\kappa_b/k_p=4$,
$c_0/(k_p\tau)=1$ and $\mu=10^{-4}$.
The dimensionless response and spectrum are consequently
\begin{equation}
 \begin{aligned}
 k_p\mathcal R_-&=
 \left[3-\mu r^2-ir+4\frac{-ir}{1-ir}\right]^{-1},\\
 \frac{k_pS_-}{k_{\rm B}T\tau}
   &=2\left[1+\frac{4}{1+r^2}\right]|k_p\mathcal R_-|^2.
 \end{aligned}
 \label{eq:benchmark-dimensionless}
\end{equation}
Here $\ell_T\simeq6.44\,\mathrm{nm}$, each particle's rms displacement
is $\sqrt{2/3}\ell_T\simeq5.25\,\mathrm{nm}$, and
$\mathsf C_u=\ell_T^2\begin{pmatrix}2/3&1/3\\1/3&2/3\end{pmatrix}$.
The particle correlation coefficient is $1/2$ in all three models.
These rms excursions are small compared with the specified clearances;
their acceptability for a real interaction additionally requires a
measured local harmonic basin. The nominal frozen elastic frequencies
are not needed to interpret the plotted relaxation crossover. At
$r\sim1$, inertia is small, and the benchmark describes strongly damped
collective motion rather than a propagating or long-lived oscillation.

The crossover is at $\omega=5\times10^4\,\mathrm{s^{-1}}$, or
$\omega/(2\pi)\simeq7.96\,\mathrm{kHz}$.
Neglecting the $10^{-4}$ inertial correction at $r=1$ gives
\begin{equation}
 \begin{array}{c|ccc}
 &\text{full branch}&\text{memory-free L}&\text{elastic E}\\ \hline
 k_p\mathcal R_-& (5-3i)^{-1}&(3-5i)^{-1}&(3-i)^{-1}\\
 |k_p\mathcal R_-|&0.1715&0.1715&0.3162\\
 \arg\mathcal R_-&30.96^\circ&59.04^\circ&18.43^\circ\\
 k_pS_-/(k_{\rm B}T\tau)&0.1765&0.2941&0.2000
 \end{array}
 \label{eq:benchmark-crossover-values}
\end{equation}
Positive phase here means displacement lag for the $e^{-i\omega t}$
convention. Magnitude alone at this frequency would scarcely distinguish
the full and low-frequency Langevin models. Their phases differ by
$28.1^\circ$, and the full fluctuation spectrum is $40\%$ lower.
Figure~\ref{fig:benchmark-predictions} uses the retained mass and the same
input table throughout; no curve has its own fitted stiffness or noise.
The plotted band $0.01\leq r\leq10$ has
$m\omega/c_0\leq10^{-3}$. An experimental application must also check
that this band lies below additional medium, structural and detector
cutoffs. Inertia may be dropped only within its stated range of accuracy.

\begin{figure}[H]
\centering\begingroup
\definecolor{vltTeal}{HTML}{176B64}
\definecolor{vltBlue}{HTML}{315D80}
\definecolor{vltGold}{HTML}{A87520}
\definecolor{vltPlum}{HTML}{725877}
\definecolor{vltRed}{HTML}{762B38}
\definecolor{vltYellow}{HTML}{D5AD3C}
\definecolor{vltNeutral}{HTML}{777D83}
\setlength{\unitlength}{1pt}
\scalebox{0.90}{%
\input{vlt-figures/figure-224.tex}%
}\endgroup
\caption{Linked predictions from one benchmark parameter set: $\mu=m/(k_p\tau^2)=10^{-4}$, $k_-/k_p=3$, $\kappa_b/k_p=4$ and $c_0/(k_p\tau)=1$. Teal, dashed burgundy and dotted blue identify the full model, low-frequency matched Langevin model L and branch-free elastic model E throughout (a)--(c). Dotted vertical guides mark $r=1$. At $r=1$, full and L nearly match in magnitude; their lags differ by $28.1^\circ$ and the full-model spectrum is $40\%$ lower. Brackets and the printed ratio mark these comparisons. All three retain the same static covariance and common-mode response. (d) The full branch alone traces its inverse-response storage/loss semicircle on equal axis scales; frequency marks use the same $\tau$. Its colored force-noise increment is fixed by $2k_BT B_b/\omega$. No stiffness, noise strength or timescale is refitted between panels. Spectral equality of integrated variance concerns the full band, beyond the displayed interval. The specified observation band is strongly damped.}
\label{fig:benchmark-predictions}
\end{figure}

\paragraph{A joint constitutive signature and its limits.}
The complex inverse-response difference cancels the shared inertia and
instantaneous resistance. Subtracting the calibrated elastic-link
contribution leaves
\begin{equation}
 \mathcal D_b=\mathcal R_-^{-1}-\mathcal R_+^{-1}-2k_l
       =\kappa_b\frac{-ir}{1-ir},\qquad
 A_b=\operatorname{Re}\mathcal D_b,\quad
 B_b=-\operatorname{Im}\mathcal D_b.
 \label{eq:benchmark-inverse-response-difference}
\end{equation}
\begin{proposition}[Linked relative-response and force-noise signature]
\label{prop:s7-benchmark-linked-signature}
For $\omega>0$, $A_b\geq0$ and $B_b\geq0$. These storage and loss
increments lie on the single-relaxation semicircle, and their force-noise
increment obeys a linked prediction:
\begin{equation}
 \boxed{\begin{aligned}
 \left(A_b-\frac{\kappa_b}{2}\right)^2+B_b^2
        &=\left(\frac{\kappa_b}{2}\right)^2,\\
 \frac{S_-}{|\mathcal R_-|^2}
       -\frac{S_+}{|\mathcal R_+|^2}
        &=\frac{2k_{\rm B}T}{\omega}B_b.
 \end{aligned}}
 \label{eq:benchmark-joint-signature}
\end{equation}
\end{proposition}
\begin{proof}
The response increment gives $A_b=\kappa_b r^2/(1+r^2)$
and $B_b=\kappa_b r/(1+r^2)$, which satisfy the circle identity.
Divide the two spectra in~\eqref{eq:benchmark-spectra} by their
respective squared responses. Their common white term cancels,
leaving $2k_{\rm B}T\kappa_b\tau/(1+r^2)
=2k_{\rm B}TB_b/\omega$.
\end{proof}

Figure~\ref{fig:benchmark-linked-response-noise-signature} pairs the semicircle with its matching force-noise spectrum at the same marked frequencies.

\begin{figure}[H]
\centering\begingroup
\definecolor{vlfGreen}{HTML}{176B64}
\definecolor{vlfDarkRed}{HTML}{762B38}
\definecolor{vlfRed}{HTML}{762B38}
\definecolor{vlfBlue}{HTML}{315D80}
\definecolor{vlfCopper}{HTML}{762B38}
\setlength{\unitlength}{1pt}
\scalebox{0.80}{%
\input{vlt-figures/figure-225.tex}%
}\endgroup
\caption{One relaxation parameter links storage/loss geometry to force noise. (a) $A_b/\kappa_b=r^2/(1+r^2)$ and $B_b/\kappa_b=r/(1+r^2)$ trace a semicircle on equal axis scales. Open endpoints denote $r\to0,\infty$, and marks identify $r=0.1,1,10$. (b) The same marks follow $\Delta S_f=2k_BT\kappa_b\tau/(1+r^2)$, obtained by dividing each displacement spectrum by its own squared response. Its finite low-frequency limit follows from $B_b\sim\kappa_b\omega\tau$. The locus, frequency trace and colored noise use one branch strength and timescale. Curves are exact predictions of the stated passive thermal model. The boxed result illustrated here is~\eqref{eq:benchmark-joint-signature}.}
\label{fig:benchmark-linked-response-noise-signature}
\end{figure}

The second identity compares modal force-noise spectra, reconstructed
from the corresponding displacement spectra and complex responses.
Dividing each displacement spectrum by its own squared response removes
that coordinate's mechanical filtering before the branch noise is
compared. It is not an equality between displacement spectra alone. The first
identity is the established single-Maxwell relaxation geometry. Its
use here is selective: the same relative branch must produce that
response increment, leave the common coordinate unchanged, preserve the
relaxed static covariance and supply precisely the associated colored
noise. A measured frequency trace must follow $r=\omega\tau$ with the
independently constrained $\tau$, not merely lie somewhere on a circle.

The held-out response, phase, fluctuations and their observation windows
therefore form one test. Agreement would support this specified passive
coordinate model over its restricted memory-free alternatives within
the resolved band. It would not identify a unique protein, envelope or
ionic mechanism: internal relaxation and projected medium memory can
have the same effective law. A discrepancy in the common control points
to a missing shared resistance, pinning or readout correction before it
can be attributed to the relative branch. Failure of the branch-specific
spectral prediction can indicate an inadequate projected constitutive
model. A genuine response--fluctuation mismatch, after matching the
coordinate and conjugate force and correcting detector effects, instead
calls for scrutiny of preparation, temperature and thermal-equilibrium
assumptions. Consistent elimination of hidden equilibrium coordinates can change
projected memory while preserving fluctuation--dissipation
\cite{Mori1965,Kubo1957}. A generalized Brownian model with the identical
kernel and thermal noise is mathematically equivalent to the benchmark.
The benchmark supplies an independently calibratable collective
prediction and a defined frequency range for testing additional
constitutive resolution. Empirical validation remains the
next step after this synthetic calculation.

\paragraph{A controlled arrangement change tests the continued state.}
The stationary benchmark can also predict a registered transition before
the new arrangement equilibrates. Keep the same two physical particles,
supports, reference centres, bath and Maxwell coordinate, and change
only the relaxed link curvature from $k_l^{\rm pre}$ to
$k_l^{\rm post}$ at $t=0$. Controlled optical-potential changes provide
an experimental precedent for testing relaxation after parameter changes
\cite{Martinez2016}; the coupled-particle calculation here uses its own
specified link and material state.

\begin{definition}[Continued link-change benchmark]
\label{def:s7-continued-link-benchmark}
Use $X=(\vct u,\vct v,y)^{\mathsf T}$ in both episodes. All of its
components are continuous at the stipulated instantaneous control
change, including $y$. The pre-event preparation is the centered joint
Gibbs law of~\eqref{eq:benchmark-storage}. Let $\mathsf Q_g$ be the
positive Hessian of that stored energy in $X$ for
$g\in\{\mathrm{pre},\mathrm{post}\}$, and put
$\mathsf\Sigma_g=k_{\rm B}T\mathsf Q_g^{-1}$. The post-event drift
$\mathsf A$ and diffusion factor $\mathsf B$ are those of
\eqref{eq:benchmark-thermal-dynamics} with $k_l=k_l^{\rm post}$.
They satisfy
$\mathsf A\mathsf\Sigma_{\rm post}+
\mathsf\Sigma_{\rm post}\mathsf A^{\mathsf T}
+\mathsf B\mathsf B^{\mathsf T}=0$.
\end{definition}

The same physical population is present before and after the event.
An association mask can select a different group without changing $X$.
The controlled connection specifies a useful continuation test for
recruitment models; a spontaneous joining experiment additionally
supplies its capture landscape, event probabilities and energy transfer.

\begin{proposition}[Post-event covariance and supplied link work]
\label{prop:s7-continued-link-covariance}
With zero drive after the change, the mean remains zero and, for
$t\geq s\geq0$,
\begin{equation}
 \begin{aligned}
 \mathsf\Sigma(t)
 &=\mathsf\Sigma_{\rm post}
    +e^{t\mathsf A}(\mathsf\Sigma_{\rm pre}
                         -\mathsf\Sigma_{\rm post})e^{t\mathsf A^{\mathsf T}},\\
 \operatorname{Cov}(X(t),X(s))
 &=e^{(t-s)\mathsf A}\mathsf\Sigma(s),\\
 \langle W_{\rm ctrl}\rangle
 &=\frac{(k_l^{\rm post}-k_l^{\rm pre})k_{\rm B}T}
                                      {k_p+2k_l^{\rm pre}}.
 \end{aligned}
 \label{eq:s7-continued-link-covariance-work}
\end{equation}
For calibrated displacement rows $\mathsf L(t)$ and $\mathsf L(s)$,
the corresponding covariance is
$\mathsf L(t)e^{(t-s)\mathsf A}\mathsf\Sigma(s)
\mathsf L(s)^{\mathsf T}$.
\end{proposition}
\begin{proof}
Continuity sets $\mathsf\Sigma(0)=\mathsf\Sigma_{\rm pre}$.
Subtract the post-event stationary Lyapunov equation from the evolving
one to obtain the first identity. Future Wiener increments are
independent of $X(s)$, giving the second. The instantaneous work is
$(k_l^{\rm post}-k_l^{\rm pre})(u_1-u_2)^2/2$.
Its pre-event Gibbs average uses
$\langle(u_1-u_2)^2\rangle=2k_{\rm B}T/(k_p+2k_l^{\rm pre})$.
Linear projection gives the detector covariance.
\end{proof}

The continued prediction retains displacement--material correlations
from the old arrangement. Resetting $y$ or sampling a new thermal
state changes that preparation and its later covariance. Controller
work and subsequent thermal exchange account for the energy change
under the independently specified post-event parameters.

An experiment can calibrate both link curvatures and the branch using
the earlier force protocols, register the control time, then hold out
the two-time displacement covariance. A response pulse sent before
the change uses the product of the two episode propagators and the
registered continuation defined earlier in this section. Exposure and membership
weights act on that complete predicted history. Before equilibration,
the covariance depends on both observation times and the stationary
spectra in~\eqref{eq:benchmark-spectra} are replaced by the finite-window
two-time prediction. This permits a direct test of retained preparation
and constitutive memory during an evolving assembly. The persistent
equilibrium limit recovers the original response--fluctuation benchmark.

Figure~\ref{fig:s7-continued-benchmark-covariance} displays the transient variance and two-time record predicted by the continued state.

\begin{figure}[H]
\centering\begingroup
\definecolor{vltTeal}{HTML}{176B64}
\definecolor{vltBlue}{HTML}{315D80}
\definecolor{vltGold}{HTML}{A87520}
\definecolor{vltPlum}{HTML}{725877}
\definecolor{vltRed}{HTML}{762B38}
\definecolor{vltYellow}{HTML}{D5AD3C}
\definecolor{vltNeutral}{HTML}{777D83}
\setlength{\unitlength}{1pt}
\scalebox{0.90}{%
\input{vlt-figures/figure-226.tex}%
}\endgroup
\caption{Continued-state prediction across a controlled link change. (a) At $t=0$, $k_l/k_p$ changes from $0.2$ to $1.1$ with the same virions, reference centres and inherited Gibbs state. Equal coil geometry registers unchanged instantaneous separation; stiffness is specified by the labels. Rulers use reference separation $d$ and origin at the left centre. Mean supplied work is $9k_{\rm B}T/14$. (b) Exact relative variances compare continuation with an artificial $y=0$ reset at identical initial particle covariance; the dotted line is the new Gibbs limit. (c) Cross-particle covariance at lag $\tau/2$ depends on waiting time $s$. Five-state matrix exponentials use $m/(k_p\tau^2)=10^{-4}$, $c_0/(k_p\tau)=1$, $\Delta k/k_p=2$ and $\ell_T^2=k_{\rm B}T/k_p$. This auxiliary protocol supplies its own controller work.}
\label{fig:s7-continued-benchmark-covariance}
\end{figure}

The benchmark makes a passive collective model testable with matched
inputs. Adding the pumped oscillator developed later in
Section~\ref{sec:phonon-quantum-active} requires independent resource
and rate constraints, its corresponding noise, and a statistic that
distinguishes its preparation. A discrepancy first tests the declared
mechanical, material and observation assumptions; those explicit
choices guide the next model refinement.

\paragraph{An angular extension with the same response--fluctuation test.}
The axial benchmark is a controlled projection of the full body mechanics.
One additional angular coordinate shows how an oriented contact enters
without changing its existing numerical comparisons. Retain
$q=q_-=(u_1-u_2)/\sqrt2$, an independently supported small libration
$\theta$, and the same length-valued material coordinate $y$. Let
$\ell_c$ be the signed contact-extension derivative with respect to that
angle, obtained from the registered site geometry. Use
\begin{equation}
 \begin{aligned}
 E_b={}&\tfrac12m\dot q^2+\tfrac12J\dot\theta^2
       +\tfrac12k_-q^2+\tfrac12k_\theta\theta^2
       +\tfrac12\Delta k(\sqrt2q+\ell_c\theta-y)^2,\\
 \dot y={}&(\sqrt2q+\ell_c\theta-y)/\tau
       +\sqrt{2k_{\rm B}T/(\Delta k\tau)}\,\xi_y .
 \end{aligned}
 \label{eq:s6patch-angular-benchmark-storage}
\end{equation}
Here $J,k_\theta,c_\theta>0$ specify angular inertia, support and drag;
$k_\theta$ has energy units and $c_\theta$ has action units. The
mechanical equations are $m\ddot q=-\partial_qE_b-c_0\dot q+f_q+\eta_q$
and $J\ddot\theta=-\partial_\theta E_b-c_\theta\dot\theta+T_\theta+\eta_\theta$,
where potential derivatives hold velocities fixed. The three independent
white noises obey the same thermal prescription as the original branch:
$\langle\eta_q(t)\eta_q(t')\rangle=2k_{\rm B}Tc_0\delta(t-t')$,
$\langle\eta_\theta(t)\eta_\theta(t')\rangle=2k_{\rm B}Tc_\theta\delta(t-t')$
and $\langle\xi_y(t)\xi_y(t')\rangle=\delta(t-t')$.

\begin{proposition}[Dynamic angular coupling with uncoupled static widths]
\label{prop:s6patch-angular-benchmark}
For this stable harmonic equilibrium model, set
$\vct w=(\sqrt2,\ell_c)^{\mathsf T}$,
$\vct z_b=(q,\theta)^{\mathsf T}$ and
$\zeta_b(\omega)=\Delta k(-i\omega\tau)/(1-i\omega\tau)$.
Eliminating $y$ gives the conjugate force/torque response and two-sided
thermal force spectrum
\begin{equation}
 \boxed{\begin{aligned}
 \mathsf D_b(\omega)&=
  \begin{pmatrix}k_--m\omega^2-i\omega c_0&0\\
              0&k_\theta-J\omega^2-i\omega c_\theta\end{pmatrix}
       +\zeta_b(\omega)\vct w\vct w^{\mathsf T},\\
 \mathsf S_{f,b}(\omega)&=2k_{\rm B}T
   \left[\begin{pmatrix}c_0&0\\0&c_\theta\end{pmatrix}
       +\frac{\Delta k\tau}{1+\omega^2\tau^2}
                           \vct w\vct w^{\mathsf T}\right].
 \end{aligned}}
 \label{eq:s6patch-angular-response-noise}
\end{equation}
The displacement/angle spectrum is
$\mathsf S_{z,b}=\mathsf D_b^{-1}\mathsf S_{f,b}\mathsf D_b^{-\dagger}$.
Nevertheless its equilibrium equal-time covariance is
\begin{equation}
 \mathsf C_b
    =k_{\rm B}T\begin{pmatrix}k_-^{-1}&0\\0&k_\theta^{-1}\end{pmatrix}.
 \label{eq:s6patch-angular-static-covariance}
\end{equation}
Here $\mathsf C_b=\operatorname{Cov}[\vct z_b,\vct z_b]$ is the
full covariance matrix of the displayed coordinate pair.
\end{proposition}
\begin{proof}
Fourier elimination of the stable scalar relaxation yields $\zeta_b$.
The eliminated branch noise acts in the direction $\vct w$ and has the
displayed Lorentzian spectrum; its off-diagonal entry is fixed by the
same coupling. The Gibbs integral over $y$ is a translation of a Gaussian
integral, independent of $q,\theta$, leaving the two diagonal supports.
\end{proof}
For $\ell_c\ne0$, the off-diagonal dynamic stiffness is
$\sqrt2\ell_c\zeta_b$ and the force--torque noises are correlated.
Thus a zero equal-time translation--angle covariance can coexist with
a measurable cross-response and lag-dependent cross-correlation. Centre
and registered-marker channels test that prediction using the same
parameters, exposure and noise law. Setting $\ell_c=0$ recovers the
original relative-coordinate response and adds an independent angular
mode. The added constants are synthetic until constrained for the chosen
VLP preparation; capsid size alone determines neither the lever arm nor
the angular support. This extends the benchmark to the orientation and
material coupling already defined in Sections~2--5 while retaining its
role as a calculation rather than experimental validation.

Figure~\ref{fig:s56-angular-response-fluctuation} compares static widths, cross-response and lagged motion for this single angular/material model.

\begin{figure}[H]
\centering\begingroup
\definecolor{vltTeal}{HTML}{176B64}
\definecolor{vltBlue}{HTML}{315D80}
\definecolor{vltGold}{HTML}{A87520}
\definecolor{vltPlum}{HTML}{725877}
\definecolor{vltRed}{HTML}{762B38}
\definecolor{vltYellow}{HTML}{D5AD3C}
\definecolor{vltNeutral}{HTML}{777D83}
\setlength{\unitlength}{1pt}
\scalebox{0.90}{%
\input{vlt-figures/figure-298.tex}%
}\endgroup
\caption{Static widths and dynamic angular coupling in the extended benchmark. Panel (a) shows the two rigid bodies and a signed contact deformation; rotation changes a surface-site separation while the centre coordinate $q$ retains its defined relative normalization. The spring denotes the retained branch, not an additional link. Panels (b--d) use one synthetic parameter set in scales $E_\star,\ell_\star,\tau_\star$: $\mathsf G_b=\diag(0.08,0.05)$, $\mathsf K_b=\diag(3,1.7)$, $\mathsf C_b^{\rm drag}=\diag(1,0.7)$, $\Delta k=2$, $\tau=1$, $\vct w=(\sqrt2,0.6)^{\mathsf T}$ and $k_BT/E_\star=1$. The upper geometry illustrates the contact map; it does not determine these assigned coefficients. The contours in (b) have squared Mahalanobis radii $1$ and $1/4$. Panel (c) plots the real and imaginary parts of the dimensionless conjugate force/torque cross-compliance from~\eqref{eq:s6patch-angular-response-noise}. Panel (d) evaluates $[e^{t\mathsf A_b}\mathsf\Sigma_b]_{q\theta}$ for the same five-state thermal model, before detector exposure. Its equal-time value vanishes although its finite-lag value does not. The signed cross quantities need not be positive.}
\label{fig:s56-angular-response-fluctuation}
\end{figure}

\subsection{A preparation-specific HPV16 application: evidence, models and prediction}
\label{sec:hpva-application}

The synthetic benchmark specifies its inputs and calculates their
consequences. An application to a measured preparation starts by
establishing those inputs and reserving an observation against which to
test the resulting prediction. Here the concrete question is whether
size, transport and surface measurements constrain the collective
interaction of one stabilized HPV16 L1 VLP preparation, and which further
record would distinguish the remaining models. The following construction
fixes the evidence, comparison hierarchy and prediction boundary. It
does not report a fit or a new experimental dataset.

\subsubsection{The selected preparation and its evidence}

\begin{definition}[Preparation-specific application]
\leavevmode\label{def:hpva-preparation}
An application specification records particle identity and structural
resolution, preparation age and history, solvent composition,
temperature, concentration convention, domain and acquisition law.
Parameters and comparisons refer to this specification. Transferring
them to another preparation requires a stated matching rule.
\end{definition}

Select the 48-hour HPV16 L1 VLP preparation studied by
Vega et al.~\cite{Vega2016}. Table~\ref{tab:hpva-provenance}
separates reported inputs from quantities requiring inference.
The nominal $55\,\mathrm{nm}$ particle in the introductory synthetic
guide remains its own geometric choice. It is not substituted for
this preparation's measured size distribution. Each retained node here
is an intact whole VLP. Population concentration may vary between
aliquots; a closed simulation assigns its own fixed enclosed pool.

\begin{table}[H]
\centering\small
\begin{tabular}{@{}p{3.15cm}p{6.2cm}p{6.1cm}@{}}
\toprule
Input & Published assignment & Provenance and role\\
\midrule
Particle preparation & Insect-cell/baculovirus HPV16 L1 VLPs; selected age $48\,\mathrm h$. & Methods and final-state measurements in~\cite{Vega2016}; structural reference.\\[4pt]
Environment & $298\,\mathrm K$, pH $6$; $0.5\,\mathrm M$ NaCl, $50\,\mathrm{mM}$ sodium citrate, $1\,\mathrm{mM}$ CaCl$_2$. & Methods; buffer identity, not a measured surface potential.\\[4pt]
Concentration & Stock $0.4\,\mathrm{mg\,mL^{-1}}$; dilution series $0.01$--$0.4\,\mathrm{mg\,mL^{-1}}$. & Methods and Fig.~5; retain each aliquot's concentration.\\[4pt]
Hydrodynamic size & Final number-average diameter about $58\pm2\,\mathrm{nm}$. & Results and S1 Table; distribution-derived size.\\[4pt]
Scattering summaries & Final $M_w=18.4\pm0.4\,\mathrm{MDa}$; quoted $B_{22}=(5.7\pm0.3)10^{-4}\,\mathrm{mol\,mL\,g^{-2}}$. & Fig.~5 analysis; coefficient convention checked below.\\[4pt]
Electrokinetics & Mobility measured at $0.1\,\mathrm{mg\,mL^{-1}}$. & Fig.~6 and S5 Fig.; numerical trace requires extraction.\\
\bottomrule
\end{tabular}
\caption{Preparation ledger for the selected literature case. Values are reported summaries, not new measurements. Source uncertainties retain their published meaning; no common confidence level is assigned.}
\label{tab:hpva-provenance}
\end{table}

The supporting inventory includes diameter values in S1 Table (DOCX)
and correlation/mobility plots in S1--S5 Figs. (TIF)~\cite{Vega2016}.
These are published numerical summaries and graphics; complete instrument
records and replicate-level covariance have not been established here.
No plotted trace has yet been digitized for the present construction.
Reported sampling includes 15-scan DLS averages, triplicate endpoint
DLS, at least 100 TEM particles per sample, and SLS error bars from
scan variability~\cite{Vega2016}. These counts distinguish technical
sampling and particle counts from independently prepared biological batches.

\begin{table}[H]
\centering\small
\begin{tabular}{@{}p{3.3cm}p{6.2cm}p{5.95cm}@{}}
\toprule
Record & Information retained & Dependency or uncertainty to preserve\\
\midrule
TEM & Particle outlines and number-weighted morphology. & Staining, selection and image measurement; a dried/stained outline is not a hydrated drag radius.\\[4pt]
DLS correlation & Field/intensity correlation, scattering vector, baseline and acquisition. & Fitted decay, inferred diffusion and inferred size share one underlying record.\\[4pt]
SLS dilution series & Excess scattering, concentration and optical calibration. & Fitted mass and virial coefficient share intercept/slope and calibration errors.\\[4pt]
Electrophoretic record & Mobility, field protocol, concentration and preparation age. & Charge conversion uses an electrokinetic model; mobility is the direct observable.\\[4pt]
Evidence status & Numeric table, published plot, digitized estimate or assigned input. & Identify scans, independent measurements and preparation batches separately.\\
\bottomrule
\end{tabular}
\caption{Observation provenance accompanying the parameter ledger. Each datum has a source locator, units and uncertainty type. Derived quantities from one record are carried jointly.}
\label{tab:hpva-observation-ledger}
\end{table}

Preparation age $t_a$ labels a slowly changing sample; correlation lag
$t_\ell$ resolves motion within a record. A correlation calculation at
fixed $t_a$ assumes that the preparation changes negligibly during that
record, or retains its change explicitly. An increase of individual
shell size or mass calls for an additional species or internal coordinate.
Recruitment of unchanged whole VLPs into a cluster is a different event.
The selected final preparation provides a starting reference; its
stationarity over any proposed measurement window remains a test.

\subsubsection{Three comparison models on one observation basis}

\begin{definition}[Application model hierarchy]
\leavevmode\label{def:hpva-hierarchy}
$\mathcal M_0$ retains the selected size distribution, an effective
isotropic pair interaction, Brownian transport and the observation map.
$\mathcal M_1$ resolves whole-body orientation and a small set of
geometry-constrained surface interaction parameters.
$\mathcal M_2$ additionally retains one responsive surface or material
coordinate, with its reciprocal mechanical force and matched thermal
forcing. The label specifies the resolution being tested, not an order
of empirical preference.
\end{definition}

For $\mathcal M_0$, retain radii as species labels and specify
$U_0=U_{\rm domain}+\sum_{i<j}W_0(r_{ij};R_i,R_j)$.
The potential includes the chosen excluded volume, finite contact law
and screened/short-range terms. For $\mathcal M_1$, replace $W_0$ by a
body-dependent pair law $U_1(\vct r_{ij},\Omega_i,\Omega_j;R_i,R_j)$
using the site geometry of Sections~2--3. Surface amplitudes are a few
constrained parameters, not one independently fitted charge at every
site. The whole-body stochastic law retains translation and rotation,
with the equilibrium-preserving mobility and noise already derived.

An unbound dilute preparation is described by encounters and scattering
statistics. A mechanically stable cluster permits the local Hessian,
mode and response reduction of the preceding sections after its full
force and torque balance is established. A harmonic well is not added
merely to make a dilute sample a lattice. Similarly, a bulk cuvette
measurement constrains bulk inputs; walls and supports in a cylindrical
application require their own geometry and calibration.

\begin{proposition}[A static orientational ambiguity]
\leavevmode\label{prop:hpva-static-ambiguity}
For a dilute equilibrium pair with fixed size labels and normalized
uniform orientation measures, define
\begin{equation}
 e^{-\beta W_0(\vct r)}=
 \left\langle e^{-\beta U_1(\vct r,\Omega_1,\Omega_2)}\right\rangle_{\Omega_1,\Omega_2}.
 \label{eq:hpva-boltzmann-average}
\end{equation}
If the pair law is rotationally covariant, $W_0$ depends only on
$r=|\vct r|$. These two descriptions give the same equilibrium
orientation-integrated pair weight and second virial coefficient.
\end{proposition}
\begin{proof}
The definition equates the integrands of the orientation-integrated
pair partition and Mayer integral. Spatial integration preserves the
equality. Rotating the separation vector can be absorbed by changing
both normalized orientation integration variables.
\end{proof}
The effective potential is a logarithmic Boltzmann average, not an
arithmetic average of energies. Equality of these static pair weights
does not fix rotational relaxation, surface-sensitive scattering or a
driven force--torque record. In a larger assembly, one body's orientation
enters several contacts; integrating it out can produce effective
many-body interactions even when the resolved law is pairwise.
The proposition matches the dilute pair weight. Eliminating orientation dynamically can
also generate memory. When no orientationally sensitive observation
distinguishes the models, an inferred detailed patch map is unsupported.

The matched comparison for $\mathcal M_2$ can be made equally explicit.
About a stable reference let $\vct x$ be length-valued retained
displacements; registered angular coordinates may first be converted to
lengths with a specified lever scale. For a dimensionless column
$\vct a$, retain one length $h$ with
\begin{equation}
 E_2=\tfrac12\vct x^{\mathsf T}\mathsf K_0\vct x
       +\tfrac\kappa2(h-\vct a^{\mathsf T}\vct x)^2,
 \qquad \mathsf K_0>0,\quad\kappa>0.
 \label{eq:hpva-matched-retained-state}
\end{equation}
Use resistances $\mathsf C_0>0$ and $c_h>0$ on $\vct x$ and $h$,
respectively, with independent generalized force noises of two-sided
spectral densities $2k_BT\mathsf C_0$ and $2k_BT c_h$.
This is the passive branch construction specialized to an application
comparison, rather than an additional force appended to an already
unmatched relaxed model.

\begin{proposition}[A matched static and dynamic comparison]
\leavevmode\label{prop:hpva-matched-state}
For the equilibrium linear overdamped model defined above, put
$\tau_h=c_h/\kappa$. Eliminating $h$ with the $e^{-i\omega t}$
convention gives
\begin{equation}
 \begin{aligned}
 \mathsf D_2(\omega)
 &=\mathsf K_0-i\omega\left[\mathsf C_0+
       \frac{\kappa\tau_h}{1-i\omega\tau_h}\vct a\vct a^{\mathsf T}\right],\\
 \mathsf S_{f,2}(\omega)
 &=2k_BT\left[\mathsf C_0+
       \frac{\kappa\tau_h}{1+\omega^2\tau_h^2}\vct a\vct a^{\mathsf T}\right],\\
 \operatorname{Cov}(\vct x)&=k_BT\mathsf K_0^{-1}.
 \end{aligned}
 \label{eq:hpva-matched-response-noise}
\end{equation}
\end{proposition}
\begin{proof}
The internal equation is
$c_h\dot h=\kappa(\vct a^{\mathsf T}\vct x-h)+\xi_h$.
Fourier substitution into the reciprocal mechanical equation gives the
first line and filters the same $\xi_h$ into the second. The equilibrium
Gaussian integral over $h-\vct a^{\mathsf T}\vct x$ is independent of
$\vct x$, yielding the covariance.
\end{proof}
The matched memory-free comparison uses $\mathsf K_0,\mathsf C_0$ and
their thermal noise. Thus equal static covariance can coexist with
different phase and lag-dependent spectra. The amplitude and time
$\tau_h$ remain application inputs to constrain; the literature size
alone assigns neither. Matching equal-time quantities isolates the
dynamic question without requiring that the models agree dynamically.

\subsubsection{Polydispersity and acquisition as controls}

At a fixed preparation age, let $p_n(R)$ be a normalized number
distribution and $F(q,R)$ the particle's optical scattering amplitude.
For independent spherical particles in an unbounded Newtonian solvent,
the single-scattering, translational reference is
\begin{equation}
 \begin{aligned}
 w(q,R)&=\frac{p_n(R)|F(q,R)|^2}
                  {\int p_n(R')|F(q,R')|^2\,\dd R'},\\
 g_1(q,t_\ell)&=\int w(q,R)e^{-q^2D(R)t_\ell}\,\dd R,
 \qquad D(R)=\frac{k_BT}{6\pi\eta R}.
 \end{aligned}
 \label{eq:hpva-polydisperse-control}
\end{equation}
The normalized weights are intensity weights, not particle counts.
The relation between field and intensity correlations, including the
optical coherence factor and baseline, is supplied by the specified
homodyne observation law. Cumulant methods characterize a limited set
of decay-distribution moments~\cite{MailerCleggPusey2015}; this reference
supports a control calculation rather than a unique inversion of a
surface-state distribution.

For a direct assigned two-species check, take intensity fractions
$w,1-w$ and diffusion coefficients $D_1,D_2>0$. Then
\begin{equation}
 \begin{aligned}
 g_1&=w e^{-q^2D_1t_\ell}+(1-w)e^{-q^2D_2t_\ell},\\
 \left.\frac{\dd^2}{\dd t_\ell^2}\ln g_1\right|_{0}
     &=q^4w(1-w)(D_1-D_2)^2.
 \end{aligned}
 \label{eq:hpva-polydispersity-curvature}
\end{equation}
Positive logarithmic curvature can therefore occur with no material
state. The parameters in this illustration are assigned, not a fit to
the selected VLP distribution. At finite concentration replace the
independent-particle expression by the appropriate collective scattering
correlation, including hydrodynamic coupling and contrast. With an
anisotropic optical marker, rotation can enter the observation even
without centre displacement. Apply the exposure and record law to the
underlying signal before comparing measured correlations; derived DLS
sizes and the correlation used to compute them are one evidence family.

\subsubsection{Calibration, unused observations and coefficient conventions}

\begin{definition}[Calibration and reserved prediction]
\leavevmode\label{def:hpva-calibration-boundary}
Before parameter adjustment, assign records to calibration
$\mathcal D_{\rm cal}$ and a reserved comparison $\mathcal D_{\rm test}$.
For a fixed model and observation specification, the admissible parameter
set $\Theta_{\rm cal}$ uses calibration records and stated external
constraints. A prediction for the reserved observable is the image
\begin{equation}
 \mathcal P_{\rm test}
   =\{\mathcal O_{\rm test}(\theta):\theta\in\Theta_{\rm cal}\}.
 \label{eq:hpva-prediction-image}
\end{equation}
This set is an admissible range. It becomes a probabilistic prediction
only after parameter distributions, observation noise and their meaning
are specified. Shared calibration errors are retained across the split.
\end{definition}

\begin{table}[H]
\centering\small
\begin{tabular}{@{}p{3.2cm}p{6.1cm}p{6.15cm}@{}}
\toprule
Route & Calibration and constraints & Reserved target and status\\
\midrule
Interaction prediction & Size/contrast, transport, mobility, independently bounded attraction and screening. & SLS interaction coefficient; an empirical test only if unused in calibration.\\[4pt]
Dynamic forecast & The same preparation, with SLS used to constrain attraction if necessary. & Lag-dependent correlation, phase or surface reporter; prospective until independently measured.\\[4pt]
Controlled transfer & Calibrated input changes and an explicit rule for transferable parameters. & Another condition or preparation; test that rule without readjusting it to its own target.\\
\bottomrule
\end{tabular}
\caption{Two immediate routes and a later transfer test. Selecting a route fixes what may count as an unused observation. A successful fit and a successful reserved prediction are reported separately.}
\label{tab:hpva-calibration-routes}
\end{table}

The static target must use the same coefficient convention as its
observation law. Write $c$ for mass concentration, $M$ for molar mass
and $\rho=N_Ac/M$ for number density. In a monodisperse dilute reference,
the equilibrium Mayer integral and osmotic expansion give
\begin{equation}
 \begin{aligned}
 B_2&=\tfrac12\int\left[1-\left\langle
                   e^{-\beta U_1(\vct r,\Omega_1,\Omega_2)}
                        \right\rangle\right]\dd^3r,\\
 \frac{\Pi}{k_BT}&=\rho+B_2\rho^2+O(\rho^3),
 \qquad A_2=\frac{N_AB_2}{M^2}.
 \end{aligned}
 \label{eq:hpva-virial-convention}
\end{equation}
Here $B_2$ has units of volume and $A_2$ has units
$\mathrm{mol}\times\mathrm{volume}\times\mathrm{mass}^{-2}$.
The zero-angle dilute optical convention is
$K_{\rm opt}c/R(0)=1/M+2A_2c+O(c^2)$,
with the optical constant and Rayleigh ratio specified
\cite{Zimm1948AngleConcentration}. It is the coefficient $A_2$ that
has the mass-concentration units quoted as $B_{22}$ in
Table~\ref{tab:hpva-provenance}. For a polydisperse sample, SLS mass is
weight averaged and the apparent interaction coefficient includes
species and optical weighting; the single-$M$ conversion is a
declared monodisperse comparison, not an exact mixture reduction.
A finite-angle record additionally needs its form factor and
concentration-dependent scattering law rather than an automatic
identification with the zero-angle expression.

\begin{example}[A dimensional audit before assigning an interaction]
\leavevmode\label{ex:hpva-virial-audit}
Use the central tabulated $M_w$ as $M$ and $58\,\mathrm{nm}$ as a
nominal hard-sphere diameter $d$ only for this consistency comparison.
Converting the quoted positive coefficient gives
\begin{equation}
 \begin{aligned}
 B_2&\simeq3.20\times10^{-19}\,\mathrm{m^3},\\
 B_2^{\rm HS}&=\frac{2\pi d^3}{3}
       \simeq4.09\times10^{-22}\,\mathrm{m^3},\qquad
 \frac{B_2}{B_2^{\rm HS}}\simeq7.8\times10^2.
 \end{aligned}
 \label{eq:hpva-virial-scale-audit}
\end{equation}
\end{example}
This large comparison makes the units, optical calibration, mixture
reduction and concentration convention important to verify before
fitting a pair potential. It neither corrects the source's reported
value nor establishes its microscopic cause. A hydrodynamic diameter
is not itself an excluded-volume contact diameter. The conversion
illustrates why a reported positive interaction coefficient cannot
simply be inserted as a spring constant.

For a resolved size mixture with number fractions $p_i$, its
thermodynamic second virial coefficient is
$B_2^{\rm mix}=\sum_{ij}p_ip_jB_{2,ij}$ when the dilute species
fractions are fixed. Optical weighting and the conversion between mass
and number distributions must accompany a comparison with SLS. At high
electrolyte concentration, use the declared ionic/charge-regulation law
within its validity range; buffer molarity alone is not a fitted charge
map. Electrophoretic mobility constrains an electrokinetic observation
model and does not by itself identify a patch potential or a unique
surface charge.

The preferred route reserves the interaction coefficient and constrains
the remaining inputs independently. If attraction cannot be bounded
without that coefficient, the second route uses it for calibration
and predicts a future dynamic or surface-sensitive record. Residuals
from the fitted dilution series assess the observation fit; they do
not independently verify the assumed microscopic surface pattern.
Holding out another preparation tests parameter transfer only when
the matching rule includes its changed chemistry and structure.

\subsubsection{An uncertainty budget attached to the observation law}
\label{sec:hpvu-budget}

The preparation ledger now becomes an uncertainty budget. Each input
records its units, source, uncertainty interpretation and dependencies
before fitting or propagation. A measured standard deviation, uncertainty
in a calibration constant, an assumed physical range and the spread
between model predictions describe different objects. Keeping them
separate makes a prediction assessable and identifies which additional
information could narrow it.

\begin{table}[H]
\centering\small
\begin{tabular}{@{}p{3.2cm}p{6.1cm}p{6.15cm}@{}}
\toprule
Source & Application examples & Treatment and interpretation\\
\midrule
Measurement and replication & Record noise, finite correlation estimates, sample-to-sample variability. & Model the replication unit and temporal dependence; distinguish variation of a record from uncertainty in its mean.\\[4pt]
Digitization and calibration & Plot coordinates, pixel scale, concentration, optical gain, baseline, exposure. & Retain shared offsets and scale errors; record extraction uncertainty separately from instrument uncertainty.\\[4pt]
Physical inputs & Size distribution, screening, interaction strength, resistance, preparation. & Use independently justified distributions or explicit bounded scenarios; retain constraints and correlations.\\[4pt]
Model assumptions & Isotropic versus patch-resolved interaction; frozen versus responsive state. & Report model-conditional predictions. Any weighting across models requires its own stated basis.\\[4pt]
Numerical calculation & Time step, finite population, relaxation solver, sampling and integration. & Check refinement and sampling stability against a declared output tolerance. Numerical precision does not measure biological uncertainty.\\
\bottomrule
\end{tabular}
\caption{Uncertainty budget for the selected HPV application. These entries are not automatically independent additive variances. A range without a probability assignment remains a sensitivity scenario.}
\label{tab:hpvu-budget}
\end{table}

For the literature preparation, dilution points share stock concentration
and optical calibration; fitted size and its underlying DLS correlation
share a record. A scan, particle image and independently prepared batch
are different replication units. Bootstrap or resampling procedures must
preserve the appropriate batch and temporal blocks. Where only summaries
are published, retain their stated errors and explore unresolved
correlations explicitly rather than reconstructing fictitious replicates.
Polydispersity and acquisition controls in
Section~\ref{sec:hpva-application} remain part of the forward model.

\begin{definition}[Joint calibration law]
\leavevmode\label{def:hpvu-joint-law}
Let $m$ identify a comparison model, $\theta$ its physical parameters
and $\nu$ its nuisance calibration and preparation variables.
After a fixed conversion of the data to dimensionless units, a Gaussian
calibration law with residual $\vct r=\vct d-\vct\mu_m(\theta,\nu)$
and positive covariance $\mathsf\Sigma_m(\theta,\nu)$ has log likelihood
\begin{equation}
 \ell_m(\theta,\nu)
 =-\tfrac12\left[\vct r^{\mathsf T}\mathsf\Sigma_m^{-1}\vct r
                       +\ln\det\mathsf\Sigma_m\right]+\mathrm{constant}.
 \label{eq:hpvu-likelihood}
\end{equation}
The law specifies which records and uncertainty sources are included.
Independent calibration data contribute their own likelihood factors;
shared nuisance variables are retained across records.
\end{definition}

The determinant term matters when the covariance is inferred or depends
on the parameters. For a linearized calibration error $\delta\nu$
and a batch effect $\vct b$, one possible marginal covariance is
\begin{equation}
 \mathsf\Sigma
 =\mathsf\Sigma_{\rm record}+\mathsf\Sigma_{\rm dig}
  +\mathsf J_\nu\mathsf\Sigma_\nu\mathsf J_\nu^{\mathsf T}
  +\mathsf Z\mathsf\Sigma_b\mathsf Z^{\mathsf T}.
 \label{eq:hpvu-shared-covariance}
\end{equation}
This expression assumes the displayed contributions are independent;
otherwise their cross covariances are retained. Treating a calibration
variable explicitly and also adding its same uncertainty as a marginal
covariance would count it twice. Nonlinear calibration uses the joint
observation law instead of this first-order approximation. Model
discrepancy needs a declared physical/statistical construction; it is
not inferred merely by enlarging every error bar until a fit succeeds.

\subsubsection{Propagation to a prediction and numerical precision}
\label{sec:hpvu-propagation}

Probability distributions for input quantities can be propagated by
Monte Carlo evaluation of the measurement model, with reported output
intervals and numerical tolerances~\cite{JCGM1012008}. Here the input
specification is joint: positive resistances, normalized fractions,
conserved populations and shared calibration variables keep their
dependencies. A literature interval does not become a Gaussian standard
deviation without an additional justification.

For a Bayesian calibration within model $m$, state the prior
$\pi_m(\theta,\nu)$ and form
\begin{equation}
 \begin{aligned}
 p_m(\theta,\nu\mid\mathcal D_{\rm cal})
   &\propto L_m(\mathcal D_{\rm cal}\mid\theta,\nu)\pi_m(\theta,\nu),\\
 p_m(y_{\rm new}\mid\mathcal D_{\rm cal})
   &=\int p_m(y_{\rm new}\mid\theta,\nu,\mathcal D_{\rm cal})
       p_m(\theta,\nu\mid\mathcal D_{\rm cal})\,\dd\theta\,\dd\nu.
 \end{aligned}
 \label{eq:hpvu-predictive-law}
\end{equation}
The first line is an inferential choice, not a consequence of a
propagation algorithm. A metrological input distribution or a bounded
sensitivity set can instead be propagated with its own stated meaning.
For a new batch or preparation, the conditional record law includes its
new variability and transfer assumptions. For a continuation of the
same material preparation, it retains the calibrated conditional state.

\begin{proposition}[Mean-response uncertainty and future-record variability]
\leavevmode\label{prop:hpvu-total-variance}
For a scalar future record $Y$ with finite second moment, a fixed model
and a specified joint input law, put
$\mu_Y(\theta,\nu)=\mathbb E[Y\mid\theta,\nu,\mathcal D_{\rm cal}]$.
Then
\begin{equation}
 \boxed{\begin{aligned}
 \operatorname{Var}(Y\mid\mathcal D_{\rm cal})
 &=\operatorname{Var}(\mu_Y\mid\mathcal D_{\rm cal})\\
 &\quad+\mathbb E\!\left[\operatorname{Var}(Y\mid\theta,\nu,
              \mathcal D_{\rm cal})\mid\mathcal D_{\rm cal}\right].
 \end{aligned}}
 \label{eq:hpvu-total-variance}
\end{equation}
\end{proposition}
\begin{proof}
Subtract the conditional mean from $Y$ and expand the squared deviation
about its unconditional mean. The cross term has zero conditional
expectation. Taking expectations yields the identity.
\end{proof}

Figure~\ref{fig:cmd-phase-design} gives a graphical specialization of this boxed relation.

The first term describes uncertainty in the conditional mean record;
the second describes variability of a future record at fixed inputs.
An interval for a mean response therefore differs from a prediction
interval for one later record. Bayesian credible intervals,
likelihood-based confidence sets and bounded sensitivity envelopes
also retain their different interpretations. Pointwise intervals over
a frequency sweep do not automatically give simultaneous coverage of
the whole curve. Phase uses a common unwrapped reference or a circular
construction when the branch cut matters.

In a Monte Carlo calculation, draw the complete input column, compute
the conditional response, and sample future records only when that is
the target. Thermal variability already represented by a finite-record
error model is not added again as an independent bath contribution.
Report the chosen probability level, interval construction and numerical
stability of its endpoints. Repeat with independent random seeds and
increase the number of draws until the stated endpoint tolerance is
resolved. Time-step, solver and population refinements assess other
numerical errors separately. More draws narrow sampling error in a
calculated interval; they do not remove uncertainty in its physical
inputs or repair an incorrect observation model.

\subsubsection{Profile likelihood for parameters and target predictions}
\label{sec:hpvu-profile}

Definitions~\ref{def:s4-response-equivalence}
and~\ref{def:s6-identifiable-prediction} distinguish parameter recovery
from prediction recovery. For the present application, examine these
questions over the full admissible family rather than only near one
optimizer output. Profile likelihood does this by readjusting the other
parameters while holding one parameter or target fixed~\cite{Raue2009}.

\begin{definition}[Profile likelihood for a declared target]
\leavevmode\label{def:hpvu-profile}
For $\psi_m(\theta,\nu)$, define
\begin{equation}
 \ell_{p,m}(a)=\sup_{\psi_m(\theta,\nu)=a}\ell_m(\theta,\nu),
 \qquad \Lambda_m(a)=2[\widehat\ell_m-\ell_{p,m}(a)],
 \label{eq:hpvu-profile}
\end{equation}
where $\widehat\ell_m$ is a finite unrestricted supremum in the same family.
A profile set is $\{a:\Lambda_m(a)\le c_\alpha\}$ with a stated threshold
and its calibration. A coordinate choice $\psi_m=\theta_j$ profiles
one parameter; a readout choice profiles the prediction itself.
\end{definition}

Refit the nuisance variables at each profile point; holding them at
their best-fit values gives a different, conditional slice. Use multiple
starts and check admissible boundaries. A broad or open profile can
reveal weak constraints even when a local Hessian suggests a small
error. A numerically flat profile may also reflect failed optimization;
an exact equivalence argument supplies stronger evidence of structural
nonidentifiability. Regular interior large-sample cases permit a
one-degree-of-freedom chi-square likelihood-ratio threshold.
Boundary cases, unidentified branch times at zero coupling and comparisons
of different model families require an appropriate separate calibration.
The threshold is not applied automatically to every model difference.

A target can be well constrained while individual microscopic parameters
remain broad. Conversely, an apparently precise fitted combination may
leave another readout uncertain. Plot the response trajectories along
the profile to identify which additional measurement could separate
those possibilities~\cite{Raue2009}. A fitted HPV dataset has not yet
been supplied here; the following calculation demonstrates the procedure
on assigned records and passive models.

\subsubsection{An explicit static ambiguity with uncertain dynamic predictions}
\label{sec:hpvu-example}

\begin{example}[A static calibration that leaves a relaxation branch unresolved]
\leavevmode\label{ex:hpvu-static-dynamic}
Use the scalar specialization of
\eqref{eq:hpva-matched-retained-state} with $a=1$ in fixed units
$k_\star,c_\star,\ell_\star$, time unit
$t_\star=c_\star/k_\star$ and thermal scale
$k_BT/(k_\star\ell_\star^2)=1$. All quantities in this example are
dimensionless unless an angle is given in degrees. The retained-state
and memory-free references have equilibrium marginal
$x\sim\mathcal N(0,1/k)$ for every positive $c,\kappa,\tau$.
Assign an independently estimated static-variance summary
$v_{\rm obs}=1$ with known Gaussian standard error $s_v=0.02$ and law
\begin{equation}
 v_{\rm obs}\mid k\sim\mathcal N(1/k,s_v^2),\qquad
 -2\ell(k,c,\kappa,\tau)
   =\frac{(v_{\rm obs}-1/k)^2}{s_v^2}+\mathrm{constant}.
 \label{eq:hpvu-static-likelihood}
\end{equation}
This is an assigned summary-error model, not the exact sampling law
of a finite Gaussian sample variance.
\end{example}

At $k=1$ the likelihood is maximized for every admissible branch time
and strength. Their static profiles are exactly flat because the
observation law contains neither. Temporal correlations would supply
additional data, so this conclusion concerns the specified static
summary, not every possible mechanical record. With the regular
scalar Gaussian likelihood-ratio cutoff $3.84146$, the profile set
for $k$ is approximately $[0.9623,1.0408]$. A uniform prior on
$k\in[0.8,1.2]$ instead gives a posterior proportional to the displayed
likelihood, with central 95\% interval approximately
$[0.9630,1.0417]$. These two intervals answer their respective
frequentist and Bayesian questions.

For the reserved target use a calibrated harmonic displacement phase
at $\omega t_\star=1$. Write
\begin{equation}
 H(\omega)=\left[k-i\omega\left(c+
                   \frac{\kappa\tau}{1-i\omega\tau}\right)\right]^{-1},
 \qquad \Phi=\arg H+\delta_\phi,\quad
 \Phi_{\rm new}=\Phi+\epsilon_\phi.
 \label{eq:hpvu-phase-target}
\end{equation}
The memory-free comparison has $\kappa=0$. In addition to the
calibrated $k$ posterior, assign independent input distributions
$c\sim\mathcal U[0.95,1.05]$, $\kappa\sim\mathcal U[1.9,2.1]$
for the retained-state cases and a shared instrument phase offset
$\delta_\phi\sim\mathcal U[-1^\circ,1^\circ]$.
Scenario A assigns $\tau\sim\mathcal U[0.2375,0.2625]$;
scenario B assigns $\tau\sim\mathcal U[1.9,2.1]$.
Their time ranges are explicit scenario priors, unchanged by the static
likelihood. An independently assigned future phase error
$\epsilon_\phi\sim\mathcal N(0,(2^\circ)^2)$ represents finite-record
measurement variability. All these input widths and the force/readout
port are synthetic specifications, not HPV calibrations.

\begin{table}[H]
\centering\small
\begin{tabular}{@{}p{2.4cm}p{3.8cm}p{4.4cm}p{4.85cm}@{}}
\toprule
Description & Nominal phase & Central 95\% interval for $\Phi$ & Central 95\% interval for $\Phi_{\rm new}$\\
\midrule
Memory-free & $45.0^\circ$ & $[42.7,47.2]^\circ$ & $[40.4,49.5]^\circ$\\[4pt]
Scenario A & $52.8^\circ$ & $[50.9,54.6]^\circ$ & $[48.4,57.0]^\circ$\\[4pt]
Scenario B & $34.7^\circ$ & $[33.0,36.4]^\circ$ & $[30.4,39.0]^\circ$\\
\bottomrule
\end{tabular}
\caption{Conditional uncertainty propagation for an assigned static calibration and a reserved phase readout. Nominal values use $k=c=1$, $\kappa=2$ and $\tau=0.25$ or $2$ for A or B. The first interval propagates input/calibration uncertainty into the conditional mean record; the second also includes the specified future-record error.}
\label{tab:hpvu-phase-predictions}
\end{table}

The static observation distributions agree across all three descriptions.
Their nominal dynamic phases differ, and the propagated future-record
intervals for A and B remain separated under the stated input laws.
This is a constructive instance of observational equivalence for one
task with different predictions for another. Nonoverlap of these
conditional central intervals is not a universally calibrated model
selection test, and does not assign either scenario to the actual VLP
preparation. The memory-free and A intervals also overlap, illustrating
why a nominal phase difference alone does not establish discrimination.

The ancillary program
\texttt{VLP\_uncertainty\_identifiability\_benchmark.py} evaluates the
calibration posterior on a refined one-dimensional grid and propagates
the stated inputs with reproducible random seeds. Four independent
batches, each with $32768$ draws, gave the intervals in
Table~\ref{tab:hpvu-phase-predictions}; the largest spread of their
estimated endpoints was about $0.086^\circ$, below the declared
$0.1^\circ$ numerical diagnostic tolerance. This batch check assesses
sampling stability, not an experimental coverage guarantee. The script
reports its grid-refinement check and keeps the earlier transport
simulation separate.

For the HPV application, replace the assigned calibration and input
laws only when the corresponding evidence is available. Until then,
report conditional scenarios, their prior sensitivity and the record
needed to distinguish them. The computation makes the uncertainty and
identifiability question explicit without claiming that the published
HPV measurements have selected a memory law.

\subsubsection{Designing the complementary measurement from surviving predictions}
\label{sec:cmd-measurement-design}

Once calibration has left several admissible descriptions, the next
measurement is chosen from their future records. A useful design names
the controlled perturbation, observed channel, timing, acquisition law
and uncertainty budget. Its purpose may be to distinguish models,
constrain a particular prediction or test whether another retained
variable is needed. These objectives can select different experiments.

\begin{definition}[Design-conditioned predictive record]
\label{def:cmd-predictive-design}
Let $d$ specify the source and detector ports, frequency or lag,
exposure, loading protocol and replication. For model $m$, propagate
the surviving calibration law to
\begin{equation}
 p_m(\vct y\mid d,\mathcal D_{\rm cal})
   =\int p_m(\vct y\mid d,\theta,\nu,\mathcal D_{\rm cal})
       p_m(\theta,\nu\mid\mathcal D_{\rm cal})\,\dd\theta\,\dd\nu.
 \label{eq:cmd-design-predictive}
\end{equation}
Shared calibration, preparation, record correlation and detector noise
are carried jointly. A bounded surviving set instead yields a
set-valued prediction with its own stated interpretation.
\end{definition}

For declared model weights $w_m>0$, $\sum_mw_m=1$, put
$\bar p_d=\sum_mw_mp_m$. An established Bayesian design utility is
\begin{equation}
 \boxed{\mathcal I(d)=\sum_mw_m\int p_m(\vct y\mid d)
                   \ln\frac{p_m(\vct y\mid d)}{\bar p_d(\vct y)}
                         \,\dd\vct y.}
 \label{eq:cmd-design-information}
\end{equation}

Figure~\ref{fig:cmd-phase-design} gives a graphical specialization of this boxed relation.

Conditioning on $\mathcal D_{\rm cal}$ is suppressed in this equation.
This weighted Jensen--Shannon divergence is the expected information
gain about the model label: Bayes' rule transforms it into the expected
log posterior-to-prior model-weight ratio. It lies between zero and
$-\sum_mw_m\ln w_m$. Design based on joint predictive distributions
has precedents in biochemical model discrimination~\cite{Vanlier2014}.
Weights and within-model input laws remain part of the design
specification. A high utility ranks a proposed experiment; it does not
establish that a model is true or provide a universal rejection threshold.

\begin{example}[Identical single-channel records and an informative joint record]
\label{ex:cmd-joint-record-information}
Two assigned reciprocal Brownian models make the value of a joint
measurement explicit. For length-valued coordinates $(x_1,x_2)$ choose
$\lambda>0$, $\ell_\star>0$ and $\rho=\sin(2\pi/5)$. Set
\begin{equation}
 \begin{aligned}
 \mathsf C_\pm&=\ell_\star^2
       \begin{pmatrix}1&\pm\rho\\\pm\rho&1\end{pmatrix},\qquad
 \mathsf K_\pm=k_BT\mathsf C_\pm^{-1},\\
 \mathsf L_\pm&=\frac{\lambda\mathsf C_\pm}{k_BT},\qquad
 \dd\vct x=-\lambda\vct x\,\dd t
                    +\sqrt{2\lambda\mathsf C_\pm}\,\dd\vct W_t.
 \end{aligned}
 \label{eq:cmd-joint-brownian-models}
\end{equation}
Here $\mathsf L_\pm$ is the reciprocal mobility.
Both storage and mobility are positive definite. The drift is
$-\mathsf L_\pm\mathsf K_\pm=-\lambda\Id$, and the diffusion obeys
the equilibrium fluctuation--dissipation relation. Supports and the
chosen interaction supply the stated force-balanced harmonic Hessian.
These are synthetic admissible models, not fitted colloidal mobilities.

In stationary equilibrium each single coordinate has the identical
complete Gaussian record law, with covariance
$\ell_\star^2e^{-\lambda|t-s|}$. Joint cross covariances have opposite
signs. For the declared binary observation $Y_i=\mathbf1_{\{x_i>0\}}$,
the single-time joint probability matrices are
\begin{equation}
 \mathsf J_+=\begin{pmatrix}0.45&0.05\\0.05&0.45\end{pmatrix},\qquad
 \mathsf J_-=\begin{pmatrix}0.05&0.45\\0.45&0.05\end{pmatrix}.
 \label{eq:cmd-joint-binary-laws}
\end{equation}
Rows and columns index $0,1$. To obtain these entries, express a
standard bivariate Gaussian as
$(Z_1,\rho Z_1+\sqrt{1-\rho^2}Z_2)$ with isotropic independent
normal $(Z_1,Z_2)$. The intersecting positive half-planes occupy angle
$\pi/2+\arcsin\rho$, giving probability
$1/4+\arcsin\rho/(2\pi)$. Reflection gives the other cells.

With equal model weights the mixture is uniform on the four outcomes.
The information utility is therefore
\begin{equation}
 \mathcal I(Y_1)=\mathcal I(Y_2)=0,\qquad
 \mathcal I(Y_1,Y_2)
     =0.9\ln1.8+0.1\ln0.2\simeq0.368064\ \mathrm{nats}.
 \label{eq:cmd-joint-information-value}
\end{equation}
The upper bound is $\ln2$. The joint record distinguishes surviving
models even though either individual trajectory has the same law.
Actual binning, exposure, localization noise and sampling dependence
are applied before evaluating an experimental predictive utility.
This example specifies distributions at a registered observation time;
repeated closely spaced pairs are not independent replications.
\end{example}

The joint record matters. Two models can predict the same one-channel
means and variances yet different cross-channel dependence. Independent
measurement noise, shared gain error and common preparation variability
then affect discrimination differently. \begin{proposition}[Mean contrast after nuisance projection]
\label{prop:cmd-nuisance-projection}
For a local Gaussian comparison
with common positive noise covariance $\mathsf\Sigma$ and unknown
unpenalized nuisance shifts $\mathsf A\vct\nu$, define
$\vct\delta=\mathsf\Sigma^{-1/2}(\vct\mu_m-\vct\mu_n)$ and
$\mathsf B=\mathsf\Sigma^{-1/2}\mathsf A$. Profiling those shifts gives
\begin{equation}
 \boxed{\min_{\vct v}\|\vct\delta-\mathsf B\vct v\|^2
  =\|\mathsf P_\perp\vct\delta\|^2,\qquad
  \mathsf P_\perp=\Id-\mathsf B\mathsf B^+.}
 \label{eq:cmd-nuisance-projection}
\end{equation}
The superscript $+$ denotes the Moore--Penrose inverse.
\end{proposition}
\begin{proof}
Split $\vct\delta$ into its orthogonal projections onto
$\operatorname{ran}\mathsf B$ and its complement. The fitted shift can
remove the first component; the second is unchanged and has the
minimum displayed squared norm.
\end{proof} A model contrast inside the
nuisance column space is lost; a calibrated additional channel can
retain a contrast outside it. Proper nuisance priors or independently
bounded shifts use their own integrated or constrained calculation.
This local score complements the full predictive utility above.

\paragraph{A reproducible frequency-and-reference comparison.}
Continue the assigned passive models of Section~\ref{sec:hpvu-example}.
Their static-variance likelihood leaves the material branch unresolved.
The program \texttt{VLP\_complementary\_measurement\_design.py}
propagates the same calibration and scenario laws into a frequency
sweep. It ranks the sweep by the smallest pairwise separation of the
mean phases divided by their combined future-record standard deviation.
This is an explicitly moment-based screening score, rather than
\eqref{eq:cmd-design-information}. Frequency sampling and Monte Carlo
stability are checked before quoting a candidate. The actual force
amplitude and observation band must remain compatible with the
calibrated linear preparation.

The same calculation compares a single phase channel with a paired
signal/reference phase. For an unbounded common phase offset, a single
channel supplies no profiled offset-independent mean contrast. The
difference of a simultaneous signal and a calibrated reference cancels
that offset, with the two record noises and their covariance retained.
More observations of the same offset-confounded channel do not create
the missing direction. A reference whose own material response is
uncertain must include that uncertainty; cancellation alone does not
make it an absolute mechanical standard.
For this assigned calculation the refined grid selects a candidate
near $\omega t_\star=5.0$, with a smallest pairwise screening score
of approximately $2.99$. Four independent batches gave a largest
spread of their peak scores of about $0.0024$; doubling the log-frequency
grid changed the selected frequency by about $1.3\%$. These are
numerical design diagnostics, not measured discrimination or biological
uncertainty estimates.
Figure~\ref{fig:cmd-phase-design} visualizes the surviving prediction
bands and the candidate frequency score. The distinction between
uncertainty of a conditional mean and variability of a future record
uses the boxed identity \eqref{eq:hpvu-total-variance}.

\begin{figure}[H]
\centering\begingroup
\setlength{\unitlength}{1pt}
\scalebox{0.84}{%
\input{vlt-figures/figure-308.tex}%
}\endgroup
\caption{An assigned complementary-measurement design using the calibration and scenario laws of Section~\ref{sec:hpvu-example}. Panel (a) shows mean phases for M0, A and B. Inner ribbons are plus/minus one standard deviation of the uncertain conditional mean; outer ribbons include the independent $2^\circ$ future-record error. This is the variance decomposition \eqref{eq:hpvu-total-variance}, not a claim of simultaneous confidence coverage. Panel (b) ranks frequencies by the minimum pairwise mean separation divided by the combined future-record standard deviation. Refinement locates a candidate near $\omega t_\star=5$ with score about $3$; this screening score is distinct from the predictive-information utility \eqref{eq:cmd-design-information}. The lower-left construction illustrates the nuisance projection \eqref{eq:cmd-nuisance-projection}: a calibrated paired reference retains an offset-independent direction while adding its own noise. The lower-right matrices are the joint binary laws of Example~\ref{ex:cmd-joint-record-information}: identical marginals give zero single-channel information, while the pair gives $0.368$ nats. All frequencies, input distributions and errors are assigned; an actual viral preparation requires a calibrated time scale and usable instrument band.}
\label{fig:cmd-phase-design}
\end{figure}

\paragraph{A structural companion to a motion record.}
For the transport application, begin with three separate candidate
mechanisms: whole-particle clustering with intact shells; reversible
shell deformation; and persistent shell damage or disassembly. An
interface-mediated mechanism supplies an additional competing load
path when the protocol exposes particles to surfaces. A centre trace
and a bulk size measurement alone can mix changes of aggregation,
shape and scattering contrast. Complementary records should target
the differing predictions: load-dependent indentation or registered
shape for shell response, particle-number/cluster statistics for
aggregation, and recovery plus integrity-sensitive observations after
unloading. The shell construction in Section~\ref{sec:cmd-shell-mechanics}
specifies how the geometric reference reaches those observables.

Use matched aliquots from one preparation for sham transport,
specified loading, and defined recovery times. Record temperature,
interfaces and delivered load; randomize or block acquisition order.
For destructive assays, the aliquots are matched samples rather than
the same tracked object. Calibrate the structural-class observation
matrix, including misclassification, against independent controls.
Then freeze the calibration and predict the reserved joint record.
Repeat-time measurements require their actual within-particle or
within-batch covariance. A new salt, pH or ligand condition carries
its changed physical inputs instead of silently sharing every parameter.

For class probabilities of registered parent particles, a calibrated
single-time classifier has nonnegative columns summing to one and
\begin{equation}
 \vct p_{\rm obs}(t)=\mathsf C_{\rm obs}\vct p(t).
 \label{eq:cmd-structural-class-readout}
\end{equation}
If $\mathsf C_{\rm obs}(\vct p_m-\vct p_n)=0$, those two populations
give identical single-time categorical observations under this law.
Additional channels are useful when their stacked map retains the
population difference. Joint-time records can contain information
absent from a single endpoint and use their full transition model.
If a shell separates into independently moving capsomers, the resolved
objects and conserved protein inventory are updated as in Section~2;
a parent-level integrity label alone does not simulate fragment motion.

For HPV, ligand-dependent AFM and structural observations provide a
precedent for linking contact changes to shell mechanics
\cite{Feng2024}; the preparation differs from the selected Vega case.
For a CCMV-inspired alternative, empty/full shell and mutation
comparisons motivate distinct mechanical inputs~\cite{Michel2006}.
Neither experiment identifies a shipping-damage mechanism. The design
uses each as a source of candidate coordinates and measurable channels.
Its next deliverable is a specified protocol with predicted separation,
uncertainty and a declared decision rule. Simulating that complete
analysis under each surviving model estimates its error rates before
an experimental claim is made. If the joint predictive laws still
overlap, the proposed record is reported as insufficient for that task.
Figure~\ref{fig:cmd-collapse-readouts} connects the competing structural
mechanisms to their complementary observations.

\begin{figure}[H]
\centering\begingroup
\setlength{\unitlength}{1pt}
\scalebox{0.84}{%
\input{vlt-figures/figure-309.tex}%
}\endgroup
\caption{Three candidate structural interpretations require complementary records. Each panel retains twelve labelled whole particles, with the same oblique viewing direction and identity colours; vertical and lateral links remain visible. Panel (a) assigns a smaller horizontal spacing, (b) flattens one shell glyph, and (c) flattens three. These are geometric mechanism illustrations, not simulated yields or fitted transport trajectories. Cluster statistics probe whole-particle association; load-dependent shape and recovery probe reversible shell deformation; a calibrated structural/integrity assay and later observation probe persistent outcomes. Misclassification enters $\vct p_{\rm obs}=\mathsf C_{\rm obs}\vct p$ with normalized nonnegative columns. Sham and loaded aliquots share a defined preparation, while destructive measurements use matched samples. The diagrams locate different target records without prescribing their biological outcome.}
\label{fig:cmd-collapse-readouts}
\end{figure}

The immediate output is therefore a preparation ledger, a specified
$\mathcal M_0$ reference, conditional $\mathcal M_1/\mathcal M_2$
comparisons, an explicit reserved observable and an uncertainty budget.
Profiles and propagated intervals quantify which predictions remain
constrained. A broad admissible range is an inference result to report. An additional state is useful
when it changes a relevant prediction and a complementary observation
can resolve that change. The existing synthetic benchmark demonstrates
this logic; the present construction identifies the evidence and
calibration needed to apply it to a selected HPV preparation.

\subsection{An STMV crystal-response benchmark with a reserved linewidth}
\label{sec:stmv-response-benchmark}

An experimentally anchored comparison can now be made for a second
species. The calculation uses the documented packing of
Section~\ref{sec:stmv-geometric-hierarchy}, the explicit passive network
of Section~\ref{sec:stmv-crystal-network}, and published Brillouin line
summaries. It first fixes a response scale, then predicts a spectral
feature excluded from that calibration. The data constrain a response
projection; assigned contact and uncertainty scenarios show what remains
unidentified.

\subsubsection{Evidence, preparation and the calibration boundary}
\label{sec:stmv-evidence-boundary}

Stephanidis et al.~\cite{Stephanidis2007} studied hydrated STMV
microcrystals with orthorhombic cells. Their polarized backscattering
spectrum separates a crystal line from a surrounding-buffer line.
Table~\ref{tab:stmv-evidence} records the numerical anchors used here.
The reported line deviations describe variation across different
crystals; replicate counts and the joint fit covariance are not
available to this calculation. The room-temperature structural packing
is a separate reference, not the same measured crystal. Dehydrated
samples and the flight-growth observations are outside this calibration.

The comparison uses unit-peak line summaries. An absolute intensity
prediction requires temperature and optical-contrast calibration.
The retained mother liquor and exact acoustic measurement temperature
must be registered for a preparation-specific material law; the
article's separate dilute-solution preparation is not silently assigned
to the wet crystal.

\begin{table}[H]
\centering\small
\caption{Published acoustic anchors and their roles in the STMV
comparison~\cite{Stephanidis2007}. Frequency widths are full widths
at half maximum of the reported Lorentzian lines. The final row is
reserved until prediction is complete.}
\label{tab:stmv-evidence}
\begin{tabular}{@{}lll@{}}
\toprule
Quantity & Reported value & Role \\
\midrule
Vacuum wavelength $\lambda_0$ & $532$ nm & Optical input \\
Crystal refractive index $n_c$ & $1.55\pm0.03$ & Optical input \\
Crystal peak $f_c$ & $11.2\pm0.2$ GHz & Elastic-scale calibration \\
Buffer peak $f_b$ & $8.00\pm0.05$ GHz & Bath-summary calibration \\
Buffer width $w_b$ & $1.33\pm0.07$ GHz & Bath-summary calibration \\
Crystal width $w_c$ & $2.69\pm0.06$ GHz & Reserved comparison \\
\bottomrule
\end{tabular}
\end{table}

The ancillary input file distinguishes these reported quantities from
assigned ones. The buffer index is varied uniformly on $[1.33,1.35]$;
this is a bounded optical scenario, not a measurement of that buffer.
The contact ratios $\vartheta=1/4,1,4$, the illustrative calibration
direction $[100]$, and independent one-percent cell-edge sensitivity
ranges are also assigned. The experiment did not register a precise
crystallographic scattering direction. Directional predictions below
therefore describe a prospective registered experiment, not a recovered
orientation from the published record.

\subsubsection{The optical channel and a passive spectral model}
\label{sec:stmv-optical-spectrum}

For internal scattering angle $\Theta$ in an effectively isotropic
optical reference, the transferred wavevector magnitude is
\begin{equation}
 q=\frac{4\pi n}{\lambda_0}\sin(\Theta/2).
 \label{eq:stmv-optical-wavevector}
\end{equation}
Backscattering gives $q_c\simeq0.03661$~nm$^{-1}$ at the nominal
crystal index. A density-sensitive channel has linearized contrast
proportional to
$-i\sum_i b_i\exp(-i\vct q\cdot\vct r_i^0)
\vct q\cdot\vct u_i$, with declared optical weights $b_i$.
Internal deformation or polarizability requires its additional
contrast derivatives. The source conjugate to that channel and the
same detector define the susceptibility. For equal acoustic mass,
the modal form is
\begin{equation}
 \chi_Y(\omega,\vct q)=
 \sum_s\frac{W_s(\vct q)}
 {m_{\rm ac}[\Omega_s^2(\vct q)-\omega^2-i\gamma_s\omega]},
 \qquad W_s\geq0.
 \label{eq:stmv-observed-susceptibility}
\end{equation}
Its classical equilibrium displacement spectrum obeys
$S_Y=2k_BT\,\operatorname{Im}\chi_Y/\omega$ in the two-sided
angular-frequency convention used above. Convolution with the
instrument response and addition of solvent, container and detector
terms give the observed optical record. An unassigned intensity scale
does not determine an absolute particle fluctuation amplitude.

The numerical benchmark matches published peak and width summaries,
rather than fitting unavailable raw spectra. For one damped oscillator
put $f_0=\Omega_0/(2\pi)$ and $g=\gamma/(2\pi)$ in the same frequency
units as $f$. Its displacement line is proportional to
\begin{equation}
 P(f)=\big[(f_0^2-f^2)^2+g^2f^2\big]^{-1},\qquad
 f_p^2=f_0^2-g^2/2.
 \label{eq:stmv-dho-peak}
\end{equation}
For an isolated positive-frequency peak whose lower half-height
crossing remains positive, its exact FWHM is
\begin{equation}
 \begin{split}
 \mathcal W(f_p,g)&=
 \sqrt{f_p^2+g\sqrt{f_p^2+g^2/4}}
 -\sqrt{f_p^2-g\sqrt{f_p^2+g^2/4}},\\
 \mathcal G(f_p,w)&=
 \sqrt{-2f_p^2+\sqrt{4f_p^4+4f_p^2w^2-w^4}}.
 \end{split}
 \label{eq:stmv-line-summary-maps}
\end{equation}
Here $g=\mathcal G(f_p,w)$ inverts the first relation.
To derive it, complete the line denominator as
$(f^2-f_p^2)^2+g^2(f_p^2+g^2/4)$ and double its minimum.
This conversion is a declared oscillator-summary surrogate for the
published Lorentzian regression. The weak-loss approximation
$\mathcal W\simeq g$ is checked separately; agreement of the two
summary conventions does not establish the full line shape.

\subsubsection{Calibration and a prediction before the crystal width is used}
\label{sec:stmv-reserved-prediction}

The simplest bath-transfer hypothesis sets the effective longitudinal
attenuation coefficient in \eqref{eq:stmv-acoustic-dynamics} equal
in buffer and crystal. This is an explicit null reference for a
hydrated assembly, not an identified material law. With
$\gamma=\nu_Lq^2$, equal wavelength and backscattering give
\begin{equation}
 \boxed{g_b=\mathcal G(f_b,w_b),\qquad
 g_c=g_b(n_c/n_b)^2,\qquad
 w_{c,\rm pred}=\mathcal W(f_c,g_c).}
 \label{eq:stmv-reserved-linewidth}
\end{equation}
The buffer line and optical inputs determine the transferred loss.
The crystal peak then
sets $f_{0c}^2=f_c^2+g_c^2/2$. For the assigned $[100]$ calibration,
\begin{equation}
 \frac{\kappa_D}{m_{\rm ac}}
 =\frac{(2\pi)^2(f_c^2+g_c^2/2)}
 {h_L(q_c\vct e_x;\vartheta)}.
 \label{eq:stmv-calibrated-scale}
\end{equation}
Frequencies in this dimensional equation are in hertz. The calibration
identifies this stiffness-to-mass ratio for each specified network;
it does not identify $\vartheta$, molecular contact energies or an
individual shell modulus. The familiar speed $f_c\lambda_0/(2n_c)$
uses the same optical and peak inputs, so it is a derived check rather
than an independent validation.

At nominal inputs the program gives $w_{c,\rm pred}\simeq1.778$~GHz.
The pole-width approximation gives $1.780$~GHz. Their small difference
here bounds that particular summary-conversion effect; uncertainty in
the material and observation model is a separate question. The
reserved value $2.69$~GHz is accessed only after this calculation.
It is broader than this reference predicts. Figure~\ref{fig:stmv-spectral-test}
shows the calibrated peak, propagated prediction and unused comparison.

\subsubsection{Uncertainty propagation and what the mismatch establishes}
\label{sec:stmv-uncertainty-assessment}

The reproducible program is
\texttt{VLP\_STMV\_crystal\_benchmark.py}, with its adjacent JSON input
ledger. It propagates Gaussian scenarios for the published marginal
deviations and the bounded buffer index. These laws are explicitly
chosen propagation scenarios. They are not fitted parameter posteriors,
standard errors of the mean, or calibrated experimental confidence
intervals. Correlations between crystal peak and index, and between
buffer peak and width, are each varied over $-0.8,0,0.8$ because their
joint covariance is unavailable.

This is a retrospective, computationally enforced feature reservation
within a published study. Independent replication and a preregistered
test remain separate experimental steps.

For $65536$ draws the central independent scenario gives a central
$95\%$ propagation interval of approximately $[1.55,2.02]$~GHz.
The envelope over the nine stated correlation scenarios has the same
rounded limits. Independent batches assess sampling error; they do
not supply missing experimental replication. The interval propagates
the stated summary and optical laws under the transfer hypothesis.
It does not cover arbitrary differences between buffer and crystal
attenuation, or all unresolved orientation and line-shape effects.
A future linewidth estimator also needs its own sampling-error model.
The displayed band carries input propagation, without adding the
reserved width's reported deviation as an advance calibration input.

As a broader optical sensitivity calculation, fix the buffer index
anywhere in $[1.30,1.45]$ and envelope the central $95\%$ input-propagation
intervals over the same correlation scenarios. Monotonicity in $n_b$
places the extremal limits at its endpoints, giving approximately
$[1.33,2.15]$~GHz. This assigned bounded comparison also remains below
the reserved mean; its index range is not an empirical confidence set.

The reported crystal width remains outside that propagated range.
This is a conditional failure of the specified attenuation-transfer
reference. The next model must allow assembly-specific loss or
additional broadening and constrain it with further records. A single
line cannot distinguish material relaxation, unresolved branches,
orientation variation and other line-shape contributions. Replacing
$\nu_L$ after inspecting the reserved width turns that feature into
calibration and requires another prediction. No formal rejection
probability is assigned without the joint sampling law. Reserving a
feature within one study is also different from an independent
experimental replication.

\subsubsection{A second, prospective test of the contact network}
\label{sec:stmv-directional-test}

The three positive contact-ratio scenarios all reproduce the same
nominal calibration peak. At equal transferred wavevector magnitude,
their calculated dominant longitudinal peaks along $[111]$ differ:
\begin{table}[H]
\centering\small
\caption{Prospective directional predictions of the assigned network.
All rows calibrate $[100]$ to $11.2$~GHz. Intervals propagate the
independent input scenario and envelope the centre and corners of
independent one-percent cell-edge ranges; these ranges are sensitivity
choices, not measured lattice uncertainties.}
\label{tab:stmv-directional-predictions}
\begin{tabular}{@{}llll@{}}
\toprule
$\vartheta$ & $[111]$ peak (GHz) & Interval (GHz) & Longitudinal weight \\
\midrule
$1/4$ & $15.59$ & $[14.71,16.51]$ & $0.986$ \\
$1$   & $11.46$ & $[10.85,12.09]$ & $0.975$ \\
$4$   & $8.73$  & $[8.26,9.21]$ & $0.912$ \\
\bottomrule
\end{tabular}
\end{table}

These separated conditional predictions illustrate what a registered
direction can add to a single unregistered line. They are not measured
directional frequencies. Such a test requires the crystal orientation,
direction-dependent optical calibration, polarization and full branch
visibility. The ancillary calculation also supplies $[010]$ and $[001]$
predictions and keeps the weaker branches in the eigenproblem.
The tabulated centres belong to the individual branch with the largest
longitudinal density weight. An unresolved measured multiplet must
use the full weighted spectrum in \eqref{eq:stmv-observed-susceptibility}.
The one-point fit leaves a family of stiffness descriptions; it does
not establish that any member is the actual microscopic STMV network.

\begin{figure}[H]
\centering\begingroup
\setlength{\unitlength}{1pt}
\scalebox{0.84}{%
\input{vlt-figures/figure-317.tex}%
}\endgroup
\caption{Calculated checks of an STMV-inspired acoustic reference. (a) Oscillator-summary curves share the calibrated peak. The green band varies the predicted linewidth over its assigned propagation envelope; the dashed red curve uses the reserved reported width only for comparison. Curves have unit peak and are not digitized spectra. (b) The predicted interval from \eqref{eq:stmv-reserved-linewidth} remains below the unused crystal width. The red range is twice the reported marginal standard deviation, not a confidence interval of a mean. (c) Three assigned contact networks match the same [100] peak but predict distinct dominant [111] peaks at equal transferred wavevector magnitude. Bars propagate the independent input scenario and envelope assigned cell-edge sensitivity ranges. The calibration direction and contact ratios are stipulated; the original acoustic record does not supply a registered crystallographic direction. Neither comparison determines a unique microscopic loss mechanism.}
\label{fig:stmv-spectral-test}
\end{figure}

The benchmark thus completes a limited, calculable application:
documented packing, explicit effective mechanics, an optical response
map, uncertainty propagation, one unused reported spectral comparison
and a further prospective measurement. Internal shell modes require
their own evidence. The same acoustic study found no firm evidence
for the searched localized particle vibrations~\cite{Stephanidis2007}.
That boundary preserves the distinction between measured assembly
acoustics and the separate viral-phonon quantum feasibility question.

The next section applies these mechanical and observation principles
to confined modes, wave packets and coupled excitations.

\clearpage

\section{Viral Phonons: Confinement, Wave Packets, and Coupling}
\label{sec:viral-phonon-mechanics}

The assembled mechanics and observation maps of Sections~2--6 now yield
concrete excitation problems: which patterns a finite ideal-virion assembly
supports, where motion concentrates, and how a specified force prepares
it. Particle identities, body orientations and prepared material states
continue through recruitment, rearrangement and breakup within the fixed
enclosed pool. Regulated charges use their surface and ionic ledgers.

\paragraph{A running HPV-inspired preparation.}
The examples below return to the empty L1 shell of
Section~\ref{sec:introduction-worked-guide}. The nominal diameter is
$55\,\mathrm{nm}$; each intact shell remains one whole-particle node.
Capsomer organization supplies the body geometry and candidate surface
locations. The interparticle Hessian, material response and detector
calibration determine the collective calculation. Larger examples retain
these roles when compatible cells assemble more particles.

\begin{definition}[Prepared HPV-inspired reference family]
\label{def:s7hpv-prepared-reference}
A member specifies the retained intact L1 VLPs from the fixed pool, their
reference centres and orientations, calibrated surface and material
states, force-balanced supports or constraints, and observation maps.
Its stable conservative Hessian defines patterns; its full causal
response defines driven motion and relaxation. The nominal protein-only
mass used here is
\begin{equation}
 m_v=20.23\times10^6\,\mathrm{Da}
       \simeq3.36\times10^{-20}\,\mathrm{kg} .
 \label{eq:s7hpv-protein-mass-reference}
\end{equation}
This is a stoichiometric reference for an ideal empty shell. A selected
hydrated preparation supplies its own inertial and hydrodynamic inputs.
\end{definition}
The mass follows the expected 72-pentamer L1-shell composition reported
by Klein et al.~\cite{Klein2025}; it adds an inertial scale to the
introductory guide, whose measured-band dynamics were overdamped.

\begin{table}[H]
\centering\small
\begin{tabular}{@{}p{3.2cm}p{5.4cm}p{6.7cm}@{}}
\toprule
Input & Running assignment & Physical role and status\\
\midrule
Whole-particle geometry & $b_v=27.5\,\mathrm{nm}$; empty L1 shell
 & HPV-inspired body and site registration.\\[3pt]
Body inertia & $m_v$ in~\eqref{eq:s7hpv-protein-mass-reference}
 & Protein-only reference; angular inertia is assigned or calibrated separately.\\[3pt]
Collective mechanics & Guide values $k_p=k_l=3\times10^{-4}\,\mathrm{N\,m^{-1}}$
 & Synthetic supported and projected interaction curvatures.\\[3pt]
Prepared medium & Guide $\Delta k=6\times10^{-4}\,\mathrm{N\,m^{-1}}$, $\tau=0.5\,\mathrm{ms}$,
 $c_0=1.5\times10^{-7}\,\mathrm{kg\,s^{-1}}$
 & Synthetic branch and resistance; not inferred from shell diameter.\\[3pt]
Surface preparation & Registered patches and retained ligand inventory
 & Binding geometry can inform coordinates; forces, populations and rates need their own inputs.\\[3pt]
Observation & Centres, registered markers and a surface-sensitive channel
 & Separate calibrated projections of one coupled physical state.\\
\bottomrule
\end{tabular}
\caption{One reference family for the Section~7 examples. The guide's
two-particle constants are reused where stated; larger arrays and angular
examples specify their additional projected coefficients. No diameter,
indentation stiffness or binding-site map alone fixes a collective mode.}
\label{tab:s7hpv-running-reference}
\end{table}

HPV-specific structure can refine the surface map. The heparin-bound
HPV16 quasivirus structure of Langley et al.~\cite{Langley2026} resolves
binding around a fivefold capsomer and associated conformational changes.
Transferring those candidate locations to an empty L1 shell is an
explicit structural idealization: that experimental particle also
contains L2 and packaged genome. The resulting site registration can
enter the patch derivatives of Section~3, while a preparation-specific
interaction law sets their strengths. This separates a tractable
geometric input from the dynamical quantities the proposed experiment
must constrain.

Figure~\ref{fig:s7hpv-prepared-array} carries the HPV-inspired node into a larger, spatially resolved assembly.

\begin{figure}[H]
\centering\begingroup
\setlength{\unitlength}{1pt}
\scalebox{0.90}{%
\input{vlt-figures/figure-299.tex}%
}\endgroup
\caption{A common geometric reference for HPV-inspired calculations. Panel (a) uses a faceted whole-body glyph for the nominal empty L1 shell; it is not a drawing of individual capsomers. Panel (b) places twelve bodies at $(x,y,z)/a\in\{-1,0,1\}\times\{-1/2,1/2\}^2$, with $a=150\,\mathrm{nm}$ and circumscribed radius $27.5\,\mathrm{nm}$. All twenty nearest-axis links, including six vertical links, are visible in one oblique projection. The coils have fixed turns, transverse amplitude and trimmed leads at this common reference geometry. They denote the selected projected coupling graph; a physical site model supplies its Hessian, prestress and truncation test. Supports balance that reference. Colours register selected labels, with the remaining bodies retained in grey. The lower chain distinguishes the geometric input, coupled mechanical/surface/material state and instrument record. This finite assembly fits inside the introductory reference cylinder.}
\label{fig:s7hpv-prepared-array}
\end{figure}

A short force pulse provides a useful physical picture. It sets a
collection of virions or resolved structural coordinates in motion.
Their restoring couplings distribute that motion among collective
patterns. Superposing patterns can concentrate a disturbance in space;
their relative phases govern its subsequent spreading and return.
The mode calculation determines these possibilities before a particular
source and readout select the observed excitation.

The biophysical extensions ask what changes when that motion couples
to another resolved process. A configuration change can shift a
mechanical equilibrium, ionic transfer can relax an electrochemical
imbalance, and membrane shape can relax through solvent motion. Each
case identifies a coordinate, its stored energy or free energy, its
evolution and its observation. This common structure connects the
confinement calculations to concrete tests of effective variables.

We begin with a stable, persistent reference and then add propagation,
mode conversion, and resolved electronic, ionic and chemical coordinates.
Each extension supplies its own coupling, evolution and energy account.
The benchmark in Section~\ref{sec:complete-benchmark} provides a
classical reference for stationary response, fluctuations and
continuation through a controlled link change before equilibration.

An \emph{excitation} is a specified departure from a reference preparation.
Classically, a mode carries continuously variable energy in its coordinate
and momentum. Under the phonon postulate, one additional occupation of
mode $s$ carries $\hbar\omega_s$. Inertia, interactions and boundary
conditions select the collective pattern and its frequency; preparation
sets the state of that oscillator. Discrete classical frequencies and
quantized energy measurements are separate properties
\cite{Misra2012,Shankar1994,FeynmanLectures1963}. Exactly harmonic quantum
modes evolve as decoupled bosonic oscillators, although their initial
occupations may be correlated or entangled. Interaction, damping and
conversion enter through the corresponding additional couplings.

Biological measurements motivate specific choices of coordinate and
readout. Atomistic capsid calculations connect collective displacements
to polarizability-based Raman signals and compare them with elastic
networks~\cite{Dykeman2010}. Orientation-sensitive terahertz measurements
resolve optical projections of vibrations in lysozyme
crystals~\cite{Acbas2014}, while single-virus ultrafast measurements probe
internal and contact-dependent motion~\cite{Zhang2025}. These distinct
preparations guide the construction of mechanical variables. Applying
them to a whole-virion array additionally requires its interparticle
curvatures and, for an occupation claim, evidence of quantization.

The medium determines the useful dynamical description. A well-separated
vibrational resonance supports an oscillatory mode over its resolved
decay and coupling window; stronger loss can produce predominantly
relaxational motion. The causal response developed in Sections~4 and~6
tests that distinction. Throughout this section, each prediction specifies
whether the measured quantity is a shape, a driven response, a fluctuation
statistic or an excitation occupation.

\subsection{Two meanings of a phonon in a well}
\label{sec:phonon-two-wells}

\subsubsection{A well in a collective coordinate}

\begin{definition}[Collective-coordinate mechanical well]
\label{def:s7-collective-coordinate-well}
A \emph{collective-coordinate well} is a restoring well in configuration
space: its energy rises
when a specified collective displacement coordinate leaves its reference
value. Choose one real positive mode with
$\vct\phi_s^{\mathsf T}\mathsf M\vct\phi_s=1$, and write its
contribution to physical displacement as $\vct u=\vct\phi_s Q_s$.
The pattern fixes which components move together, in opposite directions
or hardly at all. The scalar $Q_s$ sets the amplitude along that pattern;
it is not a location inside the cylinder. For translational coordinates
$\phi_s$ has units $\mathrm{kg}^{-1/2}$ and $Q_s$ has units
$\mathrm{kg}^{1/2}\mathrm m$.
\end{definition}

The corresponding classical energy is
$H_s=(\Pi_s^2+\omega_s^2Q_s^2)/2$, with
$\Pi_s=\dd Q_s/\dd t$. The kinetic term records collective motion and
the potential term is its restoring cost. The canonical quantum
realization is
\begin{equation}
 \widehat H_s=\frac12\bigl(\widehat\Pi_s^2+
                 \omega_s^2\widehat Q_s^2\bigr),
 \qquad
 \widehat\Pi_s=-i\hbar\frac{\dd}{\dd Q_s}.
 \label{eq:phonon-coordinate-well}
\end{equation}
This is the oscillator on $L^2(\R,\dd Q_s)$ with the self-adjoint
realization described in Section~6. Its eigenfunctions are functions
of a collective amplitude. In particular,
\begin{equation}
 \begin{aligned}
 \varphi^{\rm osc}_{0,s}(Q)
   &=\left(\frac{\omega_s}{\pi\hbar}\right)^{1/4}
                 e^{-\omega_s Q^2/(2\hbar)},\\
 \varphi^{\rm osc}_{1,s}(Q)
   &=\sqrt{\frac{2\omega_s}{\hbar}}\,Q
                              \varphi^{\rm osc}_{0,s}(Q).
 \end{aligned}
 \label{eq:phonon-oscillator-well-states}
\end{equation}
The wavefunctions have units $[Q]^{-1/2}$, making
$|\varphi^{\rm osc}_{N,s}(Q)|^2\dd Q$ a probability. The first excited
distribution has two symmetric lobes and vanishes at the reference
amplitude. Its node belongs to the collective-coordinate distribution;
the mechanical pattern $\vct\phi_s$ specifies a separate spatial shape.
Figure~\ref{fig:coordinate-well-and-collective-pattern} connects the two
through reconstructed virion displacements.

The level sequence and its adjacent increment are
\begin{equation}
 E_{N,s}=\hbar\omega_s(N+\tfrac12),\qquad
 \boxed{E_{N+1,s}-E_{N,s}=\hbar\omega_s.}
 \label{eq:phonon-mode-energy-increment}
\end{equation}
The vacuum is the oscillator ground state, with finite quantum coordinate
variance and the same physical virion population. Classical energy at this
frequency is continuous; the quantum spacing follows from canonical
quantization~\cite{Shankar1994,Misra2012}. The harmonic approximation
applies within the collision-free, small-deformation basin, so its
formal ladder must be checked against the fluctuation assessment in
Section~\ref{sec:operator-thermal-stochastic}.

\begin{figure}[H]
\centering\begingroup
\definecolor{vltTeal}{HTML}{176B64}
\definecolor{vltBlue}{HTML}{315D80}
\definecolor{vltGold}{HTML}{A87520}
\definecolor{vltPlum}{HTML}{725877}
\definecolor{vltRed}{HTML}{762B38}
\definecolor{vltYellow}{HTML}{D5AD3C}
\definecolor{vltNeutral}{HTML}{777D83}
\setlength{\unitlength}{1pt}
\scalebox{0.90}{%
\input{vlt-figures/figure-227.tex}%
}\endgroup
\caption{A collective amplitude reconstructed as whole-virion motion. (a) For $q=\sqrt{\omega_s/\hbar}Q_s$, the normalized potential is $V/(\hbar\omega_s)=q^2/2$, with physical level spacing $\hbar\omega_s$. (b) The full-line densities are $p_0=\pi^{-1/2}e^{-q^2}$ and $p_1=2q^2p_0$. (c) Both orthographic scenes use the free-end five-site mode $e_i=\sqrt{2/5}\cos[\pi(i-1/2)/5]$ and $u_i=q\sqrt{\hbar/(m\omega_s)}e_i$ at $q=\pm1$. Gray outlines register one reference; rigid bodies, enclosure and displacement gain are shared. Fixed coil turns, radius and leads make pitch follow the body-trimmed span. The configurations illustrate coordinate values; number states have zero mean displacement. The boxed result illustrated here is~\eqref{eq:phonon-mode-energy-increment}.}
\label{fig:coordinate-well-and-collective-pattern}
\end{figure}

\subsubsection{A spatial pattern and the state occupying it}

A spatial well varies the restoring structure across material locations.
Its eigenproblem selects $\omega_s$ and $\vct\phi_s$ before an occupation
is assigned. The next two subsections solve finite-boundary and
localization examples. Within their harmonic reference, changing the
oscillator preparation preserves the normalized spatial pattern.

This has a directly calculable consequence for a readout. In the number
state $\ket{N_s}$ of that one mode,
\begin{equation}
 \begin{aligned}
 \langle\widehat Q_s\rangle_{N_s}&=0,\qquad
 \langle\widehat Q_s^2\rangle_{N_s}
                         =\frac{\hbar}{\omega_s}(N_s+\tfrac12),\\
 \Delta_{N_s}\langle\widehat u_i\widehat u_j\rangle
                         &=\frac{N_s\hbar}{\omega_s}
                                           \phi_{s,i}\phi_{s,j}.
 \end{aligned}
 \label{eq:phonon-number-state-displacement-readout}
\end{equation}
Here $\Delta_{N_s}$ subtracts the vacuum contribution of this mode,
with all other modes prepared identically. A real marker
$Y=\vct\ell^{\mathsf T}\vct u$ consequently gains variance
$N_s\hbar|\vct\ell^{\mathsf T}\vct\phi_s|^2/\omega_s$.
A mode can carry an excitation while giving zero mean displacement,
or remain invisible to a marker orthogonal to its pattern. Spatial
correlation signs reflect coordinated directions of motion; they are
not probabilities of a phonon occupying individual virions.

Figure~\ref{fig:signed-phonon-covariance-and-mean} displays the corresponding signed covariance alongside its mechanical pattern, while retaining zero mean displacement.

\begin{figure}[H]
\centering\begingroup
\definecolor{vltTeal}{HTML}{176B64}
\definecolor{vltBlue}{HTML}{315D80}
\definecolor{vltGold}{HTML}{A87520}
\definecolor{vltPlum}{HTML}{725877}
\definecolor{vltRed}{HTML}{762B38}
\definecolor{vltYellow}{HTML}{D5AD3C}
\definecolor{vltNeutral}{HTML}{777D83}
\setlength{\unitlength}{1pt}
\scalebox{0.80}{%
\input{vlt-figures/figure-228.tex}%
}\endgroup
\caption{The first nonuniform equal-mass five-site path mode, $e_i=\sqrt{2/5}\cos[\pi(i-1/2)/5]$, viewed as a coefficient vector and a rank-one covariance. (a) Signed coefficients determine the relative displacement directions; the number-state mean remains zero. (b) Each cell gives $e_i e_j$, rounded to two decimals, in the vacuum-subtracted covariance $\Delta\langle u_i u_j\rangle=\hbar e_i e_j/(m\omega_s)$ for one extra occupation. Teal denotes positive entries, burgundy negative entries and gray zero; printed signs retain the distinction in grayscale. Negative cross entries express anticorrelation and coexist with the nonnegative spectrum displayed below. The neighboring figure maps this same coefficient vector onto whole-virion configurations; neither plot is a particle trajectory.}
\label{fig:signed-phonon-covariance-and-mean}
\end{figure}

For a normalized one-phonon superposition
$\ket{\Phi(0)}=\sum_s c_s\ket{1_s}$, the excitation energy is
$\sum_s\hbar\omega_s|c_s|^2$. Its vacuum-to-state displacement
matrix element is
\begin{equation}
 \bra{0}\widehat{\vct u}(t)\ket{\Phi(0)}
 =\sum_s \sqrt{\frac{\hbar}{2\omega_s}}\,
         \vct\phi_s c_s e^{-i\omega_s t}.
 \label{eq:phonon-packet-transition-readout}
\end{equation}
This amplitude contains spatial interference, although the mean
displacement of the fixed-total-one state is zero. Detecting it requires
a measurement coupling to the stated transition or to its correlations.
A coherent preparation instead has
$\langle\widehat Q_s(t)\rangle
=\sqrt{2\hbar/\omega_s}\operatorname{Re}(\alpha_s e^{-i\omega_s t})$
for initial annihilation mean $\alpha_s$, and reproduces a classical
oscillation in its first moment. A real traveling-displacement field, a coordinate probability density
and a spatial measurement effect therefore describe different
observations.

\subsubsection{What can add one quantum, and what resonance means}

An energy gap does not specify how energy enters. Let a real applied force
pattern $\vct f(t)$ couple through work
$H_{\rm d}=-\vct f^{\mathsf T}\vct u$. Its projection on the mode is
$F_s=\vct\phi_s^{\mathsf T}\vct f$, and
$H_{{\rm d},s}=-F_s(t)\widehat Q_s$.
The modal force has units $\mathrm{kg}^{1/2}\mathrm m\,\mathrm s^{-2}$.
Using $\widehat Q_s=\sqrt{\hbar/(2\omega_s)}
(\widehat b_s+\widehat b_s^\dagger)$ gives
\begin{equation}
 \bra{N'}\widehat Q_s\ket N
 =\sqrt{\frac{\hbar}{2\omega_s}}
   \left[\sqrt{N+1}\,\delta_{N',N+1}
              +\sqrt N\,\delta_{N',N-1}\right].
 \label{eq:phonon-linear-force-selection-rule}
\end{equation}
The linear coupling connects adjacent occupations, with preparation-dependent
strengths given by the square roots above. It drives a mode only when
the force has nonzero spatial overlap $F_s$ and appropriate temporal
content. Curvature modulation proportional to $\widehat Q_s^2$ instead
connects $\Delta N=0,\pm2$ at first order and draws energy from its
modulation source. The autonomous resource-dependent excitation of
Section~4 uses a separately specified source dynamics.

For weak forcing over a finite pulse $0<t<t_p$, first-order perturbation
theory yields the upward-transition probability
\begin{equation}
 P_{N\to N+1}^{(1)}
 =\frac{N+1}{2\hbar\omega_s}
       \left|\int_0^{t_p}F_s(t)e^{i\omega_s t}\,\dd t\right|^2.
 \label{eq:phonon-finite-pulse-transition}
\end{equation}
The superscript identifies the probability obtained from a first-order
transition amplitude, valid while transition probabilities remain small.
A long weak sinusoid selects a narrow interval; a short pulse has broader
spectral overlap. Damping and reservoirs introduce further timescales
when present. Thus the transition calculation combines an energy gap,
a nonzero coupling matrix element and the pulse spectrum.

Figure~\ref{fig:finite-pulse-exact-spectral-overlap} evaluates this overlap for two finite real force windows, separating pulse bandwidth from an intrinsic mode linewidth.

\begin{figure}[H]
\centering
\begingroup
\definecolor{vlfGreen}{HTML}{176B64}
\definecolor{vlfDarkRed}{HTML}{762B38}
\definecolor{vlfRed}{HTML}{762B38}
\definecolor{vlfBlue}{HTML}{315D80}
\definecolor{vlfCopper}{HTML}{762B38}

\setlength{\unitlength}{1pt}
\scalebox{0.80}{%
\input{vlt-figures/figure-229.tex}%
}%
\endgroup
\caption{Spectral overlap of a lossless oscillator with $F_s(t)=F_0\cos\Omega t$ on $0<t<t_p$. Both real-force frequency terms are included in $\mathcal P=|\int_0^{t_p}F_s(t)e^{i\omega_st}\dd t|^2/(F_0t_p/2)^2$. Equal axes compare $\omega_st_p=4\pi$ and $20\pi$. The longer window narrows the overlap. Multiplication by $(N+1)F_0^2t_p^2/(8\hbar\omega_s)$ gives the first-order upward probability, valid when it remains much smaller than one. The duration also changes deposited energy; the curve width here comes entirely from the pulse window.}
\label{fig:finite-pulse-exact-spectral-overlap}
\end{figure}

The exact harmonic result gives an additional check. From the vacuum,
the same finite linear pulse prepares a coherent displacement with
$\alpha_s(t_p)$ equal to the pulse integral in
\eqref{eq:operator-force-pulse-displacement}. Its occupation law is Poisson, as for a linearly displaced oscillator
vacuum~\cite{Glauber1963}, with
\begin{equation}
 \bar N_s=|\alpha_s(t_p)|^2
   =\frac{1}{2\hbar\omega_s}
       \left|\int_0^{t_p}F_s(t)e^{i\omega_s t}\,\dd t\right|^2,
 \qquad
 \Delta E_s=\hbar\omega_s\bar N_s.
 \label{eq:phonon-driven-excitation-energy}
\end{equation}
Repeated adjacent transitions can therefore populate many levels. In the
weak limit the one-phonon probability agrees with the preceding formula;
a finite coherent drive is not restricted to producing just one phonon.
The deposited energy also equals the classical oscillator energy from
the same pulse starting at rest. Mean response or pulse-energy transfer
alone does not discriminate those occupation descriptions.
Figure~\ref{fig:linear-force-selection-and-mean-motion} contrasts their
state and readout content.

\begin{figure}[H]
\centering\begingroup
\definecolor{vltTeal}{HTML}{176B64}
\definecolor{vltBlue}{HTML}{315D80}
\definecolor{vltGold}{HTML}{A87520}
\definecolor{vltPlum}{HTML}{725877}
\definecolor{vltRed}{HTML}{762B38}
\definecolor{vltYellow}{HTML}{D5AD3C}
\definecolor{vltNeutral}{HTML}{777D83}
\setlength{\unitlength}{1pt}
\scalebox{0.90}{%
\input{vlt-figures/figure-230.tex}%
}\endgroup
\caption{Linear-coordinate selection and the moments of a prepared oscillator. (a) The coupling $H_d=-F_sQ_s$ connects neighboring number levels. Arrow widths scale with the adjacent matrix-element magnitude; arrow directions label raising and lowering, without assigning a rate or linewidth. Repeated transitions can populate higher levels. (b) Unforced evolution of $|\alpha=1\rangle$ and $|1\rangle$ gives mean coordinates $\sqrt2\cos\tau$ and zero, respectively; their centered variances are $1/2$ and $3/2$. (c) Both states have excitation energy $\hbar\omega_s$ and total energy $3\hbar\omega_s/2$. The latter splits into mean-motion and centered-fluctuation contributions as shown. Number labels count oscillator excitations; the physical virion population stays fixed. Preparation requires its specified source and energy account.}
\label{fig:linear-force-selection-and-mean-motion}
\end{figure}

A biological excitation model therefore specifies the mechanical
coordinate, restoring frequency, force projection and measured projection.
For an optical pump, the electromechanical or polarizability coupling
determines how much supplied energy reaches that mode. The cavity and
stiffness examples below calculate the available patterns; their force
and detector maps determine preparation and visibility.

\subsubsection{A prepared quantum in a damped mode: three distinct lifetimes}
\label{sec:qfa-lifetimes}

The oscillator ladder supplies an energy unit. A bath determines how
long a preparation retains its energy, phase and counting statistics.
These questions can be calculated separately for a resolved mode.
The following reduction uses weak system--bath coupling, a bath whose
correlation time is short on the relaxation scale, and an isolated
positive carrier frequency. It is the standard thermal oscillator
master equation~\cite{Lindblad1976,Clerk2010}, applied here to a
declared collective virion coordinate. Its rates must be obtained from
that coordinate's environment.

\begin{definition}[Thermal single-mode reduction]
\label{def:qfa-thermal-mode}
Write $\widehat N=\widehat b^\dagger\widehat b$ for the selected mode,
$n_B=(e^{\hbar\omega_s/(k_{\rm B}T)}-1)^{-1}$, and
$\mathcal D[L]\varrho=L\varrho L^\dagger
-\{L^\dagger L,\varrho\}/2$. Let $\gamma_1>0$ denote occupation
relaxation and $\gamma_\phi\geq0$ pure number dephasing. The reduction is
\begin{equation}
 \begin{aligned}
 \dot\varrho={}&-i\omega_s[\widehat N,\varrho]
 +\gamma_1(n_B+1)\mathcal D[\widehat b]\varrho\\
 &+\gamma_1 n_B\mathcal D[\widehat b^\dagger]\varrho
 +2\gamma_\phi\mathcal D[\widehat N]\varrho .
 \end{aligned}
 \label{eq:qfa-thermal-generator}
\end{equation}
The specified dephasing convention makes $\gamma_\phi$ the extra
decay rate of the first amplitude moment.
\end{definition}

\begin{proposition}[Relaxation, phase loss and number-changing events]
\label{prop:qfa-three-lifetimes}
For finite required moments, \eqref{eq:qfa-thermal-generator} gives
\begin{equation}
 \begin{aligned}
 \frac{\dd}{\dd t}\langle\widehat N\rangle
   &=-\gamma_1(\langle\widehat N\rangle-n_B),\\
 \frac{\dd}{\dd t}\langle\widehat b\rangle
   &=-\bigl(i\omega_s+\gamma_1/2+\gamma_\phi\bigr)
       \langle\widehat b\rangle,\\
 R_{n,\rm jump}&=\gamma_1\bigl[n+(2n+1)n_B\bigr].
 \end{aligned}
 \label{eq:qfa-three-rates}
\end{equation}
Here $R_{n,\rm jump}$ is the first number-changing jump rate from
$\ket n$ in the unraveling with the displayed lowering and raising
channels. Starting at $\ket1$, the probability of no such bath jump
up to $t$ is $e^{-\gamma_1(1+3n_B)t}$.
\end{proposition}
\begin{proof}
Apply the adjoint dissipators to $\widehat N$ and $\widehat b$.
Their lowering and raising contributions to a number-state escape
rate are $\gamma_1(n_B+1)n$ and $\gamma_1n_B(n+1)$.
The Hamiltonian and number-dephasing term leave populations unchanged.
In the absence of a number jump the initial number state stays in
that level, so its constant escape hazard integrates exponentially.
\end{proof}

Energy above the bath's stationary mean relaxes on $\gamma_1^{-1}$;
a coherent first moment decays on
$(\gamma_1/2+\gamma_\phi)^{-1}$. A number state has zero first moment
already. Its number statistics require their own diagnostic, and
its unconditional return probability includes paths with bath jumps.
The no-jump probability is a property of the declared bath monitoring
model, rather than a direct identification with a measured coherence.

\begin{proposition}[Thermal survival window for single-phonon antibunching]
\label{prop:qfa-antibunching-window}
Start \eqref{eq:qfa-thermal-generator} in $\ket1\bra1$. Put
$r=e^{-\gamma_1t}$ and $q=n_B(1-r)$. Then
\begin{equation}
 \boxed{\begin{aligned}
 \langle\widehat N\rangle_t&=r+q,\\
 \langle\widehat N(\widehat N-1)\rangle_t&=4rq+2q^2,\\
 g^{(2)}(t)&=\frac{4rq+2q^2}{(r+q)^2}.
 \end{aligned}}
 \label{eq:qfa-single-phonon-thermal-moments}
\end{equation}
For $n_B>0$, the strict condition $g^{(2)}<1$ is equivalent to
\begin{equation}
 \boxed{\displaystyle
 t<t_{\rm ab}=\frac1{\gamma_1}
       \ln\!\left(1+\frac{\sqrt2-1}{n_B}\right).}
 \label{eq:qfa-antibunching-window}
\end{equation}
\end{proposition}
\begin{proof}
The adjoint equation for $f_2=\langle\widehat N(\widehat N-1)\rangle$
is $\dot f_2=-2\gamma_1f_2+4\gamma_1 n_B\langle\widehat N\rangle$.
Solve it with $f_2(0)=0$ and the first equation in
\eqref{eq:qfa-three-rates}. Subtracting the squared mean from
$f_2$ gives $q^2+2rq-r^2$. Since $r>0$, negativity is equivalent
to $n_B(e^{\gamma_1t}-1)<\sqrt2-1$, which gives the stated window.
\end{proof}

Figure~\ref{fig:cmd-phonon-survival-counting} gives a graphical specialization of this boxed relation.

This calculation exposes a distinction that a mean-energy record can
miss completely. At $n_B=1$, the mean occupation stays exactly one,
yet $g^{(2)}=2(1-e^{-2\gamma_1t})$ grows from zero to two.
At $\gamma_1t=0.1$ it is approximately $0.363$; it reaches one at
$\gamma_1t=(\ln2)/2\simeq0.347$ and is approximately $1.264$ at
$\gamma_1t=0.5$. These are dimensionless conditional predictions, not
HPV lifetime measurements. At $n_B=0$ ideal vacuum loss preserves
$g^{(2)}=0$ at every finite time while the mean count vanishes
asymptotically. The decreasing count rate still limits statistical
precision. Pure number dephasing leaves these population results
unchanged while suppressing phase-sensitive records.

For a proposed viral-phonon experiment, preparation, storage and readout
must be evaluated under the same bath and control protocol. If cooling
is switched off during storage, \eqref{eq:qfa-antibunching-window} uses
the storage bath's occupation and relaxation rate. A continuously cooled
mode has the combined rates of the cooling and environmental channels.
Section~\ref{sec:qfa-counting} supplies the corresponding counting map;
Section~\ref{sec:qfa-feasibility} examines which of these inputs the
selected HPV preparation actually constrains. Strongly mixed or
overdamped coordinates instead use the full causal response and quantum
bath of Section~4 before a single-mode reduction is attempted.

\begin{figure}[H]
\centering\begingroup
\setlength{\unitlength}{1pt}
\scalebox{0.84}{%
\input{vlt-figures/figure-310.tex}%
}\endgroup
\caption{Exact thermal survival and a conditional detection protocol. Panel (a) starts the weak-bath Markov oscillator in $\ket1$ with $n_B=1$. Its mean occupation remains one while $g^{(2)}=2(1-e^{-2\gamma_1t})$ rises through one at $\gamma_1t=(\ln2)/2$; shading marks the strict antibunching window. The dashed curve is the no-number-jump probability $e^{-4\gamma_1t}$ in the specified unraveling. The boxed moments and boundary are \eqref{eq:qfa-single-phonon-thermal-moments} and \eqref{eq:qfa-antibunching-window}. Panel (b) ties preparation to a finite storage delay and calibrated converted counts, using \eqref{eq:qfa-detected-count-statistic}. Pure number dephasing changes phase readouts while leaving the population moments in panel (a) intact. Times are dimensionless and no lifetime or detector coupling is assigned to HPV or CCMV.}
\label{fig:cmd-phonon-survival-counting}
\end{figure}

\FloatBarrier

\subsection{An exact finite mechanical cavity}
\label{sec:phonon-finite-cavity}

A finite chain provides an exact benchmark for end forces, mode shapes
and probe weights. Each node retains one longitudinal coordinate of a
whole ideal virion or another declared mechanical unit. Specifying its
interactions and supports makes the collective construction testable
before introducing a more detailed geometry or medium response.

\subsubsection{The end condition follows from the specified springs}

Take $n\geq2$ identical masses $m>0$, adjacent spring curvature $k_b>0$,
and displacements $u_i$ about an equilibrium reference. Choose
$\omega_{\rm pin}\geq0$ and
\begin{equation}
 U_2=\frac{k_b}{2}\sum_{i=1}^{n-1}(u_{i+1}-u_i)^2
       +\frac{m\omega_{\rm pin}^2}{2}\sum_{i=1}^n u_i^2,
 \qquad
 \mathsf D=\omega_{\rm pin}^2\Id+
                          \frac{k_b}{m}\mathsf L_{\rm path}.
 \label{eq:phonon-finite-chain-model}
\end{equation}
Here $k_b$ has newtons-per-metre units, while $\omega_{\rm pin}$ is an
angular frequency. The path Laplacian has end diagonals $1$, interior
diagonals $2$, and nearest off-diagonals $-1$. In particular the first
end obeys
$m\Dtt{u_1}=-m\omega_{\rm pin}^2u_1-k_b(u_1-u_2)+f_1$,
whereas an interior site has the restoring term
$-k_b(2u_i-u_{i-1}-u_{i+1})$.
The end has one connecting spring; its force balance introduces no
additional endpoint constraint.

The connecting network has free \emph{material} ends. Ghost equalities
$u_0=u_1$ and $u_{n+1}=u_n$ express zero extension of notional outer
bonds and allow one difference formula at every site. Pinning comes from
an independently specified conservative trap, support or restoring branch;
an inactive hard cylinder wall supplies only the enclosure. An internal
branch can also be gapped, with its own resolved inertia and readout.

Adding two springs $k_b$ to fixed external supports gives
$u_0=u_{n+1}=0$ and diagonal entries $2$ throughout. The resulting
patterns are $\sin[j\pi i/(n+1)]$, $j=1,\ldots,n$, with spring
eigenvalues $4\sin^2[j\pi/(2(n+1))]$. The attachment law therefore
selects the spectrum even for assemblies sharing one enclosure.

Figure~\ref{fig:added-end-condition-gallery} compares the first positive patterns of the two end conditions on the same displacement scale.

\begin{figure}[H]
\centering\begingroup
\definecolor{vltTeal}{HTML}{176B64}
\definecolor{vltBlue}{HTML}{315D80}
\definecolor{vltGold}{HTML}{A87520}
\definecolor{vltPlum}{HTML}{725877}
\definecolor{vltRed}{HTML}{762B38}
\definecolor{vltYellow}{HTML}{D5AD3C}
\definecolor{vltNeutral}{HTML}{777D83}
\setlength{\unitlength}{1pt}
\scalebox{0.80}{%
\input{vlt-figures/figure-231.tex}%
}\endgroup
\caption{Finite standing patterns under two attachment laws, with eight equal masses, adjacent curvature $k_b$ and $\omega_b=\sqrt{k_b/m}$. (a) Free ends give $e_{\ell i}=\sqrt{2/8}\cos[\ell\pi(i-1/2)/8]$ and $\omega_\ell/\omega_b=2\sin(\ell\pi/16)$. Common translation is omitted. (b) Two additional end springs $k_b$ give $e_{ji}=\sqrt{2/9}\sin(j\pi i/9)$ and $\omega_j/\omega_b=2\sin(j\pi/18)$. Black bars mark external supports at zero ghost coefficients, rather than retained virions. All panels share the same signed coefficient scale. Filled points are the eight physical samples; segments guide the eye. The enclosure geometry alone does not determine the attachment law.}
\label{fig:added-end-condition-gallery}
\end{figure}

\subsubsection{Exact standing modes and the translational zero coordinate}

\begin{proposition}[Spectrum of the free-connecting-end cavity]
\label{prop:s7-finite-cavity-spectrum}
For the declared path, the orthonormal cosine patterns and frequencies are
\begin{align}
 e_{\ell,i}&=\nu_\ell
    \cos\!\left[\frac{\ell\pi}{n}\left(i-\frac12\right)\right],
 &\nu_0&=n^{-1/2},\qquad \nu_{\ell>0}=(2/n)^{1/2},\nonumber\\
 \omega_\ell^2&=\omega_{\rm pin}^2+
      \frac{4k_b}{m}\sin^2\!\left(\frac{\ell\pi}{2n}\right),
 &\ell&=0,\ldots,n-1.
 \label{eq:phonon-chain-cavity-spectrum}
\end{align}
Thus $\phi_{\ell,i}=e_{\ell,i}/\sqrt m$ is the physical mass-normalized
pattern and $u_i=\sum_\ell\phi_{\ell,i}Q_\ell$.
Each positive normal coordinate has energy
$(\Pi_\ell^2+\omega_\ell^2Q_\ell^2)/2$; its optional occupation
increment is $\hbar\omega_\ell$.
\end{proposition}
\begin{proof}
Substitution into the interior difference equation gives the sine-square
eigenvalue; the half-site cosine shift satisfies both ghost equalities.
The discrete cosine orthogonality gives
$\sum_i e_{\ell,i}e_{j,i}=\delta_{\ell j}$.
\end{proof}

The $\ell=0$ mode is common translation, with frequency
$\omega_{\rm pin}$. At zero pinning it is a free canonical coordinate,
not a zero-energy oscillator whose ladder formulas can still be used.
Its Hamiltonian is $\Pi_0^2/2$ and its dynamics must be retained separately
when an experiment couples to it. The remaining $n-1$ modes form the
positive phonon sector. With positive pinning all $n$ modes are positive.
Figure~\ref{fig:finite-free-chain-modes-and-dispersion} displays a pinned
reference with free connecting ends. Its support comes from the declared
onsite curvature.

\begin{figure}[H]
\centering\begingroup
\definecolor{vltTeal}{HTML}{176B64}
\definecolor{vltBlue}{HTML}{315D80}
\definecolor{vltGold}{HTML}{A87520}
\definecolor{vltPlum}{HTML}{725877}
\definecolor{vltRed}{HTML}{762B38}
\definecolor{vltYellow}{HTML}{D5AD3C}
\definecolor{vltNeutral}{HTML}{777D83}
\setlength{\unitlength}{1pt}
\scalebox{0.90}{%
\input{vlt-figures/figure-232.tex}%
}\endgroup
\caption{An eight-site free-end path with separate onsite pinning $\omega_{\rm pin}=\omega_b/2$, $\omega_b=\sqrt{k_b/m}$. (a) Filled samples give $e_{\ell i}=\sqrt{2/8}\cos[\ell\pi(i-1/2)/8]$ for $\ell=1,3$ at $z_i/a=i-1/2$; open ghosts satisfy reflection at the dotted material ends. (b) Allowed $qa=\ell\pi/8$ sample $\omega/\omega_b=\sqrt{1/4+4\sin^2(qa/2)}$; the dashed long-wave approximation is $\sqrt{1/4+(qa)^2}$. (c) The same coefficients reconstruct rigid virions with one amplitude gain. Rulers mark the undeformed reference positions, while spring pitch follows relative displacement. Pinning and the free connecting ends are independent inputs.}
\label{fig:finite-free-chain-modes-and-dispersion}
\end{figure}

The same finite-mode construction also evolves a localized preparation. Figure~\ref{fig:added-exact-chain-spacetime} shows its signed displacement through spreading and return from the finite ends.

\begin{figure}[H]
\centering\begingroup
\definecolor{vltTeal}{HTML}{176B64}
\definecolor{vltBlue}{HTML}{315D80}
\definecolor{vltGold}{HTML}{A87520}
\definecolor{vltPlum}{HTML}{725877}
\definecolor{vltRed}{HTML}{762B38}
\definecolor{vltYellow}{HTML}{D5AD3C}
\definecolor{vltNeutral}{HTML}{777D83}
\setlength{\unitlength}{1pt}
\scalebox{0.90}{%
\input{vlt-figures/figure-233.tex}%
}\endgroup
\caption{Finite-chain evolution from $u_i(0)/A=\delta_{i8}$ and zero velocity, with fifteen free-end sites and uniform pinning. (a) Samples at $\Delta(\omega_bt)=0.5$ evaluate $u_i(t)/A=\sum_s e_{si}e_{s8}\cos\omega_st$. Teal and burgundy encode displacement sign; hatches mark values below $-0.025$ for grayscale reading. (b) Two times reconstruct the same fifteen rigid virions on one displacement gain, with gray reference outlines and fixed coil families. The matrix and conserved initial energy are printed below. The plotted field is signed displacement; its spreading and return follow conservative phase evolution.}
\label{fig:added-exact-chain-spacetime}
\end{figure}

\subsubsection{The same cavity can show one resonance or many}

Apply Section~\ref{sec:operator-probe-response} to a calibrated length readout
$Y=\sum_i\ell_i u_i$, with dimensionless real $\ell_i$.
Its conjugate force $F$ enters through $-FY$ and supplies the actual
site forces $f_i=\ell_iF$. Defining
$g_\ell=\sum_i\ell_i e_{\ell,i}/\sqrt m$, the positive-pinning response is
\begin{equation}
 \mathcal R_{YY}^{\rm ret}(z)
       =\sum_{\ell=0}^{n-1}\frac{g_\ell^2}{\omega_\ell^2-z^2},
 \qquad \sum_\ell g_\ell^2=\frac{\sum_i\ell_i^2}{m},
 \qquad\operatorname{Im}z>0.
 \label{eq:phonon-finite-chain-probe-response}
\end{equation}
Completeness of the orthonormal modes gives the second identity: a fixed
calibrated probe redistributes inverse-inertia weight among modes while
preserving its sum. Specifically, $g_\ell^2$ is the mass at
$\omega_\ell$ of the positive measure
$\dd\mu_Y(\Omega)=(2\Omega/\pi)
\operatorname{Im}\mathcal R_{YY}^{\rm ret}(\Omega+i0)\dd\Omega$
on $\Omega>0$. These integrated spectral masses describe the lossless
lines; a broadened peak requires its own damping model.

For an end probe $Y=u_1$, these weights are
\begin{equation}
 m g_0^2=\frac1n,\qquad
 m g_{\ell>0}^2=\frac2n
                   \cos^2\!\left(\frac{\ell\pi}{2n}\right).
 \label{eq:phonon-chain-end-probe-weights}
\end{equation}
Every mode is visible at the end, although high-index weights can be
small. The normalized uniform probe $Y=\sum_i u_i/\sqrt n$ instead has
$g_\ell=\delta_{\ell0}/\sqrt m$: it detects only the common mode.
Both have inverse inertia $1/m$ because both probe vectors have unit
Euclidean norm. Figure~\ref{fig:finite-cavity-probe-selection} compares
their very different spectra without changing one spring or mass.
Seeing fewer peaks therefore need not mean that fewer modes exist.

\begin{figure}[H]
\centering\begingroup
\definecolor{vlfGreen}{HTML}{176B64}
\definecolor{vlfDarkRed}{HTML}{762B38}
\definecolor{vlfRed}{HTML}{762B38}
\definecolor{vlfBlue}{HTML}{315D80}
\definecolor{vlfCopper}{HTML}{762B38}
\setlength{\unitlength}{1pt}
\scalebox{0.80}{%
\input{vlt-figures/figure-234.tex}%
}\endgroup
\caption{Two conjugate force/readout channels on the same pinned eight-site cavity. Stems give inverse-inertia masses $\mu_s=|\vct\ell^{\mathsf T}\vct e_s|^2/m$. (a) The local readout $Y=u_1$ has $m\mu_0=1/8$ and $m\mu_\ell=\tfrac14\cos^2(\ell\pi/16)$ for $\ell>0$. (b) The normalized uniform readout $Y=\sum_i u_i/\sqrt8$ has only $m\mu_0=1$, with open markers at zero weights. Both total masses equal one; note the different ordinate ranges. Frequencies are unchanged, and stems represent integrated spectral weights.}
\label{fig:finite-cavity-probe-selection}
\end{figure}

Calibration fixes the apparent inertia. The physical centre coordinate
$u_{\rm cm}=\sum_i u_i/n$ has inverse inertia $1/(nm)$; its conjugate
force is the total force, distributed as $F/n$ per site. The normalized
uniform readout is $\sqrt n\,u_{\rm cm}$ and uses a correspondingly
rescaled conjugate force. Both descriptions retain the same material mass.

Two-end response carries additional pattern information. Since
$e_{\ell,n}=(-1)^\ell e_{\ell,1}$,
$\mathcal R_{n1}^{\rm ret}(z)
=\sum_\ell(-1)^\ell e_{\ell,1}^2/[m(\omega_\ell^2-z^2)]$.
These cross-channel residues can have either sign while each diagonal
spectral measure stays nonnegative. The sign records relative end motion;
it is not negative stored energy. Calibrated force at one supported
particle and tracking at another would select this cross-response only
if their contacts and observation maps realize those channels.

For the unpinned chain, omitting the free mode removes inverse-inertia
weight $1/(mn)$ from the end probe. The uniform probe has no positive-mode
weight at all, but its free-coordinate response is
$-1/(mz^2)$, or $t/m$ after a unit impulse for $t>0$.
That response was separated, not destroyed or converted into a phonon.

\subsubsection{A controlled comparison with a continuum branch}

Assign reference spacing $a$, cell-centred positions
$z_i=(i-1/2)a$ and interval length $L=na$. The continuum ends lie a
half-step beyond the end centres; $L$ is this matching convention, not
an independently inferred cylinder dimension. With
$c=a\sqrt{k_b/m}$ and $q_\ell=\ell\pi/L$, the exact finite frequencies obey
\begin{equation}
\begin{aligned}
 \omega_\ell^2
   &=\omega_{\rm pin}^2+
       \frac{4c^2}{a^2}\sin^2(q_\ell a/2)\\
   &=\omega_{\rm pin}^2+c^2q_\ell^2
       -\frac{c^2a^2q_\ell^4}{12}
       +O(c^2a^4q_\ell^6).
 \end{aligned}
 \label{eq:phonon-chain-continuum-dispersion}
\end{equation}
The first continuum terms approximate modes with $q_\ell a\ll1$.
The Neumann cosine on $0<z<L$ has the same allowed $q_\ell$;
sampling its unit-$L^2$ pattern $\phi^{\rm cont}_\ell$ gives
$e_{\ell,i}=\sqrt a\,\phi^{\rm cont}_\ell(z_i)$.
With line mass $\mu_{\rm line}=m/a$, both physical reconstructions then
agree: $e_{\ell,i}Q_\ell/\sqrt m
=\phi^{\rm cont}_\ell(z_i)Q_\ell/\sqrt{\mu_{\rm line}}$.
Thus the limit matches normalization as well as frequency.

At fixed $L,c,\mu_{\rm line}$, numerical refinement takes
$a=L/n$, $m=\mu_{\rm line}a$ and $k_b=\mu_{\rm line}c^2/a$.
Here $m$ and $k_b$ are continuum cell parameters. For actual fixed-size
virions, a continuum model instead approximates wavelengths long relative
to the supplied spacing; the discrete model retains the high-index modes.

Without pinning the long-wave excitation energy is
$\hbar\omega\simeq\hbar c|q|$, rather than
$\hbar^2q^2/(2m)$ with the virion mass inserted into a particle-box
formula. If $\omega_{\rm pin}>0$, $qa\ll1$ and
$c|q|\ll\omega_{\rm pin}$, the near-minimum expansion instead has a
quadratic envelope dispersion. Its effective mass is a property of that
dispersion, rather than the mass of a transported virion. The derivation,
packet propagation and velocity are developed in
Section~\ref{sec:phonon-packets-conversion}.

The infinite uniform discrete chain has the finite squared-frequency
band $[\omega_{\rm pin}^2,\omega_{\rm pin}^2+4k_b/m]$.
An above-band stiff-defect mode can therefore be a lattice effect missed
by the long-wave continuum branch, which has no upper edge. The discrete
defect below makes that limitation explicit; the subsequent spatial well
uses a softer region and an exterior gap.

\subsubsection{An exact defect mode above the lattice band}
\label{sec:phonon-discrete-defect}

The upper band edge permits a complementary localization mechanism with
no continuum counterpart in the scalar long-wave model. Localized defect
vibrations are an established part of lattice dynamics
\cite{Bjork1957,Teschl2000}. Here the following elementary example fixes
the particular defect and solves its shape as well as its frequency.
Extend the uniform reference to sites $j\in\mathbb Z$, write
$g_b=k_b/m>0$, and add the conservative support energy
$\Delta k_d u_0^2/2$ at one site, with
$v_d=\Delta k_d/m>0$. The mass-weighted operator becomes
\begin{equation}
 (\mathsf D_d e)_j=(\omega_{\rm pin}^2+2g_b)e_j
       -g_b(e_{j-1}+e_{j+1})+v_d\delta_{j0}e_0.
 \label{eq:phonon-single-support-defect}
\end{equation}
This is an onsite support change. It is not a changed intersite bond,
which would change several Hessian entries and preserve common
translation in an unpinned network.

Put $e_j=C_d r_d^{|j|}$. Away from the defect the eigenvalue is
$\lambda_d=\omega_{\rm pin}^2+2g_b-g_b(r_d+r_d^{-1})$.
The equation at $j=0$ requires $v_d=g_b(r_d-r_d^{-1})$.
Choosing the root of magnitude below one gives
\begin{equation}
 \boxed{\begin{aligned}
 r_d&=\frac{v_d-\sqrt{v_d^2+4g_b^2}}{2g_b}\in(-1,0),\\
 \omega_d^2=\lambda_d
   &=\omega_{\rm pin}^2+2g_b+\sqrt{v_d^2+4g_b^2},\\
 C_d&=\sqrt{\frac{1-r_d^2}{1+r_d^2}},\qquad
       \sum_{j\in\mathbb Z}|e_j|^2=1.
 \end{aligned}}
 \label{eq:phonon-single-defect-localized-solution}
\end{equation}
Thus every positive support increment gives an eigenfrequency above the
uniform squared-frequency band. Neighboring sites alternate sign; the
envelope decays exponentially. For spacing $a$, its amplitude decay
length and small-defect separation from the band are
\begin{equation}
 \ell_d=\frac{a}{\operatorname{arsinh}(v_d/2g_b)},\qquad
 \lambda_d-(\omega_{\rm pin}^2+4g_b)
       =\frac{v_d^2}{4g_b}+O(v_d^4/g_b^3).
 \label{eq:phonon-single-defect-length-and-gap}
\end{equation}
At weak contrast the length diverges and the spectral separation
vanishes quadratically. Calling a finite-chain pattern localized then
requires enough distance to the ends relative to $\ell_d$ and a resolved
separation from its other modes. At $v_d=0$ the limiting band-edge
sequence is not square summable on the infinite chain. A finite cavity
still has its ordinary extended standing modes.

Figure~\ref{fig:phonon-discrete-defect-localization} displays this
discrete benchmark. The next calculation gives a way to assess its
finite-domain approximation without judging a mode only by its appearance.

\begin{figure}[H]
\centering\begingroup
\definecolor{vlfGreen}{HTML}{176B64}
\definecolor{vlfDarkRed}{HTML}{762B38}
\definecolor{vlfRed}{HTML}{762B38}
\definecolor{vlfBlue}{HTML}{315D80}
\definecolor{vlfCopper}{HTML}{762B38}
\setlength{\unitlength}{1pt}
\scalebox{0.80}{%
\input{vlt-figures/figure-235.tex}%
}\endgroup
\caption{A discrete support defect and its localized mode. In \eqref{eq:phonon-single-defect-localized-solution}, set $\omega_{\rm pin}^2/g_b=1$ and $v_d/g_b=2$. (a) The normalized infinite-chain coefficients are $e_j=C_d(1-\sqrt2)^{|j|}$, with alternating signs and amplitude decay length $\ell_d/a=1/\operatorname{arsinh}1\simeq1.135$. Displayed endpoints merely delimit the samples. (b) The uniform squared-frequency band is $1\le\lambda/g_b\le5$; the marked defect value is $3+\sqrt8$. Every positive defect has an eigenvalue above this band. Its open zero-defect limit is a band edge rather than a normalizable localized mode. Coefficients and squared frequencies have their stated axes; neither is an occupation probability. The extra local support is a stipulated mechanical input.}
\label{fig:phonon-discrete-defect-localization}
\end{figure}

\paragraph{Oriented bodies can mix two cavity patterns.}
The scalar cavity is one projection of the body mechanics. For an
HPV-inspired array, each whole L1 VLP can also rotate while its surface
patches load neighbouring bodies. Use the constrained body inertia of
\eqref{eq:s2patch-generalized-inertia}; first-order populations remain
separate retained states. Project the frozen conservative Hessian onto
two real inertia-orthonormal patterns: one predominantly translational,
the other predominantly librational. Their mass-normalized coordinates
$\vct Q_{\rm tr}=(Q_t,Q_r)^{\mathsf T}$ share the usual oscillator units.
An invariant pair gives an exact sector; a truncated pair also tests
discarded coupling using the residual construction below.

\begin{definition}[A two-pattern body projection]
\label{def:s7hpv-two-pattern-projection}
The declared projection has energy
$E_{\rm tr}=(\dot{\vct Q}_{\rm tr}^{\mathsf T}
\dot{\vct Q}_{\rm tr}+\vct Q_{\rm tr}^{\mathsf T}
\mathsf D_{\rm tr}\vct Q_{\rm tr})/2$, where
\begin{equation}
 \mathsf D_{\rm tr}=
 \begin{pmatrix}\omega_t^2&g_{tr}\\g_{tr}&\omega_r^2\end{pmatrix},
 \quad \omega_t^2,\omega_r^2>0,
 \quad g_{tr}^2<\omega_t^2\omega_r^2 .
 \label{eq:s7hpv-translation-libration-block}
\end{equation}
The cross coefficient $g_{tr}$ has squared-frequency units and is the
projected translation--rotation curvature of the assembled interaction.
\end{definition}

\begin{proposition}[Pattern exchange and calibrated visibility]
\label{prop:s7hpv-pattern-exchange}
Put $\bar\lambda=(\omega_t^2+\omega_r^2)/2$ and
$\rho=[(\omega_t^2-\omega_r^2)^2/4+g_{tr}^2]^{1/2}>0$.
The squared frequencies and conservative response of a real calibrated
channel $Y=\vct h^{\mathsf T}\vct Q_{\rm tr}$ are
\begin{equation}
 \boxed{\begin{aligned}
 \lambda_\pm&=\bar\lambda\pm\rho,\qquad
 \mathsf P_\pm=\tfrac12
  [\Id\pm(\mathsf D_{\rm tr}-\bar\lambda\Id)/\rho],\\
 \mathcal R_Y(\omega)&=\sum_{\sigma=\pm}
   \frac{\vct h^{\mathsf T}\mathsf P_\sigma\vct h}
                         {\lambda_\sigma-\omega^2} .
 \end{aligned}}
 \label{eq:s7hpv-avoided-crossing-readout}
\end{equation}
\end{proposition}
\begin{proof}
The traceless part of $\mathsf D_{\rm tr}$ squares to $\rho^2\Id$.
Its two projectors give both the eigenvalues and inverse response.
\end{proof}
For a synthetic sweep take $\omega_{t,r}^2=\omega_\star^2(1\mp\delta)$,
$g_{tr}=0.2\omega_\star^2$ and $|\delta|\le0.6$. At $\delta=0$ the
frequencies are $\sqrt{0.8}\omega_\star$ and $\sqrt{1.2}\omega_\star$.
A centre channel with $\vct h=(h_\star,0)$ overlaps both patterns
equally; a registered marker with calibrated $\vct h=(h_\star,h_\star)$
is dark to the lower antisymmetric pattern and bright to the upper one.
The numerator is a squared mode overlap; the denominator and acquisition
law determine the measured line or response. Changing orientation can
therefore redistribute visible motion as well as shift frequencies.
These assigned coefficients demonstrate classical pattern mixing.
The damping and memory tests below determine whether an HPV-inspired
preparation supports resolvable oscillation; phonon occupation remains
the separate quantization postulate.

Figure~\ref{fig:s7hpv-pattern-mixing} separates a change of collective pattern from a change of its visibility.

\begin{figure}[H]
\centering\begingroup
\setlength{\unitlength}{1pt}
\scalebox{0.90}{%
\input{vlt-figures/figure-300.tex}%
}\endgroup
\caption{Classical translation--libration mixing and detector selection. The projected block uses $\omega_{t,r}^2/\omega_\star^2=1\mp\delta$ and $g_{tr}/\omega_\star^2=0.2$ on $|\delta|\leq0.6$. Panel (a) diagonalizes this positive conservative matrix. Panel (b) gives the two projector weights for the calibrated marker row $h_\star(1,1)$; their sum is two, and the lower pattern has exactly zero overlap at $\delta=0$. The glyphs identify the two basis directions, not independent particle trajectories or a force-normalized displacement. The coefficients and marker row are assigned projected inputs. Damping, material response and the acquisition map determine actual records; the plotted conservative frequencies do not establish an HPV oscillatory resonance. See~\eqref{eq:s7hpv-avoided-crossing-readout}.}
\label{fig:s7hpv-pattern-mixing}
\end{figure}

\paragraph{A prepared cavity inside the fixed particle pool.}
The cavity count $n$ specifies one persistent mechanical reference.
During recruitment, the same $N_{\rm tot}$ physical particles continue
moving while the selected cavity or contact group changes. An outsider
can already have displacement, velocity and correlated material strain
before it joins. The reference states and compatible cell maps of
Sections~3--6 retain this preparation.

\begin{definition}[Prepared displacement retained by a cavity band]
\label{def:s7-cavity-prepared-retention}
Embed a chosen mass-orthonormal displacement band $\mathsf\Phi_g$
in the same full-pool laboratory coordinates for every episode $g$.
For a registered length-valued pattern $\vct U$, define
$\mathsf P_g=\mathsf\Phi_g\mathsf\Phi_g^\dagger\mathsf M$,
$\vct c_g=\mathsf\Phi_g^\dagger\mathsf M\vct U$ and
$\vct r_g=(\Id-\mathsf P_g)\vct U$. At nonzero pattern norm, its
retained displacement fraction is
$\chi_g=\|\mathsf P_g\vct U\|_M^2/\|\vct U\|_M^2$.
\end{definition}

\begin{proposition}[Incoming motion in a changed cavity band]
\label{prop:s7-cavity-incoming-pattern}
For the same physical pattern, after any reference-centre offset has
been registered, put
$\mathsf O_{hg}=\mathsf\Phi_h^\dagger\mathsf M\mathsf\Phi_g$.
Then
\begin{equation}
 \vct c_h=\mathsf O_{hg}\vct c_g+
                 \mathsf\Phi_h^\dagger\mathsf M\vct r_g,
 \qquad
 \|\vct c_h-\mathsf O_{hg}\vct c_g\|_2\le\|\vct r_g\|_M.
 \label{eq:s7-cavity-incoming-pattern}
\end{equation}
\end{proposition}
\begin{proof}
Insert $\vct U=\mathsf\Phi_g\vct c_g+\vct r_g$.
The coefficient map of a mass-orthonormal band is a contraction in
the mass norm, giving the inequality.
\end{proof}

For an explicit three-particle example, use equal masses $m$ and identical
onsite supports. The old selected pair has common pattern
$\vct\phi_g=(1,1,0)^{\mathsf T}/\sqrt{2m}$; after a connection,
the three-member common pattern is
$\vct\phi_h=(1,1,1)^{\mathsf T}/\sqrt{3m}$. For the registered
displacement $(A,A,B)^{\mathsf T}$,
\begin{equation}
 c_g=\sqrt{2m}\,A,\qquad O_{hg}=\sqrt{2/3},\qquad
 c_h=\sqrt{m/3}(2A+B).
 \label{eq:s7-cavity-three-particle-registration}
\end{equation}
The old common coefficient alone misses $\sqrt{m/3}\,B$. The new
relative patterns retain the remaining displacement, and the full
state also retains all momenta and medium variables. The example
registers a selected pattern; an actual connection additionally uses
its force law and source--medium energy account. Membership selection
by itself performs no work.

The same construction applies to a well-localized band. A large frozen
$w_s^{\rm in}$ establishes a concentrated available pattern, while
$\chi_g(t)$ tests the actual continued displacement preparation against
the chosen band. Neither is an occupation probability or an energy
fraction. Their readout consequences use the detector bounds of
Section~\ref{sec:s5-register-patterns} and the mode certificates below.
After a physical interface or contact change, propagate the registered
state before assessing these overlaps again. A replacement by a newly
relaxed state would answer a different preparation problem.

Figure~\ref{fig:s7-cavity-incoming-band-motion} resolves the incoming displacement retained by the cavity-band registration.

\begin{figure}[H]
\centering\begingroup
\definecolor{vltTeal}{HTML}{176B64}
\definecolor{vltBlue}{HTML}{315D80}
\definecolor{vltGold}{HTML}{A87520}
\definecolor{vltPlum}{HTML}{725877}
\definecolor{vltRed}{HTML}{762B38}
\definecolor{vltYellow}{HTML}{D5AD3C}
\definecolor{vltNeutral}{HTML}{777D83}
\setlength{\unitlength}{1pt}
\scalebox{0.90}{%
\input{vlt-figures/figure-236.tex}%
}\endgroup
\caption{Registering incoming displacement across a band change. (a) The same three equal masses carry $U=(A,A,B)^{\mathsf T}$, shown at $B=-A$. Shared rulers locate reference spacing $a$; arrows and matched coil families encode one displacement gain. (b) The complete new-band coefficient is $c_h=\sqrt{m/3}(2A+B)$, differing from the old-band projection by $\sqrt{m/3}B$. At the marked slice, the normalized coefficients are $1/\sqrt3$ and $2/\sqrt3$. (c) Mass-normalized patterns give the overlap and residual in \eqref{eq:s7-cavity-three-particle-registration}. These are displacement projections. The separate physical connection law determines forces and work.}
\label{fig:s7-cavity-incoming-band-motion}
\end{figure}

\subsubsection{Certifying a mode shape and its observable weight}
\label{sec:phonon-mode-certification}

A converged frequency does not by itself certify a predicted displacement
pattern. Consider a finite conservative reference with the declared
positive inertia $\mathsf M$ and real symmetric
$\mathsf D=\mathsf M^{-1/2}\mathsf K\mathsf M^{-1/2}$. A trial
displacement $\widetilde{\vct\phi}$ is normalized by
$\widetilde{\vct\phi}^{\mathsf T}\mathsf M
\widetilde{\vct\phi}=1$. Set
$\widetilde{\vct e}=\mathsf M^{1/2}\widetilde{\vct\phi}$ and compute
\begin{equation}
 \widetilde\lambda=\widetilde{\vct e}^{\mathsf T}
                         \mathsf D\widetilde{\vct e},\qquad
 \vct r_{\rm eig}=(\mathsf D-\widetilde\lambda\Id)
                         \widetilde{\vct e},\qquad
 \varepsilon_{\rm eig}=\|\vct r_{\rm eig}\|_2.
 \label{eq:phonon-mode-residual-definition}
\end{equation}
Both $\widetilde\lambda$ and $\varepsilon_{\rm eig}$ have
squared-frequency units. Expansion in the orthonormal exact modes proves
$\operatorname{dist}(\widetilde\lambda,\sigma(\mathsf D))
\le\varepsilon_{\rm eig}$: the squared residual is the sum of
$|\lambda_j-\widetilde\lambda|^2$ times the trial spectral weights.
This is a spectral residual test, with the same principle applying to a
self-adjoint continuum operator on its actual domain~\cite{Teschl2009}.

\begin{proposition}[A selected mode and its calibrated probe weight]
\label{prop:s7-mode-probe-certificate}
Suppose one simple target eigenvalue $\lambda_s$ is identified and the
remaining spectrum is separated from $\widetilde\lambda$ by
$\delta_{\rm sep}=\min_{j\ne s}|\lambda_j-\widetilde\lambda|>0$.
Projection of that same residual onto the complement gives
\begin{equation}
 \begin{aligned}
 \sqrt{1-|\vct e_s^{\mathsf T}\widetilde{\vct e}|^2}
    &\le\frac{\varepsilon_{\rm eig}}{\delta_{\rm sep}},\\
 |\widetilde\omega-\omega_s|
    &\le\frac{\varepsilon_{\rm eig}}{\widetilde\omega+\omega_s}.
 \end{aligned}
 \label{eq:phonon-mode-angle-and-frequency-bound}
\end{equation}
The frequency inequality assumes positive frequencies and that the
eigenvalue within the residual interval is the target. The angle bound
is useful only with a justified lower bound on the separating gap;
spacing between approximate eigenvalues alone need not supply it.
Residual and gap bounds are central to invariant-subspace perturbation
theory~\cite{DavisKahan1970}.

This angle controls a directly relevant observable. For the calibrated
probe $Y=\vct\ell^{\mathsf T}\vct u$, put
$\vct h=\mathsf M^{-1/2}\vct\ell$ and
$g_s=\vct h^{\mathsf T}\vct e_s$. The difference of two rank-one
orthogonal projectors has norm equal to the sine of their angle, so
\begin{equation}
 \bigl||\vct h^{\mathsf T}\widetilde{\vct e}|^2-|g_s|^2\bigr|
 \le\|\vct h\|_2^2
       \min\!\left(1,\frac{\varepsilon_{\rm eig}}{\delta_{\rm sep}}\right).
 \label{eq:phonon-mode-probe-weight-certificate}
\end{equation}
\end{proposition}
\begin{proof}
Expand the trial vector in the exact orthonormal eigenbasis.
The complementary coefficients are bounded by their residual divided
by the stated separation. The identified eigenvalue lies in its
residual interval; factoring the difference of squared frequencies
gives the frequency bound. Finally apply the rank-one projector
angle identity to the calibrated probe vector.
\end{proof}
It bounds an absolute error in spectral weight. A weak predicted line
may still have a large relative error or become effectively dark.

Near degeneracy, the stable object is often a cluster rather than a
named eigenvector. Let $\mathsf U^{\mathsf T}\mathsf U=\Id$, let
$\mathsf H_R=\mathsf U^{\mathsf T}\mathsf D\mathsf U$, and let
$\mathsf F_R=\mathsf D\mathsf U-\mathsf U\mathsf H_R$.
If an exact target projector $\mathsf P_{\mathcal C}$ has the same
rank as $\mathsf U$, and every eigenvalue of $\mathsf H_R$ is at
distance at least $\delta_{\mathcal C}>0$ from the exact complementary
spectrum, then diagonalizing $\mathsf H_R$ and summing the squared
complement residuals gives
\begin{equation}
 \begin{aligned}
 \|(\Id-\mathsf P_{\mathcal C})\mathsf U\|_F
       &\le\frac{\|\mathsf F_R\|_F}{\delta_{\mathcal C}},\\
 |\vct h^{\mathsf T}(\mathsf U\mathsf U^{\mathsf T}
             -\mathsf P_{\mathcal C})\vct h|
       &\le\|\vct h\|_2^2
           \min\!\left(1,\frac{\|\mathsf F_R\|_F}{\delta_{\mathcal C}}\right).
 \end{aligned}
 \label{eq:phonon-cluster-residual-and-readout}
\end{equation}
Here $\|\cdot\|_F$ is the Frobenius norm. Cluster-integrated probe
weight can remain reliable when individual modes rotate strongly within
the cluster. These tests quantify numerical or projection error against
a specified reference operator. They cannot validate an incorrect force
law. Comparing different biological resolutions additionally requires a
common coordinate map, inertia, boundary and readout; atomistic and
elastic-network capsid modes need not agree equally well throughout
their spectra~\cite{Dykeman2010}. Equilibrium coarse-graining and
dynamical closure also impose different requirements
\cite{Noid2008,Mori1965}.

\FloatBarrier

\subsection{A spatial stiffness well and its bound modes}
\label{sec:phonon-spatial-well}

\subsubsection{A local restoring contrast defines the problem}

The finite chain confines motion through its ends. A local stiffness
contrast can also concentrate a mode in one region while its surroundings
move. Engineered phononic structures realize this mechanism through
interaction and geometric contrasts~\cite{Eichenfield2009}. The scalar
well below makes the mechanism explicit: a biological application then
assigns the softer coordinate, its inertia and its surrounding coupling.

Take a lossless axial branch with constant effective line mass
$\mu_{\rm line}>0$ and wave speed $c>0$ on $-B<z<B$.
For a length-valued displacement $u$, set
$\xi=\sqrt{\mu_{\rm line}}\,u$. Its real quadratic energy and equation
of motion are
\begin{equation}
 \begin{aligned}
 E_2&=\frac12\int_{-B}^B
       \left[\Pdt{\xi}^2+c^2(\partial_z\xi)^2+
                           \omega_g^2(z)\xi^2\right]\dd z,\\
 \partial_t^2\xi+\mathcal D_{\rm ax}\xi&=0.
 \end{aligned}
 \label{eq:phonon-spatial-well-energy}
\end{equation}
The three terms store kinetic energy, energy of relative displacement,
and local restoring energy. The mass-rescaled field $\xi(z,t)$ records
displacement at each material location. Its coefficient $\omega_g^2(z)$
represents an onsite curvature per unit inertia or a resolved
internal-branch gap. Choose
\begin{equation}
 \begin{aligned}
 \mathcal D_{\rm ax}&=-c^2\partial_z^2+\omega_g^2(z),\\
 \omega_g(z)&=
 \begin{cases}
   \omega_{\rm in},&|z|<a_w,\\
   \omega_{\rm out},&a_w<|z|<B,
 \end{cases}
 \quad 0\leq\omega_{\rm in}<\omega_{\rm out},\quad 0<a_w<B.
 \end{aligned}
 \label{eq:phonon-spatial-well-operator}
\end{equation}
The central region costs less restoring energy for the same local
displacement. The curvature contrast and gradient energy together determine the
degree of spatial concentration. The ends obey the declared
free-traction condition $\partial_z\xi=0$; particle reflection at an
inactive hard cylindrical wall does not supply this material condition.

This continuum description retains one mechanical branch. Its widths,
penetration lengths and wavelengths must be resolved relative to the
underlying spacing; the rectangular interface replaces an unresolved
transition layer. A capsid shape, an envelope--RNP coordinate and
whole-virion array displacement each require their own effective line mass
and stiffness assignment.

Figure~\ref{fig:phonon-spatial-well} previews one fully specified curvature profile and its matched modes; the domain, eigenvalue matching and energy fractions are derived below.

\subsubsection{The domain, energy flux and positive oscillator sector}

On $L^2(-B,B)$, the operator domain is
\begin{equation}
 \mathcal D(\mathcal D_{\rm ax})=
 \{\phi\in H^2(-B,B):\phi'(-B)=\phi'(B)=0\}.
 \label{eq:phonon-well-selfadjoint-domain}
\end{equation}
Here $H^2$ requires the function and its first two weak derivatives to
be square integrable; in one dimension it also gives a continuous
function and first derivative. Thus the interface preserves $\phi$
and $\phi'$ even though the second derivative changes across the step.
Equivalently, the real closed energy form has domain $H^1(-B,B)$ and is
\begin{equation}
 \mathfrak a_{\rm ax}[\phi]=\int_{-B}^B
       [c^2|\phi'|^2+\omega_g^2|\phi|^2]\,\dd z.
 \label{eq:phonon-well-positive-form}
\end{equation}
The natural endpoint condition of this variational problem is Neumann.
The bounded real curvature step is a self-adjoint perturbation of that
Neumann Laplacian. Its finite-interval resolvent is compact, so its real
eigenvalues are discrete and the orthonormal modes form a complete
spatial basis~\cite{Teschl2009}. These separated one-dimensional boundary
conditions also give simple eigenvalues, with alternating even and odd
parity for this symmetric profile.

Integration by parts shows
$\langle\phi,\mathcal D_{\rm ax}\phi\rangle
=\mathfrak a_{\rm ax}[\phi]$.
For nonzero $\phi$, the form minus
$\omega_{\rm in}^2\|\phi\|^2$ is strictly positive: it could vanish
only for a constant function that also vanishes in the nonempty exterior.
Hence every mode satisfies $\omega_s^2>\omega_{\rm in}^2\geq0$,
including when the inner curvature itself is zero. The stiffer exterior
and the gradient term then supply the positive restoring cost. This is
a stable soft region, rather than a negative-curvature instability.

The local energy density $e$ in~\eqref{eq:phonon-spatial-well-energy}
obeys
\begin{equation}
 \partial_t e+\partial_z J_E=0,\qquad
 J_E=-c^2(\partial_t\xi)(\partial_z\xi).
 \label{eq:phonon-well-energy-flux}
\end{equation}
Continuity of displacement and traction makes this flux continuous at
the interface, and $J_E(\pm B)=0$ closes the energy account. The
exterior tail of a lossless standing mode is therefore not an escape
rate or a dissipative linewidth. If line mass or axial elastic
coefficient changes, one must return to displacement continuity and
traction continuity in the corresponding weighted eigenproblem;
matching a mass-rescaled field derivative would generally be incorrect.

Normalize $\int_{-B}^B|\phi_s|^2\dd z=1$ and expand
$\xi=\sum_s\phi_sQ_s$. Then $[\phi_s]=\mathrm m^{-1/2}$,
$[\xi]=\mathrm{kg}^{1/2}\mathrm m^{1/2}$ and
$[Q_s]=\mathrm{kg}^{1/2}\mathrm m$ as in the finite model.
The eigenproblem
$\mathcal D_{\rm ax}\phi_s=\omega_s^2\phi_s$ selects a squared
frequency. Quantizing an assigned retained mode adds
$\hbar\omega_s$ per occupation. The continuum's infinite mathematical
basis is not an assertion of infinitely many resolved biological
coordinates: a physical coarse description retains its justified band.
In particular, an absolute sum of all continuum vacuum energies requires
a cutoff or a separate regularization, whereas finite retained-mode
excitation differences are well defined.

\begin{figure}[H]
\centering
\begingroup
\definecolor{vlfGreen}{HTML}{176B64}
\definecolor{vlfDarkRed}{HTML}{762B38}
\definecolor{vlfRed}{HTML}{762B38}
\definecolor{vlfBlue}{HTML}{315D80}
\definecolor{vlfCopper}{HTML}{762B38}

\setlength{\unitlength}{1pt}
\scalebox{0.80}{%
\input{vlt-figures/figure-237.tex}%
}%
\endgroup
\caption{Finite stiffness-well modes at $B=3\ell_\star$, $a_w=\ell_\star$, $c=\ell_\star\omega_\star$, $\omega_{\rm in}=\omega_\star$ and $\omega_{\rm out}=3\omega_\star$, with free-traction ends. (a) The stepped curvature supports even and odd squared frequencies $2.3259141083$ and $5.9797605939$ in units $\omega_\star^2$. (b) Matched trigonometric interiors and Neumann $\cosh$ tails have unit full-domain norm. Central shape fractions are $0.953717,0.770383$; mechanical-energy fractions are $0.821001,0.655932$, including kinetic, strain and onsite storage. Endpoint/interface amplitude ratios are $0.011405,0.061817$. Each quantity uses its stated normalization.}
\label{fig:phonon-spatial-well}
\end{figure}

\subsubsection{Exact matching and how many subthreshold modes exist}

For $\omega_{\rm in}^2<\omega^2<\omega_{\rm out}^2$, define
\begin{equation}
 k_w=\frac{\sqrt{\omega^2-\omega_{\rm in}^2}}{c},
 \qquad
 q_w=\frac{\sqrt{\omega_{\rm out}^2-\omega^2}}{c}.
 \label{eq:phonon-well-wavenumbers}
\end{equation}
The inner solution is $A\cos(k_wz)$ for even parity and
$A\sin(k_wz)$ for odd parity. In the right exterior it is
$C\cosh[q_w(B-z)]$, which has zero derivative at $B$.
The left exterior follows by parity. Matching the value and derivative
at $a_w$ gives
\begin{align}
 k_w\tan(k_wa_w)&=q_w\tanh[q_w(B-a_w)]&&\text{(even)},\nonumber\\
 -k_w\cot(k_wa_w)&=q_w\tanh[q_w(B-a_w)]&&\text{(odd)}.
 \label{eq:phonon-finite-well-matching}
\end{align}
The matching has the form of a square-well spectral problem, with
inverse-time-squared operator units and eigenvalue $\omega^2$.
Section~\ref{sec:phonon-packets-conversion} derives the additional band
expansion that gives a particle-like envelope Hamiltonian.

\begin{proposition}[Strict subthreshold count for the finite well]
\label{prop:s7-subthreshold-count}
The matching equations also give a useful exact count. Set
$\Delta_g=\omega_{\rm out}^2-\omega_{\rm in}^2>0$ and
$\rho_w=a_w\sqrt{\Delta_g}/c$. \begin{equation}
 \boxed{N_{\rm sub}=\left\lceil\frac{2\rho_w}{\pi}\right\rceil,
      \qquad \rho_w>0,}
 \label{eq:phonon-well-subthreshold-count}
\end{equation}
where the ceiling means the smallest integer at least as large as its
argument. An equality at $\rho_w=j\pi/2$ is a threshold mode with
$q_w=0$ and is excluded from the strict subthreshold count.
There is always an even ground mode; the first odd subthreshold mode
requires $\rho_w>\pi/2$.
The count does not depend on $B-a_w>0$, although the frequencies and
shapes do. It counts existing cavity modes below the chosen exterior
threshold, not new coordinates or newly occupied phonons.
\end{proposition}
\begin{proof}
With $x=k_wa_w$, the right side,
multiplied by $a_w$, is
$\sqrt{\rho_w^2-x^2}\tanh[(B-a_w)
\sqrt{\rho_w^2-x^2}/a_w]$.
It decreases with $x$. The positive branches of $x\tan x$ and
$-x\cot x$ alternate, beginning at $x=j\pi/2$, and increase from
zero to their next pole. Each branch whose starting point is below
$\rho_w$ has exactly one crossing.
The crossings are counted by the strict inequality
$j\pi/2<\rho_w$ for integers $j\geq0$, giving the ceiling
with the stated threshold convention.
\end{proof}

Figure~\ref{fig:finite-well-matching-and-strict-count} makes the matching intersections and strict threshold endpoints explicit.

\begin{figure}[H]
\centering
\begingroup
\definecolor{vlfGreen}{HTML}{176B64}
\definecolor{vlfDarkRed}{HTML}{762B38}
\definecolor{vlfRed}{HTML}{762B38}
\definecolor{vlfBlue}{HTML}{315D80}
\definecolor{vlfCopper}{HTML}{762B38}

\setlength{\unitlength}{1pt}
\scalebox{0.80}{%
\input{vlt-figures/figure-238.tex}%
}%
\endgroup
\caption{Exact branch matching and the strict count of subthreshold modes. (a) With $\rho_w=\sqrt8$ and $(B-a_w)/a_w=2$, intersect $\sqrt{\rho_w^2-x^2}\tanh[2\sqrt{\rho_w^2-x^2}]$ with the positive branches $x\tan x$ and $-x\cot x$. The roots $1.15148344,2.23153772$ give the preceding well frequencies. Curves stop at their true domains; dotted guides locate the pole and outer cutoff. (b) In the counting variable $r_w=\rho_w/\pi>0$, $N_{\rm sub}=\lceil2r_w\rceil$. Filled/open endpoints distinguish the count at equality from the higher count immediately to its right. A threshold mode with $q_w=0$ is excluded from a strict subthreshold count, as is zero contrast. The step plot counts modes; their frequencies and attenuation lengths still vary within each step. The boxed result illustrated here is~\eqref{eq:phonon-well-subthreshold-count}.}
\label{fig:finite-well-matching-and-strict-count}
\end{figure}

\subsubsection{Existence, localization and energy concentration}

\begin{definition}[Subthreshold mode and endpoint localization diagnostic]
\label{def:s7-subthreshold-localization}
All modes of the finite interval are end-confined. Here a
\emph{subthreshold} mode means one whose exterior spatial solution is
hyperbolic rather than oscillatory. Strong central localization asks
more: its exterior must span several penetration lengths. In this model,
\begin{equation}
 \frac{|\phi(B)|}{|\phi(a_w)|}
       =\operatorname{sech}[q_w(B-a_w)].
 \label{eq:phonon-well-localization-ratio}
\end{equation}
This ratio compares two amplitudes; it is neither a probability nor
the fraction of total energy outside. Its smallness requires
$q_w(B-a_w)\gg1$.
\end{definition} In that limit the finite-end matching approaches
the decaying-exterior matching of an infinite environment.

The constant normalized trial function has Rayleigh quotient
$\omega_{\rm in}^2+(1-a_w/B)\Delta_g
<\omega_{\rm out}^2$, independently confirming a subthreshold ground
mode. Yet a weak well need not localize it. At fixed $B,a_w,c$ and
$\omega_{\rm in}$, for $\Delta_g B^2/c^2\ll1$,
\begin{equation}
 \omega_0^2=\omega_{\rm in}^2+(1-a_w/B)\Delta_g
                 +O(\Delta_g^2B^2/c^2),\qquad
 w_0^{\rm in}=a_w/B+O(\Delta_gB^2/c^2).
 \label{eq:phonon-well-weak-contrast-limit}
\end{equation}
The shape is then nearly uniform and the endpoint ratio tends to one.
Existence below a comparison threshold is insufficient evidence of
central localization. At the opposite, high-exterior-barrier limit,
the inner pattern approaches the ground mode with fixed-displacement
interfaces, giving
$\omega_0^2\to\omega_{\rm in}^2+c^2\pi^2/(4a_w^2)$.
This finite gradient cost remains even when the exterior motion is tiny.
The physical ends remain Neumann; the effective inner interface condition
arises as a limit of the coupling.

The shape weight
$w_s^{\rm in}=\int_{-a_w}^{a_w}|\phi_s|^2\dd z$ quantifies central
mass-normalized displacement concentration. Its energy interpretation
requires the other terms in the Hamiltonian. For a real single mode,
the cycle-averaged classical energy fraction is
\begin{equation}
 f_{E,s}^{\rm in}
 =\frac12 w_s^{\rm in}
  +\frac{1}{2\omega_s^2}\int_{-a_w}^{a_w}
       [c^2|\phi_s'|^2+\omega_{\rm in}^2|\phi_s|^2]\,\dd z.
 \label{eq:phonon-well-central-energy-fraction}
\end{equation}
Half the total cycle-averaged energy is kinetic; the other half includes
both local restoring and inter-region strain storage. For a subthreshold
even or odd mode, integration by parts gives
$f_{E,s}^{\rm in}=w_s^{\rm in}
-c^2q_w\tanh[q_w(B-a_w)]|\phi_s(a_w)|^2/\omega_s^2$.
The central energy fraction is consequently smaller than the shape
weight in this example. The same $f_{E,s}^{\rm in}$ is the fraction of
the vacuum-subtracted energy of a number-state excitation in this one
selected mode. An instantaneous classical fraction also depends on its
phase and need not equal the cycle average.

\begin{example}[Dimensionless confined viral-phonon modes]
Choose scales $\ell_\star,\omega_\star$ and set
$B=3\ell_\star$, $a_w=\ell_\star$,
$c=\ell_\star\omega_\star$, $\omega_{\rm in}=\omega_\star$,
and $\omega_{\rm out}=3\omega_\star$.
Then $\rho_w=\sqrt8$ lies between $\pi/2$ and $\pi$, so there are
exactly two strict subthreshold roots:
\begin{equation}
 \frac{\omega_0^2}{\omega_\star^2}\simeq2.325914,
 \qquad
 \frac{\omega_1^2}{\omega_\star^2}\simeq5.979761.
 \label{eq:phonon-well-worked-roots}
\end{equation}
They have even and odd parity respectively. Matching and normalizing
the complete inner and exterior solutions gives
\begin{center}
\small
\begin{tabular}{lccc}
\hline
Mode & $w_s^{\rm in}$ & $f_{E,s}^{\rm in}$
     & $|\phi_s(B)/\phi_s(a_w)|$\\
\hline
Even & $0.953717$ & $0.821001$ & $0.011405$\\
Odd  & $0.770383$ & $0.655932$ & $0.061817$\\
\hline
\end{tabular}
\end{center}
Thus the lowest mode has about $95\%$ of its squared shape inside,
but about $82\%$ of its cycle-averaged energy there. The tail's gradient
and larger restoring curvature account for the distinction. One
occupation adds $\hbar\omega_s$ over the full mode, with the energy
fraction just calculated. The parameters are dimensionless illustrative
inputs, not fitted viral frequencies or a measured capsid geometry.
\end{example}

Figure~\ref{fig:phonon-spatial-well} displays the exact curvature
profile, the two eigenvalues and their normalized matched shapes.

\subsubsection{Sensitivity, observation and the choice of variables}

At fixed inertia, $c$, endpoints and interface location, differentiating
a normalized simple eigenmode gives
\begin{equation}
 \frac{\partial\omega_s^2}{\partial\omega_{\rm in}^2}
       =w_s^{\rm in},\qquad
 \frac{\partial\omega_s^2}{\partial\omega_{\rm out}^2}
       =1-w_s^{\rm in}.
 \label{eq:phonon-well-curvature-sensitivity}
\end{equation}
This is the Hellmann--Feynman overlap rule of Section~6 applied to a
spatial perturbation. A shape concentrated centrally is more sensitive
to central curvature and less sensitive to exterior curvature. The
weights are not interchangeable with $f_E^{\rm in}$: frequency
sensitivity couples to the specific energy term being changed.
Figure~\ref{fig:well-depth-localization-and-sensitivity} follows the
ground mode as the exterior contrast is increased, including the
extended weak-contrast limit.

A moving interface probes a different quantity. With all other inputs
fixed and the softer interval expanded symmetrically,
\begin{equation}
 \frac{\partial\omega_s^2}{\partial a_w}
   =-\Delta_g\bigl[|\phi_s(a_w)|^2+|\phi_s(-a_w)|^2\bigr].
 \label{eq:phonon-well-interface-sensitivity}
\end{equation}
The negative sign follows because softer material replaces stiffer
material. In the weak derivative of the curvature profile,
$\partial_{a_w}\omega_g^2=-\Delta_g
[\delta(z-a_w)+\delta(z+a_w)]$; inserting these interface delta
functions into the energy-form derivative gives the displayed boundary
values. Boundary shape values control this derivative, while an
interior-curvature change uses the integrated shape weight. These
different responses give distinct tests of a proposed perturbation when
the remaining parameters are independently constrained. If the profile
is changed during motion, it also performs parameter work:
$\dd E_2/\dd t=\tfrac12\int(\partial_t\omega_g^2)\xi^2\dd z$
for fixed $c$, line mass and ends. A prescribed modulation is therefore
an energy source or sink, not autonomous self-excitation.

\begin{figure}[H]
\centering
\begingroup
\definecolor{vlfGreen}{HTML}{176B64}
\definecolor{vlfDarkRed}{HTML}{762B38}
\definecolor{vlfRed}{HTML}{762B38}
\definecolor{vlfBlue}{HTML}{315D80}
\definecolor{vlfCopper}{HTML}{762B38}

\setlength{\unitlength}{1pt}
\scalebox{0.80}{%
\input{vlt-figures/figure-239.tex}%
}%
\endgroup
\caption{Three diagnostics on the exact even ground branch, at fixed $B=3\ell_\star$, $a_w=\ell_\star$, $c=\ell_\star\omega_\star$ and $\omega_{\rm in}=\omega_\star$. Vary $\Delta=(\omega_{\rm out}^2-\omega_{\rm in}^2)/\omega_\star^2$. (a) The central shape weight and its complement sum to one; the latter equals $\partial\omega_0^2/\partial\omega_{\rm out}^2$. (b) Endpoint attenuation is $\operatorname{sech}[q_w(B-a_w)]$. Dots identify $\Delta=8$. Zero contrast tends to a uniform shape with weight $1/3$ and unit attenuation. These shape and sensitivity measures are distinct from local energy fraction.}
\label{fig:well-depth-localization-and-sensitivity}
\end{figure}

Localization and observation remain distinct. If a real force per unit
length $f(z,t)$, in newtons per metre, is applied, its conjugate term is
$-\int f u\,\dd z$ and its modal force is
$F_s(t)=\int\phi_s(z)f(z,t)\dd z/\sqrt{\mu_{\rm line}}$.
A displacement marker
$Y=\int\ell(z)u(z)\dd z$, with real $\ell$ in inverse-length units
for length-valued $Y$, has modal overlap
$L_s=\int\ell(z)\phi_s(z)\dd z/\sqrt{\mu_{\rm line}}$.
Its conservative response contains
$L_sF_s(\Omega)/(\omega_s^2-(\Omega+i0)^2)$, summed over retained
modes. A symmetric force or detector can miss the odd bound mode
altogether. A local shape maximum alone supplies neither of those
overlaps. Added environmental loss changes poles and linewidths through
its own constitutive law; the lossless tail ratio supplies neither.

The exterior gap is an essential input. On an infinite uniform acoustic
line its threshold is zero, and a nonnegative onsite curvature cannot
produce a stable positive frequency below that threshold. A negative
curvature may instead destabilize a mode. Finite cavities, real band
gaps and specified impedance structures provide other confinement
mechanisms. A discrete chain also has an upper band edge, so a sufficiently
stiff defect can support above-band localization; the scalar continuum
dispersion has no such upper edge. Each claim must be checked in its
actual finite matrix or continuum boundary problem.

\subsubsection{Heterogeneous inertia and traction-preserving interfaces}
\label{sec:phonon-heterogeneous-well}

A scalar gap contrast is only one source of heterogeneity. To vary
inertia and elastic transmission consistently, return to physical
displacement $u$ and write
\begin{equation}
 E_{\rm het}=\frac12\int_{-B}^B
       [\mu(z)|\partial_tu|^2+\mathcal T(z)|\partial_zu|^2
                                  +\kappa(z)|u|^2]\,\dd z.
 \label{eq:phonon-heterogeneous-elastic-energy}
\end{equation}
The line mass $\mu>0$ has units $\mathrm{kg\,m^{-1}}$, the elastic
coefficient $\mathcal T>0$ has force units, and the onsite coefficient
$\kappa\ge0$ has units $\mathrm{N\,m^{-2}}$. Assume bounded,
piecewise smooth coefficients, with positive lower bounds for $\mu$
and $\mathcal T$. The constant-coefficient model is recovered with
$\mu=\mu_{\rm line}$, $\mathcal T=\mu_{\rm line}c^2$ and
$\kappa=\mu_{\rm line}\omega_g^2$.

Variation of this energy gives the weighted eigenproblem and
normalization
\begin{equation}
 -\partial_z(\mathcal T\partial_z\Phi_s)+\kappa\Phi_s
       =\omega_s^2\mu\Phi_s,\qquad
 \int_{-B}^B\mu\Phi_s\Phi_t\,\dd z=\delta_{st}.
 \label{eq:phonon-heterogeneous-weighted-spectrum}
\end{equation}
At a perfectly bonded interface without a point mass or interface spring,
$\Phi_s$ and $\mathcal T\Phi_s'$ are continuous. Free ends have
$\mathcal T\Phi_s'=0$. These are the displacement and traction
conditions associated with the weighted Sturm--Liouville realization
\cite{Teschl2009}. Its form domain is $H^1(-B,B)$ in
$L^2(\mu\,\dd z)$; its operator domain additionally requires
$\mathcal T\Phi_s'$ to have a square-integrable weak derivative.
Consequently the form has no unbalanced interface force. With
$u=\sum_s\Phi_sQ_s$, the same normalization gives unit modal inertia.

This extension changes exact well matching. Let each of the inner and
outer regions have constant coefficients and gaps
$\omega_a^2=\kappa_a/\mu_a$, $a\in\{{\rm in},{\rm out}\}$.
For a subthreshold frequency define
$k_{\rm in}^2=\mu_{\rm in}(\omega^2-\omega_{\rm in}^2)/\mathcal T_{\rm in}$
and
$q_{\rm out}^2=\mu_{\rm out}(\omega_{\rm out}^2-\omega^2)/\mathcal T_{\rm out}$.
Continuity of traction now gives
\begin{equation}
 \begin{aligned}
 \mathcal T_{\rm in}k_{\rm in}\tan(k_{\rm in}a_w)
   &=\mathcal T_{\rm out}q_{\rm out}
                     \tanh[q_{\rm out}(B-a_w)]&&\text{(even)},\\
 -\mathcal T_{\rm in}k_{\rm in}\cot(k_{\rm in}a_w)
   &=\mathcal T_{\rm out}q_{\rm out}
                     \tanh[q_{\rm out}(B-a_w)]&&\text{(odd)}.
 \end{aligned}
 \label{eq:phonon-heterogeneous-well-matching}
\end{equation}
The previous equations follow when the elastic coefficients agree and
the wave speeds have their stated common value. In general one cannot
match derivatives of $\sqrt\mu\,u$ across a mass-density step:
that rescaled field can jump while physical displacement stays continuous.

For a dimensionless check, take $a_w=\ell_\star$, $B=3\ell_\star$
and a reference line mass $\mu_\star$. Choose
\begin{equation}
 \begin{aligned}
 (\mu_{\rm in},\mu_{\rm out})/\mu_\star&=(1,3/2),\\
 (\mathcal T_{\rm in},\mathcal T_{\rm out})/
              (\mu_\star\ell_\star^2\omega_\star^2)&=(2,3/4),\\
 (\kappa_{\rm in},\kappa_{\rm out})/
              (\mu_\star\omega_\star^2)&=(1,27/2).
 \end{aligned}
 \label{eq:phonon-heterogeneous-well-worked-inputs}
\end{equation}
The gaps are still $\omega_{\rm in}=\omega_\star$ and
$\omega_{\rm out}=3\omega_\star$, but the two wave speeds and
traction coefficients differ. The lowest even root of
\eqref{eq:phonon-heterogeneous-well-matching} is
$\omega_{\rm even}^2/\omega_\star^2\simeq2.797993$.
It differs from the equal-inertia, common-speed example above even
though the gap values and interface positions coincide. These are
prescribed continuum coefficients, not inferred biological parameters.

The Rayleigh quotient makes the competing effects explicit:
\begin{equation}
 \omega_s^2=
 \frac{\int[\mathcal T|\Phi_s'|^2+\kappa|\Phi_s|^2]\,\dd z}
      {\int\mu|\Phi_s|^2\,\dd z}.
 \label{eq:phonon-heterogeneous-rayleigh-quotient}
\end{equation}
For fixed interfaces and endpoints, a simple eigenvalue with unit
weighted norm has the first variation
\begin{equation}
 \delta\omega_s^2=\int_{-B}^B
   [\delta\mathcal T|\Phi_s'|^2+\delta\kappa|\Phi_s|^2
                  -\omega_s^2\delta\mu|\Phi_s|^2]\,\dd z.
 \label{eq:phonon-heterogeneous-frequency-sensitivity}
\end{equation}
This follows by differentiating the generalized eigenproblem; the
eigenfunction derivative cancels against the differentiated
normalization. Added inertia lowers the eigenvalue at fixed restoring
coefficients, while added positive elastic or onsite stiffness raises
it through different spatial overlaps. A measured shift alone cannot
distinguish these changes. This formula does not include moving-interface
shape derivatives, which require their own boundary terms. If hydration
is represented by frequency-dependent added inertia or resistance,
that dynamical reduction must replace the real, frequency-independent
$\mu$ assumed here.

For biological interpretation, the soft region might represent a
specified internal connection or supported contact-network change.
It must come from that resolution's Hessian. Atomistic capsid modeling
finds localized low-frequency motions in weakly coupled regions that
a connectivity-based coarse network can miss~\cite{Dykeman2010}.
This motivates testing retained coordinates and their force constants;
it does not identify every such region with the scalar rectangular well.
A coarse model should be compared over the same band and readout before
extra detail is treated as necessary. The confined examples distinguish allowed frequency, spatial
concentration, energy storage and excitation preparation. A membrane
application shows how related shape variables can instead relax through
a viscous medium.

\paragraph{A responsive patch makes the well frequency dependent.}
A softer region can arise from a retained surface coordinate rather than
a permanent change in support. In a fixed dimensionless body chart let
$\vct X$ describe the mechanical tangent, let $\chi$ be one local surface
perturbation, and use
\begin{equation}
 \mathcal F=\tfrac12\vct X^{\mathsf T}\mathsf H_\infty\vct X
       +b_\chi\chi\vct v^{\mathsf T}\vct X
       +\tfrac12a_\chi\chi^2,
 \quad d_\chi\dot\chi=-a_\chi\chi-b_\chi\vct v^{\mathsf T}\vct X+\eta_\chi .
 \label{eq:s7hpv-responsive-patch-storage}
\end{equation}
Here $a_\chi,d_\chi>0$, $\|\vct v\|_2=1$ in the declared chart,
and $\langle\eta_\chi(t)\eta_\chi(t')\rangle
=2k_{\rm B}Td_\chi\delta(t-t')$. The mechanical inertia is
$\mathsf G_X>0$, its instantaneous resistance is $\mathsf C_X\ge0$,
and its own thermal forcing is matched and independent of $\eta_\chi$.
The vector $\vct v$ is the
assembled derivative of the actual supported or relative patch
deformation. In a translation-invariant free assembly it annihilates
common translation; it supplies no new external pinning.

\begin{proposition}[A relaxation-controlled stiffness contrast]
\label{prop:s7hpv-responsive-well}
Set $\kappa_\chi=b_\chi^2/a_\chi$ and $\tau_\chi=d_\chi/a_\chi$,
and assume $\mathsf H_{\rm rel}=\mathsf H_\infty-
\kappa_\chi\vct v\vct v^{\mathsf T}>0$ on the supported sector.
Eliminating $\chi$ with the $e^{-i\omega t}$ convention gives
\begin{equation}
 \begin{aligned}
 \mathsf Z_X(\omega)&=\mathsf H_\infty-\omega^2\mathsf G_X
       -i\omega\mathsf C_X
       -\frac{\kappa_\chi}{1-i\omega\tau_\chi}
                              \vct v\vct v^{\mathsf T},\\
 \mathsf S_{f,X}(\omega)&=2k_{\rm B}T
       \left[\mathsf C_X+
       \frac{\kappa_\chi\tau_\chi}{1+\omega^2\tau_\chi^2}
                              \vct v\vct v^{\mathsf T}\right].
 \end{aligned}
 \label{eq:s7hpv-responsive-well-response-noise}
\end{equation}
Its low-frequency approximation uses $\mathsf H_{\rm rel}$ and
$\mathsf C_{\rm eff}=\mathsf C_X+
\kappa_\chi\tau_\chi\vct v\vct v^{\mathsf T}$. The exact remainder is
\begin{equation}
 \left\|\mathsf Z_X-
   (\mathsf H_{\rm rel}-\omega^2\mathsf G_X-i\omega\mathsf C_{\rm eff})
   \right\|_2
   =\frac{\kappa_\chi(\omega\tau_\chi)^2}
                 {\sqrt{1+\omega^2\tau_\chi^2}} .
 \label{eq:s7hpv-responsive-well-remainder}
\end{equation}
\end{proposition}
\begin{proof}
Solve the scalar relaxation in frequency space and apply the same map
to its noise. Expand $(1-ir)^{-1}=1+ir-r^2/(1-ir)$ and use the unit
norm of $\vct v\vct v^{\mathsf T}$.
\end{proof}
For example, a selected contact layer in an HPV-inspired assembly can
have a different regulated surface preparation. Its relaxed Hessian may
concentrate a pattern there, while frequencies near $\tau_\chi^{-1}$
probe a different stiffness and loss. The scalar rectangular well is
then an approximation to a declared branch over a declared band.
The existing resolvent-error test propagates the displayed remainder
to the detector. Actual localization of a damped response uses the full
poles and source/readout residues of Section~4, together with the
mixed surface channel of Section~6. Concentrated conservative
coefficients alone establish neither a long lifetime nor a large
measured signal. An angular marker and a surface-sensitive channel can
test which contribution carries the localized record.

Figure~\ref{fig:s7hpv-responsive-array} shows how a retained contact variable modifies an array response, and why its internal probability law carries additional information.

\begin{figure}[H]
\centering\begingroup
\setlength{\unitlength}{1pt}
\scalebox{0.90}{%
\input{vlt-figures/figure-301.tex}%
}\endgroup
\caption{A surface-mediated stiffness contrast and the information retained by its internal state. Panel (a) uses the twelve-body geometry of Figure~\ref{fig:s7hpv-prepared-array}; arrows define the signed relative source, and straight lines show selected topology. In dimensionless coordinates $X=u/\ell_T$, with $\ell_T^2=k_BT/k_p$, panels (b,c) prescribe $\mathsf H_\infty/E_\star=\Id+\mathsf L$, $\kappa_\chi/E_\star=2$, $E_\star=k_BT$, $\mathsf C_X/(E_\star\tau)=\Id$ and $\mathsf G_X/(E_\star\tau^2)=\mu\Id$, where $\mathsf L$ is the displayed graph Laplacian and $\mu=m_v/(k_p\tau^2)$. This is a supported projected prototype, not a fitted HPV pair potential. Panel (b) plots the real softening coefficient. Panel (c) compares the full response, its low-frequency stiffness-and-drag approximation and the frozen-surface response to the same relative source. Panel (d) evaluates the conditional Gaussian widths for $a_B=4a_A$ at $\vct X=0$, choosing $a_A=E_\star$; rescaling $b_B=2b_A,d_B=4d_A$ preserves the response within each held conformation but halves the basin partition weight at equal offsets and common physical measure, as in Example~\ref{ex:s7hpv-surface-hidden-entropy}. No linewidth or lifetime is inferred from these widths.}
\label{fig:s7hpv-responsive-array}
\end{figure}

\subsubsection{Membrane fluctuations: shape, material response and unresolved density}
\label{sec:phonon-membrane-application}

Lipid membranes connect a mechanical field directly to an image: a
fluctuating contour records shape, while its amplitudes and time
correlations constrain bending rigidity, tension and solvent response.
Helfrich curvature elasticity supplies the energy~\cite{Helfrich1973};
vesicle fluctuation theory and contour analysis connect spatial modes
to optical measurements~\cite{MilnerSafran1987,Pecreaux2004}.

For an explicit local model, consider a nearly planar, homogeneous
membrane patch with small slopes and height $h(\vct\rho,t)$, where
$\vct\rho$ is a two-dimensional reference-plane coordinate. Neglect
spontaneous curvature, composition gradients and substrate coupling.
At fixed temperature its quadratic free energy is
\begin{equation}
 \mathcal F_h=\frac12\int_{\mathcal A}
       [\kappa_{\rm bend}(\nabla^2h)^2
                         +\sigma_m|\nabla h|^2]\,\dd^2\rho,
 \qquad \kappa_{\rm bend}>0,\quad\sigma_m\ge0.
 \label{eq:phonon-membrane-height-free-energy}
\end{equation}
The bending modulus has energy units; the tension has energy-per-area
units. Use a periodic patch of area $A_{\rm p}$ only as a local
Fourier reference, with
$h_{\vct q}=A_{\rm p}^{-1/2}\int h(\vct\rho)
e^{-i\vct q\cdot\vct\rho}\,\dd^2\rho$ and $q=|\vct q|>0$.
Then $h_{\vct q}$ has area units, and
\begin{equation}
 \mathcal F_h=\frac12\sum_{\vct q\ne0}K_q|h_{\vct q}|^2,
 \qquad K_q=\kappa_{\rm bend}q^4+\sigma_mq^2,
 \qquad
 \langle|h_{\vct q}|^2\rangle=\frac{k_{\rm B}T}{K_q}.
 \label{eq:phonon-membrane-spatial-variance}
\end{equation}
The sum includes both members of each conjugate pair of the real height
field. The zero mode is fixed by the chosen reference height. Shorter
wavelengths cost more bending energy; tension dominates below the
crossover $q_\times=\sqrt{\sigma_m/\kappa_{\rm bend}}$ when tension
is positive. Measuring several resolved spatial modes can therefore
separate a tension contribution from a bending contribution more
effectively than measuring one rms height.

Relaxation asks for another physical input. In the single-sheet,
overdamped limit with identical unbounded incompressible Newtonian
fluids of viscosity $\eta_f$ on both sides, negligible membrane
inertia and surface viscosity, and no slip at the sheet, the normal
mobility is $\Lambda_q=(4\eta_fq)^{-1}$. Stokes flow transmits the
height-dependent traction into the surrounding fluid. Its thermal
completion gives
\begin{equation}
 \begin{aligned}
 \partial_th_{\vct q}&=-r_qh_{\vct q}+\zeta_{\vct q},\\
 r_q&=\Lambda_qK_q
       =\frac{\kappa_{\rm bend}q^3+\sigma_mq}{4\eta_f},\\
 \langle h_{\vct q}(t)h_{\vct q}(0)^*\rangle
       &=\frac{k_{\rm B}T}{K_q}e^{-r_q|t|},\\
 \langle\zeta_{\vct q}(t)\zeta_{\vct q'}(t')^*\rangle
       &=2k_{\rm B}T\Lambda_q\delta_{\vct q\vct q'}\delta(t-t').
 \end{aligned}
 \label{eq:phonon-membrane-height-relaxation}
\end{equation}
These shape modes relax through viscous flow. At fixed elastic parameters,
changing $\eta_f$ changes their time correlations and preserves their
equilibrium variance. Static amplitudes and relaxation thus constrain
different model ingredients. Seifert and Langer discuss this bending-only
hydrodynamic limit and its bilayer extension~\cite{SeifertLanger1993}.
Figure~\ref{fig:s7-publication-membrane-readouts} compares two wavelengths
using the same height, stiffness and rate conventions.

\begin{figure}[H]
\centering\begingroup
\definecolor{vltTeal}{HTML}{176B64}
\definecolor{vltBlue}{HTML}{315D80}
\definecolor{vltGold}{HTML}{A87520}
\definecolor{vltPlum}{HTML}{725877}
\definecolor{vltRed}{HTML}{762B38}
\definecolor{vltYellow}{HTML}{D5AD3C}
\definecolor{vltNeutral}{HTML}{777D83}
\setlength{\unitlength}{1pt}
\scalebox{0.90}{%
\input{vlt-figures/figure-240.tex}%
}\endgroup
\caption{Height, variance and relaxation of the bending-only membrane model in \eqref{eq:phonon-membrane-height-relaxation}. Set $\sigma_m=0$, $C_\star=k_{\rm B}T/(\kappa_{\rm bend}q_\star^4)$ and $t_\star=4\eta_f/(\kappa_{\rm bend}q_\star^3)$. (a) Representative cosine profiles use amplitudes $H_\star, H_\star/4$ and wavelengths $2\pi/q_\star,\pi/q_\star$ on shared axes. Their amplitude ratio follows the rms mode ratio; the profiles are illustrative configurations, with $q_\star H_\star\ll1$ for small slopes. (b) The Fourier variance decreases as $(q/q_\star)^{-4}$, reaching $1/16$ at twice the wave number. (c) The normalized correlations are $\exp[-(q/q_\star)^3t/t_\star]$, so the shorter mode relaxes eight times faster. Solid teal and dashed burgundy identify the same two scales throughout. The hydrodynamic rate assumes identical unbounded Newtonian fluids on both sides, negligible inertia and surface viscosity, and no slip. A vesicle contour uses its own surface modes and imaging projection.}
\label{fig:s7-publication-membrane-readouts}
\end{figure}

A bilayer also has a density difference between its leaflets, relaxed by
lateral flow and interleaflet slip. Coupling this coordinate to height
produces an additional mode with interleaflet-friction
dissipation~\cite{SeifertLanger1993}. When that timescale is resolved,
$(h_{\vct q},\delta\rho_{\vct q})$ carries information needed for
prediction. Eliminating the density produces history dependence of the
same form as the reciprocal internal-coordinate reductions above.

For an actual vesicle, the closed-surface geometry, volume and area
constraints select the mode basis~\cite{MilnerSafran1987}. A recorded
contour also projects the surface modes through imaging and exposure;
the flat-patch variance cannot simply be assigned to an uncorrected
contour coefficient. Refined vesicle contour analysis explicitly links
these observations to tension and bending estimates
\cite{Pecreaux2004}. The transferable application is to membrane
vesicles and resolved membrane regions whose geometry and response
justify the approximation. A supported, protein-rich or cytoskeleton-
coupled membrane requires its additional restoring and dissipative
terms. In a membrane-bearing particle, translation of its centre and
change of its surface shape are likewise distinct candidate variables.

\clearpage
\subsection{Wave packets, effective mass, and closed conversion}
\label{sec:phonon-packets-conversion}

A localized disturbance excites several of the stationary modes found
above. Their relative phases determine where the coordinated motion
concentrates and which marker detects it. This wave-packet problem
retains the same virions or structural coordinates. Coupling between modes
adds a second question: how excitation transfers between patterns.
Both can be treated within a closed mechanical model before adding loss.

\paragraph{An exact first-order wave description.}
For the real mass-normalized field $\xi$ in
\eqref{eq:phonon-spatial-well-energy}, write
$\mathsf W_{\rm ax}=\mathcal D_{\rm ax}^{1/2}$ on its positive sector.
The square root is defined spectrally: it multiplies a spatial eigenmode of
eigenvalue $\omega_s^2$ by $\omega_s$, rather than taking a pointwise square
root of the differential expression.
This construction preserves the same spatial mode basis; its eigenvalues
carry angular-frequency units instead of squared-frequency units. The complex energy amplitude
\begin{equation}
\begin{aligned}
 \psi&=\frac{\mathsf W_{\rm ax}\xi+i\partial_t\xi}{\sqrt2},\\
 i\partial_t\psi&=\mathsf W_{\rm ax}\psi,\\
 \|\psi\|_2^2&=E_2
   =\frac12\bigl(\|\partial_t\xi\|_2^2+
                         \|\mathsf W_{\rm ax}\xi\|_2^2\bigr)
 \end{aligned}
 \label{eq:phonon-packet-positive-frequency-energy}
\end{equation}
therefore contains the displacement and velocity quadratures of the same
real motion. Here $\xi$ belongs to the quadratic-form domain and
$\partial_t\xi\in L^2$; strong evolution additionally requires the
generator domain, while its unitary group supplies the finite-energy mild
evolution. With a positive lower spectral bound, reconstruction is
$\xi=\sqrt2\,\mathsf W_{\rm ax}^{-1}\operatorname{Re}\psi$ and
$\partial_t\xi=\sqrt2\operatorname{Im}\psi$. No new microscopic degree of
freedom has been introduced.

The norm is the exact quadratic mechanical energy. Because the
square-root generator is generally nonlocal, local kinetic and strain
energy are computed from the underlying real field and its energy
current. Dividing $\psi$ by $\sqrt{E_2}$ gives a normalized classical
energy amplitude. A one-phonon state uses the same spectral generator,
with coefficient probabilities $|c_s|^2$, mean excitation energy
$\hbar\sum_s\omega_s|c_s|^2$, and the displacement factors
$\omega_s^{-1/2}$ derived above. Its detector prediction follows from
that quantum readout map.

\paragraph{When a Schr\"odinger envelope is justified.}
Consider an isolated branch of an auxiliary homogeneous medium with a
smooth minimum at $k_0$. Put $q=k-k_0$ and suppose
\begin{equation}
 \omega(k_0+q)=\omega_0+\frac12\omega''(k_0)q^2+r(q),
 \qquad r(q)=o(q^2),\qquad
 \omega_0>0,\quad d_2:=\omega''(k_0)>0.
 \label{eq:phonon-band-edge-expansion}
\end{equation}
After removing the spatial carrier $e^{ik_0z}$, a sufficiently narrow-band
envelope has the leading evolution
\begin{equation}
\boxed{\begin{aligned}
 i\hbar\partial_t A
   &=\left[\hbar\omega_0-\frac{\hbar^2}{2m_{\rm eff}}\partial_z^2
                                  +V_{\rm eff}(z)\right]A,\\
 m_{\rm eff}&=\frac{\hbar}{d_2}.
 \end{aligned}}
 \label{eq:phonon-envelope-effective-mass}
\end{equation}
The constant $\hbar\omega_0$ retains the temporal carrier phase.
With $[d_2]=\mathrm{m^2\,s^{-1}}$, the effective mass has mass units
and measures branch curvature. The classical envelope equation is
independent of $\hbar$, whose factors cancel. A quantum one-phonon
envelope uses the separately postulated oscillator theory.

Figure~\ref{fig:packet-spreading-and-envelope-validity} previews a normalized spreading envelope and the distinct dispersion approximation; the formulas, domain of validity and phase-error bound are developed next.

For the gapped scalar branch
$\omega(k)=\sqrt{\omega_0^2+c^2k^2}$ near $k_0=0$,
\begin{equation}
 m_{\rm eff}=\frac{\hbar\omega_0}{c^2},\qquad
 r(q)=-\frac{c^4q^4}{8\omega_0^3}+O(q^6),\qquad
 V_{\rm eff}\simeq\frac{\hbar\,\delta\omega_g^2}{2\omega_0}.
 \label{eq:phonon-envelope-gap-expansion}
\end{equation}
The last expression requires a weak gap perturbation,
$|\delta\omega_g^2|\ll\omega_0^2$, with sufficiently slow spatial
variation for the same branch projection to remain valid. It is not a
replacement for the exact strong-contrast matching problem of
Section~\ref{sec:phonon-spatial-well}. An acoustic branch $c|k|$ at the
origin lacks a smooth positive minimum; its positive-frequency generator
is $c|{-i\partial_z}|$, and this effective-mass expansion does not apply
there. Around a smooth carrier with nonzero $\omega'(k_0)$, the first
derivative instead adds $-i\hbar\omega'(k_0)\partial_z A$ to the envelope
Hamiltonian and translates the packet with group velocity $\omega'(k_0)$;
the minimum chosen here removes this drift.

\begin{figure}[H]
\centering\begingroup
\definecolor{vltTeal}{HTML}{176B64}
\definecolor{vltBlue}{HTML}{315D80}
\definecolor{vltGold}{HTML}{A87520}
\definecolor{vltPlum}{HTML}{725877}
\definecolor{vltRed}{HTML}{762B38}
\definecolor{vltYellow}{HTML}{D5AD3C}
\definecolor{vltNeutral}{HTML}{777D83}
\setlength{\unitlength}{1pt}
\scalebox{0.80}{%
\input{vlt-figures/figure-241.tex}%
}\endgroup
\caption{An exact parabolic-envelope solution and the approximation linking it to a mechanical branch. (a) For $\tau=d_2t/(2\sigma_0^2)$, the normalized Gaussian has variance $\sigma_0^2(1+\tau^2)$ and full-line integral one. Its decreasing peak follows widening at fixed norm. (b) The exact gapped branch $\sqrt{1+\kappa^2}$ is compared with $1+\kappa^2/2$. Shading marks $|\kappa|\le1/4$, where their maximum frequency difference is $0.000473594\omega_0$. A Gaussian has spectral tails beyond any such hard band; a mechanical use therefore also controls weighted phase error, branch separation and boundary returns. The envelope density is a packet representation. A particle-position or detector record requires its own reconstruction and readout. The boxed result illustrated here is~\eqref{eq:phonon-envelope-effective-mass}.}
\label{fig:packet-spreading-and-envelope-validity}
\end{figure}

\begin{proposition}[Accumulated phase error of a fixed-branch envelope]
\label{prop:s7-packet-dispersion-error}
Narrow initial bandwidth alone does not guarantee an accurate long-time
prediction. With the unitary Fourier convention and the same initial
envelope, the unperturbed exact-branch and parabolic evolutions obey
\begin{equation}
 \|A_{\rm exact}(t)-A_{\rm par}(t)\|_2
 \le |t|\left(\int_{\R}|r(q)|^2
                         |\widetilde A(q,0)|^2\,dq\right)^{1/2}.
 \label{eq:phonon-envelope-phase-error}
\end{equation}
\end{proposition}
\begin{proof}
This follows from $|e^{-itr}-1|\le |tr|$ and makes the accumulated phase
error explicit.
\end{proof} For a hard occupied band, one may bound it by
$|t|\sup_{\rm band}|r|$ for a normalized packet. A Gaussian instead has
tails, which must be controlled rather than declared absent. In the
finite cylinder, $k$ is an auxiliary band label: the actual evolution is
a discrete mode sum. A continuum packet additionally requires many
available cavity modes across its bandwidth, wavelengths compatible with
the mechanical coarse graining, and a time interval before significant
boundary reflection or another branch invalidates the local description.

Figure~\ref{fig:phonon-envelope-phase-budget} illustrates this accumulated-error bound for a normalized packet on a prescribed spectral band.

\begin{figure}[H]
\centering\begingroup
\definecolor{vlfGreen}{HTML}{176B64}
\definecolor{vlfDarkRed}{HTML}{762B38}
\definecolor{vlfRed}{HTML}{762B38}
\definecolor{vlfBlue}{HTML}{315D80}
\definecolor{vlfCopper}{HTML}{762B38}
\setlength{\unitlength}{1pt}
\scalebox{0.80}{%
\input{vlt-figures/figure-242.tex}%
}\endgroup
\caption{A declared bandwidth gives an explicit phase-error budget. Use $\omega(q)=\sqrt{\omega_0^2+c^2q^2}$ and $\kappa=cq/\omega_0$. (a) The normalized spectral density is $\rho=2$ on $|\kappa|\le1/4$ and zero outside. (b) The parabolic remainder is $r/\omega_0=\sqrt{1+\kappa^2}-1-\kappa^2/2$, with maximum absolute value $0.000473594$. (c) Multiplication by elapsed time bounds the norm difference of exact and parabolic whole-line evolutions from that same packet. The marked time gives about $0.0947$; it is a declared error bound rather than a universal lifetime. Gaussian tails require a weighted remainder estimate, and a finite cavity also tests returns from its boundaries. The density in (a) allocates spectral norm, rather than physical virion concentration.}
\label{fig:phonon-envelope-phase-budget}
\end{figure}

\paragraph{An exactly solvable spreading problem.}
On the whole line set $V_{\rm eff}=0$ and prepare
\begin{equation}
\begin{aligned}
 A(z,0)&=(2\pi\sigma_0^2)^{-1/4}e^{-z^2/(4\sigma_0^2)},\\
 (\Delta z(t))^2&=\sigma_0^2+
                 \left(\frac{d_2t}{2\sigma_0}\right)^2,\\
 \Delta k&=\frac1{2\sigma_0}.
 \end{aligned}
 \label{eq:phonon-gaussian-envelope-spreading}
\end{equation}
The full complex solution, not just its width, is
\begin{equation}
 A(z,t)=\frac{(2\pi\sigma_0^2)^{-1/4}e^{-i\omega_0t}}
                    {\sqrt{1+i\tau}}
       \exp\!\left[-\frac{z^2}{4\sigma_0^2(1+i\tau)}\right],
 \qquad \tau=\frac{d_2t}{2\sigma_0^2}.
 \label{eq:phonon-gaussian-complex-envelope}
\end{equation}
The spatial phase is a chirp: fixed spectral weights acquire different
phases while $|A|^2$ broadens to
$\Delta z/\sigma_0=\sqrt{1+\tau^2}$.
This solution is exact for the parabolic envelope equation
\cite{Shankar1994}, subject to the mechanical conditions above.
Its variance grows quadratically in time; a diffusive variance grows
linearly. The conserved norm and phase evolution identify the mechanism.

There is also an exact envelope continuity equation,
$\partial_t|A|^2+\partial_z j_A=0$, with
$j_A=d_2\operatorname{Im}(A^*\partial_z A)$. This is the current of the
chosen envelope norm, rather than virion-number current or the exact
local mechanical energy current. For the Gaussian,
$\Delta z\,\Delta k=\tfrac12\sqrt{1+\tau^2}$ and the symmetrized
position--wave-number covariance is $\tau/2$; the covariance determinant
remains $1/4$. These are Fourier localization statements shared by
classical waves. A measurement of marker displacement still requires the
mode-to-marker overlap and its frequency weights. Figure~\ref{fig:packet-spreading-and-envelope-validity}
shows the spreading and the separate dispersion approximation.

The Gaussian tails can be included explicitly for the gapped branch.
For every $x\ge0$, Taylor's integral remainder gives
$0\le1+x/2-\sqrt{1+x}\le x^2/8$.
For the normalized Gaussian above, its Fourier intensity is a centered
Gaussian of variance $\sigma_k^2=1/(4\sigma_0^2)$ and eighth moment
$105\sigma_k^8$. Inserting these facts into
\eqref{eq:phonon-envelope-phase-error} gives the whole-line bound
\begin{equation}
 \|A_{\rm exact}(t)-A_{\rm par}(t)\|_2
 \le\min\!\left(2,
       \frac{\sqrt{105}\,c^4|t|}{128\omega_0^3\sigma_0^4}\right).
 \label{eq:phonon-gaussian-weighted-phase-error}
\end{equation}
The constant two is the distance bound between unit-norm states. This
estimate needs no artificial spectral cutoff. A useful approximation
requires the displayed error to be small; narrow packets in position
can violate that requirement quickly. It compares the two stipulated
lossless branch evolutions and does not cover branch mixing, boundaries
or environmental loss.

\paragraph{When conservative modes remain resolved in a dissipative medium.}
The packet construction also needs a lifetime assessment. As a definite
local-in-time reference, take a finite symmetric positive semidefinite resistance
$\mathsf C\ge0$ in the same physical coordinates as $\mathsf M$
and $\mathsf K$. Let $\mathsf E$ contain the real orthonormal
conservative mass-weighted modes, and put
\begin{equation}
 \begin{aligned}
 \mathsf\Gamma&=\mathsf E^{\mathsf T}\mathsf M^{-1/2}
                         \mathsf C\mathsf M^{-1/2}\mathsf E,\\
 \mathsf R_Q(\omega)&=
 [\operatorname{diag}(\omega_s^2)-\omega^2\Id-i\omega\mathsf\Gamma]^{-1}.
 \end{aligned}
 \label{eq:phonon-nonproportional-modal-resistance}
\end{equation}
$\mathsf\Gamma$ has inverse-time units. Unless it is diagonal in
this basis, damping couples the conservative coordinates. The elastic
mode shapes therefore do not automatically diagonalize the measured
response. At classical thermal equilibrium, the same resistance fixes
the modal force-noise spectrum and displacement spectrum:
\begin{equation}
 \begin{aligned}
 \mathsf S_{FF}&=2k_{\rm B}T\mathsf\Gamma,\\
 \mathsf S_{QQ}&=2k_{\rm B}T\mathsf R_Q\mathsf\Gamma\mathsf R_Q^\dagger
       =\frac{2k_{\rm B}T}{\omega}\operatorname{Im}\mathsf R_Q.
 \end{aligned}
 \label{eq:phonon-modal-resistance-matrix-fdt}
\end{equation}
Here $\operatorname{Im}\mathsf R_Q=(\mathsf R_Q-\mathsf R_Q^\dagger)/(2i)$;
the inverse is evaluated at regular real frequencies, and the last
identity follows by subtracting the inverse matrices.
Stationarity requires a compatible Gibbs preparation or decay of the
preparation transient. An undamped invariant subspace has additional
initial-state spectral lines; a noise-driven stationary formula must
not silently discard them. This is the modal form of the
Langevin and response constructions of Sections~4 and~6
\cite{Cotterill2002,CallenWelton1951}.

For an isolated weakly damped mode, perturbation of its positive pole
gives $z_s\simeq\omega_s-i\Gamma_{ss}/2$. Sufficient local separation
conditions for this approximation include
\begin{equation}
 \Gamma_{ss}\ll\omega_s,\qquad
 |\omega_s\Gamma_{st}|\ll|\omega_t^2-\omega_s^2|\quad(t\ne s).
 \label{eq:phonon-isolated-damped-mode-conditions}
\end{equation}
The second condition tests off-diagonal mixing; for a large cluster its
collective off-diagonal norm must also be controlled. A nearly
degenerate group should instead be treated as a matrix block, including
its resistance. A conservative spectral gap alone is insufficient.

When a narrow propagating packet has nonzero group velocity and nearly
constant $\Gamma_{ss}$ across its occupied band, its leading amplitude
and energy decay lengths are respectively
\begin{equation}
 \ell_{\rm amp}\simeq\frac{2|v_g|}{\Gamma_{ss}},\qquad
 \ell_E\simeq\frac{|v_g|}{\Gamma_{ss}}.
 \label{eq:phonon-packet-attenuation-lengths}
\end{equation}
These follow from amplitude decay $e^{-\Gamma_{ss}t/2}$ and energy
decay $e^{-\Gamma_{ss}t}$. They concern propagation of the selected
mode contribution; stationary thermal fluctuations remain. At the
minimum $v_g=0$, they do not replace the spreading calculation by a
ballistic distance. Dispersion error, attenuation, branch separation
and boundary-return time must all leave a common observation window.

A fluid can contribute frequency-dependent resistance, added inertia and
cross-correlations. Hydrodynamic memory can itself create Brownian
resonant features~\cite{Franosch2011}. With appreciable memory, the full
causal inverse response and matched colored noise determine the poles
and fluctuations. The strongly damped benchmark in
Section~\ref{sec:complete-benchmark} uses that response description
where a weak-damping packet approximation fails.

\paragraph{A packet window also requires a persistent arrangement.}
Choose the desired detector accuracy and a declared residence interval.
Within it, use the dispersion error above, the actual reflected signal,
the dissipative response and the registered changing-reference error
to test the same predicted record. For example, the frozen-reference
comparison in Section~\ref{sec:s5-flow-evolving-reference} is a
general state-propagator estimate after the appropriate particle-state
and detector maps are substituted. Its input includes the current
material preparation. A small geometric change can still fail that
test near a sensitive mode or nearly dark channel.

Across recruitment or breakup, continue the complete state and recompute
its band overlaps, or end the stationary packet window at the event.
A random residence time requires its preparation-conditioned survival
law; a constant-rate assumption gives the exponential special case.
When no persistent single-branch window meets the chosen accuracy,
the two-time response of Section~4 still predicts tracked correlated
motion. This provides a task-specific test of the packet as a
biophysical variable.

Figure~\ref{fig:s7-packet-residence-error-window} places the dispersion, attenuation and registered residence limits on a common accuracy window.

\begin{figure}[H]
\centering\begingroup
\definecolor{vltTeal}{HTML}{176B64}
\definecolor{vltBlue}{HTML}{315D80}
\definecolor{vltGold}{HTML}{A87520}
\definecolor{vltPlum}{HTML}{725877}
\definecolor{vltRed}{HTML}{762B38}
\definecolor{vltYellow}{HTML}{D5AD3C}
\definecolor{vltNeutral}{HTML}{777D83}
\setlength{\unitlength}{1pt}
\scalebox{0.90}{%
\input{vlt-figures/figure-243.tex}%
}\endgroup
\caption{Packet accuracy and persistence are separate inputs to one observation window. (a) The auxiliary whole-line branch is $\omega(q)=\sqrt{1+q^2}$ in units $\omega_0=c=1$, with a normalized Gaussian spectral density of standard deviation $1/4$. (b) The exact comparison includes uniform amplitude attenuation $e^{-0.006t}$ and the full dispersion, while the approximation uses the lossless parabolic branch. The rigorous bound adds $\sqrt{105}\,t/(128\sigma_0^4)$, $\sigma_0=2$, to $1-e^{-0.006t}$; direct Fourier quadrature gives the solid curve. (c) Assigning $0.05$ to each contribution gives the displayed budget times. A registered arrangement change at $t=6$ ends the earlier persistent-reference window. Continued motion then uses the full state and new band overlaps. The shaded interval is this declared pre-event comparison. A finite cavity also tests reflected contributions and its calibrated detector; the whole-line calculation supplies no universal cavity lifetime.}
\label{fig:s7-packet-residence-error-window}
\end{figure}

\paragraph{Phase mixing, interaction, and the meaning of closed conversion.}
For the exact finite harmonic reference,
$[\widehat H_0,\widehat N_s]=0$ for every full normal mode. A packet can
change shape while all these occupations remain constant. Finite
participating superpositions have arbitrarily close recurrences; exact
returns of a quantum state up to a global phase require commensurate
participating frequency differences. Return of the real displacement and
velocity instead requires the corresponding absolute modal phases to
return. Relative-phase interference, reversible energy exchange, and
irreversible equilibration are consequently different physical claims.

A nonlinear potential supplies an additional conversion mechanism. If
$U$ is $C^3$ near the reference, its leading cubic contribution is
\begin{equation}
 U_3=\frac1{3!}\sum_{s,t,v}\Lambda_{stv}Q_sQ_tQ_v,
 \qquad
 \Lambda_{stv}=\partial_{Q_s}\partial_{Q_t}\partial_{Q_v}U\big|_0.
 \label{eq:phonon-anharmonic-interaction}
\end{equation}
For the translational mass-normalized coordinates used here,
$Q_s$ has units $\mathrm{kg^{1/2}m}$ and $\Lambda_{stv}$ has units
$\mathrm{kg^{-1/2}m^{-1}s^{-2}}$. Its value is a third derivative of a
specified interaction energy, rather than a new unconstrained excitation
variable. Substituting
$Q_s=\sqrt{\hbar/(2\omega_s)}(\widehat b_s+\widehat b_s^\dagger)$
gives products whose matrix elements contain occupation square roots and
the geometric overlaps in $\Lambda$. In the interaction picture a term
such as $\widehat b_s^\dagger\widehat b_t\widehat b_v$ carries the phase
$e^{i(\omega_s-\omega_t-\omega_v)t}$, identifying a possible near-resonant
three-mode conversion. The symmetric tensor and its $1/3!$ factor must
be retained when indices repeat.

This is a local expansion of a stable potential: an isolated nonzero
cubic polynomial is not a globally stable energy. With only $C^3$
regularity the Taylor remainder is $o(\|Q\|^3)$; stabilizing higher terms
require the corresponding further physical information. Even genuine
mode interaction in a finite closed system does not by itself establish
an irreversible scattering rate. A rate description additionally needs
an appropriate receiving spectrum, preparation, and time coarse
graining~\cite{Misra2012}. None of these passive terms supplies sustained
self-excitation without an energy-bearing resource.

\paragraph{Assembling a conversion vertex from physical relative motion.}
The physical interaction derivatives can be made explicit in a scalar
longitudinal bond model. Choose one orientation
for each distinct physical interaction $e$, with linear extension
$d_e=\vct b_e^{\mathsf T}\vct u$ and cubic energy
\begin{equation}
 U_3=\frac16\sum_e\lambda_e
                       (\vct b_e^{\mathsf T}\vct u)^3,
 \qquad [\lambda_e]=\mathrm{J\,m^{-3}}.
 \label{eq:s7-cavity-bond-cubic-energy}
\end{equation}
Here $\vct b_e$ is a dimensionless signed incidence column and
$\lambda_e$ is the third derivative with respect to that declared
extension. Reversing its orientation also reverses the corresponding
cubic coefficient. Linear and quadratic reference terms retain their
separately assembled force balance and Hessian.

\begin{proposition}[Owned bond-overlap conversion tensor]
\label{prop:s7-cavity-bond-vertex}
With $\vct u=\sum_s\vct\phi_sQ_s$, the symmetric tensor in
\eqref{eq:phonon-anharmonic-interaction} is
\begin{equation}
 \Lambda_{stv}=\sum_e\lambda_e
       (\vct b_e^{\mathsf T}\vct\phi_s)
       (\vct b_e^{\mathsf T}\vct\phi_t)
       (\vct b_e^{\mathsf T}\vct\phi_v).
 \label{eq:s7-cavity-owned-bond-vertex}
\end{equation}
The same result follows from cell contributions whose weights sum to
one for each physical interaction. A common translation annihilated
by every incidence row has zero coupling through this tensor.
\end{proposition}
\begin{proof}
Insert the mode expansion into each cube and collect the ordered
products $Q_sQ_tQ_v$. Unique interaction ownership leaves each bond
once. A vanishing extension in one factor annihilates its contribution.
\end{proof}

The tensor therefore measures shared relative deformation, rather than
mere proximity of mode centres. For the three-site free path, the
normalized nonuniform patterns are
$(1,0,-1)^{\mathsf T}/\sqrt{2m}$ and
$(1,-2,1)^{\mathsf T}/\sqrt{6m}$.
With bond orientations from site~1 to~2 and from~2 to~3,
\begin{equation}
 \Lambda_{112}=\frac{\sqrt6}{4m^{3/2}}(\lambda_2-\lambda_1),
 \qquad
 \Lambda_{111}=-\frac{\lambda_1+\lambda_2}{2\sqrt2\,m^{3/2}}.
 \label{eq:s7-cavity-bond-vertex-cancellation}
\end{equation}
Equal cubic coefficients suppress the first channel by cancellation
while leaving the second generally nonzero. Breaking that equality
changes conversion strength even if the retained harmonic Hessian is
unchanged. A frequency match must still be supplied separately.

This formula is exact for the stipulated linear-extension energy.
General three-dimensional Lennard--Jones and Coulomb interactions use
the full Cartesian third derivative in
\eqref{eq:s3-phonon-anharmonic-vertex}, including changes of bond
direction and geometric terms. Their scalar third radial derivative
alone does not replace that tensor. Compatible cells still offer the
same computational advantage: assemble the physical derivative once,
then contract it with the retained collective patterns and their
measured force and readout projections.

Figure~\ref{fig:s7-cavity-owned-bond-vertex} shows how the physical bond extensions determine cancellation or addition in the owned cubic vertex.

\begin{figure}[H]
\centering\begingroup
\definecolor{vltTeal}{HTML}{176B64}
\definecolor{vltBlue}{HTML}{315D80}
\definecolor{vltGold}{HTML}{A87520}
\definecolor{vltPlum}{HTML}{725877}
\definecolor{vltRed}{HTML}{762B38}
\definecolor{vltYellow}{HTML}{D5AD3C}
\definecolor{vltNeutral}{HTML}{777D83}
\setlength{\unitlength}{1pt}
\scalebox{0.90}{%
\input{vlt-figures/figure-244.tex}%
}\endgroup
\caption{Physical bond extensions determine cubic conversion. Equal masses use the two normalized patterns printed above, $Q_s=\sqrt m\,\ell_\star$, and oriented bonds $1\to2$, $2\to3$. Rulers use reference spacing $a$ and origin at the first centre. Arrows and rigid virion shifts use one gain; fixed coil families follow the resulting spans. At equal cubic coefficients $\lambda$, bond contributions to $m^{3/2}\Lambda_{112}/\lambda$ cancel, while those to $m^{3/2}\Lambda_{111}/\lambda$ add. Bars retain signs and sums. The matrix vertex is assembled from these same owned physical extensions; it is not an additional independently assigned interaction.}
\label{fig:s7-cavity-owned-bond-vertex}
\end{figure}

\paragraph{From a cubic matrix element to a finite-time conversion probability.}
The cubic tensor can be made more predictive without introducing a
phenomenological lifetime. First take three distinct positive modes
$s,t,v$ and an initial number state. Write
$q_{{\rm zpf},j}=\sqrt{\hbar/(2\omega_j)}$ for the mass-normalized
zero-point scale. The six permutations of the distinct indices in
\eqref{eq:phonon-anharmonic-interaction} cancel its $3!$ denominator,
leaving $\Lambda_{stv}Q_sQ_tQ_v$. For the channel
$(N_s,N_t,N_v)\to(N_s-1,N_t+1,N_v+1)$ its matrix element and detuning are
\begin{equation}
 \begin{aligned}
 M_{if}&=\Lambda_{stv}q_{{\rm zpf},s}q_{{\rm zpf},t}q_{{\rm zpf},v}
           \sqrt{N_s(N_t+1)(N_v+1)},\\
 g_{if}&=M_{if}/\hbar,\qquad
 \Delta_{if}=\omega_t+\omega_v-\omega_s.
 \end{aligned}
 \label{eq:phonon-distinct-three-mode-matrix-element}
\end{equation}
$M_{if}$ is an energy and $g_{if}$ an angular frequency. A symmetry or
spatial-overlap zero in $\Lambda_{stv}$ forbids this channel at cubic
order even when its frequencies match. Thermal or prepared occupation
changes its strength through the stated oscillator matrix elements.

For a stationary weak interaction acting for time $t$, the first-order
interaction-picture amplitude gives
\begin{equation}
 P_{i\to f}^{(1)}(t)=|g_{if}|^2t^2
          \operatorname{sinc}^2(\Delta_{if}t/2),\qquad
 \operatorname{sinc}x=\frac{\sin x}{x},\quad\operatorname{sinc}0=1.
 \label{eq:phonon-finite-time-anharmonic-conversion}
\end{equation}
This follows by integrating $e^{i\Delta_{if}t'}$ from zero to $t$.
The total perturbative transition probability must remain small.
On resonance the leading growth is quadratic, not exponential; at fixed
nonzero detuning this contribution is bounded and oscillatory. For an
isolated near-resonant pair of number configurations, retaining the
resonant interaction as a two-state Hamiltonian instead gives the
reversible transfer law with frequency
$\sqrt{|g_{if}|^2+\Delta_{if}^2/4}$. That truncation additionally
requires other channels to be off resonance or unoccupied; it is not
the complete dynamics of a general cubic network.

Repeated indices change the coefficient. If the two daughter quanta
occupy the same mode $t$, the cubic contribution is
$\Lambda_{stt}Q_sQ_t^2/2$, and
\begin{equation}
 M_{if}^{(s\to2t)}=
   \frac{\Lambda_{stt}}2q_{{\rm zpf},s}q_{{\rm zpf},t}^{\,2}
       \sqrt{N_s(N_t+1)(N_t+2)}.
 \label{eq:phonon-repeated-daughter-conversion-element}
\end{equation}
Its detuning is $2\omega_t-\omega_s$. Treating this as two distinct
daughter modes would give the wrong combinatorial factor. In sums over
distinct daughter modes, each unordered pair must likewise be counted
once.

An irreversible rate requires a further limit. With many accepting
channels, a sufficiently smooth weighted density of detunings, and a
time interval that resolves that density without resolving individual
level recurrences, the distributional limit
$t\operatorname{sinc}^2(\Delta t/2)\to2\pi\delta(\Delta)$ yields
\begin{equation}
 \Gamma_{i\to\mathcal F}
       \simeq2\pi\sum_{f\in\mathcal F}|g_{if}|^2
                                             \delta(\Delta_{if}).
 \label{eq:phonon-golden-rule-receiving-spectrum}
\end{equation}
The sum denotes integration against the accepting spectral density in
the continuum approximation, not a finite delta function evaluated at
zero. A controlled weak-conversion window needs
$\tau_{\rm corr}\ll t\ll\Gamma^{-1}$ and accepting-level spacing
much smaller than $1/t$, before finite-size returns. Resumming the weak
loss into an exponential is a subsequent kinetic approximation. The
standard theory of anharmonic phonon decay supplies this type of
perturbative connection in specified solids~\cite{Klemens1966}; its
material coefficients and phase space are not transferred to virions.

Reverse channels matter in a thermal preparation. For distinct modes,
the forward and reverse occupation factors are
$N_s(N_t+1)(N_v+1)$ and $(N_s+1)N_tN_v$. Independent Bose--Gibbs
averages make them equal on $\omega_s=\omega_t+\omega_v$ because
$\bar N_j+1=e^{\beta\hbar\omega_j}\bar N_j$. Equilibrium thus has
balanced conversion rather than a net supply of excitations. These
formulas specify which derivatives, state populations and accepting
spectra are missing before a lifetime can be assigned. The cubic energy
still requires its stable higher-order completion and the same local
harmonic-basin checks stated above.

\paragraph{An exact two-channel benchmark and its energy ledger.}
Let $a,b$ now denote two uncoupled positive reference channels, for example
two resolved internal/contact motions. They are not the already diagonal
normal modes of their combined Hamiltonian. The number-conserving model,
with vacuum constants omitted, is
\begin{equation}
 \widehat H_{\rm ex}=\hbar\omega_a\widehat N_a+
  \hbar\omega_b\widehat N_b+
  \hbar J(\widehat b_a^\dagger\widehat b_b+
                         \widehat b_b^\dagger\widehat b_a),
 \qquad J\in\R,\quad J^2<\omega_a\omega_b.
 \label{eq:phonon-two-channel-hamiltonian}
\end{equation}
The positive two-by-two frequency matrix makes this quadratic operator
bounded below on bosonic Fock space; its standard self-adjoint realization
is obtained by second quantization. Total reference occupation is
conserved. Write $\delta=\omega_a-\omega_b$ and
$\omega_c=(\omega_a+\omega_b)/2$. From $\ket{1_a,0_b}$ the amplitudes are
$c_a=e^{-i\omega_ct}[\cos(\mathcal Rt)-i\delta\sin(\mathcal Rt)/(2\mathcal R)]$
and $c_b=-ie^{-i\omega_ct}J\sin(\mathcal Rt)/\mathcal R$, so
\begin{equation}
 \boxed{\displaystyle
 P_b(t)=\frac{J^2}{\mathcal R^2}\sin^2(\mathcal R t),
 \qquad
 \mathcal R=\sqrt{J^2+\frac{(\omega_a-\omega_b)^2}{4}}.}
 \label{eq:phonon-two-channel-transfer}
\end{equation}
For $J\ne0$, resonance gives complete transfer at
$t=\pi/(2|J|)$ and a subsequent return; detuning limits its maximum.
At $\mathcal R=0$ the continuous limit gives $P_b(t)=0$.
The mixed frequencies $\omega_c\pm\mathcal R$ have an avoided crossing
with minimum gap $2|J|$.

Figure~\ref{fig:phonon-closed-conversion} displays the exact transfer and the corresponding avoided crossing of this declared number-conserving model; the energy ledger and the stricter coordinate-coupling stability test follow.

\begin{figure}[H]
\centering
\begingroup
\definecolor{vlfGreen}{HTML}{176B64}
\definecolor{vlfDarkRed}{HTML}{762B38}
\definecolor{vlfRed}{HTML}{762B38}
\definecolor{vlfBlue}{HTML}{315D80}
\definecolor{vlfCopper}{HTML}{762B38}

\setlength{\unitlength}{1pt}
\scalebox{0.80}{%
\input{vlt-figures/figure-245.tex}%
}%
\endgroup
\caption{Reversible exchange and an avoided crossing from one two-channel generator. Set $J>0$, $\omega_c=100J$ and $\omega_{a,b}=\omega_c\pm\delta/2$. (a) From $|1_a,0_b\rangle$, transfer is $P_b=\sin^2(Jt)$ at resonance and $\tfrac12\sin^2(\sqrt2Jt)$ at $\delta=2J$. (b) Coupled frequencies are $\omega_\pm=\omega_c\pm\sqrt{J^2+\delta^2/4}$; the minimum physical gap is $2J$. Dashed branches are uncoupled, and both coupled frequencies stay positive over the displayed range. Reference-channel occupations exchange while occupations of the diagonal normal modes are conserved. These formulas are exact for the specified number-conserving model; a coordinate-coupling realization additionally requires a weak, near-resonant rotating-wave regime. The same amplitude matrix has a classical realization, so an exchange curve alone does not certify single-phonon preparation. The boxed result illustrated here is~\eqref{eq:phonon-two-channel-transfer}.}
\label{fig:phonon-closed-conversion}
\end{figure}

Figure~\ref{fig:added-channel-mixing-portraits} follows the corresponding eigenvector components across detuning, complementing the frequency and exchange curves.

\begin{figure}[H]
\centering\begingroup
\definecolor{vltTeal}{HTML}{176B64}
\definecolor{vltBlue}{HTML}{315D80}
\definecolor{vltGold}{HTML}{A87520}
\definecolor{vltPlum}{HTML}{725877}
\definecolor{vltRed}{HTML}{762B38}
\definecolor{vltYellow}{HTML}{D5AD3C}
\definecolor{vltNeutral}{HTML}{777D83}
\setlength{\unitlength}{1pt}
\scalebox{0.80}{%
\input{vlt-figures/figure-246.tex}%
}\endgroup
\caption{Channel composition changes through an avoided crossing. Use $H/\hbar=\omega_c\Id+J\left(\begin{smallmatrix}d/2&1\\1&-d/2\end{smallmatrix}\right)$, $J>0$, $\omega_c=10J$ and $d=(\omega_a-\omega_b)/J$. (a) Each portrait gives the lower-branch signed components on one arrow scale; the global sign is chosen so $c_a\ge0$. At resonance the vector is $(1,-1)/\sqrt2$. (b) Squaring those components gives the displayed complementary weights. Color links each component to its own curve, and the dashed curve remains identifiable in grayscale. Symbols denote reference channels, rather than individual virions or population growth. The sweep compares eigenvectors at different parameters; it is not an adiabatic time trajectory. Both eigenfrequencies stay positive over the displayed range.}
\label{fig:added-channel-mixing-portraits}
\end{figure}

The transferred occupation is not, off resonance, the fraction of bare
energy delivered to channel $b$. In the specified initial state,
\begin{equation}
\begin{aligned}
 \langle H_a+H_b\rangle&=\hbar\omega_a-\hbar\delta P_b,\\
 \langle H_{\rm int}\rangle&=\hbar\delta P_b,\\
 \langle H_{\rm ex}\rangle&=\hbar\omega_a.
 \end{aligned}
 \label{eq:phonon-two-channel-energy-ledger}
\end{equation}
In particular $\langle H_b\rangle/(\hbar\omega_a)
=(\omega_b/\omega_a)P_b$. The interaction energy accounts for the
apparent bare-energy mismatch; an unmodeled bath need not supply it.
Classical complex channel amplitudes obey the same two-by-two generator.
Their conserved squared norm measures total oscillator action in that
normalization, while their energy uses the full frequency matrix.
Consequently an oscillatory transfer curve alone does not identify
quantum occupation statistics.

The relation to a coordinate coupling must be checked separately. For
$\tfrac12\Pi_a^2+\tfrac12\omega_a^2Q_a^2+
 \tfrac12\Pi_b^2+\tfrac12\omega_b^2Q_b^2+
 \kappa_{ab}Q_aQ_b$, the coefficient $\kappa_{ab}$ has units
$\mathrm{s^{-2}}$, and stability requires
$\kappa_{ab}^2<\omega_a^2\omega_b^2$. Its exact frequencies satisfy
\begin{equation}
 \Omega_\pm^2=\frac{\omega_a^2+\omega_b^2}{2}
  \pm\sqrt{\frac{(\omega_a^2-\omega_b^2)^2}{4}+\kappa_{ab}^2},
 \qquad J=\frac{\kappa_{ab}}{2\sqrt{\omega_a\omega_b}}.
 \label{eq:phonon-full-coordinate-conversion-spectrum}
\end{equation}
The expression for $J$ follows by neglecting the pair terms
$\widehat b_a\widehat b_b+\widehat b_a^\dagger\widehat b_b^\dagger$.
This rotating-wave approximation needs weak coupling and near resonance
relative to the carrier frequencies. Its positivity condition is weaker
than the full coordinate stability condition and cannot replace it.
At resonance, for example,
$\Omega_\pm=\sqrt{\omega_c^2\pm2\omega_cJ}
=\omega_c\pm J-J^2/(2\omega_c)+O(J^3/\omega_c^2)$;
the number-conserving frequencies retain only the first two terms.
When the pair terms matter, bare total occupation need not be conserved,
although the stable full harmonic normal-mode occupations and total
energy are. This remains passive closed mechanics.

Figure~\ref{fig:closed-conversion-full-stiffness-stability} compares the two frequency constructions at resonance and identifies the different positivity boundaries.

\begin{figure}[H]
\centering
\begingroup
\definecolor{vlfGreen}{HTML}{176B64}
\definecolor{vlfDarkRed}{HTML}{762B38}
\definecolor{vlfRed}{HTML}{762B38}
\definecolor{vlfBlue}{HTML}{315D80}
\definecolor{vlfCopper}{HTML}{762B38}

\setlength{\unitlength}{1pt}
\scalebox{0.80}{%
\input{vlt-figures/figure-247.tex}%
}%
\endgroup
\caption{The rotating-wave spectrum does not replace stability of the original coordinate energy. At resonance $\omega_a=\omega_b=\omega_c>0$ and $J\geq0$, the coordinate coupling is $\kappa_{ab}=2\omega_cJ$. The exact mass-normalized stiffness eigenvalues are $\omega_c^2\pm2\omega_cJ$, giving frequencies $\Omega_\pm/\omega_c=\sqrt{1\pm2j}$ only while $0\leq j=J/\omega_c<1/2$. The dashed number-conserving branches are $\omega_\pm/\omega_c=1\pm j$, positive until $j=1$. (a) Near $j=0$ the pairs agree to first order. (b) The lower exact branch reaches zero at $j=1/2$ (open endpoint); beyond it the coordinate potential has negative curvature and is no longer the positive harmonic reference. The rotating-wave lower frequency remains positive there, showing why its weaker positivity test cannot certify the underlying coordinate model. These dimensionless parameter sweeps illustrate model domains; using a rotating-wave approximation requires weak coupling near resonance, rather than merely its own positive eigenvalues.}
\label{fig:closed-conversion-full-stiffness-stability}
\end{figure}

Figure~\ref{fig:phonon-closed-conversion} juxtaposes the exact exchange and
avoided crossing of the declared number-conserving benchmark. Frequencies,
coupling, preparation, and marker overlaps would all be required to assign
it to a particular virion system. Detuning limits reversible conversion;
added medium loss or an active biochemical reservoir changes the model
and must carry its own energy budget. The next subsection constructs a
different coupling, in which a specified electronic or configurational
state changes the mechanical equilibrium itself.

\subsection{Electronic and configurational coupling to viral phonons}
\label{sec:phonon-electronic-coupling}

Mechanical displacement changes a molecular electronic environment,
and a changed configuration exerts a reciprocal force. Expanding the
specified electronic Hamiltonian in displacement gives the coupling
derivatives and their normal-coordinate projections
\cite{Misra2012,Giustino2017}. A local chromophore, charge arrangement
or contact configuration supplies a direct finite-system application.
Periodic Bloch-state and crystal-momentum selection rules require
the corresponding extra symmetry and electronic structure.

\paragraph{What is the additional variable?}
\begin{definition}[Resolved electronic displacement coupling]
\label{def:s7-coupling-resolved-electronic}
A resolved electronic coupling specifies its retained electronic states
and the physical coordinates that change their energies. In a fixed, displacement-independent
orthonormal electronic basis $\{\ket a\}$, a local expansion is
\begin{equation}
\begin{aligned}
 \widehat H_{\rm el}(\vct u)
   &=\widehat H_{\rm el}(0)
       +\sum_s\widehat L_s Q_s+O(\|\vct u\|^2),\\
 \widehat L_s
   &=\sum_i\vct\phi_{s,i}\mathbin{\cdot}
        \nabla_{\vct u_i}\widehat H_{\rm el}\big|_0,\\
 \widehat L_s^\dagger&=\widehat L_s.
 \end{aligned}
 \label{eq:phonon-electronic-matrix-derivative}
\end{equation}
Each matrix $\widehat L_s$ describes how mode $s$ perturbs the retained
electronic problem. Diagonal entries change conditional forces;
off-diagonal entries can change the electronic state. These are different
mechanisms. A displacement-dependent electronic basis instead introduces
basis-derivative terms and requires a consistent adiabatic construction.
The mechanical reference and this electronic expansion must therefore be
chosen together~\cite{Giustino2017}.
\end{definition}

Translation invariance constrains the coupling. If an isolated object's
internal gap obeys
$\Delta\epsilon(\{\vct u_i+\vct a\})=
\Delta\epsilon(\{\vct u_i\})$ for constant $\vct a$, then
$\sum_i\nabla_{\vct u_i}\Delta\epsilon=0$.
Relative contacts, internal deformation and motion through an
inhomogeneous environment can produce nonzero derivatives. A reporter
couples to centre motion only through such a resolved physical mechanism.

A classical configurational indicator can also label conditional
mechanical surfaces. Its alternatives must describe the same enclosed
system: for example, a retained ligand is either bound or unbound, rather
than present in only one alternative. If a chemical reservoir supplies
that ligand, its chemical work belongs in the model. Conditional
isothermal free-energy surfaces can describe equilibrated hidden
coordinates, but do not by themselves define coherent electronic states
or a microscopic quantum Hamiltonian. The quantum construction below
uses specified electronic energy surfaces; the corresponding static
force-balance calculation also applies to classical conditional surfaces.

\paragraph{A configuration loads the mechanical network.}
Let $\ket g$ be the base electronic configuration and $\ket e$ the
retained second configuration, with
$\widehat P_e=\ket e\bra e$ and $\widehat P_e^2=\widehat P_e$.
Near the base equilibrium their energy difference is
\begin{equation}
\begin{aligned}
 \Delta\epsilon(\vct u)
   &=\epsilon+
              \vct g_{\rm cfg}^{\mathsf T}\vct u+O(\|\vct u\|^2),\\
 f_s&=\vct g_{\rm cfg}^{\mathsf T}\vct\phi_s,\\
 g_s&=f_s\sqrt{\frac{\hbar}{2\omega_s}}.
 \end{aligned}
 \label{eq:phonon-electronic-displacement-derivative}
\end{equation}
Here $\epsilon$ is the vertical gap at the base geometry, not yet the
gap between relaxed minima. The gradient $\vct g_{\rm cfg}$ has units
of force, and the additional physical force is
$-\widehat P_e\vct g_{\rm cfg}$. Because $Q_s$ has units
$\mathrm{kg}^{1/2}\,\mathrm m$, $f_s$ has units
$\mathrm{kg}^{1/2}\,\mathrm m\,\mathrm s^{-2}$, whereas $g_s$ is an
energy. These parameters are projections of a physical derivative,
rather than adjustable occupation labels. Retaining finitely many positive
modes and real couplings in a real mode basis gives
\begin{equation}
 \widehat H=\epsilon\widehat P_e+
       \underbrace{\sum_s\hbar\omega_s\widehat b_s^\dagger\widehat b_s}
                    _{\widehat H_{\rm ph}}+
       \underbrace{\widehat P_e\sum_sg_s
                  (\widehat b_s+\widehat b_s^\dagger)}
                    _{\widehat H_{\rm int}}.
 \label{eq:phonon-diagonal-electronic-hamiltonian}
\end{equation}
The same oscillator vacuum constants are subtracted in both electronic
sectors. Since $[\widehat H,\widehat P_e]=0$, this coupling exerts a
conditional force but never changes the configuration population. A
changing biochemical occupancy needs an additional transition mechanism.

\begin{proposition}[Conditional equilibrium shift and reorganization energy]
\label{prop:s7-coupling-reorganization}
For the finite positive-mode Hamiltonian above, define the reorganization
energy and shifted oscillators by
\begin{equation}
 \widehat B_s=\widehat b_s+
             \frac{g_s}{\hbar\omega_s}\widehat P_e,
 \qquad
 E_R=\sum_s\frac{g_s^2}{\hbar\omega_s}
      =\frac12\sum_s\frac{f_s^2}{\omega_s^2}.
 \label{eq:phonon-electronic-relaxation-energy}
\end{equation}
Then
\begin{equation}
 \boxed{\begin{aligned}
  \widehat H&=\sum_s\hbar\omega_s\widehat B_s^\dagger\widehat B_s
                            +(\epsilon-E_R)\widehat P_e,\\
  Q_s^*(\eta)&=-\eta\frac{f_s}{\omega_s^2},\qquad \eta=0,1.
 \end{aligned}}
 \label{eq:phonon-electronic-completed-square}
\end{equation}
\end{proposition}
\begin{proof}
Expand each shifted oscillator product using $\widehat P_e^2=\widehat P_e$
and its commutation with the mechanical operators. The constant term is
$E_R\widehat P_e$; minimizing each conditional quadratic gives $Q_s^*$.
\end{proof}

Figure~\ref{fig:new-s7-coupling-interior-offset} gives a graphical specialization of this boxed relation.

The reorganization energy $E_R$ is the reversible decrease from the
loaded surface at the base geometry to its own minimum. Shifted
operators retain the configuration projector. A linear force shifts
that minimum while preserving the spacings $\hbar\omega_s$; a curvature
change alters the frequency. For finitely many positive modes the
Hamiltonian stays bounded below, including when $\epsilon-E_R<0$
makes the loaded configuration energetically favoured.

Soft modes with nonzero $f_s$ give a large response, but the displaced
geometry must remain inside the collision-free harmonic basin. A force
along a free zero mechanism has no stationary harmonic displacement;
it requires restoring physics or a different reference. Moreover, a
relaxed electronic or dielectric response already included in the base
Hessian cannot be added again as an independent response without
redefining that Hessian. This avoids counting the same relaxation twice.

\subsubsection{A configuration load transmitted through compatible cells}
\label{sec:s7-coupling-cell-configuration}

A reporter or binding configuration on one virion can displace neighbouring
cells through the same assembled stiffness that carries mechanical
motion. Static condensation should therefore preserve both its force
on retained interfaces and its relaxed energy gap. These are two
independent outputs of one reduction.

\begin{definition}[A configuration load on a compatible assembly]
\label{def:s7-coupling-cell-load}
On a common constrained physical coordinate space, let $\mathsf K>0$
be the assembled stiffness and $\vct g$ the displacement derivative
of a configuration gap. For $P\in\{0,1\}$, the local potential is
$V=\vct u^{\mathsf T}\mathsf K\vct u/2+
P\vct g^{\mathsf T}\vct u+\epsilon P$.
Partition $\vct u=(\vct u_b,\vct u_i)$ into compatible interfaces
and eliminated interiors. All retained interactions, including weak
cross-cell tails, belong to $\mathsf K$.
\end{definition}

\begin{proposition}[Condensation preserves the configuration energy offset]
\label{prop:s7-coupling-cell-offset}
Define
\begin{equation}
 \begin{aligned}
 \mathsf K_S&=\mathsf K_{bb}
              -\mathsf K_{bi}\mathsf K_{ii}^{-1}\mathsf K_{ib},\\
 \vct g_S&=\vct g_b-\mathsf K_{bi}\mathsf K_{ii}^{-1}\vct g_i,
 &E_i&=\tfrac12\vct g_i^{\mathsf T}\mathsf K_{ii}^{-1}\vct g_i.
 \end{aligned}
 \label{eq:s7-coupling-condensed-load}
\end{equation}
Minimizing over $\vct u_i$ gives
\begin{equation}
 \begin{aligned}
 V_{\rm eff}(\vct u_b,P)
 &=\tfrac12\vct u_b^{\mathsf T}\mathsf K_S\vct u_b
      +P\vct g_S^{\mathsf T}\vct u_b+(\epsilon-E_i)P,\\
 E_R=\tfrac12\vct g^{\mathsf T}\mathsf K^{-1}\vct g
 &=E_i+\tfrac12\vct g_S^{\mathsf T}\mathsf K_S^{-1}\vct g_S.
 \end{aligned}
 \label{eq:s7-coupling-condensed-reorganization}
\end{equation}
\end{proposition}
\begin{proof}
Complete the interior square with minimizer
$\vct u_i^*=-\mathsf K_{ii}^{-1}
(\mathsf K_{ib}\vct u_b+P\vct g_i)$ and use $P^2=P$.
Completing the remaining interface square gives the second identity.
\end{proof}

The offset $E_i$ is the interior relaxation energy. Retaining it with
$\vct g_S$ preserves both interface forces and the relaxed gap.
Classical Gaussian integration gives the same potential, with a
configuration-independent determinant when curvature and measure are
shared. Different curvatures contribute the fluctuation terms in
\eqref{eq:phonon-configuration-harmonic-partition}. Dynamical interiors
instead use the retarded ports and induced forcing of
Subsection~\ref{sec:s4-cell-retarded-ports}.

For a fluorescence reporter coupled to a local contact deformation,
the measured equilibrium shift can constrain $\vct g_S$, while a
relaxed spectral or population gap also tests $E_i$. Both refer to the
same full-pool assembly and constraints. If a rigid zero mechanism
remains, an equilibrated load must be orthogonal to it and the inverse
acts on the specified restoring sector. Recruitment changes the
registered geometry and its matrix; it retains the actual particle,
configuration and medium state through the update. Selecting a larger
associated group does not by itself relax those states or supply the
energy represented by a new static minimum.

Figure~\ref{fig:new-s7-coupling-interior-offset} reconstructs a loaded virion chain and separates its interior and interface relaxation energies.

\begin{figure}[H]
\centering\begingroup
\definecolor{vltTeal}{HTML}{176B64}
\definecolor{vltBlue}{HTML}{315D80}
\definecolor{vltGold}{HTML}{A87520}
\definecolor{vltPlum}{HTML}{725877}
\definecolor{vltRed}{HTML}{762B38}
\definecolor{vltYellow}{HTML}{D5AD3C}
\definecolor{vltNeutral}{HTML}{777D83}
\setlength{\unitlength}{1pt}
\scalebox{0.90}{%
\input{vlt-figures/figure-248.tex}%
}\endgroup
\caption{Conditional loading of the same supported three-virion chain. Use $\mathsf K/k_\star$ with diagonal $2$, neighbouring entries $-1$, $g=k_\star u_\star(0,1,0)^{\mathsf T}$ and $\epsilon=1.5E_\star$, $E_\star=k_\star u_\star^2$. (a) Loading gives $u/u_\star=(-0.5,-1,-0.5)$; gray outlines and matched coils register the displacement. The shared ruler uses reference spacing $a$, with origin at the first centre. (b) Minimizing the interior at $u_b=(q,q)$ gives the complete potential. Omitting $-E_iP$ preserves the interface force but changes the gap. (c) Interior and interface relaxation each contribute $0.25E_\star$, leaving relaxed gap $E_\star$. Static condensation preserves both contributions without removing a physical virion. The boxed result illustrated here is~\eqref{eq:phonon-electronic-completed-square}.}
\label{fig:new-s7-coupling-interior-offset}
\end{figure}

\begin{example}[From a capsid deformation port to a collective link]
\label{ex:s7hpv-shell-contact-compliance}
Glycan-induced changes in HPV capsid indentation response motivate a
conditional shell coordinate~\cite{Feng2024}. To calculate how that
coordinate affects a whole-particle contact, let $x$ be its imposed
relative approach, $s_1,s_2$ two local shell deformations, and use
\begin{equation}
 \begin{aligned}
 E_{\rm port}&=\tfrac12 k_b(x-s_1-s_2)^2
                        +\tfrac12 k_c(s_1^2+s_2^2),\\
 k_{\rm eff}^{-1}&=k_b^{-1}+2k_c^{-1},\qquad
 \frac{k_{\rm eff}}{k_b}=\frac{1}{1+2k_b/k_c},\quad k_b,k_c>0 .
 \end{aligned}
 \label{eq:s7hpv-shell-contact-series}
\end{equation}
At equilibrium the same contact force satisfies
$F=k_b(x-s_1-s_2)=k_cs_1=k_cs_2$, proving the series law.
For assigned $k_b=3\times10^{-4}\,\mathrm{N\,m^{-1}}$, changing only
$k_c$ from $0.10$ to $0.05\,\mathrm{N\,m^{-1}}$ changes $k_{\rm eff}$
from $2.9821$ to $2.9644$ in units
$10^{-4}\,\mathrm{N\,m^{-1}}$: a decrease of about $0.59\%$.
These shell-port values are illustrative, not quoted indentation data.
\end{example}
The softer element controls this static load path. A substantial change
in one shell response can consequently have a small effect on the
collective link. A nanoindentation port and a local binding port have
different loading and boundary geometries; the former requires an
explicit transfer model before supplying $k_c$. Dynamic shell coordinates
use the full Schur response and matched noise of Sections~4--6.
This gives a practical criterion for retaining capsid deformation:
compare the predicted contact and detector corrections with the stated
band and tolerance, rather than identifying shell stiffness with
interparticle curvature.

\paragraph{A spectroscopic consequence of displaced equilibria.}
One mode makes the meaning of mechanical dressing particularly explicit.
Set $\gamma=g/(\hbar\omega)$, $S=\gamma^2$, and
$\widehat D(\alpha)=\exp(\alpha\widehat b^\dagger-
\alpha^*\widehat b)$. The relaxed excited eigenstates are
$\ket e\widehat D(-\gamma)\ket N$, with energies
$\epsilon-E_R+N\hbar\omega$. Suppose the initial state is $\ket{g,0}$,
the electronic transition operator is independent of this mechanical
coordinate, and both surfaces have the same curvature. Its mechanical
overlap factors are then
\begin{equation}
 W_N=|\bra N\widehat D(\gamma)\ket0|^2
       =e^{-S}\frac{S^N}{N!},\qquad
 E_N^{\rm tr}=\epsilon-E_R+N\hbar\omega,
 \qquad \sum_{N=0}^\infty W_NE_N^{\rm tr}=\epsilon.
 \label{eq:phonon-configuration-vibrational-overlaps}
\end{equation}
These are Franck--Condon overlap factors~\cite{Condon1928}: in the
vertical-transition approximation, the initial mechanical state projects
onto several levels of the displaced potential. The stated
coordinate-independent transition operator is the Condon approximation. When $\epsilon>E_R$ these
are positive absorption gaps from the stated ground preparation. The
lowest transition is the relaxed gap, while the weighted mean recovers
the vertical gap. The dimensionless $S$ measures the displacement in
oscillator units; it does not determine a biological rate on its own.
An observed spectrum additionally requires an allowed electronic
transition, its matrix element, linewidths, preparation, and experimental
resolution. Finite temperature or changed curvatures modifies these
weights. Figure~\ref{fig:conditional-wells-and-closed-energy} connects
the equilibrium shift to this possible spectral signature.

\begin{figure}[H]
\centering\begingroup
\definecolor{vltTeal}{HTML}{176B64}
\definecolor{vltBlue}{HTML}{315D80}
\definecolor{vltGold}{HTML}{A87520}
\definecolor{vltPlum}{HTML}{725877}
\definecolor{vltRed}{HTML}{762B38}
\definecolor{vltYellow}{HTML}{D5AD3C}
\definecolor{vltNeutral}{HTML}{777D83}
\setlength{\unitlength}{1pt}
\scalebox{0.90}{%
\input{vlt-figures/figure-249.tex}%
}\endgroup
\caption{One conditional force, its equilibrium shift and overlap spectrum. Set $\gamma=g/(\hbar\omega)=1/2$ and $\epsilon=\hbar\omega$. (a) Potentials $q^2/2$ and $1+q^2/2+\sqrt2\gamma q$ have the same curvature; the loaded minimum is $q^*=-\sqrt2\gamma$, lowered by $E_R=0.25\hbar\omega$. (b) Equal curvatures and a coordinate-independent transition operator give $W_N=e^{-S}S^N/N!$ at $E_N=\epsilon-E_R+N\hbar\omega$. The log-axis stems for $N=0,\ldots,4$ contain more than $0.999993$ of the weight; the full mean gap is $\epsilon$. Below, the auxiliary pair reconstructs $u=(-1,1)Q/\sqrt{2m}$ and loaded shifts $(0.5,-0.5)\sqrt{\hbar/(m\omega)}$. Rulers use reference separation $d$ and origin at the left reference centre; approach shortens the middle coil span and lengthens the supported end spans. A measured absorption spectrum also uses its transition matrix element and resolution.}
\label{fig:conditional-wells-and-closed-energy}
\end{figure}

\paragraph{Closed evolution after a sudden preparation.}
Prepare $\ket e\otimes\ket0$ at the undisplaced geometry. This is not
the dressed ground state. For $\gamma_s=g_s/(\hbar\omega_s)$, direct
oscillator evolution gives, up to a phase,
\begin{equation}
\begin{aligned}
 \alpha_s(t)&=-\gamma_s(1-e^{-i\omega_st}),\\
 \langle\widehat N_s\rangle_t
       &=4\gamma_s^2\sin^2(\omega_st/2),\\
 \langle\widehat Q_s\rangle_t
       &=-\frac{f_s}{\omega_s^2}(1-\cos\omega_st).
 \end{aligned}
 \label{eq:phonon-electronic-closed-dynamics}
\end{equation}
The configuration remains occupied while the mechanics oscillates about
its new equilibrium. The energy ledger is
\begin{equation}
 \langle\widehat H_{\rm ph}\rangle_t
  =2\sum_s\frac{g_s^2}{\hbar\omega_s}(1-\cos\omega_st)
  =-\langle\widehat H_{\rm int}\rangle_t,
 \qquad\langle\widehat H\rangle_t=\epsilon.
 \label{eq:phonon-electronic-closed-energy}
\end{equation}
For one mode the bare phonon energy can reach $4E_R$, although the
energy above the relaxed excited minimum is only $E_R$. The negative
interaction contribution reconciles these statements. Counting bare
quanta alone therefore does not measure energy newly created in the
complete system. This finite preparation produces reversible motion;
it supplies neither negative damping nor sustained self-excitation.

If the electronic state initially has a nonzero coherence
$\rho_{eg}(0)$, a product preparation with oscillator vacuum gives
\begin{equation}
 \frac{|\rho_{eg}(t)|}{|\rho_{eg}(0)|}
    =\exp\!\left[-\sum_s\gamma_s^2(1-\cos\omega_st)\right].
 \label{eq:phonon-electronic-coherence-overlap}
\end{equation}
The exponent measures the distinguishability of the two conditional
mechanical states: their overlap decreases when their displacements
separate. For an initially undisplaced thermal Gibbs state independent
of the electronic preparation, the displacement trace instead gives
\begin{equation}
 \frac{|\rho_{eg}(t)|}{|\rho_{eg}(0)|}
  =\exp[-\Gamma_\beta(t)],\qquad
 \Gamma_\beta(t)=\sum_s\gamma_s^2(1-\cos\omega_st)
                 \coth\!\left(\frac{\beta\hbar\omega_s}{2}\right).
 \label{eq:phonon-electronic-thermal-coherence}
\end{equation}
This thermal factor follows from
$\tr[\rho_\beta\widehat D(\alpha)]
=\exp[-|\alpha|^2\coth(\beta\hbar\omega/2)/2]$
with the present projector coupling convention~\cite{ChenChen2021}.
The full coherence, including its phase, is
\begin{equation}
 \frac{\rho_{eg}(t)}{\rho_{eg}(0)}=
 \exp\!\left[-\frac{i\epsilon t}{\hbar}
        +i\sum_s\gamma_s^2(\omega_st-\sin\omega_st)
        -\Gamma_\beta(t)\right].
 \label{eq:phonon-electronic-coherence-phase}
\end{equation}
Its initial phase rate uses the vertical gap $\epsilon$; the relaxed
gap enters together with the oscillatory phase correction. This initial phase rate checks the consistency of the energy convention.

Writing $p_e=\tr(\rho_{\rm el}\widehat P_e)$ and
$\bar N_s=(e^{\beta\hbar\omega_s}-1)^{-1}$, the same product thermal
preparation gives
$\langle\widehat N_s\rangle_t=\bar N_s+p_e|\alpha_s(t)|^2$.
Its excess bare mechanical energy is $p_e$ times
\eqref{eq:phonon-electronic-closed-energy} and is still compensated by
interaction energy. Temperature changes the initial fluctuations and
conditional phase averaging, without changing the mean force-induced
displacement at fixed parameters. A thermal state of the full dressed
Hamiltonian is a different, stationary preparation. The reduced
electronic coherence can decrease even though the total density operator
retains unit trace. In a finite set of modes there are recurrences or
near recurrences, rather than an automatic Markov decay rate
(Figure~\ref{fig:finite-mode-coherence-and-thermal-returns}). Loss of
this reduced coherence alone does not certify entanglement for an
initially mixed preparation.

These results define a shared-parameter calibration test.
For a specified configuration change, a resolved equilibrium shift
$\Delta Q_s$ predicts $f_s=-\omega_s^2\Delta Q_s$ in this model.
The same $f_s$ then fixes $E_R$, the mean transient displacement, and
the conditional quantum overlap factors. A structural readout or a
reporter spectrum must be projected onto those same coordinates and
compared at the same reference conditions. Agreement of mean motion
with the forced classical oscillator tests the mechanical coupling;
resolved quantum fluctuations or level weights would test the additional
quantized description. Neither test follows from the name of the mode.

\begin{figure}[H]
\centering
\begingroup
\definecolor{vlfGreen}{HTML}{176B64}
\definecolor{vlfDarkRed}{HTML}{762B38}
\definecolor{vlfRed}{HTML}{762B38}
\definecolor{vlfBlue}{HTML}{315D80}
\definecolor{vlfCopper}{HTML}{762B38}

\setlength{\unitlength}{1pt}
\scalebox{0.80}{%
\input{vlt-figures/figure-250.tex}%
}%
\endgroup
\caption{Energy exchange and coherence returns under the same closed diagonal coupling, with $\gamma=1/2$, $\epsilon=\hbar\omega$ and $\tau=\omega t$. (a) From $|e\rangle\otimes|0\rangle$, $\alpha=-\gamma(1-e^{-i\tau})$. Changes of bare phonon and interaction energies are $\pm2E_R(1-\cos\tau)$ and sum to zero. Configuration population stays fixed. (b) A separate product preparation with initial configuration coherence and an undisplaced thermal oscillator gives $C(t)=\exp[-\gamma^2(1-\cos\tau)\coth(b/2)]$ for $b=\infty,2,1/2$. All curves return to one at $2\pi,4\pi$; temperature changes intermediate reduction. The two panels specify different preparations, so their energy and coherence ordinates are not interchangeable. Finite commensurate modes can recur without irreversible decay; generic incommensurate sets need not share an exact period.}
\label{fig:finite-mode-coherence-and-thermal-returns}
\end{figure}

\paragraph{Two configurations can interact through shared mechanics.}
For two compatible commuting indicators $P_1,P_2$, the load on mode $s$
can be $f_{s1}P_1+f_{s2}P_2$. Eliminating its static displacement yields
\begin{equation}
 U_{\rm rel}=-\frac12\sum_s
       \frac{(f_{s1}P_1+f_{s2}P_2)^2}{\omega_s^2},\qquad
 U_{12}=-P_1P_2\sum_s\frac{f_{s1}f_{s2}}{\omega_s^2}.
 \label{eq:phonon-mediated-configuration-interaction}
\end{equation}
This is an inverse-stiffness force overlap. Equivalently, the vectors
with components $f_{sj}/\omega_s$ have a scalar product whose sign sets
the cross term: aligned loads favor joint occupation, opposed loads
penalize it, and orthogonal loads give no cross term
(Figure~\ref{fig:mechanically-mediated-configuration-overlap}). For
example, two distinct contact or binding variables could load a common
retained deformation. A possible mechanical contribution to cooperativity
then has a defined sign and magnitude, to be compared with direct
chemical interactions. Mutually exclusive states have $P_1P_2=0$ and
cannot realize this joint-occupation example. With identical conditional
curvatures, the oscillator partition-function factor is the same in all
sectors, so the static energy shift also supplies the corresponding
equilibrium free-energy difference. It still does not specify a
transition rate or a conductive electronic phase.

\begin{figure}[H]
\centering\begingroup
\definecolor{vltTeal}{HTML}{176B64}
\definecolor{vltBlue}{HTML}{315D80}
\definecolor{vltGold}{HTML}{A87520}
\definecolor{vltPlum}{HTML}{725877}
\definecolor{vltRed}{HTML}{762B38}
\definecolor{vltYellow}{HTML}{D5AD3C}
\definecolor{vltNeutral}{HTML}{777D83}
\setlength{\unitlength}{1pt}
\scalebox{0.90}{%
\input{vlt-figures/figure-251.tex}%
}\endgroup
\caption{Weighted-force overlap and a finite-network realization. (a,b) Equal-norm vectors $\vct a_j=(f_{sj}/\omega_s)_s$ give $U_{12}=-a_\star^2\cos\vartheta$; the marked $\pi/3$ point has cross energy $-a_\star^2/2$. This is a coefficient-space angle. (c) In units $a_\star^{\rm lat}$, eighteen reference sites have $x,y\in\{-1,0,1\}$ and $z=\pm0.45$. All nearest axis bonds and the positive-slope diagonal of each elementary face have equal tangent curvature. With central-extension rows $\mathsf B$ and equal onsite pins, $\mathsf K/k_\star=\Id+\mathsf B^{\mathsf T}\mathsf B$. Teal/burgundy arrows apply $+y$ forces at $(0,-1,-0.45)$ and $(0,1,0.45)$, scaled so $\vct f_j^{\mathsf T}\mathsf K^{-1}\vct f_j=E_\star$ and $\sqrt{E_\star/k_\star}=0.14a_\star^{\rm lat}$. Their joint static displacement and cross energy are computed from this same matrix. Rigid bodies, reference ghosts and face-attached helices retain their geometric meaning; onsite pins are implicit. Bond color identifies orientation, rather than curvature. This independent example uses its own weighted-force angle. Compatible joint configurations realize the cross term; mutually exclusive indicators have zero joint occupation.}
\label{fig:mechanically-mediated-configuration-overlap}
\end{figure}

\paragraph{Actual conversion changes the electronic population.}
A term off diagonal in $\ket g,\ket e$ can exchange an electronic
excitation and a mechanical quantum. A separate weak, near-resonant
model for one positive mode is
\begin{equation}
 \widehat H_{\rm tr}=\hbar\omega_{eg}\widehat P_e+
       \hbar\omega_s\widehat N_s+
       g_{eg}(\widehat\sigma_+\widehat b_s+
                           \widehat\sigma_-\widehat b_s^\dagger),
 \quad \widehat\sigma_+=\ket e\bra g,\quad
 \widehat\sigma_-=\widehat\sigma_+^\dagger,\quad g_{eg}\in\R.
 \label{eq:phonon-electronic-resonant-conversion}
\end{equation}
Here $g_{eg}$ is the transition matrix element of the displacement
coupling, with energy units; it is not the diagonal force $g_s$.
The rotating-wave approximation discards pair terms oscillating near
$\omega_{eg}+\omega_s$ and requires
$|g_{eg}|/\hbar,|\omega_{eg}-\omega_s|\ll\omega_{eg},\omega_s$.
If diagonal dressing is important it must be included consistently;
the parameters of this separate reduced transition model cannot simply
be read from an unshifted electronic gap. Its conserved excitation number
is $\widehat P_e+\widehat N_s$. In the basis
$\{\ket{e,0},\ket{g,1_s}\}$ the Hamiltonian is the two-by-two matrix
with diagonal entries $\hbar\omega_{eg},\hbar\omega_s$ and off-diagonal
entries $g_{eg}$, so, for $G=|g_{eg}|/\hbar$ and
$\delta=\omega_{eg}-\omega_s$,
\begin{equation}
 P_{g,1_s}(t)=\frac{4G^2}{\delta^2+4G^2}
       \sin^2\!\left(\frac{t}{2}\sqrt{\delta^2+4G^2}\right).
 \label{eq:phonon-electronic-conversion-probability}
\end{equation}
On resonance an initially prepared $\ket{e,0}$ becomes $\ket{g,1_s}$
at $t=\pi/(2G)$ and later returns. This is a controlled example of
single-quantum conversion, with the initial electronic excitation as its
resource. Thermally occupied initial levels give different
$G\sqrt{N+1}$ exchange frequencies and no common complete-transfer time
in general. A large optical electronic gap is not automatically resonant
with a low-frequency collective mechanical mode; a suitable transition,
its displacement derivative, preparation, and lifetime must be identified.
An irreversible decay rate additionally requires accepting modes and a
justified spectral and time coarse graining. The finite closed result
does not supply an autonomous excitation cycle.

Figure~\ref{fig:one-phonon-conversion-and-classical-drive} compares this
quantum preparation resource with the classical linear-force preparation
derived next. The two mechanisms can have related mean oscillations
while giving different phonon-number statistics. Their distinction is
therefore experimentally substantive whenever occupation statistics can
be resolved.

\begin{figure}[H]
\centering
\begingroup
\definecolor{vlfGreen}{HTML}{176B64}
\definecolor{vlfDarkRed}{HTML}{762B38}
\definecolor{vlfRed}{HTML}{762B38}
\definecolor{vlfBlue}{HTML}{315D80}
\definecolor{vlfCopper}{HTML}{762B38}

\setlength{\unitlength}{1pt}
\scalebox{0.80}{%
\input{vlt-figures/figure-252.tex}%
}%
\endgroup
\caption{Two preparation resources and their one-phonon probabilities. (a) Resonant rotating-wave conversion from $|e,0\rangle$ gives $P_{g,1_s}=\sin^2(Gt)$, with $\omega_{eg}=\omega_s=100G$ and $G=|g_{eg}|/\hbar$. It reaches one at $Gt=\pi/2$ and then reverses; the initial configuration excitation supplies the energy. (b) Fixed-frequency classical linear forcing from oscillator vacuum gives a coherent state with $P_1=\bar N e^{-\bar N}\le1/e$. The panels compare preparation families using different horizontal variables, rather than matched pulse protocols. The second bound belongs to coherent linear-drive preparation; other sources or conditioning have different constraints. Neither a mean displacement nor a response peak alone reads out these occupation probabilities. The boxed result illustrated here is~\eqref{eq:phonon-linear-drive-one-quantum-limit}.}
\label{fig:one-phonon-conversion-and-classical-drive}
\end{figure}

\subsubsection{Protein configurations: what fluctuations contribute to state populations}
\label{sec:phonon-protein-configuration-application}

The conditional surfaces have a more general biophysical use than
electronic excitation. A protein can occupy different structural
configurations, each with its own distribution of internal fluctuations.
The populations of those configurations determine which structures are
available over a measurement interval. Changes in fluctuation spectra
can alter that balance even when a mean structural coordinate changes
little. Fluctuation-mediated accounts of protein allostery provide an
established motivation for retaining this information
\cite{CooperDryden1984}; elastic-network models also connect collective
coordinates to protein fluctuation patterns~\cite{Bahar1997}.

To make the thermodynamic consequence explicit, retain two
same-composition configurations $a=A,B$, with minimum energies $E_a^0$,
the same positive inertia, and the same number $r$ of positive internal
modes. Remove the same rigid motions and impose the same external
constraints in both configurations. In a local harmonic Hamiltonian
with frequencies $\omega_{a,s}$, and basin boundaries outside the
appreciable thermal weight, the conditional free energy is
\begin{equation}
 \mathcal F_a^{\rm harm}(T)=E_a^0+
     k_{\rm B}T\sum_{s=1}^r
         \ln\!\left[2\sinh\!\left(\frac{\beta\hbar\omega_{a,s}}2\right)\right],
 \qquad \beta=(k_{\rm B}T)^{-1}.
 \label{eq:phonon-configuration-harmonic-partition}
\end{equation}
This follows by multiplying the conditional oscillator partition
functions. When curvatures differ, their zero-point offsets differ as
well and cannot both be discarded while retaining the same $E_a^0$.
For configurations whose resolved fluctuations are classical, the
Gaussian basin integral instead gives
\begin{equation}
 \begin{aligned}
 \mathcal F_B-\mathcal F_A
   &\simeq E_B^0-E_A^0
       +k_{\rm B}T\sum_s\ln\frac{\omega_{B,s}}{\omega_{A,s}}\\
   &=E_B^0-E_A^0+
       \frac{k_{\rm B}T}{2}\ln\frac{\det\mathsf K_B}{\det\mathsf K_A},\\
 \frac{p_B}{p_A}&=e^{-\beta(\mathcal F_B-\mathcal F_A)}.
 \end{aligned}
 \label{eq:phonon-configuration-curvature-populations}
\end{equation}
The determinants refer to the same constrained coordinate space and
measure. Common inertia makes its determinant cancel. The classical
formula can also be derived from conditional isothermal potentials,
provided their unresolved contributions are matched at the same
temperature and resolution. A free energy that has already integrated
out the displayed modes must not receive their partition factor again.

For example, let the minimum energies agree and let one frequency double
in configuration $B$, with all other frequencies unchanged. In the
classical limit this one mode raises $\mathcal F_B-\mathcal F_A$ by
$k_{\rm B}T\ln2$, halves $p_B/p_A$, and reduces that mode's conditional
coordinate variance by a factor of four. Both conditional mean
coordinates may remain at their respective reference values. This
specified example shows why a mean structure, a fluctuation amplitude
and a configurational population supply different information. It is
not a fitted allosteric free energy; actual communication between
binding sites requires their conditional binding free-energy changes.

\paragraph{Retaining the thermodynamic weight of a surface state.}
The same elimination also affects configurational populations. Consider
two same-composition surface conformations $c=A,B$ of an HPV-inspired
ideal VLP, each with the preceding local quadratic storage and minimum
offset $E_c^0$. Use the same constrained mechanical coordinates and the
same calibrated physical $\chi$ coordinate and integration measure in
both basins. Assume both Gaussian basins are well inside their physical
boundaries. Integrating $\chi$ gives the conditional potential of mean
force, up to one common constant,
\begin{equation}
 \mathcal F_c^{\rm eff}(\vct X)
   =E_c^0+\tfrac12\vct X^{\mathsf T}\mathsf H_{{\rm rel},c}\vct X
       +\frac{k_{\rm B}T}{2}\ln\frac{a_{\chi,c}}{a_\star},
 \qquad a_\star>0 .
 \label{eq:s7hpv-surface-condensation-free-energy}
\end{equation}
The reference $a_\star$ only makes the logarithm dimensionless. Completing
the internal square gives the relaxed mechanical curvature; its Gaussian
normalization supplies the logarithmic term. A vector surface state
replaces this ratio by $\det(\mathsf A_c\mathsf A_\star^{-1})$ on the
same allowed internal space.

\begin{example}[The same mechanical record, different basin weights]
\label{ex:s7hpv-surface-hidden-entropy}
Hold $\mathsf H_\infty,\mathsf G_X,\mathsf C_X,\vct v$ and the minimum
offset fixed. Choose
$(a_{\chi,B},b_{\chi,B},d_{\chi,B})=
(4a_{\chi,A},2b_{\chi,A},4d_{\chi,A})$.
With the conformation held during measurement, both have the same
$\kappa_\chi$, $\tau_\chi$, mechanical response and stationary
mechanical fluctuation spectrum. Their
conditional surface variances instead obey
$\operatorname{Var}(\chi\mid\vct X,c)=k_{\rm B}T/a_{\chi,c}$.
The equilibrated Gaussian reference gives $p_B/p_A=1/2$,
despite identical relaxed mechanical widths. The same physical surface
calibration is essential: a coordinate change alone would transform
the integration measure and leave the partition weight unchanged.
\end{example}
This assigned comparison concerns conformational thermodynamics, not a
measured HPV population ratio. A real ligand-binding alternative also
retains its ligand inventory or chemical-potential term. Switching
rates require the separate transition law. The example shows precisely
which free-energy information is lost if a reduced mechanical response
is used without its eliminated-state normalization.

The useful variables depend on the question. A mode coordinate $Q_s$
describes motion inside a basin. A configuration indicator $a(t)$
records which basin a molecule occupies; the probabilities $p_a$ describe
an ensemble preparation. If the internal modes equilibrate rapidly
relative to switching, their contribution can be retained through
$\mathcal F_a(T)$ while the slow configuration is explicit. If their
relaxation is resolved, selected coordinates or their history must also
be carried. Equilibrium populations constrain free-energy differences,
whereas dwell times and switching correlations constrain kinetics.
The former do not fix the latter.
The lagged-coordinate example in Section~\ref{sec:phonon-protein-slow-variables}
develops this kinetic part of the variable-selection problem.

This separation suggests complementary uses of structural, fluctuation
and population readouts. A conditional displacement shift tests a force
derivative such as $f_s$; a conditional frequency shift tests a curvature;
a change of state populations tests the complete conditional free energy.
Their joint interpretation can distinguish a shifted minimum from an
altered fluctuation entropy. It applies to protein assemblies and other
metastable biomolecular structures with well-defined conditional basins,
without requiring an electronic transition or a low-occupation quantum
preparation. The physical coordinates, constraints and matched background
free energy determine whether a proposed configuration is an adequate
biophysical variable.

\subsection{External fields: mode selection, preparation, and work}
\label{sec:phonon-electric-drive}

An applied field supplies both a force profile and a work source.
Its projection selects the loaded mechanical coordinates; the source
model determines their energy input. Electrical driving preserves the
sealed virion population while exchanging energy with its mechanics.
It therefore connects the closed conversion examples above to the
finite ionic and biochemical resources below.

\paragraph{From an electrical potential to modal force and stiffness.}
Take a prescribed quasistatic potential
$\Phi_{\rm ext}(\vct r,t)$ and effective charges $q_i$ specified for the
chosen mechanical resolution, screening conditions, and frequency range.
With $\vct E=-\nabla\Phi_{\rm ext}$, expansion gives
\begin{align}
 U_E&=\sum_iq_i\Phi_{\rm ext}(\vct R_i^0+\vct u_i,t)
   =U_E(0,t)-\sum_sF_s(t)Q_s
               +\frac12\sum_{s,v}\Delta D_{sv}(t)Q_sQ_v+\cdots,
                  \nonumber\\
 F_s&=\sum_iq_i\vct\phi_{s,i}\mathbin{\cdot}
                    \vct E(\vct R_i^0,t),\nonumber\\
 \Delta D_{sv}&=\sum_iq_i\vct\phi_{s,i}^{\mathsf T}
       [\nabla\nabla\Phi_{\rm ext}(\vct R_i^0,t)]\vct\phi_{v,i}.
 \label{eq:phonon-electric-modal-overlap}
\end{align}
The field value sets the modal force, with units
$\mathrm{kg}^{1/2}\,\mathrm m\,\mathrm s^{-2}$; its gradient sets
the curvature correction $\Delta D_{sv}$, with units $\mathrm s^{-2}$.
A static force shifts equilibrium, where the full Hessian is then
evaluated. Field gradients can mix patterns and alter stability even
when one instantaneous force projection vanishes.

The electrical boundary-value problem determines field penetration and
screening. A mechanically reflecting cylinder does not specify electrode
potentials, dielectric interfaces, or ionic charge relaxation. In a
screened solution, a charge inferred from transport or electrokinetic
measurements need not equal an unscreened net charge used in a vacuum
Coulomb model. Whichever effective coupling is chosen must reproduce the
force in its declared regime. If polarization has substantial frequency
dependence or dissipation, it needs a dynamical response description
rather than substitution of a complex charge into a conservative
instantaneous Hamiltonian.

\paragraph{A field need not excite every mechanically allowed mode.}
\begin{proposition}[Uniform-field selection for equal charge-to-mass ratios]
\label{prop:s7-coupling-uniform-field-selection}
Within the centre-charge translational model of
\eqref{eq:phonon-electric-modal-overlap}, a free network's positive
internal modes are mass-orthogonal to common translation. In a uniform
field, identical charge-to-mass ratios $q_i/m_i=a_{\rm qm}$ therefore imply
\begin{equation}
 \boxed{\sum_i m_i\vct\phi_{s,i}=0
 \quad\Longrightarrow\quad
 F_s=a_{\rm qm}\vct E\mathbin{\cdot}
                     \sum_i m_i\vct\phi_{s,i}=0.}
 \label{eq:phonon-uniform-field-dark-modes}
\end{equation}
\end{proposition}
\begin{proof}
Insert $q_i=a_{\rm qm}m_i$ into the modal force overlap and use
mass orthogonality to the translational displacement.
\end{proof}

Equal charge-to-mass ratios give equal centre accelerations, forcing
the common sector while leaving the internal patterns dark to this
drive. Identical mass-proportional pinning preserves that separation.
Unequal supports, charge heterogeneity and field gradients change the
overlaps. Visibility is thus a property of a mode together with its
specified perturbation or detector.

For example, two equal masses at $z_{1,2}=\mp a/2$, with
$q_{1,2}=q(1\pm\delta_q)$ and relative mode
$\vct\phi_r=(1,-1)/\sqrt{2m}$, see a local axial field
$E(z)=E_0+E'z$. Direct substitution gives
\begin{equation}
 F_r=\frac{\sqrt2qE_0}{\sqrt m}(\delta_q-\eta_E),
 \qquad \eta_E=\frac{aE'}{2E_0},\qquad E_0\ne0.
 \label{eq:phonon-electric-pair-overlap}
\end{equation}
Heterogeneity and the field gradient can reinforce or cancel in the
relative-force projection
(Figure~\ref{fig:external-field-relative-mode-selection}). The full
controlled model also retains common motion, induced curvature and the
shifted equilibrium. Together these predict whether a measured record
is dominated by relative displacement or translation.

Internal electrical activity requires an appropriately resolved dipole.
For a finite neutral object with total dipole
$\vct d(\{Q_s\})=\vct d_0+\sum_s\vct d_sQ_s+\cdots$,
$\vct d_s=\partial\vct d/\partial Q_s$ supplies the uniform-field
coupling $-\sum_s(\vct d_s\cdot\vct E)Q_s$.
It can be nonzero for internal deformation even when net charge is zero.
It need not equal the point-charge overlap in
\eqref{eq:phonon-electric-modal-overlap}, since retained polarization
and internal charge rearrangement also contribute. Crystal Born effective
charge tensors and their periodic normalizations cannot simply be assigned
to virion centers.

For a reversibly induced dipole with a real symmetric positive-definite
quasistatic polarizability tensor $\mathsf a_{\rm pol}$, its internal
cost and field coupling are
$U(\vct d)=\vct d^{\mathsf T}\mathsf a_{\rm pol}^{-1}\vct d/2
-\vct d\cdot\vct E$. Minimizing at fixed prescribed field gives
$\vct d=\mathsf a_{\rm pol}\vct E$ and hence
$U_{\rm ind}=-\vct E^{\mathsf T}\mathsf a_{\rm pol}\vct E/2$.
This is the reversible induced-polarization construction in that
control convention~\cite{FeynmanLectures1963}. A spatially
uniform field gives no translational force on a neutral isotropic object
with constant polarizability. Field gradients produce translation;
anisotropic polarizability can produce torque if orientation is retained;
and $\partial\mathsf a_{\rm pol}/\partial Q_s$ can force an internal
deformation even in a uniform field. These mechanisms call for different
coordinates and calibration, so neutrality alone does not imply that all
internal modes are dark.

\begin{figure}[H]
\centering\begingroup
\definecolor{vltTeal}{HTML}{176B64}
\definecolor{vltBlue}{HTML}{315D80}
\definecolor{vltGold}{HTML}{A87520}
\definecolor{vltPlum}{HTML}{725877}
\definecolor{vltRed}{HTML}{762B38}
\definecolor{vltYellow}{HTML}{D5AD3C}
\definecolor{vltNeutral}{HTML}{777D83}
\setlength{\unitlength}{1pt}
\scalebox{0.90}{%
\input{vlt-figures/figure-253.tex}%
}\endgroup
\caption{Electrical force selection from a free pair to a sixteen-virion patch. (a) In the centre-charge translation model, identical charges in a uniform field give zero relative translational overlap and equal centre acceleration. (b) At $z_{1,2}=\mp a/2$, $q_{1,2}=q(1\pm\delta_q)$ and $\eta_E=aE'/2E_0$ give $F_r/F_\star=\delta_q-\eta_E$, $F_\star=\sqrt2qE_0/\sqrt m$. (c) The independent square central-spring patch has equal masses, link curvature and onsite pinning $k_\star$. Uniform and first-row force profiles select the printed normalized vertical modes. One displacement gain and coil family are shared: row motion strains vertical links, while common translation preserves every span. This pinned example uses its own calibrated field and supports. The boxed result illustrated here is~\eqref{eq:phonon-uniform-field-dark-modes}.}
\label{fig:external-field-relative-mode-selection}
\end{figure}

\paragraph{A uniform field can select an angular pattern.}
For an HPV-inspired ideal VLP, the registered outer shape and the charge
pattern are separate inputs. A preparation with asymmetrically occupied
surface sites can have lower electrical symmetry than the ideal
icosahedron. Glycan engagement and changing HPV16 VLP surface distributions
motivate resolving such preparation states~\cite{Feng2024,Vega2016}; they
do not determine a permanent dipole or an electrical force coefficient.

\begin{definition}[Frozen-site electrical wrench]
\label{def:s7hpv-electrical-wrench}
During a specified frozen-population interval, let $Q_{ia}$ be real
chemical site charges at $\vct r_i+\vct b_{ia}$, with
$\vct b_{ia}=\mathsf O_i\vct a_{ia}$. For a prescribed local quasistatic
potential, the interaction and its work-conjugate body loads are
\begin{equation}
 \begin{aligned}
 U_E&=\sum_{ia}Q_{ia}\Phi_{\rm ext}(\vct r_i+\vct b_{ia},t),\\
 \vct F_i^E&=\sum_aQ_{ia}\vct E(\vct r_i+\vct b_{ia},t),\\
 \vct T_i^E&=\sum_a\vct b_{ia}\times
                  Q_{ia}\vct E(\vct r_i+\vct b_{ia},t).
 \end{aligned}
 \label{eq:s7hpv-site-electrical-wrench}
\end{equation}
The imposed-field and site-charge resolution must reproduce the
specified electrical boundary problem over its calibrated band.
\end{definition}

\begin{proposition}[Different translation and rotation selection rules]
\label{prop:s7hpv-field-angular-selection}
In a spatially uniform field, define $q_i=\sum_aQ_{ia}$ and
$\vct d_i=\sum_aQ_{ia}\vct b_{ia}$. Then
\begin{equation}
 \boxed{\vct F_i^E=q_i\vct E,\qquad
        \vct T_i^E=\vct d_i\times\vct E.}
 \label{eq:s7hpv-uniform-field-force-torque}
\end{equation}
Equal $q_i/m_i$ removes the direct uniform-field force overlap with a
pure internal translation as above; an angular component can still have
a nonzero torque overlap.
\end{proposition}
\begin{proof}
Vary each site position by
$\delta\vct x_{ia}=\delta\vct r_i+
\delta\vct\theta_i\times\vct b_{ia}$ in $U_E$.
The scalar triple product gives the torque. Uniformity lets the field
leave both sums, proving the stated reduction.
\end{proof}
For the dimensionless body tangent of Sections~2 and~6, the force column
is $\vct f_X=(\ell_\star\vct F_i^E,\vct T_i^E)_i$ and the modal source is
$F_s=\vct e_s^{\mathsf T}\mathsf G_X^{-1/2}\vct f_X$.
The translational scale $\ell_\star$ supplies the work-dual units.

As a synthetic example, prepare two equal-mass bodies with the same
$q_i/m_i$, reference dipoles $\vct d_1=d\vct e_x$,
$\vct d_2=-d\vct e_x$, and $\vct E=E\vct e_y$.
Their $z$ torques are $dE,-dE$. For
$\theta_-=(\theta_{1z}-\theta_{2z})/\sqrt2$, the conjugate torque is
$T_-=\sqrt2dE$, although the direct relative centre force vanishes.
A registered surface marker detects this motion through its lever arm,
as in~\eqref{eq:s6patch-marker-observation-matrix}. Mechanical coupling
can subsequently mix translation and libration; the full susceptibility
predicts that continuation. An exactly icosahedral invariant charge
pattern has $\vct d_i=0$, so this example requires its stated asymmetric
preparation. The local field, dipole, angular support and relaxation
remain independently constrained inputs. In a conducting liquid,
electrophoretic velocity uses the complete ionic and fluid ports of
Section~5; the frozen electrical wrench alone is not a mobility law.

Figure~\ref{fig:s7hpv-field-angular-selection} applies the body source map to a multi-particle preparation and its simplest relative-angular projection.

\begin{figure}[H]
\centering\begingroup
\setlength{\unitlength}{1pt}
\scalebox{0.84}{%
\input{vlt-figures/figure-302.tex}%
}\endgroup
\caption{Uniform-field force selection and angular selection are different questions. Panel (a) displays eight ideal VLPs with an alternating assigned dipole pattern in one common laboratory frame; the internal burgundy arrows denote dipoles, not displacement. Panel (b) isolates two equal-mass, equal-$q/m$ bodies with opposite $x$ dipoles in a uniform $y$ field. Their centre forces are equal, while their $z$ torques are opposite, giving zero direct relative force and conjugate relative torque $\sqrt2dE$. The curved arrows specify torque signs; angular amplitudes require the full response and supports. The dipoles, field penetration and coarse patch charges are stipulated inputs, not measured HPV values. An invariant icosahedral charge preparation has no such dipole. See~\eqref{eq:s7hpv-uniform-field-force-torque}.}
\label{fig:s7hpv-field-angular-selection}
\end{figure}

\paragraph{What state does a classical force prepare?}
For fixed positive frequencies and a real prescribed linear drive
$\widehat H_d(t)=-\sum_sF_s(t)\widehat Q_s$, the exact amplitude obeys
\begin{equation}
 \alpha_s(t)=e^{-i\omega_st}\alpha_s(0)+
     \frac{i}{\sqrt{2\hbar\omega_s}}
       \int_0^tF_s(t')e^{-i\omega_s(t-t')}\dd t'.
 \label{eq:phonon-driven-coherent-amplitude}
\end{equation}
This follows by integrating
$\partial_t\alpha_s=-i\omega_s\alpha_s+
iF_s/\sqrt{2\hbar\omega_s}$. For a coherent initial state it is its
coherent amplitude; for a general state the same affine law describes
the first moment and the propagator's displacement. After a pulse ending
at $t_p$, excitation from vacuum is set by the squared modulus of
$\int_0^{t_p}F_s(t')e^{i\omega_st'}\dd t'$: temporal frequency
matching and spatial mode overlap both matter. The excited mean motion
is a displacement wave, and the force spectrum selects which modes
participate. A measured oscillating mean does not determine the quantum
occupation statistics.

From vacuum the resulting multimode state is a product of coherent
states. Its independent Poisson counts sum to a Poisson total with mean
$\overline N=\sum_s|\alpha_s|^2$, so
\begin{equation}
 \boxed{P(N_{\rm ph,tot}=1)=\overline N e^{-\overline N}\le e^{-1}.}
 \label{eq:phonon-linear-drive-one-quantum-limit}
\end{equation}
The maximum occurs at $\overline N=1$, where vacuum and multiple-quantum
events still have nonzero probability. This linear drive prepares a coherent state with distributed
occupation, rather than a deterministic $\ket1$ state.
The resonant two-level conversion in
\eqref{eq:phonon-electronic-conversion-probability} supplies a different
preparation resource, as shown in
Figure~\ref{fig:one-phonon-conversion-and-classical-drive}. A thermal
initial state instead becomes a displaced thermal state. Quadratic
stiffness control can squeeze the state and change its number statistics,
so it lies outside this Poisson bound. Linear and quadratic Hamiltonians
acting on vacuum still produce Gaussian states, however; deterministic
single-phonon Fock preparation needs an additional non-Gaussian resource
or suitable conditioning. Changing the statistics is not by itself such
a preparation protocol.

\paragraph{Work follows from the declared Hamiltonian.}
For the fixed-frequency driven problem,
$\widehat H(t)=\widehat H_0-\sum_sF_s(t)\widehat Q_s$, the mechanical
energy and the controlled Hamiltonian obey different, compatible ledgers:
\begin{equation}
\begin{aligned}
 \frac{\dd}{\dd t}\langle\widehat H_0\rangle
     &=\sum_sF_s\langle\widehat\Pi_s\rangle,\\
 \frac{\dd}{\dd t}\langle\widehat H_0+\widehat H_d\rangle
     &=-\sum_s\frac{\dd F_s}{\dd t}\langle\widehat Q_s\rangle.
 \end{aligned}
 \label{eq:phonon-external-drive-work}
\end{equation}
Force times generalized velocity gives the first power; explicit time
dependence of the controlled Hamiltonian gives the second. Subtracting
the bare-energy power from the controlled-Hamiltonian power gives
$\dd\langle\widehat H_d\rangle/\dd t$, as required by
Hamiltonian work accounting~\cite{NeumaierWestra2011,Potts2026}.
If the force vanishes at both endpoints, the integrated work equals
the bare-energy change. Figure~\ref{fig:linear-and-parametric-control-work}
shows how their instantaneous powers differ during loading.

For clarity, stiffness control can be included in a fixed modal basis
with canonical vectors $\widehat{\vct Q},\widehat{\vct\Pi}$ and a real
symmetric positive matrix $\mathsf D_c(t)$:
\begin{equation}
 \begin{aligned}
  \widehat H_m(t)&=\tfrac12\widehat{\vct\Pi}^{\mathsf T}
                           \widehat{\vct\Pi}
       +\tfrac12\widehat{\vct Q}^{\mathsf T}\mathsf D_c(t)
                           \widehat{\vct Q},\qquad
  \widehat H(t)=\widehat H_m(t)-\vct F(t)^{\mathsf T}\widehat{\vct Q},\\
  \frac{\dd\langle\widehat H_m\rangle}{\dd t}
    &=\vct F^{\mathsf T}\langle\widehat{\vct\Pi}\rangle
       +\frac12\left\langle\widehat{\vct Q}^{\mathsf T}
            \frac{\dd\mathsf D_c}{\dd t}\widehat{\vct Q}\right\rangle,\\
  \frac{\dd\langle\widehat H\rangle}{\dd t}
    &=\frac12\left\langle\widehat{\vct Q}^{\mathsf T}
            \frac{\dd\mathsf D_c}{\dd t}\widehat{\vct Q}\right\rangle
            -\frac{\dd\vct F^{\mathsf T}}{\dd t}
                           \langle\widehat{\vct Q}\rangle .
 \end{aligned}
 \label{eq:phonon-general-control-energy-ledger}
\end{equation}
Curvature control does work on the second moment, including fluctuations;
it can act even when $\langle\widehat Q_s\rangle=0$.
A sudden one-mode change from $\omega_i$ to $\omega_f$ leaves the
state unchanged at that instant and gives
$W=(\omega_f^2-\omega_i^2)\langle\widehat Q^2\rangle/2$.
For classical initial energy $E_i$ and phase $\varphi$ with
$Q=\sqrt{2E_i}\cos\varphi/\omega_i$, this reads
$W/E_i=[(\omega_f/\omega_i)^2-1]\cos^2\varphi$.
It directly exhibits both state dependence and the sign of the work.

\begin{figure}[H]
\centering
\begingroup
\definecolor{vlfGreen}{HTML}{176B64}
\definecolor{vlfDarkRed}{HTML}{762B38}
\definecolor{vlfRed}{HTML}{762B38}
\definecolor{vlfBlue}{HTML}{315D80}
\definecolor{vlfCopper}{HTML}{762B38}

\setlength{\unitlength}{1pt}
\scalebox{0.80}{%
\input{vlt-figures/figure-254.tex}%
}%
\endgroup
\caption{Force work and curvature-control work use different terms in the ledger. (a) For an initially resting mass-normalized oscillator driven by $F_0\cos(2\omega t)$, write $q=(\cos\tau-\cos2\tau)/3$ and $p=(2\sin2\tau-\sin\tau)/3$, $\tau=\omega t$. In power units $F_0^2/\omega$, bare mechanical power is $\cos2\tau\,p$, while total controlled-Hamiltonian power is $2\sin2\tau\,q$. Their difference accounts for $H_d=-FQ$. (b) A separate instantaneous frequency quench $\omega_f=r\omega_i$ on a fixed-energy classical orbit gives $W/E_i=(r^2-1)\cos^2\varphi$, shown for $r=2,1/2$. The controller supplies or receives that phase-dependent work. Quantum mean work instead uses $\tfrac12(\omega_f^2-\omega_i^2)\langle Q^2\rangle$. Both panels specify controlled preparations and their source account.}
\label{fig:linear-and-parametric-control-work}
\end{figure}

A repeated modulation
$\omega^2(t)=\omega_0^2[1+h\cos(2\omega_0t)]$ with $|h|\ll1$
gives the Mathieu parametric-resonance problem. At leading order near
exact resonance, write
$Q=A_c(t)\cos\omega_0t+A_s(t)\sin\omega_0t$ with slowly varying
quadratures. Keeping resonant terms to first order in $h$ gives
\begin{equation}
 \frac{\dd}{\dd t}
 \begin{pmatrix}A_c\\ A_s\end{pmatrix}
 =-\frac{h\omega_0}{4}
 \begin{pmatrix}0&1\\1&0\end{pmatrix}
 \begin{pmatrix}A_c\\ A_s\end{pmatrix}.
 \label{eq:phonon-parametric-resonant-quadratures}
\end{equation}
One quadrature combination grows with amplitude exponent
$|h|\omega_0/4$ in this ideal undamped approximation; its energy grows
with twice that exponent. The prescribed
modulation supplies this energy, despite positive instantaneous
$\omega^2(t)$ when $|h|<1$. Finite detuning, damping, nonlinearities,
and departure from the local basin constrain that approximation. This
is driven amplification, distinct from the autonomous feedback and
resource dynamics needed for sustained self-excitation.

Figure~\ref{fig:parametric-quadrature-gain-and-source} resolves that leading amplification into growing and decaying quadrature directions, with the controller as its energy source.

\begin{figure}[H]
\centering
\begingroup
\definecolor{vlfGreen}{HTML}{176B64}
\definecolor{vlfDarkRed}{HTML}{762B38}
\definecolor{vlfRed}{HTML}{762B38}
\definecolor{vlfBlue}{HTML}{315D80}
\definecolor{vlfCopper}{HTML}{762B38}

\setlength{\unitlength}{1pt}
\scalebox{0.80}{%
\input{vlt-figures/figure-255.tex}%
}%
\endgroup
\caption{Phase-sensitive amplification in the leading resonant quadrature equations. With $g=h\omega_0/4$, $h>0$ and $\tau_g=gt$, $A_+=(A_c+A_s)/\sqrt2$ decays as $e^{-\tau_g}$ and $A_-=(A_c-A_s)/\sqrt2$ grows as $e^{\tau_g}$. (a) A unit initial circle maps to the displayed equal-scale ellipses at $\tau_g=0,1/2,1$. Their area is preserved. (b) The corresponding extremal cycle-averaged energy gains are $e^{\pm2\tau_g}$. These are exact solutions of the stated averaged system. Connecting them to the full parametrically driven oscillator requires small $|h|$, near resonance, an undepleted prescribed source and motion within the harmonic basin. The plotted axes are quadratures rather than a physical cylinder orbit; controller work, detuning and omitted loss remain part of the mechanical account.}
\label{fig:parametric-quadrature-gain-and-source}
\end{figure}

Electrical source conventions also change the mechanical potential.
For a mechanical coordinate $x$ changing a capacitor's capacitance
$C(x)$, fixed electrode charge $Q_{\rm el}$ gives
$U_Q=Q_{\rm el}^2/[2C(x)]$. At prescribed voltage $V$, elimination
of the voltage source gives instead the effective potential
$U_V=-C(x)V^2/2$~\cite{FeynmanLectures1963}. Both produce the
appropriate force, but their derivatives hold different electrical
variables fixed. Using $+C(x)V^2/2$ alone at fixed voltage omits source
work. The complete restoring potential must remain stable in the
chosen operating regime. A finite ionic concentration gradient supplies
a third, depletable source rather than an ideal voltage clamp; the
following subsection constructs that distinction explicitly.

\paragraph{When does a mixed field--mechanical excitation exist?}
A fully closed field extension retains the field or source degrees of
freedom, their stored energy, and the interaction. Coupling a retained
mechanical dipole to a dynamical electromagnetic mode can then yield
mixed normal modes. To call these phonon-polaritons requires a specified
electromagnetic boundary problem, polarization derivatives, dielectric
response, and consistent Coulomb or self-polarization terms. The ionic-crystal construction of Ref.~\cite{HellmanKronik2026}
illustrates these inputs at its own material resolution; its tensors
are not viral parameters. An imposed
classical field supplies a force and work protocol, whereas a dynamical
field enlarges the excitation space. This distinction identifies what
must be measured or modeled before an electromagnetic counterpart of a
viral collective excitation becomes a definite prediction.

\subsection{Finite ionic gradients and electromechanical relaxation}
\label{sec:phonon-ionic-extension}

An ionic gradient can store a finite, initially prepared work resource
inside the cylinder. Its state comprises the ionic populations, the
selective transfer pathway and the charge-separation geometry. Their
mechanical force follows by differentiating one shared free energy.
Evolving those variables then accounts for the resource as it is used.

A membrane-bearing compartment is one setting for this construction.
Reconstituted M2-containing vesicles provide a characterized example
of proton transport~\cite{Moffat2008}. Another membrane preparation
requires its own permeability, channel kinetics and mechanical coupling;
these data determine how the retained populations evolve.

\paragraph{The compartments, charge and thermodynamic description.}
Consider two well-mixed compartments, inside and outside an explicitly
introduced selective partition, both enclosed by the cylinder. Initially
hold their volumes $V_a$ fixed. For species $j$, let
$\mathcal N_{aj}$ be its particle count, $z_j\ne0$ its valence, and
$c_{aj}=\mathcal N_{aj}/V_a$ its number density. Transfer preserves
$\mathcal N_{{\rm in},j}+\mathcal N_{{\rm out},j}
=\mathcal N_{{\rm tot},j}$. These counts are different from both virion
count and phonon occupation. At a stated approximately fixed temperature
$T>0$, an ideal dilute, lumped-capacitor reference is
\begin{align}
 \mathcal F(\vct u,\mathcal N_{\rm in})
 &=U_{\rm mech}(\vct u)
   +k_BT\sum_{a,j}\mathcal N_{aj}
                [\ln(c_{aj}/c_j^\star)-1]
   +\frac{q_{\rm cap}^2}{2C(\vct u)},\nonumber\\
 q_{\rm cap}&=q_{\rm fixed}+e\sum_jz_j\mathcal N_{{\rm in},j},
 \qquad V_m=\Phi_{\rm in}-\Phi_{\rm out}
                         =\frac{q_{\rm cap}}{C(\vct u)}.
 \label{eq:phonon-ionic-common-free-energy}
\end{align}
Here $e>0$ is the elementary charge, $C>0$ is the total capacitance,
and $c_j^\star$ is the same species standard density on both sides.
Opposite outer capacitor charge and the fixed backgrounds are chosen
consistently with total charge conservation. The mixing term is the ideal configurational free energy of the ionic
populations; the capacitor term is the energy of charge separation.
The membrane's mechanics and selective pathways remain supplied inputs.
This construction follows the energy-based electrodiffusion principle
of Refs.~\cite{Mori2015,MoriLiuEisenberg2011} at a simpler retained
resolution.

Near-electroneutral bulk compartments can coexist with small excess
interfacial charges. Exact neutrality of each entire compartment,
including its interface, cannot simultaneously be imposed with an
independently nonzero $q_{\rm cap}$. The capacitor description assumes
that the unresolved charge layers and dielectric geometry admit this
reduction over the observation band. If their thickness or relaxation
is resolved, the spatial electrostatic problem is needed instead.
At fixed $T$, $\mathcal F$ is a free energy of the retained variables,
not all the internal energy of the enclosure. A finite medium can
approximate a heat reservoir over a restricted energy and time window;
an exactly isolated finite model must additionally evolve its medium
energy, entropy and temperature.

\paragraph{Nernst balance is a chemical-potential statement.}
Differentiate at fixed species totals and geometry. The transfer
electrochemical potential and the corresponding Nernst voltage are
\begin{align}
 \Delta\mu_j
 &=\frac{\partial\mathcal F}{\partial\mathcal N_{{\rm in},j}}
   =k_BT\ln\frac{c_{{\rm in},j}}{c_{{\rm out},j}}+z_jeV_m
   =z_je(V_m-E_j),\nonumber\\
 E_j&=\frac{k_BT}{z_je}
                 \ln\frac{c_{{\rm out},j}}{c_{{\rm in},j}}.
 \label{eq:phonon-nernst-transfer-potential}
\end{align}
A positive $\Delta\mu_j$ means that moving one ion outward lowers the
free energy to first order. The electrical term may oppose or assist
the concentration term. $E_j$ is the voltage that cancels their sum
for this species at its instantaneous concentrations, rather than a
separately stored battery voltage. Activities replace concentrations
for nonideal solutions, and unequal standard-state transfer energies
add their chemical-potential difference. All energies here are per
particle; a molar convention replaces $k_B,e$ by $R,F$ consistently.
For orientation, $k_BT/e\simeq25.85\,\mathrm{mV}$ at $300\,\mathrm K$,
or $59.53\,\mathrm{mV}$ per concentration decade for a monovalent ion.
These are thermal scales, not assigned viral voltages.

The finite populations also determine the restoring susceptibility.
At fixed geometry and positive populations, their Hessian is
\begin{equation}
 A_{ij}^{\rm ion}
 =\delta_{ij}k_BT\left(\frac1{\mathcal N_{{\rm in},j}}
                    +\frac1{\mathcal N_{{\rm out},j}}\right)
                    +\frac{z_iz_je^2}{C}.
 \label{eq:phonon-ionic-susceptibility}
\end{equation}
The first contribution is the chemical resistance to further depletion;
the second is a positive semidefinite charge-separation contribution.
The ideal fixed-geometry free energy is therefore strictly convex in
the independent transfer counts. A finite outside compartment is not
equivalent to an infinite bath: its concentration and hence $E_j$ change
when ions move.

For a single permeant species in equal volumes, set
$n_{\rm ion}=\mathcal N_{\rm in}/\mathcal N_{\rm tot}$ and choose
$q_{\rm fixed}=-ze\mathcal N_{\rm tot}n_{\rm ref}$.
Up to an additive constant the dimensionless free energy is
\begin{equation}
 \begin{aligned}
 \frac{\mathcal F_{\rm ion}}{\mathcal N_{\rm tot}k_BT}
   &=n_{\rm ion}\ln n_{\rm ion}
    +(1-n_{\rm ion})\ln(1-n_{\rm ion})
    +\frac{\alpha_{\rm ion}}2(n_{\rm ion}-n_{\rm ref})^2,\\
 \frac{\Delta\mu}{k_BT}
   &=\ln\frac{n_{\rm ion}}{1-n_{\rm ion}}
                   +\alpha_{\rm ion}(n_{\rm ion}-n_{\rm ref}),
 \qquad
 \alpha_{\rm ion}=\frac{(ze)^2\mathcal N_{\rm tot}}{Ck_BT}>0.
 \end{aligned}
 \label{eq:phonon-finite-ionic-resource-example}
\end{equation}
The positive second derivative
$1/n_{\rm ion}+1/(1-n_{\rm ion})+\alpha_{\rm ion}$ ensures a unique
interior equilibrium. Chemical and electrical forces can balance at
unequal concentrations. Figure~\ref{fig:finite-ionic-resource-and-equilibrium}
locates that minimum for the specified finite populations.

\begin{figure}[H]
\centering\begingroup
\definecolor{vltTeal}{HTML}{176B64}
\definecolor{vltBlue}{HTML}{315D80}
\definecolor{vltGold}{HTML}{A87520}
\definecolor{vltPlum}{HTML}{725877}
\definecolor{vltRed}{HTML}{762B38}
\definecolor{vltYellow}{HTML}{D5AD3C}
\definecolor{vltNeutral}{HTML}{777D83}
\setlength{\unitlength}{1pt}
\scalebox{0.90}{%
\input{vlt-figures/figure-256.tex}%
}\endgroup
\caption{Chemical and electrical contributions to one finite ionic resource. (a) Four inside and sixteen outside ion symbols illustrate $n_{\rm ref}=0.2$; the optional membrane-bearing object and capacitor identify the retained partition. Positive outward transfer preserves the total count. (b) For equal-volume ideal compartments with $\alpha_{\rm ion}=8$, $\psi=n\ln n+(1-n)\ln(1-n)+4(n-0.2)^2$. The curve subtracts its minimum, where $\psi'=\ln[n/(1-n)]+8(n-0.2)=0$. The count sketch is a reference preparation; the curve is the continuous population free-energy model with no independent voltage clamp.}
\label{fig:finite-ionic-resource-and-equilibrium}
\end{figure}

\paragraph{Passive transfer depletes a resource rather than maintaining it.}
With outward number transfer $J_j$ and nonnegative near-equilibrium
mobilities $\Lambda_j$, choose
\begin{equation}
 \begin{aligned}
 J_j&=\Lambda_j\Delta\mu_j,\qquad
 \frac{\dd\mathcal N_{{\rm in},j}}{\dd t}=-J_j,\\
 I_j&=z_jeJ_j=G_j(V_m-E_j),\qquad G_j=(z_je)^2\Lambda_j.
 \end{aligned}
 \label{eq:phonon-passive-ion-transfer}
\end{equation}
Here $J_j$ is a count per time, $I_j$ a current, and $G_j$ a total
conductance. The opposite population changes by $+J_j$. 

\begin{proposition}[Joint mechanical--ionic free-energy decay]
\label{prop:s7-coupling-ionic-decay}
Use the common isothermal free energy above, fixed species totals,
constant inertia and the passive transfer law
\eqref{eq:phonon-passive-ion-transfer}. For
$\mathsf M\Dtt{\vct u}=-\nabla_u\mathcal F$, their matched evolution gives
\begin{equation}
 \boxed{\frac{\dd}{\dd t}
 \left[\frac12\Dt{\vct u}^{\mathsf T}\mathsf M\Dt{\vct u}
                            +\mathcal F\right]
   =-\sum_j\Lambda_j(\Delta\mu_j)^2
   =-\sum_jG_j(V_m-E_j)^2\le0.}
 \label{eq:phonon-ionic-free-energy-dissipation}
\end{equation}
\end{proposition}
\begin{proof}
The mechanical power cancels the coordinate derivative of $\mathcal F$.
Each independent transfer count contributes
$\Delta\mu_j\,\dot{\mathcal N}_{{\rm in},j}
=-\Lambda_j(\Delta\mu_j)^2$. Summing proves the identity.
\end{proof}

Figure~\ref{fig:ionic-passive-free-energy-budget} separates stored free energy from its passive dissipation and gives a one-coordinate relaxation example.

\begin{figure}[H]
\centering\begingroup
\definecolor{vlfGreen}{HTML}{176B64}
\definecolor{vlfDarkRed}{HTML}{762B38}
\definecolor{vlfRed}{HTML}{762B38}
\definecolor{vlfBlue}{HTML}{315D80}
\definecolor{vlfCopper}{HTML}{762B38}
\setlength{\unitlength}{1pt}
\scalebox{0.80}{%
\input{vlt-figures/figure-257.tex}%
}\endgroup
\caption{A passive free-energy account for \eqref{eq:phonon-ionic-free-energy-dissipation}. (a) The stored quantity $\mathcal A$ is kinetic energy plus the isothermal free energy. With $J_j=\Lambda_j\Delta\mu_j$ and $\Lambda_j\ge0$, its decrease is $\mathcal P_{\rm dis}=\sum_jJ_j\Delta\mu_j\ge0$. The arrow denotes the decrease of available free energy; it does not identify every term as an increase of thermal internal energy. Box sizes have no quantitative meaning. (b) For one linearized ionic imbalance $y$ at fixed geometry, $\mathcal A_{\rm ex}=a_y y^2/2$, $\dot y=-\Lambda a_y y$ and $\tau=(\Lambda a_y)^{-1}$. The exact curves are the remaining fraction $e^{-2s}$ (solid) and the accumulated decrease $1-e^{-2s}$ (dashed), with $s=t/\tau$ and $\mathcal A_0=\mathcal A_{\rm ex}(0)>0$. Their sum is one. The trace is a local noiseless constitutive calculation; thermal trajectories need not decrease pointwise.}
\label{fig:ionic-passive-free-energy-budget}
\end{figure}

The decrease is available free energy, not universally an equal
increase of thermal internal energy: ideal mixing can reduce free
energy through increasing entropy. Additional mechanical resistance
adds its nonnegative dissipation to this identity. The diagonal law is
one passive choice. A coupled passive transporter may move one species
uphill using another's gradient; it is the total
$\sum_jJ_j\Delta\mu_j$ that must remain nonnegative when no other
resource is supplied~\cite{Mori2015}.

The linear law is a local approximation about positive populations.
A large-transfer constitutive model must preserve both compartments'
nonnegative counts and the relevant flux--affinity inequality. At small
counts, discrete transfer states and rates satisfying local detailed
balance are more appropriate than Gaussian population noise. In
particular, a pH specifies proton activity, not a literal macroscopic
population of unbound protons in every nanoparticle. As an ideal-dilute
scale illustration, a sphere of radius $50\,\mathrm{nm}$ has volume
$5.24\times10^{-19}\,\mathrm L$; at $10^{-7}\,\mathrm{mol\,L^{-1}}$
its mean unbound-proton count is only about $0.032$.
Buffer protonation, titratable proteins, water chemistry and other ions
may therefore be essential variables. This estimate does not assign
pH or ideal bulk behavior to an actual confined virion. It identifies
when a continuous free-proton coordinate alone would be inadequate.

Zero electrical current is also weaker than equilibrium. If
conductances and concentrations are temporarily frozen,
$\sum_jI_j=0$ gives $V_m=\sum_jG_jE_j/\sum_jG_j$ for positive total
conductance. Opposing species currents can still dissipate free energy
and deplete their gradients. Equilibrium of the independent permeation
paths instead requires $\Delta\mu_j=0$ for each permeable species.

\paragraph{How an ionic coordinate loads a mechanical mode.}
At fixed transferred charge, the capacitor supplies
\begin{equation}
 \vct f_{\rm cap}=-\nabla_u\frac{q_{\rm cap}^2}{2C(\vct u)}
                   =\frac{V_m^2}{2}\nabla_u C(\vct u).
 \label{eq:phonon-capacitive-mechanical-force}
\end{equation}
For geometry-dependent volumes the mixing terms additionally supply
osmotic forces; ideal osmotic pressure is $k_BT\sum_jc_j$ at fixed
counts. Fixed-voltage control instead requires the source-inclusive
potential $-CV_m^2/2$ and its work, as in Section~7.6. Thus a finite
gradient can shift the reference and supply transient mechanical work.
Sustained operation requires replenishment or explicit evolving fuel.

\begin{example}[An ionic coordinate gives a phonon relaxation branch]
Near a stable joint equilibrium, retain a length-valued displacement
$x$, its mass $M_x>0$, and a dimensionless number-imbalance coordinate
$y$ about that equilibrium. Write
\begin{equation}
 \mathcal F_2=\frac12k_xx^2+b_{xy}xy+\frac12a_yy^2,
 \qquad k_x>0,\quad a_y>0,\quad k_xa_y>b_{xy}^2.
 \label{eq:phonon-ion-quadratic-free-energy}
\end{equation}
Here $k_x$ has stiffness units, $b_{xy}$ force units, and $a_y$
energy units. The cross derivative $b_{xy}$ appears in both equations
because force and transfer affinity come from the same free energy.
For one transfer count in the fixed-volume capacitor model, its
capacitive contribution is
$b_{xy}=-zeV_m^*C_x^*/C^*$, where a star indicates the equilibrium
and $C_x=\partial C/\partial x$. A zero-bias reference has no such
linear cross term. Geometry-dependent chemical energies or another
declared interaction can supply additional contributions; the presence
of ions alone does not guarantee this relaxation strength.

Take
$M_x\Dtt{x}=-k_xx-b_{xy}y$ and
$\Dt{y}=-\Lambda(a_yy+b_{xy}x)$, with $\Lambda>0$.
Their energy derivative is $-\Lambda(a_yy+b_{xy}x)^2$.
For the $e^{-i\omega t}$ convention, elimination of $y$ gives
\begin{align}
 K^*(\omega)&=k_x-\frac{\Delta k}{1-i\omega\tau},
 &\Delta k&=\frac{b_{xy}^2}{a_y},\qquad\tau=(\Lambda a_y)^{-1},\nonumber\\
 K'(\omega)&=k_x-\frac{\Delta k}{1+\omega^2\tau^2},
 &K''(\omega)&=\frac{\Delta k\,\omega\tau}{1+\omega^2\tau^2}.
 \label{eq:phonon-ion-relaxation-stiffness}
\end{align}
Thus $K^*=K'-iK''$ with $K''\ge0$ at positive frequency.
Slow loading lets the ionic variable relax, giving
$K_{\rm rel}=k_x-\Delta k>0$; fast loading freezes it, giving $k_x$.
The stability condition is exactly this Schur-complement positivity.
The reciprocal ionic coordinate links the relaxed softening and loss
through the same constitutive parameters.

The equilibrium fluctuation completion at $T>0$ is
\begin{equation}
 \dd y=-\Lambda(a_yy+b_{xy}x)\,\dd t
                  +\sqrt{2k_{\rm B}T\Lambda}\,\dd\mathcal B_t,
 \qquad M_x\Dtt{x}=-k_xx-b_{xy}y,
 \label{eq:phonon-ion-thermal-completion}
\end{equation}
where $\mathcal B_t$ is a standard real Wiener process. This is a
near-equilibrium, many-count Gaussian reduction; its excursions must
remain small relative to the actual population margins. The joint
Gibbs density is proportional to
$\exp[-\beta(M_x\Dt{x}^2/2+\mathcal F_2)]$, giving
\begin{equation}
 \operatorname{Var}_\beta x=\frac{k_{\rm B}T}{K_{\rm rel}},
 \qquad K_{\rm rel}=k_x-\Delta k>0.
 \label{eq:phonon-ion-thermal-variance}
\end{equation}
For $b_{xy}\ne0$ the ionic damping reaches this mechanical coordinate.
If $b_{xy}=0$, Gibbs remains invariant but the mechanical oscillator
does not thermalize through that disconnected variable.

Elimination of the same fluctuating coordinate yields
\begin{equation}
 M_x\Dtt{x}+K_{\rm rel}x
  +\int_0^t\Gamma_y(t-t')\frac{\dd x(t')}{\dd t'}\,\dd t'
       =f_{\rm ion}(t),\qquad
 \Gamma_y(t)=\Delta k\,e^{-t/\tau}\quad(t\ge0).
 \label{eq:phonon-ion-thermal-memory}
\end{equation}
For clarity, set $z_y=y+(b_{xy}/a_y)x$. Its equation is
$\dd z_y=-z_y\dd t/\tau+(b_{xy}/a_y)\dd x
+\sqrt{2k_BT\Lambda}\dd\mathcal B_t$; the velocity convolution
produces the displayed memory. The remaining force is
\[
 f_{\rm ion}(t)=-b_{xy}\left[
 z_y(0)e^{-t/\tau}
 +\sqrt{2k_BT\Lambda}\int_0^te^{-(t-t')/\tau}\dd\mathcal B_{t'}
 \right].
\]
Conditional Gibbs preparation makes $z_y(0)$ a centered Gaussian with
variance $k_BT/a_y$, independent of $x(0),\Dt{x}(0)$. This initial
term completes the stationary colored force and gives
\begin{equation}
 \langle f_{\rm ion}(t)f_{\rm ion}(t')\rangle
                    =k_{\rm B}T\Gamma_y(|t-t'|).
 \label{eq:phonon-ion-force-fdt}
\end{equation}
For another prepared ionic branch the homogeneous force is retained
explicitly; stationarity cannot simply be assumed. This is the same
passive memory--fluctuation link used in Sections~4 and~6, now derived
from an identified population coordinate.
\end{example}

For $b_{xy}\ne0$ in this stable reference, the conjugate-force response
and joint stationary displacement spectrum are
\begin{equation}
 \begin{aligned}
 \mathcal R_{xx}(\omega)&=[K^*(\omega)-M_x\omega^2]^{-1},\\
 S_{ff}^{\rm ion}(\omega)
  &=\frac{2k_{\rm B}T\Delta k\,\tau}{1+\omega^2\tau^2},\\
 S_{xx}(\omega)&=|\mathcal R_{xx}(\omega)|^2S_{ff}^{\rm ion}(\omega)
              =\frac{2k_BT}{\omega}\operatorname{Im}\mathcal R_{xx}(\omega).
 \end{aligned}
 \label{eq:phonon-ion-observable-spectra}
\end{equation}
The last expression uses its continuous zero-frequency limit. The
two-sided angular-frequency convention is
$S(\omega)=\int_{\R}e^{i\omega t}\operatorname{Cov}(t,0)\dd t$.
Force and displacement spectra have force-squared-time and
length-squared-time units, respectively, and
$\int S_{xx}\dd\omega/(2\pi)=k_BT/K_{\rm rel}$.
Changing mobility changes the poles and distribution of spectral power,
while the integrated equilibrium variance stays fixed at the same
curvatures. The stationary displacement formula requires joint Gibbs
preparation or decay of the mechanical preparation transients, not just
conditional equilibration of $y$ at one instant.

\begin{figure}[H]
\centering\begingroup
\definecolor{vlfGreen}{HTML}{176B64}
\definecolor{vlfDarkRed}{HTML}{762B38}
\definecolor{vlfRed}{HTML}{762B38}
\definecolor{vlfBlue}{HTML}{315D80}
\definecolor{vlfCopper}{HTML}{762B38}
\setlength{\unitlength}{1pt}
\scalebox{0.80}{%
\input{vlt-figures/figure-258.tex}%
}\endgroup
\caption{One passive ionic branch changes spectral shape at fixed Gibbs variance. Use $\Delta k/k_x=1/2$, $\omega_x=\sqrt{k_x/M_x}$, $\omega_x\tau=1/4,2$ and no additional drag. (a) The force-noise ordinate is $\widetilde S_f=\omega_xS_{ff}^{\rm ion}/(2k_BT k_x)$. (b) The logarithmic displacement ordinate is $\widetilde S_x=k_x\omega_xS_{xx}/(2k_BT)$, with linear frequency $r=\omega/\omega_x$. Both use $\mathcal R=[k_x-\Delta k/(1-i\omega\tau)-M_x\omega^2]^{-1}$ and $S_{xx}=|\mathcal R|^2S_{ff}^{\rm ion}$. For each matched stationary Gibbs preparation, the full-band area is $\int_0^\infty\widetilde S_x\,\dd r=\pi$, giving $\operatorname{Var}x=2k_BT/k_x$. Mobility changes timescales and peak shapes at unchanged relaxed softness. The stated finite-ion and isothermal assumptions set the model scope.}
\label{fig:phonon-ionic-spectra}
\end{figure}

These predictions constrain the resolved relaxation branch rather than
uniquely identifying its ionic origin. An independently calibrated
ion-count or charge readout fixes the physical meaning and scale of $y$;
without it, different susceptibilities, cross derivatives and mobilities
can have the same mechanical response. Section~\ref{sec:phonon-calculable-outcomes}
makes that identification limit explicit. A new variable becomes useful
when its population or force sensitivity is separately constrained, or
when the simpler branch fails over the measured band.

\subsection{Diffusion, membrane gates, and chemical work}
\label{sec:phonon-biochemical-extensions}

A well-mixed compartment resolves a population when spatial equilibration
is fast relative to transfer and mechanical observation. Finer
resolution can require concentration fields, gate populations or fuel
counts. Each adds distinct predictive information together with its
conservation or kinetic law. The following constructions give those
couplings and energy accounts before introducing an active oscillator.

\paragraph{Spatial transport has a dissipative generator.}
For number density $c_j(\vct r,t)>0$, constant $T$, no advection and a
fixed dilute-solution standard state, write
$\widetilde\mu_j=\mu_j^\star+k_BT\ln(c_j/c_j^\star)+z_je\Phi$.
The Nernst--Planck flux and continuity law are
\begin{equation}
 \begin{aligned}
 \vct J_j&=-\frac{D_jc_j}{k_BT}\nabla\widetilde\mu_j
   =-D_j\left(\nabla c_j+\frac{z_je}{k_BT}c_j\nabla\Phi\right),\\
 \partial_tc_j+\nabla\mathbin{\cdot}\vct J_j&=0.
 \end{aligned}
 \label{eq:phonon-nernst-planck-extension}
\end{equation}
Here $D_j>0$ has area-per-time units and $\vct J_j$ is a number flux
per area. The first term mixes a concentration gradient; the second
drifts charge in the electric field. Their common mobility is
$D_jc_j/(k_BT)$, the dilute isothermal Einstein relation
\cite{Cotterill2002,Mori2015}. No-flux outer boundaries conserve each
species total. Selective membrane interfaces instead specify the
transferred flux and its opposite on the two sides.

A resolved electrostatic closure needs dielectric and boundary data,
for example
$-\nabla\cdot(\varepsilon\nabla\Phi)
=\sum_jz_jec_j+\rho_{\rm fixed}$.
For fixed geometry, fixed dielectric data and electrical boundaries
performing no work, the ideal mixing plus electrostatic free energy
obeys
\begin{equation}
 \frac{\dd\mathcal F_{\rm tr}}{\dd t}
 =-\sum_j\int\frac{D_jc_j}{k_BT}
                         |\nabla\widetilde\mu_j|^2\,\dd^3r\le0
 \label{eq:phonon-spatial-transport-dissipation}
\end{equation}
when the interfaces are also closed to transfer. Passive membrane
transfer adds its nonnegative flux--affinity dissipation; moving
boundaries add reciprocal mechanical work. Resolved charge-layer field
storage and a lumped capacitance must be matched, rather than counted
twice~\cite{MoriLiuEisenberg2011}. Prescribed electrode voltages,
advection and chemical supply introduce their corresponding boundary
or bulk work terms.

Without electric drift, an auxiliary Fourier concentration mode decays
as $e^{-D_jq^2t}$, whereas an acoustic displacement oscillates with
$\omega\simeq c|q|$. Even an electrical relaxation rate is not generally
just a diffusion rate. For a uniform ideal symmetric monovalent
electrolyte with equal diffusivities $D$ and equilibrium density $c_0$
of each species, linearizing the same flux and Poisson equations gives
\begin{equation}
 r_{\rm salt}(q)=Dq^2,\qquad
 r_{\rm charge}(q)=D(q^2+\kappa_D^2),\qquad
 \kappa_D^2=\frac{2e^2c_0}{\varepsilon k_BT}.
 \label{eq:phonon-salt-charge-relaxation}
\end{equation}
The salt fluctuation is the sum of the two densities; the charge
fluctuation is their difference. The latter creates a field that
opposes charge separation, producing the additional Debye relaxation.
These are bulk, linear, no-flow rates for nonzero admissible wave
numbers. The formal charge-relaxation intercept is not decay of the
globally conserved net charge in a sealed no-flux cavity. Actual cavity
and membrane conditions select the spatial patterns and transfer constraints.
They give concrete tests of the well-mixed and instantaneous-screening
approximations.

Figure~\ref{fig:salt-charge-resolution} compares the bulk salt and charge rates with their corresponding decay amplitudes.

\begin{figure}[H]
\centering\begingroup
\definecolor{vlfGreen}{HTML}{176B64}
\definecolor{vlfDarkRed}{HTML}{762B38}
\definecolor{vlfRed}{HTML}{762B38}
\definecolor{vlfBlue}{HTML}{315D80}
\definecolor{vlfCopper}{HTML}{762B38}
\setlength{\unitlength}{1pt}
\scalebox{0.80}{%
\input{vlt-figures/figure-259.tex}%
}\endgroup
\caption{Salt and charge are different dissipative variables. For the stipulated uniform, ideal symmetric monovalent electrolyte with equal diffusivity $D>0$, positive Debye wave number $\kappa_D$, no flow and the linear Poisson--Nernst--Planck closure, write $\eta=q/\kappa_D>0$. Panel~(a) plots the exact bulk-mode rates $r_{\rm salt}/(D\kappa_D^2)=\eta^2$ and $r_{\rm charge}/(D\kappa_D^2)=1+\eta^2$. The open point is the formal charge intercept; it is not a decaying net-charge mode in a sealed cavity. Panel~(b) uses one nonzero auxiliary mode $\eta=1$, with normalized amplitudes $e^{-s}$ and $e^{-2s}$, $s=D\kappa_D^2t$. These are relaxation amplitudes, not mechanical oscillation probabilities. Actual cavity and membrane boundary conditions select admissible spatial modes. A fast charge decay can justify eliminating a field only over a specified slower observation band.}
\label{fig:salt-charge-resolution}
\end{figure}

The thermal scale $k_BT$ sets the mobility--fluctuation relation.
Coherent forcing and mechanical inertia remain separately specified.

\subsubsection{A quantitative test of the well-mixed resolution}
\label{sec:s7-coupling-mixing-gap}

virion positions can alter the fluid paths through which an ionic imbalance
equilibrates. A mixing test therefore uses the actual accessible geometry
of a registered episode. For an auxiliary neutral diffusion component
with constant $D>0$, take a connected bounded domain $\Omega_g$ with
reflecting walls and a positive Neumann spectral gap $\lambda_{1,g}$.
Its Poincar\'e inequality is
$\|v-\bar v\|_{L^2}^2\le
\lambda_{1,g}^{-1}\|\nabla v\|_{L^2}^2$.
The domain and walls here are fixed during the test.

\begin{proposition}[Registered diffusion-gap error bound]
\label{prop:s7-coupling-mixing-bound}
Let $\partial_t c=D\Delta c+s$ with no-flux boundaries and sufficient
regularity for the energy identity. Set $\eta=c-\bar c$ and
$s_\perp=s-\bar s$, where bars are spatial means on $\Omega_g$.
Then
\begin{equation}
 \|\eta(t)\|_{L^2}
 \le e^{-\gamma_gt}\|\eta(0)\|_{L^2}
    +\int_0^t e^{-\gamma_g(t-r)}
                              \|s_\perp(r)\|_{L^2}\,\dd r,
 \qquad\gamma_g=D\lambda_{1,g}.
 \label{eq:s7-coupling-mixing-bound}
\end{equation}
In particular, a uniformly bounded nonuniform source contributes at
most $\sup_r\|s_\perp(r)\|_{L^2}/\gamma_g$.
\end{proposition}
\begin{proof}
Multiply the mean-zero diffusion equation by $\eta$, integrate by parts,
and apply Poincar\'e and Cauchy--Schwarz to obtain
$\partial_t\|\eta\|\le-\gamma_g\|\eta\|+\|s_\perp\|$.
Integrate this scalar differential inequality, with its continuous
interpretation at zero norm.
\end{proof}

The bound tests mixing against an output tolerance. Narrow passages
can lower the gap; disconnected regions retain separate conserved
means. In a forced band, $\omega_{\rm obs}/\gamma_g\ll1$ tests
timescale separation alongside the initial and source errors above.
Electrodiffusion, membrane flux and advection require their own
generator and boundary terms. Neutral diffusion supplies a baseline
test of the lumped resolution.

For computation, a conservative finite-volume discretization gives
the same test directly in matrix form. Let $\mathsf V$ contain the
positive accessible-fluid cell volumes, $\mathsf B$ their oriented
edge incidence, and $\mathsf G$ the nonnegative diffusive edge
conductances, with volume-per-time units. The concentration generator is
$-\mathsf V^{-1}\mathsf B^{\mathsf T}\mathsf G\mathsf B$.
The smallest positive eigenvalue of
$\mathsf V^{-1/2}\mathsf B^{\mathsf T}\mathsf G\mathsf B
\mathsf V^{-1/2}$ supplies $\gamma_g$ in the volume-weighted norm.
Opposite edge fluxes cancel exactly, so this calculation preserves the
species total. Its zero vectors retain each connected component's mean.
Fluid transport cells and mechanical assembly cells can have different
resolutions; the accessible passages determine their transport
conductances. An obstruction can then be tested through a changed
matrix gap before replacing a spatially resolved ionic variable by one
well-mixed population.

During rearrangement, concentration and interfacial charge continue
through transport on the evolving fluid domain, including reciprocal
boundary work. The new $\gamma_g$ tests the updated local mixing
approximation while preserving the prepared concentration field.
Recruitment can thereby change a reporter's ionic environment through
passage geometry at fixed species totals and fixed virion pool.

Figure~\ref{fig:new-s7-coupling-diffusion-gap} links a populated passage to the conserved finite-volume generator and its mixing-gap test.

\begin{figure}[H]
\centering\begingroup
\definecolor{vltTeal}{HTML}{176B64}
\definecolor{vltBlue}{HTML}{315D80}
\definecolor{vltGold}{HTML}{A87520}
\definecolor{vltPlum}{HTML}{725877}
\definecolor{vltRed}{HTML}{762B38}
\definecolor{vltYellow}{HTML}{D5AD3C}
\definecolor{vltNeutral}{HTML}{777D83}
\setlength{\unitlength}{1pt}
\scalebox{0.90}{%
\input{vlt-figures/figure-260.tex}%
}\endgroup
\caption{Accessible passages and a conservative mixing calculation. (a) Both views retain eighteen rigid virions in one cylinder; moving six central bodies narrows the passage. Conductances below are auxiliary inputs, rather than values inferred from the drawing. (b) Four fluid averages have volumes $(1,1.4,0.8,1.2)v_\star$ and edge conductances $(1,\chi,1)G_\star$. They are transport coordinates, separate from mechanical virion coordinates. (c) Evolution from $c/c_\star=3+(1,1,-1.2,-1.2)$ preserves the volume-weighted mean. For the two stated passages, gaps are $\gamma t_\star=0.52879,0.02261$, with $t_\star=v_\star/G_\star$; dotted curves bound the weighted fluctuation norm. The comparison uses neutral, source-free diffusion with reflecting boundaries.}
\label{fig:new-s7-coupling-diffusion-gap}
\end{figure}

\paragraph{A gate variable is a population with a physical transition law.}
The Hodgkin--Huxley construction illustrates dynamical conductance at
its neuronal resolution~\cite{HodgkinHuxley1952}. With inside-minus-outside voltage and outward
ionic currents, one may write
\begin{align}
 I_{\rm Na}&=\overline G_{\rm Na}m_g^3h_g(V_m-E_{\rm Na}),
 &I_{\rm K}&=\overline G_{\rm K}n_g^4(V_m-E_{\rm K}),\nonumber\\
 \Dt{g}&=\alpha_g(V_m)(1-g)-\beta_g(V_m)g,
 &g&\in\{m_g,n_g,h_g\},\nonumber\\
 \frac{\dd}{\dd t}[C(\vct u)V_m+Q_g]
   &=I_{\rm app}-I_{\rm Na}-I_{\rm K}-I_{\rm leak}.
 \label{eq:phonon-membrane-gating-extension}
\end{align}
Capacitance and conductances here are totals. The nonnegative rates
$\alpha_g,\beta_g$ have inverse-time units and preserve $0\le g\le1$.
The variables $m_g,n_g$ describe activation and $h_g$ availability or
inactivation in the stated kinetic architecture; their products model
conducting-state occupancy under additional subunit assumptions.
They are not displacements. $Q_g$ is the retained polarization or
gating-charge contribution to the charge balance; it is set to zero
only when consistently omitted. Geometry requires the derivative of
$CV_m$, not just $C\Dt{V_m}$. Evolving finite concentrations give
evolving $E_j$. The original neuronal rate functions and conductance
powers are not assigned to a virion or to the M2 example above.

A thermodynamic gate makes its force and charge explicit. For one
equilibrating two-state unit at fixed $x,V_m$, let
$\Delta\epsilon_g(x,V_m)$ be its open-minus-closed conditional free
energy. Positive forward and reverse rates obey
\begin{equation}
 \frac{\alpha_g}{\beta_g}
       =e^{-\Delta\epsilon_g/(k_BT)}.
 \label{eq:phonon-gate-detailed-balance}
\end{equation}
Local detailed balance fixes passive equilibrium odds
\cite{Seifert2012}; the independent sum $\alpha_g+\beta_g$ fixes
the relaxation rate. Equal odds can therefore accompany different
response speeds. Empirical rate fits must be tested against this
thermodynamic condition when used as a passive gate model.

For a concrete mean-population closure, take $N_g$ identical units,
open fraction $p$, a voltage-independent intrinsic gap
$\Delta\epsilon_g^0(x)$, and gating displacement charge $q_g$ per
opening. At fixed separated free charge $q_{\rm sep}$, choose
\begin{equation}
 \begin{aligned}
 \mathcal F_g={}&U_{\rm mech}(x)
  +N_g\left[p\Delta\epsilon_g^0(x)
       +k_BT\{p\ln p+(1-p)\ln(1-p)\}\right]
       +\frac{[q_{\rm sep}-N_gq_gp]^2}{2C(x)},\\
 V_m={}&\frac{q_{\rm sep}-Q_g}{C(x)},\qquad Q_g=N_gq_gp,\qquad
 \frac1{N_g}\frac{\partial\mathcal F_g}{\partial p}
       =\Delta\epsilon_g^0-q_gV_m+k_BT\ln\frac p{1-p}.
 \end{aligned}
 \label{eq:phonon-gate-charge-free-energy}
\end{equation}
The separated charge comes from the retained ionic populations or an
explicit source. The gate moves polarization charge; it does not create
net charge. This energy gives
$\Delta\epsilon_g=\Delta\epsilon_g^0-q_gV_m$ and the displayed
capacitor balance. A voltage clamp uses its source Legendre transform.
The force $-\partial_x\mathcal F_g$ includes both the gap's mechanical
derivative and the capacitance derivative, with the declared variables
held fixed. For only a few correlated gates, their joint discrete state
replaces this ideal mixing/population closure.

The population law has a useful dissipation check. At fixed $x$ and
$q_{\rm sep}$, put $j_g^+=\alpha_g(1-p)$,
$j_g^-=\beta_gp$ and $\Dt{p}=j_g^+-j_g^-$. With the local gap above,
\begin{equation}
 \left.\frac{\dd\mathcal F_g}{\dd t}\right|_{\rm gate}
 =-N_gk_BT(j_g^+-j_g^-)\ln\frac{j_g^+}{j_g^-}\le0.
 \label{eq:phonon-gate-free-energy-dissipation}
\end{equation}
The inequality follows because the logarithm increases with its argument.
It continues to hold pointwise when voltage depends on $p$ through the
same capacitor energy. Mechanical motion and electrical supply add
their own matched terms. A gate can delay a response and alter its
restoring force without becoming an autonomous energy source. Spatial
voltage propagation additionally needs axial conductance and membrane
geometry for a cable equation; it is a distinct field from an elastic
wave.

\subsubsection{A finite gate population and its charging increments}
\label{sec:s7-coupling-finite-gates}

The continuous gate fraction has a controlled finite-state counterpart.
The relevant smallness parameter is the charging increment relative to
$k_BT$, together with the number of gates; a continuous voltage alone
does not specify that approximation.

\begin{definition}[Finite charge-conserving gate ensemble]
\label{def:s7-coupling-discrete-gates}
At fixed $x$ and separated charge $q_{\rm sep}$, take $N_g$ identical
two-state units with constant displacement charge $q_g$. If $n$ are
open, define their conditional state free energy and multiplicity by
\begin{equation}
 E_n=U_{\rm mech}(x)+n\Delta\epsilon_g^0(x)
                  +\frac{(q_{\rm sep}-nq_g)^2}{2C(x)},\qquad
 d_n=\binom{N_g}{n},\qquad 0\le n\le N_g.
 \label{eq:s7-coupling-finite-gate-energy}
\end{equation}
The count description assumes identical units whose microscopic
transition law is lumpable by $n$; resolved heterogeneity uses the joint
gate states. Every state retains the same gates and ionic participants.
\end{definition}

\begin{proposition}[Exact neighbouring-state balance for finite gates]
\label{prop:s7-coupling-finite-gate-balance}
The opening free-energy increment and equilibrium count probabilities are
\begin{equation}
 \begin{aligned}
 \Delta E_n&=E_{n+1}-E_n
    =\Delta\epsilon_g^0-q_gV_n+\frac{q_g^2}{2C},
 &V_n&=\frac{q_{\rm sep}-nq_g}{C},\\
 \pi_n&=Z_g^{-1}d_ne^{-\beta E_n},
 &\frac{\lambda_n}{\mu_{n+1}}
    &=\frac{N_g-n}{n+1}e^{-\beta\Delta E_n}.
 \end{aligned}
 \label{eq:s7-coupling-finite-gate-balance}
\end{equation}
Positive birth/death rates with this ratio and reflecting count endpoints
obey detailed balance with $\pi$. For positive probabilities $P_n$ and fluxes
$J_n=\lambda_nP_n-\mu_{n+1}P_{n+1}$, the free-energy functional
$\mathcal G[P]=\sum_nP_nE_n+k_BT\sum_nP_n\ln(P_n/d_n)$ obeys
\begin{equation}
 \frac{\dd\mathcal G}{\dd t}
 =-k_BT\sum_{n=0}^{N_g-1}J_n
                  \ln\frac{\lambda_nP_n}{\mu_{n+1}P_{n+1}}\le0.
 \label{eq:s7-coupling-finite-gate-dissipation}
\end{equation}
\end{proposition}
\begin{proof}
Subtract neighbouring capacitor energies to obtain $\Delta E_n$.
The binomial ratio gives $\pi_{n+1}/\pi_n$ and hence
$\lambda_n\pi_n=\mu_{n+1}\pi_{n+1}$. Substitute
$\dot P_n=J_{n-1}-J_n$ into $\dot{\mathcal G}$ and sum by parts.
Each remaining term has the displayed nonpositive sign.
\end{proof}

The finite charging term is one half of the voltage-step contribution
to an opening. Equivalently, the electrical opening energy uses the
mean of the voltages before and after the event. Its dimensionless
size is $\beta q_g^2/(2C)$. The same pre-event voltage with a purely
differential gap omits this term when it is appreciable. A common
voltage clamp supplies a different ensemble and its source work.
For large $N_g$, Stirling's approximation to $\ln d_n$ recovers the
mixing entropy in~\eqref{eq:phonon-gate-charge-free-energy}; retaining
only a narrow distribution of $n/N_g$ gives its mean-population form.
Strong charging feedback can correlate otherwise identical gates, so
the finite distribution also supplies the occupancy variance needed
for a calibrated force-noise prediction.

At each discrete state the mechanical load remains
$-\partial_xE_n$ at fixed $n,q_{\rm sep}$. Its configuration-gap and
capacitance derivatives are therefore fixed by the same energy as the
transition ratio. Gate transitions use the stated thermal medium;
geometry changes add mechanical work. This is the
neighbouring-state local-detailed-balance construction
\cite{Seifert2012}, applied to a finite capacitor rather than inferred
from an empirical conductance fit. It permits joint tests of occupancy
odds, event rates, voltage steps and mechanical displacement while
retaining the finite ionic resource and the existing assembly mechanics.

Figure~\ref{fig:new-s7-coupling-discrete-gates} compares the exact charged gate ensemble with an independent-unit population and its finite opening increments.

\begin{figure}[H]
\centering\begingroup
\definecolor{vltTeal}{HTML}{176B64}
\definecolor{vltBlue}{HTML}{315D80}
\definecolor{vltGold}{HTML}{A87520}
\definecolor{vltPlum}{HTML}{725877}
\definecolor{vltRed}{HTML}{762B38}
\definecolor{vltYellow}{HTML}{D5AD3C}
\definecolor{vltNeutral}{HTML}{777D83}
\setlength{\unitlength}{1pt}
\scalebox{0.90}{%
\input{vlt-figures/figure-261.tex}%
}\endgroup
\caption{Finite gate populations with charge conservation. (a) Four two-state membrane units belong to one optional membrane-bearing object. Opening changes $Q_g=nq_g$ at fixed separated free charge. (b) Filled bars give $\pi_n\propto\binom4n e^{-\beta E_n}$ at $E_n/k_BT=0.2n+0.4(2-n)^2$; outline bars show an independent binomial model matched only in mean count. (c) The exact discrete opening increment exceeds the pre-event derivative by $q_g^2/(2C)=0.4k_BT$. Segments connect discrete values. Detailed-balanced rates use these finite increments and multiplicities as in \eqref{eq:s7-coupling-finite-gate-balance}; their absolute timescale is a separate input.}
\label{fig:new-s7-coupling-discrete-gates}
\end{figure}

\paragraph{Reaction rates, affinity and finite biochemical work.}
Chemical states follow the same rule: their dependence on geometry
supplies mechanical force, while a transition law specifies how their
populations change. For the elementary scheme
$E+S\rightleftharpoons ES\longrightarrow E+P$, the initial-rate
quasi-steady approximation gives
\begin{equation}
 v_0=\frac{k_{\rm cat}E_{\rm tot}[S]}{K_M+[S]},\qquad
 K_M=\frac{k_{\rm off}+k_{\rm cat}}{k_{\rm on}},\qquad
 K_D=\frac{k_{\rm off}}{k_{\rm on}}.
 \label{eq:phonon-kinetic-affinity-distinction}
\end{equation}
It follows by setting
$\partial_t[ES]=k_{\rm on}[E][S]-(k_{\rm off}+k_{\rm cat})[ES]$
approximately to zero after the binding transient, using
$[E]+[ES]=E_{\rm tot}$ while substrate changes slowly. Product backflow
has been neglected in this particular early-time scheme. $K_M$ is a
kinetic concentration scale, and equals the binding dissociation scale
only in the rapid-equilibrium limit $k_{\rm cat}\ll k_{\rm off}$
for this mechanism~\cite{Cotterill2002,IUPACMichaelis2025}. It is not
the chemical energy available per turnover. A sealed reaction cannot
retain constant substrate and negligible reverse reaction indefinitely.

For a reversible resolved reaction, use per-particle chemical potentials
and dimensionless activities:
\begin{equation}
 \Delta_rG=\sum_a\nu_a\mu_a,\qquad
 \mu_a=\mu_a^\star(\vct u,\Phi)+k_BT\ln\mathfrak a_a,
 \qquad\mathcal A=-\Delta_rG.
 \label{eq:phonon-chemical-work-affinity}
\end{equation}
Here products have positive stoichiometric coefficients $\nu_a$.
The affinity $\mathcal A$ is positive when the chosen forward event
lowers free energy. If the event also changes a retained mechanical
or electrical state, its energy belongs in these conditional potentials
or in the explicitly equivalent total state difference. A solution
reaction affinity alone need not pay an imposed load.

In an ideal well-mixed elementary reversible channel, compatible
forward and reverse progress rates $j_r^+,j_r^->0$ obey
$\ln(j_r^+/j_r^-)=\mathcal A/(k_BT)$. Consequently
\begin{equation}
 J_r=j_r^+-j_r^-,\qquad
 \mathcal A J_r=k_BT(j_r^+-j_r^-)\ln\frac{j_r^+}{j_r^-}\ge0.
 \label{eq:phonon-reaction-local-balance}
\end{equation}
This identity is the resolved-channel free-energy dissipation, not a
formula for a rate prefactor. For discrete small-population states,
local detailed balance instead compares the forward and reverse
transition propensities at the two neighboring states using their
finite free-energy change~\cite{Seifert2012}. Coarse coupled channels
must retain the total dissipation account rather than impose an
independent downhill condition on every eliminated contribution.

For example, if $F\to W$ supplies
$\Delta\mu_{\rm fuel}=\mu_F-\mu_W$ and the same forward event does
mechanical work $w_{\rm load}$, its total available affinity is
$\mathcal A=\Delta\mu_{\rm fuel}-w_{\rm load}$ in that coupled
description. The reversible bound is $w_{\rm load}\le
\Delta\mu_{\rm fuel}$ for a downhill forward operation; attaining
zero affinity removes the net forward bias of a passive reversible
channel. Catalysis changes transition barriers and timescales while retaining
the reaction's thermodynamic free-energy difference. In a finite enclosure,
$\Dt{\mathcal N_F}=-J_r$, $\Dt{\mathcal N_W}=J_r$, so the
activities and available affinity change as fuel is used. Holding them
fixed is a chemostatted approximation with chemical supply work,
not the same closed finite-resource problem.

Figure~\ref{fig:finite-chemical-resource} gives an exactly soluble, finite reversible population example in which both the affinity and excess free energy relax.

\begin{figure}[H]
\centering\begingroup
\definecolor{vltTeal}{HTML}{176B64}
\definecolor{vltBlue}{HTML}{315D80}
\definecolor{vltGold}{HTML}{A87520}
\definecolor{vltPlum}{HTML}{725877}
\definecolor{vltRed}{HTML}{762B38}
\definecolor{vltYellow}{HTML}{D5AD3C}
\definecolor{vltNeutral}{HTML}{777D83}
\setlength{\unitlength}{1pt}
\scalebox{0.80}{%
\input{vlt-figures/figure-262.tex}%
}\endgroup
\caption{Finite reversible interconversion $F\rightleftharpoons W$ at fixed total $\mathcal N$, with $k_\pm>0$, $n_{F,\rm eq}=k_-/(k_++k_-)=1/4$ and $n_F(0)=0.9$. (a) The exact fuel fraction is $n_F=1/4+0.65e^{-s}$; its complement gives the product population. The dotted level marks fuel equilibrium. (b) The same trajectory decreases the relative free energy $\psi=n_F\ln(n_F/(1/4))+(1-n_F)\ln[(1-n_F)/(3/4)]$. Its rate is $\partial_s\psi=-(n_F-1/4)\ln[3n_F/(1-n_F)]\le0$. Both pools approach positive equilibrium populations and zero affinity. This auxiliary isothermal model supplies no feed, mechanical load or permanent power source; those additions require their matched kinetics and work ledger.}
\label{fig:finite-chemical-resource}
\end{figure}

\paragraph{Configuration fluctuations become mechanical susceptibility.}
A related equilibrium calculation identifies a useful effective
variable without postulating a new force. Retain finitely many
same-composition conditional configurations with energies
$\epsilon_a(\vct u)$ and Gibbs probabilities
$p_a=e^{-\beta\epsilon_a}/\sum_b e^{-\beta\epsilon_b}$.
Each state includes the retained participants and constraints. At
fixed $T>0$, their partition gives
\begin{equation}
 \mathcal F_{\rm cfg}(\vct u)
  =U_{\rm mech}(\vct u)-k_BT\ln\sum_a e^{-\beta\epsilon_a(\vct u)}.
 \label{eq:phonon-configuration-free-energy}
\end{equation}
This is a conditional equilibrium reduction
\cite{Cotterill2002,NeumaierWestra2011}; a biochemical reservoir would
require its chemical-potential terms in the weights. 

\begin{proposition}[Configurational population contribution to curvature]
\label{prop:s7-coupling-configuration-curvature}
Write
$\vct a_a=\nabla_u\epsilon_a$ and let brackets mean the $p_a$
average. The complete force-balanced reference has curvature
\begin{equation}
 \boxed{\mathsf K_{\rm eq}
  =\mathsf K_{\rm mech}+\langle\nabla_u^2\epsilon_a\rangle
                   -\beta\operatorname{Cov}_p(\vct a_a).}
 \label{eq:phonon-configuration-equilibrium-curvature}
\end{equation}
\end{proposition}
\begin{proof}
Differentiate the conditional partition function once to obtain
$\nabla\mathcal F_{\rm cfg}=\nabla U_{\rm mech}+\langle\vct a_a\rangle$.
A second derivative uses
$\nabla_up_a=-\beta p_a(\vct a_a-\langle\vct a_a\rangle)$
and gives the covariance term.
\end{proof}

The average conditional curvature is the restoring response at fixed
populations. The negative semidefinite covariance correction comes from
population readjustment: a deformation favors the state that loads it
differently. It is a softening relative to the same populations frozen,
not a claim that all biological temperature changes soften a material.
The total curvature must still be positive on the retained restoring
sector, and an already relaxed Hessian must be rematched before these
states are restored as explicit variables.

For one physical displacement $x$, choose local surfaces
$\epsilon_0=0$, $\epsilon_1=\epsilon_c+a_{\rm cfg}x$, with slope
$a_{\rm cfg}$ in force units and frozen curvature $k_x>0$.
Include the balancing linear term $-pa_{\rm cfg}x$ in the mechanical
reference so that $x=0$ is an equilibrium of the complete free energy.
Then
\begin{equation}
 p=\frac1{1+e^{\beta\epsilon_c}},\qquad
 \Delta k_{\rm cfg}=\beta p(1-p)a_{\rm cfg}^2,\qquad
 K_{\rm cfg,rel}=k_x-\Delta k_{\rm cfg}>0.
 \label{eq:phonon-two-state-configuration-softening}
\end{equation}
At fixed $T$ and slope, the population variance is largest at equal
occupation, not at complete activation. A saturated configuration has
little capacity to redistribute, even if its static force is large.
If reversible transition rates at the reference are $r_{\rm o},r_{\rm c}$,
set $\tau_{\rm cfg}=(r_{\rm o}+r_{\rm c})^{-1}$. Linearizing the
population equation and reciprocal force gives
\begin{equation}
 \Dt{\delta p}
   =-\frac{\delta p+\beta p(1-p)a_{\rm cfg}x}{\tau_{\rm cfg}},
 \qquad K_{\rm cfg}^*(\omega)
       =k_x-\frac{\Delta k_{\rm cfg}}{1-i\omega\tau_{\rm cfg}}.
 \label{eq:phonon-two-state-configuration-relaxation}
\end{equation}
This gives another concrete realization of the ionic relaxation branch:
state variance sets its strength, transition speed its time constant.
Figure~\ref{fig:gate-population-and-reciprocal-softening} separates
these two jobs.

\begin{figure}[H]
\centering\begingroup
\definecolor{vltTeal}{HTML}{176B64}
\definecolor{vltBlue}{HTML}{315D80}
\definecolor{vltGold}{HTML}{A87520}
\definecolor{vltPlum}{HTML}{725877}
\definecolor{vltRed}{HTML}{762B38}
\definecolor{vltYellow}{HTML}{D5AD3C}
\definecolor{vltNeutral}{HTML}{777D83}
\setlength{\unitlength}{1pt}
\scalebox{0.90}{%
\input{vlt-figures/figure-263.tex}%
}\endgroup
\caption{Population susceptibility and switching kinetics in a reciprocal mechanical response. (a) Internal configurations of one retained object load coordinate $x$ through the gap derivative. (b) At gap $d$, $p=(1+e^d)^{-1}$ gives variance $p(1-p)$, maximal at equal occupancy. (c) Holding $\Delta k_{\rm cfg}/k_x=1/2$, the storage and positive loss are $K'/k_x=1-[2(1+r^2)]^{-1}$ and $K''/k_x=r/[2(1+r^2)]$. The loss peaks at $r=1$. The two abscissae vary a thermodynamic gap and a loading-rate ratio, respectively; reversible kinetics and the conjugate force connect them to a specified biophysical mechanism. The boxed result illustrated here is~\eqref{eq:phonon-configuration-equilibrium-curvature}.}
\label{fig:gate-population-and-reciprocal-softening}
\end{figure}

The corresponding noise can be checked independently. For a two-state
indicator $n_g(t)\in\{0,1\}$ with $x$ clamped at the reference,
$\operatorname{Cov}[n_g(t),n_g(0)]=p(1-p)e^{-|t|/\tau_{\rm cfg}}$.
Its fluctuating force $-a_{\rm cfg}(n_g-p)$ therefore has spectrum
\begin{equation}
 S_{ff}^{\rm cfg}(\omega)
  =\frac{2a_{\rm cfg}^2p(1-p)\tau_{\rm cfg}}
                       {1+\omega^2\tau_{\rm cfg}^2}
  =\frac{2k_BT\Delta k_{\rm cfg}\tau_{\rm cfg}}
                       {1+\omega^2\tau_{\rm cfg}^2}.
 \label{eq:phonon-gate-telegraph-fdt}
\end{equation}
Thus occupation variance, mechanical softening and equilibrium force
noise have a matched amplitude. For independent identical units their
force covariances and softening strengths add. This is the gating-force
contribution, not every source of measured ionic-current noise.
When the mechanics is released, the reciprocal near-equilibrium
linearization supplies the corresponding memory and fluctuation
completion. A conserved electronic projector in the earlier diagonal
Hamiltonian does not itself provide this population equilibration.

Population, ionic and mechanical readouts can therefore constrain a
choice of variables together: occupation odds measure a gap relative to
$k_BT$, dwell times measure transition kinetics, and calibrated force
changes measure the mechanical derivative. A measured relaxation alone
need not identify which of these mechanisms generated it. Temperature
can change both $p$ and $\beta$, and constitutive parameters can change
as well; the softening need not vary monotonically with warming.
Finally, maintaining a driven cycle requires a source account. Finite
fuel and gradients evolve, while chemostats, voltage clamps and pumps
represent supplied work or chemical free energy. That distinction is
what allows an active or quantum oscillator extension to be formulated
without turning passive thermal fluctuations into an unaccounted source.

\paragraph{A finite shared ligand pool on HPV-inspired VLPs.}
For non-enveloped L1 VLPs, a directly applicable population variable is
surface association. A retained glycan-derived ligand can be free in the
cavity or bound to a specified capsid patch. This gives a closed-pool
application of the configuration susceptibility above. The membrane and
channel constructions remain separate preparations. A glycan-induced
change in capsid indentation stiffness~\cite{Feng2024} constrains that
intracapsid response; an inter-VLP contact derivative still requires its
own interaction model or calibration.

\begin{definition}[Finite one-site ligand ensemble]
\label{def:s7hpv-finite-ligand-pool}
Let $N_S$ be the number of identical effective binding units distributed
over a fixed population of VLPs, and let $N_L$ retained ligands each occupy
at most one unit. At bound count $n$, the free count is $N_L-n$ and
$0\leq n\leq n_{\max}=\min(N_S,N_L)$.
For a well-mixed reference volume $V$, let $v_\circ$ be a standard volume.
The conditional state free energy $E_n(\vct X)$ includes mechanical,
binding and specified excess electrical terms; it excludes the ideal
free-ligand combinatorial factor below. Fixed body inertia is a reference
approximation; appreciable bound mass instead updates that inertia.
\end{definition}

\begin{proposition}[Closed-pool binding weights and reciprocal force]
\label{prop:s7hpv-finite-ligand-balance}
Under the stipulated independent-unit combinatorics, the exact equilibrium
count law at fixed geometry is
\begin{equation}
 \begin{aligned}
 \pi_n(\vct X)&=Z_L^{-1}\binom{N_S}{n}
       \frac{(V/v_\circ)^{N_L-n}}{(N_L-n)!}
                       e^{-\beta E_n(\vct X)},\\
 \frac{\lambda_n}{\mu_{n+1}}
   &=\frac{N_S-n}{n+1}\frac{(N_L-n)v_\circ}{V}
                  e^{-\beta(E_{n+1}-E_n)} .
 \end{aligned}
 \label{eq:s7hpv-finite-ligand-count-law}
\end{equation}
Positive neighbouring-state rates with this ratio and closed count
endpoints obey detailed balance with $\pi$. The mechanical load in state
$n$ is $-\nabla_XE_n$. For a fixed ligand chemical charge $z_Le$,
\begin{equation}
 Q_{\rm surface}=Q_0+z_Le n,\qquad
 Q_{\rm free}=z_Le(N_L-n),\qquad
 Q_{\rm surface}+Q_{\rm free}=Q_0+z_LeN_L .
 \label{eq:s7hpv-ligand-charge-ledger}
\end{equation}
Additional counterions or protonation events retain their own conserved
participants, as in~\eqref{eq:s2patch-proton-pool}.
\end{proposition}
\begin{proof}
Choose the occupied units and integrate the ideal free-ligand positions,
including indistinguishability. The ratio of neighbouring weights gives
the stated rate ratio. Differentiating the same conditional energy gives
the reciprocal mechanical force. Binding transfers the displayed charge
between two parts of the retained system, leaving their sum unchanged.
\end{proof}
The exact count chain supplies equilibrium force correlations and finite
association events. Its free-energy decay follows the same discrete
balance argument as~\eqref{eq:s7-coupling-finite-gate-dissipation}.
Nonidentical patches retain the joint occupations. A bridging polymer
that contacts several VLPs requires its contact graph and conformation;
the one-site law is a declared effective-ligand model.

For a single mechanical coordinate, take
$E_n(x)=\tfrac12k_xx^2-a_b\bar n\,x+n(\epsilon_b+a_bx)$,
where $a_b$ is a calibrated binding-force derivative and
$\bar n=\langle n\rangle_{x=0}$. The balancing linear term makes $x=0$
force balanced. Differentiating the count partition function gives
\begin{equation}
 \boxed{K_{\rm rel}=k_x-\beta a_b^2\operatorname{Var}_{\pi}(n)>0.}
 \label{eq:s7hpv-finite-pool-mechanical-curvature}
\end{equation}
This is an exact static susceptibility of this model. In a many-count
local Stirling approximation about an interior $n_\star$, the population
curvature and clamped variance become
\begin{equation}
 A_n\simeq k_BT\left[\frac1{n_\star}+
             \frac1{N_S-n_\star}+\frac1{N_L-n_\star}\right],
 \qquad \operatorname{Var}(n)\simeq\frac{k_BT}{A_n}.
 \label{eq:s7hpv-finite-pool-susceptibility}
\end{equation}
The final term is the free-pool depletion penalty. A chemostat removes
that term and supplies chemical exchange; it changes the physical
ensemble, even at the same mean occupation.

For a concrete synthetic preparation, let eight HPV-inspired bodies each
carry twelve effective patch units: $N_S=96$, $N_L=60$. The twelve-unit
partition is an assigned spatial resolution, not HPV capsomer or binding
stoichiometry. Choose
$\kappa_L=(v_\circ/V)e^{-\beta\epsilon_b}=1/12$.
Then $\pi_n\propto\binom{96}{n}60!\kappa_L^n/(60-n)!$ has mode $48$;
direct normalization gives $\bar n\simeq48.11$ and
$\operatorname{Var}n\simeq7.993$. An independent-site chemostat adjusted
to this same mean gives variance about $24.00$. The local estimate at
$n_\star=48$ gives $A_n=k_BT/8$ and variance $8$.
Thus shared depletion can reduce equal-time population force variance and equilibrium
softening by about a factor of three without changing the VLP count.

For example, mass-action rates
$\lambda_n=k_{\rm on}(N_L-n)(N_S-n)/V$ and
$\mu_n=k_{\rm off}n$, with
$k_{\rm on}/(Vk_{\rm off})=\kappa_L$, realize this count law at $x=0$.
Their local linearization about $48$ relaxes at $6k_{\rm off}$;
the matched fixed-concentration comparison relaxes at $2k_{\rm off}$.
A single Debye branch is this local linear resolution; the exact finite
chain generally has several decay rates. A calibrated bound-ligand
readout, centre/marker record and weak conjugate force can test the
occupancy distribution, relaxation and susceptibility with the same
inputs. Exposure is applied to those continued histories. Binding
energies, rates and $a_b$ are assigned here, while the HPV geometry
specifies the body on which the effective patches are registered.

Figure~\ref{fig:s7hpv-finite-pool} evaluates the shared-pool model with exact finite-state matrices rather than a fitted single relaxation time.

\begin{figure}[H]
\centering\begingroup
\setlength{\unitlength}{1pt}
\scalebox{0.90}{%
\input{vlt-figures/figure-303.tex}%
}\endgroup
\caption{A finite ligand pool changes both fluctuations and kinetics. The synthetic ensemble has eight HPV-inspired bodies, twelve effective one-site units per body and sixty retained ligands; the enlarged half-filled glyph illustrates one assignment at $n=48$, not HPV binding stoichiometry. Panel (b) evaluates the exact closed-pool mass function with $\kappa_L=1/12$ and an independent-site reservoir distribution adjusted to the same mean. Only counts $30$--$66$ are shown, with no renormalization; the closed law has no support above sixty. Panel (c) uses the full distributions for the variance. Panel (d) diagonalizes the reversible sixty-one-state birth--death generator at fixed geometry, with $k_{\rm off}=1$ in the rate unit and $k_{\rm on}/(Vk_{\rm off})=1/12$. It plots the two-sided clamped count spectrum $2\sum_\ell c_\ell r_\ell/(r_\ell^2+\omega^2)$, with nonnegative spectral weights summing to $\operatorname{Var}n$. The reservoir curve uses its exact independent-site relaxation rate. A binding-force readout multiplies this spectrum by $a_b^2$; released mechanics retains its joint reciprocal force law. All counts, charges and detector maps refer to the same fixed VLP population. The boxed result illustrated here is~\eqref{eq:s7hpv-finite-pool-mechanical-curvature}.}
\label{fig:s7hpv-finite-pool}
\end{figure}

\subsubsection{Protein motion: a large fluctuation need not be a slow variable}
\label{sec:phonon-protein-slow-variables}

Protein measurements illustrate how the prediction window selects
variables. Single-molecule FRET resolves structural populations in
chymotrypsin inhibitor~2~\cite{Deniz2000}. Such a distance signal is
one projection of structure, with photon statistics and label motion
in its observation model. Comparing it with mechanical amplitudes
requires both structural sensitivity and temporal resolution.

Within one stable basin, a local linearization of a calibrated reporter
can have the form $Y=\sum_j a_jx_j+\epsilon$, where the length-valued
$x_j$ are specified shape coordinates, $a_j$ are readout derivatives,
and $\epsilon$ denotes detector contributions. Consider the explicit
two-coordinate classical surrogate
\begin{equation}
 \dd x_j=-\mu_j^{\rm conf}k_j^{\rm conf}x_j\,\dd t
       +\sqrt{2k_{\rm B}T\mu_j^{\rm conf}}\,\dd B_j,
 \qquad
 r_j=\mu_j^{\rm conf}k_j^{\rm conf},\quad j\in\{f,s\}.
 \label{eq:phonon-protein-two-timescale-surrogate}
\end{equation}
The two Wiener processes are independent. Each curvature has stiffness
units and each mobility has units of velocity per force. With the
coordinates fixed in common calibrated length units, the stationary
variance is $k_{\rm B}T/k_j^{\rm conf}$, while the relaxation time is
$r_j^{-1}$. Curvature controls fluctuation size; curvature together with
mobility controls persistence.

Choose illustrative values
$k_f^{\rm conf}=k_\star$, $k_s^{\rm conf}=4k_\star$,
$\mu_f^{\rm conf}=\mu_\star$ and
$\mu_s^{\rm conf}=\mu_\star/100$. With
$t_\star=(\mu_\star k_\star)^{-1}$, the fast coordinate has four
times the variance but relaxes twenty-five times faster. A
largest-variance projection selects $x_f$. A prediction about structure
persisting over several $t_\star$ instead needs $x_s$. These numbers
define a local surrogate, not the folding kinetics of the cited protein.
A between-basin folding or opening coordinate additionally needs the
basins and transition dynamics described above.

Lagged measurements distinguish the two criteria. For centered features
$\vct X$, define
$\mathsf C_X(\tau)=\langle\vct X(t+\tau)\vct X(t)^{\mathsf T}\rangle$.
In the exact two-coordinate example,
\begin{equation}
 \mathsf C_X(\tau)\vct a_j
       =\nu_j(\tau)\mathsf C_X(0)\vct a_j,
 \qquad \nu_j(\tau)=e^{-r_j\tau},\qquad
 t_j=-\frac{\tau}{\ln\nu_j(\tau)}.
 \label{eq:phonon-biophysical-lagged-coordinate-test}
\end{equation}
The larger nonconstant eigenvalue represents the more persistent
coordinate in this reversible model. Variational approaches to slow
stochastic dynamics and time-lagged component analysis give a broader
foundation for such comparisons~\cite{NoeNuske2013,PerezHernandez2013}.
With incomplete nonlinear features, finite data or nonstationary
preparations, the generalized eigenproblem is an approximation requiring
its own checks across lag times and independent observations.

There is also a concrete closure test. Suppose a calibrated projection
reports $Y=x_f+x_s$, with detector contributions removed. Write
$R_Y(\tau)=\langle Y(t+\tau)Y(t)\rangle/\operatorname{Var}Y$ and
$w=\operatorname{Var}x_f/\operatorname{Var}Y$. Then
\begin{equation}
 \begin{aligned}
 R_Y(\tau)&=we^{-r_f\tau}+(1-w)e^{-r_s\tau},\\
 R_Y(2\tau)-R_Y(\tau)^2
       &=w(1-w)(e^{-r_f\tau}-e^{-r_s\tau})^2.
 \end{aligned}
 \label{eq:phonon-projected-protein-memory-test}
\end{equation}
A scalar Ornstein--Uhlenbeck model makes the second line zero. Two
visible distinct rates give a positive defect at nonzero lag:
matching variance alone leaves a dynamical error. A second coordinate
or a consistent memory representation restores this information.
The test concerns the stationary Gaussian linear model class; a
multi-exponential correlation alone does not exclude every nonlinear
scalar Markov model. Figure~\ref{fig:s7-publication-variance-persistence}
compares the variance, lagged persistence and projected closure tests.

\begin{figure}[H]
\centering\begingroup
\definecolor{vltTeal}{HTML}{176B64}
\definecolor{vltBlue}{HTML}{315D80}
\definecolor{vltGold}{HTML}{A87520}
\definecolor{vltPlum}{HTML}{725877}
\definecolor{vltRed}{HTML}{762B38}
\definecolor{vltYellow}{HTML}{D5AD3C}
\definecolor{vltNeutral}{HTML}{777D83}
\setlength{\unitlength}{1pt}
\scalebox{0.90}{%
\input{vlt-figures/figure-264.tex}%
}\endgroup
\caption{Variance and persistence select different coordinates in the two-mode surrogate of \eqref{eq:phonon-protein-two-timescale-surrogate}. Use $\ell_T^2=k_{\rm B}T/k_\star$, $t_\star=(\mu_\star k_\star)^{-1}$, curvatures $(1,4)k_\star$ and mobilities $(1,1/100)\mu_\star$. (a) The equal-scale unit-Mahalanobis contour has covariance $\operatorname{diag}(1,1/4)\ell_T^2$. Its horizontal fast coordinate has four times the variance. (b) Solid teal and dashed burgundy correlations decay at $1/t_\star$ and $1/(25t_\star)$, respectively. (c) For the noiseless readout $Y=x_f+x_s$, $w=4/5$ and the exact defect is $w(1-w)(e^{-\tau/t_\star}-e^{-\tau/(25t_\star)})^2$. Shading marks its positive value; the dot is its maximum. A stationary scalar Gaussian Ornstein--Uhlenbeck model gives zero. The coordinates describe local shape fluctuations in one basin; the contour and lag functions are ensemble summaries.}
\label{fig:s7-publication-variance-persistence}
\end{figure}

For biophysical interpretation, the persistent coordinate may describe
a domain arrangement or configuration that changes the availability of
a structural state during the next observation. A faster motion can
contribute much of the measured variance while averaging out over that
interval. A useful reduced variable must carry the information needed
for the intended prediction and couple to an accessible readout. The
energy-based mode construction supplies physical coordinates and
reciprocal forces; lagged statistics test which combinations remain
predictive at the chosen temporal resolution. Neither static variance
nor slow persistence alone assigns a molecular mechanism without that
structural and observational connection.

\subsection{Quantum self-excitation: pumping, saturation, and retained resources}
\label{sec:phonon-quantum-active}

Section~\ref{sec:self-excitation-passive-boundary} identifies feedback
and resources that maintain classical motion against loss. A quantum
extension specifies the transitions adding and removing quanta and
their fluctuations. The quantum Van der Pol model, developed for
engineered oscillators~\cite{LeeSadeghpour2013,WalterNunnenkampBruder2014},
provides a worked example. Applying it to a viral preparation requires
a resolved mode, a physical pump and a justified quantum regime.

\paragraph{Declare the quantum channels before assigning a gain.}
\begin{definition}[Declared active quantum oscillator]
\label{def:s7-active-channel-model}
Choose one positive mode of frequency $\omega_s$, and abbreviate its
operators by $\widehat b$, $\widehat N=\widehat b^\dagger\widehat b$ and
$\widehat H_s=\hbar\omega_s\widehat N$. Consider the time-independent
reduced master equation
\begin{equation}
 \begin{aligned}
  \partial_t\rho&=\mathcal L_{\rm a}\rho
   =-\frac{i}{\hbar}[\widehat H_s,\rho]
       +\kappa_+\mathcal D[\widehat b^\dagger]\rho
       +\kappa_-\mathcal D[\widehat b]\rho
       +\kappa_2\mathcal D[\widehat b^2]\rho,\\
  \mathcal D[\widehat L]\rho
    &=\widehat L\rho\widehat L^\dagger
         -\tfrac12\{\widehat L^\dagger\widehat L,\rho\},
  \qquad \kappa_+>0,\quad\kappa_-\geq0,\quad\kappa_2>0.
 \end{aligned}
 \label{eq:quantum-active-master-equation}
\end{equation}
All three coefficients have units $\mathrm s^{-1}$. The first jump adds
one quantum, the second removes one, and the third removes two. Their
state-dependent event rates are respectively
$\kappa_+\langle\widehat N+1\rangle$,
$\kappa_-\langle\widehat N\rangle$ and
$\kappa_2\langle\widehat N(\widehat N-1)\rangle$.
The two-quantum loss supplies saturation.
\end{definition}
The dissipator normalization is
stated because doubling its definition doubles the associated rates.
Finite occupation cutoffs give trace-preserving, completely positive
Lindblad equations~\cite{Lindblad1976}; oscillator identities below are
read on the finite-particle core and for states with the displayed finite
moments. The stationary examples use a converged cutoff limit rather
than treating the cutoff's highest level as a physical saturation process.
Frequency-resolved jumps additionally require a justified reservoir-memory
and weak-coupling hierarchy over the occupied levels; they are not an
automatic description of strongly overdamped motion.

This time-homogeneous description has no imposed periodic force, but the
mode is energetically open. The gain presupposes an available pump; the
loss channels presuppose receivers. A thermal linear bath may be included
by writing $\kappa_+=\kappa_{\rm pump}+\Gamma\bar N_T$ and
$\kappa_-=\kappa_{\rm loss}+\Gamma(\bar N_T+1)$, where
$\bar N_T=(e^{\beta\hbar\omega_s}-1)^{-1}$. Its upward transitions are
essential even when one is interested mainly in damping. The ideal
$\mathcal D[\widehat b^2]$ channel describes a cold or engineered receiver;
a finite-temperature equilibrium two-quantum bath generally also requires
reverse transitions. These choices specify distinct physical closures.

\paragraph{Excitation number and energy have exact budgets.}
\begin{proposition}[Excitation-number and mode-energy budgets]
\label{prop:s7-active-exact-budgets}
For the declared active model at fixed frequency and with the displayed
finite moments, the number and mode-energy accounts are exact.
Applying the adjoint generator to $\widehat N$ gives
\begin{equation}
 \boxed{\displaystyle
 \frac{\dd\langle\widehat N\rangle}{\dd t}
  =\kappa_+\bigl(\langle\widehat N\rangle+1\bigr)
      -\kappa_-\langle\widehat N\rangle
      -2\kappa_2\langle\widehat N(\widehat N-1)\rangle.}
 \label{eq:quantum-active-exact-number-balance}
\end{equation}
The gain's $+1$ is the oscillator matrix element
$\widehat b\widehat b^\dagger=\widehat N+1$; a pumped vacuum therefore
has initial number growth $\kappa_+$. It is not a source of energy in an
unprepared isolated vacuum. At fixed frequency the corresponding energy
currents are
\begin{equation}
 \begin{aligned}
 J_+&=\hbar\omega_s\kappa_+\langle\widehat N+1\rangle,\qquad
 J_-=\hbar\omega_s\kappa_-\langle\widehat N\rangle,\\
 J_2&=2\hbar\omega_s\kappa_2
                  \langle\widehat N(\widehat N-1)\rangle,\qquad
 \frac{\dd\langle\widehat H_s\rangle}{\dd t}=J_+-J_--J_2.
 \end{aligned}
 \label{eq:quantum-active-energy-currents}
\end{equation}
\end{proposition}
\begin{proof}
On the declared oscillator core, the three adjoint dissipators send
$\widehat N$ respectively to $\widehat N+\Id$,
$-\widehat N$ and $-2\widehat N(\widehat N-1)$.
Their weighted sum gives the number law. The Hamiltonian commutes
with number; multiplying by the fixed energy spacing
$\hbar\omega_s$ gives the mode-energy account.
\end{proof}
These are mode-energy currents in watts. A nonthermal pump contribution
cannot be called heat from an equilibrium bath without further
thermodynamic information. Figure~\ref{fig:quantum-active-jump-resource-ledger}
connects the jumps to their resource and receiver accounts.

\begin{figure}[H]
\centering\begingroup
\definecolor{vltTeal}{HTML}{176B64}
\definecolor{vltBlue}{HTML}{315D80}
\definecolor{vltGold}{HTML}{A87520}
\definecolor{vltPlum}{HTML}{725877}
\definecolor{vltRed}{HTML}{762B38}
\definecolor{vltYellow}{HTML}{D5AD3C}
\definecolor{vltNeutral}{HTML}{777D83}
\setlength{\unitlength}{1pt}
\scalebox{0.90}{%
\input{vlt-figures/figure-265.tex}%
}\endgroup
\caption{Resource and receiver channels for the declared active oscillator. (a) Open arrows identify energy transfer, with no magnitude scale. The event rates $R_+=\kappa_+\langle N+1\rangle$, $R_-=\kappa_-\langle N\rangle$, $R_2=\kappa_2\langle N(N-1)\rangle$ carry increments $+\hbar\omega_s,-\hbar\omega_s,-2\hbar\omega_s$. (b) At $\kappa_-=0$ and $\kappa_2\gg\kappa_+$, the gain path through short-lived $\ket2$ carries finite return current and yields the printed limiting mixture. The graph highlights a path in the full occupation space; boxes denote states and energy resources. The boxed result illustrated here is~\eqref{eq:quantum-active-exact-number-balance}.}
\label{fig:quantum-active-jump-resource-ledger}
\end{figure}

\paragraph{A thermal completion and a conversion-cycle test.}
The reverse of a pair-loss event absorbs two quanta from its receiver.
Adding that channel gives an explicit equilibrium benchmark against
which an engineered pump can be compared.

\begin{definition}[Paired one- and two-quantum channels]
\label{def:s7-active-paired-channels}
At the same fixed frequency and dissipator convention, let
\begin{equation}
 \mathcal L_{\rm pair}\rho
   =-i\omega_s[\widehat N,\rho]
       +\sum_{r=1}^2\left\{
          \kappa_{r,-}\mathcal D[\widehat b^r]\rho+
          \kappa_{r,+}\mathcal D[(\widehat b^\dagger)^r]\rho
                      \right\},\qquad \kappa_{r,\pm}>0.
 \label{eq:s7-active-paired-generator}
\end{equation}
The coefficients have inverse-time units and the matrix elements of
the stated jump operators supply their occupation dependence. The
oscillator-core and cutoff qualifications of
Definition~\ref{def:s7-active-channel-model} apply to this extension.
Use a finite-cutoff model, or a full-oscillator evolution whose
probability conservation and required finite moments are established.
Arbitrary positive pair-gain rates alone do not establish those
properties; the common-temperature ratios below give lower upward
than downward rates at both transition orders.
\end{definition}

\begin{proposition}[Common-temperature stationary preparation]
\label{prop:s7-active-common-temperature}
For the full oscillator at $T>0$, put
$q_T=e^{-\beta\hbar\omega_s}\in(0,1)$.
The Gibbs state with populations $p_n=(1-q_T)q_T^n$ is stationary
under the paired generator when
\begin{equation}
 \frac{\kappa_{1,+}}{\kappa_{1,-}}=q_T,\qquad
 \frac{\kappa_{2,+}}{\kappa_{2,-}}=q_T^2.
 \label{eq:s7-active-paired-thermal-ratios}
\end{equation}
Each paired channel then has zero net equilibrium energy current.
\end{proposition}
\begin{proof}
The upward rate from $n$ to $n+r$ is
$\kappa_{r,+}(n+r)!/n!$, while its reverse rate is
$\kappa_{r,-}(n+r)!/n!$. Their population-weighted fluxes agree
because $p_{n+r}=q_T^rp_n$. The Hamiltonian commutes with the Gibbs
state, so all terms vanish after summing these pairwise balances.
\end{proof}

For the population cycle $n\to n+1\to n+2\to n$, the logarithm of
the product of forward rates divided by the reverse product is
\begin{equation}
 \mathcal A_{\rm cyc}
   =\ln\frac{\kappa_{1,+}^2\kappa_{2,-}}
                   {\kappa_{1,-}^2\kappa_{2,+}}.
 \label{eq:s7-active-cycle-affinity}
\end{equation}
Oscillator factorials cancel from this dimensionless cycle affinity.
At common temperature it vanishes; a nonzero value biases conversion
between two single-quantum exchanges and one pair exchange even when
the covariance resembles a thermal value. Physical reservoirs then
set the associated energy and entropy currents~\cite{Seifert2012}.
The formula assumes positive reverse rates, with a cold pair receiver
as a one-way limit. Positive-temperature equilibrium has
$\kappa_{1,+}<\kappa_{1,-}$, so net linear gain requires a separate
source. A frequency-resolved Markov model also requires the stated
weak-coupling and memory hierarchy.

Figure~\ref{fig:s7-active-paired-thermal-cycle} evaluates the reverse-channel equilibrium test and its conversion cycle without truncating that cycle into a closed three-level system.

\begin{figure}[H]
\centering\begingroup
\definecolor{vltTeal}{HTML}{176B64}
\definecolor{vltBlue}{HTML}{315D80}
\definecolor{vltGold}{HTML}{A87520}
\definecolor{vltPlum}{HTML}{725877}
\definecolor{vltRed}{HTML}{762B38}
\definecolor{vltYellow}{HTML}{D5AD3C}
\definecolor{vltNeutral}{HTML}{777D83}
\setlength{\unitlength}{1pt}
\scalebox{0.90}{%
\input{vlt-figures/figure-266.tex}%
}\endgroup
\caption{A common-temperature check with every reverse channel retained. Set $\kappa_{1,-}t_\star=1$, $\kappa_{1,+}t_\star=1/2$, $\kappa_{2,-}t_\star=1/5$ and $\kappa_{2,+}t_\star=1/20$. Teal arrows traverse $0\to1\to2\to0$ and dashed burgundy arrows its reverse; labels are actual number-state transition rates, including oscillator matrix elements. This is one cycle within the full graph, not a three-level cutoff or a spatial particle trajectory. The forward and reverse rate products agree \eqref{eq:s7-active-cycle-affinity}. The rate ratios give $q_T=1/2$ at both transition orders \eqref{eq:s7-active-paired-thermal-ratios}. Bars show the first eight probabilities of the full normalized Gibbs law, with its omitted tail reported. Individual upward/downward energy currents are nonzero but cancel within each paired channel. No biological temperature or pump mechanism is calibrated by these units.}
\label{fig:s7-active-paired-thermal-cycle}
\end{figure}

\paragraph{The classical envelope is a controlled correspondence.}
For $\alpha=\langle\widehat b\rangle$, the exact first-moment equation is
\begin{equation}
 \partial_t\alpha=
    \left[-i\omega_s+\frac{\kappa_+-\kappa_-}{2}\right]\alpha
       -\kappa_2\langle\widehat b^\dagger\widehat b^2\rangle.
 \label{eq:quantum-active-first-moment-hierarchy}
\end{equation}
It is not closed in $\alpha$. In a large-occupation, approximately
phase-resolved regime, factorizing the last moment as $|\alpha|^2\alpha$
and removing the carrier phase gives
\begin{equation}
 \frac{\dd r}{\dd t}\simeq
    \left[\frac{\kappa_+-\kappa_-}{2}-\kappa_2r^2\right]r,
 \qquad r_*^2\simeq\frac{\kappa_+-\kappa_-}{2\kappa_2}
                  \quad(\kappa_+>\kappa_-).
 \label{eq:quantum-active-factorized-amplitude}
\end{equation}
The displacement amplitude is
$A_s=\sqrt{2\hbar/\omega_s}\,r$. Matching this drift to
\eqref{eq:self-excitation-averaged-onset} gives
$(\kappa_+-\kappa_-)/2=\gamma_{\rm a}-\gamma_{\rm p}$ and
$\kappa_2=\hbar\gamma_{\rm n}/(2\omega_sQ_c^2)$.
This identifies consistent coefficients in the weak-rate envelope
regime; it is not a unique quantization of the classical damping law.
Other reservoir models can share its leading drift while producing
different fluctuations. Moreover, the exact quantum stationary
occupation need not vanish below this factorized threshold: any
$\kappa_+>0$ also supplies fluctuations. A single finite oscillator does
not acquire an exact noiseless Hopf transition merely from this
approximate amplitude equation.

\paragraph{The deep quantum limit leaves a finite occupation.}
The populations $p_n=\bra n\rho\ket n$ obey the closed rate equation
\begin{equation}
 \begin{aligned}
 \partial_tp_n={}&\kappa_+[np_{n-1}-(n+1)p_n]
       +\kappa_-[(n+1)p_{n+1}-np_n]\\
       &+\kappa_2[(n+2)(n+1)p_{n+2}-n(n-1)p_n],
 \qquad p_{-1}=0.
 \end{aligned}
 \label{eq:quantum-active-population-equation}
\end{equation}
It follows from the jump matrix elements, rather than a classical noise
distribution. For $\kappa_2\gg\kappa_+,\kappa_-$, states above one quantum
are short lived. The path $1\to2\to0$ has a slow first step of rate
$2\kappa_+$ and a fast second step of rate $2\kappa_2$. Eliminating this
short-lived level gives the effective balance
$\kappa_+p_0=(2\kappa_++\kappa_-)p_1$, hence
\begin{equation}
 \boxed{\displaystyle
 \rho_{\rm ss}\longrightarrow
  \frac{2\kappa_++\kappa_-}{3\kappa_++\kappa_-}\ket0\bra0
   +\frac{\kappa_+}{3\kappa_++\kappa_-}\ket1\bra1.}
 \label{eq:quantum-active-deep-quantum-state}
\end{equation}

Figure~\ref{fig:strong-pair-loss-general-mixture} resolves the fast return pathway and shows how single-quantum loss changes the limiting mixture.

\begin{figure}[H]
\centering\begingroup
\definecolor{vltTeal}{HTML}{176B64}
\definecolor{vltBlue}{HTML}{315D80}
\definecolor{vltGold}{HTML}{A87520}
\definecolor{vltPlum}{HTML}{725877}
\definecolor{vltRed}{HTML}{762B38}
\definecolor{vltYellow}{HTML}{D5AD3C}
\definecolor{vltNeutral}{HTML}{777D83}
\setlength{\unitlength}{1pt}
\scalebox{0.80}{%
\input{vlt-figures/figure-267.tex}%
}\endgroup
\caption{Eliminating a fast occupation path while retaining its current. At $\kappa_2\gg\kappa_+,\kappa_-$, gain reaches $|2\rangle$ at rate $2\kappa_+$ and pair loss rapidly returns it to $|0\rangle$. This supplies an effective $1\to0$ rate $2\kappa_+$ even as the intermediate population vanishes. Balancing the two slow-state rates gives panel (b), including $(p_0,p_1)=(2/3,1/3)$ at zero single loss. The full generator retains higher states. The limiting phase-symmetric mixture has zero coherent mean. The boxed result illustrated here is~\eqref{eq:quantum-active-deep-quantum-state}.}
\label{fig:strong-pair-loss-general-mixture}
\end{figure}

For $\kappa_-=0$ the mean approaches $1/3$, while the factorized
$r_*^2=\kappa_+/(2\kappa_2)$ approaches zero. The familiar $2/3,1/3$
limit~\cite{LeeSadeghpour2013} therefore provides a sharp test of where
the classical amplitude ceases to describe quantum occupation.
This is a stationary rate-separation limit, not a prescription to make a
physical damping rate infinite. One may maintain
$\kappa_+,\kappa_-\ll\kappa_2\ll\omega_s$ within the reservoir's
Markov regime. Reducing the gain then lengthens the approach to stationarity;
taking zero gain first is a different preparation and dynamical limit.
Figure~\ref{fig:quantum-active-population-crossover} displays the
stationary crossover from the same generator. As another exact check,
$\kappa_2=\kappa_+$ and $\kappa_-=0$ give $p_n=e^{-1}/n!$.
These Poisson populations are phase diagonal, showing that Poisson
counts alone do not identify a coherent state.

The elimination is singular: $p_2\simeq(\kappa_+/\kappa_2)p_1$
vanishes, but its rapid loss retains a finite current.
For $\kappa_-=0$, $J_2\to4\hbar\omega_s\kappa_+/3=J_+$.
Projecting $\widehat b^2$ to zero in the $\{\ket0,\ket1\}$ subspace
before eliminating $\ket2$ would incorrectly remove this cycle and its
energy flow. Small occupation of a hidden state does not imply a
negligible dynamical role.

\begin{figure}[H]
\centering\begingroup
\definecolor{vltTeal}{HTML}{176B64}
\definecolor{vltBlue}{HTML}{315D80}
\definecolor{vltGold}{HTML}{A87520}
\definecolor{vltPlum}{HTML}{725877}
\definecolor{vltRed}{HTML}{762B38}
\definecolor{vltYellow}{HTML}{D5AD3C}
\definecolor{vltNeutral}{HTML}{777D83}
\setlength{\unitlength}{1pt}
\scalebox{0.80}{%
\input{vlt-figures/figure-268.tex}%
}\endgroup
\caption{Exact stationary statistics of the birth/pair-loss generator with $\kappa_-=0$. (a) Markers are integer occupations, with connecting guides. Only $N\le12$ is displayed; full-law probabilities are used without renormalization. The broad $r_\kappa=0.1$ law has omitted mass $0.0125723$ above that range. (b) Both axes are logarithmic. The exact mean tends to $1/3$ at strong pair loss, while the factorized large-occupation estimate $1/(2r_\kappa)$ tends to zero. Curves were evaluated from the normalized stationary generating function and checked against the number-state balance and exact occupation ledger. The gain resource remains explicit. The low-occupation discrepancy concerns the factorized estimate; it does not establish a nonzero phase-selected amplitude in the stationary state.}
\label{fig:quantum-active-population-crossover}
\end{figure}

\paragraph{Zero mean motion, phase-space rings, and the choice of readout.}
The unforced generator is covariant under
$\rho\mapsto e^{-i\theta\widehat N}\rho e^{i\theta\widehat N}$.
Its phase-symmetric stationary state is diagonal and has
$\langle\widehat b\rangle=0$, although its occupation and energy
throughput are nonzero. A phase-resolved preparation, a conditioned
measurement, or an added reference drive is a different question.
Temporal coherence can be tested through the ordered correlation
\begin{equation}
 C_b(t)=\langle\widehat b^\dagger(t)\widehat b(0)\rangle_{\rm ss}
     =\tr\!\left[\widehat b^\dagger
            e^{t\mathcal L_{\rm a}}(\widehat b\rho_{\rm ss})\right],
 \qquad t\geq0,\qquad C_b(0)=\langle\widehat N\rangle_{\rm ss}.
 \label{eq:quantum-active-two-time-correlation}
\end{equation}
The last equality uses quantum regression for the declared Markov model.
Its decay and Fourier spectrum contain information absent from a static
phase-space picture; a ring alone does not establish phase locking or
frequency entrainment~\cite{WalterNunnenkampBruder2014}.

The distinction is particularly clear in the zero-ordinary-loss deep
limit. In a coherent-state phase plane with dimensionless coordinate
$\alpha_c$, put $r_c=|\alpha_c|$. The same stationary density operator
has Wigner and Husimi representations. The Wigner representation
reproduces symmetrically ordered quadrature moments; the Husimi
representation is a coherent-state overlap distribution. Explicitly,
\begin{equation}
 \begin{aligned}
 W(\alpha_c)&=\frac{2}{3\pi}(1+4r_c^2)e^{-2r_c^2},\\
 Q_{\rm H}(\alpha_c)&=\frac{1}{3\pi}(2+r_c^2)e^{-r_c^2},\\
 \int W\,\dd^2\alpha_c&=\int Q_{\rm H}\,\dd^2\alpha_c=1.
 \end{aligned}
 \label{eq:quantum-active-wigner-husimi}
\end{equation}
The Wigner function peaks on $r_c=1/2$, while the Husimi function peaks
at the origin (Figure~\ref{fig:quantum-active-phase-space-representations}).
Here both are positive; a Wigner function is not generally a positive
probability density. Husimi $Q_{\rm H}=\bra{\alpha_c}\rho\ket{\alpha_c}/\pi$
uses coherent-state measurement effects and adds their resolution noise.
Their raw radius moments consequently differ:
$\int r_c^2W\,\dd^2\alpha_c=\langle N\rangle+1/2$, whereas
$\int r_c^2Q_{\rm H}\,\dd^2\alpha_c=\langle N\rangle+1$.
Neither plot is a distribution of virion positions, and the presence or
absence of a ring is not an invariant definition of self-excitation.

An even stronger readout test compares this state with a thermal Gibbs
state having the same mean $\langle N\rangle=1/3$. In canonical scaled
quadratures $q_c=(\widehat b+\widehat b^\dagger)/\sqrt2$ and
$p_c=(\widehat b-\widehat b^\dagger)/(i\sqrt2)$, both states have zero
first moments, variance $5/6$ in each quadrature, zero symmetrized
cross covariance, and determinant $\det\mathsf V=25/36$.
Nevertheless the active limiting state has
$\langle N(N-1)\rangle=0$ and
$g^{(2)}(0)=\langle N(N-1)\rangle/\langle N\rangle^2=0$;
the matched thermal state has factorial moment $2/9$ and $g^{(2)}(0)=2$.
Covariance alone therefore neither fixes occupation statistics nor
identifies a Gibbs preparation. An occupation-sensitive observable would
test a different part of the model from a displacement spectrum.

\begin{figure}[H]
\centering\begingroup
\definecolor{vltTeal}{HTML}{176B64}
\definecolor{vltBlue}{HTML}{315D80}
\definecolor{vltGold}{HTML}{A87520}
\definecolor{vltPlum}{HTML}{725877}
\definecolor{vltRed}{HTML}{762B38}
\definecolor{vltYellow}{HTML}{D5AD3C}
\definecolor{vltNeutral}{HTML}{777D83}
\setlength{\unitlength}{1pt}
\scalebox{0.90}{%
\input{vlt-figures/figure-269.tex}%
}\endgroup
\caption{Phase-space representations of one state, $\rho_*=(2|0\rangle\langle0|+|1\rangle\langle1|)/3$. (a,b) Shared coherent-amplitude axes and shading encode $\pi W$ and $\pi Q$. The Wigner maximum lies at $r=1/2$ (dashed ring); the Husimi maximum lies at zero (open centre). (c) Profiles are $\pi W=2(1+4r^2)e^{-2r^2}/3$ and $\pi Q=(2+r^2)e^{-r^2}/3$, each normalized with planar measure $2\pi r\dd r$ after restoring $1/\pi$. Their second moments differ by operator ordering. The ring belongs to this state representation; it does not specify a physical orbit.}
\label{fig:quantum-active-phase-space-representations}
\end{figure}

\paragraph{Closing the resource is an additional physical model.}
Constant pump rates assume a maintained resource or a window over which
its depletion is negligible. A finite closed construction must retain
that resource as a quantum degree of freedom. \begin{definition}[Retained resonant pump]
\label{def:s7-active-retained-pump}
For example, take $M$
resonant two-level pump units and the declared exchange Hamiltonian
\begin{equation}
 \begin{aligned}
 \widehat H_{\rm closed}
   &=\hbar\omega_s\left(\widehat N+\sum_{j=1}^M\widehat P_{e,j}\right)
      +\hbar\sum_{j=1}^M\nu_j
       (\widehat b^\dagger\widehat\sigma_{-,j}
                      +\widehat b\widehat\sigma_{+,j}),\\
 \left[\widehat H_{\rm closed},\widehat N+
                       \sum_j\widehat P_{e,j}\right]&=0.
 \end{aligned}
 \label{eq:quantum-active-finite-pump-closure}
\end{equation}
The real $\nu_j$ have frequency units; this is a specified resonant
exchange model. The commutator identifies the conserved total excitation
number of mode and pump, not a constant mechanical-mode occupation.
\end{definition}
Initial pump inversion carries the excitation energy,
and a finite initial excitation sector bounds the attainable mode
occupation. A cold receiver, if added, must also be included to make the
total system closed. Tracing out a rapidly decorrelating, maintained pump
can motivate a gain channel; a finite correlated pump does not produce
that constant-rate assumption automatically. Merely making $\kappa_+$
a function of a mean resource energy is another closure approximation,
rather than the joint quantum evolution.

Figure~\ref{fig:finite-quantum-pump-exchange} specializes this retained resource to a single resonant pump unit and displays reversible excitation exchange.

\begin{figure}[H]
\centering\begingroup
\definecolor{vltTeal}{HTML}{176B64}
\definecolor{vltBlue}{HTML}{315D80}
\definecolor{vltGold}{HTML}{A87520}
\definecolor{vltPlum}{HTML}{725877}
\definecolor{vltRed}{HTML}{762B38}
\definecolor{vltYellow}{HTML}{D5AD3C}
\definecolor{vltNeutral}{HTML}{777D83}
\setlength{\unitlength}{1pt}
\scalebox{0.80}{%
\input{vlt-figures/figure-270.tex}%
}\endgroup
\caption{Exact closed exchange with one resonant two-level pump unit, real $\nu\ne0$, no receivers and initial state $\ket{0,e}$. The conserved one-excitation sector contains the two joint basis states in (a); they share the same bare energy and are coupled by the carrier-free Hamiltonian $\hbar\nu\left(\begin{smallmatrix}0&1\\1&0\end{smallmatrix}\right)$. The arrows are reversible coupling, not transition rates. Panel (b) uses $s=|\nu|t$ and plots the exact occupations $\langle N\rangle=\sin^2s$, $\langle P_e\rangle=\cos^2s$. Their sum is one; complete transfer at $s=\pi/2$ is followed by recurrence. The carrier-free state below retains the relative phase. This exact invariant sector is not an artificial upper-level cutoff. It supplies neither a constant gain channel nor evidence of a pump mechanism in a virion.}
\label{fig:finite-quantum-pump-exchange}
\end{figure}

\paragraph{A finite pump sets a measurable throughput budget.}
Retain the joint Hamiltonian in
\eqref{eq:quantum-active-finite-pump-closure} and write
$\widehat C=\widehat N+\sum_j\widehat P_{e,j}$. Its interaction
energy also belongs in the total account. A receiver that resolves
the coupled system's transitions therefore uses the energy differences
of that full Hamiltonian, rather than assigning every event the bare
oscillator spacing by default.

\begin{proposition}[Finite-resource count and energy bounds]
\label{prop:s7-active-finite-throughput}
Let $\widehat H_{\rm closed}$ be fixed and retain a finite initial
sector $C\leq C_{\max}$. Add declared cold-receiver dissipators
$\sum_a\gamma_a\mathcal D[\widehat A_a]$, with $\gamma_a\geq0$,
whose resolved jumps satisfy
\begin{equation}
 [\widehat H_{\rm closed},\widehat A_a]
       =-\hbar\Omega_a\widehat A_a,\qquad
 [\widehat C,\widehat A_a]=-q_a\widehat A_a,
 \qquad \Omega_a>0,\quad q_a\in\{1,2\}.
 \label{eq:s7-active-resolved-receiver}
\end{equation}
There is no replenishing input. If $E_{\min}$ is a lower bound of
the retained Hamiltonian on this invariant sector, then
\begin{equation}
 \begin{aligned}
 I_C(t)&=\sum_a q_a\gamma_a
             \langle\widehat A_a^\dagger\widehat A_a\rangle_t,\\
 \int_0^t I_C(s)\,\dd s&\leq\langle\widehat C\rangle_0,\\
 P_{\rm rec}(t)&=\sum_a\hbar\Omega_a\gamma_a
             \langle\widehat A_a^\dagger\widehat A_a\rangle_t,\\
 \int_0^tP_{\rm rec}(s)\,\dd s
       &\leq\langle\widehat H_{\rm closed}\rangle_0-E_{\min}.
 \end{aligned}
 \label{eq:s7-active-finite-throughput-budget}
\end{equation}
Consequently an interval of duration $t_{\rm a}$ with outgoing mean
power at least $P_{\min}>0$ obeys
\begin{equation}
 t_{\rm a}\leq
 \frac{\langle\widehat H_{\rm closed}\rangle_0-E_{\min}}
                            {P_{\min}}.
 \label{eq:s7-active-finite-window}
\end{equation}
\end{proposition}
\begin{proof}
For each jump, the adjoint dissipator gives
$\mathcal D[\widehat A_a]^*\widehat C
=-q_a\widehat A_a^\dagger\widehat A_a$ and
$\mathcal D[\widehat A_a]^*\widehat H_{\rm closed}
=-\hbar\Omega_a\widehat A_a^\dagger\widehat A_a$.
The Hamiltonian commutes with $\widehat C$. Thus
$\partial_t\langle\widehat C\rangle=-I_C$ and
$\partial_t\langle\widehat H_{\rm closed}\rangle=-P_{\rm rec}$.
Integrate and use their lower bounds. The last inequality follows by
integrating the stipulated lower power over the interval.
\end{proof}

The count current has quanta-per-time units; $P_{\rm rec}$ has power
units. Dark residual states can keep stored resource, so these are
upper bounds rather than a claim of complete extraction. The resolved
receiver is a specified weak-bath, transition-resolving model. Its
coupled jump operators can be obtained by spectral decomposition of
the chosen system--receiver coupling. Matching those jumps to the
energy account matters in interacting systems
\cite{LevyKosloff2014}.

Local jumps $\widehat b$ and $\widehat b^2$ still give their exact
bare excitation-count budget, but they need not be eigenoperators of
$\widehat H_{\rm closed}$. Their total-energy current then also
changes interaction storage. The identification
$\hbar\Omega_a=q_a\hbar\omega_s$ belongs to a matched limit where
the exchange-induced energy shifts can be omitted at the required
accuracy. Thermal upward receiver transitions or external replenishment
add their input currents to the same ledger.

The pump-unit count $M$ is independent of $N_{\rm tot}$, the fixed
ideal-virion population. Selecting a larger cluster neither supplies new
pump inversion nor erases an existing receiver record. A physical
contact change that shifts the frequency also carries the corresponding
work and continued material state from Sections~4--6. For a proposed
capture or breakup mechanism, its actual source-to-contact coupling
must route energy through that event account. The finite-resource bound
then constrains how long a specified throughput is available; it does
not supply a universal binding threshold from phonon number alone.

Figure~\ref{fig:s7-active-dressed-receiver-budget} contrasts the count and full-Hamiltonian energy of a resolved loss event and evaluates the finite-throughput bound.

\begin{figure}[H]
\centering\begingroup
\definecolor{vltTeal}{HTML}{176B64}
\definecolor{vltBlue}{HTML}{315D80}
\definecolor{vltGold}{HTML}{A87520}
\definecolor{vltPlum}{HTML}{725877}
\definecolor{vltRed}{HTML}{762B38}
\definecolor{vltYellow}{HTML}{D5AD3C}
\definecolor{vltNeutral}{HTML}{777D83}
\setlength{\unitlength}{1pt}
\scalebox{0.90}{%
\input{vlt-figures/figure-271.tex}%
}\endgroup
\caption{Energy-resolved cold loss from one retained pump unit and mode, on the exact invariant sector $C\leq1$. Choose $\omega_st_\star=1$, $\nu t_\star=0.2$, $E_\star=\hbar/t_\star$ and $\ket\pm=(\ket{1,g}\pm\ket{0,e})/\sqrt2$. The full Hamiltonian has gaps $1.2E_\star$ and $0.8E_\star$ to $\ket{0,g}$. Each shown jump lowers excitation count by one, while its energy follows its own gap \eqref{eq:s7-active-resolved-receiver}. Take $\gamma_+t_\star=1/4$, $\gamma_-t_\star=1/10$ and initial dressed populations $3/4,1/4$. Then $p_+=3e^{-\theta/4}/4$, $p_-=e^{-\theta/10}/4$, $\langle C\rangle=p_++p_-$ and $E/E_\star=1.2p_++0.8p_-$. The integrated receiver energy is $E_0-E(t)$, with $E_0=1.1E_\star$ and $E_{\min}=0$ \eqref{eq:s7-active-finite-throughput-budget}. Shading marks the interval above the stated power threshold; its measured model duration is shorter than the resource upper bound \eqref{eq:s7-active-finite-window}. These are occupation/energy transitions, not virion trajectories.}
\label{fig:s7-active-dressed-receiver-budget}
\end{figure}

For biophysical use, these observations give a concrete hierarchy of
requirements: a resolved mode, a physical energy-conversion pathway,
measured gain and receiving processes, environmental occupation, and a
readout capable of distinguishing the proposed state. An ionic gradient
or biochemical reaction can supply classical work without establishing
the one- and two-quantum channels above. Conversely, an engineered
few-quantum realization would need to suppress or overcome its actual
thermal loading. The model specifies the additional variables, energy currents and
observables required to test an active quantum extension. An application
to a virion requires independent evidence for those inputs.

\subsection{Uncertainty relations with defined observables}
\label{sec:phonon-uncertainty-applications}

Thermal displacement, packet bandwidth, occupation fluctuations and
parameter-estimation error quantify different spreads. An uncertainty
relation answers a defined question about specified observables and
preparation. The canonical relations below constrain quantum outcome
distributions; detector calibration and inference error use their
own observation and statistical models.

\paragraph{Canonical covariance, including its orientation.}
\begin{definition}[Canonical quadrature covariance]
\label{def:s7-canonical-quadrature-covariance}
For one retained positive mode, define the dimensionless quadratures
$\widehat q_c=\sqrt{\omega_s/\hbar}\,\widehat Q_s$ and
$\widehat p_c=\widehat\Pi_s/\sqrt{\hbar\omega_s}$, so that
$[\widehat q_c,\widehat p_c]=i$. Their symmetrized covariance matrix
$\mathsf V$ has entries
$V_{ab}=\langle\{\delta\widehat r_a,\delta\widehat r_b\}\rangle/2$,
where $\widehat{\vct r}=(\widehat q_c,\widehat p_c)^{\mathsf T}$ and
$\delta\widehat r_a=\widehat r_a-\langle\widehat r_a\rangle$.
\end{definition}
\begin{proposition}[Canonical covariance area]
\label{prop:s7-canonical-covariance-area}
On the common domains and finite moments specified in Section~6,
the covariance form of Cauchy--Schwarz gives
\begin{equation}
 \det\mathsf V=V_{qq}V_{pp}-V_{qp}^2\ge\frac14,
 \qquad
 (\Delta Q_s)^2(\Delta\Pi_s)^2
       -\operatorname{Cov}(Q_s,\Pi_s)^2\ge\frac{\hbar^2}{4}.
 \label{eq:phonon-canonical-covariance-area}
\end{equation}
\end{proposition}
\begin{proof}
Apply Cauchy--Schwarz to the two centered observable vectors in a pure
state, or to their Hilbert--Schmidt counterparts for a density operator.
The imaginary part of their ordered covariance is fixed by the
commutator to $\hbar/2$. Squaring the modulus and separating its
real symmetrized part gives the physical-unit inequality; the stated
quadrature scaling gives the dimensionless form.
\end{proof}
Retaining displacement--momentum covariance gives the sharper bound
\cite{Robertson1934,NeumaierWestra2011}. Dropping its nonnegative
square gives $\Delta Q_s\Delta\Pi_s\ge\hbar/2$.
Separate repeated preparations can estimate the two distributions;
a joint-measurement protocol additionally specifies its back-action.

In a number state $\ket{N_s}$,
\begin{equation}
 (\Delta Q_s)^2=\frac{\hbar}{\omega_s}\left(N_s+\frac12\right),
 \qquad
 (\Delta\Pi_s)^2=\hbar\omega_s\left(N_s+\frac12\right),
 \qquad
 \Delta Q_s\Delta\Pi_s=\hbar\left(N_s+\frac12\right).
 \label{eq:phonon-number-state-uncertainty}
\end{equation}
A one-phonon state consequently has product $3\hbar/2$. A coherent
state can have a large mean displacement and energy while retaining
the vacuum covariance and product $\hbar/2$. A squeezed vacuum aligned
with the coordinate axes instead gives
\begin{equation}
 (\Delta Q_s)^2=\frac{\hbar}{2\omega_s}e^{-2r_{\rm sq}},\qquad
 (\Delta\Pi_s)^2=\frac{\hbar\omega_s}{2}e^{2r_{\rm sq}},\qquad
 \langle\widehat N_s\rangle=\sinh^2r_{\rm sq}.
 \label{eq:phonon-squeezing-uncertainty-cost}
\end{equation}
Squeezing reduces one variance at the cost of the conjugate variance
and preparation energy. It does not reduce the covariance determinant
below its vacuum value. With
$\mathsf R(\vartheta)=\left(
\begin{smallmatrix}\cos\vartheta&-\sin\vartheta\\
\sin\vartheta&\cos\vartheta\end{smallmatrix}\right)$,
the rotated covariance is
\begin{equation}
 \begin{aligned}
 \mathsf V(\vartheta)&=\frac12\mathsf R(\vartheta)
 \begin{pmatrix}e^{-2r_{\rm sq}}&0\\0&e^{2r_{\rm sq}}\end{pmatrix}
 \mathsf R(\vartheta)^{\mathsf T},\\
 V_{qq}V_{pp}&=\frac14+V_{qp}^2.
 \end{aligned}
 \label{eq:phonon-rotated-squeezing-covariance}
\end{equation}
Free evolution rotates the ellipse with the chosen sign convention.
Its coordinate product can increase at fixed
$\sqrt{\det\mathsf V}$
(Figure~\ref{fig:uncertainty-covariance-orientation}). The mode shapes
and frequency weights then map canonical covariance to a marker's
displacement variance. Covariance describes second moments; the pumped
and thermal preparations in Section~\ref{sec:phonon-quantum-active}
show how equal second moments can accompany different number statistics.

\begin{figure}[H]
\centering\begingroup
\definecolor{vlfGreen}{HTML}{176B64}
\definecolor{vlfDarkRed}{HTML}{762B38}
\definecolor{vlfRed}{HTML}{762B38}
\definecolor{vlfBlue}{HTML}{315D80}
\definecolor{vlfCopper}{HTML}{762B38}
\setlength{\unitlength}{1pt}
\scalebox{0.80}{%
\input{vlt-figures/figure-272.tex}%
}\endgroup
\caption{Rotation changes the standard-deviation product of a squeezed state while preserving its full covariance bound. With $q_c=\sqrt{\omega/\hbar}Q$, $p_c=\Pi/\sqrt{\hbar\omega}$ and $r_{\rm sq}=\ln2$, use $V=R(\vartheta)\operatorname{diag}(1/8,2)R(\vartheta)^{\mathsf T}$. (a) Equal-axis contours $\vct x^{\mathsf T}V^{-1}\vct x=1$ at $0,\pi/4$ retain area $\pi/2$. (b) $\Delta q_c\Delta p_c=\tfrac12\sqrt{1+\sinh^2(2r_{\rm sq})\sin^2(2\vartheta)}$, while $\sqrt{\det V}=1/2$. The difference is the coordinate--momentum covariance: $V_{qq}V_{pp}-V_{qp}^2=1/4$. Contours encode second moments; they assign no orbit or confidence probability. Physical virion variances additionally use the mode/probe overlap and physical units.}
\label{fig:uncertainty-covariance-orientation}
\end{figure}

\paragraph{Spatial localization and temporal evolution are separate bounds.}
For a whole-line normalized envelope $A\in H^1(\R)$ with
$zA\in L^2(\R)$, the unitary Fourier convention gives
\begin{equation}
 \Delta z\,\Delta k\ge\frac12.
 \label{eq:phonon-envelope-fourier-bound}
\end{equation}
This is a localization--bandwidth constraint for classical waves as
well as quantum amplitudes. The Gaussian in
\eqref{eq:phonon-gaussian-complex-envelope} starts at equality and
develops position--wave-number covariance as it spreads. A cavity's
discrete standing-wave frequencies are not automatically eigenvalues
of a self-adjoint translation generator canonically conjugate to its
bounded position coordinate. Boundary-compatible domains remain
essential. For a homogeneous whole-line branch where the commutator
is defined, $[\widehat z,\hbar\omega(\widehat k)]
=i\hbar\omega'(\widehat k)$ instead implies
$\Delta z\Delta E\ge(\hbar/2)|\langle v_g\rangle|$.
The right-hand side can vanish; localization of a stationary cavity
mode does not require a nonzero energy spread.

For a time-independent closed Hamiltonian and an observable without
explicit time dependence, the Heisenberg derivative and Robertson
bound give
\begin{equation}
 \Delta H\,\tau_A\ge\frac{\hbar}{2},\qquad
 \tau_A=\frac{\Delta A}{|\dd\langle A\rangle/\dd t|},
 \label{eq:phonon-mandelstam-tamm-time}
\end{equation}
when the derivative exists and is nonzero and both variances are finite.
It bounds the timescale on which the chosen observable changes by
its own standard deviation. This observable-dependent time is neither
a decay constant nor a detector exposure. For a pure state that actually
reaches an orthogonal state, the integrated version is
\begin{equation}
 t_\perp\ge\frac{\pi\hbar}{2\Delta H}
       =\frac{\pi}{2\Delta\omega}
          \quad\text{in the one-phonon sector}.
 \label{eq:phonon-orthogonalization-bound}
\end{equation}
Here $\Delta\omega$ is the spectral spread of the full Hamiltonian
generating the evolution~\cite{DeffnerCampbell2017}. Resonant
two-channel transfer saturates this result: the initial channel has
$\Delta H=\hbar|J|$ and reaches the other at $\pi/(2|J|)$.
Detuning leaves this initial variance unchanged but gives minimum
survival $\delta^2/(\delta^2+4J^2)>0$; the bound cannot guarantee
a transition that the dynamics forbids. An energy eigenstate has
$\Delta H=0$ and does not orthogonalize under that Hamiltonian.

Figure~\ref{fig:orthogonalization-bound-and-detuning} compares the conditional speed bound with the survival probabilities of resonant and detuned conversion.

\begin{figure}[H]
\centering\begingroup
\definecolor{vlfGreen}{HTML}{176B64}
\definecolor{vlfDarkRed}{HTML}{762B38}
\definecolor{vlfRed}{HTML}{762B38}
\definecolor{vlfBlue}{HTML}{315D80}
\definecolor{vlfCopper}{HTML}{762B38}
\setlength{\unitlength}{1pt}
\scalebox{0.80}{%
\input{vlt-figures/figure-273.tex}%
}\endgroup
\caption{A closed-system speed bound does not guarantee complete conversion. For the explicitly specified two-channel, one-excitation Hamiltonian, subtract its scalar carrier and write $H/\hbar=\left(\begin{smallmatrix}\delta/2&J\\J&-\delta/2\end{smallmatrix}\right)$ with $J>0$ and initial first channel. Its survival probability is $P_{\rm surv}=1-[4J^2/(\delta^2+4J^2)]\sin^2[\sqrt{\delta^2+4J^2}\,t/2]$. Both curves have the same initial $\Delta H=\hbar J$, hence the same Mandelstam--Tamm orthogonalization lower bound $Jt_\perp\ge\pi/2$. On resonance ($\delta=0$), the state reaches zero survival at the bound. At $\delta=2J$ the minimum is $1/2$: the state never reaches an orthogonal state, despite its nonzero energy variance. All axes are dimensionless. The vertical line is a lower bound conditional on actual orthogonalization, not a predicted transfer time for the detuned model or a dissipative lifetime.}
\label{fig:orthogonalization-bound-and-detuning}
\end{figure}

The pumped master equation of Section~\ref{sec:phonon-quantum-active}
has an additional dissipative generator. Its observable derivative is
$\dd\langle A\rangle/\dd t=\langle\mathcal L^\dagger A\rangle$,
not solely a Hamiltonian commutator, so the closed-system bound above
cannot simply be applied using the oscillator's bare $\Delta H$.
Open-system speed bounds need their generator-dependent hypotheses
\cite{DeffnerCampbell2017}. None of these relations equates lifetime,
linewidth, thermal fluctuation amplitude, or parameter-estimation error.
Likewise the full dynamical matrix commutes with its square root: it
supplies no universal positive stiffness--frequency uncertainty product.

\subsection{Temperature, phonon occupation, and biological material state}
\label{sec:phonon-temperature-comparison}

Temperature affects oscillator preparation, mechanical material properties
and hidden-variable kinetics. Comparing temperatures therefore specifies
geometry, preparation and observation band at each value. A near-zero
oscillator reference isolates quantum statistics; applying it to a
hydrated biological material also requires its actual low-temperature
structural and constitutive state.

\begin{definition}[Fixed-reference thermal comparison]
\label{def:s7-fixed-thermal-reference}
A fixed-reference comparison varies the Gibbs temperature while keeping
the physical coordinate set, inertia, restoring operator, reference
geometry and calibrated readouts fixed. Changes in contact state,
hydration or constitutive parameters instead define a registered family
of mechanical references, each with its own conditional preparation.
\end{definition}

\paragraph{First hold the mechanical reference fixed.}
\begin{proposition}[Occupation and displacement at a fixed positive frequency]
\label{prop:s7-fixed-thermal-occupation}
For a positive mode in Gibbs equilibrium,
$p_N=(1-e^{-\beta\hbar\omega_s})e^{-N\beta\hbar\omega_s}$.
Summing this geometric distribution gives
$\bar N_s=(e^{\beta\hbar\omega_s}-1)^{-1}$ and
$\langle Q_s^2\rangle=\hbar(2\bar N_s+1)/(2\omega_s)$.
The characteristic temperature $T_s^\star=\hbar\omega_s/k_{\rm B}$
compares a single quantum with thermal energy. Its limits are
\begin{equation}
 \begin{aligned}
 T\longrightarrow0:&\quad\bar N_s\longrightarrow0,
  &\langle Q_s^2\rangle&\longrightarrow\frac{\hbar}{2\omega_s},\\
 k_{\rm B}T\gg\hbar\omega_s:&\quad
  \bar N_s\simeq\frac{k_{\rm B}T}{\hbar\omega_s},
  &\langle Q_s^2\rangle&\simeq\frac{k_{\rm B}T}{\omega_s^2}.
 \end{aligned}
 \label{eq:phonon-temperature-limits}
\end{equation}
\end{proposition}
\begin{proof}
Normalize the geometric occupation distribution and differentiate its
partition sum to obtain the mean number. The diagonal oscillator
matrix element is $\langle N|\widehat Q_s^2|N\rangle=
\hbar(N+1/2)/\omega_s$. Its Gibbs average gives the variance,
and the two limits follow from the exponential expansion.
\end{proof}
Low occupation leaves zero-point fluctuations. At high temperature the
kinetic and potential terms each approach $k_{\rm B}T/2$, recovering
equipartition~\cite{FeynmanLectures1963}. The harmonic spacings do not
shift just because occupation changes: the Hamiltonian's curvature
has been held fixed. On the dimensionless scale
$\theta_T=T/T_s^\star$,
\begin{equation}
 \bar N_s=\frac1{e^{1/\theta_T}-1},\qquad
 \frac{\langle Q_s^2\rangle}{\hbar/(2\omega_s)}
       =\coth\!\left(\frac1{2\theta_T}\right),\qquad\theta_T>0.
 \label{eq:phonon-dimensionless-thermal-crossover}
\end{equation}
The corresponding classical normalized variance is $2\theta_T$.
Figure~\ref{fig:phonon-thermal-crossover} separates occupation from
the nonzero vacuum displacement scale.

\begin{figure}[H]
\centering\begingroup
\definecolor{vlfGreen}{HTML}{176B64}
\definecolor{vlfDarkRed}{HTML}{762B38}
\definecolor{vlfRed}{HTML}{762B38}
\definecolor{vlfBlue}{HTML}{315D80}
\definecolor{vlfCopper}{HTML}{762B38}
\setlength{\unitlength}{1pt}
\scalebox{0.80}{%
\input{vlt-figures/figure-274.tex}%
}\endgroup
\caption{Occupation and variance at fixed positive mode frequency, with $T^*=\hbar\omega/k_B$ and $\theta_T=T/T^*$. (a) The Gibbs occupation is $\bar N=(e^{1/\theta_T}-1)^{-1}$, compared with the classical line $\theta_T$. (b) The vacuum-normalized coordinate variance is $2\omega\langle Q^2\rangle/\hbar=\coth[1/(2\theta_T)]$, compared with $2\theta_T$. The zero-temperature points are limits: occupation vanishes while vacuum variance stays finite. Geometry, inertia and frequency are held fixed. The shared temperature axis and line styles therefore compare statistical changes under one quadratic reference; actual material changes require their own constitutive data. The panels have different ordinate normalizations.}
\label{fig:phonon-thermal-crossover}
\end{figure}

\begin{table}[H]
\centering\small
\caption{Mean Gibbs occupation at prescribed cyclic frequency
$f=\omega/(2\pi)$, with $h=2\pi\hbar$. Frequencies are illustrative,
not assigned to viral species. The elevated temperature is a comparison
point, not a clinical threshold.}
\label{tab:phonon-temperature-comparison}
\begin{tabular}{@{}lrrr@{}}
\toprule
Preparation & $T$ (K) & $f=1\,\mathrm{GHz}$ & $f=1\,\mathrm{THz}$\\
\midrule
Formal zero-temperature limit & $0$ & $0$ & $0$\\
Near-zero reference & $0.01$ & $0.00830$ & $\simeq0$\\
Room-temperature example ($25\,{}^\circ\mathrm C$) & $298.15$ & $6212$ & $5.73$\\
Body-temperature example ($37\,{}^\circ\mathrm C$) & $310.15$ & $6462$ & $5.98$\\
Elevated-temperature example ($40\,{}^\circ\mathrm C$) & $313.15$ & $6524$ & $6.04$\\
\bottomrule
\end{tabular}
\end{table}

Here $T_s^\star\simeq0.0480\,\mathrm K$ at $1\,\mathrm{GHz}$ and
$48.0\,\mathrm K$ at $1\,\mathrm{THz}$. At fixed classical curvature,
warming from $298.15$ to $310.15\,\mathrm K$ raises rms displacement
by $\sqrt{310.15/298.15}\simeq1.020$; warming from $310.15$ to
$313.15\,\mathrm K$ gives only about $1.005$.
These modest statistical factors do not bound biological changes caused
by altered interactions or structural state. A low-frequency collective
mode is highly occupied at these temperatures even if its microscopic
bonds require quantum mechanics for their description. The existence of
those bonds does not establish a low-occupation collective preparation.

\begin{example}[The thermal scale of the HPV-inspired guide]
\label{ex:s7hpv-thermal-scale-audit}
Using the nominal mass above, the guide's conservative relative coordinate
has $k_-=9\times10^{-4}\,\mathrm{N\,m^{-1}}$. Its oscillator reference
and classical equilibrium width give
\begin{equation}
 \begin{aligned}
 \omega_-^{(0)}&=\sqrt{k_-/m_v}\simeq1.64\times10^8\,\mathrm{s^{-1}},\\
 \frac{\hbar\omega_-^{(0)}}{k_{\rm B}T}
   &\simeq4.17\times10^{-6},\qquad
 \sqrt{\frac{k_{\rm B}T}{k_-}}\simeq2.15\,\mathrm{nm}
       \quad(T=300\,\mathrm K).
 \end{aligned}
 \label{eq:s7hpv-thermal-reference-scales}
\end{equation}
For that isolated oscillator Gibbs reference, the postulate would give
$\bar N\simeq2.40\times10^5$ and a zero-point length
$\sqrt{\hbar/(2m_v\omega_-^{(0)})}\simeq3.10\,\mathrm{pm}$.
The guide instead predicts measured-band motion through its damped
material response. Its stipulated constant-drag comparison has
$m_v\omega_-^{(0)}/c_0\simeq3.67\times10^{-5}$; the conservative
frequency therefore does not describe an observed oscillatory pole of
that comparison. The material rate $\tau^{-1}=2000\,\mathrm{s^{-1}}$
is a relaxation rate, not an oscillator quantum frequency.
\end{example}
These numbers locate the running example in a classical, strongly
relaxational regime over its declared observation band. An oscillatory
packet or low-occupation application uses a separately justified
preparation and full response. In particular, the guide's low-frequency
resistance is not a calibrated fluid law at $\omega_-^{(0)}$.
The conservative scale quantifies the thermal comparison; the causal
material law, preparation and sensitive readout determine whether a
quantum prediction could be distinguished experimentally.

\paragraph{An equilibrium spectral signature uses an ordered readout.}
For a fixed, undamped oscillator define
$S_{QQ}^{>}(\Omega)=\int_{\R}e^{i\Omega t}
\langle\widehat Q(t)\widehat Q(0)\rangle\,\dd t$.
Using $\widehat Q=\sqrt{\hbar/(2\omega_s)}
(\widehat b_s+\widehat b_s^\dagger)$ gives
\begin{equation}
 \begin{aligned}
 S_{QQ}^{>}(\Omega)&=\frac{\pi\hbar}{\omega_s}
  \bigl[(\bar N_s+1)\delta(\Omega-\omega_s)
                  +\bar N_s\delta(\Omega+\omega_s)\bigr],\\
 \frac{\bar N_s}{\bar N_s+1}&=e^{-\beta\hbar\omega_s}.
 \end{aligned}
 \label{eq:phonon-thermal-ordered-sidebands}
\end{equation}
The ratio compares integrated negative- and positive-frequency line
weights. A detector sensitive to the stated operator ordering can
distinguish them; a symmetrized spectrum averages the two
\cite{Clerk2010}. Its coupling, noise and back-action must be included
in the comparison. The delta lines describe the undamped reference;
a bath supplies any finite linewidth. A pumped mode requires its own
stationary distribution and cannot be assigned a thermodynamic
temperature from mean occupation alone.

Figure~\ref{fig:ordered-thermal-sideband-weights} displays the normalized line weights and their detailed-balance ratio without assigning an artificial linewidth.

\begin{figure}[H]
\centering\begingroup
\definecolor{vlfGreen}{HTML}{176B64}
\definecolor{vlfDarkRed}{HTML}{762B38}
\definecolor{vlfRed}{HTML}{762B38}
\definecolor{vlfBlue}{HTML}{315D80}
\definecolor{vlfCopper}{HTML}{762B38}
\setlength{\unitlength}{1pt}
\scalebox{0.80}{%
\input{vlt-figures/figure-275.tex}%
}\endgroup
\caption{Ordered Gibbs sideband weights for one fixed undamped oscillator. With $\theta_T=k_BT/(\hbar\omega)$, (a) the positive and negative line fractions are $w_+=1/(1+e^{-1/\theta_T})$ and $w_-=1/(1+e^{1/\theta_T})$, with sum one. (b) Their negative-to-positive ratio is $e^{-1/\theta_T}$. Zero-temperature points are limits; the horizontal guides show the high-temperature limits, reached asymptotically. These are integrated line masses in $S_{QQ}^{>}$ using the stated ordering. A readout must supply that ordering, its transduction and back-action account before these ratios can be interpreted experimentally. A classical displacement trace alone does not establish them.}
\label{fig:ordered-thermal-sideband-weights}
\end{figure}

\paragraph{Now let configurations and material properties respond.}
Even at equilibrium, the effective stiffness need not stay constant.
The two-state construction in
\eqref{eq:phonon-two-state-configuration-softening} gives, for a fixed
positive gap $\epsilon_c$ and force slope $a_{\rm cfg}$,
\begin{equation}
 \frac{\epsilon_c\Delta k_{\rm cfg}}{a_{\rm cfg}^2}
  =\frac{1}{4t_T\cosh^2[1/(2t_T)]},\qquad
 t_T=\frac{k_{\rm B}T}{\epsilon_c}>0.
 \label{eq:phonon-temperature-configuration-curvature}
\end{equation}
At low temperature the higher configuration is rarely occupied, so
there is little population response. At high temperature the explicit
$1/T$ force susceptibility becomes small. Between these limits the
softening has a maximum, unlike the monotonic Gibbs occupation of a
fixed mode (Figure~\ref{fig:temperature-population-curvature}). The
complete reference must be force balanced and stable at each temperature.
This relaxed curvature is observed only when populations can follow the
probe; faster loading measures a more frozen response. Changing rates
therefore moves the kinetic crossover as well as changing the static
population susceptibility.

\begin{figure}[H]
\centering\begingroup
\definecolor{vlfGreen}{HTML}{176B64}
\definecolor{vlfDarkRed}{HTML}{762B38}
\definecolor{vlfRed}{HTML}{762B38}
\definecolor{vlfBlue}{HTML}{315D80}
\definecolor{vlfCopper}{HTML}{762B38}
\setlength{\unitlength}{1pt}
\scalebox{0.80}{%
\input{vlt-figures/figure-276.tex}%
}\endgroup
\caption{Population-mediated softening with fixed conditional surfaces. Let $\theta_c=k_BT/\epsilon_c$ and $p=(1+e^{1/\theta_c})^{-1}$ for positive gap $\epsilon_c$. (a) The normalized softening is $\epsilon_c\Delta k_{\rm cfg}/a_{\rm cfg}^2=p(1-p)/\theta_c$. Low-temperature state selection and the high-temperature factor $\beta$ make it vanish at both limits. (b) For $a_{\rm cfg}^2/(k_x\epsilon_c)=1$, $K_{\rm rel}/k_x=1-p(1-p)/\theta_c$ stays positive despite its nonmonotone variation. Curvatures refer to a complete force-balanced reference at each temperature and equilibrated populations during slow loading. Faster loading approaches the conditional curvature $k_x$. The logarithmic axes exclude zero; changes of hydration, medium or structure are independent constitutive inputs.}
\label{fig:temperature-population-curvature}
\end{figure}

A suspension experiment also constrains hydration, medium resistance
and memory, $\mathsf K(T)$, and structural state. Contacts, envelope
organization and conformation can shift frequencies and detector
overlaps; bath kinetics sets relaxation. Comparing independently
constrained mechanical and detector parameters with the fluctuation
record separates these effects. A pumped preparation additionally
retains its resource and thermal-bath rates.

\subsubsection{A preparation-specific quantum feasibility assessment}
\label{sec:qfa-feasibility}

The quantum extension becomes an application when one physical mode,
its environment, an excitation protocol and a detector are assigned
together. This assessment uses the 48-hour HPV16 L1 preparation of
Section~\ref{sec:hpva-application} as its evidence case. Its measured
size, buffer and scattering inputs constrain the classical comparison
hierarchy. They have not yet supplied a collective oscillation frequency,
an energy-relaxation rate, a single-quantum preparation or a calibrated
quantum transduction channel~\cite{Vega2016}. Those entries remain
unassigned. The nominal $55\,\mathrm{nm}$ guide and its stipulated
stiffness and resistance remain a separate synthetic calculation.

\begin{definition}[Quantum mode specification]
\label{def:qfa-mode-specification}
A feasibility specification identifies the material coordinate and
mode shape, its readout overlap and effective inertia, the retarded
response over the proposed carrier band, bath temperature and noise,
preparation coupling and timing, and the calibrated detection statistic.
It records which entries are measured, independently bounded, fitted
or assigned. Frequencies in hertz are $f_s=\omega_s/(2\pi)$;
linewidth and relaxation conventions are stated separately.
\end{definition}

For whole-VLP translation the mode moves complete nodes and the
inertia follows the translational projection. A capsid breathing mode
instead resolves internal shell deformation and its own effective
mass and interaction with the liquid or support. The same glyph and
particle diameter do not make those two coordinates interchangeable.
The physical modal and readout projections in Sections~3 and~6
determine which coordinate a given instrument sees.

\paragraph{First establish an oscillatory response.}
For the scalar constant-drag comparison
$m\ddot y+c\dot y+ky=F$, put
$\omega_0=\sqrt{k/m}$ and $\gamma_1=c/m$. Its Laplace poles are
\begin{equation}
 s_\pm=-\frac{\gamma_1}{2}
        \pm\sqrt{\frac{\gamma_1^2}{4}-\omega_0^2}.
 \label{eq:qfa-scalar-poles}
\end{equation}
An oscillatory pair requires $\gamma_1<2\omega_0$; its oscillation
frequency is $\sqrt{\omega_0^2-\gamma_1^2/4}$. A nonzero-frequency
maximum of the displacement compliance magnitude requires the stronger
condition $\gamma_1<\sqrt2\,\omega_0$, and occurs at
$\sqrt{\omega_0^2-\gamma_1^2/2}$. These statements follow directly
from the quadratic roots and minimization of
$(k-m\Omega^2)^2+c^2\Omega^2$. Resolving a weakly damped phonon
requires a quality ratio $\omega_0/\gamma_1$ large compared with one,
separation from neighboring poles, and appreciable detector residue.
For memory or coupled modes, the full response pencil and channel
residues of Section~4 replace this scalar test. Additional phase
dephasing can broaden a line beyond the energy-relaxation width.

The guide makes the distinction calculable. Its conservative frequency
is about $26.1\,\mathrm{MHz}$, with the very small formal quality ratio
already given in Example~\ref{ex:s7hpv-thermal-scale-audit}. Retaining
its passive material branch in the overdamped observation-band model
gives
\begin{equation}
 \begin{aligned}
 D_{\rm od}(s)&=c_0s+k_-+
                    \kappa_b\frac{s\tau}{1+s\tau},\\
 c_0\tau s^2+\bigl[c_0+(k_-+\kappa_b)\tau\bigr]s+k_-&=0,
 \qquad\kappa_b=2\Delta k.
 \end{aligned}
 \label{eq:qfa-guide-material-poles}
\end{equation}
With its assigned parameters the two poles are approximately
$-7.89\times10^2$ and $-1.52\times10^4\,\mathrm{s^{-1}}$.
They describe relaxation rather than cycles. This calculation uses the
declared overdamped band; it does not extrapolate a constant liquid
resistance to the conservative megahertz reference. In this application,
phase shifts and fluctuation spectra test retained material states
without requiring a resolved quantum carrier.

\paragraph{Then compare thermal loading and the measurement timescale.}
For an independently established weakly damped mode, the bath occupation
is evaluated at that mode's frequency. For a target $n_\star>0$,
\begin{equation}
 n_B<n_\star
 \quad\Longleftrightarrow\quad
 f_s>\frac{k_{\rm B}T}{h}\ln(1+n_\star^{-1}).
 \label{eq:qfa-thermal-frequency-target}
\end{equation}
At $298\,\mathrm K$, equilibrium occupations below one and below $0.1$
require frequencies above approximately $4.30$ and $14.9\,\mathrm{THz}$,
respectively. These are thermal thresholds, not predicted HPV modes.
Cooling and conditional preparation can access different regimes, with
their actual reheating rates determining the usable window.

Ultrafast pump--probe measurements on individual lentiviral pseudoviruses
have resolved gigahertz vibrations and dephasing, demonstrating one
route to determining actual virus-mode dynamics~\cite{Zhang2025}.
Their particle structure, attachment and environment differ from the
selected aqueous HPV L1 preparation. Even an assigned
$21.8\,\mathrm{GHz}$ oscillator at $298\,\mathrm K$ would have
$n_B\simeq284$; observing a vibration at that frequency would establish
oscillatory mechanics rather than single-phonon occupation statistics.
At the guide's much lower conservative frequency, the formal relative
quantum correction to the symmetrized Gibbs displacement variance is
\begin{equation}
 \frac{x}{2}\coth\!\left(\frac{x}{2}\right)-1
     =\frac{x^2}{12}+O(x^4),\qquad
 x=\frac{\hbar\omega_0}{k_{\rm B}T}.
 \label{eq:qfa-symmetrized-quantum-correction}
\end{equation}
It is about $1.45\times10^{-12}$ for the synthetic $300\,\mathrm K$
reference. That estimate concerns its formal oscillator Gibbs law;
the actual relaxational response uses the bath model and spectral
integration of Section~4. A successful classical response--fluctuation
comparison is fully compatible with this scale separation.

\subsubsection{Preparation and readout as a conditional design}
\label{sec:qfa-preparation}

A force pulse provides a concrete coherent preparation through
\eqref{eq:phonon-driven-excitation-energy}. To test a few-quantum
counting law, one can instead examine a heralded preparation and
mode-conversion readout. This is a candidate engineered protocol for
the framework, with experimental precedents in optomechanics
\cite{Hong2017,AspelmeyerKippenbergMarquardt2014}.
For a selected optical channel $a$ and mechanical mode $b$, ideal
resonant write and read interactions have the forms
\begin{equation}
 \begin{aligned}
 H_{\rm w}&=\hbar G_{\rm w}
       (\widehat a^\dagger\widehat b^\dagger+\widehat a\widehat b),\\
 H_{\rm r}&=\hbar G_{\rm r}
       (\widehat a^\dagger\widehat b+\widehat a\widehat b^\dagger).
 \end{aligned}
 \label{eq:qfa-write-read-couplings}
\end{equation}
Starting near the two-mode vacuum, a short write pulse gives
$\ket{0,0}-i\epsilon\ket{1,1}+O(\epsilon^2)$ with
$\epsilon=G_{\rm w}t_{\rm w}\ll1$. Conditioning on the emitted
photon preferentially prepares one phonon; thermal initial occupation,
multiple pairs and false heralds alter the conditional state.
The lossless read interaction transfers an occupation with efficiency
$\sin^2(G_{\rm r}t_{\rm r})$. Its actual noisy conversion uses
Section~\ref{sec:qfa-counting} and must finish within the preparation's
statistical survival window.

An optomechanical implementation additionally determines
$g_0=|\partial\omega_c/\partial Y|Y_{\rm zpf}$ from the physical
readout coordinate $Y$ and its zero-point length. Optical linewidth,
pump photons, collection and absorption then constrain the achievable
coupling and heating~\cite{AspelmeyerKippenbergMarquardt2014}.
A whole-particle effective mass cannot be substituted for an internal
mode's inertia. An immersed VLP also requires a check that the optical
field preserves its structure and preparation. The Hamiltonians above
identify the required conversion pathways; their coefficients have not
been measured for the selected preparation.

For two independent weak Markov reservoirs, environmental relaxation
$\gamma_1$ toward $n_B$ and engineered cooling $\Gamma_c$ toward $n_c$
give the simple occupation balance
\begin{equation}
 \bar n_{\rm ss}=\frac{\gamma_1 n_B+\Gamma_c n_c}
                            {\gamma_1+\Gamma_c},\qquad
 \Gamma_c>\gamma_1\frac{n_B-n_\star}{n_\star-n_c}
 \quad(n_c<n_\star<n_B).
 \label{eq:qfa-cooling-budget}
\end{equation}
The inequality is the condition $\bar n_{\rm ss}<n_\star$ for this
model. Pump-induced heating changes the inputs and must be included;
an arbitrary gain process does not provide a cold reservoir.
After release, the thermal antibunching window in
\eqref{eq:qfa-antibunching-window} supplies the time budget. If cooling
remains on, the two raising and lowering rates combine and the
corresponding effective bath enters that calculation.

Room temperature by itself is not a universal obstruction. A heralded
single quantum of a $40\,\mathrm{THz}$ Raman-active diamond vibration
has been prepared and read out under ambient conditions
\cite{TarragoVelez2019}. This is an internal high-frequency solid-state
mode with an ultrafast protocol, rather than a collective virion
translation. It illustrates the joint role of energy spacing, lifetime
and detector speed in selecting a viable realization.

\begin{table}[H]
\centering\small
\begin{tabular}{@{}p{3.0cm}p{6.0cm}p{6.35cm}@{}}
\toprule
{\raggedright Channel\par} & {\raggedright Calculation or measurement\par} & {\raggedright Status for selected HPV preparation\par}\\
\midrule
{\raggedright Physical carrier\par} & {\raggedright Mode shape, complex pole, energy relaxation and readout residue in the actual environment.\par} & {\raggedright Unassigned; DLS/SLS and mobility constrain transport and interactions, not a resolved quantum carrier.\par}\\[4pt]
{\raggedright Thermal preparation\par} & {\raggedright $n_B$, heating and cooling rates; harmonic excursion and structural stability.\par} & {\raggedright Temperature is reported; a mode-specific frequency and bath rate remain required.\par}\\[4pt]
{\raggedright Pump--probe response\par} & {\raggedright Frequency, oscillation envelope and phase under calibrated loading.\par} & {\raggedright A possible characterization route; demonstrated virus vibrations in~\cite{Zhang2025} concern another preparation.\par}\\[4pt]
{\raggedright Ordered sidebands\par} & {\raggedright Gain-corrected ratio $n_B/(n_B+1)$ with quantum detector ordering and back-action.\par} & {\raggedright Requires a specified transducer, probe-noise controls and sufficient contrast.\par}\\[4pt]
{\raggedright Heralded counting\par} & {\raggedright Converted $g^{(2)}$, delay dependence and an upper uncertainty bound below one.\par} & {\raggedright Engineered-device precedents exist; VLP coupling, efficiency, noise and damage limits remain to be established.\par}\\
\bottomrule
\end{tabular}
\caption{Feasibility ledger for the selected literature preparation.
Rows distinguish mechanical characterization, thermal loading and
quantum state discrimination. No assigned guide parameter fills an
unmeasured entry.}
\label{tab:qfa-feasibility-ledger}
\end{table}

For sideband thermometry, detector ordering, gain imbalance, optical
back-action and classical laser noise belong in the same inference
model. Laser phase noise can bias or mimic sideband asymmetry
\cite{SafaviNaeini2013}. A symmetrized displacement spectrum on its own
can be reproduced by an appropriate classical noise model. A heralded
counting experiment instead tests the normally ordered statistic and
classical bound of Section~\ref{sec:qfa-counting}, with its own
sampling and detector uncertainty. The desired witness therefore
determines which complementary record should be collected.

For the present evidence case, the defensible outcome is a classical
application plus an explicit quantum assessment with unfilled
experimental inputs. The synthetic guide predicts relaxational motion
in its stated band and offers no resolved single-phonon realization.
A future candidate is advanced by measuring its complex response,
assigning the bath and physical preparation coupling, and propagating
their uncertainties into a predicted detector statistic and required
record length. That procedure can reveal a viable regime or identify
which measurement would rule one out, while leaving the classical
interaction and variable-selection tests intact. The companion program
\texttt{VLP\_quantum\_feasibility\_audit.py} reproduces the assigned
scale comparisons, material poles, thermal survival example and ideal
background-count calculation; it performs no fit to measured HPV modes.

\subsection{Transport protocols, structural transitions, and delivery records}
\label{sec:s7transport-delivery}

A transport experiment asks which particles retain their intended
structure after a specified loading history. The HPV-inspired particle
provides the initial shell and surface preparation; its subsequent
normal, deformed and damaged classes are additional resolved states.
Mechanical-fatigue experiments on adenovirus and nonlinear capsid
deformation models demonstrate physical routes to structural failure
under stated loads~\cite{OrtegaEsteban2013,Kononova2016}. Applying such a
mechanism to transported HPV L1 VLPs requires its own force history,
barrier parameters and structural readouts. The calculation below
formulates that hypothesis and a direct model comparison.

\subsubsection{A structural state basis and its mechanical coupling}

\begin{definition}[Transport integrity states]
\label{def:s7transport-integrity-states}
For each retained labelled particle use the probability basis
$\ket{\mathrm N},\ket{\mathrm R},\ket{\mathrm D}$ for its declared normal,
reversibly deformed and long-lived damaged basins. The damaged class
denotes altered morphology or mechanical integrity; genetic change is a
different variable, absent from the empty L1-shell reference.
For $n$ tracked material objects the basis is
$\ket{\vct c}=\bigotimes_{i=1}^n\ket{c_i}$. A classical preparation is
\begin{equation}
 \ket{p(t)}=\sum_{\vct c}p_{\vct c}(t)\ket{\vct c},\qquad
 p_{\vct c}\geq0,\qquad
 \bra{\vct1}\ket{p}=1,\quad
 \bra{\vct1}=\sum_{\vct c}\bra{\vct c}.
 \label{eq:s7transport-probability-kets}
\end{equation}
The coefficients are probabilities, with the normalization above.
These structural labels are distinct from harmonic phonon occupations.
Each basin supplies its conditional energy/free energy, mechanics and
observation map. Disassembly into separately moving capsomers instead
changes the resolved objects and uses a conserved protein-building-block
description, as in Section~2.
\end{definition}

Combining two structural preparations with fractions $r$ and $1-r$
gives the classical population $r\vct p^{(1)}+(1-r)\vct p^{(2)}$,
$0\leq r\leq1$. Its entries are probabilities and have no adjustable
quantum relative phase. With the same specified joint generator, their
probability laws evolve by the same convex combination. Each conditional
mechanical motion can still use linear superposition over its valid
reference. The mixture of structural classes and the addition of those
motions answer different questions, as in Table~\ref{tab:sup-coefficient-meanings}.

The fixed-particle construction specializes the arrangement sectors of
the author's open-system path-integral formulation~\cite{KleessOpenSystem2025}.
Here a retained state $X$ includes centre and angular motion, any shell
deformation coordinate needed to resolve failure, and prepared medium
and surface variables. With a registered protocol $\Lambda(t)$,
the joint probability densities obey
\begin{equation}
 \partial_t P_{\vct c}
   =\mathcal L_{\vct c,\Lambda(t)}^*P_{\vct c}
    +\sum_{\vct c'\ne\vct c}
       [w_{\vct c\leftarrow\vct c'}(X,t)P_{\vct c'}
              -w_{\vct c'\leftarrow\vct c}(X,t)P_{\vct c}] .
 \label{eq:s7transport-joint-forward-law}
\end{equation}
Its continuous generators use the conditional mechanical force and
matched thermal forcing. A state transition continues the registered
physical coordinates and material preparation; a changed reference uses
the maps of Section~4. If transitions exchange only heat with an
isothermal bath at fixed control, their positive reverse pairs satisfy
$w_{c'\leftarrow c}/w_{c\leftarrow c'}=
\exp[-\beta(\mathcal F_{c'}-\mathcal F_c)]$ at the common retained state.
Driven chemistry or an eliminated nonequilibrium bath carries its
specified additional work and entropy terms.

Transport enters through physical ports: container acceleration relative
to the suspended bodies, spatial flow gradients, particle or wall
contacts, and interfacial or osmotic loading. Uniform co-moving
translation preserves relative deformation. The source power is the
force--velocity and torque--angular-velocity pairing of Section~5,
together with any pressure or chemical ports. Thus motion excites a
failure coordinate through a declared load path, rather than through
laboratory speed alone. A stable small-motion Hessian supplies the
within-basin response. Transition barriers require the extended energy
landscape or a separately calibrated kinetic law.

For example, an overdamped shell reaction coordinate $s$ can obey
\begin{equation}
 \gamma_s\dd s=
   [-\partial_s\mathcal F_c(s,h)+f_{\rm tr}(t)]\dd t
                       +\sqrt{2k_BT\gamma_s}\,\dd W_t .
 \label{eq:s7transport-deformation-coordinate}
\end{equation}
The retained strain $h$ has its own coupled evolution and noise.
In a metastable, thermally activated regime, a local barrier approximation
is $\Delta G_{c'c}(f)=\Delta G_{c'c}(0)-f x_{c'c}^{\ddagger}$ and
$w_{c'\leftarrow c}\simeq\nu_{c'c}(f)
e^{-\beta\Delta G_{c'c}(f)}$~\cite{Kramers1940,HanggiTalknerBorkovec1990}.
The coordinate, activation distance, prefactor and barrier validity
range are specified inputs. This converts a predicted local loading
history into a structural-transition probability.

If an additional resolved external load is Gaussian at the relevant
quasistatic barrier timescale, with mean $\bar f$ and variance
$\sigma_f^2$, a constant-prefactor version gives
\begin{equation}
 \langle w(f)\rangle
 =w(\bar f)\exp\!\left[
           \tfrac12\beta^2(x^{\ddagger})^2\sigma_f^2\right].
 \label{eq:s7transport-fluctuating-barrier-load}
\end{equation}
This conditional averaging concerns the additional load after the
thermal barrier bath has been specified. Counting the same thermal
fluctuation again would change the model. When loading and structural
relaxation occur on comparable timescales, the joint law above replaces
this quasistatic average. A mode can transfer mechanical energy toward
the reaction coordinate; a specific single-phonon failure mechanism
would require an additional resolved coupling and quantum evidence.

\paragraph{Superposed loads can change a nonlinear transition rate.}
\label{sec:sup-activated-phase-loading}
The transition model gives a direct biophysical use of phase-resolved
mechanics. Suppose two resolved drives act on the same shell reaction
coordinate with local force
$f(t)=\bar f+F_1\cos\Omega t+F_2\cos(\Omega t+\theta)$,
where $F_1,F_2\geq0$. These are the forces at that physical port,
including the mechanical transfer phases. Use the constant-prefactor
barrier approximation already stated, so that
$w(f)=w(\bar f)\exp[\beta x^\ddagger(f-\bar f)]$.
Assume intrabasin relaxation is rapid compared with a forcing period,
escape is rare within a period, and the barrier approximation remains
valid throughout the combined load.

\begin{proposition}[Cycle-averaged rate under two phase-related loads]
\label{prop:sup-activated-phase-loading}
For the fixed-basin, quasistatic rate above, the cycle average is
\begin{equation}
 \begin{aligned}
 F_{\rm eff}^2&=F_1^2+F_2^2+2F_1F_2\cos\theta,\\
 \overline w&=w(\bar f)I_0(\beta x^\ddagger F_{\rm eff}),\qquad
 I_0(z)=\frac1{2\pi}\int_0^{2\pi}e^{z\cos v}\,\dd v.
 \end{aligned}
 \label{eq:sup-activated-phase-loading}
\end{equation}
\end{proposition}
\begin{proof}
Adding the two force phasors gives a single sinusoid of amplitude
$F_{\rm eff}$. Integrating its exponential rate over a period gives
the displayed modified Bessel function~\cite{DLMFSuperposition}.
\end{proof}
For assigned $F_1=F_2=F$ and $\beta x^\ddagger F=1/2$, in-phase
loading gives $\overline w/w(\bar f)=I_0(1)\simeq1.2661$;
opposite phases cancel the oscillatory port load and give one.
The mean force agrees in both cases, while the rates differ. This is a
synthetic consequence of the barrier law, rather than an HPV transport
measurement. It illustrates why linear addition of forces does not imply
linear addition of their failure probabilities. Cancellation at a camera
readout need not cancel this shell load; its force and work-conjugate
coordinate must be calibrated separately. Finite changes of preparation
or escape probability within a cycle instead require the joint
time-dependent generator, and retained thermal forcing is not counted
again as an additional imposed load.

\subsubsection{A finite shipment and delayed structural loss}

\begin{example}[Normal and damaged particles after one delivery]
\label{ex:s7transport-three-state-delivery}
Consider identical independent particles under a prescribed protocol,
with reduced rates $a$ for $\mathrm N\to\mathrm R$, $b$ for
$\mathrm R\to\mathrm N$ and $d$ for $\mathrm R\to\mathrm D$.
Treat $\mathrm D$ as absorbing only over the observation window. The
column-generator convention gives
\begin{equation}
 \dot{\ket p}=\mathsf W\ket p,\qquad
 \mathsf W=
 \begin{pmatrix}-a&b&0\\a&-(b+d)&0\\0&d&0\end{pmatrix},
 \qquad \ket{p(0)}=\ket{\mathrm N} .
 \label{eq:s7transport-three-state-generator}
\end{equation}
In an assigned time unit $\tau_T$, let transport last $2\tau_T$ with
$(a,b,d)\tau_T=(1.2,0.4,0.1)$. At delivery,
$\vct p\simeq(0.26494,0.63703,0.09804)^{\mathsf T}$.
After unloading set $a=0$ and $(b,d)\tau_T=(0.6,0.3)$.
For elapsed holding time $u\geq0$,
\begin{equation}
 \boxed{\begin{aligned}
 p_R(T+u)&=p_R(T)e^{-(b+d)u},\\
 p_D(T+u)&=p_D(T)+\frac{d}{b+d}p_R(T)[1-e^{-(b+d)u}],\\
 p_N(T+u)&=p_N(T)+\frac{b}{b+d}p_R(T)[1-e^{-(b+d)u}].
 \end{aligned}}
 \label{eq:s7transport-delayed-delivery-fractions}
\end{equation}
At $u=3\tau_T$ these fractions are approximately
$(0.66108,0.04281,0.29611)$. Some deformed particles recover while others
enter the damaged basin after the applied transport drive has stopped.
These assigned rates illustrate a metastable preparation retained at
delivery; they are not a fit to HPV shipment data.
\end{example}
For twelve independent objects the class counts have the corresponding
multinomial law and mean $12\vct p$. Coupled contacts, shared medium
fluctuations and common uncertain loading can correlate outcomes, so the
joint basis rather than independent counts is then retained. The rates
in this reduced example represent a declared kinetic resolution.
Connecting them to the guide's mechanical parameters requires the
reaction-coordinate calibration above. The example predicts waiting-time
dependence in addition to an endpoint damaged fraction.

Structural readout supplies another map. For calibrated probabilities
$C_{ac}$ of assigning observed class $a$ to physical class $c$,
$\vct y_{\rm class}=\mathsf C\vct p$ gives expected class fractions.
An explicit missed-detection class completes the probability ledger.
Morphology imaging, particle counts, a mechanical compliance and a
surface/antigenic assay can have different $C_{ac}$ or readout weights.
A normal centre track alone therefore need not determine shell integrity.
Comparing stationary controls, loading histories and several post-delivery
waiting times tests the transport mechanism and the observation model
separately.

Figure~\ref{fig:s7transport-delivery-states} gives an executable finite-state transport benchmark with both recovery and delayed structural loss.

\begin{figure}[H]
\centering\begingroup
\setlength{\unitlength}{1pt}
\scalebox{0.90}{%
\input{vlt-figures/figure-304.tex}%
}\endgroup
\caption{Transport, retained deformation and delayed damage in the synthetic three-state example. The upper glyphs define normal, reversibly deformed and damaged structural classes; their shapes are schematic and are not a computed capsid deformation field. The same twelve labelled material objects appear at $t/\tau_T=2$ and $5$, using exact continuous-time Gillespie paths with seed $20261102$. Colours encode structural class, not a new particle identity. No interparticle links are drawn because these sample paths use the stipulated independent-object approximation. The lower curves are exact ensemble fractions from~\eqref{eq:s7transport-three-state-generator}, with rates $(a,b,d)\tau_T=(1.2,0.4,0.1)$ until delivery and $(0,0.6,0.3)$ thereafter. The damaged fraction rises from $0.0980$ to $0.2961$ during the shown holding interval while some deformed particles recover. Rates and the time unit are assigned; a mechanochemical HPV model must constrain the loading-to-rate map and structural classifier. The continued state at delivery is not reset. The boxed result illustrated here is~\eqref{eq:s7transport-delayed-delivery-fractions}.}
\label{fig:s7transport-delivery-states}
\end{figure}

\subsubsection{Path survival and nonequilibrium steady operation}

For a conservative safe-state process $Z_t$ with backward generator
$\mathcal L_t$ and failure hazard $\lambda_D(Z_t,t)\geq0$, define
\begin{equation}
 \boxed{\begin{aligned}
 S(z,t;T)&=
 \mathbb E_{z,t}\!\left[
   \exp\!\left(-\int_t^T\lambda_D(Z_u,u)\,\dd u\right)\right],\\
 \partial_tS+\mathcal L_tS-\lambda_DS&=0,\qquad S(z,T;T)=1 .
 \end{aligned}}
 \label{eq:s7transport-survival-feynman-kac}
\end{equation}
This Feynman--Kac representation applies to a well-posed safe-state law
with the required finite survival functional. It is the killed-process
survival probability; failure is an event rather than an extra arbitrary
reweighting. A specified terminal readout replaces the final value $1$.
Initial-state averaging gives shipment survival, while the full
generator gives conditional final integrity classes. Retained material
history can make two histories with the same endpoint positions have
different survival or later collapse. This is the path-space connection
to the companion formulation~\cite{KleessOpenSystem2025}.

For sustained autonomous operation, a stationary state instead satisfies
$\mathsf W\vct\pi=0$. Its edge currents are
$J_{c'c}=W_{c'c}\pi_c-W_{cc'}\pi_{c'}$.
Stationary nonzero currents distinguish a driven steady state from
detailed balance~\cite{Seifert2012}. A transient current can also occur
under an equilibrium generator started away from equilibrium; finite
transport is consequently solved as an initial-value protocol.
An absorbing damaged class has no sustained nontrivial steady cycle
without a specified return, repair or replacement channel.

For an explicit steady-state comparison, take three reusable
configuration classes on a ring, clockwise rate $r_+>0$ and
counterclockwise rate $r_->0$. Direct substitution gives
\begin{equation}
 \pi_c=\tfrac13,\quad J=\frac{r_+-r_-}{3},\quad
 \mathcal A=3\ln\frac{r_+}{r_-},\quad
 \dot S_{\rm tot}=k_B(r_+-r_-)\ln\frac{r_+}{r_-}\geq0 .
 \label{eq:s7transport-driven-ring-current}
\end{equation}
The maintained work or chemical port supplies the cycle affinity. These
reusable classes are distinct from the absorbing-damage approximation
above. Periodic shaking generally produces a periodic long-time law;
stationary stochastic agitation may admit a stationary driven joint law.
Both retain their source work, state-dependent kinetics and measured
outputs. Stopping at delivery changes the generator and its preparation;
it does not prepare a new stationary state instantaneously.

Figure~\ref{fig:s7transport-path-and-steady-current} contrasts a path-survival calculation with a source-maintained steady current, using explicit generators in both cases.

\begin{figure}[H]
\centering\begingroup
\setlength{\unitlength}{1pt}
\scalebox{0.90}{%
\input{vlt-figures/figure-305.tex}%
}\endgroup
\caption{Finite delivery and sustained nonequilibrium operation answer different questions. Panel (a) prescribes two safe-state hazards, $\lambda_A\tau_T=0.2$ and $\lambda_B\tau_T=0.2[1+4\sin^2(\pi t/(2\tau_T))]$, on $0\leq t/\tau_T\leq2$. The hazards agree at both endpoints, but their integrated survival probabilities differ as printed; these are exact deterministic instances of~\eqref{eq:s7transport-survival-feynman-kac}. Panels (b--d) use a separate autonomous three-class ring with clockwise rate $r_+$ and reverse rate $r_-$. All classes have stationary weight $1/3$, while directed currents persist when the rates differ. The nonnegative entropy-production rate follows~\eqref{eq:s7transport-driven-ring-current} and requires the stated maintained work or chemical source. The ring classes are reusable configurations; they are not the absorbing damaged state of the finite shipment. No finite-time failure rate or stationary circulation is inferred from a static image alone.}
\label{fig:s7transport-path-and-steady-current}
\end{figure}

\subsection{Biophysical applications and selection of effective variables}
\label{sec:phonon-calculable-outcomes}

An observation map can itself reshape a spectrum. Figure~\ref{fig:added-exposure-transfer} isolates finite exposure before the following model comparisons are interpreted.

The common question is which variables predict calibrated response and
fluctuations under the stated preparation. Each useful variable has a
physical assignment, an evolution law and a measured projection.
Ionic imbalance, configuration population, material relaxation and
phonon occupation supply different information. Their distinguishing
tests appear in Table~\ref{tab:phonon-model-comparison}.

\begin{table}[H]
\centering\small
\begin{tabular}{@{}p{3.0cm}p{4.4cm}p{7.3cm}@{}}
\toprule
\raggedright Mechanism & \raggedright Predicted readout
& \raggedright Required specification \tabularnewline
\midrule
\raggedright Spatial stiffness well
& \raggedright Frequencies, shape and penetration length
& \raggedright Curvature profile, inertia, end conditions and probe overlap
\tabularnewline\addlinespace[3pt]
\raggedright Coupled channels
& \raggedright Mixed frequencies and exchange contrast
& \raggedright Channel assignment, coupling, detuning, preparation and loss
\tabularnewline\addlinespace[3pt]
\raggedright Configuration loading
& \raggedright Conditional displacement, population softening and relaxation
& \raggedright Same-system surfaces, force derivatives, kinetics and temperature
\tabularnewline\addlinespace[3pt]
\raggedright Ionic relaxation
& \raggedright Compliance, colored force noise and relaxed variance
& \raggedright Finite populations, capacitance, mobility and conjugate force
\tabularnewline\addlinespace[3pt]
\raggedright Pumped quantum mode
& \raggedright Count distribution, number correlations and two-time coherence
& \raggedright Gain/loss channels, resource, bath, preparation and sensitive readout
\tabularnewline\addlinespace[3pt]
\raggedright Quantized preparation
& \raggedright Canonical covariance, occupation moments and a defined witness
& \raggedright State, subsystem division and calibrated quantum transduction
\tabularnewline
\bottomrule
\end{tabular}
\caption{Each mechanism defines a distinct comparison. Structural data
motivate possible coordinates; dynamics, preparation and transduction
determine which prediction is tested. These rows are modeling options,
not a claim that every mechanism occurs in every viral system.}
\label{tab:phonon-model-comparison}
\end{table}

\begin{figure}[H]
\centering\begingroup
\definecolor{vlfGreen}{HTML}{176B64}
\definecolor{vlfDarkRed}{HTML}{762B38}
\definecolor{vlfRed}{HTML}{762B38}
\definecolor{vlfBlue}{HTML}{315D80}
\definecolor{vlfCopper}{HTML}{762B38}
\setlength{\unitlength}{1pt}
\scalebox{0.80}{%
\input{vlt-figures/figure-277.tex}%
}\endgroup
\caption{Finite exposure filters a fixed motion. (a) A centred half-period exposure multiplies a sinusoidal displacement by $2/\pi$. (b) The general amplitude transfer is $H=\operatorname{sinc}(\omega t_{\rm exp}/2)$; its square gives power transfer. Solid and dashed curves distinguish signed amplitude from nonnegative power, with zeros at the indicated exposures. The preparation and mechanical frequency are unchanged. An experimental spectrum uses this detector filter together with its other transfer and noise contributions.}
\label{fig:added-exposure-transfer}
\end{figure}

\paragraph{What a paired mechanical measurement identifies.}
Take the single-coordinate ionic model after equilibration at known
$T$, with a calibrated physical displacement $x$ and its conjugate
force. Independently constrained inertia and the measured complex
response give
\begin{equation}
 K^*(\omega)=\mathcal R_{xx}(\omega)^{-1}+M_x\omega^2,
 \qquad K_{\rm rel}=\frac{k_{\rm B}T}{\operatorname{Var}x}.
 \label{eq:phonon-response-fluctuation-inference}
\end{equation}
These are statements for this scalar model and its thermal preparation,
after detector covariance, exposure and transfer-function corrections.
A resonance in displacement alone is not a measurement of frozen
stiffness: inertia and loss also enter the response. For a nonzero branch $\Delta k>0$, if the resolved
band contains both plateaus of the stipulated single relaxation,
\begin{equation}
 \boxed{\Delta k=k_x-K_{\rm rel},\qquad
 \tau=\omega_{\rm loss}^{-1},\qquad
 K''_{\max}=\frac{\Delta k}{2}.}
 \label{eq:phonon-relaxation-parameter-inference}
\end{equation}

Figure~\ref{fig:relaxation-plateaus-and-loss-inference} identifies the two plateaus and the loss maximum used to infer the branch strength and relaxation time.

\begin{figure}[H]
\centering\begingroup
\definecolor{vlfGreen}{HTML}{176B64}
\definecolor{vlfDarkRed}{HTML}{762B38}
\definecolor{vlfRed}{HTML}{762B38}
\definecolor{vlfBlue}{HTML}{315D80}
\definecolor{vlfCopper}{HTML}{762B38}
\setlength{\unitlength}{1pt}
\scalebox{0.80}{%
\input{vlt-figures/figure-278.tex}%
}\endgroup
\caption{Relaxed and frozen plateaus and the loss maximum of one stable relaxation branch, as in \eqref{eq:phonon-relaxation-parameter-inference}. Set $\Delta k/k_x=0.6$, so $K_{\rm rel}/k_x=0.4>0$, and use logarithmic frequency $r=\omega\tau>0$. (a) The exact storage curve $K^\prime/k_x=1-0.6/(1+r^2)$ joins the relaxed and frozen plateaus; their difference is $\Delta k$. (b) The exact loss curve $K^{\prime\prime}/k_x=0.6r/(1+r^2)$ peaks at $r=1$, with height $0.3$. The horizontal frequency scale and the plateau difference determine $\tau$ and $\Delta k$, respectively. These are constitutive stiffness curves, not displacement spectra: the maximum of a displacement spectrum can be elsewhere because inertia and mechanical filtering also enter. Inference requires a nonzero branch and a resolved band that covers both plateaus below omitted cutoffs.}
\label{fig:relaxation-plateaus-and-loss-inference}
\end{figure}

Here $\omega_{\rm loss}$ is the maximum of the loss stiffness, not
necessarily the displacement-spectrum peak. The frozen plateau needs
$\tau^{-1}\ll\omega$ below other structural, medium and detector
cutoffs. If that interval is absent, an apparent single relaxation
can be poorly constrained even when a fit looks smooth.

\begin{proposition}[Mechanical-channel parameter equivalence]
\label{prop:s7-mechanical-parameter-equivalence}
Use the stipulated stable single-relaxation model, its stationary
thermal preparation and only the calibrated mechanical force/readout.
Assume the nonzero branch and resolved-response band specified above.
The mechanical record identifies $\Delta k=b_{xy}^2/a_y$ and
$\tau=(\Lambda a_y)^{-1}$, rather than three separate hidden
parameters. Indeed, for any dimensionless $\lambda>0$,
\begin{equation}
 a_y^{(\lambda)}=\lambda^2a_y,\qquad
 b_{xy}^{(\lambda)}=\lambda b_{xy},\qquad
 \Lambda^{(\lambda)}=\Lambda/\lambda^2
 \quad\Longrightarrow\quad
 K^{*(\lambda)}(\omega)=K^*(\omega).
 \label{eq:phonon-hidden-parameter-equivalence}
\end{equation}
\end{proposition}
\begin{proof}
The transformation leaves $b_{xy}^2/a_y$ and $\Lambda a_y$
unchanged. Substitution into the same scalar constitutive law
therefore leaves its complex response unchanged. The matched
stationary thermal fluctuation prediction is unchanged as well.
\end{proof}
Reversing the sign of $b_{xy}$ also leaves this mechanical response
unchanged. These are distinct allowed constitutive parameters for a
hidden coordinate with a fixed physical meaning and unit; the ambiguity
does not require changing that unit. Their equilibrium conditional
variance $k_{\rm B}T/a_y$ and conditional slope $-b_{xy}/a_y$ differ,
so an additional ionic or configurational measurement can break the
degeneracy (Figure~\ref{fig:mechanical-hidden-variable-nonidentifiability}).
For example, a calibrated population fluctuation or charge readout,
combined with the mechanical response, supplies information absent
from $x$ alone. This is structural nonidentifiability: an ideal noiseless mechanical
spectrum still leaves equivalent parameter sets~\cite{Raue2009}.
Longer averaging can reduce sampling error but cannot remove this
equivalence.

\begin{figure}[H]
\centering\begingroup
\definecolor{vltTeal}{HTML}{176B64}
\definecolor{vltBlue}{HTML}{315D80}
\definecolor{vltGold}{HTML}{A87520}
\definecolor{vltPlum}{HTML}{725877}
\definecolor{vltRed}{HTML}{762B38}
\definecolor{vltYellow}{HTML}{D5AD3C}
\definecolor{vltNeutral}{HTML}{777D83}
\setlength{\unitlength}{1pt}
\scalebox{0.80}{%
\input{vlt-figures/figure-279.tex}%
}\endgroup
\caption{Mechanical parameter equivalence and an additional discriminating channel. All three models use the same physical $x,y$ coordinates and $k_x=2$. (a) Equal-scale contours show $\mathcal F_2=(2x^2+2bxy+a_yy^2)/2=1/2$. (b) Their common combinations $\Delta k=b^2/a_y=1$ and $\tau=(\Lambda a_y)^{-1}=1$ give identical $K^*(\omega)=2-(1-i\omega)^{-1}$, hence identical mechanical response and Gibbs variance at shared inertia. Solid and dashed black curves give storage and loss. (c) With no independently applied hidden force, harmonic continuation gives $y/x=-[b/a_y]/(1-i\omega\tau)$. Its real part separates both the hidden scale and coupling sign when that coordinate has an independently calibrated readout. Colors and line styles join each contour to its hidden-channel curve. This comparison changes constitutive parameters at fixed coordinate units; it assigns no unique molecular mechanism from the mechanical channel alone.}
\label{fig:mechanical-hidden-variable-nonidentifiability}
\end{figure}

Environmental reconstruction uses a specified particle, interaction,
boundary and detector model. Tensorial microrheology relates probe
translation and rotation to material response under hydrodynamic
assumptions~\cite{SquiresMason2010}. Independent motions can test those
assumptions once geometry, surface coupling and usable bandwidth are
constrained. Otherwise support compliance, internal deformation and
medium memory can compensate in the inferred response.

\subsubsection{Two-particle microrheology: elastic coupling and shared solvent motion}
\label{sec:phonon-two-probe-application}

Two-point microrheology measures how a soft material transmits stress
between tracked locations. Correlations have been used to characterize
inhomogeneous soft media~\cite{Crocker2000}, giving cross-response a
direct spatial interpretation. Probe geometry, interface conditions,
separation and constitutive bandwidth determine that interpretation.

An exact low-inertia comparison separates hydrodynamic coupling from an
elastic link. Take two rigid spherical probes with small longitudinal
displacements about separate harmonic traps, both of stiffness
$k_{\rm tr}>0$, and no direct elastic link. In a Newtonian reference,
use the constant symmetric mobility
\begin{equation}
 \begin{aligned}
 \mathsf\mu_{\rm pair}&=
   \begin{pmatrix}\mu_0&\mu_c\\\mu_c&\mu_0\end{pmatrix},
 \qquad 0<\mu_c<\mu_0,\\
 \dd\vct u&=-k_{\rm tr}\mathsf\mu_{\rm pair}\vct u\,\dd t
       +\sqrt{2k_{\rm B}T\mathsf\mu_{\rm pair}}\,\dd\vct B.
 \end{aligned}
 \label{eq:phonon-hydrodynamic-pair-reference}
\end{equation}
For equal no-slip spheres of radius $a$ in an unbounded liquid of
viscosity $\eta_f$, the leading far-field longitudinal values are
$\mu_0=(6\pi\eta_fa)^{-1}$ and
$\mu_c=(4\pi\eta_fd)^{-1}$ at centre separation $d\gg a$.
The frozen-separation approximation requires excursions small compared
with $d$. Fluid inertia, nearby walls and viscoelastic memory are
excluded from this reference. Hydrodynamic anticorrelations between
trapped particles have been measured and explained using this type of
coupling~\cite{MeinersQuake1999}.

The orthonormal common and relative coordinates
$u_\pm=(u_1\pm u_2)/\sqrt2$ now have different mobilities,
$\mu_\pm=\mu_0\pm\mu_c$, while their trap curvatures are equal.
Their conjugate-force responses and particle equilibrium covariance are
\begin{equation}
 \mathcal R_\pm(\omega)
       =\left[k_{\rm tr}-\frac{i\omega}{\mu_\pm}\right]^{-1},
 \qquad
 \mathsf C_u(0)=\frac{k_{\rm B}T}{k_{\rm tr}}\Id_2.
 \label{eq:phonon-hydrodynamic-pair-response-and-variance}
\end{equation}
Thus the two positions are uncorrelated at equal time in this harmonic
equilibrium model, even though their thermal increments are correlated
and their relaxation rates differ. Reconstructing the site covariance
from the common and relative decays gives
\begin{equation}
 C_{12}(t)=\frac{k_{\rm B}T}{2k_{\rm tr}}
       [e^{-k_{\rm tr}\mu_+|t|}-e^{-k_{\rm tr}\mu_-|t|}].
 \label{eq:phonon-hydrodynamic-lagged-anticorrelation}
\end{equation}
The negative cross-covariance at nonzero lag records relaxation through
the shared solvent and returns to zero. Its correlated thermal noise
is fixed by the off-diagonal mobility in
\eqref{eq:phonon-hydrodynamic-pair-reference}; together they preserve
the uncoupled equilibrium trap energy.
Figure~\ref{fig:biophysical-two-probe-lagged-correlation} displays this
testable distinction between equal-time and lagged correlations.

\begin{figure}[H]
\centering\begingroup
\definecolor{vltTeal}{HTML}{176B64}
\definecolor{vltBlue}{HTML}{315D80}
\definecolor{vltGold}{HTML}{A87520}
\definecolor{vltPlum}{HTML}{725877}
\definecolor{vltRed}{HTML}{762B38}
\definecolor{vltYellow}{HTML}{D5AD3C}
\definecolor{vltNeutral}{HTML}{777D83}
\setlength{\unitlength}{1pt}
\scalebox{0.90}{%
\input{vlt-figures/figure-280.tex}%
}\endgroup
\caption{Cross-correlations can reveal shared-solvent motion with no equilibrium cross-spring. (a) The stipulated spherical probes have equal independent traps and symmetric positive mobility, $0<\epsilon_\mu=\mu_c/\mu_0<1$. The paired vertical marks indicate traps, and arrows define positive displacement. Correlated thermal increments use the same mobility that generates drift. (b) The exact covariance in \eqref{eq:phonon-hydrodynamic-lagged-anticorrelation} is negative at nonzero lag even though the equal-time position covariance is diagonal. (c) For $\epsilon_\mu=0.2$, normalized common and relative covariances decay as $e^{-1.2\theta}$ and $e^{-0.8\theta}$; their half-difference reconstructs the cross record. The unbounded, no-slip, far-field sphere reference sets this mobility approximation. Probe dimensions are schematic; the positive-gap condition, fixed separation and absence of walls or fluid memory retain the scope stated in the text. No link spring is assigned to these probes.}
\label{fig:biophysical-two-probe-lagged-correlation}
\end{figure}

This comparison sharpens the interpretation of the benchmark in
Section~\ref{sec:complete-benchmark}.
An elastic link changes the equilibrium covariance through its energy;
instantaneous hydrodynamic coupling can change the lagged covariance
while leaving that equilibrium energy unchanged. Their common and
relative driven responses distinguish the two contributions. A medium
with resolved relaxation additionally changes their frequency
dependence and matched force noise. The data can therefore establish
whether a mobility tensor is sufficient or whether a further material
coordinate or memory kernel is needed. A single positional correlation
cannot decide among these possibilities.

For hydrated polymer or protein networks, local probe binding, network
deformation and solvent motion may contribute over different spatial
and temporal ranges. Comparing self- and cross-responses over the same
band tests that separation~\cite{Crocker2000,SquiresMason2010}.
The probe's radius and surface coupling are constitutive inputs;
$\vct u$ is its retained state, and a measured cross-correlation is an
observable of that state. A compliant biological particle can require
internal deformation variables as well. This is how the same framework
can be used both for a particle assembly and for a particle probing a
surrounding biomaterial, with different physical assignments of the
couplings.

\paragraph{Separate calibration from the prediction being tested.}
One can estimate $K_{\rm rel}$ from equilibrium variance, but then
that variance is a calibration input. To test a fluctuation prediction,
constrain the stiffness using independent force--displacement data and
hold out the fluctuation record. Parameter uncertainties should propagate
into a band of predicted responses and spectra, rather than be removed
by fitting each observable separately. A held-out comparison means
distinct calibration and test data, not necessarily distinct instruments.
The pair benchmark in Section~\ref{sec:complete-benchmark} makes this
division concrete.

For a stable equilibrium classical scalar coordinate with a matched
response and force, define
\begin{equation}
 \mathcal E_{\rm FDT}(\omega)
  =S_{xx}(\omega)-\frac{2k_{\rm B}T}{\omega}
                         \operatorname{Im}\mathcal R_{xx}(\omega),
 \qquad \omega\ne0.
 \label{eq:phonon-equilibrium-consistency-residual}
\end{equation}
The equilibrium prediction is zero, within the model and calibration
uncertainties. Projecting out equilibrium coordinates changes the
effective memory but preserves the appropriate equilibrium
fluctuation--dissipation relation. Thus a nonzero corrected residual
cannot be explained merely by naming another hidden equilibrium mode.
Preparation drift, temperature mismatch, an incorrect conjugate force,
detector effects and genuine activity are different possibilities that
must be distinguished. In the quantum regime one must instead compare
the appropriate operator ordering and quantum thermal factor. The
classical residual is not a universal test for quantum coherence or
for the identity of a chemical pump.

\paragraph{A concrete HPV-inspired comparison sequence.}
The running family supports a compact implementation. First select
intact L1 VLPs, register their body and reporter geometry, and constrain
the supported interparticle force/torque response in the selected band.
Capsid indentation, surface binding and centre mobility enter through
their respective ports. The source map then specifies which pattern
is loaded, and the observation map includes exposure, localization and
surface transduction before fitting a record.

Next compare a centre-only reference with one that retains angular
motion, and then with a model carrying the finite surface pool and
prepared material coordinates. A controlled change of ligand supply or
surface preparation can leave centre geometry fixed while changing
marker, surface or later force records. For a closed-pool protocol,
redistribution preserves the bound-plus-free ligand and charge ledgers;
a wash or reagent addition is a registered open chemical port. After a
geometry change, the actual retained state is continued rather than
sampled anew. The same fitted parameters predict the held-out
cross-response, fluctuation spectrum and preparation-age dependence.

Three specific discrepancies guide the next retained variable. A marker
response absent from the calibrated centre prediction tests angular or
shape resolution. A surface-population record that distinguishes
mechanically equivalent preparations tests the hidden-state assignment.
A post-contact response that depends on preparation age tests the
retained material history. The angular benchmark of Section~6, the
finite-pool calculation below and the surface-basin comparison above
make these alternatives calculable. Their discrimination remains a
task-specific result with uncertainty and bandwidth; choosing the next
measurement is part of the construction.

\subsubsection{Establishing variables at a specified biophysical resolution}
\label{sec:phonon-establishing-biophysical-variables}

The applications give an operational meaning to establishing variables
in biophysics: select physical information sufficient for a stated
prediction. The choice depends on spatial resolution, forcing,
preparation and observation time.

\begin{definition}[Effective variables for a biophysical prediction task]
\label{def:s7-task-effective-variables}
Fix a family of admissible preparations and controls, calibrated target
observables, a time window and an error tolerance $\epsilon$ in declared
physical units and norms. A retained-variable description specifies
$Z=\mathcal B(X,g)$, its evolution or retained-history law, its
preparation and its target-readout maps. It is effective for that task
when those maps predict the targeted observables of the declared
mechanical model within $\epsilon$ throughout that family and window.
The exact case has $\epsilon=0$; statistical tasks specify the moments
or probability laws being compared.
\end{definition}

The pointwise and event-closure criterion of
Proposition~\ref{prop:s6-history-closed-variables} tests exact linear
state sufficiency. Definition~\ref{def:s6-identifiable-prediction}
separately tests which targets an observation task can determine.
A description can pass one test while needing more state or more data
for the other. Projection theories and equilibrium coarse-graining
supply complementary foundations for controlled reductions
\cite{Mori1965,Noid2008}.

Figure~\ref{fig:s7-effective-variable-task-window} tests the retained state against a specified preparation family, target and error window.

\begin{figure}[H]
\centering\begingroup
\definecolor{vltTeal}{HTML}{176B64}
\definecolor{vltBlue}{HTML}{315D80}
\definecolor{vltGold}{HTML}{A87520}
\definecolor{vltPlum}{HTML}{725877}
\definecolor{vltRed}{HTML}{762B38}
\definecolor{vltYellow}{HTML}{D5AD3C}
\definecolor{vltNeutral}{HTML}{777D83}
\setlength{\unitlength}{1pt}
\scalebox{0.90}{%
\input{vlt-figures/figure-281.tex}%
}\endgroup
\caption{Prepared material history changes a future displacement record, illustrating Definition~\ref{def:s7-task-effective-variables}. The scaled passive model is $\dot u=v$, $\dot v=-u-0.2v-z$, $\dot z=v-z$, with storage $(u^2+v^2+z^2)/2$. (a,b) Two preparations share the particle position and velocity but have $z_0=\pm1$. Only the Maxwell branch is drawn; the onsite stiffness and instantaneous drag remain in the stated dynamics. The material dot has $y_0=-z_0$; the same natural spring span, six turns and $12\,\mathrm{pt}$ per unit extension make the positive-$z$ spring longer. Burgundy arrows show the opposite branch forces. (c) For $u_0=v_0=0$, $|z_0|\leq1$, zero input and target displacement on $[0,T]$, omitting $z_0$ gives minimax error $\mathcal E(T)=\max_{t\leq T}|h(t)|$, where $h(t)=(e^{t\mathsf A})_{13}$. The pointwise error can fall while this window error remains large. The tolerance is in the declared scaled displacement unit. Retaining the material state closes this model; determining unique microscopic parameters is the separate observation problem developed above. The glyphs show alternative preparations of one ideal virion, not two interacting virions.}
\label{fig:s7-effective-variable-task-window}
\end{figure}

Three roles organize the variable-selection problem. A state variable
evolves, such as probe displacement, membrane height or a configuration
indicator. A constitutive parameter, such as bending modulus or mobility,
specifies that evolution within a reference. An observable is a
calibrated function or temporal projection of the state. A parameter
becomes an additional state when its resolved evolution is needed for
prediction. Table~\ref{tab:biophysical-variable-establishment} applies
these roles to the worked examples.

\begin{table}[H]
\centering\small
\begin{tabular}{@{}p{2.8cm}p{5.5cm}p{6.7cm}@{}}
\toprule
Biophysical setting & State and constitutive inputs & Distinguishing observation\\
\midrule
\raggedright Tracked probes in a soft medium
& \raggedright State: centre displacements and, if needed, material relaxation coordinates. Inputs: trap curvature, mobility and interface coupling.
& \raggedright Equal-time covariance, common/relative response and lagged cross-correlation separate stored elastic coupling from shared drag.\tabularnewline\addlinespace[4pt]
\raggedright Fluctuating membrane
& \raggedright State: resolved height modes; leaflet density difference when its relaxation is resolved. Inputs: bending modulus, tension and dissipative coefficients.
& \raggedright Spatial fluctuation amplitudes constrain restoring terms; time correlations test mobility and the need for another relaxing field.\tabularnewline\addlinespace[4pt]
\raggedright Protein within a structural basin
& \raggedright State: calibrated shape coordinates or their predictive combinations. Inputs: conditional curvature, mobility and reporter derivatives.
& \raggedright Variance and lagged persistence select different information; additional readouts test structurally hidden motion.\tabularnewline\addlinespace[4pt]
\raggedright Switching biomolecular configurations
& \raggedright State: the configuration indicator, or ensemble populations and resolved internal coordinates. Inputs: conditional free energies and transition laws.
& \raggedright Population ratios constrain free-energy differences; conditional fluctuations and switching statistics distinguish internal entropy from kinetics.\tabularnewline
\bottomrule
\end{tabular}
\caption{Physical roles in establishing effective variables. These
examples use different resolutions of biomolecular or soft-material
motion. Their variables are selected for a defined prediction and do
not identify a universal microscopic coordinate set.}
\label{tab:biophysical-variable-establishment}
\end{table}

Several checks then work together. The coordinate must have a physical
definition, units and compatible constraints, so that changing it has a
defined energy or free-energy cost and a work-conjugate force. Its
evolution must close to the required accuracy: preparations with the
same retained state should give compatible conditional predictions,
or their relevant preparation history must be retained. The protein
two-timescale example shows explicitly how projection can fail this
test while preserving static variance. The membrane example identifies
a missing density coordinate through the dynamics it carries. The
probe example shows that adding instantaneous cross-mobility can
sometimes suffice without adding another independent state.

Observability and identifiability impose a different check. A real
coordinate can be dark to a chosen reporter, and several physical
mechanisms can generate the same projected memory. A second readout or
independently constrained force sensitivity can distinguish models that
one record cannot, as in~\eqref{eq:phonon-hidden-parameter-equivalence}.
In a near-degenerate structural cluster, the stable observable may be
its combined projector weight rather than a uniquely named eigenvector;
the residual bounds of Section~\ref{sec:phonon-mode-certification}
quantify that distinction. Statistical persistence identifies candidates,
while geometry, energetics and calibrated couplings give them physical
meaning~\cite{NoeNuske2013,PerezHernandez2013}.

A reduction is useful when it preserves the target response, covariance
or population statistics over the declared band. Fast internal modes
can contribute conditional free energy; slower populations can produce
memory and matched fluctuations. These contributions determine whether
an additional state improves prediction at the required accuracy.
Physical assignment, closure and observable tests jointly establish
effective variables for the biophysical question.

\clearpage

\section{Conclusion}
\label{sec:conclusion}

Viral Lattice Theory connects a specified particle assembly to
predictions of motion and measurement. Geometry, interactions, inertia
and constraints determine force balance and local restoring response.
Compatible cells assemble larger systems. Material coordinates retain
relaxation and preparation; eliminating them produces causal memory and
preparation-dependent forcing, with matching fluctuations under the stated
thermal-equilibrium assumptions. Registered changes continue the physical
preparation through arrangement and coordinate maps.

The mechanics, coarse-graining, stochastic dynamics and response methods
are established. The contribution is their organization for collective
virion motion and the conditional results and comparisons developed from
it. A generalized Brownian description with the same mechanical inputs,
memory, preparation and thermal noise gives the same passive predictions.
The formulation makes the required state, physical inputs and observation
map explicit, so that a researcher can calculate what a record
distinguishes and which further information would change a prediction.

Four tasks give this organization concrete uses. The joint frame--state
calculation accounts for motion during camera exposure and its
correlation with the next state, predicting acquisition-dependent records
at fixed mechanics. Two-probe comparisons separate elastic coupling,
which changes equilibrium energy, from shared-fluid mobility, which can
change lagged correlations at fixed energy. A constructed parameter
family has identical mechanical response and thermal fluctuation records
but different hidden properties; a complementary calibrated channel
distinguishes those alternatives. Retained-state criteria test whether
particle coordinates suffice or whether additional material variables or
history are needed for a chosen preparation family, record and time
window. State sufficiency and parameter identifiability ask different
questions, even when they use the same measured trajectories.

The passive-pair benchmark makes these distinctions reproducible. One
synthetic input set determines complex response, fluctuation spectra and
covariance. The common channel provides a control and relative motion
senses an additional relaxation branch. In the overdamped introductory
guide, the full and matched memory-free models share the static
covariance and response magnitude at the selected crossover, yet differ
by $28.1^\circ$ in phase and $40\%$ in spectral density there.
The finite-inertia benchmark in Section~\ref{sec:complete-benchmark}
retains the same discrimination, with nearly equal crossover magnitudes.
Finite-rise forcing tests a calibration requirement; a controlled link
change tests continuation of prepared material state before the new
reference equilibrates. These are calculated consequences of supplied
models and a protocol for comparing complementary measurements.

The STMV crystal example adds a published-summary-anchored test.
The documented reference packing and a specified positive acoustic
network distinguish internal shell geometry from whole-virion crystal
response. A bath-transfer reference calibrated without the crystal
linewidth predicts about $1.78$~GHz, with an assigned propagation range
of about $[1.55,2.02]$~GHz; the reserved reported width is
$2.69$~GHz. This identifies a failure of that transfer assumption,
without assigning its cause to a unique microscopic mechanism.
Three contact networks fit the same illustrative calibration peak yet
predict distinct registered-direction responses. These are conditional
comparisons with explicit preparation, optical and structural inputs;
they neither recover all crystal elastic constants nor validate an
internal shell vibration or a crystal-growth mechanism.

HPV-inspired ideal VLPs give the worked examples a common physical
reference. Each empty L1 shell is one whole-particle node. Capsid geometry
and candidate binding locations help assign coordinates; collective
interaction, material and observation parameters require their own
constraints. The later examples distinguish shell compliance from link
stiffness, translation from libration, and mechanical response from
surface-population information. A finite ligand pool changes occupancy
fluctuations and kinetics through shared depletion. These calculations
show how additional variables enter when a prediction depends on them.
At centre-coordinate resolution, the same equations can also describe
other interacting colloids. Biological specificity comes from the actual
preparation, surface physics, medium and readout.

The transport extension applies this discipline to structural stability.
Classical probabilities label normal, reversibly deformed and long-lived
damaged objects, with conditional mechanics, transition rates and
structural readouts. The synthetic shipment continues its delivered
preparation and predicts both recovery and delayed loss after unloading.
The guide's unchanged-hazard comparison shows that delayed damage can
persist without increasing the damage rate, while its reset comparison
shows the error from discarding the deformed population. These results
follow from assigned kinetics; they do not establish an HPV transport
mechanism. A biological test must constrain the load paths, structural
barriers or rates, and class-observation map. Several waiting times and
an integrity-sensitive measurement provide tests beyond an endpoint
centre trajectory. Path survival records finite loading history; a
separate source-maintained cycle illustrates nonequilibrium steady
circulation.

Superposition connects the spectral construction to these experiments.
For a fixed linear mechanical law, motion contributions add with their
phases; prepared response is affine and is compared about the same
unforced trajectory. Local cancellation can coexist with motion and
storage in other channels. The phase-dependent barrier example further
shows how a superposed load can feed a nonlinear structural rate.
Classical structural mixtures and quantum occupation superpositions keep
their distinct probability and coherence rules.

Confinement, localization, wave-packet dispersion and coupled channels
extend the collective mechanical description. A separate viral-phonon
quantization postulate supports occupation statistics, quantum covariance
bounds, entanglement-witness and quantum self-excitation calculations.
Their biological interpretation requires an identified mode and
preparation, consistent damping and thermal conditions, a resource
account, and a readout that distinguishes the proposed statistics from
classical alternatives. The preparation-specific assessment in
Section~\ref{sec:qfa-feasibility} leaves unmeasured carrier and detector
inputs explicit. The guide has relaxational poles in its declared band;
a separate conditional single-phonon calculation predicts loss of
antibunching even when mean occupation stays fixed. Its thermal and
counting budgets give testable requirements for a future realization.
The classical motion, inference and delivery comparisons stand on their
own stated models.

The next empirical step is a specified realization with independently
constrained geometry, interactions, surface and material response, and
acquisition characteristics. Calibrate on designated records, propagate
input uncertainty, then use the same parameters to predict complementary
driven, fluctuation and preparation-dependent measurements. For transport,
include structural outcomes across loading protocols and post-delivery
times. Such comparisons test which retained variables and couplings
suffice for a particular biophysical question and which additional state
or measurement is warranted.

\clearpage
\phantomsection


\addcontentsline{toc}{section}{References}

\begin{thebibliography}{999}
\fancyhead[R]{\small References}
\setlength{\itemsep}{3pt}

\bibitem{FeynmanLectures1963}
R.~P.~Feynman, R.~B.~Leighton, and M.~Sands,
\emph{The Feynman Lectures on Physics}, Vols.~I--III,
Addison-Wesley (1963--1965), especially Vol.~I, Chapters~41 and~47--49,
Vol.~II, Chapters~8, 10, and~19, and Vol.~III, Chapters~3 and~8.
\href{https://www.feynmanlectures.caltech.edu/}{Official Caltech online edition}.

\bibitem{Leake2016}
M.~C.~Leake, \emph{Biophysics: Tools and Techniques}, CRC Press (2016),
especially Chapter~1 and Sections~2.1--2.2.6.
ISBN 978-1-4987-0243-0.

\bibitem{Liljeroos2011}
L.~Liljeroos, J.~T.~Huiskonen, A.~Ora, P.~Susi, and S.~J.~Butcher,
``Electron cryotomography of measles virus reveals how matrix protein coats
the ribonucleocapsid within intact virions,''
\emph{Proceedings of the National Academy of Sciences} \textbf{108}(44),
18085--18090 (2011).
\href{https://doi.org/10.1073/pnas.1105770108}{doi:10.1073/pnas.1105770108}.

\bibitem{Ke2018}
Z.~Ke, J.~D.~Strauss, C.~M.~Hampton, M.~A.~Brindley, R.~S.~Dillard,
F.~Leon, K.~M.~Lamb, R.~K.~Plemper, and E.~R.~Wright,
``Promotion of virus assembly and organization by the measles virus matrix
protein,'' \emph{Nature Communications} \textbf{9}, 1736 (2018).
\href{https://doi.org/10.1038/s41467-018-04058-2}{doi:10.1038/s41467-018-04058-2}.

\bibitem{Zhang2025}
Y.~Zhang, R.~Wu, M.~Shahjahan, C.~Yang, D.~Pyeon, and E.~Harel,
``Nanoscopic acoustic vibrational dynamics of a single virus captured by
ultrafast spectroscopy,''
\emph{Proceedings of the National Academy of Sciences} \textbf{122}(4),
e2420428122 (2025).
\href{https://doi.org/10.1073/pnas.2420428122}{doi:10.1073/pnas.2420428122}.

\bibitem{CasparKlug1962}
D.~L.~D.~Caspar and A.~Klug,
``Physical principles in the construction of regular viruses,''
\emph{Cold Spring Harbor Symposia on Quantitative Biology}
\textbf{27}, 1--24 (1962).
\href{https://doi.org/10.1101/SQB.1962.027.001.005}{doi:10.1101/SQB.1962.027.001.005}.

\bibitem{Lidmar2003}
J.~Lidmar, L.~Mirny, and D.~R.~Nelson,
``Virus shapes and buckling transitions in spherical shells,''
\emph{Physical Review E} \textbf{68}, 051910 (2003).
\href{https://doi.org/10.1103/PhysRevE.68.051910}{doi:10.1103/PhysRevE.68.051910}.

\bibitem{Sirohi2016}
D.~Sirohi, Z.~Chen, L.~Sun, T.~Klose, T.~C.~Pierson,
M.~G.~Rossmann, and R.~J.~Kuhn,
``The 3.8~\AA\ resolution cryo-EM structure of Zika virus,''
\emph{Science} \textbf{352}(6284), 467--470 (2016).
\href{https://doi.org/10.1126/science.aaf5316}{doi:10.1126/science.aaf5316}.

\bibitem{Carrasco2006}
C.~Carrasco, A.~Carreira, I.~A.~T.~Schaap, P.~A.~Serena,
J.~G\'omez-Herrero, M.~G.~Mateu, and P.~J.~de Pablo,
``DNA-mediated anisotropic mechanical reinforcement of a virus,''
\emph{Proceedings of the National Academy of Sciences}
\textbf{103}(37), 13706--13711 (2006).
\href{https://doi.org/10.1073/pnas.0601881103}{doi:10.1073/pnas.0601881103}.

\bibitem{Israelachvili2011}
J.~N.~Israelachvili, \emph{Intermolecular and Surface Forces}, 3rd ed.,
Academic Press (2011).
\href{https://doi.org/10.1016/C2009-0-21560-1}{doi:10.1016/C2009-0-21560-1}.

\bibitem{Cotterill2002}
R.~M.~J.~Cotterill, \emph{Biophysics: An Introduction},
John Wiley \& Sons (2002), especially Sections~3.5, 4.3, and Chapter~5.
ISBN 978-0-471-48538-4.

\bibitem{Coimbra2004}
C.~F.~M.~Coimbra, D.~L'Esperance, R.~A.~Lambert, J.~D.~Trolinger,
and R.~H.~Rangel,
\textquotedblleft An experimental study on stationary history effects
in high-frequency Stokes flows,\textquotedblright
\emph{Journal of Fluid Mechanics} \textbf{504}, 353--363 (2004).
\href{https://doi.org/10.1017/S002211200400789X}{doi:10.1017/S002211200400789X}.

\bibitem{Tatsuo2000}
H.~Tatsuo, N.~Ono, K.~Tanaka, and Y.~Yanagi,
\textquotedblleft SLAM (CDw150) is a cellular receptor for measles virus,\textquotedblright
\emph{Nature} \textbf{406}, 893--897 (2000).
\href{https://doi.org/10.1038/35022579}{doi:10.1038/35022579}.

\bibitem{Muhlebach2011}
M.~D.~M\"uhlebach et al.,
\textquotedblleft Adherens junction protein nectin-4 is the epithelial receptor
for measles virus,\textquotedblright
\emph{Nature} \textbf{480}, 530--533 (2011).
\href{https://doi.org/10.1038/nature10639}{doi:10.1038/nature10639}.

\bibitem{Brindley2012}
M.~A.~Brindley, M.~Takeda, P.~Plattet, and R.~K.~Plemper,
\textquotedblleft Triggering the measles virus membrane fusion machinery,\textquotedblright
\emph{Proceedings of the National Academy of Sciences}
\textbf{109}, E3018--E3027 (2012).
\href{https://doi.org/10.1073/pnas.1210925109}{doi:10.1073/pnas.1210925109}.

\bibitem{Misra2012}
P.~K.~Misra, \emph{Physics of Condensed Matter}, Academic Press
(2012 copyright edition), especially Chapters~1--2 and Section~8.11.
ISBN 978-0-12-384954-0.

\bibitem{Shankar1994}
R.~Shankar, \emph{Principles of Quantum Mechanics}, 2nd ed.,
Plenum Press (1994), especially Chapters~1 and~4--7.
\href{https://doi.org/10.1007/978-1-4757-0576-8}{doi:10.1007/978-1-4757-0576-8}.

\bibitem{Loewe2019}
D.~Loewe, J.~H\"aussler, T.~A.~Grein, H.~Dieken, T.~Weidner,
D.~Salzig, and P.~Czermak,
``Forced degradation studies to identify critical process parameters for
the purification of infectious measles virus,''
\emph{Viruses} \textbf{11}(8), 725 (2019).
\href{https://doi.org/10.3390/v11080725}{doi:10.3390/v11080725}.

\bibitem{Deriglazov2010}
A.~Deriglazov, \emph{Classical Mechanics}, Springer (2010),
especially Section~2.1.2, printed pp.~79--83.
ISBN 978-3-642-14036-5.

\bibitem{VinjanampathyAnders2016}
S.~Vinjanampathy and J.~Anders,
\textquotedblleft Quantum thermodynamics,\textquotedblright
\emph{Contemporary Physics} \textbf{57}(4), 545--579 (2016).
\href{https://doi.org/10.1080/00107514.2016.1201896}{doi:10.1080/00107514.2016.1201896}.

\bibitem{Potts2026}
P.~P.~Potts,
\textquotedblleft Quantum Thermodynamics,\textquotedblright
\emph{SciPost Physics Lecture Notes} \textbf{126} (2026).
\href{https://doi.org/10.21468/SciPostPhysLectNotes.126}{doi:10.21468/SciPostPhysLectNotes.126}.
Author version:
\href{https://arxiv.org/abs/2406.19206v3}{arXiv:2406.19206v3},
especially Sections~2.1.1, 3.3, and~8.3.

\bibitem{TalknerHanggi2020}
P.~Talkner and P.~H\"anggi,
\textquotedblleft Colloquium: Statistical mechanics and thermodynamics
at strong coupling: Quantum and classical,\textquotedblright
\emph{Reviews of Modern Physics} \textbf{92}, 041002 (2020),
especially Section~III.A.
\href{https://doi.org/10.1103/RevModPhys.92.041002}{doi:10.1103/RevModPhys.92.041002}.

\bibitem{Michel2006}
J.~P.~Michel, I.~L.~Ivanovska, M.~M.~Gibbons, W.~S.~Klug,
C.~M.~Knobler, G.~J.~L.~Wuite, and C.~F.~Schmidt,
``Nanoindentation studies of full and empty viral capsids and the effects
of capsid protein mutations on elasticity and strength,''
\emph{Proceedings of the National Academy of Sciences}
\textbf{103}(16), 6184--6189 (2006).
\href{https://doi.org/10.1073/pnas.0601744103}{doi:10.1073/pnas.0601744103}.

\bibitem{Lakes2009}
\begingroup\raggedright
R.~S.~Lakes, \emph{Viscoelastic Materials}, Cambridge University Press (2009).
\href{https://doi.org/10.1017/CBO9780511626722}{doi:10.1017/CBO9780511626722}.
\par\endgroup

\bibitem{Landau1986}
L.~D.~Landau and E.~M.~Lifshitz, \emph{Theory of Elasticity},
3rd ed., \emph{Course of Theoretical Physics}, Vol.~7 (1986),
revised by E.~M.~Lifshitz, A.~M.~Kosevich, and L.~P.~Pitaevskii.
\href{https://doi.org/10.1016/C2009-0-25521-8}{doi:10.1016/C2009-0-25521-8}.

\bibitem{Lubensky2015}
T.~C.~Lubensky, C.~L.~Kane, X.~Mao, A.~Souslov, and K.~Sun,
\textquotedblleft Phonons and elasticity in critically coordinated lattices,\textquotedblright
\emph{Reports on Progress in Physics} \textbf{78}, 073901 (2015),
\href{https://doi.org/10.1088/0034-4885/78/7/073901}{doi:10.1088/0034-4885/78/7/073901}.
Author version, especially Section~II:
\href{https://arxiv.org/abs/1503.01324}{arXiv:1503.01324}.

\bibitem{Keim2004}
P.~Keim, G.~Maret, U.~Herz, and H.~H.~von Gr\"unberg,
``Harmonic lattice behavior of two-dimensional colloidal crystals,''
\emph{Physical Review Letters} \textbf{92}(21), 215504 (2004).
\href{https://doi.org/10.1103/PhysRevLett.92.215504}{doi:10.1103/PhysRevLett.92.215504}.

\bibitem{HornJohnson2013}
R.~A.~Horn and C.~R.~Johnson, \emph{Matrix Analysis}, 2nd ed.,
Cambridge University Press (2013), especially Chapter~6.
\href{https://doi.org/10.1017/CBO9781139020411}{doi:10.1017/CBO9781139020411}.

\bibitem{Clerk2010}
A.~A.~Clerk, M.~H.~Devoret, S.~M.~Girvin, F.~Marquardt, and R.~J.~Schoelkopf,
\textquotedblleft Introduction to quantum noise, measurement, and amplification,\textquotedblright
\emph{Reviews of Modern Physics} \textbf{82}, 1155--1208 (2010),
especially Sections~II.A--II.B, IV.A, and Appendix~E.2.
\href{https://doi.org/10.1103/RevModPhys.82.1155}{doi:10.1103/RevModPhys.82.1155}.

\bibitem{Tomaiuolo2016}
G.~Tomaiuolo, A.~Carciati, S.~Caserta, and S.~Guido,
``Blood linear viscoelasticity by small amplitude oscillatory flow,''
\emph{Rheologica Acta} \textbf{55}(6), 485--495 (2016).
\href{https://doi.org/10.1007/s00397-015-0894-3}{doi:10.1007/s00397-015-0894-3}.

\bibitem{GuzmanSepulveda2022}
J.~R.~Guzman-Sepulveda, M.~Batarseh, R.~Wu, W.~M.~DeCampli, and A.~Dogariu,
``Passive high-frequency microrheology of blood,''
\emph{Soft Matter} \textbf{18}(12), 2452--2461 (2022).
\href{https://doi.org/10.1039/D1SM01726H}{doi:10.1039/D1SM01726H}.

\bibitem{Wolf2011}
M.~Wolf, R.~Gulich, P.~Lunkenheimer, and A.~Loidl,
``Broadband dielectric spectroscopy on human blood,''
\emph{Biochimica et Biophysica Acta---General Subjects}
\textbf{1810}(8), 727--740 (2011).
\href{https://doi.org/10.1016/j.bbagen.2011.05.012}{doi:10.1016/j.bbagen.2011.05.012}.

\bibitem{Brust2013}
M.~Brust, C.~Schaefer, R.~Doerr, L.~Pan, M.~Garcia, P.~E.~Arratia, and C.~Wagner,
``Rheology of Human Blood Plasma: Viscoelastic Versus Newtonian Behavior,''
\emph{Physical Review Letters} \textbf{110}(7), 078305 (2013).
\href{https://doi.org/10.1103/PhysRevLett.110.078305}{doi:10.1103/PhysRevLett.110.078305}.

\bibitem{Grimm2012}
M.~Grimm, T.~Franosch, and S.~Jeney,
\textquotedblleft High-resolution detection of Brownian motion for quantitative
optical tweezers experiments,\textquotedblright
\emph{Physical Review E} \textbf{86}, 021912 (2012).
\href{https://doi.org/10.1103/PhysRevE.86.021912}{doi:10.1103/PhysRevE.86.021912}.
Author version: \href{https://arxiv.org/abs/1208.1114}{arXiv:1208.1114}.

\bibitem{GrimmMaxwell2011}
M.~Grimm, S.~Jeney, and T.~Franosch,
\textquotedblleft Brownian motion in a Maxwell fluid,\textquotedblright
\emph{Soft Matter} \textbf{7}, 2076--2084 (2011).
\href{https://doi.org/10.1039/C0SM00636J}{doi:10.1039/C0SM00636J}.

\bibitem{Roylance2001}
D.~Roylance, \emph{Engineering Viscoelasticity}, Department of Materials
Science and Engineering, Massachusetts Institute of Technology (2001),
especially Sections~4--5.
\href{https://web.mit.edu/course/3/3.11/www/modules/visco.pdf}{MIT lecture notes}.

\bibitem{Hanyga2013}
A.~Hanyga,
\textquotedblleft Wave propagation in linear viscoelastic media with completely
monotonic relaxation moduli,\textquotedblright
\emph{Wave Motion} \textbf{50}(5), 909--928 (2013).
\href{https://doi.org/10.1016/j.wavemoti.2013.03.002}{doi:10.1016/j.wavemoti.2013.03.002}.
Author version: \href{https://arxiv.org/abs/1302.0402v2}{arXiv:1302.0402v2}.

\bibitem{Zwanzig1973}
R.~Zwanzig, ``Nonlinear generalized Langevin equations,''
\emph{Journal of Statistical Physics} \textbf{9}, 215--220 (1973).
\href{https://doi.org/10.1007/BF01008729}{doi:10.1007/BF01008729}.

\bibitem{Tokmakoff2009}
A.~Tokmakoff, ``Linear response theory,'' Lecture~7,
\emph{Introductory Quantum Mechanics II}, Massachusetts Institute of
Technology OpenCourseWare (Spring 2009).
\href{https://ocw.mit.edu/courses/5-74-introductory-quantum-mechanics-ii-spring-2009/resources/mit5_74s09_lec07/}{MIT course notes}.

\bibitem{Lindblad1976}
G.~Lindblad, ``On the generators of quantum dynamical semigroups,''
\emph{Communications in Mathematical Physics} \textbf{48}, 119--130 (1976).
\href{https://doi.org/10.1007/BF01608499}{doi:10.1007/BF01608499}.

\bibitem{Ford1988}
G.~W.~Ford, J.~T.~Lewis, and R.~F.~O'Connell, ``Quantum Langevin equation,''
\emph{Physical Review A} \textbf{37}, 4419--4428 (1988).
\href{https://doi.org/10.1103/PhysRevA.37.4419}{doi:10.1103/PhysRevA.37.4419}.

\bibitem{Georgi1992}
H.~Georgi, \emph{The Physics of Waves}, Benjamin Cummings (1992),
especially Chapter~1.
\href{https://ocw.mit.edu/courses/8-03sc-physics-iii-vibrations-and-waves-fall-2016/8eb46d081c5ba4728cbbb369a03ba45e_MIT8_03SCF16_Text_Ch1.pdf}{MIT-hosted textbook chapter}.

\bibitem{Wunsch2005}
C.~Wunsch, \emph{Frequency Domain Formulation}, Chapter~1 of
MIT~12.864, \emph{Inference from Data and Models}, Spring~2005,
especially Sections~1 and~18.
\href{https://ocw.mit.edu/courses/12-864-inference-from-data-and-models-spring-2005/ace73f7c12b91019a0ad145dc5313a1d_tsamsfmt_chap1.pdf}{Primary lecture notes}.

\bibitem{Teschl2009}
G.~Teschl, \emph{Mathematical Methods in Quantum Mechanics:
With Applications to Schr\"odinger Operators}, Graduate Studies in
Mathematics, Vol.~99, American Mathematical Society (2009),
especially Sections~1.1--1.3, 2.2--2.3, 3.1, 5.1, 8.1, 8.3, and~9.2.
\href{https://www.mat.univie.ac.at/~gerald/ftp/book-schroe/schroe.pdf}{Author-hosted online edition}.

\bibitem{Zwiebach2013}
B.~Zwiebach, \emph{Quantum Dynamics}, lecture notes for MIT~8.05,
\emph{Quantum Physics II}, Fall~2013, dated 4~November~2013,
especially Section~4.5.
\href{https://ocw.mit.edu/courses/8-05-quantum-physics-ii-fall-2013/79a8091ef4f18d8e3ff76a097e8db33c_MIT8_05F13_Chap_06.pdf}{Primary lecture notes}.

\bibitem{AtkinsFriedman2005}
P.~Atkins and R.~Friedman, \emph{Molecular Quantum Mechanics},
4th ed., Oxford University Press (2005), especially Chapter~1.
ISBN 978-0-19-927498-7.

\bibitem{NeumaierWestra2011}
A.~Neumaier and D.~Westra, \emph{Classical and Quantum Mechanics
via Lie algebras}, draft book, 14 April 2011,
arXiv:0810.1019v2, especially Sections~5.2, 8.2, 8.4, 11.5, 15.3, and~20.5.
\href{https://arxiv.org/abs/0810.1019v2}{Author-submitted manuscript}.

\bibitem{Sternberg2004}
S.~Sternberg, \emph{Lie algebras}, lecture notes, 23 April 2004,
especially Section~2.1.
\href{https://people.math.harvard.edu/~shlomo/docs/lie_algebras.pdf}{Harvard author-hosted notes}.

\bibitem{Ottobre2015}
\begingroup\raggedright
M.~Ottobre, G.~A.~Pavliotis, and K.~Pravda-Starov,
\textquotedblleft Some remarks on degenerate hypoelliptic
Ornstein--Uhlenbeck operators,\textquotedblright
\emph{Journal of Mathematical Analysis and Applications}
\textbf{429}(2), 676--712 (2015), especially Introduction,
Eqs.~(1.1)--(1.8).
\href{https://doi.org/10.1016/j.jmaa.2015.04.019}{doi:10.1016/j.jmaa.2015.04.019}.
\par\endgroup

\bibitem{Guff2022}
\begingroup\raggedright
T.~Guff, N.~Boulle, and I.~Pikovski,
\textquotedblleft Optimal fidelity witnesses for gravitational entanglement,\textquotedblright
\emph{Physical Review A} \textbf{105}, 022444 (2022),
especially Section~II~A.
\href{https://doi.org/10.1103/PhysRevA.105.022444}{doi:10.1103/PhysRevA.105.022444}.
\par\endgroup

\bibitem{Simon2000}
R.~Simon,
\textquotedblleft Peres--Horodecki separability criterion for
continuous variable systems,\textquotedblright
\emph{Physical Review Letters} \textbf{84}, 2726--2729 (2000),
especially Eqs.~(12), (14), (21), and the two-mode Gaussian theorem.
\href{https://doi.org/10.1103/PhysRevLett.84.2726}{doi:10.1103/PhysRevLett.84.2726}.

\bibitem{PeetersTaormina2009}
K.~Peeters and A.~Taormina,
``Group theory of icosahedral virus capsid vibrations: A top-down approach,''
\emph{Journal of Theoretical Biology} \textbf{256}(4), 607--624 (2009),
especially Sections~2.1 and Appendix~A.2.
\href{https://doi.org/10.1016/j.jtbi.2008.10.019}{doi:10.1016/j.jtbi.2008.10.019}.

\bibitem{Canestrari2025}
N.~Canestrari, D.~Nelli, and R.~Ferrando,
``General theory for packing icosahedral shells into multi-component aggregates,''
\emph{Nature Communications} \textbf{16}, 1655 (2025),
especially Introduction and the individual-shell construction.
\href{https://doi.org/10.1038/s41467-025-56952-1}{doi:10.1038/s41467-025-56952-1}.
Author version: \href{https://arxiv.org/abs/2406.07137v3}{arXiv:2406.07137v3}.

\bibitem{Eichenfield2009}
M.~Eichenfield, J.~Chan, R.~M.~Camacho, K.~J.~Vahala, and O.~Painter,
\textquotedblleft Optomechanical crystals,\textquotedblright
\emph{Nature} \textbf{462}, 78--82 (2009).
\href{https://doi.org/10.1038/nature08524}{doi:10.1038/nature08524}.
Author version: \href{https://arxiv.org/abs/0906.1236}{arXiv:0906.1236}.

\bibitem{Giustino2017}
F.~Giustino,
\textquotedblleft Electron--phonon interactions from first principles,\textquotedblright
\emph{Reviews of Modern Physics} \textbf{89}, 015003 (2017).
\href{https://doi.org/10.1103/RevModPhys.89.015003}{doi:10.1103/RevModPhys.89.015003}.
Author version: \href{https://arxiv.org/abs/1603.06965v2}{arXiv:1603.06965v2}.

\bibitem{ChenChen2021}
\begingroup\raggedright
H.-B.~Chen and Y.-N.~Chen,
\textquotedblleft Canonical Hamiltonian ensemble representation of
dephasing dynamics and the impact of thermal fluctuations on
quantum-to-classical transition,\textquotedblright
\emph{Scientific Reports} \textbf{11}, 10046 (2021),
especially Methods, Eqs.~(41)--(45).
\href{https://doi.org/10.1038/s41598-021-89400-3}{doi:10.1038/s41598-021-89400-3}.
\par\endgroup

\bibitem{HellmanKronik2026}
O.~Hellman and L.~Kronik,
\textquotedblleft Microscopic theory of phonon polaritons and long wavelength
dielectric response,\textquotedblright
arXiv:2505.03915v2, revised 12 January 2026.
\href{https://arxiv.org/abs/2505.03915v2}{Author-submitted manuscript}.

\bibitem{Mori2015}
Y.~Mori,
\textquotedblleft A multidomain model for ionic electrodiffusion and osmosis
with an application to cortical spreading depression,\textquotedblright
\emph{Physica D: Nonlinear Phenomena} \textbf{308}, 94--108 (2015).
\href{https://doi.org/10.1016/j.physd.2015.06.008}{doi:10.1016/j.physd.2015.06.008}.
Author version: \href{https://arxiv.org/abs/1410.8391}{arXiv:1410.8391}.

\bibitem{MoriLiuEisenberg2011}
Y.~Mori, C.~Liu, and R.~S.~Eisenberg,
\textquotedblleft A model of electrodiffusion and osmotic water flow and its
energetic structure,\textquotedblright
\emph{Physica D: Nonlinear Phenomena} \textbf{240}(22), 1835--1852 (2011).
\href{https://doi.org/10.1016/j.physd.2011.08.010}{doi:10.1016/j.physd.2011.08.010}.
Author version: \href{https://arxiv.org/abs/1101.5193}{arXiv:1101.5193}.

\bibitem{HodgkinHuxley1952}
A.~L.~Hodgkin and A.~F.~Huxley,
\textquotedblleft A quantitative description of membrane current and its
application to conduction and excitation in nerve,\textquotedblright
\emph{Journal of Physiology} \textbf{117}(4), 500--544 (1952).
\href{https://doi.org/10.1113/jphysiol.1952.sp004764}{doi:10.1113/jphysiol.1952.sp004764}.

\bibitem{IUPACMichaelis2025}
\begingroup\raggedright
International Union of Pure and Applied Chemistry,
\textquotedblleft Michaelis--Menten equation,\textquotedblright
\emph{Compendium of Chemical Terminology}, 5th ed., online version 5.0.0
(2025).
\href{https://doi.org/10.1351/goldbook.11546}{doi:10.1351/goldbook.11546}.
\par\endgroup

\bibitem{DeffnerCampbell2017}
S.~Deffner and S.~Campbell,
\textquotedblleft Quantum speed limits: from Heisenberg's uncertainty
principle to optimal quantum control,\textquotedblright
\emph{Journal of Physics A: Mathematical and Theoretical} \textbf{50},
453001 (2017).
\href{https://doi.org/10.1088/1751-8121/aa86c6}{doi:10.1088/1751-8121/aa86c6}.
Author version: \href{https://arxiv.org/abs/1705.08023}{arXiv:1705.08023}.
\bibitem{Hough2002}
L.~A.~Hough and H.~D.~Ou-Yang,
``Correlated motions of two hydrodynamically coupled particles confined in
separate quadratic potential wells,''
\emph{Physical Review E} \textbf{65}, 021906 (2002).
\href{https://doi.org/10.1103/PhysRevE.65.021906}{doi:10.1103/PhysRevE.65.021906}.

\bibitem{Paul2021}
S.~Paul, N.~Narinder, A.~Banerjee, K.~R.~Nayak, J.~Steindl,
and C.~Bechinger,
``Bayesian inference of the viscoelastic properties of a Jeffrey's fluid
using optical tweezers,''
\emph{Scientific Reports} \textbf{11}, 2023 (2021).
\href{https://doi.org/10.1038/s41598-021-81094-x}{doi:10.1038/s41598-021-81094-x}.

\bibitem{Simon2011}
S.~H.~Simon,
\emph{Lecture Notes for Solid State Physics}, Oxford University,
Hilary Term 2011, dated January 8, 2011, especially Sec.~5.2.3,
pp.~62--64.
\href{https://www-thphys.physics.ox.ac.uk/people/SteveSimon/condmat2011/LectureNotes.pdf}{Author-hosted course text}.

\bibitem{Harrow2016}
A.~Harrow,
\emph{Time-dependent approximation methods}, Chapter~2 of
MIT 8.06 Quantum Physics III, Spring 2016, revised March 12, 2016,
especially Secs.~1.2 and 1.4, pp.~3, 5--7.
\href{https://ocw.mit.edu/courses/8-06-quantum-physics-iii-spring-2016/0c27511c09675d8d385577023328248b_MIT8_06S16_chap2.pdf}{MIT course text}.

\bibitem{Stewart2014}
I.~Stewart,
\emph{Classical Mechanics III (8.09)}, MIT OpenCourseWare,
Fall 2014, Chapter~7, especially Secs.~7.4--7.4.2, pp.~172--177.
\href{https://ocw.mit.edu/courses/8-09-classical-mechanics-iii-fall-2014/a3b977a17091d225ee1870d100c7a308_MIT8_09F14_Chapter_7.pdf}{MIT course text}.

\bibitem{Tedrake2023}
R.~Tedrake,
\emph{Underactuated Robotics: Algorithms for Walking, Running, Swimming,
Flying, and Manipulation}, course notes for MIT 6.832 (2023),
Chapters~4, 9, and 16, accessed October 10, 2026.
\href{https://underactuated.mit.edu/simple_legs.html}{Limit cycles and the Van der Pol oscillator};
\href{https://underactuated.mit.edu/limit_cycles.html}{Orbital stability and transverse coordinates};
\href{https://underactuated.mit.edu/lyapunov.html}{Lyapunov analysis and the invariance principle}.

\bibitem{JulicherProst1997}
F.~J{\"u}licher and J.~Prost,
``Spontaneous oscillations of collective molecular motors,''
\emph{Physical Review Letters} \textbf{78}(23), 4510--4513 (1997).
\href{https://doi.org/10.1103/PhysRevLett.78.4510}{doi:10.1103/PhysRevLett.78.4510}.

\bibitem{Seifert2012}
U.~Seifert,
``Stochastic thermodynamics, fluctuation theorems and molecular machines,''
\emph{Reports on Progress in Physics} \textbf{75}(12), 126001 (2012).
\href{https://doi.org/10.1088/0034-4885/75/12/126001}{doi:10.1088/0034-4885/75/12/126001}.


\bibitem{Dhar2008}
A.~Dhar, \textquotedblleft Heat transport in low-dimensional systems,\textquotedblright
\emph{Advances in Physics} \textbf{57}(5), 457--537 (2008), especially Section~2.
\href{https://doi.org/10.1080/00018730802538522}{doi:10.1080/00018730802538522}.


\bibitem{Trefethen1997}
L.~N.~Trefethen,
\textquotedblleft Pseudospectra of linear operators,\textquotedblright
\emph{SIAM Review} \textbf{39}(3), 383--406 (1997),
especially Sections~1--2.
\href{https://doi.org/10.1137/S0036144595295284}{doi:10.1137/S0036144595295284}.


\bibitem{Dykeman2010}
E.~C.~Dykeman and O.~F.~Sankey,
``Atomistic modeling of the low-frequency mechanical modes and Raman spectra of icosahedral virus capsids,''
\emph{Physical Review E} \textbf{81}, 021918 (2010).
\href{https://doi.org/10.1103/PhysRevE.81.021918}{doi:10.1103/PhysRevE.81.021918}.

\bibitem{Acbas2014}
G.~Acbas, K.~A.~Niessen, E.~H.~Snell, and A.~G.~Markelz,
``Optical measurements of long-range protein vibrations,''
\emph{Nature Communications} \textbf{5}, 3076 (2014).
\href{https://doi.org/10.1038/ncomms4076}{doi:10.1038/ncomms4076}.


\bibitem{SquiresMason2010}
T.~M.~Squires and T.~G.~Mason,
``Tensorial generalized Stokes--Einstein relation for anisotropic probe
microrheology,'' \emph{Rheologica Acta} \textbf{49}, 1165--1177 (2010).
\href{https://doi.org/10.1007/s00397-010-0490-5}{doi:10.1007/s00397-010-0490-5}.

\bibitem{LeeSadeghpour2013}
T.~E.~Lee and H.~R.~Sadeghpour, ``Quantum synchronization of quantum
van der Pol oscillators with trapped ions,'' \emph{Physical Review Letters}
\textbf{111}, 234101 (2013),
\href{https://doi.org/10.1103/PhysRevLett.111.234101}{doi:10.1103/PhysRevLett.111.234101};
author manuscript \href{https://arxiv.org/abs/1306.6359}{arXiv:1306.6359}.

\bibitem{WalterNunnenkampBruder2014}
S.~Walter, A.~Nunnenkamp and C.~Bruder, ``Quantum synchronization of a
driven self-sustained oscillator,'' \emph{Physical Review Letters}
\textbf{112}, 094102 (2014),
\href{https://doi.org/10.1103/PhysRevLett.112.094102}{doi:10.1103/PhysRevLett.112.094102};
author manuscript \href{https://arxiv.org/abs/1307.7044}{arXiv:1307.7044}.

\bibitem{Moffat2008}
J.~C.~Moffat, V.~Vijayvergiya, P.~F.~Gao, T.~A.~Cross,
D.~J.~Woodbury, and D.~D.~Busath,
\textquotedblleft Proton transport through influenza A virus M2 protein reconstituted in vesicles,\textquotedblright
\emph{Biophysical Journal} \textbf{94}(2), 434--445 (2008).
\href{https://doi.org/10.1529/biophysj.107.109082}{doi:10.1529/biophysj.107.109082}.


\bibitem{Ivanovska2004}
I.~L.~Ivanovska, P.~J.~de Pablo, B.~Ibarra, G.~Sgalari, F.~C.~MacKintosh,
J.~L.~Carrascosa, C.~F.~Schmidt, and G.~J.~L.~Wuite,
``Bacteriophage capsids: Tough nanoshells with complex elastic properties,''
\emph{Proceedings of the National Academy of Sciences} \textbf{101}(20),
7600--7605 (2004).
\href{https://doi.org/10.1073/pnas.0308198101}{doi:10.1073/pnas.0308198101}.

\bibitem{Franosch2011}
T.~Franosch, M.~Grimm, M.~Belushkin, F.~M.~Mor, G.~Foffi, L.~Forr\'o,
and S.~Jeney,
``Resonances arising from hydrodynamic memory in Brownian motion,''
\emph{Nature} \textbf{478}, 85--88 (2011).
\href{https://doi.org/10.1038/nature10498}{doi:10.1038/nature10498}.

\bibitem{Crocker2000}
J.~C.~Crocker, M.~T.~Valentine, E.~R.~Weeks, T.~Gisler, P.~D.~Kaplan,
A.~G.~Yodh, and D.~A.~Weitz,
``Two-point microrheology of inhomogeneous soft materials,''
\emph{Physical Review Letters} \textbf{85}, 888--891 (2000).
\href{https://doi.org/10.1103/PhysRevLett.85.888}{doi:10.1103/PhysRevLett.85.888}.

\bibitem{CallenWelton1951}
H.~B.~Callen and T.~A.~Welton,
``Irreversibility and generalized noise,''
\emph{Physical Review} \textbf{83}, 34--40 (1951).
\href{https://doi.org/10.1103/PhysRev.83.34}{doi:10.1103/PhysRev.83.34}.

\bibitem{Raue2009}
A.~Raue, C.~Kreutz, T.~Maiwald, J.~Bachmann, M.~Schilling,
U.~Klingm\"uller, and J.~Timmer,
``Structural and practical identifiability analysis of partially observed
dynamical models by exploiting the profile likelihood,''
\emph{Bioinformatics} \textbf{25}(15), 1923--1929 (2009).
\href{https://doi.org/10.1093/bioinformatics/btp358}{doi:10.1093/bioinformatics/btp358}.

\bibitem{Noid2008}
W.~G.~Noid, J.-W.~Chu, G.~S.~Ayton, V.~Krishna, S.~Izvekov,
G.~A.~Voth, A.~Das, and H.~C.~Andersen,
``The multiscale coarse-graining method. I. A rigorous bridge between
atomistic and coarse-grained models,''
\emph{The Journal of Chemical Physics} \textbf{128}, 244114 (2008).
\href{https://doi.org/10.1063/1.2938860}{doi:10.1063/1.2938860}.

\bibitem{Mori1965}
H.~Mori, ``Transport, collective motion, and Brownian motion,''
\emph{Progress of Theoretical Physics} \textbf{33}(3), 423--455 (1965).
\href{https://doi.org/10.1143/PTP.33.423}{doi:10.1143/PTP.33.423}.

\bibitem{MasonWeitz1995}
T.~G.~Mason and D.~A.~Weitz,
``Optical measurements of frequency-dependent linear viscoelastic moduli
of complex fluids,''
\emph{Physical Review Letters} \textbf{74}, 1250--1253 (1995).
\href{https://doi.org/10.1103/PhysRevLett.74.1250}{doi:10.1103/PhysRevLett.74.1250}.


\bibitem{Welch1967}
P.~D.~Welch,
``The use of fast Fourier transform for the estimation of power spectra:
A method based on time averaging over short, modified periodograms,''
\emph{IEEE Transactions on Audio and Electroacoustics} \textbf{15}(2),
70--73 (1967).
\href{https://doi.org/10.1109/TAU.1967.1161901}{doi:10.1109/TAU.1967.1161901}.

\bibitem{Kalman1963}
R.~E.~Kalman, ``Mathematical description of linear dynamical systems,''
\emph{Journal of the Society for Industrial and Applied Mathematics,
Series A: Control} \textbf{1}(2), 152--192 (1963).
\href{https://doi.org/10.1137/0301010}{doi:10.1137/0301010}.

\bibitem{Berglund2010}
A.~J.~Berglund, ``Statistics of camera-based single-particle tracking,''
\emph{Physical Review E} \textbf{82}, 011917 (2010).
\href{https://doi.org/10.1103/PhysRevE.82.011917}{doi:10.1103/PhysRevE.82.011917}.

\bibitem{Glauber1963}
R.~J.~Glauber, ``Coherent and incoherent states of the radiation field,''
\emph{Physical Review} \textbf{131}, 2766--2788 (1963).
\href{https://doi.org/10.1103/PhysRev.131.2766}{doi:10.1103/PhysRev.131.2766}.

\bibitem{Robertson1929}
H.~P.~Robertson, ``The uncertainty principle,''
\emph{Physical Review} \textbf{34}, 163--164 (1929).
\href{https://doi.org/10.1103/PhysRev.34.163}{doi:10.1103/PhysRev.34.163}.

\bibitem{Robertson1934}
H.~P.~Robertson, ``An indeterminacy relation for several observables and
its classical interpretation,''
\emph{Physical Review} \textbf{46}, 794--801 (1934).
\href{https://doi.org/10.1103/PhysRev.46.794}{doi:10.1103/PhysRev.46.794}.

\bibitem{Kubo1957}
R.~Kubo, ``Statistical-mechanical theory of irreversible processes. I.
General theory and simple applications to magnetic and conduction problems,''
\emph{Journal of the Physical Society of Japan} \textbf{12}(6),
570--586 (1957).
\href{https://doi.org/10.1143/JPSJ.12.570}{doi:10.1143/JPSJ.12.570}.

\bibitem{VanHove1954}
L.~Van Hove, ``Correlations in space and time and Born approximation
scattering in systems of interacting particles,''
\emph{Physical Review} \textbf{95}, 249--262 (1954).
\href{https://doi.org/10.1103/PhysRev.95.249}{doi:10.1103/PhysRev.95.249}.

\bibitem{Peres1996}
A.~Peres, ``Separability criterion for density matrices,''
\emph{Physical Review Letters} \textbf{77}, 1413--1415 (1996).
\href{https://doi.org/10.1103/PhysRevLett.77.1413}{doi:10.1103/PhysRevLett.77.1413}.

\bibitem{Condon1928}
E.~U.~Condon, ``Nuclear motions associated with electron transitions in
diatomic molecules,''
\emph{Physical Review} \textbf{32}, 858--872 (1928).
\href{https://doi.org/10.1103/PhysRev.32.858}{doi:10.1103/PhysRev.32.858}.

\bibitem{Bahar1997}
I.~Bahar, A.~R.~Atilgan, and B.~Erman,
``Direct evaluation of thermal fluctuations in proteins using a
single-parameter harmonic potential,''
\emph{Folding \& Design} \textbf{2}(3), 173--181 (1997).
\href{https://doi.org/10.1016/S1359-0278(97)00024-2}{doi:10.1016/S1359-0278(97)00024-2}.

\bibitem{Teschl2000}
G.~Teschl, \emph{Jacobi Operators and Completely Integrable Nonlinear
Lattices}, Mathematical Surveys and Monographs, Vol.~72,
American Mathematical Society (2000), especially Chapter~1.
ISBN 978-0-8218-1940-1.
\href{https://www.mat.univie.ac.at/~gerald/ftp/book-jac/}{Author's book page}.

\bibitem{Bjork1957}
R.~L.~Bjork, ``Impurity-induced localized modes of lattice vibration in a
diatomic chain,'' \emph{Physical Review} \textbf{105}, 456 (1957).
\href{https://doi.org/10.1103/PhysRev.105.456}{doi:10.1103/PhysRev.105.456}.

\bibitem{DavisKahan1970}
C.~Davis and W.~M.~Kahan, ``The rotation of eigenvectors by a perturbation.
III,'' \emph{SIAM Journal on Numerical Analysis} \textbf{7}(1),
1--46 (1970).
\href{https://doi.org/10.1137/0707001}{doi:10.1137/0707001}.

\bibitem{Klemens1966}
P.~G.~Klemens, ``Anharmonic decay of optical phonons,''
\emph{Physical Review} \textbf{148}, 845--848 (1966).
\href{https://doi.org/10.1103/PhysRev.148.845}{doi:10.1103/PhysRev.148.845}.


\bibitem{Helfrich1973}
W.~Helfrich, ``Elastic properties of lipid bilayers: Theory and possible
experiments,'' \emph{Zeitschrift f\"ur Naturforschung C} \textbf{28},
693--703 (1973).
\href{https://doi.org/10.1515/znc-1973-11-1209}{doi:10.1515/znc-1973-11-1209}.

\bibitem{MilnerSafran1987}
S.~T.~Milner and S.~A.~Safran, ``Dynamical fluctuations of droplet
microemulsions and vesicles,'' \emph{Physical Review A} \textbf{36},
4371--4379 (1987).
\href{https://doi.org/10.1103/PhysRevA.36.4371}{doi:10.1103/PhysRevA.36.4371}.

\bibitem{SeifertLanger1993}
U.~Seifert and S.~A.~Langer, ``Viscous modes of fluid bilayer membranes,''
\emph{Europhysics Letters} \textbf{23}(1), 71--76 (1993).
\href{https://doi.org/10.1209/0295-5075/23/1/012}{doi:10.1209/0295-5075/23/1/012}.

\bibitem{Pecreaux2004}
J.~P\'ecr\'eaux, H.-G.~D\"obereiner, J.~Prost, J.-F.~Joanny, and
P.~Bassereau, ``Refined contour analysis of giant unilamellar vesicles,''
\emph{The European Physical Journal E} \textbf{13}, 277--290 (2004).
\href{https://doi.org/10.1140/epje/i2004-10001-9}{doi:10.1140/epje/i2004-10001-9}.

\bibitem{CooperDryden1984}
A.~Cooper and D.~T.~F.~Dryden, ``Allostery without conformational change:
A plausible model,'' \emph{European Biophysics Journal} \textbf{11},
103--109 (1984).
\href{https://doi.org/10.1007/BF00276625}{doi:10.1007/BF00276625}.

\bibitem{Deniz2000}
A.~A.~Deniz, T.~A.~Laurence, G.~S.~Beligere, M.~Dahan, A.~B.~Martin,
D.~S.~Chemla, P.~E.~Dawson, P.~G.~Schultz, and S.~Weiss,
``Single-molecule protein folding: Diffusion fluorescence resonance energy
transfer studies of the denaturation of chymotrypsin inhibitor~2,''
\emph{Proceedings of the National Academy of Sciences} \textbf{97}(10),
5179--5184 (2000).
\href{https://doi.org/10.1073/pnas.090104997}{doi:10.1073/pnas.090104997}.

\bibitem{NoeNuske2013}
F.~No\'e and F.~N\"uske, ``A variational approach to modeling slow
processes in stochastic dynamical systems,'' \emph{Multiscale Modeling
\& Simulation} \textbf{11}(2), 635--655 (2013).
\href{https://doi.org/10.1137/110858616}{doi:10.1137/110858616}.

\bibitem{PerezHernandez2013}
G.~P\'erez-Hern\'andez, F.~Paul, T.~Giorgino, G.~De~Fabritiis, and
F.~No\'e, ``Identification of slow molecular order parameters for Markov
model construction,'' \emph{The Journal of Chemical Physics}
\textbf{139}, 015102 (2013).
\href{https://doi.org/10.1063/1.4811489}{doi:10.1063/1.4811489}.

\bibitem{MeinersQuake1999}
J.-C.~Meiners and S.~R.~Quake, ``Direct measurement of hydrodynamic cross
correlations between two particles in an external potential,''
\emph{Physical Review Letters} \textbf{82}, 2211--2214 (1999).
\href{https://doi.org/10.1103/PhysRevLett.82.2211}{doi:10.1103/PhysRevLett.82.2211}.


\bibitem{HeadLevineMacKintosh2003}
D.~A.~Head, A.~J.~Levine, and F.~C.~MacKintosh,
``Deformation of cross-linked semiflexible polymer networks,''
\emph{Physical Review Letters} \textbf{91}(10), 108102 (2003).
\href{https://doi.org/10.1103/PhysRevLett.91.108102}{doi:10.1103/PhysRevLett.91.108102}.
Author-hosted full text:
\href{https://www.nat.vu.nl/~fcm/Papers/HeadPRL.pdf}{published article}.

\bibitem{Broedersz2010}
C.~P.~Broedersz, M.~Depken, N.~Y.~Yao, M.~R.~Pollak,
D.~A.~Weitz, and F.~C.~MacKintosh,
``Cross-link-governed dynamics of biopolymer networks,''
\emph{Physical Review Letters} \textbf{105}(23), 238101 (2010).
\href{https://doi.org/10.1103/PhysRevLett.105.238101}{doi:10.1103/PhysRevLett.105.238101}.
Author-hosted full text:
\href{https://www.nat.vu.nl/~fcm/Papers/PhysRevLett.105.238101.pdf}{published article}.

\bibitem{Dove2011}
M.~T.~Dove, ``Introduction to the theory of lattice dynamics,\textquotedblright
\emph{Collection SFN} \textbf{12}, 123--159 (2011).
\href{https://doi.org/10.1051/sfn/201112007}{doi:10.1051/sfn/201112007}.

\bibitem{WuVanderbiltHamann2005}
X.~Wu, D.~Vanderbilt, and D.~R.~Hamann,
``Systematic treatment of displacements, strains, and electric fields
in density-functional perturbation theory,''
\emph{Physical Review B} \textbf{72}, 035105 (2005).
\href{https://doi.org/10.1103/PhysRevB.72.035105}{doi:10.1103/PhysRevB.72.035105}.

\bibitem{GuazzelliMorris2012}
\'E.~Guazzelli and J.~F.~Morris,
\emph{A Physical Introduction to Suspension Dynamics},
Cambridge Texts in Applied Mathematics, Vol.~45,
Cambridge University Press (2012), especially Chapter~1 and Section~3.5.
\href{https://doi.org/10.1017/CBO9780511894671}{doi:10.1017/CBO9780511894671}.

\bibitem{LevineLubensky2001}
A.~J.~Levine and T.~C.~Lubensky,
``Response function of a sphere in a viscoelastic two-fluid medium,''
\emph{Physical Review E} \textbf{63}, 041510 (2001).
\href{https://doi.org/10.1103/PhysRevE.63.041510}{doi:10.1103/PhysRevE.63.041510}.
Author version: \href{https://arxiv.org/abs/cond-mat/0010027}{arXiv:cond-mat/0010027}.

\bibitem{CeriottiBussiParrinello2010}
M.~Ceriotti, G.~Bussi, and M.~Parrinello,
``Colored-noise thermostats \`a la carte,''
\emph{Journal of Chemical Theory and Computation} \textbf{6}(4),
1170--1180 (2010), especially Section~I.1, Eq.~(6), and Appendix~A.
\href{https://doi.org/10.1021/ct900563s}{doi:10.1021/ct900563s}.
Author version: \href{https://arxiv.org/abs/1204.0822}{arXiv:1204.0822}.


\bibitem{Beyn2012}
W.-J.~Beyn,
``An integral method for solving nonlinear eigenvalue problems,''
\emph{Linear Algebra and its Applications} \textbf{436}(10), 3839--3863 (2012),
especially Section~2, Theorem~2.4.
\href{https://doi.org/10.1016/j.laa.2011.03.030}{doi:10.1016/j.laa.2011.03.030}.
Author version: \href{https://arxiv.org/abs/1003.1580}{arXiv:1003.1580}.

\bibitem{GuettelTisseur2017}
S.~G\"uttel and F.~Tisseur,
``The nonlinear eigenvalue problem,''
\emph{Acta Numerica} \textbf{26}, 1--94 (2017),
especially Sections~2.2 and~2.6, and Theorem~2.20.
\href{https://doi.org/10.1017/S0962492917000034}{doi:10.1017/S0962492917000034}.
\href{https://pure.manchester.ac.uk/ws/files/55114004/GT17_final.pdf}{Accepted author manuscript}.

\bibitem{Onsager1931}
L.~Onsager, ``Reciprocal relations in irreversible processes. I,''
\emph{Physical Review} \textbf{37}(4), 405--426 (1931).
\href{https://doi.org/10.1103/PhysRev.37.405}{doi:10.1103/PhysRev.37.405}.

\bibitem{Doi2011}
M.~Doi, ``Onsager's variational principle in soft matter,''
\emph{Journal of Physics: Condensed Matter} \textbf{23}(28), 284118 (2011).
\href{https://doi.org/10.1088/0953-8984/23/28/284118}{doi:10.1088/0953-8984/23/28/284118}.

\bibitem{Willems1972}
J.~C.~Willems, ``Dissipative dynamical systems part I: General theory,''
\emph{Archive for Rational Mechanics and Analysis} \textbf{45},
321--351 (1972).
\href{https://doi.org/10.1007/BF00276493}{doi:10.1007/BF00276493}.

\bibitem{Khalil2002}
H.~K.~Khalil,
\emph{Nonlinear Systems}, 3rd ed. (Prentice Hall, Upper Saddle River, NJ,
2002), Section~4.2 (the invariance principle).
\href{https://www.pearson.com/en-ca/subject-catalog/p/nonlinear-systems/P200000003306/9780130673893}{Publisher record};
\href{https://www.egr.msu.edu/~khalil/NonlinearSystems/}{Author's book page}.

\bibitem{Villaverde2016}
A.~F.~Villaverde, A.~Barreiro, and A.~Papachristodoulou,
``Structural identifiability of dynamic systems biology models,''
\emph{PLOS Computational Biology} \textbf{12}(10), e1005153 (2016).
\href{https://doi.org/10.1371/journal.pcbi.1005153}{doi:10.1371/journal.pcbi.1005153}.

\bibitem{SquiresMason2010Review}
T.~M.~Squires and T.~G.~Mason,
``Fluid mechanics of microrheology,''
\emph{Annual Review of Fluid Mechanics} \textbf{42}, 413--438 (2010).
\href{https://doi.org/10.1146/annurev-fluid-121108-145608}{doi:10.1146/annurev-fluid-121108-145608}.

\bibitem{Wunsch2005BasicMachinery}
C.~Wunsch, \emph{Basic Machinery}, Chapter~2 of
MIT~12.864, \emph{Inference from Data and Models}, Spring~2005,
especially Section~2.4.
\href{https://ocw.mit.edu/courses/12-864-inference-from-data-and-models-spring-2005/7b21edbfc62016cc3575242c344ea9a4_chap2lg.pdf}{Primary lecture notes}.

\bibitem{Picinbono1996}
B.~Picinbono, ``Second-order complex random vectors and normal distributions,''
\emph{IEEE Transactions on Signal Processing} \textbf{44}(10),
2637--2640 (1996).
\href{https://doi.org/10.1109/78.539051}{doi:10.1109/78.539051}.

\bibitem{Weedbrook2012}
C.~Weedbrook, S.~Pirandola, R.~Garc\'ia-Patr\'on, N.~J.~Cerf,
T.~C.~Ralph, J.~H.~Shapiro, and S.~Lloyd,
``Gaussian quantum information,''
\emph{Reviews of Modern Physics} \textbf{84}, 621--669 (2012),
especially Section~II.A.2.
\href{https://doi.org/10.1103/RevModPhys.84.621}{doi:10.1103/RevModPhys.84.621};
\href{https://arxiv.org/abs/1110.3234}{Author-submitted manuscript}.

\bibitem{ClermontZenker2015}
G.~Clermont and S.~Zenker, ``The inverse problem in mathematical biology,''
\emph{Mathematical Biosciences} \textbf{260}, 11--15 (2015).
\href{https://doi.org/10.1016/j.mbs.2014.09.001}{doi:10.1016/j.mbs.2014.09.001}.

\bibitem{Ercole2016}
L.~Ercole, A.~Marcolongo, P.~Umari, and S.~Baroni,
``Gauge invariance of thermal transport coefficients,''
\emph{Journal of Low Temperature Physics} \textbf{185}, 79--86 (2016).
\href{https://doi.org/10.1007/s10909-016-1617-6}{doi:10.1007/s10909-016-1617-6};
\href{https://arxiv.org/abs/1601.00158}{Author-submitted manuscript}.

\bibitem{Willems1972Linear}
J.~C.~Willems, ``Dissipative dynamical systems Part II: Linear systems
with quadratic supply rates,''
\emph{Archive for Rational Mechanics and Analysis} \textbf{45},
352--393 (1972).
\href{https://doi.org/10.1007/BF00276494}{doi:10.1007/BF00276494}.

\bibitem{Womersley1955}
J.~R.~Womersley, ``Method for the calculation of velocity, rate of flow
and viscous drag in arteries when the pressure gradient is known,''
\emph{The Journal of Physiology} \textbf{127}(3), 553--563 (1955).
\href{https://doi.org/10.1113/jphysiol.1955.sp005276}{doi:10.1113/jphysiol.1955.sp005276}.

\bibitem{CasanellasOrtin2011}
L.~Casanellas and J.~Ort\'in, ``Laminar oscillatory flow of Maxwell and
Oldroyd-B fluids: Theoretical analysis,''
\emph{Journal of Non-Newtonian Fluid Mechanics} \textbf{166}(23--24),
1315--1326 (2011).
\href{https://doi.org/10.1016/j.jnnfm.2011.08.010}{doi:10.1016/j.jnnfm.2011.08.010}.

\bibitem{Verwulgen2018}
S.~Verwulgen, K.~Beyers, T.~Van~Mulder, T.~Peeters, S.~Truijen,
F.~Dams, and V.~Vankerckhoven,
``Assessment of forces in intradermal injection devices: Hydrodynamic
versus human factors,''
\emph{Pharmaceutical Research} \textbf{35}(6), 120 (2018).
\href{https://doi.org/10.1007/s11095-018-2397-2}{doi:10.1007/s11095-018-2397-2}.

\bibitem{Thueer2018}
T.~Thueer, L.~Birkhaeuer, and D.~Reilly,
``Development of an advanced injection time model for an autoinjector,''
\emph{Medical Devices: Evidence and Research} \textbf{11}, 215--224 (2018).
\href{https://doi.org/10.2147/MDER.S151727}{doi:10.2147/MDER.S151727}.

\bibitem{Staelin2011}
D.~H.~Staelin, \emph{Electromagnetics and Applications}, MIT course
notes (2011), Sections~2.7 and~13.1--13.3.
\url{https://ocw.mit.edu/courses/6-013-electromagnetics-and-applications-spring-2009/d3be4ea78b036a6362230fb41780cf54_MIT6_013S09_notes.pdf}.

\bibitem{Silva2014AbsorbingParticle}
G.~T.~Silva, ``Acoustic radiation force and torque on an absorbing
compressible particle in an inviscid fluid,'' \emph{J. Acoust. Soc. Am.}
\textbf{136}, 2405--2413 (2014).
\url{https://doi.org/10.1121/1.4895691}.

\bibitem{FitzpatrickSphericalWaves}
R.~Fitzpatrick, ``Spherical Waves,'' in \emph{Waves and Oscillations},
University of Texas at Austin, online lecture text.
\url{https://farside.ph.utexas.edu/teaching/315/Waves/node55.html}.

\bibitem{EVandenEijnden2006}
W.~E and E.~Vanden-Eijnden,
``Towards a theory of transition paths,''
\emph{Journal of Statistical Physics} \textbf{123}, 503--523 (2006).
\url{https://doi.org/10.1007/s10955-005-9003-9}.

\bibitem{SzaboSchultenSchulten1980}
A.~Szabo, K.~Schulten and Z.~Schulten,
``First passage time approach to diffusion controlled reactions,''
\emph{The Journal of Chemical Physics} \textbf{72}, 4350--4357 (1980).
\url{https://doi.org/10.1063/1.439715}.

\bibitem{Kramers1940}
H.~A.~Kramers,
``Brownian motion in a field of force and the diffusion model of chemical reactions,''
\emph{Physica} \textbf{7}, 284--304 (1940).
\url{https://doi.org/10.1016/S0031-8914(40)90098-2}.

\bibitem{HanggiTalknerBorkovec1990}
P.~H\"anggi, P.~Talkner and M.~Borkovec,
``Reaction-rate theory: fifty years after Kramers,''
\emph{Reviews of Modern Physics} \textbf{62}, 251--341 (1990).
\url{https://doi.org/10.1103/RevModPhys.62.251}.

\bibitem{DudkoHummerSzabo2006}
O.~K.~Dudko, G.~Hummer and A.~Szabo,
``Intrinsic rates and activation free energies from single-molecule pulling experiments,''
\emph{Physical Review Letters} \textbf{96}, 108101 (2006).
\url{https://doi.org/10.1103/PhysRevLett.96.108101}.

\bibitem{RoylanceFEA2001}
D.~Roylance, \emph{Finite Element Analysis}, Department of Materials
Science and Engineering, Massachusetts Institute of Technology
(28 February 2001).
\href{https://ocw.mit.edu/courses/3-11-mechanics-of-materials-fall-1999/812a432c72099369788f78fb344eb047_MIT3_11F99_fea.pdf}{MIT lecture notes}.

\bibitem{CraigBampton1968}
R.~R.~Craig, Jr., and M.~C.~C.~Bampton,
\textquotedblleft Coupling of substructures for dynamic analyses,\textquotedblright
\emph{AIAA Journal} \textbf{6}(7), 1313--1319 (1968).
\href{https://doi.org/10.2514/3.4741}{doi:10.2514/3.4741}.

\bibitem{BatheDong2014}
K.-J.~Bathe and J.~Dong,
\textquotedblleft Component mode synthesis with subspace iterations for
controlled accuracy of frequency and mode shape solutions,\textquotedblright
\emph{Computers \& Structures} \textbf{139}, 28--32 (2014).
\href{https://doi.org/10.1016/j.compstruc.2014.03.003}{doi:10.1016/j.compstruc.2014.03.003}.

\bibitem{TongSolidState2017}
D.~Tong, \emph{Solid State Physics}, University of Cambridge lecture notes
(2017), Chapter~4.
\href{https://davidtong.org/teaching/solid-state-physics/}{Author's course and notes}.

\bibitem{Stone2000}
M.~Stone, \textquotedblleft Phonons and Forces: Momentum versus
Pseudomomentum in Moving Fluids,\textquotedblright
arXiv:cond-mat/0012316 (2000).
\href{https://arxiv.org/abs/cond-mat/0012316}{Author manuscript}.

\bibitem{EspositoKrichevskyNicolis2019}
A.~Esposito, R.~Krichevsky, and A.~Nicolis,
\textquotedblleft Gravitational Mass Carried by Sound Waves,\textquotedblright
\emph{Physical Review Letters} \textbf{122}, 084501 (2019).
\href{https://doi.org/10.1103/PhysRevLett.122.084501}{doi:10.1103/PhysRevLett.122.084501}.

\bibitem{Soper2020}
D.~E.~Soper, \textquotedblleft Sound waves move matter,\textquotedblright
\emph{Physical Review Research} \textbf{2}, 013258 (2020).
\href{https://doi.org/10.1103/PhysRevResearch.2.013258}{doi:10.1103/PhysRevResearch.2.013258}.

\bibitem{AspelmeyerKippenbergMarquardt2014}
M.~Aspelmeyer, T.~J.~Kippenberg, and F.~Marquardt,
\textquotedblleft Cavity optomechanics,\textquotedblright
\emph{Reviews of Modern Physics} \textbf{86}, 1391--1452 (2014).
\href{https://doi.org/10.1103/RevModPhys.86.1391}{doi:10.1103/RevModPhys.86.1391}.

\bibitem{WangWangLu2008}
J.-S.~Wang, J.~Wang, and J.-T.~L{\"u},
\textquotedblleft Quantum thermal transport in nanostructures,\textquotedblright
\emph{European Physical Journal B} \textbf{62}, 381--404 (2008).
\href{https://doi.org/10.1140/epjb/e2008-00195-8}{doi:10.1140/epjb/e2008-00195-8}.

\bibitem{Tamura1983}
S.-i.~Tamura, \textquotedblleft Isotope scattering of dispersive phonons
in Ge,\textquotedblright
\emph{Physical Review B} \textbf{27}, 858--866 (1983).
\href{https://doi.org/10.1103/PhysRevB.27.858}{doi:10.1103/PhysRevB.27.858}.

\bibitem{Anderson1954}
P.~W.~Anderson,
``A Mathematical Model for the Narrowing of Spectral Lines by Exchange or Motion,''
\emph{Journal of the Physical Society of Japan} \textbf{9}, 316--339 (1954).
\href{https://doi.org/10.1143/JPSJ.9.316}{doi:10.1143/JPSJ.9.316}.

\bibitem{MengPritchardTerentjev2016}
F.~Meng, R.~H.~Pritchard, and E.~M.~Terentjev,
``Stress relaxation, dynamics, and plasticity of transient polymer networks,''
\emph{Macromolecules} \textbf{49}(7), 2843--2852 (2016).
\href{https://doi.org/10.1021/acs.macromol.5b02667}{doi:10.1021/acs.macromol.5b02667}.

\bibitem{Higham1986}
N.~J.~Higham, ``Computing the polar decomposition---with applications,''
\emph{SIAM Journal on Scientific and Statistical Computing}
\textbf{7}(4), 1160--1174 (1986).
\href{https://doi.org/10.1137/0907079}{doi:10.1137/0907079}.

\bibitem{SchreierScharf2003}
P.~J.~Schreier and L.~L.~Scharf,
``Second-order analysis of improper complex random vectors and processes,''
\emph{IEEE Transactions on Signal Processing} \textbf{51}(3),
714--725 (2003).
\href{https://doi.org/10.1109/TSP.2002.808085}{doi:10.1109/TSP.2002.808085}.

\bibitem{Liberzon2003}
D.~Liberzon, \emph{Switching in Systems and Control},
Systems \& Control: Foundations \& Applications, Birkh\"auser (2003).
\href{https://doi.org/10.1007/978-1-4612-0017-8}{doi:10.1007/978-1-4612-0017-8}.

\bibitem{LevyKosloff2014}
A.~Levy and R.~Kosloff, ``The local approach to quantum transport may
violate the second law of thermodynamics,'' \emph{EPL}
\textbf{107}(2), 20004 (2014).
\href{https://doi.org/10.1209/0295-5075/107/20004}{doi:10.1209/0295-5075/107/20004}.

\bibitem{Martinez2016}
I.~A.~Mart\'inez, A.~Petrosyan, D.~Gu\'ery-Odelin, E.~Trizac,
and S.~Ciliberto, ``Engineered swift equilibration of a Brownian particle,''
\emph{Nature Physics} \textbf{12}, 843--846 (2016).
\href{https://doi.org/10.1038/nphys3758}{doi:10.1038/nphys3758}.

\bibitem{SarkkaSolin2019} S.~S\"arkk\"a and A.~Solin,
\emph{Applied Stochastic Differential Equations},
Cambridge University Press (2019), Chapters~6 and~10.
\href{https://doi.org/10.1017/9781108186735}{doi:10.1017/9781108186735}.

\bibitem{Calderon2016} C.~P.~Calderon,
``Motion blur filtering: A statistical approach for extracting confinement
forces and diffusivity from a single blurred trajectory,''
\emph{Physical Review E} \textbf{93}, 053303 (2016).
\href{https://doi.org/10.1103/PhysRevE.93.053303}{doi:10.1103/PhysRevE.93.053303}.

\bibitem{Goetschius2021}
D.~J.~Goetschius, S.~R.~Hartmann, S.~Subramanian, C.~M.~Bator,
N.~D.~Christensen, and S.~L.~Hafenstein,
``High resolution cryo EM analysis of HPV16 identifies minor structural
protein L2 and describes capsid flexibility,''
\emph{Scientific Reports} \textbf{11}, 3498 (2021).
\href{https://doi.org/10.1038/s41598-021-83076-5}{doi:10.1038/s41598-021-83076-5}.

\bibitem{Zhao2014}
Q.~Zhao, C.~S.~Potter, B.~Carragher, et al.,
``Characterization of virus-like particles in GARDASIL by cryo transmission
electron microscopy,''
\emph{Human Vaccines \& Immunotherapeutics} \textbf{10}(3), 734--739 (2014).
\href{https://doi.org/10.4161/hv.27316}{doi:10.4161/hv.27316}.

\bibitem{Vega2016}
J.~F.~Vega, E.~Vicente-Alique, R.~N\'u\~nez-Ram\'irez,
Y.~Wang, and J.~Mart\'inez-Salazar,
``Evidences of changes in surface electrostatic charge distribution
during stabilization of HPV16 virus-like particles,''
\emph{PLoS ONE} \textbf{11}(2), e0149009 (2016).
\href{https://doi.org/10.1371/journal.pone.0149009}{doi:10.1371/journal.pone.0149009}.

\bibitem{Feng2024}
Y.~Feng, D.~van Bodegraven, A.~K\'adek, I.~L.~B.~Munguira,
L.~Soria-Martinez, S.~Nentwich, S.~Saha, F.~Chardon, D.~Kavan,
C.~Uetrecht, M.~Schelhaas, and W.~H.~Roos,
``Glycan-induced structural activation softens the human papillomavirus
capsid for entry through reduction of intercapsomere flexibility,''
\emph{Nature Communications} \textbf{15}, 10076 (2024).
\href{https://doi.org/10.1038/s41467-024-54373-0}{doi:10.1038/s41467-024-54373-0}.

\bibitem{Hamaker1937}
H.~C.~Hamaker,
``The London--van der Waals attraction between spherical particles,''
\emph{Physica} \textbf{4}(10), 1058--1072 (1937).
\href{https://doi.org/10.1016/S0031-8914(37)80203-7}{doi:10.1016/S0031-8914(37)80203-7}.

\bibitem{HoggHealyFuerstenau1966}
R.~Hogg, T.~W.~Healy, and D.~W.~Fuerstenau,
``Mutual coagulation of colloidal dispersions,''
\emph{Transactions of the Faraday Society} \textbf{62}, 1638--1651 (1966).
\href{https://doi.org/10.1039/TF9666201638}{doi:10.1039/TF9666201638}.

\bibitem{NinhamParsegian1971}
B.~W.~Ninham and V.~A.~Parsegian,
``Electrostatic potential between surfaces bearing ionizable groups in
ionic equilibrium with physiologic saline solution,''
\emph{Journal of Theoretical Biology} \textbf{31}(3), 405--428 (1971).
\href{https://doi.org/10.1016/0022-5193(71)90019-1}{doi:10.1016/0022-5193(71)90019-1}.

\bibitem{BozicPodgornik2013}
A.~Lo\v{s}dorfer Bo\v{z}i\v{c} and R.~Podgornik,
``Symmetry effects in electrostatic interactions between two arbitrarily
charged spherical shells in the Debye--H\"uckel approximation,''
\emph{The Journal of Chemical Physics} \textbf{138}, 074902 (2013).
\href{https://doi.org/10.1063/1.4790576}{doi:10.1063/1.4790576}.

\bibitem{BozicPodgornik2018}
A.~Lo\v{s}dorfer Bo\v{z}i\v{c} and R.~Podgornik,
``Anomalous multipole expansion: Charge regulation of patchy
inhomogeneously charged spherical particles,''
\emph{The Journal of Chemical Physics} \textbf{149}, 163307 (2018).
\href{https://doi.org/10.1063/1.5037044}{doi:10.1063/1.5037044}.

\bibitem{Gnidovec2025}
A.~Gnidovec, E.~Locatelli, S.~\v{C}opar, A.~Bo\v{z}i\v{c}, and E.~Bianchi,
``Anisotropic DLVO-like interaction for charge patchiness in colloids
and proteins,''
\emph{Nature Communications} \textbf{16}, 4277 (2025).
\href{https://doi.org/10.1038/s41467-025-58991-0}{doi:10.1038/s41467-025-58991-0}.

\bibitem{AsakuraOosawa1958}
S.~Asakura and F.~Oosawa,
``Interaction between particles suspended in solutions of macromolecules,''
\emph{Journal of Polymer Science} \textbf{33}(126), 183--192 (1958).
\href{https://doi.org/10.1002/pol.1958.1203312618}{doi:10.1002/pol.1958.1203312618}.

\bibitem{DelongUsabiagaDonev2015}
S.~Delong, F.~Balboa Usabiaga, and A.~Donev,
``Brownian dynamics of confined rigid bodies,''
\emph{The Journal of Chemical Physics} \textbf{143}, 144107 (2015).
\href{https://doi.org/10.1063/1.4932062}{doi:10.1063/1.4932062}.

\bibitem{Eringen1966}
A.~C.~Eringen, ``Linear theory of micropolar elasticity,''
\emph{Journal of Mathematics and Mechanics} \textbf{15}(6), 909--923 (1966).
\href{https://doi.org/10.1512/iumj.1966.15.15060}{doi:10.1512/iumj.1966.15.15060}.

\bibitem{Eringen1968NASA}
A.~C.~Eringen, \emph{Compatibility Conditions of the Theory of
Micromorphic Elastic Solids}, NASA Contractor Report CR-1236 (1968),
Section~4.
\href{https://ntrs.nasa.gov/citations/19690001486}{NASA Technical Reports Server}.

\bibitem{LinPonceMarzari2022}
C.~Lin, S.~Ponc\'e, and N.~Marzari,
``General invariance and equilibrium conditions for lattice dynamics in
1D, 2D, and 3D materials,'' \emph{npj Computational Materials}
\textbf{8}, 236 (2022).
\href{https://doi.org/10.1038/s41524-022-00920-6}{doi:10.1038/s41524-022-00920-6}.

\bibitem{ShoupLipariSzabo1981}
D.~Shoup, G.~Lipari, and A.~Szabo,
``Diffusion-controlled bimolecular reaction rates. The effect of rotational
diffusion and orientation constraints,'' \emph{Biophysical Journal}
\textbf{36}(3), 697--714 (1981).
\href{https://doi.org/10.1016/S0006-3495(81)84759-5}{doi:10.1016/S0006-3495(81)84759-5}.

\bibitem{Langer1969}
J.~S.~Langer, ``Statistical theory of the decay of metastable states,''
\emph{Annals of Physics} \textbf{54}(2), 258--275 (1969).
\href{https://doi.org/10.1016/0003-4916(69)90153-5}{doi:10.1016/0003-4916(69)90153-5}.

\bibitem{GroteHynes1980}
R.~F.~Grote and J.~T.~Hynes, ``The stable states picture of chemical
reactions. II. Rate constants for condensed and gas phase reaction
models,'' \emph{The Journal of Chemical Physics} \textbf{73}(6),
2715--2732 (1980).
\href{https://doi.org/10.1063/1.440485}{doi:10.1063/1.440485}.

\bibitem{HaganElradJack2011}
M.~F.~Hagan, O.~M.~Elrad, and R.~L.~Jack,
``Mechanisms of kinetic trapping in self-assembly and phase transformation,''
\emph{The Journal of Chemical Physics} \textbf{135}, 104115 (2011).
\href{https://doi.org/10.1063/1.3635775}{doi:10.1063/1.3635775}.

\bibitem{SzermerOlearnik2017}
B.~Szermer-Olearnik, M.~Drab, M.~M\k{a}kosa, M.~Zembala,
J.~Barbasz, K.~D\k{a}browska, and J.~Boraty\'nski,
``Aggregation/dispersion transitions of T4 phage triggered by
environmental ion availability,'' \emph{Journal of Nanobiotechnology}
\textbf{15}, 32 (2017).
\href{https://doi.org/10.1186/s12951-017-0266-5}{doi:10.1186/s12951-017-0266-5}.

\bibitem{KimKarrila1991}
S.~Kim and S.~J.~Karrila, \emph{Microhydrodynamics: Principles and
Selected Applications}, Butterworth--Heinemann (1991), Chapters~2--3.
\href{https://shop.elsevier.com/books/microhydrodynamics/kim/978-0-7506-9173-4}{Publisher record, ISBN 978-0-7506-9173-4}.

\bibitem{MortensenOlesenBruus2006}
N.~A.~Mortensen, L.~H.~Olesen, and H.~Bruus,
``Transport coefficients for electrolytes in arbitrarily shaped nano-
and microfluidic channels,'' \emph{New Journal of Physics} \textbf{8},
37 (2006).
\href{https://doi.org/10.1088/1367-2630/8/3/037}{doi:10.1088/1367-2630/8/3/037}.

\bibitem{OhgaItoKolchinsky2023}
N.~Ohga, S.~Ito, and A.~Kolchinsky,
``Thermodynamic Bound on the Asymmetry of Cross-Correlations,''
\emph{Physical Review Letters} \textbf{131}, 077101 (2023).
\href{https://doi.org/10.1103/PhysRevLett.131.077101}{doi:10.1103/PhysRevLett.131.077101}.

\bibitem{Backlund2012}
M.~P.~Backlund, M.~D.~Lew, A.~S.~Backer, S.~J.~Sahl, G.~Grover,
A.~Agrawal, R.~Piestun, and W.~E.~Moerner,
``Simultaneous, accurate measurement of the 3D position and orientation
of single molecules,'' \emph{Proceedings of the National Academy of
Sciences USA} \textbf{109}(47), 19087--19092 (2012).
\href{https://doi.org/10.1073/pnas.1216687109}{doi:10.1073/pnas.1216687109}.

\bibitem{SvergunBarberatoKoch1995}
D.~Svergun, C.~Barberato, and M.~H.~J.~Koch,
``CRYSOL---a Program to Evaluate X-ray Solution Scattering of Biological
Macromolecules from Atomic Coordinates,''
\emph{Journal of Applied Crystallography} \textbf{28}, 768--773 (1995).
\href{https://doi.org/10.1107/S0021889895007047}{doi:10.1107/S0021889895007047}.

\bibitem{StoperaMorales2022}
C.~Stopera and J.~A.~Morales,
``Temporally stable rotational coherent states for molecular simulations.
II. Symmetric rotor case,'' \emph{Journal of Chemical Physics}
\textbf{156}, 224114 (2022).
\href{https://doi.org/10.1063/5.0092167}{doi:10.1063/5.0092167}.

\bibitem{BrownErnstSheppWolpert2014}
L.~Brown, P.~Ernst, L.~Shepp, and R.~Wolpert,
``The General Stationary Gaussian Markov Process,''
arXiv:1401.0251 [math.PR] (2014).
\href{https://doi.org/10.48550/arXiv.1401.0251}{doi:10.48550/arXiv.1401.0251}.

\bibitem{TegmarkTaylorHeavens1997}
M.~Tegmark, A.~N.~Taylor, and A.~F.~Heavens,
``Karhunen--Lo\`eve Eigenvalue Problems in Cosmology: How Should We
Tackle Large Data Sets?'' \emph{Astrophysical Journal} \textbf{480},
22--35 (1997).
\href{https://doi.org/10.1086/303939}{doi:10.1086/303939}.

\bibitem{Klein2025}
S.~M.~Klein, A.~Patterson, K.~Young, M.~P.~Biever, L.~M.~Miller,
A.~Zlotnick, S.~C.~Jacobson, and M.~F.~Jarrold,
``Speed Matters: Directed Assembly of Icosahedral HPV Virus-Like
Particles,'' \emph{Journal of the American Chemical Society}
\textbf{147}(28), 24950--24957 (2025).
\href{https://doi.org/10.1021/jacs.5c07472}{doi:10.1021/jacs.5c07472}.

\bibitem{Langley2026}
C.~H.~Langley, D.~J.~Goetschius, S.~Antol\'inez, E.~P.~Orji,
C.~M.~Bator, S.~A.~Brendle, N.~D.~Christensen,
J.~A.~Hadden-Perilla, and S.~L.~Hafenstein,
``Atomic resolution structure of human papillomavirus bound to heparin,''
\emph{Nature Communications} \textbf{17}, 10239 (2026).
\href{https://doi.org/10.1038/s41467-026-77019-9}{doi:10.1038/s41467-026-77019-9}.

\bibitem{Kononova2016}
O.~Kononova, J.~Snijder, Y.~Kholodov, K.~A.~Marx,
G.~J.~L.~Wuite, W.~H.~Roos, and V.~Barsegov,
``Fluctuating Nonlinear Spring Model of Mechanical Deformation of
Biological Particles,'' \emph{PLoS Computational Biology}
\textbf{12}(1), e1004729 (2016).
\href{https://doi.org/10.1371/journal.pcbi.1004729}{doi:10.1371/journal.pcbi.1004729}.

\bibitem{OrtegaEsteban2013}
A.~Ortega-Esteban, A.~J.~P\'erez-Bern\'a, R.~Men\'endez-Conejero,
S.~J.~Flint, C.~San Mart\'in, and P.~J.~de Pablo,
``Monitoring dynamics of human adenovirus disassembly induced by
mechanical fatigue,'' \emph{Scientific Reports} \textbf{3}, 1434 (2013).
\href{https://doi.org/10.1038/srep01434}{doi:10.1038/srep01434}.

\bibitem{KleessOpenSystem2025}
L.~St.~Kleess,
``Open-System Virus Particle Physics: A Path-Integral Viral Lattice Theory
Using Non-Self Adjoint Stochastic PDEs and Fock-Space Formalism,''
arXiv:2502.12318 [physics.bio-ph] (2025).
\href{https://doi.org/10.48550/arXiv.2502.12318}{doi:10.48550/arXiv.2502.12318}.

\bibitem{DLMFSuperposition}
National Institute of Standards and Technology,
\emph{Digital Library of Mathematical Functions},
Section~10.32(i), Eq.~(10.32.1), integral representation of $I_0$.
\href{https://dlmf.nist.gov/10.32.E1}{DLMF 10.32.1}.

\bibitem{MailerCleggPusey2015}
A.~G.~Mailer, P.~S.~Clegg, and P.~N.~Pusey,
``Particle sizing by dynamic light scattering: non-linear cumulant analysis,''
\emph{Journal of Physics: Condensed Matter} \textbf{27}(14), 145102 (2015).
\href{https://doi.org/10.1088/0953-8984/27/14/145102}{doi:10.1088/0953-8984/27/14/145102};
\href{https://arxiv.org/abs/1504.06502}{arXiv:1504.06502}.

\bibitem{Zimm1948AngleConcentration}
B.~H.~Zimm,
``The dependence of the scattering of light on angle and concentration in linear polymer solutions,''
\emph{Journal of Physical and Colloid Chemistry} \textbf{52}(1), 260--267 (1948).
\href{https://doi.org/10.1021/j150457a022}{doi:10.1021/j150457a022}.

\bibitem{JCGM1012008}
Joint Committee for Guides in Metrology,
\emph{Evaluation of measurement data---Supplement 1 to the Guide to the expression of uncertainty in measurement---Propagation of distributions using a Monte Carlo method},
JCGM 101:2008 (2008), especially Sections~5--7.
\href{https://doi.org/10.59161/JCGM101-2008}{doi:10.59161/JCGM101-2008}.

\bibitem{Hong2017}
S.~Hong, R.~Riedinger, I.~Marinkovi\'c, A.~Wallucks, S.~G.~Hofer,
R.~A.~Norte, M.~Aspelmeyer, and S.~Gr\"oblacher,
``Hanbury Brown and Twiss interferometry of single phonons from an
optomechanical resonator,'' \emph{Science} \textbf{358}, 203--206 (2017).
\href{https://doi.org/10.1126/science.aan7939}{doi:10.1126/science.aan7939}.

\bibitem{SafaviNaeini2013}
A.~H.~Safavi-Naeini, J.~Chan, J.~T.~Hill, S.~Gr\"oblacher, H.~Miao,
Y.~Chen, M.~Aspelmeyer, and O.~Painter,
``Laser noise in cavity-optomechanical cooling and thermometry,''
\emph{New Journal of Physics} \textbf{15}, 035007 (2013).
\href{https://doi.org/10.1088/1367-2630/15/3/035007}{doi:10.1088/1367-2630/15/3/035007}.

\bibitem{TarragoVelez2019}
S.~Tarrago Velez, K.~Seibold, N.~Kipfer, M.~D.~Anderson,
V.~Sudhir, and C.~Galland,
``Preparation and decay of a single quantum of vibration at ambient
conditions,'' \emph{Physical Review X} \textbf{9}, 041007 (2019).
\href{https://doi.org/10.1103/PhysRevX.9.041007}{doi:10.1103/PhysRevX.9.041007}.

\bibitem{KeefTaorminaTwarock2005}
T.~Keef, A.~Taormina, and R.~Twarock,
``Assembly models for Papovaviridae based on tiling theory,''
\emph{Physical Biology} \textbf{2}, 175--188 (2005).
\href{https://doi.org/10.1088/1478-3975/2/3/005}{doi:10.1088/1478-3975/2/3/005}.

\bibitem{Vanlier2014}
J.~Vanlier, C.~A.~Tiemann, P.~A.~J.~Hilbers, and N.~A.~W.~van Riel,
``Optimal experiment design for model selection in biochemical networks,''
\emph{BMC Systems Biology} \textbf{8}, 20 (2014).
\href{https://doi.org/10.1186/1752-0509-8-20}{doi:10.1186/1752-0509-8-20}.

\bibitem{LarsonSTMV2014}
S.~B.~Larson, J.~S.~Day, and A.~McPherson,
``Satellite tobacco mosaic virus refined to 1.4~\AA\ resolution,''
\emph{Acta Crystallographica D} \textbf{70}, 2316--2330 (2014).
\href{https://doi.org/10.1107/S1399004714013789}{doi:10.1107/S1399004714013789}.

\bibitem{Stephanidis2007}
B.~Stephanidis, S.~Adichtchev, P.~Gouet, A.~McPherson, and A.~Mermet,
``Elastic properties of viruses,''
\emph{Biophysical Journal} \textbf{93}, 1354--1359 (2007).
\href{https://doi.org/10.1529/biophysj.107.109033}{doi:10.1529/biophysj.107.109033}.

\end{thebibliography}
\end{document}